  \documentclass{PT1}
  \usepackage{natbib}
  \usepackage[figuresright]{rotating}
  \usepackage{floatpag}
    \rotfloatpagestyle{empty}

  \usepackage{amsthm}

\usepackage{type1cm}         

\usepackage{makeidx}         
\usepackage{graphicx}        
\usepackage{multicol}        
\usepackage[bottom]{footmisc}

\usepackage{newtxtext}       %
\usepackage[varvw]{newtxmath}

\usepackage{color}
\usepackage{xcolor}
\usepackage{colortbl}
\usepackage{tikz}
\usepackage{adjustbox}
\usepackage{soul}
\usepackage{listings}
\usepackage{sidecap}
\usepackage{todonotes}

\usepackage[]{graphicx}
\usepackage{latexsym}
\usepackage{natbib}
\usepackage{amsmath}
\usepackage[caption=false]{subfig}
\usepackage{multirow}
\usepackage{mathtools}
\usepackage{algorithm, algpseudocode}
\usepackage{longtable}
\usepackage{textcomp}
\usepackage{rotating}
\usetikzlibrary{arrows}
\usetikzlibrary{quotes}
\usetikzlibrary{calc,topaths,hobby}
\usetikzlibrary{arrows.meta}
\usepackage{tabularx}
\usepackage{comment}
\newcommand{\commentAlt}[1]{\ignorespaces}
\newcommand{\commentLongAlt}[1]{\ignorespaces}
\usepackage{threeparttable} 
\usepackage{booktabs}

\definecolor{myEcoliColor}{rgb}{1, 0.58, 0.58}
\definecolor{myBsubColor}{rgb}{0.93, 0.77, 0.54}
\usepackage{array}
\usepackage{longtable}
\newcolumntype{H}{>{\setbox0=\hbox\bgroup}c<{\egroup}@{}}

\PassOptionsToPackage{hyphens}{url}
\usepackage[hyphens]{url}
\usepackage[breaklinks=true]{hyperref}

\usepackage[final]{feynmp}
\newcommand{\cS}{\mathcal{S}}

\newcommand{\hd}{\mathbf{h}}
\newcommand{\tl}{\mathbf{t}}

\newcommand{\Matrixr}[1]{\ensuremath{\left(\begin{array}{ccccccccccccccccccccccccr} #1 \end{array}\right)}}
\newcommand{\CC}{\mathcal{C}}

\newcommand{\II}{\mathcal{I}}
\newcommand{\beqn}{\begin{eqnarray*}}
\newcommand{\eeqn}{\end{eqnarray*}}
\newcommand{\eps}{\varepsilon}
\newcommand{\BB}{\mathcal{B}}
\newcommand{\Z}{\mathbb{Z}}
\newcommand{\R}{\mathbb{R}}
\newcommand{\Matrix}[1]{\ensuremath{\left[\begin{array}%
{cccccccccccccccccccccccc} #1 \end{array}\right]}}

\newcommand{\GG}{{\mathcal G}}

\newcommand{\Aut}{\mathrm{Aut}}
\newcommand{\ignore}[1]{}

\newcommand{\shf}{\mbox{\footnotesize $\frac{1}{2}$}}
\newcommand{\ii}{{\mathrm{i}}}
\newcommand{\sot}{\mbox{\footnotesize $\frac{1}{3}$}}

\makeatletter
\newcommand{\Letter}[1]{\@Alph{#1}}
\makeatother
\newcolumntype{C}[1]{>{\centering\arraybackslash}p{#1}}
\usetikzlibrary{math}

\usepackage{mdframed}
\usepackage{framed}
\usepackage{xcolor}

\mdfdefinestyle{MyShadeQuoteStyle}{%
    leftmargin=15pt,
    rightmargin=15pt,
    backgroundcolor=gray!25,
    linewidth=0pt,
    skipbelow=\topskip,
    skipabove=\topskip
}

\newenvironment{keyquote}[1][]{%
    \ignorespaces%
    \begin{mdframed}[style=MyShadeQuoteStyle,#1]%
}{%
    \end{mdframed}%
    \ignorespacesafterend%
}%

\makeatletter
\tikzset{
    @pos/.style={@pos1={#1},@pos2={#1}},
    @ratio/.style={@ratio1={#1},@ratio2={#1}},
    @delta/.style={@delta1={#1},@delta2={#1}},
    @edge/.style={@@edge/.append style={#1}},
    @edge 0/.style={@@edge 0/.append style={#1}},
    @edge 1/.style={@@edge 1/.append style={#1}},
    @edge 2/.style={@@edge 2/.append style={#1}},
    @edge 3/.style={@@edge 3/.append style={#1}},
    @edge 4/.style={@@edge 4/.append style={#1}},
    @pos1/.store in=\qrr@posA,
    @pos2/.store in=\qrr@posB,
    @ratio1/.store in=\qrr@ratioA,
    @ratio2/.store in=\qrr@ratioB,
    @delta1/.store in=\qrr@deltaA,
    @delta2/.store in=\qrr@deltaB,
    @pos=.5,
    @ratio=.5,
    @delta=.1,
}
\newcommand*{\connectThree}[4][]{
    \begingroup
    \tikzset{#1}
    \coordinate (@aux1) at ($(#2)!\qrr@ratioA!(#3)$);
    \coordinate (@aux2) at ($(#4)!\qrr@posA!(@aux1)$);
    \path (@aux2) edge[@@edge/.try, @@edge 0/.try, @@edge 3/.try] (#4);
    \draw[@@edge/.try, @@edge 1/.try] (@aux2) .. controls ($(#4)!\qrr@posA+\qrr@deltaA!(@aux1)$) .. (#2);
    \draw[@@edge/.try, @@edge 2/.try] (@aux2) .. controls ($(#4)!\qrr@posA+\qrr@deltaA!(@aux1)$) .. (#3);
    \endgroup
}

\newcommand*{\connectFour}[5][]{
    \begingroup
    \tikzset{#1}
    \coordinate (@aux1a) at ($(#2)!\qrr@ratioA!(#3)$);
    \coordinate (@aux1b) at ($(#4)!\qrr@ratioB!(#5)$);
    \coordinate (@aux2a) at ($(@aux1b)!\qrr@posA!(@aux1a)$);
    \coordinate (@aux2b) at ($(@aux1a)!\qrr@posB!(@aux1b)$);
    \path (@aux2a) edge[@@edge/.try,@@edge 0/.try] (@aux2b);
    \draw[@@edge/.try,@@edge 1/.try] (@aux2a) .. controls ($(@aux1b)!\qrr@posA+\qrr@deltaA!(@aux1a)$) .. (#2);
    \draw[@@edge/.try,@@edge 2/.try] (@aux2a) .. controls ($(@aux1b)!\qrr@posA+\qrr@deltaA!(@aux1a)$) .. (#3);
    \draw[@@edge/.try,@@edge 3/.try] (@aux2b) .. controls ($(@aux1a)!\qrr@posB+\qrr@deltaB!(@aux1b)$) .. (#4);
    \draw[@@edge/.try,@@edge 4/.try] (@aux2b) .. controls ($(@aux1a)!\qrr@posB+\qrr@deltaB!(@aux1b)$) .. (#5);
    \endgroup
}

 \makeindex

  \theoremstyle{plain}
  \newtheorem{theorem}{Theorem}[chapter]

    \newtheorem{example}{Example}
  \theoremstyle{definition}
  \newtheorem{definition}[theorem]{Definition}
  
  \newtheorem{example-norules}[theorem]{Example}%
  \newtheorem{proposition}[theorem]{Proposition}

  \theoremstyle{remark}
  \newtheorem*{remark}{Remark}

\begin{document}

  \title[Symmetry Fibrations and Synchronization in Biological
    Networks]{Symmetries of Living Systems}

    \author{Hern\'an A. Makse, Paolo Boldi,\\Francesco Sorrentino, Ian Stewart}

    \details{Cambridge University Press, 2025}


  \frontmatter
  \maketitle
%
%


\chapter*{Preface}

A symmetry is a `change without change'. As simple as it sounds, this
concept is the fundamental cornerstone that unifies all branches of 
theoretical physics. Virtually all physical laws---ranging from
classical mechanics and electrodynamics to relativity, quantum
mechanics, and the standard model---can be expressed in terms of symmetry
invariances. In this book, we explore whether the same
principle can also explain the emergent laws of biological
systems.

This book is an attempt to describe, in simple terms, the branches of
theoretical physics and mathematics that are relevant to the
application of symmetries to biology and to make these ideas
accessible to a wider audience of researchers. We provide details for
numerous biological networks of current interest.

\subsubsection{Organization of the book} 

Part I introduces the theoretical concepts needed to understand symmetry fibrations. We do this
through simple examples, either mathematical or biological. From
Chapter \ref{chap:definitions} onwards, we also present formal definitions
of these concepts. The aim is to make the material accessible to
non-mathematicians while avoiding over-simplification.
Part II demonstrates, in much greater detail, 
exactly how these concepts apply to a variety of biological networks ranging from genetic networks to the brain.

Part I begins with Chapter \ref{chap:intro}, which poses the main question of this book: can the
same symmetry principles that describe physical systems be applied to
biological systems? We argue that a more general symmetry notion is required
to take account of the flexibility found in biology.

Chapter \ref{chap:nutshell} gives an informal overview of the 
essential ideas involved in this more general kind of symmetry: 
symmetry fibrations, input trees,
balanced colorings, and synchrony patterns. We also discuss admissible equations, which are the most general ordinary differential equations
that are compatible with the network topology and, therefore, include
all possible model equations for the network concerned. 
We explain the difference between the global group-theoretic symmetries used in
physics and the local symmetries of fibrations, which we suggest are more suitable for biology.

Chapter \ref{chap:definitions} starts by defining basic concepts such as
graphs, networks, partitions, complete versus cluster synchronization,
and elaborates on a recurrent topic in this book: the
structure-function relation.
We discuss how
symmetry fibrations guarantee the synchronization of network node
activity. The sets of synchronized nodes are called fibers, and they
represent nodes in the network that have the same dynamical
state, i.e., they are synchronized. This chapter includes the mathematical
proof of how symmetry fibrations lead to synchronization in fibers.

Chapter ~\ref{chap:group} describes automorphisms and their relation to
cluster synchronization.  These symmetries describe all physical
systems but few biological ones.  However, they serve as an introduction to the
biological symmetry fibrations to be discussed in
Chapters \ref{chap:fibration_1} and \ref{chap:fibration_2}.

Chapter \ref{chap:fibration_1} describes the graph fibration formalism,
including the definition of input-tree isomorphisms and minimal
graph fibrations, leading to symmetry fibrations, balanced colorings, 
and the equivalent concept of an equitable partition. These are the
natural symmetries of biological networks. We discuss necessary conditions for
synchrony to be possible and relate these to input trees.

Chapter \ref{chap:fibration_2} expands on the notion of a 
fibration symmetry, introduced informally in Chapter \ref{chap:nutshell}. We begin with the more general
notion of a graph homomorphism, and then specialize to fibrations.
We relate fibrations to equitable partitions, and emphasize
the `lifting property', which distinguishes fibrations from
homomorphisms. The connections between fibrations, synchrony,
balanced colorings, and equitable partitions are examined in
greater depth. 

Chapter \ref{chap:groupoid} is mainly about the groupoid formalism,
a mathematical context suited to the local symmetries that arise
from fibrations. A groupoid is like a group, except that
a groupoid need not be closed under the operation of composition.
We give a brief, concrete discussion of groupoids that avoids most technicalities. 

Chapter \ref{chap:stability} focuses on the important issue of stability.
A dynamical state is stable if it remains essentially unchanged 
after a sufficiently small perturbation is applied. 
There are several stability notions, and we focus on the main ones.
We discuss the stability of equilibria and apply the theory to some
biological examples. The chapter ends with a brief introduction to 
bifurcations, where the stability of a state changes as parameters
are varied, leading to radical changes in the dynamics.

Chapter \ref{chap:hypergraph} generalizes the setting 
from graphs to related structures, such as multi-layered 
and weighted graphs. In particular, we discuss
hypergraphs, which formalize specific types of
many-body interaction. These types of
network are important for a more complete description of biological
networks across the spectrum of biological interactions from gene
regulation, protein-protein interactions, protein-metabolite
interactions and metabolism.

Chapter \ref{chap:bundles} discusses the relation between topological
fibrations, graph fibrations and fiber bundles. Fiber bundles are at
the foundations of particle physics as the formalism used to describe
the symmetries of the standard model. This chapter attempts to
make a connection between the formalism of fibrations in the broader
realm of theoretical physics.

Chapter \ref{chap:robustness} discusses how fibrations provide a
structure for biological networks that is robust to failure of its
components, e.g., under gene mutations. A comparison with group
symmetry structures shows that networks with fibrations are less
vulnerable to such failures than those with group symmetries. This suggests an
evolutionary drive for fibrations in biological networks. We discuss how
structure can constrain behavior (phenotype).

Chapter \ref{chap:geometrization} proposes a geometrization of
biological networks, by analogy with the geometrization of physics, by
following a parallel between fibrations in biology and fiber bundles
in physics. Using a financial analogue proposed by Maldacena we
show that the concepts of curvature and connection in
physical fiber bundles have analogues in biological fibrations, such as those in
genetic networks and the brain. 

Chapter \ref{chap:algorithms} discusses algorithms to find fibers in
networks by exploiting the link between fibers and balanced
colorings. A set of algorithms and their implementations is provided to
 calculate fibrations and fiber building
blocks easily for any network. Online sources for algorithms used in this book are listed in Appendix \ref{sec:list-software}.

Chapter \ref{chap:hierarchy_1} starts Part II. We show how symmetry fibrations
describe a hierarchy of fiber circuits across biological networks, especially
transcription regulatory networks, but also more generally.
We describe how these fibers define the building blocks of
these networks and  provide a broad classification of building blocks.

Chapter \ref{chap:hierarchy_2} applies the ideas of Chapter \ref{chap:hierarchy_1} to specific simple fibration building blocks of biological 
significance. We return to some of these networks later, in greater detail and
with more realistic models. The circuits discussed include the autoregulation loop,
several types of feed-forward fiber, Fibonacci fibers, and binary and $n$-ary trees.
We discuss operons and regulons, which can be viewed as trivial examples of fibers,
and analyze how building blocks vary across different species.

Chapter \ref{chap:complex} considers metabolic networks and enzyme networks,
where more complex building blocks arise. We focus on {\em E. coli}
where, in particular, complex composite Fibonacci building blocks appear.
These circuits involve longer-range feed-forward and feedback loops.

Chapter \ref{chap:motif} compares the building blocks obtained from
fibrations with network motifs and modules, the popular ways to
identify building blocks of biological networks. We define $p$-values and $Z$-scores,
which quantify the concepts involved. A key difference between motifs and 
fibration building blocks is that motifs are defined statistically, whereas
fibration building blocks are defined dynamically.

Chapter \ref{chap:breaking} discusses the cell as a computational
device. It shows how fibration symmetry breaking can identify genetic
circuits analogous to toggle switches and memory storage flip-flops
that build the computational logic machinery of genetic networks.
This leads to the identification of the function of each gene in a
genetic network of bacteria as a part of a computational logic
structure of the network, which is considered as a finite state
machine of computation.

Chapter \ref{chap:minimal} concerns the key issue of complexity reduction.
Most real biological systems are inordinately complex; to understand them
we must somehow reduce the complexity. We show that
the application of fibrations to
genetic networks leads to a systematic method for 
uncovering a minimal functional network, thus
reducing biological complexity to its minimal form. We analyze 
some biological examples in detail.

Chapter \ref{chap:synchronization} describes applications of
fibrations to biological systems displaying cluster synchronization
such as gene coexpression and neural synchronization. The main aim
is to investigate the structure $\rightsquigarrow$ function relation.
We also describe
the types of systems of admissible dynamical equations that can be
studied by fibrations.

Chapter \ref{chap:alive} asks how living organisms can exist when
everything in biology conspires against synchrony. We attempt to answer this
question in terms of weak symmetry breaking, in which idealized symmetric
networks are modified by natural selection to break symmetry while
preserving approximate synchrony. In essence, this chapter is about the
relationship between idealized mathematical models and realistic biology.

Chapter \ref{chap:function} addresses the issue of reconstructing 
idealized networks from incomplete biological information. This is the
converse of the structure $\rightsquigarrow$ function relation: the function
 $\rightsquigarrow$ structure relation.

Chapter \ref{chap:brain1} asks whether it is realistic to seek
organizing principles for brains. In this chapter, we consider circuits in
the {\em C. elegans} connectome. We apply the methods developed in
previous chapters to use fibrations to classify cell types and consider
locomotion in {\em C. elegans} from this point of view.

Chapter \ref{chap:brain2} continues the analysis of brains
in the more complex case of the mouse. Here, we focus on
memory storage and the notion of an engram.

Chapter \ref{chap:brain3} pushes ahead into the even more complex
structures of the human brain, with an emphasis on language. The role
of synchrony is emphasized, and we discuss fibration symmetry breaking in the human language connectome.


Chapter \ref{chap:outlook} ends the book with an overview of its main messages and an outlook for future work.

Appendix \ref{sec:list-software} then provides the code and data to perform fibration data analysis and reproduce all the results of the book.

\subsubsection{How to read this book}

This book aims to provide insights across various levels of complexity
to inspire interdisciplinary collaboration:

\begin{itemize}
\item For readers new to the concepts of symmetries in
  networks and seek a quick yet informative overview, we recommend
  focusing on Chapters \ref{chap:intro} and \ref{chap:nutshell} only. These
  chapters lay the foundation and discuss all concepts needed for the application of symmetries in biology.

\item If you are interested in the mathematically rigorous formulation
  of fibrations, we encourage you to read Chapters
  \ref{chap:definitions} through \ref{chap:geometrization}, which
  provide a detailed examination of the mathematical principles that
  underpin the fibration formalism.

\item For those with an interest in the practical implications of
  fibrations in both biological networks and artificial intelligence,
  transitioning to Part II after completing the first two introductory
  chapters is advisable. Each chapter includes self-contained
  real-world applications relevant to practical scenarios.

\item Similarly, if your goal is to perform a fibration/symmetry
  analysis of complex networks, you can directly proceed to Chapter
  \ref{chap:algorithms} following your review of the first two
  introductory chapters. Here, you will find the essential algorithms,
  along with the accompanying codes provided in the Appendix
  \ref{sec:list-software}, to facilitate practical applications.

\end{itemize}

\subsubsection{Acknowledgment and fibration companion website}

We are deeply indebted to the many enlightening discussions we have shared with our collaborators. Our journey into the fascinating world of fibrations in biology began with inspiring conversations with Flaviano Morone 
as we sought to develop a theory that bridges the gap between structure and function in biological networks. Throughout this journey, we have been fortunate to collaborate with an exceptional group of individuals who have deepened our understanding of symmetries in biology, including Luis \'Alvarez-Garc\'ia, Francesca Arese-Lucini, Pedro Augusto, Bryant \'Avila,
Jos\'e Soares Andrade Jr,   Pablo Balenzuela, Guido Caldarelli, Santiago Canals, Sof\'ia Morena del Pozo, Andrea Gabrielli, Raquel Garc\'ia-Hern\'andez, Tommaso Gili, Martin Golubitsky, Alireza Hashimi, Cecilia Ishida,
Andrei  Holodny,  Ian Leifer, Wolfram Liebermeister,  Higor Monteiro, Amir Nazerian,
Lucas Parra, David Phillips,  Saulo Reis, Nastassia Samadzelkava, Matteo Serafino, Mariano Sigman,  Silvina Tomassone, Osvaldo Velarde, Sebastiano Vigna,  Stefan Wuchty, and Manuel Zimmer.

The companion website of the book
\url{https://fibration.org} includes resources, codes, and
datasets discussed in this book.  We would love to hear from our
readers. Please send any comments or requests
to \href{mailto:hmakse@ccny.cuny.edu}{hmakse@ccny.cuny.edu}.





\vspace{\baselineskip}
\begin{flushright}\noindent
New York, USA\hfill {\it Hern\'an A. Makse}\\
Milan, Italy\hfill {\it Paolo Boldi}\\
Albuquerque, USA\hfill {\it Francesco Sorrentino}\\
Coventry, UK\hfill {\it Ian Stewart}\\
February 2025\\
\end{flushright}

  \tableofcontents

  
  \mainmatter

  \partquote{Part I introduces the concepts of symmetry fibration, balanced coloring, and cluster synchronization in networks, comparing these ideas to alternative frameworks based on symmetry groups. We demonstrate that the global group symmetries typically used to describe physical systems are too restrictive to fully represent the complexity of biological networks. In contrast, a more flexible local symmetry---specifically fibration symmetry---can effectively capture the structures within biological networks that lead to cluster synchronization. In Part II, we explore the applications of fibration symmetry to various biological networks, including genetic and metabolic networks as well as neural networks in the brain.}
  
\part{Theory: Symmetry Groups in Physics; Symmetry Fibrations in Biology}


\chapter[Biological Symmetry]{\bf\textsf{Biological Symmetry}}
\label{chap:intro}

\begin{chapterquote}
  This chapter presents the central question we address in this book: the role of symmetry in our understanding of nature. Symmetry is fundamental in physics at all levels, from the building blocks of matter and the shapes of snowflakes to the fabric of spacetime. The effectiveness of symmetries in explaining the physical world leads us to question why these same concepts cannot account for emergent properties in biological systems. Given that the same physical laws apply to both living and non-living systems, we must ask: if life is an emergent property of physics, why cannot the symmetry principles of physics also explain the organization of life?
  We propose a more general notion of symmetry, fibration symmetry, that is better suited for this purpose, and explain how it relates
to the important biological function of synchrony.
\end{chapterquote}

\section{Symmetry in physics and biology}
\label{sec:symmetry}

The mathematical scaffold of modern physics is built around principles
of symmetry \citep{gellmann1995,weinberg1995,wilczek2016beautiful}.  Since
nature seems to favor symmetry at a very fundamental level, symmetry
has become a powerful tool for discovering new laws of nature.
Physicists traditionally think of the notion of symmetry
as being synonymous with the theory of groups\index{group } and their actions, a
beautiful and extremely powerful branch of mathematics.
The theory of symmetry groups is the cornerstone of modern theoretical
physics, on which our understanding of particles, forces, and matter
has been built, via Lorentz invariance,\index{Lorentz invariance }\index{Lorentz, Hendrik } Einstein's equivalence principle,\index{Einstein, Albert }\index{Einstein's equivalence principle }
gauge invariance,\index{gauge invariance } Noether's theorem,\index{Noether's theorem }\index{Noether, Emmy } Bloch's theorem,\index{Bloch's theorem }\index{Bloch, Felix } Landau's theory\index{phase transition }\index{Landau, Lev }
of phase transitions, and so on
\citep{weinberg1995,landau1977quantum,georgi2018,dixon1996,stewart2003,stewart2005}.
  
In particular, symmetries and group theory lie at the core of the theoretical
formulation of the Standard Model,\index{Standard Model } which provides a unified
description of all fundamental particles and their interactions
(with the exception of gravity). This description is formalized by continuous (Lie)
symmetry groups of the Lagrangian \citep{weinberg1995}. 

In crystallography, point symmetry groups systematically classify 
crystal structures \citep{vainshtein1994}.  Molecular symmetry groups in
chemistry classify the three-dimensional arrangements of atoms and
their chemical bonds that constitute all molecules
\citep{landau1977quantum}.  Furthermore, the unifying mechanism of
spontaneous symmetry breaking explains pattern formation
in numerous areas of science, the
existence of phase transitions (e.g., ferromagnetism, superconductivity, and
superfluidity), and the emergence of universal critical behavior
unifying condensed matter and high-energy physics
\citep{weinberg1995}.

Symmetries are significant in theoretical physics, and this encourages exploring whether they can help us organize the vast amounts of information present in biological systems. These systems consist of interacting components arranged in networks, such as the billions of neurons in the human brain or the tens of thousands of genes, proteins, metabolites, and other biomolecules that form genetic, protein, metabolic, and signaling networks within an organism. 

A key question is whether the structure of these biological networks has developed according to the same symmetry principles that govern molecules and matter. Furthermore, we need to investigate whether these symmetries can assist in identifying the functional building blocks of living networks, much like they help identify the fundamental building blocks in particle physics. If this proves to be the case, it could lead to a systematic organization of biological complexity, with significant implications for theoretical biology, similar to the role of symmetries in theoretical physics.

However, this program has not yet been done, strongly suggesting that the group-theoretic 
notion of symmetry so prevalent in physics is not an appropriate fundamental framework for biology.
We aim to bring the concept of symmetry into the age of genomics and systems biology by exploring whether a more general notion of symmetry can describe the structure of biological graphs, including transcriptomes, proteomes, metabolomes, and connectomes. The new theory of fibration symmetries presented in this book extends the concept of global symmetry groups, transitioning into more flexible and local symmetries.  It demonstrates that the fibration formalism serves as the ideal mathematical framework for describing biological networks.

\section{Group-theoretic symmetries in biology}

\begin{figure}
  \centering \includegraphics[width=.8\linewidth]{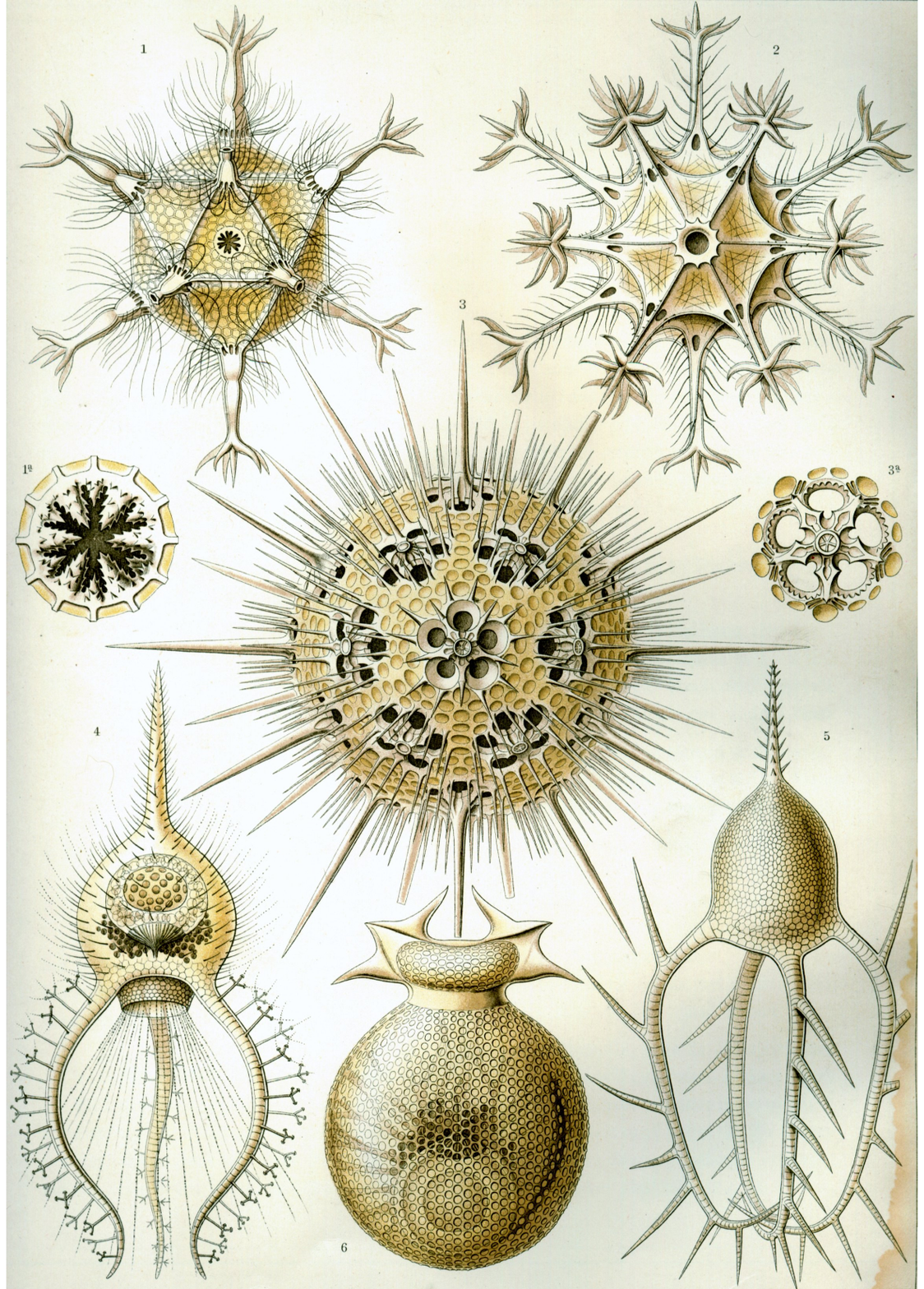}
   \caption{\textbf{The group symmetries of growth and forms in living systems are fascinating, yet they do not describe biological graphs.} In his influential book, \cite{darcy} examines numerous examples of what we now call group symmetries within simple biological systems, particularly highlighted by the polyhedral forms of radiolaria that Ernst Haeckel illustrated in 1904, shown here. 
   Unfortunately, these group symmetries cannot be translated to explain genetic graphs underlying life. Reprinted from
Wikipedia
\url{https://en.wikipedia.org/wiki/Radiolaria}.
} 
\label{fig:darcy}
\commentAlt{Figure~\ref{fig:darcy}: 
Illustration only. Beautiful engravings of radiolaria, many of which have
group-theoretic symmetry. Two are elaborate shapes with the symmetries 
of the icosahedron.
}
\end{figure}

We are not suggesting that the `global' notion of symmetry formalized in group theory is irrelevant to biology.
A notable early example is
the influential 1917 book {\it On Growth and Form} by D'Arcy Thompson\index{Thompson, D'Arcy }
\citep{darcy}. In this all-time favorite, Thompson meticulously documented
 the symmetries found in the structural
patterns and forms of simple living systems. He explored various examples, ranging from the shapes of viruses and the
icosahedral structure of radiolaria\index{radiolaria } to the left-right symmetry of
bilaterians and the spirals of mollusks, as well as the Fibonacci
phyllotaxis\index{phyllotaxis } observed in sunflowers. 
Figure \ref{fig:darcy} shows one of Ernst Haeckel's\index{Haeckel, Ernst } famous drawings of radiolaria
 from the {\em Challenger} expedition. Modern
research has shown that Haeckel often idealized the symmetry \citep{jungck2019}, but the figure is a dramatic illustration of the spirit
of Thompson's ideas. 
The body of literature surrounding Thompson's concepts is huge and diverse; \citep{chaplain1999} provides a balanced assessment and modern update.

Thompson's book was published prior to the discovery of DNA structure and the subsequent big-data 'omics' era in systems biology. 
While Thompson
enumerates these beautiful geometric invariants,  he does not offer 
 insights into the molecular origins of such symmetries at
the level of networks involving genes, proteins, and other biomolecules, which are of interest here. Not surprisingly,
given the era in which he worked, he focuses on symmetries
of the {\em organism}, not on the underlying molecular biological {\em graphs}. 
Additionally, he overlooks 
the principles of Darwinian evolution,\index{evolution } 
being somewhat antipathetic to it,
on the grounds that it ignores physical constraints on form.
                                                
Thompson's\index{Thompson, D'Arcy } book inspired Alan Turing's\index{Turing, Alan } groundbreaking work on
mathematical biology \citep{turing}.  Turing aimed to explain how symmetric pattern
formation in living organisms emerges from a uniform state through the process of 
morphogenesis in catalytic chemical reactions. These patterns, known as 
`Turing patterns'\index{Turing pattern } include examples such as the stripes of a zebra and the spirals
in the skin pigmentation of a giant pufferfish. Turing's theory fell out of favor with the huge advances in
molecular biology, but more recently it has been revived with
the identification of specific molecules as `morphogens'\index{morphogen }
and modern experimental techniques \citep{murray1989,meinhardt1995,kondo1995,sick2006,economou2012,sheth2012}.

The search for symmetries in biology re-emerged during the
Eleventh Nobel Symposium, chaired by Jacques Monod\index{Monod, Jacques } in 1968, and was
documented in its Proceedings \citep{monod1970symmetry}. During the Symposium, scientists discussed how the functions of macromolecules may arise from the symmetric arrangements of specific biomolecules. They explored how a protein's function could be encoded in the three-dimensional (quasi-) symmetric shape of its quaternary structure or how the functionality of an enzyme is determined by the highly symmetric coupling that allows a metabolite to bind at its active site. Additionally, it was acknowledged that various types of symmetry are observed at every level of the physical structure of biomolecules. Examples range from quantum mechanical atomic orbitals and the shapes of individual molecules to genetic sequences, the configurations of amino acids, and the alpha helices and beta sheets in proteins. Moreover, structures such as pores and proton pumps in cell membranes often exhibit symmetry, typically belonging to either a cyclic or dihedral group.

On the other hand, Monod\index{Monod, Jacques } and others also noted
that biomolecular symmetries seldom extend beyond the level of
quaternary structure of proteins. They play a role in early biological development; for instance, the symmetries of the fertilized
egg or the blastula break to create important structure in
the growing organism---but genetic effects quickly come into play
as well.
Beyond this level, the group symmetries of physics go silent
when explaining biological function or even the gross shape of
an organism. 

\begin{figure}
  \centering \includegraphics[width=\linewidth]{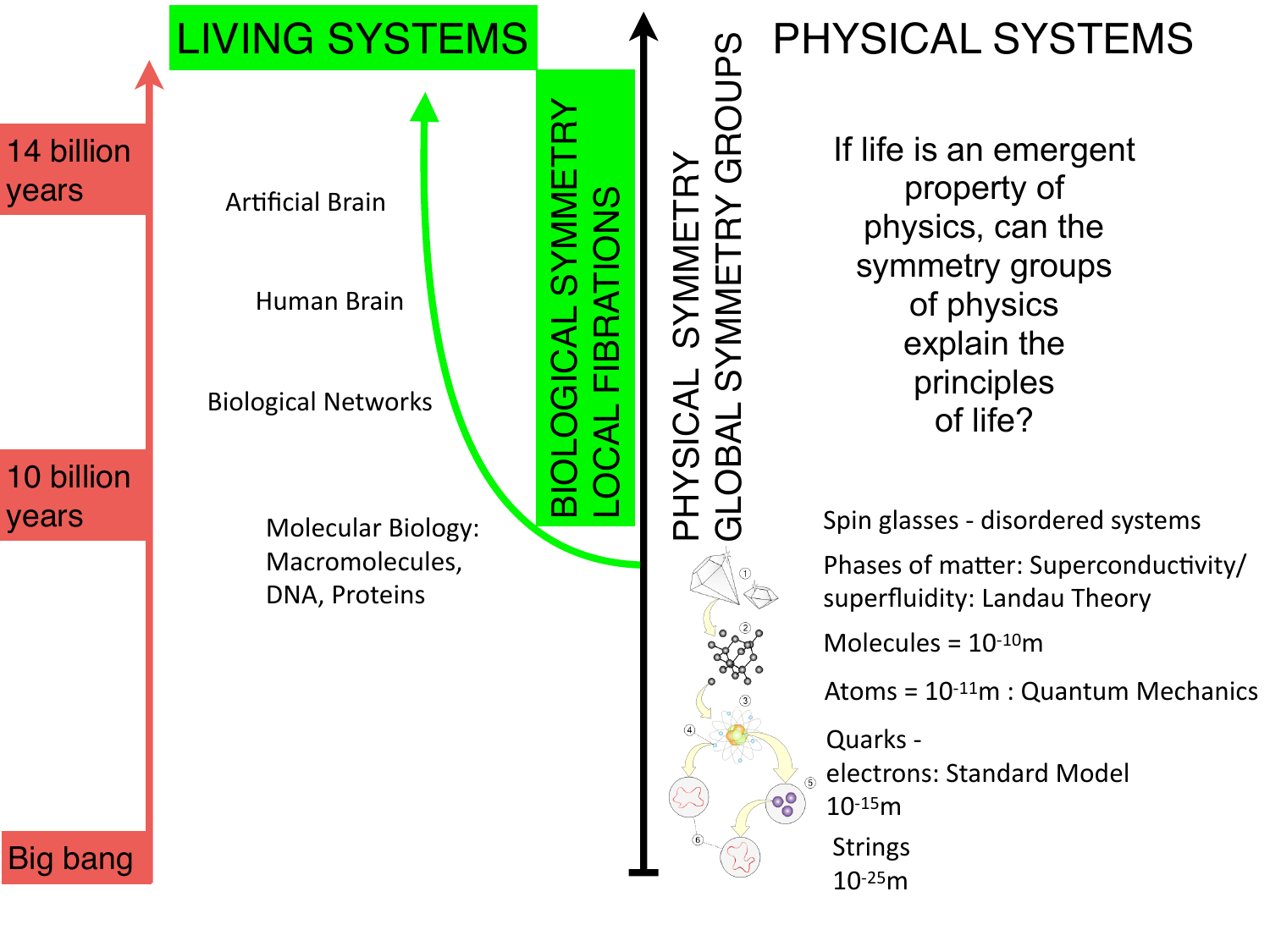}
   \caption{\textbf{Group symmetry in physics and fibration symmetry in biology.} The leitmotiv of this book is to explore the question posed by \cite{morone2020fibration}: `If life is an emergent property of physics---why cannot the same symmetry principles that explain physical phenomena also explain the organizing principle of life?' The success of group symmetries in elucidating the physical world---from general relativity to the standard model of particle physics and all phases of matter---raises an important question about their applicability in understanding the emergent properties of biological systems. In this book, we argue that a relaxed form of local symmetry, called symmetry fibration, explains the structure and synchronization observed in biological networks. 
 } 
\label{continuity}
\commentAlt{Figure~\ref{continuity}: 
Left column - time scale with labels 14 billon years, 10 billion years, Big Bang.
Middle column - List of living systems: artificial brain, human brain, biological networks,
molecular biology. Right column -  Physical systems. "If life is an emergent
property of physics, can the symmetry groups of physics explain the principles of life?"
Spin glasses, phases of matter, molecules, atoms, quarks, strings.
}
\end{figure}

Two fundamental questions were
raised at the Symposium and they remain unanswered \citep{monod1970symmetry,dualism}:

\vspace{5pt}
\noindent
{\bf Q1.}   \textit{'Is there continuity in kind between the
  symmetries seen at the levels of the molecule, the organelles, the
  virus, the cell, the organism?'} (Fig. \ref{continuity}).
  
\vspace{5pt}
\noindent  
{\bf Q2.}   \textit{'Are these symmetrical arrangements really
  useful only structurally, or might they be more directly related to
  dynamic function?'}
  
\vspace{5pt}
\noindent
Despite raising the right questions at the symposium, the hypothesis that biological
function could emerge from symmetry quickly fell into oblivion afterward. 
This may have happened because J. D. Bernal---another towering figure in molecular biology---gave an emphatic `no' in response to {\bf Q1}. He pointed out that\index{Bernal, John Desmond (J.D.) } `the symmetry of a sea star does not
  arise from self-assembly of the arms' \citep{dualism}, and further
stated \citep{bernal}: \textit{`There is radical difference between
  symmetry as observable in organisms and that in
  crystals'.}

More recent efforts have focused on applying group symmetries to dynamical systems within the field of systems biology. These studies encompass areas such as locomotion, evolution, and visual hallucinations \citep{stewart2003book, stewart2015}. A survey examining symmetry in neuronal circuits from the perspective of `form and function' can be found in \citep{stewart2022}. We shall not provide 
a comprehensive review of such work here. However, these applications tend to be specialized, often rely on idealized models, and do not adequately tackle the large data sets currently available, like gene regulatory networks\index{network !gene regulatory } and connectomes.\index{connectome } To date, these individual areas have
not had the impact on biology that group-theoretic symmetry has had on physics.

\section{A more general notion of symmetry}
\label{sec:MGNS}
 It therefore seems worthwhile to seek
a more general concept of symmetry, better adapted to
the needs of `big data' biology.
The present book aims at resuming the line of research started by
Thompson,\index{Thompson, D'Arcy } Turing\index{Turing, Alan } and Monod,\index{Monod, Jacques } by bringing new symmetry principles, not
based on groups, into the
era of big-data `omics'.  Bernal,\index{Bernal, John Desmond (J.D.) } like other historical figures of his
time, did not have access to the vast amount of large-scale
high-throughput data on all branches of functional genomics that is available
to present-day systems biologists. We will show how the principle of network
symmetry, formulated in an appropriately flexible and local manner, augments modern graph theory. This provides a theoretical method for
identifying functional building blocks of biological synchronization and for
understanding how the structure of a network determines cellular function.

This more general notion of symmetry does not displace or compete
with the existing group-theoretic one: instead, it complements it. Its mathematical origins go back a century
to \cite{brandt1927}, who introduced the notion of a `groupoid' and established many basic properties; it has been employed increasingly
in pure mathematics ever since, especially in algebraic topology \citep{B06}. Its applications in science are more
recent; in the biological context just described, it turns
out to be a natural way to approach key biological issues
of form and function.

Traditionally, the symmetries of a network have been formalized as
permutation symmetries of the nodes of the graph that leave its
adjacency matrix invariant. In simple terms, these
symmetries are those permutations of nodes that preserve the global
connectivity of the network.  Each of these permutation symmetries
imposes strong conditions on the dynamics of the nodes in the graph
because symmetry permutations leave invariant any `admissible'
 equation---one that is compatible with the network topology---that describes the dynamics of the system. Nodes
that are permuted by the symmetries are divided into the `orbits' of the symmetry group acting on the graph, and
these orbits form clusters of synchronized nodes. This phenomenon is known
as `cluster synchronization'.\index{synchronization !cluster }

These permutation symmetries are called automorphisms. They form a
group called the symmetry group of the graph: each symmetry has an inverse, their composition is associative, and the
identity function is a symmetry permutation. Symmetry
groups are the theoretical basis of all symmetries in physics, yet
they have found comparatively few applications in biology. At the same time,
cluster synchronization is ubiquitous in biological networks since the
collective dynamics of synchronized units has biological significance
for their function.

\begin{floatingbox}[h!]
    \processfloatingbox{The conundrum of this book} {How can we
      rationalize the widespread existence of cluster synchronization found in
      all biological systems in the absence of symmetry groups in
      their underlying biological networks?}
\end{floatingbox}

The main contention of this book is that there is a more general form of
network symmetry known as `fibration symmetry'\index{fibration !symmetry }
\citep{morone2020fibration,leifer2020circuits}. This new type of symmetry is local, allowing for a broader range of synchronous clusters compared to global group symmetries.
Mathematically, fibration symmetry is both necessary
and sufficient for robust cluster synchrony \citep{GS2023}.
Here `robust' means that such clusters occur for
any admissible ODE. However, {\it existence does not guarantee stability}, which depends on the specific model ODE under consideration. On the other hand, any
steady-state pattern of synchronous clusters that arises from
a fibration is stable for {\em some} admissible ODE.

Fibration symmetry, therefore, explains the widespread
existence of cluster synchronization in biology in the absence of
traditional symmetry groups. (We use both terms `symmetry fibration' and `fibration symmetry'. They are essentially synonymous, but the former focuses
on the type of fibration, while the latter focuses
on the type of symmetry.)

\begin{floatingbox}[h!]
\processfloatingbox{Our working hypothesis} 
    {The function of biological systems is engraved in the fibration symmetries of the underlying biological networks that make up the genome, transcriptome, proteome, metabolome, and the brain connectome, which define life at the system level.}
  \end{floatingbox}

\begin{figure}
	\centering
	\includegraphics[width=\linewidth]{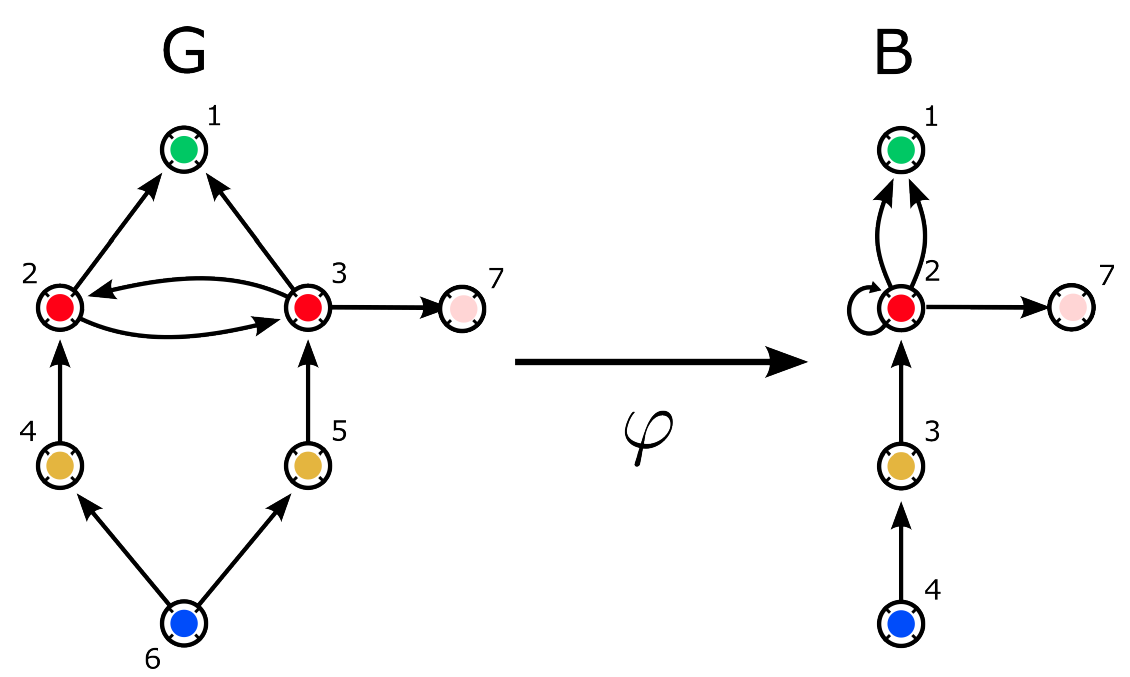}
	\caption{\textbf{A symmetry fibration $\varphi$ from a graph $G$ to
          its base $B$.} The graph $G$ can represent any biological network, such as a set of regulating genes or interacting neurons. It has no automorphism (no group symmetry) except for the
          identity. Yet, it possesses a rich fibration symmetry. The
          colored nodes demonstrate a balanced coloring of the graph. Equal-colored nodes receive the same 'amount' of colors from their in-neighbors, ensuring cluster synchronization among nodes of the same colors in the absence of a symmetry group. The fibration symmetry $\varphi$ collapses these clusters (called fibers) while preserving the dynamics of graph $G$ into the base $B$. Edges $(2, 3)$ and $(3, 2)$ are collapsed into the
          loop $(2, 2)$ and two edges $(2, 1)$ and $(3, 1)$
          become a multi-edge $(2, 1)$. This book elaborates on these concepts and their application to biological networks.
          }
	\label{fig:synchronyfibers}
\commentAlt{Figure~\ref{fig:synchronyfibers}: 
Left: graph with vertices 1-7. Colors are: 1 green, 2 and 3 red, 4 and 5 yellow, 6 blue, 7 pink.
Arrows go from 2 to 1 and 3; 3 to 1, 2, and 7; 4 to 2; 5 to 3; 6 to 4 and 5.
Right: `Collapsed graph' with vertices 1-4 and 7. Colors are: 1 green, 2  red, 3 yellow, 4 blue, 7 pink.
Arrows go from 2 to 1 (two arrows), 2 to 2, 2 to 7, 3 to 2, 4 to 3. In between is a large\
arrow labeled `phi' to show the fibration.
}
\end{figure}

Symmetry fibrations offer a more general and less restrictive form of symmetry
than automorphisms. They are not associated with symmetry groups but
with symmetry groupoids and are formalized by graph
fibrations \citep{grothendieck1959,boldi2002fibrations,stewart2006,GS2023}.
Graph fibrations are homomorphisms---structure-preserving 
maps---between graphs, subject to some simple but vital
extra conditions. These transformations do not preserve the adjacency matrix of the network, as automorphisms do, but they
define a natural way to `collapse' sets of nodes into clusters. Automorphisms are homomorphisms but with extra restrictions. Graph fibrations can
be of many types: injective, surjective, bijective, or none of those, and
minimal or not minimal. Not all of these possibilities are relevant to biological
systems. Only one particular type of graph fibration has a profound biological
meaning because it can reduce the network to its fundamental
building blocks for cluster synchronization, as shown in Fig. \ref{fig:synchronyfibers}.  These graph fibrations
are surjective and minimal, and they were named by \cite{morone2020fibration} as {\it symmetry fibrations}. They are closely related to balanced colorings
and initially we use this term informally through examples, without defining it.

\section{Cluster synchronization and fibrations}

The previous discussion implies that the existence of synchronized clusters of nodes in biological
networks is not accidental but follows from the presence of fibration
symmetries in the network, which apply to any
compatible dynamical system.
These clusters play a fundamental role in understanding living
systems because synchronization is an essential component of
biological processes: cluster synchronization of biological units
(proteins, neurons, etc.) indicates their participation in a common
function.  Cluster synchronization appears in synchronized gene
expression patterns \citep{eisen1998,langfelder2008} and the
synchronized oscillatory dynamics of ensembles of neurons in the
brain \citep{singer1999,uhlhass2009}.  Synchronization and oscillations further appear in circadian
rhythms, the cell-division cycle, the sleep-wake cycle, respiration,
locomotion, and cardiac rhythms, to name just a
few~\citep{strogatz2018,stewart2006}.

Graph fibrations are analogous to, and inspired by, 
topological fibrations, which were introduced by mystic genius
\cite{grothendieck1959} within category theory. The graph-theoretic
analog was developed by \cite{boldi2002fibrations} in the context
of distributed systems. In successively more general form,
\cite{stewart2003,stewart2005} and 
\cite{GS2023} developed an alternative formalism to fibrations
based on balanced colorings of a graph and its symmetry groupoids. Balanced colorings of 
graphs, also called `equitable partitions', are an alternative,
mathematically equivalent way to describe
fibers in the fibration formalism (Fig. \ref{fig:synchronyfibers}). 
They therefore lead to the same robust cluster
synchrony patterns. Thus, both fibers and balanced colorings describe,
in two different ways, the same synchronous dynamics of clusters of nodes.  

\begin{figure}
	\centering
        \includegraphics[width=.9\linewidth]{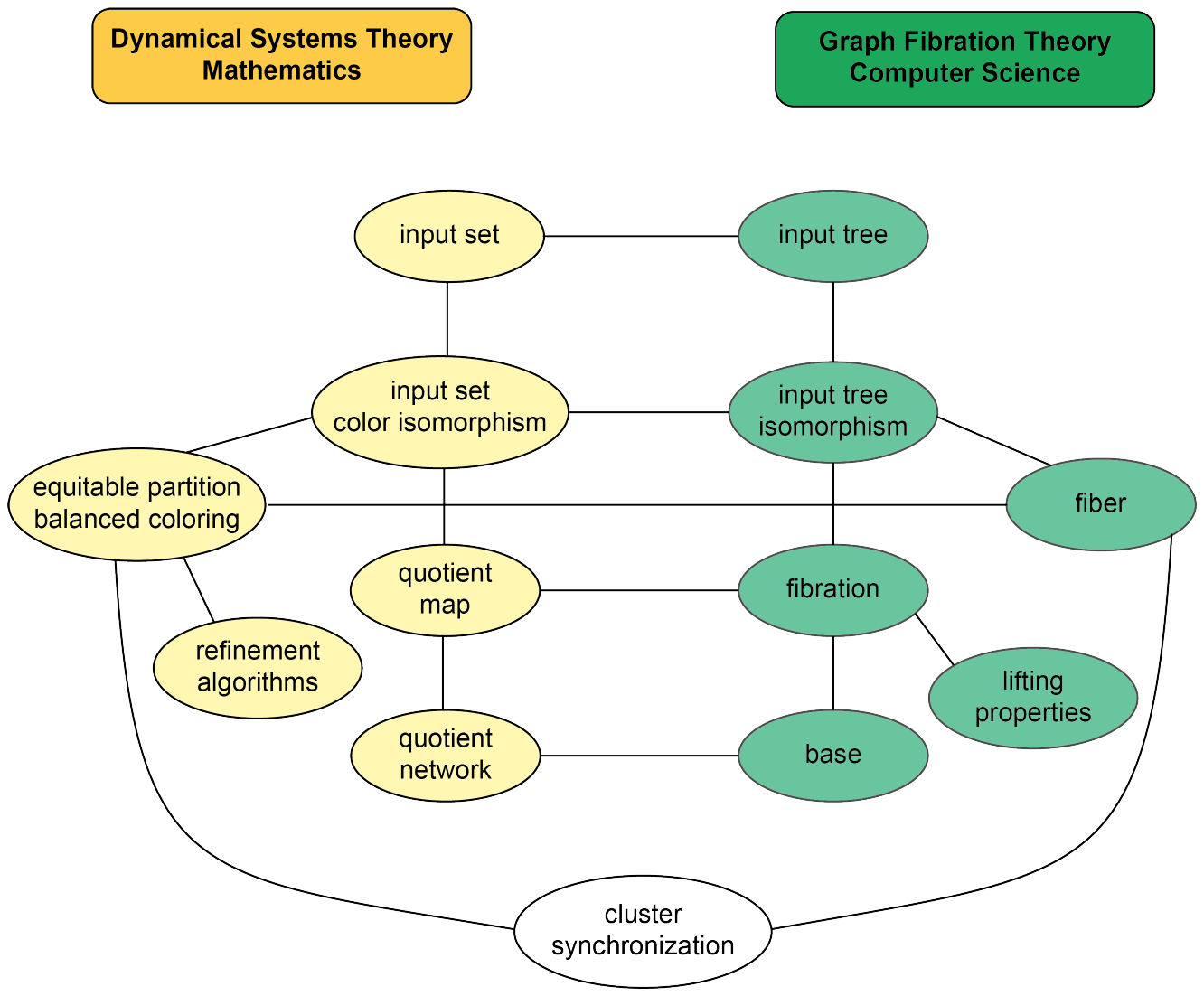}
	\caption{\textbf{Main mathematical concepts}. In Part I of
          this book, we examine key mathematical concepts as they are
          understood by various communities involved in the
          development of these theories. Mathematicians, computer
          scientists, and dynamical system theorists often use
          different terminology for the same concepts which are all
          related as shown here. We aim to unify these diverse
          groups within a common framework as shown in the graph,
          which is further developed to enhance the understanding of
          biological systems.}
	\label{Fig1-theory}
\commentAlt{Figure~\ref{Fig1-theory}:  Left: concepts from dynamical systems theory
(Mathematics): input set, input set color isomorphism, equitable partition/balanced coloring,
quotient map, refinement algorithms, quotient network. Right: input tree, input tree isomorphism, fibration, fiber,  base, lifting properties. Bottom: cluster synchronization.
}
\end{figure}

Fibrations and balanced colorings were shown to be
equivalent by \cite{aldis2008}.
The equivalence between fibration
symmetry and groupoid symmetry was demonstrated by
\cite{rink2013,lerman2015b} and discussed in
\citep{lerman2015,nijholt2016}. An approach
from an engineering viewpoint was investigated by \citep{field2004} and developed in \citep{agarwal2010a,agarwal2010b}.
The
connection between these approaches and the nomenclature relating the
formalisms of fibrations, groupoids, and balanced colorings are shown
in Fig.~\ref{Fig1-theory}, and discussed in detail in the rest of the book. 

Figure \ref{fig:synchronyfibers} shows an example of a graph $G$ with
no automorphisms (except for the identity), yet with a
symmetry fibration that determines fibers, which are the balanced coloring shown
by the differently colored nodes.
Given any admissible system of dynamical equations
associated with
the graph, it becomes possible for these fibers to synchronize their activity.
This synchrony pattern cannot be captured by
automorphisms. The symmetry fibration collapses the graph $G$ onto its
base $B$ as shown. In
succeeding chapters we discuss such fibrations extensively, and characterize admissible dynamics.

In this book we show that symmetry fibrations define the key features of the structure of
biological networks. We show that many biological networks have
fibration symmetries and display dynamical invariances, despite
the absence of automorphisms.  We
further show that the requirement of integration of functionally
segregated modular units in biological systems strongly suppresses the
applicability of automorphisms to biology.

 Applications of networks are not restricted to biology but also
include social networks, financial networks, the Internet, infrastructure
networks, and ecosystems, among others. 
Studying each of these networks
separately is a fascinating problem, but having a unified approach to
study the dynamics of all such networks from a symmetry point of view,
without needing to know how to describe the dynamics of each node,
is very beneficial.
Groupoids and fibrations describe the symmetries of
any network, where each node of the network defines an interacting
unit, and an edge between the nodes represents the transmission of
information---a signal of physical contact from one node to the other,
realized as the influence of one node on the dynamics of another.
 
\cite{morone2020fibration} and \cite{leifer2020circuits} have applied fibration symmetries to
information-processing networks to discover functional building blocks
of biological networks and to study their functions.  This book describes
these discoveries in a coherent fashion, as well as the theoretical
machinery behind the
fibration approach that has been developed over the last few decades in the fields of dynamical
systems.
This is done in Part I in a simple yet rigorous way to make the concepts and methods available
to a wider audience in the fields of graph theory, complex networks, and biology, as well as to connect the dots between those fields for more
mathematically inclined readers. Part II of this book deals with concrete applications to biological systems ranging from genetics networks to the brain.


\chapter[Symmetry Fibrations in a Nutshell]{\bf\textsf{Symmetry Fibrations in a Nutshell}}
\label{chap:nutshell}
\begin{chapterquote}
If all you have is fifteen minutes and want to grasp the essence of fibrations, this chapter is for you. We explain, in
simple terms, how fibrations generalize the concept of automorphisms---that is,
symmetry groups---and why it is difficult to realize these symmetry groups in biological networks. The central point of this book is that fibration symmetry, 
a local and less rigid form of symmetry, 
represents the structure of biological networks better than global symmetry
groups do. In particular, it controls the important feature of
cluster synchronization.
 We introduce the minimal concepts needed to understand
fibrations and their relation to cluster synchronization, concepts that are elaborated later in the book.
\end{chapterquote}

\section{Informal introduction to fibrations}
\label{sec:informal}

In this chapter, we introduce, in an informal manner, a small
number of key mathematical concepts that determine and describe
fibrations and cluster synchronization in networks. We do this in the context of biological
networks, specifically simple gene regulatory networks (GRNs), because
the primary target audience of the book is the biological community. A
secondary aim is to introduce the biological and mathematical concepts
to mathematicians, physicists, computer scientists, and engineers, who
are interested in applications to biology (or, indeed, to other areas
of `real world' science).

Because of this diverse audience, we avoid technical definitions in
this chapter. Instead, we use simple but typical examples to
illustrate the main ideas, which are treated in more detail in
subsequent chapters. The mathematical concepts can be defined, and
analyzed, in much greater generality, but we leave these extensions for later. The ideas presented here are repeated later in Part I in
greater depth and through a richer set of examples. Part II focuses on applications of fibrations in biology and artificial intelligence.

Research into biological networks has focused on the importance of
synchronization (also referred to as synchrony). Nodes in a network
are {\it synchronized} if they display identical dynamics.  
A significant amount of research in dynamical systems has focused on understanding \textit{complete} synchronization in networks. Complete synchronization occurs when all nodes in the network are synchronized in the same state. This research primarily examines dynamics described by the Laplacian matrix: $L=D-A$, where $D$ is the degree matrix, and $A$ is the adjacency matrix of the graph \citep{PC90}. 
 Laplacian dynamics describe networked systems interacting with diffusive couplings (like the Kuramoto model, see Section  \ref{sec:laplacian}).
 
 However, complete synchronization is rarely observed in biological networks.
Cluster
synchronization, which occurs when nodes split into disjoint subsets,
such that they synchronize within each subset but do not synchronize
between different subsets, is mathematically more common and also
biologically more significant. 

Examples are abundant in biology. In a GRN, the genes concerned are
active in clusters at the same time and in the same manner, and this cluster synchronization
constitutes a basic type of functionality
\citep{leifer2020circuits}.
Hebbian learning \citep{hebb1949} is based on the principle `neurons
that fire together, wire together', and 
the neurons concerned fire (or do not fire) at the same
time in neuronal assemblies. 
  In any network, cluster synchronization reveals
functional building blocks of the network structure
\citep{morone2019symmetry,morone2020fibration,leifer2020circuits}.

These biological systems are not described by the diffuse couplings of Laplacian dynamics. Instead, the dynamics in the biological systems treated in this book are defined in terms of the adjacency matrix of the network and their input trees, whose symmetries explain the cluster synchronization. 
 
The area of network dynamics is becoming increasingly important for
several reasons. Scientists (and indeed other researchers) have come
to realize that many real-world systems are naturally structured as
networks: GRNs, neuronal networks in the brain or elsewhere,
predator-prey networks, evolutionary trees, epidemic networks, the
Internet, social networks, transportation networks, economic
networks$\ldots$ All of these networks benefit from the understanding of
symmetry and synchrony.

\section{Fibrations, input trees, balanced colorings, and synchrony}

The notions discussed in this section can be formulated in three
distinct but equivalent ways: as (a) \textit{ synchrony patterns}, (b) \textit{fibrations, fibers} and \textit{input trees}, and (c) 
\textit{balanced colorings} with their associated {\em quotient networks} or {\em bases}.  All three
formulations occur in the literature, and it is important to
understand how they are related.  

Synchrony between nodes is
defined as nodes that exhibit identical
time-series and synchrony clusters can be obtained
using {\em graph  fibrations}.
The synchrony clusters form a partition of the network that is balanced colored (nodes with the same colors receive the same colors from their neighbors, Figs. \ref{fig:synchronyfibers} and  \ref{F:Uxur}) and nodes in a synchrony cluster (also called fiber) have isomorphic input trees (Fig. \ref{F:Uxur_input_tree}). 

If $G$ and $B$ are graphs, a fibration $\varphi: G \to B$ is a map that
sends nodes to nodes, arrows to arrows, and preserves input sets (where the input set of a node consists of all
of its input arrows).\index{input set }
Since fibrations preserve input sets, they also preserve input trees.
Figure  \ref{F:Uxur_input_tree} is a simple example. Nodes 2 and 3
belong to the same fiber (pink), and they have isomorphic input trees,
shown there to level  3, but the isomorphism is valid to all levels. 

We elaborate on these definitions next.

\section{Symmetry fibration in {\it Escherichia coli}}
\label{sec:SFEC}

\begin{figure}[h!]
\centerline{%
\includegraphics[width=0.6\textwidth]{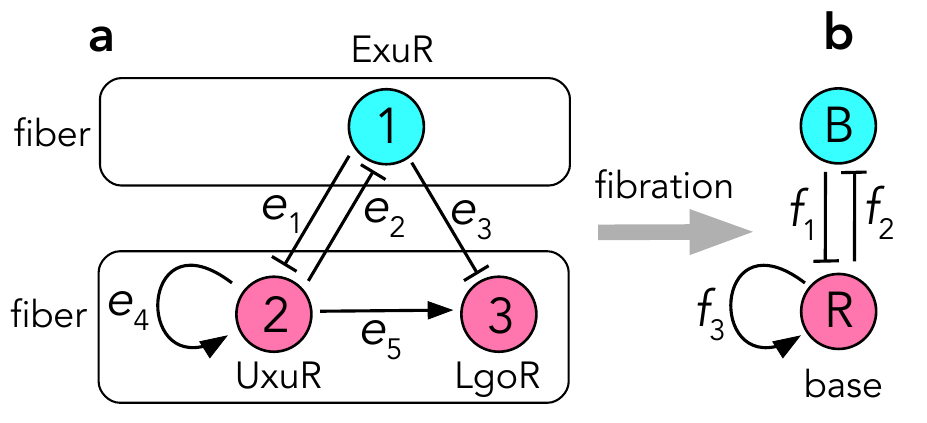}
}
\caption{\textbf{Sample network to define the main concepts.}  (\textbf{a}) Subnetwork of GRN of {\it E. coli}, extracted from the {\it E. coli} TRN (Transcription Regulatory Network) controlling the gene {\it Uxur}. Colors show a synchronization
  pattern. (\textbf{b}) The corresponding base in which nodes of the same
  color are `collapsed' to a single node. The gray arrow indicates a
  fibration: a map from the GRN to its base that preserves the
  directed edges that input to any given node. Curved rectangles outline the two fibers.
  Sharp arrow =
  activator, barred arrow = repressor.}
\label{F:Uxur}
\commentAlt{Figure~\ref{F:Uxur}: 
Left: graph with vertices 1 (blue), 2, 3 (red). Labels correspond to genes:
1 = ExuR, 2 = UxuR, 3 = lgoR. Rectangle encloses vertex 1, labeled `fiber'.
Rectangle encloses vertices 2 and 3, labeled `fiber'.
Arrows link the vertices: e1 from 1 to 2 (barred head
indicates repression),
e2 from 2 to 1 (barred head), e3 from 1 to 3 (barred head), e4 from 2 to 2 
(sharp head indicates activation), 
e5 from 2 to 3 (sharp head). Right: Quotient by coloring. Labeled `base'. Vertices B (blue)
and R (red). Arrows link the vertices: f1 from B to R (barred head),
f2 from R to B (barred head), f3 from R to R (sharp head). 
In between: gray arrow labeled `fibration'.
}

\end{figure}

We begin by analyzing a simple example of synchronization in a tiny
part of the genome of the bacterium {\it Escherichia coli}, one of the
standard model organisms in genomics.
Figure \ref{F:Uxur}a illustrates a typical simple example of this
situation: it shows a small subnetwork of the gene regulatory network
of {\it E. coli}. The transcriptional regulators ExuR, UxuR, and LgoR
control d-galacturonate and d-glucuronate metabolism
\citep{tutukina2016, shimada2018}.

For simplicity we assume that the subnetwork concerned has no external
inputs; that is, we consider only the dynamics of this network on its
own.  We also assume that nodes $1$, $2$ and $3$ have the same
internal dynamics: they react to similar stimuli in similar ways.
Edges $e_1, e_2, e_3$ determine identical repressor couplings, and
edges $e_4, e_5$ determine identical activator couplings.

This 3-node network has no group symmetry. More formally, there is only a trivial group symmetry: there is no automorphism (permutation) of the network that preserves its topology except for the identity. 

However, there exists a synchrony pattern in which nodes
2 and 3 are synchronized (see Section \ref{S:admissible_eq}). (Node 1 typically is not synchronized with
any other node.) How can this cluster of synchrony arise in the absence of automorphisms? The reason is that, just as for a group symmetry in the
usual sense, if nodes 2 and 3 are synchronized, then they both receive
the same input signals, hence remaining synchronized for all future
time. To verify this claim, we can ignore node 1 since it is not
expected to synchronize with any other node. Node 2 receives inputs
from edges $e_1$ and $e_4$, whereas node 3 receives inputs from edges
$e_3$ and $e_5$. However, the signals along edges $e_1$ and $e_3$ are
identical because they both emanate from node 1 and are both
repressors; the signals along edges $e_4$ and $e_5$ are identical
because they both emanate from the same node
and are both activators.

We now formalize this intuition and explain the details included in Fig. \ref{F:Uxur}, which relate
to a general explanation of this type of cluster synchronization.

In the example network of Fig. \ref{F:Uxur}, edges have a specific type (in the example,
the types are `repressor' and `activator'). When there is more than
one edge type involved, the types must be taken into consideration
when looking at the symmetry.

The colors `red' and `blue' partition the nodes into subsets $\{1\}$
and $\{2,3\}$, which we call `fibers'.  Subfigure {\bf (b)} shows a
simpler network, the `base'. It has only two nodes B (blue) and R (red), one for
each fiber (hence for each color), repressor edges $f_1$ and $f_2$,
and an activator edge $f_3$.

The large gray arrow in Fig. \ref{F:Uxur} 
indicates that subfigures {\bf (a)} and {\bf (b)} are
related by a `fibration'. This is a map that collapses fibers to
single nodes while retaining the set of input edges. More precisely, it is
a map $\varphi$ from the GRN to its base, which, among other properties
described immediately below, maps nodes to nodes and edges to edges,
while preserving sources and targets and also preserving the types of
edges (here, activator/repressor).  It does so according to the following scheme for nodes and edges, where B = blue, R = red:
\begin{eqnarray*}
1 \mapsto {\rm B} & 2,3 \mapsto {\rm R} \\
e_1, e_3 \mapsto f_1 & \qquad e_2 \mapsto f_2 \qquad e_4,e_5 \mapsto f_3
\end{eqnarray*}

In general, the `fiber' of a node in the original network is the
set of all nodes that map to a given node in the base. In Fig. \ref{F:Uxur}
the fibers are $\{1\}$, the only node that maps to $B$, and
$\{2,3\}$, the two nodes that map to $R$. The fibers correspond to
the colors and determine the clusters in the corresponding synchrony pattern,
which exists provided the map satisfies two additional properties,
as we now explain.

Any map of this kind is called a `graph homomorphism'. However,
$\varphi$ has three further properties that make it a fibration:
\begin{enumerate}
\item $\varphi$ is `surjective': every node in the base
is the image of some node in the GRN.
\item $\varphi$ preserves the structure of `input sets', the set of all
  edges with a given target node. 
\item $\varphi$ preserves the colors. 
\end{enumerate}

The above conditions mean that (under general assumptions) 
fibrations preserve synchronous dynamics. That is,
when nodes of the same color are
synchronized, the dynamics in the GRN is the same as the dynamics in the base. 

We verify these two conditions for this example.
For condition 1:
\begin{eqnarray*}
&& \mbox{Node 1 has input set}\ \{e_2\} \\
&& \mbox{Node B has input set}\ \{f_2\} 
\end{eqnarray*}
and these match each other, both in type (one repressor)
and in the color of the source nodes (blue).
Similarly
\begin{eqnarray*}
&& \mbox{Node 2 has input set}\ \{e_1,e_4\} \\
&& \mbox{Node 3 has input set}\ \{e_3, e_5\} \\
&& \mbox{Node R has input set}\ \{f_1,f_3\} 
\end{eqnarray*}
and these match each other, both in type (one repressor and one activator)
and in the color of the source nodes (blue and red respectively).

These conditions guarantee that if nodes 2 and 3 are synchronized initially, then they receive synchronized signals, hence
remain synchronized in the future.

Translated into the terminology of colorings, these statements
show that:
\begin{keyquote}
    Nodes of the same color have inputs that match each other, as regards both the type of edge and the color of the source of that edge. That is, nodes of the same colors `receive' the same colors from in-neighbors.
\end{keyquote}
Such a coloring is said to be a \textit{ balanced coloring}. 

Mathematically,
the notions of a fibration and a balanced coloring are equivalent,
and the fibers correspond exactly to the colored clusters, that is,
the sets of nodes that synchronize according to the pattern of colors.

\subsection{Input trees}

In the fibration formalism, the behavior of each node is encoded in its input
tree, which is a rooted tree that contains all the paths in the
network that terminate at the node \citep{morone2020fibration}.  This encoding is more
convenient than the typical graph representation of the adjacency matrix, because it allows
for the separation of the whole network into parts so that the
symmetries that directly preserve the dynamics can be analyzed
separately.

\begin{figure}[h!]
  \centering
  \includegraphics[width=0.8\linewidth]{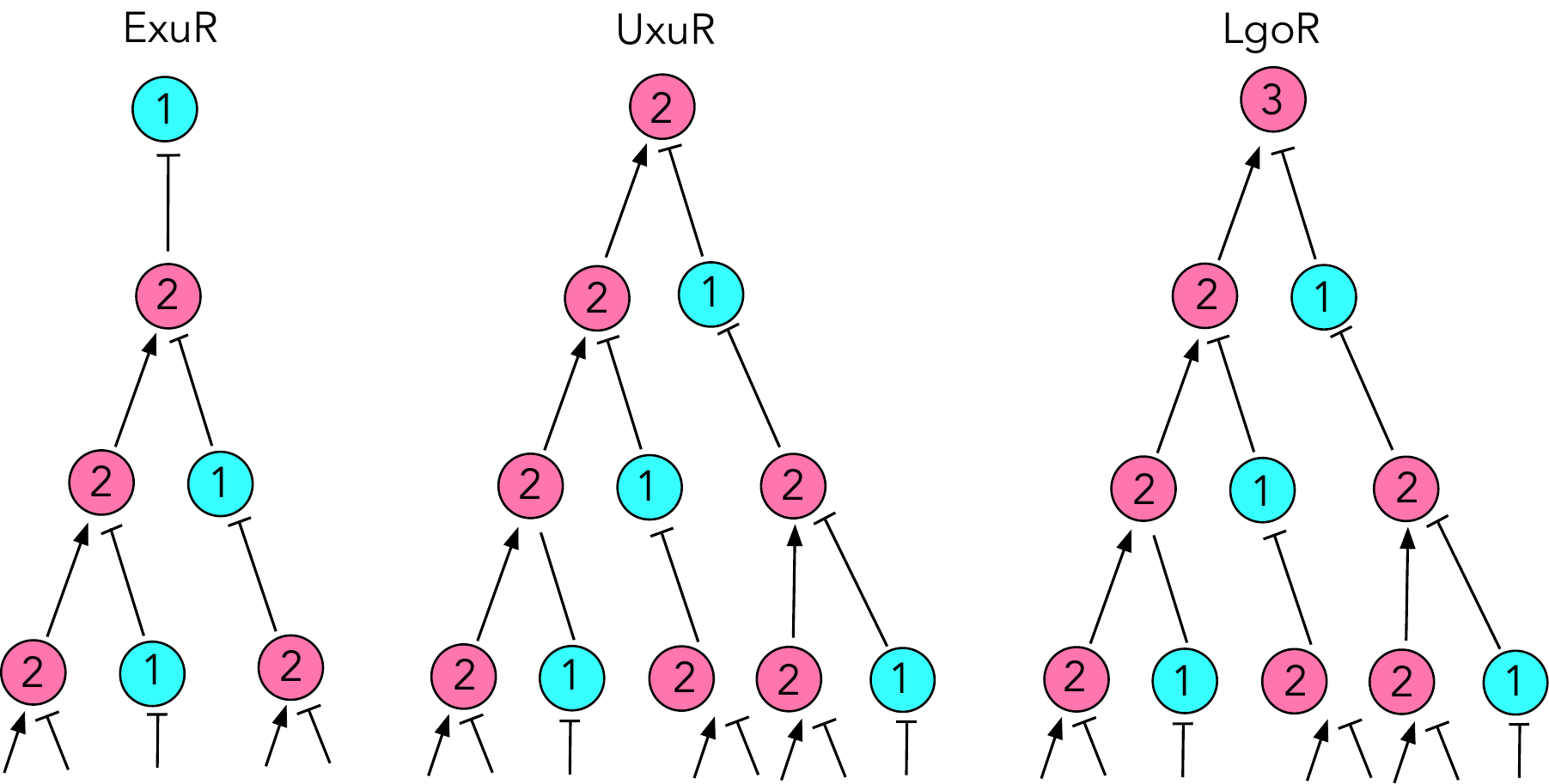}
  \caption{\textbf{Input tree examples.}
    Input trees for the {\it E. coli} network of Fig. \ref{F:Uxur}.
    Nodes 2 and 3 have isomorphic input trees, whereas that of node 1
    is different. Thus, nodes 2 and 3 can synchronize, but node 1
    cannot synchronize with either node 2 or node 3. Colors show the
    synchrony pattern but are not part of the definition of an input
    tree. }
\label{F:Uxur_input_tree}
\commentAlt{Figure~\ref{F:Uxur_input_tree}: 
Left: Input tree for vertex 1. Middle: Input tree for vertex 2. right: Input tree for vertex 3.
The middle and right trees are isomorphic; the left one is not. Further description in text.
}
\end{figure}

Since fibrations preserve input sets, they also preserve input trees in the following sense: If $\varphi:G \to B$ is
a fibration, then the input tree of any node $i$ of $G$
is isomorphic to the input tree of its image $\varphi(i)\in B$, where $B$ is the base.
If $\varphi$ is not surjective, there could be nodes in
$B$ with input sets (hence also input trees) that
differ from those in $G$.
Figure \ref{F:Uxur_input_tree} is a simple example. Nodes 2 and 3
belong to the same fiber (pink), and they have isomorphic input trees,
shown in Fig. \ref{F:Uxur_input_tree} to level 3 (the isomorphism is valid to all levels) 

A \textit{symmetry fibration} is a surjective fibration that transforms the graph into its
{\it minimal} base by collapsing {\it all} the nodes with isomorphic input trees. That is, the symmetry fibration produces the minimal base (with minimal colors) 
by collecting all the symmetries of the graph. The
collapsed nodes belong to a fiber of the graph and, through the
application of the symmetry fibration, the fibers are collapsed into
one representative node at the base of the graph.

For example, Fig. \ref{F:Uxur_input_tree} shows the three input trees
for the graph in Fig. \ref{F:Uxur}. The input trees for nodes 2 and 3 have the same topology, while the input tree for node 1 has a different topology.
Indeed, the colors on the input trees for nodes 2 and 3 correspond
as well. Clearly, such an isomorphism of input trees
is necessary for a fibration to exist. In fact, this condition is also sufficient:
isomorphism of input trees determines the {\it minimal fibration}---the
one with the fewest colors.

Invariance of the input tree under a symmetry fibration preserves the
dynamics of the system. Intuitively, this happens because directed paths in the graph correspond
to the flow of information through the network. Section \ref{S:admissible_eq}
makes this idea precise.
This invariance has
important consequences for the dynamics of the nodes,
since it predicts robust cluster synchronization of nodes in the same
fiber. Nodes in the same fiber synchronize their activity via the
fibration.

A fibration symmetry is not a group symmetry of the graph: it
is a symmetry that relates input trees. It represents invariances of the transmission of information
through the network, which explains its widespread appearance in
information processing networks in biology. In this book, we show that
the synchronized fibers identified by symmetry fibrations are the
fundamental building blocks from which biological networks are built.

\section{Admissible equations and cluster synchronization}
\label{S:admissible_eq}
To illustrate the mathematics in a biological context,
we consider a `toy' model of the GRN of Fig. \ref{F:Uxur}, employing
Hill functions and linear interactions to represent gene dynamics.
For simplicity, we assume a `lumped' model in which the protein and mRNA
concentrations are combined into a single real variable:
\begin{equation}
\label{E:Ecoli_model}
\begin{array}{rcl}
\dot x_1 &=& -ax_1+ b\frac{1}{1+x_2^2}\\
\dot x_2 &=& -cx_2 + b\frac{1}{1+x_1^2} + d\frac{x_2^2}{1+x_2^2}\\
\dot x_3 &=& -cx_3+  b\frac{1}{1+x_1^2} +d\frac{x_2^2}{1+x_2^2}
\end{array}
\end{equation}
Here $x_i$ is a measure of the activity of node $i$.  The constants
$a,b,c,d > 0$ are parameters whose values determine the amount of
activation or repression and degradation rates. The minus signs on $a,
c$ indicate degradation. These terms represent the internal dynamics of
each node.  The Hill function $\frac{1}{1+x^2}$ corresponds to
repression, while $\frac{x^2}{1+x^2}$ corresponds to activation.

To build the ODE model of the graph Fig. \ref{F:Uxur}, we have made several important hidden assumptions of the graph representation. For instance, we consider 
each edge to determine only the contribution of one input to the input function of the node and that these input edges are additive in the input function.
Thus, individual arrows in the graph represent individual terms in the ODE. They can be set up that way for additive interactions.
Also, the same parameters $b,c,d$ appear several times because
arrows of the same type model identical couplings.
These assumptions and parameter choices will be discussed in more detail in Section 
\ref{sec:hypergraph-metabolic}
and Chapter \ref{chap:alive}.

To investigate synchronized solutions with the given coloring,
we set $x_1(t) = x(t)$ and $x_2(t)= x_3(t)=y(t)$. Equations
\eqref{E:Ecoli_model} become:
\begin{equation}
\label{E:Ecoli_model2}
\begin{array}{rcl}
\dot x &=& -ax + b\frac{1}{1+y^2}\\
\dot y &=& -cy + b\frac{1}{1+x^2} + d\frac{y^2}{1+y^2}\\
\dot y &=& -cy + b\frac{1}{1+x^2} +d\frac{y^2}{1+y^2}
\end{array}
\end{equation}
Such a system would be `overdetermined' in general; that is, there
would be more equations than unknowns. However, in this case, the repeated equation (for $\dot y$)
is the same on both occasions, so no contradiction arises. Deleting the repeated equation for $\dot y$ implies that any solution of the
{\em  restricted equation}
\begin{equation}
\label{E:Ecoli_model3}
\begin{array}{rcl}
\dot x &=& -ax + b\frac{1}{1+y^2}\\
\dot y &=& -cy + b\frac{1}{1+y^2} + d\frac{y^2}{1+y^2}
\end{array}
\end{equation}
corresponds to a synchronized solution of \eqref{E:Ecoli_model}
by setting
\[
x_1(t) = x(t) \qquad x_2(t)=x_3(t) = y(t)
\]
and conversely. Moreover, \eqref{E:Ecoli_model3}
represents an analogous model for the base network.

Equations \eqref{E:Ecoli_model2} are not the most general ones
that are consistent with the network topology. They
assume that interaction terms are additive, with specific forms,
and that each arrow or node symbol determines one such term. The general theory of
network dynamics shows that persistence of the synchrony pattern 
is not just an accident 
of this type of model. It holds for {\it any} differential 
(or difference or delay differential) equation model 
that respects the network topology and our basic modeling
assumptions. Such a model is said to be {\em admissible}
for the network topology.

The relation between \eqref{E:Ecoli_model} and
\eqref{E:Ecoli_model3} implies that the
synchrony pattern described by that equation is valid for all times $t$. However, it
is important to understand that this statement does not
imply that the synchrony pattern concerned is dynamically {\em stable}---a property that is necessary for a given synchrony pattern to occur in reality. In practice the synchrony pattern persists, as time passes, only when
small perturbations do not have a significant effect on the dynamics. 
If such perturbations preserve synchrony, or they break it but die down, then the synchrony pattern is stable.
But if perturbations that break the synchrony tend to grow, the synchronized
state is unstable. 

Conditions for stability are not determined by the network
topology alone but also by details of whichever admissible ODE
is used as a model. However, there are
some general principles that characterize how the topology constrains stability. We return to this crucial point in Chapter \ref{chap:stability}.

\section{Key principle}
\label{sec:key_principle}

We now state a key principle for the existence of (not necessarily stable) synchrony patterns:

\begin{keyquote}
    Synchronized nodes must have the same internal dynamics and must receive synchronized signals.
\end{keyquote}

This principle is highly intuitive since differences in
internal dynamics or incoming signals are likely to break the
synchronization.
This principle turns out (subject to mild technical conditions
that exclude `accidental' synchronies permitted by special kinds of coupling) 
to be equivalent to any of the following statements:
\begin{itemize}
\item Synchrony clusters determine a fibration.
\item Synchrony clusters determine a balanced coloring.
\item The subspace of all possible synchronized states is invariant under all admissible equations (i.e., those that respect the network architecture).
\end{itemize}

In a sense, everything else in this book follows from the
above principle or expands on its implications and applications.

\section{Global symmetry of automorphisms versus local symmetry of fibrations}
\label{SS:global_sym}

We now discuss the two different notions of `symmetry'.
The first is the global group-theoretic symmetry
prevalent in physics. The second is the fibration symmetry which is local and usually more suitable for biology.

\subsubsection{Global symmetries}

\begin{figure}[b]
\centerline{%
\includegraphics[width=0.85\textwidth]{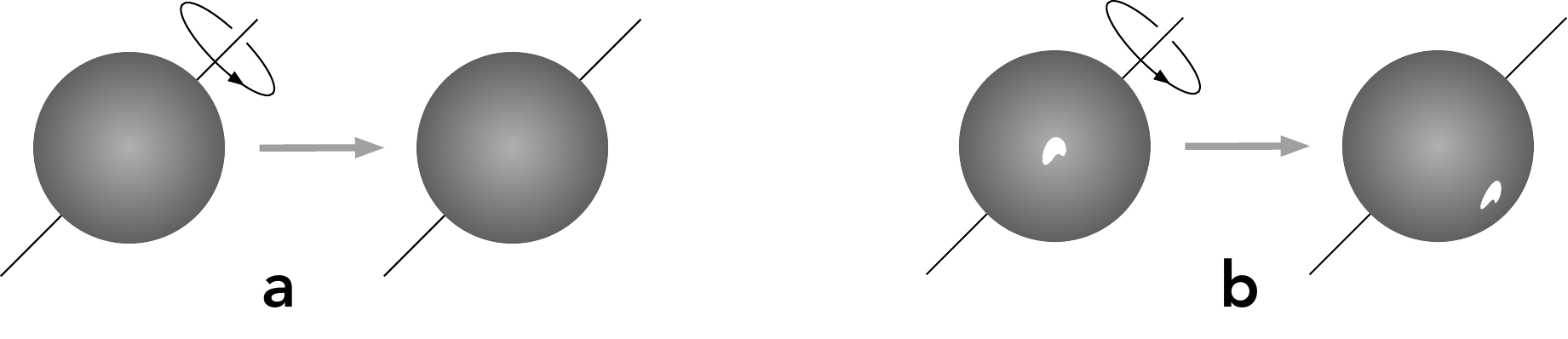}}
\caption{ \textbf{Global symmetry of a sphere. }(\textbf{a}) Rotational symmetry of a sphere. (\textbf{b}) One small hole in the sphere destroys the rotational symmetry.}
\label{F:sphere_sym}
\commentAlt{Figure~\ref{F:sphere_sym}: No alt-text required.
}
\end{figure}

Recall that in mathematics and physics, a `symmetry' of some object normally means
a transformation that preserves specified structural features of that
object. A sphere has `spherical symmetry', meaning that
it looks the same if we rotate it about its center (or
reflect it in a plane through its center); see Fig. \ref{F:sphere_sym}a. 

\begin{figure}[h!]
\centerline{%
\includegraphics[width=0.8\textwidth]{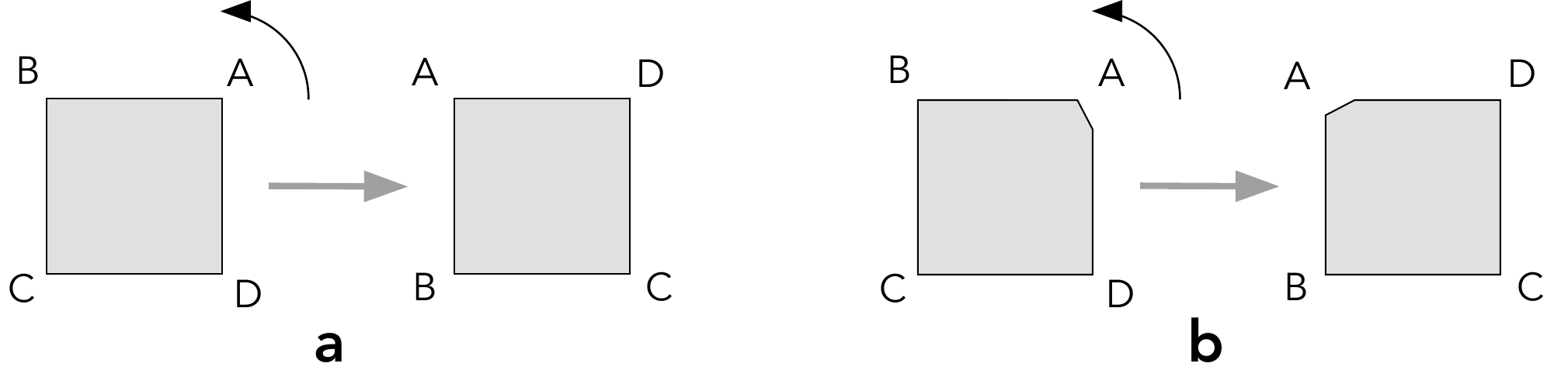}}
\caption{\textbf{Global symmetry of a square.} (\textbf{a}) Rotational symmetry of a square though one right angle. (\textbf{b}) One small change to the square destroys the rotational symmetry.}
\label{F:square_sym}
\commentAlt{Figure~\ref{F:square_sym}: No alt-text required.
}
\end{figure}

Similarly, a square has
`square symmetry', meaning that
it looks the same if we rotate it about its center through
one or more right angles, (or
reflect it in a diagonal or a line bisecting opposite sides); see Fig. \ref{F:square_sym}a. 

The symmetries of any object, in this sense, form an algebraic structure, which, as previously mentioned, is called a `group'.
One key property
of groups is that if two symmetry transformations
are composed---performed one after the other---the result is also a symmetry. This is the {\em closure property}.
For example, if we rotate a sphere and rotate it again, 
the combined effect is the same as some other rotation.
The same goes for a square.

Figures \ref{F:sphere_sym} and \ref{F:square_sym} involve {\em geometric} symmetries, and the symmetry transformations concerned are global motions in space. These global transformations preserve the global structure of the objects.
But the concept of a global symmetry is more general.
In particular,
the corresponding concept for a graph/network is an {\it automorphism}. This is a {\em permutation}
of nodes that preserves the global graph connectivity. It follows that any two automorphisms compose to give
an automorphism---the closure property again---so the set of such permutations forms a group.

\begin{figure}[h!]
\centerline{%
\includegraphics[width=0.8\textwidth]{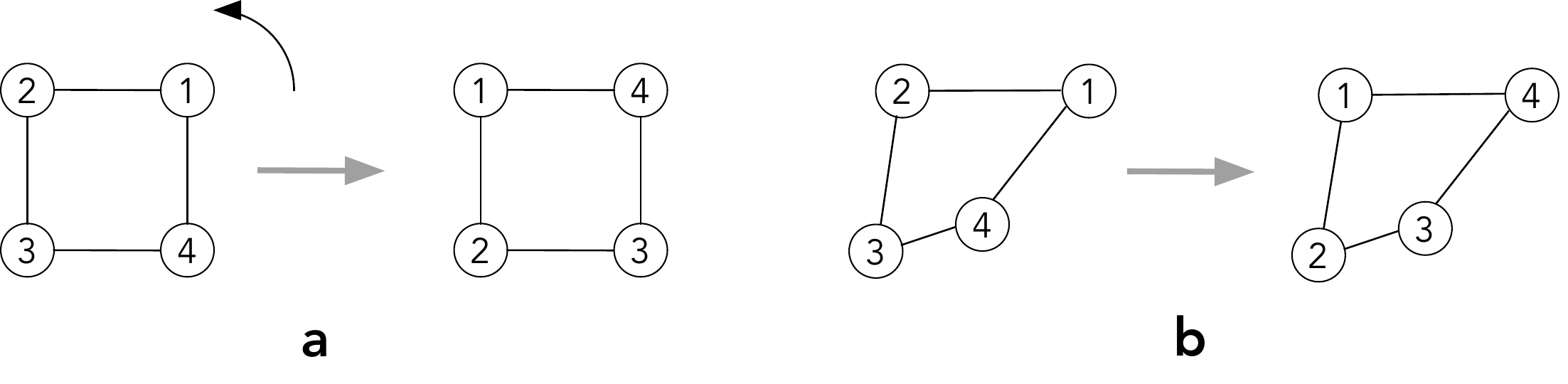}}
\caption{ \textbf{Global symmetry in a graph.} (\textbf{a}) This 4-node graph has the same symmetries as a square. (\textbf{b}) Any graph with the same topology also has the same symmetries as a square,
despite being drawn as a different shape.}
\label{F:square_sym_graph}
\commentAlt{Figure~\ref{F:square_sym_graph}: No alt-text required.
}
\end{figure}

Figure \ref{F:square_sym_graph} illustrates this idea
using a graph with four nodes. In (\textbf{a})
we draw it as a square, which makes the analogy with
a geometric square obvious. In (\textbf{b})
we draw a graph with exactly the same
topology, which no longer looks like a geometric square.
Because the two drawings have the same
topology, they have the same automorphisms.
For example, the geometric right-angle rotation
corresponds to the permutation
\[
\sigma = \left(\begin{array}{llll}
    1 & 2 & 3 & 4\\
    \downarrow & \downarrow & \downarrow & \downarrow \\
    2 & 3 & 4 & 1
\end{array}\right)
\]
For either drawing, this permutation `preserves' connecting edges---that is, it
preserves the graph topology---in the sense that node $i$
is connected to node $j$ if and only if $\sigma(i)$
is connected to $\sigma(j)$. For example, 1 is connected to 2,
and $\sigma(1) = 2$ is connected to $\sigma(2) = 3$.
It is easy to check that the four connections, and only those,
are preserved.

\subsubsection{Local symmetries}

Figs. \ref{F:sphere_sym} and \ref{F:square_sym} show that
even a very small change to a shape can destroy its symmetry. For example, if we cut a tiny hole in the sphere, as in
Fig. \ref{F:sphere_sym}b, it no longer looks
the same after it is rotated. If we cut a tiny piece off a square, as in
Fig. \ref{F:square_sym}b, it no longer looks
the same after it is rotated.
These examples show that a symmetry, in this sense, is a {\em global} property of the entire object. We have seen that in mathematics and physics, global symmetries are formalized using the notion of a {\em group}. 

Although the group notion of symmetry is widely used in physics
and mathematics, there
are features that it fails to capture. For example, most
regions of the sphere in 
Fig. \ref{F:sphere_sym}b
look the same. And three corners of the square in
Fig. \ref{F:square_sym}b are identical right angles. The rigid, global notion of a group symmetry does not include these resemblances. 
So there seem to be `local symmetries', not
captured by the usual notion of symmetry. (One might consider the concept of pseudo-symmetry in this case, as discussed in Section \ref{sec:pseudosymmetry}; however, this alone will go nowhere.)

A major point of this book is that in biology the more flexible notion
of fibration symmetry is more natural and more useful. A fibration can be viewed 
as a system of `local' symmetries, relating nodes in any given fiber. The closure property no longer holds, so these new kinds of symmetries do not form a group. Instead, the fibration
symmetries form a more general structure called a `groupoid'. We develop this viewpoint in more
detail as the book progresses.

\begin{figure}[h!]
\centerline{%
\includegraphics[width=0.7\textwidth]{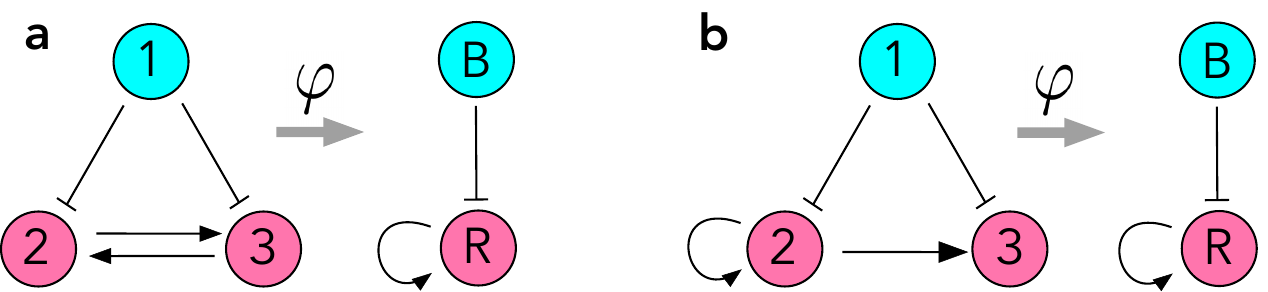}}
\caption{ \textbf{Global and local symmetries.} (\textbf{a}) This graph has global left-right symmetry. Assigning the same color to symmetrically related nodes defines a fibration $\varphi$ to a graph with two nodes. 
(\textbf{b}) This graph has no (nontrivial) global symmetry. However, assigning colors as shown also defines a fibration $\varphi$ to a graph with two nodes.}
\label{F:global_local}
\commentAlt{Figure~\ref{F:global_local}: Left: graph with vertices
1 (blue), 2 and 3 (red). Barred arrows 1 to 2, and 1 to 3.
Sharp arrows 2 to 3, and 3 to 2. Grey arrow labeled `phi' points to graph with
vertices B(blue) and R (red). Barred arrow B to R. Sharp arrow R to R.
Right: graph with vertices
1 (blue), 2 and 3 (red). Barred arrows 1 to 2, and 1 to 3.
Sharp arrows 2 to 2, and 2 to 3. Grey arrow labeled `phi' points to graph with
vertices B(blue) and R (red). Barred arrow B to R. Sharp arrow R to R.
}
\end{figure}

We can illustrate the difference between global
and local symmetries using a graph resembling
Fig. \ref{F:Uxur}, with one edge removed to
simplify the construction. This graph is shown in
Fig. \ref{F:global_local}b. We compare and
contrast it with Fig. \ref{F:global_local}a.

Figure  \ref{F:global_local}a has a global
permutation symmetry
\[
\rho = \left(\begin{array}{lll}
    1 & 2 & 3 \\
    \downarrow & \downarrow & \downarrow  \\
    1 & 3 & 2
\end{array}\right)
\]
which leaves node 1 fixed and interchanges 2 and 3.
This permutation is a global symmetry of the entire network. Geometrically, it reflects the graph left-right.
If we color symmetrically related nodes with the same 
color, node 1 is colored blue while nodes 2 and 3 are colored red. There is now a fibration $\varphi$
to a 2-node base, as shown. 

Figure \ref{F:global_local}b has no global
permutation symmetry (aside from a trivial symmetry:
the identity permutation). If we apply the permutation
$\rho$, we obtain a different graph, so $\rho$ is not
a global symmetry. Despite that, there is a fibration
with the same coloring as in (\textbf{a}). The reason
is that there is a local symmetry: nodes 2 and 3 receive
exactly the same sets of input arrows, of the same types 
and with the same colors for
source nodes. That is, the coloring is balanced. This is a local condition since it affects only the in-neighborhood of nodes 2 and 3, but not the rest of the graph, like node 1 and the outputs of 2 and 3.

The same phenomenon occurs in Fig. \ref{F:Uxur},
which is why that graph has a nontrivial fibration symmetry, 
despite the lack of any global symmetry.

These examples illustrate two key points. First,
we see that:

\begin{keyquote}
Balanced colorings (equivalently, symmetry fibrations) need not be determined by global (group) symmetries.
\end{keyquote}

In full generality, it can be proved that:

\begin{keyquote}
    Every balanced coloring is given by the fibers of a fibration symmetry, and conversely.
\end{keyquote}

Since we have already seen that robust synchrony patterns 
correspond precisely to balanced colorings, it is clear that
fibration symmetry is {\it the} central concept in network synchronization (here, robust, as defined in Section \ref{sec:robust}, is the key word):

\begin{keyquote}
    Every {\it robust} cluster synchrony pattern is given by the fibers of a fibration symmetry, and conversely.
\end{keyquote}

\section{Physics and biology}

The previous sections have introduced, in an informal manner, several
basic concepts related to synchronization in networks.  We end this
chapter by summarizing some of the most important implications for
biological networks. The reasoning behind these statements will be
explained in more detail as the book progresses.

In the physical sciences, the deepest features of the laws of physics
are their symmetries. These symmetries are global, defined by a
symmetry group; they occur because the laws of physics are the same
everywhere and are compatible with group symmetries, which are rigid.
For example, electrons are the same in all molecules; they are
interchangeable and identical.

Biological sciences do not work like that. It is not even clear
whether analogous `laws of biology' exist, but if they do, they must
explain how variability and flexibility lead to robust (in a non-technical sense) biological processes, in spite of various
external and internal changes.  For example, DNA is in every single
cell and works in the same way, yet each species, even each organism,
has a different genome.  All electrons are the same, but every
elephant is different \citep{wuppuluri}.  They belong of the same
species (more precisely, several closely related ones), but they are
not identical.

This suggests that the group concept of symmetry used so successfully
in physics is not entirely appropriate for biology. Group symmetries
are too rigid. In particular, they are easily destroyed by any
perturbation of network structure. It is reasonable to ask whether
there might be some more general and more flexible notion of symmetry
that is appropriate for biology. Unlike electrons, elephants are not
identical, but they are easily distinguishable from mice or cats. This
is some kind of symmetry statement, in a metaphorical sense, at least.

Can the metaphor be formalized? If so, is it useful?

In this book we argue that fibration symmetries provide just such a
notion, at least for biological networks. The appropriate algebraic
analog of a group turns out to be a groupoid, which can be viewed as a
coherent system of local groups.  Group symmetry preserves the global
structure of a physical system. Groupoid symmetry does not preserve
all relevant properties of a system, but it does preserve the system's
synchronized dynamics.  A key feature of the groupoid structure is the
associated notion of a fibration, so we also use the term `fibration
symmetry'.  Fibration symmetries are flexible and robust. And biology
needs to be flexible and robust. It is therefore natural to seek to relate fibration symmetries to biology.

In this book, we use symmetry in this more general sense.  The
`correct' notion of symmetry for biology, we contend, is fibration
symmetry. The distinction is important, but there are some useful
analogies, which help to motivate and organize the theory. For instance, in a network with group symmetry, symmetrically related
objects are in the same orbit of the group. In a network with fibration
symmetry, symmetrically related objects are in the same fiber.
A group symmetry is always a fibration symmetry (but not the opposite).
Every group of automorphisms has a corresponding fibration symmetry,
whose fibers are the group orbits, but the converse need not hold.
Thus fibration symmetries are strict generalizations of automorphisms.
Their local nature makes them more flexible and more robust.  Small
changes to a network can destroy group symmetry completely.  But many
of the fibration symmetries remain.

Part I of the book is mainly directed towards setting up and
explaining the corresponding mathematical machinery, although we use
biological examples when we can.
All very well, but `the proof of the pudding is in the eating'.  The
proof that fibration symmetries are useful in biological networks
requires us to put all this machinery into action, to solve
interesting biological problems. Part II of this book is an attempt to
do exactly that.


\chapter[Graphs, Networks, and Synchronization]{\bf\textsf{Graphs, Networks, and Synchronization}}

\label{chap:definitions}

\begin{chapterquote}
  We now begin to supply  a little more detail and to define concepts
  in precise mathematical terms. We continue to illustrate the
  formal concepts with examples, and avoid mathematical
  complexities where possible.
  This chapter defines the basic concepts that will be used in this
  book, such as the difference between a graph\index{graph } and a network\index{network } (and the
  admissible dynamical systems of equations\index{admissible !system } defining its dynamics),
  the notion of partitions\index{partition } of a graph, and the difference between
  complete\index{synchronization !complete } and cluster synchronization.\index{synchronization !cluster } We also elaborate on an
  important problem treated in this book: the structure-function
  relation in networks.\index{structure-function relation }
  
  \end{chapterquote}

\section{Basic definitions}

To set the scene, we consider a general biological
system composed of $N$ interacting units, such as neurons or genes.
The dynamical behavior of each biological unit is described by a
continuous variable $x_i(t)$, which measures the biological activity of
unit $i$ at time $t$, such as the firing of a neuron \citep{hodgkin1952} or the
gene expression level of mRNA and proteins
\citep{klipp2016book,alon2019}. We use a network representation of
biological systems~\citep{buchanan2010}, where biological units are
represented as nodes, and their interactions as directed links between
nodes. The nodes and links form a {\it biological graph},\index{graph !biological } and when we
attach a dynamical system to every node they form a
{\it biological network (system)}.\index{network !biological }

\begin{definition}
    \textbf{Directed graph}\index{graph !directed } \citep{harary1969,BerGH}.
    A {\em directed graph}\index{graph !directed } $G=(V,E)$ is defined by a set $V = \lbrace
    1,\dots,N\rbrace$ of $N=|V|$ nodes (vertices, cells, compartments, units\ldots)\index{node }\index{vertex }\index{cell }\index{compartment }\index{unit } and a set
    $E\subseteq V\times V$ of $M=|E|$ edges (arcs, links, arrows).\index{edge }\index{arc }\index{link }\index{arrow }
    The set $E$ determines the connections among the nodes. If $E = \{e_1, e_2,
    \dots, e_M\}$, we can write each edge in the form
    $e_\ell = (i, j),$ with $\ell \in
         [1, M]$. This pair of nodes represents a directed connection from node $i \in V$ (the {\em source}\index{source } of the edge) to node $j \in V$ (the {\em target}\index{target } of the edge).
         The source is denoted by $s(e_\ell)$ and the target is
        denoted by $t(e_\ell)$.
\label{graph}
\end{definition}

\begin{remark}
In various parts of the literature, 
 the source (also called \emph{tail})\index{tail } is denoted by ${\bf t}(e_\ell)$ or $\mathcal{T}(e_\ell)$. The target (also called \emph{head})\index{head } is denoted by ${\bf h}(e_\ell)$
    or $\mathcal{H}(e_\ell)$.  We do not use these alternative notations. 

The terms `node' and `link'  are usually used in the complex network
literature \citep{buchanan2010}, while `vertex' and `edge/arc' are employed
in the graph theory literature \citep{harary1969,bollobas2012} especially when referring to undirected networks (i.e., networks where arc direction is irrelevant). The terms `node' and `arrow' are used in \citep{GS2023} and related articles.
In this book, we mainly use `node' and `edge/link', but `vertex' and `arc' are also used, depending on context.
\end{remark}

\begin{definition}
    \textbf{Adjacency matrix of a directed graph.}\index{adjacency matrix !directed graph } Let $G=(V,E)$ be a directed graph. The {\em adjacency matrix} $A$
    of the graph $G$ is a $N \times N$ matrix
    ($N=\left|{V}\right|$) \index{adjacency matrix } with elements
    $A_{ij}$ representing the connectivity of the graph:
    \begin{equation}
        A = 
        \begin{cases}
            A_{ij} = 1 \,\,\, \textnormal{if there exists an edge }\,\,e_\ell = (i, j)\\
            A_{ij} = 0 \,\,\,\,\,\, \textnormal{otherwise}.
        \end{cases}
    \end{equation}
The edge $(i,j)$ represents an {\em output}\index{output } from node $i$ and an
{\em input}\index{input } to node $j$.

    We write $G(A)$ for the graph determined by the
    adjacency matrix $A$.
\end{definition}

\begin{definition}
    \textbf{Undirected graph}.
    A graph $G=(V,E)$ is {\em undirected}\index{graph !undirected } if and only if $(i,j) \in E$ implies $(j,i) \in E$.
    The pair of arcs $(i,j), (j,i)$ is interpreted as a single undirected connection between $i$ and $j$, which is called an {\em edge}.\index{edge } In this context, nodes are often referred to as {\em vertices}.\index{vertex }
\label{graph_undirected}
\end{definition}

In other words, an undirected graph can be thought of as a special type of directed graph,
in which those pairs of nodes that are connected 
are linked by two directed edges, one in each direction.

We shall use the directed and undirected terminologies interchangeably, and the context will usually be enough to distinguish whether we are talking about a directed or an undirected graph. In particular we often use `edge' rather than `arc' for directed graphs.

Undirected graphs are also called `symmetric' graphs, but we do not use this name to avoid confusion.

A graph may be weighted or unweighted.\index{graph !weighted }\index{graph !unweighted } In a {\em weighted} graph each edge $e_\ell$ is
associated with the {\em weight} $\omega_\ell \in \R$.  For an unweighted graph all
interactions are weighted equally and $\omega_\ell = 1$. 

The adjacency matrix of a weighted (directed or undirected) graph
is defined in the same way as that of an unweighted graph, but now we set $A_{ij} = \omega_{ij}$, the
weight of the corresponding edge from $i$ to $j$.

Sometimes, edges can have a type\index{type, of edge} (like in the example of Fig.~\ref{F:Uxur}, where we have repressor and activator edges). Similarly to weights, types provide extra information about
the nature of the corresponding edges. Also nodes can have a type\index{type, of node}: this is especially important for networks that include nodes with different classes (for instance, proteins, genes, metabolites etc.). 

\begin{definition}
    \textbf{Node/edge typing}.
    A \emph{typed graph}\index{graph!typed } is a graph $G=(V,E)$
     endowed with a set $T$ of types, and two functions $\nu: V \to T$ and $\eta: E \to T$ that assigns a type to every node and edge respectively. 
\label{graph_typed}
\end{definition}

Although we are using one single set of types, normally the types of nodes and edges are different. In the following, we can assume that all nodes have the same type, and the same for edges, unless otherwise specified. Some more systematical examples of graphs with different types of nodes and edges is postponed to Section~\ref{sec:heterogeneous}.

{\em Warning}: 
Some authors, such as \citep{GS2023}, define the adjacency matrix
so that $A_{ij}$ represents edges from node $j$ to node $i$; that is,
they use the transpose of the adjacency matrix defined above. This
modification is slightly more natural when constructing admissible equations,
but to avoid possible confusion we do not use it in this book.

It is also
important to define subgraph types since they are useful for
defining building blocks.

\begin{definition}
  \textbf{Subgraph} \citep{harary1969}. A {\em subgraph}\index{subgraph }
  $G'=(N',E')$ of a graph $G=(N,E)$ is a graph such that $N'\subseteq N$ and
  $E'\subseteq E$, and if $e \in E'$ then $s(e), t(e) \in N'$.
  
   That is, $G'$ is a subgraph of $G$ if $G'$ consists of
  some nodes and some edges of the nodes and edges of $G$, and the
  sources and targets of edges are among the nodes concerned. 
  
  \label{subgraph}
  \end{definition}

\begin{definition}
  \textbf{Induced subgraph}\index{subgraph !induced } \citep{harary1969}.  An {\em induced
  subgraph} $G'=(N',E')$ of $G=(N,E)$, induced by the vertex set $N' \subseteq
  N$, is the graph $G'=(N', E')$ such that $E'=\{ e = (n_1,n_2) \in E
  \, | \, n_1,n_2 \in N'\}$. The subgraph of $G$ induced by
  $N' \subseteq N$ is denoted by $G[N']$ and contains {\em all} of
  the edges in $E$ that connect nodes in $N'$.
\label{induced}
\end{definition}

The difference between an induced subgraph and a subgraph is that an
induced subgraph includes all the edges connecting the nodes in it,
while a subgraph can omit some edges. This definition will be useful
when defining the building blocks of the network in Part II, since such a
circuit can be studied either in isolation from the original graph or
embedded on it.

We end this section with two further concepts that we use throughout the book:

\begin{definition} \label{def:path}{\bf Directed path.}
A {\em (directed) path}\index{path } in a directed graph is 
any sequence of directed edges such that the target/head node of each edge is the source/tail node of the next. Repetitions of nodes or edges are permitted. 
\end{definition}

In graph theory the term `walk' is often used for this concept, a `trail' is a walk with no repeated edges, and a `path' is a walk with no repeated vertices. In this book, we use `path' rather than `walk'. SCCs will be used in Part II in the analysis of biological networks.

\begin{definition} \label{def:SCC}
{\bf Strongly connected component (SCC).}
A directed graph is {\it strongly connected}\index{strongly connected } if there is a (directed) path between all
pairs of vertices.  A {\it strongly connected component}\index{strongly connected component } (SCC)\index{SCC } of a directed
graph is a maximal induced subgraph that is strongly connected.
\end{definition}

For brevity, we often omit the adjective `strongly'
when this does not cause confusion, and talk of `connected components' of
a directed graph. In the graph theory literature this term usually refers to an
analogous concept based on undirected paths, but we do not need
that concept in this book since it is not related to network dynamics in any useful manner.

\section{Graphs, multigraphs, and hypergraphs}
\label{SS:GMH}
   \index{graph }\index{multigraph }\index{hypergraph }
Sometimes we need to allow graphs to have multiple edges between 
a given pair of nodes, in part because in some cases a fibration can collapse a graph without multiple links to
a base that has multiple links \citep[Sections 8.10, 10.6]{GS2023}.
This is already apparent in Fig. \ref{fig:synchronyfibers}. The graph
$G$ has no loops\index{loop } or multiple edges,\index{multiple edge } but the base $B$ has both.
These features are important to make the admissible maps for the
base correspond precisely to synchronized restrictions of admissible maps
for the graph.

To allow for multiple edges\index{multiple edge } we need a slightly more general notion of graph:

\begin{definition}
  \textbf{Multigraph} \citep{berge-hypergraph}. A {\em multigraph}\index{multigraph } $G=(V,E,s,t)$
  consists of a set of nodes $V$ and a set of multiedges\index{multiedge } $E$ allowing for multiple
  edges (parallel edges) between two nodes (while a graph allows only
  one edge connecting any two given nodes). The two functions $s,t: E \to V$ describe the source and the target node of each edge: the directed edge $e\in E$ has source $s(e)$ and target $t(e)$.
  \label{def:multigraph}
\end{definition}
A graph can always be seen as a multigraph in which each edge has
multiplicity 1, so the notion of a multigraph is in fact an extension of Definition~\ref{graph}.  The notions of edge weight and/or type carry over to multigraphs.

In the following, it is convenient to use the term `graph' for either a graph in the strict sense or a multigraph, and we do this when it is clear which concept is intended; in this case, we say \emph{simple graph}\index{graph !simple } to mean a graph according to Definition~\ref{graph}. 

In a simple graph, two nodes uniquely
define one edge, so we can use the simplified notation $e_\ell = (i, j)\in
E$, which is equivalent to letting $s(e) = i$ and $t(e) = j$. This notation is
unambiguous because in a simple graph the source
and target nodes uniquely define the edge, which is not the case for a
multigraph.

It is sometimes asserted that since the network formalism of \citep{GS2023}
is based on graphs and multigraphs, it
describes only two-body interactions.\index{interaction !two-body } On the contrary, 
the dynamic of a node depends on {\em all} of its input nodes,
and can be any function of their states; it is not limited to
a linear combination of pairwise interactions.
This misconception may have arisen through confusion with standard `linear interaction' models in which
the interaction term is additive in the input variables and the interaction strengths
are prescribed by an adjacency matrix.

The main issue here is not whether many-body interactions\index{interaction !many-body } are permitted,
but what extra constraints can be imposed on them. A systematic way to
do this is to use hypergraphs\index{hypergraph } \citep{berge-hypergraph, gracht2023}, which we discuss in
Chapter \ref{chap:hypergraph}.

Most of the examples in the book are treated with
simple graphs and multigraphs unless otherwise noted. Multigraphs are important since
the application of a fibration to a graph usually leads to a graph
with multiple parallel edges. Hypergraphs are treated in Section
\ref{S:hypergraph}.  Examples of the different types of graphs can
be seen in Fig.~\ref{fig:example}.

\begin{figure}[ht]
	\centering
        \begin{tabular}{ccccc}
	\begin{adjustbox}{valign=c}
		\begin{tikzpicture}[every node/.style={draw, circle}, scale=.6, ->, >=latex]
			\node (A1) at (1,4) {$1$};
			\node (A2) at (0,2) {$2$};
			\node (A3) at (2,2) {$3$};
			\node (A4) at (0,0) {$4$};
			\node (A5) at (2,0) {$5$};
			\draw (A2) -- (A1);
			\draw (A1) -- (A3);
			\draw (A2) -- (A4);
			\draw (A5) -- (A4);
			\draw (A5) -- (A3);
			\draw (A2) to[bend left] (A3);
			\draw (A3) to[bend left] (A2);
			\path (A5) edge[loop right] (A5);
		\end{tikzpicture}
	\end{adjustbox}
&
	\begin{adjustbox}{valign=c}
		\begin{tikzpicture}[every node/.style={draw, circle}, scale=.6, ->, >=latex]
			\node (A1) at (1,4) {$1$};
			\node (A2) at (0,2) {$2$};
			\node (A3) at (2,2) {$3$};
			\draw (A2) -- (A1);
			\draw (A1) -- (A3);
			\draw (A2) to[bend left] (A3);
		\end{tikzpicture}
	\end{adjustbox}
&
	\begin{adjustbox}{valign=c}
		\begin{tikzpicture}[every node/.style={draw, circle}, scale=.6, ->, >=latex]
			\node (A1) at (1,4) {$1$};
			\node (A2) at (0,2) {$2$};
			\node (A3) at (2,2) {$3$};
			\draw (A2) -- (A1);
			\draw (A1) -- (A3);
			\draw (A2) to[bend left] (A3);
			\draw (A3) to[bend left] (A2);
		\end{tikzpicture}
	\end{adjustbox}
&
	\begin{adjustbox}{valign=c}
		\begin{tikzpicture}[every node/.style={draw, circle}, scale=.6, ->, >=latex]
			\node (A1) at (0,3) {$1$};
			\node (A2) at (0,0) {$2$};
			\node (A3) at (2,0) {$3$};
			\draw (A2) to[bend right=20] (A1);
			\draw (A1) to[bend right=20] (A2);
			\draw (A1) to[bend right=40] (A2);
			\draw (A2) -- (A3);
			\path (A3) edge[loop right] (A3);
			\path (A3) edge[loop above] (A3);
		\end{tikzpicture}
	\end{adjustbox}
&
	\begin{adjustbox}{valign=c}
		\begin{tikzpicture}[every node/.style={draw, circle}, scale=.6, every loop/.style={}]
			\node (A1) at (1,4) {$1$};
			\node (A2) at (0,2) {$2$};
			\node (A3) at (2,2) {$3$};
			\node (A4) at (0,0) {$4$};
			\node (A5) at (2,0) {$5$};
			\draw (A2) -- (A1);
			\draw (A1) -- (A3);
			\draw (A2) -- (A4);
			\draw (A5) -- (A4);
			\draw (A5) -- (A3);
			\draw (A2) to[bend left] (A3);
			\draw (A3) to[bend left] (A2);
			\path (A5) edge[loop right] (A5);
		\end{tikzpicture}
	\end{adjustbox}
\\[5ex]
(a) & (b) & (c) & (d) & (e)
\\[5ex]
\end{tabular}
	\caption{\textbf{Simple examples of graphs.} 
        (\textbf{a}) A graph $G$ with 5 nodes and 8 edges (one of them is a self-loop on node $5$).
        (\textbf{b}) A subgraph of $G$.
        (\textbf{c}) An induced subgraph of $G$ (the subgraph induced by $\{1,2,3\}$).
        (\textbf{d}) A multigraph. Node $3$ has two loops, and there are two parallel arcs from $1$ to $2$.
        (\textbf{e}) An undirected multigraph (every edge should be interpreted as a pair of arcs in opposite directions; the loop itself is just a loop).
        }
	\label{fig:example}
\commentAlt{Figure~\ref{fig:example}: (a) Graph with vertices 1-5.
Arrows: 1 to 3; 2 to 1,3,4; 3 to 2; 5 to 3,4,5. (b) Graph with vertices 1-3.
Arrows 1 to 3, 2 to 1 and 3. (c) Graph with vertices 1-3.
Arrows 1 to 3, 2 to 1 and 3, 3 to 2. (d) Graph with vertices 1-3.
Arrows 1 to 2 (two arrows), 2 to 1 and 3, 3 to 3 (two arrows).
a) Graph with vertices 1-5. Undirected edges: 1 to 2 and 3; 2 to 3 and 4; 
3 to 5; 4 to 5; 5 to 5.
}
\end{figure}

\section{Setting up model equations}
\label{SS:SMEq}
\index{model equation, construction of }
In many areas of science, networks are used to set up model equations.
Typically, nodes represent state variables and edges represent
couplings.  The network structure is a modeling assumption, which
generally represents a simplification of a more complex reality.
As we said at the beginning of this section,  each node $i$ in the graph comes with its state variable
$x_i(t)$ (the state of node $i$ at time $t$), which is an element of some node space $M_i$. 

The model equations are often selected from a small catalog of
equations that are standard in each specialized area
\citep{buchanan2010}---for example, in neuroscience, typical model
neurons are governed by
FitzHugh--Nagumo~\citep{fitzhugh1961,nagumo1962},\index{FitzHugh--Nagumo equation }
Morris--Lecar~\citep{morris1981},\index{Morris--Lecar equation } or Hodgkin--Huxley~\citep{hodgkin1952}\index{Hodgkin--Huxley equation }
equations. In systems biology, Hill functions\index{Hill function } are used to model
binding interactions between protein and DNA in gene regulatory
networks \citep{alon2019}, while Michaelis--Menten kinetics\index{Michaelis--Menten kinetics } describe
biochemical reactions in metabolic networks
\citep{klipp2016book}. Couplings are also standardized, for example as
additive voltage coupling or multiplicative gates in gene regulation
\citep{alon2019}.

The manner in which the network embodies the model equations
often differs between areas of application. An example is
the distinction between activator and repressor edges,
with (fairly) standard symbols for each. In molecular reaction
networks,\index{network !molecular reaction } edges may point to other edges, not nodes.
When conservation laws\index{conservation law } such as mass balance\index{mass balance } apply, an edge
sometimes indicates an output from its source as well as an
input to its target. These conventions must be borne in mind
when comparing network diagrams in different areas.
The network diagram may then
take on a different appearance from that assumed in 
this book, which represents the structure of
the {\em influence network}:\index{network !influence } which
node variables affect which other node variables.

\section{Admissible ODEs}\index{admissible !ODE }
\label{sec:adODEs}
Although the models are standard in a given area, they differ from one
area to another. Moreover, the model may be an ordinary differential
equation (ODE), a delay differential equation, a stochastic
differential equation, or have other (for instance, probabilistic) features. Further,
the dynamic can be continuous time (differential equation) or discrete time
(iterated map). A general theory has to encompass this variety of
models, but it also has to avoid being {\em too} general.  We
therefore, restrict attention mostly to ODE models, although iterative
maps will also be treated in few occasions. Because of the variety of models encountered in
different areas, a key feature of the general formalism is that a
graph determines an entire {\em class} of differential equations: 

\begin{definition}
\label{D:admissible_ODE_def}{\bf Admissible differential equations} \citep{stewart2003,stewart2006,GS2023}.
An ODE is {\em admissible}\index{admissible !ODE } for a graph $G$ if it respects the graph structure
of $G$. That is, the component of the ODE for node $c$ has the form
\begin{equation}
\label{E:admissible_ODE_general}  
\dot x_c = f_c(x_c, x_{i_1}, \ldots, x_{i_k})
\end{equation}
where the $i_j$ are the source nodes of the input edges to $c$.
The function $f_c$ is called the {\em input function} at node $c$, and it specifies how the inputs $(x_{i_1}, \ldots, x_{i_k})$ are combined to determine the dynamics at node $c$. The first variable in $f_c$ is the node $c$, specifying the internal dynamics of the node.
Moreover, if nodes $c$ and $d$ have isomorphic input sets
then $f_c=f_d$ (the ordering of the input edges needs to be
consistent with the isomorphism). This means that $f_c$ and $f_d$ represent the same input functions (same type and same coefficients defining the interaction edge).
\label{admissible}
\end{definition}

The admissible ODEs are not specific equations; they define the class of
{\em all} ODEs that respect the graph structure.
The specific equations that are standard in any given area belong to the class of admissible equations---provided they are ODEs---so general results can be specialized to
those models. However, special models can have additional features,
not present in every admissible equation\index{admissible !equation }. Examples of such features 
are the mass-balance
condition in chemical kinetics, diffusive or generalized diffusive coupling,
and the widespread use of linear coupling. 

The definition of admissible equations generalizes readily to other kinds of dynamic:
discrete time dynamics (iterated functions), finite-state automata, Boolean
networks, and so on, although a general theory is currently less well developed in these contexts.

\subsection{From graph to admissible ODEs}
\label{S:Uxur_admiss}

The concept of admissible equations assumes that the equations of the
dynamical system respect the structure of the graph. In particular,
the symmetries (group or fibration) of the graph are respected by the dynamical system
too. 

The admissible equations are a set of all models compatible with the graph. However, most of the time, we are interested in a particular ODE that represents the biological system of study. In this case, there is a one-to-one relation between the graph or the hypergraph
and the ODE: we can write the form of the
dynamical equations using only the information contained in the
graph. Each edge in the graph represents one type of coupling or
interaction variable in the dynamical system, and the information
contained in each edge is enough to specify the form of the dynamical
equation (modulo some general considerations) It is also necessary to choose the node spaces $M_i$:
the structural form of admissible ODEs is the same for all such choices,
but the meaning of the node coordinate $x_i$ differs. Examples of these are discussed in Section \ref{sec:hypergraph-metabolic}.

\begin{definition}{\bf Network dynamical system} \citep{stewart2003,stewart2006}.
    We define a {\it network dynamical system}\index{network !dynamical system } corresponding to
   the graph $G=(V,E)$ by attaching to every node $i\in G$ a
    {\em state}\index{state } determined by a variable $x_{i}(t) \in M_i$ taking values in
    a finite-dimensional vector space $M_i$. In most cases treated here
    we use $M_i=\mathbb{R}$. Moreover, the state $x_i(t)$ evolves following the
    set of admissible equations for the graph under the influence of
    the nodes specified by the graph.  
\label{network}
\end{definition}

Examples of biological networks defined on an underlying graph,
multigraph or hypergraph are: 
\begin{itemize}
\item
The transcriptome,\index{transcriptome } which
encompasses the activity of all RNA transcripts interacting on an
underlying transcriptional regulatory graph. 
\item
The proteome,\index{proteome } which
encompasses the activity of all proteins expressed in the cell
interacting on an underlying protein interaction network. 
\item
The metabolome,\index{metabolome } made of small molecule chemicals and metabolites
interacting in a metabolic graph.
\item The neural dynamical
network of brain activity produced by neurons interacting in the
underlying graph connectome.\index{connectome } 
\end{itemize}
These networks are coupled via other
biological processes, such as signaling pathways and the environment. The
functionality of the cell then emerges at the system level through the
integration of these basic `omics' networks.

Conversely, the form of admissible ODEs specifies the
connections and distinguishes their types, with two caveats.
First, in the general network theory we can say when two edges have 
the same types, or different types, but not what those types are.
All we know is when they are the same and when they differ. 
Two nodes whose input sets are in a one-to-one correspondence
that preserves types are `input isomorphic',
\index{input! isomorphic } and the
correspondence is an `input isomorphism'.\index{input! isomorphism}
Second, if two nodes are {\it not}
input isomorphic, there is no formal relation between the
input edges of one node and those of the other. The next example
illustrates these statements.
Definition \ref{def:input_iso} provides a formal definition.

\begin{example}\em
Consider the network of Fig. \ref{F:Uxur}. Admissible ODEs have the form
\begin{equation}
\label{E:Ecoli_model4dup}
\begin{array}{rcl}
\dot x_1 &=& f(x_1,x_2)\\
\dot x_2 &=& g(x_2,x_1,x_2)\\
\dot x_3 &=& g(x_3,x_1,x_2)
\end{array}
\end{equation}
for arbitrary input functions $f,g$ with suitable domain and codomain.
\end{example}

We now explain why that form follows from \eqref{E:admissible_ODE_general}, i.e., it is admissible for the graph.
As discussed in Section \ref{sec:SFEC}, the input sets of the nodes are:
\begin{equation}
\label{E:ISexample}
\begin{array}{rcl}
\mbox{Node}\ 1: & & \{e_2\}.\\
\mbox{Node}\ 2: & & \{e_1,e_4\}.\\
\mbox{Node}\ 3: & & \{e_3,e_5\}.
\end{array}
\end{equation}
We see that node $1$ is not input isomorphic to the other two nodes, 
because its input set consists of a single edge,
whereas nodes 2 and 3 have two input edges.
Nodes
$2$ and $3$ are input isomorphic to each other:
the correspondence
\[
e_1 \leftrightarrow e_3 \quad e_4 \leftrightarrow e_5
\]
is one-to-one and corresponding edges have the same type.
In this case, $e_1$ and $e_3$ are inhibitors, while $e_4$
and $e_5$ are activators.

Therefore we use a function
$f_1=f$ for the $\dot x_1$ equation, and a {\em different} function
$f_2=f_3=g$ for the $\dot x_2$ and $\dot x_3$ equations.
The fact that $f_2=f_3=g$ is important. We can see from the graph that they both represent an input function receiving one inhibitor edge (from node 1) and one activator edge (from node 2). Thus, $f_2$ and $f_3$  are the same input functions.
However, a further (somehow hidden) assumption is that these input functions have the same coefficients defining the interaction terms. We will come back to this issue below and in Chapter \ref{chap:alive}.

The behavior of node $1$ depends on its own state $x_1$ (this is indicated by the first variable in $f(x_1,x_2)$), and on that of
the only node that is the source of its input edge $e_2$, which is node $2$ (indicated by the second variable in $f(x_1,x_2)$).
In the $\dot x_1$ equation, we apply $f$ to these node variables to obtain:
\[
\dot x_1 = f(x_1,x_2).
\]
The behavior of node $2$ depends on its own state $x_2$, and on that of
the two nodes that are the sources of its input edges $e_1,e_4$, which are nodes 1 and 2 respectively. 
In the $\dot x_2$ equation, we apply $g$ to these node variables:
\[
\dot x_2 = g(x_2,x_1,x_2).
\]
The behavior of node $3$ depends on its own state $x_3$, and on that of
the two nodes that are the sources of its input edges $e_3,e_5$, which are nodes $1$ and $2$ respectively. 
In the $\dot x_3$ equation, we apply $g$ to these node variables:
\[
\dot x_3 = g(x_3,x_1,x_2).
\]
Here, it is important to list the variables for nodes $2$ and $3$ in the
order that gives the input isomorphism, which, in this
case is: node, repressor, activator.

The result is the form of admissible ODEs stated in \eqref{E:Ecoli_model4dup}.

\subsection{From admissible ODEs to graph}

We can also `reverse-engineer' the network from the form
of admissible ODEs,\index{admissible !ODE } subject to some technical remarks below. 
Since there are three equations, there are
three corresponding nodes. The variables inside the function on the right-hand side specify the source nodes of input edges. By convention,  the
first variable represents the node itself. For node $1$ there is
only one input edge, with source $2$, so there is an edge from $2$ to $1$.
Similarly, the variables in the equations for nodes $2$ and $3$ list
the sources of their inputs in corresponding order. One
subtlety of the general theory is that when a node has several input edges
of the same type, the function is symmetric under all permutations of those edges. (These permutations are input {\it automorphisms}; that is, 
isomorphisms of the input set with itself; the form of the ODE
must be invariant under all input isomorphisms, and that includes
the input automorphisms. Another term is {\em vertex symmetry}.) Since we have not specified any such symmetries
for the function $g$, the two input edges are of different types.

A further subtlety is that although admissible ODEs distinguish
different types of edge, these types do not themselves tell us which are 
repressors and which are activators. This level of detail comes into
play only when we specialize the ODE to more specific models.
Similarly, because node $1$ is not input isomorphic to nodes $2$ and $3$,
the admissible ODE does not tell us that node $e_2$ is `the same as'
nodes $e_1$ and $e_3$, so it does not specify that this edge is a repressor.
Again, that information must be provided by specializing  the ODE model.

It is common to associate individual {\it terms}
in a model ODE with the edges (and with the nodes). Often, inputs
are combined by adding them (or, in some cases, multiplying them). In such cases,
the model also contains information on the type of edge---say
using a positive coefficient for an activator and a negative one for a repressor. However, in other areas, this kind of additive structure for the inputs
is not appropriate, which is why the general theory does not assume it. In general, inputs can be combined in any form, representing a multi-body interaction that can be described in more detail by a hypergraph if desired (Section \ref{S:hypergraph}).  The general form \eqref{E:admissible_ODE_general} is valid for graphs and hypergraphs. 

If there exist different types of interactions between nodes,
this should be reflected in the graph. For instance,
genetic networks are composed of activators and
repressors, and neural networks contain excitatory and
inhibitory interactions, so two classes of edges are required to
describe these networks. When the network is unweighted and untyped, all
edges are the same. However, when the edges are weighted/typed, representing,
for instance, different strength/type of interactions, then each edge
should, in principle, come with its specific weight/type. We treat these cases in detail
in Chapter \ref{chap:alive}. 

The question arises: why do we not use a hypergraph to describe the network in Fig. \ref{F:Uxur}? After all, nodes 2 and 3 receive two inputs each, and we should specify how these inputs are combined through input functions specified by a hypergraph. However, the key assumption is that both nodes combine their inputs in the same way, i.e., they are input isomorphic. This is why we use the same function $g$ to specify the input functions at both nodes. Additionally, it is biologically 
plausible---though not true in general---that edges $e_1$ and $e_3$ represent the same interaction terms in the ODE (not only the same type, like activator/inhibitor but also with the same coefficients defining the input function, eg, the same Hill function with the same coefficient and combined in the same way) since they originate from the same source node (the same applies to edges $e_4$ and $e_5$). Therefore, the complexity of the hypergraph representation can be effectively simplified to a standard graph structure. 

However, if the input functions at these nodes differ, then the same function $g$ cannot be used for 2 and 3, and we may need to employ a hypergraph to accurately describe how the inputs are combined at each of these nodes. This discussion may sound a bit cryptic at this point, but it will make sense when we apply the concepts to specific biological networks in Part II; we dedicate Section  \ref{sec:hypergraph-metabolic} to an analysis of this  issue.

\section{Partition of a graph}
\label{sec:partition}

To understand important concepts in this book such as cluster
synchronization, orbits, fibers and balanced colorings/equitable
partitions, it is useful to introduce the concepts of a partition and
the corresponding graph coloring. Formally:
\begin{definition}
  \textbf{Partition of a graph.}\index{partition } Let $G=(V,E)$ be a graph with
$N=|V|$ nodes. Let $c_1, \ldots, c_k$ be
subsets of nodes. Then 
  $\mathcal{P} = \{ c_1, \ldots , c_K \}$ is a \emph{partition} of the graph (and the sets $c_k$ are the \emph{parts} or \emph{clusters}\index{cluster } of the partition $\mathcal{P}$)
  if $\bigcup_{k=1}^K c_k=V$, $c_k \neq \emptyset$, $c_k \cap c_l =\emptyset $ for $k \neq
  l$. Clearly, $N= \sum_{k=1}^K |c_k|$ where $|c_k|$ is the number of
  nodes in cluster $c_k$.
  \label{partition}
\end{definition}

In other words, $\mathcal{P}$ partitions the graph into non-overlapping (or disjoint)
non-empty sets
of nodes, $c_k$, which are the clusters (or parts) of the partition. If
$|c_k|=1$, then the cluster is  {\it trivial}; {\it
  nontrivial} clusters contain more than one node. A useful way to
visualize a partition is to assign a color to each cluster,
creating a colored graph.

\begin{definition} {\bf Colored graph.}
    A \emph{(node-)colored graph}\index{graph !colored } with $K$ colors 
    $\{\phi_1, \phi_2, \dots, \phi_K\}$ is a pair $(G, \phi)$ comprising a graph
    $G = (V, E)$ and a function $\phi : V \rightarrow
    \{\phi_1, \phi_2, \dots, \phi_K\}$ that
    maps nodes of $G$ to their colors.
\label{def:color}
\end{definition}

Intuitively, this definition provides a convenient way to visualize the partition. 
Clearly, the two notions are equivalent: every partition into $K$ clusters defines a coloring\index{coloring } (using $K$ colors, and coloring two nodes with the same color if and only if they belong to the same cluster); and
every coloring can be interpreted as a partition (which gathers all nodes with the same color in the same cluster). We are here assuming, without loss of generality, that $\phi$ is surjective (i.e., that all colors are used). 

We have not imposed any restrictions or constraints on the assignment of
colors.  Imposing constraints creates different types of {\it coloring
  problems}, like the balanced colorings/equitable partition to be
treated in Chapter \ref{chap:fibration_1}, or the famous (and unrelated
to balanced coloring) {\it vertex coloring problem} which consists of
coloring the vertices of an undirected graph in such a way that no two adjacent vertices
have the same color.  In what follows, we consider two main types
of partitions/colorings of the graph: orbits, discussed in section
\ref{sec:orbit}, and balanced colorings/equitable partitions/fibers, which are equivalent
to each other and are discussed in Chapter
\ref{chap:fibration_1}.  These partitions are related to each other,
and each leads to cluster synchronization in its own way.

\section{Complete synchronization and cluster synchronization}
\label{sec:cluster}

Drawing a conclusion about  network dynamics, and in particular
 synchronization, based on the symmetries of the graph is a wide
field of research. The problem of generic synchronization can be
studied for the general set of equations (\ref{E:admissible_ODE_general}). It is important
to distinguish between complete synchronization and cluster
synchronization.

\begin{definition}
\label{def:complete}
\textbf{Complete synchronization.} \label{complete} {\em Complete
synchronization}\index{synchronization !complete } for the system of equations \eqref{E:admissible_ODE_general}, admissible for a graph
$G=(V,E)$ with $N=|V|$ nodes, arises when
\begin{equation}
    x_1(t)= x_2(t)= \cdots = x_N(t) \ .
\end{equation}
Such a state is {\em completely synchronous} or
{\em completely synchronized}.

A network is {\em completely synchronizable}\index{synchronizable, completely } if some 
open set of initial conditions converges to a completely synchronous state. That is, there is a dynamically stable, completely synchronous state.
\end{definition}
Complete synchronization refers to the {\it existence} of completely synchronous solutions, while complete synchronizability refers to the ability of the network to achieve a {\it stable} completely synchronous state \citep{pecoraMSF}.

Synchronization and synchronizability have been extensively
studied in the engineering, nonlinear dynamics and physics communities
\citep{Belykh2008,Do2012,Dahms2012,Fu2013,Kanter2011,pecoraMSF,Rosin2013,Sorrentino2007,strogatz2018}. Some of these papers consider the more general phenomenon of
cluster synchronization, which is  a collective behavior of dynamical systems
in which only certain subsets (modules, clusters) of nodes undergo the same
evolution. Following
\citep{stewart2003,lerman2015b,nijholt2016,pecora2016b,siddique2018symmetry} 
we define cluster synchrony as
follows:

\begin{definition}
  \textbf{Cluster synchronization.} {\em Cluster synchronization}\index{synchronization !cluster } for the
  system of equations \eqref{E:admissible_ODE_general}, admissible for a graph $G=(V,E)$, arises when
  there exists a graph partition $\mathcal{P} = \{ c_1, \ldots , c_K \}$ such that
\begin{equation}
  x_i(t)= x_j(t)\,\,\, \mbox{for} \,\, i,j
  \in c_k, \,\, k=1, \dotso , K \, . 
  \end{equation}
\label{cluster}
\end{definition}
In the language of \citep{stewart2003}, cluster synchronization defines
a {\em synchrony subspace}\index{synchrony !subspace } or {\em polydiagonal subspace}\index{polydiagonal subspace}
\[
\Delta_\phi = \{x : x_c=x_d\ \mbox{whenever}\ c\ \mbox{and}\ d\ \mbox{are synchronized}\}.
\]
Another term is {\em cluster synchronization manifold}.\index{cluster synchronization manifold }
The key feature is that for balanced colorings,\index{coloring !balanced } and only those colorings,
this subspace is invariant under any admissible ODE. 
Intuitively, this means that
nodes that initially lie in the same cluster remain in the same cluster as time passes. (However, they may not do so stably.
Perturbations may destroy the synchrony pattern.)
The dynamics of the full network, when restricted to the synchrony subspace, represent
the dynamics of the clusters.

\subsection{Laplacian systems}
\label{sec:laplacian} 

In the graph theory literature, the Laplacian matrix of a graph
 is at least as important as that of the adjacency matrix. 
Indeed, a huge amount of work on synchronization in networks has been carried out for Laplacian systems. 
 We, therefore, give a brief and highly simplified description of
 this approach, and then explain why we do not
 adopt it in this book.
 
 The {\em Laplacian (matrix)} $L$ of a graph 
 (simple, undirected, and unweighted) is defined by:
 \[
 L_{ij} = \left\{ \begin{array}{ccl}
 	{\rm deg}(i) &\mbox{if} & i = j \\
	-1 &\mbox{if} & i\neq j\ \mbox{and}\ i\ \mbox{is adjacent to}\ j
 \end{array}\right.
 \]
 where $i,j$ runs through the nodes. Here, deg is the in-degree, which equals the number of adjacent nodes. The Laplacian can also be written as $L=D-A$, with $D$ the degree matrix and $A$ the adjacency matrix.

Suppose that the dynamic of a network is defined by diffusive couplings, meaning  that all
interactions are additive and of the form $H(x_i)-H(x_j)$,
for some possibly nonlinear function $H$, with the
same $H$ for every node. 
Then
the interaction between synchronous nodes, where $x_i=x_j$,
is $H(x_i)-H(x_i)$, which is zero. That is,
synchronous pairs of nodes {\em decouple}. Typically, the
linear part of the model ODE is then determined by the Laplacian.

Assume that all nodes have the same state space,
and that the internal dynamics of
all nodes are identical: $\dot x_i = F(x_i)$ for a fixed $F$.
Then, the model ODE takes the form
\begin{equation}
\label{E:MSFdup}
\dot x_i = F(x_i) - S \sum_{j\neq i} A_{ij}(H(x_i)-H(x_j)), \quad i=1,...,N,
\end{equation}
where $A$ is the adjacency matrix and $S$ a coupling constant.

Assuming $H(x)$ is linear, we obtain `Laplacian' dynamics
\begin{equation}
\label{eq:laplacian}
\dot x_i = F(x_i) - S \sum_{j\neq i} L_{ij} H(x_j), \quad i=1,...,N.
\end{equation}
(See Section \ref{S:MSFSI} for more detail.)
Interactions of the form (\ref{E:MSFdup}) and (\ref{eq:laplacian}) imply that a completely
synchronized state {\em always} exists, no matter what the
topology of the graph may be \citep{pecoraMSF}. 
The reason is that synchronous nodes decouple dynamically
for this kind of interaction, unlike more general couplings.
To see why, start with a completely
synchronous initial condition $x_i(0)=x_j(0)$. Then
each node satisfies the same ODE $\dot x_i = F(x_i)$
with the same initial condition on $x_i(0)$, so
$x_i(t)=x_j(t)$ for all times $t$. 
In other words, the {\em existence} of a completely synchronous state
is built into the model from the start.
From the general viewpoint, the coloring with all nodes having the same
color is not balanced. This can be dealt with by introducing a notion of balance tailored to such models: ignore input arrows whose
source and target have the same color.

Whether this synchronous state is {\em stable} is another issue, and
much work has been done on that question;
see Section \ref{S:MSFSI}, which explains how the Laplacian
enters into the stability analysis. 
Similar remarks apply to another widely used type of
model, the Kuramoto model \citep{K84}.

In this book, we assume more general interactions than those described by diffusive couplings and Laplacians.
The existence of a {\em completely}
synchronous state then depends on the topology of the network 
and often does not exist. Cluster synchrony may occur even when
complete synchrony does not, which is why we focus on clusters.

Moreover, as discussed in section \ref{sec:informal}, few systems in
biology have interaction terms like $H(x_i) - H(x_j)$ 
(a notable exception being Turing-type reaction-diffusion models of pattern formation
and development), and complete synchronization is not common (and deleterious) in biological networks. In particular, none of
the biological systems studied in this book is diffusive. Thus, these systems are not described by the Laplacian $L$ but by the adjacency matrix $A$. This remark includes genetic, protein, metabolic, signaling, ecological, and brain networks, as well as artificial neural networks.

For Laplacian models, the link between synchrony and fibrations changes dramatically. For example, a different
notion of a balanced coloring is required to characterize
robust synchrony patterns, which presumably 
requires a different notion of a fibration. The main reason
for this difference is that in the Laplacian case, the dynamics
of a given node is unaffected by any nodes that are synchronous with it. So inputs to such a node from any synchronous nodes
should be ignored when seeking robust synchrony patterns.
Moreover, which nodes should be ignored depends on the pattern.

For this reason, despite their importance in other
areas, we do not address Laplacian-based models again, except 
in Section \ref{S:MSFSI}.

\subsection{Cluster synchronization and cellular function}
\label{S:finding_clusters}

An important question addressed in this book is: which cluster synchronization\index{synchronization !cluster } and
partitions are supported by the different symmetries of the graph? In
general, there may be many such partitions, and enumerating them all is
computationally challenging \citep{kamei2013}. We discuss cluster
synchronization by automorphisms in Chapter \ref{chap:group}, and by
symmetry fibrations in Chapter \ref{chap:fibration_2}. Algorithms to find fibrations, hence clusters, are described in Chapter \ref{chap:algorithms}.

Cluster synchronization is important because
different synchronized clusters of units in a biological network
perform different biological functions.  In this scenario, biological
structures can be pictured as an orchestra\index{orchestra } in which each instrument is a
node. (However, unlike a real orchestra, there is no conductor---unless
that role is played by the flow of time.) 
When the instruments play coherently, with synchronized,
rhythmical or structured temporal patterns, the system becomes
functional.  Here, we concentrate on the simplest temporal
organization, one in which some units (instruments) act synchronously
in time in clusters (modules), a ubiquitous pattern observed in all
biological systems.

Biological systems are constrained by the requirements of modularity
and integration~\citep{hartwell1999}. Cellular functions are performed
by modules: segregated molecular units synchronize to perform a
specific function in the cell. Modules ought to be sufficiently
independent to guarantee functional specialization and, at the same
time sufficiently connected to bind multiple units for efficient
global integration that guarantees higher-level emergent behavior of a
unified cellular machine~\citep{hartwell1999}. How to realize the
conflicting requirements of modularity in the presence of integration
is a conundrum that has been largely studied in neuroscience
~\citep{tononi1994}, but it applies equally to any information-processing 
living system, from genes to proteins to organisms~\citep{hartwell1999}.

Modules can perform independent synchronized functions, yet
they can still be integrated in the network, providing segregation
of function in the presence of global integration.  Biological cluster
synchronization allows functional modules of biomolecules, proteins,
metabolites, neurons and other features to work together in coherent
dynamics.  This includes clusters of co-expression patterns of genes
and the cluster of synchronized neural ensembles in the brain
\citep{strogatz2018,klipp2016book}.  Thus, cluster synchronization is
required by a functional biological network, while complete
synchronization is not.

For weighted (and unweighted) networks,\index{network !weighted } with {\it linear} coupling, the
way to obtain complete synchronization from a system of equations
is to ensure that the rows of the underlying adjacency
matrix $A$ sum to a constant, which is analogous to using diffusive couplings in the linear regime, see  \eqref{eq:laplacian}. This leads to the use of the Laplacian
matrix to describe the dynamics \citep{pecoraMSF} explained in section \ref{sec:laplacian}. As discussed there,  this assumption
is not realistic for biological systems since interactions are not always diffusive, so they are not captured correctly by Laplacian matrices. 

For a general network, complete synchronization\index{synchronization !complete } is  possible if
each adjacency matrix\index{adjacency matrix } has constant row-sums, because this
condition means that all nodes have the same number of input edges
of any given type.
In this book we do not assume that there is a single adjacency matrix $A$ 
whose rows all sum to zero. Removing this
assumption (which effectively is the condition that $f(0)= 0$ in the model equation
$\dot x = f(x)$) implies that complete synchronization may not be achievable.
However, it is still possible for several synchronized clusters or
modules to occur, as opposed to only one cluster corresponding to
complete synchronization in the entire network. These clusters arise
through fibrations, the main subject of this book.

\section{From graph to network dynamics -- from structure to function}
\label{sec:from}

In this book, we often distinguish between the terms `graph' and `network'.\index{graph }\index{network }
`Graph' is straightforward: a set of nodes and (usually directed)
edges that satisfies Definition \ref{graph}. In part of the literature, a `network' is
a graph whose nodes and edges can be split into distinct types
\citep[Chapter 9]{GS2023}. Graph theorists call such graphs `colored graphs', where the colors represent different types, but in this book we associate colors with clusters. We therefore avoid this terminology. A graph can be viewed as a network with only one type of edge, so the term `network' includes `graph'.

When a graph is equipped with state variables and dynamical equations,
it technically becomes a `network system' of ODEs. However, we sometimes abbreviate this
term to `network' when it is clear that we are referring to
an admissible equation. The point of view is that a graph is
just a combinatorial structure, whereas a network is a
graph with extra features, intended to be used to define 
the structure of the associated differential equations.

To complicate matters, the terms `graph' and `network'
are often used interchangeably in parts of the
literature, especially in the field of complex networks, and we anticipate
that we may also interchange them here from time to time.  However,
separating the graph from the network can be important, since this
difference underlies a recurrent problem that will come up in this
book many times: {\it how structure determines function} in biological systems.\index{structure-function relation } By structure we mean the structure of the graph, and by
function we mean the function of the biological system given by the
solution of an admissible dynamical system of equations of the
corresponding network.

Clearly specific features of the dynamics, such as the quantitative value of a state variable or the frequency of
an oscillation, depend on precise details of the model equations.
We focus on more general features, which can occur for broad classes of
models. Synchronization is the prime example of such a feature.

The types of questions we ask are:

\begin{itemize}
  \item What can we say about the dynamics and function of a network based
    solely on the graph structure?
    \item Can we predict the function of a
network or parts of it by knowing the structure of the graph alone,
independently of the dynamical equations?
\item What are the features of the
  graph that determine the function/dynamics of the network?
  \item Under what conditions can we decouple the dynamics from the
    structure?
  \item To be more concrete, can we predict anything about the
    dynamics, for instance, cluster synchronization, by knowing only the graph?
\end{itemize}

These are admittedly difficult questions at the core of systems science. They
 appear systematically in the study of biological networks like gene
regulation and brain networks, and there are no easy answers for most
of them. When the graph has no structure---think for instance of an
Erd\H{o}s--Rényi graph\index{graph !Erd\H{o}s--Rényi } where nodes are connected at random---only limited conclusions can be drawn about its function. Away from randomness, many structures have
been found in graphs, specially in biological graphs, such as modules,
motifs, hubs, small worlds, fractality and others. However, most of these
structures say little about the function of the network.

We will show that  when the structure of the graph is determined by
symmetries, in particular fibration symmetries, then important aspects of the relation
between structure (graph symmetries and broken symmetries) and
dynamics in the network (cluster synchronization and bifurcations)
become clear. We develop this line of thinking throughout the book.


\chapter[Automorphisms and Symmetry Groups]{\bf\textsf{Automorphisms and Symmetry \hspace{5pt} Groups}}
\label{chap:group}\index{automorphism }\index{symmetry !group }

\begin{chapterquote}
  This book is about symmetry fibrations. But in order to
  understand these symmetries, we start with more familiar notions of automorphisms
  and symmetry groups. Traditionally, graph symmetries have been 
  studied in terms of graph automorphisms,\index{automorphism } which are permutations of
  nodes that preserve the graph connectivity. For the
  didactic purpose of understanding fibrations\index{fibration } in the context of
  known symmetries, we  first discuss classical
  symmetries given by automorphism groups. We review these ideas as a warm-up,
  before introducing fibration symmetries and groupoids as the relevant
  symmetries in biology, as discussed in subsequent chapters. The
  reader familiar with group symmetries is invited to jump directly to
  Chapters \ref{chap:fibration_1} and \ref{chap:fibration_2} on graph fibrations.
\end{chapterquote}

\section{Automorphisms and the symmetry group of a graph}
\label{sec:automorphism}

Historically, the concept of a group traces back to Joseph-Louis Lagrange (1766--1769),\index{Lagrange, Joseph-Louis }
Paolo Ruffini (1765--1822),\index{Ruffini, Paolo } Niels Henrik Abel (1802--1829),\index{Abel, Niels Henrik } and especially 
\'Evariste Galois (1811--1832),\index{Galois, \'Evariste } who developed it in connection
with solutions of polynomial equations. It was further developed by Camille Jordan (1838-1922)\index{Jordan, Camille }
and others, and rapidly spread to all main areas of mathematics, a development heavily influenced by Emmy Noether (1882--1935).
The concepts of groupoids\index{groupoid } and fibrations\index{fibration } are of more recent vintage:
groupoids were introduced by \cite{brandt1927}, and fibrations
by \cite{grothendieck1959}.\index{Grothendieck, Alexander } Their main uses have been in
algebraic geometry and algebraic topology.

Consider a dynamical system (system of ODEs)
\begin{equation}
\label{E:system_of_ODEs}
\dot x = F(x)
\end{equation}
which in coordinates becomes
\begin{equation}
\label{E:system_of_ODEs_2}
\begin{array}{rcl}
\dot x_1 &=& f_1(x_1, \ldots, x_N) \\
\dot x_2 &=& f_2(x_1, \ldots, x_N) \\
& \vdots & \\
\dot x_N &=& f_N(x_1, \ldots, x_N) 
\end{array}
\end{equation}

When the dynamical system \eqref{E:system_of_ODEs_2} is admissible
for a graph, it inherits the structural
symmetries of the graph. These symmetries
are particularly important because they define synchronized clusters
that do not depend on the details of the dynamics, i.e., on the form
of the inputs $f_i(x_1,\dots, x_N)$.  

Symmetries of networks have traditionally been described using group
theory. Recall that the  symmetry group of a network is the set
of permutations of nodes preserving the adjacency structure of the network~\citep{bollobas2012}. A permutation with
this property is called an automorphism.\index{automorphism } Any two
automorphisms compose to give an automorphism, a key feature of the
group concept.

Suppose that a permutation maps node $i$ to node
$j$ in \eqref{E:system_of_ODEs}. If this permutation does not change the
dynamical equations, which should be the case if it is an automorphism,
then $x_i$ and $x_j$ should play the same role in
the dynamical system, evolve in the same way, and thus be able to
synchronize.  On the graph, an interchange of variables corresponds to
an interchange of nodes, and symmetries of the dynamical system
\eqref{E:system_of_ODEs} correspond to symmetries of the graph,
that is, automorphisms. An automorphism implies that
many features of the network are invariant when its nodes are interchanged. An example
of such a feature is adjacency of nodes, encoded in the
adjacency matrix $A = (a_{ij})$. In
this case, a symmetry is a permutation of nodes that does not change
the global connectivity of the network
\citep{pecora2014,pecora2016a,bollobas2012,stewart2003book}:
 if any two nodes are connected before the permutation, they stay
connected after the permutation. 

Of course, features that refer to specific nodes, such as `node 1 connects to node 2', may not be invariant. The node labels must be permuted according to the automorphism to preserve connectivity.

\subsection{Formal definitions}

We now make these statements precise and give examples.

\begin{definition}
    \textbf{Permutation of a graph}.\index{permutation }
A {\em permutation} $\pi$ of a graph $G(V,E)$ is a
bijective map from the set of nodes $V=\{1,\dots,N\}$ to itself,
$\pi:V\to V$. It permutes the node labels.
\label{permutation}
\end{definition}

\begin{figure}[htb]
\centerline{%
 \includegraphics[width=.6\textwidth]{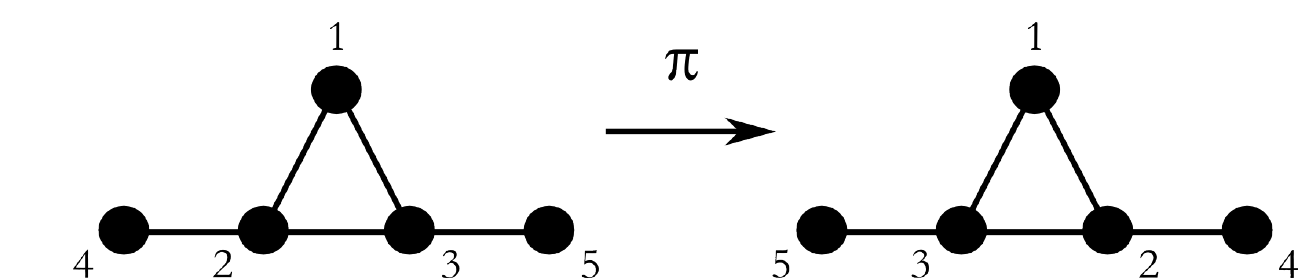}
\qquad 
\includegraphics[width=.28\textwidth]{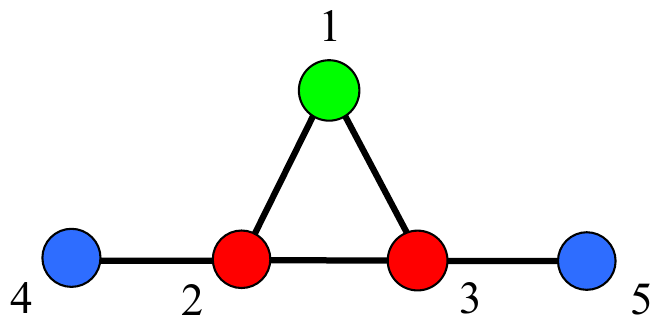}
}
\caption{\textbf{Symmetric permutation}.  {\em Left}:
Permutation \eqref{eq:legal_perm1} is an example of 
symmetric permutation preserving adjacency of nodes.
{\em Right}: Corresponding orbital partition. Nodes that are interchanged are assigned the same color.}
\label{fig5}
\commentAlt{Figure~\ref{fig5}: Left: graph with vertices 1-5. Edges 
1 to 2,3; 2 to 3,4; 3 to 5. Then horizontal arrow labeled `pi'. 
Middle: Same graph with labels 2, 3 ,4, 5 replaced by 3, 2, 5, 4.
Right: Same graph with colored vertices: 1 (green), 2 and 3 (red), 4 and 5 (blue).
}
\end{figure}

For example, the permutation $\pi$ shown in the graph of
Fig.~\ref{fig5} maps 1 to 1, 2 to 3, 3 to 2, 4 to 5, and 5 to 4. It is
denoted by:
\begin{equation}
\pi\ =\ \left(
\begin{matrix}
  1 & 2 & 3 & 4 & 5 \\
    1 & 3 & 2 & 5 & 4 
\end{matrix}
\right)\ ,
\label{eq:legal_perm1}
\end{equation}
in Cauchy's two-line notation which consists of a first
  row with the  original labels of nodes, and a second row with the permuted labels. Sometimes, for clarity,
  vertical arrows are inserted between the two rows.
  A compact alternative is 
   cycle notation: $\pi = (2\,3) (4\,5)$, meaning that $2 \to 3 \to 2$ and
   $4 \to 5 \to 4$.

A {\it permutation matrix}\index{permutation !matrix } $P$ is an $N \times N$ matrix that is
obtained from the identity matrix by permuting both rows and columns
according to $\pi$.
In the case of permutation (\ref{eq:legal_perm1}), this is:
\begin{equation}
    P = 
    \begin{pmatrix}
        1 & 0 & 0 & 0 & 0 \\
        0 & 0 & 1 & 0 & 0 \\
        0 & 1 & 0 & 0 & 0 \\
        0 & 0 & 0 & 0 & 1 \\
        0 & 0 & 0 & 1 & 0
    \end{pmatrix}
    .
\end{equation}

\begin{definition}{\bf Automorphism (group symmetry) of a graph} \citep{harary1994}.
  \label{def:automorphism}
  An {\it automorphism}\index{automorphism } or {\it group symmetry}\index{group symmetry } of a graph $G = (V,E)$ is a
  permutation $\pi : V \to V$ of the vertex set $V$, such that the
  pair of vertices $i$ and $j$ forms an edge $(i, j)$ if and only if
  ($\pi(i) , \pi(j)$) also forms a edge.
\end{definition}

That is, an automorphism preserves adjacency and non-adjacency in the
graph; two edges are adjacent after the permutation if and only if they were
adjacent before the permutation.  The set of automorphisms of a graph
with adjacency matrix $A$, with the operation of composition, forms the
automorphism group, or symmetry group, of the
graph. We use the following notation:  

\begin{definition} {\bf Symmetry/automorphism group of a graph.}
Given a graph $G(V,E)$, we define:
\begin{equation}
{\rm Aut}(G) = \{\pi\ |\ \pi {\rm\ is\ a\ symmetry\ permutation} \}\ 
\label{eq:group}
\end{equation} 
to be the set of all symmetry permutations. Composition preserves the network structure, so the set
${\rm Aut}(G)$ is a group under composition and defines the {\it
  symmetry group} or {\em automorphism group} of the graph. It is a subgroup of $\mathbb{S}_N$.
\end{definition}

An automorphism of a graph is a permutation of nodes 
that preserves the adjacency
matrix \citep{babai}. Any permutation matrix of a permutation $\pi$
transforms the adjacency matrix into another $A'$ as $A' = P A
P^{-1}$. If $P$ represents an automorphism, then $A'=A$, so
\begin{equation}
\label{E:perm_auto}
    A = P A P^{-1}.
\end{equation}
Since the group consist of matrices we can state this condition
in another way. Equation \eqref{E:perm_auto} holds if and only if
the matrix $P$ commutes with $A$, i.e., $PA=AP$. Equivalently, 
the {\em commutator}\index{commutator } is zero:
\begin{equation}
  [P, A] = P A - A P = \textbf{0} .
\end{equation}
The permutation \eqref {eq:legal_perm1} is a
symmetry of
 the network in
Fig.~\ref{fig5}.  For a network with $N$ nodes there are $N!$ possible permutations forming the symmetric group.

\begin{definition}
 {\bf Symmetric group}. The set of all permutations of the nodes of a graph $G(V,E)$ forms a
group $\mathbb{S}_N = \{P_1, \dots P_K\}$ where $K = N!$. It is called the {\it
  symmetric group}\index{group !symmetric } (not to be confused with a symmetry group).
      \label{def:symm_gp}
\end{definition}

Some of these permutations are symmetries, and the rest are not. The example network
in Fig.~\ref{fig5} has $5!=120$ permutations, 2 of which are
symmetries ($\pi$ defined in \eqref {eq:legal_perm1} and the
identity) and 118 are not.

The symmetry group of a network
can be calculated from the adjacency matrix $A$ using widely available
discrete algebra routines \citep{Stein,GAP4}. The problem of finding
all automorphisms of a graph (or the generators of the group) is known
as the Graph Automorphism Problem.\index{Graph Automorphism Problem }  Although in the worst case
there is no polynomial-time algorithm to solve this problem, 
several algorithms work quite well for
large graphs in practical applications. A popular algorithm to find automorphisms of a graph is
McKay's {\it Nauty}\index{Nauty algorithm } algorithm~\citep{nauty}, which is based on the well-known problem of
testing isomorphism between graphs (see Section \ref{sec:testing}). Other
algorithms based on Nauty are {\it bliss} \citep{junttilla2007} and {\it
  saucy} \citep{darga2008}, which is particularly good for sparse graphs.

The {\em identity permutation},\index{identity permutation } denoted by {\it id}, is the transformation
that does not move any nodes in the graph. The identity is called a {\it trivial}
automorphism. All graphs have a symmetry
group that is composed at least of the identity.  We will find that in most real
biological graphs,  the identity is the only
automorphism of the graph.

\begin{definition}   A graph with no nontrivial automorphisms
  has only the identity permutation as a symmetry.  Such a
  graph is said to be \emph{rigid}.\index{rigid } We also refer to this kind of
  graph as a graph with `no' group symmetries (`nontrivial' is 
  understood here: the identity map is always a symmetry). 
  
   However, we repeat that in this book the word `symmetry'\index{symmetry } has a more general meaning
  than `automorphism': it also refers to fibration symmetry.
Therefore  we do not call a rigid graph 
  an {\it asymmetric} graph, as is common in the mathematical literature.
\label{def:rigid}
\end{definition}

\section{Cluster synchronization by automorphisms: orbital partition}
\label{sec:orbit}

Next, we discuss \textit{orbital partitions}
\index{partition !orbital }
arising from automorphisms of graphs. This idea is
already present in \citep{gol1985}, applied to Hopf bifurcation for rings of oscillators, and is a special case of results in \citep{golubitsky1988} for general symmetric dynamical systems.
Pecora, Sorrentino, and others
studied orbital partitions and cluster synchronization\index{synchronization !cluster } employing group
symmetries and automorphisms in a series of papers \citep{pecora2014,
  pecora2016b, pecora2019, nishikawa2016,schaub2016}. The connection
between symmetry and synchronization in networks with
group symmetry is explained in detail in
\cite[Chapter 16]{GS2023}. Other authors have studied group symmetries in real
networks \citep{macarthur,garcia2020} as well as in multilayer graphs \cite{della2020symmetries} and hypergraphs
\citep{hypergraph}.  

Group symmetries, either by the whole automorphism group of a subgroup, naturally induce certain kinds of synchrony clusters.
The set of automorphisms of the graph permutes certain subsets of nodes
among each other.  When the symmetry group ${\rm Aut}(G)$ acts on the
network, a given node $i$ is moved by the permutations of the group to
various other nodes $j$.  In the language of group theory, the set of
all nodes to which $i$ can be moved defines the {\it orbit}\index{orbit } of
node $i$.
Two nodes $i$ and $j$ are in the same orbit
if there is an automorphism that maps $i$ to $j$.
This in turn defines the orbital partition\index{partition !orbital } of the network:

\begin{definition} {\bf Orbit of a node and orbital partition.}
  Given a graph $G(V,E)$, the {\it orbit}\index{orbit } of a node $i \in V$ for a subgroup
 $H$ of ${\rm Aut}(G)$ is:
    \begin{equation}
        \mathcal{S}_H(i) = \{j \in V \,|\, \exists \, \pi \in H \,\, s.t.\,\,
        \pi(i) = j \}.
    \end{equation}
    \label{orbits}
    If $H={\rm Aut}(G)$ we write $\mathcal{S}(i)$ rather than $ \mathcal{S}_H(i)$. 
    It can easily be proved that two orbits $\mathcal{S}_H(i)$ and $\mathcal{S}_H(j)$
are either equal or disjoint, and the union of all
orbits is equal to $V$.  Therefore the set of all orbits induces a
partition of the nodes into mutually disjoint clusters.
This set of all orbits forms the {\em $H$-orbital
partition}.\index{partition !orbital } When $H={\rm Aut}(G)$ we call this the {\em orbital partition}: it is the  $H$-orbital
partition into the fewest clusters. 
\end{definition}

For examples, in the network of Fig. \ref{fig5} the orbits are $\{1\}$, colored green; $\{2,3\}$, colored red; and $\{4,5\}$, colored blue.

There is a similar concept of an $H$-orbital partition for a subgroup $H$ of the automorphism group.

\subsection{Orbits define clusters}

The orbits of subgroups of the symmetry group determine clusters of nodes that
can synchronize. In other words, these orbits guarantee that the
 synchrony subspace of the cluster is flow-invariant
\citep[Lemma XIII.2.1]{golubitsky1988},\citep{pecora2014}. 

\begin{definition} {\bf Cluster synchronization in orbital partitions.}
All nodes in the orbit $\mathcal{S}_H(i)$ of node $i$ 
 have the same dynamical state as node $i$, i.e.,
\begin{equation}
  x_i(t)=x_j(t) \,\,\mbox{ if}\,\, j \in \mathcal{S}_H(i) , 
\end{equation}
forming a synchronization cluster of an admissible system of equations to the graph.
\end{definition}

A different question is whether the  
corresponding synchronous solution is stable; stability\index{stability } here is referred to as
{\em synchronizability} and treated in detail in chapter \ref{chap:stability}. \index{synchronizability } A third question is whether any particular
cluster pattern occurs for a specific admissible differential equation,
stably or not.

\subsection{Orbital clusters and synchrony}

To exemplify the connection between the orbits of the symmetry group
${\rm Aut}(G)$ and cluster synchronization, we use the network in
Fig.~\ref{fig5}.  A simple example of an admissible dynamical system compatible with this
network (not the most general, see the Remark below)
is of the form
\begin{equation}
\dot{x_i}\ = \ f(x_i) + \sum_{j=1}^5A_{ij}g(x_i, x_j)\ , 
\end{equation}
or explicitly
\begin{equation}
\begin{aligned}
&{\dot x}_1\ =\ f(x_1)\ + g(x_1, x_2) + g(x_1, x_3)\ ,\\
&{\dot x}_2\ =\ f(x_2)\ + g(x_2, x_1) + g(x_2, x_3) +  g(x_2, x_4)\ ,\\
&{\dot x}_3\ =\ f(x_3)\ + g(x_3, x_1) + g(x_3, x_2)+  g(x_3, x_5)\ ,\\
&{\dot x}_4\ =\ f(x_4)\ + g(x_4, x_2) \ ,\\
&{\dot x}_5\ =\ f(x_5)\ + g(x_5, x_3)\ .
\end{aligned}
\label{eq:dynamics_group}
\end{equation}

Notice that we use the same function $g(\cdot)$ in all equations. This
makes the system admissible for the graph, since the graph has all
edges of the same type.
\begin{remark}
\label{R:general}
In this example, specific functions have been assigned to edges,
and interactions are assumed to be additive. These assumptions are common
in biological models, and are often appropriate, but they are special cases.

In the general formalism for network dynamics used in this book,
couplings need not be additive. In the example above, the most general admissible system has the form
\begin{equation}
\begin{aligned}
&{\dot x}_1\ =\ f(x_1,\overline{x_2, x_3}),\\
&{\dot x}_2\ =\ g(x_2, \overline{x_1,x_3,x_4}),\\
&{\dot x}_3\ =\ g(x_3, \overline{x_1,x_2,x_5}),\\
&{\dot x}_4\ =\ h(x_4, x_2),\\
&{\dot x}_5\ =\ h(x_5, x_3).
\end{aligned}
\label{eq:dynamics_group_general}
\end{equation}
for three distinct and arbitrary functions $f,g,h$ with appropriate
domains and ranges.
Here the overlines indicate that the functions are symmetric under all permutations of those variables.

This level of generality is required in the theory 
of admissible ODEs because
many models do not assign coupling functions
in an additive (or multiplicative) manner.
\end{remark}

The symmetry group of this network is ${\rm Aut}(G)=\{ \pi, id\}$, where
$\pi=(2\ 3)(4\ 5)$, as in \eqref{eq:legal_perm1}, and $id$ is the
identity.  The orbits of the network are obtained as follow. We take
node 2 and apply $\pi$ to get node 3 and {\it vice versa}. This defines the
first orbit $\mathcal{S}_1=\{2, 3\}$. We repeat with node 4 and obtain
$\mathcal{S}_2=\{4, 5\}$. Node 1 just leads to a trivial cluster
$\{1\}$.

We now show that nodes in these orbits synchronize.  Applying $\pi$ as in
\eqref{eq:legal_perm1} to both sides of
\eqref{eq:dynamics_group} we get the transformed dynamical system:
\begin{equation}
\begin{aligned}
&{\dot x}_1\ =\ f(x_1)\ + g(x_1, x_3) + g(x_1, x_2)\ ,\\
&{\dot x}_3\ =\ f(x_3)\ + g(x_3, x_1) + g(x_3, x_2) +  g(x_3, x_5)\ ,\\
&{\dot x}_2\ =\ f(x_2)\ + g(x_2, x_1) + g(x_2, x_3)+  g(x_2, x_4)\ ,\\
&{\dot x}_5\ =\ f(x_5)\ + g(x_5, x_3) \ ,\\
&{\dot x}_4\ =\ f(x_4)\ + g(x_4, x_2)\ ,
\end{aligned}
\label{eq:dynamics_group2}
\end{equation}
which is identical to the original dynamical system
\eqref{eq:dynamics_group} before the permutation (though written in a different order).  The invariance
of \eqref{eq:dynamics_group} under $\pi$ implies that if
$x=(x_1, x_2, x_3, x_4, x_5)$ is a solution
to \eqref{eq:dynamics_group}, then  $\pi(x)=(x_1, x_3, x_2, x_5,
x_4)$ is also a solution, since it solves the same equations.  

Moreover, the fixed points of the
automorphism, given by $\pi(x) = x$, are obtained by substituting
$x_2(t)=x_3(t)$ and $x_4(t)=x_5(t)$ in the equations, and
the resulting
system is consistent. Indeed, the ODE reduces to
just three components, one for each cluster:
\begin{equation}
\begin{aligned}
&{\dot x}_1\ =\ f(x_1,\overline{x_2, x_2}),\\
&{\dot x}_2\ =\ g(x_2, \overline{x_1,x_2,x_4}),\\
&{\dot x}_4\ =\ h(x_4, x_2).
\end{aligned}
\label{eq:dynamics_group3}
\end{equation}
After the substitution, the equation for $\dot x_3$ is the same as that for $\dot x_2$,
and the equation for $\dot x_5$ is the same as that for $\dot x_4$.

\begin{definition}{\bf Minimal orbital partition.}\index{partition !minimal orbital } For a given network there
  may be several orbital partitions.  Of all the orbital partitions,
  there is exactly one partition with the smallest number of
  orbits. We refer to this as the {\it minimal orbital
    partition}.
\end{definition}

Unless the symmetry group is trivial or has few subgroups, there
are also non-minimal orbital partitions.
These are generated by applying Definition \ref{orbits}
with a given subgroup $H$ of the symmetry group.  Furthermore, 
as we discuss in Chapter \ref{chap:fibration_1}, there are
other partitions that lead to cluster synchronization without
automorphisms, and generally a graph may have many more symmetries than
those captured by the orbits. These symmetries correspond to more
general partitions, called `equitable';\index{partition !equitable } equivalently to balanced colorings,\index{coloring !balanced } which can also be viewed as fibers,\index{fiber }
which respect fibration symmetries.\index{fibration !symmetry }

To summarize, the orbits define synchronous clusters, but there can be
other synchronized clusters that are not orbits. Orbital
partitions are equitable, but not all equitable partitions are orbits
(more on this in section \ref{minimal_partition}).  Permutation
symmetries form a symmetry group with the same algebraic structure as
groups in physics, including those groups associated with quantum
mechanics and particle physics
\citep{landau1977quantum,weinberg1995}. 

While symmetry groups are
all-pervasive in physics, permutation symmetry groups are rare in
biological networks. Instead, the more general symmetry of fibrations\index{fibration }
captures biological symmetry.
Among the few examples of  biological networks with nontrivial
automorphisms are the forward and backward locomotion
networks of the
nematode {\it C. elegans},\index{C. elegans@{\it C. elegans} } which are subgraphs of the neural network  which contains only 302 neurons and has
been fully mapped: see section \ref{sec:celegans}.  
In general, automorphisms seldom occur in biological networks,
but fibrations are common.

\subsection{Two types of quotient graph}
\label{sec:two_quots}\index{quotient !graph }

The graph-theoretic literature includes a construction for the
quotient of a graph by its automorphism group---or more generally by any group of automorphisms; see \citep{hahn}. This construction is natural in that context.
The nodes of this quotient graph correspond to
the orbits of the automorphism group (or subgroup),
and therefore determine a balanced coloring, the `orbit coloring'
\citep[Section 16.6]{GS2023}. The dynamics of this
synchronous state can be found from the model ODE
concerned.

However, to avoid confusion, we emphasize 
that it is not the most appropriate notion 
of a quotient to represent synchronous dynamics.
The reason is that although the nodes represent
the clusters, the edges of this quotient graph
do not represent the couplings inherited from the original graph in a manner that preserves the dynamics of all
admissible ODEs.
We expand on this point below in Example \ref{ex:wrong_quot}.

Suppose that $G$ is a graph (directed or undirected)
with automorphism group ${\rm Aut}(G)$. There is 
a balanced coloring in which all nodes in an orbit of
${\rm Aut}(G)$ are assigned the same color.
The quotient as defined in \citep{hahn} collapses
the nodes in each orbit to form a single node,
and defines edges between pairs of orbits in terms of edges between
pairs of nodes, one in each orbit:

\begin{definition}{\textbf{Quotient graph by an automorphism group in the sense of \cite{hahn}.}}
\label{def:aut_quot}
 Let $G$ be a graph and let $\mathcal{S}(i)$ be the orbits of the
automorphism group ${\rm Aut}(G)$ of the graph.  The {\it quotient graph}\index{quotient !graph } of the symmetry group acting on graph $G$
 is a graph whose vertices are the orbits $\mathcal{S}(i)$:
 \begin{equation}
  G /{\rm Aut}(G) = \{ \mathcal{S}(i)\ : \ i\in G \ \}
\end{equation}
with a unique edge $\mathcal{S}(j) \to \mathcal{S}(i)$ whenever 
some (hence any) node
in $\mathcal{S}(j)$ is connected to some node in $\mathcal{S}(i)$.

The mapping
 $\pi_{\mathcal{P}} : V(G) \to V(G/\mathcal{P})$ defined by
 $\pi_{\mathcal{P}}(u) = V_i$ such that $u \in V_i$, is the {\it natural
 map} for $\mathcal{P}$.
\end{definition}

Although this notion of a quotient graph is well-known,
it is not entirely appropriate
for our purposes. The reason is that nodes in $G/{\rm Aut}(G)$ are connected either by a unique edge, or none. This implies that
in general the natural map is not a fibration: it
fails to preserve input sets or input trees. Therefore
the admissible ODEs for this notion of a quotient graph\index{quotient !graph } 
are not the same as the restrictions of the admissible
ODEs for $G$ to the synchrony subspace that corresponds to the clusters.

The next example illustrates this point and clarifies what we mean.

\begin{example}\em
\label{ex:wrong_quot}

The admissible ODEs for the graph $G$ of Fig. \ref{fig:6node_orbital_quot}a are:
\begin{equation}
\label{E:6node_orbital_quot_a}
\begin{array}{lcl}
\dot x_1 &=& f(x_1,x_6) \\
\dot x_2 &=& f(x_2,x_1) \\
\dot x_3 &=& g(x_3,x_2,x_5) \\
\dot x_4 &=& f(x_4,x_3) \\
\dot x_5 &=& f(x_5,x_4) \\
\dot x_6 &=& g(x_6,x_2,x_5) 
\end{array}
\end{equation}
where $g$ is symmetric in the second and third variables.

\begin{figure}[b!]
\centerline{%
\includegraphics[width=0.75\textwidth]{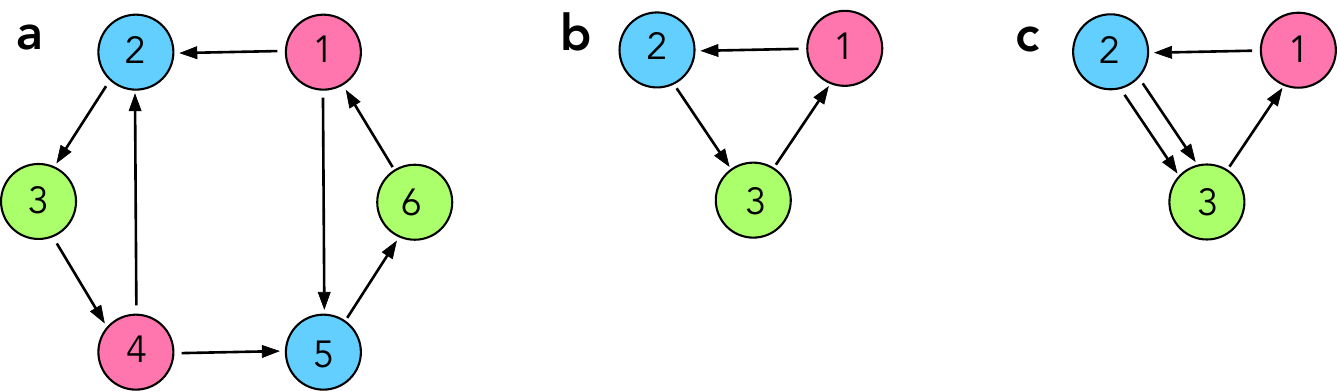}}
\caption{\textbf{Quotient graph.} Difference between the quotient graph as defined in \citep{hahn} and the
quotient network\index{quotient !network } for the corresponding fibration.
(\textbf{a}) A directed graph
with six nodes and automorphism group $\Z_2$ generated by
$180^\circ$ rotation $(1\ 4)(2\ 5)(3\ 6)$.
The coloring is the orbital coloring. (\textbf{b}) Quotient graph in the sense of 
\citep{hahn}. This is not the base of a fibration because
the input arrow to node 2 does not lift uniquely, see chapter \ref{chap:fibration_2}.
(\textbf{c}) Quotient network defined by the fibration
corresponding to the colors has an extra arrow from
node 2 to node 3, and is (necessarily) a multigraph.
}
\label{fig:6node_orbital_quot}
\commentAlt{Figure~\ref{fig:6node_orbital_quot}:  (a) graph with nodes 1-6.
Arrows 1 to 2 and 5; 2 to 3; 3 to 4; 4 to 2 and 5; 5 to 6; 6 to 1. Colors
1 and 4 (red), 2 and 5 (blue), 3 and 6 (green).
(b) Graph with nodes 1 (red), 2 (blue), 3 (green). Arrows 1 to 2; 2 to 3; 3 to 1.
(c) like (a) but with a second arrow 2 to 3.
}
\end{figure}

When restricted to the synchrony subspace corresponding to the orbits of ${\rm Aut}(G)$, equation \ref{E:6node_orbital_quot_a} takes the form
\begin{equation}
\label{E:6node_orbital_quot_c}
\begin{array}{lcl}
\dot x_1 &=& f(x_1,x_3) \\
\dot x_2 &=& f(x_2,x_1) \\
\dot x_3 &=& g(x_3,x_2,x_2) 
\end{array}
\end{equation}
This is the general admissible ODE for Fig. \ref{fig:6node_orbital_quot}c, because of the double arrow.
This network is the quotient network of
\ref{fig:6node_orbital_quot}a determined by the
coloring, in the sense we use in this book: see Definition \ref{def:quot_net}
for a general construction.

In contrast, the general admissible ODE for $G/{\rm Aut}(G)$,
namely Fig. \ref{fig:6node_orbital_quot}b, is:
\begin{equation}
\label{E:6node_orbital_quot_b}
\begin{array}{lcl}
\dot x_1 &=& f(x_1,x_3) \\
\dot x_2 &=& f(x_2,x_1) \\
\dot x_3 &=& f(x_3,x_2) 
\end{array}
\end{equation}
Equation \ref{E:6node_orbital_quot_b} differs
from \ref{E:6node_orbital_quot_c}.
In particular, \ref{E:6node_orbital_quot_b} involves
a single function $a$ and has a $\Z_3$ automorphism group.
This
leads to generic dynamics that would not be generic
for \ref{E:6node_orbital_quot_c}. The latter involves a
second function $g$, and has trivial automorphism group.
Its dynamics reduces to \ref{E:6node_orbital_quot_b} if
we choose $g$ so that $g(u,v,w)=f(u,v)$, so the 
quotient network in the sense of this book (Definition
\ref{def:quot_net}) has more general dynamics.
\end{example}


\chapter[Input Trees, Synchrony, Balanced Colorings, and Equitable Partitions]{\bf\textsf{Input Trees, Synchrony, and Equitable Partitions}}
\label{chap:fibration_1}

\begin{chapterquote}

Robust synchrony patterns in networks correspond precisely to
fibration symmetries, or equivalently equitable partitions of the nodes, also called balanced colorings. We motivate these notions using two 
simple biological networks, the metabolator\index{metabolator } and the Smolen oscillator.\index{Smolen oscillator }
Each has two nodes, which can synchronize. For the metabolator, the explanation is symmetry in the group-theoretic sense. For the Smolen oscillator, however, the symmetry
group is trivial, and the explanation of synchrony is a fibration.\index{fibration }
We discuss the  key difference between global group-theoretic symmetries and local fibration symmetries. We state a series of modeling assumptions that can
naturally create synchrony, and show how they lead to the fibration concept.
A crucial mathematical point is that fibrations are determined by
the structure of the input trees of nodes, which represent the flow
of information through the network. The minimal equitable partition\index{partition !minimal equitable } 
(largest clusters of synchronous nodes) is determined by isomorphism of input trees.\index{input tree }
\end{chapterquote}

\section{Intuition and motivation}
\label{intuition}

\begin{figure}[htb]
\centerline{
\includegraphics[width=.3\linewidth]{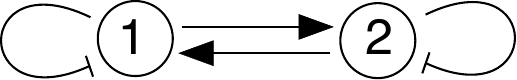} \qquad\qquad
\includegraphics[width=.3\linewidth]{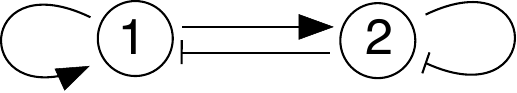}
}
\caption{{\bf Synchronization without group symmetry.} {\em Left}:
  Metabolator network. {\em Right}: Smolen oscillator
  network. Arrowheads indicate activator (pointed) and repressor
  (bar). In both graphs, nodes 1 and 2 are found to synchronize. In the
  metabolator, this synchronization can be explained by the group
  symmetry of the graph (and also by the fibration symmetry since it
  includes automorphisms). In the Smolen oscillator the
  synchronization is not explained by the group symmetry since this
  graph has no nontrivial automorphisms. Instead, the fibration
  symmetry captures the synchronization.}
\label{F:meta_smol1}
\commentAlt{Figure~\ref{F:meta_smol1}: Left: graph with nodes 1 and 2. 
Barred arrows 1 to 1, 2 to 2. Sharp arrows 1 to 2; 2 to 1.
Right: graph with nodes 1 and 2. 
Barred arrows 2 to 1, 2 to 2. Sharp arrows 1 to 1; 1 to 2.
}
\end{figure}

Even though automorphisms of a graph impose a rigid type of symmetry with strict
conditions on the graph structure, it seems to be widely believed that these
symmetries determine all possible synchrony patterns in a
network. Simple examples show that this is false. We already saw this for the UxuR GRN in Fig. \ref{F:Uxur}. We now compare two even
simpler networks, which we revisit to illustrate fibrations
in Chapter \ref{chap:fibration_2}. Each has two nodes; one is symmetric under
interchange of the nodes, and the other is not.
Figure ~\ref{F:meta_smol1} shows these networks, the metabolator\index{metabolator } on the left
and the Smolen oscillator\index{Smolen oscillator } on the right, which occur in synthetic
genetic oscillators \citep{purcell2010}.
Here, two genes regulate each other, and also self-regulate themselves as
activators (pointed) or repressors (bar).

Both of these networks support synchronous states, where both nodes
have identical states. For the metabolator, this is explained by its
$\Z_2$ group symmetry, which permutes the two nodes. Admissible ODEs
have the form
\begin{eqnarray}
    \dot x_1 &=& f(x_1,x_2) \\
    \dot x_2 &=& f(x_2,x_1) 
\end{eqnarray}
which both reduce to $\dot x = f(x,x)$ when $x_1=x_2=x$.
So this equation describes the synchronous states.
However, the
Smolen oscillator has trivial symmetry group (only the identity), yet it also supports the same synchronous state. Now admissible ODEs
have the form
\begin{eqnarray}
    \dot x_1 &=& f(x_1,x_2) \\
    \dot x_2 &=& f(x_1,x_2) 
\end{eqnarray}
and again, both reduce to $\dot x = f(x,x)$ when $x_1=x_2=x$.
That is, node 1 synchronizes with node 2 in both
graphs. How can we explain this synchronization in the absence of nontrivial group
symmetry?

\begin{figure}[h!]
\centerline{
\includegraphics[width=.4\textwidth]{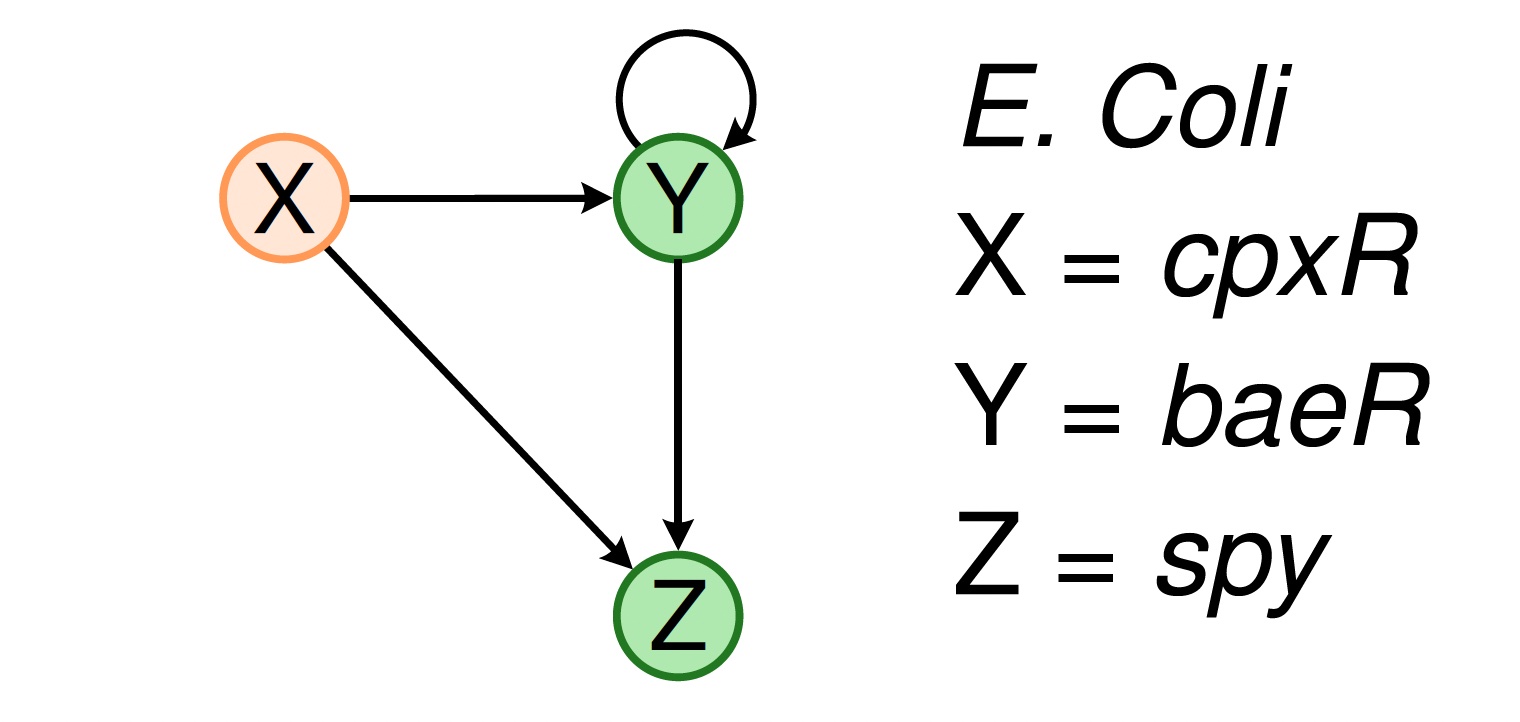}
\qquad \includegraphics[width=.5\textwidth]{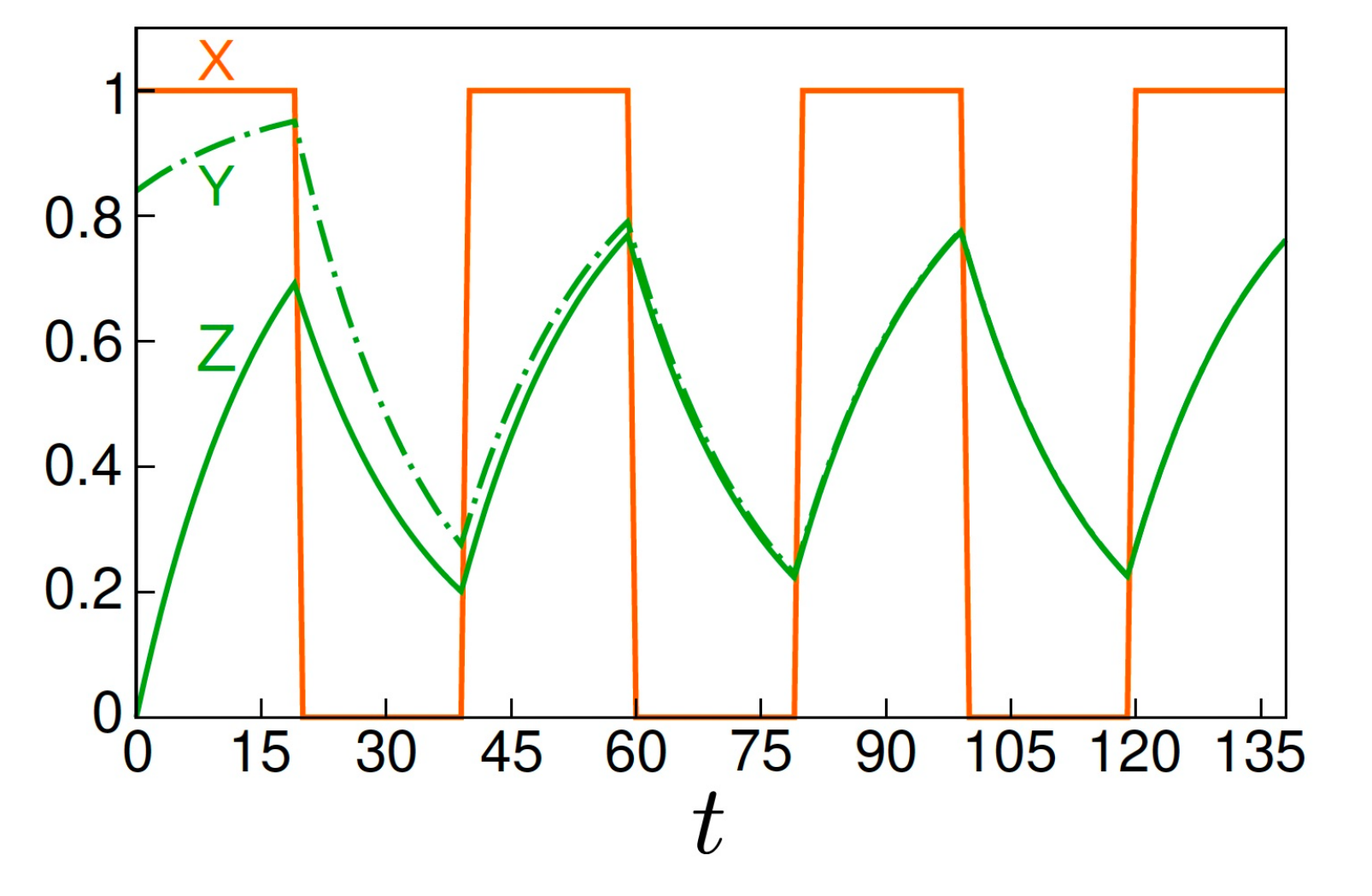}
}
\caption{ \textbf{Synchronization without a symmetry group in a gene
 regulatory circuit.}  As an example, we consider a building block
  of the gene regulatory network of {\it E. coli} called the
  feed-forward fiber \citep{leifer2020circuits}, composed of genes {\it cpxR} = X, {\it baeR} = Y,
  and {\it spy} = Z.  The circuit has no group symmetry, yet a
  numerical solution of the dynamics shows synchronization of the
  expression levels of genes Y and Z. }
\label{fig1-saulo}
\commentAlt{Figure~\ref{fig1-saulo}:  Left: Gene network, nodes X, Y, Z.
Sharp arrows  X to Y and Z, Y to Y and Z. Legend: E.coli. X= cpxR, Y=baeR, Z=spy.
Right: Graph of X against t is a square wave (orange). Graphs of Y and Z against t  
give sawtooth waves (green) which superpose exactly as t increases.
}
\end{figure}

We can further illustrate this conundrum with the circuit depicted in
Fig. \ref{fig1-saulo} left, which is called a feed-forward fiber
(FFF).\index{feed-forward fiber }\index{FFF } This circuit is a
fundamental fibration building block of gene regulatory networks. It is found
abundantly in {\it E. coli} as well as in other biological networks
\citep{morone2020fibration,leifer2020circuits}, and will be discussed
in detail in Chapter \ref{chap:hierarchy_2}. It is similar to the UxuR
circuit of Fig. \ref{F:Uxur}, but now all edges have the same type.
The FFF circuit consists of a feed-forward loop motif as found in
\citep{milo2002network} between genes X, Y and Z, plus an
autoregulation\index{autoregulation } loop at Y. Again, this circuit
has no nontrivial group symmetries. However, numerical simulations shown in Fig. \ref{fig1-saulo} right and analytical solutions by
\cite{leifer2020circuits} confirm the existence of cluster
synchronization of genes Y and Z. We use the model equations \beqn
y_{t+1}-y_t &=& -\alpha y_t +
\gamma_x\theta(x_t-k_x)\gamma_y\theta(y_t-k_y) \\ z_{t+1}-z_t &=&
-\alpha z_t + \gamma_x\theta(x_t-k_x)\gamma_y\theta(y_t-k_y) \eeqn
with parameters $\alpha = 0.06$, $\gamma_x = 0.775$, $\gamma_y=0.775$,
$k_x = 0.5$, $k_y = 0.1$, $y_0 = 0.85$, $z_0 = 0.0$. Notice that the chosen parameters make this model system admissible for the graph.

\section{Global symmetries versus local symmetries}
\label{global}

\begin{figure*}
\includegraphics[width=\textwidth]{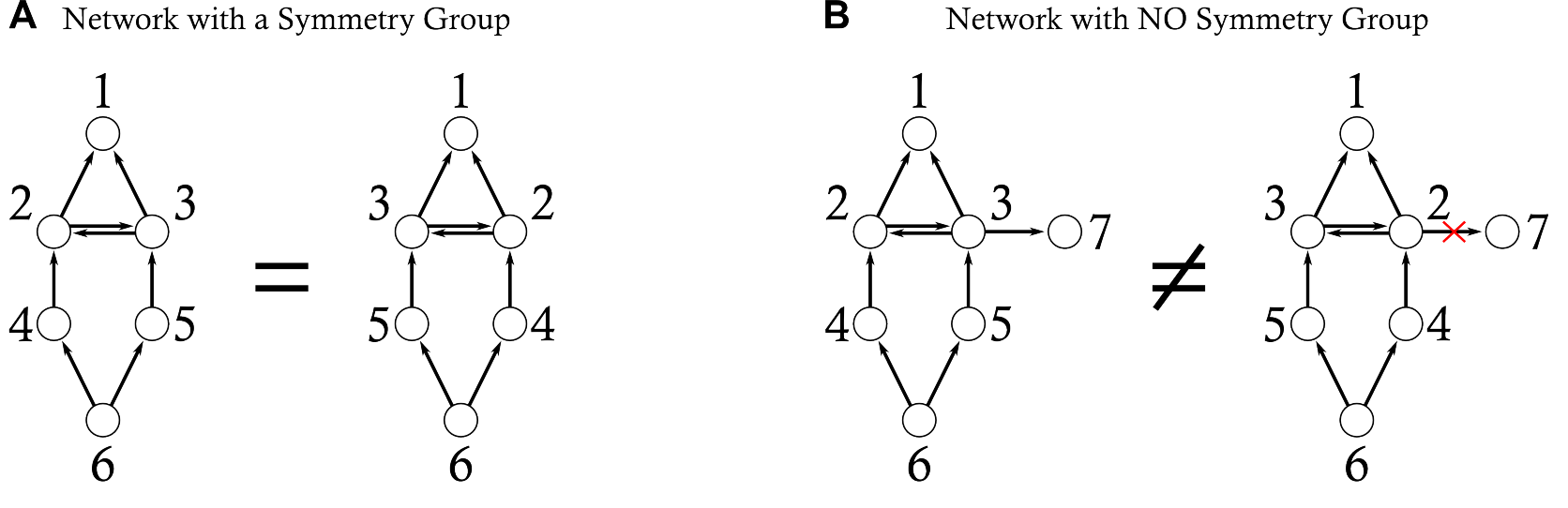} \centering
\caption {\textbf{Group symmetries and fibrations}.  (\textbf{a}) Example
  of a network with a (nontrivial) symmetry group.  The automorphism shown maps the
  network to itself leaving invariant the connectivity of every node
  in the network. (\textbf{b}) A graph without (nontrivial) automorphisms
  but with a fibration symmetry.\index{fibration !symmetry } The addition of a single out-going
  edge from 3 $\to 7$ breaks the global group symmetry of the graph
  rendering the graph with only a trivial group symmetry (the
  identity). We will see that, even though this graph has no group
  symmetry, it is still symmetric under a symmetry fibration: see Fig. \ref{fig:minimal}.}
\label{fig: example} 
\commentAlt{Figure~\ref{fig: example}: A: graph with nodes 1-6.
Sharp arrows 2 to 1 and 3, 3 to 1 and 2, 4 to 2, 5 to 3, 6 to 4 and 5.
`Equals' sign, then same graph flipped left/right.
B: Same graph with extra node 7 and arrow 2 to 7. `Not equals' sign;
then same graph flipped except for node 7,
with arrow from 7 to 2 crossed out in red.
}
\end{figure*}

Fibrations are more relevant to biology than automorphisms
because they are more robust, often surviving changes to
the network. Automorphisms are far more delicate. To
see why, we compare fibrations and automorphisms using the graph in Fig. \ref{fig:synchronyfibers},
which is replotted in Fig.~\ref{fig: example}b together with a slight
variation in Fig.~\ref{fig: example}a.  The graph of Fig.~\ref{fig: example}a
exhibits a global left-right group symmetry defined by the
automorphism:
\begin{equation}
\pi\ =\ \left(
\begin{matrix}
    1 & 2 & 3 & 4 & 5 & 6 \\
    1 & 3 & 2 & 5 & 4 & 6 \\  
\end{matrix}
\right)\ ,
\label{eq:legal_perm2}
\end{equation}
or in cycle notation $\pi=(2\,3) (4\,5)$; nodes are connected in
exactly the same way before and after applying $\pi$. This
automorphism preserves the connectivity of the entire graph and
represents a global symmetry of the network. In particular, it
preserves both the inputs and the outputs of the nodes, as specified by
the adjacency matrix. This automorphism leads to a partition into four
orbits, as defined in Definition \ref{orbits}; the nontrivial ones are composed of nodes
$\mathcal{S}(2) = \{2, 3\}$ and $\mathcal{S}(4) = \{4, 5 \}$. Nodes in
these orbits can synchronize in clusters. {\em Both} orbits must
synchronize simultaneously to create such a state, which
satisfies
\[
x_2 = x_3 \qquad x_4=x_5.
\]

Next, consider the slightly modified graph depicted on the left of
Fig.~\ref{fig: example}b, which differs from the network in
Fig.~\ref{fig: example}a by one extra out-going edge from node 3 to 7. In
this graph, the permutation of nodes $\pi = (2,3) (4,5)$ in
\eqref{eq:legal_perm2} is no longer an automorphism, because it
does not preserve the in and out connectivities of nodes 2, 3 and
7. For instance, node 3 is connected to 7 in the original graph but it
is not after the permutation is applied (Fig.~\ref{fig: example}b
right). Likewise, the connectivity of nodes 2 and 7 is different
before and after the permutation. There are no other symmetries in
this graph.

We see how fragile group symmetries are: by adding just one outgoing
link from the symmetric graph (the edge 3 $\to$ 7), the global symmetry is
broken. The automorphism group of the network in Fig.~\ref{fig: example}b
now contains only the trivial identity permutation.
This fragility occurs because automorphisms require very strict global
arrangements of nodes and links to preserve the global
structure of the network.

The breaking of global symmetry by adding edge 3 $\to$ 7 changes the automorphism group of the graph, and also changes its orbits. The predicted absence
of an orbital partition and concomitant cluster synchronization is
significant, because it represents a global vulnerability to the
perturbation. That is, the addition of outgoing link 3 $\to$ 7, which
in principle should not affect the dynamics of node 3 and much less
that of nodes 4 or 5, breaks the synchronization of not only the orbit
$\mathcal{S}(2)= \{3, 4 \}$ but also of the other orbit $\mathcal{S}(4)=
\{4, 5 \}$, even though the added link does not directly connect to
this orbit nor emanate from it. This exemplifies the nonlocal
vulnerability of group symmetry. It also shows that (group) symmetry breaking\index{symmetry breaking } is different from synchrony-breaking.\index{synchrony-breaking }

\subsection{Biology is robust}\index{biology, robustness of }
\label{sec:bio_robust}

This example shows that an orbital partition based on a global
symmetry cannot properly capture cluster synchronization: a
change in any part of the network can produce a loss of
synchronization in an orbit located in another part of the network
that is not even connected with the site of this change. Such a
vulnerable biological network would not survive under evolutionary
pressure. A resilient evolving network requires a more robust type of symmetry
that preserves local synchronization under many small changes to the network, or the equations modeling its dynamics. Such a
local symmetry could guarantee robustness of the synchronization,
while at the same time providing flexibility to changes in the
network structure for evolvability\index{evolvability } of new phenotypes under natural
selection.

This example raises the following question: although there are no
(nontrivial) automorphisms, are there extra symmetries in the network
shown in Fig.~\ref{fig: example}b that could lead to robust
synchronization? The answer to this question is, `yes' in this case: there are
fibration symmetries. Fibration symmetries, by
not being global symmetries of the network, allow
more robust patterns of synchrony under random addition or deletion of
nodes. This fact allows the network to evolve more freely than one that is
constrained by symmetry groups would do, and can still preserve  synchronization. 

The same example shows that synchronization is usually not a property of a single cluster. Instead, it is a collective property of a set of clusters, corresponding to the partition (coloring) determined by the clusters. So when we say that certain nodes `can synchronize', this statement is always made in the context of a set of clusters, and
usually requires other sets of nodes to synchronize as well.

Automorphisms always preserve outputs as well as inputs.
Fibrations preserve only inputs. This is one reason why they are more common.
In undirected graphs every input is also an output, so fibrations are 
likely to be more common in directed networks. Most biological networks
are directed. That said, most directed networks, constructed randomly, 
have only the trivial fibration symmetry.
Moreover, reversing the extra arrow $2\to 7$ in Fig. \ref{fig: example}b
to get $7 \to 2$ also destroys the group symmetry, so
the input/output distinction is not the only reason for
the loss of symmetry.

In summary: automorphisms describe invariances in all physical systems (elementary
particles, forces, atoms and molecules) but are rare in biological
networks, perhaps due to their inherent vulnerabilities.  Next, we
will see how fibration symmetries---with their emphasis on preserving
only input trees locally but not globally---offer a less constrained
and more robust alternative to symmetry groups, still producing the
necessary synchronization required for biological function.

\section{Modeling assumptions leading to synchrony}
\label{sec:introfib}

As discussed in section \ref{sec:from}, one of our aims is to decipher the
structure $\rightsquigarrow$ function 
relation\index{structure-function relation } in biology. This means understanding how
the graph structure determines the dynamics of the biological system
and, in turn, the biological functions of its components. Our purpose is to decouple the
dynamics from the structure of the graph, making predictions of
 synchronization based on graph properties alone. The way to
achieve this is via the fibration formalism. To make theoretical
progress in this direction we start with a series of modeling
assumptions about conditions that permit synchrony. In
Chapter \ref{chap:fibration_2} we show how these assumptions
let us apply fibrations to biological graphs. Later, we relax these
assumptions by switching to more
realistic models of biological networks, tested by experimental data on
biological synchronization, but the same
basic principles continue to hold.

We propose that processes in cellular networks can be
seen as computations. Initially, for simplicity, we think of a
biological network or, more generally, an information-processing
network, as a Von Neumann cellular automaton\index{cellular automaton
} or `deterministic finite-state machine':\index{finite-state machine
} a computational machine made of units passing signals or `messages'
through directed connections to perform some kind of computation based
on their previous states \citep{leifer2020circuits}. In such models,
the phase spaces of nodes are discrete (indeed, finite).  More complex
dynamics such as ODEs are analogous, and the same metaphors apply, subject to suitable interpretations.

These signals could represent
neuronal spiking, forming the basis for information transfer in the
brain, or `message passing' from source gene to target gene in a
transcriptional regulatory network, where a transcription factor
expressed by the source gene diffuses in the cell until it binds to
the promoter DNA site of the target gene to turn it on and off.
Our aim is to capture key aspects of the more complex reality in a
simple model, and we start with the simplest setting.

In a directed network, nodes pass signals or messages from source to target
along the directed links. In particular, nodes can
receive signals from other nodes. This information flow respects the
direction of edges: if there is an edge from node $i$ to node $j$,
that connection can be used to send information \emph{from} $i$
\emph{to} $j$, but not the other way around.  Based on the signals
they receive, nodes compute and then broadcast further signals through
their outgoing edges, and so on.
The nodes of such a network are autonomous processing entities,
behaving according to well-defined rules. More precisely,
we start from the following three assumptions. Later we add two more.

\begin{itemize}
    \item {\bf States:}\index{state } Every node, at every moment in time, finds
      itself in a certain state.
    \item {\bf Determinism:}\index{determinism } Its future behavior depends
      deterministically (i.e., in a uniquely prescribed way)
      only on its current state and on the inputs it receives from
      incoming signals.
    \item {\bf Broadcast: }\index{broadcast } When a node sends out a signal, it sends
      the same signal through all of its outgoing edges of the same
      type.
\end{itemize}

While the first two assumptions are quite natural (and are the essence
of a Von Neumann cellular automaton\index{cellular automaton } or deterministic finite-state
machine),\index{finite-state machine } the last one is more puzzling. What we are saying is that,
for example, when node $6$ in the network of Fig. \ref{fig: example}b
communicates a signal, it will communicate the same piece of
information both to node $4$ and to node $5$ (the only two nodes it
can communicate to). This postulate is not as restrictive as it might
seem, and can be circumvented using typed edges. For example, to enable
node $6$ to communicate differently with nodes $4$ and $5$, we 
give the edges $6 \to 4$ and $6 \to 5$ different types. For instance, one edge could be an activator and the other a
repressor. Thus edge labels (often indicated in diagrams by different
forms for the arrows)
can capture different types of
biological interactions, such as activators and repressors in a gene
regulatory network, or excitatory/inhibitory connections in the brain. Thus the third constraint above can be relaxed
even further by requiring the broadcast to be uniform only for the
messages sent through a given type of edge.  We will get back
to this later, but for the moment, and for the sake of simplicity, we
 maintain the assumption above and refrain from using different types of edges.

Based on these assumptions, for instance, since node $6$ of Fig. \ref{fig: example}
has no incoming
edges, it receives no signals, so its future behavior depends
only on its initial state. In particular, this observation entails
that the signals that node $6$ sends to $4$ and $5$ depend only on
its initial state, and nothing else: it can be considered as an
external parameter of the system.

Now, consider nodes $5$ and $4$. They receive 
information only from node $6$;
moreover, they receive the same information
because of the broadcast postulate. So if
they start in the same state, they will remain in the same
state because of the determinism postulate.  Therefore they
behave in synchrony, even though there is no graph
automorphism that exchanges them.

In this chapter we are concerned only with showing that these
synchronous states {\em exist}, as a consequence of fibration symmetries of the
graph. However, we repeat that the existence of cluster synchronization does not
guarantee that the solutions concerned are dynamically stable. 
As already mentioned, the study of the stability\index{stability }
and attractiveness of these synchronous states is referred to as
synchronizability.\index{synchronizability }  The stability of the synchronized solution is 
a separate issue from its existence. A
synchronous solution predicted by fibration symmetries or any other
symmetry may result in an unstable state, which would not be biologically meaningful.
We discuss stability in Chapter \ref{chap:stability}.

\subsection{Necessary conditions for synchrony}\index{synchrony !necessary conditions }
\label{sec:necessary_conditions}

To make one  further step towards understanding the  nature of biological networks, we continue to study the example of Fig. \ref{fig: example}. We consider a subset $S$ of nodes, and ask what conditions
are required for $S$ to be a synchronized cluster. We say that such a set can `potentially' synchronize.

One obvious condition is that the concept of synchrony must make
sense for the nodes concerned. This is not the case if, for example,
one node has phase space $\R$ (the real numbers) and the other has phase space ${\bf S}^1$ (the circle, that is, angles), since it is not legitimate to compare a number with an angle.

In some types of model the number $0$ plays a special role,
indicating `no activity', and such a comparison might be
considered legitimate for some purposes, but the topology of the real line differs from that of the circle. In the theoretical parts of this book we do not assume that $0$ is special;
in particular we do not assume that the functions defining the dynamics in an admissible ODE vanish at their origin---an assumption that is often made in the literature. So, in general, there is no
automatic complete synchronization state like those given by Laplacian dynamics in Section \ref{sec:laplacian}. Instead, the {\it structure
of the network} is what makes synchronized states possible 
without building them into the model from the start.

Since nodes in a synchronized cluster are synchronized for all time,
we avoid breaking synchrony via unsynchronized initial conditions by
assuming that all nodes that potentially could synchronize are initially set to the 
same state. This condition assumes that `same state'
can meaningfully be defined for all nodes, which need not be the case
in general, but it must make sense for nodes in a synchronized cluster,
because it is the key feature defining synchrony.
For this reason we add a fourth assumption:
\begin{itemize}
    \item {\bf Anonymity: }\index{anonymity } When seeking synchrony patterns, we may assume that all nodes that can potentially synchronize start from the same initial state.
\end{itemize}

Suppose we apply this reasoning to nodes $2$ and $3$. Node $2$
receives signals from node $4$ and from node $3$, whereas node $3$
receives signals from node $5$ and from node $2$. As we said, node $4$
and node $5$ behave in the same way, 
so node $2$ and node $3$ will
receive the same signals along one of the edges (whenever some piece of
information is sent along the edge $4 \to 2$, the same piece of
information is sent along the edge $5 \to 3$). Nodes $2$
and $3$ also receive signals from each other, but again this 
cannot break their synchrony if they start in the same state.

The anonymity\index{anonymity } assumption might look very restrictive; indeed,
no real biological system can satisfy it, since the initial
states are constantly changing and the cell needs to adapt to
continuously changing signals from the environment. Mathematically, it represents the principle that 
any meaningful synchrony pattern should persist over time.
Another way to say this is that the initial state lies in the appropriate synchrony 
subspace.\index{synchrony !subspace }
In line with our
purpose of decoupling the message-passing dynamics from the
graph structure, this assumption is necessary at this stage. 
However, this restriction can easily be relaxed afterward.
In particular, it does not imply that nodes that start in different states cannot tend toward synchrony as time passes.
This condition, in fact, avoids the problem of synchronizability to be treated in Chapter \ref{chap:stability}.

To avoid `accidental' synchrony patterns, it is convenient to add a fifth assumption:
\begin{itemize}
    \item {\bf No Fragility: }\index{fragility } The synchrony pattern is not {\em fragile }, i.e., it does not depend on
    special features of the admissible equations. In particular,
    it is not destroyed by small admissible changes to the equations.
\end{itemize}
For example, consider Fig. \ref{fig5}. In an extreme case, suppose that
the function $g$ that represents the coupling is identically zero,
so the equations decouple. Then, all five nodes obey the same
equation $\dot x_i = f(x_i)$, so complete synchronization is possible.
However, complete synchronization is {\it not} possible in this
network with more general choices of $g$, because for many $g$ the components $\dot x_1$ and
$\dot x_2$ must satisfy different equations. This makes the equations logically inconsistent. In this case, what below we call the
`minimal equitable partition'---the synchronization pattern with fewest
clusters---is the pattern with clusters $\{1\},\{2,3\},\{4,5\}$.

\begin{remark}
More seriously, the same problem arises if the coupling function $g$
vanishes whenever the source and target nodes are synchronized;
that is, if $g(y,y) = 0$ for all $y$. Again, the equations
decouple and complete synchronization becomes possible. A common
example of this kind of coupling is `diffusive coupling'.
Since this is a common modeling assumption in some areas, the definition of
a fibration or balanced coloring should be modified. This can be done
\citep[Section 8.9]{GS2023}, see also \cite{kamei2013computation,diaz2017evaluating,pecora2016b,siddique2018symmetry}.
Another widespread context in which coupling between synchronous nodes is always zero occurs in models based on the graph Laplacian. See Sections \ref{sec:robust} and \ref{S:MSFSI}.
\end{remark}

\section{Input sets and input trees}
\label{sec:input_sets_trees}

We now study the consequences of the observations listed in the previous section,
starting from the notions of input set and input tree.  There are several definitions of the input set in the
literature. We define it in a way that is most consistent
with information-processing networks. Simply stated, the input set\index{input set } of
a node is the set of incoming {\it edges}, and the number of them is the
{\it in-degree}\index{in-degree } (or {\em valence})\index{valence } of that node. 

Using only edges may seem unnatural or pedantic, but
it makes the mathematical formalism work better, and is necessary
for multigraphs. Related information, such as source or target nodes of an edge, can
be specified separately as additional structure. We also think of the
node itself as a special type of edge, and include it in the
input set. (In \citep{GS2023} this version is called
the `extended input set'. The node is included in the formalism as a distinguished variable in model equations.)

\begin{definition} {\bf Input set of a node.} \index{input set }
  The \textit{input set} of the node $i \in V$ in a graph $G=(V,E)$ is
  the set $I_i = \{i\} \cup \{e \mid e \in E, t(e) = i\}$ that
  consists of the node itself and all of its incoming edges (which can
  be self-loops, if the node has any).
  \label{inputset}
\end{definition}

We denote the input set of a node $i\in V$ with in-degree $k_{\rm in}$
as $I_i=(i: e_1, \dotso, e_{k_{\rm in}})$, where $e_j\in E$
are the incoming edges. This also defines the {\it local
  in-neighborhood} of the node $i$ in the terminology of graph
fibrations \citep{boldi2002fibrations}.  An analogous notation also
employed in the literature on symmetry groupoids is in terms of the
nodes connecting to $i$ from incoming edges, thus $I_i = (i :
1, \dotso, k_{\rm in})$, where $(1, \dotso, k_{\rm in}) \in V$ are the
in-neighborhood nodes of $i$ (see Fig. \ref{fig:inputsets}): here, we are assuming that the graph itself is simple.

Conceptually, the input set of $i$ is the set of `objects' that can
influence $i$'s behavior: it includes $i$ itself (because the current
state of the node will of course, influence its own future behavior) as
well as all its incoming edges (because it is through those edges that
input signals come, and input signals influence the node's
behavior). We deliberately prefer to use incoming edges
instead of their source nodes because multigraphs can have
several edges with the same source and the same target. This
influences the behavior differently compared to a single
edge (even when they all carry the same signal). However, it is the
state of the source node that generates that signal: we do
not equip edges with states.

That said, when drawing input sets, it is often convenient to include
their source nodes as well as their target nodes. The source nodes are not part of the input set, but they add useful information for purposes such as the construction of admissible ODEs. In such drawings, distinct edges are drawn with
distinct sources; however, if the nodes are numbered, some source nodes may correspond to the same number in the case of multigraphs. That is, the sources
of multiple edges are split apart but numbered in a way that shows
how to identify them.

An input set is naturally represented by a graph. Consider
Fig.~\ref{fig:example-inputsets}, where we show again the network of
Fig.~\ref{fig: example}b left, but with names on the edges so that they can
be identified; Fig.~\ref{fig:inputsets} shows the input sets
of nodes $1$, $2$, $3$ and $4$.

\begin{figure}
    \centering
    \scalebox{1}{
        \begin{tikzpicture}
        \tikzset{vertex/.style = {shape=circle,draw}}
        \tikzset{edge/.style = {->,> = latex'}}
        \node[vertex] (v6) at  (1,0) {$6$};
        \node[vertex] (v4) at  (0,1) {$4$};
        \node[vertex] (v5) at  (2,1) {$5$};
        \node[vertex] (v2) at  (0,3) {$2$};
        \node[vertex] (v3) at  (2,3) {$3$};
        \node[vertex] (v7) at  (4,3) {$7$};
        \node[vertex] (v1) at  (1,4) {$1$};
        \draw[edge] (v6) to node [below] {$g$} (v4);
        \draw[edge] (v6) to node [below] {$h$} (v5);
        \draw[edge] (v4) to node [left] {$e$} (v2);
        \draw[edge] (v5) to node [right] {$f$}(v3);
        \draw[edge] (v2) to[bend left] node [below] {$c$} (v3);
        \draw[edge] (v3) to[bend left] node [below] {$d$}(v2);
        \draw[edge] (v3) to node [above] {$i$} (v7);
        \draw[edge] (v2) to node [above] {$a$} (v1);
        \draw[edge] (v3) to node [above] {$b$} (v1);
        \end{tikzpicture}
    }
    \caption{\textbf{A network with no automorphisms.} The same as in
      Fig.~\ref{fig: example}b (left) but with labeled edges. Usually
      nodes are labeled with numbers and edges with letters.}
    \label{fig:example-inputsets}
\commentAlt{Figure~\ref{fig:example-inputsets}:  Same graph as 
Figure~\ref{fig: example} A. Arrows labeled 
a (2 to 1), b (3 to 1), c (2 to 3), d (3 to 2), e (4 to 2), f (5 to 4),
g (6 to 4), h (6 to 5), i (3 to 7).
}
\end{figure}

\begin{figure}
    \centering
    \begin{tabular}{cccc}
    \scalebox{1}{
        \begin{tikzpicture}[baseline=(current bounding box.center)]
        \tikzset{vertex/.style = {shape=circle,draw}}
        \tikzset{edge/.style = {->,> = latex'}}
        \node[vertex] (v1) at  (1,1) {$1$};
        \node[vertex] (v2) at  (0,0) {$2$};
        \node[vertex] (v3) at  (2,0) {$3$};
        \draw[edge] (v2) to node [above] {$a$}(v1);
        \draw[edge] (v3) to node [above] {$b$}(v1);
        \end{tikzpicture}
    }
    &
    \scalebox{1}{
        \begin{tikzpicture}[baseline=(current bounding box.center)]
        \tikzset{vertex/.style = {shape=circle,draw}}
        \tikzset{edge/.style = {->,> = latex'}}
        \node[vertex] (v2) at  (1,1) {$2$};
        \node[vertex] (v3) at  (0,0) {$3$};
        \node[vertex] (v4) at  (2,0) {$4$};
        \draw[edge] (v3) to node [above] {$d$}(v2);
        \draw[edge] (v4) to node [above] {$e$} (v2);
        \end{tikzpicture}
    }
    &
    \scalebox{1}{
        \begin{tikzpicture}[baseline=(current bounding box.center)]
        \tikzset{vertex/.style = {shape=circle,draw}}
        \tikzset{edge/.style = {->,> = latex'}}
        \node[vertex] (v3) at  (1,1) {$3$};
        \node[vertex] (v5) at  (0,0) {$5$};
        \node[vertex] (v2) at  (2,0) {$2$};
        \draw[edge] (v5) to node [above] {$f$}(v3);
        \draw[edge] (v2) to node [above] {$c$} (v3);
        \end{tikzpicture}
    }
    & 
    \scalebox{1}{
        \begin{tikzpicture}[baseline=(current bounding box.center)]
        \tikzset{vertex/.style = {shape=circle,draw}}
        \tikzset{edge/.style = {->,> = latex'}}
        \node[vertex] (v4) at  (0,1) {$4$};
        \node[vertex] (v6) at  (0,0) {$6$};
        \draw[edge] (v6) to node [left] {$g$}(v4);
        \end{tikzpicture}
    }
    \end{tabular} 
    \caption{\textbf{Definition of input set.} Input sets for the nodes
      $1$, $2$, $3$ and $4$ of the network of
      Fig.~\ref{fig:example-inputsets}, together with their source and target nodes. We denote the input set of,
      for instance, node 1 as: $I_1 = (1 : a, b)$.}
    \label{fig:inputsets}
\commentAlt{Figure~\ref{fig:inputsets}:  From  left to right:
Input set for node 1: nodes 1, 2, 3 and edges a, b.
Input set for node 2: nodes 2, 3, 4 and edges d,e.
Input set for node 3: nodes 2, 3, 5 and edges f,c.
Input set for node 4: nodes 4, 6 and edge g.
}
\end{figure}

Input sets can be put together inductively to form {\it input trees},\index{input tree }
following ~\citep{morone2020fibration}. Input trees are referred to as an
`infinite depth input tree' in \citep{aldis2008, stewart2007} and as a `universal
total graph' in~\citep{boldi2002fibrations}.
Technically, the tree graph is a `universal cover' of the original graph.

\begin{definition}{\bf Input tree of a node.} \index{input tree }
We inductively define the \emph{input tree} $T_i$ of a node $i \in G$ as
follows:
\begin{itemize}
    \item if $I_i=\{i\}$, then the input tree $T_i$ is the root-only tree;
    \item if $I_i= ( i : e_1,\dots,e_{k_{\rm in}} )$, then the input
      tree $T_i$ is formed by a root with $k_{\rm in}$ incoming edges from its
      $k_{\rm in}$ input nodes, plus the subtrees rooted at the input nodes
      $T_{s(e_1)}, \dots, T_{s(e_{k_{\rm in}})}$ in no particular order.
\end{itemize}
\end{definition}

Again, only the edges are required, but it is convenient to add
the target and sources to help specify paths through the graph.

Input trees can be finite or infinite, but they are infinite in most
practical cases (as soon as there is a cycle in the graph). More
precisely, $T_i$ is infinite if and only if there is a node $j$ that has a
path to $i$ and is involved in a cycle. (This cycle could be an
`autoregulation' loop from $j$ to itself.)

\begin{figure}
    \centering
    \scalebox{1}{
        \begin{tikzpicture}
        \tikzset{vertex/.style = {shape=circle,draw}}
        \tikzset{edge/.style = {->,> = latex'}}
        \node[vertex] (v2) at  (4,5) {$2$};
        \node[vertex] (v32) at  (2,4) {$3$};
        \node[vertex] (v42) at  (7,4) {$4$};
        \node[vertex] (v532) at  (0,3) {$5$};
        \node[vertex] (v232) at  (3,3) {$2$};
        \node[vertex] (v642) at  (7,3) {$6$};
        \node[vertex] (v6532) at  (0,2) {$6$};
        \node[vertex] (v3232) at  (2,2) {$3$};
        \node[vertex] (v4232) at  (5,2) {$4$};
        \node[vertex] (v53232) at  (1,1) {$5$};        
        \node[vertex] (v23232) at  (3,1) {$2$};        
        \node[vertex] (v64232) at  (5,1) {$6$};        
        \node (vd53232) at  (1,0) {$\vdots$};        
        \node (vd23232) at  (3,0) {$\vdots$};                
        \draw[edge] (v32) to node [below] {$d$} (v2);
        \draw[edge] (v42) to node [below] {$e$} (v2);        
        \draw[edge] (v532) to node [below] {$f$} (v32);
        \draw[edge] (v232) to node [below] {$c$} (v32);
        \draw[edge] (v642) to node [left] {$g$} (v42);
        \draw[edge] (v6532) to node [left] {$h$} (v532);
        \draw[edge] (v3232) to node [below] {$d$} (v232);
        \draw[edge] (v4232) to node [below] {$e$} (v232);
        \draw[edge] (v53232) to node [below] {$f$} (v3232);
        \draw[edge] (v23232) to node [below] {$c$} (v3232);
        \draw[edge] (v64232) to node [left] {$g$} (v4232);
        \draw[edge] (vd53232) to  (v53232);
        \draw[edge] (vd23232) to  (v23232);
        \end{tikzpicture}
    }
    \caption{\textbf{Example of an input tree.} The input tree $T_2$ of node
      $2$ in the network of Fig.~\ref{fig:example-inputsets}. The tree
      is infinite (we show only its upper levels).}
    \label{fig:example-tree2}
\commentAlt{Figure~\ref{fig:example-tree2}: Tree graph for Figure~\ref{fig:inputsets}.
Nodes and edges connected as for that figure.
Level 0: node 2.
Level -1: nodes 3 and 4; edges d and e.
Level -2: nodes 5, 2, 6; edges f, c, g.
Level -3: Nodes 6, 3, 4; edges h, d, e.
Level -4: Nodes 5, 2, 5; edges f ,c, g.
}
\end{figure}

As we said, input trees can be built by starting from input sets and
stitching them together, and here the node labels are useful.  In Fig.~\ref{fig:example-tree2} we see the
input tree $T_2$ for node $2$ of the network of
Fig.~\ref{fig:example-inputsets}. Observe that the same node can
(and usually does) occur repeatedly in different positions 
in the tree, and we do not
identify these nodes in the tree diagram.  So these node
and edge labels are there only
for clarity. An input tree can be imagined as the set
of ways in which signals (or, if you prefer, states) can reach a given
node. That is, the directed {\it paths}\index{path }
through the graph. For instance, node $2$ (the root of the tree) can receive
information from node $4$ (through the edge named $e$), whose state
can depend on the information it receives from node $6$ (through the
edge named $g$). This is equivalent to saying that the state of node
$6$ influences the state of node $4$, which in turn influences the
state of node $2$. Paths through the tree correspond to paths through the original graph, but in the tree, distinct paths do not overlap or share common edges, as they may do in the graph.

Input trees can be better understood if we consider them as unfoldings
of network behavior along the axis of time. 
This way of reasoning
requires a further assumption:
\begin{itemize}
    \item {\bf Synchronous update:}\index{synchronous update } all nodes change their state at
      the same time.
\end{itemize}

This quite strong assumption can be thought of as yet another form of
worst-case scenario: once more, we wish to look at the graph structure
as the only source of asymmetry, so we keep all other
sources of asymmetry (initial states of the node, communication time,
broadcasting, etc) constant. These constraints can be relaxed later.

\begin{remark}
Synchrony between nodes can be broken based on 
communication delays;\index{delay } that is, two otherwise symmetric nodes can receive the
same signal with a different delay, which may cause asynchronous
dynamics. In this case the edges of the network, which represent
{\em couplings} between nodes, 
have different types, but a symmetry must preserve edge types.
This chapter concerns the kind of symmetry
breaking in which states of the system have less symmetry than
that of the network. This is possible because nonlinear systems
can support several distinct states simultaneously, selected by initial conditions, and a fully symmetric state may be unstable.
Symmetry breaking\index{symmetry breaking !caused by delays } caused by different delays,
together with the question of the dynamics of
networks with non-fixed architecture, are outside of the scope of
this book.   
\end{remark}

Synchronous update implies that there is some global clock\index{clock } that keeps
ticking, and at every tick, akin to a computer, simultaneously all
nodes receive their incoming signals and change their state.
This is `discrete' dynamics.\index{dynamics !discrete } It is also possible to model
the dynamics in continuous time\index{dynamics !continuous } using differential equations,
and here the use of a common time variable for all nodes effectively
updates their states synchronously. Again, there is a global clock,
but it `ticks' continuously.

\begin{figure}
    \centering
    \scalebox{1}{
        \begin{tikzpicture}
        \tikzset{vertex/.style = {shape=circle,draw}}
        \tikzset{edge/.style = {->,> = latex'}}
        \node[vertex] (v2) at  (4,5) {$2$};
        \node[vertex] (v32) at  (2,4) {$3$};
        \node[vertex] (v42) at  (7,4) {$4$};
        \node[vertex] (v532) at  (0,3) {$5$};
        \node[vertex] (v232) at  (3,3) {$2$};
        \node[vertex] (v642) at  (7,3) {$6$};
        \node[vertex] (v6532) at  (0,2) {$6$};
        \node[vertex] (v3232) at  (2,2) {$3$};
        \node[vertex] (v4232) at  (5,2) {$4$};
        \node[vertex] (v53232) at  (1,1) {$5$};        
        \node[vertex] (v23232) at  (3,1) {$2$};        
        \node[vertex] (v64232) at  (5,1) {$6$};        
        \node (vd53232) at  (1,0) {$\vdots$};        
        \node (vd23232) at  (3,0) {$\vdots$};                
        \draw[edge] (v32) to node [below] {$d$} (v2);
        \draw[edge] (v42) to node [below] {$e$} (v2);        
        \draw[edge] (v532) to node [below] {$f$} (v32);
        \draw[edge] (v232) to node [below] {$c$} (v32);
        \draw[edge] (v642) to node [left] {$g$} (v42);
        \draw[edge] (v6532) to node [left] {$h$} (v532);
        \draw[edge] (v3232) to node [below] {$d$} (v232);
        \draw[edge] (v4232) to node [below] {$e$} (v232);
        \draw[edge] (v53232) to node [below] {$f$} (v3232);
        \draw[edge] (v23232) to node [below] {$c$} (v3232);
        \draw[edge] (v64232) to node [left] {$g$} (v4232);
        \draw[edge] (vd53232) to  (v53232);
        \draw[edge] (vd23232) to  (v23232);
        \draw (-0.55,5) -- (v2) -- (7.5,5) [dashed];
        \draw (-0.55,4) -- (v32) -- (v42) -- (7.5,4) [dashed];
        \draw (-0.55,3) -- (v532) -- (v232) -- (v642) -- (7.5,3) [dashed];
        \draw (-0.55,2) -- (v6532) -- (v3232) -- (v4232) -- (7.5,2) [dashed];
        \draw (-0.55,1) -- (v53232) -- (v23232) -- (7.5,1) [dashed];
        \draw[->,line width=1.8pt]  (-0.5,0) to (-0.5,5.5);
        \node at (-0.8,5) {$0$};
        \node at (-0.9,4) {$-1$};
        \node at (-0.9,3) {$-2$};
        \node at (-0.9,2) {$-3$};
        \node at (-0.9,1) {$-4$};
        \end{tikzpicture}
    }
    \caption{\textbf{Dynamical interpretation of the input tree.} The
      input tree $T_2$ of node $2$ in the network of
      Fig.~\ref{fig:example-inputsets}, as a representation of the
      information flow represented as an unfolding in time of the
      signals that node $2$ receives.}
    \label{fig:example-tree2-time}
\commentAlt{Figure~\ref{fig:example-tree2-time}: 
Same graph as Figure~\ref{fig:example-tree2} with levels
marked by horizontal dashed lines.
}
\end{figure}

Figure ~\ref{fig:example-tree2-time} shows the same input tree $T_2$,
but on a timeline: the axis of time is shown on the left, and nodes of
the tree are aligned with time ticks. 
Thus, the state of node $2$ at
time $0$ depends on the states of nodes $3$ and $4$ at times $-1$. In
turn, the state of node $3$ at time $-1$ depends on the states of
nodes $5$ and $2$ at time $-2$ and so on.

Here we have assumed discrete dynamics. For continuous dynamics, the {\em whole tree}
updates simultaneously.

The input tree\index{input tree !as set of paths } of a node $i$
can be thought of as the complete set of all pathways in the network
that terminate at $i$
~\citep{morone2020fibration,boldi2002fibrations,aldis2008} (the node itself
can be a part of that path if it has a self-loop). Therefore, an input tree
contains all the pathways that can contribute to the dynamics of the
node, showing all the `information' about the network that is available
to the node, hence the only form of \emph{asymmetry} that the node
can use to differentiate itself from the other nodes in the network.

\begin{figure}
    \centering
    \scalebox{1}{
        \begin{tikzpicture}
        \tikzset{vertex/.style = {shape=circle,draw}}
        \tikzset{edge/.style = {->,> = latex'}}
        \node[vertex] (v3) at  (4,5) {$3$};
        \node[vertex] (v23) at  (2,4) {$2$};
        \node[vertex] (v53) at  (7,4) {$5$};
        \node[vertex] (v423) at  (0,3) {$4$};
        \node[vertex] (v323) at  (3,3) {$3$};
        \node[vertex] (v653) at  (7,3) {$6$};
        \node[vertex] (v6423) at  (0,2) {$6$};
        \node[vertex] (v2323) at  (2,2) {$2$};
        \node[vertex] (v5323) at  (5,2) {$5$};
        \node[vertex] (v42323) at  (1,1) {$4$};        
        \node[vertex] (v32323) at  (3,1) {$3$};        
        \node[vertex] (v65323) at  (5,1) {$6$};        
        \node (vd42323) at  (1,0) {$\vdots$};        
        \node (vd32323) at  (3,0) {$\vdots$};                
        \draw[edge] (v23) to node [below] {$c$} (v3);
        \draw[edge] (v53) to node [below] {$f$} (v3);        
        \draw[edge] (v423) to node [below] {$e$} (v23);
        \draw[edge] (v323) to node [below] {$d$} (v23);
        \draw[edge] (v653) to node [left] {$h$} (v53);
        \draw[edge] (v6423) to node [left] {$g$} (v423);
        \draw[edge] (v2323) to node [below] {$c$} (v323);
        \draw[edge] (v5323) to node [below] {$f$} (v323);
        \draw[edge] (v42323) to node [below] {$e$} (v2323);
        \draw[edge] (v32323) to node [below] {$d$} (v2323);
        \draw[edge] (v65323) to node [left] {$h$} (v5323);
        \draw[edge] (vd42323) to  (v42323);
        \draw[edge] (vd32323) to  (v32323);
        \end{tikzpicture}
    }
    \caption{\textbf{Example of input tree.} The input tree $T_3$ of node
      $3$ in the network of Fig.~\ref{fig:example-inputsets}. The tree
      is infinite (only its upper levels are shown).}
    \label{fig:example-tree3}
\commentAlt{Figure~\ref{fig:example-tree3}: Same figure
as Figure~\ref{fig:example-tree2} but drawn for node 3.
Level 0: node 3.
Level -1: nodes 2 and 5; edges c and f.
Level -2: nodes 4, 3, 6; edges e, d, h.
Level -3: Nodes 6, 2, 5; edges g, c, f.
Level -4: Nodes 4, 3, 6; edges e, d, h.
}
\end{figure}

\section{Examples of input trees}
\label{sec:ex_input_tree}

\begin{figure}
	\centering
	\includegraphics[width=0.7\linewidth]{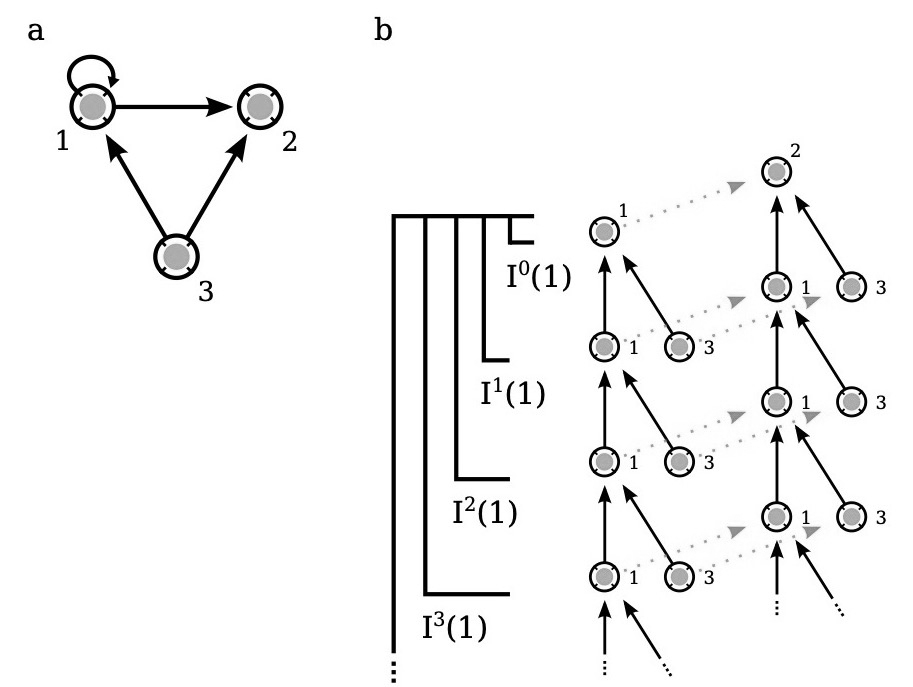}
	\caption{\textbf{Example of the input tree of the FFF.}
          (\textbf{a}) The feed-forward fiber (FFF) circuit.
          (\textbf{b}) Input trees of the three nodes in the
          graph. The input set of node $1$ consists of nodes $1$ and
          $3$ and the next layer of the input tree is the input set of
          $1$ attached to node $1$.  The input set of node $3$ is only
          the node itself. The procedure is repeated. The red dotted
          arrow shows the bijection between input trees $I_1$ and
          $I_2$ illustrating the definition
          \ref{input tree-isomorphism} of graph isomorphism.}
	\label{fig:inputtree}
\commentAlt{Figure~\ref{fig:inputtree}: No alt-text required.
}
\end{figure}

Input trees are defined inductively as input sets of input sets \ldots
taken an infinite number of times. We previewed one example in 
Fig. \ref{F:Uxur_input_tree}. We now discuss a biological example, shown in
Fig.~\ref{fig:inputtree}: the input tree of the FFF circuit, already
discussed in Fig. \ref{fig1-saulo}. Figure ~\ref{fig:inputtree} shows
the process of constructing an infinite depth input tree for the FFF
circuit. For instance, consider node 1. The root of the tree is
the node itself. The first layer is the input set which contains nodes
3 and 1, since there is a self-loop at node 1 and 3 connects to 1.
The next layer is constructed by attaching the input set of $1$ to the
leaf $1$ and the input set of $3$ to the leaf $3$.  The procedure is
repeated {\it ad infinitum} to represent the full input tree.  Even if
the graph is finite, the input tree can be finite or infinite.  If the
graph has no cycles and self-loops, then all input trees are finite
\citep{aldis2008}. As soon as any node in the input tree belongs to a
cycle (even just a self-loop), the input tree has an infinite number of
layers. See also \citep[Section 13.3]{GS2023}.

\section{Input tree isomorphism}
\label{sec:isom}

Figure ~\ref{fig:example-tree2} and Fig.~\ref{fig:example-tree3} show
the input trees of node $2$ and $3$ (respectively) in the network of
Fig.~\ref{fig:example-inputsets}: although the node and edge labels
appearing there are different, the shape is the same. A more precise
statement would be that the two trees are isomorphic, which means
`equal shape'. This is why nodes $2$ and $3$ can be exchanged
by a fibration symmetry. 
We now formalize 
equivalence of the `shape' of two graphs, while ignoring their labeling
and the way they are drawn.

\begin{definition} {\bf Graph isomorphism} \citep{harary1969}. 
\index{graph isomorphism }
    Two graphs $G = \{N_G, E_G\}$ and $H = \{N_H, E_H\}$ are {\em isomorphic} if there exists a bijection  $\alpha_N:N_G\rightarrow
    N_H$ between nodes and another bijection  $\alpha_E:E_G\rightarrow E_H$ between edges, such that $s(\alpha_E(e)) = \alpha_N(s(e))$ and
    $t(\alpha_E(e)) = \alpha_N(t(e))$ for all $e\in E_G$. 
      We denote isomorphism of $G$ and $H$ by
    \begin{equation}
  G \simeq H\, .
\end{equation}
\label{def:isomorphism} 

Usually we just write $\alpha$ instead of
    $\alpha_N$ and $\alpha_E$, unless this would cause confusion.  We say that $\alpha$ is a {\em graph isomorphism}.
\end{definition}

\begin{figure}[htb]
\centerline{
\includegraphics[width=0.6\textwidth]{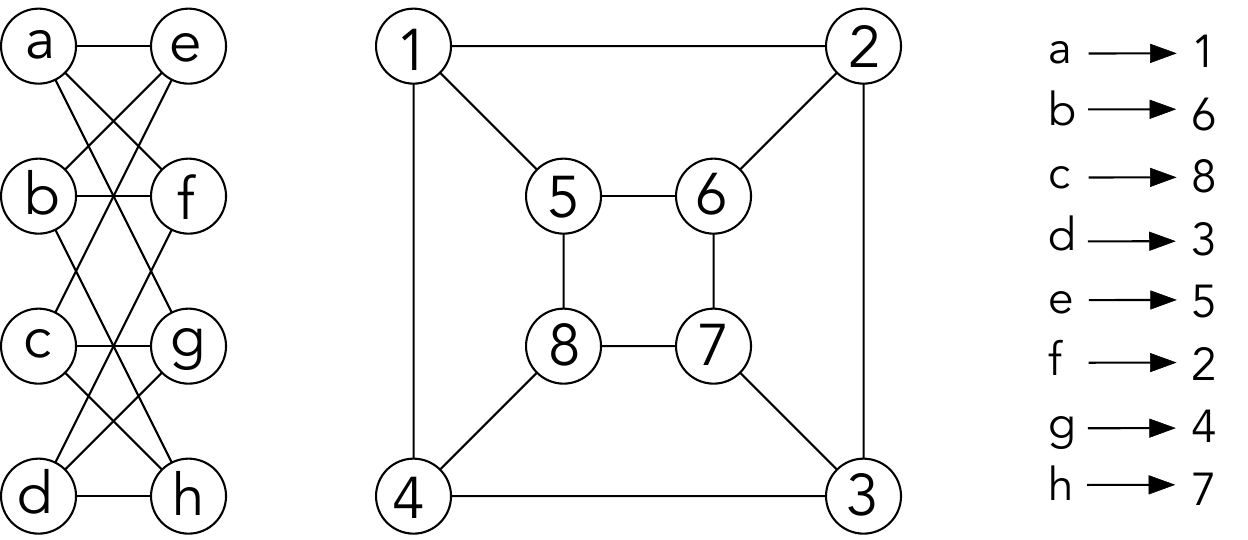} 
}
\caption{\textbf{Graph isomorphism.}  Two graphs $G$ and $H$ are
  isomorphic ($G \simeq H$) despite their different drawings and labels,
  meaning that when we relabel one of the graphs with the isomorphism
  shown on the right, the graphs are identical. Here labels are just
  numbers and letters, but in a biological network labels are genes (or proteins, neurons, and so on),
  and the graph isomorphism acquires an important dimension.}
\label{fig:iso}
\commentAlt{Figure~\ref{fig:iso}: Left: Graph with nodes
a, b, c, d, e, f, g, h. Edges connect ae, af ag; be, bf, bh;
ce, cg, ch; df, dg, dh. Middle: Graph with nodes
1, 2, 3, 4, 5, 6, 7, 8. Edges connect
12, 14, 15, 23, 26, 34, 37, 48, 56, 58, 67, 78.
Right: list of corresponding nodes:
(a, b, c, d, e, f, g, h) map to (1, 6, 8, 3, 5, 2, 4, 7).
}
\end{figure}

\begin{figure}[h!]
\centerline{
\includegraphics[width=\linewidth]{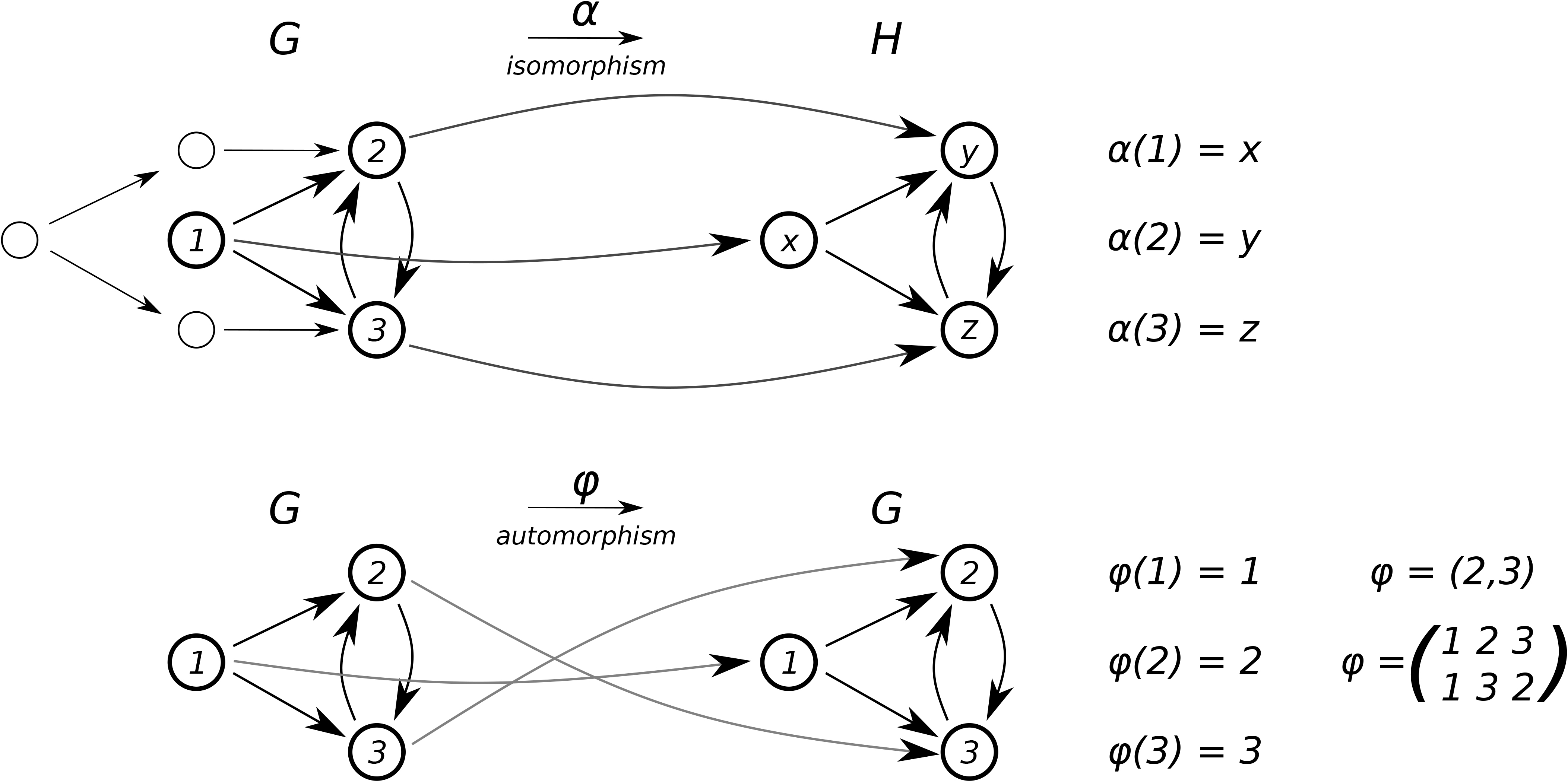} \qquad
}
\caption{\textbf{Isomorphism and automorphism.} Example of an isomorphism of a subgraph and an automorphism of the whole graph.}
\label{fig:iso-auto}
\commentAlt{Figure~\ref{fig:iso-auto}: 
Top: Graph with nodes 1, 2, 3 and arrows 1 to 2 and 3, 2 to 3, 3 to 2. 
Blue arrows (labeled `alpha', isomorphism) link this to graph with nodes x, y, z and arrows x to y and z, y to z, z to y,
showing map 1 to x, 2 to y, 3 to z.
Right: Two copies of graph with nodes 1,2,3. Red arrows (labeled `phi', automorphism)
link this to itself
but swap 2 and 3.
}
\end{figure}

Simply speaking, an isomorphism $\alpha$ maps nodes and edges of
$G$ to nodes and edges of $H$ in such a way that the image of the edge
connects the image of the source of this
edge to the image of the target, as in Figs. \ref{fig:iso} and 
 \ref{fig:iso-auto}).

A graph automorphism is an isomorphism that maps a graph onto itself,
i.e., $\alpha:G\rightarrow G$, see Fig. \ref{fig:iso-auto}. An
automorphism of a graph is a graph isomorphism with itself, i.e., a
mapping from $G$ back to $G$ such that the resulting graph is
isomorphic with $G$. The set of all automorphisms of $G$ is a group under functional composition, and it is precisely the symmetry group described in Chapter~\ref{chap:group}.

In general, determining whether two graphs are isomorphic is a difficult problem. There exists no known polynomial-time algorithm for graph isomorphism testing, although the problem has not been shown to
be NP-complete. It is believed that the problem belongs to a low hierarchy of the class NP, but its exact standing is unknown and it is itself an intensely active field of research~\citep{grohe2020}. However, we observe below that
when the original graph is finite, there are
efficient algorithms to test for isomorphisms between its input trees.

Since input trees are special instances of (possibly infinite) graphs,
we can apply the same notion of isomorphism to compare trees with the
same shape:

\begin{definition} {\bf Input tree isomorphism}. 
\index{input tree !isomorphism }
  Two input trees $T_i$ and $T_j$ are {\it isomorphic} when their
  graph representations are isomorphic:
\begin{equation}
  T_i \simeq T_j \, .
\end{equation}
That is, there exists an isomorphism $\alpha : T_i \to T_j$ that maps
one input tree to the other one. We call $\alpha$ an {\it input tree
isomorphism}.
\label{input tree-isomorphism}
\end{definition}

Looking back at Fig.~\ref{fig:example-tree2} and
Fig.~\ref{fig:example-tree3}, we see that $T_2 \simeq T_3$, see also
the example of Fig. \ref{fig:inputtree}. As observed above, if
$T_i\simeq T_j$ then $i$ and $j$ are indistinguishable, in the sense
that if they start in the same state, then at all times they will always have the same state. This means that nodes with
isomorphic input trees are synchronized, provided this pattern is applied to all nodes. This fact can be seen
immediately, looking at how $T_i$ was defined:
\begin{itemize}
    \item Suppose that $T_i \simeq T_j$ and that they are both a
      one-node tree. Assume these nodes have the same internal dynamic (which is always necessary for synchronization). Then $I_i=\{i\}$, and $I_j=\{j\}$, that is,
      neither $i$ nor $j$ have any incoming edge: by the anonymity
      postulate, they start in the same state, so they will remain in
      the same state forever, because they cannot receive any external
      stimulus. They are synchronous (although, of course, this is a trivial synchronization).
    \item Now, suppose that $T_i \simeq T_j$ and that the roots of these
      trees have $k>0$ children. The number of children must be the
      same, because the trees are isomorphic. Then
      $I_i=\{ i , e_1,\dots,e_k\}$ and $I_j=\{j , e_1',\dots,e_k'\}$;
      moreover, the input trees $T_{s(e_1)},\dots,T_{s(e_k)}$ are
      isomorphic to the input trees $T_{s(e_1')},\dots,T_{s(e_k')}$,
      suitably permuted.  For the sake of simplicity, let us assume
      that $T_{s(e_1)}\simeq T_{s(e_1')}$, \dots, $T_{s(e_1)}\simeq
      T_{s(e_k')}$. By induction, $s(e_l)$ and $s(e_l')$ will be
      forever in the same state at the same time, and so will $i$ and $j$ (because $i$
      receives stimuli through $e_1,e_2,\dots$, whereas $j$ receives
      stimuli through $e_1',e_2',\dots$).
\end{itemize}

Figure ~\ref{fig:example-alltrees} shows all input
trees for the network of Fig. \ref{fig:example-inputsets}, grouping
together those that are isomorphic to each other.  

\begin{figure}
    \centering
        \begin{tabular}{c|c}
    \hline
    $T_1$ &
    \scalebox{0.6}{
        \begin{tikzpicture}
        \tikzset{vertex/.style = {shape=circle,draw}}
        \tikzset{>={Latex[width=2mm,length=2mm]}}
        \node[] () at (6,6.3) {}; 
        \node[vertex] (v) at (7.2,6.6) {};
        \node[vertex] (v3) at  (4,5) {};
        \node[vertex] (v23) at  (2,4) {};
        \node[vertex] (v53) at  (7,4) {};
        \node[vertex] (v423) at  (0,3) {};
        \node[vertex] (v323) at  (3,3) {};
        \node[vertex] (v653) at  (7,3) {};
        \node[vertex] (v6423) at  (0,2) {};
        \node[vertex] (v2323) at  (2,2) {};
        \node[vertex] (v5323) at  (4,2) {};
        \node[vertex] (v42323) at  (1,1) {};        
        \node[vertex] (v32323) at  (3,1) {};        
        \node[vertex] (v65323) at  (4,1) {};        
        \node (vd42323) at  (1,0) {$\vdots$};        
        \node (vd32323) at  (3,0) {$\vdots$};    
        \node[vertex] (v2) at  (12,5) {};
        \node[vertex] (v32) at  (10,4) {};
        \node[vertex] (v42) at  (15,4) {};
        \node[vertex] (v532) at  (8,3) {};
        \node[vertex] (v232) at  (11,3) {};
        \node[vertex] (v642) at  (15,3) {};
        \node[vertex] (v6532) at  (8,2) {};
        \node[vertex] (v3232) at  (10,2) {};
        \node[vertex] (v4232) at  (12,2) {};
        \node[vertex] (v53232) at  (9,1) {};  
        \node[vertex] (v23232) at  (11,1) {};        
        \node[vertex] (v64232) at  (13,1) {};        
        \node (vd53232) at  (9,0) {$\vdots$};        
        \node (vd23232) at  (11,0) {$\vdots$};                
        \node [above left = 0mm and 0mm of v]() {$1$};
        \node [above left = 0mm and 0mm of v3]() {$3$};
        \node [above left = 0mm and 0mm of v23]() {$2$};
        \node [below left = 0mm and 0mm of v53]() {$5$};
        \node [above left = 0mm and 0mm of v423]() {$4$};
        \node [above right = 0mm and 0mm of v323]() {$3$};
        \node [below left = 0mm and 0mm of v653]() {$6$};
        \node [above left = 0mm and 0mm of v6423]() {$6$};
        \node [above left = 0mm and 0mm of v2323]() {$2$};
        \node [above right = 0mm and 0mm of v5323]() {$5$};
        \node [above left = 0mm and 0mm of v42323]() {$4$};
        \node [above right = 0mm and 0mm of v32323]() {$3$};
        \node [above right = 0mm and 0mm of v65323]() {$6$};
        \node [above right = 0mm and 0mm of v2]() {$2$};
        \node [above right = 0mm and 0mm of v32]() {$3$};
        \node [above right = 0mm and 0mm of v42]() {$4$};
        \node [above left = 0mm and 0mm of v532]() {$5$};
        \node [above right = 0mm and 0mm of v232]() {$2$};
        \node [above right = 0mm and 0mm of v642]() {$6$};
        \node [above left = 0mm and 0mm of v6532]() {$6$};
        \node [above left = 0mm and 0mm of v3232]() {$3$};
        \node [above right = 0mm and 0mm of v4232]() {$4$};
        \node [above left = 0mm and 0mm of v53232]() {$5$};
        \node [above right = 0mm and 0mm of v23232]() {$2$};
        \node [above right = 0mm and 0mm of v64232]() {$6$};
        \draw[->] (v2)--(v);
        \draw[->] (v2) to (v);
        \draw[->] (v3) to (v);
        \draw[->] (v23) to (v3);
        \draw[->] (v53) to (v3);        
        \draw[->] (v423) to (v23);
        \draw[->] (v323) to (v23);
        \draw[->] (v653) to (v53);
        \draw[->] (v6423) to (v423);
        \draw[->] (v2323) to (v323);
        \draw[->] (v5323) to (v323);
        \draw[->] (v42323) to (v2323);
        \draw[->] (v32323) to (v2323);
        \draw[->] (v65323) to (v5323);
        \draw[->] (vd42323) to  (v42323);
        \draw[->] (vd32323) to  (v32323);
        \draw[->] (v32) to (v2);
        \draw[->] (v42) to (v2);        
        \draw[->] (v532) to (v32);
        \draw[->] (v232) to (v32);
        \draw[->] (v642) to (v42);
        \draw[->] (v6532) to (v532);
        \draw[->] (v3232) to (v232);
        \draw[->] (v4232) to (v232);
        \draw[->] (v53232) to (v3232);
        \draw[->] (v23232) to (v3232);
        \draw[->] (v64232) to (v4232);
        \draw[->] (vd53232) to  (v53232);
        \draw[->] (vd23232) to  (v23232);
        \end{tikzpicture}}
    \\
    \hline
    $T_2 \simeq T_3$ &
    \scalebox{.6}{
        \begin{tikzpicture}
        \tikzset{vertex/.style = {shape=circle,draw}}
        \tikzset{>={Latex[width=2mm,length=2mm]}}
        \node[] () at (4,5.3) {}; 
        \node[vertex] (v3) at  (4,5) {};
        \node[vertex] (v23) at  (2,4) {};
        \node[vertex] (v53) at  (6,4) {};
        \node[vertex] (v423) at  (0,3) {};
        \node[vertex] (v323) at  (3,3) {};
        \node[vertex] (v653) at  (6,3) {};
        \node[vertex] (v6423) at  (0,2) {};
        \node[vertex] (v2323) at  (2,2) {};
        \node[vertex] (v5323) at  (4,2) {};
        \node[vertex] (v42323) at  (1,1) {};        
        \node[vertex] (v32323) at  (3,1) {};        
        \node[vertex] (v65323) at  (4,1) {};        
        \node (vd42323) at  (1,0) {$\vdots$};        
        \node (vd32323) at  (3,0) {$\vdots$};                
        \node [above left = 0mm and 0mm of v3]() {$2$};
        \node [above left = 0mm and 0mm of v23]() {$3$};
        \node [above right = 0mm and 0mm of v53]() {$4$};
        \node [above left = 0mm and 0mm of v423]() {$5$};
        \node [above right = 0mm and 0mm of v323]() {$2$};
        \node [above right = 0mm and 0mm of v653]() {$6$};
        \node [above left = 0mm and 0mm of v6423]() {$6$};
        \node [above left = 0mm and 0mm of v2323]() {$3$};
        \node [above right = 0mm and 0mm of v5323]() {$4$};
        \node [above left = 0mm and 0mm of v42323]() {$5$};
        \node [above right = 0mm and 0mm of v32323]() {$2$};
        \node [above right = 0mm and 0mm of v65323]() {$6$};
        \draw[->] (v23) to (v3);
        \draw[->] (v53) to (v3);        
        \draw[->] (v423) to (v23);
        \draw[->] (v323) to (v23);
        \draw[->] (v653) to (v53);
        \draw[->] (v6423) to (v423);
        \draw[->] (v2323) to (v323);
        \draw[->] (v5323) to (v323);
        \draw[->] (v42323) to (v2323);
        \draw[->] (v32323) to (v2323);
        \draw[->] (v65323) to (v5323);
        \draw[->] (vd42323) to  (v42323);
        \draw[->] (vd32323) to  (v32323);
        \end{tikzpicture}
    }
    \raisebox{1cm}{$\qquad\simeq\qquad$}
    \scalebox{.6}{
        \begin{tikzpicture}
        \tikzset{vertex/.style = {shape=circle,draw}}
        \tikzset{>={Latex[width=2mm,length=2mm]}}
        \node[] () at (4,5.3) {}; 
        \node[vertex] (v3) at  (4,5) {};
        \node[vertex] (v23) at  (2,4) {};
        \node[vertex] (v53) at  (6,4) {};
        \node[vertex] (v423) at  (0,3) {};
        \node[vertex] (v323) at  (3,3) {};
        \node[vertex] (v653) at  (6,3) {};
        \node[vertex] (v6423) at  (0,2) {};
        \node[vertex] (v2323) at  (2,2) {};
        \node[vertex] (v5323) at  (4,2) {};
        \node[vertex] (v42323) at  (1,1) {};        
        \node[vertex] (v32323) at  (3,1) {};        
        \node[vertex] (v65323) at  (4,1) {};        
        \node (vd42323) at  (1,0) {$\vdots$};        
        \node (vd32323) at  (3,0) {$\vdots$};                
        \node [above left = 0mm and 0mm of v3]() {$3$};
        \node [above left = 0mm and 0mm of v23]() {$2$};
        \node [above right = 0mm and 0mm of v53]() {$5$};
        \node [above left = 0mm and 0mm of v423]() {$4$};
        \node [above right = 0mm and 0mm of v323]() {$3$};
        \node [above right = 0mm and 0mm of v653]() {$6$};
        \node [above left = 0mm and 0mm of v6423]() {$6$};
        \node [above left = 0mm and 0mm of v2323]() {$2$};
        \node [above right = 0mm and 0mm of v5323]() {$5$};
        \node [above left = 0mm and 0mm of v42323]() {$4$};
        \node [above right = 0mm and 0mm of v32323]() {$3$};
        \node [above right = 0mm and 0mm of v65323]() {$6$};
        \draw[->] (v23) to (v3);
        \draw[->] (v53) to (v3);        
        \draw[->] (v423) to (v23);
        \draw[->] (v323) to (v23);
        \draw[->] (v653) to (v53);
        \draw[->] (v6423) to (v423);
        \draw[->] (v2323) to (v323);
        \draw[->] (v5323) to (v323);
        \draw[->] (v42323) to (v2323);
        \draw[->] (v32323) to (v2323);
        \draw[->] (v65323) to (v5323);
        \draw[->] (vd42323) to  (v42323);
        \draw[->] (vd32323) to  (v32323);
        \end{tikzpicture}
    }
    \\
    \hline
    $T_4 \simeq T_5$ &
    \scalebox{.6}{
        \begin{tikzpicture}
        \tikzset{vertex/.style = {shape=circle,draw}}
        \tikzset{>={Latex[width=2mm,length=2mm]}}
        \node[] () at (0,1.3) {}; 
        \node[vertex] (v4) at  (0,1) {};
        \node[vertex] (v64) at  (0,0) {};
        \node [above left = 0mm and 0mm of v4]() {$4$};
        \node [above left = 0mm and 0mm of v64]() {$6$};
        \draw[->] (v64) to (v4);
        \end{tikzpicture}
    }
    \raisebox{0.3cm}{$\qquad\simeq\qquad$}
    \scalebox{.6}{
        \begin{tikzpicture}
        \tikzset{vertex/.style = {shape=circle,draw}}
        \tikzset{>={Latex[width=2mm,length=2mm]}}
        \node[] () at (0,1.3) {}; 
        \node[vertex] (v4) at  (0,1) {};
        \node[vertex] (v64) at  (0,0) {};
        \node [above left = 0mm and 0mm of v4]() {$5$};
        \node [above left = 0mm and 0mm of v64]() {$6$};
        \draw[->] (v64) to (v4);
        \end{tikzpicture}
    }
    \\
        \hline
    $T_6$ &
    \scalebox{.6}{
        \begin{tikzpicture}
        \tikzset{vertex/.style = {shape=circle,draw}}
        \tikzset{>={Latex[width=2mm,length=2mm]}}
        \node[] () at (0,0.3) {}; 
        \node[vertex] (v6) at  (0,0) {};
        \end{tikzpicture}
    }
    \\
    \hline
    $T_7$ &
    \scalebox{.6}{
        \begin{tikzpicture}
        \tikzset{vertex/.style = {shape=circle,draw}}
        \tikzset{>={Latex[width=2mm,length=2mm]}}
        \node[] () at (4,6.3) {}; 
        \node[vertex] (v7) at (4,6) {};
        \node[vertex] (v3) at  (4,5) {};
        \node[vertex] (v23) at  (2,4) {};
        \node[vertex] (v53) at  (6,4) {};
        \node[vertex] (v423) at  (0,3) {};
        \node[vertex] (v323) at  (3,3) {};
        \node[vertex] (v653) at  (6,3) {};
        \node[vertex] (v6423) at  (0,2) {};
        \node[vertex] (v2323) at  (2,2) {};
        \node[vertex] (v5323) at  (4,2) {};
        \node[vertex] (v42323) at  (1,1) {};        
        \node[vertex] (v32323) at  (3,1) {};        
        \node[vertex] (v65323) at  (4,1) {};        
        \node (vd42323) at  (1,0) {$\vdots$};        
        \node (vd32323) at  (3,0) {$\vdots$}; 
        \node [above left = 0mm and 0mm of v7]() {$7$};
        \node [above left = 0mm and 0mm of v3]() {$3$};               
        \node [above left = 0mm and 0mm of v23]() {$2$};               
        \node [above right = 0mm and 0mm of v53]() {$5$};               
        \node [above left = 0mm and 0mm of v423]() {$4$};               
        \node [above right = 0mm and 0mm of v323]() {$3$};               
        \node [above right = 0mm and 0mm of v653]() {$6$};               
        \node [above left = 0mm and 0mm of v6423]() {$6$};               
        \node [above left = 0mm and 0mm of v2323]() {$2$};               
        \node [above right = 0mm and 0mm of v5323]() {$5$};               
        \node [above left = 0mm and 0mm of v42323]() {$4$};               
        \node [above right = 0mm and 0mm of v32323]() {$3$};               
        \node [above right = 0mm and 0mm of v65323]() {$6$};               
        \draw[->] (v3) to (v7);
        \draw[->] (v23) to (v3);
        \draw[->] (v53) to (v3);        
        \draw[->] (v423) to (v23);
        \draw[->] (v323) to (v23);
        \draw[->] (v653) to (v53);
        \draw[->] (v6423) to (v423);
        \draw[->] (v2323) to (v323);
        \draw[->] (v5323) to (v323);
        \draw[->] (v42323) to (v2323);
        \draw[->] (v32323) to (v2323);
        \draw[->] (v65323) to (v5323);
        \draw[->] (vd42323) to  (v42323);
        \draw[->] (vd32323) to  (v32323);
        \end{tikzpicture}
    }
    \\\hline
    \end{tabular}
    \caption{\textbf{Examples of input tree isomorphisms.} The input trees $T_i$ of
      all the nodes $i$ of the network of
      Fig.~\ref{fig:example-inputsets} grouping those that are isomorphic.}
    \label{fig:example-alltrees}
\end{figure}

\section{The graph isomorphism problem: testing input tree isomorphism}
\label{sec:testing}

The graph isomorphism (GI) problem\index{Graph Isomorphism Problem } is the problem of determining whether
two finite graphs are isomorphic, that is, whether the graphs are identical
despite different drawing and/or labeling (Definition
\ref{def:isomorphism}).  Algorithms for the solution of this problem have many
practical applications, ranging from chemistry (identification of
chemical compounds), computer vision and pattern recognition, to
electronic circuit design. The exact complexity of this problem
remains one of the most important
unsolved problems in theoretical computer science
\citep{grohe2020}.

While the exact complexity standing of GI is unknown, it has not been proved to be NP-complete\index{NP-complete } (and it is unlikely to be NP-complete, according to most researchers).
There has been recent breakthroughs by Babai
\citep{babai2016,cho2015,grohe2020} who proved   that GI is almost
efficiently solvable, i.e., it is solvable in quasipolynomial time,
meaning that Babai's algorithm\index{Babai's algorithm } runs in time as $N^{p(\log N)}$, for
some polynomial $p(x)$.
Its generalization, the Subgraph Isomorphism Problem,\index{Subgraph Isomorphism Problem } consists of determining whether a graph $G$ has a subgraph that is isomorphic to $H$,
has been proved NP-complete\index{NP-complete } \citep{Cook71}.

A breakthrough in practical algorithms for isomorphism testing in
large graphs was McKay's Practical Graph Isomorphism\index{Practical Graph Isomorphism }
~\citep{mckay1981practical,mckay2014}, which is quite efficient in practice. It is the basis for McKay's {\it Nauty} algorithm\index{Nauty algorithm } to find graph
automorphisms \citep{nauty}.  Indeed, the graph isomorphism problem is
computationally equivalent to computing the automorphism group of a
graph \citep{luks1993}.

Despite GI's complexity being unknown, in many realistic applications or for sufficiently small instances, the problem is easy to solve.
For instance, \cite{luks1982} shows that isomorphism can be
decided in polynomial time for graphs with bounded in-degree, and this condition is easily satisfied in
all biological networks.  Furthermore, here we are 
concerned only with isomorphisms between input trees, and for trees, GI has
efficient, polynomial-time solutions \citep{kelly1957}.

For our needs, we therefore limit ourselves to isomorphism of trees,
which is a polynomial-time solvable problem. Furthermore, in practice, testing for symmetry fibrations does  not require testing for
input tree isomorphisms. Input trees are a useful theoretical concept, but in most practical applications, we 
use algorithms to find balanced colorings to obtain the fibers of the
symmetry fibration without using the actual input trees.
However, input trees are useful conceptually to classify
building blocks, see Section \ref{sec:definition-building}.

As an example, consider again the FFF in Fig. \ref{fig:inputtree}.
Figure  \ref{fig:inputtree}b shows the input trees, and those of nodes 1 and 2
are isomorphic. In this case the isomorphism $\alpha : T_1\to T_2$ is
a map of nodes as indicated in Fig. \ref{fig:inputtree}b by the dotted
red lines. At the root: $\alpha(1) = 2$; in the first layer:
$\alpha(1) = 1$, $\alpha(3)= 3$, and so on. The edges are also mapped
correspondingly: $e_{T_1} = \{1, 1\}$ in the first layer is mapped to the edge
$e_{T_1} = \{\alpha(1), \alpha(1)\} = \{1, 2\}$, and so on. The map
$\alpha$ is a bijection, because the unique edge connects the image of
its source to the image of its target. Therefore, the input trees of
$Y_1$ and $Y_2$ are isomorphic (i.e., isomorphic as graphs).

These input trees are represented by a graph of infinite size.  Do we
need to check every layer in the input trees?  Obviously, it is not practical to check every edge
in an infinite set of edges. However, an important
result of \cite{angluin1980local} and \cite{norris1995} proves that if the first $N -
1$ levels of the input trees are isomorphic ($N$ is the number of
nodes in the graph $G$), then the (infinite) input trees are isomorphic.
For example, in the FFF circuit, $N = 3$, so the first two levels of
the input tree are enough to show that an isomorphism exists.

Norris's theorem\index{Norris's theorem } is a fundamental result for fibrations, since it lets us
 define isomorphism for any input tree, infinite or not,
in a finite number of steps. It also has important consequences for
interpreting the dynamics, since it limits the number of steps that we
need to go back in time to determine the state of the root node in the
input tree. If we interpret the input tree as in
Fig.~\ref{fig:example-inputsets}, representing how the
information flow of the signals
that the root node receives unfold in time, then Norris's theorem implies that the
state of the root nodes that are connected to a cycle in the graph is
affected for the first time by a bounded number of steps along the cycle. This
number of steps is the number of nodes in the 
circuit. However, a cycle implies a positive feedback on the nodes involved: after the first time a certain node $x$ can impact on the behavior of node $y$ (which depends only on the length of the shortest path from $x$ to $y$), $x$ will have further influence on $y$, if $x$ and $y$ are involved together in one (or more) cycles.

Norris's theorem also determines how we define fiber
building blocks when the input trees are infinite and determined by
cycles in the graph (section \ref{sec:definition-building}).

\section{Minimal equitable partition and balanced coloring partition}
\label{minimal_partition}

To take one further step in our analysis, we consider graph
partitions, Definition \ref{partition}.  We recall: a
partition is a way to group nodes into non-overlapping clusters.  We
visualize a partition by coloring the nodes in each cluster with the
same color, using different colors for different clusters.

Of course, there are many possible ways to partition the nodes of a
graph. Here, we are exclusively interested in {\it equitable
  partitions}, also called {\it balanced coloring partitions}.
Within these partitions we mainly focus on {\it minimal}
equitable and balanced coloring partitions. These correspond to maximal
synchronization---fewest clusters.

\begin{definition}{\bf Equitable Partition.}
Let $\mathcal{S} = {S_1, \cdot \cdot \cdot ,S_K}$ be a partition of
the nodes of a network $G=(V,E)$, where $K$ is the total number of
parts or clusters in the partition (Definition \ref{partition}). The
partition is {\it equitable}\index{partition !equitable } if and only if each node in cluster $\mathcal{S}_\mu$ has the same number $k_{\nu\mu}$ of incoming edges
from nodes in cluster
$\mathcal{S}_\nu$, for $1\leq \mu, \nu \leq K$.
\label{equitable}
\end{definition}

Equitable partitions and (robust) cluster synchronization are logically
equivalent concepts \cite[Theorem 10.18]{GS2023}.  We think of an equitable partition as a way to
represent a cluster of synchronous states of the system. Nodes in the
same partition have the same (local) state, i.e., they synchronize in
the cluster. Nodes that are in different clusters may have different
(local) states or different synchronous states. Equitable partitions
are defined by input tree isomorphisms.  We use colors to visualize
these partitions: colors represent local states, and the clusters of
the partition are the groups of nodes with the same color.

In general there are many equitable partitions, and enumerating them
all is computationally challenging. 
However, several algorithms to do this exist, for
example \citep{kamei2013, belykh2011}; see Chapter \ref{chap:algorithms}.
But one partition
is of particular interest here because it collects all the symmetries
that the graph has. This is the {\it minimal equitable partition}
or {\it coarsest balanced coloring}, and it
 partitions the graph into the smallest number of clusters.
  
\begin{definition}{\bf Minimal equitable partition.}
  The {\it
    minimal} equitable partition\index{partition !minimal equitable } is the (unique) equitable partition with the 
  minimal number of clusters (colors) or the \emph{coarsest} equitable
  partition that can be found.
  It can be proved to be the equitable partition where two nodes are in the same
  cluster if and only if they have isomorphic input trees.  
\label{def:minimal}
\end{definition}

\begin{figure*}
\includegraphics[width=\textwidth]{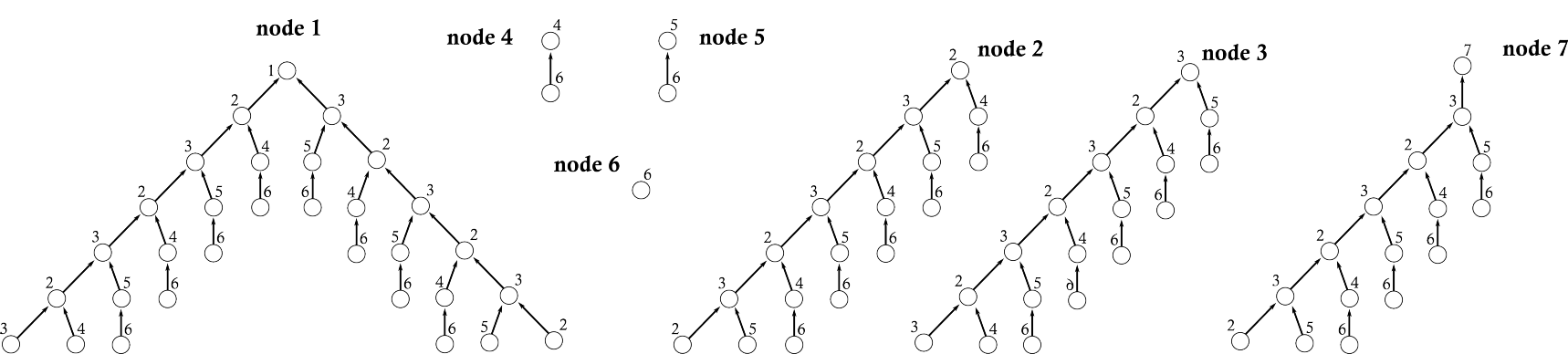}
\centering
\caption {\textbf{Isomorphic input trees for the minimal equitable
    partition}.  Input trees of the graph shown in
  Fig.~\ref{fig:example-inputsets}.  We find: $T_2 \simeq T_3$ and
  $T_4 \simeq T_5$. Grouping these two clusters lead to the minimal
  equitable partition of Fig. \ref{fig:minimal}. If
  we do not group all isomorphic input trees but just some of them,
  then the resulting partition is still equitable but not minimal, see
  Fig. \ref{fig:non-minimal}. Grouping nodes that are not isomorphic
  results in a partition that is not equitable, see
  Fig. \ref{fig:non-equitable}.}
\label{fig:isomorphic}
\commentAlt{Figure~\ref{fig:isomorphic}:  No alt-text required.
}
\end{figure*}

\begin{figure}
    \centering
    \scalebox{1}{
        \begin{tikzpicture}
        \tikzset{vertex/.style = {shape=circle,draw}}
        \tikzset{edge/.style = {->,> = latex'}}
        \node[vertex, fill=blue] (v6) at  (1,0) {$6$};
        \node[vertex, fill=red] (v4) at  (0,1) {$4$};
        \node[vertex, fill=red] (v5) at  (2,1) {$5$};
        \node[vertex, fill=yellow] (v2) at  (0,3) {$2$};
        \node[vertex, fill=yellow] (v3) at  (2,3) {$3$};
        \node[vertex, fill=green] (v7) at  (4,3) {$7$};
        \node[vertex, fill=magenta] (v1) at  (1,4) {$1$};
        \draw[edge] (v6) to node [below] {$g$} (v4);
        \draw[edge] (v6) to node [below] {$h$} (v5);
        \draw[edge] (v4) to node [left] {$e$} (v2);
        \draw[edge] (v5) to node [right] {$f$}(v3);
        \draw[edge] (v2) to[bend left] node [below] {$c$} (v3);
        \draw[edge] (v3) to[bend left] node [below] {$d$}(v2);
        \draw[edge] (v3) to node [above] {$i$} (v7);
        \draw[edge] (v2) to node [above] {$a$} (v1);
        \draw[edge] (v3) to node [above] {$b$} (v1);
        \end{tikzpicture}
    }
    \caption{\textbf{Minimal equitable partition.} The graph of
      Fig. \ref{fig:example-inputsets} with nodes colored according to
      their input trees. Two nodes have the same color if and only if
      their input trees are isomorphic, which specifies the minimal equitable
      partition.}
    \label{fig:minimal}
\commentAlt{Figure~\ref{fig:minimal}: 
Graph of Figure~\ref{fig:example-inputsets} with colored nodes: 
1 (magenta), 2 and 3 (yellow), 4 and 5 (red), 6 (blue), 7 (green).
}
    \end{figure}

Figure  \ref{fig:isomorphic} shows the input tree isomorphisms of the
graph of Fig. \ref{fig:example-inputsets}.  Input trees of node 2 and
3 are isomorphic, $T_2 \simeq T_3$, and nodes 4 and 5 too, $T_4 \simeq
T_5$. There are no more isomorphisms as shown by the rest of the input
trees. If we group {\it all} isomorphic nodes into clusters we
obtain the minimal equitable partition with 5 colors shown in
Fig. \ref{fig:minimal}.  This equitable partition is
minimal because all nodes with isomorphic trees are in the same cluster.

If we break one of these clusters and assign two different colors, for
instance, to nodes 2 and 3, but do not break synchrony of nodes 4 and 5, 
the partition is still equitable but not
minimal, since it has 6 colors rather than 5, see
Fig. \ref{fig:non-minimal}. Note, however, that in general if we take one cluster from an equitable partition and break it arbitrarily, the resulting partition may not be equitable. For instance, if we break the synchrony between 4 and 5 in Fig. \ref{fig:minimal},
then also the synchrony between 2 and 3 ceases to exist. On the other hand, if we break \emph{both} clusters we obtain again an equitable partition (a trivial one in which each cluster contains just one node).

\begin{figure}
    \centering
    \scalebox{1}{
        \begin{tikzpicture}
        \tikzset{vertex/.style = {shape=circle,draw}}
        \tikzset{edge/.style = {->,> = latex'}}
        \node[vertex, fill=blue] (v6) at  (1,0) {$6$};
        \node[vertex, fill=red] (v4) at  (0,1) {$4$};
        \node[vertex, fill=red] (v5) at  (2,1) {$5$};
        \node[vertex, fill=yellow] (v2) at  (0,3) {$2$};
        \node[vertex, fill=brown] (v3) at  (2,3) {$3$};
        \node[vertex, fill=green] (v7) at  (4,3) {$7$};
        \node[vertex, fill=magenta] (v1) at  (1,4) {$1$};
        \draw[edge] (v6) to node [below] {$g$} (v4);
        \draw[edge] (v6) to node [below] {$h$} (v5);
        \draw[edge] (v4) to node [left] {$e$} (v2);
        \draw[edge] (v5) to node [right] {$f$}(v3);
        \draw[edge] (v2) to[bend left] node [below] {$c$} (v3);
        \draw[edge] (v3) to[bend left] node [below] {$d$}(v2);
        \draw[edge] (v3) to node [above] {$i$} (v7);
        \draw[edge] (v2) to node [above] {$a$} (v1);
        \draw[edge] (v3) to node [above] {$b$} (v1);
        \end{tikzpicture}
    }
    \caption{\textbf{Equitable partition that is not minimal.} When we do not
      group all the nodes with isomorphic input trees with the same
      color, for instance, node 2 and 3 are isomorphic but we assign
      them different colors, we end up with an equitable partition
      that is not the minimal one, since we are using two colors to color
     what previously was one cluster.}
    \label{fig:non-minimal}
\commentAlt{Figure~\ref{fig:non-minimal}: 
Graph of Figure~\ref{fig:example-inputsets} with colored nodes: 
1 (magenta), 2 (yellow), 3 (brown) , 4 and 5 (red), 6 (blue), 7 (green).
}
\end{figure}

\begin{figure}
    \centering
    \scalebox{1}{
        \begin{tikzpicture}
        \tikzset{vertex/.style = {shape=circle,draw}}
        \tikzset{edge/.style = {->,> = latex'}}
        \node[vertex, fill=blue] (v6) at  (1,0) {$6$};
        \node[vertex, fill=red] (v4) at  (0,1) {$4$};
        \node[vertex, fill=yellow] (v5) at  (2,1) {$5$};
        \node[vertex, fill=yellow] (v2) at  (0,3) {$2$};
        \node[vertex, fill=yellow] (v3) at  (2,3) {$3$};
        \node[vertex, fill=green] (v7) at  (4,3) {$7$};
        \node[vertex, fill=magenta] (v1) at  (1,4) {$1$};
        \draw[edge] (v6) to node [below] {$g$} (v4);
        \draw[edge] (v6) to node [below] {$h$} (v5);
        \draw[edge] (v4) to node [left] {$e$} (v2);
        \draw[edge] (v5) to node [right] {$f$}(v3);
        \draw[edge] (v2) to[bend left] node [below] {$c$} (v3);
        \draw[edge] (v3) to[bend left] node [below] {$d$}(v2);
        \draw[edge] (v3) to node [above] {$i$} (v7);
        \draw[edge] (v2) to node [above] {$a$} (v1);
        \draw[edge] (v3) to node [above] {$b$} (v1);
        \end{tikzpicture}
    }
    \caption{\textbf{Partition that is not equitable.} If we color the
      graph of Fig.~\ref{fig:example-inputsets} by grouping nodes with
      the same color that do not have isomorphic input trees, for
      instance yellow for node 5 which is not isomorphic with 2 and 3,
      the resulting partition is not equitable. The coloring
      represents an non-equitable partition (nodes with the same color
      have different color-isomorphic input sets).}  
    \label{fig:non-equitable}
\commentAlt{Figure~\ref{fig:non-equitable}: 
Graph of Figure~\ref{fig:example-inputsets} with colored nodes: 
1 (magenta), 2,3,5 (yellow), 4 (red), 6 (blue), 7 (green).
}
\end{figure}

We have established that all equitable partitions (not just the
minimal one) group nodes with isomorphic input trees.  An equitable
partition has another special property, which not all partitions share.
If you look at the \textit{input sets} (not the input trees)  of all nodes, and color them according
to the equitable partition, one important property holds true: if two
nodes have the same color, then their input sets are color-isomorphic\index{color-isomorphic }
(that is, isomorphic and with an isomorphism that respects the
colors).  This means that nodes with the same color receive the same
colors from their respective input nodes.  For instance, in
Fig. \ref{fig:minimal}, nodes 2 receives one red color and one yellow
color, and node 3 as well, although it receives a red from a different
node. Then, nodes 2 and 3 are colored the same. Color-isomorphism
between input sets is the main concept that connects the fibration
formalism with the groupoid formalism of Golubitsky and Stewart
\citep{stewart2006}, which is treated in detail in Chapter
\ref{chap:groupoid}.

\begin{figure}
    \centering
    \begin{tabular}{cccc}
    \scalebox{1}{
        \begin{tikzpicture}[baseline=(current bounding box.center)]
        \tikzset{vertex/.style = {shape=circle,draw}}
        \tikzset{edge/.style = {->,> = latex'}}
        \node[vertex,fill=magenta] (v1) at  (1,1) {$1$};
        \node[vertex,fill=yellow] (v2) at  (0,0) {$2$};
        \node[vertex,fill=yellow] (v3) at  (2,0) {$3$};
        \draw[edge] (v2) to node [above] {$a$}(v1);
        \draw[edge] (v3) to node [above] {$b$}(v1);
        \end{tikzpicture}
    }
    &
    \scalebox{1}{
        \begin{tikzpicture}[baseline=(current bounding box.center)]
        \tikzset{vertex/.style = {shape=circle,draw}}
        \tikzset{edge/.style = {->,> = latex'}}
        \node[vertex,fill=yellow] (v2) at  (1,1) {$2$};
        \node[vertex,fill=yellow] (v3) at  (0,0) {$3$};
        \node[vertex,fill=red] (v4) at  (2,0) {$4$};
        \draw[edge] (v3) to node [above] {$d$}(v2);
        \draw[edge] (v4) to node [above] {$e$} (v2);
        \end{tikzpicture}
    }
    &
    \scalebox{1}{
        \begin{tikzpicture}[baseline=(current bounding box.center)]
        \tikzset{vertex/.style = {shape=circle,draw}}
        \tikzset{edge/.style = {->,> = latex'}}
        \node[vertex,fill=yellow] (v3) at  (1,1) {$3$};
        \node[vertex,fill=red] (v5) at  (0,0) {$5$};
        \node[vertex,fill=yellow] (v2) at  (2,0) {$2$};
        \draw[edge] (v5) to node [above] {$f$}(v3);
        \draw[edge] (v2) to node [above] {$c$} (v3);
        \end{tikzpicture}
    }
    & 
    \scalebox{1}{
        \begin{tikzpicture}[baseline=(current bounding box.center)]
        \tikzset{vertex/.style = {shape=circle,draw}}
        \tikzset{edge/.style = {->,> = latex'}}
        \node[vertex,fill=red] (v4) at  (0,1) {$4$};
        \node[vertex,fill=blue] (v6) at  (0,0) {$6$};
        \draw[edge] (v6) to node [left] {$g$}(v4);
        \end{tikzpicture}
    }
    \end{tabular} 
    \vspace{10pt}
    \caption{\textbf{Color-isomorphic input sets in the equitable
        partition.} The input sets of Fig.~\ref{fig:example-inputsets}
      (shown in Fig.~ \ref{fig:inputsets}) but, here, colored
      according to the coloring of Fig.
      \ref{fig:minimal}: the coloring represents an
      equitable partition (nodes with the same color have
      color-isomorphic input sets). }
    \label{fig:inputsets-colors}
\commentAlt{Figure~\ref{fig:inputsets-colors}: Input sets for nodes 1, 2, 3, 4.
Node 1: 1 (magenta), 2 (yellow) with arrow a to 1, 3 (yellow) with arrow b to 1.
Node 2: 2 (yellow), 3 (yellow)  with arrow d to 21, 4 (red) with arrow e to 2.
Node 3: 3 (yellow), 2 (yellow) with arrow c to 3, 5 (red) with arrow f to 3.
Node 4: 4 (red), 6 (blue) with arrow g to 4.
}
\end{figure}

This fact is clear in Fig.~\ref{fig:minimal}. The
input sets $I_1$, $I_2$ and $I_3$ are all isomorphic, because nodes
$1$, $2$, $3$ have all exactly two incoming edges, see
Fig. \ref{fig:inputsets}. However, after coloring we see that only $I_2$ and
$I_3$ are color isomorphic, see
Fig. \ref{fig:inputsets-colors}.

This observation leads to an alternative way to define an equitable
partition without using the full input tree, but just using its first
layer, i.e. the input set. This may seem an improvement, since we
do not need to consider (possible) infinite trees, but the drawback is
that we need to know the colors first.
For this definition to work, we must first find the equitable
partition of the nodes, and then test color-isomorphism of the input
sets. This problem
disappears if we use the input tree to define the equitable
partition, Definition \ref{def:minimal}. However, we obtain only the
minimal equitable partition in this manner.

\subsubsection{Balanced coloring partition}
Let us formalize these statements. An equitable partition (Definition \ref{equitable}) is an
equivalent concept to a balanced coloring partition, defined
below. These terms are used in different areas, and
have a slightly different emphasis. The usual definition in the
formalism of \citep{GS2023} is:

\begin{definition}{\bf Balanced coloring partition.}
    A {\it coloring}\index{coloring } $\kappa$ of a graph $G(V,E)$ assigns a color $\kappa(c)$ to each node $c$,
    where $\kappa(c)$ belongs to a set $\mathcal{K}$, called the set of {\em colors}.
    The coloring $\kappa$ is {\em balanced}\index{coloring !balanced } if $\kappa(c) = \kappa(d)$
    implies that $I(c)$ and $I(d)$ are color-isomorphic.\index{color-isomorphic }
    
    Equivalently, a
    balanced coloring is a coloring such that each node
    with color $\mu$ in cluster $S_\mu$ is connected (by arrows of the same type) to the same
    number of nodes $k_{\mu\nu}$ with color $\nu$, for $1\leq \mu, \nu
    \leq K$ where $K$ is the number of colors.
\label{balance-coloring}
\end{definition}

This definition is essentially the same as that of an equitable partition,\index{partition !equitable } Definition \ref{equitable}, but in different terminology.
There is also a natural minimal balanced coloring partition, which is
the same as the minimal equitable partition \citep[Theorem 13.16]{GS2023}:

\begin{definition}{\bf Minimal balanced coloring partition.}\index{partition !minimal balanced coloring }
  This is a balanced coloring of a graph with the minimal number of
  colors. It is the same as the minimal equitable partition.
  \label{minimal-balance-coloring}
\end{definition}

In terms of synchronization,\index{synchronization } nodes inside the same subset $\mathcal{S}_i$ of the
balanced coloring
partition can synchronize, since they receive the same color inputs
from the same synchronized nodes. Thus, color balance is another way
to say that the partition is equitable. 

An equitable partition represents a global state that has a robust synchrony pattern.
In fact, equitability means that if two nodes (say $i$
and $j$) are currently in the same state, they read inputs from
sources in the same state, so they will never diverge from each
other. Conversely, in non-equitable partitions this behavior does not take
place: in the global state of Fig.~\ref{fig:non-equitable}, nodes $2$
and $3$ are both in the state yellow, but one receives inputs from two
yellow nodes, and the other receives inputs from one yellow and one
red, so $2$ and $3$ may well change state.
For a rigorous proof, see \citep[Proposition 10.20]{GS2023}.

And what about our old friend, the orbital partition from Definition
\ref{orbits}?  As discussed in section \ref{sec:orbit}, the orbital
partition also produces a partition of synchronized nodes, and it is,
by definition, equitable and color-balanced. The origin of
synchronization in the orbital partition is the existence of
automorphisms in the graphs. Automorphisms determine certain special
fibrations, and are the best-known way to create
balanced colorings. But, as we have already seen,
not all equitable partitions are
orbits of subgroups of automorphisms of the network,
because there can be additional fibrations. In particular, the minimal
orbital partition need not be the minimal equitable partition.

In fact, throughout this book, we  deal almost
exclusively with equitable partitions that are not orbital, that is,
 equitable partitions that do not arise from automorphisms of the
network, but from a symmetry fibration.

The main difference is that balanced and equitable partitions are
about fibrations and groupoid symmetry, which are local concepts.
Orbits are about
global group symmetry. In general, there is very little reason for a local symmetry
to extend to a global one, which opens the door to a
much richer world of symmetries than those represented by orbits and
automorphisms.

\subsection{Examples of balanced coloring partitions}
\index{partition !equitable, examples }

We can continue our journey by providing some examples.  We are mainly
interested in the minimal partition, which captures most of the
symmetries. 

\begin{example}
\label{ex:minimal_1}\em
Figure \ref{fig:orbital} shows a case where the orbital
partition is indeed the minimal equitable partition. That is, all the
symmetries of the graph are captured by its automorphisms, and fibration
symmetries do not add much to explain this structure. We studied this
graph in Fig. \ref{fig: example}a and concluded that it has left-right
symmetry. This induces the orbital partition shown in Fig.
\ref{fig:orbital}. This is an equitable partition and indeed the
minimal one, as the input trees show.
\end{example}

\begin{figure}
    \centering
    \scalebox{1}{
        \begin{tikzpicture}
        \tikzset{vertex/.style = {shape=circle,draw}}
        \tikzset{edge/.style = {->,> = latex'}}
        \node[vertex, fill=blue] (v6) at  (1,0) {$6$};
        \node[vertex, fill=red] (v4) at  (0,1) {$4$};
        \node[vertex, fill=red] (v5) at  (2,1) {$5$};
        \node[vertex, fill=yellow] (v2) at  (0,3) {$2$};
        \node[vertex, fill=yellow] (v3) at  (2,3) {$3$};
        \node[vertex, fill=magenta] (v1) at  (1,4) {$1$};
        \draw[edge] (v6) to node [below] {$g$} (v4);
        \draw[edge] (v6) to node [below] {$h$} (v5);
        \draw[edge] (v4) to node [left] {$e$} (v2);
        \draw[edge] (v5) to node [right] {$f$}(v3);
        \draw[edge] (v2) to[bend left] node [below] {$c$} (v3);
        \draw[edge] (v3) to[bend left] node [below] {$d$}(v2);
        \draw[edge] (v2) to node [above] {$a$} (v1);
        \draw[edge] (v3) to node [above] {$b$} (v1);
        \end{tikzpicture}
    }
    \caption{\textbf{Example of orbital partition that is also the
        minimal equitable partition.} A graph with a left-right
      automorphism that induces the shown orbital partition. There are
      no extra symmetries in this graph, so the orbital partition is
      the most symmetric one. The orbital partition is the minimal
      equitable partition and fibrations do not add much information
      to this analysis.}
    \label{fig:orbital}
\commentAlt{Figure~\ref{fig:orbital}: 
Graph with nodes
1 (magenta), 2, 3 (yellow), 4,5 (red), 6 (blue).  Edges
a (from 2 to 1), b (3 to 1), c (2 to 3), d (3 to 2), e (4 to 2),
f (5 to 3), g (6 to 4), h (6 to 5).
}
\end{figure}

\begin{example}
\label{ex:minimal_2}\em
However, we have shown that the addition of a single outgoing edge
from node 3 to node 7 as in Fig. \ref{fig:minimal} destroys this global
automorphism, and the only automorphism of the modified graph is the
identity, so the graph is rigid. Consequently, the orbital
partition is now reduced to the most trivial one (see
Fig. \ref{fig:trivial}): the identity partition where each node has
a different color. This partition is also equitable: in this case the
definition of equitable (if two nodes have the same state then\dots)
is trivially true because no two nodes have the same
state. This trivial partition requires seven colors (one color per
node), and it is the least coarse (the \emph{finest}), the one with maximal number of colors, and
farthest from the minimal equitable partition of five colors shown in
Fig. \ref{fig:minimal}.
\end{example}

\begin{figure}
    \centering
    \scalebox{1}{
        \begin{tikzpicture}
        \tikzset{vertex/.style = {shape=circle,draw}}
        \tikzset{edge/.style = {->,> = latex'}}
        \node[vertex, fill=blue] (v6) at  (1,0) {$6$};
        \node[vertex, fill=orange] (v4) at  (0,1) {$4$};
        \node[vertex, fill=red] (v5) at  (2,1) {$5$};
        \node[vertex, fill=yellow] (v2) at  (0,3) {$2$};
        \node[vertex, fill=brown] (v3) at  (2,3) {$3$};
        \node[vertex, fill=green] (v7) at  (4,3) {$7$};
        \node[vertex, fill=magenta] (v1) at  (1,4) {$1$};
        \draw[edge] (v6) to node [below] {$g$} (v4);
        \draw[edge] (v6) to node [below] {$h$} (v5);
        \draw[edge] (v4) to node [left] {$e$} (v2);
        \draw[edge] (v5) to node [right] {$f$}(v3);
        \draw[edge] (v2) to[bend left] node [below] {$c$} (v3);
        \draw[edge] (v3) to[bend left] node [below] {$d$}(v2);
        \draw[edge] (v3) to node [above] {$i$} (v7);
        \draw[edge] (v2) to node [above] {$a$} (v1);
        \draw[edge] (v3) to node [above] {$b$} (v1);
        \end{tikzpicture}
    }
    \caption{\textbf{Example of trivial orbital partition in a graph with
        nontrivial minimal equitable partition.} Adding the single
      edge to node 7 from Fig. \ref{fig:orbital} reduces the orbital
      partition to the trivial one: one color per node.  This trivial
      orbital partition is far from the minimal equitable partition
      for this graph shown in Fig. \ref{fig:minimal}. This represents
      an example where fibration symmetries are necessary to capture
      the symmetries of the graph. It is ubiquitous in biological
      networks.}
    \label{fig:trivial}
\commentAlt{Figure~\ref{fig:trivial}: Graph of Figure~\ref{fig:orbital}
with extra node 7. nodes
1 (magenta), 2, (yellow),3 (brown), 4 (orange), 5 (red), 6 (blue). 7 (green).  Edges
a (from 2 to 1), b (3 to 1), c (2 to 3), d (3 to 2), e (4 to 2),
f (5 to 3), g (6 to 4), h (6 to 5), i (3 to 7).
}
\end{figure}

\begin{figure}[h!]
  \centering{
  \includegraphics[width=0.6\textwidth]{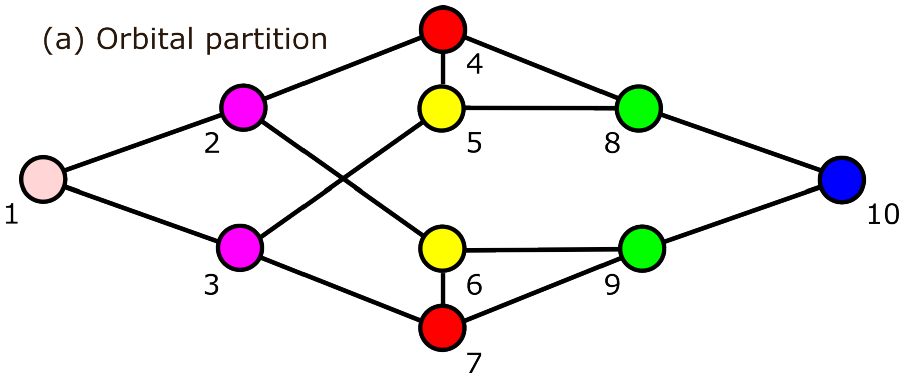}
  \includegraphics[width=0.6\textwidth]{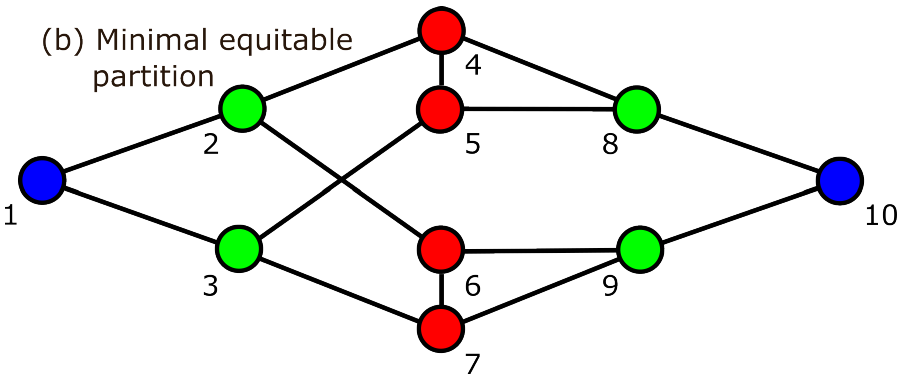}}  
  \caption {\textbf{Example of a graph
      whose orbital partition is not the minimal equitable
      partition.}  {\bf (a)} Orbital partition of the graph with six
    colors arising from the up-down automorphisms of the graph.  {\bf
      (b)} Minimal equitable partition with three colors, showing
    extra symmetries not captured by the orbits. In fact, the graph
    has no left-right symmetry, yet the green and blue clusters are
    still fibration symmetric in their local topology.}
\label{fig:intermediate}
\commentAlt{Figure~\ref{fig:intermediate}:  (a) Orbital partition.
Graph with nodes 1 (pink); 2, 3 (magenta); 4, 7 (red); 5, 6 (yellow); 8, 9 (green); 10 (blue).
Edges 1-2, 1-3, 2-4, 2-6, 3-5, 3-7, 4-5, 6-7, 5-8, 6-9, 8-10, 9-10.
(b) Minimal equitable partition: The same, except that nodes 5 and 6 are red.
}
\end{figure}

\begin{example}
\label{ex:minimal_3}\em
An intermediate example of an undirected graph  
with coexistence of an orbital partition that is not a
minimal partition is shown Fig. \ref{fig:intermediate}.  The graph has
only a global up-down symmetry, creating the six orbits in pairs of
nodes as shown in Fig. \ref{fig:intermediate}a. However, these are not
the only symmetries of this graph, and the orbits do not capture all
the clusters of synchronous states.  While the graph has no global
left-right symmetry, locally, there are extra left-right symmetries
creating extra synchronous states by the minimal equitable partition
shown in Fig. \ref{fig:intermediate}b.

All green nodes are connected to two red nodes and one
blue node. All red nodes are connected to two green nodes and one red
node, and all blue nodes are connected to two green nodes, creating a
more symmetric minimal balanced coloring than the orbital partition
with just three colors. The green nodes (and blue nodes) are symmetric
in terms of their local topology despite the lack of global left-right
symmetry. The coloring represents a left-right symmetry through the
conservation of color in a local neighborhood of each node, i.e., the
left-right mirror transformation conserves the balanced coloring. The
symmetry in local topology captured by the minimal balanced coloring
is less restrictive (and more perspicuous) than the global one.
\end{example}
    
\begin{example}
\label{ex:minimal_4}\em
{\bf Frucht Graph}.

A notable example, and a more extreme one in some sense, is the Frucht
graph\index{graph !Frucht } described by Robert Frucht\index{Frucht, Robert } in 1939 \citep{frucht-wiki} and shown
in Fig. \ref{F:frucht_coloring_bal}.
Frucht is better known for his influential
theorem stating that any group can be realized as the group of
symmetries of a graph. Even more interesting is his subsequent theorem
stating that all groups can be represented as the automorphism
group of a 3-regular graph (or cubic graph, a graph in which all
nodes have degree three) \citep{frucht1949}.

The Frucht graph is an undirected graph with
all degrees equal to 3 (undirected edges are interpreted as
bidirectional, i.e., each edge can be seen as a pair of arrows in
opposite directions).  It is famous because, despite the apparent
regularity of its structure, it has no nontrivial automorphisms: its
only automorphism is the identity, hence the only orbit coloring is
the one in Fig. \ref{F:frucht_coloring_bal}e, where all nodes have
a different color.

Although its automorphism group is trivial, the Frucht graph  has a minimal equitable
partition with just one color that represents the most symmetric state
(Fig. \ref{F:frucht_coloring_bal}a). This is possible because the graph is {\em regular}: all nodes have
the same degree and receive the same unique color from adjacent nodes.
The Frucht graph can therefore display complete synchronization under any admissible dynamics.

\begin{figure}[h!]
\centerline{%
\includegraphics[width=\textwidth]{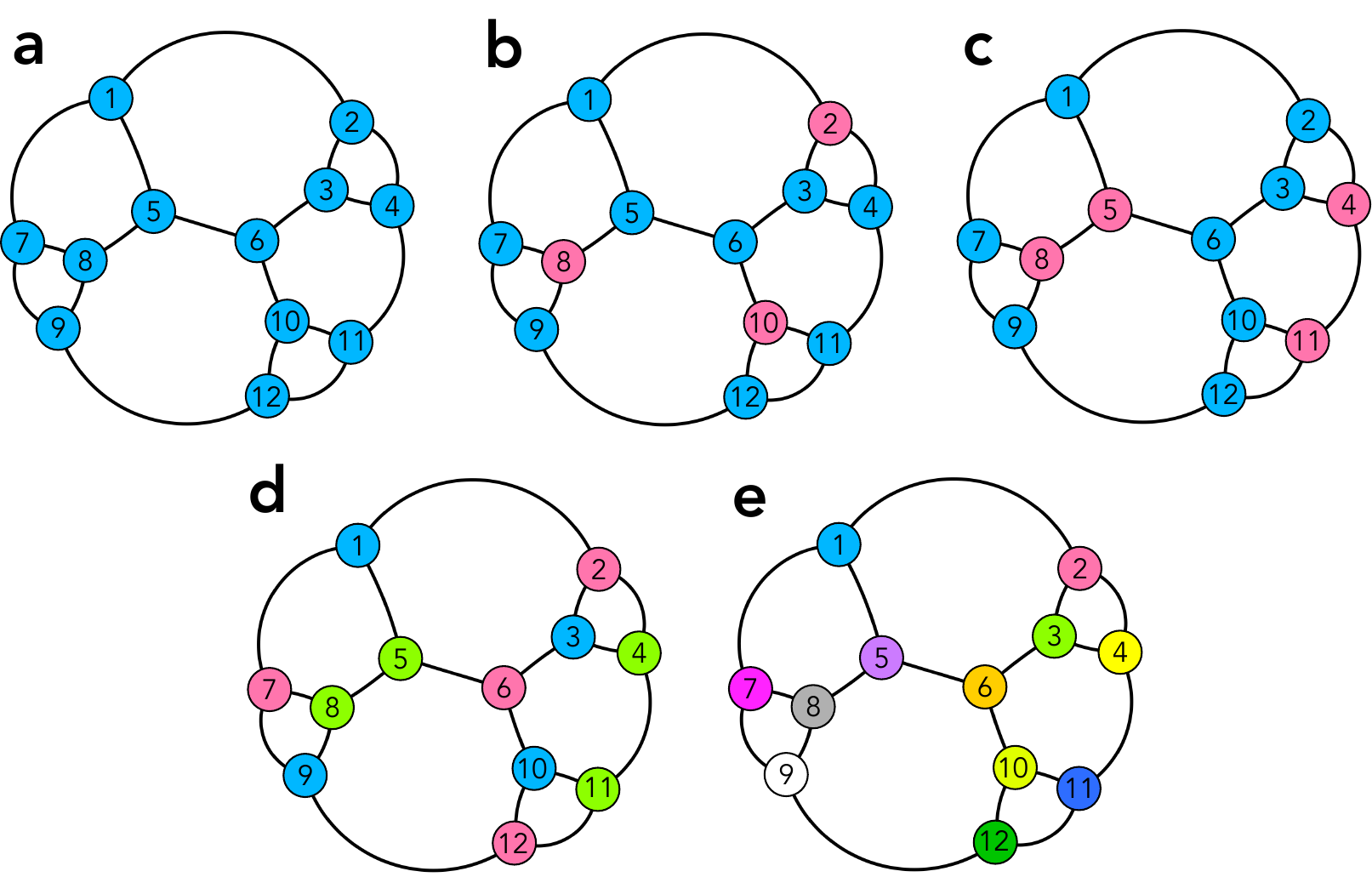}
}
\caption{\textbf{The amazing hidden symmetries of the Frucht graph}. Complete list of balanced colorings of the Frucht graph, assuming all edges
  are bidirectional. (\textbf{a}) Full symmetry: All nodes have identical colors, leading to complete
  synchrony. (\textbf{b}) Two colors: each red node has three blue inputs;
  each blue node has one red and two blue inputs. (\textbf{c}) Two colors:
  each red node has one red and two blue inputs; each blue node has
  one red and two blue inputs. (\textbf{d}) Three colors: each blue node
  has two red and one green input, each red node has two blue and one
  green input, and each green node has one red, one blue, and one
  green input.  (\textbf{e}) All nodes are different: trivial coloring.}
\label{F:frucht_coloring_bal}
\commentAlt{Figure~\ref{F:frucht_coloring_bal}: All five frames show the same
graph with nodes 1-12. Undirected edges connect 1 to 2, 5, 7; 2 to 3, 4; 3 to 4,6;
4 to 11; 5 to 6, 8; 6 to 10; 7 to 8, 9; 8 to 9; 9 to 12; 10 to 11, 12; 11 to 12. 
Frame 1: All nodes blue.
Frame 2: All nodes blue except for 2, 8, 10: red.
Frame 3: All nodes blue except for 4, 5, 8, 11: red.
Frame 4: Nodes 1, 3, 9, 10 blue. Nodes 2, 6, 7, 12 red. Nodes 4, 5, 8, 11 green.
Frame 5: All nodes different colors.
}
\end{figure}

To study whether the Frucht graph has any further nontrivial balanced
colorings, we can use the combinatorial characterization of a balanced
coloring.  This requires nodes of the same color to have
correspondingly colored inputs, for suitable permutations of the input
arrows. The analysis leads to precisely five different balanced
colorings, shown in Fig. \ref{F:frucht_coloring_bal}. Only coloring (e) is given by the symmetry group, which is trivial.
The other four balanced colorings do not arise from automorphisms. 
Coloring (a) is obvious from the regularity of the graph, and it
constitutes the minimal equitable partition. The other
three are far from obvious, but can be found by various methods
that we do not enter into here.
The algorithms discussed in Chapter \ref{chap:algorithms} show that these colorings are
the only balanced colorings of the Frucht graph. 

\end{example}

Any $n$-node 3-regular graph has $6n^2$ fibration symmetries.
Here $n=12$, giving 864 fibration symmetries, compared to 1 group symmetry. Nevertheless, these fibration symmetries lead to only
5 equitable partitions. This shows that the relation between
fibration symmetries and equitable partitions is subtle.

\begin{example}\em
\label{ex:minimal_5}
{\bf Regular graphs.}\index{graph !regular }
More generally, any $k$-regular graph (i.e., any undirected graph whose
nodes have exactly $k$ neighbors) with $N$ nodes can be colored
with one single color (we will see later that it can be fibered over a
$k$-bouquet: a one-node graph with $k$ loops).  This coloring
corresponds to the highest symmetry of the
graph: all nodes are the same. This contrasts with the conclusion that
would arise by considering only automorphism symmetries. 
\end{example}

From the
point of view of automorphisms, it is known that for $k\geq 3$ almost
all $k$-regular graphs are rigid (have no automorphisms except for the
identity). `Almost all' here means that the fraction of all
$k$-regular graphs with $N$ nodes that are rigid tends to 1 as $N\to\infty$. In other words, the Frucht graph is just an example of a more general and almost unavoidable phenomenon.  Although
these graphs have no nontrivial automorphisms, and in principle every node can be
distinguished topologically from every other node, these graphs have
the most symmetric balanced coloring possible: all nodes are the same,
and in fact, they can synchronize in unison under any
admissible dynamics. Moreover, any synchrony pattern determined by a balanced 
coloring can be realized as a {\em stable} equilibrium of {\em some} admissible ODE
\citep{GS2023}.

To move forward we need to abandon the notion of an automorphism as the
only useful symmetry notion, and embrace the more general form of symmetry of
equitable partitions and fibrations.  Only then we will be able to
describe relevant symmetries of biological networks.

\begin{remark}
The balanced coloring problem should not be confused with the `vertex
coloring problem', sometimes referred to as just the `coloring
problem' or `graph coloring'.  The vertex coloring problem attempts to find a coloring of
the graph such that no two adjacent nodes have the same color. The balance condition is different. These two problems are
unrelated.
\end{remark}

\section{Robust versus fragile synchrony}
\label{sec:robust}
In Section \ref{sec:key_principle}, we stated a key principle for this book:
\begin{equation}
\label{E:key_principle}
\begin{array}{c}
\mbox{Synchronized nodes must have the same internal dynamics} \\
\mbox{and must receive synchronized signals.}
\end{array}
\end{equation}
This principle, which is highly intuitive, led to the conclusion
that synchrony clusters determine a fibration/balanced coloring. 
However, in Section \ref{sec:necessary_conditions} we remarked that this deduction
requires an extra technical condition to be mathematically
valid.  In this section we expand on this remark. This discussion is a mathematical fine point and can be skipped.

\begin{figure}[htb]
\centerline{%
\includegraphics[width=\textwidth]{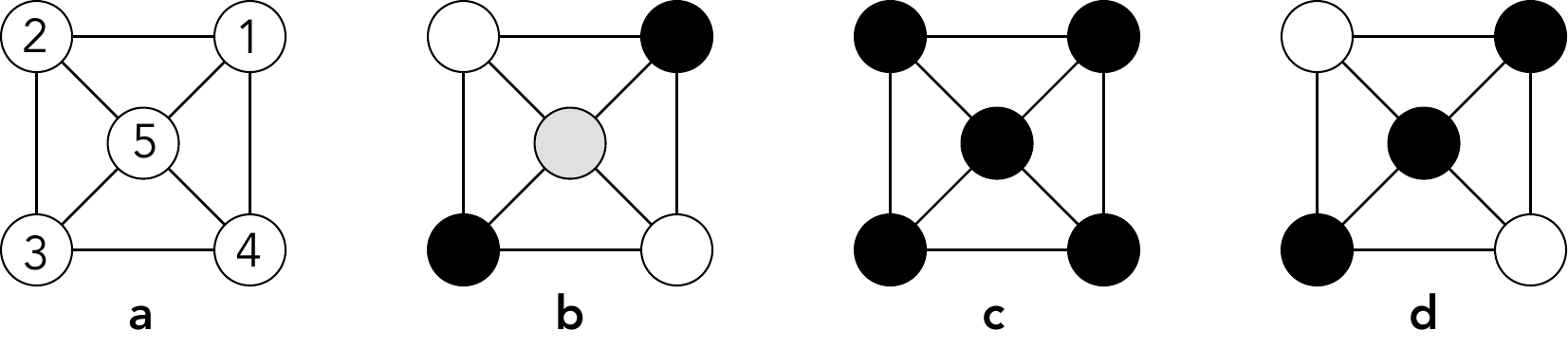} 
}
\caption{ \textbf{Colorings of a 5-node undirected graph.}  ({\textbf a}) The original graph. ({\textbf b}) A balanced coloring.
({\textbf c}) The coloring for complete synchrony is not balanced. ({\textbf d}) This coloring is also not balanced. However, all three colorings correspond to cluster synchrony in the model ODE \eqref{E:0.1}.}
\label{F:diffusive_unbal}
\commentAlt{Figure~\ref{F:diffusive_unbal}: All frames show the same 5-node graph.
Node 1 connects by undirected edges  to 2, 4, 5. Node 2 connects to 3, 5. Node 3 connects to 4, 5. Node 4 connects to 5.
Frame 1: The graph.
Frame 2: Nodes 1, 3 (black), 2, 4 (white), 5 (gray).
Frame 3: All nodes black.
Frame 2: Nodes 1, 3, 5 (black), 2, 4 (white).
}
\end{figure}

We begin with an example to illustrate the mathematical phenomenon
involved.
\begin{example}\em
\label{ex:not_robust}
Consider the $5$-node bidirectional graph in Fig. \ref{F:diffusive_unbal}a.
Here nodes 1, 2, 3, and 4 are input isomorphic, with in-degree (or valence) 3.
Node 5 has in-degree 4, so it is not input isomorphic to the other four nodes.
The coloring in Fig. \ref{F:diffusive_unbal}b is balanced. However, the colorings in  Fig. \ref{F:diffusive_unbal}c, d
are not balanced, because nodes with the same color must be input isomorphic
in any balanced coloring.
\end{example}

Now consider a discrete reaction-diffusion equation\index{reaction-diffusion equation }\eqref{E:0.1}, where $f$
is any reaction function and $a$ is a diffusion coefficient. 
\begin{equation}
\label{E:0.1}
\begin{array}{lll}
\dot x_1 &=&f(x_1) + a(x_1-x_2)+ a(x_1-x_4)+ a(x_1-x_5)\\
\dot x_2 &=&f(x_2) + a(x_2-x_1)+ a(x_2-x_3)+ a(x_2-x_5)\\
\dot x_3 &=&f(x_3) + a(x_3-x_2)+ a(x_3-x_4)+ a(x_3-x_5)\\
\dot x_4 &=&f(x_4) + a(x_4-x_1)+ a(x_4-x_3)+ a(x_4-x_5)\\
\dot x_5 &=&f(x_5) + a(x_5-x_1)+ a(x_5-x_2)+ a(x_5-x_3)+ a(x_5-x_4).
\end{array}
\end{equation}
This is an admissible ODE for the graph. By general theory, coloring (b) gives a
consistent equation for cluster synchrony, as can be checked directly
by setting $x_1=x_3=x, x_2=x_4=y, x_5=z$.
Although coloring (c) is  not balanced, substituting $x_1=x_2=x_3=x_4=x_5 = x$
leads to five copies of the same equation $\dot x = f(x)$. There is no inconsistency,
and any solution of that equation yields a solution of \eqref{E:0.1}
in which all five nodes are synchronous. Similarly,
although coloring (d) is  not balanced, substituting $x_1=x_3=x_5=x$ and $x_2=x_4 = y$
leads to
\beqn
\dot x &=& f(x) + 2a(x-y)\\
\dot y &=& f(y) + 3a(y-x)
\eeqn
and again, there is no inconsistency

At first sight,  the cluster synchrony patterns (c) and (d) 
contradict our claim that principle \eqref{E:key_principle} implies that
every synchrony pattern corresponds to a balanced coloring.
It does not contradict the principle itself: nodes
$x$ and $y$ do indeed receive synchronized inputs,
{\em for this ODE}. However,
as Example \ref{ex:not_robust} shows,
the inference of a balanced coloring can have exceptions.
In this case, the exceptional patterns arise from the special nature of
the coupling terms, which vanish when the nodes concerned are synchronous.
These interactions are, therefore, `invisible' to the dynamics.
In fact, these types of couplings are the diffusive couplings leading to the Laplacian dynamics described in Section \ref{sec:laplacian}. They are a popular way to describe complete synchronization patterns, but we see here that in terms of cluster synchronization, they are fragile patterns of the admissible ODEs. 

To avoid such exceptions (which are `rare' in a suitable context, 
here admissible ODEs) we can modify \eqref{E:0.1} by imposing the extra condition of 
{\em  robustness} \citep{stewart2003}. This was mentioned in
Section \ref{sec:necessary_conditions}
and is defined as follows:
\begin{definition}
\label{D:robust_pattern}
A cluster synchrony pattern for a given graph and admissible ODE,
is {\em robust}\index{robust } if that pattern, when it is imposed upon {\em any}
admissible ODE, 
leads to a consistent ODE for the cluster dynamics.

A cluster synchrony pattern that is not robust is {\em fragile}.\index{fragile }
\end{definition}

It is easy to see that adding an extra small term such
as $\eps x_5$ to the component of \eqref{E:0.1} for $\dot x_5$, which preserves admissibility, leads to two conflicting equations for $\dot x$ in patterns (c) and (d). Thus these patterns
are fragile. Of course, if diffusive coupling is a sensible
modeling assumption, such fragility would not be
relevant for biological purposes.

The term `robust' should not be confused with less formal uses of the word. In this section we use it in this technical sense, 
which goes back to \cite{stewart2003}, in order to clarify when
the relation between cluster synchrony and balanced colorings
holds. Robustness in this technical sense
carries no other connotations that might be associated with that word. 
In particular, it does not imply the stability of the cluster state 
concerned, only its existence. Stability depends on details of the model ODE, and it is discussed in Chapter
\ref{chap:stability}. Section 
\ref{sec:structural_stability} further discusses this notion of robustness and compares it with others in the literature. The implication of the robustness Definition \ref{D:robust_pattern} for the existence of synchrony and the representation of the system is further discussed in Section \ref{sec:hypergraph-metabolic}.

The central point of this example, and others like it, is that
input isomorphisms are properties of the {\em equation}, not just of some synchronous {\em solution}.

\begin{remark}
In fact, a less restrictive  
condition also ensures that cluster synchrony corresponds 
precisely to fibrations/balanced colorings. The
Rigid Equilibrium Theorem\index{Rigid Equilibrium Theorem } \citep[Theorem 14.9]{GS2023}
shows that if clusters of an equilibrium state persist after
any {\em sufficiently small} admissible perturbation, then the corresponding coloring must be
balanced. This formulation
is highly relevant to biology. It tells us that if any
{\em small} change to the model equations that is consistent with the graph structure does not change the clusters---which
is a reasonable requirement, since most biological models lack the precision of
physical ones---then the clusters must arise from a fibration.

A similar result is conjectured for all networks
when the dynamics is time-periodic. This has been proved for
a wide variety of networks, and for all networks under stronger
hypotheses: see \citep[Chapter 15]{GS2023}. Something similar should usually apply to chaotic states as well,
but the issues involved are outside the scope of this book.
\end{remark}

The key point is that
fibrations/balanced colorings are fundamental to the 
existence of meaningful cluster states, {\em unless}
the area of application uses specific modeling assumptions 
that allow unbalanced
colorings\index{unbalanced coloring, conditions for } to occur naturally.

\section{Comments on equivalent definitions in different disciplines}

Now that we have established that `equitable partition = balanced
coloring', it is worth asking why we are 
providing two definitions for the same concept. The answer is
that the literature has grown from several quasi-independent 
sources, in different disciplines. Anyone consulting these sources
needs to be aware of differences in viewpoint and terminology.

The
term `equitable partition' is typically used in the computer science
literature where the graph fibration formalism was originally developed by
\cite{boldi2002fibrations}. On the other hand, the name `balanced
coloring' is used in the dynamical systems and mathematics communities where
the groupoid formalism was originally developed in \citep{stewart2003,stewart2005}.

Both concepts are mathematically equivalent. The overlap of these concepts across
various subject areas from computer science, mathematics, dynamical
systems, chaos and physics makes this topic multidisciplinary. This
introduces a problem of communication between these disparate
disciplines. Unfortunately, the fibration literature from computer
science and the balanced coloring literature from dynamical systems
hardly overlap, even though they speak about the same concepts.  One
of the goals of this book is to bring these communities together by
providing common definitions and concepts. Figure \ref{Fig1-theory} is
an attempt to do just that. Sometimes, it is just a matter of
language, as when one community calls an equitable partition what the other
calls a balanced coloring. Sometimes, however, concepts need to be related in
less obvious ways.

To make things more complicated, we are targeting another community with
its own jargon: physicists. In physics, symmetries are synonymous with
symmetry groups. In particular, particle physics and general
relativity are built on fiber bundles (discussed in Chapter
\ref{chap:bundles}). Fibrations are generalizations of fiber bundles,
and following physics jargon, we refer to the synchronous clusters in
equitable/colored partitions as the {\it fibers} of the graph (see
section \ref{fiber}) in analogy with fiber bundles in physics.

We may finally wonder: where does biology enter into this picture? It
has not entered anywhere so far, because fibrations have not previously been
discussed in biology, so there is no biological jargon related to fibrations
to worry
about. However, `fiber' has other meanings
in biology; this should not cause confusion.
This book is an attempt to combine all previous results to
build a new language for the application of fibrations to biological
networks.


\chapter[Graph Fibration Formalism]{\bf\textsf{Graph Fibration Formalism}}
\label{chap:fibration_2}

\begin{chapterquote}

  We now focus on the central concept of this book: that of a
  {\em symmetry fibration}\index{symmetry !fibration } or {\em fibration symmetry}.\index{fibration !symmetry }
 We define (graph) homomorphisms, which `collapse' nodes and edges in
  a consistent manner, and (graph) fibrations, which are homomorphisms
  that also preserve input sets. We relate fibrations
  to equitable partitions, which are equivalent to balanced colorings of the nodes. We discuss the key property of
  a fibration, the lifting property. We emphasize the role of
  minimal equitable partitions (minimal balanced colorings),
  which employ the smallest number of colors and group the
  nodes into the fewest clusters. Finally, we consider non-minimal colorings.
\end{chapterquote}

So far, we have established that equitable partitions describe robust
symmetries between the nodes of a network. A more perspicuous way of
describing equitable partitions is by means of a graph fibration,
which is a special form of graph homomorphism.  Next, we define a graph
homomorphism, and then we show that the minimal equitable partition
emerges from a symmetry fibration of the graph.

\section{Graph homomorphisms}
\label{sec:morphism}

Since equitable partitions, balanced colorings and cluster
synchronization are all about colored source nodes of incoming edges captured by the
input trees, it is natural that the maps that define these
partitions should preserve certain incidence relations among the
nodes. Maps that preserve incidence relations are called
homomorphisms, and a fibration is a particular kind of graph
homomorphism that further preserves a particular kind of incidence, the
lifting property. To understand fibrations we must first understand
homomorphisms
\citep{boldi2002fibrations}. They are similar to graph
isomorphisms, Definition \ref{def:isomorphism}, but the
map concerned need not be a bijection.

Recall that an undirected graph can be viewed as a directed graph in which
all arrows come in pairs, pointing in opposite directions.
This lets us focus on directed graphs without losing generality.
It is also a natural convention for admissible dynamics.

A graph homomorphism is a map from nodes to nodes and edges to edges
that preserves source and target of every arc~\citep{harary1969,cameron,hahn,hell}; in other words, if it sends some arc $a$ of $G$ to some other arc $a'$ of $H$, then it must also send the source (target) of $a$ in $G$ to the source (target) of $a'$ in $H$. Formally:

\begin{definition}{\bf Graph homomorphism}.
  Given two graphs $G = (N_G, E_G)$ and $H = (N_H, E_H)$, a
  \emph{graph homomorphism}\index{graph !homomorphism } $\alpha: G \to H$ is a map from $G$ to $H$
  such that if $i$ and $j$ are adjacent in $G$ via an edge, then $\alpha(i)$ and
  $\alpha(j)$ are adjacent in $H$ by an edge in the same direction.
\label{morphism}
\end{definition}

In other words, a graph homomorphism $\alpha$ maps nodes of $G$ to
nodes of $H$ and edges of $G$ to edges of $H$ in such a way that the
image of the edge connects the image of the source and the image of
the target of this edge.  
This requirement is very minimal, and the existence of a homomorphism from $G$ to $H$ does in general not imply that $G$ and $H$ are in any way similar.

In particular, a homomorphism need not be an isomorphism; it is a more general
concept that encompasses also automorphisms and fibrations, as well as other
maps between graphs that are not fibrations. A homomorphism is a weak form of incidence preservation: If there is more than one edge between $i$ and $j$, then it requires at least one between $\alpha(i)$ and $\alpha(j)$.  A fibration is a stricter form of homomorphism since not only preserves incidence but it also preserves the input trees (via the lifting property) and therefore preserves the dynamics between $G$ and $H$, which is not necessarily preserved by the homomorphism.

Multigraphs (with parallel edges, Definition \ref{def:multigraph})
require a generalization:

\begin{definition}{\bf Multigraph homomorphism}.
\label{def:multihomo}
A {\it multigraph homomorphism}\index{multigraph !homomorphism } is a pair of functions
$\alpha_N: N_G \to N_H$ and $\alpha_E: E_G \to E_H$ such that:
\begin{equation} s_H(\alpha_E(a))=\alpha_N(s_G(a)) \,\,\,\, \mbox{\rm and} \,\,\,\,s_H(\alpha_E(a))=\alpha_N(s_G(a)).
\end{equation}
\label{def:multigraph-morphism}
\end{definition}

For typed (multi)graphs, it is required that types of nodes and edges are respected: if a node $i$ (or an arc $a$) has a certain type, then its image $\alpha_N(i)$ ($\alpha_E(a)$, respectively) must have the same type.

In various branches of abstract algebra, the analogs of graph homomorphisms are called just 
  homomorphisms, which are structure-preserving maps between two
algebraic structures or sets of the same type (group, ring, module $\ldots$). 
Vector space homomorphisms are called linear maps, for historical reasons. 
This notion is then
further generalized in category theory, where they are called {\it
  morphisms}. Homomorphisms are important since they encompass
the definitions of isomorphisms, automorphisms and fibrations as
special cases. Despite their similar names, homomorphisms are
different from {\it homeomorphisms}, which arise in topology and   preserve
topological structures, like deforming a coffee mug into a donut
(torus).

\begin{figure}[htb]
\centerline{
\includegraphics[width=0.9\linewidth]{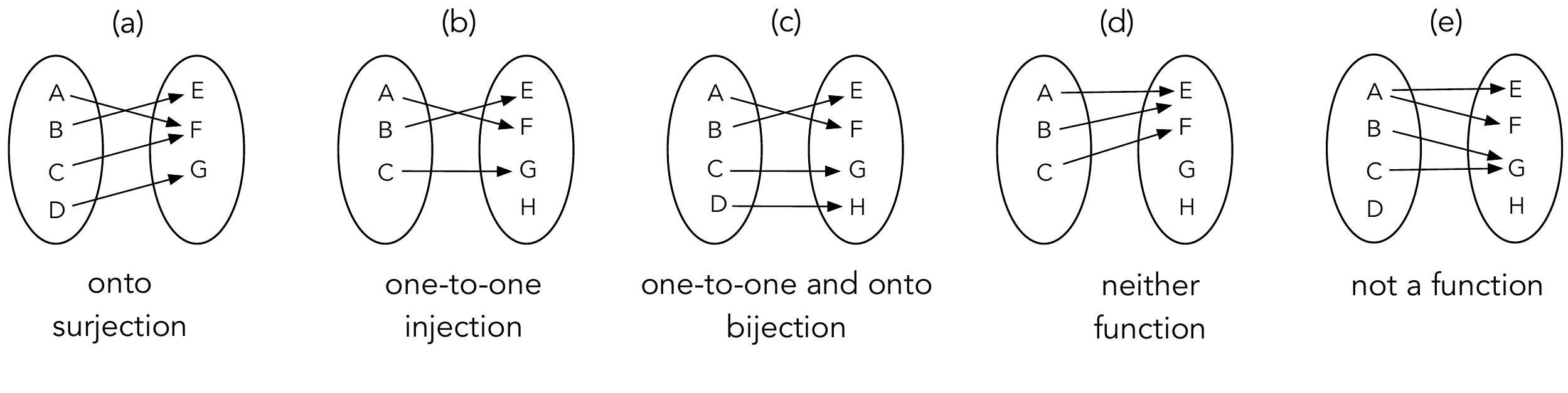} \qquad
}
\caption{\textbf{Types of maps}. (\textbf{a})
  Injective. (\textbf{b}) Surjective. (\textbf{c}) Bijective. (\textbf{d}) Neither injective nor surjective. (\textbf{e}) Not a function.}
\label{maps}
\commentAlt{Figure~\ref{maps}: 
(a) Left: oval containing A, B, C, D. Right: oval containing E, F, G. Arrows AF, BE, CF, DG.
Label: onto, surjection.
(b) Left: oval containing A, B, C. Right: oval containing E, F, G, H. Arrows AF, BE, CG.
Label: one-to-one, injection.
(c) Left: oval containing A, B, C, D. Right: oval containing E, F, G, H. Arrows AF, BE, CG, DH.
Label: one-to-one and onto, bijection.
(d) Left: oval containing A, B, C. Right: oval containing E, F, G, H. Arrows AE, BE, CF.
Label: neither, function.
(e) Left: oval containing A, B, C, D. Right: oval containing E, F, G, H. Arrows AE, AF, BG, CG.
Label: not a function.
}
\end{figure}

Graph homomorphisms are maps, and any map can be injective,
surjective, bijective (injective and surjective), or neither (see
Fig. \ref{maps}). We recall these definitions since they will be
important for the types of fibrations that we use to describe biological
graphs. For more details see \citep{ST2015}.

\begin{definition}{\bf Injective map.}
A map is {\em injective}\index{injective } if it maps distinct elements of its domain to
distinct elements of its codomain (Fig. \ref{maps}a).
\label{injective}
\end{definition}

Therefore, every element of the codomain is the
image of at most one element of its domain. An injective map is also
called a `one-to-one map'. An injective homomorphism (i.e., one that maps injectively both nodes and edges) can map a graph to a bigger
graph. Even though the map is one-to-one, there could be elements in
the codomain that are not images of any elements in the
domain. Thus, a one-to-one map need not be a correspondence (as in a
bijection), and an injective map need not be invertible. 

When dealing with (multigraph) homomorphisms (Definition~\ref{def:multihomo}), injectivity is required both for the node map and for the edge map. More precisely, a multigraph homomorphism $\alpha: G \to H$ is injective if and only if both $\alpha_N$ and $\alpha_E$ are injective.

\begin{definition}{\bf Surjective map.}
A map $\alpha : G \to H$ is {\em surjective}\index{surjective } if for every element of
the codomain of $y\in H$, there is an element in the domain $x\in G$
such that $\alpha(x) = y$ (Fig. \ref{maps}b).
\label{surjective}
\end{definition}

That is, in a surjective map every element of the codomain is the
image of at least one element in the domain. For this reason,
a surjective map is also called
`onto'. Such a map may lead to a reduction of the domain through the
`collapse' of some elements, and will be the main type of map studied
here.  

Again, when dealing with (multigraph) homomorphisms (Definition~\ref{def:multihomo}), surjectivity is required both for the node map and for the edge map. More precisely, a multigraph homomorphism $\alpha: G \to H$ is surjective if and only if both $\alpha_N$ and $\alpha_E$ are surjective.
Surjective fibrations, which collapse nodes or edges, are the
most common fibrations discussed in this book, since they describe
cluster synchronization and dimension reduction in biological models.

\begin{definition}{\bf Bijective map.}
A {\em bijection}\index{bijection } is a map that is both injective and surjective (Fig. \ref{maps}c).
\label{bijection}
\end{definition}

A bijection is a map such that every element in the codomain is an
image of exactly one element in the domain. Thus, a bijection is both
injective and surjective. A bijection is a `one-to-one correspondence'
between two sets, and it is invertible. 

When dealing with (multigraph) homomorphisms (Definition~\ref{def:multihomo}), bijectivity is required both for the node map and for the edge map, as usual.

When the map preserves some kind of algebraic structure, 
an injection is often called
a `monomorphism', a surjection is an `epimorphism', and a bijection
is an `isomorphism'. Over the years, different communities have developed
different terms for the same concept, and mathematical fashions have changed.

We have seen already two examples of bijective homomorphisms, namely
isomorphisms and automorphisms, which are compared to each other in
Fig. \ref{fig:iso}. Injective homomorphisms are not very relevant for
biological networks since they lead to larger graphs (although an
inverse injective fibration will be used in Chapter \ref{chap:minimal}
to reduce the complexity of the network). The relevant types of
homomorphism for biological applications are the surjections, and
within those the surjective fibrations. Fibrations are a very special
class of homomorphism with an additional condition: preserving 
color-isomorphic input sets by satisfying the lifting property.  We
 discuss this concept next.

\section{Graph fibrations}
\label{sec:fibmin}

A fibration\index{fibration } is a homomorphism that preserves more adjacency
structure than the simple incidence relation: the exact definition is
related to the so-called lifting property, and we postpone it to Section~\ref{lifting}.
The basic idea of fibrations, though, is that they can 
collapse only nodes in synchronous fibers, i.e., nodes in equitable
partitions. Here `collapse' means `map to the same thing', that is,
`identify'. Equivalently, they collapse nodes with isomorphic input trees or with
color-isomorphic input sets.  This is why we believe that fibrations are the right type of graph
homomorphism to describe biological networks.

\section{Examples of fibrations}

Before giving the general definition of a graph fibration we
consider two examples.

\begin{example}\em
\label{ex:meta+smol_fib}

{\bf Metabolator and Smolen Circuits revisited}.
We return to the metabolator and Smolen networks of Section
\ref{intuition}, Fig. \ref{F:meta_smol1}, each of which has two nodes and two arrow types---biologically, activation and repression. We saw that
both circuits support completely synchronous states,
in which nodes 1 and 2 synchronize; see Fig. \ref{fig:metab+smo_fibrations} (top).

For the metabolator,\index{metabolator }
this state is the minimal orbital partition for its automorphism group $\Z_2$. The Smolen circuit\index{Smolen circuit } has trivial automorphism group, and its
minimal orbital partition is trivial, with parts $\{1\}$ and $\{2\}$.
The minimal equitable partition has a single part $\{1,2\}$.
This occurs because the nodes have color-isomorphic input sets:
each pink node receives one pink input of each arrow-type.

\begin{figure}[b]
\centerline{%
\includegraphics[width=0.7\textwidth]{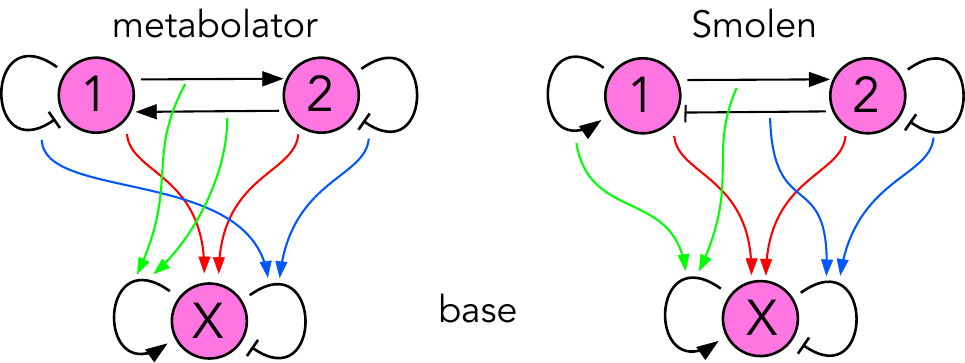}}
\caption{\textbf{Fibrations for the minimal equitable partition.} The 
metabolator and Smolen networks.}
\label{fig:metab+smo_fibrations}
\commentAlt{Figure~\ref{fig:metab+smo_fibrations}: 
Left: Label `metabolator'. Top: graph with nodes 1, 2. Barred arrows join 11, 22.
Sharp arrows 12, 21. Bottom: graph with node X. Label `base'. Barred arrow XX, sharp arrow XX.
Red arrows run from top nodes to bottom one.
Blue arrows run from top barred arrows to bottom one.
Green arrows run from top sharp arrows to bottom one.
Right: Label `Smolen'. Top: graph with nodes 1, 2. Barred arrows join 21, 22.
Sharp arrows 11, 12. Bottom: graph with node X. Label `base'. Barred arrow XX, sharp arrow XX.
Red arrows run from top nodes to bottom one.
Blue arrows run from top barred arrows to bottom one.
Green arrows run from top sharp arrows to bottom one.
}
\end{figure}

We now recast this analysis in the language of fibrations.\index{fibration }
Both networks have fibrations to the {\em base}\index{base } illustrated in
Fig. \ref{fig:metab+smo_fibrations} (bottom). The base, which is the
same for both circuits, has
a single node, and receives one input of each arrow type from itself.

The colored arrows indicate the corresponding fibrations. Red arrows
show how the nodes map (both nodes 1 and 2 map to node X in the base).
Blue arrows show how the activator arrows map; green arrows
show how the repressor arrows map. The key feature that makes this map a fibration is that it preserves input sets. Each node, be it 1, 2, or X, has one activator input and one repressor input.

There is a unique fiber in each case, namely the set $\{1,2\}$ of all nodes.
\end{example}

\begin{figure}[h!]
	\centering
        \includegraphics[width=\textwidth]{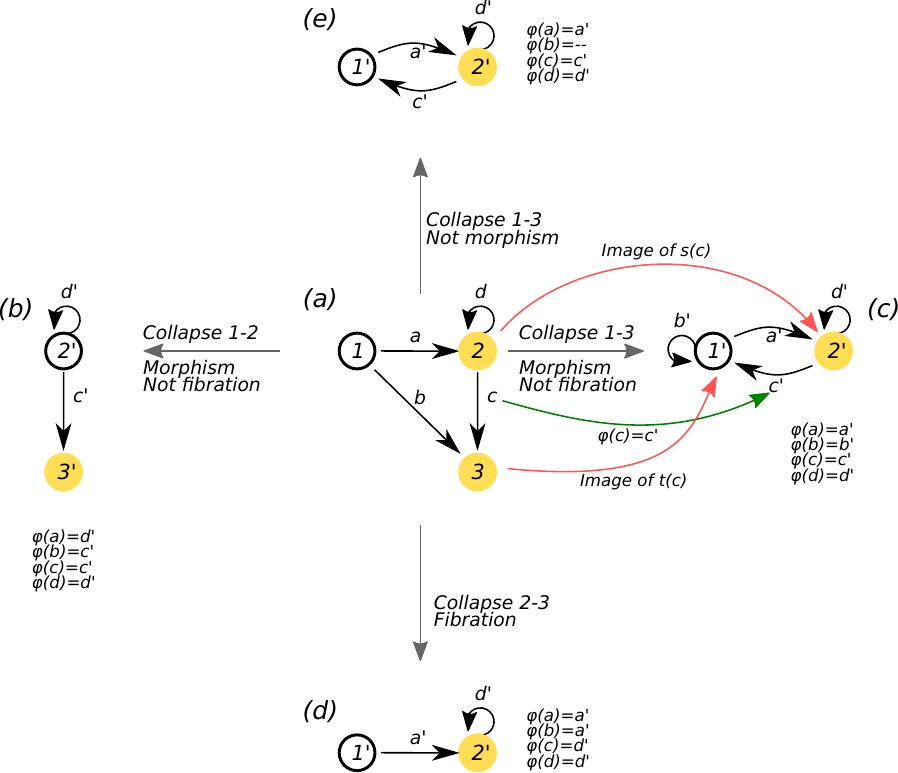}
	\caption{\textbf{Types of fibrations and homomorphisms in the FFF
            circuit.} (\textbf{a}) The FFF circuit is the graph $G$ and
          we map this to four graphs $H$ as shown. 
          The node component is described in this caption, whereas the edge component is described in the figure.
          Collapsing different nodes gives rise to different
          maps $\varphi: G \to H$: these can be either homomorphisms, fibrations
          or just maps. Maps contain
          homomorphisms, which contain fibrations. (\textbf{b})
          Collapsing 1 and 2 (both mapped to node $2'$) with the edge map $\varphi$ shown in the
          figure provides a homomorphism but not a fibration.  (\textbf{c}) Collapsing 1 and 3 (both mapped to node $1'$) with the edge map $\varphi$ shown is a
          homomorphism, but not a fibration because $b'$ cannot be
          lifted to any edge in $G$ (however, $a'$, $c'$ and $d'$ can be lifted
          to an edge in $G$). (\textbf{d}) Only when we collapse the
          fiber 2 and 3 (mapping them both to node $2'$) do we obtain a fibration. All edges in $B$ can be lifted to edges in
          $G$. (\textbf{e}) Collapsing 1 and 3 without mapping edge $b$
          is not even a homomorphism. It is just a map among the node sets.}
    \label{morphism-fibration}
\commentAlt{Figure~\ref{morphism-fibration}: 
No alt-text required.
}
\end{figure}

\begin{example}\em
\label{ex:FFF_fib}

Next, we consider a less trivial example of a
fibration using the FFF\index{FFF } circuit depicted in
Fig. \ref{morphism-fibration} (center). 
We show in yellow nodes 2 and 3, which belong to the same part of an equitable partition:
they have isomorphic input sets, or equivalently isomorphic input trees.  We apply to
the FFF graph (Fig. \ref{morphism-fibration}a) three surjective
homomorphisms: Fig. \ref{morphism-fibration}b, a surjective homomorphism
collapsing nodes of 1 and 2 (both mapped to node $3'$); Fig. \ref{morphism-fibration}c, a
surjective homomorphism collapsing nodes of 1 and 3 (both mapped to node $1'$); and
Fig. \ref{morphism-fibration}d, a surjective fibration collapsing
nodes 2 and 3 (both mapped to $2'$).  We also show in Fig. \ref{morphism-fibration}e a
collapsed graph that is obtained with a surjective node map but not a homomorphism.

A homomorphism preserves adjacency: edges are mapped in such a way that the
image of the edge connects the image of the source to the image of the
target, transferring the incidence relation from source to target node in
$G$ to their respective images in $H$. 

In a homomorphism, it can happen that nodes are collapsed (i.e., they are mapped to the same image); because of its very definition, the homomorphism `drags along' the edges of the collapsed nodes to $H$, enough to have at least one edge
satisfying the preservation of adjacency.

In practice this means that when we collapse nodes 1 and 3 into the single node 1'
($\varphi(1)=\varphi(3)=1'$) in the surjective homomorphism of Fig.
\ref{morphism-fibration}c, node 3 drags its incoming edges into the
collapsed node 1': this is achieved by mapping $c \to c'$ and $b \to
b'$, such that both incoming edges of $3$ are transferred to $1'$.
Thus, through the edge map $\varphi(c)=c'$, we see that the image
($c'$) of the edge ($c$) connects the image of the source
$\varphi(s(c))$ and the image of the target $\varphi(t(c))$ of the
source edge $c$, thus preserving incidence. The same occurs for the
edge $b$, that is mapped to $b'$ ($\varphi(b)=b'$).
\end{example}

Homomorphisms preserve the source and target incidence relations
between nodes and edges, but they are not required to preserve any
other structure. Fibrations also preserve input sets, isomorphically,
moreover, they collapse only nodes in the same equitable
partition; that is, in synchronous fibers with color-isomorphic input sets. Thus Fig. \ref{morphism-fibration}b and
Fig. \ref{morphism-fibration}c are homomorphisms but not fibrations.
The map of Fig. \ref{morphism-fibration}d, besides being
a homomorphism, is quite special and different
from the previous two. It is the only one of these examples that
collapses only nodes in the same equitable partition. Therefore this
homomorphism has special significance since the collapsed graph inherits the
synchronous  dynamics of $G$. A homomorphism that collapses nodes in
equitable partitions constitutes a graph fibration. Dynamical
invariance is not satisfied by the other homomorphisms in
Fig. \ref{morphism-fibration}, which are not fibrations.

In a more intuitive description, a graph fibration is a homomorphism in which every edge
targeting an image node is an image of a {\em unique} edge, so multiple
edges targeting the same node cannot be collapsed to fewer edges (as
we did in Fig. \ref{morphism-fibration}b collapsing $b$ and $c$ into
$c'$). Neither can new edges targeting the image of a node be added
(as we did in Fig. \ref{morphism-fibration}c by creating $b'$). The
only way to avoid these two conditions is to collapse equitable
partitions. Thus, a fibration implies a stronger condition than the simple
preservation of the incidence relation by homomorphisms. This extra
conditions is captured by the lifting property, Section \ref{lifting} below.

The definition of a fibration through the collapse of equitable nodes
 applies only to surjective fibrations. A more general definition of
fibration, valid for all fibrations (surjective or not), uses the
lifting property. We now elaborate on both definitions in turn.

\section{Graph fibrations through equitable partitions}
\label{fiber}

We now generalize these observations to give a general definition of a fibration, and discuss how fibrations lead to equitable partitions
in this general context.

There are many equivalent ways to define graph fibrations. 
Normally they are defined using the `lifting property' (see Section~\ref{lifting}), but we can start by providing an alternative and possibly more intuitive definition based on input isomorphisms, which
are local concepts.

\begin{definition}{\bf Definition of graph fibration as a local input isomorphism.}
   \label{def:fiblocalin}
   \index{graph fibration } A graph homomorphism $\varphi:
  G \to B$ is a graph \emph{fibration}\index{fibration } if and only if, for all vertices $x \in V_G$,
  $\varphi$ induces an isomorphism between the set of edges of $G$ having $x$ as target and the set of edges of $B$ having $\varphi(x)$ as target. The graph $G$ is the \emph{total graph} of the fibration, $B$ is called the \emph{base graph},\index{base } and for every vertex $x$ of $B$, the set of vertices of $G$ that are mapped to $x$ by $\varphi$ is the \emph{fiber (over $x$)}. \index{fiber }
\end{definition}

The condition says that, when looked at locally (i.e., at a specific vertex of the graph $G$ and at its image in the graph $B$) the map provides an isomorphism between incoming edges. There is no condition on \emph{outgoing} edges.

Based on this definition we can immediately state some properties that fibrations enjoy, relating them to equitable partitions. 

\begin{proposition}
\label{prop:surjfib-eqp}
If $\varphi: G \to B$ is a fibration, its fibers define an equitable partition of $G$, the equitable partition \emph{associated to $\varphi$}.   
\end{proposition}

The converse of Proposition~\ref{prop:surjfib-eqp} is true:

\begin{proposition}
\label{prop:eqp-surjfib}
For every equitable partition of the nodes of $G$, there exists a graph $B$ and a surjective graph fibration $\varphi: G \to B$ such that two nodes are in the same cluster of the equitable partition if and only if they are mapped by $\varphi$
to the same node of $B$. 
\end{proposition}

In other words, the fibers of $\varphi$ are exactly the clusters\index{cluster } of the equitable partition, so fibrations
are homomorphisms that collapse nodes belonging to the same cluster in an
equitable partition. 

Instead of starting with the fibration and considering its base,
we can construct the base from the corresponding
equitable partition.\index{partition !equitable }

\begin{definition}
\label{def:quot_net}
{\bf Quotient network}.
Let $G$ be a directed graph with an equitable partition $P$. 
The corresponding {\em quotient network}\index{quotient !network } $G/P$ is the graph
whose nodes correspond to the elements of $P$ (that is,
the colors in the equitable partition). To construct the input edges
to a given node we choose any node of that color and consider
its input set. For each edge in the input set we define an
edge in $G/P$ by connecting nodes of the corresponding colors.

The map from $G$ to $G/P$ that assigns to each node of $G$ its
color in $P$ is now a fibration, because each edge in
$G/P$ is the image of a unique edge in $G$; that is, the lifting
property holds. Moreover, the base of this fibration is $G/P$.
\end{definition}

For instance, the identity map $id: G \to G$ (which maps each node and
edge to itself) is a fibration, and the partition it induces
(the identity partition) is equitable; in this case, no nodes are actually collapsed, though.  More generally, every
automorphism is, trivially, a fibration; hence all group symmetries
are \emph{a fortiori} equitable partitions. So fibrations generalize
automorphism in a proper mathematical sense.

The notion of a fibration can easily be extended to graphs with various
types of edges, such as a gene regulatory network which has both
activation and repressor edges, or a brain network with excitatory and
inhibitory synaptic connections. In this case, what we have to do is to restrict the notion of homomorphism to send nodes or edges of a certain type to nodes or edges of the same type.

\begin{figure}
    \centering
    \scalebox{1}{
        \begin{tikzpicture}
        \tikzset{vertex/.style = {shape=circle,draw}}
        \tikzset{edge/.style = {->,> = latex'}}
        \node[vertex] (v6) at  (1,0) {$6$};
        \node[vertex] (v4) at  (0,1) {$4$};
        \node[vertex] (v5) at  (2,1) {$5$};
        \node[vertex] (v2) at  (0,3) {$2$};
        \node[vertex] (v3) at  (2,3) {$3$};
        \node[vertex] (v7) at  (4,3) {$7$};
        \node[vertex] (v1) at  (1,4) {$1$};
        \draw[edge] (v6) to node [below] {$g$} (v4);
        \draw[edge] (v6) to node [below] {$h$} (v5);
        \draw[edge] (v4) to node [left] {$e$} (v2);
        \draw[edge] (v5) to node [right] {$f$}(v3);
        \draw[edge] (v2) to[bend left] node [below] {$c$} (v3);
        \draw[edge] (v3) to[bend left] node [below] {$d$}(v2);
        \draw[edge] (v3) to node [above] {$i$} (v7);
        \draw[edge] (v2) to node [above] {$a$} (v1);
        \draw[edge] (v3) to node [above] {$b$} (v1);
        \node[vertex] (b6) at (7,0) {$b_6$};
        \node[vertex] (b45) at (7,1) {$b_{45}$};
        \node[vertex] (b2) at (6,3) {$b_2$};
        \node[vertex] (b3) at (8,3) {$b_3$};
        \node[vertex] (b7) at (10,3) {$b_7$};
        \node[vertex] (b1) at (7,4) {$b_1$};
        \draw[edge] (b6) to node [left] {$g'$} (b45);
        \draw[edge] (b45) to node [left] {$e'$} (b2);
        \draw[edge] (b45) to node [right] {$f'$} (b3);
        \draw[edge] (b2) to[bend left] node [below] {$c'$} (b3);
        \draw[edge] (b3) to[bend left] node [below] {$d'$} (b2);
        \draw[edge] (b3) to node [above] {$i'$} (b7);
        \draw[edge] (b2) to node [above] {$a'$} (b1);
        \draw[edge] (b3) to node [above] {$b'$} (b1);
        \end{tikzpicture}
    }
    \caption{\textbf{Example of a fibration that is not minimal.} The
      network $G$ of Fig.~\ref{fig:example-inputsets} (left)
      and a nontrivial fibration $\varphi: G \to B$ to another
      network (right). Edges $a$, $b$, \dots are mapped to
      $a'$, $b'$, \dots respectively, with the exception of $h$, which
      is mapped to $g'$ (like $g$); node $i$ is mapped (injectively)
      to node $b_i$ for all $i$, except for nodes $4$ and $5$, which are
      both mapped to $b_{45}$.}
    \label{fig:example-fibration}
\commentAlt{Figure~\ref{fig:example-fibration}: 
Left: A graph with nodes 1-7. Directed edges: arrows from 2 to 1 (label a),
from 3 to 1 (label b), from 2 to 3 (label c), from 3 to 2 (label d),
from 4 to 2 (label e), from 5 to 7 (label f), from 6 to 4 (label g),
from 6 to 5 (label h), from 3 to 7 (label i).
Right: A graph with nodes called b1, b2, b3, b45, b6, b7. Directed edges: arrows
from b2 to b1 (label a'), from b3 to b1 (label b'), from b2 to b3 (label c'),
from b3 to b2 (label d'), from b45 to b2 (label e'), from b45 to b3 (label f'),
from b6 to b45 (label g'), from b3 to b7 (label i').
}
\end{figure}

Since nodes in the same equitable partition have isomorphic input trees, this is true also for nodes in the same fiber. More precisely:

\begin{proposition}
    \label{prop:fibinputtree}
    If $\varphi: G \to B$ is a fibration, and $\varphi(x)=\varphi(y)$, then $T_x$ is isomorphic to $T_y$. Conversely, if $x$ and $y$ are two nodes of a graph $G$ with isomorphic input trees (i.e., such that $T_x \sim T_y$), then there exists some graph $B$ and fibration $\varphi: G \to B$ such that $\varphi(x)=\varphi(y)$.
\end{proposition}

As we have said in Section~\ref{minimal_partition}, the minimal equitable partition\index{partition !minimal equitable }  
is the \emph{coarsest} equitable partition that can be found.  That
is, it is the partition that collects all the symmetries that the
network has, and partitions the network into the smallest number of
fibers. Proposition~\ref{prop:eqp-surjfib} now implies:

\begin{proposition}
  \label{prop:minimal}
  For every graph $G$, there is a graph $\hat G$, the \emph{minimum base}\index{base !minimum } of $G$ and a surjective fibration $\mu: G \to \hat G$ whose associated fibers correspond to the minimal equitable partition. Because of the properties of the minimal equitable partition, $\mu(x)=\mu(y)$ if and
  only if $x$ and $y$ have isomorphic input trees (i.e., $T_x \simeq
  T_y$). The fibration $\mu$ is the \emph{minimal fibration}\index{fibration !minimal } of $G$.
\end{proposition}

It turns out that this graph $\hat G$ is unique (up to
isomorphism), whereas the fibration $\mu: G \to \hat G$ is not
necessarily unique, but it is unique on nodes.
  
Figure ~\ref{fig:example-minimum} shows a graph fibration that collapses
the total graph $G$ of Fig.~\ref{fig:example-inputsets} into the
minimum base $\hat G$ of $G$, and a minimal fibration $\mu: G \to \hat
B$ that maps the total graph to the minimal base by collapsing all the
fibers of the network.

To summarize this section, a graph fibration $G \to B$
maps the nodes of a graph $G$ so that nodes that are mapped to the same node of $B$ form an equitable partition. 
When we are looking only at surjective fibrations, 
every node in $B$ has a preimage in
$G$: therefore, except in trivial cases (i.e., when the map is a bijection),
some distinct nodes in $G$ are `collapsed' into a
single node.
Nodes that are collapsed together by the fibration belong to the same
\textit{fiber}, and the size of the fiber is the number of collapsed
nodes.  Nodes that belong to
the same fiber of surjective fibrations can synchronize their dynamics, which explains why fibrations are the natural way to describe synchronization.

All nodes in the same fiber have isomorphic input trees. Similarly, given an equitable partition of $G$, nodes in the same cluster have isomorphic input trees, and there is a surjective fibration $G \to B$ whose fibers precisely correspond to the clusters of the equitable partition.

The minimal surjective graph fibration 
is the fibration associated to the minimal (i.e., coarsest) equitable partition; in this case, two nodes $v_1, v_2 \in G$ belong
to the same fiber $\iff$ their input trees are
isomorphic. $\Rightarrow$ follows from the second part of Theorem 2 in
\citep{boldi2002fibrations}, and $\Leftarrow$ is Proposition 23 in
\citep{boldi2002fibrations}. See also~\citep{stewart2007}.   In other words, a surjective fibration is minimal if it
collapses all nodes that can be collapsed.

\begin{figure}
    \centering
    \scalebox{1}{
        \begin{tikzpicture}
        \tikzset{vertex/.style = {shape=circle,draw}}
        \tikzset{edge/.style = {->,> = latex'}}
        \node[vertex] (v6) at  (1,0) {$6$};
        \node[vertex] (v4) at  (0,1) {$4$};
        \node[vertex] (v5) at  (2,1) {$5$};
        \node[vertex] (v2) at  (0,3) {$2$};
        \node[vertex] (v3) at  (2,3) {$3$};
        \node[vertex] (v7) at  (4,3) {$7$};
        \node[vertex] (v1) at  (1,4) {$1$};
        \draw[edge] (v6) to node [below] {$g$} (v4);
        \draw[edge] (v6) to node [below] {$h$} (v5);
        \draw[edge] (v4) to node [left] {$e$} (v2);
        \draw[edge] (v5) to node [right] {$f$}(v3);
        \draw[edge] (v2) to[bend left] node [below] {$c$} (v3);
        \draw[edge] (v3) to[bend left] node [below] {$d$}(v2);
        \draw[edge] (v3) to node [above] {$i$} (v7);
        \draw[edge] (v2) to node [above] {$a$} (v1);
        \draw[edge] (v3) to node [above] {$b$} (v1);
        \node[vertex] (b6) at (7,0) {$b_6$};
        \node[vertex] (b45) at (7,1) {$b_{45}$};
        \node[vertex] (b23) at (7,3) {$b_{23}$};
        \node[vertex] (b7) at (10,3) {$b_7$};
        \node[vertex] (b1) at (7,4) {$b_1$};
        \draw[edge] (b6) to node [left] {$g'$} (b45);
        \draw[edge] (b45) to node [left] {$e'$} (b23);
        \draw[edge] (b23) to[loop left] node [left] {$c'$} (b23);
        \draw[edge] (b23) to node [above] {$i'$} (b7);
        \draw[edge] (b23) to[bend left] node [left] {$a'$} (b1);
        \draw[edge] (b23) to[bend right] node [right] {$b'$} (b1);
        \end{tikzpicture}
    }
    \caption{\textbf{Minimal fibration and minimal base.} The network $G$ of Fig.~\ref{fig:example-inputsets} (left) and its minimum base $\hat G$ (right). A minimal fibration
      $\mu: G \to \hat G$ maps node $i$ to node $b_i$ except $4$ and
      $5$ (which are both mapped to $b_{45}$) and nodes $2$ and $3$
      (which are both mapped to $b_{23})$. Edges $a$, $b$, \dots are
      mapped to $a'$, $b'$, \dots respectively, with the following
      exceptions: $d \mapsto c'$, $f \mapsto e'$, $h \mapsto g'$. The
      only other minimal fibration is exactly the same but maps $a
      \mapsto b'$ and $b \mapsto a'$.}
    \label{fig:example-minimum}
\commentAlt{Figure~\ref{fig:example-minimum}: 
Left: A graph with nodes 1-7. Directed edges: arrows from 2 to 1 (label a),
from 3 to 1 (label b), from 2 to 3 (label c), from 3 to 2 (label d),
from 4 to 2 (label e), from 5 to 7 (label f), from 6 to 4 (label g),
from 6 to 5 (label h), from 3 to 7 (label i).
Right: A graph with nodes called b1, b23, b45, b6, b7. Directed edges: arrows
from b23 to b1 (label a'), from b23 to b1 (label b'), 
from b23 to itself (label c'), from b45 to b23 (label e'), 
from b6 to b45 (label g'), from b23 to b7 (label i').
}
\end{figure}

\section{The lifting property}
\label{lifting}\index{lifting property }

For practical purposes, as well as to study
fibrations in biological networks as in Chapters
\ref{chap:hierarchy_1} and \ref{chap:hierarchy_2}, Propositions~\ref{prop:fibinputtree} and \ref{prop:minimal} (specifying how fibrations and input-tree isomorphisms are related) are
enough to understand fibration symmetries.
That is, the search for fibration symmetries in real networks is
performed by searching for the fibers of networks that correspond to
the coarsest universal equitable partition or balanced coloring. The minimal base of the
graph is then obtained by collapsing the fibers into a common
node at the base. Algorithms to find the fibers of the network are
based on finding the coarsest balanced coloring partition, and are
discussed in Chapter \ref{chap:algorithms}.

In \index{lifting } an alternative way (totally equivalent to Definition~\ref{def:fiblocalin}), graph fibrations
can be defined independently using the so-called
{\it lifting property}, which rigorously distinguishes
a graph fibration from a generic homomorphism \citep{boldi2002fibrations}:
 
\begin{definition}
  {\bf Definition of graph fibration via the lifting
    property}. \index{graph fibration } A graph homomorphism $\varphi:
  G \to B$ is a {\em graph fibration} if and only if, for all edges $e' \in
  E_B$ and for all vertices $i \in N_G$, if $\varphi(i)=t_B(e')$ then
  there exists exactly one edge $e \in E_G$ such that $t_G(e)=i$ and
  $\varphi(e)=e'$. This unique edge $e$ is the {\it lift} of
    $e'$ at $i$.

    In particular, every edge targeting $i'=\varphi(i)$ can be
\emph{uniquely} lifted to an edge in $G$ targeting $i$. This is 
the {\it lifting property}.\index{lifting property }
\label{def:fibration}
\end{definition}

The relation between Definition~\ref{def:fiblocalin} and Definition~\ref{def:fibration} should be immediately clear: Definition~\ref{def:fibration} postulates the existence of an isomorphism between the edges entering into $\varphi(i)$ and the edges entering into $i$ (existence and uniqueness, together, exactly define a one-to-one relation). In turn, this means that if $\varphi(i)=\varphi(j)$, then the set of edges entering into $i$ and $j$ must be isomorphic (because they are both isomorphic to the set of edges entering into $\varphi(i)=\varphi(j)$), which is what  Definition~\ref{def:fiblocalin} requires.

\begin{figure}
	\centering
                \includegraphics[width=\textwidth]{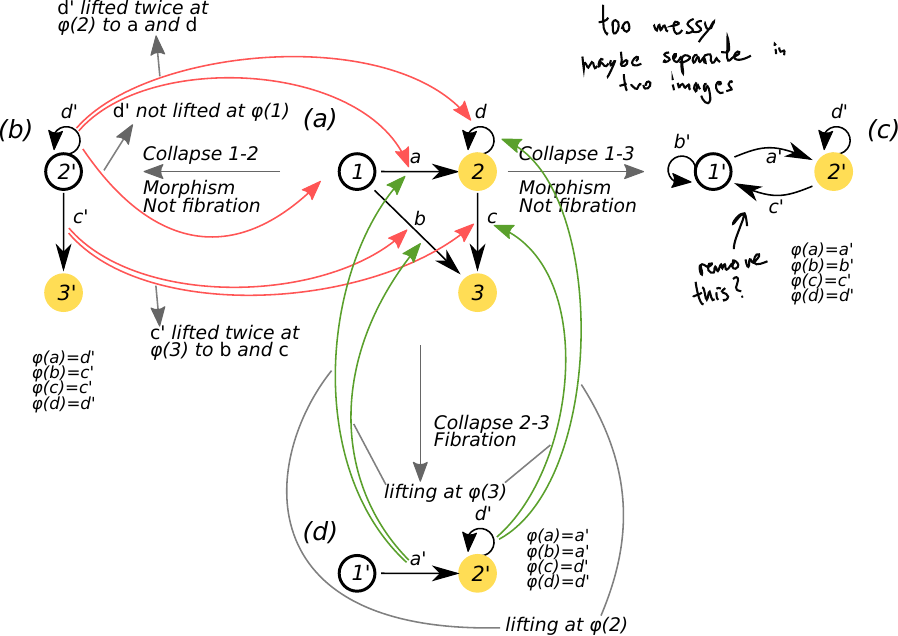}
	\caption{\textbf{The amazing world of the  lifting property.} Example of the lifting property for the fibration of
          Fig. \ref{morphism-fibration}d, and an example of too many
          lifts for the homomorphism of
          Fig. \ref{morphism-fibration}b and lack of a lift to
          node 1 in \ref{morphism-fibration}c.}
    \label{fig:lifting}
\end{figure}

This definition means that multiple edges, targeting the same node in $G$, cannot be
collapsed to fewer edges in $B$, unlike a homomorphism; neither can new
edges targeting the image of a node be added. This is exemplified in
Fig. \ref{morphism-fibration}d. This definition still allows
for reduction of the network, because nodes with same number of inputs
can be mapped to the same image.
 
The homomorphism of
Fig. \ref{morphism-fibration}d satisfies the lifting property and hence it
is a fibration, while the homomorphisms of
Figs. \ref{morphism-fibration}b and \ref{morphism-fibration}c do not, so
they are not fibrations.

To understand more deeply the lifting property,\index{lifting property }
 we
examine why these homomorphisms fail to be fibrations.
To do so, we test the lifting property.  The fibration of
Fig. \ref{morphism-fibration}d $\varphi: G \to B$ is defined by the
following map between nodes: $\varphi(1)=1'$, $\varphi(2)=2'$,
$\varphi(3)=2'$, and by the map between edges: $\varphi(a)=a'$,
$\varphi(b)=a'$, $\varphi(c)=d'$, and $\varphi(d)=d'$, as shown in the
figure.  We test lifting: for edge $a'$ pointing to $2'$ in $B$, there
are two nodes $x=2$ and $x=3$ in the fiber over $2'$.  For each of these nodes in $G$ we must
 check that there is a unique edge (the lift) pointing to them,
so that upon fibration we get $a'$. Indeed, $\varphi$ maps two edges to $a'$, namely $a$ and $b$: one targets $2$ and one targets $3$. The same procedure demonstrates the existence and uniqueness of the
lifts of $d'$ at $2$ and  $3$ proving that
$\varphi$ is a fibration.  The checks are shown in Table \ref{lift1}.

On the other hand, the homomorphism in Fig. \ref{morphism-fibration}b
fails uniqueness of lifting for the edge $d'$ at $2$ (both $a$ and $d$ target $2$ and are mapped to $d'$), and the same goes for an attempt at lifting $c'$ at 3 (both $b$ and $c$ target $3$ and are mapped to $c'$); moreover, $d'$ cannot be lifted at $1$. Again the tests are summarized in Table \ref{lift2}.  Therefore the
homomorphism is not a fibration. The failure of the lifting property
in this homomorphism is somewhat subtle. Sometimes existence fails, sometimes uniqueness fails. 

Finally, the homomorphism of Fig. \ref{morphism-fibration}c is also
not a fibration as shown in Table \ref{lift3}. In principle, every edge in $B$ can be
lifted to an edge in $G$ as indicated in the figure. However, there
are lifts missing for node $x=1$, so the homomorphism fails to be a
fibration.  This happens because these homomorphisms collapse two
nodes ($1$ and $3$) that are not input tree isomorphic: recall that a
fibration can collapse only nodes with
isomorphic input trees.  Welcome to the world
of fibrations, where these subtle differences are all relevant and
important!

While the lifting property might look threatening, it is just another
way to say that we cannot place any nodes with non-isomorphic input
trees in the same fiber. For practical purposes, then, we first find the fibers
using algorithms from Chapter \ref{chap:algorithms}, 
based on finding a balanced coloring, and then we apply the notion of fibration to
reduce the network (i.e., to decide how \emph{edges} can be mapped). Not the other way around, since we are not aware of
any algorithm that can systematically find fibrations using the
lifting property alone.  Nevertheless, the lifting property is helpful when
thinking in terms of a model of evolution of biological networks via
duplication and speciation, discussed in Chapter \ref{chap:robustness}.

\begin{table*}[t!]
\centering
\begin{threeparttable}
  {\setlength\doublerulesep{ 2pt}
	\begin{tabular}{|c|c|c|c||c|c||c|c|c|}
          \hline
          Fibration & Lifting & $i\in G$  & $e'\in B$  & $t_B(e')$ & $\varphi(i)$ & $e\in G$ & $t_G(e)$  & $\varphi(e)$  \\
          &  Property & \textsuperscript{*} & \textsuperscript{**} & & & &  \textsuperscript{*} &  \textsuperscript{**}\\ 
          \hline\hline
          $\surd$ & $a$ lift $a'$ at 2 & 2  & $a'$ & 2' & 2' & $a$ & 2  & $a'$  \\ \hline
          $\surd$ & $d$ lift $d'$ at 2 & 2  & $d'$ & 2' & 2' & $d$ & 2  & $d'$  \\ \hline
          $\surd$ & $b$ lift $a'$ at 3 & 3  & $a'$ & 2' & 2' & $b$ & 3  & $a'$  \\ \hline
          $\surd$ & $c$ lift $d'$ at 3 & 3  & $d'$ & 2' & 2' & $c$ & 3  & $d'$  \\ \hline
        \end{tabular}
}
\label{lift1}
\begin{tablenotes}
\item[*] \footnotesize{These two columns should be the same for fibrations.}
\item[**] \footnotesize{These two columns should be the same for fibrations.}
\end{tablenotes}
\end{threeparttable}
\vspace{10pt}
\caption{Testing the lifting property for
  the fibration in Fig. \ref{morphism-fibration}d. All edges in $B$
  can be lifted uniquely, so the homomorphism is a fibration.}
\end{table*}

\begin{table*}[t!]
\centering

\begin{threeparttable}
  {\setlength\doublerulesep{ 2pt}
	\begin{tabular}{|c|c|c|c||c|c||c|c|c|}
          \hline
          Fibration & Lifting & $i\in G$  & $e'\in B$  & $t_B(e')$ & $\varphi(i)$ & $e\in G$ & $t_G(e)$ & $\varphi(e)$ \\
          &  Property & \textsuperscript{*} & \textsuperscript{**} & & & &  \textsuperscript{*} &  \textsuperscript{**}\\ 
          \hline\hline
          $\times$ & none &            1  & $d'$ & 2' & 2' & none & none  & none  \\ \hline
          $\times$ & $>$ one &            2  & $d'$ & 2' & 2' & $a$ or $d$ & 2  & $d'$  \\ \hline          
          $\times$ & $>$  one &            3  & $c'$ & 3' & 3' & $b$ or $c$ & $3$  & $c'$  \\ \hline
        \end{tabular}
}
\label{lift2}
\begin{tablenotes}
\item[*] \footnotesize{These two columns should be the same for fibrations.}
\item[**] \footnotesize{These two columns should be the same for fibrations.}
\end{tablenotes}
\end{threeparttable}
\vspace{10pt}
\caption{Testing the lifting property for the homomorphism in
  Fig. \ref{morphism-fibration}b. Edge $d'$ cannot be lifted at $1$, but it has two possible lifts at $2$; edge $c'$ also has two possible lifts at $3$. Therefore the homomorphism is not a fibration.}
\end{table*}

\begin{table*}[t!]
\centering
\begin{threeparttable}
  {\setlength\doublerulesep{ 2pt}
	\begin{tabular}{|c|c|c|c||c|c||c|c|c|}
          \hline
          Fibration & Lifting & $i\in G$ & $e'\in B$  & $t_B(e')$ & $\varphi(i)$ & $e\in G$ & $t_G(e)$  & $\varphi(e)$  \\
            &  Property &  \textsuperscript{*} & \textsuperscript{**}  &  &  &  & \textsuperscript{*} &  \textsuperscript{**} \\
           \hline\hline
          $\times$ & none &            1  & $b'$ & 1' & 1' & none & none  & none  \\ \hline
          $\times$ & none &            1  & $c'$ & 1' & 1' & none & none  & none  \\ \hline
          $\surd$ & $b$ lift $b'$ & 3  & $b'$ & 1' & 1' & $b$ & 3  & $b'$  \\ \hline
          $\surd$ & $c$ lift $c'$ & 3  & $c'$ & 1' & 1' & $c$ & 3  & $c'$  \\ \hline
          $\surd$ & $a$ lift $a'$ & 2  & $a'$ & 2' & 2' & $a$ & 2  & $a'$  \\ \hline
          $\surd$ & $d$ lift $d'$ & 2  & $d'$ & 2' & 2' & $d$ & 2  & $d'$  \\ \hline
        \end{tabular}
}
\label{lift3}
\begin{tablenotes}
\item[*] \footnotesize{These two columns should be the same for fibrations.}
\item[**] \footnotesize{These two columns should be the same for fibrations.}
\end{tablenotes}
\end{threeparttable}
\vspace{10pt}
\caption{Testing the lifting property for the homomorphism in
  Fig. \ref{morphism-fibration}c. Neither $b'$ nor $c'$ can be lifted at $1$, therefore the homomorphism is not a fibration.}
\end{table*}

A fibration can be surjective or not, injective or not \citep{lerman2015b} (see Fig. \ref{maps}). Surjectivity is mainly a technical requirement: we can make any homomorphism (and in particular, any fibration) $\varphi: G \to B$ surjective by modifying $B$, removing the nodes and edges that are not in the image of $\varphi$; for this reason we focus mainly on surjective fibrations. On the other hand, we are mainly interested in non-injective fibrations, because those are the fibrations that collapse nodes 
and thus identify nontrivial clusters of synchronous nodes.

\begin{example}\em
\label{wrong_quot_2}
We return to Example \ref{ex:wrong_quot}
and Fig. \ref{fig:6node_orbital_quot}, which illustrates
the key difference between graph homomorphisms and fibrations.
Let $G$ be the graph of Fig. \ref{fig:6node_orbital_quot}a.
The quotient in the sense of \citep{hahn}, that is, $G/{\rm Aut}(G)$,
is shown in Fig. \ref{fig:6node_orbital_quot}b.
The natural map from $G$ to $G/{\rm Aut}(G)$ maps
each node of $G$ to the node with the corresponding color in 
$G/{\rm Aut}(G)$. This map is a graph homomorphism since it has at least an edge between collapsed nodes in the quotient if there was an edge between any nodes of the collapsed clusters in the original graph, thus preserving incidence under the collapse.
However, it is not a fibration, because the edges 
$(1,5)$ and $(4,5)$ in Fig. \ref{fig:6node_orbital_quot}a
both map to the same edge $(1,2)$ in Fig. \ref{fig:6node_orbital_quot}b.

In contrast, denote the equitable partition/coloring
by $P$. Then the quotient network $G/P$ is shown in Fig. \ref{fig:6node_orbital_quot}c. Because the arrow
from node 2 to node 3 is doubled, the lifting property
holds. Now the map that assigns to each node of $G$ 
the node of the corresponding color in Fig. \ref{fig:6node_orbital_quot}c
is a fibration. Input sets are preserved, and
the admissible ODEs for the quotient network
determine, precisely, the possible cluster dynamics
for the coloring.

We also see that homomorphisms and fibrations
have different effects on input trees. Homomorphisms
can collapse them, but fibrations always preserve them.

Because the natural map from $G$ to $G/{\rm Aut}(G)$ is not
a fibration but merely a graph homomorphism, 
input sets and input trees in the base are not isomorphic to
input sets and input trees in the original graph. 
Since the map from $G$ to $G/P$ is a fibration, 
input trees in the base {\em are} isomorphic to
input sets and input trees in the original graph.
Figure  \ref{fig:6node_input_tree} illustrates this point.

\begin{figure}[h!]
\centerline{%
\includegraphics[width=0.43\textwidth]{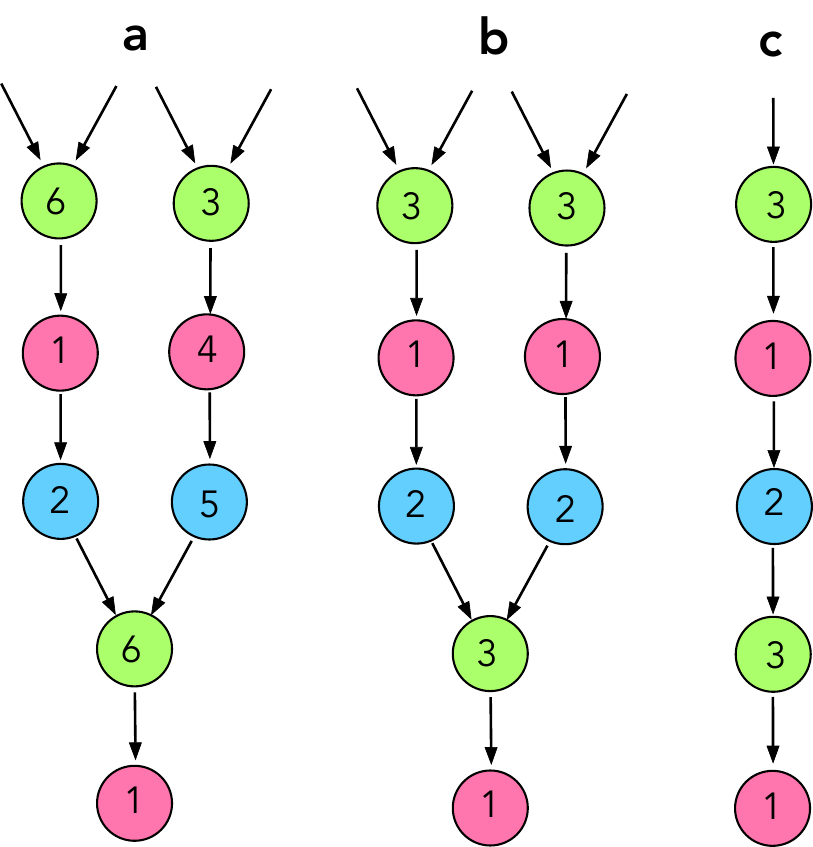}}
\caption{\textbf{Differences in input trees between homomorphisms and fibrations.} (\textbf{a}) Input tree of node 1 in 
Fig. \ref{fig:6node_orbital_quot}a.
(\textbf{b}) Input tree of node 1 in 
Fig. \ref{fig:6node_orbital_quot}c is isomorphic to
that in (\textbf{a}), as it should be for a fibration.
(\textbf{c}) Input tree of node 1 in 
Fig. \ref{fig:6node_orbital_quot}b is not isomorphic to
that in (\textbf{a}).
}
\label{fig:6node_input_tree}
\commentAlt{Figure~\ref{fig:6node_input_tree}: 
Three downwards-oriented trees, with labels a, b, c.
All trees have colored and labeled nodes
as explained in Long alt-text.
}

\commentLongAlt{Figure~\ref{fig:6node_input_tree}: 
Tree a (described from its topmost level down). 
Level 1: nodes 6 and 3 (both green), with two incoming
arrows from above (not connected to anything).
Level 2: node 1 and 4 (both red), with arrows from 6 (of level 1) and
from 3 (of level 1), respectively.
Level 3: node 2 and 5 (both blue), with arrows from 1 (of level 2) and
from 4 (of level 2), respectively.
Level 4: node 6 (green), with arrows from 2 and 5 (of level 3).
Level 5: node 1 (red), with arrow from 6 (level 4).
Tree b (described from its topmost level down). 
Level 1: nodes 3 and 3 (both green), with two incoming
arrows from above (not connected to anything).
Level 2: nodes 1 and 1 (both red), with arrows from leftmost 3 (of level 1)
and rightmost 3 (of level 1), respectively.
Level 3: nodes 2 and 2 (both blue), with arrows from leftmost 1 (of level 2)
and rightmost 1 (of level 2), respectively.
Level 4: node 3 (green), with arrows from the two nodes label 2 (of level 3).
Level 5: node 1 (red), with arrow from 3 (level 4).
Tree c (described from its topmost level down). 
Level 1: node 3 (green), with one incoming
arrow from above (not connected to anything).
Level 2: node 1 (red), with arrow from 3 (level 1).
Level 3: node 2 (blue), with arrow from 1 (level 2).
Level 4: node 3 (green), with arrow from 2 (level 3).
Level 5: node 1 (red), with arrow from 3 (level 4).
}
\end{figure}
\end{example}

\section{Hierarchy of symmetries
 from homomorphisms to fibrations (op-fibrations) to coverings to automorphisms}
\label{sec:fibcovaut}

Having defined the main tool that we use in this book---fibrations---we revisit our initial purpose: to describe symmetry in graphs and synchronization in networks. The aim of this section is
to summarize the two axes along which automorphisms and fibrations differ  (local vs. global, in vs. in/out). To do this we introduce  other local intermediate symmetries  between these two (local+in = fibrations, local+out = op-fibrations, local+in+out = coverings) forming a hierarchy of symmetries from less to more strict:
homomorphism $\to$ fibration (and op-fibration) $\to$ covering $\to$ automorphism.

As discussed, automorphisms capture a global type of symmetry: a bijective homomorphism from the structure to itself that preserves the \emph{whole} structure. We can think of this as a \emph{change of label} that does not change the shape. Applying this notion to graphs means that we consider graph homomorphisms $\varphi: G \to G$ that are bijections. In particular, such bijections preserve outgoing edges as well as incoming edges.
Automorphisms (for graphs and in general) have a nice algebraic structure: 
\begin{itemize}
    \item They can composed (as functions) and composition is \emph{associative}, that is, $(\varphi_1 \circ \varphi_2) \circ \varphi_3 = \varphi_1 \circ (\varphi_2 \circ \varphi_3)$: in other words, we can get rid of all parentheses when composing automorphisms.
    \item There is an \emph{identity} map $1: G \to G$, which maps each node/edge to itself. (In group theory the identity map is
    usually denoted by $1$, not {\em id}.) This map is an automorphism, and it is the identity for composition, because $\varphi \circ 1=1 \circ \varphi=\varphi$.
    \item Every automorphism is \emph{invertible}, that is, for every automorphism $\varphi$ there exists an automorphism $\varphi^{-1}$ such that $\varphi \circ \varphi^{-1}=\varphi^{-1} \circ \varphi = 1$.
\end{itemize}
These three properties together make what mathematicians call a \emph{group};\index{group } that is, the set $\Aut(G)$ of all automorphisms of $G$ is a group with respect to function composition. This nice, simple algebraic structure gives rise to a whole rich theory. 
There is a quotient graph $G/\Aut(G)$ (Definition \ref{def:aut_quot}),
but this is not suitable for describing synchrony patterns
or the dynamics of clusters.

A key point, 
previously discussed, is that automorphisms are hardly the right way to describe symmetries of biology. The restrictions imposed by the global essence of an automorphism make it very unlikely that a real-world graph should possess any nontrivial automorphism (any automorphism other than the identity). So $G/\Aut(G)$ is usually the same as $G$.

Fibrations overcome this obstacle. They offer another way to look at symmetry, with much humbler requirements. First, fibrations require that structure be preserved only locally and not globally. Second, they ask that only incoming edges are preserved, whereas outgoing edges are not taken into account. This is precisely what Definition~\ref{def:fiblocalin} requires. All automorphisms are of course fibrations, but there are typically many more fibrations than automorphisms. Every fibration $\varphi: G \to B$ captures some of the fibration symmetry of the graph $G$. Proposition~\ref{prop:fibinputtree} tells us that the input trees are the same in every fiber; that is, $T_x \simeq T_y$ if $\varphi(x)=\varphi(y)$. This is important, because as long as we are interested only in the symmetry of vertices, we can simply look at their input trees.
The analog of $G/\Aut(G)$ for fibrations is the minimum base $\hat G$ and the minimal fibration $\mu: G \to \hat G$. Again, looking only at the effect of $\mu$ on vertices, $\mu(x)=\mu(y)$ if and only if $T_x \simeq T_y$, as explained in Proposition~\ref{prop:fibinputtree}.

\subsection{Preserving outputs instead of inputs: op-fibrations}
A brief conceptual stop is needed at this point. The duality global vs.~local (automorphisms are global symmetries, fibrations are local symmetries) is clear. Less clear is why we look 
only at incoming edges and ignore outgoing edges. This is
philosophically challenging. 

The choice, here, comes from the meaning of directed edges in nature: a directed edge $x \to y$ represents a form of communication \emph{from} $x$ \emph{to} $y$; the exact type of communication is not relevant, here (it will be different from case to case) but the idea is always that there is a flow of information going from one node ($x$) to another node ($y$). It is in the very nature of information that the recipient of a piece of information can change its state based on the information it obtained; the sender does not. This (trivial) observation explains why incoming edges are more important in practice than outgoing edges, and this is why fibrations are defined in that way.

One caveat: in some areas where network diagrams are commonly used, outputs can change the state of a node.
The mass balance condition in biochemical networks imposes
such a restriction, for instance. We can (and do) sidestep this issue by working with
the `influence network',\index{network !influence }. This has extra {\em input} edges to a given node, which represent
the effect of these outputs on the dynamics of that node. We have already discussed this point in subsection \ref{SS:SMEq}, but we repeat the message here. The component of an admissible ODE for node $c$
has the form $\dot x_c = f_c(x_c,x_{i_1}, x_{i_2}, \ldots$
for a general map $f_c$ and `input' variables $x_{i_1}, x_{i_2}, \ldots$. The equation specifies how node $c$
{\em processes} the listed variables to affect the dynamics
of its own variable $x_c$. In some areas of application, these variables
may include outputs a well as inputs, but in the
general theory we can represent their
effects using the influence network, in which case we refer to them as inputs.

Of course, from a purely mathematical standpoint, we could consider a notion that is absolutely analogous to that of fibration, but where outgoing edges are considered instead and the output tree is preserved rather than the input tree. This notion is called \emph{op-fibration}\index{op-fibration } (`op' stands for opposite, and it is the standard categorical way to name things when you `reverse' them). Now, op-fibrations are no different than fibrations: everything we know about fibrations applies \emph{mutatis mutandis}. This time what counts is \emph{output trees}\index{output trees } (the trees of outgoing paths), $\check T_x$. We can imagine the op-fibration equivalent of Proposition~\ref{prop:fibinputtree}: given an op-fibration $\varphi: G \to B$ and $\varphi(x)=\varphi(y)$, the output trees will be isomorphic, i.e., $\check T_x \simeq \check T_y$. 

And again there will be a minimum op-fibration base $\check G$\index{minimum op-fibration base } and a minimal op-fibration $\check \mu: G \to \check G$. 
Figure  \ref{fig:main_fig} shows an example of output trees.

Op-fibrations are not more informative or useful mathematically than fibrations: the two concepts are dual to each other. However, given that input trees are more relevant for dynamical systems in biology, we can put, in principle, op-fibrations in the cabinet of curiosities and leave them there. Except for a very important application: In chapter \ref{chap:ai} we will show that op-fibrations and their output trees describe the symmetries in the learning process through backpropagation in deep neural networks, while fibrations and input trees describe the inference process.

\subsection{A graph having different fibrations and op-fibrations}

To appreciate the difference between fibrations and op-fibrations, look at the graph $G$ of Fig. \ref{fig:example-fibopfib}. This graph has non-trivial fibrations and op-fibrations, but the two are different from each other.

\begin{figure}
  \centering
  \begin{tikzpicture}
      \tikzset{vertex/.style = {shape=circle,draw,minimum size=4mm}}
      \tikzset{>={Latex[width=2mm,length=2mm]}}
      \node[vertex] (v1) at (-2,0) {$1$};
      \node[vertex] (v2) at (-2,-2) {$2$};
      \node[vertex] (v3) at (0,0) {$3$};
      \node[vertex] (v4) at (0,-2) {$4$};
      \draw[->] (v1) to (v3);
      \draw[->] (v1) to[bend left] (v2);
      \draw[->] (v2) to[bend left] (v1);
      \draw[->] (v3) to[bend left] (v4);
      \draw[->] (v4) to[bend left] (v3);
  \end{tikzpicture}
  \caption{{\bf Fibrations and op-fibrations.} This graph $G$ has different non-trivial fibrations and op-fibrations. It has, however, neither non-trivial coverings nor non-trivial automorphisms.\label{fig:example-fibopfib}}
\commentAlt{Figure~\ref{fig:example-fibopfib}: 
A graph with nodes labeled 1-4.
Directed edges: arrows from 1 to 3, from 1 to 2, from 2 to 1, from 3 to 4,
from 4 to 3.
}
\end{figure}

To see why this fact happens, let us build the input trees and output trees of each node (see Fig. \ref{fig:example-fibopfib-minfib}). As you can see, tree isomorphism clusters the nodes of the graph $G$ in two different ways, depending on whether you are looking at input trees or output trees.

From the viewpoint of \emph{inputs}, node 1 and 2 are totally indistinguishable because 1 only receives input from 2, and 2 only receives input from 1; 3 and 4, on the other hand, are totally different from each other, and also from 1 and 2, because 3 receives two inputs (it is in fact the only node in the graph with two incoming arcs), and 4 receive input from 3. 

\begin{figure}
  \centering
  \begin{tabular}{cc}
  {
  \begin{tabular}{c|c}
  \hline
  $T_1\simeq T_2$ &
  \scalebox{0.6}{
      \begin{tikzpicture}
      \tikzset{vertex/.style = {shape=circle,draw}}
      \tikzset{>={Latex[width=2mm,length=2mm]}}
      \node[] () at (0,6.3) {}; 
      \node[vertex] (v1) at (0,6) {};
      \node[vertex] (v21) at  (0,5) {};
      \node[vertex] (v121) at  (0,4) {};
      \node[vertex] (v2121) at  (0,3) {};
      \node (vd2121) at  (0,2) {$\vdots$};        
      \node [above left = 0mm and 0mm of v1]() {$1$};
      \node [above left = 0mm and 0mm of v21]() {$2$};
      \node [above left = 0mm and 0mm of v121]() {$1$};
      \node [above left = 0mm and 0mm of v2121]() {$2$};
      \draw[->] (v21)--(v1);
      \draw[->] (v121) to (v21);
      \draw[->] (v2121) to (v121);
      \draw[->] (vd2121) to (v2121);
      \end{tikzpicture}
  } 
  \raisebox{1cm}{$\qquad\simeq\qquad$}
  \scalebox{0.6}{
      \begin{tikzpicture}
      \tikzset{vertex/.style = {shape=circle,draw}}
      \tikzset{>={Latex[width=2mm,length=2mm]}}
      \node[] () at (0,6.3) {}; 
      \node[vertex] (v1) at (0,6) {};
      \node[vertex] (v21) at  (0,5) {};
      \node[vertex] (v121) at  (0,4) {};
      \node[vertex] (v2121) at  (0,3) {};
      \node (vd2121) at  (0,2) {$\vdots$};        
      \node [above left = 0mm and 0mm of v1]() {$2$};
      \node [above left = 0mm and 0mm of v21]() {$1$};
      \node [above left = 0mm and 0mm of v121]() {$2$};
      \node [above left = 0mm and 0mm of v2121]() {$1$};
      \draw[->] (v21)--(v1);
      \draw[->] (v121) to (v21);
      \draw[->] (v2121) to (v121);
      \draw[->] (vd2121) to (v2121);
      \end{tikzpicture}
  } 
  \\
  \hline
  $T_3$ &
  \scalebox{0.6}{
      \begin{tikzpicture}
      \tikzset{vertex/.style = {shape=circle,draw}}
      \tikzset{>={Latex[width=2mm,length=2mm]}}
      \node[] () at (0,6.3) {}; 
      \node[vertex] (v3) at (0,6) {};
      \node[vertex] (v13) at  (-2,5) {};
      \node[vertex] (v213) at  (-2,4) {};
      \node[vertex] (v1213) at  (-2,3) {};
      \node[vertex] (v21213) at  (-2,2) {};
      \node[vertex] (v121213) at  (-2,1) {};
      \node (vd121213) at  (-2,0) {$\vdots$};        
      \node[vertex] (v43) at  (2,5) {};
      \node[vertex] (v343) at  (2,4) {};
      \node[vertex] (v1343) at  (0,3) {};
      \node[vertex] (v21343) at  (0,2) {};
      \node[vertex] (v121343) at  (0,1) {};
      \node (vd121343) at  (0,0) {$\vdots$};        
      \node[vertex] (v4343) at (3,3) {};
      \node[vertex] (v34343) at (3,2) {};
      \node[vertex] (v134343) at (2,1) {};
      \node(vd134343) at (2,0) {$\vdots$};
      \node[vertex] (v434343) at (4,1) {};
      \node(vd434343) at (4,0) {$\vdots$};
      \node [above left = 0mm and 0mm of v3]() {$3$};
      \node [above left = 0mm and 0mm of v13]() {$1$};
      \node [above left = 0mm and 0mm of v213]() {$2$};
      \node [above left = 0mm and 0mm of v1213]() {$1$};
      \node [above left = 0mm and 0mm of v21213]() {$2$};
      \node [above left = 0mm and 0mm of v121213]() {$1$};
      \node [above left = 0mm and 0mm of v43]() {$4$};
      \node [above left = 0mm and 0mm of v343]() {$3$};
      \node [above left = 0mm and 0mm of v1343]() {$1$};
      \node [above left = 0mm and 0mm of v21343]() {$2$};
      \node [above left = 0mm and 0mm of v121343]() {$1$};
      \node [above right = 0mm and 0mm of v4343]() {$4$};
      \node [above left = 0mm and 0mm of v34343]() {$3$};
      \node [above left = 0mm and 0mm of v134343]() {$1$};
      \node [above right = 0mm and 0mm of v434343]() {$4$};
      \draw[->] (v13)--(v3);
      \draw[->] (v213) to (v13);
      \draw[->] (v1213) to (v213);
      \draw[->] (v21213) to (v1213);
      \draw[->] (v121213) to (v21213);
      \draw[->] (vd121213) to (v121213);
      \draw[->] (v43)--(v3);
      \draw[->] (v343) to (v43);
      \draw[->] (v1343) to (v343);
      \draw[->] (v21343) to (v1343);
      \draw[->] (v121343) to (v21343);
      \draw[->] (vd121343) to (v121343);
      \draw[->] (v4343) to (v343);
      \draw[->] (v34343) to (v4343);
      \draw[->] (v134343) to (v34343);
      \draw[->] (vd134343) to (v134343);
      \draw[->] (v434343) to (v34343);
      \draw[->] (vd434343) to (v434343);
      \end{tikzpicture}
  } 
  \\
  \hline
  $T_4$ &
  \scalebox{0.6}{
      \begin{tikzpicture}
      \tikzset{vertex/.style = {shape=circle,draw}}
      \tikzset{>={Latex[width=2mm,length=2mm]}}
      \node[] () at (0,7.3) {}; 
      \node[vertex] (v) at (0,7) {};
      \node[vertex] (v3) at (0,6) {};
      \node[vertex] (v13) at  (-2,5) {};
      \node[vertex] (v213) at  (-2,4) {};
      \node[vertex] (v1213) at  (-2,3) {};
      \node[vertex] (v21213) at  (-2,2) {};
      \node[vertex] (v121213) at  (-2,1) {};
      \node (vd121213) at  (-2,0) {$\vdots$};        
      \node[vertex] (v43) at  (2,5) {};
      \node[vertex] (v343) at  (2,4) {};
      \node[vertex] (v1343) at  (0,3) {};
      \node[vertex] (v21343) at  (0,2) {};
      \node[vertex] (v121343) at  (0,1) {};
      \node (vd121343) at  (0,0) {$\vdots$};        
      \node[vertex] (v4343) at (3,3) {};
      \node[vertex] (v34343) at (3,2) {};
      \node[vertex] (v134343) at (2,1) {};
      \node(vd134343) at (2,0) {$\vdots$};
      \node[vertex] (v434343) at (4,1) {};
      \node(vd434343) at (4,0) {$\vdots$};
      \node [above left = 0mm and 0mm of v]() {$4$};
      \node [above left = 0mm and 0mm of v3]() {$3$};
      \node [above left = 0mm and 0mm of v13]() {$1$};
      \node [above left = 0mm and 0mm of v213]() {$2$};
      \node [above left = 0mm and 0mm of v1213]() {$1$};
      \node [above left = 0mm and 0mm of v21213]() {$2$};
      \node [above left = 0mm and 0mm of v121213]() {$1$};
      \node [above left = 0mm and 0mm of v43]() {$4$};
      \node [above left = 0mm and 0mm of v343]() {$3$};
      \node [above left = 0mm and 0mm of v1343]() {$1$};
      \node [above left = 0mm and 0mm of v21343]() {$2$};
      \node [above left = 0mm and 0mm of v121343]() {$1$};
      \node [above right = 0mm and 0mm of v4343]() {$4$};
      \node [above left = 0mm and 0mm of v34343]() {$3$};
      \node [above left = 0mm and 0mm of v134343]() {$1$};
      \node [above right = 0mm and 0mm of v434343]() {$4$};
      \draw[->] (v3) to (v);
      \draw[->] (v13)--(v3);
      \draw[->] (v213) to (v13);
      \draw[->] (v1213) to (v213);
      \draw[->] (v21213) to (v1213);
      \draw[->] (v121213) to (v21213);
      \draw[->] (vd121213) to (v121213);
      \draw[->] (v43)--(v3);
      \draw[->] (v343) to (v43);
      \draw[->] (v1343) to (v343);
      \draw[->] (v21343) to (v1343);
      \draw[->] (v121343) to (v21343);
      \draw[->] (vd121343) to (v121343);
      \draw[->] (v4343) to (v343);
      \draw[->] (v34343) to (v4343);
      \draw[->] (v134343) to (v34343);
      \draw[->] (vd134343) to (v134343);
      \draw[->] (v434343) to (v34343);
      \draw[->] (vd434343) to (v434343);
      \end{tikzpicture}
  } 
  \end{tabular}
  } &
  {\begin{tabular}{c|c}
      \hline
      $\check T_3 \simeq \check T_4$ &
      \scalebox{0.6}{
          \begin{tikzpicture}
          \tikzset{vertex/.style = {shape=circle,draw}}
          \tikzset{>={Latex[width=2mm,length=2mm]}}
          \node[] () at (0,6.3) {}; 
          \node[vertex] (v1) at (0,6) {};
          \node[vertex] (v21) at  (0,5) {};
          \node[vertex] (v121) at  (0,4) {};
          \node[vertex] (v2121) at  (0,3) {};
          \node (vd2121) at  (0,2) {$\vdots$};        
          \node [above left = 0mm and 0mm of v1]() {$3$};
          \node [above left = 0mm and 0mm of v21]() {$4$};
          \node [above left = 0mm and 0mm of v121]() {$3$};
          \node [above left = 0mm and 0mm of v2121]() {$4$};
          \draw[<-] (v21)--(v1);
          \draw[<-] (v121) to (v21);
          \draw[<-] (v2121) to (v121);
          \draw[<-] (vd2121) to (v2121);
          \end{tikzpicture}
      } 
      \raisebox{1cm}{$\qquad\simeq\qquad$}
      \scalebox{0.6}{
          \begin{tikzpicture}
          \tikzset{vertex/.style = {shape=circle,draw}}
          \tikzset{>={Latex[width=2mm,length=2mm]}}
          \node[] () at (0,6.3) {}; 
          \node[vertex] (v1) at (0,6) {};
          \node[vertex] (v21) at  (0,5) {};
          \node[vertex] (v121) at  (0,4) {};
          \node[vertex] (v2121) at  (0,3) {};
          \node (vd2121) at  (0,2) {$\vdots$};        
          \node [above left = 0mm and 0mm of v1]() {$4$};
          \node [above left = 0mm and 0mm of v21]() {$3$};
          \node [above left = 0mm and 0mm of v121]() {$4$};
          \node [above left = 0mm and 0mm of v2121]() {$3$};
          \draw[<-] (v21)--(v1);
          \draw[<-] (v121) to (v21);
          \draw[<-] (v2121) to (v121);
          \draw[<-] (vd2121) to (v2121);
          \end{tikzpicture}
      } 
      \\
      \hline
      $\check T_1$ &
      \scalebox{0.6}{
          \begin{tikzpicture}
          \tikzset{vertex/.style = {shape=circle,draw}}
          \tikzset{>={Latex[width=2mm,length=2mm]}}
          \node[] () at (0,6.3) {}; 
          \node[vertex] (v3) at (0,6) {};
          \node[vertex] (v13) at  (-2,5) {};
          \node[vertex] (v213) at  (-2,4) {};
          \node[vertex] (v1213) at  (-2,3) {};
          \node[vertex] (v21213) at  (-2,2) {};
          \node[vertex] (v121213) at  (-2,1) {};
          \node (vd121213) at  (-2,0) {$\vdots$};        
          \node[vertex] (v43) at  (2,5) {};
          \node[vertex] (v343) at  (2,4) {};
          \node[vertex] (v1343) at  (0,3) {};
          \node[vertex] (v21343) at  (0,2) {};
          \node[vertex] (v121343) at  (0,1) {};
          \node (vd121343) at  (0,0) {$\vdots$};        
          \node[vertex] (v4343) at (3,3) {};
          \node[vertex] (v34343) at (3,2) {};
          \node[vertex] (v134343) at (2,1) {};
          \node(vd134343) at (2,0) {$\vdots$};
          \node[vertex] (v434343) at (4,1) {};
          \node(vd434343) at (4,0) {$\vdots$};
          \node [above left = 0mm and 0mm of v3]() {$1$};
          \node [above left = 0mm and 0mm of v13]() {$3$};
          \node [above left = 0mm and 0mm of v213]() {$4$};
          \node [above left = 0mm and 0mm of v1213]() {$3$};
          \node [above left = 0mm and 0mm of v21213]() {$4$};
          \node [above left = 0mm and 0mm of v121213]() {$3$};
          \node [above left = 0mm and 0mm of v43]() {$2$};
          \node [above left = 0mm and 0mm of v343]() {$1$};
          \node [above left = 0mm and 0mm of v1343]() {$3$};
          \node [above left = 0mm and 0mm of v21343]() {$4$};
          \node [above left = 0mm and 0mm of v121343]() {$3$};
          \node [above right = 0mm and 0mm of v4343]() {$2$};
          \node [above left = 0mm and 0mm of v34343]() {$1$};
          \node [above left = 0mm and 0mm of v134343]() {$3$};
          \node [above right = 0mm and 0mm of v434343]() {$2$};
          \draw[<-] (v13)--(v3);
          \draw[<-] (v213) to (v13);
          \draw[<-] (v1213) to (v213);
          \draw[<-] (v21213) to (v1213);
          \draw[<-] (v121213) to (v21213);
          \draw[<-] (vd121213) to (v121213);
          \draw[<-] (v43)--(v3);
          \draw[<-] (v343) to (v43);
          \draw[<-] (v1343) to (v343);
          \draw[<-] (v21343) to (v1343);
          \draw[<-] (v121343) to (v21343);
          \draw[<-] (vd121343) to (v121343);
          \draw[<-] (v4343) to (v343);
          \draw[<-] (v34343) to (v4343);
          \draw[<-] (v134343) to (v34343);
          \draw[<-] (vd134343) to (v134343);
          \draw[<-] (v434343) to (v34343);
          \draw[<-] (vd434343) to (v434343);
          \end{tikzpicture}
      } 
      \\
      \hline
      $\check T_2$ &
      \scalebox{0.6}{
          \begin{tikzpicture}
          \tikzset{vertex/.style = {shape=circle,draw}}
          \tikzset{>={Latex[width=2mm,length=2mm]}}
          \node[] () at (0,7.3) {}; 
          \node[vertex] (v) at (0,7) {};
          \node[vertex] (v3) at (0,6) {};
          \node[vertex] (v13) at  (-2,5) {};
          \node[vertex] (v213) at  (-2,4) {};
          \node[vertex] (v1213) at  (-2,3) {};
          \node[vertex] (v21213) at  (-2,2) {};
          \node[vertex] (v121213) at  (-2,1) {};
          \node (vd121213) at  (-2,0) {$\vdots$};        
          \node[vertex] (v43) at  (2,5) {};
          \node[vertex] (v343) at  (2,4) {};
          \node[vertex] (v1343) at  (0,3) {};
          \node[vertex] (v21343) at  (0,2) {};
          \node[vertex] (v121343) at  (0,1) {};
          \node (vd121343) at  (0,0) {$\vdots$};        
          \node[vertex] (v4343) at (3,3) {};
          \node[vertex] (v34343) at (3,2) {};
          \node[vertex] (v134343) at (2,1) {};
          \node(vd134343) at (2,0) {$\vdots$};
          \node[vertex] (v434343) at (4,1) {};
          \node(vd434343) at (4,0) {$\vdots$};
          \node [above left = 0mm and 0mm of v]() {$2$};
          \node [above left = 0mm and 0mm of v3]() {$1$};
          \node [above left = 0mm and 0mm of v13]() {$3$};
          \node [above left = 0mm and 0mm of v213]() {$4$};
          \node [above left = 0mm and 0mm of v1213]() {$3$};
          \node [above left = 0mm and 0mm of v21213]() {$4$};
          \node [above left = 0mm and 0mm of v121213]() {$3$};
          \node [above left = 0mm and 0mm of v43]() {$2$};
          \node [above left = 0mm and 0mm of v343]() {$1$};
          \node [above left = 0mm and 0mm of v1343]() {$3$};
          \node [above left = 0mm and 0mm of v21343]() {$4$};
          \node [above left = 0mm and 0mm of v121343]() {$3$};
          \node [above right = 0mm and 0mm of v4343]() {$2$};
          \node [above left = 0mm and 0mm of v34343]() {$1$};
          \node [above left = 0mm and 0mm of v134343]() {$3$};
          \node [above right = 0mm and 0mm of v434343]() {$2$};
          \draw[<-] (v3) to (v);
          \draw[<-] (v13) to (v3);
          \draw[<-] (v213) to (v13);
          \draw[<-] (v1213) to (v213);
          \draw[<-] (v21213) to (v1213);
          \draw[<-] (v121213) to (v21213);
          \draw[<-] (vd121213) to (v121213);
          \draw[<-] (v43)--(v3);
          \draw[<-] (v343) to (v43);
          \draw[<-] (v1343) to (v343);
          \draw[<-] (v21343) to (v1343);
          \draw[<-] (v121343) to (v21343);
          \draw[<-] (vd121343) to (v121343);
          \draw[<-] (v4343) to (v343);
          \draw[<-] (v34343) to (v4343);
          \draw[<-] (v134343) to (v34343);
          \draw[<-] (vd134343) to (v134343);
          \draw[<-] (v434343) to (v34343);
          \draw[<-] (vd434343) to (v434343);
          \end{tikzpicture}
      } 
      \end{tabular}
      }
  \end{tabular}
  \caption{{\bf Input and output tree isomorphisms.} The input trees $T_i$ and the output trees $\check T_i$ of
    all the nodes $i$ of the network of
    Fig.~\ref{fig:example-fibopfib}, grouping those that are isomorphic.}
  \label{fig:fibopfib-in}

\end{figure}

In fact, the minimum base $\hat G$ over which $G$ is fibered in shown in Fig. \ref{fig:example-fibopfib-minfib}: nodes 1 and 2 belong to the same fiber, whereas 3 and 4 are in two different fibers.
\begin{figure}
  \centering
  \begin{tikzpicture}
      \tikzset{vertex/.style = {shape=circle,draw,minimum size=4mm}}
      \tikzset{>={Latex[width=2mm,length=2mm]}}
      \node[vertex] (bv12) at (-8,-5) {$12$};
      \node[vertex] (bv3) at (-6,-5) {$3$};
      \node[vertex] (bv4) at (-6,-7) {$4$};
      \draw[->] (bv12) to[loop left] (bv12); 
      \draw[->] (bv12) to (bv3);
      \draw[->] (bv3) to[bend left] (bv4);
      \draw[->] (bv4) to[bend left] (bv3);
  \end{tikzpicture}
  \caption{{\bf Minimum fibration base.} The minimum fibration base $\hat G$ on which the graph $G$ of Fig. \ref{fig:example-fibopfib} is fibered; nodes $1$ and $2$ of $G$ are both mapped to node $12$ of $\hat G$, whereas $3\mapsto 3$ and $4 \mapsto 4$.}
  \label{fig:example-fibopfib-minfib}
\commentAlt{Figure~\ref{fig:example-fibopfib-minfib}: 
A graph with nodes labeled 12, 3, 4. Directed edges: arrows
from 12 to itself, from 12 to 3, from 3 to 4, from 4 to 3.
}
\end{figure}

Now, if we look at \emph{outputs} instead the situation changes dramatically: this time it is 3 and 4 that are alike, because 3 only provides output to 4, and 4 only provides output to 3. Conversely, 1 is special in that it is the only node with two outgoing arcs, and 2 is special because it provides output to 1. 
The minimum op-fibration base $\check G$ is shown in Fig. \ref{fig:example-fibopfib-minfib2}: this time, 3 and 4  belong to the same op-fiber, whereas 1 and 2 live in separate op-fibers.
 
\begin{figure}
  \centering
  \begin{tikzpicture}
      \tikzset{vertex/.style = {shape=circle,draw,minimum size=4mm}}
      \tikzset{>={Latex[width=2mm,length=2mm]}}
      \node[vertex] (gv1) at (4,-5) {$1$};
      \node[vertex] (gv2) at (4,-7) {$2$};
      \node[vertex] (gv34) at (6,-5) {$34$};
      \draw[->] (gv34) to[loop right] (gv34); 
      \draw[->] (gv1) to (gv34);
      \draw[->] (gv1) to[bend left] (gv2);
      \draw[->] (gv2) to[bend left] (gv1);
  \end{tikzpicture}
  \caption{{\bf Minimum op-fibration base.} The minimum op-fibration base $\check G$ on which the graph $G$ of Fig. \ref{fig:example-fibopfib} is op-fibered; nodes $3$ and $4$ of $G$ are both mapped to node $34$ of $\hat G$, whereas $1\mapsto 1$ and $2 \mapsto 2$.}
  \label{fig:example-fibopfib-minfib2}
\commentAlt{Figure~\ref{fig:example-fibopfib-minfib2}: 
A graph with nodes labeled 1, 2, 34. Directed edges: arrows
from 1 to 34, from 1 to 2, from 2 to 1, from 34 to itself.
}
\end{figure}

\subsection{Preserving input and output at the same time: coverings}

The existence of two notions of local symmetry, one based on inputs and one based on outputs, raises a natural mathematical question: what if we want to preserve \emph{both} input and output? Is there a natural notion that puts together the definition of fibration and that of op-fibration at the same time? more importantly: is it the same thing as an automorphism?

This notion exists, and it is called \emph{(graph) covering}\index{graph !covering } which is a symmetry of the input and output trees and applies to the learning and inference process in neural networks (Chapter \ref{chap:ai}). 

A graph covering is a homomorphism $\varphi: G \to B$ such that, for every node $x$ of $G$:
\begin{itemize} 
    \item $\varphi$ induces a bijection between the edges of $G$ incoming in $x$ and the edges of $B$ incoming in $\varphi(x)$ (like a fibration does);
    \item $\varphi$ induces a bijection between the edges of $G$ outgoing from $x$ and the edges of $B$ outgoing from $\varphi(x)$ (like an op-fibration does).
\end{itemize}

If we look again at the example of Fig. \ref{fig:example-fibopfib}, while it has both non-trivial fibrations and non-trivial op-fibrations, it has no non-trivial coverings, simply because there do not exist two nodes that have at the same time isomorphic input trees and isomorphic output trees.
In other words, the graph of Fig. \ref{fig:example-fibopfib} has no non-trivial coverings, and it is easy to see that this graph is also rigid, i.e., it has no non-trivial automorphisms.

\begin{figure}
  \centering
  \begin{tikzpicture}
      \tikzset{vertex/.style = {shape=circle,draw,minimum size=4mm}}
      \tikzset{>={Latex[width=2mm,length=2mm]}}
      \node[vertex] (v1) at (-2,0) {$1$};
      \node[vertex] (v1p) at (-1,0) {$1'$};
      \node[vertex] (v2) at (-2,-2) {$2$};
      \node[vertex] (v2p) at (-1,-2) {$2'$};
      \node[vertex] (v3) at (3,0) {$3$};
      \node[vertex] (v3p) at (4,0) {$3'$};
      \node[vertex] (v4) at (3,-2) {$4$};
      \node[vertex] (v4p) at (4,-2) {$4'$};
      \draw[->] (v1p) to (v3);
      \draw[->] (v1) to[bend left] (v3p);
      \draw[->] (v1p) to[bend left] (v2p);
      \draw[->] (v1) to[bend right] (v2);
      \draw[->] (v2p) to (v1);
      \draw[->] (v2) to (v1p);
      \draw[->] (v3p) to[bend left] (v4p);
      \draw[->] (v3) to[bend right] (v4);
      \draw[->] (v4p) to (v3p);
      \draw[->] (v4) to (v3);
  \end{tikzpicture}
  \caption{{\bf Coverings.} This graph covers the graph $G$ of Fig. \ref{fig:example-fibopfib} mapping $1$ and $1'$ to $1$, $2$ and $2'$ to $2$ and so on.}
  \label{fig:example-cov}
\commentAlt{Figure~\ref{fig:example-cov}: 
A graph with nodes labeled 1-4 and 1'-4'. Directed edges: arrows
from 1 to 2 and 3', from 2 to 1', from 3 to 4, from 4 to 3, 
from 1' to 2' and 3, from 2' to 1, from 3' to 4', from 4' to 3'.
}
\end{figure}

Yet, it is possible to find graphs that have non-trivial coverings, but have no non-trivial automorphisms: Figure~\ref{fig:example-cov} shows a graph that covers the graph of Figure~\ref{fig:example-fibopfib}, but it is nevertheless rigid. This is because the notion of covering is \emph{local}, albeit taking both inputs and outputs into account simultaneously.

\subsection{Levels of symmetry}

\begin{figure}[ht]
\centering
\includegraphics[scale=0.14]{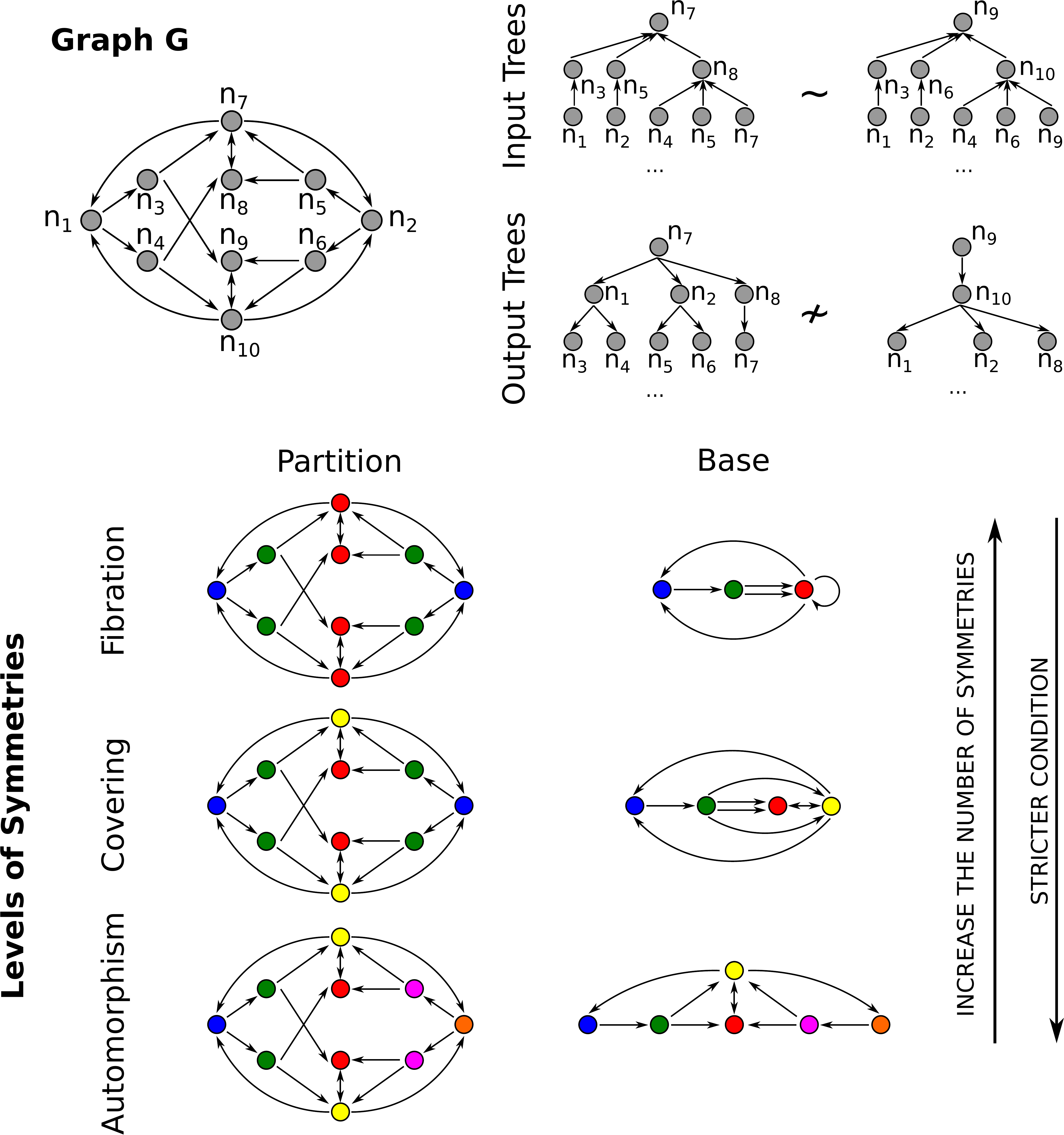}
\caption{\textbf{Level of symmetries of a graph}. For each level of
  symmetry (i.e., fibrations, coverings, automorphisms), there exists
  a partition of the nodes of a graph $G$ induced by the definition of
  the level. The original graph can be compressed into a smaller graph
  (its \textit{base} $B$). In the \textit{base} $B$, nodes
  represent equivalence classes (indicated with colors).}
\label{fig:main_fig}
\end{figure}

As we saw, a covering offers a more rigid kind of requirement than both fibrations and op-fibrations separately, but still far less rigid than that of an automorphism, because  the bijection is  required \emph{locally} and not at a global level. This means that there are interesting examples of networks with different colorings according to the level of symmetry. In Fig. \ref{fig:main_fig}, we show a graph and the corresponding symmetries at the level of fibrations, covering and automorphisms. Figure \ref{fig:main_fig} shows a network with three balanced colorings. One coloring arises from the strictest symmetry of automorphisms, preserving the whole structure (6 colors). This symmetry preserves the input and output of every node globally. An intermediate symmetry of the covering preserves the input and output trees of every node locally and produces 4 colors. It captures more symmetries than automorphisms, yet less than fibration.
The fibration produces the highest symmetry with 3 colors in the base as shown. This intriguing example shows the hierarchy of symmetries at its best.
The various types of graph symmetries are summarized in Table~\ref{tab:graphsym}.

\begin{table*}[t!]
    \centering
    \begin{tabular}{|l|c|c|c|}
    \hline
    & {\bf Local input } & {\bf Local output } & {\bf Global }\\
    & {\bf  isomorphism} & {\bf  isomorphism} & {\bf  isomorphism}\\
    \hline\hline
    fibration & $\surd$ & $\times$ & $\times$\\
    op-fibration & $\times$& $\surd$ &$\times$\\
    covering & $\surd$ & $\surd$ &$\times$\\
    automorphism & $\surd$ & $\surd$ & $\surd$ \\
    \hline
    \end{tabular}
    \vspace{10pt}
        \caption{Types of graph symmetries.\label{tab:graphsym}}
\end{table*}

The level of automorphisms is unique in preserving global properties,
whereas the other two levels preserve only certain local
properties. This distinction between global and local symmetries is
evident in other disciplines. In quantum field theory, a global
symmetry implies that fields undergo transformations that remain
constant across all space-time. An example
is temporal invariance, which implies the conservation of energy. On
the other hand, a local symmetry implies that fields can undergo local
transformations that vary in space and time, such as gauge
transformations \citep{jackson_classical_2012}. For example, all four
fundamental interactions (i.e. gravity, electromagnetism, weak
interaction, and strong interaction) are described by gauge local symmetries
that keep the Lagrangians invariant \citep{berghofer_gauge_2023}, although the kind of locality that occurs in gauge theories is not analogous to the local condition of a fibration; a fibration is a less strict local condition than the gauge symmetries of the fiber bundle. More on this in Chapter \ref{chap:bundles}.

\section{Further thoughts on global versus local symmetries}
\label{S:IOR}

Figure \ref{fig: example} shows that adding a single output edge to
a network with global group symmetry can destroy that symmetry
completely, even though the new edge does not affect the dynamics
of any node in the original network. An extra output of this type
preserves all existing fibration symmetries, although the new node
forms an extra fiber.

This example suggests, correctly, that fibration symmetries are more
robust than group symmetries to changes in the network topology, but
in some ways it is also misleading, because it tends to suggest that
the important difference is between inputs and outputs.  Fibrations
are defined in terms of input isomorphisms; outputs as such are irrelevant
in this definition. (Outputs do affect the dynamics, but they 
are represented in the theory as inputs to the nodes
that are influenced by their signals.) In
contrast, automorphisms and coverings also preserve output sets (defined in the
obvious way). Therefore the definition of an automorphism and coverings requires the
preservation of structure that is irrelevant to synchronization.

\begin{figure}[h!]
\centerline{%
\includegraphics[width=0.8\textwidth]{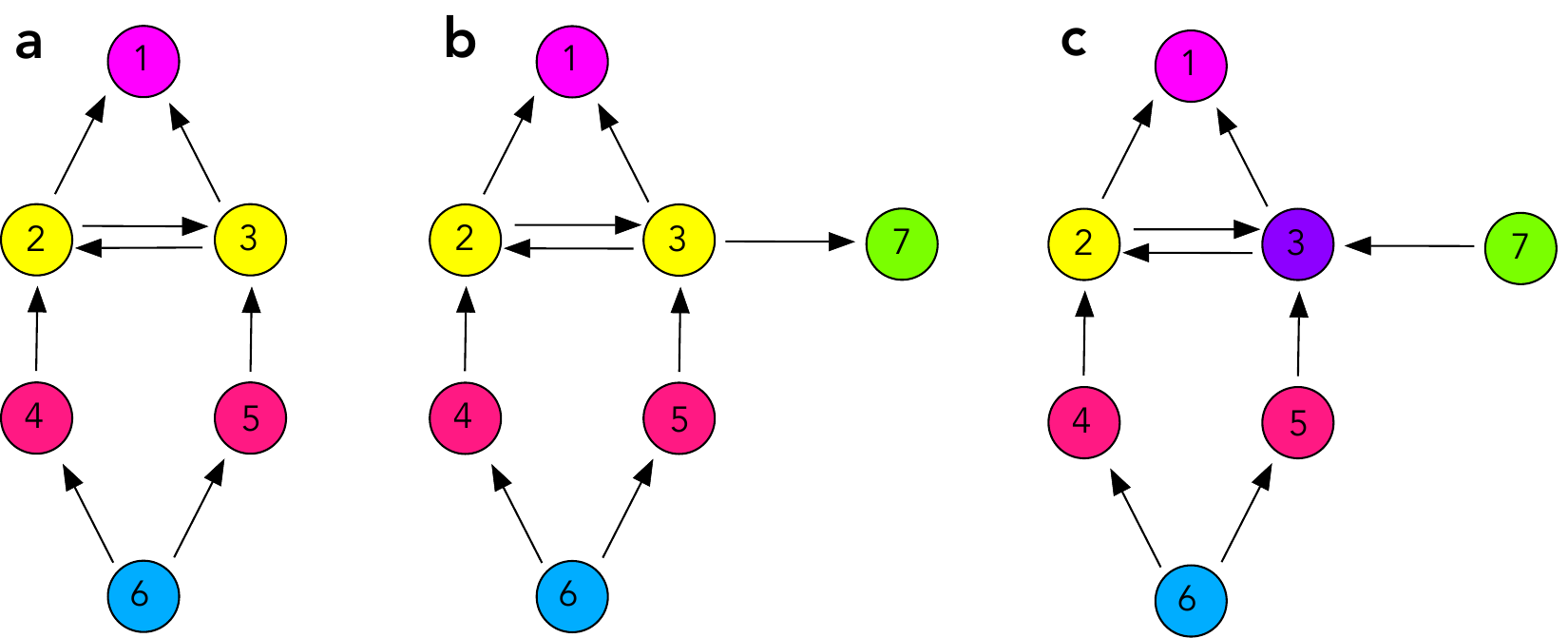}
}
\caption{ \textbf{Global versus local.} (\textbf{a}) The network of Fig. \ref{fig: example} with its
  minimal balanced coloring, defined by the automorphism group. (\textbf{b}) The effect of adding a new output edge and target node 7. There
  are now no automorphisms except the identity. The fibration
  symmetries remain and the coloring remains balanced (with a new
  color for node 7).  (\textbf{c}) The effect of adding a new input edge
  and source node 7. There are now no automorphisms except the
  identity. The fibration-induced synchrony of nodes 2 and 3 is
  destroyed because node 3 now has an extra input. However, the
  fibration-induced synchrony of nodes 4 and 5 persists.}
\label{F:6node_example}
\commentAlt{Figure~\ref{F:6node_example}: 
Three graphs called a, b, c (from left to right).
Graph a has nodes 1 (magenta), 2 and 3 (yellow), 4 and 5 (red),
6 (blue). Directed edges: arrows from 2 to 1 and 3, from 3 to 1 and 2,
from 4 to 2, from 5 to 3, from 6 to 4 and 5.
Graph b has nodes 1 (magenta), 2 and 3 (yellow), 4 and 5 (red),
6 (blue), 7 (green). Directed edges: arrows from 2 to 1 and 3,
from 3 to 1, 2 and 7, from 4 to 2, from 5 to 3, from 6 to 4 and 5.
Graph c has nodes 1 (magenta), 2 (yellow), 3 (purple), 4 and 5 (red),
6 (blue), 7 (green). Directed edges: arrows from 2 to 1 and 3,
from 3 to 1 and 2, from 4 to 2, from 5 to 3, from 6 to 4 and 5,
from 7 to 3.
}
\end{figure}

It is
tempting to trace the destruction of symmetry to this difference
between inputs and outputs.
However, the actual situation is subtler. For a start, if the
extra edge in Fig. \ref{fig: example} is reversed, this {\it also} destroys
the group automorphism, as in Fig. \ref{F:6node_example}.
In this case, the new edge also destroys {\it part of} the fibration
symmetry, by breaking up the fiber $\{2,3\}$. However, the other
nontrivial fiber $\{4,5\}$ remains untouched.

What this example tells us is that the most important distinction
between automorphisms and fibrations is that the former are global
properties of the network, while the latter are local. Any change
anywhere in the network, be it a new input or a new output, can
therefore affect the automorphism group, often in a fairly drastic
way. (It is easy to concoct examples where an extra edge actually {\it
  increases} the size of the automorphism group. Indeed, this happens
to Fig. \ref{fig:forwardgap}b when it is idealized to Fig.
\ref{fig:forwardgap}d. However, the most likely effect of a change is to make
the automorphism group smaller.)

For fibration symmetries the distinction between inputs and outputs
has more relevance. New outputs do not change the input set of the
node impacted by the addition, so such a fiber remains unchanged.
What does change is the fiber of the node to which the output goes,
since it is an input to that node.

If a new edge is added, it changes the input set of its target node,
splitting that node off from its previous fiber. All other
input sets remain unchanged, so the symmetry  becomes
smaller, but much of it usually survives. Ironically, it is possible
for the node that is split off in this manner to end up in
some other fiber for a suitable fibration.

Despite all of the above caveats, two basic statements remain largely
valid: 
\begin{itemize}
    \item 
Fibration symmetries are generally more robust to small changes in
network topology than automorphism (group) symmetries.
\item Changes to the output edges of a node have no effect on its
  input set. Changes to the input edges of course do. Which fibers
  survive such a change depends on where the outputs go.
\end{itemize}

\section{Example of fibrations, synchrony, colorings, and balance}
\label{coloring}

We saw in Chapter \ref{chap:fibration_2} that graph symmetry theory can
be presented in two distinct ways, which are mathematically equivalent \citep{lerman2015b,lerman2015}
but emphasize different aspects of the structure: balanced
colorings and graph
fibrations. The two are linked by the
concept of an admissible ODE---model equations that reflect the
network topology---and quotient networks, a construction that collapses
synchronous nodes together, into a single node in a collapsed network that is
analogous to the base of the fibration. Fibrations focus on the map
between the nodes of the network and those of the quotient network;
balanced colorings focus on the clustering of nodes induced by
identifying those that are synchronous.  The quotient network focuses
on the interactions between the clusters; admissible ODEs focus on the
consequent dynamics and synchronization.

Below, we exemplify these ideas using two common networks
from biology and two modified networks whose mathematical features
shed light on the general theory.

\begin{figure}[htb]
\centerline{
\includegraphics[width=0.7\textwidth]{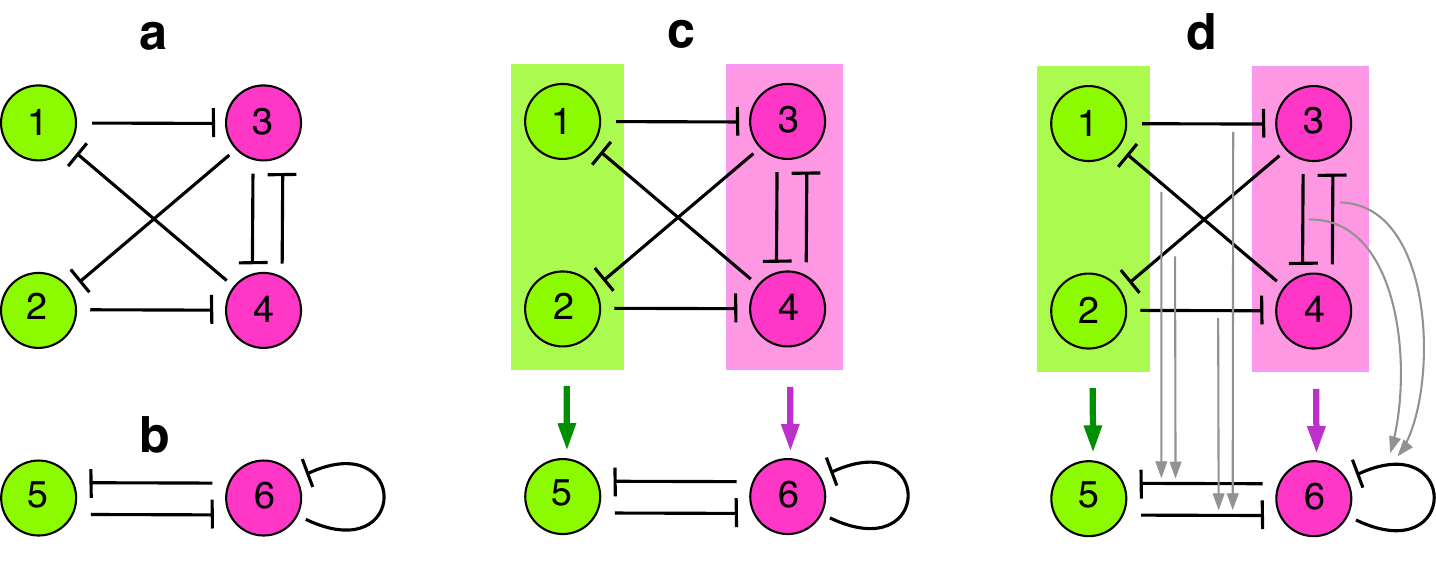}}
\caption{\textbf{A quotient network and the corresponding fibration.} (\textbf{a})
  Replica Fibonacci circuit with nodes classified into two disjoint
  subsets (colors).  (\textbf{b}) Base Fibonacci circuit. (\textbf{c}) Fibration
  map on nodes identifies all nodes of the same color; the base is the
  quotient network of the replica for the coloring shown.  (\textbf{d})
  Fibration map on arrows maps each arrow in the replica to the arrow
  in the base with the same colors for its head and tail nodes.}
\label{F:circuits_1}
\commentAlt{Figure~\ref{F:circuits_1}: 
Three almost identical graphs in the top row, called a, c, d (from left to right).
Three almost identical graphs in the bottom row, the first is called b and the other two are unnamed.
}

\commentLongAlt{Figure~\ref{F:circuits_1}: 
All the graphs in the first row have nodes named 1-4, with 1 and 2 on the left, 3 and 4 on the right. 
In all graphs 1 and 2 are green, 3 and 4 are magenta.
The arcs are all directed inhibition arcs, 
and are the following: arrows from 1 to 3, from 2 to 4, from 3 to 2 and 4, from 4 to 1 and 3.
All the graphs in the second row have nodes named 5 (left, green) and 6 (right, magenta).
The arcs are all directed inhibition arcs,
and are the following: arrow from 5 to 6, from 6 to 5 and 6.
The graphs named c and d have the two left green nodes grouped in a green box, and the two
right magenta nodes grouped in a magenta box. A green arrow connects the green box to the
green node of the graph below, and a magenta arrow connects the magenta box to the
magenta node of the graph below.
The graph named d has some further arrows connecting the inhibition arrow of the graph itself
with the inhibition arrow of the graph below, as follows: 1 to 3 and 2 to 4 are both connected to the arrow 5 to 6,
3 to 2 and 4 to 1 are both connected to the arrow 6 to 5, the remaining arrows are connected to the arrow 6 to 6.
}

\end{figure}

We begin with two networks from~\cite[Fig. 4]{leifer2020circuits},
there called replica\index{Fibonacci circuit !replica } and base
Fibonacci circuits,\index{Fibonacci circuit !base } shown respectively
in Fig.~\ref{F:circuits_1}a and b. All nodes have the
same type, and also all arrows have the same type
(repressor/inhibitory).  These two networks yield two distinct classes
of models.  Network Fig.~\ref{F:circuits_1}a has symmetry
fibration (which is  also a group $\Z_2$, which swaps nodes 1 and 2, and also swaps 3 and 4). Network Fig.~\ref{F:circuits_1}b has trivial group symmetry (i.e. the identity): in particular, node 5
has one input but node 6 has two.

Nodes 1 and 2, colored green, each receive one input arrow. Nodes 3
and 4, colored red, each receive two input arrows. Synchrony in
particular requires nodes to have the same number of input arrows (of
each type); that is, the same input tree to level 1, up to
isomorphism. Therefore nodes 1, 2 cannot synchronize with nodes 3,
4. However, it might be possible for nodes 1 and 2 to synchronize,
and/or for nodes 3 and 4 to synchronize. 

If nodes 1 and 2 synchronize then so must their inputs,
which implies that nodes 3 and 4 must also synchronize with each other
(though not necessarily with nodes 1 and 2). Then the inputs to
nodes 3 and 4 must synchronize (because in this case the only inputs
are from a single node, namely nodes 4 and 3 respectively).
Thus we can work backwards along the chain of inputs, which is
why synchronization can be described using input trees.

Since synchrony requires
nodes to have isomorphic input trees to all levels, we could use this
condition as a test for possible synchrony patterns. Alternatively, we
can take a short cut using the colors. In Fig.~\ref{F:circuits_1}a each green node receives one input from a red node, and no other
inputs. Similarly, each red node receives two inputs, both from a
green node, and no other inputs. Thus the coupling between nodes
preserves their colors. That is, it is a  balanced coloring as in Definition
\ref{balance-coloring}.  As we explained above, this condition ensures
the existence of synchronous states in the dynamics.

\section{Fibers and minimal balanced colorings}

The fiber terminology of Section \ref{fiber} arises because there is
an alternative way to describe the quotient network construction,
illustrated in Fig.~\ref{F:circuits_1}cd, which is the
graph fibration introduced in Section \ref{sec:fibmin}.

A graph fibration\index{fibration } is a map from network {\bf (a)} to network
{\bf (b)} that preserves certain aspects of the graph structure
(Definition \ref{def:fibration}). More precisely, it is a pair
of maps: one acting on nodes, the other on arrows.  In {\bf (c)} we show
the map acting on nodes (thick colored arrows). Nodes 1 and 2 both map
to node 5; nodes 3 and 4 map to node 6. In other words, the fibration
preserves the colors of the nodes. The map on arrows is depicted by
the gray arrows in {\bf (d)}. It preserves the colors of head and tail
nodes; for example the arrow from node 4 to node 1 has red head and
green tail, so it maps to the arrow from 6 to 5.

In order to be a fibration, such a (pair of) map(s) must also preserve
the input sets of nodes (level 1 of the input tree). For example,
nodes 1 and 2 both have one input arrow; so does node 5; moreover, the
sets of input arrows in {\bf (a)} and {\bf (b)} correspond under the
fibration. Similarly nodes 3, 4, and 6 have two input arrows, and
again they correspond under the fibration.

This final condition is the analog of balance from the point of view
of fibrations.  More precisely, given a coloring---balanced or not
---we can define a new network by identifying all nodes of a given
color. We can also map nodes of the original network to the
corresponding node of the new one, by mapping a node to its
color. However, this map may not extend consistently to a map of
arrows, so the new network does not have a well-defined topology. If a
consistent map of arrows exists, we have a homomorphism, but this need
not preserve input sets.  However, if the coloring is balanced, the
map of the nodes does extend consistently to a map of arrows {\em and}
how they are arranged in input sets.  Now we obtain a well-defined
quotient network, and the maps of nodes and arrows form a fibration.

Passing from the fibration viewpoint to the balance/quotient viewpoint
is straightforward. Given a fibration, the fiber of a node in the
image network is the set of all nodes that map to it. Then we assign a
separate color to each fiber. The fibration condition implies that
this coloring is balanced, and the image is the quotient network for
that coloring. Conversely, given a balanced coloring, there is a
natural map from the original network to the quotient.  This maps each
node to its color (thought of as a node in the quotient), and maps
arrows to preserve the input set and the colors, so it determines a
fibration.

Figure  \ref{F:circuits_1}c is in fact the {\em minimal fibration}
or {\em coarsest balanced coloring} as in Definition
\ref{def:minimal}. That is, the number of synchrony classes (clusters,
fibers, colors) is as small as possible. Every network has a unique
minimal fibration~\citep{stewart2007}.

\begin{figure}[htb]
\centerline{
\includegraphics[width=0.7\textwidth]{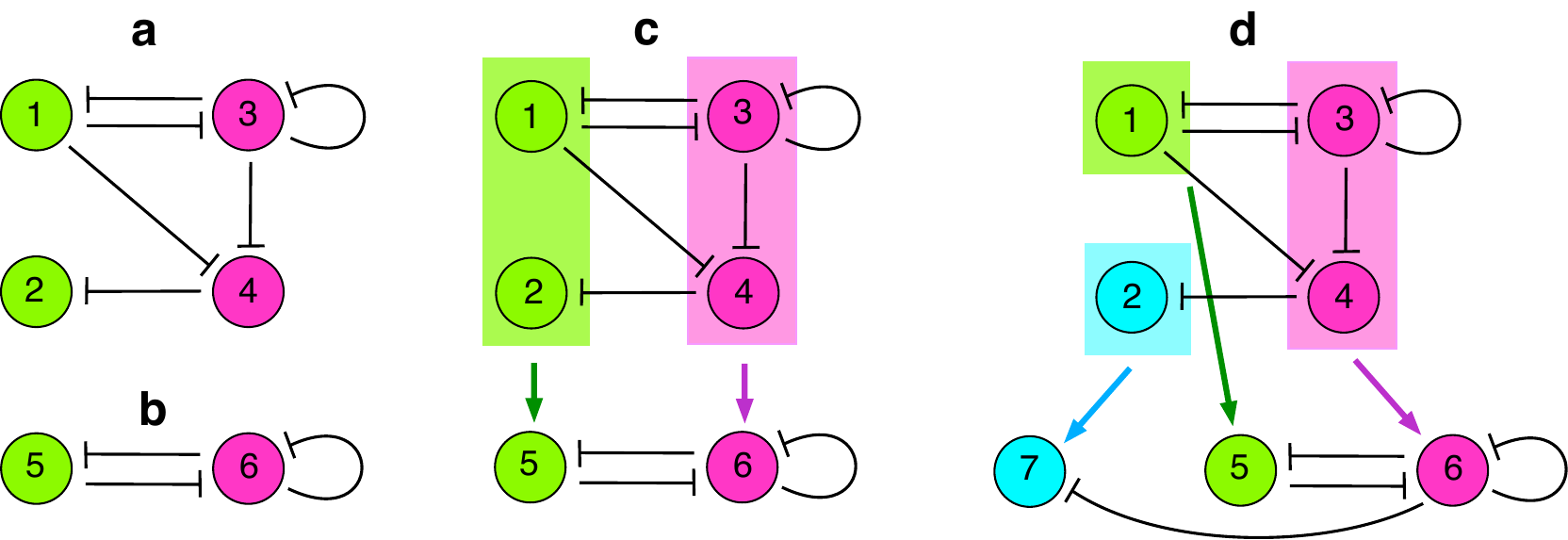}}
\caption{\textbf{Example of fibration. (a)} A rewiring of the replica Fibonacci circuit, with
  nodes classified into the same two colors.  (\textbf{b}) Fibonacci
  circuit, base, unchanged. (\textbf{c}) Fibration on nodes identifies all
  nodes of the same color. Arrows are mapped (map not shown) according
  to their head and tail colors as before; the quotient network of
  network (\textbf{a}) for this coloring is network (\textbf{b}).
  (\textbf{a}) A different fibration of (\textbf{a}) with three colors, leading to a quotient network with three nodes.}
\label{F:circuits_2}
\end{figure}

We now modify Fig. \ref{F:circuits_1}a.  We do not claim that
the resulting network arises naturally in biology: it is introduced to
illustrate some important mathematical features.
Figure ~\ref{F:circuits_2}a rewires the arrows of
Fig.~\ref{F:circuits_1}a, but retains the numbers of input
arrows and the colors of their heads and tails.  Accordingly, the
quotient network {\bf (b)} is the same as before. There is a
corresponding fibration, shown in {\bf (c)} for nodes only. Again this
is the minimal fibration (we cannot make node 1 red as well, while
retaining balance, because it has only one input arrow).
 
However, now there is another balanced coloring, hence a corresponding
fibration, shown in {\bf (d)}.  This fibration is not minimal. It has
three colors, with a new color blue for node 2. Unlike the Fibonacci
network, this coloring is balanced, because both red nodes receive
inputs from node 1. The quotient network, with three nodes, is also
shown in {\bf (d)}.  Thus, although the minimal fibration is unique,
there can be other fibrations, hence other balanced colorings, that
use more colors.

\section{ Beyond the minimal balanced coloring}
\label{S:LBC}

A network may have many different colorings, in addition to its
minimal coloring. The set of all possible balanced colorings has its
own structure as a partially ordered set.\index{partially ordered set }  Indeed, in \cite[Chapter
  13]{GS2023} and \cite{A10,stewart2007} it is proved that the set of
all colorings of a network forms a {\em lattice},\index{lattice } in the sense of a
partially ordered set that satisfies certain axioms; see \citep{DP90}.
The set of all balanced colorings is a sublattice of this.

The partial order is refinement:

\begin{definition}
\label{D:fine_coarse}
A coloring $\kappa_1$ is {\em finer}\index{finer } than, or {\em refines},\index{refine } a coloring $\kappa_2$  if 
$\kappa_1(c)=\kappa_1(d)$ implies $\kappa_2(c)=\kappa_2(d)$, for all pairs of nodes $c,d$ from the set $\mathcal{C}$ of nodes.
We denote this relation by
$
\kappa_1 \preceq \kappa_2
$.
The coloring $\kappa_1$ is {\em coarser}\index{coarser } than $\kappa_2$ if 
$\kappa_2$ is finer than $\kappa_1$. We denote this by
$
\kappa_1 \succeq \kappa_2
$.
\end{definition}

A finer coloring has more colors, and these are obtained by subdividing
color classes of the coarser coloring. Conversely, a coarser coloring arises by amalgamating
color classes of a finer one. 
The finest coloring occurs when all
nodes have different colors; trivially this coloring is balanced.
The coarsest coloring is the 1-coloring in which all
nodes have the same color.  It is balanced if and only if
the network is homogeneous;\index{network !homogeneous } that is, all nodes are input isomorphic.

Several algorithms exist to determine some or all balanced colorings of
a given network: see Chapter \ref{chap:algorithms}.

\begin{figure}[htb]
  \centerline{%
   \includegraphics[width=.2\textwidth]{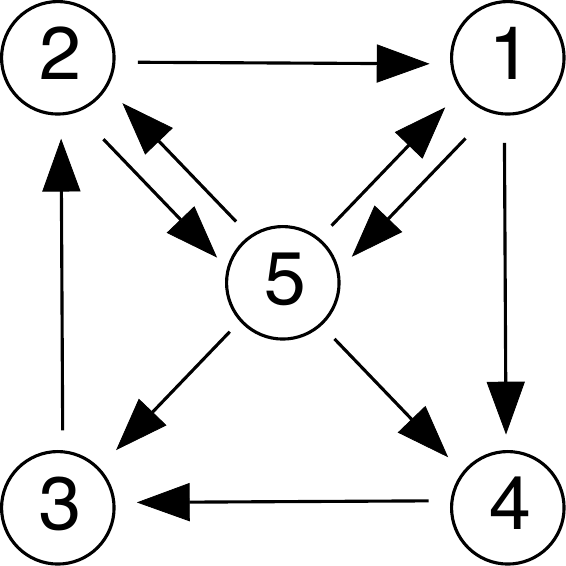}\qquad \qquad
   \includegraphics[width=.6\textwidth]{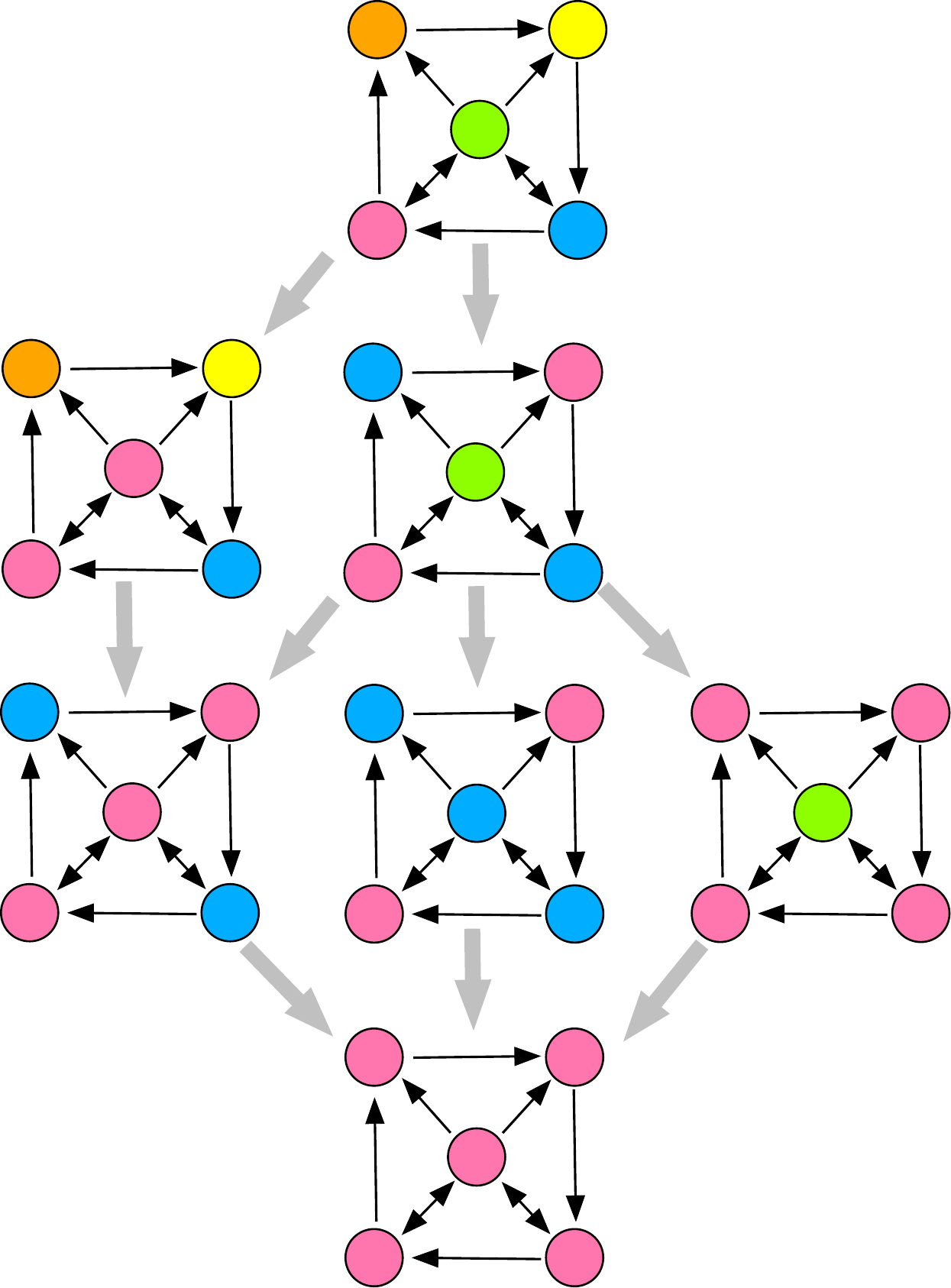}
  }
   \caption{\textbf{Lattice of balanced colorings.} {\em Left}: A five-node network with seven balanced colorings and a minimal fibration which fibers to a single node with two loops. {\em Right}: Lattice of balanced colorings for this 5-node network.
Grey arrows represent coarsening of the coloring (head is coarser than tail). Compositions of grey arrows not shown. Each grey arrow also corresponds to a fibration in which the coloring becomes coarser. Figure  \ref{F:5lat_fibration} shows the corresponding bases.}
  \label{F:fivecell}
\commentAlt{Figure~\ref{F:fivecell}: 
On the left, there is a graph with nodes named 1-5. Directed edges: arrows
from 1 to 4 and 5, from 2 to 1 and 5, from 3 to 2, from 4 to 3, from 5 to 1, 2, 3 and 4.
On the right, the same graph is redrawn seven times, with different colorings of its nodes.
The seven versions are arranged in a diagram with 4 levels.
}

\commentLongAlt{Figure~\ref{F:fivecell}: 
Level 1 contains only one version.
Level 2 contains two versions, and there are arcs from the version on level 1 to both versions of level 2.
Level 3 contains three versions, and there are arcs from both versions of level 2 to the first version of level 3, 
and arcs from the second version of level 2 to all versions of level 3.
Level 4 contains only one version, and there are arcs from all the versions of level 3 to the only version of level 4.
Here follows a description of the versions. Each version is specified by indicating the color of each node (recall
that the graph is the one on the left described above).
Level 1, only version: 1 yellow, 2 orange, 3 magenta, 4 blue, 5 green.
Level 2, first version: 1 yellow, 2 orange, 3 magenta, 4 blue, 5 magenta.
Level 2, second version: 1 magenta, 2 blue, 3 magenta, 4 blue, 5 green.
Level 3, first version: 1 magenta, 2 blue, 3 magenta, 4 blue, 5 magenta.
Level 3, second version: 1 magenta, 2 blue, 3 magenta, 4 blue, 5 blue.
Level 3, third version: 1 magenta, 2 magenta, 3 magenta, 4 magenta, 5 green.
Level 4, only version: 1 magenta, 2 magenta, 3 magenta, 4 magenta, 5 magenta.
}
\end{figure}

\begin{example}\em
\label{ex:5node_lattice}

Consider the $5$-node network of Fig.~\ref{F:fivecell} (left). 
Assume all nodes have the same node-type
with state space $\R^k$, and all arrows have the same arrow-type.
\end{example}

The algorithm of \cite{kamei2013} is able to calculate all balanced colorings of a graph and confirms that
the network of Fig. \ref{F:fivecell} 
has exactly $7$ balanced colorings.
The colorings and the lattice that they form under refinement are shown in Fig.~\ref{F:fivecell} (right).

\begin{figure}[htb]
\centerline{
\includegraphics[width=.8\textwidth]{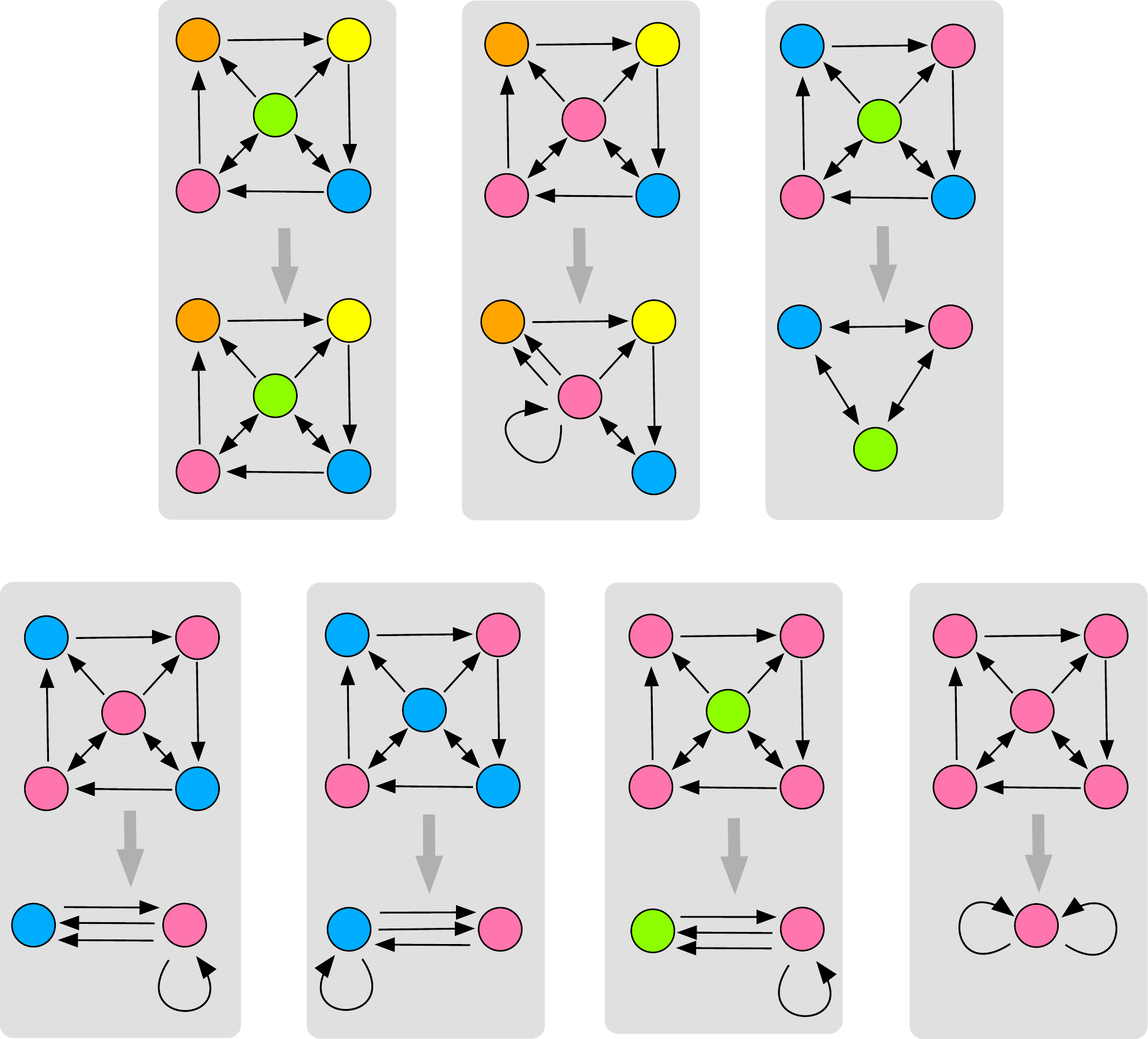}}
\caption{\textbf{Bases for the fibrations corresponding to
the balanced colorings of the 5-node network in Fig.~\ref{F:fivecell} (right).}
Each grey arrow represents the fibration that maps nodes of
a given color in the network to the node of the same color in the base. The associated fiber is the set of nodes of that color; that is, the corresponding cluster. Multiple edges and self-loops are common.
}
\label{F:5lat_fibration}
\end{figure}

The lattice of balanced colorings can also be considered in  terms
of fibrations. The relation of coarsening, which is the reverse of refinement, corresponds to a fibration:
the coarser coloring fibers over the finer one. 
The bases of the fibrations are  shown in Fig. \ref{F:5lat_fibration}.

Some of the bases in this figure also fiber over other
bases with fewer colors. For example, they all fiber over
the minimal base at the lower right.

\chapter[Symmetry Groupoid Formalism]{\bf\textsf{Symmetry Groupoid Formalism}}
\label{chap:groupoid}

\begin{chapterquote}
  The graph fibration formalism for synchronization and the associated
  balanced coloring and fibers can be cast into the groupoid formalism
  developed in
  \citep{stewart2003,stewart2005,stewart2006,stewart2016,GS2023}.
  This chapter provides a brief, informal introduction to the main
  concepts of the formal theory of groupoids in networks described in
  the cited sources. We relate
  the groupoid framework to the fibration  and
  balanced coloring frameworks. The
  discussion is restricted to basic topics that are relevant to cluster synchrony. This chapter need not be read in detail.
    \end{chapterquote}

\section{Formulation in terms of symmetry groupoids}
\label{S:FTSG}
In this chapter we reinterpret parts of Chapter \ref{chap:fibration_2}
from a different but closely related point of view,
emphasizing algebraic structure alongside graph topology.
We reformulate the problem of finding balanced colorings in a
theoretical manner using an algebraic tool called the \emph{symmetry groupoid}.\index{symmetry !groupoid }
We describe this concept in
detail in the next section, after recalling some background. 

Every balanced coloring induces a local isomorphism between nodes with
the same color: this local isomorphism is, concretely, a bijection
between the input sets of the two nodes, i.e., between the sets of
their incoming edges.  The symmetry groupoid is a formal way to
represent a collection of input isomorphisms: every element of the
groupoid contains the bijections between input sets of all pairs of
nodes with the same indegree. Bijections are functions, so when they
can be composed and their composition is associative. Moreover, any
bijection has an inverse, and here this is also an input isomorphism.
However, input isomorphisms cannot always be composed, because each
bijection is local: its domain is the input set of a node, and its
codomain is the input set of another node. Thus the natural algebraic
structure of the set of all input isomorphisms---the `local
symmetries' of the network---is that of a groupoid.  Informally, we
can think of this as a group whose composition operation is defined
only partially.

In marked contrast to the automorphism group (which, as we said,
contains only one element), the symmetry groupoid of the Frucht graph\index{graph !Frucht !symmetry groupoid }
contains 864 elements, see Example \ref{ex:minimal_4}.  It has one element for every possible
isomorphism between incoming edges of nodes with the same indegree,
and in the Frucht graph all nodes have the same indegree, which is 3.
The symmetry groupoid consists of all bijections between each pair of
input sets.  There are 12 nodes, so there are $12 \times 12=144$ pairs
of nodes; for each pair of nodes $(c,d)$, the input sets of $c$ and
$d$ contain three edges each, so we can choose $3!=6$ possible
bijections, therefore we have $144 \times 6 = 864$ input isomorphisms.
Every balanced coloring induces a set of input isomorphisms, but not
all sets of input isomorphisms correspond to balanced colorings.

\begin{figure}[h!]
\centerline{%
\includegraphics[width=0.75\textwidth]{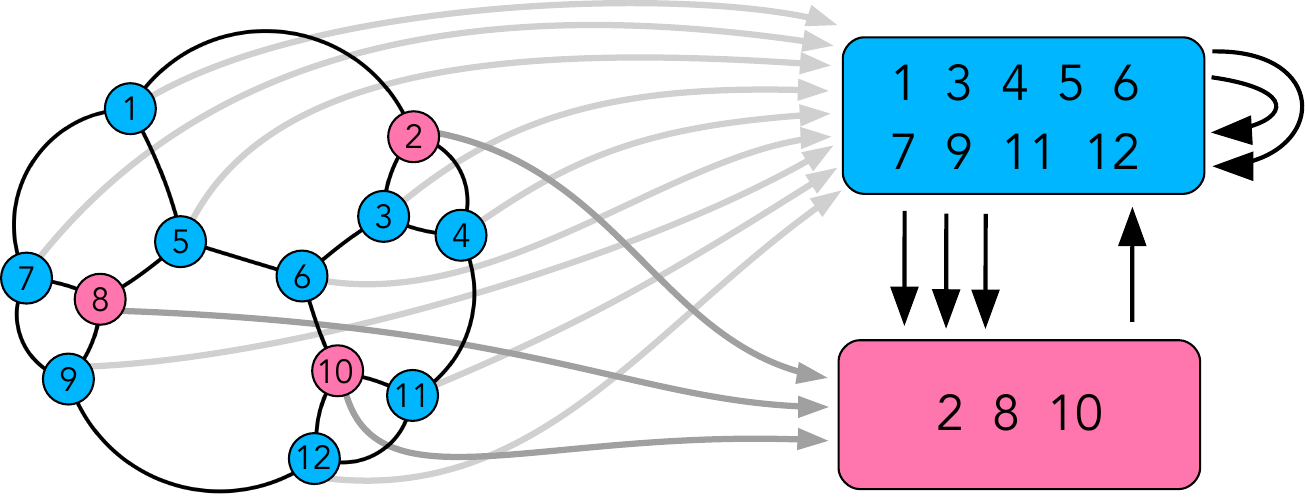}
}
\caption{\textbf{Fibration (gray arrows) corresponding to balanced coloring
  (\textbf{b}) of the Frucht graph.} Edges of main graph are bidirectional;
  those of quotient are directed, shown by black arrows.}
\label{F:frucht_coloring_quot}
\commentAlt{Figure~\ref{F:frucht_coloring_quot}: 
On the left, the Frucht graph. It is an undirected graph with nodes named 1-12 (all blue, except 2, 8 and 10 
whose color is magenta), and the following edges: 1-2, 1-5, 1-7, 2-3, 2-4, 3-4, 3-6, 4-11, 5-6, 5-8,
6-10, 7-8, 7-9, 8-9, 9-12, 10-11, 10-12, 11-12.
On the right, there is a graph with two nodes (represented as large boxes). The node above is blue and
contains the numbers 1, 3, 4, 5, 6, 7, 9, 11, 12; the node below is magenta and contains the numbers
2, 8, 10. There are three arrows from blue to magenta, one arrow from magenta to blue and two arrows from
blue to itself.
The nodes of the Frucht graph are connected with gray arrows to the nodes of the right graph, following the rule
that all blue nodes are connected to the blue node, and all magenta nodes are connected to the magenta node.
}
\end{figure}

As already remarked, the algorithm of \cite{kamei2013} shows that the
Frucht graph has exactly five balanced colorings,\index{coloring !balanced, of Frucht graph } shown in Fig.
\ref{F:frucht_coloring_bal}. In coloring (a), all nodes have the same
color; this coloring is balanced because all nodes have the same
in-degree: it represents complete synchrony. Coloring (e) is trivial:
all nodes have different colors; here no pairs of nodes
synchronize. Coloring (b) and (c) are more interesting: they both have
two colors, and represents a state in which nodes group into two
synchronized clusters: in one case 
(Fig. \ref{F:frucht_coloring_bal}b), $\{1,3,4,5,6,7,9,11,12\}$ and $\{2,8,10\}$; in the other case
(Fig. \ref{F:frucht_coloring_bal}c) $\{1,2,3,6,7,9,10,12\}$ and
$\{4,5,7,11\}$. In both cases, the two clusters are the fibers of the
fibration obtained by mapping each node to its color; Fig.
\ref{F:frucht_coloring_quot} shows the fibration for
Fig. \ref{F:frucht_coloring_bal}b. Edges are mapped according to
their source and target colors.

Finally, coloring (d) uses three colors and is perhaps the most
surprising one: in this final case, there are three clusters of the
same size (four nodes in each cluster). Observe that (d) is a
refinement of (c): nodes that are in the same cluster with respect to
(d) are also in the same cluster with respect to (c); more precisely,
the blue cluster of (c) (with 8 nodes) has split into two equally
sized clusters.

As observed in~\citep{stewart2007,kamei2013}, the balanced colorings
form a lattice according to refinement, 
see Section \ref{S:LBC}.

\begin{figure}[h!]
  \begin{tikzpicture}[node distance=3cm]
    \node (A) at (0,3)  {(a)};
    \node (B) at (-1,2) {(b)};
    \node (C) at (1,2)  {(c)};
    \node (D) at (1,1)  {(d)};
    \node (E) at (0,0)  {(e)};
    \draw[->] (A) -- (B);
    \draw[->] (A) -- (C);
    \draw[->] (C) -- (D);
    \draw[->] (B) -- (E);
    \draw[->] (D) -- (E);
  \end{tikzpicture}
\caption{\textbf{The lattice of balanced colorings of the Frucht graph}.  Every arrow
  represents a refinement (head is finer than tail). The
  letters refer to Fig.~\ref{F:frucht_coloring_bal}.}
\label{F:frucht_lattice}
\commentAlt{Figure~\ref{F:frucht_lattice}: 
A diagram with five elements named (a)-(e).
(a) is on top, with two arrows below, to (b) and (c).
(b) and (c) are on the same level. (b) has an arrow to (e), and (c) has an arrow to (d).
(d) is alone on a level, with an arrow to (e).
(e) is alone on the bottommost level.
}
\end{figure}

\section{Groupoids}

As we mentioned, although the input isomorphisms of a network seldom
form a group, they do have a lot of algebraic structure. Indeed, their
natural structure is that of a `groupoid'.\index{groupoid } Groupoids are of more
recent vintage than groups, and were introduced by \citep{brandt1927}. They resemble
groups, except that sometimes two elements cannot be composed.  For completeness, we
provide the formal definition of a groupoid in
Section \ref{sec:FDG}; however, for present purposes, examples suffice to convey the
key ideas.  This section also shows that the groupoid formalism is of
some use in the study of network synchronization, but its role is
mainly to provide an abstract algebraic setting analogous to the way
groups codify automorphisms.  For the moment, we content
ourselves with one example: Fig. \ref{F:Uxur}a.  This network
has precisely five input isomorphisms. Expressed as bijections of
input sets (which consist of all {\it edges} entering in a
node). These bijections are:
\begin{eqnarray*}
&& \varepsilon_1 = \left(\begin{array}{c}e_2 \\ e_2 \end{array}\right) \qquad
\varepsilon_2 = \left(\begin{array}{cc}e_1 & e_4 \\ e_1 & e_4         \end{array}\right) \qquad
\varepsilon_3 = \left(\begin{array}{cc}e_3 & e_5 \\ e_3 & e_5  \end{array}\right) \\
&& \tau = \left(\begin{array}{cc}e_1 & e_4 \\ e_3 & e_5         \end{array}\right) \qquad
\tau^{-1} = \left(\begin{array}{cc}e_3 & e_5 \\ e_1 & e_4         \end{array}\right)
\end{eqnarray*}
Here the notation for the bijections is a variant of the usual
notation for permutations. The top row shows elements of the domain;
the bottom row shows their images in the codomain.

\begin{figure}[h!]
\centerline{%
\includegraphics[width=0.5\textwidth]{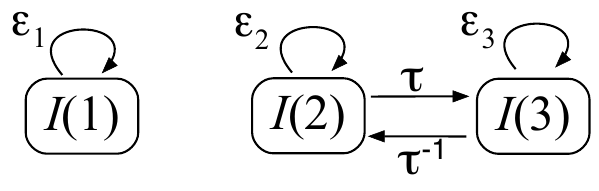}
}
\caption{\textbf{Symmetry groupoid.} Domains and codomains of the five bijections that form the symmetry groupoid of the network in Fig. \ref{F:Uxur}a.}
\label{F:Ecoli_groupoid}
\commentAlt{Figure~\ref{F:Ecoli_groupoid}: 
A graph with three nodes named I(1), I(2), I(3).
Arcs are arrows: from I(1) to itself (labeled epsilon1),
from I(2) to itself (labeled epsilon2),
from I(3) to itself (labeled epsilon3),
from I(2) to I(3) (labeled tau),
from I(3) to I(2) (labeled tau^-1).
}
\end{figure}

Input set $I(1)$ is disconnected from the other two in this figure.
This happens because it is not input isomorphic to $I(2)$ or $I(3)$
(node $1$ has only one incoming edge, whereas both $2$ and $3$ have
two incoming edges).  The maps $\varepsilon_1, \varepsilon_2,
\varepsilon_3$ are distinct `identity maps', because when they can
legally compose with any other map, the result is that map. For
example, $\varepsilon_3\tau = \tau$. We also have $\tau\tau^{-1} =
\varepsilon_3$ and $\tau^{-1}\tau = \varepsilon_2$. On the other hand,
the composition $\varepsilon_2\varepsilon_3$ is not defined because
the codomain of $\varepsilon_3$ is $I(3)$, and this is not the domain
$I(2)$ of $\varepsilon_2$.  We consider only isomorphisms that respect
edge type (e.g., we cannot map a repressor to an activator, or the other way
round).

We can calculate the `multiplication table' of this groupoid, Table
\ref{T:mult_groupoid}. Here the entry in row $\alpha$ and column
$\beta$ is the composition $\alpha\beta$, which is obtained by
applying first $\beta$, then $\alpha$.  Diagrams like Fig.
\ref{F:Ecoli_groupoid} or tables like Table~\ref{T:mult_groupoid} are
useful when contemplating groupoid structure. But even then, groupoids
are something of an acquired taste.

\begin{table*}[t!]
\centering
\caption{Multiplication table of the groupoid.}
\vspace{10pt}
\begin{tabular}{|l||l|l|l|l|l|}
\hline
  & $\varepsilon_1$ & $\varepsilon_2$ & $\varepsilon_3$ & $\tau$ & $\tau^{-1}$\\
\hline
\hline
$\varepsilon_1$ & $\varepsilon_1$ & $-$ & $-$ & $-$ & $-$\\
$\varepsilon_2$ & $-$ & $\varepsilon_2$ & $-$ & $-$ & $\tau^{-1}$ \\
$\varepsilon_3$ &  $-$ & $-$ &$\varepsilon_3$ & $\tau$ &   $-$\\
$\tau$ & $-$ & $\tau$  & $-$ & $-$ & $\varepsilon_3$ \\
$\tau^{-1}$ &  $-$ & $-$ &$\tau^{-1}$ & $\varepsilon_2$ & $-$ \\
\hline
\end{tabular}
\label{T:mult_groupoid}
\end{table*}

\section{Input isomorphisms between input sets}

As shown in Section \ref{sec:orbit}, a group symmetry gives rise to an
orbital partition (Definition \ref{orbits}) which we can show is
also a balanced coloring, the orbit coloring.
An orbit coloring is always balanced.
However, a balanced coloring may not necessarily represent an orbit
(Section \ref{sec:orbit}).
More generally, every balanced coloring is determined by some fibration.
Now the colors are given by the fibers of the fibration,
and balance is ensured by the definition of a fibration.

Any orbit coloring, being balanced, corresponds to a fibration in
which nodes and arrows map to their orbits. However, not all fibrations 
 arise in this manner, leading to the generalization from groups to
groupoids to be discussed in this chapter. The role of
groupoids is theoretical, and in practice somewhat peripheral; in
contrast, the
concept of balance/fibration is essential.

We have seen so far that cluster synchronization is captured by the
existence of a symmetry fibration emerging from isomorphic input
trees. The input trees need not necessarily be colored to be
isomorphic. The theory of groupoids is based on the input sets rather
than on input trees; the symmetry groupoid\index{symmetry !groupoid } $\mathcal{B}$ of the
network is the set of all {\it input isomorphisms} between its input
sets, with natural composition operations. For these isomorphisms to
be relevant to cluster synchronization, the input sets are given a
balanced coloring, and the isomorphisms are required to respect this
coloring.  These `color-preserving input isomorphisms' form the
`symmetry groupoid' of the coloring, and this is a subgroupoid of
$\mathcal{B}$. In this chapter we formalize this notion and compare it
with symmetry fibrations and symmetry groups.

Since input isomorphisms are bijections between input sets, to lay out
the formal definition of symmetry groupoid we first recall the
definition of the input set of a node (Definition \ref{inputset}) and
then proceed introducing the notion of an `input isomorphism'.

According to Definition \ref{inputset}, the input set of a node
consists of the node itself, together with its incoming {\it edges}. In multigraphs (i.e., graphs that allow for parallel edges) the node
may possess different edges coming from the same source. Such `parallel'
edges are uncommon in biological networks, though they sometimes
occur. To keep the discussion (and notation) simple, we assume that
the network under consideration has no multiple edges: in that case,
we can identify an edge with the pair $(j,i)$ of nodes, where $i$ is
the target node and $j$ the source node.  (Some authors use the
reverse convention, replacing $(j,i)$ by $(i,j)$.)

Furthermore, when discussing the input set of node $i$, we can
identify the input edge with its source node $j$. We make this simplification
in this section. It assumes that there are no multiple edges between
the same pair of nodes, and no self-loops. 
Without this simplifying assumption, 
we must also label the edges with
their own symbols, complicating the notation without improving
understanding.

\begin{remark}
The general theory of \citep{GS2023} permits these features; indeed,
they are mathematically necessary because even when a network has
no such features, they can arise in a quotient network by a balanced coloring; equivalently, in the base of a fibration. The lifting property---restricted ODEs on the synchrony
space are {\em precisely} the admissible ODEs for the quotient network---works only when networks with multiple edges and self-loops 
are permitted. Self-loops are common in biology (as `autoregulation' 
in genetic networks, for instance). Multiple edges
are less common, in part because they can often be considered
as weighted edges with larger weights.
\end{remark}

Given a node $i$, we denote its input set by:
\[
{\rm In}(i) = \{i:
{\rm input}_1,\dots, {\rm input}_{k_{\rm in}}\}.
\]
(Alternative
notations are $\partial^{\rm in}_i$ and $I(i)$.)
This notation emphasizes that there is a `base' node $i$ with in-degree
$k_{\rm in}$ that receives input edges from nodes ${\rm input}_1,\ldots,{\rm input}_{k_{\rm in}}$. 
For example, the input sets
of nodes in the network of Fig.~\ref{fig:groupoid_formal}a are:
\begin{equation}
\begin{aligned}
{\rm In}(1)\ &= \{1 : 2, 3\}\  & \ \ \ \ {\rm In}(2)\ &= \{2 : 3,
4\}\  \\ {\rm In}(3)\ &= \{3 : 3, 5\}\  & \ \ \ \ {\rm In}(4)\ &= \{4
: 6\}\ \\ {\rm In}(5)\ &= \{5 : 6\}\  & \ \ \ \ {\rm In}(6)\ &= \{6 :
\empty\}\  \\ {\rm In}(7)\ &= \{7 : 3\}\ ,  & \\
\end{aligned}
\label{eq:input-sets}
\end{equation}
and are depicted in Fig.~\ref{fig:groupoid_formal}b.

\begin{definition} {\bf Input isomorphism.}
\label{def:input_iso}
  An {\em input isomorphism}\index{input !isomorphism } between nodes $i$ and $j$ is a bijective map
  \[
  \beta: {\rm In}(i) \to {\rm In}(j)
  \]
If such a map exists, the input sets are {\em input isomorphic}:
\index{input !isomorphic }
 \begin{equation}
   {\rm In}(i) \simeq {\rm In}(j) \, .
\end{equation}
\end{definition}

For example, an input isomorphism between ${\rm In}(1)$ and ${\rm
  In}(2)$ in Fig.~\ref{fig:groupoid_formal}b is given by the map
$\tau_{1\to 2}:{\rm In}(1)\to{\rm In}(2)$, defined as follows:
\begin{equation}
\tau_{1\to 2}\ =\ \left(
\begin{matrix}
    1: & 2 & 3  \\
    \downarrow & \downarrow & \downarrow\\
    2: & 3 & 4  
\end{matrix}
\right)\ ,
\label{eq:tau}
\end{equation}
which maps base node 1 to base node 2, and their respective inputs: 2
to 3, and 3 to 4 (see Fig.~\ref{fig:groupoid_formal}a). The only
condition for this map to exist is that it should be bijective,
that is, possess a well-defined inverse. In practical terms, the only
condition for an input isomorphism in a given network is that it
maps two input sets that have the same in-degree $k_{\rm
  in}$.

  \begin{remark}
 In the more general setting of \cite[Sections 7.6, 9.2]{GS2023}, where edges can have different
  types, an input isomorphism must also preserve the types of edges and of the base nodes.
  That is, if $e \in {\rm In}(i)$ then $\beta(e) \in {\rm In}(j)$
  must have the same type as $e$. Now the interpretation is that
  nodes $i$ and $j$ have the same number of input edges for each type. Moreover, multiple arrows (having the same target and the same source, but not necessarily the same type) are not only permitted, but necessary, in order for possible synchronous dynamics to be determined precisely by the admissible ODEs for the quotient network. 
\end{remark}

The example network of Fig.~\ref{fig:groupoid_formal}a has in total 28
input isomorphisms, which can be obtained by inspection. These constitute
its symmetry groupoid. We saw that the symmetry group defines
orbits, which give balanced colorings, or equivalently fibrations,
but not all colorings are balanced. Analogously, the symmetry groupoid
defines colorings via fibrations, and these are balanced, but other colorings
need not be.

Specifically, two input sets may be input isomorphic, yet fail to
synchronize. The relation of input isomorphism is in general not balanced.
This happens because the input set\index{input set }
captures only the first layer of the input tree.\index{input tree } To transform this
truncated input set into a carrier of a fibration symmetry, we must color
it with a balanced coloring.  Then, to capture the symmetry, the
input isomorphism must preserve the balanced coloring of the
nodes. Thus, if we first color the input sets with a balanced coloring
and then look for the subset of input isomorphisms that are also color
preserving, then we have captured the isomorphisms that characterize
the equitable partition without using the full input tree that is
needed for the symmetry fibration.  These isomorphisms are {\it
  symmetry isomorphisms} and form groupoids, not groups. We discuss
this case next.

\section{Symmetry isomorphisms and symmetry groupoid}
\label{subgroupoid}

In the same way that only some permutations can be permutation
symmetries (i.e. automorphisms) of a network, only some input
isomorphisms are symmetry isomorphisms. The set of
all input
isomorphisms has a natural algebraic structure: it is a groupoid.
To explain what this means and why it happens, we first 
generalize the notion
to color-preserving isomorphisms (which are more important for our purposes):

\begin{definition}
  {\bf Symmetry-isomorphism for colorings.} 
  \label{def:symiso}
Consider a network equipped with a balanced coloring. Then
  a {\em symmetry isomorphism}\index{symmetry !isomorphism } is a
  color-preserving input isomorphism. That is, it preserves input sets and
  maps nodes of a given color
  to nodes of the same color.
This set is the {\em coloring
symmetry groupoid}\index{groupoid !coloring
symmetry }
 $\mathcal{B}_\phi$ of the network, for the coloring $\phi$.
\end{definition}

The set $\mathcal{B}_\phi$ is a groupoid because not all input isomorphisms
can be composed. Two such isomorphisms $\alpha,\beta$  can be composed 
to define $\alpha\beta$ only when the codomain of $\beta$ is the same as the
domain of $\alpha$. However, when color-preserving
isomorphisms can be composed, the result is also color-preserving.
Moreover, composition is associative when defined,
and there are identity maps and inverses.

\begin{figure}[t!]
   \includegraphics[width=0.8\textwidth]{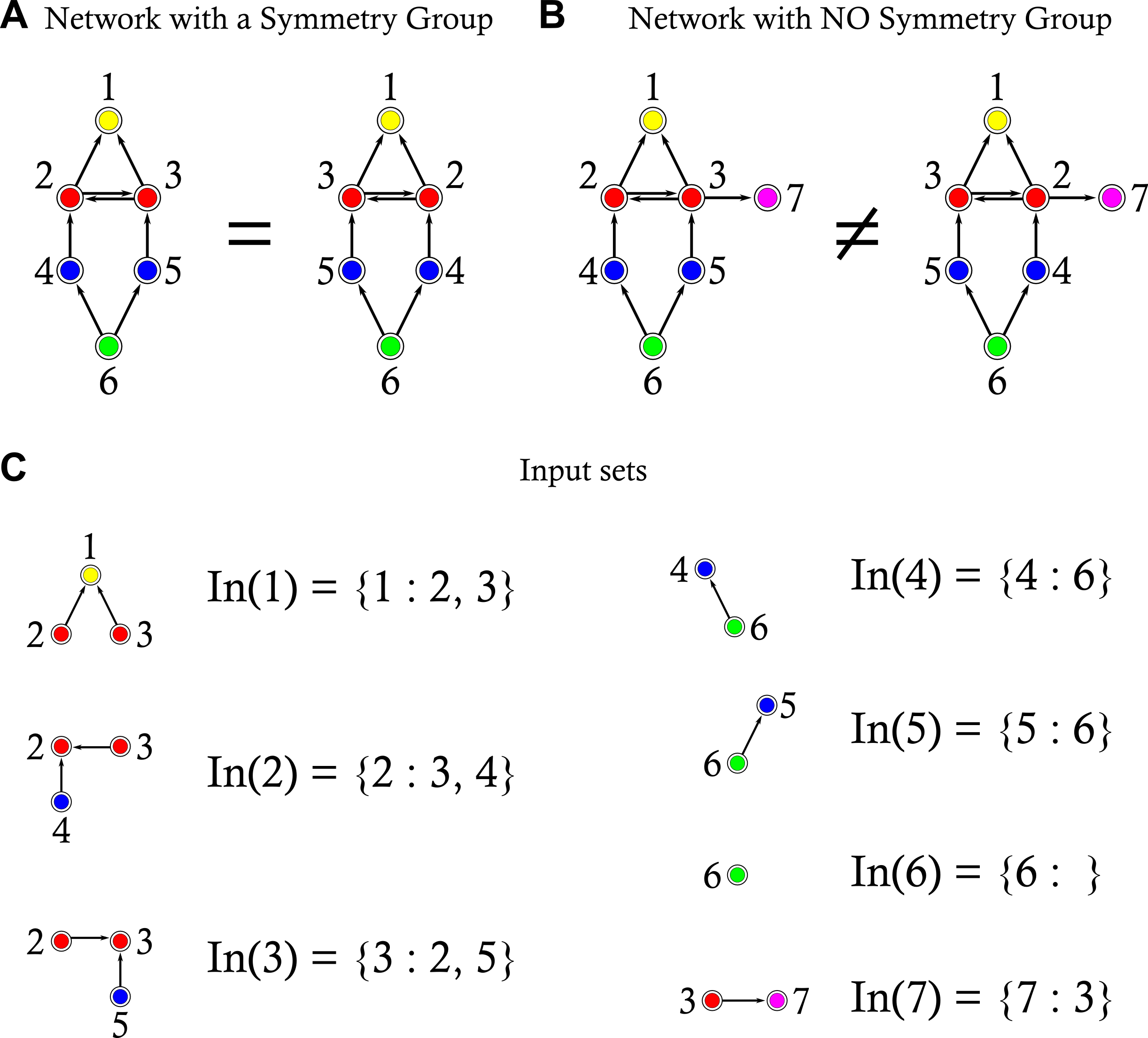} 
\centering
\caption {\textbf{From input sets to input isomorphisms and symmetry
    groupoids}.  (\textbf{a}) Example of a directed network. (\textbf{b}) Input
  sets of nodes of the network in (\textbf{a}), drawn with added source and target 
  nodes for clarity. The input set ${\rm In}(i)$ of  node $i$ is
  the set of all edges $j\to i$. For example, the
  input set of node 1 includes nodes 1, 2, and 3, i.e., ${\rm
    In}(1)=\{1 : 2,\ 3\}$. (\textbf{c}) Input isomorphisms defining the
  groupoid of the network. (\textbf{d}) Input set symmetry isomorphisms
  defining the symmetry groupoid of the network. }
\label{fig:groupoid_formal}
\end{figure}

For example, in the sample network Fig.~\ref{fig:groupoid_formal}a, with the given coloring,
the input isomorphism $\tau_{2\to 3}$ is a symmetry isomorphism:
\begin{equation}
\label{tau_sym}
  \tau_{2\to 3} = \left(
\begin{matrix}
    2: & 3 & 4 \\
    \downarrow & \downarrow & \downarrow \\
    3 & 2 & 5 
\end{matrix}
\right)\ ,
\end{equation}
because it maps red nodes to red nodes and a blue node to a blue node.
Among the 28 input isomorphisms of the network in
Fig.~\ref{fig:groupoid_formal}a only 10 are symmetry-isomorphisms, and
they are given by the following maps:
\begin{equation}
\label{eq:symmetry_isomorphism}
\begin{array}{ll}

\tau_{1\to 1} = \left(
\begin{array}{lll}
    1: & 2 & 3 \\
    \downarrow & \downarrow & \downarrow \\
    1 & 2 & 3 
\end{array}
\right)\ , \qquad

\tau_{2\to 2} = \left(
\begin{array}{lll}
    2: & 3 & 4 \\
    \downarrow & \downarrow & \downarrow \\
    2 & 3 & 4 
\end{array}
\right)\ , \\

\tau_{3\to 3} = \left(
\begin{array}{lll}
    3: & 2 & 5 \\
    \downarrow & \downarrow & \downarrow \\
    3 & 3 & 5 
\end{array}
\right)\ ,\qquad

\tau_{4\to 4} = \left(
\begin{array}{lll}
    4: & 6 \\
    \downarrow & \downarrow \\
    4 & 6  
\end{array}
\right)\ , \\ 

\tau'_{1\to 1} = \left(
\begin{array}{lll}
    1: & 2 & 3 \\
    \downarrow & \downarrow & \downarrow \\
    1 & 3 & 2 
\end{array}
\right)\ ,\qquad

\tau_{2\to 3} = \left(
\begin{array}{lll}
    2: & 3 & 4 \\
    \downarrow & \downarrow & \downarrow \\
    3 & 2 & 5 
\end{array}
\right)\ ,\\

\tau_{3\to 2} = \left(
\begin{array}{lll}
    3: & 2 & 5 \\
    \downarrow & \downarrow & \downarrow \\
    2 & 3 & 4 
\end{array}
\right)\ ,\qquad

\tau_{5\to 5} = \left(
\begin{array}{lll}
    5: & 6 \\
    \downarrow & \downarrow \\
    5 & 6  
\end{array}
\right)\ , \\

\tau_{4\to 5} = \left(
\begin{array}{lll}
    4: & 6 \\
    \downarrow & \downarrow \\
    5 & 6  
\end{array}
\right)\ , \qquad\quad\ 

\tau_{5\to 4} = \left(
\begin{array}{lll}
    5: & 6 \\
    \downarrow & \downarrow \\
    4 & 6  
\end{array}
\right)\ , \\

\tau_{7\to 7} = \left(
\begin{array}{lll}
    7: & 3 \\
    \downarrow & \downarrow \\
    7 & 3  
\end{array}
\right)\ ,\qquad\quad\ 

\tau_{6\to 6} = \left(
\begin{array}{lll}
    6: \\
    \downarrow \\
    6:  
\end{array}
\right)\ .
\end{array}
\end{equation}

Now we consider the full set of input isomorphisms:

\begin{definition}{\bf Symmetry groupoid.}
The {\em symmetry groupoid}\index{symmetry !groupoid } of the network is defined as the set of
all symmetry isomorphisms of the network~\citep{stewart2006}:
\begin{equation}
\begin{array}{l}
{\rm Symmetry \,\, groupoid}=\\ \{\tau_{i\to j}\ |\ \tau_{i\to
  j}\ {\rm is\ a\ symmetry \ isomorphism \ from}\ i \to j \}\ .
\end{array}
\end{equation}
\end{definition}

For example, the symmetry groupoid of the network in
Fig.~\ref{fig:groupoid_formal}a is given by the isomorphisms of
(\ref{eq:symmetry_isomorphism}):
\begin{equation}
\begin{array}{l}
{\rm Symmetry \,\, groupoid}\ =\\
\{\tau_{1\to 1}, \tau'_{1\to 1}, 
\tau_{2\to 2},  \tau_{2\to 3},
\tau_{3\to 3},  \tau_{3\to 2}, 
\tau_{4\to 4},  \tau_{4\to 5}, 
\tau_{5\to 5},  \tau_{5\to 4},
\tau_{6\to 6},  \tau_{7\to 7}\}\ .
\end{array}
\end{equation}

We think of symmetry-isomorphisms as the fibration analogs of group
automorphisms.  Then we can summarize the analogies between groups and
groupoids via the following correspondences:
\begin{equation}
\begin{aligned}
{\rm Groups}\ &\longrightarrow\ {\rm Groupoids}\ \\
{\rm Permutations}\ &\longrightarrow\ {\rm Input \  isomorphisms}\ \\
{\rm Automorphisms}\ &\longrightarrow\ {\rm Symmetry  \ isomorphisms}\  
\end{aligned}
\end{equation}

Symmetry-isomorphisms formalize the concept that nodes of the same
color have input-sets of the same color~\citep{stewart2006}.  Thus, in
this definition of equivalence, nodes in the input sets lose their
identity since which node sends the
information is no longer important. What matters is that the information received by symmetrically related nodes is
the same.  In a balanced coloring, if two nodes $i$ and $j$ have the
same color then their input sets ${\rm In}(i)$ and ${\rm In}(j)$
are isomorphic by a bijection that preserves colors of source nodes
(and edge-types when edges can have more than one type).

In practice, the notion of a balanced coloring is primary,
and its interpretation in groupoid language is included here to
show that the groupoid formalism is equivalent
to both the fibration viewpoint and that of balanced colorings.
This also reinforces the analogy between the global symmetries
captured by groups, and local symmetries captured by groupoids.
Moreover, it establishes a `philosophical' distinction between
the group symmetries that are central to physics and the 
more general groupoid symmetries that we argue are natural for biology.

For this reason the mathematical formalism of
groupoids, while theoretically important, has no practical use when
finding balanced colorings and equitable partitions. In contrast, the
fibration formalism, based on the full input tree, 
does not require a previous coloring to find the fibers,
and this is the formalism that we will use in the remainder of this
book in practical applications. The input tree determines only
the minimal balanced coloring, but this is the most important coloring
when determining building blocks of biological circuits.

An example of a balanced coloring is
given for the network shown in Fig.~\ref{fig:groupoid_formal}d, and we
have already checked that nodes of the same color receive inputs from nodes of
the same colors.

Some groupoid symmetries are of a simple type, in which the input sets of the
nodes concerned are exactly the same. Thus, they depend on the inbound
neighboring list of nodes. However, more elaborated groupoid
symmetries can occur for nodes that do not have exactly the same input set, but
instead have equivalent input sets. The identification of these
groupoid symmetries requires the input tree formalism of fibrations and
iterative algorithms discussed in Chapter
\ref{chap:algorithms}.

\section{Example of groupoids and cluster synchronization}

Fibrations can be viewed as a generalization of the classical notion
of group symmetry. To compare the two, we recall
Fig.~\ref{F:meta_smol1} showing two simple networks, which in
particular occur as synthetic genetic oscillators~\citep{purcell2010}.
The metabolator\index{metabolator } network has a global symmetry, which swaps the two
nodes; this gives rise to a balanced coloring with an associated
fibration.  The Smolen oscillator\index{Smolen oscillator } has no group symmetry, but it does have a
fibration.  We show below that the synchronous dynamics of these two
networks obeys the same equations, but the synchrony-breaking dynamics
is typically very different in the two cases. This further exemplifies the difference between a symmetry group and fibration.

First, we write down the admissible ODEs for Fig.~\ref{F:meta_smol1}
(left). We list variables in the order: node, source of repressor
connection, source of activator connection.

The admissible ODEs then have the form:
\begin{equation}
\label{E:Z2net1}
\begin{array}{rcl}
\dot{x}_1 &=& f(x_1,x_1,x_2) \\
\dot{x}_2 &=& f(x_2,x_2,x_1) \ .
\end{array}
\end{equation}

Nodes 1 and 2 are synchronous if $x_1(t) = x_2(t)$ for all times
$t$. Setting both equal to $y(t)$ the two equations become:
\begin{equation}
\label{E:Z2net2}
\begin{array}{rcl}
\dot{y} &=& f(y,y,y) \\
\dot{y} &=& f(y,y,y) \ .
\end{array}
\end{equation}
Any solution of the common equation $\dot{y} = f(y,y,y)$ corresponds
to a synchronous solution $(y(t)$ of\eqref{E:Z2net1}, and {\em vice
  versa}.

For comparison, consider Fig.~\ref{F:meta_smol1} (right). This is not
symmetric. Listing variables in the same order, admissible ODEs have
the form:
\begin{equation}
\label{E:Asymnet1}
\begin{array}{rcl}
\dot{x}_1 &=& f(x_1,x_2,x_1) \\
\dot{x}_2 &=& f(x_2,x_2,x_1) \ .
\end{array}
\end{equation}
Again, nodes 1 and 2 are synchronous if if $x_1(t) = x_2(t)$ for all
times $t$. Setting both equal to $y(t)$ the two equations once more
reduce to\eqref{E:Z2net2}, and any solution of the common equation
$\dot{y} = f(y,y,y)$ corresponds to a synchronous solution
$(y(t),y(t))$ of\eqref{E:Z2net1}, and {\em vice versa}.

\begin{figure}[htb]
\centerline{
\includegraphics[width=0.4\textwidth]{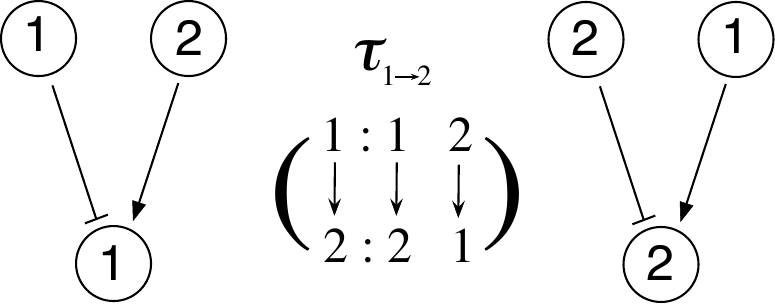} \qquad
\includegraphics[width=0.4\textwidth]{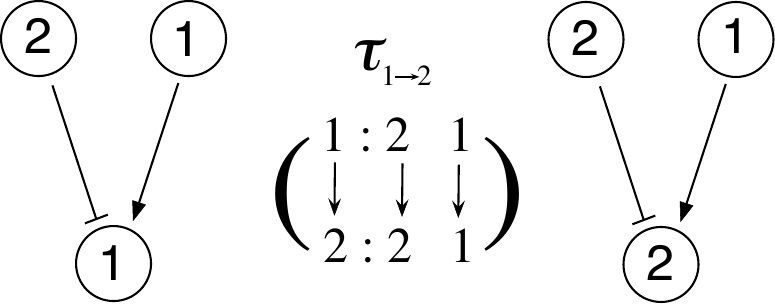}
}
\caption{\textbf{Input sets and input isomorphisms.} {\em Left}: Input sets for the metabolator network, and the
  unique input isomorphism between them. {\em Right}: Input sets for
  the Smolen oscillator network, and the unique input isomorphism
  between them.}
\label{F:meta_smol_input}
\commentAlt{Figure~\ref{F:meta_smol_input}: 
There are two subfigures, each containing an input tree, followed by a permutation, and another input tree.
All input trees have some directed arrows and some inhibitor directed arrow.
First subfigure, first input tree: 1 inhibitor-to 1, 2 to 1. 
First subfigure, permutation: name tau_1to2, upper row 1:1 2, lower row 2:2 1 (arrows between numbers in the same horizontal position).
First subfigure, second input tree: 2 inhibitor-to 2, 1 to 2.
Second subfigure, first input tree: 2 inhibitor-to 1, 1 to 1.
Second subfigure, permutation: name tau_1to2, upper row 1:2 1, lower row 2:2 1 (arrows between numbers in the same horizontal position).
Second subfigure, second input tree: 2 inhibitor-to 2, 1 to 2.
}
\end{figure}

Figure ~\ref{F:meta_smol_input} shows the input sets and input
isomorphisms for these two networks. They appear very similar, but
there are subtle differences. In particular, for the metabolator the
map $\tau_{1 \rightarrow 2}$ swaps nodes 1 and 2 in all three columns,
because this is a global symmetry of the network. In the Smolen
oscillator, the map $\tau_{1 \rightarrow 2}$ swaps nodes 1 and 2 in
the first column, but fixes both in the second.

We anticipate that the groupoid for both oscillators consists of an
identity map for each input set, together with the corresponding map
$\tau_{1 \rightarrow 2}$ and its inverse $\tau_{2 \rightarrow 1}$.

The existence of synchronous states for Fig. \ref{F:meta_smol1}
(right) is governed by the analogous fibration symmetry (Chapter
\ref{chap:fibration_2}), balanced coloring (Chapter \ref{chap:nutshell}
and this Chapter), or {\em input symmetry} described below in this
Chapter. The crucial property here is that when certain nodes are
required to have identical dynamics (synchrony) the entire system of
admissible ODEs leads to a {\em consistent} set of equations for the
synchrony classes (clusters).

For either oscillator in Fig. \ref{F:meta_smol_input}, the base of the
fibration is a network with a single node with two self-loops: one for
each type of arrow. The fibration maps both nodes to the node of the
base, and maps arrows to preserve their type.

If the functions $f$ in\eqref{E:Z2net1} and \eqref{E:Asymnet1} are
the same, then the ODE \eqref{E:Z2net2} for synchronous states is the
same for both networks. Since symmetry groups and fibrations capture
the same synchronized state in the metabolator and the Smolen
oscillator, the question arises: What is the main difference captured
by the fibration that is not captured by the symmetry group? The
difference shows up when synchrony is broken, as we show for
stability of equilibria in Section \ref{sec:saulo}.

\section{Algebraic formulation: lack of composition in groupoids}
\label{lack}

The composition law (totality or closure axiom in the axiomatic
definition of abstract algebraic structures) is a fundamental axiom of groups. It
states that if two transformations $\pi$ and $\tau$ permute the nodes,
then the composition $\pi \circ \tau$ 
is also some permutation of the nodes.  This fundamental
property of permutations, the closure axiom, is not always valid for
the set of input isomorphisms defined in a given network.

Input isomorphisms in the groupoid\index{groupoid } formalism are the analogs of
permutations in the theory of groups. However, whilst
permutations can always be composed, input isomorphisms cannot, and
for this reason they generate a groupoid rather than a group.

\begin{figure}[h!]
\includegraphics[width= \textwidth]{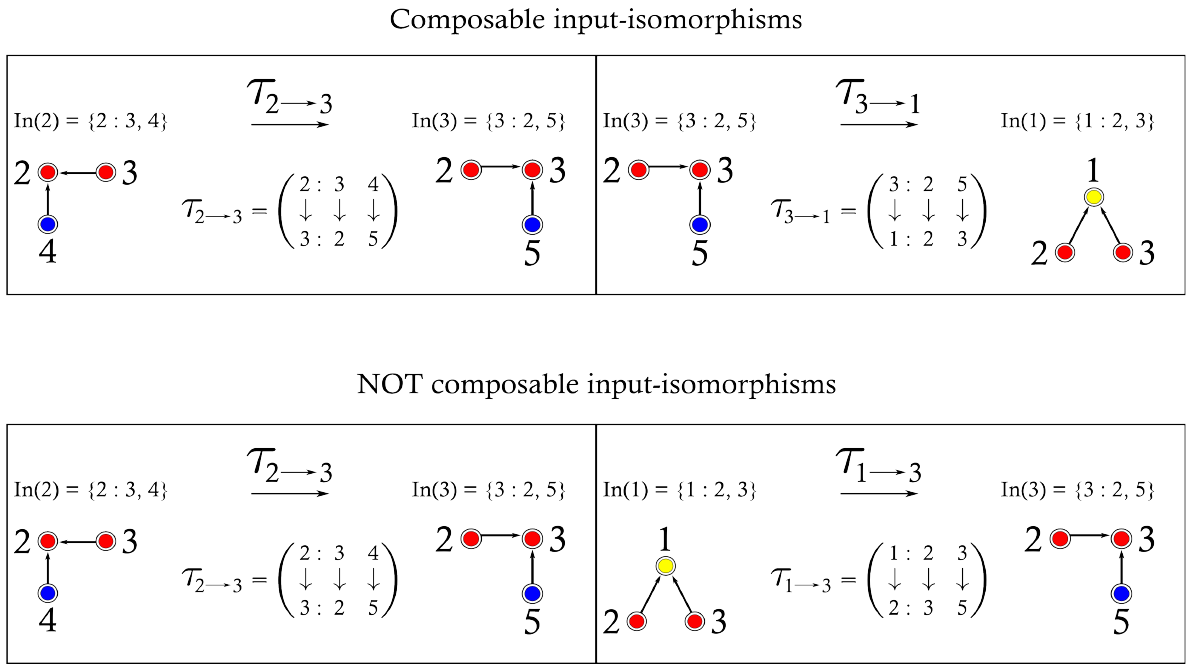}
\centering
\caption {\textbf{Lack of composition law for input isomorphisms}.  {\em Top}: Example of composition of two input isomorphisms $\tau_{2\to
    3}$ and $\tau_{3\to 1}$ of the network shown in
  Fig.~\ref{fig:groupoid_formal}b. Composition is possible if we apply
  first $\tau_{2\to 3}$ and then $\tau_{3\to 1}$, but not in the
  opposite order. More generally, for composition to be possible, the
  codomain of the first isomorphism must coincide with the domain of
  the second isomorphism. This condition is satisfied only if
  $\tau_{2\to 3}$ is applied before $\tau_{3\to 1}$, but not if they
  are applied in the reverse order.  {\em Bottom}: Example of non composable
  input isomorphisms. In this case the isomorphisms $\tau_{2\to 3}$
  and $\tau_{1\to 3}$ are never composable, since there is no match
  between the codomain of one and the domain of the other, no matter
  in which order they are applied.  }
\label{lack-composition}
\end{figure}

It is easy to verify that input isomorphisms cannot always be
composed.  For instance,  $\tau_{2\to 3}$ and $\tau_{3\to 1}$, depicted in
Fig.~\ref{lack-composition}a, are input isomorphisms of the
network concerned. The figure shows that the
composition $\tau_{2\to 3} \circ \tau_{3\to 1}$ is well-defined, so 
this is an example of a legal composition.  However, if we consider
the same $\tau_{2\to 3}$ but we try to compose it with the inverse
input isomorphism $\tau_{1\to 3}$ (Fig.~\ref{lack-composition}b), we
immediately see that this composition is not possible, since the
codomain of $\tau_{2\to 3}$ is node 3 and its input set and the domain
of $\tau_{1\to 3}$ is node 1. Thus, $\tau_{2\to 3}$ and $\tau_{1\to
  3}$ cannot be composed.

In general, for two input isomorphisms to be composed, the codomain of
the first and the domain domain and second input isomorphisms,
respectively, need to be the same. That is, two isomorphisms cannot be
applied in succession $\sigma \circ \tau$ if the codomain of $\tau$
does not coincide with the domain of $\sigma$.

In a group of permutations of a set $X$, all domains and codomains are
the same, namely $X$. This is why any pair of permutations can be
composed. The groupoid\index{groupoid } structure arises when domains and codomains can
differ, and this ultimately relates to the presence of special `base
points' for the input sets, namely the nodes that constitute the
common targets, which lead to different domains and codomains.

Because the composition law need not hold, the
set of input isomorphisms is not a group. Instead, these maps are
groupoid transformations.  However, other properties of a group
persist. Groupoid\index{groupoid } transformations still have a well-defined inverse,
an identity (indeed, often more than one identity), and composition
satisfies the associative law {\it when composition is possible}.
In summary, there are two groupoids associated with any network that
is equipped with a coloring.  One groupoid, say $\mathcal{G}$, consists of all
input isomorphisms for all pairs of nodes. The other, say $\mathcal{H}$,
consists of all input isomorphisms that map every node to a node of
the same color.  Here $\mathcal{G}$ depends only on the graph, whereas $\mathcal{H}$
depends on the coloring concerned.  These coloring groupoids are all
subgroupoids\index{subgroupoid } of the main groupoid.

The coloring symmetry groupoid, for a given coloring, lists the
elements of $\mathcal{H}$. The groupoid $\mathcal{G}$ contains extra maps. For example, in
Fig.~\ref{fig:groupoid_formal}, nodes 1 and 2 are input isomorphic,
because they each have two input arrows. So there is an input
isomorphism $\tau_{1\rightarrow 2}$. In fact, there are two of these,
since we can swap the two arrows. Node 3 is input isomorphic to both 1
and 2, giving further input isomorphisms. So $\mathcal{G}$ is larger than $\mathcal{H}$,
or equivalently, $\mathcal{H}$ is a subset of $\mathcal{G}$. However, in addition, $\mathcal{H}$ is
a groupoid in its own right, because composing two maps the preserve
colors also preserves colors. Technically speaking, $\mathcal{H}$ is a {\em
  subgroupoid} of $\mathcal{G}$. Every coloring of the network gives rise to a
subgroupoid in this manner.

Conversely, we can recover the coloring from the subgroupoid $\mathcal{H}$: two
nodes $i, j$ have the same color if some element of $\mathcal{H}$ maps the input
set of $i$ to that of $j$.  This construction defines a coloring of
the nodes for any subgroupoid $\mathcal{H}$, but in general the resulting
coloring need not be balanced. In order for the coloring thus defined
to be balanced, the subgroupoid $\mathcal{H}$ must satisfy extra
conditions. These are somewhat technical, and will not be stated here,
see \citep{stewart2006}; they restate in groupoid terminology the
condition that the coloring reconstructed from $\mathcal{H}$ must be balanced.

\section{Relations between properties of fibrations, groups, and groupoids}

We have associated a variety of mathematical structures with a graph,
all related to concepts of symmetry;
notably their symmetry fibrations, symmetry groups, and symmetry groupoids.
There are numerous analogies between these fibrations, groups, and groupoids.
Although the technical definitions of these terms differ
considerably, these analogies let us `translate' concepts from one
structure to another. Table \ref{table:difference2} presents
an informal `dictionary' showing how some of the basic concepts relate to each other. As with languages, subtle details can become lost in
translation: that is, we should not expect every property of
one concept to transfer unchanged to an analogous one. But
bearing this dictionary in mind helps to organize important
features and provides some motivation for them.

\begin{table*}[h!]
\centering
\begin{tabular}{| c | c |  c | c |}
\hline 
Property & Fibration &\ \ Group\ \ &\ \ Groupoid\ \ \\
\hline \hline 
preserved  & input tree & adjacency & colored  \\
equivalence relation &  &  & input set \\
\hline 
transformation & input tree  & permutation & input set   \\
 & isomorphism &  & isomorphism  \\
\hline 
symmetry  & symmetry input tree  & automorphism & symmetry  \\
 transformation & isomorphism &  &  isomorphism \\
\hline
symmetry projection & symmetry fibration & quotient map & symmetry groupoid  \\
\hline
cluster synchronization & fibers & orbits & balanced coloring \\
\hline 
axiom of closure & not needed & needed &  not needed \\
\hline 
symmetry type  & local & global & local\\
\hline 
application & biology/AI  & physics & biology \\
\hline 
\end{tabular}
\vspace{10pt}
\caption{\textbf{Informal Dictionary.} Properties of symmetry group, symmetry groupoid and symmetry
  fibration of a graph, indicating analogies between them and areas of science to which they usually apply.}
\label{table:difference2}
\end{table*}

\section{Formal definition of a groupoid}
\label{sec:FDG}

Until now we have discussed groupoid notions informally, without defining their structure precisely. 
For completeness we now give a formal definition  \cite{B06, H71,IR20}, and
formalize the relation between fibrations and groupoids.
 
This section can be skipped by anyone who does not want to see a formal treatment.

\begin{definition}
\label{D:groupoid}
A {\em groupoid}\index{groupoid } $A$ consists of:

 (a) A set $\mathcal{O}$ of {\em objects}\index{object } $i,j,k,\ldots $.

(b) For each $(i,j) \in \mathcal{O} \times \mathcal{O}$, a set $\Theta(i,j)$ of {\em morphisms}.\index{morphism } This set
may be empty.

Instead of writing $\theta \in \Theta(i,j)$, we often write $i \stackrel{\theta}{\to}j$,
which is more intuitive.

The $\Theta(i,j)$ are assumed to be disjoint, or if necessary are made to be
disjoint by redefining them. Thus the set of all morphisms is
$\Theta = \bigsqcup \Theta(i,j)$ where the `squarecup' indicates disjoint union.

(c) For each $i \in \mathcal{O}$, the set $\Theta(i,i)$ contains a distinguished
morphism $\eps_i$. In particular, $\Theta(i,i) \neq \emptyset$ for all $i \in \II$.

(d) There is a composition rule $(\theta,\phi) \mapsto \phi\theta$ whenever
$i \stackrel{\theta}{\to} j\stackrel{\phi}{\to}k$, and $i \stackrel{\phi\theta}{\to}k$.

\noindent
The following axioms must be satisfied:

(e) {\em Identities}: If $i \stackrel{\theta}{\to}j$ then $\theta\eps_i = \theta = \eps_j\theta$.

(f) {\em Inverses}: If $i \stackrel{\theta}{\to}j$ there exists $j \stackrel{\phi}{\to}i$
such that $\phi\theta = \eps_i$ and $\theta\phi=\eps_j$. We write $\phi = \theta^{-1}$.

(g) {\em Associativity}: If $i \stackrel{\theta}{\to}j\stackrel{\phi}{\to}k \stackrel{\psi}{\to}l$
then $(\psi\phi)\theta = \psi(\phi\theta)$.
\end{definition}

Writing products of subsets $X, Y$ of $\Theta$ as $XY$, it is straightforward
to deduce that
\begin{equation}
\label{E:Theta_prods}
\Theta(j,k)\Theta(i,j) = \Theta(i,k)
\end{equation}

If $\theta \in Theta(i,j)$ then $i$ is the {\em source} of $\theta$ and $j$ is the
{\em target} of $\theta$.

Unlike a group, a groupoid has a different identity element
for each object. It
is a group if and only if it has a single object, or equivalently a unique identity element.

\begin{example}\em
\label{ex:groupoid_ex_1}
Let the set of objects be $\mathcal{O} = \{1,2\}$. Define the sets of morphisms by:
\[
\Theta(1,1) = \{\eps_1\} \qquad
\Theta(2,2) = \{\eps_2\} \qquad
\Theta(1,2) = \{\alpha\} \qquad
\Theta(2,1) =  \{\beta\}
\]
with composition defined by Table 1.

\begin{table}[h!]
\label{T:comp_table}
\begin{center}
Table 1. Composition table of groupoid in Example \ref{ex:groupoid_ex_1}. 
Entry in row $a$, column $b$ is $ab$.

\vspace{.2in}
\begin{tabular}{|c|cccc|}
\hline
 & $\eps_1$ & $\eps_2$ & $\alpha$ & $\beta$ \\
\hline 
$\eps_1$ & $\eps_1$ & -- & -- & $\beta$ \\
$\eps_2$ & -- & $\eps_2$ & $\alpha$ & -- \\
$\alpha$ & $\alpha$ & -- & -- & $\eps_2$ \\
$\beta$ & -- & $\beta$ & $\eps_1$ & -- \\
\hline
\end{tabular}
\end{center}
\end{table}

Figure \ref{F:2node_groupoid} is a schematic diagram,
showing these four morphisms and their relation to the objects $1$ and $2$.
Composition is defined only when the corresponding arrows connect head to tail,
forming a directed path.
The table is then easy to construct. The identities are $\eps_1, \eps_2$, and 
$\beta = \alpha^{-1}$.

\begin{figure}[h!]
\centerline{%
\includegraphics[width=0.4\textwidth]{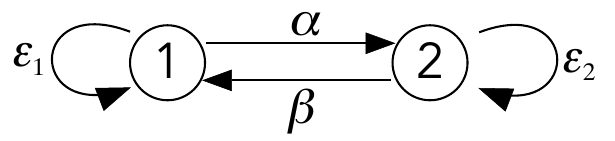}
}
\caption{\textbf{Groupoid.} Schematic diagram of a groupoid with objects $1$, $2$ and morphisms $\eps_1, \eps_2,\alpha, \beta$ composing
only when the corresponding arrows connect head to tail.}
\label{F:2node_groupoid}
\commentAlt{Figure~\ref{F:2node_groupoid}: 
A groupoid with two elements named 1 and 2. An arc from 1 to itself (labeled epsilon1), an arc from 2 to itself (labeled epsilon2), and two
further arcs: from 1 to 2 (labeled alpha), from 2 to 1 (labeled beta).
}
\end{figure}

\end{example}

\subsubsection{Network groupoids}

We now state the connection between networks
and groupoid theory in more formal terms.

\begin{definition}
\label{D:gpoid_of_ntwk}
Let $G$ be a network, with $\CC$ its set of nodes.
The {\em network groupoid}\index{network !groupoid } $\BB_G$ is defined as follows:

The {\em objects} $\mathcal{O}$ of $\BB_G$ are the nodes $c \in \CC$,
so $\mathcal{O}=\CC$.
(Alternatively the objects can be taken as the input sets $I(c)$ of 
these nodes, which are indexed by $c \in \CC$.)

The {\em morphisms} of $\BB_G$ in $\Theta(c,d)$ are the input isomorphisms
$\beta: I(c) \to I(d)$. The usual notation in network dynamics is $B(c,d)$, but for this section we continue to
use $\Theta(c,d)$.

The {\em network groupoid} $\BB_G$ {\em of} $G$ is the disjoint union
\[
\BB_G = \bigsqcup_{c,d} \Theta(c,d)
\]
with the operation of composition (where possible). 
\end{definition}

The {\em identity elements} of $\BB_G$ are the identity maps
$\eps_i:I(i) \to I(i)$.

The inverse $\alpha^{-1}$ of $\alpha:I(i) \to I(j)$
is its inverse as a mapping: $\alpha^{-1}:I(j) \to I(i)$.

It is easy to verify the groupoid axioms.

\subsubsection{Coloring subgroupoids}

A subgroupoid $G'$ of a groupoid $G$ is defined by taking
a subset  $\mathcal{O}' \subseteq \mathcal{O}$ and a subset of 
morphisms $\Theta'(i,j) \subseteq \Theta(i,j)$, with the condition
that these subsets are chosen so that
the groupoid axioms remain valid.

We can relate a fibration of a network $G$
to a subgroupoid of its network groupoid $\BB_G$ by considering the
corresponding balanced coloring (where colors $\leftrightarrow$ fibers).
Indeed, any coloring of $G$,
balanced or not, is associated with a subgroupoid of $\BB_G$:

\begin{definition}
\label{D:color_preserving}
Let $G$ be a network with a coloring $\kappa$. 
An input isomorphism $\beta \in \Theta(c,d)$ is {\em color-preserving} if
for all $e\in I(c)$,
\[
\kappa(\hd (\beta(e))) = \kappa(\hd(e)) \quad \mbox{and}\quad \kappa(\tl(\beta(e))) = \kappa(\tl( e))
\]
Write $\Theta^\kappa(c,d)$ for the set of all color-preserving
elements of $\Theta(c,d)$.
\end{definition}

A coloring is balanced if and only if
\begin{equation}
\label{E:groupoid_bal_con}
\kappa(c) = \kappa(d) \implies \Theta^\kappa(c,d) \neq \emptyset
\end{equation}

\begin{definition}
\label{D:coloring_subgroupoid}
Let $G$ be a network with a coloring $\kappa$. 
The {\em coloring subgroupoid}\index{coloring subgroupoid } $\BB_G^\kappa$ of $\kappa$ is the set of
all color-preserving input isomorphisms for $\kappa$. We have
\[
\BB_G^\kappa = \bigsqcup \Theta^\kappa(c,d)
\]
\end{definition}

The set $\BB_G^\kappa$ is a 'full subgroupoid' of $\BB_G$; that is,
it contains all of the identity elements $\eps_i$ for $i \in \CC$.  

\begin{example}\em
\label{ex:5node_color}

\begin{figure}[h!]
\centerline{%
\includegraphics[width=0.25\textwidth]{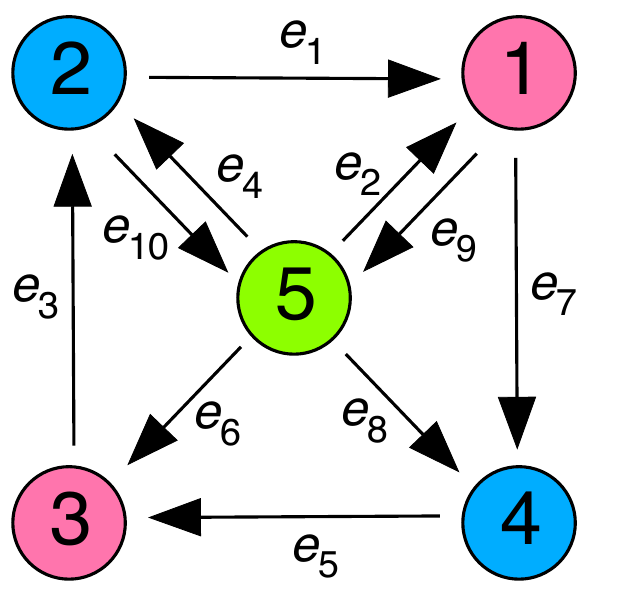}\qquad
\includegraphics[width=0.6\textwidth]{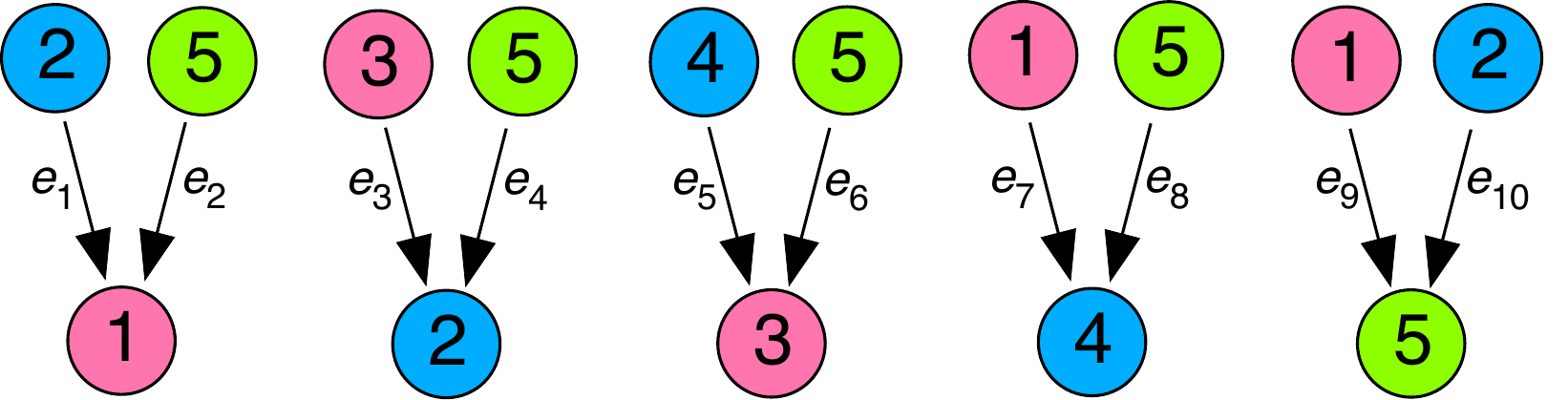}}
\caption{\textbf{Example of colored input sets.} {\em Left}: Network with balanced 
coloring. {\em Right}:
Input sets including head and tail nodes and their colors.}
\label{F:5node_groupoid_col}
\commentAlt{Figure~\ref{F:5node_groupoid_col}: 
One graph and five input sets.
}

\commentLongAlt{Figure~\ref{F:5node_groupoid_col}: 
The graph has nodes named 1 (magenta), 2 (blue), 3 (magenta), 4 (blue), 5 (green). Directed edges; arrows
from 1 to 4 (label e7), from 1 to 5 (label e9), from 2 to 1 (label e1), from 2 to 5 (label e10), 
from 3 to 2 (label e3), from 4 to 3 (label e5),
from 5 to 1 (label e2), from 5 to 2 (label e4), from 5 to 3 (label e6), from 5 to 4 (label e8).
All input trees have one node on the bottom, and two nodes on top; the top nodes are connected with an arrow to the bottom node.
All nodes in the input trees have the same color as the corresponding node in the graph, and all arcs in the input trees
have the same label as the corresponding arc in the graph.
We hereafter describe the five input trees.
First: top 2 and 5, bottom 1.
Second: top 3 and 5, bottom 2.
Third: top 4 and 5, bottom 3.
Fourth: top 1 and 5, bottom 4.
Fifth: top 1 and 2, bottom 5.
}
\end{figure}

Consider the 5-node network of Fig. \ref{F:5node_groupoid_col} (left). 
The groupoid $\BB_G$ has 50 elements,
the input isomorphisms $\beta$. We now consider which maps $\beta$ preserve the colors
of head and tail nodes of arrows, in the sense that $\kappa(\hd(e)) = \kappa (\beta(\hd(e)))$
and $\kappa(\tl(e)) = \kappa (\beta(\tl(e)))$.
Clearly each identity element $\eps_1, \ldots,\eps_5$ does so. There
are precisely four others, in subsets $B(c,d)$ where $c \neq d$ and
$\kappa(c) = \kappa(d)$:
\[
\Matrixr{e_1 & e_2 \\\downarrow & \downarrow \\ e_5& e_6} \qquad \Matrixr{e_5 & e_6 \\ \downarrow & \downarrow \\e_1& e_2} \qquad \Matrixr{e_3 & e_4 \\ \downarrow & \downarrow \\e_7& e_8}
	\qquad \Matrixr{ e_7 & e_8 \\ \downarrow & \downarrow \\e_3& e_4}
\]
\end{example}

Many results about balanced colorings can be interpreted
using coloring subgroupoids, but the theory is technical.
See \citep{stewart2024groupoids}.

\section{Hierarchy of algebraic structures}
\label{axiomatic}

Algebraic structures like groups and groupoids can be defined in an
axiomatic way. Four axioms defines the algebraic structures: closure
(law of composition), invertibility, identity, and associativity, to
which we add a fifth, commutativity, for completeness \citep{dixon1996, hamermesh2012}.
In the present study we consider the following algebraic structures
which satisfy a number of axioms and are not required to satisfy
others, see Table~\ref{table:difference}.
The table shows a hierarchy of increasingly
general algebraic structures, in which the axioms assumed become
successively less restrictive.

\begin{table*}[h!]
\centering
\begin{tabular}{| c | c | c | c | c | c |}
  \hline Algebraic structure & Commutative & Closure & Inverses
  & Identity & Associative \\ \hline \hline Abelian group & $\surd$ &
  $\surd$ & $\surd$ & $\surd$ & $\surd$\\ Non-Abelian group & $\bigtimes$ &
  $\surd$ & $\surd$ & $\surd$ & $\surd$ \\ Groupoid & $\bigtimes$ & $\bigtimes$ & $\surd$ & $\surd$ & $\surd$ \\ Small Category & $\bigtimes$ &
  $\bigtimes$ & $\bigtimes$ & $\surd$ & $\surd$ \\ Semigroupoid & $\bigtimes$ &$\bigtimes$ & $\bigtimes$ & $\bigtimes$ & $\surd$\\ \hline
\end{tabular}
\vspace{10pt}
\caption{Hierarchy of algebraic structures considered in the present
  study. There are other algebraic structures not listed here.
  $\surd$: required axiom. 
  $\bigtimes$: not required.}
  \label{table:difference}
\end{table*}

Starting from a commutative (Abelian)
group which satisfies all five axioms, we drop one axiom at a time to
create less restricted structures, resulting in the
nested hierarchy:

Abelian groups $\to$ Non-abelian groups $\to$ Groupoids $\to$
Small Categories $\to$ Semigroupoids.

The most restrictive algebraic structure satisfies all the axioms listed:
this is the Abelian
Group. An example in physics is quantum electrodynamics (QED), with an
Abelian gauge symmetry group $U(1)$. Abelian groups include graphs of
atomic bonds and molecules. 

A non-Abelian group is also called a non-commutative group. Examples in physics are the
non-Abelian gauge theory of weak interactions, $SU(2)$, and the strong
interaction, $SU(3)$. Physics uses other structures, such as
Lie algebras and `quantum groups'---which despite the name are not groups---but
there is a strong emphasis on groups and related structures.
Permutations forming a symmetry group satisfy the axiom of
closure, since they can be always composed and the result of this
composition is again a permutation in the group.

Dropping the composition axiom, we obtain groupoid
symmetries and fibrations,
which describe many biological systems. The objects of the groupoid
are (or correspond to) the nodes of the network. The morphisms
are isomorphisms of input sets. These are `local symmetries',
indicating  that certain {\em parts} of the network---the input sets of the nodes concerned---have the same structure. If there are input isomorphisms from node $i$ to node $j$ and from node $j$ to node $k$:
\[
i \stackrel{\alpha}{\longrightarrow} j\qquad j \stackrel{\beta}{\longrightarrow} k
\]
then these compose to give an input isomorphism from node $i$ to node $k$:
\[
i \stackrel{\beta\alpha}{\longrightarrow} k
\]
Otherwise, there is no natural way to compose input isomorphisms,
which is why we obtain a groupoid, not a group.

The list of algebraic structures is completed with further structures
obtained by removing one axiom at a time: a small category which has
an identity and satisfies associativity, and finally a semigroupoid
which satisfies only the axiom of associativity. We do not make use of these last two, but groupoids are often viewed as categories
in which every morphism has an inverse, and many groupoid structures
generalize naturally to categories.

The transition from groups to groupoids is
where life begins; biology is modeled by more general
and more flexible structures than physics. 
We elaborate on this quasi-philosophical idea after 
Chapter \ref{chap:bundles}, which reviews how groups describe physics with fiber
bundles. These are similar to fibrations, but more restrictive.


\chapter[Stability and Synchronizability]{\bf\textsf{Stability and Synchronizability}}
\label{chap:stability}

\begin{chapterquote}
  Biological systems need more than the {\em existence} of cluster synchrony,
because these systems must continue to function in a variable environment.
The synchronous state must therefore be {\em stable}.
In this chapter we discuss the stability of states of network
ODEs, focusing on equilibria with cluster synchrony induced by  fibrations. We  
begin with a quick review of asymptotic and linear stability. An equilibrium is
asymptotically stable if nearby initial conditions converge to it
under the flow of the ODE. It is linearly stable if all eigenvalues
of the derivative (linearization, Jacobian matrix) have negative real part.  
Cluster synchrony has an important feature: the Jacobian decomposes into a block 
triangular matrix, whose eigenvalues are those of the two diagonal blocks.
Thus the eigenvalues split into two subsets: synchronous eigenvalues, which  correspond to synchrony-preserving
perturbations, and transverse eigenvalues, which 
correspond to synchrony-breaking perturbations. 
We illustrate these concepts
for a variety of genetic circuits using several types of models. 
Finally, we
discuss the `master stability function' of \cite{pecoraMSF}, initially developed to study stability of complete synchronization in Laplacian systems, and its generalizations to the problem of cluster synchronization in networks.
  
  \end{chapterquote}

\section{Biology needs stability}
  
  So far, we have concentrated on finding 
 conditions for the {\it existence} of cluster synchrony.\index{synchrony !cluster } These conditions
  can be stated in four equivalent ways: a fibration symmetry,
  a balanced coloring, an equitable partition of nodes, the
  existence of a flow-invariant subspace for {\it any} admissible ODE.
  (A fifth viewpoint, groupoid symmetry, is not needed here.)
  For example, Fig. \ref{F:Uxur} shows a 3-node subnetwork of the GRN
  of {\it E. coli}. The fibration concerned is the map from
  this network to the 2-node network on the right hand side of the figure.
  The balanced coloring is given by the blue and pink colors. The
  equitable partition has two parts: $\{1\}$ and $\{2,3\}$. The
  synchrony subspace is
  \[
  \Delta = \{(u,v,v) : u,v \in P^3 \}
  \]
where each node has state space $P$. 

All four formulations convey the same
information with different emphasis.
The key property for cluster synchrony is flow-invariance of $\Delta$.
This means that, 
dynamically, if we set the initial condition of the system to
respect the fibers---that is, if the nodes in each fiber initially have the same
state---then the same clusters remain synchronous for all forward time. 
However, this
property alone is irrelevant to biology, because biological systems must
constantly adapt to a changing environment. What matters for biology
is {\it stable} cluster synchronization; that is, {\it synchronizability}.\index{synchronizability }
Even when initial conditions are not synchronous, the dynamic trajectory
should converge to the cluster state for all initial conditions in
some neighborhood of that state.

That said, it is not possible to study the stability of a state unless
that state exists. Moreover, the network topology does have one 
useful implication for stability analysis.
The flow-invariance property associated with any
cluster defined by a fibration splits the stability analysis
into two parts: {\it synchrony-preserving} stability and
{\it synchrony-breaking} stability. The first refers to stability to
perturbations within the synchrony subspace; the second
to stability transverse to the synchrony subspace. See Section \ref{S:STEE}.

The remarkable robustness of biological organisms and systems
is a big question, and stability\index{stability } of an equilibrium (or a more complex
dynamical state) is only part of the answer, since this refers to
a specific model. A subtler issue is how the state of the system changes
if the model itself is perturbed. This is `structural stability', the
main motivation for the topological approach to nonlinear dynamics
arising from the work of \cite{A89, S67, thom} from the 1960s. 

For a given model, the concept of `stability' has been given many different mathematical formulations,
some of which we discuss in Section \ref{S:stab_notions}. They all have much the same intuitive
implication: if the system is in an equilibrium state (in particular cluster synchrony)
and it is then subjected to a small disturbance, it will settle back down
towards either the original state or one that is very close to it---and in
the network case, a state that has the same clusters, subject to the usual
approximations in modeling assumptions.
The main stability notions are {\em local}: they refer to sufficiently small
perturbations of initial conditions; that is, to neighborhoods of the equilibrium.
A more global indication of stability is the largest such neighborhood,
the {\em basin of attraction}.\index{basin of attraction } For an equilibrium, or a more complex 
state---an `attractor'---the basin consist of all points whose dynamic trajectories limit on that attractor
in forward time. The larger the basin, the `more stable' the state becomes.
Thus, for cluster synchrony, the size of the basin is one way to
quantify synchronizability.

The existence of any given pattern of cluster synchrony depends only
on the topology of the network; it is independent of the (admissible)
dynamics because fibrations are purely graph-theoretic. However, the dynamics of the clusters depends on the choice of admissible ODE. The same goes for
stability, as illustrated below. Stability is therefore a more complex issue than fibrations.
There are two natural questions:

Q1: Is a given cluster synchronous state stable or unstable?

Q2: If it is stable, how big is its basin of attraction?

\noindent
These questions need to be investigated more
deeply to create a solid foundation for developing research applied to
real networks. 

Much research on Q1 has focused on complete synchrony,
in which all nodes are synchronous: see Sections \ref{S:EWSCS} and \ref{S:MSFSI}.
References \citep{pecora2016b,siddique2018symmetry,panahi2021cluster,lodi2021one} address these questions
for nontrivial clusters arising from fibrations, although the problem is better studied for
automorphisms \citep{golubitsky1988,GS2023,pecora2014} where methods from group representation
theory apply. We mention some articles that apply to fibrations 
in Section \ref{S:EWSCS}.
Q1 is local: it concerns only an arbitrarily small neighborhood
of the state concerned. Q2 is harder because the basin
of attraction is a global object: in particular, it can be the entire state space,
in which case the cluster state is globally synchronizable.

However, existing results already have major consequences for
information-processing networks, and show that synchronizability is
possible under certain conditions. The existence of cluster synchrony
gives a better understanding of the function of some
information-processing networks. Synchronization reveals functional
building blocks in biological networks \cite{morone2020fibration},
which in turn shows that some circuits in genetic networks perform
core logic computations \cite{leifer2020circuits}.

\paragraph{Outline of Chapter}

We first explain general concepts of stability with a specific focus on cluster synchrony of equilibria and illustrate them with examples. We show how
flow-invariance of a synchrony space decomposes the eigenvalues of the Jacobian,
which determine linear stability,
into synchrony-preserving and synchrony-breaking parts.

We then apply these ideas to analyze
particular examples of stability in some small genetic circuits,
including bistability of the toggle-switch. This circuit has,
for suitable parameters, two coexisting stable equilibria,
and it can be made switch between them. In this respect
it behaves just like an electronic flip-flop circuit.
   
We provide a short discussion of basic bifurcation theory, illustrated
for the pattern of $\sot$-period phase shifts that occurs in oscillatory states
of the symmetric repressilator, a synthetic genetic circuit.
Finally, we discuss the `master stability function' of \cite{pecoraMSF}
to study stability of the complete synchronization for networks of coupled dynamical systems and its generalization to cluster synchronization developed in \citep{pecora2014, pecora2016b,siddique2018symmetry,panahi2021cluster,lodi2021one}.

\subsection{Existing work on stability of cluster states}
\label{S:EWSCS}

Some general theories have been developed with regard to {\it complete} synchronization under Laplacian dynamics.
We briefly discuss two of them. 

Stability of synchronous states has been widely studied for
special models, such as the Kuramoto model\index{Kuramoto model } \citep{K84,MBBP19}.
Here the nodes are phase oscillators: the state
space of a node is a topological circle, whose points define the phase.
Nodes are considered to be synchronous if their states
are approximately the same, and asynchronous if their states
differ significantly. 
The `Kuramoto order parameter' has been proposed to quantify
synchronization \citep{nykamp}. We do not discuss phase oscillators in this book, but they have many
applications in  both physical and biological science.

The second approach is the `master stability function'\index{master stability function } of \citep{pecoraMSF},
which in its original formulation applies to complete synchronization and relies on a 
specific form of the admissible ODE, namely Laplacian dynamics. It has since been generalized to cluster synchrony; see Sections \ref{S:MSFSI}.
and \ref{sec:MSFACS}.

Complete synchrony is rare for most biological problems, 
and when it occurs it is often deleterious for
an organism. For example, complete synchrony has been associated with epileptic seizures. In contrast, cluster synchronization is common in biology---and important.
In this chapter we deal with cluster synchronization by basing our analysis on general stability notions
from nonlinear dynamics, which we explain through simple biological examples.
In general, the stability issue is very complex and is highly sensitive to
the choice of model and its parameters. Even regular 3-node networks
illustrate these complexities \citep{leite2006}. 

\cite{boldi1997}
study a network of processors that execute the
same algorithm, showing the existence of self-stabilizing
(decaying towards a synchronous state) algorithms and how they can be
constructed in some cases. 
It is also known that for any given network, any cluster corresponding to a
 fibration/balanced coloring can occur as a stable equilibrium for {\it some} admissible
 ODE. This does not guarantee stability for biologically plausible
 model ODEs, but it does show that a full classification of fibrations
 is useful. See \citep[Theorem 14.20]{GS2023}.

\cite{belykh2011, belykh2015} study synchronization in neural networks
using algorithms that produce groupoids (which are equivalent to
fibrations in networks, see Chapters \ref{chap:nutshell} and
\ref{chap:groupoid}). They find synchronous solutions and investigate
the stability of these solutions.  In their models, synchronous
solutions are locally stable if the coupling term is big
enough. \cite{leifer2020circuits} observe that synchronous solutions
for genetic networks simulated by first order ODEs and first order
DDEs are stable and have a very wide basin of attraction. In fact,
solutions synchronize for any initial conditions given sufficient
time; that is, if $x(t)$ and $y(t)$ simulate synchronous genes, then
$\lim_{t\to\infty} (x(t)- y(t)) = 0$.

The stability of
cluster synchronous steady states to synchrony-breaking
perturbations is discussed in \citep[Section 14.9]{GS2023}. Some 
general results for networks with a certain type of feed-forward
structure are obtained in \citep{SW2024a} for both equilibria and periodic states,
with applications to animal locomotion \citep{SW2025loco} and neuron models \citep{SW2025neuron}.
Other approaches and related results can be found in \citep{BPP00,PC90,PK17}. 

The study of stability of the cluster synchronous solution, either periodic or chaotic, has been investigated in \citep{pecora2016b,siddique2018symmetry,panahi2021cluster,lodi2021one}. An extension of these studies to oscillators of different types can be found in \citep{Sorrentino2007,blaha2019cluster,della2020symmetries}. The case of cluster synchronization originating from the presence of different delays over the network edges is described in \citep{nathe2020delays}. Finally, cluster synchronization in adaptive networks is investigated in \citep{lodi2024patterns}.

\section{Notions of stability}
\label{S:stab_notions}

Recall that `synchronizability' means the existence of a {\em stable} cluster synchrony pattern. The ultimate goal of this chapter is
to provide methods that help to 
understand synchronizability  of cluster synchronous states. 

However, this issue is best approached by through general features of the stability of
equilibria for any system of ODEs, because the
stability analysis for clusters depends on these results. 

In nonlinear dynamics, the term `stable'
has many technically different meanings \citep{BS70}.
Intuitively, a solution of an ODE is stable if it persists after any
sufficiently small perturbation to initial conditions. `Persists'
means that the solution converges back towards the previous state as
time tends to infinity, or, less strongly, tends towards something
close to the original solution. For simplicity we focus
on the stability of equilibria, which is the main concern
in many biological applications.

\begin{figure}[h!]
\centerline{%
\includegraphics[width=0.6\textwidth]{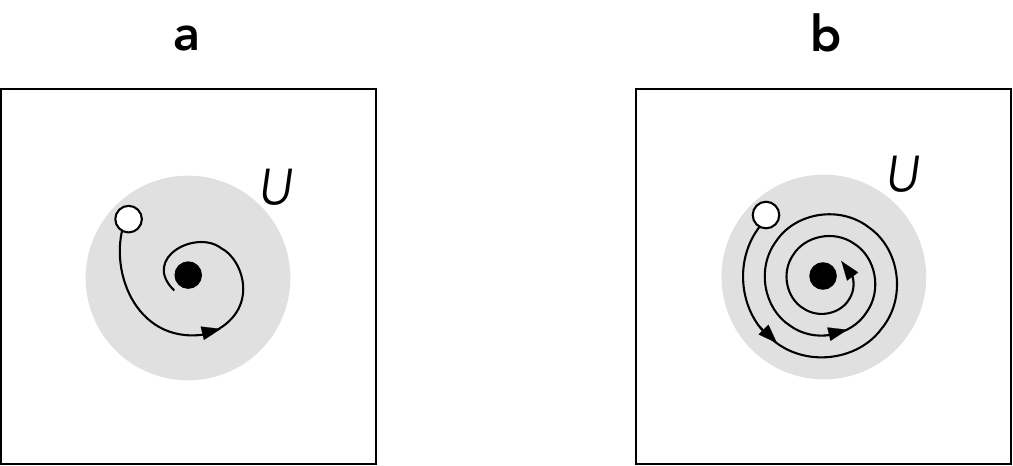}}
\caption{
\textbf{Notions of stability for equilibria.} (\textbf{a}) Exponential (= Linear) stability: There exists a ball $U$ centered on the equilibrium, such that all trajectories starting inside $U$ converge exponentially to the equilibrium.
(\textbf{b}) Asymptotic stability: There exists a ball $U$ centered on the equilibrium, such that all trajectories starting inside $U$ converge to the equilibrium---possibly more slowly than exponential convergence.
}
\label{F:stabilities}
\commentAlt{Figure~\ref{F:stabilities}: 
Two rectangles named a and b.
In both there is a gray spot called U with a black dot in the center and a white ball in the periphery.
In figure a the ball moves towards the dot with a spiral that rapidly approaches the center.
In figure b the ball moves towards the dot with a spiral with many more concentric circles.
}
\end{figure}

To set the analysis in context, we define
three standard notions of stability for equilibria, namely
linear, exponential, and asymptotic stability.
The first two of these are equivalent; the third is weaker. Linear and asymptotic stability are illustrated in Fig. \ref{F:stabilities}. 
There are other stability notions; a common one is Liapunov stability, but we do not use this notion in this book.
For further information see \citep[Chapter 4]{MLS93}
and \citep{L64,LL61,L92}.  

To define these stability notions, consider the ODE $\dot x = f(x)$ for $x \in \R^n$,
where $f:\R^n \to \R^n$. Let $x^*$ be an equilibrium point; that is, $f(x^*)=0$.
Recall that $\|x\|$ denotes the norm of a vector $x$ in Euclidean space 
$\R^n$, defined as
$\|x\| = \sqrt{x_1^2 + \cdots + x_n^2}$ when $x = (x_1, \ldots, x_n)$.

 \paragraph{Linear Stability}\index{stability !linear }
The equilibrium $x^*$ is {\em linearly stable} if all eigenvalues of the Jacobian 
$\mathrm{D}f$ evaluated at $x^*$ have negative real part.
\vspace{.1in}

\paragraph{Exponential Stability}\index{stability !exponential }
The equilibrium $x^*$ is {\em exponentially stable} if 
the rate of convergence is exponential. That is, there is a
neighborhood $U$ of $x^*$ and constants $K, \alpha > 0$ such that
$\|x(t)-x^*\| < K e^{-\alpha t}$ for all $x(0) \in U$.  

\paragraph{Asymptotic Stability}\index{stability !asymptotic }
The equilibrium $x^*$ is {\em asymptotically stable}  if 
all solutions with initial conditions sufficiently close
to $x^*$ converge to $x^*$ in forward time. That is, there is a
neighborhood $U$ of $x^*$ such that $\|x(t)-x^*\|
\to 0$ as $t \to +\infty$ for all $x(0) \in U$. Unlike exponential
stability, the rate of convergence to $x^*$ can be very slow.

\vspace{5pt}
Exponential stability is equivalent to linear stability,
and either of them implies asymptotic stability, but the converse is 
not valid in general. Linear stability
is equivalent to exponential stability.

`Liapunov stability',\index{stability !Liapunov } is a weaker stability notion still, in which solutions that start
sufficiently close to $x^*$ remain close to it in forward
time. We do not give the formal definition since we
avoid this concept of stability in favor of stronger ones.
This notion goes back to \citep{L92} and is the
central topic of \citep{LL61}.
It becomes useful when there is a
suitable `Liapunov function', which guarantees Liapunov stability. Such functions
can sometimes be constructed when other stability notions
are difficult to establish.
See \citep{HS74,SW2024a} for further information.

Stability theory for periodic orbits is analogous, and reduces to
the theory for an equilibrium by taking a `Poincar\'e section'\index{Poincar\'e section }
transverse to the periodic orbit and considering the corresponding
`Poincar\'e map',\index{Poincar\'e map } which maps a point in the section to the first
point at which its trajectory returns to the section. Linear stability
then becomes what classically is known as `Floquet theory'.\index{Floquet theory }
Nonlinear flows can also produce chaotic  oscillations. Chaotic trajectories show sensitive dependence on the initial condition, i.e., a very small change in the initial condition results in a completely different trajectory for large enough time. Despite this characterization, chaotic oscillators can synchronize and linear stability can be used to assess whether synchronous chaotic oscillations are stable or not \citep{pecoraMSF}. This topic is discussed at the end of this Chapter.

\section{Stability of equilibria of linear ODEs}

In what follows we consider only linear stability of an equilibrium. This is the strongest notion of stability,
and it implies exponential, asymptotic, and Liapunov stability. An advantage of linear stability is
that it is determined by the eigenvalues of the Jacobian (derivative,
Fr\'echet derivative) at the equilibrium point, and so reduces to
linear algebra. Excellent algorithms exist for numerical computations.
  
 We begin with the simplest case: equilibria of linear ODEs. This case provides
 motivation for the general theory and leads naturally to stability conditions for
 more complicated systems. Consider the following example, in which
 the equations `decouple':
 \[
 \dot x = ax \qquad
 \dot y = by
\]
 where $a,b \neq 0$ are constants. There is an explicit solution:
 \[
 x(t) = e^{at}x_0 \qquad y(t) = e^{bt}y_0
 \]
 where $x_0, y_0$ are initial conditions at $t=0$. That is, $x(0)=x_0, y(0)=y_0$.
 
\begin{figure}[h!]
\centerline{%
\includegraphics[width=0.8\textwidth]{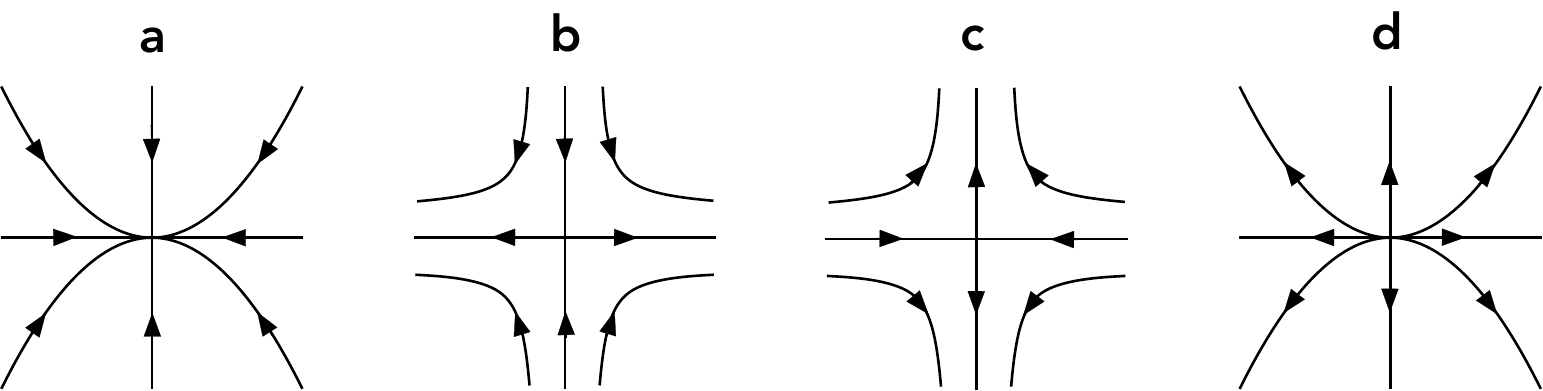}}
\caption{\textbf{Phase portraits for a decoupled linear ODE in $\R^2$.} (\textbf{a}) $a <0, b < 0$ (sink).
(\textbf{b}) $a >0, b < 0$ (saddle). (\textbf{c}) $a <0, b >0$ (saddle). (\textbf{d}) $a >0, b > 0$ (source).}
\label{F:2Dstable}
\commentAlt{Figure~\ref{F:2Dstable}: 
Four phase portraits, named a,b,c,d.
Portrait a: the four axes are directed towards the origin; two parabolas with vertex in the origin, one upward and one downward, all oriented towards the origin.
Portrait b: the Y axis is directed towards the origin, the X axis is directed from the origin. Four rectangular hyperbolas, one in each quadrant.
They are all directed towards the branch that narrows the X axis at infinity.
Portrait c: the X axis is directed towards the origin, the Y axis is directed from the origin. Four rectangular hyperbolas, one in each quadrant.
They are all directed towards the branch that narrows the Y axis at infinity.
Portrait d: the four axes are directed from the origin; two parabolas with vertex in the origin, one upward and one downward, all oriented from the origin.
}
\end{figure}

 The unique equilibrium is the origin $(0,0)$. If $a,b < 0$, all solutions tend towards the origin as
 $t$ tends to infinity. If either of $a,b > 0$, most solutions tend to infinity.
 So the condition for $(0,0)$ to be a stable equilibrium is that $a,b< 0$.
 We call the origin a {\em sink} in this case.
  Otherwise the origin is a {\em saddle} or a {\em source}, as in Fig. \ref{F:2Dstable}.
 
 We recast this analysis in matrix form. Let
 \[
 X = \Matrix{x \\y} \qquad A = \Matrix{a & 0 \\ 0 & b} .
 \]
 Then the equation becomes
 \[
 \dot X = A X
 \]
so that
 \[
X = \Matrix{x \\ y} = \Matrix{e^{at} & 0 \\ 0 & e^{bt}} \Matrix{x_0 \\ y_0} = e^{At}X_0.
 \]
 
We recognize the constants $a,b$ as the eigenvalues of $A$. More generally,
if we replace $A$ by any diagonalizable matrix, a linear change of coordinates
brings it to the form just analyzed, and again stability depends on the signs of
the eigenvalues. More precisely, it depends on the signs of their real parts, because
eigenvalues can occur in complex pairs $c\pm \ii d$, and
$e^{c\pm \ii d} = e^c e^{\pm \ii d}$.
The term $e^{\pm \ii d}$ corresponds to an oscillation of constant amplitude,
and the factor $e^c$ damps the amplitude down to zero when $c < 0$, but blows it
up to infinity if $c > 0$. The characterization in terms of the signs
of the real parts of the eigenvalues applies even when the matrix is not 
diagonalizable, using Jordan normal form.

A general linear ODE on $\R^n$ has the form
\[
\dot X = AX
\]
 where now $X = (x_1, \ldots, x_n) \in \R^n$ is thought of as a column matrix
 (equal to $[x_1, \ldots, x_n]^\mathrm{T}$ where T indicates the transpose)
 and $A = [a_{ij}]$ is an $n \times n$ matrix. The general solution is
 \begin{equation}
 \label{E:gen_lin_ODE}
 X(t) = e^{At}X_0. 
 \end{equation}
 Considering the eigenvalues leads to:
 
 \begin{theorem}
 \label{T:stable_eigen}
 The origin is a stable equilibrium of \eqref{E:gen_lin_ODE}
 if all eigenvalues of $A$ have negative real part. It is unstable if
 any eigenvalue of $A$ has positive real part.
 \end{theorem}
 If any eigenvalue of $A$ has zero real part, the stability is not
 determined. This case is `non-generic' in an individual ODE; that is, it typically does not occur. In Section \ref{S:bifurcations} we see that in a nonlinear ODE, this case
 indicates the possibility of a `bifurcation', and higher-order terms
 are needed to determine stability at the bifurcation point. 

\section{Stability of equilibria of nonlinear ODEs} 

In general, linear models are too simple
and artificial to capture genuine biological behavior, so most biological 
models are nonlinear. However, linear stability
theory for nonlinear ODEs is---as the name suggests---modeled on the linear case.
Let $X = (x_1, \ldots, x_n) \in \R^n$, and consider the ODE 
\begin{equation}
\label{E:nonlinODE}
\dot X = F(X)
\end{equation}
where $F:\R^n \to \R^n$. In coordinates, this takes the form
\beqn
\dot x_1 &=& f_1(x_1, \ldots, x_n) \\
\dot x_1 &=& f_2(x_1, \ldots, x_n) \\
&\vdots & \\
\dot x_n &=& f_n(x_1, \ldots, x_n) 
\eeqn
where $F = (f_1, \ldots, f_n)$.

Assume that \eqref{E:nonlinODE} has an equilibrium $X_0 \in \R^n$,
so that $F(X_0) = 0$. (Here we use $0$ for the zero element of $\R^n$,
which in coordinates is $(0,0,\ldots, 0)$.) Form the Taylor series of $F$ to order 1:
\begin{equation}
\label{E:Taylor_O(2)}
F(X_0+X) = F(X_0) + JX + O(2)
\end{equation}
where $O(2)$ indicates an error term of order $\|X\|^2$, and
$J$ is the {\em Jacobian matrix}\index{Jacobian matrix } (or (Fr\'echet derivative) evaluated at $X_0$:
\[
J = {\rm D}F|_{X_0} .
\]
In components, $J$ is the matrix of partial derivatives of components $f_i$ of $F$:
\[
J= ({\rm D}f|_{X_0})_{ij} = \frac{\partial f_i}{\partial x_j}(X_0) .
\]

Intuitively, the $O(2)$ term in \eqref{E:Taylor_O(2)} becomes arbitrarily small
compared to $JX$ as $\|X\| \to 0$, so we expect solution trajectories $X(t)$
near $X_0$ to resemble those of the {\it linearization}
\[
\dot X = JX.
\]
This can be proved to be the case provided the equilibrium is  {\it hyperbolic}:\index{hyperbolic }
$J$ has no eigenvalues on the imaginary axis. The Kupka--Smale Theorem\index{Kupka--Smale Theorem }
states that hyperbolicity is a `generic' (typical)
property in a general dynamical system \citep{kupka1963,kupka1964,smale1963,peixoto1966}.

Bearing Theorem \ref{T:stable_eigen} in mind, we state:
\begin{definition}\em
\label{D:lin_stab_general}
The equilibrium $X_0$ of \eqref{E:nonlinODE} is {\it linearly stable}\index{stability !linear } if all eigenvalues
of the Jacobian matrix, evaluated at $X_0$, have negative real part.
\end{definition}
Linear stability implies asymptotic stability for nonlinear ODEs. 
We therefore take linear stability
as the most convenient stability notion for equilibria---in particular,
for cluster synchronous equilibria.

\section{Two examples}
\label{chap:stability_examples}

We now illustrate the use of the eigenvalues of the Jacobian
to establish stability (or not) of equilibria for two simple
networks: the Goodwin circuit\index{Goodwin circuit } and a circuit with two different
fibrations. First, we set up suitable notation for
the second of these examples, Example \ref{ex:artificial_3node}. We use the same notation later in this chapter.

\paragraph{Notation for partial derivatives}

In the multigraph setting it is sometimes necessary
to use notation for partial derivatives that
differs from the familiar $\partial f/\partial x_j$ notation.\index{partial derivative, notation for }
Let $f:\R^m \to \R$ be one component of an ODE on $\R^m$.
We denote the partial derivative of $f$ with respect to
the $i$th variable by $f_i$. (An alternative, often used, is $\partial_i f$.)

We explain why we use this notation instead of the more usual
$\partial f/\partial x_j$. When different input arrows have different
tail nodes, $f_i$ is equal to $\partial f/\partial x_i |
_{(x_0,y_0)}$, as usual. However, this notation is ambiguous when the same tail node
occurs for more than one arrow. For example, suppose that one
component of the admissible ODE is $\dot x = f(x,y,y)$. Then $\partial
f/\partial y$ could refer either to the second or third
variable. Technically we should define $f$ using dummy variables; for
example let $f(u,v,w)= u^2+v+w$. Then the three partial derivatives
$\partial f/\partial u$, $\partial f/\partial v$, and $\partial
f/\partial w$ are uniquely defined: they are $2x, 1, 1$
respectively. When we substitute $u=x$ and $v = w = y$ we get
$f(x,y,y) = x^2+2y$. Now
\[
\frac{\partial f(x,y,y)}{\partial y} = \frac{\partial (x^2+2y)}{\partial y} = 2
\]
is uniquely defined, but $\partial f/\partial y$ is not, since $f$
depends on $u,v,w$, not on $y$.  The problem is that $f$ is doing
double duty: as the name of the function and as the result of making
the substitution.

Indeed, if all we know is that, say,  $f(x,y,y) = x^2+2y$, then
$f$ is not unique. Perhaps $f(u,v,w) = u^2+2v$, or
$f(u,v,w) = u^2+3v-w$.

Using subscripts to denote partial derivatives of $f(u,v,w)$
avoids this issue
by specifying which variable of $f$ is concerned. In detail,
\[
\frac{\partial f(x,y,y)}{\partial y} = f_2(x,y,y)+f_3(x,y,y)
\]
by the chain rule for partial derivatives. In the above example 
$f_2=f_3=1$, so 
this formula gives $1+1$, which equals $2$, as before.

This notation should not be confused with $f_i$ for the $i$th component
of a map $F = (f_1, \ldots, f_i)$. In that case the partial derivatives
would be denoted $f_{i,j}$, or for greater clarity,
$\partial_j f_i$.

\begin{example}\em
\label{ex:goodwin}

Perhaps the simplest example is the Goodwin circuit, often called
the Goodwin oscillator because suitable models generate oscillatory behavior
\citep{goodwin1963}. This `circuit' comprises a single node that autoregulates
via a repressor arrow. It is also one of the network motifs---called negative auto-regulation loop---identified in bacterial genetic networks in \citep{alon2019}. Suitable models generate oscillatory behavior,
hence the name, but here we consider equilibria. 

We assume a model in which the state of each node is determined by two variables:
the mRNA concentration $x^R$ and the protein concentration $x^P$.
The most general admissible ODE is then:
\[
\dot x^R = f(x^R,x^P) \qquad
\dot x^P = g(x^P,x^R).
\]
Assume there is an equilibrium state $X_0=(x_0^R,x_0^P)$.
The Jacobian at $X_0$ is
\[
J = \Matrix{\frac{\partial f}{\partial x^R} & \frac{\partial f}{\partial x^P} \\ \frac{\partial g}{\partial x^R} & \frac{\partial g}{\partial x^P}}
\]
which must be evaluated at $X_0$ to specify the linear stability of $X_0$.

For suitable $f$ and $g$ this can be any $2 \times 2$ matrix.
So we cannot say much about it without specifying $f$ and $g$.
To make progress, consider the `special model' 
\[
\dot x^R = -\delta x^R + \frac{1}{1+(x^P)^2}  \qquad
\dot x^P = -\alpha x^P + \beta x^R.
\]
Here the term $-\delta x^R$ represents mRNA degradation, 
 $-\alpha x^R$ represents protein degradation, $\beta$
 represents the rate at which mRNA produces protein, and 
$\frac{1}{1+(x^P)^2}$ is a `Hill function', see Section \ref{S:hill_fn}.
The constants $\alpha,\beta,\delta $ are all positive for this circuit.

It is convenient to simplify notation by setting
\[
x^R = x \qquad x^P = y
\]
so the equations become
\[
\dot x = -\delta x + \frac{1}{1+y^2} \qquad
\dot y = -\alpha y + \beta x.
\]
Equilibria therefore satisfy
\[
0 = -\delta x + \frac{1}{1+y^2} \qquad
0 = -\alpha y + \beta x.
\]
The second equation gives  $y = \frac{\beta}{\alpha}x$. Let $\gamma = \frac{\beta}{\alpha}$
and write this relation as $y = \gamma x$ to keep the formulas simple. Then
the first equation becomes
\[
\delta x = \frac{1}{1+\gamma^2x^2}
\]
which simplifies to a cubic equation for $x$:
\[
\gamma^2\delta x^3 + \delta x = 1.
\]
Since $\gamma, \delta > 0$ this equation has a unique real solution $x_0$,
and it is positive. Positivity is a biological requirement since $x$ is the RNA concentration.
The equilibrium is therefore at $(x_0,y_0)$ where $y_0 = \gamma x_0$.

The Jacobian is
\[
J = \Matrix{-\delta & \frac{-2y}{(1+y^2)^2} \\ \beta & -\alpha}\left|_{(x_0,y_0)}\right. .
\]
A standard result about $2\times 2$ matrices states that
the eigenvalues have negative real parts if and only if the trace is
negative and the determinant is positive. Thus stability requires
\[
-\delta -\alpha < 0 \qquad
\alpha\delta + \frac{2y_0}{(1+y_0^2)^2} > 0 .
\]
Since $\alpha, \delta < 0$ the first inequality holds.
Moreover, $y_0 = \gamma x_0$, so $y_0 > 0$; also $\alpha\delta > 0$.
Therefore the second inequality holds.
We conclude that there is always a unique equilibrium, which is positive,
and it is linearly stable for all parameters $\alpha,\beta,\delta > 0$. 
In more complex models, this state can become unstable,
creating oscillations.
\end{example}

Even for this simple model, we run into a cubic equation. 
For more complex networks and models, explicit solutions
are very unlikely, so numerical methods must be used.
However, sometimes partial information can be obtained
analytically. 

\begin{example}\em
\label{ex:artificial_3node}

Next we consider Fig.~\ref{F:two_fib}, an artificial network introduced to
illustrate the mathematics for cluster synchrony. There are two possible divisions of this
graph into fibers: one is minimal and one is not.

\begin{figure}[h!]
\centerline{%
\includegraphics[width=0.7\textwidth]{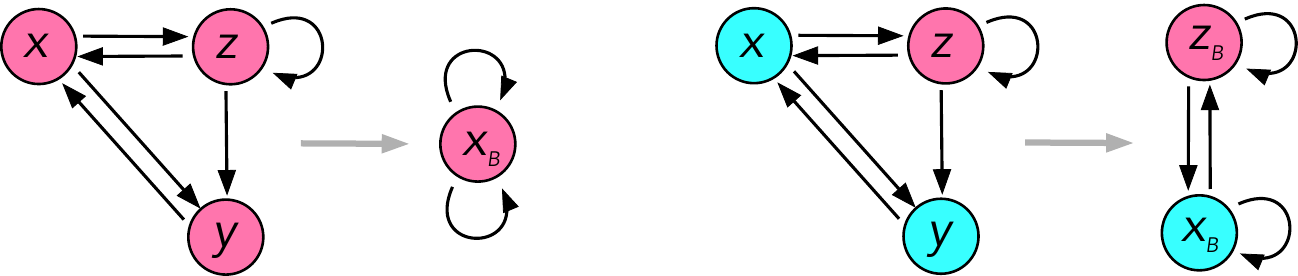}}
\caption{\textbf{Two fibrations of a 3-node network.} {\em Left}: Minimal base: one color; base has one node.
{\em Right}: Two colors; base has two nodes. }
\label{F:two_fib}
\commentAlt{Figure~\ref{F:two_fib}: 
Two graphs each with a minimum base.
Left graph, with nodes called x,y,z, all of them magenta; arrows x to y and z, z to x, y and itself, y to x.
Left base, one node called xB, magenta; two arrows from xB to itself.
Right graph, with nodes called x,y (blue) and z (magenta); arrows x to y and z, z to x, y and itself, y to x.
Right base, two nodes called xB (blue) and zB (magenta); arrows zB to xB and itself, xB to zB and itself.
}
\end{figure}

In the minimal case, all nodes are synchronous, $x(t) =  y(t) =
z(t) = x_B(t)$, on the left of the figure. On the right is a non-minimal cluster state,
$x(t) =  y(t) = x_B(t), z(t) = z_B(t)$. For illustrative purposes,
suppose that these states are equilibria, and assume that the state of each
node is determined by a single variable; that is, $x,y,z \in \R$. Admissible
ODEs are:
\beqn
\dot x &=& f(x,\overline{y,z})\\
\dot y &=& f(y,\overline{x,z})\\
\dot z &=& f(z,\overline{x,z})
\eeqn
where the overline indicates that $f$ is symmetric in its second and third variables. 

The restriction of the ODE to the synchrony subspace is
$\dot u = f(u,u,u)$.
A fully synchronous equilibrium $X_0=(u,u,u)$ is a solution of
$f(u,u,u)=0$. The Jacobian of the restricted ODE on the synchrony subspace 
$\Delta = (u,u,u): u \in \R$  is
\[
J|_\Delta = f_1+f_2+f_3
\]
where, as discussed above, $f_i$ is the partial derivative of $f$ with respect to the $i$th variable.
(We cannot use the notation $\partial f/\partial u$ since the same variable
$u$ occurs three times. See section \ref{S:PDJAM}.) Thus the completely synchronous
state is linearly stable to {\em synchrony-preserving} perturbations
if and only if $f_1(X_0)+f_2(X_0)+f_3(X_0) < 0$. The symmetry condition
on $f$ implies that $f_2(X_0)=f_3(X_0)$ because $X_0$ is fixed by
the symmetry, so this becomes $f_1(X_0)+2f_2(X_0) < 0$.

The Jacobian of the full ODE on $\R^3$ is
\[
J = \Matrix{f_1 & f_2 & f_3 \\ f_2 & f_1 & f_3 \\ f_2 & 0 & f_1+f_3} .
\]
Again $f_2(X_0)=f_3(X_0)$, so this simplifies to
\[
J = \Matrix{f_1 & f_2 & f_2 \\ f_2 & f_1 & f_2 \\ f_2 & 0 & f_1+f_2}.
\]
The eigenvalues are
\[
f_1+2f_2, f_1, f_1-f_2
\]
all evaluated at $X_0$. 

It is easy to see that either of $f_1$ or $f_1-f_2$ (or both) can be positive,
while keeping $f_1+2f_2$ negative. Thus the fully synchronous state
$X_0$ can be stable with respect to synchrony-preserving perturbations,
yet unstable to synchrony-breaking perturbations.

On the other hand, all three eigenvalues can be negative,
in which case $X_0$ is stable for the full ODE on $\R^3$. 
\end{example}

This calculation shows that:
\begin{itemize}
\item[\rm (a)] The stability of an equilibrium depends on the details of the model.
\item[\rm (b)] Stability with respect to synchrony-preserving perturbations,
although necessary for stability of the cluster state, need not be sufficient.
\end{itemize}

The cluster state with two colors can be analyzed in the same manner.
Now $X_0 = (u,u,v)$ and $\Delta = \{u,u,v): u,v \in \R\}$, which is
2-dimensional. There are two equilibrium conditions for $X_0$,
namely $f(u,u,v) = 0$ and $f(v,u,v)=0$.
Since the second and third coordinates of $X_0$
are different, we no longer have $f_2(X_0)=f_3(X_0)$.
The eigenvalues of the Jacobian $J$ for the full ODE on $\R^3$ turn out to be
\[
f_1+f_2+f_3 \qquad f_1 \qquad f_1-f_3
\]
and these must be evaluated at $X_0$.
We recognize the first two as the synchronous eigenvalues---those of the restricted Jacobian $J|_\Delta$. These eigenvalues determine stability with respect to
synchrony-preserving perturbations.
The third is the transverse eigenvalue, determining 
 stability with respect to
synchrony-breaking perturbations.
Again the stability
depends heavily on the model equations, and statements (a) and (b) above apply.

Both cluster states for this network can be stable for the
same model and parameters. To see this, choose $f$ so that $a,b,c < 0$,
$a<b$ and $a<c$. It is also possible for either state to be stable while the other is unstable, or for both to be unstable.

\subsection{Synchronous and transverse eigenvalues and eigenvectors}
\label{S:STEE}

Example \ref{ex:artificial_3node} illustrates a more general theorem,
in which the stability of a cluster equilibrium is decomposed into two parts: 
stability with respect to synchrony-preserving perturbations, and
stability with respect to synchrony-breaking perturbations.
This decomposition arises from the 
flow-invariance of the synchrony subspace $\Delta$. It
simplifies the calculation of eigenvalues and eigenvectors by
decomposing state space $P$ into two parts: $\Delta$ itself, and the
quotient space $P/\Delta$. In general $\Delta$ need not have a flow-invariant
complement, but the eigenvalues that do not correspond to eigenvectors lying
in the synchrony subspace can be found from the vector space quotient.

Because $\Delta$ is flow-invariant, the Jacobian can be placed in
block-triangular form by choosing a basis for $\Delta$ and then extending
it to a basis for $P$. Then
\[
J = \Matrix{A & 0 \\B & C}
\]
for suitable matrices $A,B,C$. The eigenvalues of $J$
are therefore those of $A$ and those of $C$, which are smaller matrices
than $J$, aiding computations.

The matrix $A$  is the restriction
$J|_\Delta$ of $J$ to the synchrony space. The matrix $C=\vec{J}$
is that for the action of $J$ on the quotient space $P/\Delta$.
Stability with respect to synchrony-preserving\index{synchrony-preserving } perturbations\index{perturbation }
is determined by the eigenvalues\index{eigenvalue } of $A=J|_\Delta$. The remaining
eigenvalues of $J$ are those of $C=\vec{J}$, and
determine the stability with respect to synchrony-breaking\index{synchrony-breaking }
perturbations.

\begin{definition}\em
\label{D:ADGW3.4}
The transverse Jacobian\index{transverse Jacobian } at $x_0$ is
the linear map $\vec{J}|_{x_0}$ induced by $J|_{x_0}$ on the vector space
quotient $P/\Delta$. 

The transverse eigenvalues\index{eigenvalue !transverse } of $J|_{x_0}$ are precisely
those of $\vec{J}|_{x_0}$.

The  synchronous eigenvalues\index{eigenvalue !synchronous } of $J|_{x_0}$ are precisely
those of $J|_\Delta$.
\end{definition}

\begin{theorem}
\label{T:synch+trans_ev}
The equilibrium $x_0$ is linearly stable if and only if all
synchronous eigenvalues have negative real parts and all
transverse eigenvalues have negative real parts.
\end{theorem}

For nontrivial clusters, both of the matrices  $J|_{x_0}$
and $\vec{J}|_{x_0}$ are smaller than 
$J|_{x_0}$, so this decomposition simplifies the calculation
of the eigenvalues, hence the linear stability of the equilibrium.
A simple formula for $\vec{J}|_{x_0}$ is stated and proved in \citep[Theorem 14.23]{GS2023},
but we do not use this here.

\section{Some small genetic circuits}
\label{S:6GC}

\begin{figure}[b!]
\centering \includegraphics[width=0.9\textwidth]{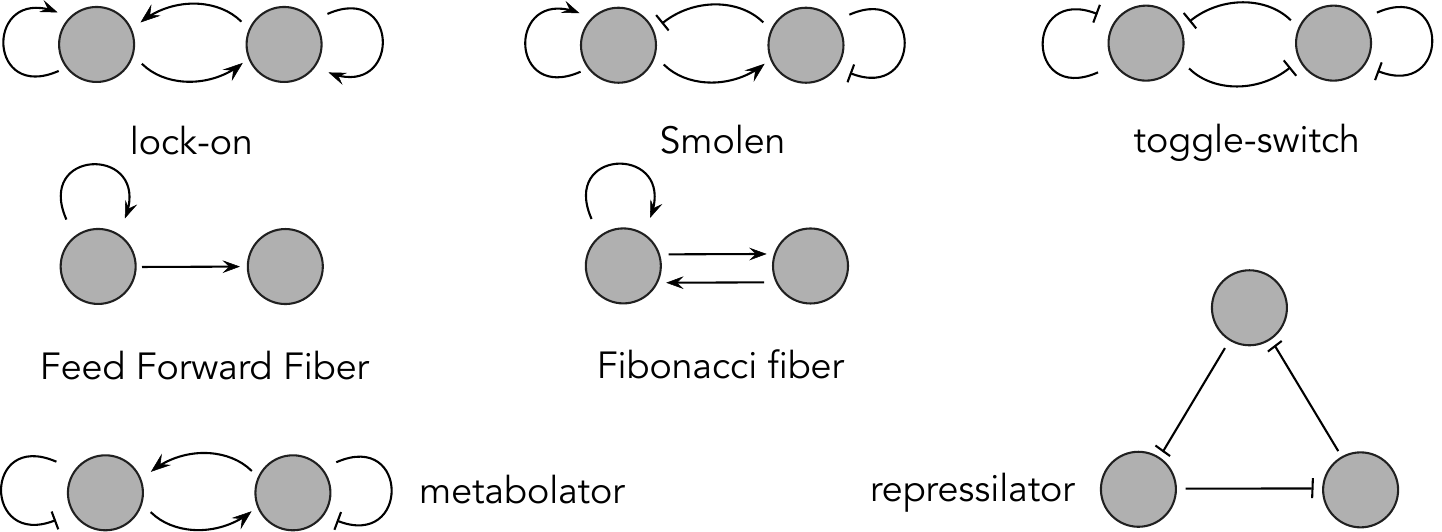}
\caption{\textbf{Seven small circuits that occur as
functional and synchronous building blocks in GRNs}. The lock-on,\index{lock-on } Smolen,\index{Smolen circuit }  toggle-switch,\index{toggle-switch } feed-forward fiber,\index{feed-forward fiber } and Fibonacci fiber\index{Fibonacci fiber } 
are bases for circuits found in the {\em E. coli} GRN. The repressilator\index{repressilator } is a synthetic GRN that sustains periodic oscillations. The metabolator\index{metabolator } is a GRN that includes metabolic effects.}
 \label{fig:stability_circuits}
\commentAlt{Figure~\ref{fig:stability_circuits}: 
Seven circuits, each with a name, two or three nodes, and some directed or inhibition arc.
Graph named lock-on; two nodes, say 1,2; arcs 1 to itself and 2, 2 to itself and 1.
Graph named Smolen; two nodes, say 1,2; arcs 1 to itself and 2, 2 inhibition-to itself and 1.
Graph named toggle-switch; two nodes, say 1,2; arcs 1 inhibition-to itself and 2, 2 inhibition-to itself and 1.
Graph named Feed Forward Fiber; two nodes, say 1,2; arcs 1 to itself and 2.
Graph named Fibonacci fiber; two nodes, say 1,2; arcs 1 to itself and 2, 2 to 1.
Graph named Metabolator; two nodes, say 1,2; arcs 1 inhibition-to itself, 2 inhibition-to itself, 1 to 2, 2 to 1.
Graph named Repressilator; three nodes, say 1,2,3; arcs 1 inhibition-to 2, 2 inhibition-to 3, 3 inhibition-to 1.
}
\end{figure}

In Part II we will identify building blocks of gene regulatory
networks through the fibration symmetries of the underlying biological
graph. In \citep{stewart2024dynamics} we analyze analytically the
cluster synchronized solutions, 
and their stability, for seven of these
`circuits', shown in Fig. \ref{fig:stability_circuits}.
In this chapter we summarize the results for five of these
circuits, omitting the feed-forward fiber and Fibonacci fiber
circuits. Their biological significance, and in particular
their appearance in the {\em E. coli} GRN, is discussed in 
Chapter \ref{chap:hierarchy_2}. \cite{arese2020diversity} discusses stability in ecosystems.

All of these circuits occur as functional and synchronous building
blocks in real or synthetic genetic networks. The lock-on,\index{lock-on } toggle-switch,\index{toggle-switch }
Smolen,\index{Smolen circuit } feed-forward fiber,\index{feed-forward fiber } and Fibonacci fiber\index{Fibonacci fiber } circuits occur
in living organisms; they were first observed in {\em E. coli} in
\citep{morone2020fibration} and later in other organisms from yeast to
humans \citep{leifer2020circuits}. The sixth is the metabolator,\index{metabolator }
a GRN incorporating metabolic effects \citep{fung2005asynthetic}, and the seventh is
the synthetic GRN
repressilator\index{repressilator } \citep{elowitz2000}. 

In Section
\ref{sec:saulo} we discuss the lock-on circuit as a typical example.
In Section \ref{S:metab+smol} we compare the Smolen circuit with the
very similar metabolator circuit \citep{fung2005asynthetic,purcell2010}. Section \ref{S:bi_tog} discusses bistability of the
toggle-switch. The repressilator's oscillatory states are discussed in
Section \ref{S:repressilator}. The
feed-forward fiber and Fibonacci fiber circuits are analyzed from
the same viewpoint in \citep{stewart2024dynamics}.

For each of these circuits, the network
topology admits a fibration symmetry, leading to the existence of
synchronous cluster states. These states can be steady, periodic, or chaotic,
depending on the model. 
In Section \ref{sec:saulo} we show how
the network framework permits a systematic analysis of certain
aspects of dynamics and bifurcations in these GRN circuits.
Similar methods apply to other small circuits. The main aim in
this chapter is to illustrate several mathematical ideas 
for analyzing circuit dynamics on examples of biological relevance.
We do not aim at a systematic analysis of all the dynamic behavior
that can occur in these circuits, which in any case depends on the choice
of model. We focus on general principles that give insight into
the stability of equilibria, and emphasize
the distinction between synchrony-preserving and synchrony-breaking
stability. We show how fibration symmetry simplifies the
calculation of the eigenvalues of the Jacobian at a cluster
equilibrium.

In Section \ref{S:bifurcations} we recall some basics of bifurcation theory.\index{bifurcation }
A bifurcation occurs when varying a parameter causes a stable state to
become unstable, in  which case the system typically 
moves to some other stable state. 
The two main types of local bifurcation from an equilibrium are
steady-state bifurcation,\index{bifurcation !steady-state } resulting in new equilibria,
and Hopf bifurcation,\index{bifurcation !Hopf } resulting in a periodic oscillation. Bifurcations can either
preserve or break cluster synchrony, depending on the corresponding critical eigenvectors.
We apply these ideas to some of the six circuits in this section,
after we have set up the necessary mathematical background.
Similar methods apply to any small circuit with fibration symmetries,
but the calculations may be more complicated.

\section{Protein/mRNA network structure}
\label{sec:PRN}

A gene regulatory network (GRN)\index{network !gene regulatory }\index{GRN } is often referred to as a transcriptional regulatory network (TRN), although the former is more general than the latter.\index{network !transcriptional regulatory }\index{TRN } We use both
terms here.
The usual format for GRN diagrams in the
biological literature uses one node per gene. However, for
some mathematical purposes it can be more convenient to work with
a modified representation of gene regulatory networks,
so that standard models occur as admissible ODEs in a manner that
makes the different roles of transcription and translation explicit.

To do this we use a protein-mRNA network (PRN)\index{network
  !protein-mRNA }\index{PRN } representation
\citep{stewart2024dynamics}, which splits each gene into two nodes:
one for mRNA and one for protein.  The PRN format effectively gives
each gene node a 2-dimensional node space with a restriction on the
node dynamic: the arrow from the mRNA component to the protein
component must be inside that node.  In contrast, arrows from the
protein component can connect to mRNA nodes of other genes via
regulation.

This unorthodox representation is, of course, implicit in the simpler networks
commonly used in biology, but there are mathematical 
advantages in making it explicit. An arbitrary 2-dimensional node
state space permits more general ODEs that fail to incorporate important
biological information: namely, that translation of mRNA to protein occurs within
a single gene, but transcription of DNA to mRNA, regulated
by a protein, can link different genes. Subsection \ref{S:BS}
draws useful conclusions from this structure.

\subsection{Hill functions}
\label{S:hill_fn}

We will illustrate some stability calculations with simple, explicit models
based on Hill functions.
A {\em Hill function}\index{Hill function }  $H:\R^{\geq 0} \to \R$ is any function of the form
\begin{equation}
\label{E:hill}
H^-(z) = \frac{1}{1+z^n} \quad 1 \leq n \leq \infty .
\end{equation}
(Since $z$ is a concentration we require $z \geq 0$.)  This function has a `sigmoid' shape
and is monotonic decreasing for all $n$.  We have $H^{-}(0) = 1$ and
$H^-(z) \to 0$ as $z \to \infty$.  Therefore this form represents
    {\it repression} of gene $j$ on gene $i$ when used in the context
    of $f^R_i(x^P_j)$.  The minus sign in $H^-$ is a reminder of this.
    For {\it activation} we use
\begin{equation}
H^+(z) =1-H^-(z) = \frac{z^n}{1+z^n}.
\end{equation}
This function also has a sigmoid shape, but now it
is monotonically increasing for all $n$. 
We have $H^+(0) = 0$ and $H^+(z) \to 1$ as $z \to \infty$.
This form represents {\it activation} 
of gene $j$ on gene $i$, with the plus sign as a reminder.

The function $H^\pm$ can be modified in several ways: change the value
of $n$; scale the variable $z$ considering $H^\pm(\mu z)$ for a
parameter $\mu > 0$; scale the {\it value} of $H^\pm$ by using $\nu
H^\pm$; move the origin by translating $z$ to $z+z_0$ for some
$z_0$. Usually we take $n=2$, both for simplicity and because it is
relatively typical.

\subsection{Bipartite structure}
\label{S:BS}

\begin{definition}\em
A graph is {\it bipartite}\index{bipartite } if its nodes split into two components, so that the 
 head and tail of any arrow lie in different components.
\end{definition}

Every PRN is bipartite. The two components are the  mRNA nodes
and the protein nodes.
The mRNA nodes connect to proteins only via internal translation arrows, which we draw with wavy lines,
while proteins connect to mRNA either within a single gene or between
distinct genes via transcription arrows, which we draw with solid lines, either
activator or repressor. (The PRN structure does not consider other types of interaction between genes, but modified
network structures can represent these.)

We distinguish two types of model:
\begin{itemize}
\item {\em General model}:\index{general model } the most general ODEs that are compatible
  with the network structure; that is, the admissible ODEs for the PRN.
\item {\em Special model}:\index{special model } this common type of model uses a Hill
  function and a linear degradation term for mRNA concentration, and a
  linear function and linear degradation term for protein
  concentration.  Activators\index{activator } and repressors\index{repressor } are represented by the
  signs of coefficients and the choice of Hill functions. 
  Other models, many not based on Hill functions, are also possible.
\end{itemize}

\subsection{Relation between GRN and PRN}

\begin{figure}[htb]
\centerline{%
  \includegraphics[width=.45
  \textwidth]{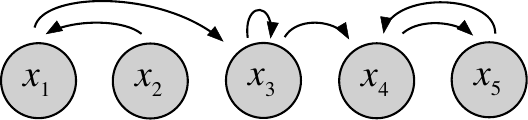} }
\caption{\textbf{Example GRN with 5 nodes.} There is an autoregulation loop at node 3.}
 \label{F:bipGRNgeneformat}
\commentAlt{Figure~\ref{F:bipGRNgeneformat}: 
A graph with nodes named x1-x5. Directed edges: arrows from x1 to x5, from x2 to x1, from x3 to itself and x4,
from x4 to x5, from x5 to x4.
}
\end{figure} 

\begin{figure}[htb]
\centerline{%
\includegraphics[width=.45\textwidth]{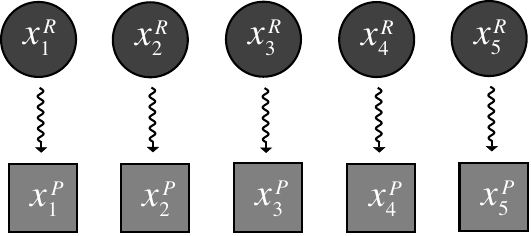} \qquad\qquad
\includegraphics[width=.45\textwidth]{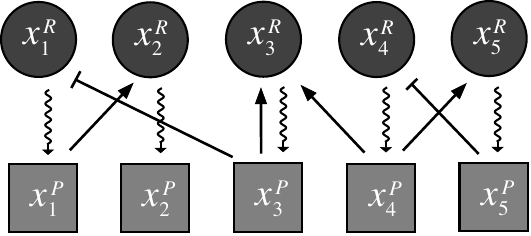} 
}
\caption{\textbf{Bipartite structure of PRN for Fig. \ref{F:bipGRNgeneformat}.} {\em Left}: translation
  arrows (wavy) only.  {\em Right}: Typical example with added
  transcription arrows (solid).}
 \label{F:bipGRN}
\commentAlt{Figure~\ref{F:bipGRN}: 
Two graphs. On the left, a graph with five blue circular nodes named x1R-x5R (top row), and five red rectangular nodes named x1P-x5P (bottom row).
Every blue node is connected with the corresponding red node through a squiggly arrow.
On the right, a graph with the same nodes, but different arcs. Squiggly arrows connect x1R to x1P, x2R to x2P, x3R to x3P, x4R to x4P, x5R to x5P;
inhibition arrows connect x3P to x1R and x5P to x4R. Directed arrows connect x1P to x2R, x3P to x3R, x4P to x3R, x4P to x5R.
}
\end{figure}

Figure \ref{F:bipGRNgeneformat} shows a GRN in the usual biological form,
with one node per gene.
Figure ~\ref{F:bipGRN} (left) shows the corresponding PRN
with just the internal translation arrows from R to P (wavy arrows), which are
common to all PRNs.  Figure ~\ref{F:bipGRN} (right) shows 
the transcription arrows (solid) as well. The
autoregulation loop on node 3 can be seen as the pair of 
arrows between nodes $x_3^R$ and $x_3^P$.  

\subsection{Relation between Jacobian and adjacency matrices}
\label{S:PDJAM}

In preparation for the next section, we point out one feature of the relation between
adjacency matrices and the Jacobian that might cause confusion.
This is a technical issue that arises because different disciplines
prefer different conventions. It is basically straightforward, but can cause confusion
if it is not borne in mind when reading literature from diverse sources.

The issue is that the natural structure of the Jacobian, and the standard definition in graph theory
of an adjacency matrix,\index{adjacency matrix, convention for } do not correspond as well as we might wish. 
There is one
adjacency matrix for each node- or arrow-type. The admissible linear maps
are linear combinations of the {\em transposes} of these matrices. The transpose arises
because of different conventions for Jacobian and adjacency matrices.

To explain what we mean, consider the Smolen network of Fig. \ref{fig:stability_circuits}.
The admissible ODEs are
\begin{equation}
\label{E:smolen_admiss_partial}
\dot x_1 = f(x_1,x_1,x_2) \qquad
\dot x_2 = f(x_2,x_1,x_2).
\end{equation}
Recall that, following most graph theory texts, we adopt the
convention that row $i$ of an adjacency matrix represents nodes to which node $i$
sends an {\em output} arrow.
There are three adjacency matrices: one ($N$) corresponding to the nodes,
a second ($A$) to the activator arrows, and a third ($R$) to the repressor arrows.
They are:
\[
N = \Matrix{1 & 0 \\ 0 & 1} \qquad  A = \Matrix{1 & 1 \\ 0 & 0} \qquad R = \Matrix{0 & 0 \\ 1 & 1} .
\]
The Jacobian for \eqref{E:smolen_admiss_partial} has the form
\[
J = \Matrix{f_1+f_2 & f_3 \\ f_2 & f_1+f_3}
\]
where, as explained above, $f_k$ is the partial derivative of $f$ with
 respect to the $k$th variable. Now
\beqn
J &=& f_1 \Matrix{1 & 0 \\ 0 & 1}  +f_2 \Matrix{1 & 0 \\ 1 & 0} +f_3 \Matrix{0 & 1 \\ 0 & 1}\\
&=& f_1N^\mathrm{T} + f_2 A^\mathrm{T} + f_3 R^\mathrm{T}.
\eeqn
Thus the Jacobian corresponds naturally
to the sum, over the three `arrow' types (node, activator, repressor)
of the corresponding partial derivatives of $f$ multiplied by the
transpose of the corresponding adjacency matrix.
Similar results hold for any network \citep[Theorem 11.4]{GS2023}.

In practice this clash of conventions
causes no serious issues, because the Jacobian and its transpose have
the same eigenvalues. However, they do not have the same eigenvectors,
and it is the eigenvectors of the Jacobian that correspond to 
important features of the dynamics such as cluster patterns.

The arrangement of entries for the Jacobian is natural because 
it corresponds to the arrangement of variables in the components of the ODE.
The component for $\dot x_c$ depends on the variables listed inside the corresponding 
$f$, which appear as a row. 
To avoid the need for a transpose, we can either redefine the Jacobian in an unnatural manner,
or redefine the adjacency matrices so that entry $ij$ corresponds to the
tails nodes of input arrow to node $i$, not outputs. Indeed, this is what is done
in sources such as \citep{GS2023,pecoraMSF,pecora2016b}. 

\subsection{Adjacency matrix and Jacobian}
\label{S:AMJ}

The eigenvalues of the Jacobian at a cluster synchronous equilibrium of a
PRN can be deduced from those of the corresponding GRN. This 
simplifies some calculations. The key point is that 
every PRN is bipartite.\index{PRN }\index{bipartite } The two parts are the mRNA nodes (R) and the
protein nodes (P). Arrows from R to P are internal to the combined
gene node.  Assume there are $n$ genes; then the network has $n$ R
nodes and $n$ P nodes.  

In this subsection it is convenient to list the nodes with the R nodes as
$1, \ldots, n$ and the P nodes as $n+1, \ldots, 2n$.

The transpose of the weighted adjacency matrix for the network
$\GG$, and the Jacobian at a general point, have the form
\begin{equation}
\label{E:matrix_M}
M = \Matrix{P & Q \\ R & S}.
\end{equation}
partitioned into blocks according to the bipartite structure.
The blocks $P,R,S$ are diagonal; say
\[
P = \Matrix{p_1 & 0 & \cdots & 0 \\ 0 &p_2 & \cdots & 0 \\ \vdots &
  \vdots & \ddots & \vdots \\ 0 & 0 & \cdots &p_n} \qquad R =
\Matrix{r_1 & 0 & \cdots & 0 \\ 0 & r_2 & \cdots & 0 \\ \vdots &
  \vdots & \ddots & \vdots \\ 0 & 0 & \cdots & r_n} \qquad S =
\Matrix{s_1 & 0 & \cdots & 0 \\ 0 & s_2 & \cdots & 0 \\ \vdots &
  \vdots & \ddots & \vdots \\ 0 & 0 & \cdots & s_n}
\]
In contrast, the block $Q$ is arbitrary, because proteins produced by
one gene can regulate mRNA in other genes.  In the example of
Fig.~\ref{F:bipGRN} (right),
\[
Q = \Matrix{0 & q_1 & 0 & 0 & 0\\ 0 & 0 & 0 & 0 & 0 \\ q_2 & 0 & q_3 &
0 & 0 \\ 0 & 0 & q_4 & q_5 & 0 \\ 0 & 0 &0 & 0 & q_6}.
\]
The interpretation of these blocks is:

\quad $P$: internal mRNA dynamic (usually degradation of mRNA).

\quad $Q$: {\em interaction matrix} $Q_{ij}$ denoting the binding of the transcription factor expressed by gene $i$ to the regulatory domain of gene $j$ to regulate (turn on and off) its expression of mRNA .

\quad $R$: translation of mRNA to protein.

\quad $S$: internal protein dynamic (usually degradation of the protein).

\noindent
The structure of the matrix $M$ in \eqref{E:matrix_M} does not constrain its
eigenvalues in any simple manner.  In
Section~\ref{S:HC} we specialize to cases where the eigenvalues of $M$
can be determined directly from those of $Q$, see Theorem
\ref{T:MeigCeig}.  The matrix $Q$ is the {\em transpose} of the adjacency 
matrix for the GRN
version of the network, weighted by connection strengths if appropriate.

\subsubsection{Homogeneous case}
\label{S:HC}

To make progress we now assume the network is homogeneous. 
That is, all nodes are input isomorphic. This implies that $p_i
= p, r_i = r, s_i =s$ for $1 \leq i \leq n$. Here $p,r,s$ are
constants.  (For the usual models $p, s$ represent degradation of mRNA
and protein concentrations, so are negative, and $r$ represents the
rate of translation, so is positive. However, at this stage we allow
arbitrary signs.)  Now
\begin{equation}
\label{E:Meqn}
M = \Matrix{pI_n & Q  \\  rI_n & sI_n}
\end{equation}
where $I_n$ is the $n \times n$ identity matrix.  The eigenvalues of
$M$ are related to those of $Q$, which in general are simpler to
compute. We prove:

\begin{theorem}
\label{T:MeigCeig}
If the eigenvalues of $Q$ are $\mu_i$, then those of $M$ are
\begin{equation}
\label{E:MeigCeig}
\lambda^\pm_i =  \shf\left(p+s \pm \sqrt{(p-s)^2+4r\mu_i} \right) .
\end{equation}
\end{theorem}
Each $\mu_i$ gives two values of $\lambda_i$, in
accordance with doubling the size of the matrix.
\begin{proof}
Suppose that $\Matrix{u\\v}$ is an eigenvector of $M$ with eigenvalue $\lambda$.
Then
\[
\Matrix{\lambda u\\ \lambda v} =  \Matrix{p I_n & Q \\ rI_n & sI_n}\Matrix{u\\v} =
	\Matrix{pu+Qv\\ru + sv} .
\]
Therefore
\[
(\lambda -p) u = Qv \qquad
(\lambda -s)v = ru
\]
so
\[
u = \frac{\lambda -s}{r} v \qquad
\frac{(\lambda -p)(\lambda -s)}{r} v = Qv.
\]
Thus $v$ is an eigenvector of $Q$ with eigenvalue 
\[
\mu = \frac{(\lambda -p)(\lambda -s)}{r}.
\]
Rewrite this as
\begin{equation}
\label{E:lambda_mu}
(\lambda -p)(\lambda -s)-r \mu = 0
\end{equation}
whose solutions are
\eqref{E:MeigCeig}. Now let $\lambda$ run through all eigenvalues of $M$.
\end{proof}

\begin{remark}\rm
This calculation does not use homogeneity as such---which 
requires all nodes to be input isomorphic. All it needs is
that $p_i = p, r_i = r, s_i = s$ for $1 \leq i \leq n$.  That is,
degradation and transcription rates are independent of the node, which (unlike homogeneity) does not impose conditions on the block matrix $Q$. This
assumption is often realistic and is common in the literature, in
particular in {\it E. coli} regulation dynamics.
\end{remark}

\section{Stability analysis of biologically relevant circuits}
\label{sec:saulo}

We now consider the GRN circuits in Fig. \ref{fig:stability_circuits}.
A full discussion can be found in \citep{stewart2024dynamics}, so we
present a representative selection of results and refer to that paper
for a more extensive analysis. We focus on linear stability of cluster
states induced by a fibration.

We also compare the Smolen circuit
with the metabolator circuit, which has the same base,
hence the same synchronous stability, but different
transverse stability.

\subsection{Stability analysis for lock-on}
\label{S:lock-on}

\begin{figure}[htb]
\centerline{
\includegraphics[width=.32\textwidth]{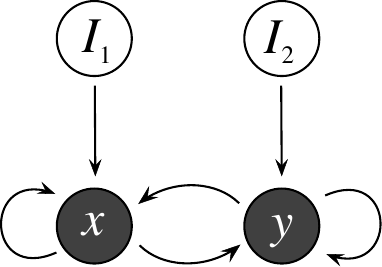} \qquad 
\includegraphics[width=.24\textwidth]{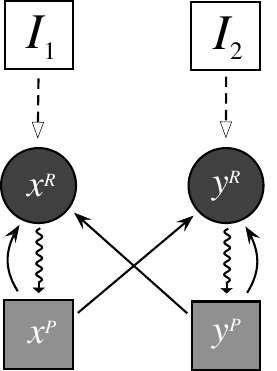} \qquad 
\includegraphics[width=.32\textwidth]{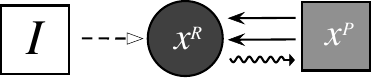}
}
\caption{\textbf{Lock-on circuit, with inputs.} 
{\em Left}: GRN. {\em Middle}: PRN. {\em Right}: Base of fibration when $I_1=I_2=I$,
shown by colors in Middle and Right figures.}
\label{F:LO}
\commentAlt{Figure~\ref{F:LO}: 
Three graphs.
On the left, a graph with two circular white nodes called I1 and I2, and two circular blue nodes called x and y. Directed arrows
connect I1 to x, I2 to y, x to itself and y, y to itself and x.
In the center, a graph with two rectangular white nodes called I1 and I2, two circular blue nodes called xR and yR, two rectangular red nodes called xP and yP.
Squiggly arrows connect xR to xP and yR to yP; directed arrows connect xP to xR and yR, yP to xR and yR; directed dashed arrows connect I1 to xR and I2 to yR.
On the right, a graph with one rectangular white node called I, one circular blue node called xR and one rectangular red node called xP.
A squiggly arrow connects the blue node to the red node; two directed arrows go from the red node to the blue node; one directed dashed arrow goes from I to the blue node.
}
\end{figure}

We begin with the {\em lock-on} circuit,\index{lock-on }
Fig.~\ref{F:LO}.  The figure includes {\em inputs} $I_1, I_2$.  For
a synchronous state to exist we require $I_1 = I_2 = I$, and we make
this assumption.  The network then has a group-theoretic symmetry that
swaps $x \leftrightarrow y$. Together with the identity permutation,
this forms the cyclic group $\Z_2$ of order 2.  

\subsubsection{Lock-on: general PRN model}

For consistency, we order the component equations as in Section~\ref{S:BS}; that
is, mRNA nodes first, followed by the corresponding protein nodes.
Within each equation, the arguments of the function concerned are listed
in an order that respects input isomorphisms. A bar over a set of variables indicates 
a vertex symmetry: the function is invariant under any permutation of those variables.

Admissible ODEs for the lock-on PRN are:
\begin{equation}
\label{E:LO_genRP}
\begin{array}{rcl}
\dot{x}^R & = & f(x^R, \overline{x^P,y^P},I) \\
\dot{y}^R & = &  f(y^R, \overline{x^P,y^P},I)\\
\dot{x}^P & = & g(x^P,x^R) \\
\dot{y}^P & = & g(y^P,y^R)
\end{array}
\end{equation}
in the coordinates $X= (x^R, y^R,x^P, y^P) \in \R^4$.  We treat $I \in
\R$ as a fixed but arbitrary parameter.  We call this the {\em general
  model} for the lock-on circuit. In \eqref{E:LO_genRP}
the first argument of functions $f$ and $g$ is the respective
degradation variable.

The Jacobian for \eqref{E:LO_genRP} is
\[
J = \Matrix{f_1 & 0 & f_2 & f_3 \\	
	0 & f_1 & f_2 & f_3 \\  g_2 & 0 & g_1 & 0 \\ 0 & g_2& 0 & g_1
	}
\]
where, as explained in Section \ref{S:PDJAM}, subscripts 
$f_i, g_i$ on $f$ and $g$ indicate the partial
derivative with respect to the $i$th variable.

The lock-on circuit has a nontrivial (group-theoretic) fibration indicated by the
colors in Fig.~\ref{F:LO}, which shows the orbit coloring 
 for the $\Z_2$ symmetry.
The corresponding synchrony subspace is
\[
\Delta = \{(x^R,y^R,x^P,y^P): x^R=y^R, x^P=y^P)\} = \{(u,u,v,v): u,v \in \R\} .
\]
The space $\Delta$ can be
identified with the state space for the base of the fibration, Fig. \ref{F:LO} (right).
We take the nodes of the base to be the two colors (blue, red)
and identify each point $(u,v)$ in its state space 
with $(u,u,v,v)$ in the state space of the lock-on PRN. The restricted ODE 
on $\Delta$ is admissible for the base.

The linear stability of the synchronous state is determined by the eigenvalues of
the Jacobian when it is evaluated at a synchronous equilibrium
\[
x^R=y^R = u_0 \qquad x^P=y^P = v_0
\]
which we denote by
\begin{equation}
\label{e:synceq}
X_0 = (u_0,u_0,v_0,v_0).
\end{equation}

At such a synchronous equilibrium, vertex symmetry implies that
$f_2=f_3$. This follows from the Taylor series expansion about the
equilibrium point concerned.  Consider the first term in equation
\ref{E:LO_genRP}. The linear part of the Taylor series is
$f_1x^R+f_2x^P+f_3y^P$.  However, the function is symmetric when we
interchange the second and third variables, which gives
$f_1x^R+f_2y^P+f_3x^P$.  These expressions must be equal, so $f_2
=f_3$ when evaluated at the synchronous state $X_0$. (This observation
is a special case of a more general result, which we do not address here.)

To simplify the notation further, let
\[
a = f_1 \quad b = f_2=f_3 \quad c = g_1 \quad d = g_2.
\]
We emphasize that the entries $a,b,c,d$ are not constants: they are functions of the synchronous equilibrium~\eqref{e:synceq}. The Jacobian is
\begin{equation}
\label{E:LOJ}
J = \Matrix{aI & Q \\  dI & bI} 
	\qquad \mbox{where} \qquad Q = \Matrix{b & b \\ b & b}
\end{equation}
in accordance with \eqref{E:Meqn}. (Here $I$ is the $2 \times 2$ identity 
matrix, not an input parameter.)
The eigenvalues of $Q$
are $\mu = 0, 2b$. By Theorem~\ref{T:MeigCeig} the eigenvalues of $J$ are:
\begin{equation}
\label{E:LO_eigen}
\begin{array}{rcl}
\lambda_1 &=& a \\
\lambda_2 &=& c \\
\lambda_3^\pm &=& \shf \left(a + c \pm \sqrt{K} \right) 
\end{array}
\end{equation}
where
\begin{equation}
\label{E:Q}
K = (a-c)^2 + 8 bd.
\end{equation}
Here $\lambda_1$ and $\lambda_2$ are real, but $\lambda_3^\pm$ can be
complex, depending on the sign of $K$. They are real if
$K\geq 0$ and a complex conjugate pair if $K<0$.

Computing the eigenvectors for the eigenvalues $\lambda_3^\pm$ shows that
these eigenvectors lie in the invariant synchrony subspace
$\Delta$. Therefore these are the synchronous
  eigenvalues. In contrast, $\lambda_1,\lambda_2$ have
eigenvectors in a complement to $\Delta$ and are 
 the transverse eigenvalues. 
  
We claim that the equilibrium $X_0$ is linearly stable if and only if,
when evaluated at $X_0$, the partial derivatives $a,b,c$ satisfy one
of the following conditions:
\begin{itemize}
\item[\rm (1)] $a < 0, c < 0, a+c+\sqrt{(a-c)^2+8bd} > 0, (a-c)^2+8bd > 0$,
\item[\rm (2)] $a < 0, c < 0, (a-c)^2+8bd < 0$.
\end{itemize}
The proof is a short series of simple calculations. Let
$K=(a-c)^2+8bg$ as in \eqref{E:Q}. Linear stability is equivalent to
all eigenvalues $\lambda_1, \lambda_2,\lambda_3^+, \lambda_3^-$
having negative real parts. Therefore $a<0$ and $c<0$. 

If $K >0$ then $\lambda_3^+$ and$  \lambda_3^-$ are real.
Since $\lambda_3^+> \lambda_3^-$, stability requires $\lambda_3^+< 0$.
(The factor $\shf$ in $\lambda_3^\pm$ does not affect the sign.)

If $K <0$ then $\lambda_3^+$ and $\lambda_3^-$ are complex conjugate,
with real part $\shf(a+c)$. Since $a,c <0$ this is automatically negative.
This proves the claim.

The values of $a,b,c,d$ that can occur in a specific model depend on 
the model.

\subsubsection{Lock-on: special PRN model}

For biological applications, the functional forms of
the functions $f, g$ in \eqref{E:LO_genRP}
are normally chosen from a standard range. A common choice,
used here because it is simple and illustrates the analysis, leads to
the special model:
\begin{equation}
\label{E:LO_specRP}
\begin{array}{rcl}
\dot{x}^R & = & -\delta x^R + H^+(x^P) + H^+(y^P) + I \\
\dot{y}^R & = &  -\delta y^R + H^+(x^P) + H^+(y^P) + I\\
\dot{x}^P & = & -\alpha x^P + \beta x^R \\
\dot{y}^P & = & -\alpha y^P + \beta y^R.
\end{array}
\end{equation}
The parameters $\alpha, \delta > 0$ represent degradation of mRNA and
protein respectively. The parameter $\beta > 0$ is the rate at which
the mRNA produces protein. We use the same Hill function $H^+$ 
for both activator arrows,
since these give identical couplings. This reflects the vertex symmetry
of $f$ indicated by the overline.

Consider the synchronous equilibrium $X_0$ in \eqref{e:synceq},
for which $x^R=y^R=u_0, x^P=y^P=v_0$.
The quantities $u_0$ and $v_0$ depend on the input $I$, so $I$ can
act as a bifurcation parameter.

For \eqref{E:LO_specRP} we have
\[
a = -\delta < 0 \qquad b = H^{+\prime}(v_0) > 0 \qquad c = -\alpha < 0 \qquad d = \beta > 0.
\]
where $H^{+\prime}$ indicates the derivative of $H^{+}$.
The sign of $b$ is always positive because $H^+$ is monotonic increasing,
reflecting the activation arrows.

The transverse eigenvalues are $-\delta, -\alpha < 0$, in accordance with
the stability criteria (1) and (2) above.
Thus the dynamic transverse to the synchrony subspace
is attracting; this is caused by the degradation terms. Indeed, here the transverse dynamic is {\em constant}: independent of the
equilibrium point in the synchrony space. 
The quantity $K =  (a-c)^2+8bd$ is always positive, so we apply 
criterion (1): stability also requires $a+c+\sqrt{K} < 0$. This is
equivalent to $\sqrt{K} < -a-c$, that is, $K < (a-c)^2$.
Therefore $bd <0$, but this does not occur. 

We conclude that the synchronous equilibrium is unstable
for this choice of model.

\subsection{Bistability of the toggle-switch}
\label{S:bi_tog}

\begin{figure}[h!]
\centerline{%
\includegraphics[width=0.3\textwidth]{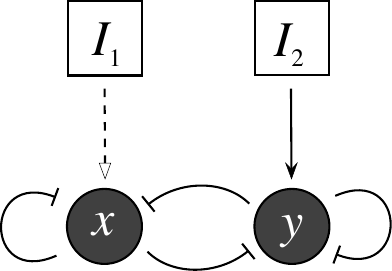} \qquad
\includegraphics[width=0.23\textwidth]{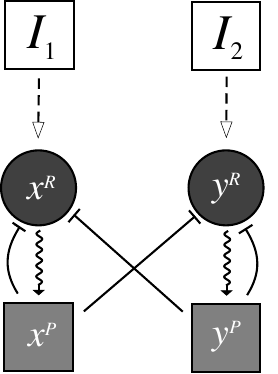} \qquad
\includegraphics[width=0.32\textwidth]{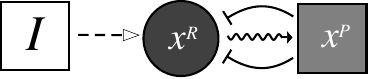}
}
\caption{\textbf{Toggle-switch.} {\em Left:} GRN with inputs. {\em Middle}: PRN with inputs. {\em Right}: Base.}
\label{F:toggle_bio_math_input}
\commentAlt{Figure~\ref{F:toggle_bio_math_input}: 
Three graphs.
On the left, a graph with two circular white nodes called I1 and I2, and two circular blue nodes called xR and yR. Directed dashed arrows
connect I1 to xR and I2 to yR; inhibition xR to itself and yR, yR to itself and xR.
In the center, a graph with two rectangular white nodes called I1 and I2, two circular blue nodes called xR and yR, two rectangular red nodes called xP and yP.
Squiggly arrows connect xR to xP and yR to yP; inhibition arrows connect xP to xR and yR, yP to xR and yR; directed dashed arrows connect I1 to xR and I2 to yR.
On the right, a graph with one rectangular white node called I, one unnamed circular blue node and one unnamed rectangular red node.
A squiggly arrow connects the blue node to the red node; two inhibition arrows go from the red node to the blue node; one directed dashed arrow goes from I to the blue node.
}
\end{figure}

The toggle-switch\index{toggle-switch } is shown in Fig. \ref{F:toggle_bio_math_input}.
The general model for the toggle-switch is the same as for the
lock-on circuit, because the network formalism used to determine admissible
ODEs does not distinguish activator arrows from repressor arrows;
and each network has only one type of arrow.

Rather than repeating the analysis using a special model analogous
to that for the toggle-switch but with $H^+$ replaced by $H^-$, we focus on
the most interesting feature of the toggle-switch, and the reason for its name.
It  is {\em bistable}, and can be switched between two stable equilibria. 
This is discussed thoroughly in \citep{gardner2000construction}, using the following dimensionless model:
\begin{equation}
\label{E:gardner}
\dot u = \displaystyle \frac{\alpha_1}{1+v^\beta} - u +I_1\qquad 
\dot v =\displaystyle \frac{\alpha_2}{1+u^\gamma} - v+I_2 .
\end{equation}
Here $u, v$ are the concentrations of the two repressors, 
the $\alpha_i$ are lumped parameters representing their effective synthesis rates, $\beta$ is
the cooperativity of repression for promoter 2, and $\alpha$ is
the cooperativity of repression for promoter 1. The terms $I_1,I_2$ are
external inputs.

We analyze their model in the case where $\beta=\gamma=2$ and 
$I_1=I_2=0$. For notational simplicity we set 
$\alpha_1=a, \alpha_2=b$. 
There is a fibration symmetry if and only if $\alpha_1 = \alpha_2$
and $\beta=\gamma$, when the symmetry group is $\Z_2$.
When $\alpha_1 \neq \alpha_2$ we have `induced' symmetry breaking. 

The ODE becomes
\begin{equation}
\label{E:gardner2}
\dot u = \displaystyle \frac{a}{1+v^2} - u \qquad
\dot v =\displaystyle \frac{b}{1+u^2} - v .
\end{equation}

\subsubsection{Restriction to synchrony subspace}

We first analyze \eqref{E:gardner2} on the synchrony space where $u=v, a=b$,
but considering also transverse eigenvalues.
The equation becomes
\begin{equation}
\label{E:gardner_synch}
\dot u = \displaystyle \frac{a}{1+u^2} - u \quad 
	\mbox{with equilibria when}  \quad u^3+u=a.
\end{equation}
Since the cubic $u^3+u$ is monotone increasing and passes through
the origin, there is a unique solution $u_0(a)$ for each value of $a$. This solution
is positive when $a>0$, and $u_0(a)$ increases monotonically with $a$.
We write it as $u_0$, remembering that this varies with $a$.

The Jacobian for \eqref{E:gardner_synch} at $u_0$ is the $1 \times 1$ matrix
\[
\hat J = \Matrix{-1-\frac{2au_0}{(1+u_0^2)^2}} = -1-\frac{2u_0^3}{a}
\]
since $(1+u_0)^2 = \frac{a}{u_0} $ by \eqref{E:gardner_synch}.
When $a >0$ this is always negative, so the equilibrium $u_0$
is stable with respect to synchrony-preserving perturbations, for all
values of $a$.

The Jacobian for \eqref{E:gardner2} is 
\[
J = \Matrix{-1 & \frac{-2av}{(1+v^2)^2} \\ \frac{-2bu}{(1+u^2)^2} &-1}
\]
Evaluated at a synchronous equilibrium $(u_0,u_0)$ when $a=b$ this becomes
\[
J|_{(u_0,u_0)} = \Matrix{-1 & \frac{-2au_0}{(1+u_0^2)^2} \\ \frac{-2au_0}{(1+u_0^2)^2} &-1}
\]
Being a symmetric matrix, this has real eigenvalues. To compute them, we
observe that $(1+u_0)^2 = \frac{a}{u_0} $ by \eqref{E:gardner_synch}, so
\[
J|_{(u_0,u_0)} = \Matrix{-1 & \frac{-2u_0^3}{a} \\  \frac{-2u_0^3}{a} &-1}
\]
The eigenvalues are
\[
\lambda_1 = -1+\frac{2u_0^3}{a} \qquad \lambda_2 = -1-\frac{2u_0^3}{a}
\]
with corresponding eigenvectors $[1,-1]^{\rm T}$ and  $[1,1]^{\rm T}$.
Therefore the transverse eigenvalue is $\lambda_1$ and the synchronous
 eigenvalue is $\lambda_2$. Since $\lambda_1>\lambda_2$, the first eigenvalue to change sign
 as $u_0$ increases is $\lambda_1$, so the synchronous state
 loses stability when $2u_0^3/a = 1$. Now 
 $a = 2u_0^3 = 2(a-u_0)$, so $u_0=a/2$, which implies that 
$(a/2)^3+(a/2)=a$, whence $a = \pm 2$. The positive value is $a=2$,
and this implies that $u_0=1$. Thus the
synchronous state is stable in the full state space if $a<2$,
but becomes unstable to synchrony-breaking perturbations 
when $a>2$.

\subsubsection{Full state space}

\begin{figure}[h!]
\centerline{%
\includegraphics[width=0.2\textwidth]{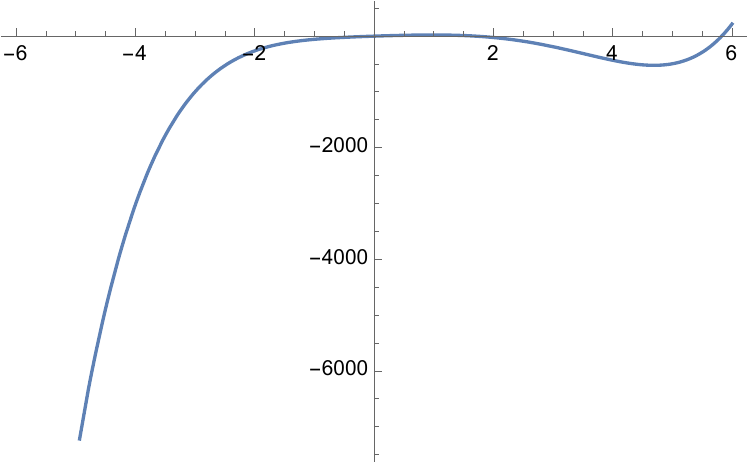} \qquad
\includegraphics[width=0.2\textwidth]{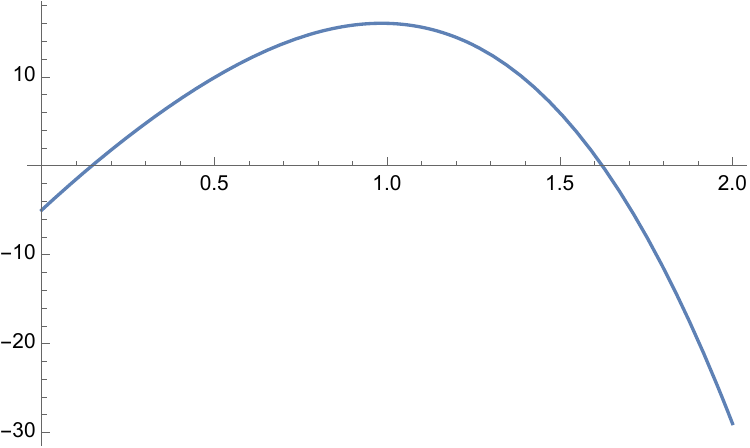} \qquad
\includegraphics[width=0.2\textwidth]{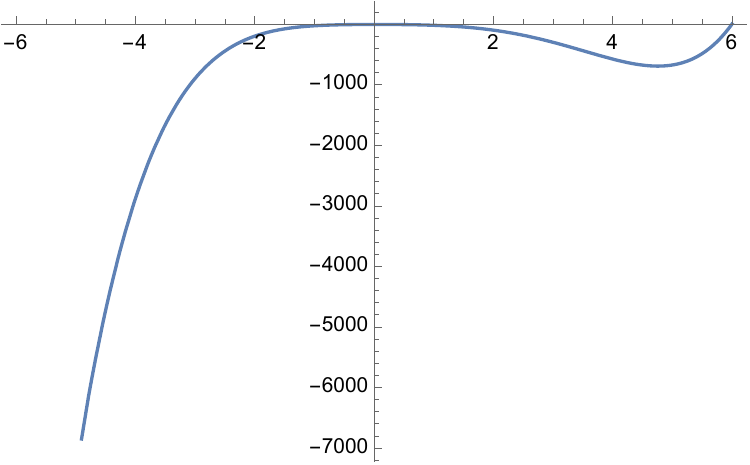} \qquad
\includegraphics[width=0.2\textwidth]{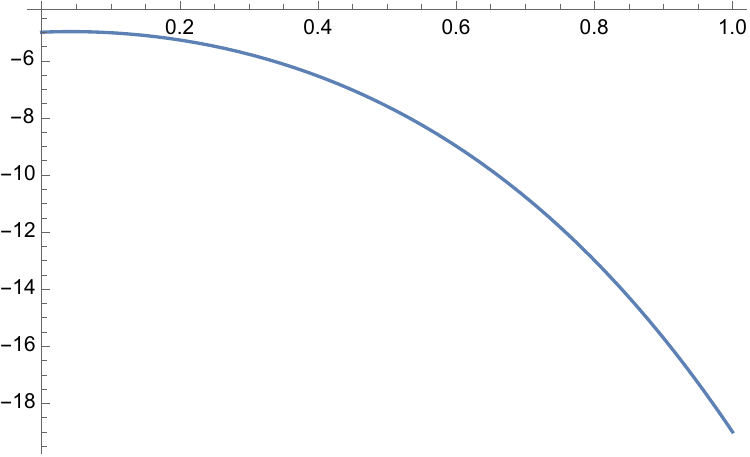} 
}
\caption{\textbf{Plots of stability.} {\em From left to right}: Graph of $p(u)$ when $a=b=6$. Blowup near $u=0$
to show two of the three zeros. Graph of $p(u)$ when $a=6, b=1$. Blowup near $u=0$
to show there are no zeros in that region.}
\label{F:quintics}
\commentAlt{Figure~\ref{F:quintics}: 
Four plots of curves on the Cartesian plane. All are graphs of the quintic polynomial
p(u) = \ref{E:toggle_cusp1} for different values of parameters a, b as stated in caption.
Plot 1: The curve increases from negative values on the left. It crosses the
x-axis up and then down at two nearby points, then increase again to cut it a third time.
So there are three zeros.
Plot 2: This is a blow-up of the graph showing behavior near the two points that
are close together.
Plot 3: The curve increases from negative values on the left. It comes close to the
x-axis but does not meet it, then increases again to cut it for the first time.
So there is one zero.
Plot 4: This is a blow-up of the graph showing behavior near the x-axis, 
where the curve comes close but does not cross  the x-axis.
}
\end{figure}

We now extend the analysis of \eqref{E:gardner2} to the full state space
and characterize the equilibria and their stabilities. At equilibrium,
\[
u = \frac{a}{1+v^2} \qquad v = \frac{b}{1+u^2}
\]
so
\[
u(1+v^2) = a \qquad v(1+u^2)= b.
\]
Eliminate $v$ using the second equation: the first becomes
\[
u\left( 1+\left[ \frac{b^2}{(1+u^2)^2}\right]\right) = a.
\]
Multiply throughout by $(1+u^2)^2$:
\[
u(1+u^2)^2+ub^2 = a(1+u^2)^2
\]
which expands to give the 5th degree polynomial equation
\begin{equation}
\label{E:toggle_cusp1}
p(u) = u^5-a u^4 + 2u^3 - 2a u^2 + (b^2+1)u - a = 0.
\end{equation}
Plotting the polynomial $p(u)$ numerically for various values of $a,b$ 
we find that sometimes it has one real zero, and sometimes three real zeros.
Fig. \ref{F:quintics} shows typical plots. Since the curves are close to the 
horizontal axis near the origin, blow-ups near $u=0$ clarify the number of zeros.

Applying a standard viewpoint in singularity/catastrophe theory, we
plot the bifurcation variety in $(a,b)$-space. This
is the set of values of $(a,b)$ for which the polynomial
has a multiple zero. It divides $(a,b)$-space into regions
where the number of roots is 3 or 1. The condition for a multiple zero is that
$u$ is also a zero of the derivative:
\begin{equation}
\label{E:toggle_cusp2}
 p'(u) = 5u^4 -4a u^3 - 6 u^2 -4a u + (b^2+1) = 0.
\end{equation}
Rewrite equations \eqref{E:toggle_cusp1} and \eqref{E:toggle_cusp2} as
\beqn
a(u^4+2u^2+1) &=& u^5+2u^3+(b^2+1)u, \\
a(4u^3+4u) &=& 5u^4+6u^2+(b^2+1) .
\eeqn
These are linear in $a$ and $\beta^2$, so we solve
explicitly and take the square root of $b^2$ to get
\begin{equation}
\label{E:toggle_cusp3}
a = \frac{4u^3}{3u^2-1},\qquad
b = \sqrt{\frac{(u^2+1)^3}{3u^2-1}}.
\end{equation}

\begin{figure}[h!]
\centerline{%
\includegraphics[width=0.35\textwidth]{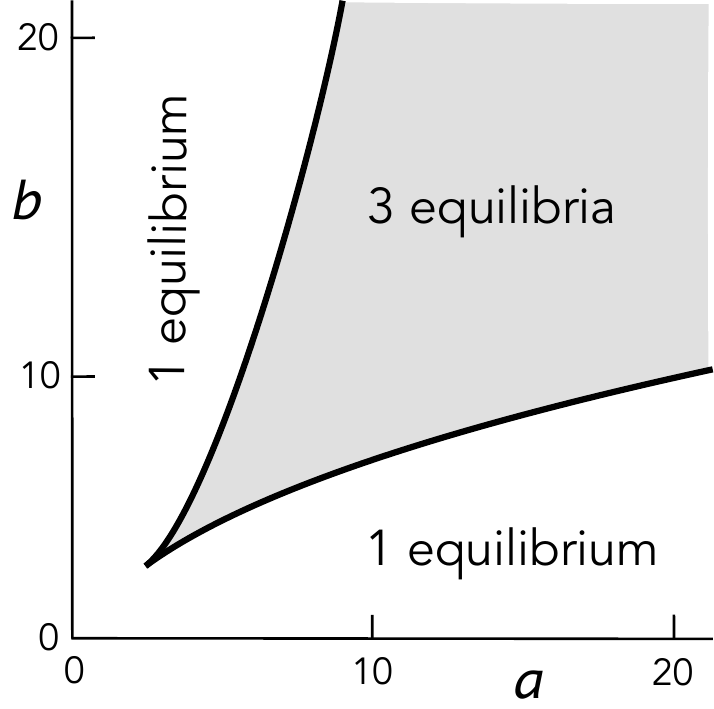} \qquad\qquad
\includegraphics[width=0.3\textwidth]{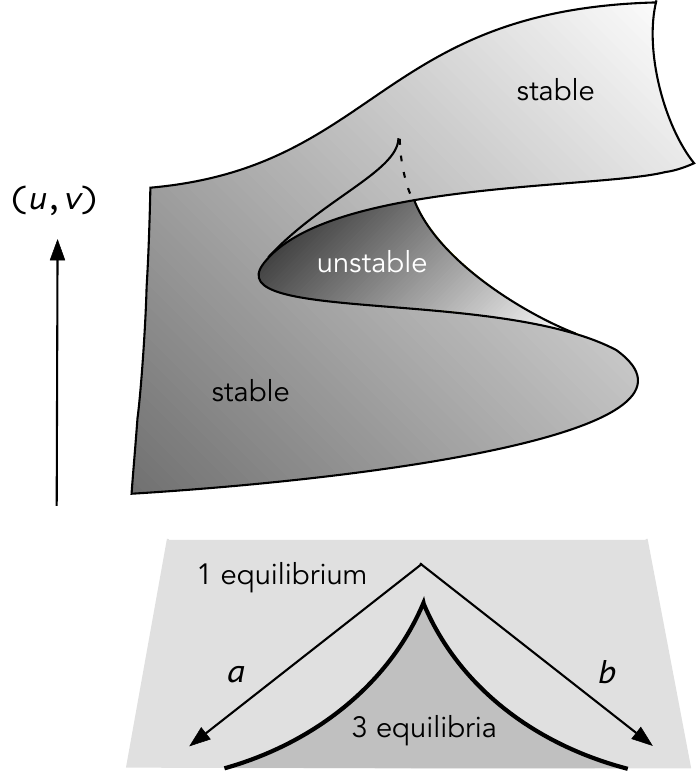}
}
\caption{\textbf{Singularity/catastrophe theory.} {\em Left}: Bifurcation variety in $(a,b)$-space. {\em Right}: Cusp catastrophe surface showing how equilibria vary with parameters $(a,b)$. (Schematic to show topology: the vertical axis is 2-dimensional.) }
\label{F:bif_var}
\commentAlt{Figure~\ref{F:bif_var}: 
Left image: Plane with coordinates (a,b). A shaded V-shaped region with a sharp cusp
is outlined by formed by two symmetrically related curves either side of the diagonal.
Inside this region is label `3 equilibria'. Outside `1 equilibrium'.
right image: a 3-dimensional plot of a surface. Below is the plane of the
left image.  Above is a surface which slopes upwards form left to right at the rear,
but is S-shaped at the front. The overall effect in like a pleat in a piece of cloth.
Three layers of the surface lie above points in the shaded region; only
one layer above points outside it.
}
\end{figure}

Now we can consider \eqref{E:toggle_cusp3}
as a parametric representation of the {\em bifurcation variety} in $(a,b)$-space, using $u$ as parameter---a convenient way to plot the curves.
Figure \ref{F:bif_var} (left) shows the result, where we retain only positive values
of $a$ and $b$.  \cite{gardner2000construction} obtain similar cusped curves for
other values of $\beta,\gamma$ in \eqref{E:gardner}.
This figure is symmetric about the diagonal $a=b$, due to the
`parameter symmetry' $u \leftrightarrow v$, $\alpha_1 \leftrightarrow \alpha_2$,
$I_1 \leftrightarrow I_2$, $\beta \leftrightarrow \gamma$
of \eqref{E:gardner}.

The shaded region inside the cusped curve corresponds to
a bistable state with two stable equilibria and one unstable one.
Outside the cusped curve there is a unique equilibrium, which is stable.
The geometry  is that of the cusp catastrophe 
 \citep{stewart2014, zeeman1977}; see Fig. \ref{F:bif_var} (right). The cusped curve
indicates transitional values of the parameters $(\alpha_1,\alpha_2)$
at which a stable equilibrium merges with an unstable one. 
With quasi-static dynamics (the equilibrium state changes continuously as parameters
vary, except for discontinuous jumps when that state disappears) the system
jumps between the upper and lower sheets of the surface
at the folds, and thus exhibits
 hysteresis, leading to the characteristic toggle-switch behavior
of this circuit.

The intersection of the diagonal line $\alpha_1 = \alpha_2$ with the cusp catastrophe
surface gives a pitchfork bifurcation: see Section \ref{S:bifurcations}. 
This is symmetric under the map
$(u,v) \mapsto (v,u)$  but appears asymmetric if projected onto $u$ or $v$.
There are intervals of uni- and bi-stability
but no hysteresis. Hysteresis requires asymmetric parameter values.

\subsection{Comparison of metabolator and Smolen}
\label{S:metab+smol}

\begin{figure}[htb!]
\centerline{%
\includegraphics[width=0.3\textwidth]{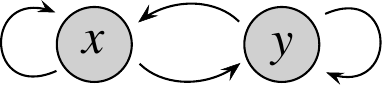}\qquad\qquad
\includegraphics[width=0.3\textwidth]{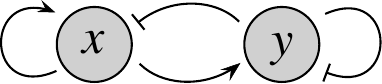}
}
\caption{\textbf{Sample circuits with the same base.} {\em Left}: Metabolator network. {\em Right}: Smolen
circuit.}
\label{F:metab+smolen}
\commentAlt{Figure~\ref{F:metab+smolen}: 
Two graphs, both with two blue nodes called x1 and x2.
The graph on the left has two inhibition arrows from each node to itself, and two directed arrows connecting x1 to x2 and x2 to x1.
The graph on the right has two inhibition arrows from x1 to itself and from x1 to x2, and two directed arrows from x2 to itself and from x2 to x1.
}
\end{figure}

To illustrate how the transverse eigenvalues affect the stability of a synchronous
state, when this state is stable on the synchrony subspace, we consider
the Smolen circuit\index{Smolen circuit } and the similar but subtly different metabolator\index{metabolator }
circuit, Fig. \ref{F:metab+smolen}.

In this section we use a `lumped' model with a single variable for each node, 
rather than the two variables of a PRN model, since this illustrates the main
point in a simple manner. The node variables are then $x_1,x_2 \in \R$ for each circuit.

The admissible ODEs for the metabolator are
\begin{equation}
\label{E:Z2net1metab}
\begin{array}{rcl}
\dot{x}_1 &=& f(x_1,x_1,x_2) \\
\dot{x}_2 &=& f(x_2,x_2,x_1)
\end{array}
\end{equation}
where the function $f$ is arbitrary. The variables inside $f$ are
taken in the order: node, repressor arrow, activator arrow.

For the Smolen circuit, the admissible ODEs are
\begin{equation}
\label{E:Asymnet1smo}
\begin{array}{rcl}
\dot{x}_1 &=& f(x_1,x_1,x_2), \\
\dot{x}_2 &=& f(x_2,x_1,x_2).
\end{array}
\end{equation}
If we set $x_1=x_2=x$, a balanced synchrony pattern, both
ODEs reduce to
\[
\dot x = f(x,x,x).
\]
Using the same $f$ is both models, they have identical synchronous
dynamics. However, the stability of
the synchronous state to synchrony-breaking perturbations---transverse
stability---can be different.

Assume that \eqref{E:Z2net1metab} has a synchronous equilibrium point at $x^*=(x_0,x_0)$
for some $y_0 \in \R$. Then \eqref{E:Asymnet1smo} also has a synchronous equilibrium point at $x^*$,
since the equations are identical. Such an equilibrium is linearly stable if all eigenvalues of
the Jacobian at $x^*$ have negative real parts. By the chain rule, 
the Jacobians $J_M$ for the metabolator and $J_S$ for the Smolen network are
\begin{eqnarray*}
J_M &=& \Matrix{\partial_1 f+\partial_2 f & \partial_3 f \\
	\partial_3 f & \partial_1 f+\partial_2 f}_{x^*} 
		\quad= \quad \Matrix{\alpha+\beta & \gamma \\ \gamma & \alpha+\beta}
\\
J_S &=& \Matrix{\partial_1 f+\partial_2 f & \partial_3 f \\
	\partial_2 f & \partial_1 f+\partial_3 f}_{x^*} 
		\quad= \quad \Matrix{\alpha+\beta & \gamma \\ \beta & \alpha+\gamma}
\end{eqnarray*}
(say), 
 where the subscript indicates evaluation at $x^*$. The corresponding eigenvalues
 are easily seen to be
\begin{eqnarray*}
\mbox{metabolator:} &:& \alpha + \beta + \gamma \qquad \alpha + \beta - \gamma, \\
\mbox{Smolen:} &:& \alpha + \beta + \gamma \qquad \alpha .
\end{eqnarray*}
  
The condition for stability to synchrony-preserving perturbations is 
$\alpha + \beta + \gamma < 0$, which is the same for both networks.
This is the synchronous eigenvalue.

\begin{figure}[h!]
\centerline{%
\includegraphics[width=0.3\textwidth]{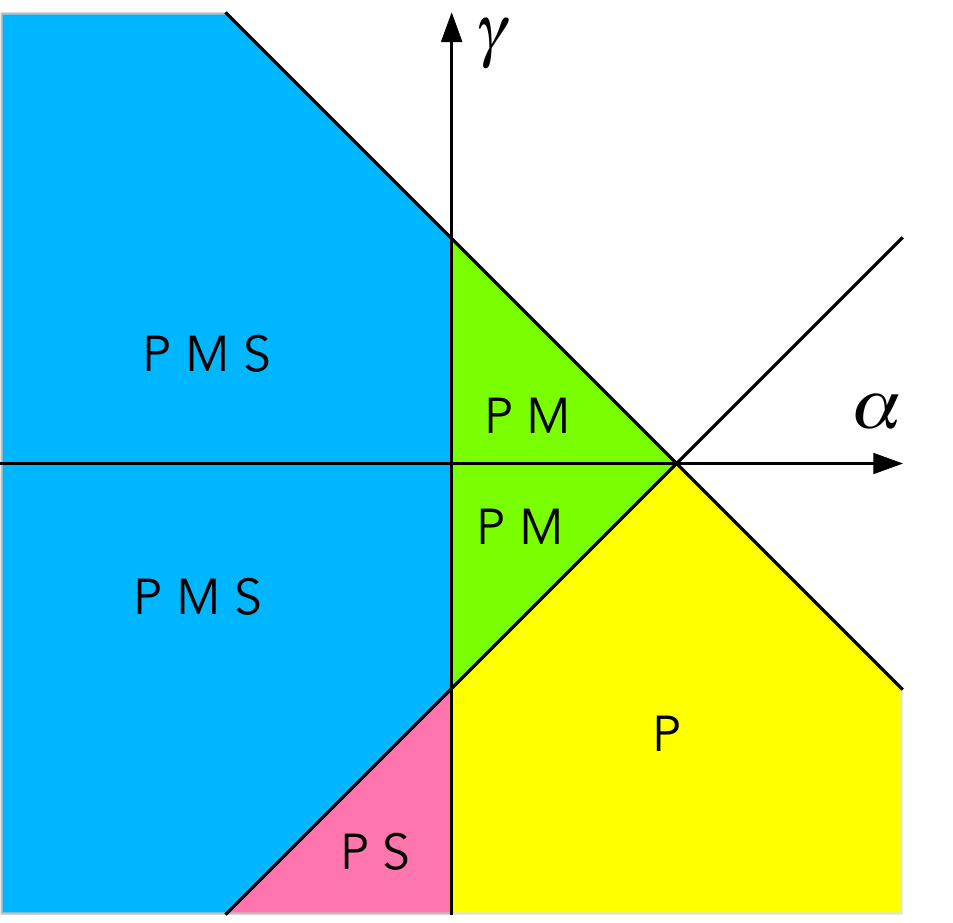}
\quad
\includegraphics[width=0.3\textwidth]{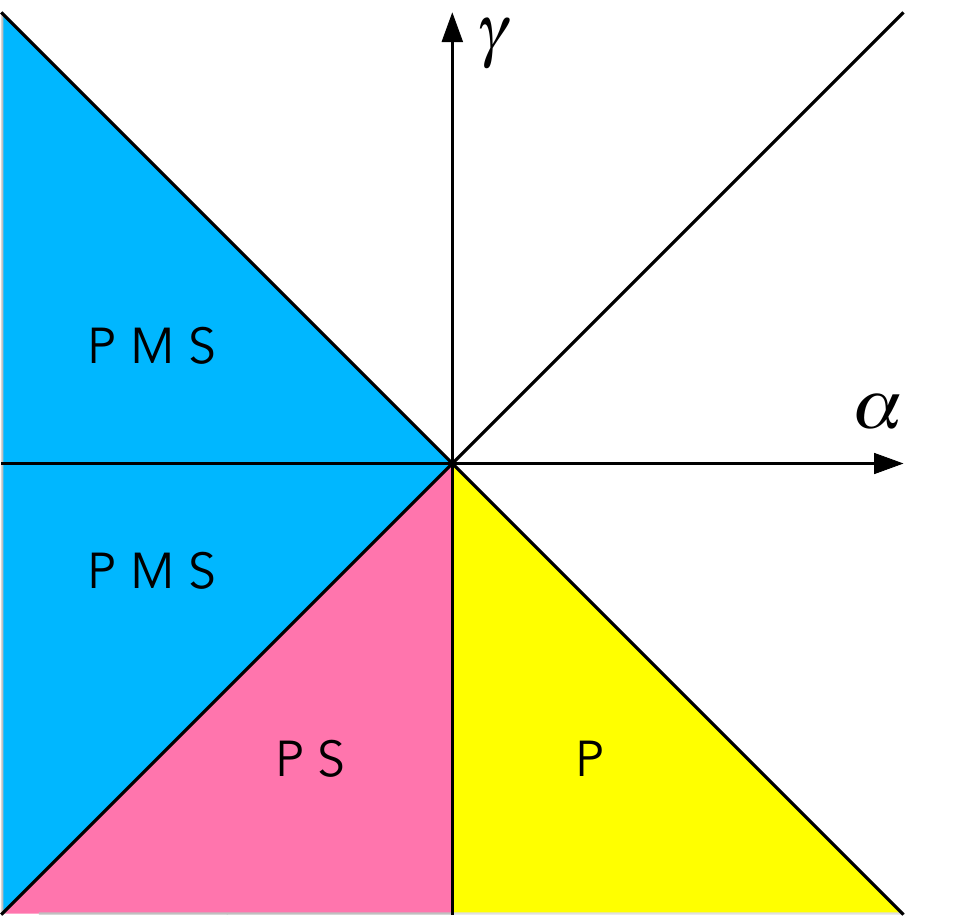}
\quad
\includegraphics[width=0.3\textwidth]{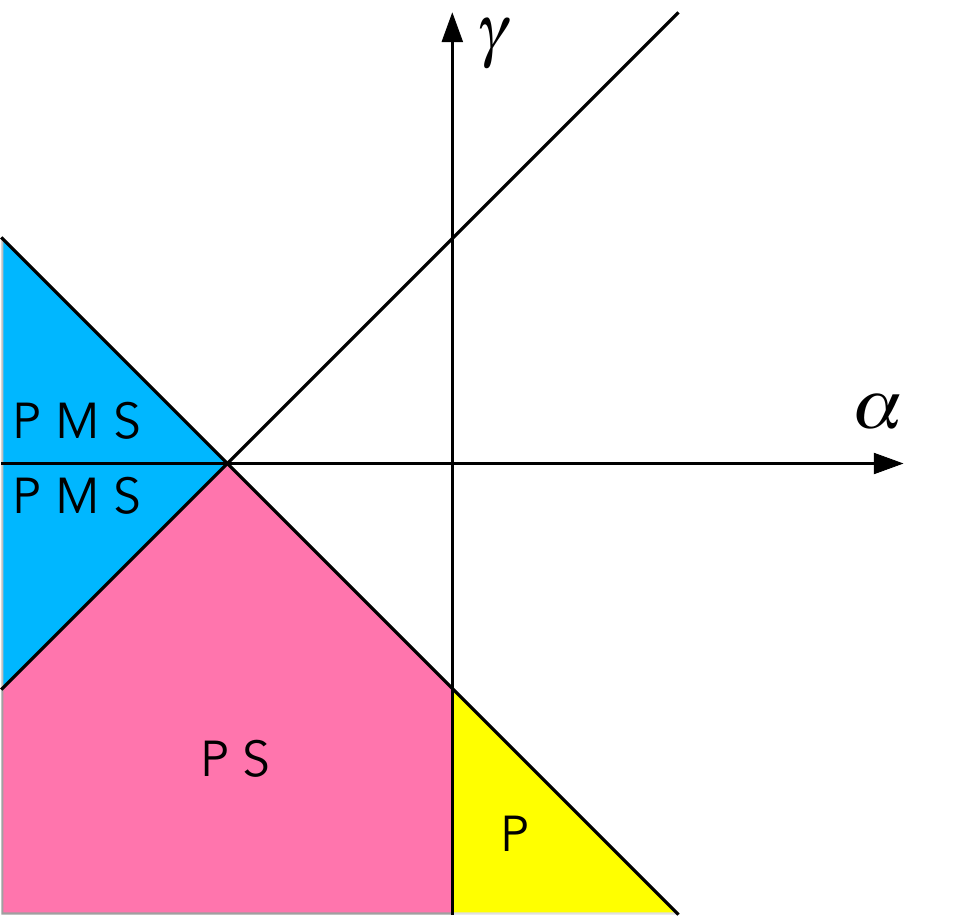}
}
\caption{\textbf{Stability regions in the $(\alpha,\gamma)$ plane.} P = both
  metabolator and Smolen stable to synchrony-preserving perturbations;
  M = metabolator also stable to synchrony-breaking perturbations; S =
  Smolen network also stable to synchrony-breaking perturbations.
  {\em Left}: $\beta < 0$. {\em Middle}: $\beta = 0$. {\em Right}:
  $\beta > 0$.}
\label{F:MSstab}
\commentAlt{Figure~\ref{F:MSstab}: 
Three two-dimensional plots with coordinates alpha, gamma.
showing which combinations of these parameters give stable 
equilibria for the metabolator and Smolen networks.
}

\commentLongAlt{Figure~\ref{F:MSstab}: 
Plot 1: Divided by three lines: y-axis, 45 degree angle sloping upward
passing through a point on the alpha-axis to the right of the origin;
45 degree angle sloping downward
passing through the same point on the alpha-axis. These lines divide
the plane into five regions. From left to right and top to bottom these
are labeled  `empty' (white), PMS (blue), PS (pink), PM (green), P (yellow).
Plot 2: Again divided by three lines, but now the sloping ones pass through the origin.
The green PM region disappears,
so only four regions remain.
These regions have labels and colors as before
from left to right and top to bottom these
are labeled  `empty' (white), PMS (blue), PS (pink), P (yellow).
Plot 3: Again divided by three lines, but now the sloping ones pass through
a point on the alpha-axis to the left of the origin. Only four regions remain:
from left to right and top to bottom these
are labeled  `empty' (white), PMS (blue), PS (pink), P (yellow).
}
\end{figure}

Assuming stability to synchrony-preserving perturbations, 
an additional condition for stability to synchrony-breaking perturbations is required.
This is the transverse eigenvalue
\begin{eqnarray*}
\mbox{metabolator:} &:&  \alpha + \beta - \gamma < 0,\\
\mbox{Smolen:} &:& \alpha < 0 ,
\end{eqnarray*}
and these conditions are different. Figure  \ref{F:MSstab} sketches the
regions of the $(\alpha,\gamma)$ plane in which various stability conditions
hold, for $\beta < 0$, $\beta = 0$, and $\beta > 0$. The regions are as follows:

P: Stable to synchrony-preserving perturbations (same condition for
metabolator and Smolen).

M: Satisfying P, and also stable to synchrony-breaking perturbations
(hence stable) for metabolator.

S: Satisfying P, and also stable to synchrony-breaking perturbations
(hence stable) for Smolen.

For suitable choices of $\alpha, \beta, \gamma$, it is possible for
both networks, either one, or neither to be stable. Both can be stable
or both can be unstable to synchrony-preserving perturbations. It is
not possible for one to be stable to synchrony-preserving
perturbations while the other is unstable to synchrony-preserving
perturbations, since the restricted ODE is the same in both cases.

In general the stability of synchrony patterns depends in complicated
ways on the model ODE, especially for non-equilibrium states. There
are few useful general principles, although some results exist for
regular, symmetric, and feed-forward networks \citep[Chapter
  18]{GS2023}, or with mild extra assumptions on the model ODE
\citep{HCLP09}.

\section{Bifurcations---a short sketch and examples}
\label{S:bifurcations}

We now move on from specific networks and models to 
give a brief summary of the commonest bifurcations.\index{bifurcation } These are the
`local' bifurcations, where every important change happens in
any sufficiently small neighborhood of the bifurcation point.

In a one-parameter family of ODEs, which is common in applications,
it is possible to track how a given
state changes as the parameter $\lambda$ varies. For `most' parameter values, this state
usually varies continuously with the parameter and retains its dynamical
character---steady, periodic, chaotic, and so on.
However, there can exist {\em bifurcation points} $\lambda=\lambda_0$
at which the type of the state, or the number of possible states, changes.
The system then adopts a different state.  Such a change is called a {\em bifurcation}.\index{bifurcation }
The main cause of
bifurcation is {\em loss of stability} of the state being tracked.
Typically, the values of $\lambda$ at which a bifurcation occurs are isolated.

Bifurcations provide a powerful method for analyzing stabilities,
because they split the parameter values into intervals along which
the stability remains the same. (When there is more than one parameter,
similar ideas apply, but now bifurcations typically occur on
`codimension 1' subsets of parameter space---that is on curves
for two parameters, surfaces for three, and dimension $d-1$ submanifolds
for $d$ parameters.) In Fig. \ref{F:MSstab} the straight lines bounding the
various regions are sections of planes,
and these planes define values of $(\alpha,\beta,\gamma)$
at which bifurcations from synchronous equilibrium occur. In general
these `bifurcation varieties' can be curved.

In dynamical systems theory there are many standard `generic' bifurcation scenarios
 \citep{guckenheimer1983}. Networks introduce new complexities,
because behavior that is generic in a general dynamical system need not
be generic in a network dynamical system. More precisely, the
structure of an admissible ODE for a network has a strong influence on
the possible bifurcations. 
The relation between the possible bifurcations and the graph topology
is, in general, subtle and complex \citep{leite2006}. However, a few useful general principles
apply \citep[Chapters 3,6, 18--22]{GS2023}.

\begin{definition}{\bf Bifurcation.} 
A {\em bifurcation}\index{bifurcation } occurs in a $\lambda$-parametrized family of ODEs
$\dot x = f(x,\lambda)$ 
when the set of attractors changes its topological type as the parameter $\lambda
\in \R^n$ varies. The parameter can be multidimensional ($n > 1$) but
is commonly 1-dimensional ($\lambda\in\R$).  

{\em Local bifurcation}\index{bifurcation !local }
occurs when a parametrized family of equilibria loses stability,
leading to one or more branches of new solutions.

An eigenvalue is {\em critical}\index{eigenvalue !critical } if it has zero real part.
\end{definition}

There are two kinds of generic local bifurcation:
\begin{itemize}
\item[\rm (a)] {\em Steady-state} bifurcation:\index{bifurcation !steady-state } a real eigenvalue of
  the Jacobian (linearization) passes through $0$ as $\lambda$ passes through
  some value $\lambda_0$.  The generic
  bifurcation is saddle-node\index{bifurcation !saddle-node } for a general dynamical system,
  transcritical\index{bifurcation !transcritical } if the family has a branch of trivial solutions, and
  pitchfork\index{bifurcation !pitchfork } if the system has a symmetry of order 2. Pitchforks and higher singularities can occur generically if $n>1$,
  and symmetry is not essential~\citep{golubitsky1988}.

\item[\rm (b)] {\em Hopf} bifurcation:\index{bifurcation !Hopf } a pair of complex conjugate
  eigenvalues of the Jacobian crosses the imaginary axis as $\lambda$ passes through
  some value $\lambda_0$. Typically
  this causes a new branch of time-periodic states, with frequency
  tending to $\frac{2\pi}{\omega}$ at the bifurcation point, where the
  imaginary eigenvalues are $\pm \omega i$.  See~\citep{hassard1981}.
\end{itemize}
In either case, $\lambda_0$ is the {\em bifurcation point} concerned.

\begin{figure}[h!]
\centerline{%
\includegraphics[width=.8\textwidth]{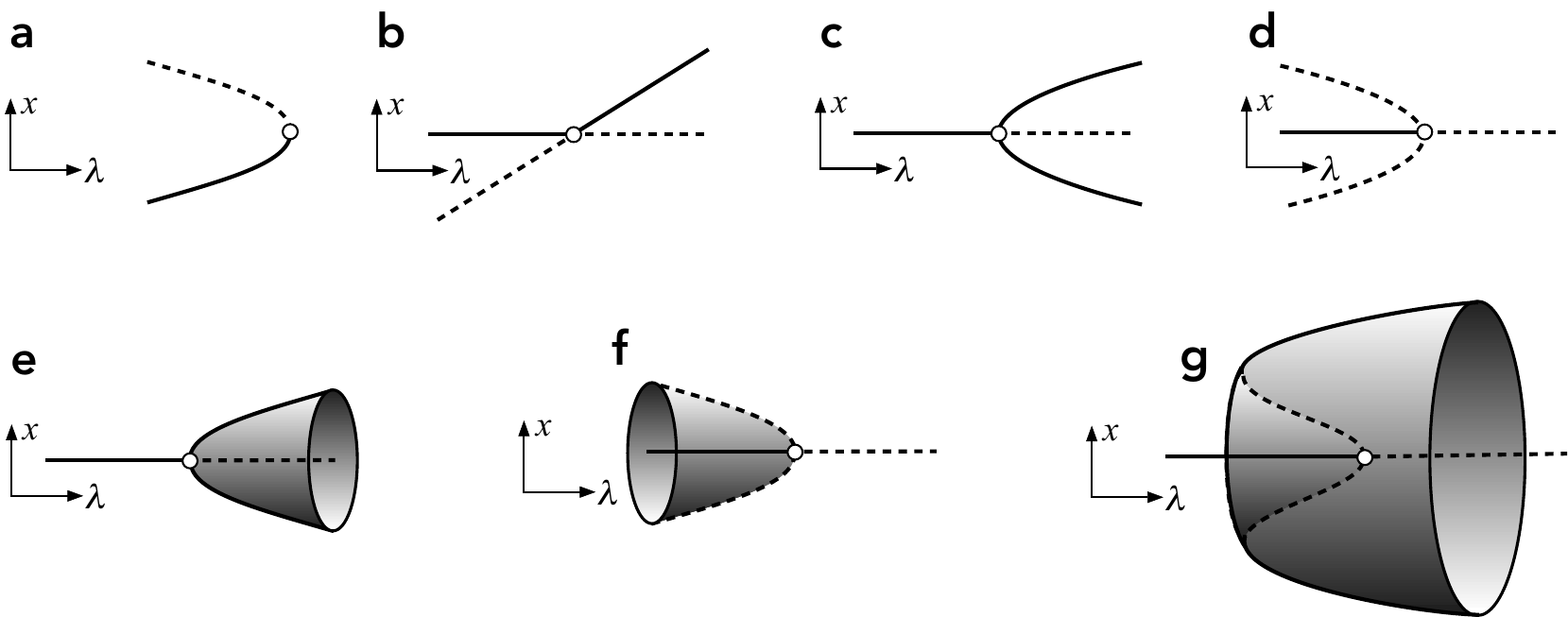}}
\caption{\textbf{Standard types of bifurcation.} Solid line: stable. Dashed line: unstable. White dot: bifurcation point.
 (\textbf{a}) Saddle-node. (\textbf{b}) Transcritical,
generic when $f(0, \lambda)  = 0$ for all $\lambda$. (\textbf{c}) Supercritical pitchfork,
generic when $f(x, \lambda)$ is even in $x$. (\textbf{d}) Subcritical pitchfork,
generic when $f(x, \lambda)$ is even in $x$. 
(\textbf{e}) Supercritical Hopf bifurcation. (\textbf{f}) Subcritical Hopf bifurcation.
(\textbf{g}) Subcritical Hopf bifurcation `turning around' to create a stable
branch, not local to the origin.
}
\label{F:bifurcation_list}
\commentAlt{Figure~\ref{F:bifurcation_list}: 
Caption gives short description.
}

\commentLongAlt{Figure~\ref{F:bifurcation_list}: 
Seven subfigures. All have coordinates lambda, x.
(a) Parabola with horizontal axis opening to left: top half dashed, bottom half solid.
(b) Horizontal axis crossed by sloping line. Top segments solid, lower ones dashed.
(c) Horizontal lambda-axis plus parabola with horizontal axis opening to right. All lines
solid except for right half of lambda-axis, which is dashed.
(d) Horizontal lambda-axis plus parabola with horizontal axis opening to left. All lines
dashed except for left half of lambda-axis, which is solid.
(e) Horizontal lambda-axis plus paraboloid of revolution with horizontal axis opening to right. All lines
solid except for right half of lambda-axis, which is dashed.
(f) Horizontal lambda-axis plus paraboloid of revolution with horizontal axis opening to left. All lines
dashed except for left half of lambda-axis, which is solid.
(g) Like (e) except that the left-hand end of the paraboloid is
pushed towards the right, folding in under the rest of the surface like
a `Mexican hat' surface.
}
\end{figure}

Figure \ref{F:bifurcation_list} shows bifurcation diagrams for these two types of local
bifurcation from a branch of stable equilibria, and variants of them under various extra conditions.
(a) For arbitrary $f$ the only generic local bifurcation to an equilibrium state
from a branch of equilibria is
the saddle-node, where the stable equilibrium merges with an unstable one
and no equilibria exist for $\lambda > 0$. (b) If $f(0, \lambda) = 0$ for all $\lambda$,
a common assumption in network models, a transcritical bifurcation (`exchange of stability') is generic. (c,d) If $f(x, \lambda)$ is even in $x$ the generic
bifurcation is a pitchfork, which can be supercritical or subcritical. 
(e,f) For arbitrary $f$ the other generic local bifurcation from a branch of equilibria is
Hopf bifurcation to a branch of periodic states, which can be supercritical or subcritical.
(g) A subcritical Hopf branch can `turn around' and become stable.

In a network with group symmetry, bifurcations can either preserve symmetry
or break it. {\it Symmetry breaking} bifurcations are of special
interest because they lead to a change of symmetry, which manifests as
pattern formation.  Similarly,  in a network with fibration symmetry, bifurcations
can either preserve synchrony or break it. {\it Synchrony-breaking}
bifurcations are of special interest because they lead to a change of
the synchrony pattern, which again can manifest as pattern formation.
The two bifurcations set-ups are analogous, and conform to the group
symmetry/fibration distinction.

An informative example occurs for the toggle-switch circuit, Section \ref{S:bi_tog},
which has $\Z_2$ symmetry swapping the two nodes. We observed
that the synchronous state loses stability to synchrony-breaking
perturbations (but not synchrony-preserving ones) when the
parameters satisfy $a=b=2$. These values correspond to the cusp
point in Fig. \ref{F:bif_var}. The cross-section of the cusp catastrophe surface
along the diagonal $a=b$ is a pitchfork bifurcation in $(u,v)$-space.

Hopf bifurcation for the Smolen and metabolator circuits is discussed
in \citep{stewart2024dynamics}.

\subsection{First bifurcation}

One useful general stability principle is the notion of `first bifurcation'.\index{bifurcation !first }
For an ODE on $\R^k$ where $k>1$, the Jacobian $J$ has
more than one eigenvalue. Eigenvalues move continuously as
parameters vary continuously \citep{lancaster1985}. Bifurcations occur when one of these
eigenvalues crosses the imaginary axis.

A standard way to analyze bifurcations is to
arrange for the bifurcation parameter
$\lambda$ appear in the ODE as the family $\dot x = F(x)+\lambda x$.
In this case, all eigenvalues move towards the right in $\mathbb{C}$ as $\lambda$
increases. Therefore stability of the equilibrium concerned is first lost
when the eigenvalue with largest real part crosses the imaginary axis;
we call this the {\em first bifurcation}.
A similar remark applies if the bifurcation parameter appears as
an added input, such as $I_1, I_2$ in \eqref{E:gardner}, since these 
have n effect in the eigenvalues because
the form of the Jacobian does not change. However, the point at 
which it is evaluated depends on the value(s) of the input(s).

If the path of equilibria persists for larger values of $\lambda$ another
bifurcation from that path occurs if another eigenvalue crosses the imaginary axis.
However, the path of equilibria leading up to that bifurcation is
unstable, because at least one eigenvalue now has positive real part, 
so the bifurcating branch initially describes unstable states. 
Therefore the most significant bifurcation for applications is
the first one.

That said, other
choices of bifurcation parameter can change the ordering of
the real parts of eigenvalues, because the form $\dot x = F(x)+\lambda x$
is not the only possible choice. The family of ODEs must remain admissible,
but $\mu x$ can be replaced by an independent linear term for each
type of node and each type of arrow, or by nonlinear expressions. We omit details.

\cite{stewart2024dynamics} tabulates the conditions for first bifurcation for the
six circuits featured in  Fig. \ref{fig:stability_circuits}, for
the general admissible PRN model, on the assumption that the
bifurcation parameter arises as an added term $\mu x$ or as an
added input. We omit details here because they depend
on various partial derivatives, and the notation used depends on the
circuit. Table \ref{T:first_bif_summary} provides a summary stating the type of the first 
bifurcation, again under the stated assumption on how the bifurcation
parameter appears in the model.

\begin{table*}[h!]
  \centering
\begin{tabular}{|l|l|l|l|}
\hline 
circuit & possible type of first bifurcation\\
\hline
\hline
lock-on & SB steady \quad SP steady \\
toggle-switch & SB steady \quad SP steady \\
Smolen oscillator & SB steady \quad SP steady \\
feed-forward fiber (FFF)& SB steady \quad SP steady \\
Fibonacci fiber & SB steady \quad SP steady \quad SP Hopf \\
Repressilator &  SB Hopf $\sot$-period phase relation rotating wave\\
\hline
\end{tabular}
\vspace{10pt}
\caption{Possible first bifurcations from a stable synchronous 
equilibrium in the PRN model, for the six circuits. SB = synchrony-breaking, SP= synchrony-preserving.}
\label{T:first_bif_summary}
\end{table*}

\begin{example}\em
\label{ex:first_LO}
{\bf First bifurcation for the lock-on circuit}
As an example of how Table \ref{T:first_bif_summary}
 is derived, we study the conditions for first bifurcation
in the lock-on circuit.\index{lock-on }

Routine calculations using \eqref{E:LO_eigen} determine which 
eigenvalues can occur as the first bifurcation. The idea is to compare
the real parts of pairs of eigenvalues and determine when they are
equal: this happens when they change order.  Let $K = (a-c)^2 + 8bd$. 
For $\lambda_3^\pm$ we
must again distinguish two cases: $K<0$, where $\lambda_3^\pm$ are complex
conjugate, so Hopf bifurcation may occur, and $K>0$, where $\lambda_3^\pm$
are real, so only steady-state bifurcation is possible.  We 
 summarize the results in Table~\ref{T:LO_first}. 

\begin{table*}[h!]
  \centering
\begin{tabular}{|c|c|c|l|}
\hline 
eigenvalue & $K<0$ & $K\geq0$ & type of bifurcation\\
\hline
\hline
$\mu_1$ & $a=0, c<0$ & $a=0,c<0,bd<0$ & SB steady $|$  SB steady\\
$\mu_2$ & $a<0, c=0$ & $a<0,c=0,bd<0$ & SB steady $|$  SB steady\\
$\mu_3$ & impossible& $a<0,c<0,ac=2bd$ & SP Hopf $|$  SP steady\\
\hline
\end{tabular}
\vspace{10pt}
\caption{Conditions for first bifurcation from a synchronous
  equilibrium for the general PRN lock-on and toggle-switch models. SB
  = synchrony-breaking, SP= synchrony-preserving.}
  \label{T:LO_first}
\end{table*}

{\em Case 1}: $K<0$.

{\em Subcase 1}: $\lambda_1$ critical.

Here $a=0$. We have
\[
\lambda_2 = c \qquad \lambda_3^\pm = \shf(c\pm\sqrt{c^2+8bd})
\]
All three of these have real part with the same sign as $c$. Therefore when $K<0$,
the first bifurcation occurs at $\mu_1$ if and only if $a=0, c < 0$.

{\em Subcase 2}: $\lambda_2$ critical.

Here $c=0$. We have
\[
\lambda_1 = a \qquad \lambda_3^\pm = \shf(a\pm\sqrt{a^2+8bd})
\]
All three of these have real part with the same sign as $a$. Therefore when $K<0$,
the first bifurcation occurs at $\mu_2$ if and only if $a<0, c= 0$.

{\em Subcase 3}: $\mu_3^\pm$ critical. (Being complex conjugates, both eigenvalues
are critical simultaneously.)

Here $a+c=0$ so $c=-a$. Now
\[
\lambda_1 = a \qquad \lambda_2 = -a
\]
which cannot both be negative. Therefore $\mu_3^\pm$ cannot be the first bifurcation.

{\em Case 2}: $K>0$.

This condition is equivalent to
\begin{equation}
\label{bd_sign}
bd > -\frac{(a-c)^2}{8}
\end{equation}

{\em Subcase 1}: $\mu_1$ critical.

Now $a=0$ and
\[
\lambda_2 = c \qquad \lambda_3^\pm = \shf(c\pm\sqrt{c^2+8bd})
\]
Thus $c < 0$ and $c+\sqrt{c^2+8bd} < 0$. Since $c < 0$ this is easily seen to be
equivalent to $bd < 0$.

{\em Subcase 2}: $\lambda_2$ critical.

Now $c =0$, and 
\[
\lambda_2 = c \qquad \lambda_3^\pm = \shf(a\pm\sqrt{a^2+8bd})
\]
Therefore $a <0$.
A similar calculation again leads to $bd < 0$.

{\em Subcase 3}: $\lambda_3^\pm$ critical. Since $\lambda_3^- < \lambda_3^+$,
first bifurcation must happen for $\lambda_3^+$, if at all.

We have $a+c+\sqrt{K} = 0$, which by a simple calculation is equivalent to
\[
ac = 2bd
\]
Now we want $\lambda_1 = a < 0$ and $\lambda_2 = c < 0$.
Therefore $ac > 0$, so $bd > \shf ac$. This is stronger than \eqref{bd_sign}.
\end{example}

\begin{example}\em
In some cases, certain eigenvalues cannot occur as the first bifurcation,
no matter how the bifurcation parameter is chosen. An example
occurs for the Smolen circuit, whose general model has the form
\[
\begin{array}{rcl}
\dot{x}^R & = & f(x^R, x^P,y^P,I) \\
\dot{y}^R & = &  f(y^R, x^P,y^P,I)\\
\dot{x}^P & = & g(x^P,x^R) \\
\dot{y}^P & = & g(y^P,y^R)
\end{array}
\]
There is a
nontrivial fibration symmetry $x=y$, giving a flow-invariant synchrony
subspace.  At a synchronous equilibrium define
\[
a = f_1 \quad b = f_2  \quad c = g_1 \quad d = g_2 \quad e = f_3
\]
The Jacobian is
\[
J = \Matrix{aI & Q \\ dI & cI}\qquad \mbox{where} \qquad Q = \Matrix{b & e \\ b & e}
\]
The eigenvalues of $Q$ are $\mu = 0, b+e$.
By Theorem~\ref{T:MeigCeig} the eigenvalues of $J$ are
\[
\lambda_1=a \qquad \lambda_2=c \qquad \lambda_3^\pm=\shf\left(a + c \pm \sqrt{K}\right)
\]
where $K = (a-c)^2 + 4d(b+e)$.
Of these, $\lambda_1,\lambda_2$  are synchrony-breaking. The $\lambda_3^\pm$ pair is synchrony-preserving. Hopf bifurcation can occur only for
 $\lambda_3^\pm$, when $K < 0$. The real part of  $\lambda_3^\pm$ is
 $\shf(a+c)$, the average of the other two eigenvalues $a$ and $c$.
 A Hopf branch can arise only when $\shf(a+c)=0$, but a
 necessary condition for this branch to be stable is that the
 other two eigenvalues $a$ and $c$ are negative. This combination
 is impossible.

Thus the Smolen circuit with fibration symmetry cannot exhibit stable oscillations
created at a stable synchrony-breaking Hopf bifurcation. However,
stable oscillations can occur through other dynamic mechanisms.
\cite{hasty2002} and \cite{hasty2008} analyze the Smolen circuit using
different mathematical models, and find oscillations that
arise by Hopf bifurcation---but not by a synchrony-breaking bifurcation
from a synchronous state. Indeed, those models do not have
a synchronous state; they employ asymmetric parameter values and input values.

We conclude that more complex models than those discussed above are required to generate sustained oscillations in the Smolen
circuit. Indeed, even in the cited models, the parameter regions for
Hopf bifurcation are relatively small and delicate. Thus the presence
of oscillations in models of the Smolen circuit is highly sensitive to
the functions used.
\end{example}

\subsection{Hopf bifurcation in the symmetric repressilator}
\label{S:repressilator}

We illustrate the effect of symmetric network topology
on Hopf bifurcation with a discussion of the {\em repressilator}\index{repressilator }~\citep{elowitz2000}.  The repressilator does not occur naturally in
{\it E. coli}, but its base is in the class 0-FFF as shown in
Fig. \ref{fig:stability_circuits}.  In \citep{elowitz2000} it is modeled both in the symmetric case
(where it has $\Z_3$ symmetry) and in an asymmetric
generalization. The symmetric case exhibits sustained periodic
oscillations in which successive genes are $\sot$-period out of
phase. In the asymmetric case the oscillations persist, with phase
shifts near $\sot$-period, but varying amplitudes (partly due to
including stochastic noise in the model).

\begin{figure}[b]
\centerline{
\includegraphics[width=.24\textwidth]{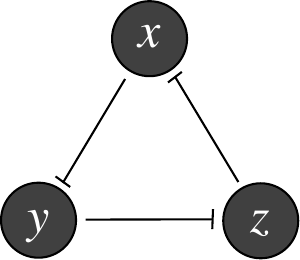} \qquad 
\includegraphics[width=.44\textwidth]{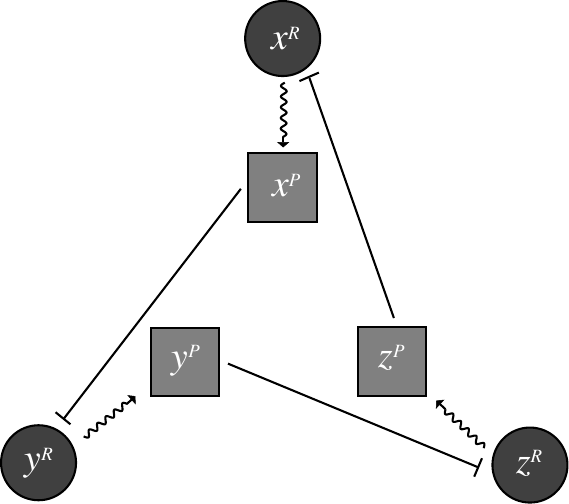} \qquad
\includegraphics[width=.18\textwidth]{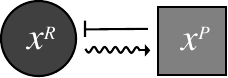}
}
\caption{\textbf{Repressilator circuit.} {\em Left}:  GRN. {\em Middle}: PRN, drawn to 
show $\Z_3$ symmetry. {\em Right}:  Base of PRN for (blue, red) fibration symmetry.}
\label{F:Z3rep}
\commentAlt{Figure~\ref{F:Z3rep}: 
Three graphs. 
On the left: a graph with three circular blue nodes called x, y, z. Three inhibition arrows connect x to y, y to z, z to x.
In the center: a graph with three circular blue nodes called xR, yR, zR and three rectangular red nodes called xP, yP, zP.
Inhibition arrows connect xP to yR, yP to zR, zP to xR. Squiggly arrows connect xR to xP, yR to yP, zR to zP.
On the right: a graph with one circular blue node called xR and one rectangular red node called xP. Inhibition arrow from xP to xR.
Squiggly arrow from xR to xP.
}
\end{figure}

Here we derive analytic results for the general
PRN model in the $\Z_3$-symmetric case, Fig.~\ref{F:Z3rep}. We show
that the $\sot$-period phase shifts occur through a symmetry breaking Hopf bifurcation,\index{bifurcation !Hopf, symmetry breaking }
and that this must be the first bifurcation.  We also perform simulations 
that illustrate this phase pattern in a specific model.

The mRNA and protein variables are
 $x = (x^R,x^P), y = (y^R,y^P), z = (z^R,z^P)$. Admissible ODEs are:
\beqn
\dot{x}^R &=& f(x^R,z^P)\\
\dot{x}^P &=& g(x^P,x^R)\\
\dot{y}^R &=& f(y^R,x^P) \\
\dot{y}^P &=& g(y^P,y^R)\\
\dot{z}^R &=& f(x^R,y^P)\\
\dot{z}^P &=& g(z^P,z^R)
\eeqn
The Jacobian $J$ is block-circulant, because of $\Z_3$ symmetry. Let
\[
A = \Matrix{a & 0 \\ c & d} \qquad B = \Matrix{0 & b \\ 0 & 0}
\]
where
\[
a= f_1  \quad b= f_2  \quad c= g_2 \quad d= g_1 
\]
Then in block matrix notation
\[
J = \Matrix{A & 0 & B \\ B & A & 0 \\ 0 & B & A}
\]
Using representation theory and equivariant dynamics \citep{golubitsky1988}, or
by direct calculation,
the eigenvalues and form of the eigenvectors of $J$ are:
\beqn
\mbox{eigenvalues of}\ A+B &:&   [u,u,u]^{\mathrm{T}}\\
\mbox{eigenvalues of}\ A+\zeta^2B &:& [u,\zeta u,\zeta^2 u]^{\mathrm{T}} \\
\mbox{eigenvalues of}\ A+\zeta B &:& [u,\zeta^2 u,\zeta u]^{\mathrm{T}}
\eeqn
where $u \in \R^2$ is an eigenvector of the $2 \times 2$ matrix listed
and $\zeta = e^{2\pi\ii/3}$ is a primitive cube root of unity.
Here $A+B$ is symmetry-preserving and the others are symmetry breaking.
Now
\[
A+B = \Matrix{a & b \\ c & d} \qquad A+\zeta B = \Matrix{a & \zeta b \\ c & d} 
	\qquad A+\zeta^2 B = \Matrix{a & \zeta^2 b \\ c & d} 
\]
For Hopf bifurcation we seek purely imaginary eigenvalues of these 
matrices.

We claim that if the first bifurcation is Hopf, it must be
symmetry breaking. That is, the nodes do not oscillate in synchrony, with the same phases. To see why,
suppose, on the contrary, that the bifurcation is symmetry-preserving. Then it
comes from a pair of imaginary eigenvalues of $A+B$, so the trace
$a+d=0$ and $d=-a$.  The eigenvalues of $A+\zeta B$ sum to give the
trace of $A+\zeta B$, but again this is $a+d=0$. Therefore the real parts of
these eigenvalues also sum to zero, so one is positive and the other
negative.  Therefore the first bifurcation cannot arise from $A+B$.

Thus it must be a symmetry breaking Hopf bifurcation, arising
either from $A+\zeta B$ or $A+\zeta^2 B$. Since these are complex
conjugates, we can consider just $A+\zeta B$.
The conditions for $A+\zeta B$ to have purely imaginary eigenvalues can be derived without
difficulty, and are:
\begin{eqnarray}
\label{E:hopfcond1}
ad+\shf bc &>& 0 \\ 
\label{E:hopfcond2}
(a+d)^2(ad+\shf bc) &=& \frac{3}{4}b^2c^2
\end{eqnarray}

The way $\zeta$ appears in the eigenvectors of $J$ exhibits the $\sot$-period phase
shift pattern, since multiplication by $\zeta$ rotates the complex plane
through $120^\circ$ counterclockwise.
The Equivariant Hopf Theorem \citep[Chapter XVI Theorem 4.1]{golubitsky1988} proves
that this pattern persists, exactly, in the bifurcating periodic solution
of the nonlinear ODE. Thus a
symmetry breaking Hopf bifurcation corresponds to the repressilator's
rotating wave state with $\sot$-period phase shifts, simulated below in Fig. \ref{F:xRPosc}.

\begin{figure}[h!]
\centerline{
\includegraphics[width=.5\textwidth]{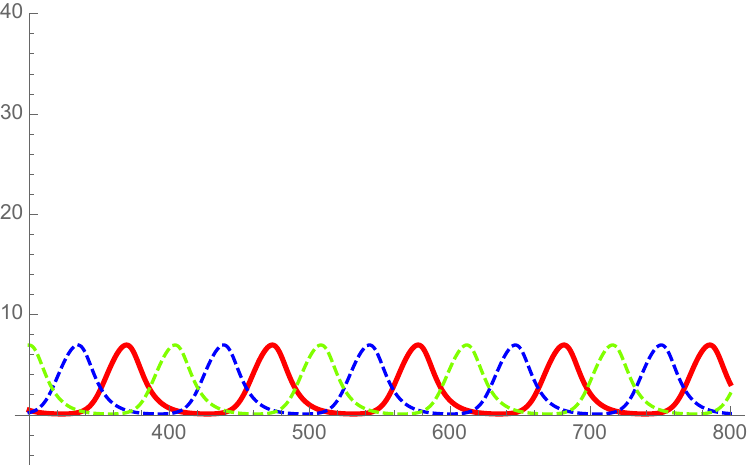} \qquad
\includegraphics[width=.5\textwidth]{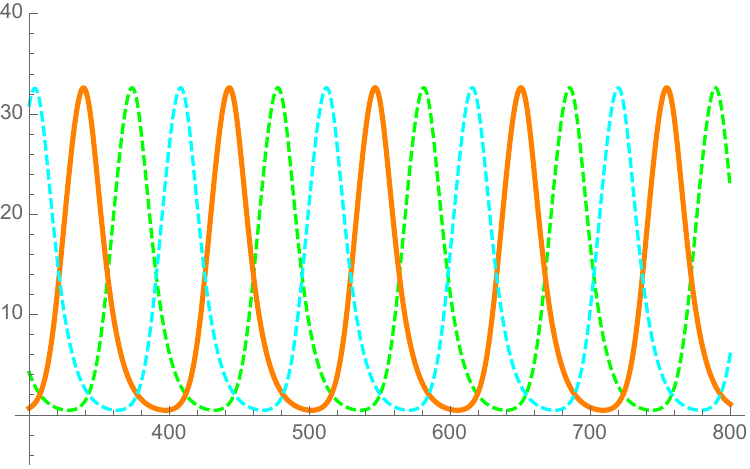}
}
\caption{\textbf{Simulations of special PRN model for repressilator}. We use $S(x) = 1/(1+x^2)$ and 
parameters: $\alpha = 0.2$, $\beta = 1$, $\delta = 0.1$.
{\em Left}:
Time series of $x^R,y^R,z^R$ superposed. {\em Right}:
Time series of $x^P,y^P,z^P$ superposed.}
\label{F:xRPosc}
\commentAlt{Figure~\ref{F:xRPosc}: 
Left: Three oscillatory time series superposed: all have the same shape but successive
ones are phase-shifted by one third of the period. Low amplitude.
Left: Three oscillatory time series superposed: all have the same shape but successive
ones are phase-shifted by one third of the period. High amplitude.
}
\end{figure}

We consider a numerical example for the following model:
\[
\begin{array}{rcl}
f(x^R,x^P) & = & -\delta x^R + H^-(x^P)\\
g(x^P,x^R) & = & -\alpha x^P + \beta x^R
\end{array}
\]
Figure ~\ref{F:xRPosc} shows typical time series for all six variables, superposed
in threes:
mRNA on the left, protein on the right. It can be checked that
this state arises by symmetry breaking Hopf bifurcation.
The oscillations are very similar to those in~\citep{elowitz2000},
having the $\sot$-period phase shift
rotating wave form predicted by equivariant bifurcation theory
\citep{golubitsky1988}.

\subsection{Discussion and Conclusions}

In the previous sections we analyzed six circuits that serve as functional building blocks in natural or synthetic gene regulatory networks, classified into classes based on their fibration symmetry. We emphasized analytic conditions for local bifurcation from synchronous states in each circuit, focusing on
fibration symmetries and synchrony-breaking local bifurcations.
The main results are summarized in Table \ref{T:first_bif_summary}.

The lock-on, toggle-switch, Smolen circuit, feed-forward fiber, and Fibonacci fiber support both steady symmetry
breaking and symmetry preserving bifurcations. Additionally, the
Repressilator supports robust oscillatory states via symmetry-breaking
Hopf bifurcation, taking the form of rotating waves with $\sot$-period
phase shifts.  The Fibonacci fiber circuit supports damped oscillatory
states. Oscillations in the Smolen circuit cannot arise by
synchrony-breaking Hopf bifurcation in a GRN model because of sign
constraints, but can occur in neuron models and through other
mechanisms. This analysis enhances the understanding of gene
regulatory networks and highlights the importance of both
group-theoretic and fibration symmetry for their dynamics.   Overall, this
study contributes to a systematic characterization of the building
blocks of bacterial regulatory networks. It complements previous
studies based only on topological properties of the circuits \citep{leifer2020circuits,morone2020fibration}
 by extending the analysis to the stability
properties of the dynamical solutions observed for biologically
inspired ODEs that are admissible for the graphs.

\section{Master stability function for complete synchronization}
\label{S:MSFSI}

\cite{pecoraMSF} introduced the `master stability function'\index{master stability function }
as a way to study synchronization of a completely synchronous
state. They work with Laplacian-based models,\index{Laplacian } which we
discussed briefly in Section \ref{sec:laplacian}.
Their approach is described next. Its main disadvantage is that it assumes specific types of couplings (diffusive), which precludes its applicability to many biological networks. The fact that the complete synchronous state is built into the model from the start is also not realistic in biological situations since this state is deleterious for biological functionality. However, when the form of equations that it assumes
is an appropriate model it
has three fundamental advantages: 

(1) It allows the study of arbitrary synchronous solutions, possibly periodic or chaotic.\index{chaos } 

(2) It permits the analysis of stability of the complete synchronous solution for arbitrary networks described by possibly weighted adjacency matrices.

(3) It leads to a dimensionality reduction of the stability problem into lower dimensional `modes'.\index{mode } 

The latter point is of relevance to networks with a large number of nodes. It also leads to another indirect benefit: the stability problem can be parametrized in a convenient form, as we see in what follows.

In this approach to stability, the
model ODE is assumed to have a specific structure,
 which, among other things, 
permits the existence of a completely synchronous state for any structure of the network, whether it has symmetries or not (see section \ref{sec:laplacian}). The assumption is:
\begin{equation}
\label{E:MSF}
\dot x_i = F(x_i) -S \sum_{j\neq i} A_{ij}(H(x_i)-H(x_j)), \quad i=1,...,N
\end{equation}
for suitable functions $F,H$, possibly weighted adjacency matrix $A=\{A_{ij}\geq 0\}$, and coupling strength parameter $S>0$. The state of node $i$ is $n\geq 1$ dimensional, and the number of nodes $N$ is at least $2$.
The same function $F$ is used for all variables; that is, the internal dynamic
is assumed to be the same for all nodes. The coupling terms are of
generalized diffusive type;\index{generalized diffusive coupling } in particular, the coupling between two nodes
vanishes whenever they are in the same state. However, it need not 
vanish when node states differ. 

Equation \eqref{E:MSF} can be transformed into a convenient form
by defining the {\em network Laplacian} $L=\{L_{ij}\}$, with entries $L_{ij}=\delta_{ij} (\sum_j A_{ij})-A_{ij}$, where $\delta_{ij}$ is the Kronecker delta. In Laplacian\index{Laplacian } notation, Equation \eqref{E:MSF} can be rewritten as follows,
\begin{equation}
\label{E:MSFrepeat}
\dot x_i = F(x_i) -S \sum_{j\neq i} L_{ij} H(x_j) \quad i=1,..,N.
\end{equation}

By construction, all the rows of the Laplacian\index{Laplacian } matrix $L=\{L_{ij}\}$ sum to zero, which implies 
that $[1, 1,..., 1]$ is an eigenvector with 
eigenvalue $\lambda_1=0$. This ensures the existence of a completely synchronous solution 
\begin{equation}\label{completesynch}
{x}_1(t)={x}_2(t)=...={x}_N(t)= {x}_s(t)
\end{equation} 
for all $t$. This solution obeys the equation
\begin{equation} \label{dotxs}
\dot{x}_s(t)=F(x_s(t)).
\end{equation}
Typically we are interested in those models \eqref{dotxs} that produce either periodic or chaotic oscillations.

From the viewpoint of modern dynamical systems,
an equilibrium is stable if and only if all eigenvalues of the
Jacobian have negative real part at the equilibrium point,
see Definition \ref{D:lin_stab_general}. In the literature
on the master stability function, this leads to a
more specific description, which can also be used to study
stability of more complex dynamical states. We therefore
present the ideas involved in that form.

To study stability we consider small perturbations ${\delta x}_i(t) = ({x}_i(t)-{x}_s(t))$ about the synchronous solution, each of which evolves according to the variational equation
\begin{equation} \label{Eq_deltaxi}
\delta  \dot{x}_i(t)= {\rm D}F({x}_s(t)) {\delta x}_i(t) -S \sum_j L_{ij} {\rm D}H({x}_s(t)) {\delta x}_j(t) \quad i=1,..,N.
\end{equation}
Here ${\rm D}F(x_s(t))$ is the Jacobian of $F$ and ${\rm D}H(x_s(t))$ is the Jacobian of $H$, both evaluated at the synchronous solution.
Equation \eqref{Eq_deltaxi} can also be rewritten in vector form
\begin{equation}\label{net:varW}
    \delta  \dot{X}(t)= [I_N \otimes {\rm D}F({x}_s(t)) -S L \otimes {\rm D}H({x}_s(t))] {\delta X}(t)
\end{equation}
where ${\delta X}(t)=[\delta x_1^T(t),\delta x_2^T(t),...,\delta x_N^T(t)]^T$ is an
 $N \times n$ dimensional vector, and
the symbol $\otimes$ indicates the Kronecker product\index{Kronecker product } (or tensor product) of matrices.

If $L$ is diagonalizable, the linearization \eqref{net:varW}
decomposes into components along the eigenvectors of $L$, and the
corresponding solutions are the following decoupled `modes'.

To see this, we apply the transformation   $V^{-1} \otimes I_n$ to \eqref{net:varW}, where $V$ is the matrix whose columns are the eigenvectors of $L$, $V^{-1} L V= \Lambda$,
 $\Lambda$ is a diagonal matrix that has the eigenvalues of $L$ on the main diagonal, and $I_n$ is the identity matrix of dimension $n$. Now
\begin{equation}
    \dot{Y}(t)= [I_N \otimes {\rm D}F({x}_s(t)) -S \Lambda \otimes DH({x}_s(t))] {Y}(t)
\end{equation}
where $Y(t)=V^{-1} \otimes I_n \delta X(t)$. This equation decouples as
\begin{equation} \label{ind}
     \dot{\eta}_k(t)= [ {\rm D}F({x}_s(t)) - S \lambda_k  {\rm D}H({x}_s(t))] {\eta}_k(t)
\end{equation}
$k=1,...,N$, where ${\eta}_k(t)$ is the eigenmode\index{eigenmode } associated with the eigenvalue $\lambda_k$ of $L$, $k=1,...,N$.
The modes \eqref{ind} are independent of one another. One mode (the one associated with the eigenvalue $\lambda_1=0$ of $L$) corresponds to a perturbation inside the synchrony subspace  with associated eigenvector $[1,1,\ldots,1]$. Such a mode is synchrony-preserving.\index{synchrony-preserving } The remaining modes associated with the `transverse eigenvalues'\index{eigenvalue !transverse } of $L$, namely $\lambda_2,\lambda_3,...,\lambda_N$, correspond to perturbations in the transverse subspace. These modes are synchrony-breaking.\index{synchrony-breaking } We emphasize that based on the assumption that $A_{ij}\geq 0$, the minus sign on $S$ means that the
real parts $\rho$ of the eigenvalues $\lambda_k = \rho_k + \ii \sigma_k$ must all be positive.

We can rewrite \eqref{ind} in parametric form
\begin{equation} \label{param}
     \dot{\eta}_k(t)= [ {\rm D}F({x}_s)(t) + \xi  {\rm D}H({x}_s)(t)] {\eta}_k(t)
\end{equation}
where the parameter $\xi$ may be complex.  Equation \eqref{param} is called the \textit{master stability equation}.\index{master stability equation }  We then introduce the  \textit{master stability function} (MSF):\index{master stability function }
\begin{equation} \label{MSF0}
\mathcal{M}(\xi)
\end{equation}
which associates the Maximum Liapunov Exponent (MLE) of  \eqref{param} to any value of $\xi$. Maximum Liapunov Exponents\index{maximum Liapunov exponent } \citep{ott2002chaos} are well-defined measures of the asymptotic growth rate of a perturbation about a trajectory (in this case $x_s(t)$). They are negative when the perturbation asymptotically decays back to the trajectory and positive when it asymptotically departs from the trajectory.
They should not be confused with eigenvalues of the Jacobian evaluated at the solution.
We then have:
\begin{theorem}
\label{T:MSFstability}
The totally synchronous solution is stable if 
\begin{equation}
\label{E:MSFstability}
\mathcal{M}(-S \lambda_k)<0
\end{equation}
for all transverse eigenvalues $\lambda_k$, $k=2,...,N$. 
\end{theorem}
This condition must be satisfied for all transverse eigenvalues, not just one.

It is important to emphasize that \eqref{param} has dimension $n$, as well as \eqref{dotxs}. Therefore, with respect to the original problem \eqref{net:varW} we have achieved a dimensional reduction from $N \times n$ to $2 \times n$, where the number $2$ comes from counting both the dynamics of the synchronous evolution \eqref{dotxs} as well as the dynamics of the decoupled modes \eqref{ind}. This is especially significant when the number of network nodes $N$ is very large.

There is also another important implication of the parametrization \eqref{param}. Namely,
for a given choice of the functions $F$ and $H$, we can compute {\it a priori} the region of the complex plane $\Omega$ within which $\mathcal{M}(\xi)<0$, and then test whether the transverse eigenvalues of a given Laplacian\index{Laplacian } matrix $L$ belong to $\Omega$. Accordingly, the condition for stability of the synchronous solution becomes
\begin{equation}
    -S \lambda_k \in \Omega, \quad k=2,...,N.
\end{equation}
This has the significant advantage that computing $\Omega$  needs to be done only once. For a known $\Omega$, assessing  stability of the complete synchronous solution for a given network topology simply requires checking whether the transverse eigenvalues of the Laplacian multiplied by $-S$ fall into the region $\Omega$. This gives a quick stability check for many different network topologies at once. 

As a useful rule of thumb, it is suggested that when $\Omega$ is bounded, the more widely the transverse eigenvalues are spread,
the less likely it is for all of them to lie in $\Omega$, so stability is less
likely. One plausible measure of the spread is the standard deviation\index{standard deviation } of the eigenvalues.
\cite{barahona2002synchronization, nishikawa2010} propose this as a synchronizability index\index{synchronizability index } of a given network topology.

The main limitations of the master stability function approach described in this section are the assumptions of
identical node dynamics and generalized diffusive coupling\index{generalized diffusive coupling } in \eqref{E:MSFrepeat} leading to the Laplacian dynamics which is not widespread in biology, as well as the impossibility of applying this analysis to cluster synchronization, which is of fundamental importance for biological systems.

\section{A master stability function approach for cluster synchronization}
\label{sec:MSFACS}

A large body of work  by Pecora, Sorrentino and others, e.g. \citep{Sorrentino2007,pecora2016b, siddique2018symmetry,blaha2019cluster,della2020symmetries,panahi2021cluster,lodi2021one,lodi2024patterns}, has focused on extending the master stability function\index{master stability function !for clusters } approach to cluster synchronization.
\index{cluster synchronization } The first paper to accomplish this goal for an arbitrary network topology was \citep{pecora2014}, which,  however, focused on the case of symmetries produced by the graph automorphism group, and on orbital partitions.  Here we are mostly interested in equitable partitions arising from the more general fibration symmetries, a case considered in \citep{panahi2021cluster,lodi2021one}. The general set of dynamical equations considered is:
\begin{equation}
\label{E:MSF:A}
\dot x_i = F(x_i) +S \sum_{j\neq i} A_{ij} H(x_j), \quad i=1,...,N.
\end{equation}
These equations are  
 not in the Laplacian form of \eqref{E:MSF}. The case when the adjacency matrix is symmetric is studied in \citep{panahi2021cluster} and the general case
 is studied in \citep{lodi2021one}.
The method presents similar advantages to those described above, namely: 

(i) It can be applied to study stability of either periodic or chaotic cluster synchronous solutions. 

(ii) It can be applied to any given network and to any equitable partition (minimal or not). 

(iii) It leads to a dimensional reduction of the stability problem into subproblems of smaller dimension. 

However, it is important to stress that, unlike \citep{pecoraMSF}, the extent of the dimension reduction varies from network to network, and can be limited.

For a given adjacency matrix $A$ we partition the set of network nodes $V$ into $C \geq 1$ equitable clusters, $\mathcal{C}_1,\mathcal{C}_2,..,\mathcal{C}_C$, so that $\cup_{k=1}^C \mathcal{C}_k={V}$ and $\mathcal{C}_k \cap \mathcal{C}_\ell=\emptyset$ for $k \neq \ell$, with the condition of balanced coloring:
\begin{equation} \label{ecp}
  \sum_{h \in \mathcal{C}_{\ell}} A_{ih} = \sum_{h \in \mathcal{C}_{\ell}} A_{jh}, \quad  \forall i,j \in \mathcal{C}_k, \quad \forall \mathcal{C}_k,\mathcal{C}_{\ell} \subset {V}. 
\end{equation}
\noindent We let $|\mathcal{C}_k|=n_k$ be the number of nodes in cluster $k=1,...,C$, so that $\sum_{k=1}^C n_k=N$. Given such an equitable partition we can assume, without loss of generality, that the network nodes are ordered so that the first $n_1$ nodes are those in cluster $\mathcal{C}_1$, followed by the $n_2$ nodes in cluster $\mathcal{C}_2$, and so on.
Then we construct {\em indicator matrices} $E_k$:
\begin{equation}\label{E_i}
    E_1=\begin{pmatrix}
  I_{n_1} & 0\\ 
  0 & {0}_{N-n_{1}}
    \end{pmatrix}
    \quad
        E_2=\begin{pmatrix}
  {0}_{n_{1}} & 0 & 0\\ 
  0 & I_{n_2} & 0\\
  0 & 0 & {0}_{N-(n_{1}+n_{2})}
    \end{pmatrix}
    \quad
    \cdots
    \quad
    E_C=\begin{pmatrix}
  {0}_{N-n_{C}} & 0\\ 
  0 & I_{n_C}
    \end{pmatrix}
\end{equation}
where $I_{n_i}$ is the identity matrix of size $n_{i}$, and $0_{n_i}$ is the zero matrix of size $n_{i}$. 

  We have seen that an equitable partition defines an invariant subspace for \eqref{E:MSF:A}, the synchrony subspace of the cluster (also called the cluster synchronization manifold).\index{cluster synchronization manifold } The dynamics on this subspace 
gives the time evolution of the clusters \citep{golubitsky2005patterns} $\lbrace s_{1}(t), s_{2}(t), \cdots, s_{C}(t)\rbrace$, where $s_{1}(t)$ is the synchronous solution for all nodes in cluster $\mathcal{C}_1$, $s_{2}(t)$ is the synchronous solution for nodes in cluster $\mathcal{C}_2$, and so on.
       
The $C \times C$ quotient matrix $Q$ is defined so that, for each pair of equitable clusters
$\mathcal{C}_k$ and $\mathcal{C}_l$,
\begin{equation}\label{quo1}
    Q_{kl}=\sum_{j \in \mathcal{C}_{l}}A_{ij}  \quad i \in \mathcal{C}_{k}.
    \end{equation}
      The quotient matrix describes a quotient network, in which all nodes in each equitable cluster collapse to a single quotient node. Assuming that \eqref{E:MSF:A} evolves on the synchrony subspace, and averaging over all the nodes in each cluster, we can derive the equations for the time evolution of the quotient network:
\begin{equation}\label{quo2}
    \dot{s}_{k}(t) = F(s_{k}(t)) + S \sum_{l=1}^{C}Q_{kl}H(s_{v}(t)),
\quad
    k,l = 1, 2, \cdots, C
\end{equation}
where the $n$-dimensional vector $s_{k}(t)$ represents the state of the quotient network node $k=1,...,C$. 

To investigate the stability of the cluster synchronous solution, we consider a small perturbation $\delta x_i=(x_i - s_k)$, $i \in \mathcal{C}_k$. Linearizing  \eqref{E:MSF:A} about \eqref{quo2}, we can write
\begin{equation}\label{z}
\delta \dot{x}(t) = \left[ \sum_{c=1}^{C}E_{c}\otimes {\rm D}F(s_{c}(t))+S \sum_{c=1}^{C}A E_{c}\otimes {\rm D}H(s_{c}(t))\right] \delta x(t)
\end{equation}
in the $nN$- dimensional vector $\delta {x}(t)=[\delta {x}_1^T(t),\delta {x}_2^T(t),...,\delta {x}_N^T(t)]^T$ and the indicator matrices defined in \eqref{E_i}. 

We are interested in the possibility that the stability problem for the $Nn$-dimensional system \eqref{z} can be decoupled into a set of lower-dimensional equations. Equation \eqref{z} is analogous to \eqref{net:varW}, introduced in the previous section for the case of complete synchronization. However, decoupling \eqref{z} into a set of lower dimensional equations is typically a much more complex
task. The challenge is to find a transformation matrix $T$ that transforms \eqref{z} into a set of lower dimensional equations of lowest dimension. (In the case of \eqref{net:varW} this matrix $T$ was easily constructed as the matrix of the eigenvectors of the Laplacian matrix $L$.)

\cite{panahi2021cluster} and \cite{lodi2021one} differ for the specific technique used to construct the matrix $T$, where the approach of \citep{panahi2021cluster} is faster but works only when the matrix $A$ is symmetric, while the approach of \citep{lodi2021one} is slower but can be applied to the general case. Details of the construction of either transformation
matrix $T$ are in \citep{panahi2021cluster} and \citep{lodi2021one}. Here we take the approach of \citep{panahi2021cluster} and directly present the final result: the linearized system \eqref{z} is transformed
into 
\begin{equation}
     \dot{{\eta}}(t)= \left[\sum_{k=1}^C J_{k} \otimes {\rm D}F(s_{k}(t)) + \sum_{k=1}^C B J_{k} \otimes {\rm D}H(s_{k}(t))\right]{\pmb \eta}(t).
     \label{Eq11}
 \end{equation}
Here 
$B$ is the transform of $A$, that is, $B=T^{-1} A T$, and the matrices $J_k$ are the transforms of $E_k$, so $J_k=T^{-1} E_k T$, where $k=1,...,C$. The matrix $B$ is block-diagonal in $r \geq 2$ blocks and the matrices $J_k$ are diagonal, so the dimensional reduction is determined by the structure of the $r$ blocks into which $B$ is decomposed. Thus \eqref{Eq11} decouples into $r \geq 2$ independent equations of smaller dimension, in the $r$ independent blocks of the block-diagonal matrices $J_k$ and $B J_k$. The structure of these blocks conveys important information about key properties of the stability problem, and in particular about the possible breakings of the clusters that may lead to the emergence of other synchronization patterns.

As already stated, there is always one $C$-dimensional symmetry preserving block, which we label $k=1$; this is associated with dynamics parallel to the synchrony subspace. 
The remaining $k=2,...,r$ symmetry-breaking blocks are associated with dynamics transverse to the synchrony subspace. This is analogous to what we saw earlier in this chapter for  equilibria, the difference being that for equilibria we had symmetry preserving and symmetry breaking eigenvalues and eigenvectors, and in the case of synchronous oscillations we have symmetry preserving and symmetry breaking modes.
The condition for stability of the cluster synchronous solution is that all the maximum Liapunov exponents associated with the symmetry breaking blocks $k=2,...,r$ are negative.\index{maximum Liapunov exponent }

\subsection{Examples of cluster stability analysis}

\begin{figure}
\centering
  \includegraphics[scale=.32]{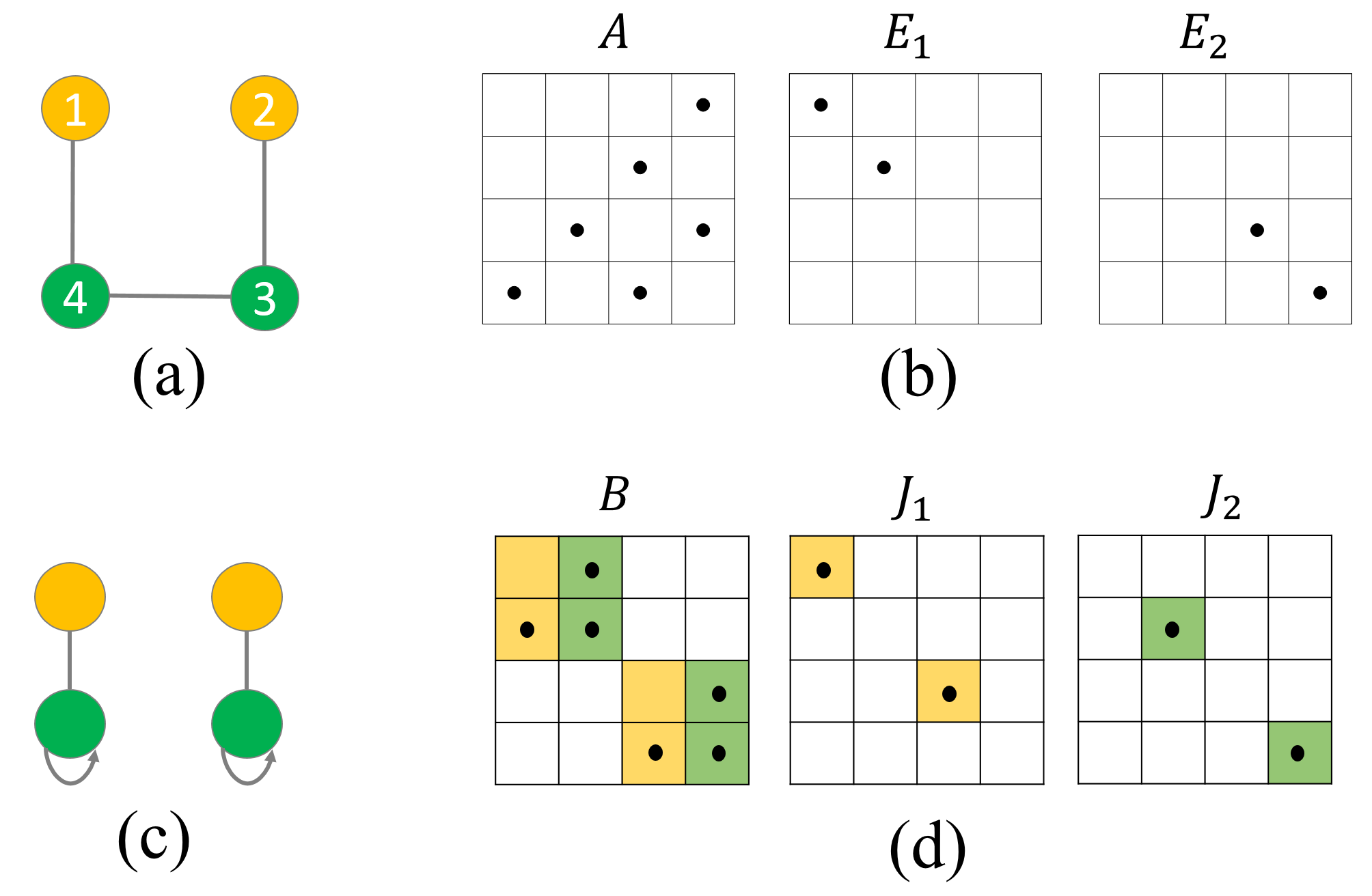}
\caption{\textbf{Stability
of the cluster synchronous solution.} (\textbf{a}) An $N=4$ node network with $C=2$ clusters. Nodes are color-coded according to the clusters to which they belong. (\textbf{b}) The adjacency matrix of the network ($A$) and the two cluster indicator matrices ($(E_1,E_2)$) are shown graphically; a nonzero entry is indicated by a black dot. Panel (\textbf{c}) shows the quotient and transverse subnetworks after applying $T$. Subnetwork nodes are colored according to the cluster with which they are associated.    
(\textbf{d}) The set of matrices $\{ B, J_1, J_2 \}$ is
obtained using the transformation $T$. The background  color of each matrix entry indicates the cluster about which the dynamics is linearized.}
\label{4n}
\commentAlt{Figure~\ref{4n}: 
Two graphs on the left (called (a) and (c)), with three matrices on the right-hand side of each graph (called (b) and (d)). All matrices are 4x4.
}

\commentLongAlt{Figure~\ref{4n}: 
Graph (a) is undirected, and has nodes 1 and 2 (yellow), and 3 and 4 (green). Undirected edges connect 1 to 4, 4 to 3, 3 to 2.
Graph (c) contains directed and undirected arcs, and has two yellow nodes (say 1,2) and two green nodes (say 3,4). Two undirected edges
connect 1 to 3 and 2 to 4. Two directed edges connect 3 to itself and 4 to itself.
The group of matrices called (b) is made of three matrices, called A, E1, E2. 
A contains dots in the entries (1,4), (2,3), (3,2), (3,4), (4,1), (4,3).
E1 contains dots in the entries (1,1), (2,2).
E2 contains dots in the entries (3,3), (4,4).
The group of matrices called (d) is made of three matrices, called B, J1, J2.
B contains dots in the entries (1,2), (2,1), (2,2), (3,4), (4,3), (4,4). Moreover the first two entries of the first column and the last two entries of the third column are yellow;
the first two entries of the second column and the last two entries of the last column are green.
J1 contains dots in the entries (1,1), (3,3). Dotted entries are yellow.
J2 contains dots in the entries (2,2), (4,4). Dotted entries are green.
}
\end{figure}

Next, we present two examples to illustrate how this procedure works. 
First, we consider the $N=4$-dimensional network with $C=2$ clusters shown in Fig.~\ref{4n}(a).  The nodes are colored according to the cluster to which they belong. The synchrony subspace is $\Delta= \{x_1, x_1, x_2, x_2 \} : x_1,x_2 \in \mathbb{R}^n$.

 Figure \ref{4n} shows that applying the transformation $T$ from \citep{panahi2021cluster} to the matrices $A$, $E_1$, and $E_2$ reduces the $4n$-dimensional stability problem of \eqref{z} into $r=2$ separate blocks: a $2n$-dimensional equation corresponding to the quotient dynamics (symmetry preserving block) and a $2n$-dimensional equation corresponding to the transverse dynamics (symmetry breaking block). Only the latter is responsible for stability of the cluster synchronous solution, i.e., it is the maximum Liapunov exponent associated with this latter block that determines whether the cluster synchronous pattern in (a) is stable. 
 Figure \ref{4n}b represents the adjacency matrix $A$ and the indicator matrices $E_1$ and $E_2$, where a black dot corresponds to an entry equal to $1$ and the absence of a dot corresponds to a zero entry.
 Figure \ref{4n}c represents the two decoupled `networks' obtained after applying the transformation, and this is consistent with the block structure of the matrices $B$, $J_1$, and $J_2$ in (d). 
 
 We may wonder why there is only one symmetry breaking block and not two. In fact, since the equitable partition in (a) has $C=2$ clusters, we might expect that stability of each  of the two clusters corresponds to a separate symmetry breaking condition. The reason why only one symmetry breaking condition occurs is that the two clusters are `intertwined' \citep{pecora2014}.That is, it is not possible to break the yellow cluster without also breaking the green cluster and {\it vice versa}. Thus either the two clusters are simultaneously synchronized, or they are simultaneously desynchronized. As a result, stability of the cluster pattern (a) depends only on one maximum Liapunov exponent, which corresponds to simultaneous breaking of the yellow and green clusters. There is no other way in which the cluster synchronous solution can be broken.

\begin{figure}
\centering
\includegraphics[width=1\linewidth]{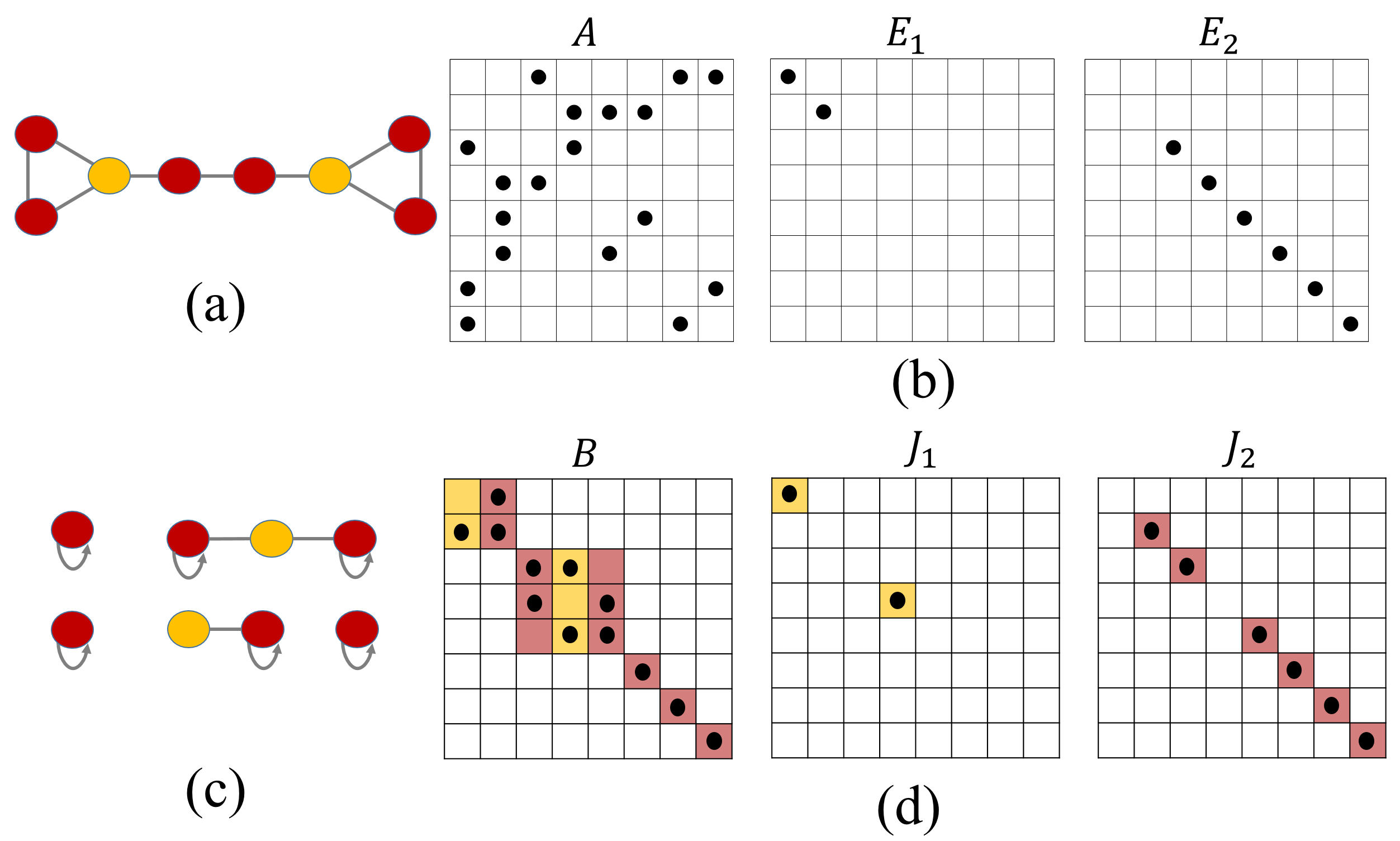}
  \caption{\textbf{The papillon network with minimal balanced coloring/equitable partition}. (\textbf{a}) An $N=8$-dimensional network with $C=2$ equitable clusters. Nodes are color-coded according to the clusters to which they belong. (\textbf{b}) The adjacency matrix of the network ($A$) and the two cluster indicator matrices $E_1,E_2$ are shown graphically, in which a nonzero entry is indicated by a black dot. (\textbf{c}) shows the quotient and transverse subnetworks after applying the transformation $T$. Subnetwork nodes are colored based on the cluster to which they are associated. 
(\textbf{d}) The set of matrices $\{B, J_1, J_2 \}$  obtained by using the transformation $T$. The background color of each matrix entry indicates the cluster about which the dynamics is linearized.}
\label{82}
\commentAlt{Figure~\ref{82}: 
Two graphs on the left (called (a) and (c)), with three matrices on the right-hand side of each graph (called (b) and (d)). All matrices are 8x8.
}

\commentLongAlt{Figure~\ref{82}: 
Graph (a) is undirected, and has nodes 1-8. Node 1,2,4,5,7,8 are red, nodes 3,6 are yellow. Undirected edges connect
1 with 2 and 3, 2 with 3, 3 with 4, 4 with 5, 5 with 6, 6 with 7 and 8, 7 with 8.
Graph (c) contains directed and undirected arcs, and has nodes 1-8. Nodes 1,2,4,5,7,8 are red, nodes 3,6 are yellow.
Undirected edges connect 3 with 2 and 4, 6 with 7. Nodes 1,2,4,5,7,8 have a directed arrow to themselves.
The group of matrices called (b) is made of three matrices, called A, E1, E2. 
A contains dots in the entries (1,3), (1,7), (1,8), (2,4), (2,5), (2,6), (3,1), (3,4), (4,2), (4,3), (5,2), (5,6), (6,2), (6,5), (7,1), (7,8), (8,1), (8,7).
E1 contains dots in the entries (1,1), (2,2).
E2 contains dots in the entries (3,3), (4,4), ..., (8,8).
The group of matrices called (d) is made of three matrices, called B, J1, J2.
B contains dots in the entries (1,2), (2,1), (2,2), (3,3), (3,4), (4,3), (4,5), (5,4), (5,5), (6,6), (7,7), (8,8).
Furthermore the first two rows of the first column, and the rows 3-5 of column 4 are yellow; rows 1-2 of column 2, rows 3-5 of columns 3 and 5, and the diagonal elements (6,6), (7,7), (8,8) are orange.
J1 contains dots in the entries (1,1), (4,4). Both are yellow.
J2 contains dots in all the diagonal entries except (1,1) and (4,4). All dotted entries are orange.
}
\end{figure}

\begin{figure}[t!]
\centering
\includegraphics[width=1\linewidth]{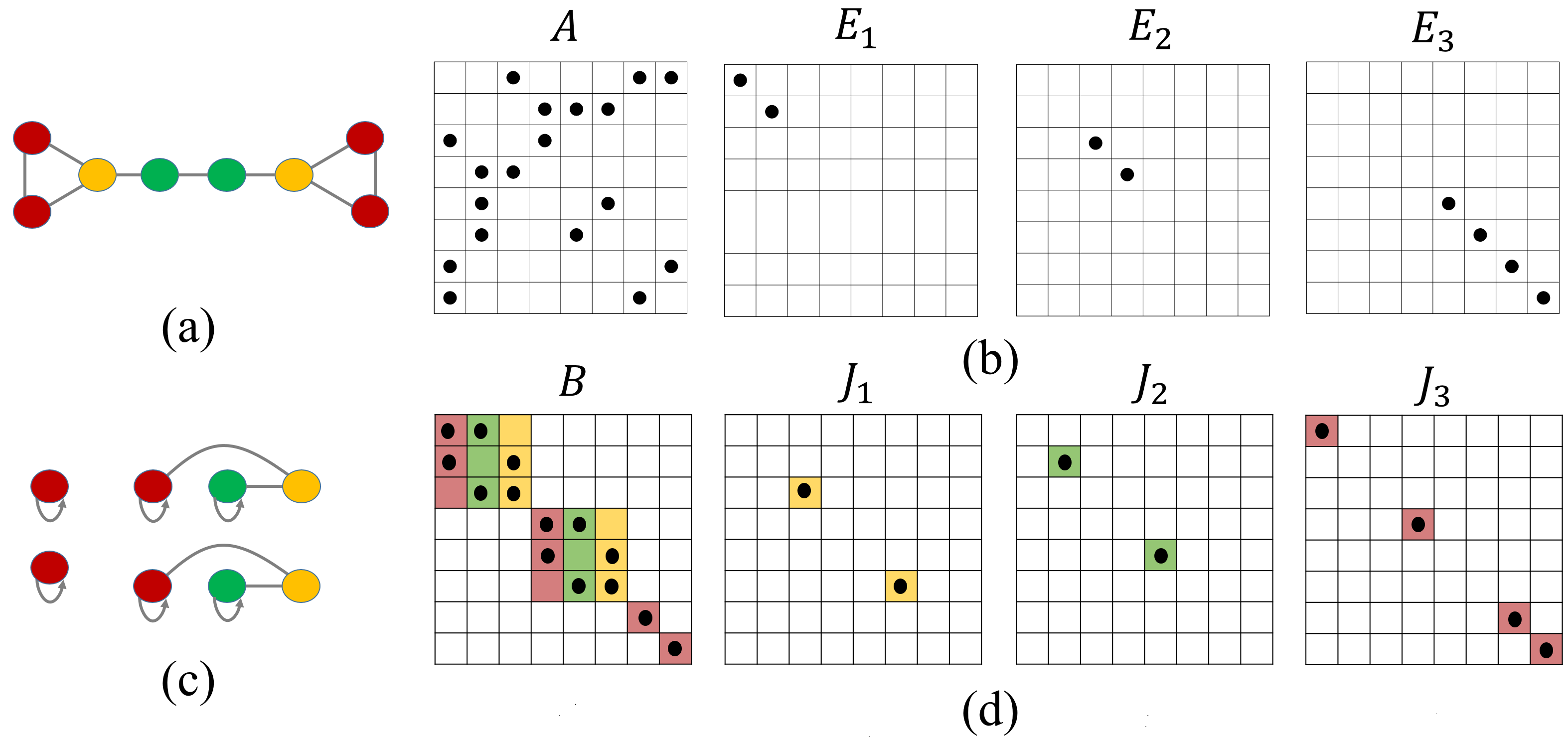}
  \caption{\textbf{The papillon network with minimal orbital partition.} (\textbf{a}) An $N=8$-dimensional network with $C=3$ orbital clusters. Nodes are color-coded according to the clusters to which they belong. (\textbf{b}) The adjacency matrix of the network ($A$) and the three cluster indicator matrices $E_1,E_2,E_3$ are shown graphically; a nonzero entry is indicated by a black dot. (\textbf{c}) The quotient and transverse subnetworks after applying the transformation $T$. Subnetwork nodes are colored based on the cluster to which they are associated.   
(\textbf{d}) The set of matrices $\{B, J_1, J_2, J_3 \}$  obtained by using the transformations $T$ and $\tilde{T}$, respectively. The background color of each matrix entry indicates the cluster about which the dynamics is linearized. }
\label{83}
\commentAlt{Figure~\ref{83}: 
Two graphs on the left (called (a) and (c)), with four matrices on the right-hand side of each graph (called (b) and (d)). All matrices are 8x8.
}

\commentLongAlt{Figure~\ref{83}:
Graph (a) is undirected, and has nodes 1-8. Node 1,2,7,8 are red, nodes 3,6 are yellow, nodes 4,5 are green. Undirected edges connect
1 with 2 and 3, 2 with 3, 3 with 4, 4 with 5, 5 with 6, 6 with 7 and 8, 7 with 8.
Graph (c) contains directed and undirected arcs, and has nodes 1-8. Nodes 1,2,5,6 are red, nodes 4,8 are yellow, nodes 3,7 are green.
Undirected edges connect 2 with 4, 3 with 4, 6 with 8, 7 with 8. Nodes 1,2,3,5,6,7 have a directed arrow to themselves.
The group of matrices called (b) is made of three matrices, called A, E1, E2, E3, E4.
A contains dots in the entries (1,3), (1,7), (1,8), (2,4), (2,5), (2,6), (3,1), (3,4), (4,2), (4,3), (5,2), (5,6), (6,2), (6,5), (7,1), (7,8), (8,1), (8,7).
E1 contains dots in the entries (1,1), (2,2).
E2 contains dots in the entries (3,3), (4,4), ..., (8,8).
The group of matrices called (d) is made of three matrices, called B, J1, J2.
B contains dots in the entries (1,2), (2,1), (2,2), (3,3), (3,4), (4,3), (4,5), (5,4), (5,5), (6,6), (7,7), (8,8).
Furthermore the first two rows of the first column, and the rows 3-5 of column 4 are yellow; rows 1-2 of column 2, rows 3-5 of columns 3 and 5, and the diagonal elements (6,6), (7,7), (8,8) are orange.
J1 contains dots in the entries (1,1), (4,4). Both are yellow.
J2 contains dots in all the diagonal entries except (1,1) and (4,4). All dotted entries are orange.
}
\end{figure}

  We now apply our approach to a network for which the clusters of the minimal equitable partition and the clusters of the minimal orbital partition do not coincide. This has $N=8$ nodes, and is taken from \citep{kudose2009}; we call it the {\em papillon network}.  
  Figure \ref{82} 
 presents the case of the network minimal equitable partition with $C=2$ clusters, and Fig. \ref{83} 
 presents the case of the network minimal orbital partition with $C=3$ clusters. Both figures are similar in content and style to Fig. \ref{4n}. 
 
 Figure \ref{82} 
 shows that applying the transformation $T$  leads to a total of $5$ blocks. The first block is $2$-dimensional and corresponds to the quotient network (symmetry preserving block). The remaining four blocks, with dimensions 3, 1, 1, 1, correspond to the transverse dynamics (symmetry breaking blocks). The quotient block is the only $2$-dimensional block. Stability of the cluster synchronous solution depends on the maximum Liapunov exponents associated to the four symmetry breaking blocks. Each of these symmetry breaking blocks has a nice geometric interpretation in terms of possible breakings of the cluster synchronous solution in (a), which may lead to the emergence of a different (non-minimal) equitable partition of the network nodes. 
 
 The $3$-dimensional transverse block corresponds to a left/right symmetry breaking in which the cluster pattern in (a) is broken along the vertical symmetry line that divides the network into two halves of four nodes each. This may happen without breaking synchronization of each of the two pairs of red nodes on the left and on the right of the papillon in (a). In addition, the are three other symmetry breakings of the red cluster alone that may disrupt the cluster synchronization pattern in (a). One is for the two center red nodes to depart from the four outer red nodes (the corresponding new cluster synchronization pattern is shown in panel (a) of Fig. \ref{83}). The remaining two correspond to the two possible ways in which the four outer red nodes can be broken into two halves of two nodes each, either along the top-bottom horizontal symmetry line or along the top left-bottom right diagonal symmetry line. In order for the cluster synchronous pattern in (a) to be stable, all the four maximum Liapunov exponents corresponding to these symmetry-breaking blocks should be negative, thus preventing any of these possible breakings from happening.  

Figure \ref{83} 
 shows the same analysis applied to the papillon network but for the minimal orbital partition shown in panel (a). In this case, applying the transformation $T$ leads to a total of $4$ blocks: one $3$-dimensional block corresponding to the quotient network (symmetry preserving block), 
 and  three blocks with dimensions 3, 1, 1 corresponding to the transverse dynamics
(symmetry breaking blocks). 
  Stability of the cluster synchronous solution depends only on the maximum Liapunov exponents associated to the three symmetry breaking blocks, each of which has a nice geometric interpretation in terms of possible breakings of the cluster synchronous pattern (a). 
  
  Similarly to  Fig. \ref{82}, the $3$-dimensional transverse block corresponds to a left/ right symmetry breaking in which the cluster pattern in (a) is broken along the vertical symmetry line that divides the network into two halves of four nodes each. This may happen without breaking synchronization of the two pairs of red nodes on the left and on the right of the papillon in (a). In addition, the are other two symmetry breakings of the red cluster that may disrupt the cluster synchronization pattern in (a). These correspond to the two possible ways in which the four outer red nodes can be broken into two halves of two nodes each, without breaking the green and yellow clusters: either along the top-bottom horizontal symmetry line or along the top left-bottom right diagonal symmetry line.
  
  In order for the cluster synchronous pattern in (a) to be stable, all the three maximum Liapunov exponents corresponding to these symmetry-breaking blocks should be negative. The main difference from Fig. \ref{82} is that the $6$-node red cluster in Fig. \ref{82} already appears to be broken into two clusters in Fig. \ref{83}a. This corresponds to an increase in the dimension of the quotient network (which has now $C=3$ nodes) and a corresponding decrease in the dimension of the transverse blocks.
Hence, there are situations for which the cluster synchronization pattern in \ref{83}a is stable but the cluster synchronization pattern in \ref{83}b is not, because this requires stability of one additional maximum Liapunov exponent.

\section{Structural stability: changes to equations, parameters, and graphs}
\label{sec:structural_stability}

`Stability'\index{stability } is a word with numerous different meanings.
Several other important properties of network models
have an air of stability as well as robustness about them, and it is important not to
confuse them with stability as discussed so far in this chapter.
All of them are important to biological modeling, so for completeness we mention them here and make a few pertinent remarks.
They include:

(1) \textbf{Synchronizability:} {\em Stability against small changes to initial conditions.} This is the stability concept described in previous sections of this chapter, and it is perhaps the most basic. A solution of an ODE is stable in this sense
if a small change in initial conditions leads to a
solution that is the same as, or at least `very close to' the original solution. It is called synchronizability when the state is synchronous.  This is important for biology because
the environment is constantly changing, and such changes
should not cause radical differences in the behavior predicted
by a model.

(2) \textbf{Hard structural robustness:} {\em Stability against (small) changes in the model equations},
when these remain consistent with the same underlying graph. This
is a form of {\em structural stability}\index{structural stability } in the sense of 
\cite{smale1963}, tailored to the graph setting. 
There it is known as {\em robustness}\index{robust ! technical meaning }, see Definition \ref{D:robust_pattern}.
An apparently weaker condition, which turns out to be equivalent for
equilibria (and probably for more complex dynamical states) is 
{\em rigidity},\index{rigid }. This notion of stability is independent of (1). An unstable state can
nonetheless be structurally stable if it persists after
small perturbations to the model ODE. It will still exist but remain unstable. This notion is important for biology because
model equations are seldom known exactly; they are usually
phenomenological models involving functions and parameters
that have been chosen to offer useful insights but do
not reflect reality with high precision. This form
of stability preserves the underlying graph representing known
biological interactions but allows minor changes to the functions
in the model. 

(3) \textbf{Soft structural robustness}: {\em Stability against changes to parameters}.\index{parameter changes } 
This is a special form of (2): the allowed perturbations to the model
are already built in as numerical parameters, which may not be known
accurately. Changing these parameters changes the equations,
which is why (2) applies, but the range of changes considered is more restrictive. The relevant mathematical concepts are
hyperbolicity\index{hyperbolicity } \citep{guckenheimer1983}, which ensures that
in typical circumstances,
sufficiently small changes to a model do not destroy equilibria or
periodic states (even if these are unstable in the sense of (1) above). This stability notion is important for biology because
biology is disordered and random; for example, the binding of molecules depends on many things, and it cannot be assumed to be uniform in biological systems. See Chapter \ref{chap:alive}.

(4) \textbf{Hard structural robustness:}\index{robust ! structural }
{\em Stability against graph modifications}.  \index{graph!
  modification } This is an interesting but relatively new area of
mathematics, which becomes relevant when we are uncertain whether the
graph being assumed is an accurate representation of all relevant
biological interactions.  Even when this is the case, biological
events such as evolution and mutations may change the graph. Again,
hyperbolicity\index{hyperbolicity } can indicate this kind of
stability as long as the graph modifications lead to small changes in
the ODE.  Simulations are often the only way to study this type of
question.

In this book, we reserve the word `stable' for (1), unless otherwise indicated.


\chapter[Extending Fibrations From Graphs to Hypergraphs]{\bf\textsf{Extending Fibrations Beyond Graphs to Hypergraphs}}
\label{chap:hypergraph}

\begin{chapterquote}
 Up to this point, we have discussed systems that can be described by
  directed (multi)graphs with two-body interactions
   associated with graph edges.
  Modeling biological
  networks often requires a more specific type of structure 
  that prescribes the many-body interactions among the
  biological units. These kinds of interactions are widespread in
  biology and are not described by graphs when (as is often assumed)
  each edge corresponds to a specific function of a single input variable. A popular way to impose such constraints is to
  use a hypergraph. Hypergraph equations are admissible
  but satisfy extra conditions on the terms that appear in the ODE.
 In this chapter, we describe how the techniques we have
  previously discussed can be applied, \emph{mutatis
  mutandis}, in this more general setting. We also discuss other
  generalizations of fibrations to weighted, multiplex, and
  multilayer networks.
\end{chapterquote}

\section{Introduction}
\label{sec:hyper_intro}

So far, we have discussed many different types of systems that can all be described in the form of directed graphs (networks): a set of entities (protein, genes, neurons, cells, chemical compounds, etc.) with directed edges between them representing some form of binary action (control, relation, enhancement, activation, repression, excitation, inhibition, etc.). 
The notions of symmetries that we have taken into consideration mainly target this simple (yet rich) kind of structure. We should, however, notice that in many cases (most probably all) biological networks are more complex, and describing them merely as graphs would be an oversimplification.
In this chapter, we want to take some of these complex structures into consideration; the most relevant to biology are hypergraphs and weighted networks. We discuss in each case whether and how the techniques described in the book can still be applied to them. These applications remain open for additional theoretical developments.

\section{Hypergraphs}\index{hypergraph }
\label{S:hypergraph}

While graphs serve as a powerful tool for modeling complex systems, they are not always the best choice, because in many cases
interactions between components are not binary but can involve more than two components at the same time. As discussed in Section \ref{SS:GMH}, a general
admissible ODE for an ordinary (multi)graph is not limited to binary interactions:  
if a node has three inputs, for instance, then
{\em all possible functions} of the three tail nodes can occur in an admissible ODE.
If the input set has $m$ edges, then $m$-ary interactions are, in principle, permitted;
$m+1$ if we count the node variable.

The problem is that in many areas of application, standard models
involve specific functional forms for interactions. The resulting 
admissible ODEs include all possible admissible equations,
many of which do not have the required 
specific form used by modelers in the area concerned. 
While the theory of admissible equations is important from the mathematical point of view, leading to general theorems about admissible dynamics that apply to {\em all} models (see Section \ref{sec:adODEs} and discussion in Section \ref{sec:robust}), a modeler 
in a specific discipline would be more interested in a representation of the ODE that describes the dynamics of interest in a particular practical application.

Hypergraphs~\citep{BerGH} are a popular solution to this problem and are also adopted in data modeling of higher-order interactions
\citep{battiston2020networks,millantopology2025}. Hypergraphs provide a systematic way to represent specific {\em restrictions} on the form of model ODEs, which are not consequences of the usual network structure.

For instance, in social network analysis, a hypergraph can represent a group of people (vertices) participating in an event (hyperedge), a situation that cannot be accurately represented by a simple graph.
However, hypergraphs also find a very natural use as models for biological networks, such as protein-protein interaction networks, gene regulatory networks, and metabolic networks. In all of these networks, complex interactions often involve more than two entities, making hypergraphs a suitable modeling tool.

\begin{figure*}[htbp]
  \centering
  \includegraphics[scale=0.3]{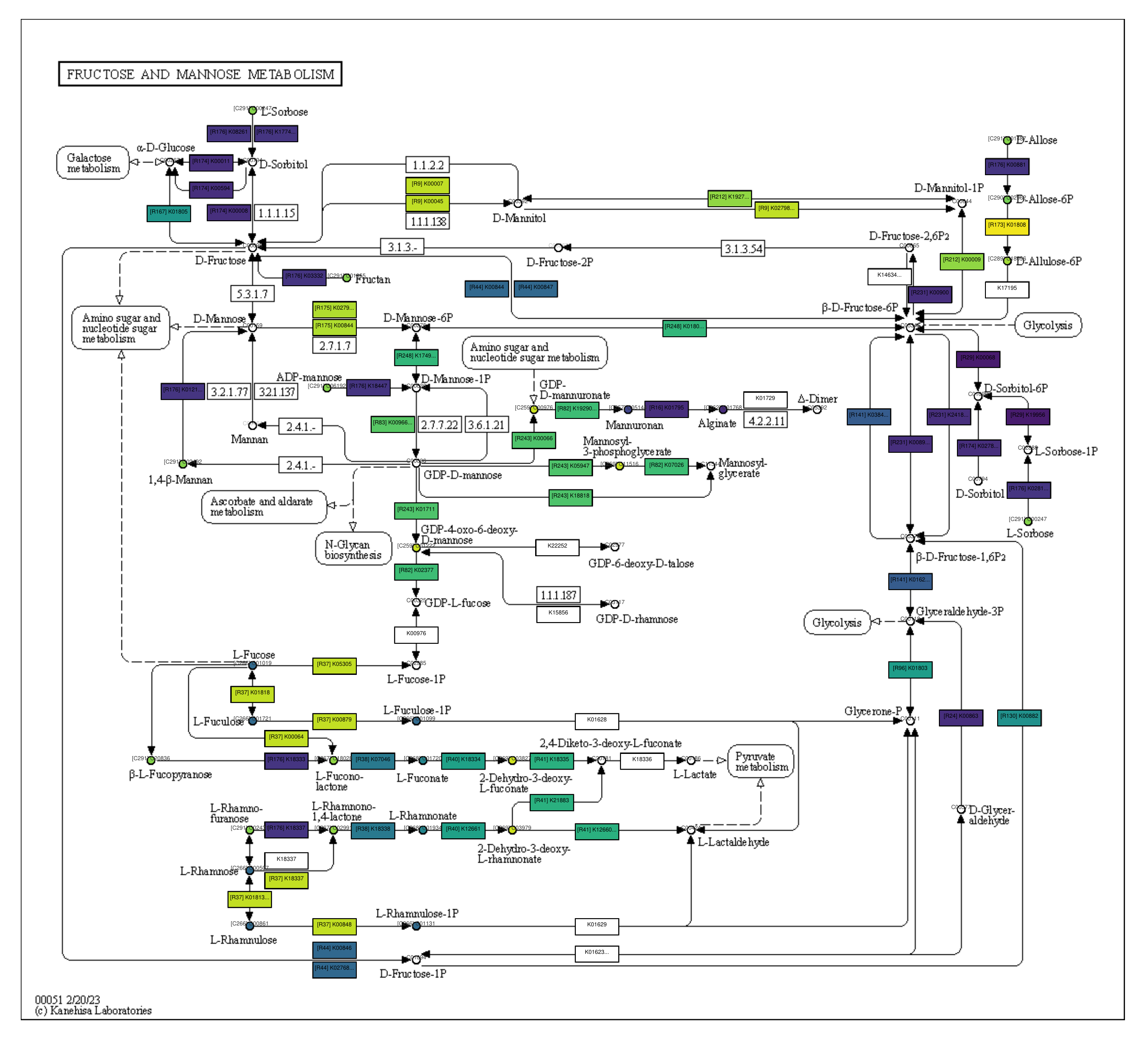}
  \caption{ \textbf{Example of a complex hypergraph in biology.} The metabolic pathway describing fructose and mannose metabolism from KEGG. Figure reproduced from \citep{kanehisa2004kegg}.  Copyright \copyright ~2004, Oxford University Press.}
  \label{fig:kegg}
\commentAlt{Figure~\ref{fig:kegg}: 
No alt-text required.
}
\end{figure*}

A biological example is shown in Fig. \ref{fig:kegg}, taken from the KEGG Or\-tho\-logy da\-ta\-base~\citep{kanehisa2004kegg}; it represents the metabolic pathway of fructose and mannose metabolism. A metabolic pathway like this is a complex network connecting metabolic reactions, genes, and compounds. Reactions typically have many inputs (reactants) and/or outputs (products). 
A general admissible ODE for this graph would permit numerous models
(infinitely many) that do not adequately capture the biological meaning.
Some of these models are realistic representations, but some are not. The modeler is more interested on a particular representation with particular types of interactions.

\smallskip
Generally speaking, hypergraphs\index{hypergraph } are generalizations of graphs where each edge (called a `hyperedge' or
`hyperarc')\index{hyperedge }\index{hyperarc } involves an arbitrary number of nodes instead of just two. This idea can take a number of flavors, depending on the actual situation we want to model. For instance, Fig. \ref{fig:und-hypergraph} is a standard depiction of an \emph{undirected hypergraph}\index{hypergraph !undirected }: hyperedges here are just sets of nodes; while some edges contain exactly two nodes (like $h_2$ in the picture), which is what happens in a standard undirected hypergraph, there are also hyperedges containing more ($h_1$ and $h_3$, in the example) or less ($h_2$) nodes.

Let us focus on another type of hypergraph that is particularly common in the biological context: in biological systems, hypergraphs are typically directed: a \emph{directed hypergraph}\index{hypergraph !directed } has hyperedges\index{hypergraph !hyperedge } identified with pairs $(S,T)$ where $S$ and $T$ are nonempty sets of nodes: the set $S$ represents the source (also called `domain' or `tail') of the hyperedge, and the set $T$ represents its target (also called `codomain' or `head''). Figure~\ref{fig:dir-hypergraph} (left) shows an example of a directed hypergraph: in this example, all hyperedges have exactly one target.

\begin{figure}[t]
  \centering
  \begin{tikzpicture}[use Hobby shortcut]
    \node (v1) at (0,2) {};
    \node (v2) at (1.5,3) {};
    \node (v3) at (4,2.5) {};
    \node (v4) at (1,0) {};
    \node (v5) at (2,4.5) {};
    \node (v6) at (3.5,4.5) {};
    \node (v7) at (3.5,0) {};

    \begin{scope}[fill opacity=0.8]
    \filldraw[fill=red!70] ($(v1)+(-0.5,0)$) 
        to[out=90,in=180] ($(v2) + (0,0.5)$) 
        to[out=0,in=90] ($(v3) + (1,0)$)
        to[out=270,in=0] ($(v2) + (1,-0.8)$)
        to[out=180,in=270] ($(v1)+(-0.5,0)$);
    \filldraw[fill=yellow!70] ($(v4)+(-0.5,0.2)$)
        to[out=90,in=180] ($(v4)+(0,1)$)
        to[out=0,in=90] ($(v4)+(0.6,0.3)$)
        to[out=270,in=0] ($(v4)+(0,-0.6)$)
        to[out=180,in=270] ($(v4)+(-0.5,0.2)$);
    \filldraw[fill=blue!80,closed] ($(v5)+(-0.5,0)$)
        .. ($(v5)+(0,1)$)
        .. ($(v6)+(0,1)$)
        .. ($(v3)+(0.5,0)$)
        .. ($(v3)+(0,-1)$)
        .. ($(v3)+(-0.5,0)$)
        .. ($(v5)+(0,-1)$)
        ;
    \filldraw[fill=green!70] ($(v2)+(-0.5,-0.2)$) 
        to[out=90,in=180] ($(v2) + (0.2,0.4)$) 
        to[out=0,in=180] ($(v3) + (0,0.3)$)
        to[out=0,in=90] ($(v3) + (0.3,-0.1)$)
        to[out=270,in=0] ($(v3) + (0,-0.3)$)
        to[out=180,in=0] ($(v3) + (-1.3,0)$)
        to[out=180,in=270] ($(v2)+(-0.5,-0.2)$);
    \end{scope}

    \foreach \v in {1,2,...,7} {
        \fill (v\v) circle (0.1);
    }

    \fill (v1) circle (0.1) node [right] {$x_4$};
    \fill (v2) circle (0.1) node [below left] {$x_7$};
    \fill (v3) circle (0.1) node [left] {$x_1$};
    \fill (v4) circle (0.1) node [below] {$x_2$};
    \fill (v5) circle (0.1) node [below right] {$x_3$};
    \fill (v6) circle (0.1) node [below left] {$x_6$};
    \fill (v7) circle (0.1) node [below right] {$x_5$};

    \node at (0.2,2.8) {$h_1$};
    \node at (2.3,3) {$h_2$};
    \node at (2.8,5) {$h_3$};
    \node at (1.1,0.7) {$h_4$};
\end{tikzpicture}
  \caption{\textbf{Example of undirected hypergraph.} It contains seven nodes and four hyperedges.}
  \label{fig:und-hypergraph}
\commentAlt{Figure~\ref{fig:und-hypergraph}: 
An undirected hypergraph: nodes are named x1-x6, hyperarcs are named h1-h4 and are represented as colored blobs. 
Hyperedges are as follows: h1 (red) includes x1, x4, x7, h2 (green) includes x1,x7, h3 (blue) includes x1, x3, x6, h4 (yellow) includes x2.
}
\end{figure}

\begin{figure}[b]
\centerline{%
        \begin{tikzpicture}
  \node [circle,draw] (v1) at (0,0)   { $v_1$ };
  \node [circle,draw] (v2) at (-1,-3) { $v_2$ };
  \node [circle,draw] (v3) at (2,-2)  { $v_3$ };
  \node [circle,draw] (v4) at (1,-4)  { $v_4$ };
  \node [circle,draw] (v5) at (3,1)   { $v_5$ };
  \node [circle,draw] (v6) at (5,-1)  { $v_6$ };
  \node [circle,draw] (v7) at (4,-5)  { $v_7$ };
        \connectThree[
          @edge 3=->
        ]{v1}{v3}{v5}
        \connectThree[
          @ratio=0,
          @edge 1=->
        ]{v6}{v7}{v5}
        \connectThree[
          @edge 1=->
        ]{v3}{v4}{v7}
        \connectFour[
          @ratio=.4,
          @pos1=.7,
          @pos2=.5,
          @edge 3=->
        ]{v1}{v2}{v3}{v4}
    \node[font=\small] (h1) at (0.9,-1.9) {$h_1$};
    \node[font=\small] (h2) at (2.1,-0.4) {$h_2$};
    \node[font=\small] (h3) at (2.9,-3.8) {$h_3$};
    \node[font=\small] (h4) at (4.2,0.1) {$h_4$};
\end{tikzpicture} 
        $\quad$
        \begin{tikzpicture}
  \tikzset{>={Latex[width=2mm,length=2mm]}}
  \node [circle,draw] (v1) at (0,0)   { $v_1$ };
  \node [circle,draw] (v2) at (-1,-3) { $v_2$ };
  \node [circle,draw] (v3) at (2,-2)  { $v_3$ };
  \node [circle,draw] (v4) at (1,-4)  { $v_4$ };
  \node [circle,draw] (v5) at (3,1)   { $v_5$ };
  \node [circle,draw] (v6) at (5,-1)  { $v_6$ };
  \node [circle,draw] (v7) at (4,-5)  { $v_7$ };
  \node [rectangle,draw] (h1) at (0.8, -2) { $h_1$ };
  \node [rectangle,draw] (h2) at (2.1, -0.4) { $h_2$ };
  \node [rectangle,draw] (h3) at (2.9, -3.8) { $h_3$ };
  \node [rectangle,draw] (h4) at (4.2, 0.1) { $h_4$ };
  \draw [->] (v1) -- (h1);
  \draw [->] (v2) -- (h1);
  \draw [->] (v4) -- (h1);
  \draw [->] (h1) -- (v3);
  \draw [->] (v1) -- (h2);
  \draw [->] (v3) -- (h2);
  \draw [->] (h2) -- (v5);
  \draw [->] (v4) -- (h3);
  \draw [->] (v7) -- (h3);
  \draw [->] (h3) -- (v3);
  \draw [->] (v5) -- (h4);
  \draw [->] (v7) -- (h4);
  \draw [->] (h4) -- (v6);
\end{tikzpicture} 
}
\caption{\textbf{Example of a directed hypergraph and a bipartite representation as a factor graph.} {\em Left}: A directed hypergraph. For instance, nodes $v_1, v_2, v_4$ are inputs to the hyperedge $h_1$ that is the input function for node $v_3$. {\em Right}: Its representation as a bipartite graph or factor graph. The square nodes are the hyperedges also called the factors. They represent the way to combine the inputs into a multi-body interaction in the input function of the output node. This represents the multi-body interaction in the ODE.}
\label{fig:dir-hypergraph}
\commentAlt{Figure~\ref{fig:dir-hypergraph}: 
Two representations of the same directed hypergraph. 
On the left, an actual hypergraph with nodes named v1-v7, and hyperarcs named h1-h4. Each hyperarc has many sources and one single target; more precisely
h1 has sources v1, v2, v4 and target v3; h2 has sources v1, v3 and target v5; h3 has sources v4, v7 and target v3; h4 has sources v5, v7 and target v6.
On the right, the same hypergraph is represented, but this time its nodes are of two types: nodes v1-v7 are circular, nodes h1-h4 are rectangles. The connections
from circular nodes to rectangles and from rectangles to circular nodes correspond to the hypergraph on the left.
For instance h1 has three incoming arrows from v1, v2, v4 and one outgoing arrow to v3.
}
\end{figure}

The easiest way to approach the problem of extending the notion of fibration to hypergraphs, as also suggested in~\citep{preti2024higher}, is to represent hypergraphs in the form of a \emph{factor graph}\index{hypergraph !factor graph }, a bipartite directed graph: we turn every node and hyperedge of the hypergraph $H$ into a node of the factor graph $[H]$, and use edges in $[H]$ to represent sources and targets.
Figure~\ref{fig:dir-hypergraph} (right) shows the factor graph representation $[H] $of the hypergraph $H$ appearing above in the same figure: the seven nodes of $H$ ($v_1$, \dots, $v_7$) are also nodes of the factor graph $[H]$, but in the factor graph we also have one extra node for every hyperedge. These extra nodes are called factors and are represented as rectangles (to highlight the fact that the factor graph is bipartite). 
Each factor is linked to its sources and targets by incoming and outgoing edges in the factor graph: for instance, hyperedge $h_1$ has source $\{v_1,v_2,v_4\}$ and target $\{v_3\}$. In the ODE, the factor represents the input function of the factor's output node.

Now, besides being convenient, it can be formally shown~\citep{boldi2023emergence} that graph fibrations from $[H]$ are equivalent to hypergraph fibrations\index{hypergraph !fibration } from $H$, provided that typing is taken into account. Typing here is
needed because we do not want to send nodes to factors or factors to nodes, 
something that we may end up doing without typing: recall that both factors and nodes of $H$ are nodes of $[H]$, so a homomorphism that does not consider types may freely mix factors and nodes, a behavior that we need to avoid.

Hence the blueprint for hypergraph fibration symmetries would be to use the map $[-]$ (technically, an equivalence functor between the category of hypergraphs and the category of typed graphs) to get back to the world of typed graphs, and then use the machinery we have already set up to discuss symmetries in that context.

A study of synchrony patterns in hypergraphs with identical nodes, which
also considers their stability,
can be found in \citep{aguiar2023}.

\section{Graph, hypergraph, and admissible graph representations of ODEs}
\label{sec:hypergraph-metabolic}

The representation of the ODE system is crucial to the application of
fibrations. Here, we review the three representations discussed so far
and explain when they are useful and when they are not.

The purpose of representing an ODE as a graph is to enable the
calculation of balanced colorings. These balanced colorings must align
with the cluster synchrony present in the ODE. This alignment is
essential for the practical application of fibrations in typical
scenarios. However, there is more to discuss beyond this initial
explanation.

As remarked before, each area of application has its own rules for
turning graphs (or multigraphs or hypergraphs) into model
equations. These rules may differ from those used in this book, and
how the graph is drawn can differ from one area to another.  We
further illustrate this issue by discussing the graph, hypergraph and
admissible graph representations using a genetic and metabolic
network,\index{network !metabolic } an important hypergraph to which we
return in Chapter \ref{chap:complex}.

\subsection{Graph representation}

To review the assumption of the graph representation, we return to the
UxuR-LgoR fiber in {\it E. coli} exemplified in Fig. \ref{F:Uxur} and
whose ODE is represented in (\ref{E:Ecoli_model}), as an example of a
system with two-body interactions. There are some hidden assumptions
that allow the graph to be a valid representation of the ODE and vice
versa (see also Chapter \ref{chap:alive}).  If couplings are additive
and two-body, for example---as we have often assumed in this book---then each edge in the graph determines only the contribution of one
input to the input function of the node.

Each edge-type in the graph Fig. \ref{F:Uxur} represents either an
activator of the form $d\frac{x^2}{(1+x^2)}$ or a repressor
$b\frac{1}{(1+x^2)}$.  Additionally, in the ODE (\ref{E:Ecoli_model})
we use the same constants $b$ and $d$ in $x_2$ and $x_3$ interaction
terms, leading to the same link in the graph. That is, $e_1=e_3$ and
$e_4=e_5$.  This consistency is necessary to generate the fibration;
if the parameters were different, the fiber would be gone (effectively
'bye-bye' to the fiber due to the symmetry breaking discussed in
Chapter \ref{chap:alive}). This may seem disappointing, as we dedicate
an entire chapter to explore this aspect.  Furthermore, the internal
degradation dynamics are assumed to be the same for nodes 2 and 3
since we use the same constant $c$ in the ODE. The equal internal
dynamics is represented in the graph by the same symbol for nodes 2
and 3.

There is an extra assumption for the graph to be a valid
representation of the ODE: we need to specify how different input
links are combined in a node's input function. In the assumption
presented in (\ref{E:Ecoli_model}), the combination is additive;
however, any other form of input function can also be valid. The key
point of the graph representation is that the type of input function
must remain the same for every node. So, all the links are combined in
the same way for every node's input function. This condition includes
two-body interactions but also many-body interactions. That is, a
many-body system with the same type of input function for all nodes
still admits a graph representation. Otherwise, a factor graph
(hypergraph) is needed to specify the different input functions (even in
the pair-wise case).



Thus, when more specific functions are required, the graph
representation must be augmented. In such scenarios, a hypergraph and
its corresponding factor graph would be appropriate to represent the
meaning of the different input functions, even for two-body
interactions.  Consequently, the entire book should be rewritten to
focus on hypergraphs. Fortunately, many of the results can be adapted
from graphs to hypergraphs, \emph{mutatis mutandis}, although
hypergraph fibrations remain an exciting and open area of research.

On the contrary, the more general theory of admissible ODEs from
Section \ref{sec:adODEs} is not to assign each term in a formula to a
corresponding arrow. Instead, the input isomorphism class of the input
set determines which components of the admissible ODE use the same
function (with corresponding variables assigned according to the
relevant input isomorphisms) but otherwise, this function is not
specified. Further discussed in this section, this representation is
the {\em admissible graph}.\index{admissible graph }

The extra structure imposed by passing from an admissible graph to a
hypergraph should not be viewed as {\em extending} the range of
admissible models, but as {\em restricting} it to a smaller class of
models whose form corresponds more closely to the biology and the
particular type of ODE describing the system.

\subsection{Hypergraph representation}

\begin{figure}[htb]
\centerline{%
\includegraphics[width=\textwidth]{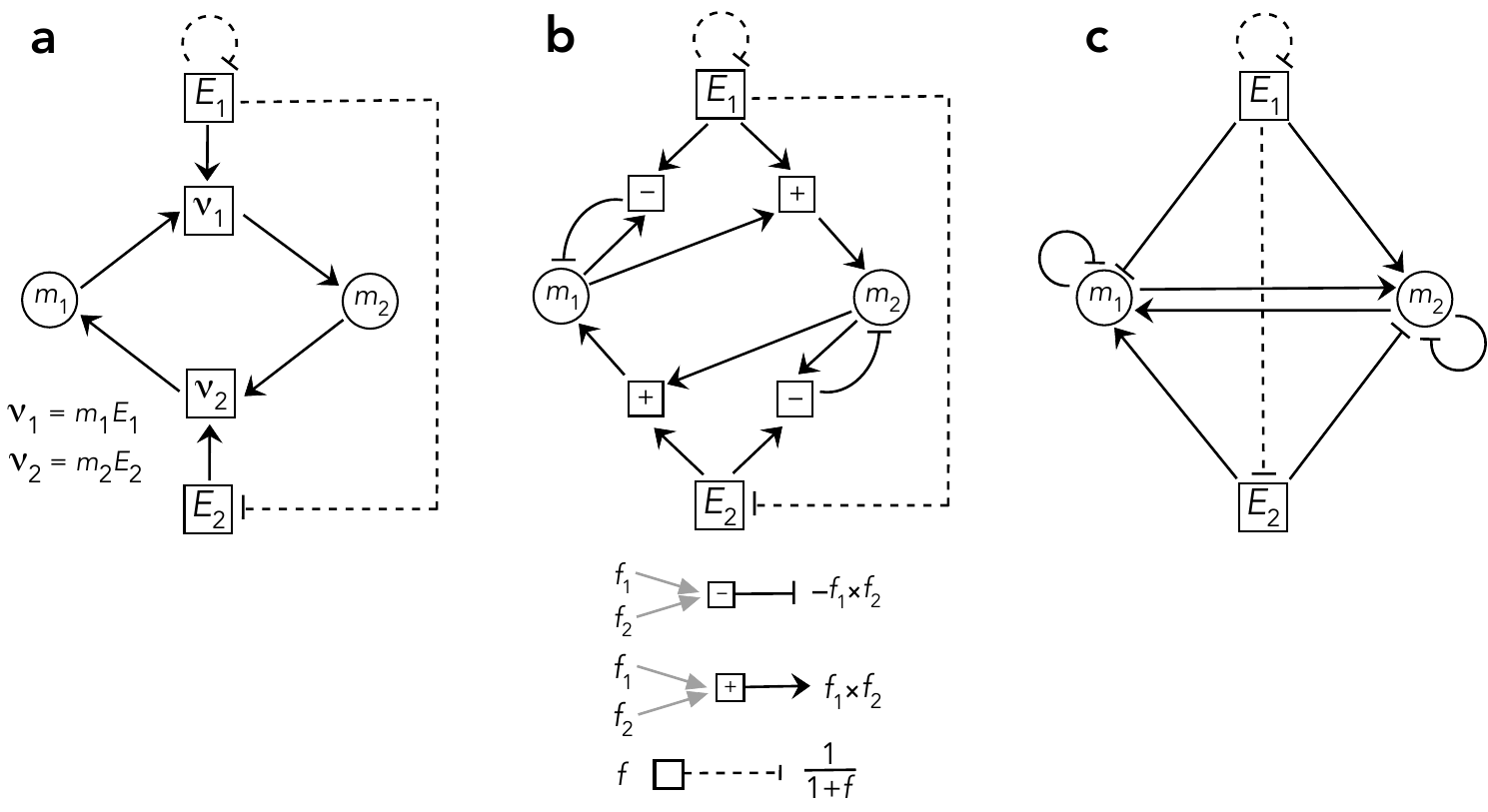} 
}
\caption{\textbf {Comparison of different graphical representations of an ODE.} (\textbf{a}) Biochemical representation
for \eqref{E:metabolic_network}.
\textbf{b}) Hypergraph representation.
(\textbf{c})  Admissible graph representation from Section \ref{sec:adODEs} following the conventions of \citep{GS2023}.}
\label{F:metabolic_hypergraph}
\end{figure}

As a typical example, we consider two metabolites $m_1$ and $m_2$ in a bidirectional biochemical  reaction metabolized by enzymes $E_1$ and $E_2$, which are, in turn, repressed, as in the ODE: 
\begin{equation}
\label{E:metabolic_network}
\begin{array}{rcl}
\dot m_1 &=& -E_1m_1+E_2m_2 \\
\dot m_2 &=& E_1m_1-E_2m_2 \\
\dot E_1 &=& -\alpha E_1 + \frac{1}{1+E_1} \\
\dot E_2 &=& -\alpha E_2 + \frac{1}{1+E_1} \\
\end{array}
\end{equation}
This ODE can be represented by a standard biochemical representation
as in Fig. \ref{F:metabolic_hypergraph}a; by
a hypergraph, Fig. \ref{F:metabolic_hypergraph}b;
or by a network following the conventions of the admissible graph framework in Section \ref{sec:adODEs} from \citep{GS2023}),
Fig. \ref{F:metabolic_hypergraph}c.
The hypergraph\index{hypergraph } in (b)  conveys the information to reproduce exactly the form of the ODE, including the detailed functions involved
in the ODE.
In the figures, the node variables
$m_1, m_2, E_1, E_2$
are written inside their node symbols.

We discuss now how these different graphical representations relate to
the ODE \eqref{E:metabolic_network}.  In the biochemical
representation Fig. \ref{F:metabolic_hypergraph}a, the variables $E_1,
E_2$ are replaced by $\nu_1=m_1E_1$ and $\nu_2=m_2E_2$. Both inputs
and outputs contribute to the terms of the ODE. The mass balance
condition implies that an outgoing arrow contributes a negative
term. For example node $m_1$ has an output to $\nu_1$ representing
$-m_1E_1$, and an input from $\nu_2$ representing $+m_2E_2$. The rules
for converting the graph to the ODE are standard in the area of
metabolic networks.  However, this is neither a graph representation
nor a hypergraph. It is just a convenient representation of the
biochemical pathways. Thus, it cannot be used in a fibration analysis
to calculate the balanced coloring. It does not represent the ODE.

In the hypergraph representation of Fig. \ref{F:metabolic_hypergraph}b, separate
terms in the ODE, and the variables they combine, are
drawn as smaller squares that contain $-, +$.
These represent the factors, which are the input functions of the output node of the factor, as shown on the right.
Their source nodes are combined using the appropriate function,
and the output is sent to the target of the output arrow.
For example, the term $-E_1m_1$ in the $\dot m_1$ component of
the ODE is represented by the two arrows targeting the top left-hand 
square marked `$-$', and the output arrow to node $m_1$ indicates
that this term appears in the $\dot m_1$ component.
It is assumed, by convention, that these terms are added
together in the model ODE.

\begin{figure}[htb]
\centerline{%
\includegraphics[width=\textwidth]{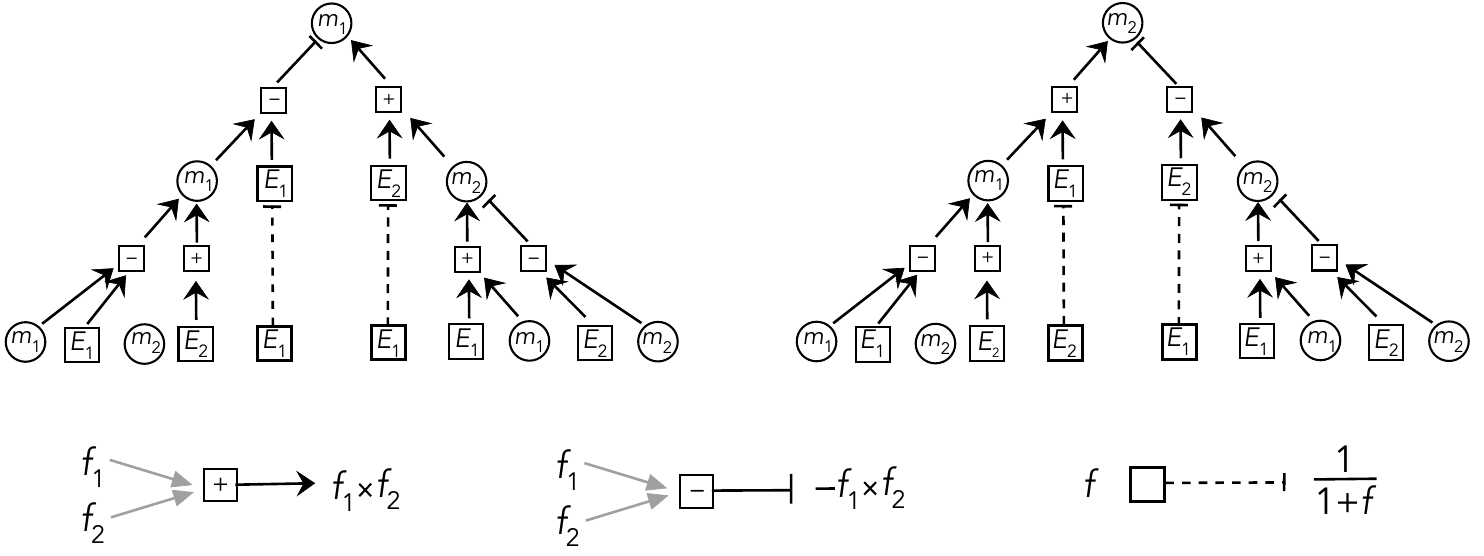} 
}
\caption{\textbf{Input trees in the hypergraph.} Input trees for nodes 1 and 2 in Fig. \ref{F:metabolic_hypergraph} are isomorphic, as would
be expected from a fibration.}
\label{F:metabolic_hypergraph_trees}
\end{figure}

The ODE \eqref{E:metabolic_network} does not have $\Z_2$ symmetry,
because the same term $\frac{1}{1+E_1}$ occurs in the last two
components. The situation is similar to that of the metabolator and
Smolen circuits, see Section \ref{S:metab+smol}, and we seek a
fibration rather than an automorphism.  Inspecting for input
isomorphisms, we see that there is a fibration in which nodes $m_1$
and $m_2$ are synchronized and nodes $E_1$ and $E_2$ are
synchronized. To support this claim, we show the input trees (to level
2) of nodes 1 and 2 in Fig. \ref{F:metabolic_hypergraph_trees}. The
input trees of $E_1$ and $E_2$ are standard AR loops.

This is the correct representation of the ODE that yields a balanced
coloring in agreement with the synchronous dynamics. In practical
applications, either a graph (provided the aforementioned conditions
are met) or a hypergraph should be utilized. We then ask: what is the
purpose of the admissible graph representation (Section
\ref{sec:adODEs}) in Fig. \ref{F:metabolic_hypergraph}c?

\subsection{Admissible graph representation}

We recall that the admissible graph is constructed as a graph of 'influence' between the nodes. Thus, an edge from $i$ to $j$  appears in the admissible graph  when there is an interaction term in the ODE of $j$ where $i$ appears.

In this book we normally use a multigraph to
represent admissible graphs---but, as previously explained,
its edges do not represent single-body interactions, as
is the case in the graph representation. Instead, our conventions on admissible ODEs, which focus on
input isomorphisms, effectively treat the multigraph as
a special kind of hypergraph. The entire input set of arrows (together with their source nodes)
is treated as a single hyperedge\index{hyperedge } representing the collective interactions
created by the source nodes. No other hyperedges occur.

Unlike a hypergraph, there is one extra feature: if
the input set includes more than one arrow of the same type, 
the function $f_c$ is symmetric in the corresponding entries. 
The intuition is that the node concerned `does not know' which nodes send which signal. It just processes the entire set of signals
according to $f_c$.
These constraints are formalized by `vertex symmetries'\index{vertex symmetry }, which permute
source node variables for arrows of the same type. An example 
occurs in equation \eqref{E:LO_genRP} for the lock-on\index{lock-on }
circuit in Section \ref{S:lock-on}.

In the admissible graph representation Fig. \ref{F:metabolic_hypergraph}c, the small square nodes representing functional
combinations no longer appear. 
Input arrows from node $j$ to node $i$
show that $\dot x_i$ depends on $x_j$. The node symbol itself
shows that $\dot x_i$ depends on $x_i$. Thus
$\dot x_i = f_i(x_i, x_{j_1}, \ldots, x_{j_k})$, where
$x_{j_1}, \ldots, x_{j_k}$ run through the source nodes
of the input arrows. Arrows drawn in the same style (solid, dashed, \ldots) indicate how these variables correspond in
the component functions; however, this information applies 
{\em only} to input isomorphic nodes. The graph
then encodes not a specific ODE, but the set of all
admissible ODEs. In this case, the admissible ODEs
have the form
\begin{equation}
\label{E:metab_admiss}
\begin{array}{rcl}
\dot m_1 &=& F(m_1,E_1,m_2,E_2) \\
\dot m_2 &=& F(m_2,E_2,m_1,E_1) \\
\dot E_1 &=& G(E_1,E_1) \\
\dot E_2 &=& G(E_2,E_1)
\end{array}
\end{equation}
for {\em arbitrary} functions $F,G$. 

\begin{remark}
We can also require the function $F$
to be symmetric under interchange of the first two
components, and of the last two, because $m_1E_1 = E_1m_1$
and $m_2E_2 = E_2m_2$.
But this feature is rather incidental, and we ignore it here.

In each component, the first variable is the node variable;
for example in the $\dot m_1$ component is the function
$F(m_1,E_1,m_2,E_2)$. In \eqref{E:metabolic_network}, there is no explicit
term involving $m_1$ on its own, so in an intuitive sense, the
`internal dynamic' is zero. But, as previously explained,
the formalism of admissible graphs does not associate
individual arrows with individual terms in the ODE.)
\end{remark}

If in \eqref{E:metab_admiss} we set
\[
F(u,v,w,x) = -uv+wx \qquad\quad G(y,z) = 
 -\alpha y + \frac{1}{1+z}
\]
(where $u,v,w,x,y,z$ are `dummy' variables used to define the form of
$F$ and $G$), then we recover \eqref{E:metabolic_network}.  However,
infinitely many other choices of $F$ and $G$ are possible in
general. Which choices make biological sense depends on the modeling
assumptions.  Crucially, this graph has a fibration, corresponding to
the cluster synchrony in which $m_1=m_2$ and $E_1=E_2$.  We deduce
that this cluster pattern is consistent with the ODE for {\em any}
choices of $F$ and $G$.  This is a much stronger statement than its
consistency with \eqref{E:metabolic_network}.  It implies that the
synchrony state is robust in the sense of Section \ref{sec:robust}.

But the price we pay for a stronger statement is that the graph does
not tell us how to write down a specific model ODE; just the class of
all admissible ones. So (c) is more useful for some theoretical
purposes, such as classifying possible cluster synchrony patterns, but
less so for modeling specific biological systems, especially in a
quantitative manner.

\subsection{Which representation do we use?}
\label{sec:which_rep}

The answer to this question is subjective and depends on the perspective of the practitioner. In most cases, we recommend using graph representations when the conditions for their validity are satisfied; otherwise, the more general hypergraph should be used. In practical situations, a practitioner---such as a bioinformatician or a physicist---may be interested in a specific system, like the regulatory network of prostate cancer. This system is described by particular ODEs but not by the entire set of admissible ones. Therefore, the theory of admissible representations would be inappropriate in this context. For any practical application, the practitioner should always use a hypergraph, which, as shown in the above cases, always gives the correct balanced coloring that will be reproduced in experiments.

The admissible graph representation is more useful for mathematicians attempting to prove general theorems. A good example of this is the case of robustness under general modifications of the admissible equations (Section \ref{sec:robust}). For a robust system (in this technical sense), we can claim that the synchronous pattern arises from a fibration, which is a fundamental theorem in the fibration framework. This is a very significant result: {\it fibration implies synchrony, and synchrony implies fibrations.} This means that every time we find synchrony in a system, we can use fibrations to describe it. This is the major theoretical underpinning for the applications in Chapters \ref{chap:brain1}, \ref{chap:brain2}, and \ref{chap:brain3}, where we will use fibrations to infer brain networks from synchronization data. 
But the theorem relies on the assumption of robustness for all general admissible models. It is not valid for balanced colorings on the hypergraph that do not satisfy the robustness condition.
However, such colorings may still exist experimentally, which is what matters in all applications.

Robustness is, of course, an important issue for biology too. So
results ought not to depend on the precise formula in the model.
However, network practitioners do not always appreciate this. In their
support, a hypergraph can reflect known interactions, restricting the
type of model more than by just assuming a fully general one.

Overall, the issues discussed in this section are fascinating, as they
explore the challenges of transitioning from ideal mathematical
theories to the realities of the physics of life.  The core argument
of this book is that once this tension is resolved, biology will
become more akin to physics, with theoretical predictions leading to
experiments rather than the other way around.

\section{Weighted networks}
\label{sec:weighted}

Another very common class of networks, not only in the context of
biology, is that of weighted graphs: a {\em weighted
  graph}\index{network !weighted }\index{graph !weighted } is a
network where each edge has a {\em weight}\index{weight }, a real
number that represents the strength of the influence between the two
nodes concerned.  While this situation is extremely common, and the
necessity to model it does not require any further discussion, weights
are difficult to deal with in a purely combinatorial way, and at base
of our approach is a definition of symmetry which is exclusively
combinatorial (i.e., discrete) in nature. The same limitation also
holds for automorphisms. Similarly, fibrations do not lend themselves
easily to the weighted world.

A possible, quite brutal, way to treat weights is by \emph{thresholding}:\index{thresholding } we establish a threshold $\theta$, and just throw away all edges with weights smaller than $\theta$, whereas the remaining edges are retained and their weight is discarded.
This solution is arbitrary, and the results we obtain depend on the choice of $\theta$. Thus, studies concentrate on varying $\theta$ and choosing an appropriate threshold where connectivity of the network arises. This naturally leads to a percolation problem.  Such a thresholding process is popular in the analysis of networks in biology, specially in functional brain networks from fMRI data as treated in Chapter \ref{chap:brain3}.

An alternative, more elegant, solution is \emph{discretization}:\index{discretization } we forget about fine-grained differences between weights, and consider edges with significantly different weights as if they represent two different types of interaction.
In other words, we turn weights into types, provided that the weights are first suitably coalesced (or `coarse-grained').
For a concrete example, consider the weighted network of Figure~\ref{fig:weighted}.

\begin{figure}[htbp]
\begin{tikzpicture}[main/.style = {draw, circle}, scale=.6, font=\tiny]

\def \n {12}
\def \radius {3cm}
\def \margin {8} 

\node[draw,circle,font=\footnotesize] at (0,0) (one) {$1$};
\node[draw,circle,font=\footnotesize] at (4.5,-2) (two) {$2$};
\node[draw,circle,font=\footnotesize] at (1.5,-2) (six) {$6$};
\node[draw,circle,font=\footnotesize] at (3,0) (three) {$3$};
\node[draw,circle,font=\footnotesize] at (6,0) (four) {$4$};
\node[draw,circle,font=\footnotesize] at (7,-2) (five) {$5$};
\node[draw,circle,font=\footnotesize] at (9,0) (seven) {$7$};

\draw[->, >=latex] (three) -- (one) node[midway,above] {$0.1$};
\draw[->, >=latex] (two) -- (three) node[midway,left] {$0.2$};
\draw[->, >=latex] (two) -- (four) node[midway,right] {$0.3$};
\draw[->, >=latex] (three) to[bend left, "0.7"] (four);
\draw[->, >=latex] (three) --(six) node[midway,left] {$0.1$};
\draw[->, >=latex] (four) to[bend left, "0.8"] (three);
\draw[->, >=latex] (five) --(two) node[midway,below] {$0.8$};
\draw[->, >=latex] (one) --(six) node[midway,left] {$0.3$};
\draw[->, >=latex] (two) --(six) node[midway,below] {$0.7$};
\draw[->, >=latex] (four) --(seven) node[midway,above] {$0.9$};
\draw[->, >=latex] (four) --(five) node[midway,right] {$0.8$};
\draw[->, >=latex] (seven) --(five) node[midway,right] {$0.4$};

\end{tikzpicture}
  \caption{\textbf{A weighted network.} Each edge is labeled by its weight. }
  \label{fig:weighted}
\commentAlt{Figure~\ref{fig:weighted}: 
A directed graph with weights on its arcs. Nodes are named 1-7.
Directed edges (weights in parenthesis) are: 1 to 6 (0.3), 2 to 3 (0.2), 2 to 4 (0.3), 2 to 6 (0.7), 3 to 1 (0.1), 3 to 4 (0.7), 3 to 6 (0.1),
4 to 3 (0.8), 4 to 5 (0.8), 4 to 7 (0.9), 5 to 2 (0.8), 7 to 5 (0.4).
}
\end{figure}

In Fig.~\ref{fig:weighted-discr}, we show three different ways to discretize the weights by turning them into types. For readability, types are displayed as edge labels, and are integers, but they should not be interpreted as weights any more, but as labels for types.

\begin{figure}[b!]
\begin{tabular}{ccp{0.5\textwidth}}
(a) & $\quad$ & \begin{tikzpicture}[main/.style = {draw, circle}, scale=.6, font=\tiny, baseline=(current bounding box.center)]

\def \n {12}
\def \radius {3cm}
\def \margin {8} 

\node[draw,circle,fill=yellow,font=\footnotesize] at (0,0) (one) {$1$};
\node[draw,circle,fill=yellow,font=\footnotesize] at (4.5,-2) (two) {$2$};
\node[draw,circle,fill=green,font=\footnotesize] at (1.5,-2) (six) {$6$};
\node[draw,circle,fill=blue,font=\footnotesize] at (3,0) (three) {$3$};
\node[draw,circle,fill=blue,font=\footnotesize] at (6,0) (four) {$4$};
\node[draw,circle,fill=blue,font=\footnotesize] at (7,-2) (five) {$5$};
\node[draw,circle,fill=yellow,font=\footnotesize] at (9,0) (seven) {$7$};

\draw[->, >=latex] (three) -- (one) node[midway,above] {$1$};
\draw[->, >=latex] (two) -- (three) node[midway,left] {$1$};
\draw[->, >=latex] (two) -- (four) node[midway,right] {$1$};
\draw[->, >=latex] (three) to[bend left, "1"] (four);
\draw[->, >=latex] (three) --(six) node[midway,left] {$1$};
\draw[->, >=latex] (four) to[bend left, "1"] (three);
\draw[->, >=latex] (five) --(two) node[midway,below] {$1$};
\draw[->, >=latex] (one) --(six) node[midway,left] {$1$};
\draw[->, >=latex] (two) --(six) node[midway,below] {$1$};
\draw[->, >=latex] (four) --(seven) node[midway,above] {$1$};
\draw[->, >=latex] (four) --(five) node[midway,right] {$1$};
\draw[->, >=latex] (seven) --(five) node[midway,right] {$1$};

\end{tikzpicture} \\[5em]
(b) & $\quad$ & \begin{tikzpicture}[main/.style = {draw, circle}, scale=.6, font=\tiny, baseline=(current bounding box.center)]

\def \n {12}
\def \radius {3cm}
\def \margin {8} 

\node[draw,circle,fill=purple,font=\footnotesize] at (0,0) (one) {$1$};
\node[draw,circle,fill=yellow,font=\footnotesize] at (4.5,-2) (two) {$2$};
\node[draw,circle,fill=green,font=\footnotesize] at (1.5,-2) (six) {$6$};
\node[draw,circle,fill=blue,font=\footnotesize] at (3,0) (three) {$3$};
\node[draw,circle,fill=blue,font=\footnotesize] at (6,0) (four) {$4$};
\node[draw,circle,fill=blue,font=\footnotesize] at (7,-2) (five) {$5$};
\node[draw,circle,fill=yellow,font=\footnotesize] at (9,0) (seven) {$7$};

\draw[->, >=latex] (three) -- (one) node[midway,above] {$1$};
\draw[->, >=latex] (two) -- (three) node[midway,left] {$1$};
\draw[->, >=latex] (two) -- (four) node[midway,right] {$1$};
\draw[->, >=latex] (three) to[bend left, "2"] (four);
\draw[->, >=latex] (three) --(six) node[midway,left] {$1$};
\draw[->, >=latex] (four) to[bend left, "2"] (three);
\draw[->, >=latex] (five) --(two) node[midway,below] {$2$};
\draw[->, >=latex] (one) --(six) node[midway,left] {$1$};
\draw[->, >=latex] (two) --(six) node[midway,below] {$2$};
\draw[->, >=latex] (four) --(seven) node[midway,above] {$2$};
\draw[->, >=latex] (four) --(five) node[midway,right] {$2$};
\draw[->, >=latex] (seven) --(five) node[midway,right] {$1$};

\end{tikzpicture} \\[5em]
(c) & $\quad$ & \begin{tikzpicture}[main/.style = {draw, circle}, scale=.6, font=\tiny,baseline=(current bounding box.center)]

\def \n {12}
\def \radius {3cm}
\def \margin {8} 

\node[draw,circle,fill=purple,font=\footnotesize] at (0,0) (one) {$1$};
\node[draw,circle,fill=yellow,font=\footnotesize] at (4.5,-2) (two) {$2$};
\node[draw,circle,fill=green,font=\footnotesize] at (1.5,-2) (six) {$6$};
\node[draw,circle,fill=blue,font=\footnotesize] at (3,0) (three) {$3$};
\node[draw,circle,fill=red,font=\footnotesize] at (6,0) (four) {$4$};
\node[draw,circle,fill=brown,font=\footnotesize] at (7,-2) (five) {$5$};
\node[draw,circle,fill=pink,font=\footnotesize] at (9,0) (seven) {$7$};

\draw[->, >=latex] (three) -- (one) node[midway,above] {$1$};
\draw[->, >=latex] (two) -- (three) node[midway,left] {$2$};
\draw[->, >=latex] (two) -- (four) node[midway,right] {$3$};
\draw[->, >=latex] (three) to[bend left, "1"] (four);
\draw[->, >=latex] (three) --(six) node[midway,left] {$1$};
\draw[->, >=latex] (four) to[bend left, "8"] (three);
\draw[->, >=latex] (five) --(two) node[midway,below] {$8$};
\draw[->, >=latex] (one) --(six) node[midway,left] {$3$};
\draw[->, >=latex] (two) --(six) node[midway,below] {$7$};
\draw[->, >=latex] (four) --(seven) node[midway,above] {$9$};
\draw[->, >=latex] (four) --(five) node[midway,right] {$8$};
\draw[->, >=latex] (seven) --(five) node[midway,right] {$4$};

\end{tikzpicture} 
\end{tabular}
  \caption{\textbf{Possible discretizations of a weighted graph.} Three typings obtained as different discretizations of the weighted network in Fig.~\ref{fig:weighted}, with the corresponding coarsest equitable partitions (balanced colorings).}
  \label{fig:weighted-discr}
\commentAlt{Figure~\ref{fig:weighted-discr}: 
This figure shows three graphs identical to the one described in~\ref{fig:weighted}, but with different weights and colored nodes.
Graph (a): all weights are 1; nodes 1,2,7 are yellow, nodes 3,4,5 are blue, node 6 is green.
Graph (b): weights of the original graph that were above 0.5 are now 2, and the other weights are 1; node 1 is red, nodes 2,7 are yellow, nodes 3,4,5 are blue, node 6 is green.
Graph (c): weights are the same as in the original graph but multiplied by 10; node 1 is dark red, node 2 is yellow, node 3 is blue, node 4 is light red, node 5 is brown, node 6 is green, node 7 is pink.
}

\end{figure}

The three discretizations in Figure~\ref{fig:weighted-discr} are obtained as follows:
\begin{itemize}
    \item (a) The type of edge is the same ($1$) regardless of the weight, i.e., $\eta(a)=1$.
    \item (b) We distinguish between weights smaller than 0.5 and weights larger than or equal to 0.5, i.e., $\eta(a)=\lfloor 1+2w(a)\rfloor$.
    \item (c) We use a more fine-grained discretization, letting $\eta(a)=\lfloor 1+10w(a)\rfloor$.
\end{itemize}
In practice, the coarsest equitable partition (a) does not take weights into account \emph{at all}: the symmetries we find are simply due to the graph structure. For instance, blue nodes all have one incoming edge from blue and one from yellow; whereas yellow all have just one incoming edge from blue; the green node is the only one with indegree three.

Now, when we consider (b), we see that blue nodes are still together (they all receive a $2$-edge from blue and a $1$-edge from yellow), but the three yellow nodes are now not together in the same cluster any more: while $2$ and $7$ both receive a $2$-edge from blue, $1$ receives a $1$-edge from blue, so they get different inputs. Intuitively, the different weights in the connecting edges make the two recipients different.

The more fine-grained the discretization becomes, the finer is the equivalence relation we obtain at the end. In (c) almost all edge types are different, and the few that are not are irrelevant to induce a symmetry.

Of course, even if in our example we use a linear form of discretization (in fact, translating weights into types by using their magnitude and deciding where to put boundaries), other forms of discretizations are entirely possible (e.g., logarithmic) and can make sense in some contexts.

This proposal has a number of evident shortcomings:
\begin{itemize}
    \item Discretization is arbitrary, and how we discretize weights influences strongly the result.
    \item Every form of discretization is by its very nature discontinuous; hence, for weights that are `close to the border' we must somehow decide which type they belong to, and how we take this decision may lead to largely different outcomes. 
    \item Closely related to the previous item is the fact that there can be no overlap between types, and that each edge has but one type.
    \item There is no inherent form of `arithmetic' between weights: receiving two edges of weights $w_1$ and $w_2$ from nodes of the same class is totally different from receiving one single edge with weight $w_1+w_2$ from a node of the same class. This happens because lifting is a discrete property, that must hold for every single edge, so a node with two incoming edges will never ever be matched with a node with just one incoming edge, independently of how weights are discretized. 
    \item The previous objection can be avoided to some extent by redefining the notion of a fibration, but only if the admissible ODEs have a suitable additive structure such as linear coupling. This is treated in \citep{aguiar2018, aguiar2020,sequeira2021,sequeira2022}.
\end{itemize}

\subsection{Balanced coloring for a weighted network}
\label{sec:balanced}

The discretization and thresholding techniques discussed in the previous section enable the transformation of a weighted network into a binary network, which can then be analyzed using fibrations. Alternatively, we can redefine balanced coloring directly for the weighted network without filtering. While this approach does not yield a fibration, it often suffices in many real-world applications where the primary goal is to identify the balanced coloring of the network. Therefore, generalizing balanced colorings from unweighted to weighted networks can be adequate for numerous applications.

We can reinterpret the condition of balanced coloring for an unweighted network $G=(V,E)$ in matrix form. Let the 
adjacency matrix be $A_{ki}=\{0,1\}$, and for simplicity assume that all edges have the
same type. Then $i$ and $j$ are balanced in clusters $C \in \mathcal{C}$ if, for
all $C \in \mathcal{C} $, all pairs of distinct nodes $i,j \in C$, and all $D
\in \mathcal{C}$, we have:
\begin{equation}
  i \sim j \iff \sum_{k \in D: ki \in E} A_{ki} = \sum_{r \in D: rj \in E} A_{rj} .
  \label{eq:balanced_unweighted}
\end{equation}
Here $i \sim j$ means that $i$ and $j$ are in the same cluster.

The weighted case
requires new variables and a modification of constraints. The adjacency matrix is modified using $w_{ij}$ to
denote the weight of edge $ij \in E$, and the condition of balanced coloring becomes:
\begin{equation}
  i \sim j \iff \sum_{k \in D: ki \in E} w_{ki} = \sum_{r \in D: rj \in E} w_{rj} .
  \label{eq:balanced_weighted}
\end{equation}

We acknowledge that this is one possible solution, which assumes that the colors are balanced by the sum of the weights. This assumption is reasonable if the system of ODEs represented by the weighted graph is additive in the weights of various input terms within the input function that defines the dynamics of each node. 
However, this assumption may not hold true in all cases. Each application should take into account different rules to achieve a balance in the weights. 
We revisit the issue of weighted networks in Chapter \ref{chap:alive}.

\section{Multiplex and multilayer graphs
\label{sec:heterogeneous}}

More often than not, biological systems have a heterogeneous structure. This heterogeneity mainly happens in two different ways~\citep{hammoud2020multilayer,della2020symmetries}.

\begin{figure}[b!]
  \centering 
  \includegraphics[width=0.4\textwidth]{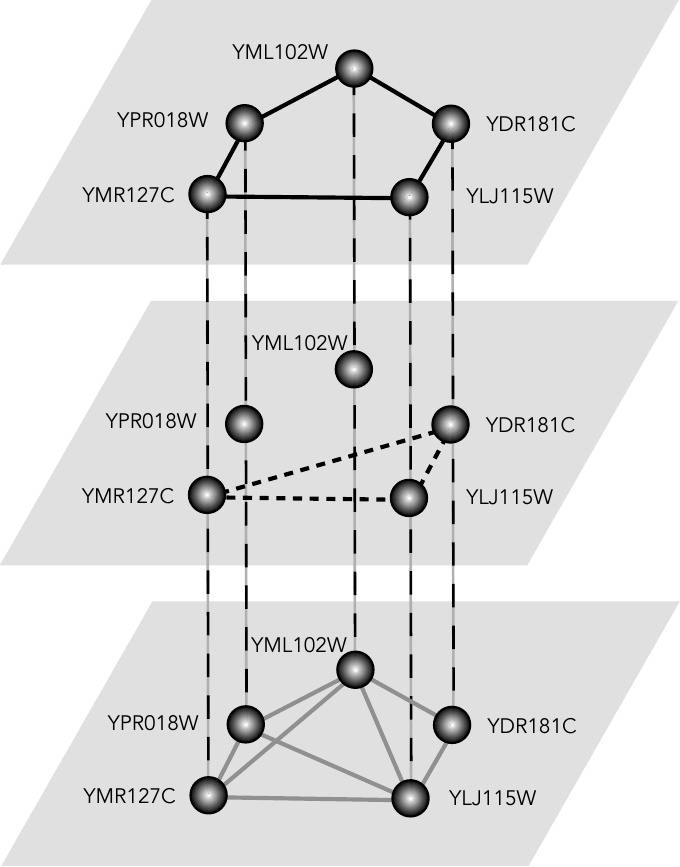} \qquad
  \includegraphics[width=0.47\textwidth]{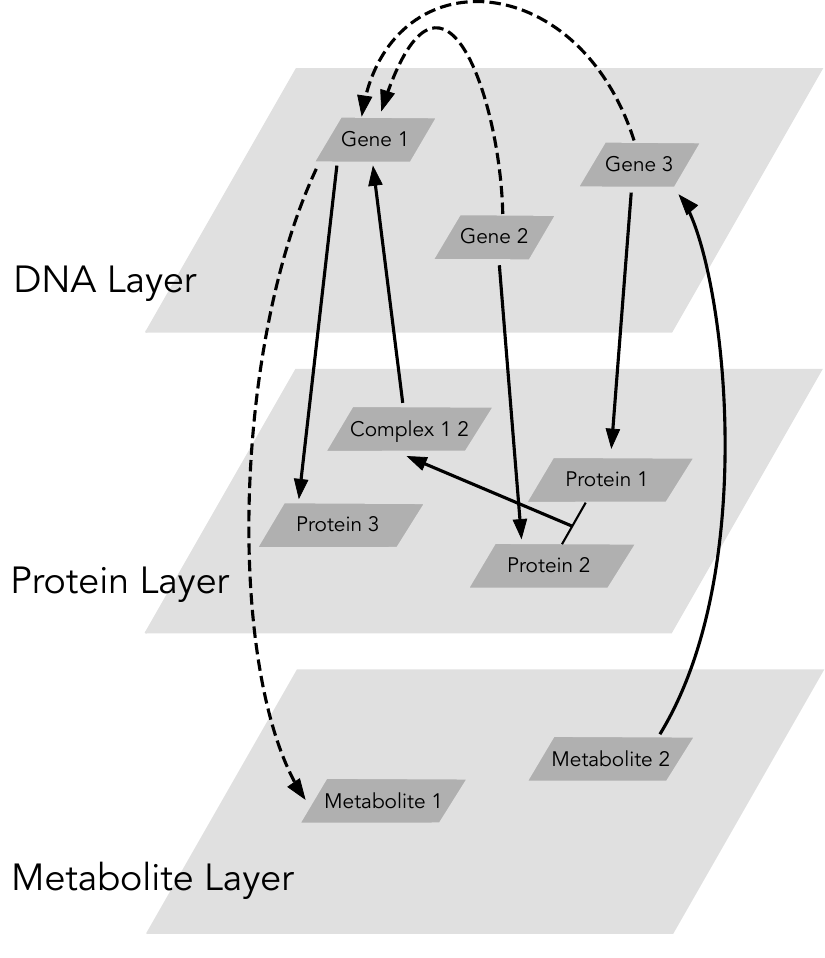}
   \caption{\textbf{Heterogeneous graphs.} {\em Left}: A multiplex, arc-heterogeneous (undirected) network of proteins: different layers (edge types) correspond to different types of interaction. {\em Right}: A metabolic network with multiple types of nodes forming a multilayer network. Figure reproduced from \citep{hammoud2020multilayer}.}
    \label{fig:multilayered}
\commentAlt{Figure~\ref{fig:multilayered}: 
Two pictures, both show a tridimensional projection of three grey layers, with networks on each layer and connection between nodes in the layers.
}

\commentLongAlt{Figure~\ref{fig:multilayered}: 
Picture on the left.
Upper layer: an undirected graph with nodes called YML102W, YPR018W, YMR127C, YLJ115W, YDR181C, connected in a five-loop (a pentagon).
Middle layer: an undirected graph with nodes having the same names as in the upper layer, and the same positions; connections are dashed and 
are only between YMR127C, YLJ115W, YDR181C, connected in a three-loop (a triangle).
Lower layer: an undirected graph with nodes having the same names as in the upper layer, and the same positions; this time besides the same connections as 
in the upper layer there are also the connections YMR127C-YML102W, YPR018W-YLJ115W and YML102W-YLJ115W.
Nodes with the same name are connected across the layers.
Picture on the right.
Upper layer (DNA Layer): three boxes named Gene 1, Gene 2, Gene 3.
Middle layer (Protein Layer): four boxes named Protein 1, Protein 2, Protein 3, Complex 1 2.
Third layer (Metabolite Layer): two boxes named Metabolite 1 and Metabolite 2.
Directed connections: Gene 1 to Protein 3, Gene 2 to Protein 2, Gene 3 to Protein 1, Complex 1 2 to Gene 1, Metabolite 2 to Gene 3.
Directed dashed connections: Gene 1 to Metabolite 1, Gene 2 to Gene 1, Gene 3 to Gene 1.
Hyperarc directed connection: sources Protein 1 and Protein 2, target Complex 1 2.
}
\end{figure}

\begin{itemize}
    \item Edges of the network may bear different meanings: we already presented examples of networks with activation and inhibition edges, but the number of `edge types' is not limited to two. Networks with different types of edges  are often referred to as 
{\emph{arc-heterogeneous} 
or \emph{multiplex}\index{network !multi-layered }. Since these networks present different types of interactions (edges) between the same set of actors (nodes), they are conveniently represented by replicating the same set of nodes over different layers and assigning to each layer all edges  of a given type.} 
    Figure~\ref{fig:multilayered} (left) shows an example of a network of five proteins with three different layers. The example in this figure is undirected, but directed graphs, or networks mixing directed and undirected layers, can also occur.
    \item Nodes of the network can also have different types; in this case the network is often said to be \emph{node-heterogeneous} 
 \emph{or multi-layered}. For instance, in a metabolic network\index{network !metabolic } some nodes may represent genes, while others represent proteins or metabolites.  Figure~\ref{fig:multilayered}  (right) shows an example, {for which each layer now contains all nodes of the same type. 
    An important difference is that the nodes in different layers of the multilayer network of Fig.~\ref{fig:multilayered}  (right) are  different objects (genes, proteins, metabolites), while  the nodes in different layers of the multiplex network in Fig.~\ref{fig:multilayered}  (left)  are replicas of the same objects across the layers. 
    }
\end{itemize}

Clearly, both classes of heterogeneity can be present at the same time. For instance, there may be different types of nodes (proteins, genes, metabolites, cells, \dots) and different types of relations between them (inhibition, activation, repression, expression, \dots). In most cases, a given type of relation can  occur only between two specific types of nodes. But, given two types of nodes, there can be many (or no) types of relation connecting them. Again, some relations can be directed, others can be undirected. All these interactions can be of many-body types leading to hypergraphs.

As mentioned earlier, the fibration framework can easily and uniformly deal with this kind of scenario, that is, with graphs that have different types of nodes and/or edges. In the general fibration formalism, 
notions of node\index{node type } and arrow types\index{arrow type } are built into the definitions from
the beginning. 

More often than not, types are referred to as `colors' in the graph-theoretic literature, and heterogeneous graphs are hence called (and represented as) node-colored and/or edge-colored graphs.\index{graph !colored } We prefer to refrain from using the term `color' here, because we are already using colors to refer to node equitable partitions, and having two types of colors would  be confusing.
Clearly non-heterogeneous graphs are a special case of heterogeneous graphs, where the set $T$ contains only one element (all nodes and all edges have the same type).

What a type represents depends on the context, but nodes and edges with different types cannot be mixed by a symmetry: a gene cannot have the same role as a metabolite, an inhibition relation cannot have the same role as an activation. This `hard barrier' imposed by typing in a heterogeneous world is formalized by requiring the relevant graph homomorphisms to preserve types.

Recall that, if $G$ and $H$ are heterogeneous, a graph homomorphism $f: G \to H$ is a graph homomorphism which further satisfies
\[
    \nu(x)=\nu(f(x)) \qquad \eta(a)=\eta(f(a))
\]
for all nodes $x \in V_G$ and edges $a \in E_G$.
The idea behind this condition is quite obvious: typing provides an impassable wall 
between nodes and edges. Two nodes of different types cannot be mapped to each other, because they have a different nature. Similarly, two edges of different types will represent different types of interactions, so they cannot be related by a symmetry.

Once the notion of homomorphism has been generalized to the world of typed graphs, we can extend all the notions of fibrations, minimal fibration, etc. Everything holds \emph{mutatis mutandis} in the heterogeneous world: we will have heterogeneous equitable partitions, the heterogeneous coarsest equitable partition, and algorithms to find such partitions just extend to the heterogeneous case by taking  node and edges types into account at all stages. 
We discuss this further in Chapter \ref{chap:algorithms} when talking about algorithms to find fibers.

\subsection{Balanced coloring for a multiplex/multilayer weighted network}
\label{sec:multiplex}

As an example, we generalize the balanced coloring to a brain multiplex.
Neurons can communicate with each other via different signaling
mechanisms and over a range of different timescales.
The result is a multiplex, multilayer weighted hypergraph of  neuronal communication, which combines all the existing heterogeneities into one system. Understanding the symmetries of such a system poses
significant challenges to standard graph theoretical 
approaches.

\begin{figure}
  \centering 
  \includegraphics[width=\textwidth]{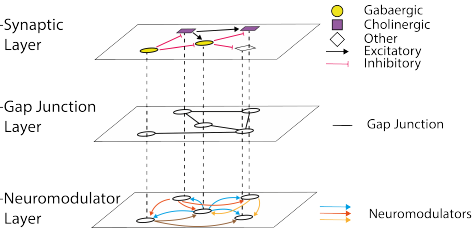} 
   \caption{\textbf{Multiplex/multilayer weighted brain hypergraph.} The complexity of the brain network requires a heterogeneous model combining different types.}
    \label{fig:multiplex}
\commentAlt{Figure~\ref{fig:multiplex}: 
A picture showing a tridimensional projection of three layers, with networks on each layer and connection between nodes in the layers.
Upper layer (Synaptic layer): a graph with two nodes representing Cholinergic (say C1, C2), two nodes representing Gabaergic (say G1, G2) and one node representing Other (say O).
There are Excitatory arcs G1 to G2 and G1 to C2.
There are Inhibitory arcs C1 to G1, C1 to C2, C2 to G2, C2 to O.
Middle layer (Gap Junction Layer): nodes in the same quantity and positions as in the upper layer, but this time they all have the same circular shape and no name. There are undirected connections
between (using the names of the upper layer) C1-C2, C2-O, O-G2, G2-C1, O-G1.
Lower layer (Neuromodulator Layer): nodes in the same quantity and positions as in the upper layer, but this time they all have the same circular shape and no name. They are freely connected by colored
arcs with different colors, representing neuromodulators.
}
\end{figure}

To account for the
diversity of neuronal and synaptic properties in biological neuronal
circuits, different types of links and nodes can be considered (Fig. \ref{fig:multiplex}).  Typically, a
neuron can simultaneously communicate via three routes: (1)
chemical signaling via neurotransmitters, (2) electrical signaling via
gap junctions, and (3) neuroendocrine wireless signaling via
neuropeptides. Therefore, we consider three main classes of
graphs $c$: $c=1$ chemical synapses graph, $c=2$ gap-junctions graph,
and $c=3$ neuromodulatory graph made of neuropeptide-mediated
interactions. Each of these graphs will have different types of nodes
and links, which reflect the diversity of the neuronal types, as well
as of neuropeptides identities, gap junction properties, and diverse neuropeptide
interactions. 

For instance, the $c=1$ synaptic class 
is subdivided into types $t$ as excitatory, inhibitory, and
neurotransmitter (NT) identity, i.e., glutamate, acetylcholine (ACh)
and Gamma-Aminobutyric Acid (GABA). The wireless $c=3$ graph is made of neuropeptide modulators.
There is a large number of neuropeptide precursors encoded in the
genome, potentially producing over 300 individual neuropeptides, and
each neuron releases a unique cocktail of them.

Simply collapsing these interactions onto a single binary layer would result
in an almost fully connected graph. At the other extreme, considering
the combinatorial identity of each pairwise connection in this matrix
would make nearly all of its elements unique. Both extremes limit any
analysis of symmetrical structure or other graph theoretical measures. So, a symmetry analysis needs to be performed with care.

The model for balanced coloring which takes all
this information into account is the following.
The different multilayer graphs and strengths are taken into account
by generalizing \eqref{eq:balanced_weighted} to a multilayer/multiplex weighted
matrix $w_{ki}^{c,t}$ which defines the strength of an edge
$ki$ of class $c=\{$chemical synapses, gap-junctions, neuropeptide
connections$\}$ and type $t$ as defined above.  Each of these classes
and types are balanced as:
\begin{equation}
i\sim j \iff   \sum_{k \in D: ki \in E} w_{ki}^{c,t} = \sum_{r \in D: rj \in E}
   w_{rj}^{c,t}  \,\,\,\,\,\,\,\,\, \mbox{\textrm for each class $c$ and type
     $t$ .}
  \label{eq:balanced2}
\end{equation}

\section{Using typing to restrict the class of homomorphisms}
\label{sec:treatment}

Heterogeneity, expressed by typing, is often inherent in the nature of the system we are modeling. We can exploit heterogeneity to restrict the kind of symmetries we want to consider. Indeed, this is sometimes necessary to capture properties that otherwise would fail to be represented \citep{morone2020fibration}. As an example, we now consider a relevant problem related to nodes with no inputs.\index{node !with no inputs }

\begin{figure}[htbp]
  \begin{tikzpicture}[main/.style = {draw, circle}, scale=.6]

\def \n {12}
\def \radius {3cm}
\def \margin {8} 

\node[draw,circle] at (0,0) (one) {$1$};
\node[draw,circle] at (4.5,-2) (two) {$2$};
\node[draw,circle] at (1.5,-2) (six) {$6$};
\node[draw,circle] at (2.25,-1) (seven) {$7$};
\node[draw,circle] at (3,0) (three) {$3$};
\node[draw,circle] at (6,0) (four) {$4$};
\node[draw,circle] at (9,0) (five) {$5$};

\draw[->, >=latex] (one) -- (three);
\draw[->, >=latex] (two) -- (three);
\draw[->, >=latex] (two) -- (four);
\draw[->, >=latex] (three) to[bend left] (four);
\draw[->, >=latex] (four) to[bend left] (three);
\draw[->, >=latex] (five) to (four);
\draw[->, >=latex] (one) to (six);
\draw[->, >=latex] (two) to (six);
\draw[->, >=latex] (one) to (seven);
\draw[->, >=latex] (two) to (seven);

\end{tikzpicture}
  \caption{\textbf{Typing. } A graph with many nodes having no inputs ($1$, $2$ and $5$).}
  \label{fig:many-noinputs}
\commentAlt{Figure~\ref{fig:many-noinputs}: 
A directed graph with nodes named 1-7. Directed edges connect 1 to 3, 7, 6; 2 to 3, 4, 7, 6; 3 to 4; 5 to 4. 
}
\end{figure}
Consider as an example the graph of Fig.~\ref{fig:many-noinputs}. Here each node has either one or two inputs. Its coarsest equitable partition is the one shown in Fig.~\ref{fig:many-noinputs-bc}.

\begin{figure}[htbp]
\begin{tikzpicture}[main/.style = {draw, circle}, scale=.6]

\def \n {12}
\def \radius {3cm}
\def \margin {8} 

\node[draw,circle,fill=green] at (0,0) (one) {$1$};
\node[draw,circle,fill=green] at (4.5,-2) (two) {$2$};
\node[draw,circle,fill=red] at (1.5,-2) (six) {$6$};
\node[draw,circle,fill=red] at (2.25,-1) (seven) {$7$};
\node[draw,circle,fill=blue] at (3,0) (three) {$3$};
\node[draw,circle,fill=blue] at (6,0) (four) {$4$};
\node[draw,circle,fill=green] at (9,0) (five) {$5$};

\draw[->, >=latex] (one) -- (three);
\draw[->, >=latex] (two) -- (three);
\draw[->, >=latex] (two) -- (four);
\draw[->, >=latex] (three) to[bend left] (four);
\draw[->, >=latex] (four) to[bend left] (three);
\draw[->, >=latex] (five) to (four);
\draw[->, >=latex] (one) to (six);
\draw[->, >=latex] (two) to (six);
\draw[->, >=latex] (one) to (seven);
\draw[->, >=latex] (two) to (seven);

\end{tikzpicture}
  \caption{\textbf{Minimal coloring.} The coarsest equitable partition of the graph in Fig.~\ref{fig:many-noinputs}.}
  \label{fig:many-noinputs-bc}
\commentAlt{Figure~\ref{fig:many-noinputs-bc}: 
Same graph as in Figure~\ref{fig:many-noinputs}, but with nodes colored as follows: 1,2,5 are green, 3,4 are blue, 6,7 are red.
}
\end{figure}

While this coloring can make sense in some contexts, in most cases we cannot simply postulate that all nodes with no inputs ($1$, $2$ and $5$ in this example) belong to the same class. Think, for instance, of a very large network with two such nodes very far apart, or even belonging to different components. 
In many situations it is much more natural to suppose that different nodes with no inputs all belong to different classes.  One way to take care of this observation is to modify the color refinement algorithm in the following way (as suggested in \citep{morone2020fibration}):
\begin{enumerate}
    \item Identify all nodes with no inputs (nodes that receive signals only from themselves) from other nodes, and assign to each of them a different initial color.
    \item Identify all nodes with no inputs from any node (including itself) and assign each of them its own separate color during every iteration of the algorithm.
\end{enumerate}
While this solution certainly solves the problem, we can try to tackle the situation in a more principled and general way.

We can take the graph $G=(V,E)$ under consideration and suitably add types to its nodes as follows: nodes with no inputs are all assigned a different type, whereas nodes with one or more inputs are assigned the same type. This typing corresponds to the idea that nodes with no input will always be different from one another (and different from all the other nodes of the network, anyway), whereas we seek symmetries in the remaining nodes.

Formally, we use the following typing function for nodes:
\[
    \nu(x) = \begin{cases}
            x & \text{if $x$ has no inputs}\\
            * & \text{otherwise.}
            \end{cases}
\]
In this way, all nodes with one or more inputs have the same type, whereas nodes with no inputs all have different types. 
When executing any algorithm to compute balanced partitions on the heterogeneous graph obtained in this way, nodes with no inputs will be distinct. 

Now, if we determine the coarsest equitable partition of this heterogeneous graph (see Section~\ref{sec:alghet}) we will obtain the coloring of Figure~\ref{fig:many-noinputs-bcfix-pre}:
now nodes $1$, $2$ and $5$ have different colors (because we imposed them to have types). A consequence of this is that also $3$ and $4$ are no longer symmetrical to each other, because each of them receives an input from a different node ($3$ receives inputs from $1$, and $4$ receives inputs from $5$). Nonetheless, some symmetry remains ($6$ and $7$ have the same color, because they do receive inputs from symmetrical nodes).
\begin{figure}[htbp]
\begin{tikzpicture}[main/.style = {draw, circle}, scale=.6]

\def \n {12}
\def \radius {3cm}
\def \margin {8} 

\node[draw,circle,fill=green] at (0,0) (one) {$1$};
\node[draw,circle,fill=yellow] at (4.5,-2) (two) {$2$};
\node[draw,circle,fill=red] at (1.5,-2) (six) {$6$};
\node[draw,circle,fill=red] at (2.25,-1) (seven) {$7$};
\node[draw,circle,fill=blue] at (3,0) (three) {$3$};
\node[draw,circle,fill=blue!20] at (6,0) (four) {$4$};
\node[draw,circle,fill=brown] at (9,0) (five) {$5$};

\draw[->, >=latex] (one) -- (three);
\draw[->, >=latex] (two) -- (three);
\draw[->, >=latex] (two) -- (four);
\draw[->, >=latex] (three) to[bend left] (four);
\draw[->, >=latex] (four) to[bend left] (three);
\draw[->, >=latex] (five) to (four);
\draw[->, >=latex] (one) to (six);
\draw[->, >=latex] (two) to (six);
\draw[->, >=latex] (one) to (seven);
\draw[->, >=latex] (two) to (seven);

\end{tikzpicture}
  \caption{\textbf{How to deal with nodes with no inputs.} The coarsest equitable partition of the graph in Fig.~\ref{fig:many-noinputs} after preprocessing the nodes with no inputs. In many realistic cases, i.e., biological networks where nodes with no inputs corresponds to external sources, it makes more sense to consider them with different colors, rather than with the same colors as in the original definition of fibration.}
  \label{fig:many-noinputs-bcfix-pre}
\commentAlt{Figure~\ref{fig:many-noinputs-bcfix-pre}: 
Same graph as in Figure~\ref{fig:many-noinputs}, but with nodes colored as follows: 1 is green, 2 is yellow, 3 is dark blue, 4 is light blue, 5 is brown, 6, 7 are red.
}
\end{figure}

We have discussed, for simplicity, only the case of a homogeneous graph on which we impose a typing that takes the no-input case into account. This idea can be easily extended to the typed case in a straightforward way. Nodes can be assigned the same color only if they have the same node type and the same number of
input edges of each type. Not all such colorings are balanced---correspond to fibrations---but all balanced colorings must have this property since
fibrations have the lifting property, and in this generalization, they must also preserve node and edge types.


\chapter[Fiber Bundles for Physics --- Fibrations for Biology]{\bf\textsf{Fiber Bundles for Physics---Fibrations for Biology}}
\label{chap:bundles}

\begin{chapterquote}
  The notions of fiber bundles, general fibrations, and graph
  fibrations are closely related, and further insight into fibrations is
  obtained when we compare them. In this
  chapter we elaborate on the links between these mathematical
  notions, seeking unification between the distinct areas in physics
  (where fiber bundles are abundant) and biology (where fibrations are
  abundant).  Fiber bundles and fibrations
  are related but arise in different contexts: fibrations in homotopy
  theory and fiber bundles in topological geometry.  The aim of the
  chapter is to sketch how these concepts are related, without
  providing many formal definitions.  We give references for more
  formal treatments of these topics.
  \end{chapterquote}

\section{Fiber bundles}
\label{fiber_bundles}

Fiber bundles\index{fiber bundle } go back to \citep{seifert1933}, in a special
case that contains the main ideas; it was later developed
by numerous mathematicians, with the first general definition being given by \cite{whitney1935}. `Fibration’ as a term
in topology was introduced (with a slightly different name) by \cite{serre1951}.

Fibrations \index{fibration } became established as a general concept in the 1960s 
through the work of
Serre\index{Serre, Jean-Pierre } and \cite{grothendieck1959}\index{Grothendieck, Alexander } in algebraic geometry\index{algebraic geometry }
and category theory,\index{category theory } inspired by the topological concept with the same
name. Fibrations are maps similar to fiber bundles but more general ~\citep{bundles1,bundles2}.  
In this usage, the
structure of the fibers in fiber bundles is constrained: often all
fibers should be equivalent topologically with local Cartesian product
structure. Moreover, each fiber can be equipped with
a `structure group' $G$ that acts on it (in the same way for each fiber), in which case the fiber bundle is called a `$G$-bundle'. Instead, in graph fibrations, the fibers can differ from
one base point to another, which makes them more suitable for describing the great complexity of  biology.

Fiber bundles\index{fiber bundle } and
group theory provide a unified theoretical framework
for modern physics. Fiber bundles describe the fundamental
interactions in nature in a unified manner, forming the basis of
the standard model of particle physics through gauge invariance of
the principal bundle. Likewise, the requirement of invariance under
local rotations in curved spacetime implies that general relativity is
also a gauge invariant theory described by a fiber bundle, thus
unifying all forces under the common notion of bundles.

To compare bundles and fibrations, it is convenient to first describe
fiber bundles\index{fiber bundle !over manifold } over manifolds and then generalize to the discrete case
of fiber bundles and fibrations for graphs.\index{fibration !graph } Fiber bundles over
manifolds appear when building a new manifold from
the topological product of two given manifolds. Products of topological
spaces are very common in mathematics, since they allow the
construction of larger spaces from smaller spaces.

Similarly to fibrations, fiber bundles consist of a base\index{base } $B$ and a
fiber\index{fiber } $F$ which is copied to every point in the base to form the total
space $E$. Both fiber bundles and fibrations capture the idea of one
manifold (the fiber) being indexed or parametrized over another
(the base) to form the total space of the bundle. Thus, the fiber of
the bundle is a topological space which is parametrized by the
base. From this point of view, fiber bundles appear to be similar to
fibrations. Differences are discussed in the next subsection.

Fiber bundles can be seen as generalization of Cartesian products.\index{Cartesian product } We
start by defining these product spaces, called trivial bundles;\index{bundle !trivial } then
we show how nontrivial bundles emerge by adding a `twist' to the
Cartesian product. We then repeat the same sequence of
definitions for graphs, beginning with Cartesian graph products,
generalizing to
graph bundles, and ending with graph fibrations. With these definitions
to hand, we then speculate on why fibrations, being a less
restricted form of symmetry than groups, describe the natural symmetries of
biological networks. We elaborate on why groups describe the inanimate
world, while fibrations and groupoids describe the living world.

\subsection{Cartesian products and trivial fiber bundles}

The simplest (trivial) way to obtain a fiber bundle is by
`multiplying' two spaces using the Cartesian product.

\begin{definition}{\bf Cartesian product.}
In set theory, the {\em Cartesian product}\index{Cartesian product } of two sets $A$ and $B$ is the
set of all ordered pairs $(a, b)$: $ A\times B=\{(a,b)\mid a\in
A\ {\mbox{ and }}\ b\in B\}.$ 
\end{definition}

If $A$ and $B$ are endowed with a topology, their product comes also equipped with
a natural topology~\citep{kelley1955}.

\begin{figure*}
    	\centering \includegraphics[width=\linewidth]{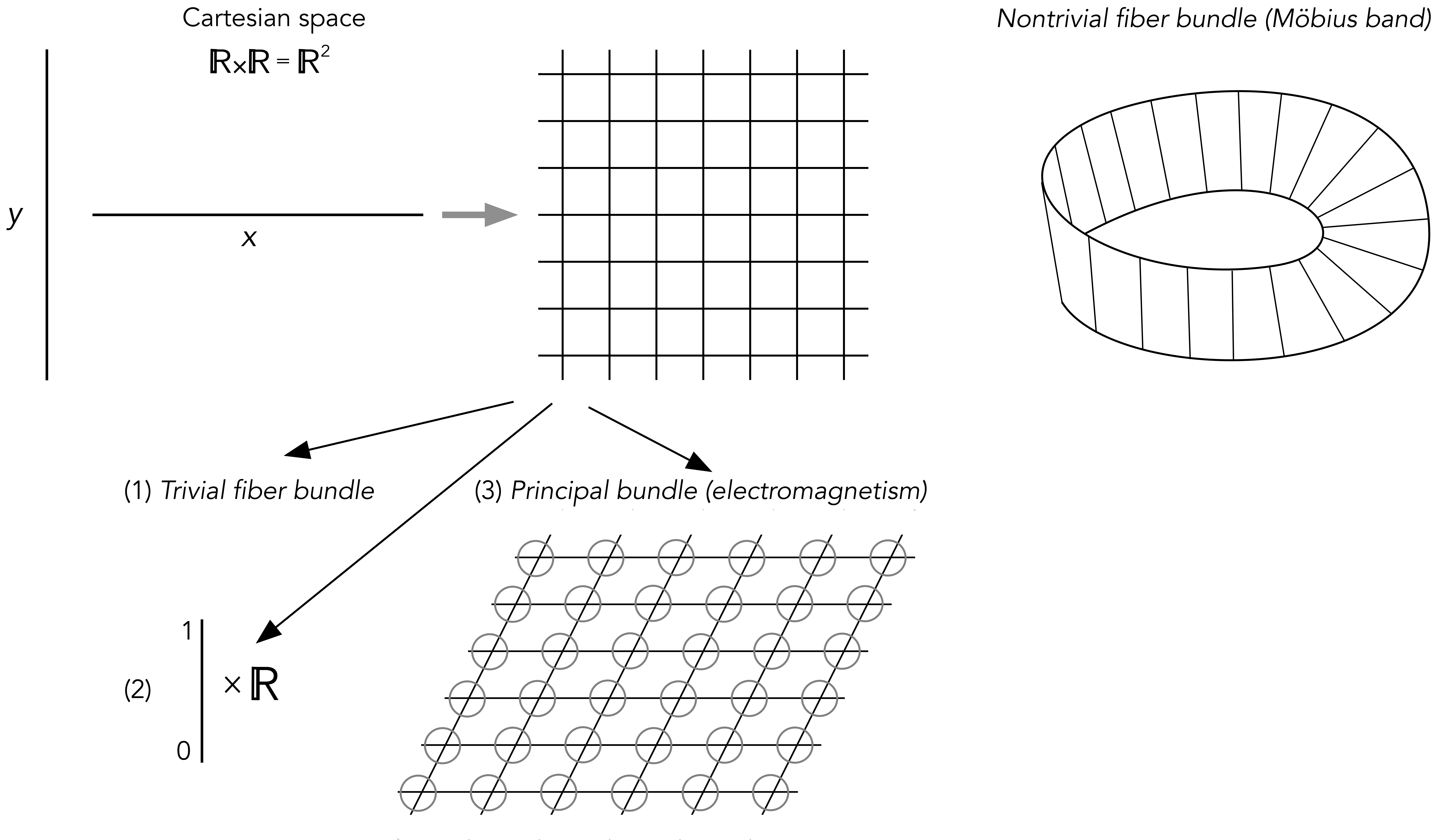}
	\caption{\textbf{Fiber bundles.} {\it Top left}: the plane as
          an $\R$-bundle over $\R$. {\it Top right}: the M\"obius band as
          an $\R$-bundle over the circle. {\it Bottom}: Electromagnetism as a circle bundle over $R^2$.}
	\label{bundlestofibrations}
\commentAlt{Figure~\ref{bundlestofibrations}: 
Four subfigures in two rows of two. Top left: Labeled `Cartesian space R x R = R^2'.
Vertical line labeled y, horizontal line labeled x. arrow points to square grid.
Top right: Labeled `nontrivial fiber bundle (M\"obius band).
Images of M\"obius band divided by a series of lines at right angles to its edges.
Below these: three arrows. First points to label `(1) trivial fiber bundle'.
Second labeled (2)  points to bottom left figure.
Third labeled `(3) Principal bundle (electromagnetism)' points to bottom right figure.
Bottom left: Vertical line labeled 0 at bottom, 1 at top, followed by x R.
Bottom right: square grid drawn in perspective as a horizontal plane.
Attached to each grid point is a small circle.
}
\end{figure*}

A simple example of a Cartesian product is $\mathbb{R}^2 = \mathbb{R}
\times \mathbb{R}$, which takes a line $\mathbb{R}$ and places
another line, also $\mathbb{R}$, at each point of the first line, to form a
plane (in this case $\mathbb{R}$ is endowed with a topology, which carries over to
$\mathbb{R}^2$). Thus the two-dimensional space consists of two identical
spaces $\mathbb{R}$ that are combined to create the total
space $\mathbb{R}^2$, see Fig. \ref{bundlestofibrations} (top left). Likewise, we
could consider the product of a line segment $(0,1)$ and a circle
(1-sphere $\mathbb{S}^1$) to form a cylinder.

These Cartesian products\index{Cartesian product } are special examples of fiber bundles: a
fiber (line segment) is combined with the base ($\mathbb{S}^1$) to
form the total space (essentially, a cylinder). The total space can be
written as $E=B\times F$: here the total space is not just `locally a
product': it is {\em globally} a product.  Cartesian
products are {\it trivial} fiber bundles. In both cases
$\mathbb{R}^2$ and the cylinder, a set of global coordinates is enough
to specify any position in the total space. {\it Nontrivial}
fiber bundles are generalizations of the global Cartesian product that
look like a direct product locally, but not globally. Nontrivial
fiber bundles are `twisted' Cartesian products\index{Cartesian product !twisted } like the M\"obius band,\index{Mobius band @M\"obius band } Fig. \ref{bundlestofibrations} (top right).

Before discussing nontrivial fiber bundles we make a detour to
explain the importance of trivial fiber bundles in modern theoretical
physics, by showing how trivial bundles describe all fundamental
interactions between elementary particles. Then we explain what we can
learn from this detour to describe the biological world.

\subsection{Principal bundles in physics}
     
Fiber bundles can combine all kinds of topological spaces and structures. They
are classified in terms of the kind of fiber space that is attached to
the base manifold. For instance, when the fiber is a vector space, the
resulting bundle is a vector bundle.\index{bundle !vector } A tangent bundle\index{bundle !tangent } consists of the
tangent spaces of a manifold. These fibers are important in the
Lagrangian formulation of classical mechanics \citep{kibble2004}. When the fiber is a Lie
group we have a principal bundle \citep{bundles1}.\index{bundle !principal }

Principal bundles play a fundamental role in physics by enabling the
construction of a geometric theory of the fundamental forces.
Interactions can be seen as fiber bundles over Minkowski spacetime,\index{Minkowski spacetime }
with a continuous Lie group as the fiber describing the particular
gauge symmetry\index{gauge symmetry } of the interaction. For instance, the quantum
mechanical description of electromagnetism\index{electromagnetism } is understood as a fiber
bundle over spacetime whose fibers have the gauge symmetry of
rotations of a circle \citep{weyl2}. Figure
\ref{bundlestofibrations} (bottom) pictures this fiber bundle by assigning a
circle to every point of a two-dimensional spacetime base. The extra
dimension given by the angle on the circle specifies the phase factor
of the complex-valued wave function $\psi(x,t)$ describing
electrons or charged particles moving in the manifold.

Gauge symmetry\index{gauge symmetry !in quantum mechanics } in quantum mechanics arises because observables are
real numbers and therefore depend on the wave function according to the form
$\psi^*\psi$. This creates a gauge freedom in the definition of the
wave function, since it can be rescaled by
\begin{equation}
  \psi \to \tilde{\psi} = e^{i \alpha} \, \psi,
\label{u1}
\end{equation}
without any consequence for the observables. Indeed,
\[
(e^{i\alpha}\psi)^*(e^{i\alpha}\psi) = e^{-i\alpha}e^{i\alpha}\psi^*\psi
= \psi^*\psi
\]

Mathematically, we can see these rotations as the action of the
symmetry group $U(1)$ by rotations $e^{i \alpha}$ in the complex plane
through an arbitrary angle $\alpha$. The elements of this group are
the fibers of the fiber bundle. The electromagnetic interaction
between electrons arises when the phase factor is allowed to change
along the base $\alpha(x)$ with $x\in B$. The gauge field that `glues'
the different circles together, while keeping the dynamics invariant
under the local shifts of the phase factors, is the {\it connection}
of the fiber bundle, which in this case is the electromagnetic
potential $A_\mu$. This gauge field determines how the position of an
electron in the circle changes from a point in spacetime to its
neighbor. The observed electromagnetic field or magnetic flux arises
as the non-zero curvature\index{curvature } of this trivial fiber bundle along closed
trajectories (loops) in spacetime \citep{maldacena}.

The symmetries of gauge theories are local in the sense that the
parameter $\alpha(x)$ of the Lie group depends on the position $x$ in
the base. Global symmetry in this context is a gauge transformation
with constant parameter $\alpha$ at every spacetime point, and it is
analogous to a rigid rotation of the geometric coordinate system. A
gauge transformation with a local parameter $\alpha(x)$ generates a
local symmetry, and the gauge field of the physical interaction.
However this `locality' does not change the structure group defining
the fiber bundle, which remains a trivial Cartesian product of the
groups for the spacetime manifold and the manifold of the gauge
group. Thus, even though the gauge symmetry is a local symmetry, the
underlying fiber bundle remains of a trivial kind as the Cartesian
product.

The mathematics of gauge theories was discovered by mathematicians as
fiber bundles.  Generalizing the gauge theory of electromagnetism, the
theory of electroweak interactions was later discovered by considering
the fibers as elements of the symmetry group $SU(2)$. Geometrically,
this is described by a 3-sphere $\mathbb{S}^3$ whose elements are
$2\times 2$ matrices with unit determinant, and corresponds to the
surface of a four-dimensional ball describing isospin
space \citep{glashow1961}. In similar fashion, the strong
interaction of $SU(3)$ describes the color space of quantum
chromodynamics \citep{greiner2007}.\index{quantum
chromodynamics }

Going beyond the standard model, it is interesting that fibrations
appeared in physics in the late 1980s through the quantum geometry of
string theory,\index{string theory } since these theories require the dimension of spacetime
to be 10, while we observe only 4. A string theory vacuum decomposes
the 10-dimensional space as $M_{10} = M_4 \times X$, where $M_4$ is
Minkowski 4-dimensional space and $X$ provides the extra 6 dimensions
in a tightly curled-up manifold called a `Calabi--Yau fibration'
\citep{calabi}.\index{fibration !Calabi--Yau }

\subsection{Nontrivial fiber bundles}

Fiber bundles admit more interesting products than the trivial ones
described above. Nontrivial fiber bundles generalize the Cartesian
direct product of spaces to `twisted' direct products by describing
structures that look like a trivial direct product locally, but not
globally. This is analogous to a manifold, which generalizes Euclidean
space by being locally, but not globally, Euclidean. In general, a
nontrivial bundle is a local product of the base with the fiber.

The standard example of a nontrivial fiber bundle is the M\"obius
band,\index{Mobius band @M\"obius band } which is a `twisted' Cartesian product of a line segment and a
circle, Fig. \ref{bundlestofibrations} (top right). The M\"obius band is created
by attaching a copy of the line segment $[0,1]$ to the circle
$\mathbb{S}^1$ but in a nontrivial manner, by adding a 180$^\circ$
twist as we go around one revolution. The band can still be mapped
onto the base $\mathbb{S}^1$ and, locally, still looks like the
trivial cylinder. However, a point in the band cannot be described by
global coordinates (the angle in $\mathbb{S}^1$ and the position in
the line interval) as for the trivial global parametrization of the
cylinder. Indeed, the M\"obius band is not topologically equivalent to
the cylinder; in particular it has a single boundary component and is
non-orientable. The cyclic group $G=\Z_2$ is an example of the structure group
of a fibration. It acts on each fiber
by flipping it end to end, $x \mapsto 1-x$. Traversing once round the base transforms the fiber in this manner.

     \begin{figure}[b!]   
  	\centering 
   \includegraphics[width=0.6\textwidth]{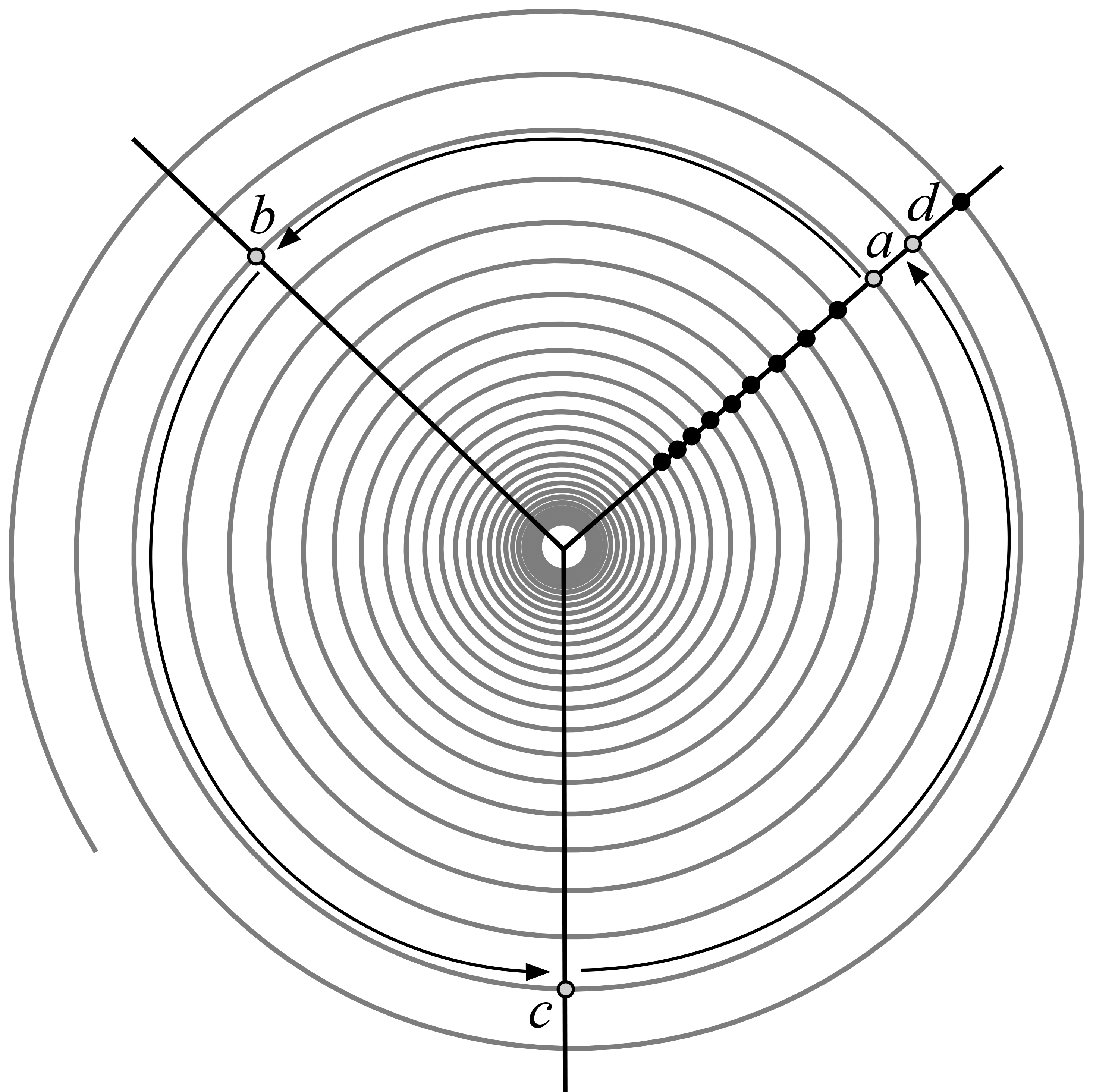}
	\caption{\textbf{Logarithmic spiral.} A fiber bundle with base $\mathbb{S}^1$ and fiber $\Z$ (black dots). One counterclockwise rotation around the base moves a typical point $a$ on the fiber, via $b$ and $c$,
 returning to the original fiber at $d= a+1$. The entire fiber shifts one step
 outwards, from $n$ to $n+1$.}
	\label{fibration_spiral}
\commentAlt{Figure~\ref{fibration_spiral}: 
A blue spiral with many coils becoming denser and denser as they get close to the center.
Three axes are drawn from the center going out, leaving 120 degrees between each axis and the following. A single coil is highlighted in dark, with arrows going outwards.
The connection points with the axes are called a, b, c, d (d is on the same axis as a, but on the immediately outer coil of the spiral).
}
\end{figure}

Another example is a logarithmic spiral
of Fig. \ref{fibration_spiral},\index{logarithmic spiral } which
can be projected radially onto the base $\mathbb{S}^1$.
For the full spiral, infinite in both directions, this
is a fibration with discrete fibres homeomorphic to the integers $\Z$
(the intersection of a radius with the spiral). Going once round
the circle shifts the fiber by $n \mapsto n\pm1$ depending on the direction
(clockwise or counterclockwise), so the structure group is $\Z$
under addition. 
Globally, the spiral and $\mathbb{S}^1$ are different.
Without the shift up or down by 1, the figure would be an infinite set
of concentric circles, a trivial $\Z$-bundle over $\mathbb{S}^1$.

\subsubsection{Connections}

In terms of fiber bundles, a {\it connection}\index{connection } tells us how movement in
the total space induces change along the fiber.  A connection
describes how to move from one fiber to another by describing how a
quantity associated with a manifold changes as we move from one point
to another by connecting neighboring fiber spaces.  As discussed
above, in the principal bundles of the standard model the connections
are the gauge fields mediating the interactions, i.e., the fundamental
forces.  We  come back to connections in Chapter
\ref{chap:geometrization} in the context of a financial model
and its analogs for genetic and brain networks.

\section{Fibrations in algebraic topology and category theory}

In algebraic geometry, the notion of fibration\index{fibration !in category theory } was introduced by
\cite{grothendieck1959} in the context of category
theory, inspired by the topological concept with the same name.  But
the topological idea, which developed into what arose in gauge
theory, predates Grothendieck. In algebraic geometry, a fibration\index{fibration !in algebraic geometry } is
a map from an algebraic variety to a lower-dimensional variety having some
reasonable properties like lifting or surjectivity. In topology,
a fibration is a further generalization of the concept of a fiber
bundle. Fibrations in topology are denoted by $F\to E\to B$.

The common feature of bundles and fibrations is that both maps
project a big space to a smaller one, and certain features transverse
to the fiber are preserved. So each fiber is embedded in the big space
in a nice way. These features are local neighborhoods for topology and
 input sets for graphs.

A distinctive feature of fiber bundles is that all fibers are
homeomorphic to each other.  In simpler terms, a fiber bundle is a map
between fibers to the base, where all fibers are topologically equivalent to the same space.
In a fibration, however, fibers need not necessarily be the
same space. They still parametrize a topological space by a family
of spaces, one for each point of the base, but there is a certain amount of freedom to choose
different fibers along the base.  In particular, in graph fibrations,
the input tree needs not vary in any nice way from one fiber to its
neighbors in the base.

Certainly, the condition that the fibers preserve local
in-neighborhoods (that is, input sets) is the same for both fibrations and fibre bundles,
although the technical implications are even more interesting since
homeomorphisms do not have a direction, whereas combinatorial
fibrations do not preserve out-neighborhoods, but only
in-neighborhoods. This is the main difference that lets us apply
fibrations to biological networks, which are usually directed
networks.

\begin{figure*}
  	\centering 
        \includegraphics[width=.6\linewidth]{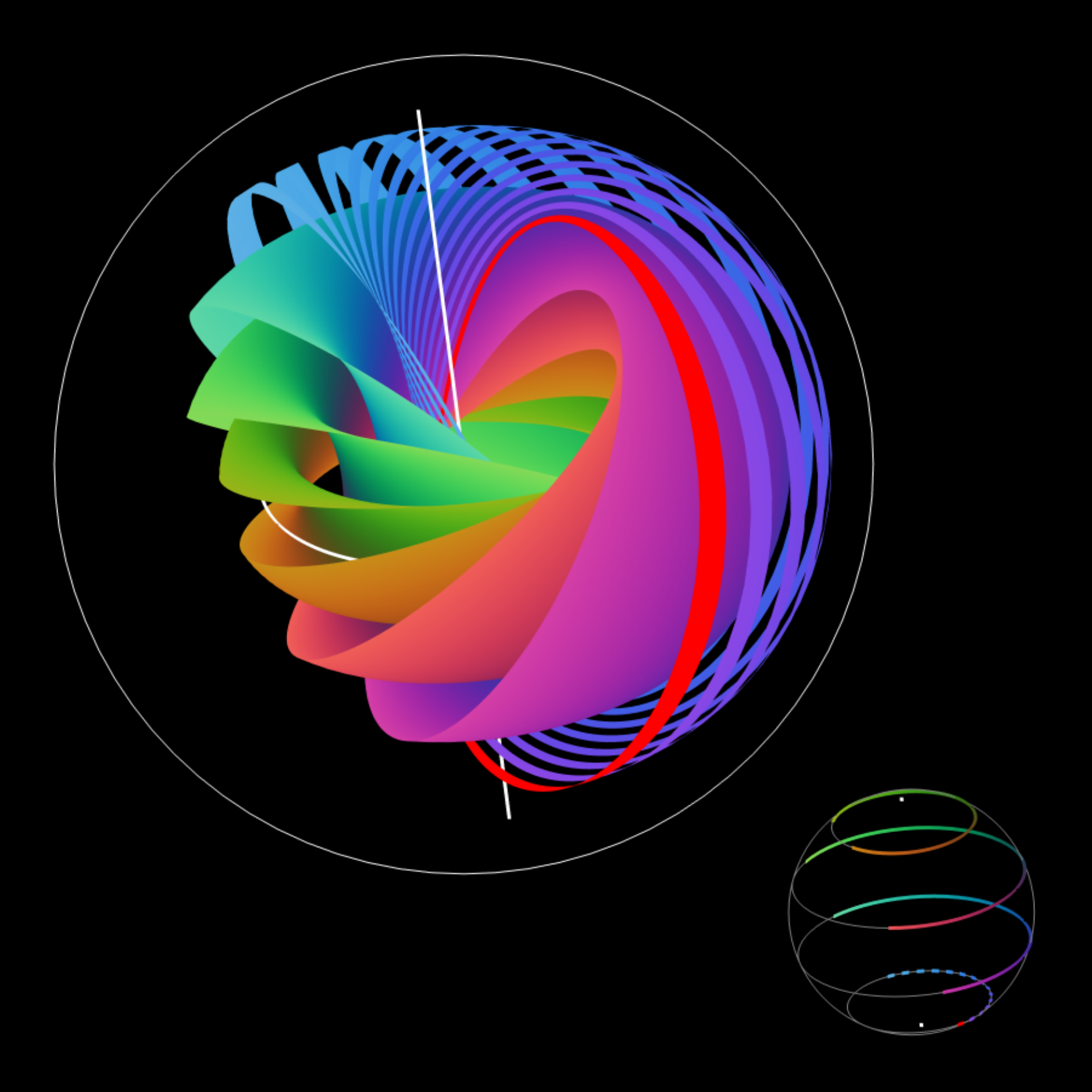}
	\caption{\textbf{Hopf fibration.} Ken Shoemake's\index{Shoemake, Ken }
          picture of the Hopf fibration donated to the Hopf topology
          archive at purdue.pagaloo.com/d/edu/hopf.math \citep{niles}. }
	\label{F:bundlestofibrations2}
\commentAlt{Figure~\ref{F:bundlestofibrations2}: 
Illustrative only. The Hopf fibration  is shown as a set of circles sweeping out surfaces.
The circles are great circles on a 3-sphere, and each circle corresponds to a point
on a 2-sphere. So the 3-sphere is a circle bundle with base the 2-sphere.
}
\end{figure*}

One of the earliest examples of fibration in topology is the Hopf
fibration,\index{fibration !Hopf } Fig. \ref{F:bundlestofibrations2}. This a locally trivial fibration since it is locally a product space. Thus, this fibration is
technically a fiber bundle, referred to as a fibration in the
literature for historical reasons.  The Hopf fibration structure is denoted:
$\mathbb{S}^{1}\hookrightarrow \mathbb{S}^{3}{\xrightarrow {\ p\,}}\mathbb{S}^{2}$. It has a
local product structure since every point of $\mathbb{S}^2$ has some neighborhood $U$
whose inverse image in $\mathbb{S}^3$ can be identified with the product of $U$
and a circle: $p^{-1}(U) \cong U \times \mathbb{S}^1$. While this fibration is
locally trivial, it is not a trivial fiber bundle, i.e., $\mathbb{S}^3$ is not
a global product of $\mathbb{S}^2$ and $\mathbb{S}^1$ although locally it is
indistinguishable from it. See Fig. \ref{F:bundlestofibrations2}.

\subsubsection{Definitions}

For the mathematically inclined reader, we end this section with
topological definitions of fiber bundles and fibrations following
\citep{cohen}. This is one of the few expositions in the literature where
both concepts are directly compared.

\begin{definition}{\bf Fiber bundle in topology.}
 A fiber bundle\index{fiber bundle } $F \to E: \pi \to B$ consists of a
  base space\index{base !space } $B$, fiber space $F$, and total space $E$, together with
  a projection map $\pi: E \rightarrow B$.  The total space $E$
  locally looks like the product of the base and fiber
  spaces. More precisely, every point $x \in B$ is contained in an
  open set $U \subset B$ such that there is a homeomorphism $\phi:
  \pi^{-1}(U) \to U \times F$. Further, $\pi \circ \phi^{-1}$ gives
  the projection onto the first factor of $U \times F$.
\end{definition}

In classical homotopy theory,\index{homotopy theory }
 a fibration\index{fibration !in homotopy theory }
 $p : E \to B$ is a
continuous map between topological spaces such that the `homotopy
lifting property'\index{homotopy
lifting property } holds with respect to all fiber spaces.  The most basic
property is that given a point $e \in E$ and a path $[0,1] \to B$ in
$B$ starting at $p(e)$, the path can be lifted to a path in $E$
starting at $e$.  A generalization of the notion of a fiber bundle due
to Serre is simply a map that satisfies this type of lifting property.
That is, a Serre fibration\index{fibration !Serre } is a map between topological spaces $ \varphi : E
\to B$ that satisfies the `homotopy lifting property'.
More generally, a Hurewicz fibration\index{fibration !Hurewicz } is a surjective,
  continuous map $p : E \to B$ that satisfies the homotopy lifting
  property for all spaces.

\section{Graph products}

We now generalize the topological concepts of Cartesian product, fiber
bundle and fibrations to their combinatorial counterparts defined,
respectively, as graph products, graph bundles, and graph fibrations. We
 explain the relations among these concepts, and what we can learn
from them to describe biological networks. We use graph terminology,
namely `vertex' rather than `node'. Table \ref{comparison} provides
a short `dictionary' for this analogy.

\begin{table*}
\centering
\begin{tabular}{| l  c  r |}
\hline 
 Topology & $\longrightarrow$  & Combinatorics \\ \hline \hline
  Cartesian product & $\longrightarrow$ & Cartesian graph product \\
  Fiber bundle & $\longrightarrow$ & Cartesian graph bundle \\
  Fibration & $\longrightarrow$ & Graph fibration\\
\hline 
\end{tabular}
\vspace{10pt}
\caption{From topology to combinatorics.}
\label{comparison}
\end{table*}

Generalizing the Cartesian product of sets, {\it graph products}\index{graph product } are
graphs whose vertex set is the Cartesian product of the vertex sets of
its factors. In addition, they satisfy some condition on adjacency
of vertices in the product, which depends only on adjacency in the
factors. Graph products have been studied in discrete mathematics
\citep{hammak}, computer science \citep{heine} and even applied in
biology \citep{fontana}.  

\begin{figure}[h!]
	\centering
        \includegraphics[width=\linewidth]{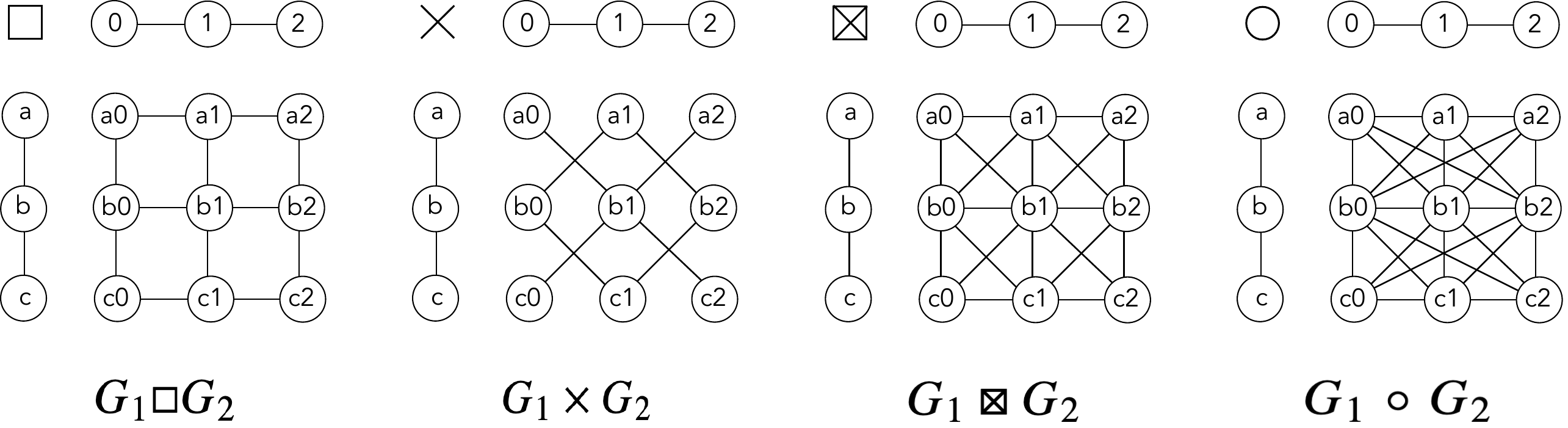}
	\caption{\textbf{Graph products.} The four standard graph products.}
	\label{graph-products}
\commentAlt{Figure~\ref{graph-products}: 
Four graphs representing each one type of product (in order: Cartesian, Direct, Strong, Lexicographic) between the path 0-1-2 and the path a-b-c.
The resulting products is always a graph with nodes a0, a1, a2, b0, b1, b2, c0, c1, c2 but with different undirected connections. The a? are on the first row, the b? on the second, the c? on the third.
Cartesian product: connections are only on rows and columns (a0-a1-a2, b0-b1-b2, ..., a0-b0-c0, ...)
Direct product: connections are on the diagonals: a0-b1-c2, a1-b2, b0-c1, a1-b0, a2-b1-c0, b2-c1.
Strong product: all the connections appearing in either Cartesian or Direct product.
Lexicographic product: each element of each row is connected to all elements of the following row.
}
\end{figure}

Graphs can be `multiplied' in a variety of
ways. In a 1975 paper \citep{imrich}, it was shown that there are
exactly 256 possible products where adjacency in the product graph
depends only on the adjacency properties of its factors.
From these 256 products, only four standard products
\citep{hammak} are typically used in the literature (Fig. \ref{graph-products}). Below we
follow the exposition in \citep{ostermeiter,hammak}.  Assuming two
graphs $G_1=(N_1, E_1)$ and $G_2=(N_2, E_2)$ (in all cases $V (G_1
\times G_2) = V (G_1) \times V (G_2)$), two vertices $(x_1, x_2),
(y_1, y_2)$ are adjacent (Fig. \ref{graph-products}) in the following ways:

\begin{definition}{\bf Four standard graph products:}
\begin{itemize}
\item {\it Cartesian product}:\index{graph product !Cartesian } $G_1 \square G_2$. Vertices are adjacent if (i) $(x_1, y_1) \in
E(G_1)$ and $x_2 = y_2$, or (ii) $(x_2,y_2) \in E(G_2)$ and $x_1 =
y_1$.

\item  {\it Direct product}:\index{graph product !direct } $G_1 \times G_2$. Vertices are adjacent if (iii) $(x_1, y_1) \in
  E(G_1)$ and $(x_2, y_2) \in E(G_2)$.

\item  {\it Strong product}:\index{graph product !strong } $G_1 \boxtimes G_2$ has edge set $E(G_1
  \square G_2) \cup E(G_1 \times G_2)$, i.e., it is the union of the
  Cartesian product and the direct product.

  \item  {\it Lexicographic product}:\index{graph product !lexicographic } $G_1 \circ G_2$. Vertices are adjacent if they
satisfy (ii) or if (iv) $(x_1, y_1) \in E(G_1)$.
\end{itemize}
\end{definition}

\section{Cartesian graph bundles}

Graph bundles\index{graph bundle } are a generalization of both Cartesian products and
covering graphs.  They can be seen as analogs for graphs of nontrivial
fiber bundles like the M\"{o}bius band.  While other graph
products have been generalized to bundles, the most common one is the
Cartesian product, so most  graph bundles are Cartesian
graph bundles.

\begin{definition}{\bf Cartesian graph bundle.}
A {\it Cartesian graph bundle}\index{graph bundle !Cartesian } is a graph $G$ composed of a {\it fiber graph} $F$
and a {\it base graph} $B$, together with a {\it graph map} $p: G \to B$ such that
$p^{-1}(v) \subset F$ for each vertex $v \in V(B)$, and $p^{-1}(e)
\subset K_2\square F$ for each edge $e \in E(B)$. Here $K_n$ is the
complete graph with $n$ vertices.
\end{definition}

\begin{figure}[h!]
	\centering
	\includegraphics[width=.9\linewidth]{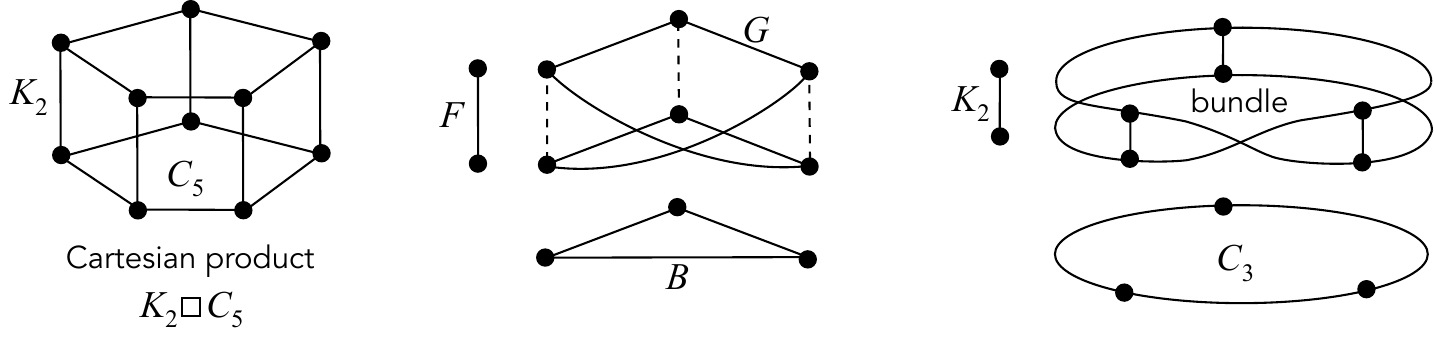}
	        \caption{\textbf{Cartesian products.}  We show a graph bundle $K_2\square
                  C_5$ and a Cartesian graph bundle $G$
                  over base graph $B =K_3$ with fiber $F = K_2$. The
                  non-degenerate edges are highlighted with
                  solid lines; the degenerate edges, which
                  belong to the copies of the fiber, are depicted with
                  dashed lines. 
                  }
	\label{graph-bundles}
\commentAlt{Figure~\ref{graph-bundles}: 
Three subfigures.
Left: ten nodes forming the vertices of a pentagonal prism. Label: `Cartesian product 
K2 box C5'.
Middle:Six-node graph: vertices of a triangular prism, except that the ends of two
vertical edges are joined as if one has been twisted through 180 degrees. Below,
three nodes in a triangle form the base of the fibration. 
Right: The same as the middle but drawn to show the resemblance to a
discrete version of a M\"obius band.
}
\end{figure}

Figure \ref{graph-bundles} shows an example of the Cartesian graph
bundle over the base graph $K_3$ with fiber $K_2$, which is a graph
representation of a M\"obius band.

The concept of graph bundle is quite interesting since it has the properties of a fiber bundle and at the same time
is discrete. Graph bundles were introduced in a seminal
preprint by \cite{pisansky}, which remains
unpublished. Despite the importance of graph bundles as a bridge between topology and
graphs, this preprint has enjoyed only mild attention from
mathematicians, gathering only 36 citations in almost four decades.

\section{Graph fibrations---history}
\label{sec:litfibrations}

The history of fibrations\index{fibration !history of } and their ramifications in various fields of
algebraic geometry, topology, and other fields of mathematics, and
beyond to computer science and systems theory is very rich and complex:
more details can be found in Sebastiano Vigna's\index{Vigna, Sebastiano }
\emph{Graph-Fibrations Home Page}
({\small\url{http://vigna.di.unimi.it/fibrations/}}). In the
following, we just want to give an idea of the main lines of research
involved.

The combinatorial definition of graph fibration used here was
formalized in~\citep{boldi2002fibrations} as a
special case of Grothendieck's\index{Grothendieck, Alexander } notion, obtained by looking at the free
categories generated by the graphs~\citep{mac2013categories}. More
precisely, a graph morphism $\varphi: G \to H$ is a fibration if and
only if the corresponding covariant functor $\varphi^*: G^* \to H^*$
is a Grothendieck fibration,\index{fibration !Grothendieck } where $L^*$ is the free category
generated by the graph $L$, also called a quiver\index{quiver } (its objects and arrows are the nodes of
and the paths of $L$, respectively).

Special instances of graph fibrations were rediscovered many times,
under different names, in the literature before the notion was finally
systematized in~\citep{boldi2002fibrations}:
\begin{itemize}
    \item In the context of spectral graph theory, rear
      divisors~\citep{cvetkovic1998spectra} induce a graph fibration,
      thus providing a connection between fibrations, eigenvalues,
      eigenvectors and factorizations of the characteristic
      polynomial.
    \item A parallel line of research, also related to the
      computation of the characteristic polynomial, led to the
      introduction of equitable partitions~\citep{schwenk1974computing},
      although apparently the connection with rear divisors went
      unnoticed.
    \item In the search for practical (albeit non-polynomial)
      techniques for the graph isomorphism problem, some efficient
      algorithms for the construction of the minimum base were
      discovered~\citep{corneil1970efficient,cardon1982partitioning}, or even
      earlier \citep{unger1964git}. \cite{mckay1981practical} made it explicit
      that the algorithm 
       computes the coarsest (in our wording, minimal)
      equitable partition.
    \item In the computer science literature, the use of
      homomorphisms between graphs representing networks as a tool to
      prove impossibility results started
      with~\citep{angluin1980local}; the necessity of extending those
      results to the directed case justified the shift from local
      isomorphisms (coverings) to fibrations.
\end{itemize}

Graph fibrations were introduced in computer science to prove
(im)possility results in distributed
systems~\citep{boldi2001,boldi2002universal}.  This study  was
inspired (see chapter 2.5 in \citep{boldi2002fibrations}) by an
analysis of a network of processors (each node of this network
executes the same algorithm). If all processors start in the same
initial state, then existence of the fibration will cause all
processors in the same fiber to stay in the same state, or in other
words, to be synchronized. This is why fibrations have
important consequences for information-processing networks and in computer science.

Fibrations have been also applied to study
self-stabilization~\citep{boldi2002universal} and to understand some
topological properties of Google's ranking algorithm
PageRank~\citep{boldi2006}.  Finally, in the context of system
dynamics, fibrations were proposed to study interconnections
between subsystems and to generalize the group of
symmetries ~\citep{lerman2015b}.

Armed with the analogies between fibrations, groupoids and fiber
bundles in physics discussed above, in the next chapter we discuss the
general applicability of fibrations to biology. Part II  then provides empirical evidence of fibrations in
biological networks.


\chapter[Symmetry and Robustness in Biological Networks]{\bf\textsf{Symmetry and Robustness in Biological Networks}}
\label{chap:robustness}

\begin{chapterquote}
    In Section \ref{S:IOR} we saw that fibration symmetries are more robust
    to changes in the network than automorphism symmetries,
    mainly because fibrations are local but automorphisms are global.
    We also saw that (in suitable circumstances) new outputs have 
    less effect than new inputs.
In this chapter we consider some biological examples where this principle
  applies: gene duplication and speciation. We
  argue that symmetry fibrations in biological
  networks may have been favored over symmetry groups by evolution due
  to their increased robustness (in an informal sense) and resilience under mutation events.
  \end{chapterquote}

\section{Homomorphisms for biological networks}

We have seen that the concept of homomorphism (Definition
\ref{morphism}) encompasses all the main maps that are important to
characterize biological networks.
We recall the definitions:
\begin{enumerate}
\item A (graph) homomorphism\index{homomorphism } is a map that preserves the adjacency relation(s) for
each node- and edge-type.

\item When the homomorphism is a bijection (injective and surjective)
  then it is an isomorphism. Isomorphisms induce an equivalence
  between input trees in fibers.
  
\item When the homomorphism is an isomorphism of the graph to itself,
  then it is an automorphism, i.e., a symmetry permutation of the
  graph.

\item When the homomorphism satisfies the lifting property, that is,
  when it collapses only nodes with isomorphic input trees, then it is
  a graph fibration.\index{fibration !graph }

\item Two nodes in the same fiber of a fibration always have isomorphic input trees.\index{input tree }

\item 
  A graph fibration that collapses all the available fibers into the
  minimal base of the graph is a minimal fibration or symmetry
  fibration. In a minimal fibration two nodes are in the same fiber \emph{if and only if} they have isomorphic input trees.
\end{enumerate}

A homomorphism can collapse any nodes in the network provided it
preserves adjacency of nodes. A quotient map of a group of automorphisms
collapses all nodes in the same orbit. A fibration collapses only nodes within each fiber
(color classes for balanced colorings or nodes with isomorphic input trees). Every
automorphism is a fibration, so all orbits are fibers, but not the
converse.

To understand how these features of homomorphisms come together to
explain the structure of biological networks, we use the example of
graph $G$ in Fig. \ref{summary}a, which consists of the FFF\index{FFF } to which we
add a two-layer multilayer subgraph with a symmetry group. This
graph contains four fibers (three are trivial and contain a single node, the other contains two nodes) and two orbits, as indicated by the coloring. 
Orbits of the symmetry group are also fibers, but we want to make a distinction
between the two types. Because of the automorphism $(4\,5) (6\,7)$, a quotient map
(Definition \ref{def:aut_quot}) can be applied to this network to reduce
it to $G/$Aut$(G)$ as in the figure.

From this quotient network,\index{quotient !network }
a fibration $\varphi$ can be applied to reduce the fiber (in yellow) to
its base as indicated. A direct fibration from $G$ to $B$
can be also applied since every automorphism is a fibration as
shown (fibrations compose to give fibrations). We indicate this minimal fibration in the figure by $\varphi: 
G\to B$ collapsing the green, blue and yellow fibers.

Both orbits and fibers synchronize. However, we  show in the next section that
fibers may be favored over orbits under evolutionary pressure, because
orbital synchrony is more vulnerable to duplication events and mutations.

\begin{remark}
We can use $G/$Aut$(G)$ here because it
gives the same quotient network as Definition \ref{def:quot_net}.
In general, however, $G/$Aut$(G)$ gives a different quotient, even for fibrations induced by automorphisms. See
Section \ref{sec:two_quots}.
\end{remark}

\section{Duplication events break a
  group symmetry but not a fibration}
\label{sec:duplication}

\begin{figure}[t!]
  
   \includegraphics[width=.8\textwidth]{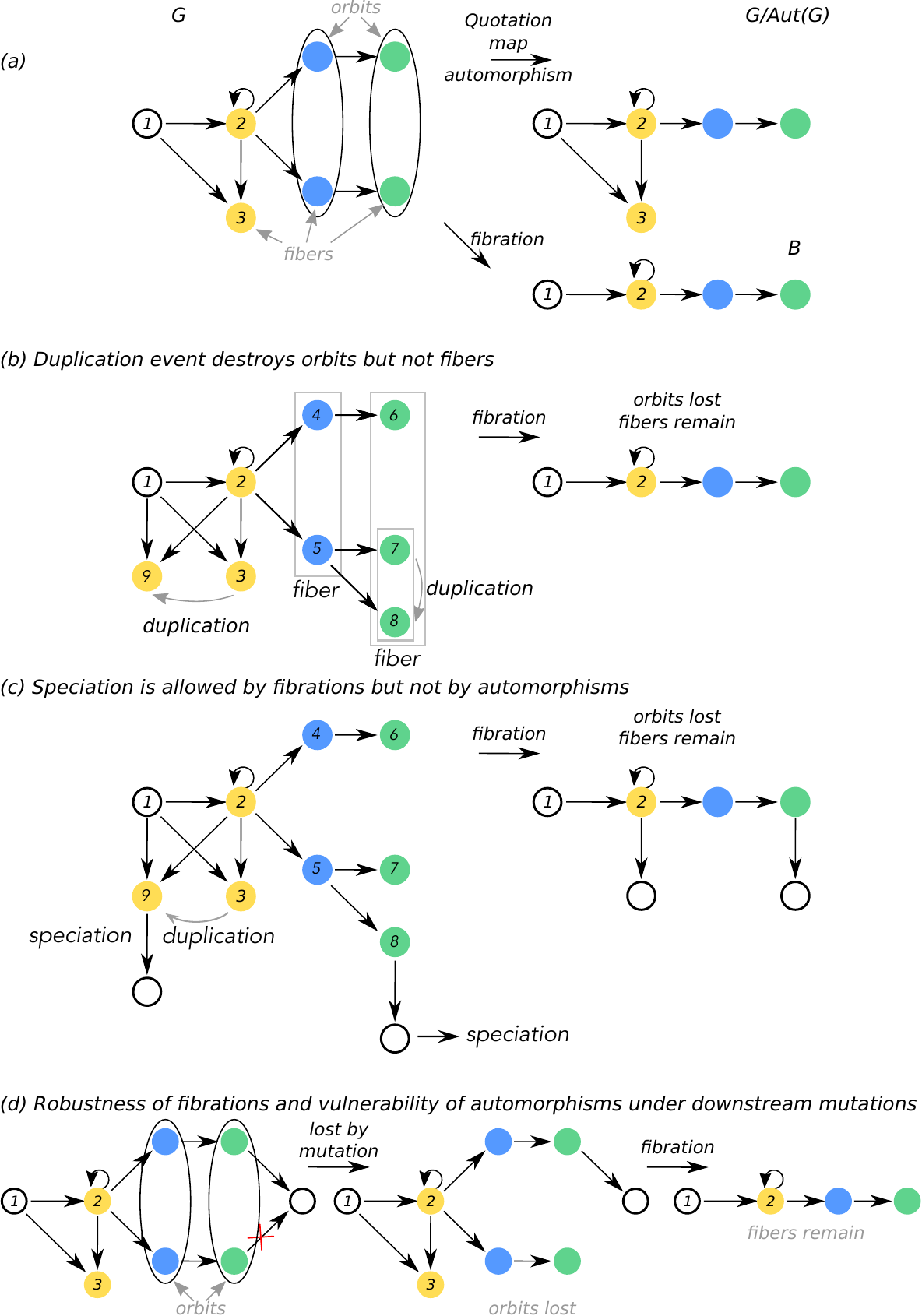}
\caption{ \textbf{Why fibrations are favored over automorphisms in
    biological networks.} Fibration robustness versus automorphism
  vulnerability in biological networks.  (\textbf{a}) A network and it fibration and quotient map under automorphism. (\textbf{b}) A generic gene duplication event destroys the automorphisms but fibration remains.
  (\textbf{c}) Fibrations permit speciation, automorphisms do not. \textbf{(d)} Fibrations are robust to
  downstream mutations; automorphisms are vulnerable.
  }
\label{summary} 
\end{figure}

Genetic evolution\index{evolution } is in part driven by duplication of parts of the
genome. Gene duplication\index{gene duplication } events are shown schematically in the genetic
network of Fig. \ref{summary}b, where gene 9 is a duplicate of gene 3
and gene 8 is a duplication of gene 7. Both events duplicate the
original gene, together with its expressed protein and promoter region.
Therefore, the duplication event duplicates also the incoming link of
the gene and the child gene  
inherits the full input tree of the
parent gene.  Thus, by definition, the duplication event preserves 
synchronization of the fiber.  This demonstrates 
robustness of the fibration symmetry.

\begin{remark}\em
Here and in the rest of the book, the word `robust'\index{robust !informal usage } is used informally to indicate resilience in changing conditions.
It is {\em not} used in the technical sense of Section \ref{sec:robust}.
\end{remark}

On the other hand, duplication of gene 7 into 8 breaks not only
the orbit $\{6, 7\}$, but also the upstream
 orbit $\{4, 5\}$. Thus the entire symmetry
group is removed by this single perturbation. This happens because of the
global character of the symmetry group, which creates vulnerabilities
that might propagate across the network structure.

In contrast, a fibration symmetry is more robust. Considering
the fibration of the circuit, we observe synchronization of genes
$\{4, 5\}$ and $\{6, 7, 8\}$. In this last case, the duplicated gene
is added to the fiber, showing how the fibration not only preserves
the synchronization state, but  supports the addition of a new
duplicated gene to a previous synchronized state without 
synchrony-breaking. This is shown in the base of Fig. \ref{summary}b after the
application of the fibration. On the other hand, the automorphism
group predicts the appearance of a new partial synchrony in $\{7, 8\}$, but not
 $\{6, 7, 8\}$.

This vulnerability of automorphisms\index{automorphism, vulnerability of } is also exemplified in the globally
symmetric network example of Fig. \ref{fig: example}a, which show how
fragile group symmetries are: if the network evolves by adding just
one extra node to the network, as in Fig.~\ref{fig: example}b, the
global symmetry is completely broken. The automorphism group of the network
in Fig.~\ref{fig: example}b now contains only the identity permutation. This occurs because
automorphisms require strict arrangements of nodes and links to
preserve the global structure of the network.  Thus the
addition of the outgoing link from node 3 breaks the synchronization,
not only of the orbit $\{2, 3\}$, but also of $\{4, 5 \}$, even though the
added link does not connect to or emanate from this orbit.

These two cases exemplify the vulnerability of a synchronization state based on a global symmetry: a small perturbation in a part of the network
can change the possible cluster synchrony patterns
of the entire network, which
often become very different. Potential synchronies
can be destroyed---or even, sometimes, created. These changes can
affect nodes located in
another part of the network, not connected to the site
of the added (or removed) edge or node.

A biological network with this global vulnerability cannot survive
under evolutionary pressure. A robust evolving network requires a
locally-symmetric fibration based on the input trees of symmetrically related
nodes. Robustness induced by local symmetries
guarantees stability of the phenotype, while at the same time
providing flexibility to changes in network structure, suitable
for evolvability of new phenotypes under natural selection.

\subsection{Speciation events break a group symmetry but not
  a fibration}

After a duplication event, speciation may occur by the addition of extra
links, creating a new function for the duplicated gene, which no longer needs to
be synchronized with the parent gene. This is exemplified in
the network of Fig. \ref{summary}c by the addition of node 10 to the
duplicated gene 8 and the addition of gene 11 to the duplicated gene 9
in the fiber.

Again, the extra link to gene 10 breaks synchronization in
the orbit $\{7, 8\}$, due to the new outgoing edge of node 8. Yet the
synchronization predicted by the fiber $\{6, 7, 8\}$ remains intact
after the speciation event, because the input trees of these nodes are not
affected by the extra outgoing edge. Likewise, the addition of node 11
does not break the symmetry in the yellow fiber. The fibration ignores
the break of synchronization predicted by the broken group orbits, and
predicts the correct synchrony as observed in the base of the
fibration, as in Fig. \ref{summary}c.  Because the sole condition
for a network to be coherent and functional is symmetry in the inputs,
but not in the outputs, there is complete freedom for the set of genes that
connect the fiber to the external world to evolve, without altering the already
functional fiber. These results show how a structure characterized by
fibrations produces increased robustness of synchronization patterns compared to automorphisms, under
duplication and divergence in the evolution of the genome.

\subsection{Vulnerability of automorphisms under mutations}

Figure \ref{summary}d shows how a downstream mutation\index{mutation } causing the loss of an edge
breaks synchronization in both orbits. Again, in contrast, the fibration
preserves synchronization in both fibers, since in this case the fibration does
not depend on the outgoing edges. The loss of synchrony in the group
orbits is severe, since it happens upstream of the mutation, affecting
genes that are not connected to the lost edge. This is another
expression of the vulnerability of a synchrony space based on a global
symmetry.  The fibration shown in the figure indicates that, instead,
it is robust against this mutation.

\section{Understanding `structure $\rightsquigarrow$ phenotype'}
\label{sec:structure_to_function}
\index{structure-function relation }\index{phenotype }
In biological development, the `big data' networks of molecular biology
eventually manifest themselves in the phenotype of the organism.
Early hopes that the DNA sequence of the genome could be `read' to
reveal the phenotype proved overoptimistic, and the way phenotype emerges
from genotype remains enigmatic, despite numerous advances. It is also
one of the central problems in biology.

In describing the arrow $\rightsquigarrow$:
\begin{equation}
{\rm structure}\  \rightsquigarrow \ {\rm phenotype}
  \end{equation}
we consider the structure to be given by the biological network, and the
phenotype by its function and consider the simplest form of
coherent functionality which is achieved by synchronization of the
activity of the biological units: genes, metabolites, proteins,
or neurons.
We represent the space of biological functions as the quotient or base
of the phenotype. In particular we evaluate the redundancy of the
map $\rightsquigarrow$ associated with the cardinalities of the fibers
\citep{gromov}.

Thus the existence of robust fibration symmetries has important
consequences for the correlated dynamics of biological systems. The
synchronization of neural and gene activity, as a consequence of
network symmetry, is observed in patterns of correlated activity in
biological networks, and this synchronization is largely independent of
the dynamical model used to describe the activity of the nodes.
Indeed, this is necessary, because precise models that match real
biology with high accuracy are uncommon, unlike the situation in physics.
Most biological models are phenomenological, and even those that
include a lot of detail involve unknown parameters such as reaction
rates and connection strengths.

\subsection{Segregation versus integration of function}

The collapse of the fibers by the symmetry fibration determines a
partition of genes into segregated `sectors'\index{sector } with a well defined functional
interpretation. This suggests that the segregation of function in
genetic networks could be a direct consequence of the symmetry
fibration.  This segregation of function occurs despite the
integration of genes into a globally connected network. Therefore, the
existence of fibers provides a possible mechanism to explain the
conundrum of how segregation of functions can coexist with an
integrated biological network. Thus, while a gene may be highly
integrated by connections to other genes, the precise arrangement of
links by fibrations still allows the genes to perform different
functions according to the fiber controls, while remaining integrated 
into the whole network for coherent dynamics.

Thus, fibrations can explain not only the existence of modular structure
in biological networks \citep{hartwell}, but also their robustness and
evolvability.  Loss of function by breaking the symmetry
of a fiber need not propagate to other fibers, because
the fibration acts locally on the fibers. As a consequence, symmetry
breaking in a fiber does not affect the symmetry of another fiber.  If
we associate each symmetry fiber with a function, then the failure of a
given function does not affect the functions of other sectors.  This
theoretical prediction opens the possibility of experimental testing,
by manipulating the
activity of genes chosen to break the symmetry of a given fiber,
and monitoring the resulting activity inside and outside the chosen fiber.

\section{Axiomatization of biology}
\label{S:AOB}
\index{biology, axiomatization of }

It is interesting to elaborate on how the biological symmetry of
fibrations and groupoids emerges from a mathematical perspective in the
hierarchy of fundamental algebraic structures. The axiomatic
representation of the relevant abstract algebras involves four axioms: (i) Identity, (ii)
Invertibility, (iii) Closure or composition law, and (iv)
Associativity.  A group is an algebraic structure whose elements
satisfy the four axioms.  In addition, we could add a fifth condition
which is not considered to be a group axiom but is interesting in the context
of earlier theories of quantum electrodynamics: (v)
Commutativity, which leads to an Abelian group.\index{group !Abelian }

Elementary particles and their fundamental interactions are described
by group symmetries following the four axioms in the non-Abelian gauge
theory of the Standard Model.\index{Standard Model } In addition the Abelian gauge theory of
quantum electrodynamics also obeys commutativity \citep{weinberg1995}.

Automorphisms\index{automorphism } represent global symmetries since they involve
permutations of all nodes. In contrast, input isomorphisms and
fibrations\index{fibration } involve only local transformations that act on a subset of
nodes and their input-sets. These
isomorphisms are not necessarily closed
under composition, and form a groupoid\index{groupoid }, which does not satisfy the
axiom of closure. This suggests that a groupoid
cannot be a symmetry of physics, since spacetime is composable
\citep{weinberg1995}.  Removing one axiom from the algebraic structure of the
universe brings huge freedom to the system---like
looking at an entire new universe by dropping one fundamental axiom in
the groups that are common in physics.

Groupoids are widely used in algebraic topology, because various topological
constructions require specifying a base point in a topological space,
and comparing what happens relative to that point with
what happens relative to another point. The natural structure that
results is a groupoid. 
Groupoids might be relevant in some places in physics,
but probably not on the deep philosophical level of fundamental laws.

Outside their uses in certain branches of mathematics, groupoids
are seldom encountered. They are not widely used in the way that 
groups are used in physics, with powerful tools such as group
representations, irreducible representations, and so on. The groupoid
formalism does not provide a similarly 
powerful mathematical technique that can be widely
applied. Instead, the most important information that can be
extracted from groupoids is the concepts they lead to.
In this case, these are things like balanced colorings and fibrations. 

With Einstein's hat on: the big group symmetries of physics arise
because the laws of physics are the same at every point in spacetime.
If the laws of physics changed from one
region to another, there would not be a symmetry group for those
laws... but there could be a symmetry groupoid.\index{groupoid } This symmetry groupoid
would encode the answer to the following question: if two physicists were at different
places, would they observe the same laws of physics? If the laws
of physics are consistent with a symmetry group, the answer 
 would be `yes'.  But if the laws of physics are consistent with a symmetry groupoid, then
 physicists would see the same physics only when they are in the
same connected component of the groupoid (in the sense of composition), with the same local 
vertex symmetries. 

We argue that the axiom of closure may be an `accidental' constraint
rather than a fundamental one, and thus unnecessary for a living
system. We call it accidental because this axiom holds only for
special symmetries of the system, i.e. automorphisms, whereas in general it is not valid for the majority 
of symmetries, which constitute the symmetry groupoid.

Symmetry groupoid transformations identify clusters of synchronized
balanced-colored nodes.
Symmetry groupoid transformations do not impose any
restriction on where the outputs are sent. That is, nodes in a
synchronized groupoid cluster are free to send their outputs to any
node in the network that does not lie in that cluster. (This can
break up other clusters, however.)
The outputs of nodes
in a group orbit---a synchronized group cluster---are
restricted by the global structure of the whole network.

In Section \ref{S:IOR} we compared the rigidity of automorphisms
with the flexibility of fibration symmetries. We concluded that
although the situation is subtle, the focus on input sets rather
than outputs as well, combined with the local nature of fibration symmetries, groupoid\index{fibration !symmetry }
allows biological networks to undergo changes such as gene duplication
and speciation without much change in synchrony patterns.
The distinction between inputs and outputs requires a directed
graph,
such as a genetic or neuronal network, where the distinction
between the input and output sets of a node is meaningful. This might be
 why, in many biological systems, directionality of an
effect has been favored. This may have given a groupoid circuit a subtle
evolutionary advantage, allowing it to emerge more commonly than group circuits. In an undirected network, like the gap junction connections in the neural network of {\it C. elegans} studied in Section
\ref{sec:celegans-synchronization}, the evolutionary advantage of fibration over automorphisms is diminished. Indeed, these gap junctions form a group like the physical bonds in
molecules. 

The ability of different nodes in the
circuit to perform the same function---may give a groupoid structure
an evolutionary advantage over group structures with a concomitant
increase of robustness. If matter is a group and life is a groupoid, so to speak, then a key feature of
genetic networks might be that their biological information is encoded
in the symmetries of the directed genetic signals that genes receive
from each other, rather than in the identity of the gene itself. 

\subsection{Speculations on axioms for biology and the Erlangen Program}
\label{sec:speculations}

\index{biology, axiomatization of }

\begin{figure*}
  \includegraphics[width=.55\textwidth]{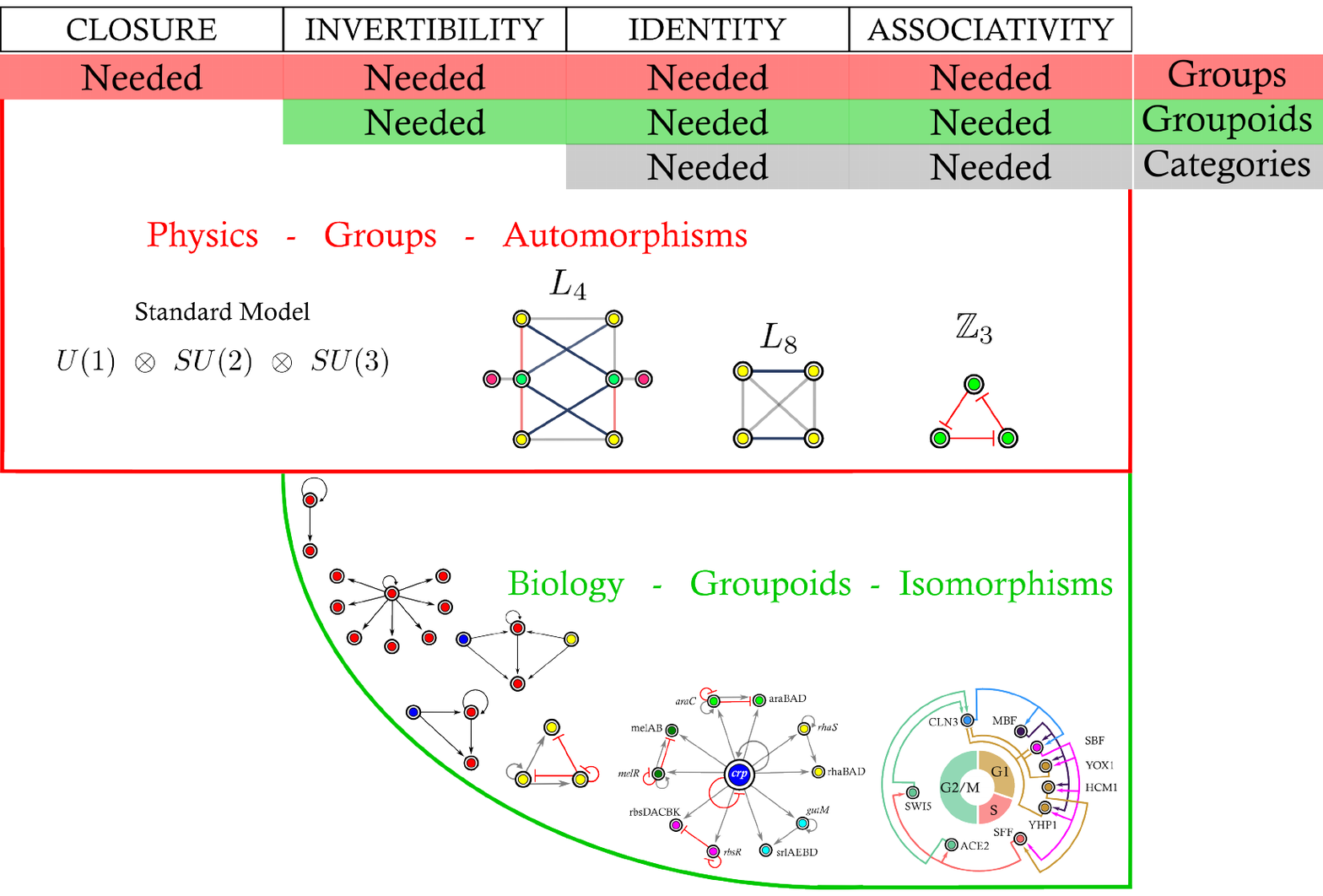}
  \includegraphics[width=.4\textwidth]{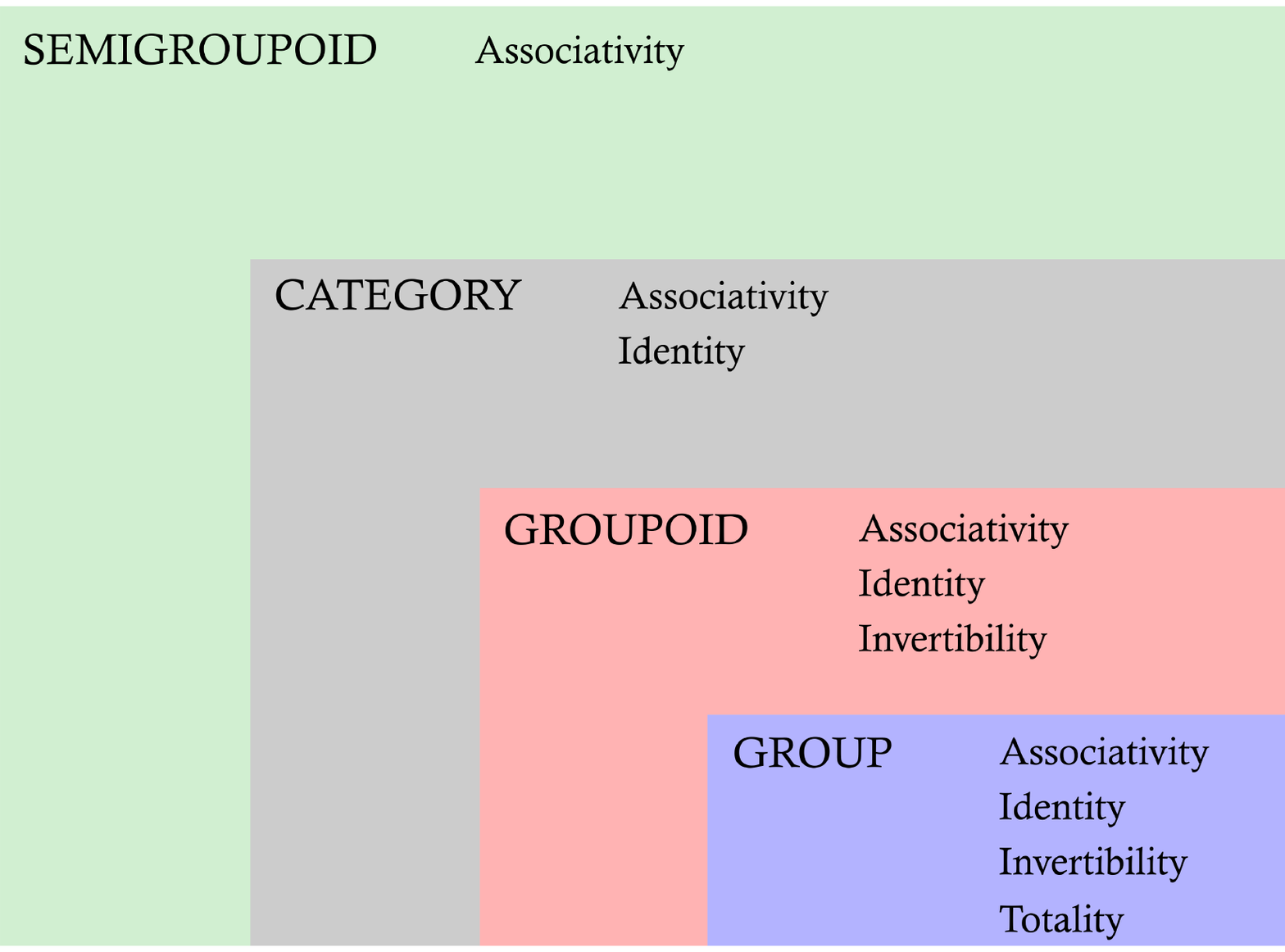} 
\caption{  \textbf{Evolution of groupoids in the hierarchy of algebraic
    structures.}  {\em Left}: Biological networks may have emerged by an 'algebra
  breaking' event by relaxing the axiom of closure (law of composition)
  from the more stringent structures characterized by physical
  symmetry groups. Symmetry groups shown are $U(1)\otimes SU(2)
  \otimes SU(3)$ of the Standard Model \citep{weinberg1995}, the locomotors
  $L_4$ and $L_8$ (Section
\ref{sec:celegans-synchronization}), and repressilator $\mathbb{Z}_3$
  ~\citep{elowitz2000}. The symmetry groupoid branch represents the
  building blocks and composite groupoids found in biological networks.
  {\em Right}: Nested hierarchy of algebraic structures considered in this
    book. Starting from the most restrictive algebra, Abelian
  Groups, we relax one axiom at a time to obtain: Abelian groups $\to$
  Non-Abelian groups $\to$ Groupoids $\to$ Categories $\to$
  Semigroupoids.}
\label{axioms} 
\commentAlt{Figure~\ref{axioms}: 
Left: Table with five columns, labels `closure', `invertibility', `identity', `associativity', and empty.
Row 1: needed, needed, needed, needed, groups.
Row 2: ---, needed, needed, needed, groupoids.
Row 3: ---, needed, ---, needed, categories.
Plus decorative images : 
one set illustrating Physics - groups-automorphisms; the other 
illustrating Biology - groupoids- isomorphisms.
Right: Nested series of rectangles. In order from outermost to innermost:
1: Semigroupoid - associativity.
2: Category - associativity, identity.
3: Groupoid - associativity, identity, invertibility.
4: Group - associativity, identity, invertibility, totality.
}
\end{figure*}

Nearly all physical systems have emerged from higher to lower
symmetries following a series of spontaneous `symmetry breaking' events
\citep{weinberg1995,georgi2018,landau1977quantum}, without
abandoning the four
algebraic axioms of non-abelian groups which were hard-wired in the
fabric of spacetime from its origin. From a very speculative viewpoint: the emergence of biological
groupoids and life may have involved a more drastic change in the appropriate algebraic
structure itself, through a process that we might
metaphorically describe as `algebra breaking' from
physical groups by abandoning the axiom of composition to obtain the
less restrictive algebraic structure of groupoids
(Fig.~\ref{axioms}).  If evolution from matter to life can be seen as
an algebra-breaking event, from non-abelian groups to groupoids by relaxing
one axiom, it is natural to consider what other algebraic structures could
emerge from groupoids.  Relaxing the axiom of invertibility leads to a
category of irreversible processes~\citep{categories} which satisfies
only associativity and the existence of the identity and further
completes a series of transitions from Groups $\to$ Groupoids $\to$ Categories.

Of course, there are more ways to generalize
  an algebraic structure than by dropping individual axioms, and
alternative axiom systems can define the same structure.
But the hierarchy of structures here is especially significant.
Biology, due to its complexity, necessitates less
  constrained laws than physics does. Physics is static, so groups can
  describe all molecules, but biology is evolving, so it needs less
  rigid algebras. There is an analogy with the physics of phase
  transitions: a system can have full symmetry, like a liquid, but
  after a phase transition, a lower symmetry state is found in the
  ground state. Thus a solid breaks the symmetry of the liquid and
  brings up a new state of matter. All symmetries in physics are
  eventually broken.
  To go from biology to physics, a more fundamental property is
  broken: the fundamental algebraic structure that describes biology
  as opposed to physics. 

Such an algebraic view of natural processes may lead to a
`geometrization' or `axiomatization' of biology,\index{biology, axiomatization of } analogous to
Felix Klein's\index{Klein, Felix } 1928 `Erlangen Program'\index{Erlangen Program } \citep{klein1872}, that could
successfully organize scattered biological knowledge in a systematic
way. 

A geometrization program of biology (more details in Chapter \ref{chap:geometrization}) would follow the systematic
application of symmetry principles to all biological networks.  This
program could provide a classification of the elementary building
blocks of living networks from first principles to open the way to
predict function and dynamics from the symmetries of transcriptomes,
proteomes, metabolomes, connectomes and beyond.  A geometrization of
biology would imply a systematic classification of all building blocks
of biological networks from the bottom up using first principles based
on symmetries (in analogy to geometry). We will attempt the first steps of such a program in Part II of this book.

It is as though this axiomatization of symmetry in biology might be as
fundamental in biology as in physics, but with a twist. As noted by
\cite{bourbaki1994} p.47, the law of composition is the most
primitive notion in the axiomatization of algebra, to the extent that
it has been inseparable from the evolution of algebra since the first
calculations with whole numbers. By liberating natural systems from
the strong requirements of the composition law imposed by physics and
its symmetry groups, a gentler symmetry has emerged, and with it, the
circuits that sustain life.

\chapter[Geometrization of Biology]{\bf\textsf{Geometrization of Biology}}
\label{chap:geometrization}
\index{biology, geometrization of }

\begin{chapterquote}
  In this chapter, we sketch a
  geometrization of biological networks (fibrations) by analogy with the geometrization of physics (fiber bundles). There, geometrization is performed by introducing the
  curvature of a fiber bundle, using the concept of parallel
  transport along closed loops. In physics, this procedure leads to
  the association of the local curvature (e.g., of spacetime) with the
  fundamental forces of nature (e.g., gravity) via the connection of
  the fiber bundle. We first introduce the concept of curvature
  from a classical geometric point of view and generalize the
  concept to the modern view of gauge theory in theoretical physics.  
  We then describe genetic networks as a
  fibration in a closed loop with nonzero curvature.  
  As it
  turns out, an analogous gauge theory for biological networks is the
  dilation symmetry group used by \cite{weyl1} 
  in his early (but failed) attempt to unify electromagnetism and general relativity. 
  Weyl gauge theory has been applied to a toy model of
  financial markets by \cite{ilinski} and \cite{young}, and
  reviewed by \cite{maldacena}. Borrowing heavily from Maldacena's essay, we follow this simple
  analogy between gauge theories in physics and financial markets
   to introduce geometry into
  biology through the more general framework of fibrations.
  \end{chapterquote}

\section{Curvature and connection in geometry}

 Section \ref{S:AOB} was speculative and informal.
  In this chapter, we attempt to tighten the reasoning and clarify
  some of the ideas it leads to, though this discussion remains speculative. One such idea is curvature.

\begin{figure}
	\centering \includegraphics[width=\linewidth]{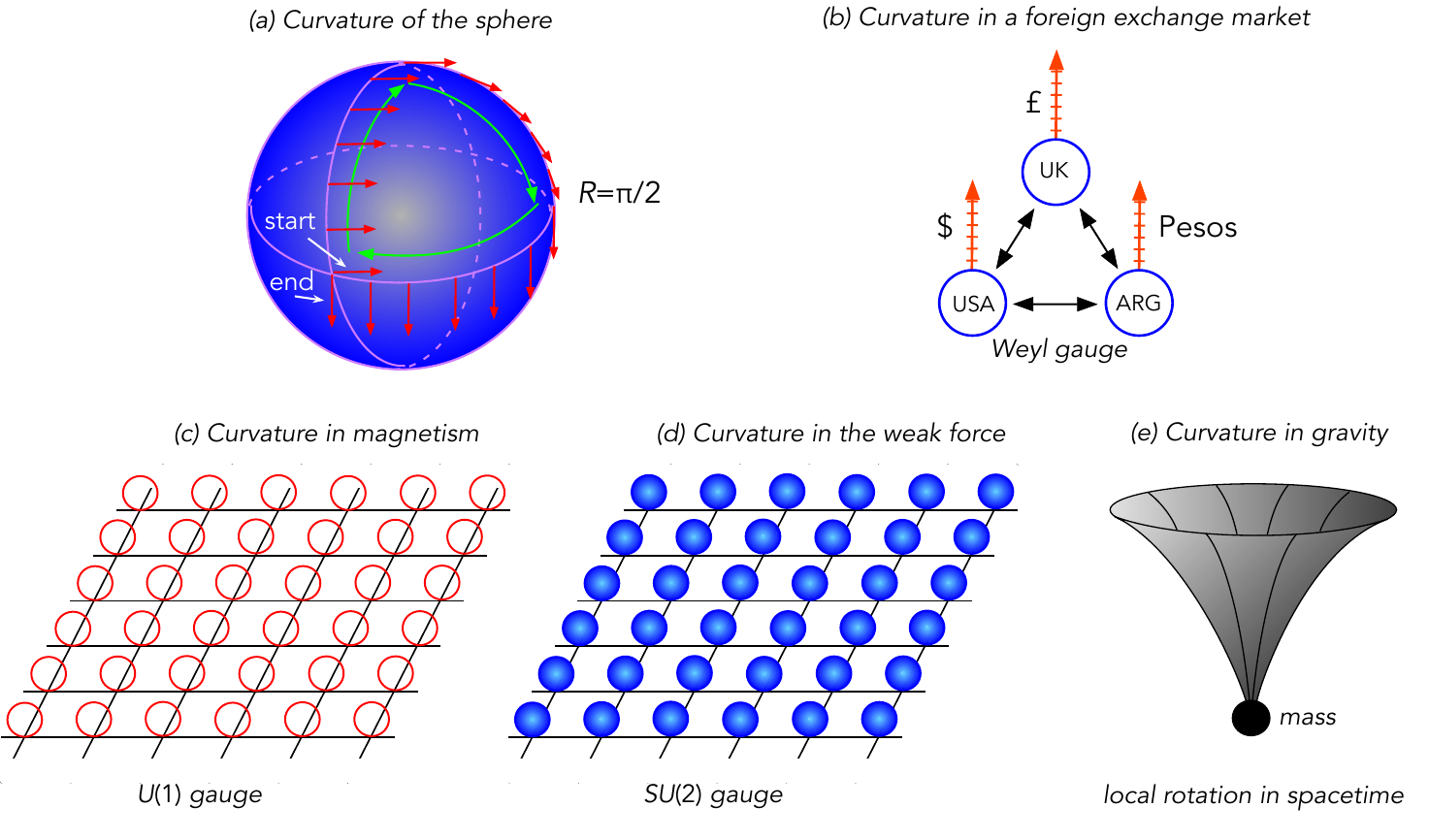}
	\caption{\textbf{Curvature and connections.}\ (\textbf{a}) Curvature of the sphere is measured by the parallel transport of a vector in a closed circuit (green arrows from `start' to `end'). (\textbf{b}) Curvature in a toy model of an exchange
          market. (\textbf{c}) Curvature in QED.  (\textbf{d}) Curvature in electroweak
          fiber bundle.  (\textbf{e}) The curvature of spacetime in general
          relativity.} 
       	\label{sphere}
\commentAlt{Figure~\ref{sphere}: 
Five images illustrating curvature in various contexts.
(a) Curvature of the sphere.
(b) Curvature in a foreign exchange  market.
(c) Curvature  in magnetism.
(d) Curvature in the weak force.
(e) Curvature in gravity.
}

\commentLongAlt{Figure~\ref{sphere}: 
(a) Curvature of the sphere. A sphere on which three great circles define 
a spherical triangle with all three angles right-angles. A vector starts
at one vertex of the spherical triangle, lower left, on the equator. It is transported
parallel to itself up a line of longitude to the north pole. Then down a line of longitude
to the equator, now tangent to that line. Finally, it is transported
back along the equator, pointing south at right angles to the equator.
The start direction and end direction are at right angles.
(b) Curvature in a foreign exchange  market. Three nodes connected
in a triangle (bidirectional links). Nodes labeled UK, ARG, USA.
Each has a vertical red arrow attached, labeled 
respectively `pound', `pesos', `dollar'. Beneath is `Weyl gauge'.
(c) Curvature in magnetism. Square grid in perspective. Attached 
to each grid point is a circle. Beneath is `U(1) gauge'.
(d) Curvature in the weak force. Square grid in perspective. Attached 
to each grid point is a sphere. Beneath is `SU(2) gauge'.
(e) Curvature in gravity. Funnel-shaped surface narrows towards the bottom,
where a small sphere is attached, labeled `mass'.
Beneath is `local rotation in spacetime'.
}
\end{figure}

The curvature\index{curvature } of a surface is measured by transporting a stick 
(a vector in a vector
field) along a closed loop on the surface while attempting
to keep the stick straight along the trajectory. When the orientation
of the stick at the end of the transport across the loop is different
from the initial orientation, the surface has non-zero
curvature. This construction is realized by the rules of parallel
transport\index{parallel transport } of the stick, and can be easily visualized on the 2-sphere
$\mathbb{S}^2$ (Fig. \ref{sphere}a).  Parallel transport of a vector tangent to
the sphere occurs when we transport the vector along a great circle of
the sphere (i.e., the intersection of the sphere and a plane that
passes through the center of the sphere), in such a way that the
vector tangent to the great circle moves always tangent towards
another point in the great circle. Intuitively, we transport a stick
while keeping it always parallel to the local tangent plane of the
surface. The definition of parallel transport also applies to any
path, irrespective of whether it follows a great circle or not, by
decomposing the path into a series of small steps along great circles.

In this setting, curvature arises when the vector moving under
parallel transport along a closed loop returns to the initial location
pointing in a different direction. This difference in direction 
is proportional to the curvature of the surface.

For instance, the parallel transport of a vector along a loop composed
of the spherical triangle defined by three great circles of the
2-sphere depicted in Fig. \ref{sphere}a results in a rotation by
$\pi/2$ compared with the initial vector. This is a consequence of the
curvature of the sphere, and the difference in orientation along a
closed loop encodes this curvature. In contrast, when we parallel
transport a vector on a flat surface (which can be viewed,
heuristically, as a sphere with infinite
radius, hence zero curvature) the initial and final orientations of
the vectors are the same.

This idea gave rise to the 
concept of a connection,\index{connection }, which arose historically as a
way to define differentiation on a curved spherical surface.
Here the comparison of two arbitrary tangent vectors is not
well-defined, because the coordinate system changes from point to point
along the surface. To capture this change of coordinate system along a
curve in the surface, a `connection' field is defined, which produces
parallel transport along the curve. The notion of a connection then
captures this change of coordinate systems, which lets us compare two
vectors on the sphere, and it makes precise the idea of transporting a
vector in a parallel and consistent manner along the curved surface.
When two vectors at two points along the curved surface are compared
by the connection field, then, provided there is no difference between the
vectors, those vectors are the result of parallel transport along the
surface, and the covariant difference along the curve is zero. On the
contrary, if the vectors are not connected by parallel transport,
there is a non-zero covariant difference along the curve.

The area enclosed by the curve of the closed loop, $S$, times the
curvature associated with the connection, $\mathcal{R}$, is the
differential between the vectors as they are parallel-transported along
the loop. In symbols,
\begin{equation}
\Delta = \mathcal{R} S.
\end{equation}
For the closed loop on the sphere in Fig. \ref{sphere}a,
the curvature of the sphere of radius $R$ is $\mathcal{R} = R^{-2}$,
and the area enclosed by the loop is $S = \pi R^2/2$.  Therefore, the
difference is $\Delta = \pi/2$. This agrees with the previous discussion.

\section{Curvature and connection in physics}
\label{curve1phys}
\index{connection !in physics }\index{curvature !in physics }

Curvature can be defined locally (near a given point), and there are
several distinct notions of curvature. Gaussian curvature is
local: at a given point, it is the product of the two principal curvatures.
Local curvature is a metric property, preserved by isometries
(transformations that preserve distances {\it within the surface}.
Thus a cylinder has zero local curvature at every point, just like a plane,
and indeed the cylinder can be unrolled to form a plane without distorting distances.)

In contrast, total curvature is global: it is the integral of the Gaussian curvature over the entire surface. 
It is more general than the actual physical
geometry of the object, characterized by such quantities as angle, metric, or length. There is no need for a metric to define the geometry because
curvature is a property of the topological class of the shape. Thus if
two surfaces are homeomorphic to one another (same topology), they will
have the same total curvature, even if they have different 
shapes. This is the Gauss-Bonnet Theorem, and it is valid for compact
surfaces without boundary \citep{docarmo1992}.

Likewise, curvature can be defined for a fiber bundle,
characterizing its local geometry through a connection.  Indeed, the
2-sphere forms the base of a fiber bundle with two-dimensional tangent
planes (fibers) attached to every point of the base to form the total
space. The curvature of parallel transport across a closed loop in a
fiber bundle characterizes its geometry in the same way as the
curvature of the 2-sphere.

The geometrization of physics is achieved through the curvature of the
fiber bundle, and fiber bundle geometry can be extended to
biological fibrations.  To set up a geometrization of
biology, we 
begin with an intriguing analog of the geometrization of physics in the financial sector.
Then we develop this analogy more specifically for the
 salient features of the
geometrization of modern physics in general relativity and the
standard model. Finally, we extend the analogy to biology.

\section{Curvature and connection in a financial fiber bundle}
\label{curve1}
\index{connection !in finance }\index{curvature !in finance }

Physical fiber bundles consist of Minkowski spacetime
as the base together with different fibers associated with gauge invariant
groups that represent the fundamental forces of nature:
gravity, electroweak and strong forces, as given by general relativity
and the standard model of physics, respectively.  These forces correspond to
local rotational invariances of general relativity and the quantum
gauge theories with fiber symmetry groups $U(1) \times SU(2)
\times SU(3)$.
 However, trying to understand biology through an analogy with the
gauge invariances of gravity and quantum mechanics of the standard
model could be quite traumatic and threatening. 

Fortunately, this is
not necessary. The main ingredients for understanding curvature are
already present in the simplest gauge-invariant theory, represented by
the multiplicative group of dilation introduced by \cite{weyl1} to describe
electromagnetism. 
Gauge invariance is the
term used by physicists to talk about what mathematicians call the structure group
of the fiber bundle \citep{weyl2}.  Thus, gauge theory and fiber bundles are two
  ways to talk about the same theory, the first for physicists, the
  second for mathematicians.  
  Modern quantum gauge theories are
  Yang-Mills theories from the 1950s, but the term `gauge invariance'
  originated in a seminal (but failed) attempt of
  \cite{weyl1,weyl0,weyl0g,weyl2} to unify gravity and electromagnetism, using the
  dilation group as the gauge invariance.  
 
It turns out that the dilation group describes also
the gauge symmetry of a toy model of a foreign exchange market which
maps to classical electromagnetism. This happens because the value of
one currency in terms of another depends on the exchange rate, which is
the ratio of the numerical amounts concerned. If a currency is
revalued by government intervention (in the example below, by defining
a new currency with three zeros removed from the denomination) and all exchange rates are adjusted
to reflect this change (multiply or divide the relevant exchange rate by 1000) 
then the underlying financial transaction are effectively unchanged.

This analogy between foreign
exchange and gauge theory was developed by \cite{ilinski} and
\cite{young}, and popularized in \citep{maldacena,jakob}.  We
follow the exposition of Maldacena,\index{Maldacena, Juan } where it was shown
that Weyl\index{Weyl, Hermann } gauge invariance describes an economic toy model of
speculators exchanging currencies between different countries. This is
a very simple example of a gauge theory, and it is a genuine analog of
classical electromagnetism\index{electromagnetism } in the absence of charges. It turns out
that curvature is the key concept that unifies all these disparate
systems.

Gauge invariance\index{gauge invariance } is present in the economic model of foreign exchange\index{foreign exchange } of
\cite{ilinski}.  Imagine a toy model of foreign exchange among three
countries with their own currency, USA (USD), UK (£), and
Argentina (Pesos).\index{Argentina } Speculators can freely exchange currency among
these three countries according to the exchange rates defined by the
respective central banks. The base of the fiber bundle is the network
of three countries shown in Fig. \ref{sphere}b.  The network has
undirected links since speculators can exchange money either way.  The
system is a fiber bundle with gauge symmetry, and we will show that it
displays curvature analogous to magnetic flux.

We attach a fiber representing the money space $F = \R_{+} = [0,
  +\infty)$ to each country in the base. The fiber measures the amount
  of money in that particular country, or the price of any good in the
  currency of that country. This means that an element of $f\in F$ can
  be any real positive number; it expresses the value of an asset or
  good in the currency of the country. This fiber has gauge
  symmetry: the units of the currency can be changed arbitrarily without
  any consequence for the economy. For example, an apple in Argentina
  can cost 1 Peso or 1 trillion Pesos according to the units of the
  currency set by the central bank, and, in principle, nobody would
  notice the difference. Using either currency would just imply
  changing the numbers of zeros on the banknotes, with no real-life
  consequences.

\begin{figure}
	\centering  
       \includegraphics[width=0.6\linewidth]{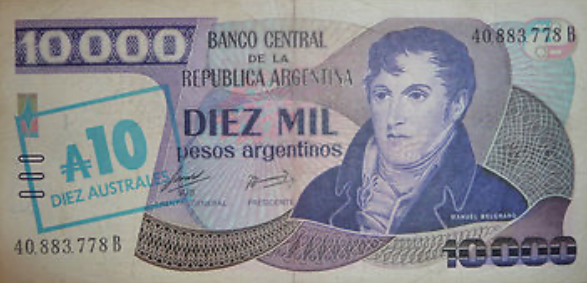}
	\caption{\textbf{Argentinian monetary gauge symmetry in
          action} \citep{maldacena}.  In the 1980s the
          Argentina Central Bank removed three zeros from the currency
          denomination and called the new currency `Australes'. This
          change had no noticeable impact on economics, since the
          prices of all goods adjusted to the new denominations
          instantly, revealing a gauge symmetry in the market. In
          fact, the Central Bank did not even find it necessary to
          print new notes, and a simple stamp sufficed to apply the
          gauge transformation. However, this change in currency led
          to an adjustment of the exchange rates with other countries,
          leading to a `force' (connection) with those countries. When
          these connections are non-zero, a curvature emerges in the
          market, which is a manifestation of arbitrage opportunities
          round a closed circuit, analogous to magnetic flux in
          electromagnetism.}
	\label{australes}
\commentAlt{Figure~\ref{australes}: 
No alt-text required.
}
\end{figure}

In fact, because of recurrent systemic inflation over the last 60 years,
Argentina has changed the unit of currency over time by removing zeros
from its bills (not to be confused with devaluing the currency).
Since the 1960s this Argentinian gauge symmetry has been applied many
times, removing a total of 12 zeros.  One Peso of
today is equivalent to 1 trillion Pesos of the 1960s. One of these
gauge transformations is shown in Fig. \ref{australes}. It occurred in
1985 when the Central Bank removed three zeros from the Pesos bill to
transform it into a new currency called `Australes' at a rate of 1
Austral = 1,000 Pesos. 

Only a toy model of the economy that neglects
psychological effects or transaction costs can claim that removing
zeros from the currency has {\it no} consequences for the
markets, so this gauge symmetry is never perfect in reality. However, it is a
useful didactic example of a gauge transformation that leaves the system
(approximately) invariant.

Such a transformation corresponds to the action of the dilation group,
$GL^+(1,\R_+)$: this is the one-dimensional general linear group over the real
numbers (with positive determinant), and it consists of
all multiplications by any positive real number.  This group is the
structure group $G$ of the fiber bundle. In this case, $G$ is the
group of functions $g: \R_+ \to \R_+$ that acts by multiplying $x\in
\R_+$ by a constant $\lambda(g) \in \R_+$: $g(x) = \lambda(g) x$.

While the change of currency unit has no consequence for the internal
economy, there is a catch: the Central Banks and foreign exchanges
need to update their exchange rates with Argentina to keep the
comparison of prices of goods intact, i.e., the dynamics of
speculators exchanging money among the counties should remain
invariant under the gauge transformation.

\begin{figure*}
    	\centering \includegraphics[width=0.8\linewidth]{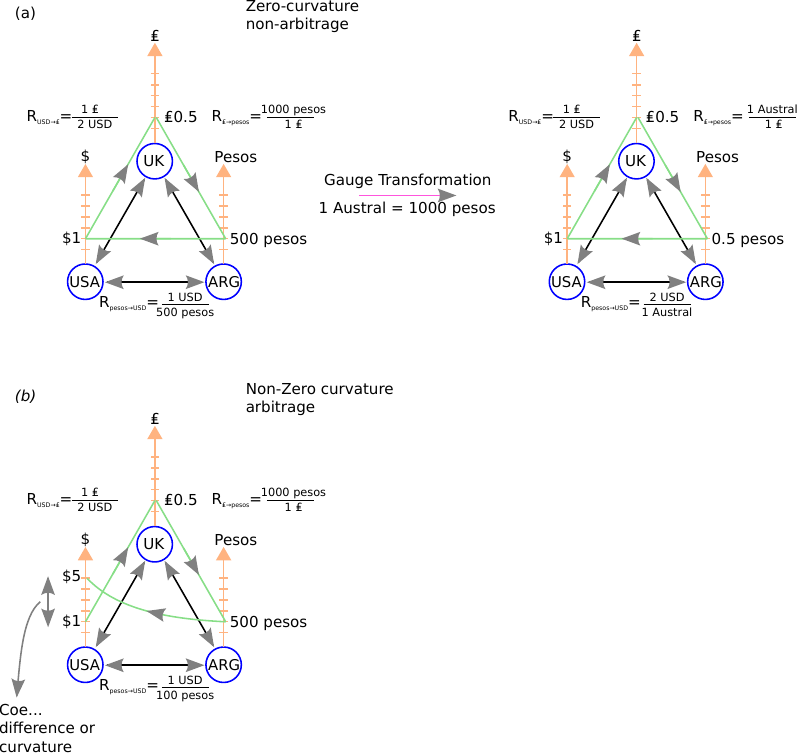}
	\caption{\textbf{Financial curvature.} (\textbf{a}) Zero curvature in an equilibrium market with no
          arbitrage. (\textbf{b}) Non-zero curvature in a financial market with
          arbitrage opportunities to make money out of thin air by
          just going around the loop.}
\label{financial}
\end{figure*}

For instance, imagine that before the adjustment to Australes, \mbox{£}1
 was equivalent to 1,000 pesos, i.e., exchange rate
$R_{\mbox{£}\to{\rm pesos}} = 1,000$ pesos/\mbox{£}1, and the exchange
rate to dollars was $R_{{\rm pesos}\to {\rm USD}} =  500
 {\rm pesos}/1{\rm USD}$. After the transformation of the Argentinian currency
1,000 Pesos $ \to 1$ Australes, the exchange rates with neighboring
countries need to be adjusted by the central banks to $1 \mbox{£} = 1 $
Australes and 1 Australes = 2 USD, so that the dynamics of money
exchange across the countries remains unaffected by the change, and we
can safely (covariantly) compared the value of goods in all currencies
(Fig. \ref{financial}).

These exchange rates are exactly the connections of the financial
fiber bundle,\index{fiber bundle !financial } and they are the glue that lets us compare how much a
currency is worth in terms of another (e.g., how much gold is
worth in one country with respect to another one) in the same way that
the covariant derivative in the sphere lets us compare vectors
moving along the curved surface of the sphere.  Thus the connection
provides the rule for adjusting the coordinate systems in fibers
associated with different neighboring points of the base.

To define the curvature of the fiber bundle we perform a parallel
transport\index{parallel
transport } of the currencies crossing from country to country along a
closed loop of the base, e.g., from USA $\to$ UK $\to$ ARG $\to$ USA,
Fig. \ref{financial}. This is the financial analogous of the parallel transport in the sphere of Fig. \ref{sphere}a. 

For this, we define the third connection
between USA and UK with a rate $R_{{\rm USD}\to \mbox{£}} = 1 \mbox{£} /
2 {\rm USD} $. A speculator going around the loop could take one
dollar in USA and parallel transport that dollar across the loop and
obtain around the circuit: 1 USD $\to 0.5 \mbox{£} \to 500$ pesos $\to$
1 USD. In detail: 1 USD $ \times R_{{\rm USD}\to \mbox{£}} \times R_{ \mbox{£} \to
  {\rm pesos} } \times R_{{\rm pesos}\to {\rm USD} } = 1\ {\rm USD}
\times  \mbox{£}1 / 2 {\rm USD} \times 1,000\ {\rm pesos}/ \mbox{£}
\times 1 {\rm pesos} / 500\ {\rm USD} = 1$ USD. This is exactly the
same dollar at the end of the loop when the speculator is back in the USA,
Fig. \ref{financial}a. That is, the gain or {\it arbitrage} around the loop
is zero. This means that the curvature of the fiber bundle defined by
these particular connection exchange rates is zero: it is a `flat
space', as flat as a plane, and the speculators would gain no money by
traversing this financial circuit.

In practice, speculators go around financial circuits to find opportunities for
arbitrage: an opportunity to make money out of thin
air (risk-free profit) from fluctuations in the market. In the
case of currency arbitrage, it is the opportunity to make money from a
mis-adjustment in exchange rates. For instance, imagine that the
Argentinian Central Bank is not quick enough to adjust their exchange
rate after the adjustment of currency and deviate (even slightly) from
the zero curvature connection discussed above.  In doing so, instead
they set the exchange rate with USA as $R_{{\rm pesos}\to {\rm USD} }
= 1 {\rm USD} / 100 {\rm USD}$. Then, going around parallel
transporting money in the same circuit starting with 1 USD in USA, the
speculator will take advantage end up with a 5$\times$ gain,
Fig. \ref{financial}b. This gain corresponds, geometrically, to the
curvature of the `surface' enclosed by the circuit, by analogy with the
sphere. Thus, whenever we move in a loop and end up in a
different state from the initial one, we say that the space of the
base is curved, and this curvature defines the geometry of the fiber
bundle.

\begin{figure}
  \centering \includegraphics[width=0.4\linewidth]{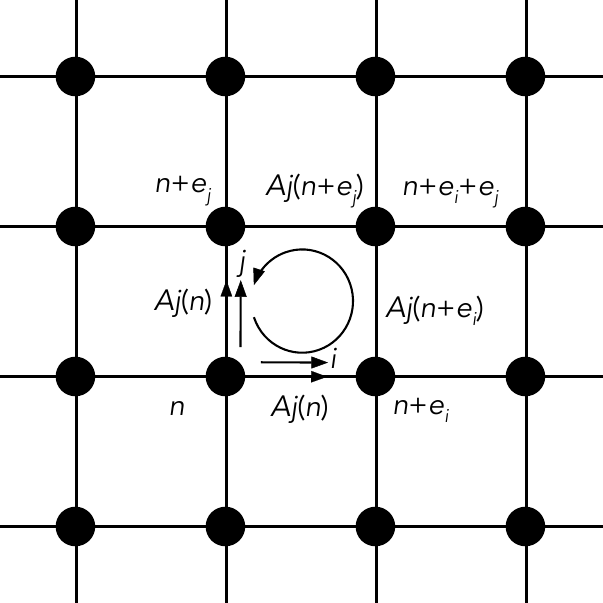}
  \caption{\textbf{Lattice gauge theory.} Example for a
    financial market  and electromagnetism.}
\label{lattice}
\commentAlt{Figure~\ref{lattice}: 
A bidimensional grid with 4 x 4 connection points (the grid borders continue past the last connection point on each direction).
The central square is labeled both on the dots and on the connecting edge. Reading the labels from the upper left corner and in clockwise direction:
n+ej, Aj(n+ej), n+ei+ej, Aj(n+ei), n+ei, Aj(n), n, Aj(n).
The two edges stemming from the lower left corner are arrows directed outwards.
Two further arrows, parallel to these two edges read i (horizontally), j (vertically).
There is a counter-clockwise arrow in the inner part of the square.
}
\end{figure}

\subsection{From finance to physics}

To calculate the
curvature, we identify countries as points in a lattice,\index{lattice } which for
convenience we consider to be a two-dimensional lattice as 
in Fig. \ref{lattice}.  Lattice gauge theories in physics
\citep{creutz} are generalizations of continuum four-dimensional
spacetime in physics to a lattice base.  Lattice gauge theories\index{lattice !gauge theory } were
introduced by \cite{wilson} to describe non-Abelian gauge
theories of the strong interaction on a space-time lattice
\citep{creutz}.  A lattice formulation of the fiber bundle is shown in
the discrete spaces of the financial market (Fig. \ref{financial}) and
electromagnetism (Fig. \ref{lattice}).

We now consider the lattice model of Fig. \ref{lattice}, following
\citep{maldacena}. In physics, the logarithm is introduced to work with
the Lie algebra rather than the Lie group, which is a more complicated
object. We therefore work with the logarithm of the exchange rates,
which plays the role of the magnetic potential $A_i(\vec{n})$, at
position $\vec{n}$ in a two dimensional lattice $d=2$, and $i=1,
\cdots, d$, according to the formula
\begin{equation}
  R_i(\vec{n})= e^{A_i(\vec{n})}.
\end{equation}

The gauge transformation at position $\vec{n}$ multiplies the currency
of country at $\vec{n}$ and can be written as $f(\vec{n}) =
e^{\varepsilon(\vec{n})}$. Under the action of this gauge, the
connections with neighboring countries at direction $\vec{e}_i$ (see
Fig. \ref{lattice}) are changed as:
\begin{equation}
  R_i(\vec{n}) \longrightarrow \frac{f(\vec{n}+\vec{e}_i)}{f(\vec{n})}
  R_i(\vec{n}),
\end{equation}
or
\begin{equation}
  A_i(\vec{n}) \longrightarrow
  A_i(\vec{n}) + \varepsilon_i(\vec{n}+\vec{e}_i) -\varepsilon_i(\vec{n}).
\label{ai}
\end{equation}
The curvature $F_{ij}$ is then obtained as the traders go around the
elementary circuit shown in Fig. \ref{lattice} that starts and
finishes at $(i,j)$. The gain (or loss) of money in such an elementary
financial circuit is:
\begin{equation}
  {\rm gain} = R_i(\vec{n}) R_j(\vec{n}+\vec{e}_i) \frac{1}{
    R_i(\vec{n}+\vec{e}_j)} \frac{1}{ R_j(\vec{n})} = e^{F_{ij}(\vec{n})},
\end{equation}
or
\begin{equation}
  F_{ij}(\vec{n}) = A_j(\vec{n}+\vec{e}_i) - A_j(\vec{n}) -
  [ A_i(\vec{n}+\vec{e}_j) - A_i(\vec{n}) ].
  \end{equation}
Extending the base space to four-dimensional spacetime with elements
$x = x^\mu$, with $\mu=0, 1, 2, 3$, and taking the continuum limit
(lattice spacing in Fig. \ref{lattice} to zero, see \citep{maldacena}),
we obtain the curvature tensor characterizing the financial market
geometry:
\begin{equation}
  \mathcal{F}_{\mu\nu}(x) = \frac{\partial A_\nu}{\partial x^\mu} - \frac{\partial A_\mu}{\partial x^\nu},
  \label{fij}
\end{equation}
with the transformation of the connection (\ref{ai}) given by
\begin{equation}
  A_\mu(x) \longrightarrow
  A_\mu(x) + \frac{\partial}{\partial x^\mu}\varepsilon(x).
  \label{ai2}
\end{equation}
This notion of curvature\index{curvature } is analogous to magnetic flux\index{magnetic flux } in electromagnetism.
It is invariant under the change of currency, and provides the
gain or loss of money around the elementary speculative circuit.

\section{Curvature and connection in the electromagnetic fiber bundle}
\label{curve2}

We have seen that speculators can make (or lose) money by traversing a closed
financial circuit if there is positive (negative) curvature; \index{curvature } moreover,
this curvature is analogous to the manifestation of magnetic
flux in a gauge theory of electromagnetism. This analogy can be
further extended by incorporating the transport of goods, like silver
or gold, analogous to moving electrons. This now introduces a particle
field, in addition to the gauge field of the interaction, but we stop
the analogy here since we already have enough elements to move to the
physical fiber bundle of electromagnetism,\index{electromagnetism } and then to jump to describe
curvature in a genetic fibration.

In the electrodynamic analogy, the exchange rates are the magnetic
potentials defined along the edges in the base space. The speculators
are the electrons, and in the presence of non-zero magnetic fields
they move in circles, like speculators in search of arbitrage
opportunities.  The total gain obtained from the curvature of the
financial circuit is now the total magnetic flux enclosed by the
closed loop. Thus an electron moving in a loop because of the magnetic field
is analogous to a speculator moving in a financial circuit, getting
richer because of the non-zero curvature.

This gauge invariance in classical electromagnetism is a manifestation
of the non-observability of $A_\mu$; that is, only the electric and
magnetic fields can produce observable effects (modulo the Aharonov--Bohm
effect, see below).\index{Aharonov--Bohm effect }  Briefly, the electromagnetic four-potential is
$A^{\mu}= (\phi/c, {\bf A})$, where $\phi$ is the scalar potential and
${\bf A}$ is the vector potential. Maxwell's equations are invariant
under the transformation:
\begin{equation}
  {\bf A} \longrightarrow {\bf A} + \nabla \Lambda \,, \,\,\,\,\, \phi
  \longrightarrow \phi + \frac{1}{c} \frac{\partial \Lambda}{\partial
    t},
\end{equation}
where $\Lambda$ is an arbitrary function  \citep{weyl2}.
In terms of the four-potential:
\begin{equation}
  A_\mu(\vec{x}) \longrightarrow A_\mu(\vec{x}) + \frac{\partial}{\partial x^\mu}\Lambda(x),
  \label{ai3}
\end{equation}
which is the same transformation as in the financial model
(\ref{ai2}). In other words, the connection field in Maxwell's equations\index{Maxwell's equations } 
$A_\mu$ transforms under a gauge transformation in a fiber bundle with
a multiplicative group, offering the same geometric analogy as in
the financial model. Furthermore, in electromagnetism the curvature
tensor $ \mathcal{F}_{\mu\nu}$ is the electromagnetic field tensor,
whose components are expressed in terms of the electric and magnetic
fields, meaning that the observable electric and magnetic fields define
the curvature\index{curvature } of the bundle. We see then that a complete geometric
theory is possible for both the financial model and classical
electromagnetism.  The physical electric and magnetic fields are the
elements of the curvature tensor, the four-potential is the connection
and `spiritus' of the interaction, and the dynamical equations
(Maxwell's) are invariant under the symmetry transformation.

The dilation gauge invariance employed in (\ref{ai3}) is the
invariance employed by Weyl\index{Weyl, Hermann } in his failed attempt to unify
electromagnetism and gravity by enlarging the gauge group of
general relativity of local rotational symmetries in spacetime 
(Lorentz transformations) to 
include also dilation symmetry of vectors.  Indeed,
since the invariance in electromagnetism (\ref{ai3}) and the
dilation group (\ref{ai2}) are the same, it was natural to
interpret the four-potential as the connection of the dilation gauge
symmetry of electromagnetism.  

However, as was pointed out by Einstein and
others, this symmetry is incompatible with special
relativity. Scale invariance of the dilation group implies that
there is no well defined scale of physical vectors. However, the
length of the four-momentum is $\| p \| = - E^2/c^2 + {\bf p} \cdot
{\bf p} = -m^2 c^2$, so the mass of a particle breaks
dilation symmetry. 
Local gauge invariance survived, and 18 years after Weyl, a quantum
framework showed that the correct gauge invariance of electromagnetism
is not the dilation group, but the one-parameter $U(1)$ symmetry
group of rotations in the complex plane. This symmetry group is also a
multiplicative group. Indeed, there are only two options to choose
from if we are interested in a one-parameter multiplication group:
$U(1)$ and multiplication by positive real numbers (dilation group)
\citep{ilinski}.

Electromagnetism\index{electromagnetism } is based on the gauge invariance\index{gauge invariance } specified by the
symmetry group $U(1)$, which corresponds to the symmetry of rotations of a
circle. This group is associated with redundancy of the phase factor in the wave
function of the electron, Eq. (\ref{u1}). This is captured by a fiber
bundle of spacetime where at every spacetime point a circle is
attached that defines the phase factor (Fig. \ref{sphere}c).  That is, the complex plane
with coordinates $z=x + i y$ is the fiber, and the structure group is
the group of multiplications by a complex number $e^{i \alpha}$ of
unit modulus: $U(1)$. Thus, the phase $\alpha$ is the variable to be
gauged.  As the electron moves along the base of the fiber bundle, the
magnetic field, defined by the connection or magnetic potential
$A_\mu$ determines how the phase factor changes from circle
to circle for two neighboring points in the base, Fig. \ref{sphere}c.

Gauge theories are relevant to all the forces in theoretical physics.
The weak force arises from the symmetry of a sphere at each point of
 spacetime, Fig. \ref{sphere}d.  The dynamical field or wave
function $\Psi$ is an isospin doublet where the two components refer
to the quantum amplitudes measuring the state of a particle that is
either a proton or a neutron, rather than a single complex number. This is approximate since it assumes that we turn off electromagnetism,
so we cannot distinguish a proton from a neutron by the electric charge;
it also ignores small mass differences.  These isospin states are
indistinguishable: the doublet can be transformed as $\Psi(x)
\longrightarrow \Psi'(x) = T \Psi(x)$ where $T$ is a $2\times 2$
matrix for isospin and it is an element of a Lie group. In general,
the groups are non-Abelian, as opposed to the Abelian group $U(1)$.
When the gauge is $U(1)\times SU(2)$ we obtain the QED theory
of the electroweak interaction\index{electroweak interaction } \citep{weinberg1995}, while adding the gauge group $SU(3)$
completes the standard model with the strong interaction \citep{glashow1961,greiner2007}.

\section{Curvature and connection in gravity}

In the {\it Scholium Generale} of the {\it Principia Mathematica}\index{Principia Mathematica@{\it Principia Mathematica} }
\citep{newton}, Isaac Newton\index{Newton, Isaac } laments his own failure: {\it `Hactenus ph\ae
  nomena c\ae lorum \& maris nostri per vim gravitatis exposui, sed
  causam gravitatis nondum assignavi.}'  (`Thus far I have explained
the phenomena of the heavens and our sea by the force of gravity, but
I have not yet assigned a cause to gravity.')  In the {\it `Hypotheses
  non fingo'} paragraph that has attracted much scholarly attention,
Newton concludes, regarding the origin of gravity: {\it `Sed h\ae c
  paucis exponi non possunt; neque adest sufficiens copia
  experimentorum, quibus leges actionum hujus spiritus accurate
  determinari \& monstari debent.'}  (`But these things cannot be
explained in a few words; furthermore, there is not a sufficient
number of experiments to determine and demonstrate accurately the laws
governing the actions of this spirit.')

Humanity had to wait 249 years until Einstein\index{Einstein, Albert } read the last chapter of
Newton's {\it Principia} and disentangled the meaning of this {\it
  `spiritus'} through his geometric theory of gravitation, showing
that massive objects influence each other through the curvature of empty
spacetime, which is itself caused by those objects. 
This concept of curvature was then
applied to all fundamental forces and particles in the standard model
through the curvature of the fiber bundle, providing the intellectual
framework to think about them all with the same kind of concept 
underlying all interactions. In the next section, we generalize curvature and
connection in terms of biological fibrations, thus extending Einstein's
geometric view to genetic (and other biological) interactions.

We have seen that the core concept of force arises from seemingly
superfluous redundancies in gauging the fibers of the system, like
changing the currency units in economics or the phase factor of the
wave function in quantum mechanics. Yet these redundancies are highly
nontrivial, because they trigger changes in the interactions between
neighboring elements in the base through the connection. The
interaction is then the adjustment of the connection that is needed to compare
elements along a path in the base.

The parallel transport of elements in the fiber along two different
paths with the same initial and end points can be different. The
curvature of the fiber bundle captures this difference, and non-zero
curvature indicates the existence of a physical force in nature, or the
possibility of risk-free financial gain or arbitrage in exchange
markets. This non-zero curvature then triggers the motion around the
loop of fundamental particles (with charge) and speculators,
respectively. The non-zero curvature is the `spiritus' of the force;
thus answering Newton's final scholium question. We now show that this 'spiritus' is also present in the transfer of genetic information in the fibration formalism.

\section{Curvature and connection in biology}
\label{connections}
After this lengthy but informative motivation, we reach the main
point of this chapter: to
 generalize the curvature\index{curvature !genetic fibration } concept to genetic fibrations.

\begin{figure*}[h!]
    	\centering \includegraphics[width=0.5\textwidth]{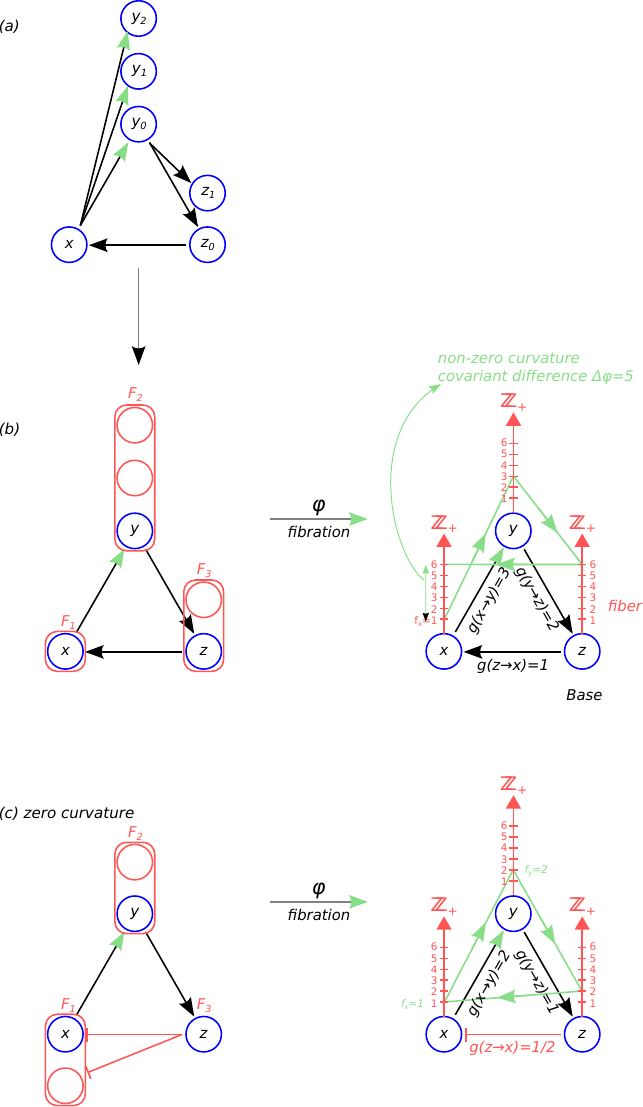}
  \caption{\textbf{Curvature in a genetic fibration.} (\textbf{a}) Genetic network with a closed loop to define the
    genetic curvature. (\textbf{b}) The fibers represents redundancy in the
    network.  These redundancies can be eliminated by collapsing
    all genes in a fiber to its base gene of the fibration. We show
    the connections along the edges (gauge fields) and the elements in
    the fibers. The cross-sections going around the loop gives rise to
    a positive curvature. The TFs going around the circuit
    get `richer', as in the financial model with arbitrage. (\textbf{c}) Zero
    curvature in a cycle with repressor links. The curvature of this
    graph is zero since the cross-sections going around the loop end
    up with the same value as the original one. The curvature is zero,
    and the TFs do not get `richer' or `poorer', by analogy with an
    equilibrium economic model with no arbitrage, or the absence of gravity
    in a flat spacetime.}
\label{genetic}
\end{figure*}

\subsection{Flow of information in a genetic network: the  genetic `spiritus'}

We consider a transcriptional regulatory network (genetic network)
with a closed loop as shown in Fig. \ref{genetic}a.  We explain the
concepts in this simple fibration structure with three fibers, $F_1$,
$F_2$ and $F_3$ (Fig. \ref{genetic}b), forming a $\Z_3$-symmetric ring at the
base. The ring is the closed loop or cycle necessary to develop
curvature.  We will develop the concept of genetic curvature
using this simple example, which can then be generalized to more
complex structures cycles.

\begin{figure}
    	\centering \includegraphics[width=\linewidth]{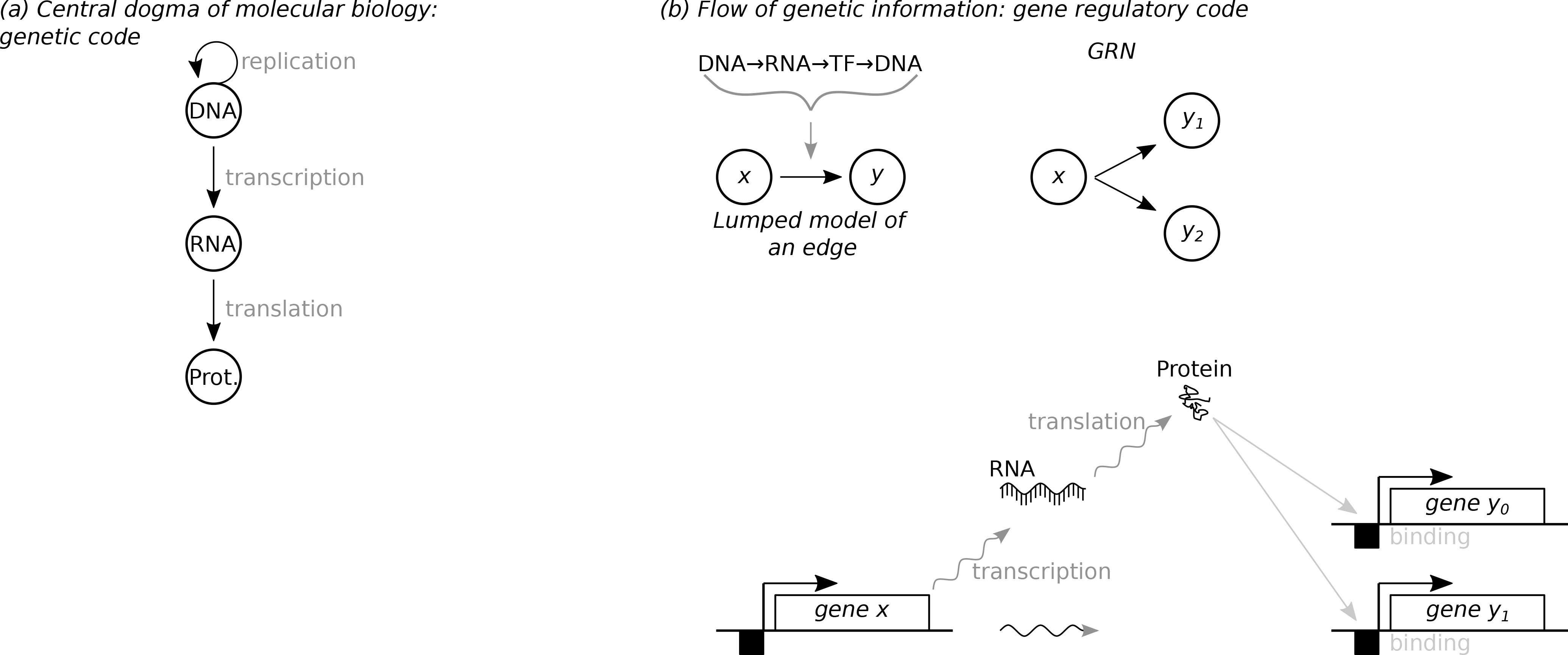}
 \caption{ \textbf{Flow of information in a genetic network: the  genetic `spiritus'}. (\textbf{a}) Central dogma of molecular biology \citep{crick2}. Rare
   processes are excluded. (\textbf{b}) Interpretation of the information flow
   in the genetic network. A gene is expressed (first transcribed
   into a long mRNA molecule, which is then translated into a protein
   (transcription factor, TF), which, upon folding, acquires a structure
   that allows the TF to bind to the promoter site, controlling the
   expression of another gene. Thus, the source
   gene sends a `TF messenger' to activate or repress the target gene,
   which represents the transmission of genetic information across
   the network.}
\label{dogma}
\commentAlt{Figure~\ref{dogma}: 
On the left, labeled by (a), there is a diagram with title: Central dogma of molecular biology: genetic code.
The diagram is a graph with three nodes with names DNA, RNA, Prot.
There is an arrow from DNA to itself (label: replication), from DNA to RNA (label: transcription), from RNA to Prot. (label: translation).
On the right, labeled by (b), there is a diagram with title: Flow of genetic information: gene regulatory code.
}

\commentLongAlt{Figure~\ref{dogma}:
The right-hand diagram is made of two graphs. 
The one on the left has two nodes called x,y with an arc from x to y reading: Lumped model of an edge.
The arc is further labeled with the sequence DNA to RNA to TF to DNA.
The one on the right has three nodes called x,y1,y2 with arcs from x to y1 and y2. This graph is labeled GRN.
On the bottom, there is a picture representing three genes (called Gene x, Gene y0, Gene y1).
From Gene x there is a squiggly arrow with red tip going to the other genes.
There is also a red squiggly arrow from Gene x to the drawing of a piece of RNA (labeled RNA); the arrow is labeled: transcription.
From RNA there is a further red squiggly arrow to the drawing of a folded protein (labeled Protein); the arrow is labeled: translation.
From the protein there are two green arrows towards Gene y0 and Gene y1, that connect to a place labeled binding.
}
\end{figure}

To facilitate the analogy with financial and electromagnetic fiber
bundles, we interpret the genetic network in terms of message-passing 
along its edges.  A link in a transcriptional network
represents regulatory messages that are dynamically sent from a
source gene to a target gene via a transcription factor, which acts as the
messenger, Fig. \ref{dogma}. Each arrow in the network
represents a message sent from a source gene to a target gene, by
analogy with links between countries when speculators exchange
 currencies, or spacetime links in a lattice model of
electromagnetism or gravity. Here we simplify the biological
regulation process to the case of bacteria. Eukaryotes have more
complex forms of regulation, e.g., splicing and TF binding to distance
enhancers that regulate genes via DNA looping, which also need to be taken
into consideration; however, they do not alter the main picture.

This flow of information\index{information flow} in biology is understood in terms of Francis
Crick's\index{Crick, Francis } `central dogma' of molecular biology: `DNA makes RNA
(transcription), and RNA makes protein (translation)';
Fig. \ref{dogma}a ~\citep{crick1,crick2,crick3}. Information about
the function of the protein is encoded in the sequence of amino acid
residues in the protein, which in turn is encoded in the sequence of
nucleotides in DNA. The dogma is the framework for understanding the
transfer of this information from DNA, encoded in the A, G, T, C
sequence, to the structure and function of the protein.

The DNA sequence\index{DNA sequence } of the source gene is first read by an RNA
polymerase to transcribe the gene's A, G, T, C code into RNA
codes A, G, U, C, where the RNA polymerase binds the target DNA at
the promoter region at the start of the gene.  In doing so, the
segment of DNA representing the source gene is copied to RNA. The
RNA is then read by the ribosome and translated into an amino acid
chain (or polypeptide), which then turns spontaneously into a folded
protein whose structure determines its function in the cell. The
encoding of each amino acid of the protein is done through the genetic code in groups of three
nucleotides, called codons. The combined
transcription $\to$ translation $\to$ folding process is called gene
expression. To this pipeline, a loop DNA $\to$ DNA is added to
represent the self-replication process necessary for life
(Fig. \ref{dogma}a).

However, the transfer of information does not stop with the folded
protein. To realize its function, the protein binds with other
biopolymers and molecules, thus extending the information transmission
through the network of binding among proteins, DNA, metabolites and
other small molecules. Recall that a transcription factor (TF) is a protein that binds to another DNA segment.\index{transcription factor } It thus acts as a messenger that controls the
transcription rate of the target gene by binding to its promoter
region (Fig. \ref{dogma}b).

This information flow\index{information flow} is not restricted to two interacting genes, but
it is transferred to different regions within the network that are
accessible through the connected pathways.  Likewise, a link can
represent any kind of regulation, either repression or activation of
any strength, directed or bidirectional, and can describe any
dynamical law of genetic regulation, such as a Hill input function. This
genetic information travels through the network and contains the whole
history of information transmitted through all possible pathways that
reach a given gene.  This process of communication of information
between different genes is formalized by the input tree of the gene.

Thus, a link in a transcriptional network does not represent a static
physical link, like a molecular bond. Links in a genetic network
represent regulatory messages that are sent dynamically from a source
gene to a target gene via the transcription factor which acts as a
messenger of genetic information by binding to the promoter region of
the target gene to turn its activity on and off.  The causal
interaction from source $\to$ target extends the flow of genetic
information of the central dogma from DNA$_{\rm source} \to$ mRNA
$\to$ protein (TF) $\to$ DNA$_{\rm target}$, by adding the last step
where the protein is a transcription factor that regulates the
transcription rate of the target gene (Fig. \ref{dogma}b).
This flow of genetic information\index{information flow} is the biological analogous of Newton's 'spiritus'.

\subsection{Curvature and connection in a genetic network}
\label{genetic-connection}

We have extended the causal flow of genetic information from source
gene $\to$ target gene using the transcription factor\index{transcription factor } as the messenger.
In this framework, DNA makes RNA, RNA makes proteins (TFs), TF
regulates DNA to make fibrations, and fibrations make the phenotype.

We now define this genetic interaction in terms of a
connection between the fibers attached to the base of the
fibration. We follow the definition of a connection in the fiber bundle,
defined in Sections \ref{curve1} and \ref{curve2} for the
financial and electromagnetic model, using a discrete base.
These lattices can be generalized to a graph to describe genetic
interactions.  We follow the generalization of the continuum space to a
discrete base from \citep{ilinski}, which is particularly suited
to describe curvature and connection in graph fibrations.

The discrete base of pointlike countries in Fig. \ref{financial}
now becomes a graph of genes interacting according to Fig. \ref{genetic}a.
In this case we have three genes in the base arranged in a $\Z_3$
ring: $B = \{ x, y, z \}$.  To each gene in the base we assign a fiber
$F_x, F_y, F_z$ with a given number of genes $|F_x|=1$, $|F_y|=3$,
$|F_z|=2$.  The genes inside each fiber are all equivalent, since they
have the same input tree, and are therefore redundant from a dynamical
point of view.  The analogous fiber bundle is now represented as in
Fig. \ref{genetic}b, where we exemplify the base graph and the fibers
attached to each node of the base. The fibration collapses this total space to give the base. 

By repeated duplication,\index{gene duplication } we can add any number of genes to the fibers without changing the
dynamics, and this represents the gauge invariance of the system
analogous to changing the units of the currency in the financial
model, or the phase factor in quantum mechanics. From this
perspective, the fiber is defined as the number of genes associated to
each gene at the base. An element $f$ of the fiber $F_i$ indicates
that there are $f$ redundant genes at the base $i$. Thus the fiber is
the space $\Z_+$, the positive integers:
\begin{equation}
  F_i = \Z_+ \equiv [1, +\infty)\,\,\,\, (i=x, y, z) .
    \end{equation}
The number $|F_i|$ is fixed for a given genetic realization of the cell, but
it can be changed arbitrarily with no changes to the dynamics of the
genes (aside from changes to some connection strengths, see below). 
An element $f$ of the fiber can increase by one by an
evolutionary duplication event (this will be treated in more detail in Section \ref{sec:duplication-lifting},  see Fig. \ref{fig:duplication} and Section \ref{sec:duplication}) by creating
one gene in the fiber $F_i$.  Likewise, a mutation or speciation can
remove a gene from a fiber. These changes are performed by a
transformation of the structure group $G$ associated to the fiber, in
similar way as the $GL(1,\R_+)$ group generates the changes of
arbitrary units of currency in the financial fiber bundle, or the
transformations of $U(1)$ generate gauge rotations in the phase factor
of the wave function in the electromagnetic fiber bundle.

The structure group acting on the fibers is the dilation (or
contraction) group over $\Z_+$ corresponding to maps $g$ that multiply
the number of genes in the fiber by an arbitrary number. For a fixed
$f\in F_i$ this is a `one-parameter semigroup' (with identity, that is, a `monoid') ${\bf G} = \mathbb{Q}_+
=\{q \in \mathbb{Q}: q > 0 \}$. This semigroup acts as
transformations of the
fiber, with an action ${\bf G} \times F \to F$ defined by:
\begin{equation}
\label{eq:Glambda}
(\lambda, f) \mapsto f' = \lceil \lambda
  f \rceil \qquad (\lambda \in {\bf G}).
\end{equation}
(This formula is not strictly an action, because the `ceiling' function does not preserve addition. However, it can be viewed as an `approximate action' and the resemblance justifies the terminology in the present analogy.)
The bundle resulting from this structure semigroup {\bf G} (analogous to a structure
group but lacking inverses) is depicted in
Fig. \ref{genetic}b, c right where we plot the genes in the base forming a ring
to which we attached a space $\Z_+$ denoting the number of genes $f$
in each fiber which can be changed by the action of ${\bf G}$.

The connection associated with edge $x\to y$ in the base of this graph
is a map $g_{x\to y}$ of the fiber $F_x$ associated with gene $x$ of
the base to the fiber $F_y$ above the gene $y$ that defines the
parallel transport of an element $f$ of the fiber along the edge. That
is, the expression $g_{x\to y} \circ f \in F_y$ is the result of the
parallel transport of $f \in F_x$ to $y\in B$.

The gauge transformation in the fiber is induced by the function of
the base $\lambda(x) : B \to {\bf G}$ defined by: $F_x \to \lceil
\lambda(x) F_x \rceil$, for any $x\in B$. That is, this is a gauge
transformation that changes the number of genes in the fiber.

While the action of ${\bf G}$ over the fibers changes the
number of genes in the fiber, it is also required to leave the
dynamics invariant, since the expression activity in the base does not
change. Thus, the redundant creation or deletion of a gene in the
fiber needs to be accompanied by a rescaling in the interaction with
the input genes to the fiber \citep[Section 8.8]{GS2023}. This is analogous to the rescaling of
the exchange rates in the financial model when the currency is changed
arbitrarily. In the genetic model, the addition of a node in, for
instance, fiber $F_y$ creates a new regulatory edge from gene $x$ to
the newly created gene in fiber $F_y$. This change in the interaction
is reflected by the action of the group ${\bf G}$ on the
connection. The change in the connection is necessary to keep the
dynamics in the graph invariant under the arbitrary increase/decrease
in the fiber.

\subsubsection{Genetic connection}

We define a genetic connection,\index{connection !genetic} which provides the rule of parallel
transport\index{parallel transport } of an element $f$ in a fiber by analogy with the financial
fiber bundle. We associate an independent element of ${\bf G}$
with each directed edge in the graph connecting a pair of genes in the
base.  The connection between gene $x$ and gene $y$ in the base is an
operator $g_{x\to y} \in {\bf G}$ that belongs to the structure
group ${\bf G}$ and acts from a fiber $F_x$ to a neighboring
fiber $F_y$ along the base:
\begin{equation}
  g_{x\to y}  : F_x \to F_y .
\end{equation}
This connection measures the necessary adjustment in the number of interactions
of the gene at base $x$ due to a gauge change in the number of genes
at base $y$.

Using $ g_{x\to y} $ we can define parallel transport of an
element $f$ of the fiber $F_x$ along the path $x\to y$ by $g_{x\to y}
f \in F_y$.  The result of a parallel transport of an element of a
fiber along two paths with the same initial and end points can be
different.  The curvature of the graph quantifies this difference
which is a measurement of the 'spiritus' or `force' in the system.

We recall that, in the financial model, the parallel transport
operator on the edge $x\to y$ connecting two countries in the base is
a factor by which we convert an asset in the currency of $x$ to the
currency in $y$. This is the factor by which a given currency needs to
be multiplied as a result of the operation represented by the edge
from $x\to y$ in the base. Analogously, the parallel transport
operator in the genetic base from gene $x$ to gene $y$ is the number
of genes that a single gene $x$ in the base regulates in the fiber at
position $y$ in the base.  For instance, in the genetic network in
Fig. \ref{genetic}b we have the connections:
\begin{equation}
  g_{x\to y}=3   \,, \,\,\,\, g_{y\to z}=2 \, , \,\,\,\, g_{z\to x}=1 \, ,
  \end{equation}
since a base gene at $x$ regulates 3 genes in the fiber $F_y$ at $y$,
a base gene at $y$ regulates 2 genes in $F_z$ and the base at $z$
regulates one gene in $F_x$.

The operator of parallel transport along the cycle from $x\to x$ in
Fig. \ref{genetic}b is defined as the product of the operators of
parallel transport along the edges that constitute the cycle
\citep{ilinski}:
\begin{equation}
  g_{x\to x} = g_{x\to y} \times g_{y\to z} \times g_{z\to x} .
\end{equation}

A parallel transport of one unit of expression starting at gene $x$
will result into $3$ units at $y$ and then $6=3 \times 2$ units at
base $y$ and back to 6 units at base $x$. The ratio between the
initial unit of expression and the final after traversing the cycle in
the graph is a measurement of the curvature tensor $ \mathcal{R}$ of
the fiber bundle, which is defined as \citep{ilinski}:
\begin{equation}
  \mathcal{R} = g_{x\to y} \times g_{y\to z} \times g_{z\to x} - 1 .
\label{curvature}
\end{equation}

\subsubsection{Curvature}

In this particular case, the curvature\index{curvature } of the bundle in
Fig. \ref{genetic}b is $\mathcal{R}=5$, a positive
curvature that indicates the existence of a `genetic force' or
interaction. This positive curvature makes the messenger TFs go around
the cycle in an analogous way to how arbitrage triggers speculators to
move currency between countries in a speculative circuit; it is also
analogous to electrons moving in cycles under the influence of a
non-zero magnetic flux. All of these examples can be captured by the
concept of non-zero curvature in the fiber bundle
\citep{maldacena}. This definition of curvature is motivated by the similar
definition for a lattice \citep{ilinski}, and gives the correct
curvature for electromagnetism in the continuum limit when the lattice
size shrinks to zero.

We see that the curvature of the circuit of Fig. \ref{genetic}b is
positive. In the language of financial models, starting at $x$,
the speculators (TFs) in the circuit can go around the financial
circuit ($\Z_3$ ring) and take advantage of the `arbitrage
opportunities'  (connections) in exchange rates, and get richer (gain
in genetic activity) when they return to $x$. Thus a non-zero
positive curvature measures the excess return (increase in expression
activity) of the financial operation (activation of the genetic
circuit).

Zero or negative curvature in a genetic circuit requires the
existence of repression interactions. A repressor edge gives rise to a
connection that is smaller than $1$. It is analogous to the exchange
rate from Argentina to USA in the financial model of
Fig. \ref{financial}a with $R_{{\rm pesos}\to {\rm USD}} = 
 500 {\rm pesos}/1{\rm USD} $, since the dollar is stronger than the peso; and also to
the exchange rate from USA to UK with $R_{{\rm USD}\to
  {\mbox{£}}} = 1{\mbox{£}} / 2 {\rm USD}$, since the pound is stronger
than the dollar.

A circuit with repression is shown in Fig. \ref{genetic}c, where the
edge from $z$ to $x$ is now a repressor of activity of gene $x$. The
connections in this case are:
\begin{equation}
  g_{x\to y}=2   \,, \,\,\,\, g_{y\to z}=1 \, , \,\,\,\, g_{z\to x} =1/2 \, ,
\end{equation}
where the repressor $z\to x$ is represented by the 
connection $g_{z\to x}=1/2$, which is smaller than $1$.  
Going around the circuit, the parallel
transport operator is $g_{x\to x}=1$ and the curvature is $\mathcal{R}
= 0$ indicating that there is no excess gain of expression activity
(or arbitrage in the financial model). Larger repression strength in
$g(z\to x)$ would lead to smaller connection and negative curvature,
analogous to a loss of money for the speculators in the financial
circuit.

\subsubsection{Generalization to cycles of any length}

The framework can be generalized to discrete cycles\index{cycle } of length $n$ in
the base defined as $\Gamma = \{ x_1, x_2, \cdots, x_n, x_1\}$, where
the gene $x_1$ is the initial and end point of the cycle, and the
points in the cycle are ordered \citep{ilinski}. The connection along
the cycle $\Gamma$ is defined by the products of the structure group
elements corresponding to each of the connections at the edges in the
path $g_{x_i\to x_{i+1}}$:
\begin{equation}
  g_\Gamma = g_{x_1\to x_2} g_{x_2\to x_3} \cdots g_{x_{n-1}\to x_n}
   g_{x_{n}\to x_1}.
\end{equation}
The graph curvature is obtained by parallel transport around the
cycle $\Gamma$ in the graph with a starting point $x_1$, and it is
a generalization of (\ref{curvature}):
\begin{equation}
  \mathcal{R}_\Gamma = g_\Gamma - 1 .
\end{equation}

\begin{figure}
         	\centering \includegraphics[width=0.35\linewidth]{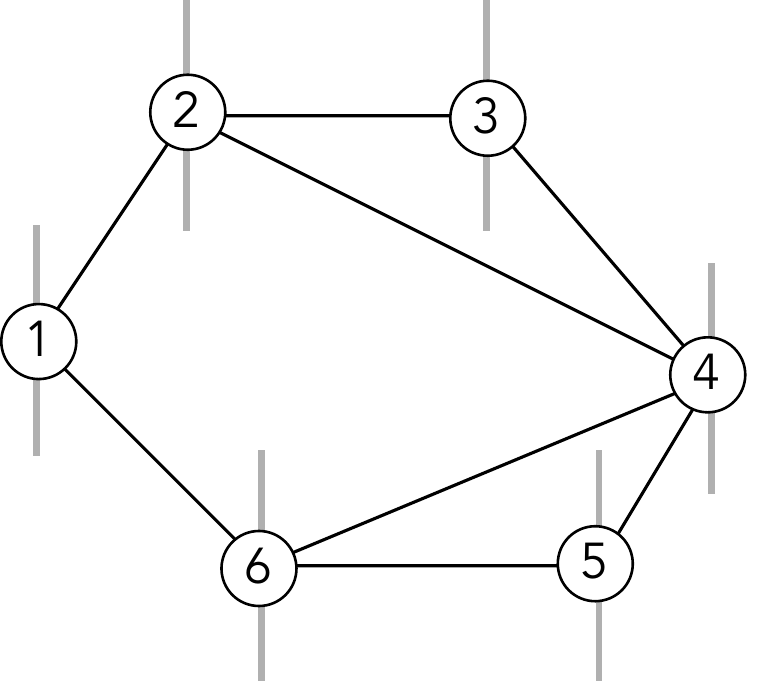}
  \caption{ \textbf{Elementary cycles}.\ The cycle 2342 is elementary
    while 124561 is not, since here we can consider the edge between 4
    and 6 to obtain a pair of cycles 12461 and 4564. The operator of parallel transport along 124561
    is the product of the operators along the elementary cycles 12461
    and 4564.}
\label{cycles}
\commentAlt{Figure~\ref{cycles}: 
An undirected graph with nodes named 1-6; nodes appear as circles with two rays going up and down.
The undirected connection closes the cycle 1-2-3-4-5-6-1, and there are further connections 2-4 and 4-6.
}
\end{figure}

When there are multiple cycles in the network, the curvature is
restricted to the elementary cycles in the base, as in
Fig. \ref{cycles}. Elementary cycles are cycles that cannot be reduced
to a pair of other cycles by the 
consideration of an existing link in
the base; see \citep{ilinski} and Fig. \ref{cycles}. The curvature is then:
\begin{equation}
  \mathcal{R}_{\Gamma_{\rm elem}} = g_{\Gamma_{\rm elem}} - 1 ,
\end{equation}
where $\Gamma_{\rm elem}$ is the set of all elementary cycles in
the base.

The gauge transformation in the fiber: $f_y \to \lambda(y) f_y = f'_y$
with $\lambda(y)\in {\bf G}$ transforms the connection as:
$g_{x\to y} \to \lceil \lambda(y) g_{x\to y} \rceil = g'_{x\to y}$.
Additionally, we can define the cross-section as a map from the base
to the total graph, $\psi: B \to E$ such that $\psi(x)$ is an element
of $F_x$ for each $x$.  The cross-section defines the covariant
difference between two connected points in the base as:
\begin{equation}
\Delta \psi =
\psi(y) - g_{x\to y} \psi(x) .
\label{covariant}
\end{equation}
When ${\bf G}$ is a Lie group, this covariant difference
transforms into the covariant derivative.
By definition, when the value $\psi(y)$ is the solution of the problem
of parallel transport of an element $f = \psi(x)$ along the edge $x\to
y$, then the covariant difference of $\psi$ is zero; in symbols, $\Delta \psi = 0$.

At a given point in time, a given cross-section is realized in the
network. For instance, in the graph of Fig. \ref{genetic}b the cross
sections are $\psi_x=1$, $\psi_y=3$, $\psi_z=2$, which represent the
biological realization of the redundant genes in the fibers at a given
time in evolution. This cross-sections give rise to a curvature.
Evolutionary changes can occur by deletion or additions of genes to
the fibers according to the gauge transformation. At a given time, a
given cross-section is realized. The space of all cross-sections is
the space of all possible scenarios under biological evolution, and
corresponds to all possible trajectories in the language of Feynman's
path integral\index{Feynman
path integral } in quantum mechanics \citep{ilinski,creutz,weyl2}.

\section{Curvature and connection in the {\it E. coli}
  genetic network}

\begin{figure}[h!]
      	\centering \includegraphics[width=0.6\linewidth]{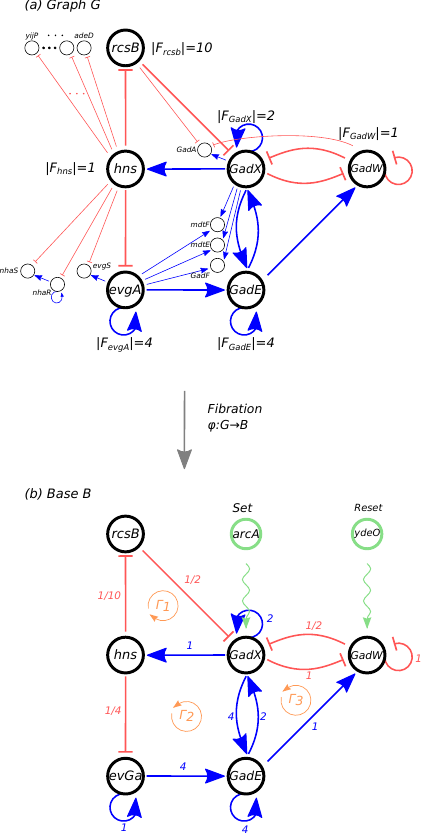}
  \caption{\textbf{ Curvature in a genetic fibration occurring in {\it E. coli}.} (\textbf{a}) Graph $G$. (\textbf{b}) Base $B$.}
\label{curvature-ecoli}
\end{figure}

An example of such a geometric model of a genetic network is realized in {\it
  E. coli}.\index{E. coli @{\em E. coli} } This network will be studied in detail in Part II. Here we
consider a sample of the strongly connected component (SCC)\index{strongly connected component }\index{SCC } of the
genetic network that controls the pH response of the bacterium,
through maintenance of pH homeostasis and control of the principal
acid resistance system, as an example of genetic fibration with
geometric curvature. We consider a part of the SCC in the acid
response regulation circuit shown in Fig. \ref{curvature-ecoli}a. The
SCC contains genes that are connected by a path between any two pair
of genes in the component. This path may consists of activators or
repressors, indistinctly. In other words, for every gene in the SCC,
there is at least a cycle that starts and ends at that gene. An input
tree for a gene in the SCC is, by definition, infinite, unless the
SCC contains just one gene with no autoregulation.

The circuit consists of two fibers $F_{\it rscB}$ and $F_{\it evgA}$
associated with the genes {\it rscB} and {\it evgA},
respectively. These fibers are studied in detail in Chapter
\ref{chap:hierarchy_2}; they are examples of the star fiber and FFF.

The full SCC of the pH response has a total of 53 elementary cycles.
For the sake of simplicity, to show the concept of curvature in a
simplified real setting, we consider three such cycles in
Fig. \ref{curvature-ecoli}a. These cycles run through both fibers and
pass through other genes in the pH network, starting at the master
regulator {\it hns}, and then going to genes identified with the acid
response system such as {\it gadX}, {\it gadW}, and {\it gadE}.

While at least a gene in the fiber must belong to the SCC, not all the
genes in the fiber do. For instance, $F_{\it rscB}$ consists of 10
genes, but only {\it rscB} belongs to the SCC. The rest of the genes
in the fiber from {\it adcD} to {\it yjjP}  receive only a repressor
edge from {\it hns}, but have no outgoing edge, and therefore do not
belong to the SCC. The same situation is observed at $F_{\rm evgA}$;
only {\it evgA} sends information back to the SCC while the gene {\it
  nhaR} belongs to the fiber but not to the SCC. Thus, only one gene
in each of these fibers sends back information to the SCC, and that is the
only gene that belongs to the SCC. We call these two fibers by the name
of that gene.

The fibration collapses the network into the base shown in
Fig. \ref{curvature-ecoli}b, where all genes in the base belong to
the SCC. We observe three elementary cycles of path length larger than
2, namely $\Gamma_1, \Gamma_2, \Gamma_3$. For simplicity, we do not analyze
the feedback loops of path length 2, nor the auto-regulation loop of
path length 1. Along the edges, we assign the connections.  In order
to define  the connections properly it is necessary to count all the
genes associated with a given fiber, including the genes inside an
operon. An {\it operon} is a set of contiguous genes that are transcribed in
a row by the same RNA polymerase; they are trivially
synchronized and they form a trivial fibration. 
In all of this analysis we consider the genes
inside operons to be a single unit. For example, the genes {\it evgA} and
{\it evgS} are considered as a single unit in the operon {\it
  evgAS} (denoted {\it
  evgA} in the figure). 
  
  However, to  assign the connections we need to take
into account the multiplicity of the operon. In this case, the operon
{\it evgAS} has to be counted as two genes, in order to 
assign the connection properly. Therefore, in the network of
Fig. \ref{curvature-ecoli}a, genes in an operon are considered
separate for the purpose of counting the connections.  However, when
we count fibers (Chapter \ref{chap:hierarchy_2}) we do not count operons
as fibers since they are trivial. In some studies, {\it gadAXW} is
considered to be a whole operon, but this cannot be so, since {\it gadY}
sits between {\it gadW} and {\it gadX} and these genes have their own
promoters and regulate different sets of genes. Thus we consider {\it
  gadW} to be separated from the operon {\it gadXA}, which is a true
operon.

An example of a connection is the edge {\it hns}$\to${\it rscB} with a
connection $g({\it hns} \to {\it rscB}) = 1/10$ since {\it hns} is a
repressor gene and $|F_{\it rscB}|=10$.  The activator edge {\it
  evgA}$\to${\it gadE} has a connection $g({\it evgA} \to {\it gadE})
= 4$ since $|F_{\it gadE}|=4$. The four genes of this fiber are {\it
  gadE, gadF} and {\it mdtE, mdtF}, which form an operon, but they are
considered separate only for the purpose of defining the connection,
as discussed above. We also assign connections to the self-loop as
indicated in the figure.

When we consider the elementary cycle $\Gamma_1 = \{{\it hns}\to {\it
  rscB} \to {\it gadX} \to {\it hns}\}$ (excluding the self-loop at
{\it gadX} in Fig. \ref{curvature-ecoli}b) we obtain a negative
curvature:
  \begin{equation}
    \mathcal{R}_{\Gamma_1} = \frac{1}{10} \times \frac{1}{2} \times 1 - 1 = - 0.95
    \end{equation}
  indicating a loss of expression activity along the cycle.  The
  curvature along $\Gamma_2 = \{ {\it hns} \to {\it evgA} \to {\it
    gadE} \to {\it gadX} \to {\it hns} \}$ is positive:
  \begin{equation}
    \mathcal{R}_{\Gamma_2} = \frac{1}{4} \times 4 \times 2 \times 1 - 1 = 1 , 
    \end{equation}
  indicating a gain of expression activity along this cycle.  Finally,
  the cycle $\Gamma_3 = \{ {\it gadX} \to {\it gadE} \to {\it gadW}\to
  {\it gadX} \}$ has a positive curvature too:
  \begin{equation}
    \mathcal{R}_{\Gamma_3} = 4 \times 1 \times \frac{1}{2} - 1 = 1 .
    \end{equation}

  This last circuit is interesting since it adds another form of
  regulation to the activity along the cycles. The genes {\it gadX}
  and {\it gadW} form a toggle-switch,\index{toggle-switch } like the circuit shown in
  Fig. \ref{fig:stability_circuits}, through their repressor feedback loop.  This configuration is analogous to
  an electronic flip-flop, a device used in computers to store a bit
  of memory. This circuit is ubiquitous in {\it E. coli} and also in
  eukaryotes, and it has been well-studied in synthetic circuits
  \citep{gardner2000construction}. It will be investigated in detail in Chapter
  \ref{chap:breaking}. This flip-flop can store one bit of information by the
      {\it set} gene, in this case by the activation of gene {\it
        arcA} (see Fig. \ref{curvature-ecoli}b). Gene {\it arcA} can
      turn the toggle switch on by storing a bit in {\it gadX} and
      turning off {\it gadW}, thus storing a memory in the activation
      of {\it gadX}. This state will remain even when {\it arcA} is
      turned off. Thus, the toggle switch works as a memory device
      storing one bit of information even when we turn off {\it arcA}.
      By this activation, the cycles $\Gamma_1$ and $\Gamma_2$ are
      activated and their negative and positive curvature would lead
      to repression and activation circuits, respectively. At this
      point, the cycle $\Gamma_3$ will not be activated, since the
      toggle switch stored a 0 at gene {\it gadW}. This state will
      last even if {\it arcA} is switched off.

  Only when the {\it reset} gene {\it ydeO} is on, is the flip-flop
  reset. Now {\it gad X} is off and {\it gadW} is on. At this point,
  the cycles $\Gamma_1$ and $\Gamma_2$ cease to operate, and they
   remain so even if the reset gene {\it ydeO} is turned off
  again. Thus, again, the flip-flop\index{flip-flop } device remembers the state. Only
  when the system is reset by {\it arcA} will the $\Gamma_1$ and
  $\Gamma_2$ cycles be activated again, and the process can be
  repeated between the set and reset states.

  These circuits represent examples of how fibrations, curvature and
  memory determine the computational logic underlying the function of
  the regulatory network. Altogether, the symmetries of the fibers,
  the geometry of the cycles, and the memory devices turning 
  the cycles in the network on and off, contribute to the computational logic of
  the gene regulatory code. These structures will be studied in detail
  in Part II of this book. We will show that they characterize the
  structure of biological networks and contribute to the phenotype by
  determining the functionality of genes, proteins and metabolites in
  the cell.


\chapter[Software and Algorithms to Perform Fibration Analysis]{\bf\textsf{Software and Algorithms to Perform Fibration Analysis}}
\label{chap:algorithms}

\begin{chapterquote}
  In the previous chapters, we have seen that fibers of fibrations  predict synchronization in networks.
  However, we have not yet discussed how such symmetries can actually be found.
  This chapter answers these questions in two ways. On the one hand, we consider the problem
  from an algorithmic viewpoint: How difficult is it to discover symmetries? Is it easier or harder to find fibration symmetries as opposed to automorphism symmetries? What methods do we currently know to find them? 
	And what do these questions mean exactly? 
  On the other hand, we provide a list of pointers to actual software tools that 
  are typically used in the context of bioinformatics and dynamical systems to find symmetries. Readers who are not interested in delving into the deeper algorithmic aspects can safely skip to Section~\ref{sec:software}.
\end{chapterquote}

\section{Finding symmetries}

To summarize in a nutshell what we have seen so far, biological
structures and dynamical systems can be seen as networks of
objects showing various degrees of symmetry, which in turn determine
synchronization in the behavior of the objects composing the
network. The types of symmetry observed in such systems depend on the
deep nature of what we are looking at: classical physical systems tend
to display a more rigid form of symmetry, captured by the group of
automorphisms of the network; biological systems tend to
display symmetries that are better described by fibrations.
This chapter explains the algorithms to find these symmetries.

The algorithms\index{algorithm } we describe take as input a graph, and produce as output a partition of its nodes.
The questions outlined above can therefore be rephrased more precisely as
follows: if we have a graph, how can we find the node partition induced by a
certain kind of symmetry? This is an algorithmic questions that implies, also, a
question of complexity: how difficult is it to find symmetries? We discuss both algorithms and complexity in this chapter, focusing
on fibrations\index{fibration } (in particular, symmetry fibrations). We start
with automorphisms:\index{automorphism } a pleasant and
important fact is that finding automorphism symmetries is apparently harder than
finding fibration symmetries.

\section{Finding the automorphisms of a graph}
\label{sec:autom}

Before embarking on a search for fibration symmetries, we discuss the more classical problem of finding automorphisms.\index{automorphism } How difficult is it to find the automorphisms of a graph? Recall that an automorphism of a graph $G$ is a bijective graph homomorphism $\varphi: G \to G$; i.e., a function (in fact, a pair of functions, one for the vertices and one for the nodes, though for graphs with no parallel edges the former uniquely determines the latter). 
This search turns out to be equivalent from a computational viewpoint to another, which is subtly different but has the same answer: how difficult is it to decide whether two graphs are isomorphic? Given two graphs $G$ and $H$, does there exists an isomorphism (i.e., a bijective homomorphism) $\varphi: G \to H$? 

This is the \emph{graph isomorphism problem}, and it is one of the most famous problems in computer science. Apart from its many applications, it is also interesting from a theoretical viewpoint, as it is one of the few real-world problems that are currently not known to be either polynomial-time solvable or NP-hard\index{NP-hard} (i.e., unlikely to be solvable in polynomial time), see~\citep{babai,babai2016}.

To give an idea of why this problem is difficult, consider the three graphs in Fig.~\ref{fig:isodraw}. Can the reader tell which of these graphs are isomorphic, if any? 

\begin{figure}
  \centering 
  \begin{tabular}{ccc}
  \begin{tikzpicture}[main/.style = {draw, fill, circle}, scale=.7, baseline=(current bounding box.center)] 
\node[main] (1)  at (0,0) {}; 
\node[main] (2)  at (5,0) {};
\node[main] (3)  at (2,-1) {};
\node[main] (4)  at (3,-1) {};
\node[main] (5)  at (1,-2) {};
\node[main] (6)  at (2.5,-2) {};
\node[main] (7)  at (4,-2) {};
\node[main] (8)  at (2.5,-2.5) {};
\node[main] (9)  at (2,-3) {};
\node[main] (10) at (3,-3) {};
\node[main] (11) at (0,-4) {};
\node[main] (12) at (5,-4) {};

\draw (1) -- (2);
\draw (1) -- (3);
\draw (1) -- (11);
\draw (2) -- (4);
\draw (2) -- (12);
\draw (3) -- (6);
\draw (3) -- (7);
\draw (4) -- (5);
\draw (4) -- (6);
\draw (5) -- (9);
\draw (5) -- (11);
\draw (6) -- (8);
\draw (7) -- (10);
\draw (7) -- (12);
\draw (8) -- (9);
\draw (8) -- (10);
\draw (9) -- (10);
\draw (11) -- (12);

\end{tikzpicture}  & 
  \begin{tikzpicture}[main/.style = {draw, fill, circle}, scale=.7, baseline=(current bounding box.center)]
\node[main] (1)  at (2.5,0) {}; 
\node[main] (2)  at (0,-1) {};
\node[main] (3)  at (5,-1) {};
\node[main] (4)  at (2.5,-1.2) {};
\node[main] (5)  at (2,-1.6) {};
\node[main] (6)  at (2.5,-2) {};
\node[main] (7)  at (2,-2.8) {};
\node[main] (8)  at (0,-3) {};
\node[main] (9)  at (5,-3) {};
\node[main] (10) at (2.5,-3.5) {};
\node[main] (11) at (1,-4) {};
\node[main] (12) at (4,-4) {};

\draw (1) -- (2);
\draw (1) -- (3);
\draw (1) -- (5);
\draw (2) -- (4);
\draw (2) -- (8);
\draw (3) -- (4);
\draw (3) -- (9);
\draw (4) -- (6);
\draw (5) -- (6);
\draw (5) -- (7);
\draw (6) -- (10);
\draw (7) -- (8);
\draw (7) -- (9);
\draw (8) -- (11);
\draw (9) -- (12);
\draw (10) -- (11);
\draw (10) -- (12);
\draw (11) -- (12);

\end{tikzpicture}  & 
  \begin{tikzpicture}[main/.style = {draw, fill, circle},  scale=.7, baseline=(current bounding box.center)] 
\node[main] (1)  at (1,0) {}; 
\node[main] (2)  at (4,0) {};
\node[main] (3)  at (2,-1) {};
\node[main] (4)  at (3,-1) {};
\node[main] (5)  at (2.5,-1.5) {};
\node[main] (6)  at (0,-2) {};
\node[main] (7)  at (5,-2) {};
\node[main] (8)  at (2.5,-2.5) {};
\node[main] (9)  at (0,-3) {};
\node[main] (10) at (3,-3) {};
\node[main] (11) at (5,-3) {};
\node[main] (12) at (2.5,-4) {};

\draw (1) -- (3);
\draw (1) -- (4);
\draw (1) -- (6);
\draw (2) -- (3);
\draw (2) -- (4);
\draw (2) -- (7);
\draw (3) -- (5);
\draw (4) -- (5);
\draw (5) -- (8);
\draw (6) -- (7);
\draw (6) -- (9);
\draw (7) -- (11);
\draw (8) -- (9);
\draw (8) -- (10);
\draw (9) -- (12);
\draw (10) -- (11);
\draw (10) -- (12);
\draw (11) -- (12);

\end{tikzpicture} 
  \end{tabular}
\caption{\textbf{Graph isomorphism problem.} Are any of these three undirected graphs isomorphic?\label{fig:isodraw}}
\commentAlt{Figure~\ref{fig:isodraw}: 
Three undirected graphs with many nodes, intentionally drawn so that it is hard to tell if they are or not different representations of the same graph.
}
\end{figure}

We might object that drawings are not how graphs are normally provided as input to algorithms, which is true. A more standard way would be to specify them as a list of edges, or using adjacency matrices. 
Fig.~\ref{fig:isomat} shows three adjacency matrices, and Fig.~\ref{fig:isoedges} shows the same graphs but in the form of a list of edges. Are any of these graphs isomorphic?

\begin{figure}
  \centering 
  \small
  \begingroup
  \setlength\arraycolsep{3pt}
  \begin{tabular}{ccc}
    $\left(\begin{matrix}
       0 & 0 & 0 & 1 & 0 & 0 & 0 & 0 & 0 & 0 \\
       0 & 0 & 1 & 0 & 0 & 0 & 1 & 0 & 0 & 0 \\
       0 & 1 & 0 & 1 & 0 & 1 & 0 & 0 & 1 & 1 \\
       1 & 0 & 1 & 0 & 1 & 0 & 0 & 0 & 0 & 0 \\
       0 & 0 & 0 & 1 & 0 & 0 & 0 & 1 & 0 & 1 \\
       0 & 0 & 1 & 0 & 0 & 0 & 0 & 0 & 0 & 1 \\
       0 & 1 & 0 & 0 & 0 & 0 & 0 & 0 & 1 & 1 \\
       0 & 0 & 0 & 0 & 1 & 0 & 0 & 0 & 0 & 0 \\
       0 & 0 & 1 & 0 & 0 & 0 & 1 & 0 & 0 & 1 \\
       0 & 0 & 1 & 0 & 1 & 1 & 1 & 0 & 1 & 0 
    \end{matrix}\right)$  &
      $\left(\begin{matrix}
         0 & 0 & 0 & 0 & 0 & 0 & 0 & 0 & 1 & 0 \\
         0 & 0 & 1 & 0 & 1 & 0 & 1 & 0 & 0 & 0 \\
         0 & 1 & 0 & 0 & 1 & 1 & 0 & 1 & 1 & 0 \\
         0 & 0 & 0 & 0 & 1 & 0 & 0 & 0 & 1 & 1 \\
         0 & 1 & 1 & 1 & 0 & 1 & 1 & 0 & 0 & 0 \\
         0 & 0 & 1 & 0 & 1 & 0 & 0 & 0 & 0 & 0 \\
         0 & 1 & 0 & 0 & 1 & 0 & 0 & 1 & 0 & 0 \\
         0 & 0 & 1 & 0 & 0 & 0 & 1 & 0 & 0 & 0 \\
         1 & 0 & 1 & 1 & 0 & 0 & 0 & 0 & 0 & 0 \\
         0 & 0 & 0 & 1 & 0 & 0 & 0 & 0 & 0 & 0 
        \end{matrix}\right)$ &
        $\left(\begin{matrix}
           0 & 0 & 0 & 0 & 1 & 0 & 1 & 1 & 0 & 0 \\
           0 & 0 & 1 & 0 & 0 & 1 & 0 & 0 & 0 & 1 \\
           0 & 1 & 0 & 0 & 0 & 0 & 0 & 0 & 0 & 0 \\
           0 & 0 & 0 & 0 & 0 & 1 & 0 & 0 & 0 & 0 \\
           1 & 0 & 0 & 0 & 0 & 0 & 0 & 0 & 0 & 1 \\
           0 & 1 & 0 & 1 & 0 & 0 & 0 & 1 & 0 & 0 \\
           1 & 0 & 0 & 0 & 0 & 0 & 0 & 1 & 0 & 1 \\
           1 & 0 & 0 & 0 & 0 & 1 & 1 & 0 & 1 & 1 \\
           0 & 0 & 0 & 0 & 0 & 0 & 0 & 1 & 0 & 1 \\
           0 & 1 & 0 & 0 & 1 & 0 & 1 & 1 & 1 & 0 
          \end{matrix}\right)$        
  \end{tabular}
  \endgroup
    \vspace{10pt}
\caption{\textbf{Graph isomorphism problem.} Are any of these three graphs, represented by adjacency matrices, isomorphic?\label{fig:isomat}}
\commentAlt{Figure~\ref{fig:isomat}: 
The adjacency matrices corresponding to the graphs of Fig.~\ref{fig:isodraw}. 
}
\end{figure}

\begin{figure}
  \centering 
  \small
  \begin{tabular}{ccccc}
    \begin{tabular}{cc}
      $0$&$3$\\
      $1$&$2$\\
      $1$&$6$\\
      $2$&$1$\\
      $2$&$3$\\
      $2$&$5$\\
      $2$&$8$\\
      $2$&$9$\\
      $3$&$0$\\
      $3$&$2$\\
      $3$&$4$\\
      $4$&$3$\\
      $4$&$7$\\
      $4$&$9$\\
      $5$&$2$\\
      $5$&$9$\\
      $6$&$1$\\
      $6$&$8$\\
      $6$&$9$\\
      $7$&$4$\\
      $8$&$2$\\
      $8$&$6$\\
      $8$&$9$\\
      $9$&$2$\\
      $9$&$4$\\
      $9$&$5$\\
      $9$&$6$\\
      $9$&$8$\\
      \end{tabular}
& $\qquad$ &
\begin{tabular}{cc}
  $0$&$8$\\
  $1$&$2$\\
  $1$&$4$\\
  $1$&$6$\\
  $2$&$1$\\
  $2$&$4$\\
  $2$&$5$\\
  $2$&$7$\\
  $2$&$8$\\
  $3$&$4$\\
  $3$&$8$\\
  $3$&$9$\\
  $4$&$1$\\
  $4$&$2$\\
  $4$&$3$\\
  $4$&$5$\\
  $4$&$6$\\
  $5$&$2$\\
  $5$&$4$\\
  $6$&$1$\\
  $6$&$4$\\
  $6$&$7$\\
  $7$&$2$\\
  $7$&$6$\\
  $8$&$0$\\
  $8$&$2$\\
  $8$&$3$\\
  $9$&$3$\\
  \end{tabular}
  & $\qquad$ &
  \begin{tabular}{cc}
    $0$&$4$\\
    $0$&$6$\\
    $0$&$7$\\
    $1$&$2$\\
    $1$&$5$\\
    $1$&$9$\\
    $2$&$1$\\
    $3$&$5$\\
    $4$&$0$\\
    $4$&$9$\\
    $5$&$1$\\
    $5$&$3$\\
    $5$&$7$\\
    $6$&$0$\\
    $6$&$7$\\
    $6$&$9$\\
    $7$&$0$\\
    $7$&$5$\\
    $7$&$6$\\
    $7$&$8$\\
    $7$&$9$\\
    $8$&$7$\\
    $8$&$9$\\
    $9$&$1$\\
    $9$&$4$\\
    $9$&$6$\\
    $9$&$7$\\
    $9$&$8$\\
    \end{tabular}
  \end{tabular}
  \vspace{10pt}
\caption{\textbf{Graph isomorphism problem.} Are these three graphs, represented as lists of edges, isomorphic?\label{fig:isoedges}}
\commentAlt{Figure~\ref{fig:isoedges}: 
Three lists of edges of large graphs: again, it is not immediately clear if they are the same graph (up to renaming) or not.
}
\end{figure}

These examples already suggest that finding whether two graphs are isomorphic (or, equivalently, looking for the automorphisms of a graph) 
is not an easy task. As we said above, it falls into an unusual category: no known algorithm solves it in polynomial-time, but there is no proof that it is NP-hard---a property that most computer scientists would interpret as `not having any polynomial-time solution'~\citep{GareyJ79}. 

Regardless of the actual complexity of this problem, though, it is often easy to give a \emph{negative} answer to the graph isomorphism question: for instance, if two graphs have a different number of vertices or a different number of edges, they \emph{cannot} be isomorphic. This is an example of a \emph{(polynomial) graph isomorphism test}, or GIT:\index{graph isomorphism test } it is something that we can be computed easily (`polynomial' here means `in polynomial time')
and that sometimes lets us rule out the possibility of two graphs being isomorphic.

There is a number of such tests, and each test, if it succeeds, provides a negative answer to whether the two graphs are isomorphic.
However, if all such tests succeed, 
whether the graphs are isomorphic remains unknown. 
 Although this conundrum is still unsolved, there are algorithms that are very efficient in practice, and that can solve the problem for many examples of interest. These algorithms
do not work in polynomial time, so we should expect them to run fast typically only on small graphs, but to become slower and slower  as the graph size increases.
The most famous of these algorithms is the Nauty algorithm\index{Nauty algorithm } of Brendan McKay~\citep{nauty,mckay1981practical,mckay2014}.\index{McKay, Brendon } 

Algorithms to find the coarsest equitable partitions,\index{equitable partition !coarsest } which we cover in the next section, 
were originally studied (starting from the late 1960s) in the context
of the graph isomorphism problem, i.e., precisely as polynomial GITs. We must also mention the
most famous of all graph isomorphism tests, the\index{Weisfeiler--Lehman test }
\emph{Weisfeiler--Lehman test}~\citep{weisfeiler1968reduction}, also
called the WL-test.\index{WL-test } The WL-test is in fact a hierarchy of graph
isomorphism tests depending on a dimension $k$, and it turns
out~\citep{huang2021short} that the color refinement algorithm (which, we shall see, is the simplest way to find the coarsest equitable partition)
is equivalent to the Weisfeiler--Lehman test of
dimension $1$, also called $1$-WL.  WL tests have been revived
recently because it was discovered that they are a quite powerful tool
to analyze the expressivity (i.e., computational capability) of graph
neural networks~\citep{xu2018powerful}. We discuss the WL test and graph fibrations further in Chapter \ref{chap:ai}.

  \section{Finding the fibration symmetries of a graph}
\label{sec:autom_fib}

Recall from Definition~\ref{equitable} that an equitable partition\index{equitable partition }
is a partition $\mathcal{S} = \{S_1, \cdot \cdot \cdot ,S_K\}$ of the
set $V$ of nodes of a network $G=(V,E)$, such that each node in
cluster $\mathcal{S}_\mu$ has the same number $k_{\nu\mu}$ of incoming
edges from nodes in cluster $\mathcal{S}_\nu$, for all $1\leq \mu, \nu
\leq K$.  In other words, equitable partitions are `special'
partitions of the set $V$ of nodes of a graph $G$, whose structure is
related to the structure of the graph $G$. These partitions are the balanced colorings or fibers of the network.

For obvious reasons, the singleton partition (all nodes in different clusters) is equitable, whereas the
discrete partition (all nodes in the same cluster) is equitable only for some graphs (precisely: only
if all nodes have the same number of incoming edges). But, in the
general case, is there a \emph{coarsest} equitable partition?\index{equitable partition !coarsest } We already know that
the answer is positive, as we learned in Chapter~\ref{chap:fibration_1}, but the problem
by itself is far from trivial.
Historically, the first formal proof of the existence of a
coarsest equitable partition was obtained by~\cite{cardon1982partitioning}, but
the intuition behind that dates back to~\cite{unger1964git}. While we
are not interested in discussing their proof here, it is useful to
recall what we already know about equitable partitions (balanced colorings, fibers),  fibrations and
input trees. We will get back to Cardon--Crochemore\index{Cardon--Crochemore algorithm } and to algorithms
to find equitable partitions in subsequent sections.

From Section~\ref{fiber} we know that equitable partitions of a graph
are the same as fibers of surjective fibrations. In other words,
if we have a surjective fibration $\varphi: G \to B$, then the fibers
of $\varphi$ are an equitable partition of $G$; on the other hand, if
we have an equitable partition $\mathcal{P}$ of $G$, there exists a
graph $B$ and a surjective fibration $\varphi: G \to B$ such that the clusters of 
$\mathcal{P}$ are the fibers of $\varphi$.  The base $B$ may need to be
a multigraph, even if $G$ is just a simple graph.

So, in our quest for the coarsest equitable partition of $G$, i.e., the partition with the minimal number of colors (fibers),  we may try to find a
surjective fibration of $G$ onto some `smallest' base $B$. Here we are
implying not only that $B$ should have the minimal possible number of nodes, but
also that for every surjective fibration $\varphi: G \to H$ (with $H
\neq B$) there should be another fibration $\psi: H \to
B$. (Actually, this property holds automatically because
the set of equitable partitions forms a lattice under refinement, see Section \ref{S:LBC}.)
The latter fibration may be an isomorphism, if $H$ is already `smallest'
from the viewpoint of its number of nodes; otherwise (i.,e., if $H$
has more nodes than $B$) $\psi$ will actually collapse some nodes.

We can now use another result that has previously been established: if $\varphi: G \to B$ is a fibration, then for every fiber the
input trees of the nodes of $G$ in that fiber are isomorphic, and they
are also all isomorphic to the input tree of their image in $B$. In
other words, if $\varphi(x_1)=\varphi(x_2)=y$ then the input trees of
$x_1$ and $x_2$ in $G$ are isomorphic, and they are isomorphic to the
input tree of $y$ in $B$.

This observation suggests how to construct the minimal base $\hat G$:
take the graph $G$ and collapse all the nodes that have
isomorphic input trees. Edges are defined in $\hat G$ in such a way as to
preserve input trees, but if we are interested only in finding the
coarsest equitable partition of $G$, we do not need to construct the
edges of $\hat G$.

\begin{algorithm}[H]
  \begin{flushleft}
    \hspace*{\algorithmicindent} \textbf{Input:} A graph $G=(V,E)$.\\
    \hspace*{\algorithmicindent} \textbf{Output:} A coloring of the
    nodes representing the coarsest equitable partition of $G$.
  \end{flushleft}
  \begin{algorithmic}[1]
    \State For every $x \in V$ build the input tree $T(x)$ truncated
    at the first $|V|-1$  levels
    \State Arbitrarily assign a number $\upsilon(T)$ to
    each of the non-isomorphic input trees $T$ \State Let $\kappa$ be
    defined by $\kappa(x)=\upsilon(T(x))$
  \end{algorithmic}
  \caption{Finding the coarsest equitable partition by building truncated input trees.}
  \label{algo:treenaive}
\end{algorithm}

This conceptual construction is not very useful, even though Norris's theorem\index{Norris's theorem } \citep{norris1995}, see Section \ref{sec:testing}, 
implies that we do not need to look at the whole
(typically: infinite) input tree to check for isomorphism---only the
first $|V|-1$ levels suffice.  So this construction is a genuine
algorithm, but a very inefficient one. It is, nonetheless, formally
described in Algorithm~\ref{algo:treenaive}.

In the following Section, we describe more efficient algorithms\index{algorithm } based on finding a minimal balanced coloring with refinement algorithms, and
also discuss the algorithmic problem in historical perspective.

\subsection{Coarsest equitable partition and tests for graph isomorphism}

In Section~\ref{sec:autom} we saw that a polynomial graph isomorphism test,
or GIT, is a polynomial-time algorithm that, given two graphs $G_1$ and
$G_2$, either outputs that they are not isomorphic ($G_1 \not\cong
G_2$), or it concludes that they \emph{may be isomorphic} (meaning
that it was unable to prove the opposite). In the latter case, we must
either try with some other GIT, or we can try to see if the graphs are
actually isomorphic in a brute force way. Sometimes, when the GIT is
unable to conclude that $G_1 \not\cong G_2$ it can provide some
additional information that can be used to speed up the brute-force
search. Nauty~\citep{nauty}\index{Nauty algorithm } typically applies a certain number of GITs in order,
and uses brute force only if all GITs fail to conclude that $G_1
\not\cong G_2$.

To understand the relation between GIT's and equitable
partitions, suppose that we want to determine whether $G_1 \cong G_2$. 
Instead of looking at the two graphs separately, we look at their
disjoint union $H=G_1 \cup G_2$: this is the graph obtained by placing
$G_1$ and $G_2$ `side by side' without adding any other edge.

\begin{figure}[h!]
  \begin{tikzpicture}[main/.style = {draw, circle}, scale=.6, font=\fontsize{6}{6}\selectfont,baseline]

	\node[draw,circle] at (0,0) (n1) {$1$};
	\node[draw,circle] at (3,3) (n2) {$2$};
	\node[draw,circle] at (3,0) (n5) {$5$};
	\node[draw,circle] at (3,-3) (n4) {$4$};
	\node[draw,circle] at (7,0) (n3) {$3$};

	\node[draw,circle] at (11,0) (nA) {$A$};
	\node[draw,circle] at (14,3) (nB) {$B$};
	\node[draw,circle] at (17,1) (nC) {$C$};
	\node[draw,circle] at (15,-3) (nE) {$E$};
	\node[draw,circle] at (18,-1) (nD) {$D$};

	\draw[->,>=latex] (n1) -- (n2);
	\draw[->,>=latex] (n2) -- (n3);
	\draw[->,>=latex] (n2) -- (n5);
	\draw[->,>=latex] (n3) -- (n4);
	\draw[->,>=latex] (n4) -- (n1);
	\draw[->,>=latex] (n4) -- (n5);
	\draw[->,>=latex] (n5) -- (n1);
	\draw[->,>=latex] (n5) -- (n3);

	\draw[->,>=latex] (nA) -- (nB);
	\draw[->,>=latex] (nA) -- (nC);
	\draw[->,>=latex] (nB) -- (nC);
	\draw[->,>=latex] (nB) -- (nE);
	\draw[->,>=latex] (nC) -- (nD);
	\draw[->,>=latex] (nD) -- (nB);
	\draw[->,>=latex] (nD) -- (nE);
	\draw[->,>=latex] (nE) -- (nA);
	
\end{tikzpicture}
  \caption{\textbf{The disjoint union of two graphs.} A graph $H$ obtained as the disjoint union of two graphs $G_1$
    and $G_2$, drawn on the left and right,
    respectively: a disjoint union does not add any further arc, it is just a juxtaposition of the two graphs. This example is taken from~\citep{unger1964git}.}
  \label{fig:git-example-uncolored}
\commentAlt{Figure~\ref{fig:git-example-uncolored}: 
The disjoint union of two directed graphs (since it is a disjoint union, the graphs appear juxtaposed).
Their nodes are called 1-4 and A-D.
The directed connections are 1 to 2, 2 to 3 and 5, 3 to 4, 4 to 1 and 5, 5 to 1 and 3,
A to B and C, B to C and E, C to D, D to B and E, E to A.
}
\end{figure}

Take the coarsest equitable partition\index{equitable partition !coarsest } $\mathcal{P}$ of $H$. If the two
graphs are isomorphic (that is, essentially, if they are the same
graph up to node identity), then every cluster of $\mathcal{P}$ will
contain the same number of nodes in $G_1$ and in $G_2$. On the other
hand, if we find that there is some cluster containing a different
number of nodes from $G_1$ and $G_2$, then for sure $G_1 \not\cong
G_2$.

\begin{figure}[h!]
  \begin{tikzpicture}[main/.style = {draw, circle}, scale=.6, font=\fontsize{6}{6}\selectfont,baseline]

	\node[draw,circle,fill=yellow!70] at (0,0) (n1) {$1$};
	\node[draw,circle,fill=red!70] at (3,3) (n2) {$2$};
	\node[draw,circle,fill=blue!70] at (3,0) (n5) {$5$};
	\node[draw,circle,fill=red!70] at (3,-3) (n4) {$4$};
	\node[draw,circle,fill=yellow!70] at (7,0) (n3) {$3$};

	\node[draw,circle,fill=red!70] at (11,0) (nA) {$A$};
	\node[draw,circle,fill=blue!70] at (14,3) (nB) {$B$};
	\node[draw,circle,fill=yellow!70] at (17,1) (nC) {$C$};
	\node[draw,circle,fill=yellow!70] at (15,-3) (nE) {$E$};
	\node[draw,circle,fill=red!70] at (18,-1) (nD) {$D$};

	\draw[->,>=latex] (n1) -- (n2);
	\draw[->,>=latex] (n2) -- (n3);
	\draw[->,>=latex] (n2) -- (n5);
	\draw[->,>=latex] (n3) -- (n4);
	\draw[->,>=latex] (n4) -- (n1);
	\draw[->,>=latex] (n4) -- (n5);
	\draw[->,>=latex] (n5) -- (n1);
	\draw[->,>=latex] (n5) -- (n3);

	\draw[->,>=latex] (nA) -- (nB);
	\draw[->,>=latex] (nA) -- (nC);
	\draw[->,>=latex] (nB) -- (nC);
	\draw[->,>=latex] (nB) -- (nE);
	\draw[->,>=latex] (nC) -- (nD);
	\draw[->,>=latex] (nD) -- (nB);
	\draw[->,>=latex] (nD) -- (nE);
	\draw[->,>=latex] (nE) -- (nA);
	
\end{tikzpicture}
    \vspace{10pt}
  \caption{\textbf{Computing the coarsest equitable partition of the disjoint union.} The coarsest equitable partition of the graph $H$ in Fig.~\ref{fig:git-example-uncolored}.}  
  \label{fig:git-example}
\commentAlt{Figure~\ref{fig:git-example}: 
Same graph as in Fig.~ref{fig:git-example-uncolored} but colored as follows: 5 and B are blue; 1,3,C,E are yellow; 2,4,A,D are red.
}
\end{figure}

If we look at the coarsest equitable partition of the graph in
Fig.~\ref{fig:git-example-uncolored} we obtain the partition shown in
Fig.~\ref{fig:git-example}. This partition has three clusters (red, yellow and blue). Both $G_1$ and
$G_2$ contain two yellow nodes, two red nodes and one blue node. So we
can conclude that $G_1$ \emph{may} be isomorphic to $G_2$. So this GIT
is inconclusive.

However, any isomorphism $\alpha: G_1 \to G_2$, if one exists,
must respect the colors of the coarsest partition.  For instance,
$\alpha(5)=B$ necessarily. Intuitively, this constraint can be
explained as follows: in both graphs there are only three nodes with
two incoming edges: in $G_1$ they are $1$, $3$ and $5$, whereas in
$G_2$ they are $B$, $C$, $E$. Since $\alpha(5)$ must be a node of
$G_2$ with two incoming edges, it must be one of $B$, $C$ or $E$.  On
the other hand, the two incoming edges of $5$ both come from nodes with
just \emph{one} incoming edge, whereas one of the two incoming edges of
$C$ has source $B$, which has two incoming edges.  The same is true of
$E$. Hence the only possibility is that $\alpha(5)=B$.  This is
probably better appreciated by looking at the minimal base $\hat H$
(Fig.~\ref{fig:git-example-base}): both the yellow and the blue node
have in-degree two, but only the blue node has two incoming edges both
arriving from a node with in-degree one (the red node).

\begin{figure}[h!]
  \begin{tikzpicture}[main/.style = {draw, circle}, scale=.7, font=\fontsize{6}{6}\selectfont,baseline]

	\node[draw,circle,fill=red!70] at (0,0) (red) {$\vphantom{A}$};
	\node[draw,circle,fill=yellow!70] at (3,0) (yellow) {$\vphantom{A}$};
	\node[draw,circle,fill=blue!70] at (6,0) (blue) {$\vphantom{A}$};

	\draw[->,>=latex] (red) to [out=60,in=120] (blue);
	\draw[->,>=latex] (red) to [out=-60,in=-120] (blue);
	\draw[->,>=latex] (blue) to (yellow);
	\draw[->,>=latex] (red) to[out=20,in=160] (yellow);
	\draw[->,>=latex] (yellow) to[out=-160,in=-20] (red);
\end{tikzpicture}
    \vspace{10pt}
  \caption{\textbf{Minimal base.} The minimal base $\hat H$ of the graph $H$ in Fig.~\ref{fig:git-example}: the colors of the nodes indicate the fibers of the fibration symmetry.}  
  \label{fig:git-example-base}
\commentAlt{Figure~\ref{fig:git-example-base}: 
A directed graph with three unnamed nodes (red, yellow, blue) and the following arrows: red to blue (twice), blue to yellow, yellow to red, red to yellow.
}
\end{figure}

So we conclude that $G_1$ and $G_2$ may be isomorphic: in other words, our GIT could not exclude the two graphs from being isomorphic. In fact, we can check (e.g., by brute force) that they are.

\subsection{The color refinement algorithm}

The first algorithm to find the coarsest equitable partition\index{equitable partition !coarsest }
 of a
graph was originally proposed as a GIT by \cite{unger1964git}. The
algorithm that Unger\index{Unger, Stephen } provided to find the coarsest equitable partition
of a graph $G$ is today called the \emph{color refinement algorithm}\index{color refinement algorithm }
(or `naive vertex classification'). It is described in
Algorithm~\ref{algo:naive}. Building on Unger's method, a further
improvement was proposed (always starting from the coarsest equitable
partition) in~\citep{corneil1970efficient}.

\begin{algorithm}[H]
  \begin{flushleft}
    \hspace*{\algorithmicindent} \textbf{Input:} A graph $G=(V,E)$\\
    \hspace*{\algorithmicindent} \textbf{Output:} A coloring of the nodes representing the coarsest equitable partition of $G$
  \end{flushleft}
  \begin{algorithmic}[1]
    \State Let $\kappa_0$ be a coloring such that $\kappa_0(x)=0$ for all $x \in V$
    \State Let $c_0=1$  (the number of colors used by $\kappa_0$)
    \State Let $i=0$ (the number of iterations so far)  
    \While{true}
      \State For each node $x$ of $V$ consider the multiset $M_{i+1}(x)=\{\!\!\{\kappa_i(s(e)) \mid e \in E, t(e)=x\}\!\!\}$ 
      \State Let $c_{i+1}$ be the number of distinct multisets $M_{i+1}(x)$ (for $x \in V$)
      \State Arbitrarily assign a number $\upsilon(A)$ to each of the $c_{i+1}$ multisets $A$
      \State Let $\kappa_{i+1}$ be a coloring such that $\kappa_{i+1}(x)=\upsilon(M_{i+1}(x))$ for all $x \in V$
      \State Stop if $c_{i+1}=c_i$ (the number of colors did not change in the last iteration)
      \State $i=i+1$
    \EndWhile
    \State Output $\kappa_i$
  \end{algorithmic}
  \caption{Color refinement algorithm~\citep{unger1964git}.}
  \label{algo:naive}
\end{algorithm}

It is called a `refinement' algorithm because it starts with the
coarsest partition (the one in which all nodes have the
same color), $\kappa_0$, and then refines it until an
equitable partition is obtained. \cite{cardon1982partitioning} proved that the process does
eventually stop at the coarsest equitable partition.

At every step, we compute the multiset\index{multiset } of colors $M(x)$ of the incoming edges
of each node, and we assign a different color to each distinct
multiset. The algorithm stops when the number of colors does not
change any more. A multiset allows for multiple instances for each of its elements, for example $\{a, a, b\}$ where the element $a$ has multiplicity 2. $M(x)$ is a multiset, because $x$
may be the target of many edges coming from nodes of the same color
(either different nodes of the same color, or even the same node, if
$G$ is itself a multigraph).

The step where the algorithm turns a multiset into numbers (arbitrarily)
is not strictly necessary, but it is useful if we want to be sure that
we are always using `numbers' as colors, and nothing more exotic
(like multisets).

The running time complexity of the color refinement
Algorithm~\ref{algo:naive} is $O(|E|\cdot |V|^2)$~\citep{cardon1982partitioning},
which is $O(|V|^3)$ for simple dense graphs, and $O(|V|^2)$ for simple
sparse graphs. This is not very efficient, but it is still polynomial.
This method can be applied to small graphs, but scales poorly on
larger graphs. 

\cite{belykh2011} present a simple algorithm to find the 
coarsest equitable
partition, based on the same general principles.

\subsection{The Cardon--Crochemore algorithm and canonical colorings}

Starting from the color refinement algorithm, \cite{cardon1982partitioning}
proposed a more efficient algorithm to find the coarsest equitable
partition of a graph or minimal balanced coloring. The algorithm is based on Hopcroft's\index{Hopcroft, John } technique
for minimizing finite-state automata~\citep{hopcroft1971}.\index{finite-state automaton } A simpler
version of the Cardon--Crochemore algorithm\index{Cardon--Crochemore algorithm } was presented
in~\citep{paige1987three}, and later a similar algorithm with the same
running time was discussed in~\citep{junttilla2007}.
Another variant of the same ideas was proposed
in~\citep{berkholz2017tight}, and the latter is presented in
Algorithm~\ref{algo:cc}.  

\begin{algorithm}[b!]
  \begin{flushleft}
    \hspace*{\algorithmicindent} \textbf{Input:} A graph $G=(V,E)$\\
    \hspace*{\algorithmicindent} \textbf{Output:} A coloring of the nodes representing the coarsest equitable partition of $G$
  \end{flushleft}
  \begin{algorithmic}[1]
    \State Let $C_1=V$
    \State Let $k=1$ (the number of colors)
    \State Let $S_R$ be the stack containing only $1$ (the only color used)
    \While{$S_R$ is not empty}
      \State Let $r=\mathrm{pop}(S_R)$ (the topmost color on the stack) \label{algo:cc-r}
      \State For every $x \in V$, let $d^-_r(v)=|\{e \mid t(e)=v, s(e)\in C_r\}|$
      \State Let $C_S=\{c\in\{1,\dots,k\} \mid \exists x,y\in C_c, d^-_r(x)\neq d^-_r(y)\}$\label{algo:cc-cs}
      \For{$s \in C_s$ (in increasing order)} \label{algo:cc-fors}
        \State Let $m=\max_{v \in C_s} d_r^-(v)$
        \State For $i=0,\dots,m$, let $n_i=|\{v \in C_s \mid d_r^-(v)=i\}|$
        \State Let $D=\{i=0,\dots,m \mid n_i >0\}$
        \State Let $I=\{s\} \cup \{k+1, \dots, k+|D|-1\}$
        \State Construct a bijection $f: D \to I$ such that $i<j$ implies $f(i)<f(j)$
        \For{$v \in C_s$}
          \If{$f(d_r^-(v)) \neq s$}
            \State Remove $v$ from $C_s$
            \State Add $v$ to $C_{f(d_r^-(v))}$
          \EndIf
        \EndFor
        \If{$s \in S_R$}
          \For{$c \in I \setminus\{s\}$ (in increasing order)}
            \State $\mathrm{push}(S_R, c)$
          \EndFor
        \Else 
          \State Let $b=\min\{i=0,\dots,m \mid \forall j n_i \geq n_j\}$
          \For{$c \in I$ (in increasing order)}
            \State If $c \neq b$, then $\mathrm{push}(S_R, c)$
          \EndFor
        \EndIf
        \State Let $k=k+|I|-1$
      \EndFor
    \EndWhile
    \State Output the coloring mapping $v$ to the index $c$ such that $v \in C_c$
  \end{algorithmic}
  \caption{The algorithm of~\citep{berkholz2017tight}, based on the same techniques (and with the same running time as) as 
  Cardon--Crochemore's~\citep{cardon1982partitioning}.}
  \label{algo:cc}
\end{algorithm}

The algorithm is very technical, and we do
 discuss it in detail here.  Essentially, in
Line~\ref{algo:cc-cs} we try to understand what are the colors
(clusters) that can be broken based on $r$ (i.e., that contain at
least two nodes that have a different number of incoming edges from the
color $r$). These clusters are analyzed in increasing index order
(Line~\ref{algo:cc-fors}), and for each cluster we group the vertices
that differ in the number of incoming edges from $r$: we leave in the
original cluster only those with the smallest number of incoming edges,
and we move those with more incoming edges to newly created clusters
(in increasing order of number of incoming edges).

The important point, for our discussion, is that the Cardon--Crochemore
algorithm\index{Cardon--Crochemore algorithm } has complexity $O((|E|+|V|) \log |V|)$, which is $O(|V|^2
\log |V|)$ for simple dense graphs and $O(|V| \log |V|)$ for simple
sparse graph.

It was recently proved~\citep{berkholz2017tight} that no algorithm can
be more efficient than Cardon--Crochemore, so $O((|E|+|V|) \log |V|)$
is the best possible bound for finding the coarsest equitable
partition of a graph. (However, the result
of~\citep{berkholz2017tight} is not for the standard general RAM
computation, but only for the special class of refinement-based
algorithms.)

Now Cardon--Crochemore, like all the algorithms described so far,
produces an equitable partition (in fact: the coarsest equitable
partition) in the form of a coloring. In other words, it takes as
input a graph $G=(V,E)$, and outputs a function $\kappa: V \to C$ over
some fixed set of colors $C$ such that two nodes are in the same part
of the partition if and only if they have the same color.

But this coloring is not guaranteed to be `canonical'. An algorithm taking as input a graph $G$ produces a \emph{canonical coloring}\index{coloring!canonical } if the output coloring remains the same when the algorithm is presented with an isomorphic graph $G'$.
In other words, if the same algorithm is provided another graph $G'=(V',E')$
that is isomorphic to $G$ (say, via $\alpha: G \to G'$), it must output
a coloring $\kappa': V' \to C$ such that
$\kappa'(\alpha(x))=\kappa(x)$.  

There are many situations where we wish to use algorithms producing canonical
colorings, and Algorithm~\ref{algo:cc} does in fact produce a canonical
coloring. Nonetheless, the guarantee that the coloring produced is canonical  is irrelevant for our purposes, and for the discussion so far, because we never look at the actual coloring, but rather at the node partition it induces.

\section{Algorithms to find all equitable partitions}
\label{sec:kamei}

In~\citep{kamei2013}, the authors studied the problem of finding not only
the coarsest equitable partition, but the {\em whole lattice} of possible equitable partitions.
We start discussing this problem with a specific concrete example.

Consider the graph $C_k$ (the directed
$k$-cycle), consisting of nodes $\{0,1,\dots,k-1\}$ and edges $0 \to
1$, $1 \to 2$, \dots, $k-2 \to k-1$, $k-1 \to 0$; in
Fig.~\ref{fig:cycle12-uncolored} we show the $12$-cycle $C_{12}$.  What are the equitable partitions of this graph? An
equitable partition\index{equitable partition, algorithm } is a partition of the set
$\{0,1,\dots,k-1\}$: there are overall $4,213,597$ partitions of a
set of 12 elements (the 12-th Bell number), but not all of them are
equitable.

\begin{figure}[h!]
    \begin{tikzpicture}[main/.style = {draw, circle}, scale=.7, font=\fontsize{6}{6}\selectfont,baseline]

\colorlet{color min rgb}[rgb]{red}
\colorlet{color max rgb}[rgb]{lime}

\def \n {12}
\def \d {1}
\def \radius {3cm}
\def \margin {8} 

\foreach \s in {1,...,\n}
{
	\tikzmath{
		int \v, \modlcass;
		\v = \s-1;
		\modclass = int(\v-\d*int(\v/\d));
	}
	\pgfmathsetmacro\myvalue{\modclass};
	\pgfmathtruncatemacro\lambda{\myvalue/\d*100};
	\colorlet{my color rgb}[rgb]{color min rgb!\lambda!color max rgb};
	\node[draw, circle] at ({360/\n * (\s - 1)}:\radius) {\v};
	\draw[->, >=latex] ({360/\n * (\s - 1)+\margin}:\radius) arc ({360/\n * (\s - 1)+\margin}:{360/\n * (\s)-\margin}:\radius);
}
\end{tikzpicture}
    \caption{\textbf{Finding all equitable partitions of a directed cycle.} The directed 12-cycle $C_{12}$.}
  \label{fig:cycle12-uncolored}
\commentAlt{Figure~\ref{fig:cycle12-uncolored}: 
A directed cycle with nodes labeled 0-11, and arcs connecting (counterclockwise) 0 to 1, 1 to 2, ..., 10 to 11, 11 to 0.
}
\end{figure}

In fact, it turns out that the equitable partitions of a $k$-cycle are
only as many as the number of divisors of $k$. More precisely, for
every integer $d$ that is a divisor of $k$, there is an equitable
partition $\mathcal{P}_d$ into $d$ clusters: the partition
$\mathcal{P}_d$ puts in the same cluster all the nodes $i$ such that
$i \equiv j \mod d$, that is to say, if $i$ and $j$ give the same
remainder when divided by $d$.

For instance, let $k=12$; one of the divisors of $12$ is $d=2$. The
equitable partition $\mathcal{P}_2$ contains exactly two clusters: one
includes the even-numbered nodes $\{0,2,4,6,8,10\}$, and the other
contains the odd-numbered nodes $\{1,3,5,7,9,11\}$. For each node of
each of the two clusters, the number of incoming edges from the same
cluster is zero, and the number of incoming edges from the other
cluster is one, so this is indeed an equitable partition.

If we look at the nodes in the order of the cycle, we see that
$\mathcal{P}_2$ alternates between two colors.  There is one such
partition for every divisor of $12$, so we have 6 equitable partitions
(one for each divisor of $12$, that is, $1$, $2$, $3$, $4$, $6$ and
$12$). The equitable partitions, represented as colorings, are shown
in Fig.~\ref{fig:cycle12}.

\begin{figure}[h!]
  \begin{tikzpicture}[main/.style = {draw, circle}, scale=.5, font=\fontsize{5}{5}\selectfont,baseline]

\colorlet{color min rgb}[rgb]{blue!80}
\colorlet{color max rgb}[rgb]{yellow!80}

\def \n {12}
\def \radius {3cm}
\def \margin {8} 

\foreach \d in {1,2,3,4,6,12}
{
	{
	\ifnum \d = 1
		\tikzmath {
			\sx = 0;
			\sy = 4;
		}
	\else \ifnum \d = 2
		\tikzmath {
			\sx = 4;
			\sy = 4;
		}
	\else \ifnum \d = 3 
		\tikzmath {
			\sx = 8;
			\sy = 4;
		}
	\else \ifnum \d = 4 
		\tikzmath {
			\sx = 0;
			\sy = 0;
		}
	\else \ifnum \d = 6
		\tikzmath {
			\sx = 4;
			\sy = 0;
		}
	\else 
		\tikzmath {
			\sx = 8;
			\sy = 0;
		}
	\fi\fi\fi\fi\fi
	\begin{scope}[shift={(2*\sx,2*\sy)}]
		\node at (-\radius,\radius) {$\mathcal{P}_{\d}$};
		\foreach \s in {1,...,\n}
		{
			\tikzmath{
				int \v, \modlcass;
				\v = \s-1;
				\modclass = int(\v-\d*int(\v/\d));
			}
			\pgfmathsetmacro\myvalue{\modclass};
			\pgfmathtruncatemacro\lambda{\myvalue/\d*100};
			\colorlet{my color rgb}[rgb]{color min rgb!\lambda!color max rgb};
			\node[draw, circle, fill=my color rgb] at ({360/\n * (\s - 1)}:\radius) {\v};
			\draw[->, >=latex] ({360/\n * (\s - 1)+\margin}:\radius) arc ({360/\n * (\s - 1)+\margin}:{360/\n * (\s)-\margin}:\radius);
			
		}
	\end{scope}
	}
}
\end{tikzpicture}
  \caption{\textbf{The equitable partitions of $C_{12}$}.  The partition
    $\mathcal{P}_d$ puts in the same cluster two nodes $i$ and $j$
    such that $i \equiv j \mod d$. For instance, $\mathcal{P}_2$ has
    two clusters: one for the odd-numbered nodes and one for the
    even-numbered nodes.}
\label{fig:cycle12}
\commentAlt{Figure~\ref{fig:cycle12}: 
Six replicas of the graph of Fig.~\ref{fig:cycle12-uncolored}, labeled P1, P2, P3, P4, P6, P12.
The colors repeat cyclically every 1, 2, 3, 4, 6, 12 nodes respectively.
So for instance: all nodes in P1 have the same color; the nodes in P2 have two alternate colors; the nodes in P3 have a pattern of three different consecutive colors repeating; etc.
}
\end{figure}

\begin{figure}[h!]
  \begin{tikzpicture}[main/.style = {draw, circle}, scale=1.2, font=\fontsize{6}{6}\selectfont,baseline]
	\node at (2,0) (p1) {$\mathcal{P}_{1}$};
	\node at (0,1) (p2) {$\mathcal{P}_{2}$};
	\node at (3,1) (p3) {$\mathcal{P}_{3}$};
	\node at (2,2) (p6) {$\mathcal{P}_{6}$};
	\node at (0,2) (p4) {$\mathcal{P}_{4}$};
	\node at (1,3) (p12) {$\mathcal{P}_{12}$};
	\draw[<-] (p12) -- (p4);
	\draw[<-] (p12) -- (p6);
	\draw[<-] (p4) -- (p2);
	\draw[<-] (p6) -- (p2);
	\draw[<-] (p6) -- (p3);
	\draw[<-] (p2) -- (p1);
	\draw[<-] (p3) -- (p1);
\end{tikzpicture}
  \caption{\textbf{The equitable partitions of $C_{12}$, ordered by
    refinement.} A partition $\mathcal{P}$ is finer than $\mathcal{P}'$
    if there is a directed path of arrows from $\mathcal{P}'$ to
    $\mathcal{P}$. For instance $\mathcal{P}_1$ is coarser than all
    the other partitions; $\mathcal{P}_2$ is coarser than
    $\mathcal{P}_4$, $\mathcal{P}_6$ and $\mathcal{P}_{12}$, finer
    than $\mathcal{P}_{1}$ but incomparable with (i.e., neither coarser nor finer than)  $\mathcal{P}_3$.  }
  \label{fig:cycle12-hasse}
\commentAlt{Figure~\ref{fig:cycle12-hasse}: 
A diagram with the elements P1, P2, P3, P4, P6, P12 drawn by levels (P12 at the top, P4 and P6 on the second level, P2 and P3 on the third level, P1 at the bottom)
and arrows P1 to P2, P1 to P3, P2 to P6, P3 to P6, P2 to P4, P4 to P12, P6 to P12.
}
\end{figure}

It is possible to check that these are the \emph{only} possible
equitable partitions of $C_{12}$, and they are all orbital. (See for example \cite[Chapter 26]{GS2023}, which in particular
classifies all equitable partitions for uni- and bi-directional rings with
nearest-neighbor connections.) If we look at them from the
viewpoint of refinement, we obtain the scenario depicted in
Fig.~\ref{fig:cycle12-hasse}, called the \emph{Hasse diagram} or {\em lattice} of equitable partitions under refinement. The elements shown in the diagram
are the equitable partitions of $C_{12}$, and they are
represented in such a way that it is possible to understand which
relations are finer and which are coarser. Finer relations appear
above, coarser relations below, and arrows determine which
relations are refinements of which. As expected, there is a
\emph{finest} equitable partition $\mathcal{P}_{12}$: this is 
the finest possible partition, the singleton partition that puts all
nodes into different clusters. But there is also a
coarsest equitable partition: in this example, it is the partition
$\mathcal{P}_1$ which is actually the trivial partition, the coarsest
of all.

\begin{figure}[t!]
  \begin{tikzpicture}[main/.style = {draw, circle}, scale=.6, font=\fontsize{6}{6}\selectfont,baseline]

\colorlet{color min rgb}[rgb]{blue!80}
\colorlet{color max rgb}[rgb]{yellow!80}

\def \n {12}
\def \radius {3cm}
\def \margin {8} 

\foreach \d [count=\i] in {1,3,6} 
{
	{
	\begin{scope}[shift={(0,8*\i)}]
		\node at (-\radius,\radius) {$\mathcal{P}_{\d}$};
		\foreach \s in {1,...,\n}
		{
			\tikzmath{
				int \v, \modlcass;
				\v = \s-1;
				\modclass = int(\v-\d*int(\v/\d));
			}
			\pgfmathsetmacro\myvalue{\modclass};
			\pgfmathtruncatemacro\lambda{\myvalue/\d*100};
			\colorlet{my color rgb}[rgb]{color min rgb!\lambda!color max rgb};
			\node[draw, circle, fill=my color rgb] at ({360/\n * (\s - 1)}:\radius) {\v};
			\draw[->, >=latex] ({360/\n * (\s - 1)+\margin}:\radius) arc ({360/\n * (\s - 1)+\margin}:{360/\n * (\s)-\margin}:\radius);
			
		}
		\draw[->,>=latex] (\radius+1cm,0) -- (\radius+2cm,0) node[midway, above] {$\varphi_{\d}$};
		\begin{scope}[shift={(3*\radius,0)}]
			\foreach \s in {1,...,\d}
			{
				\tikzmath{
					int \v, \modlcass;
					real \rradius;
					\v = \s-1;
					\modclass = \v;
					\rradius = \radius;
				}
				\pgfmathsetmacro\myvalue{\modclass};
				\pgfmathtruncatemacro\lambda{\myvalue/\d*100};
				\colorlet{my color rgb}[rgb]{color min rgb!\lambda!color max rgb};
				\node[draw, circle, fill=my color rgb] at ({360/\d * (\s - 1)}:\radius) {\v};
				\draw[->, >=latex] ({360/\d * (\s - 1)+\margin}:\radius) arc ({360/\d * (\s - 1)+\margin}:{360/\d * (\s)-\margin}:\radius);
			}
		\end{scope}
	\end{scope}
	}
}
\end{tikzpicture}
  \caption{\textbf{Fibrations of $C_{12}$.} Some of the equitable partitions of $C_{12}$, shown as
    fibers of surjective fibrations over some base graph (on the
    right). The fibration acts on the nodes as suggested by colors,
    whereas the fibration on edges is the only possible one.}
  \label{fig:cycle12-fib}
\commentAlt{Figure~\ref{fig:cycle12-fib}: 
On the left the colored graphs that were labeled as P6, P3, P1 in Fig.~\ref{fig:cycle12} are redrawn.
On the right there are three graphs at the same level of P6, P3, P1; between them there is an arrow with text phi6, phi3, phi1, respectively.
The graph on the right of P6 has 6 nodes named 0-5, closed in a cycle 0 to 1, 1 to 2, ..., 4 to 5, 5 to 0 and with the same pattern of colors as the graph on its left.
The graph on the right of P3 has 3 nodes named 0-2, closed in a cycle 0 to 1, 1 to 2, 2 to 0 and with the same pattern of colors as the graph on its left.
The graph on the right of P1 has 1 node named 0, closed in a cycle 0 to 0 and with the color as all the nodes in the graph on its left.
}
\end{figure}

In Fig.~\ref{fig:cycle12-fib} we look at some of the equitable
partitions of $C_{12}$ as fibrations over some graph. All the base
graphs, in this case, are themselves cycles. More precisely, the
equitable partition $\mathcal{P}_d$ is the fiber of a fibration
$\varphi_d: C_{12} \to C_{d}$, which maps node $i$ to node $i \mod d$.

Both \citep{kamei2013} and \citep{aguiar2014lattice} provide a general algorithm to
determine the full lattice of all equitable partitions, as we did for the example of
$C_{12}$. While they prove that the problem is intractable in general, their
algorithm can deal with networks of up to 15 nodes---even more if the synchrony patterns are large.

\section{How to extend fibration algorithms to take noise into consideration}
\label{algo-pseudo}

The presence of imperfections in real data is a major hurdle in biological
network analysis: not all the systems in biology present the perfect
symmetries that we have discussed so far.
Indeed, biological data are inherently noisy and disordered, and it
would be naive and unrealistic if we were to expect that genomes and
connectomes would give us the perfect symmetries discussed
theoretically. Analogously, biological systems will display  approximate cluster synchronization, as opposed to exact cluster synchronization, something that was first pointed out in \citep{sorrentino2016approximate}.

\subsection{Pseudosymmetries}
\label{sec:pseudosymmetry}

This observation motivates the development of a theory of approximate symmetries.\index{symmetry !approximate }  
The general idea is to take a notion of symmetry (e.g., automorphism symmetry or fibration symmetry) and to relax the conditions it imposes to allow for a limited number of \emph{exceptions}. 
A first step in this direction was proposed in~\citep{morone2019symmetry}. 
with the concept of an $\eps$-pseudosymmetry;\index{pseudosymmetry } they also observe that
any $\eps$-pseudosymmetry can be converted into a genuine symmetry
of a `repaired' graph.

We include this development for completeness; however, we do not recommend it as the best framework for dealing with noisy environments. First, group symmetries are not particularly useful in biology. Second, fibrations are more general and encompass group symmetries as well, making the theory of quasi-fibrations and pseudo-balanced coloring more applicable. These topics will be discussed in the following two sections. 

Even the pseudo-balanced colorings addressed in Section \ref{sec:repair} and the associated repair algorithm for restoring symmetry may not be the most effective approach. This limitation arises because the algorithm does not account for the most common scenario, which involves assisting network repair with some form of dynamical data, such as information about cluster synchronization within the system. This more useful case will be explored in Chapter \ref{chap:function}, along with its application to brain network inference in subsequent chapters.

Recall that a permutation matrix is a matrix of zeros and ones containing exactly one 1 in every row and column. 
If $A$ is the adjacency matrix of a graph $G$, then an automorphism of $G$ 
is a permutation matrix $P$ such that $PAP^{-1}=A$. (The matrix $PAP^{-1}$ permutes both the rows and columns of $A$ according to $P$, so this
equation states that the permuted adjacency matrix is equal to the original one. This is the property required for an automorphism.)

Equivalently $PA=AP$, which we rewrite as
\begin{equation}
\label{eqn:autperm}
    [A,P]={\mathbf 0}
\end{equation}
where $[A,P]=AP-PA$ and $\mathbf 0$ is the matrix whose entries are all zeros. 

A natural extension to allow for errors is the following. Define the
{\em $L^1$ norm}\index{L1 norm @{\em $L^1$ norm} } of a matrix $M=(m_{ij})$ to be
\[
\|M\| = \sum_{i,j} |m_{ij}|
\]
This norm has several useful properties.

\noindent
(a)\quad
If $M$ has integer entries, as is the case for an adjacency matrix,\index{adjacency matrix } then $\|M\|$ is a nonnegative integer. 

\noindent
(b)\quad
If the entries of $M$
are always 0 or 1, as is also the case for an adjacency matrix,
then $\|M\|$ is the number of nonzero entries of $M$.

\noindent
(c)\quad
Therefore if $M,N$ are two matrices, whose entries are always 0 or 1,
the quantity $\|M-N\|$ is equal to the number of entries at
which $M$ and $N$ differ.

\noindent
(d)\quad
If $Q, R$ are permutation matrices\index{permutation !matrix } then $\|QMR\|=\|M\|$. This
holds because $M$ and $QMR$ have the same set of entries: the permutation matrices only permute their positions.

Fix $A$ and suppose that $Q$ is any permutation matrix. If $Q$ is an automorphism
then, by (\ref{eqn:autperm}), $[Q,A]={\mathbf 0}$, so $\|[A,Q]\|=0$. We replace this by
the (milder) condition 
\begin{equation}
\label{E:pseudo-symm2}
 \|[A,Q]\| \leq \eps.
\end{equation}
By property (d) above this is equivalent to either of
\begin{equation}
\label{E:pseudo-symm1}
\|AQ-QA\|\leq \eps \qquad \|A-QAQ^{-1}\|\leq \eps.
\end{equation}
The case of interest is when $\eps > 0$ is small compared to the number of entries in the matrix, but initially we consider any $\eps > 0$. 

\begin{definition}
\label{D:pseudosymm}
The matrix $Q$ is an $\eps$-{\em pseudosymmetry}\index{eps- pseudosymmetry @$\eps$-pseudosymmetry }\index{pseudosymmetry } of $A$ if
\eqref{E:pseudo-symm2} holds, or equivalently \eqref{E:pseudo-symm1} holds.
\end{definition}

For a given $\varepsilon\geq 0$, consider the set 
\[
    {\mathcal P}_\varepsilon=\{\,Q\mid  \|[A,Q]\|\leq \varepsilon\,\}.
\]
This is the set of all permutation matrices $Q$ such that $[A,Q]$ has a 
small $L^1$ norm  rather than being zero. These permutations are the pseudosymmetries of $G$. 

`Small' here means `small relative to the size of the network',
because an integer-valued norm 
is never `small' in any intuitive sense---unless it is zero. 
Here, the smallest nonzero value of $\eps$ is 1. However, if
we measure the network size by the number $n$ of nodes, say,
then $\eps/n$ is relatively small, and we get a
good approximation when $\eps/n$ is small enough.
Alternatively, we might divide $\eps$ by $n^2$, the number
of matrix entries, or by the number of edges,
or by some other measure of network complexity. 

To understand better what pseudosymmetries are, consider a fixed
but arbitrary pseudosymmetry $Q \in {\mathcal P}_\varepsilon$. It need not represent an automorphism of $G$, 
but by (c) its norm is precisely the number of entries at which
$A$ and $QAQ^{-1}$ are different. Then one of them has a 0 entry where the other has a 1 entry. We can `repair' the graph by making all such
entries of $A$ equal to 1. (Alternatively, to 0.) Let $A_\varepsilon$ be the resulting adjacency matrix, bearing in mind that this matrix
depends on $Q$ as well as on $\eps$.

Then $[A_\varepsilon,Q] = 0$, so $Q$ is an automorphism of the repaired graph with
adjacency matrix $A_\varepsilon$, obtained by adding or removing some edges. 
Call this graph $G_\eps$, again bearing in mind that this also
depends on $Q$. Now:
\[
    \varepsilon \geq \|[A,Q]\| =
                \|AQ-Q A\| 
                = \|A-Q A Q^{-1}\|,
\]
where in the last equality, we use property (d). Adding and subtracting $A_\varepsilon$ we obtain
\[
    \varepsilon \geq \|A-A_\varepsilon+A_\varepsilon-Q A Q^{-1}\| 
                = \|A-A_\varepsilon+Q A_\varepsilon Q^{-1}-Q A Q^{-1}\|.
\]
In the last equality we used the fact that $[A_\varepsilon, Q]={\mathbf 0}$, that is, $A_\varepsilon Q=Q A_\varepsilon$, hence $A_\varepsilon=Q A_\varepsilon Q^{-1}$.
Hence
\[
    \varepsilon \geq \|(A-A_\varepsilon)-Q (A-A_\varepsilon)Q^{-1}\| \geq \|A-A_\varepsilon\|+\|Q (A-A_\varepsilon)Q^{-1}\| = 2\|A-A_\varepsilon\|.
\]
The latter term ($\|A-A_\varepsilon\|$) is exactly the number of edges that have been added or removed from $G$, and the inequality shows how this number is related to $\varepsilon$. 

In other words, for a fixed value of $\varepsilon$ we have an upper bound on the number of edges that can be changed to convert a pseudosymmetry\index{pseudosymmetry } $Q$ of $G$
into a symmetry $Q$ of the repaired graph $G_\eps$.

\begin{example}\em
\label{ex:pseudo_6node}
Let $G$ be as in Fig. \ref{fig:pseudo_6node} (top).
The adjacency matrix is
\[
A = \Matrix{0&1&0&0&0&0\\0&0&0&0&0&0\\0&0&0&1&0&0\\0&0&0&0&1&0\\0&0&0&0&0&1\\0&1&0&0&0&0}
\]
This graph is obtained by changing one connection in
a unidirectional ring with six nodes and nearest-neighbor arrows.
It can also be seen as a unidirectional ring with five nodes 2 -- 6, plus
an extra input node 1. 

\begin{figure}[h!]
\centerline{%
\includegraphics[width=0.4\textwidth]{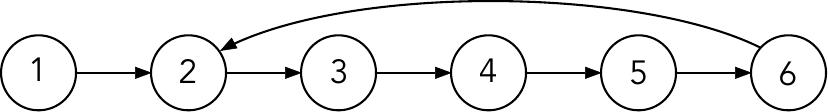}}
\vspace{20pt}
\centerline{%
\includegraphics[width=0.4\textwidth]{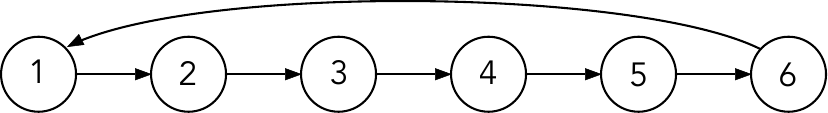} \qquad
\includegraphics[width=0.4\textwidth]{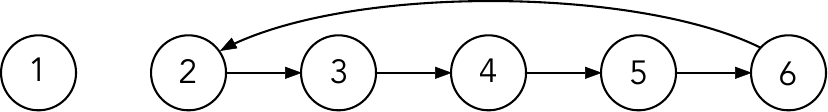}}
\caption{\textbf{Pseudosymmetries.} {\em Top}: Graph with 6 nodes, exemplifying pseudosymmetries.
{\em Bottom}: Two potential repairs. {\em Left}: Symmetry group
$\Z_6.$ {\em Right}: Symmetry group
$\Z_5.$}
\label{fig:pseudo_6node}
\commentAlt{Figure~\ref{fig:pseudo_6node}: 
Three directed graphs with six nodes each, named 1-6.
The first graph (Top) has directed connections: 1 to 2, 2 to 3, ..., 5 to 6, 6 to 2.
The second graph (Left) has directed connections: 1 to 2, 2 to 3, ..., 5 to 6, 6 to 1.
The third graph (Right) has directed connections: 2 to 3, 3 to 4, 4 to 5, 5 to 6, 6 to 2 (node 1 is not connected to anything).
}
\end{figure}

Let $Q$ be the matrix of the 6-cycle $(123456)$, so that
\[
Q = \Matrix{0&0&0&0&0&1\\1&0&0&0&0&0\\0&1&0&0&0&0\\0&0&1&0&0&0\\0&0&0&1&0&0\\0&0&0&0&1&0}
\]
Then
\[
[A,Q] = \Matrix{1&-1&0&0&0&0\\0&0&0&0&0&0\\0&0&0&0&0&0\\0&0&0&0&0&0\\0&0&0&0&0&0\\1&0&0&0&0&-1}
\]
and $\|[A,Q\|=4$. Therefore $Q$ is an $\eps$-pseudosymmetry with $\eps=4$.
The maximum value that the norm can take is 36, so this gives a proportion
$1/9$ of differing entries.

Alternatively, we might try the 5-cycle $(23456)$. Now
\[
Q = \Matrix{0&0&0&0&0&1\\1&0&0&0&0&0\\0&1&0&0&0&0\\0&0&1&0&0&0\\0&0&0&1&0&0\\0&0&0&0&1&0}
\]
so that
\[
[A,Q] = \Matrix{0&-1&0&0&0&1\\0&-1&1&0&0&0\\0&0&0&0&0&0\\0&0&0&0&0&0\\0&0&0&0&0&0\\0&-1&0&0&0&1}
\]
and $\|[A,Q\|=6$. Therefore $Q$ is an $\eps$-pseudosymmetry with $\eps=6$,
which is a bigger error than 4.

In the first case the natural repaired network is a unidirectional
ring on all six nodes, Fig. \ref{fig:pseudo_6node} (bottom left). In the second case, it is a unidirectional
ring on nodes 2 -- 6, with 1 as an isolated node, Fig. \ref{fig:pseudo_6node} (bottom right).
\end{example}

\subsection{Quasifibrations}
\label{sec:quasifibrations}

A related issue is that of statistical noise in observations.
When moving to the world of fibrations,\index{fibration } we can approach the problem of considering noise in a number of ways. One possibility, described in~\citep{leifer2022symmetry-driven}, is to define a relaxed notion of equitable partition that explicitly takes statistical noise into account. In terms of the adjacency matrix $A$ of a graph $G$, a node partition ${\mathcal S}$ is equitable if and only if, for all clusters $C,D$ of ${\mathcal S}$ and for all pairs of nodes $p,q \in C$, the number of edges entering in $p$ and $q$ and coming from $D$ are the same, that is
\[
    \sum_{j \in D, (j,p) \in E} A_{jp} = \sum_{j \in D, (j,q) \in E} A_{jq}.
\]
If we let $\Phi$ denote the perturbation matrix\index{perturbation !matrix } representing the noise, we can require that
\[
    \sum_{j \in D, (j,p) \in E} \left(A_{jp}+\Phi_{jp}\right) = 
    \sum_{j \in D, (j,q) \in E} \left(A_{jq}+\Phi_{jq}\right).
\]
In this case ${\mathcal S}$ is called \emph{quasi-equitable},\index{quasi-equitable } up to the noise matrix\index{noise matrix } $\Phi$. Besides $\Phi$ (which contains only information on how \emph{existing} edges should be modified), the full formulation also allows  extra edges to be added, which requires a second noise matrix $\Omega$: this case is presented in the next subsection.

Here there are two factors to play with: the number of clusters $|{\mathcal S}|$ 
(in particular, the number of clusters of the coarsest equitable partition) and the norm of the perturbation matrix $\|\Phi\|$. Of course, we expect these two values to be negatively correlated: if we allow for more noise, we can get a more symmetrical structure, reducing the number of clusters.  \cite{leifer2022symmetry-driven} prove that for a fixed value of $|{\mathcal S}|$, minimizing $\|\Phi\|$ is NP-hard. However, they formulate an integer programming optimization which can be used as a heuristic to find an approximate solution for the coarsest quasi-equitable partition.

An alternative approach to noisy fibration symmetries is described in~\citep{boldi2021}.
Recall that a fibration $\varphi: G \to B$ is a homomorphism
that satisfies the lifting property: in other words, for every edge $a$
of $B$ and every node $x$ of $G$ in the fiber of the target of $a$,
there must be \emph{exactly one} edge of $G$ with target $x$ that
$\varphi$ maps to $a$.  In a way, though, any homomorphism $f: G \to B$ can be seen as
a `fibration with errors'. Errors can be of two types:
\emph{deficiencies} (i.e., edges that have \emph{no} lift at a
certain target) and \emph{excesses} (i.e., edges that have \emph{more
  than one} lift at a certain target). The overall number of
deficiencies and excesses can be used as a measure of how much a
given homomorphism deviates from being a fibration. Such a homomorphism is called a {\em quasifibration}.\index{quasifibration }

Suppose, for a moment, that we know what an equitable partition
${\mathcal S}$ of $G$ looks like. That is, that we know how the nodes
of $G$ can be divided to obtain an equitable partition. Then we
can consider a homomorphisms $f: G \to B$, whose fibers are
exactly the clusters of $\mathcal{P}$, to be a fibration with errors. The
question is whether we can find, among all these homomorphisms, one that
is as close as possible to being a fibration, i.e., one that is a
quasifibration\index{quasifibration } with the smallest number of deficiencies and
excesses. This problem turns out to be polynomial-time
solvable in an exact way \citep{boldi2021}.

This result is only partially satisfactory, for at least two
reasons. The first is that we need to have some idea of an equitable
partition to start with; this equitable partition can be obtained
using various heuristics, but it is certainly a limitation that must
somehow be overcome (in a heuristic way, necessarily, for otherwise the optimization problem in~\citep{leifer2022symmetry-driven} would not be NP-hard). The second limitation is that the algorithm makes a
very specific assumption on the metric that measures the
distance from the space of fibrations; using other metrics can be
difficult (and, in particular, it is proved in~\citep{boldi2021} that the general problem for arbitrary metrics is NP-hard).

\subsection{Pseudo-balanced colorings and repair algorithm without knowledge of balanced coloring}\index{pseudo-balanced coloring }
\label{sec:repair}

\cite{leifer2022symmetry-driven} explore an alternative approach to finding fibration symmetry of a graph in the presence of noise.\index{noise } 

The algorithm reconstructs a network by optimally selecting the minimal number of modifications needed to achieve a fiber-symmetric network. In this algorithm, the balanced coloring is not known {\it a priori}, and we do not know the number of colors either. Therefore, the algorithm is initially run using a fixed number of colors, and the network is reconstructed accordingly.

Afterward, the algorithm can be executed with different numbers of colors, and the best solution should be chosen based on heuristic metrics, as discussed in \citep{leifer2022symmetry-driven}. A simpler and more effective algorithm is developed in Chapter \ref{chap:function}, which assumes that we know the balanced coloring from synchronization experiments.

We do not consider unweighted multigraphs, but rather weighted simple graphs.\index{graph! weighted } The difference is that between each source/target pair there can be only one edge, but that edge is weighted. Although the definition of fibrations is not well suited for weighted graphs (as discussed in Section~\ref{sec:weighted}), balanced colorings can easily be extended to this case. Instead of requiring that each pair of nodes with the same color $c$ receive the same number of edges from each color class $c'$, we require that the \emph{sum of the weights} of edges coming from $c'$ should be the same.
Incidentally, if all weights are integers and the weighted simple graph is interpreted as an unweighted multigraph (with weights corresponding to multiplicities), then the two notions of balanced colorings coincide.

While the scope of \citep{leifer2022symmetry-driven} is broad, next we discuss the particular approach they develop to repair a given unweighted undirected graph $G(V,E)$ with adjacency matrix $A$, aiming to produce a pseudobalanced coloring\index{pseudo-balanced coloring } with $k$ colors. This is relevant to all situations where imperfect knowledge of a given graph may be affected by missing links. The underlying assumption is that the original graph $G$ may lack some of the true fibration symmetries and, as a result, produce balanced colorings with an exceedingly large number of trivial colors.

\cite{leifer2022symmetry-driven} introduce the perturbation matrix $\Omega=\{\omega_{ij} \in \{0,1\} \}$. Then they formulate the pseudo-balanced coloring integer program\index{pseudo-balanced coloring !integer program } (PBCIP)\index{PBCIP } in the quantities $\omega_{ij}$, which minimizes the total number of added links:
\begin{equation}
\min \Biggl\{ \sum_{i,j \in V, ij \notin E} \omega_{ij} \Biggr\},
\end{equation}
This is done in a manner that makes the repaired graph $G'$, with adjacency matrix $A'=A+\Omega$, have a balanced coloring with $k$ colors. This allows only adding links
that are not already present, not removing them.
Solving this PBCIP finds the minimal number of edges to add to a given graph to ensure a balanced coloring for a fixed number of colors. The overall repair method presented in \citep{leifer2022symmetry-driven}  proceeds according to the following steps:

\begin{enumerate}
    \item Pre-processing: compute the minimal balanced coloring for the original graph $G$ and identify the initial non-trivial set of colors. Let $C$ denote the number of non-trivial colors and $T$ the number of trivial colors in this initial coloring.
    \item  for all $k$ from $C$ to $C+T$ do the following:
    \begin{enumerate}
\item[a] Solve the (PBCIC) for $k$ colors, with added constraints to fix the nodes that were already assigned to one of the $C$ non-trivial colors.
\item[b] Construct the resulting graph $G'$ with the added edges dictated by (PBCIP)\index{PBCIP } and color the nodes according to the minimal balanced coloring of $G'$.
\end{enumerate}
    \item Evaluate the resulting $T$ graphs to identify the number of colors of the \textit{best} repaired graph. In general this involves both a quantitative and a qualitative assessment, which may be based on available information about the graph structure and functions.  
\end{enumerate}

Then~\cite{leifer2022symmetry-driven} proceed by considering an optimization problem that looks for some balanced coloring with $K$ colors in a given weighted graph $G$, which allows for the presence of noise in the following sense:
\begin{itemize}
    \item We can introduce some (positive or negative) noise into the weights of existing edges $\Phi=(\phi_{ij})_{ij \in E}$.
    \item We can add some new edges $E'$ with some weights on them: $\Omega=(\omega_{ij})_{ij \in E'}$.
\end{itemize}
The noise\index{noise } on existing edges and the weights of new edges can be restricted to a certain class $\mathcal R$, i.e., $(\Phi, \Omega) \in \mathcal R$. Also, the search space (the set of allowed colorings) is restricted by providing a coarser coloring $\mathcal F$ (in the sense that we are not allowed to mix color classes that are distinct in $\mathcal F$).

This framework yields an optimization problem\index{optimization } whose objective function\index{objective function } is the overall noise
\[
    \sum_{ij \in E} \left|\phi_{ij}\right| + \sum_{ij \in E'} \left|\omega_{ij}\right|,
\]
and whose constraints are the graph $G$, the number of colors $K$, the type of noise we can introduce $\mathcal R$, and the constraint $\mathcal F$ on colors.
While the most general version of this optimization problem is known to be computationally hard, the authors discuss an integer formulation that is at the same time amenable to polynomial-time approximation and, in practice, provides very competitive results.

\section{How to extend fibration algorithms to take types into consideration}
\label{sec:alghet}

As discussed in Section~\ref{sec:heterogeneous}, our algorithms should be extended to take heterogeneity into consideration. 
We limit ourselves to showing in Algorithm~\ref{algo:naive-heterogeneous} how the naive color refinement algorithm (Algorithm~\ref{algo:naive}) is adapted to the heterogeneous case.
Essentially:
\begin{itemize}
    \item Every time a color is assigned to some node $x$, both during initialization (line 1) and during iterations (line 8), the color takes the node type $\nu(x)$ into account (so that nodes of different types are always assigned different colors);
    \item Every time we look at the input set of a node $x$ (line 5) we consider the colors of edges connecting $x$ to its in-neighbors.
\end{itemize}

\begin{algorithm}[H]
  \begin{flushleft}
    \hspace*{\algorithmicindent} \textbf{Input:} A heterogeneous graph $G=(V,E)$, with $\nu: V \to T$ and $\eta: E \to T$ being the maps specifying the types of nodes and edges\\
    \hspace*{\algorithmicindent} \textbf{Output:} A coloring of the nodes representing the coarsest equitable partition of $G$
  \end{flushleft}
  \begin{algorithmic}[1]
    \State Let $\kappa_0$ be a coloring such that $\kappa_0(x)=\nu(x)$ for all $x \in V$
    \State Let $c_0=|\nu^{-1}(V)|$  (the number of colors used by $\kappa_0$)
    \State Let $i=0$ (the number of iterations so far)  
    \While{true}
      \State For each node $x$ of $V$ consider the multiset $M_{i+1}(x)=\{\!\!\{\langle \eta(e),\kappa_i(s(e))\rangle \mid e \in E, t(e)=x\}\!\!\}$ 
      \State Let $c_{i+1}$ be the number of distinct pairs $\langle \nu(x),M_{i+1}(x)\rangle$ (for $x \in V$)
      \State Arbitrarily assign a number $\upsilon(\langle t,A\rangle)$ to each of the $c_{i+1}$ pairs $\langle t,A\rangle$
      \State Let $\kappa_{i+1}$ be a coloring such that $\kappa_{i+1}(x)=\upsilon(\langle \nu(x),M_{i+1}(x)\rangle)$ for all $x \in V$
      \State Stop if $c_{i+1}=c_i$ (the number of colors did not change in the last iteration)
      \State $i=i+1$
    \EndWhile
    \State Output $\kappa_i$
  \end{algorithmic}
  \caption{Color refinement algorithm (Algorithm~\ref{algo:naive}) modified for heterogeneous graphs.}
  \label{algo:naive-heterogeneous}
\end{algorithm}

We exemplify the execution of Algorithm~\ref{algo:naive-heterogeneous} on the graph of Fig.~\ref{fig:many-noinputs}.
Let us first try to find its coarsest equitable partition \emph{in absence of typing}, for instance, using color refinement (Algorithm~\ref{algo:naive}): this algorithm starts with all nodes having the same color, and then refines the coloring based on the color of in-neighborhoods. In the example of Fig.~\ref{fig:many-noinputs}, at the beginning all nodes are assigned the same color (call it $\ell_0$), and at the first iteration of the algorithm 
\begin{itemize}
    \item nodes with no inputs ($1$, $2$ and $5$) all see the empty multiset
    $\{\!\!\{\}\!\!\}$ of colors 
    \item nodes $3$ and $4$ both see a multiset with three elements (because they have exactly three incoming edges): the multiset is 
    $\{\!\!\{\ell_0,\ell_0,\ell_0\}\!\!\}$
    \item finally nodes $6$ and $7$ both see the multiset $\{\!\!\{\ell_0,\ell_0\}\!\!\}$.
\end{itemize}
Let $\ell_1$ be the color assigned to the first multiset, $\ell_2$ the color assigned to the second multiset, and $\ell_3$ the color assigned to the last multiset. At the next iteration
\begin{itemize}
    \item nodes with no inputs ($1$, $2$ and $5$) all see the empty multiset
    $\{\!\!\{\}\!\!\}$ of colors 
    \item nodes $3$ and $4$ all see a multiset with three elements (because they have exactly three incoming edges): the multiset is 
    $\{\!\!\{\ell_1,\ell_1,\ell_2\}\!\!\}$
    \item finally nodes $6$ and $7$ sees the multiset is 
    $\{\!\!\{\ell_1,\ell_1\}\!\!\}$.
\end{itemize}
Since the number of colors did not change with respect to the previous iteration, the algorithm stops and the final color (corresponding to the coarsest equitable partition, i.e., to the fibration symmetry) is the one shown in Fig. \ref{fig:many-noinputs-bc}.


Now, let us add a typing on top of the graph, as discussed in Section~\ref{sec:treatment}, so as to take into consideration in a proper way the nodes that have no input. Explicitly, the type function $\nu(x)$ is defined by:
\[
    \nu(x)=\begin{cases}
            x & \text{if $x$ has no inputs}\\
            * & \text{otherwise.}
            \end{cases}
\]
The renumbering at each step (function $\upsilon$ in Algorithm~\ref{algo:naive-heterogeneous} is arbitrary; here we use lexicographic ordering of the pair $\langle \nu(x), M_i(x)\rangle$).
\begin{table}
    \centering
    \begin{tabular}{c|c|c|c|c|c|c}
        $x$ & $\kappa_0(x)$ & $\langle \nu(x), M_1(x) \rangle$ & $\kappa_1(x)$ & $\langle \nu(x), M_2(x) \rangle$ & $\kappa_2(x)$\\
        \hline
         1 & 1 & $\langle 1, \{\!\!\{\}\!\!\}\rangle$ & 3 & $\langle 1, \{\!\!\{\}\!\!\}\rangle$ & 3\\
         2 & 2 & $\langle 2, \{\!\!\{\}\!\!\}\rangle$ & 4 & $\langle 2, \{\!\!\{\}\!\!\}\rangle$ & 4\\
         3 & $*$ & $\langle *, \{\!\!\{1,2,*\}\!\!\}\rangle$ & 1 & $\langle *, \{\!\!\{2,3,4\}\!\!\}\rangle$ & 1\\
         4 & $*$ & $\langle *, \{\!\!\{2,5,*\}\!\!\}\rangle$ & 2 & $\langle *, \{\!\!\{1,4,5\}\!\!\}\rangle$ & 0\\
         5 & 5 & $\langle 5, \{\!\!\{\}\!\!\}\rangle$ & 5 & $\langle 5, \{\!\!\{\}\!\!\}\rangle$ & 5\\
         6 & $*$ & $\langle *, \{\!\!\{1,2\}\!\!\}\rangle$ & 0 & $\langle *, \{\!\!\{3,4\}\!\!\}\rangle$ & 2\\
         7 & $*$ & $\langle *, \{\!\!\{1,2\}\!\!\}\rangle$ & 0 & $\langle *, \{\!\!\{3,4\}\!\!\}\rangle$ & 2\\
        \hline
        & $c_0=4$ & & $c_1=6$ & & $c_2=6$
    \end{tabular}
    \label{tab:my_label}
\end{table}

The symmetry fibration obtained is shown in Fig.
\ref{fig:many-noinputs-bcfix-pre}.
Not only do all nodes with no inputs belong to different classes (this is trivial, because they do not have the same type), but also nodes $3$ and $4$ are different because they receive inputs from different nodes (albeit nodes with no input themselves).
Here the only nontrivial fiber is the one formed by nodes $6$ and $7$.


An implementation of Algorithm~\ref{algo:naive-heterogeneous} with the preprocessing required for nodes with no input is available at {\small\url{https://github.com/MakseLab/FibrationSymmetries}} in the form of an R package.

\section{Software used in this book to find minimal balanced colorings }
\label{sec:software}
\index{software implementation }

All fibration analysis done in this book has been performed using a version of the refinement algorithm developed by \cite{leifer2022thesis}, see also the supplementary information of \citep{morone2020fibration} and \citep{monteiroAlgorithm}. The particular implementation of the refinement algorithm explained in this section is available in R at {\small\url{https://github.com/MakseLab/FibrationSymmetries}}.
This algorithm takes into account nodes with no inputs from the very beginning, implicitly assigning them different types (like in Section~\ref{sec:alghet}). 
We explain this implementation below. The reader interested in reproducing all results presented in this book can download the codes and data to perform the fibration analysis from the repository of the book at {\small\url{https://github.com/MakseLab/}} and {\small\url{https://osf.io/4ern8/}}. Appendix \ref{sec:list-software} lists all source code and data used in this book.

The analysis identifies fibers by searching for the minimal balanced coloring, i.e., the network's coarsest equitable partition (minimal base). The most efficient algorithm is the one developed by \cite{paige1987three}, and has time complexity $O(M \log N)$ and space complexity $O(M+N)$. Implementation of this algorithm, however, is fairly complicated, and in the interest of giving a concise explanation, we present our implementation of the algorithm developed in \citep{leifer2022thesis} and \citep{morone2020fibration} based on \citep{kamei2013}. This algorithm has runtime complexity $O(N^2 \log N)$ and space complexity is $O(N^2)$.

\cite{leifer2022thesis} defines the \textit{Input Set Color Vector (ISCV)}.  
 Let $G = (H, \chi)$ be a colored graph with $k$ colors of $H = (N, E)$ and $\chi:N \rightarrow (\chi_1, \chi_2, \dots, \chi_k), k \in \mathbb{N}_{<|N|}$ be a function assigning each node from $N$ a color. Let $\mathcal{P} = \{c_1, c_2, \dots, c_k\}$ be the partition corresponding to $\chi$.
\begin{definition}
    The {\em ISCV (Input Set Color Vector)}\index{ISCV }\index{input set color vector } of node $n$ is a $k$-dimensional vector $ISCV(n) = (ISCV_1(n), ISCV_2(n), \dots ISCV_k(n))$ in which $ISCV_i(n) = | (N_{I(n)} \setminus \{n\}) \bigcap c_i|$.
\end{definition}
That is, $ISCV(n)$ is a $k$-dimensional vector with $i$-th entry equal to the number of nodes of $i$-th color in the input set of $n$ (excluding the node itself).

\begin{figure}[ht!]
    \centering
    \includegraphics[width=0.7\textwidth]{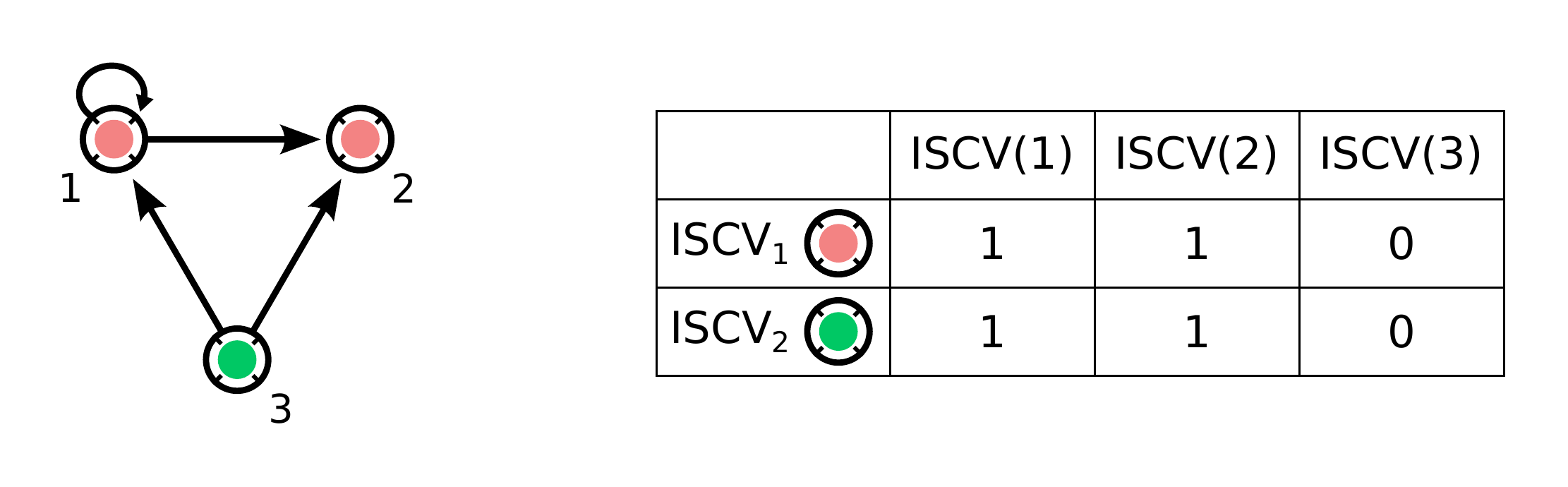} 
    \caption{\textbf{Input Set Color Vector}. An example of an ISCV is found on the graph nodes with minimal balanced coloring.}
    \label{Fig:ISCVExample}
\commentAlt{Figure~\ref{Fig:ISCVExample}: 
On the left, a directed graph with nodes named 1-3 (1 and 2 are red, 3 is green), and directed connections 1 to itself, 1 to 2, 3 to 1, 3 to 2.
On the right, a table with columns ISCV(1), ISCV(2), ISCV(3) and rows ISCV1(red), ISCV2(green).
The entries (top to bottom, left to right) are: 1, 1, 0, 1, 1, 0.
}
\end{figure}

A coloring is balanced and minimal when nodes of the same color have the same ISCVs and nodes of different colors have different ISCVs. Figure ~\ref{Fig:ISCVExample} shows an example of a minimal balanced coloring. The graph in Fig.~\ref{Fig:ISCVExample} has 2 colors. Therefore, the ISCV is a 2-dimensional vector. Node $1$ receives input from itself (red color) and from node $3$ (green color); therefore, $ISCV(1)=(1, 1)$. Similarly, $ISCV(2)=(1, 1)$ and $ISCV(3)=(0, 0)$.

The algorithm for balanced coloring is based on the iterative procedure of partition refinement. The minimal balanced coloring is unique and can be obtained by refining the unit partition until no further refinement is possible.

\begin{enumerate}
    \itemsep0em
    \item Assign all nodes the same color. Set $d = 1$.
    \item Find the ISCV of all nodes (In the first iteration, the ISCV is a 1-dimensional vector equal to the in-degree of the node).
    \item Let $m$ be the number of unique ISCVs.
    \item If $m$ is equal to $d$, stop.
    \item Assign each of the $m$ unique ISCVs a new color and recolor the graph accordingly. Set $d = m$.
    \item Repeat steps 2-5 until the condition in step 4 is satisfied.
\end{enumerate}

\begin{figure}[ht!]
    \centering
    \includegraphics[width=.8\textwidth]{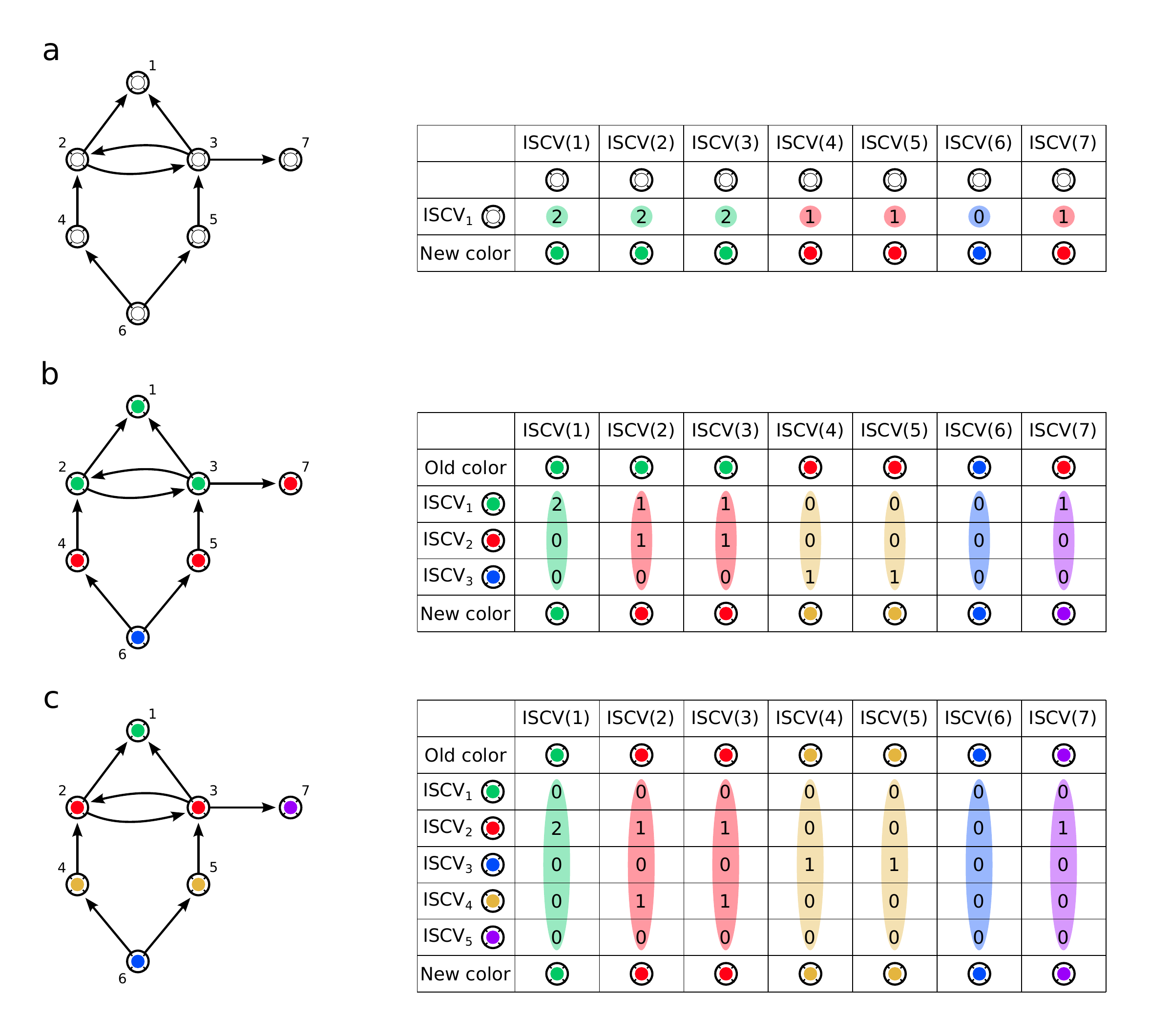} 
    \caption{\textbf{Example of the step-by-step execution of software to find the minimal balanced coloring}. (\textbf{a}) All nodes are assigned the color white. The ISCV corresponds to the in-degree of each node. Three unique ISCV values, $(0)$, $(1)$, and $(2)$, define three new colors: green, red, and blue. (\textbf{b}) A new coloring is defined by three unique ISCVs in (\textbf{a}). The ISCV is now a 3-dimensional vector. There are five unique ISCVs, so there are five new colors. (\textbf{c}) A new coloring is defined by five unique ISCVs in (\textbf{b}). The ISCV is now a 5-dimensional vector. There are five unique values of the ISCV, so the algorithm stops.}
    \label{Fig:AlgorithmExample}
\end{figure}

Figure ~\ref{Fig:AlgorithmExample} shows an example of the step-by-step execution of the software implementation. Both examples provided in this chapter involve unweighted networks, but the software can be applied to undirected and weighted networks as well.

As explained in the previous section, real applications in biology require the solution of a slightly modified problem: finding an almost minimal coloring.
This modified balanced coloring algorithm essentially assigns a different color to each node that has no inputs. 

The symmetry fibration obtained by this procedure is not minimal but rather `as minimal as possible' given that nodes without inputs belong to different fibers. \cite{morone2020fibration} offer a modification of the algorithm that obtains symmetry fibers and takes into account the biological fact that nodes without inputs are not synchronized, even though in the fibration theory, they would be put into the same fiber, since their input tree, which is a null tree, would be mathematically isomorphic. 
However, common sense indicates that such nodes should be put into different fibers since they do not receive the same inputs; thus, they will never synchronize. 

In the analysis of real biological networks, it is rare for a situation to arise where a gene receives no inputs. If this occurs, it typically indicates that the data set is incomplete. In a regulatory network, for example, all genes must receive inputs in order to be activated or repressed. If a gene has no input, it suggests that the network is not fully complete.
Regarding small molecules and metabolites, it is possible for these nodes to have no inputs when they are part of the external environment. However, in such cases, it is unrealistic to assume that all external molecules are synchronized, as the environment is constantly changing.

Thus, we modify the refinement algorithm to assume that nodes with no inputs are in different fibers, as discussed in Section \ref{sec:treatment}. \cite{leifer2022thesis} developed the implementation explained below, which takes this modification into account, producing a coloring that is not exactly minimal but differs from the minimal coloring by the number of colors assigned to the nodes with zero inputs.

 This modification is a heuristic that works for all networks considered by the authors in the analysis of all data included in this book. However, in some specific examples, it may not work. We first describe the heuristic, then give some examples that can be problematic and discuss alternative approaches. The heuristic is based on two modifications applied to the algorithm:
\begin{enumerate}
    \itemsep0em
    \item Identify all nodes with no inputs from other nodes (nodes that receive only from themselves) and assign each of them a different initial color.
    \item Identify all nodes with no inputs from any node (including themselves) and assign each of them its color during every iteration of the algorithm.
\end{enumerate}

\begin{figure}[ht!]
    \centering
    \includegraphics[width=\textwidth]{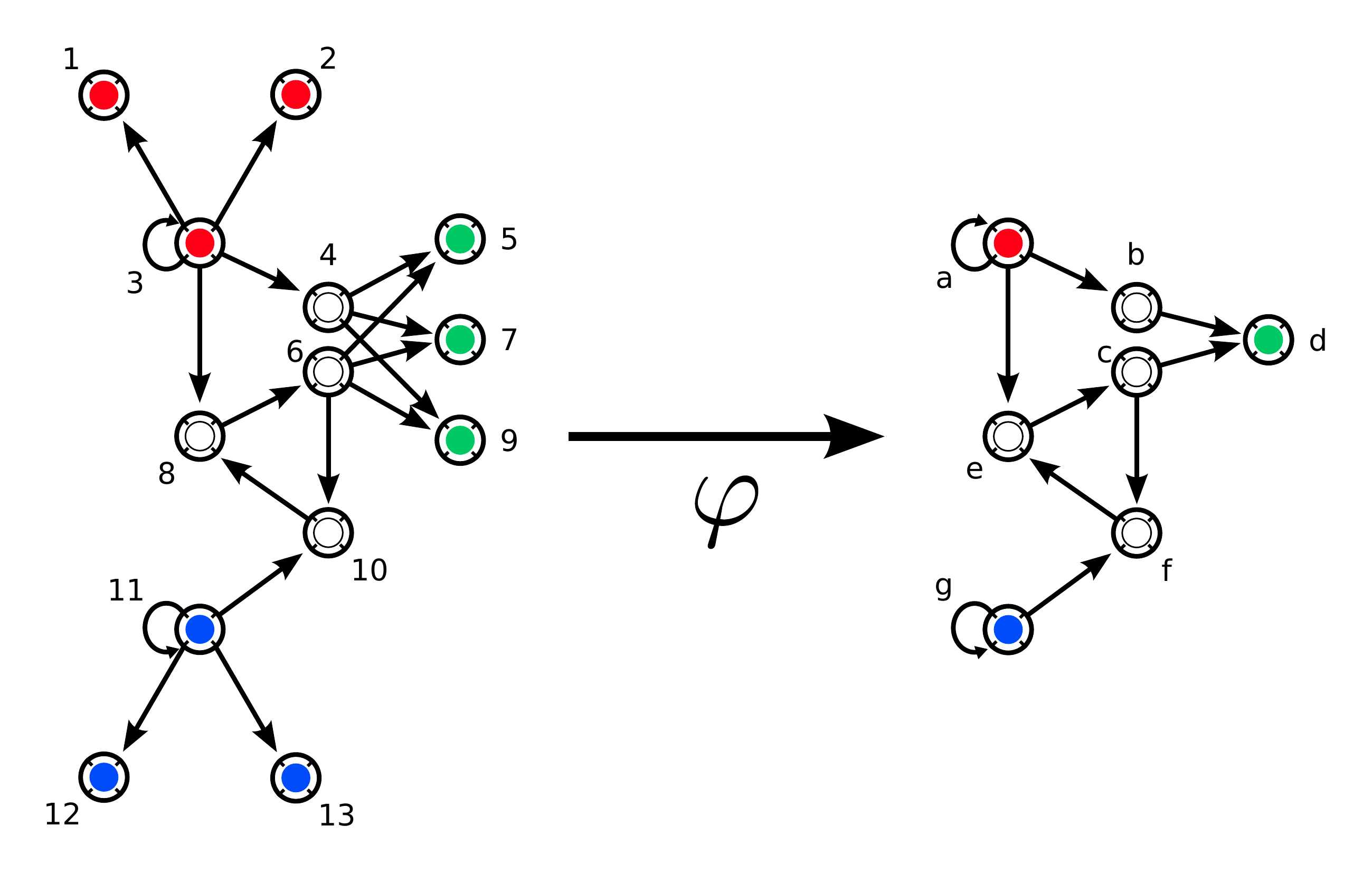} 
    \caption{\textbf{Special cases of fibers.} Nodes are colored according to the symmetry fiber partition. White color is used to show the nodes in trivial (consisting of only one node) fibers. The input trees of nodes $3$ and $11$ are isomorphic, however they are separated into two different fibers in order to avoid a fiber partition in which nodes without a common input are grouped together. This is a biological requirement.}
    \label{Fig:SymmetryFibration}
\commentAlt{Figure~\ref{Fig:SymmetryFibration}: 
Two directed graphs with an arrow connecting them labeled phi.
The graph on the left has nodes named 1-13. 1,2,3 are red, 5,7,9 are green, 11,12,13 are blue, the remaining ones are white.
Directed connections: 3 to 1, 3 to 2, 3 to itself, 3 to 4, 3 to 8, 4 to 5, 4 to 7, 4 to 9, 6 to 5, 6 to 7, 6 to 9, 6 to 10,
8 to 6, 10 to 6, 11 to 10, 11 to itself, 11 to 12, 11 to 13.
The graph on the right has nodes named a-g. a is red, d is green, g is blue, the remaining ones are white.
Directed connections: a to itself, a to b, a to e, b to d, c to d, c to f, e to f, f to e, g to f, g to itself.
}
\end{figure}

The first modification is aimed at dealing with the situation shown in Fig.~\ref{Fig:SymmetryFibration}. The ISCV of a node that receives inputs only from itself consists of one non-zero entry corresponding to the color of this node. The ISCV of any node that receives inputs only from this node is equal to the ISCV of the node, which creates fibers like those shown in blue and red in Fig.~\ref{Fig:SymmetryFibration} while keeping blue and red fibers separate from each other. 

\begin{figure}[ht!]
    \centering
    \includegraphics[width=0.4\textwidth]{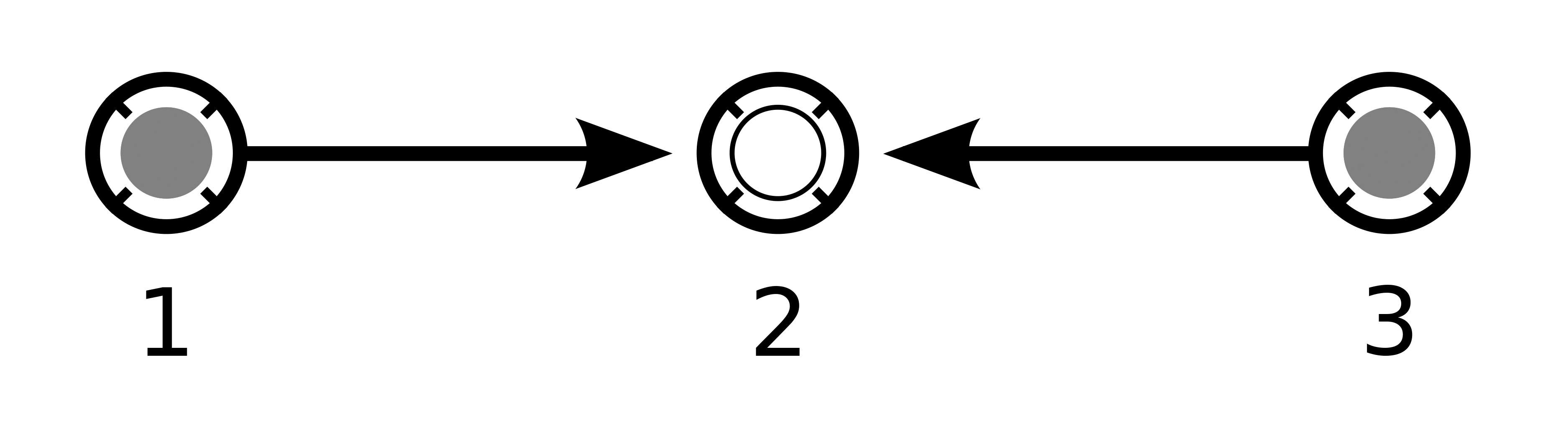} 
    \caption{\textbf{Special cases of fibers.} An example of the network in which fibrations considered in \citep{morone2020fibration} are not minimal. Nodes $1$ and $3$ have no inputs and therefore belong to the same fiber of the minimal fibration by definition. However \cite{morone2020fibration} put them in separate fibers to fiber partitions in which nodes with no common input are put in the same fiber, since they are unlikely to synchronize in a real network.}
    \label{Fig:MinFibersExample}
\commentAlt{Figure~\ref{Fig:MinFibersExample}: 
A directed graph with nodes named 1-3; 1 and 3 are red, 2 is white.
Directed connections: 1 to 2, 3 to 2.
}
\end{figure}

The second modification deals with the more trivial situation in Fig.~\ref{Fig:MinFibersExample}. The ISCVs of all nodes with no inputs are equal to the zero vector at every step of the algorithm; therefore, assigning different initial colors is not enough, and repeated assignment during every iteration is needed to maintain the partition.

\begin{algorithm}[H]
    \begin{flushleft}
      \hspace*{\algorithmicindent} \textbf{Input:} A graph $G=(V,E)$ with $V=\{1,\dots,N\}$\\
      \hspace*{\algorithmicindent} \textbf{Output:} Colors of the symmetry fibers $\chi_1, \chi_2, \dots, \chi_N$, $N = |V|$
    \end{flushleft}
    \begin{algorithmic}[1]
      \State For every node $j$, we denote with $I^j$ the ISCV of $j$, and with $I^j_i$ its $i$-th component
      \State We let $d(j,i)$ be the number of nodes of color $i$ adjacent to node $j$
      \If {$G$ is directed}
        \State Let $z_1,\dots,z_\ell$ be the nodes with no inputs
        \State Let $x_1,\dots,x_m$ be the nodes that receive input only from themselves
      \Else
        \State Let $\ell=m=0$
      \EndIf
      \ForAll {$j \in \{1,\dots,N\}$}
        \State Let $\chi_j=1$
      \EndFor
      \ForAll {$j \in \{1,\dots,\ell\}$}
        \State Let $\chi_{z_j}=j$
      \EndFor
      \ForAll {$j \in \{1,\dots,m\}$}
        \State Let $\chi_{x_j}=\ell+j$
      \EndFor
      \If {$\ell+m \neq N$}
        \State Let $d=\ell+m+1$
      \Else 
        \State {\bf return} $\chi_1,\dots,\chi_N$
      \EndIf 
      \While {true}
        \ForAll {$i \in \{1,\dots,d\}$ and $j \in \{1,\dots,N\}$}
            \State Let $I^j_i=d(j,i)$
        \EndFor
        \State Assume that there are $k$ distinct ISCV's, say $H^1,\dots,H^k$ (numbering is arbitrary)
        \ForAll {$i \in \{1,\dots,k\}$ and $j \in \{1,\dots,N\}$}
            \If {$I^j=H^i$}
                \State Let $\chi_j=\ell+i$
            \EndIf
        \EndFor
        \ForAll {$j \in \{1,\dots,\ell\}$}
            \State Let $\chi_{z_j}=j$
        \EndFor
        \If {$d=\ell+k$}
            \State {\bf return} $\chi_1,\dots,\chi_N$
        \Else
            \State $d=\ell+k$
        \EndIf
      \EndWhile
    \end{algorithmic}
    \caption{Implementation of the refinement algorithm developed by \cite{leifer2022thesis} to find minimal balanced colorings in graphs. This implementation is used to obtain all results displayed in this book. See also  \citep{morone2020fibration}.}
    \label{Algo:SymmFibers}
\end{algorithm}

\begin{figure}[ht!]
    \centering
    \includegraphics[width=.4\textwidth]{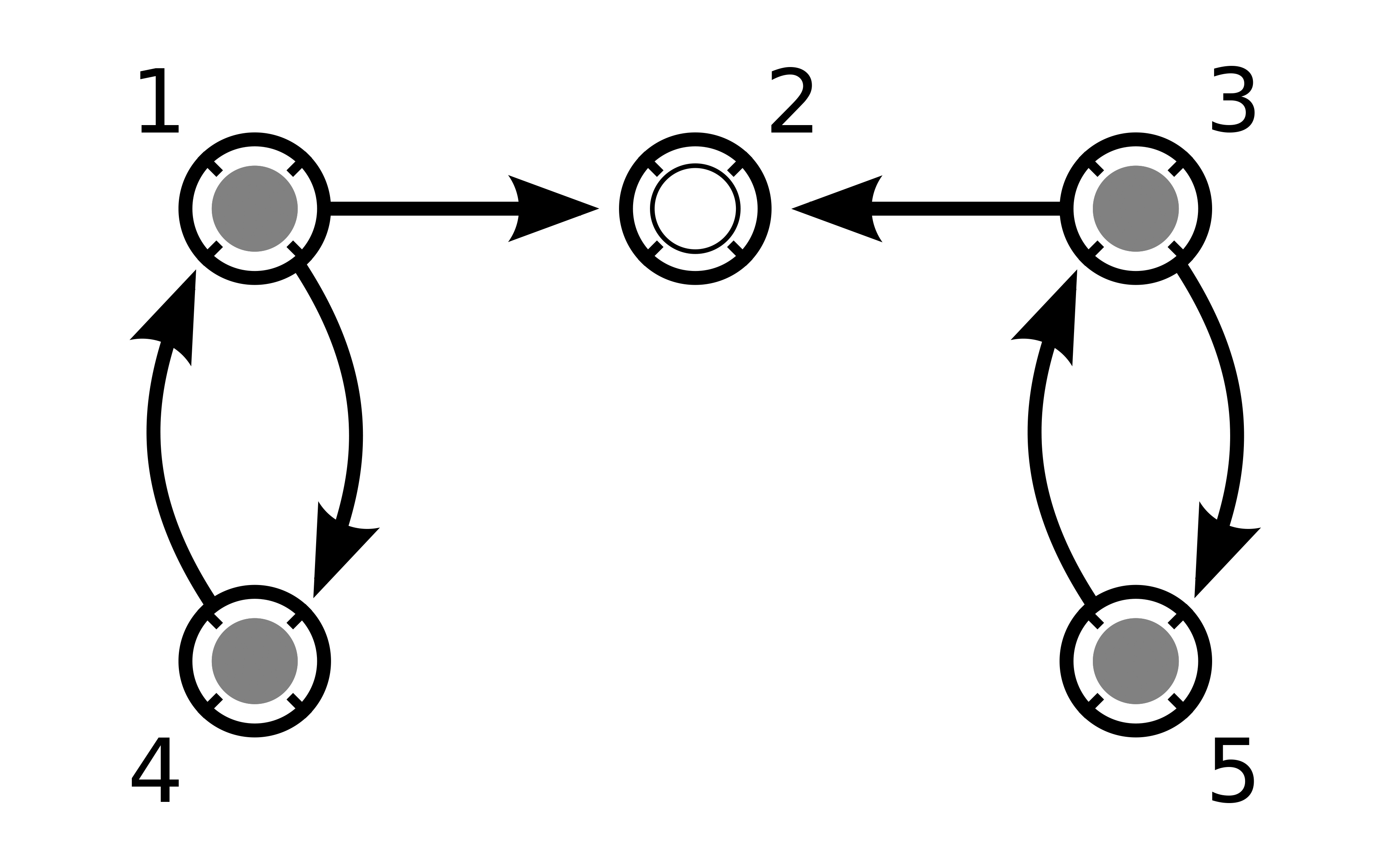} 
    \caption{\textbf{Special cases of fibers.} Example of a graph in which the algorithm does not produce the symmetry fiber partition. Nodes $1$, $3$, $4$, and $5$ have the same initial color and the same ISCVs. However, nodes $1$ and $4$ have no common inputs with nodes $3$ and $5$ and biology requires the two clusters to be of different colors.}
    \label{Fig:SCCHeuristicCounterexample}
\commentAlt{Figure~\ref{Fig:SCCHeuristicCounterexample}: 
A directed graph with nodes named 1-5; all nodes are red except for 2 (white).
Directed connections: 1 to 2, 1 to 4, 3 to 2, 3 to 5, 4 to 1, 5 to 3.
}
\end{figure}

Consider now the graph in Fig.~\ref{Fig:SCCHeuristicCounterexample}. It is easy to see that the coloring produced by the algorithm assigns nodes $1$, $3$, $4$, and $5$ the same color. Indeed, all four nodes are assigned the same initial color, and corresponding ISCVs are the same. However, nodes $1$ and $4$ have no inputs in common with $3$ and $5$. Therefore, this coloring is not a realistic fiber partition from a biological viewpoint. The situation described here can be repaired using a further modification of the initial coloring. Let $y_1, y_2, \dots, y_k$ be the set of sets of nodes belonging to the strongly connected components of the graph that receive no input, i.e., $y_i = j_1, j_2, \dots, j_{k_1}$, where $j_m$ are nodes that belong to the strongly connected component that receives no input with $id = i$. Then, the problem described here can be negated by assigning $y_1, y_2, \dots, y_k$ different initial colors. This case appeared only once in the 373 networks studied in \citep{leifer2020circuits}. Therefore, we take it as a special case.

All these special biological cases are considered in  Algorithm~\ref{Algo:SymmFibers} showing the pseudocode employing these heuristics. An implementation of this algorithm in the form of an R package is available at {\small\url{https://github.com/makselab/fibrationSymmetries}}.
See Appendix \ref{sec:list-software} for a comprehensive list of all codes for fibration analysis.
We suggest the reader use this implementation in any study of a real-world biological network. However, a mathematician should adhere to the original minimal balanced coloring definition and ignore these special cases, i.e., the colorings of Fig.~\ref{Fig:MinFibersExample} and Fig.~\ref{Fig:SCCHeuristicCounterexample} should be considered the minimal.

As was mentioned at the beginning of this section, the Kamei--Cock algorithm is slower than the Paige--Tarjan algorithm. A faster implementation of the Paige--Tarjan algorithm to find symmetry fibers, called Fast Fiber Partitioning (FFP), is presented in \citep{monteiroAlgorithm}.

\partquote{
Part I developed the theoretical basis for the application of
symmetries to biological networks. We saw that symmetries of the
network come in different flavors. The more stringent symmetries are
automorphisms with their related orbital partitions of synchronized
nodes.  
Fibration symmetries and their associated clusters of
synchronized fibers impose weaker conditions on the structure of
the network since they preserve only input trees.
Synchronized fibers are determined by the balanced
colorings of the network. 
Part I provided theoretical evidence that fibration
symmetry is necessary and sufficient to ensure cluster synchronization
in the network.  A primary objective of systems science is to
decompose the network into independently functioning parts and
scrutinize the emergence of the network dynamics from the
interaction of these parts. Fibration symmetries provide a framework
for this decomposition.

\quad Part II focuses on biological applications of these
theoretical methods.  We show how symmetry fibrations describe 
symmetries and synchronization in biological networks including
genetic networks, metabolic networks, signaling networks, and
connectomes across species.
We first consider two popular model organisms, {\it E. coli} and
      {\it C. elegans}, for which the most complete reconstructions of
      genetic, protein, and metabolic networks and the connectome,
      respectively, are known. Later we generalize our results to more
      complex organisms, including humans. We show that
      symmetries can directly predict gene synchronization activity of
      coexpression patterns observed experimentally, as well as neural
      synchronization in the brain. Furthermore, symmetry breaking of
      the synchronized fibers uncovers further building building
      blocks with functionalities resembling memory devices of
      computation. The concept of a fibration can thus provide a full
      characterization of all genes in the genetic network, based on
      their symmetries and symmetry breaking.  This provides a general
      organizing principle to understand the structure-function relation in
      biological networks. 
  }

\part{Applications: from Genes to the Brain}


\chapter[Fibration Analysis of Biological Networks]{\bf\textsf{Fibration Analysis of Biological Networks}}
\label{chap:hierarchy_1}

\begin{chapterquote}
  Part I shows that cluster synchronization emerges from a symmetry
  fibration. This chapter shows that genetic networks exhibit symmetry
  fibrations, which group the genes in fibers with synchronized
  behavior, as shown in subsequent chapters. This creates a hierarchy
  of fibration building blocks of biological networks.

\end{chapterquote}

\section{Geometric biology}

Inspired by the symmetry theory of Part I,  we now ask whether a
geometrization process based on symmetry fibrations\index{fibration } can be invoked to
tame biological complexity, and if so, precisely how. We start by
reviewing applications of symmetries to understand the structure of
gene regulatory networks. Gene expression\index{gene expression } and its regulation have been
studied since the time of the bacterium operon\index{operon } model by \cite{jacob1961genetic,jacob1977}. Gene products that cooperate
to accomplish a certain task, for example, lactose utilization in the
case of the Lac operon,\index{Lac operon } tend to be coexpressed, i.e., differentially
expressed in a synchronized manner. This genetic synchronization
largely determines the function of the biological circuit. This leads to a hierarchy of fibration building blocks describing the geometry of the system.

\section{Transcriptional regulatory networks}

Systems biology relies on the construction of different types of
biological networks including transcriptional, metabolic, signaling,
protein-protein interactions and neural networks. The basic network
that we analyze here is the transcriptional regulatory network\index{network !transcriptional regulatory } (TRN)\index{TRN }
and the metabolic network of {\it E. coli}. This model bacterium
\emph{E.~coli} possesses a genome with 4,690 genes, as compiled by
RegulonDB's aggregate of results to the date of publication of this
book \citep{tierrafria2022regulondb}. Of these, 1,843 are known to have
regulatory functions in the transcriptional network.

The transcriptional regulatory network is responsible for performing
the decision-making process of how to change its gene expression
levels according to the prevailing circumstances.  For instance,
organisms need to regulate the expression of their genes according to
their (possibly changing) environment. To do so they must process
information from transient environmental signals as inputs, and, along
with the current state of expression levels, compute an appropriate
response as an output, in the form of a new state of expression
levels: namely, to decide which genes to turn off and which genes to
turn on. In this manner they tune the expression level continuously as circumstances
require.

Genes are regulated by transcription factors (TFs)\index{transcription factor }\index{TF } that bind to their
promoter region, which either activate the transcription of the target
gene, `turning on' the gene, or repress the transcription from
happening, thus preventing expression `turning off' the gene. The TFs
are proteins coded by genes, so this amounts to a pairwise relation
between genes, albeit indirectly: a gene codes for a TF, which then
regulates other genes' expression. The entirety of transcriptional
regulations constitutes the transcriptional regulatory network (TRN)
of the cell, which is in charge of deciding which genes to
express. This endows the cell with the emergent property of being able
to compute the next state of expression levels for the cell. The TFs
are, in turn, usually activated or inhibited by the presence of
specific small molecules, which function as input signals from the
environment or the cell's internal state. These small molecules, or
metabolites, can either come from the environment or be the product of
a metabolic pathway.\index{metabolic pathway } Eukaryotes, as opposed to bacteria,  have other forms of more complex regulations as well.

\begin{figure*}[!t] 
    \centering \includegraphics[width=\textwidth]{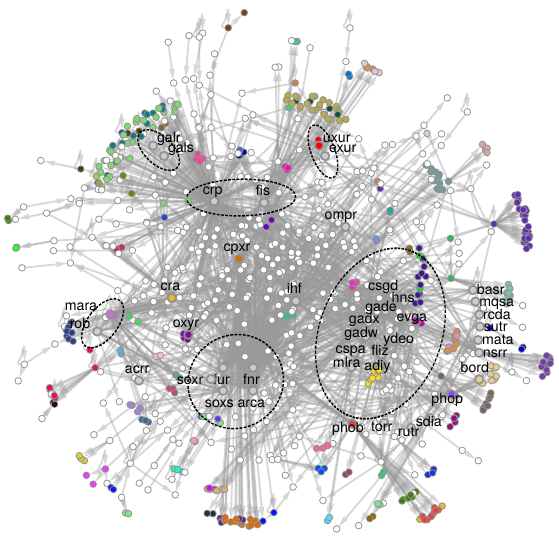}
\caption{ \textbf{The complexity of the \emph{E.~coli} TRN in its full
    glory.} We show the weakly connected component of the 
  operon-TRN for \emph{E.~coli} with 879 nodes. Small disconnected pieces of
  the network are not shown since they do not play a role in the
  network's dynamics.  Genes in the same fibers are colored the
  same. Strongly connected components are marked by ellipses with
  their gene names. In this chapter, we  unravel the origin of
  this complex structure one piece at a time. Figure reproduced from
  \citep{alvarez2024fibration}.}
\label{fig:main_trn}
\commentAlt{Figure~\ref{fig:main_trn}: 
Mainly illustrative. A very complicated network with hundreds of nodes and edges.
Six regions are surrounded by dotted ovals. Clusters of genes in same fiber
are given same color. Details not relevant. Subsequent text describes
how the fibers are found.
}
\end{figure*}

To see how signals propagate\index{signal propagation } from gene to gene (in time or as a causal
chain determining steady states), we need to consider the entire gene
regulatory network. For simplicity, and due to limited knowledge about
some regulation steps, we simplify the network and consider directed
regulation edges between transcription factors and their target genes
(all of them being described by protein\index{protein } levels). Neglecting many
specific regulation steps (e.g., in transcription termination,
ribosome binding, and mRNA\index{mRNA } or protein degradation),\index{degradation } we describe the
expression of a target gene by a simple \textit{input function}, with the TF
expression levels as function arguments.

Indeed, every transcriptional interaction between two genes (edge in
the TRN) requires multiple parameters for a precise mathematical
description of the genes' expression dynamics \citep{klipp2016book}.  These
effective parameters capture the gene regulatory process from
transcription, translation, protein folding, to binding and unbinding
of the transcription factor to DNA, plus ribosome and polymerase
binding, mRNA and protein lifetimes, to mRNA and protein degradation. This set of events is lumped into one single edge
representing the whole process in the TRN, yet, this edge encapsulates
a large number of parameters which are, in principle, unknown to the
modeler trying to understand the regulatory process.

Since most of these parameters are unknown, structural graph analyses of these
lumped networks\index{network !lumped } have been performed by ignoring these heterogeneities
and assuming that all edges are the same, i.e., each edge represents
the same set of parameters, modulo some important differences such as
repression and activation. (Chapter \ref{chap:alive} examines in detail this assumption and offers alternatives.)  Such an approach has led to considerable
insights, in particular, the discovery of network motifs \citep{shen2002network,milo2002network}. 
Motifs are building blocks of the network that are identified by
statistical over-representation of small circuits in the network, as
compared to randomized networks. Their existence suggests
that, as expected, the network structure has most likely been shaped
by evolutionary pressures rather than being randomly connected. Even though the individual dynamics of some motifs is
more or less understood  \citep{shen2002network, milo2002network},
given that they are in essence, local structures, they do little to
unravel the global topology of the network or the global
dynamics. Also, statistical significance alone is not enough to
explain function. 

We construct a directed network for this process, representing all the
transcriptional regulatory signals. A directed edge in the network
represents a transcriptional signal from one gene to another, where
the source of the edge represents the gene coding for the
transcription factor that binds to the gene represented by the target
node of the edge. The type of edge corresponds to the type of
regulation the transcription factor performs, either activation or
repression.  (Further distinctions, such as connection strengths or
weights, can be made, but we ignore these refinements at this stage, see Chapter \ref{chap:alive}.)
The full complexity of the {\it E. coli} TRN\index{E.~coli TRN @{\it E.~coli} TRN } is shown
in Fig. \ref{fig:main_trn} highlighting the structures---strongly
connected components,\index{strongly
connected component } fibers\index{fiber } and building blocks\index{building block }---that we 
unravel in this chapter.

The process of sending regulatory messages across genes defines the
information flow in the transcriptional network, which we can think of
as an ’information package’ or ’message passing’ from the source gene
to the target gene. Thus, we interpret a link in a transcriptional
network as representing regulatory `messages' that are dynamically sent
from a source gene to a target gene via the transcription factor. The
transcription factor acts as a `messenger' to repress or activate the
transcription rate of the target gene. This information flow is not
restricted to two interacting genes, but is transferred to different
regions within the network that are accessible through the connecting
pathways. The information arriving at a gene contains the entire
history transmitted through all pathways that reach this gene, which is
formalized by the input tree of the gene.

On a large scale, it has been proposed that the {\it E. coli} TRN\index{E.~coli TRN @{\it E.~coli} TRN }
has a feed-forward structure \citep{ma2004hierarchical,
  martinez2008functional, dobrin2004aggregation,shen2002network},
where signals flow unidirectionally from a core of sensors and master
regulators through a series of parallel layers down to an outer
periphery in a feed-forward manner \citep{dobrin2004aggregation,
  shen2002network, milo2002network}.  However, there also exist
feedback loops that complicate this picture. Thus, many questions
remain answered: 
\begin{enumerate}
\item What is the
relevant structure of this system that determines its function?  \item What are the network's building blocks? 
\item What
is the computational core structure in charge of decision-making in
the cell? 
\item How does this structure control the rest of the network? 
\item Is
there a minimal structure that explains the function of the TRN is a
simplified manner? 
\end{enumerate}
Fibration symmetries offer possible answers to
these questions.

\section{Fibration analysis of the TRN}
\label{sec:fibration-analysis}

We search for symmetries in the {\it E. coli} transcriptional network
using the compilation TF interactions at RegulonDB
 \citep{tierrafria2022regulondb}. Nodes are genes; a directed link
represents an activator, repressor (or sometimes a dual)
transcriptional regulation mediated by a transcription factor.
\begin{figure}[h!]
  \centering
    \includegraphics[width=.75\textwidth]{fig14.2.pdf}
\caption {\textbf{Example of input tree, fibration, fiber, and base in a
    simple {\it E. coli} circuit}.  (\textbf{a}) The circuit controlled by
  the {\it cpxR} gene regulates a series of fibers, as shown by the
  different colored genes.  (\textbf{b}) The input tree
  of representative genes involved in the {\it cpxR} circuit shows
  the isomorphisms that define the fibers. (\textbf{c})
  Isomorphism between the input trees of {\it baeR} and {\it spy}.
  (\textbf{d}) Fibration $\psi$ transforms the {\it cpxR} circuit $G$
  into its base $B$ by collapsing the genes in the fibers into one.
  (\textbf{e}) Fibration of the {\it fadR} circuit and (\textbf{f}) its
  isomorphic input trees.  (\textbf{g}) Symmetric genes in the fiber
  receive the same information through the pathways in the network
  and, therefore, synchronize their activity to produce activity
  levels leading to a common cellular function. Figure reproduced from
  \citep{morone2020fibration}. Copyright \copyright ~2020, National Academy of
  Sciences U.S.A.}
\label{cpxr}
\end{figure}

We start our discussion by analyzing the symmetries in a sample sub-circuit
extracted from the TRN of {\it E. coli}.
\index{E.~coli TRN @{\it E.~coli} TRN } This circuit is regulated by
gene {\it cpxR} which regulates its own expression (via an
autoregulation positive loop) and regulates other genes as shown in
Fig. \ref{cpxr}a. (The circuit regulates more genes,
  represented by the dotted lines, which are not displayed for
  simplicity and do not change the results.)  Gene {\it cpxR} is not regulated by any other
transcription factor, so it does not receive `information' from the
main network (we say that this gene forms its own `strongly connected
component', see below). Therefore, it is an ideal simple circuit to
exemplify the concept of graph fibration in the TRN. 

The input tree of gene {\it
  spy} (Fig. \ref{cpxr}b) starts with {\it spy} at the root (first
layer). Since this gene is upregulated by {\it baeR} and {\it cpxR},
the second layer of the input tree contains these two pathways
of length one starting at both genes. Gene {\it baeR} is further
upregulated by {\it cpxR} and by itself through the autoregulation
loop, and {\it cpxR} is also autoregulated. Thus, the input tree
continues to the third layer, taking into account these three possible
pathways of length 2: one starting at {\it baeR}, and two starting at
{\it cpxR}. The procedure now continues, and since there are loops in
the circuit---the autoregulation loops at {\it baeR} and {\it cpxR}---the input tree has an infinite number of layers.

As discussed in Section \ref{sec:isom}, the theorem of \cite{norris1995} allows us to prove the
equivalence of two input trees\index{input tree } with an infinite number of
levels (or layers).
It suffices to find an isomorphism up to $N-1$ levels,
where $N$ is the number of nodes in the circuit. Since the ribose
system contains three genes, only input trees up to the second
level are needed to prove the existence of an isomorphism. In fact, the
input trees of {\it rbsR} and rbsDACBK are isomorphic, but not with
{\it crp}.

The input trees in the {\it cpxR} circuit are displayed in
Fig.~\ref{cpxr}b. The input trees of {\it baeR} and {\it spy}
(Fig.~\ref{cpxr}c) are isomorphic and define the {\it baeR}-{\it spy}
fiber with an isomorphism:
\begin{equation}
  \begin{array}{ll}
  \tau = ({\it baeR}, {\it baeR}, {\it cpxR}, {\it baeR}, {\it cpxR},
       {\it cpxR}, \dots) \\
       \qquad \to ({\it spy}, {\it baeR}, {\it cpxR}, {\it
         baeR}, {\it cpxR}, {\it cpxR}, \dots).
  \end{array}
\end{equation}

Clearly, the input tree of {\it cpxR} is not isomorphic to either {\it
  baeR} or {\it spy}, and therefore {\it cpxR} is not symmetric with
these genes, but it is isomorphic to {\it bacA, slt} and {\it yebE}
forming another fiber.  Likewise, {\it ung, tsr} and {\it psd} are all
isomorphic forming their corresponding fiber
(Fig. \ref{cpxr}b). Figure \ref{cpxr}d shows the fibration $\psi: G\to
B$ that collapses the genes in the fibers to the base $B$. Figure
\ref{cpxr}e shows another example (out of many) of a single connected
component, {\it fadR}, and its corresponding isomorphic input trees
(Fig. \ref{cpxr}f), fibers and base.
 
In Fig. \ref{cpxr}g we use the
  mathematical model of gene regulatory dynamics of Boolean kinetics\index{Boolean kinetics }
  from \citep{shen2002network} to show synchronization inside
  the fiber {\it baeR-spy} when the fiber is activated by its
  regulator {\it cpxR}. Notice that {\it cpxR} does not synchronize
  with the fiber. Sigmoidal interactions lead to
  qualitatively similar results.

\begin{figure}
\centering
  \includegraphics[width=0.9\textwidth]{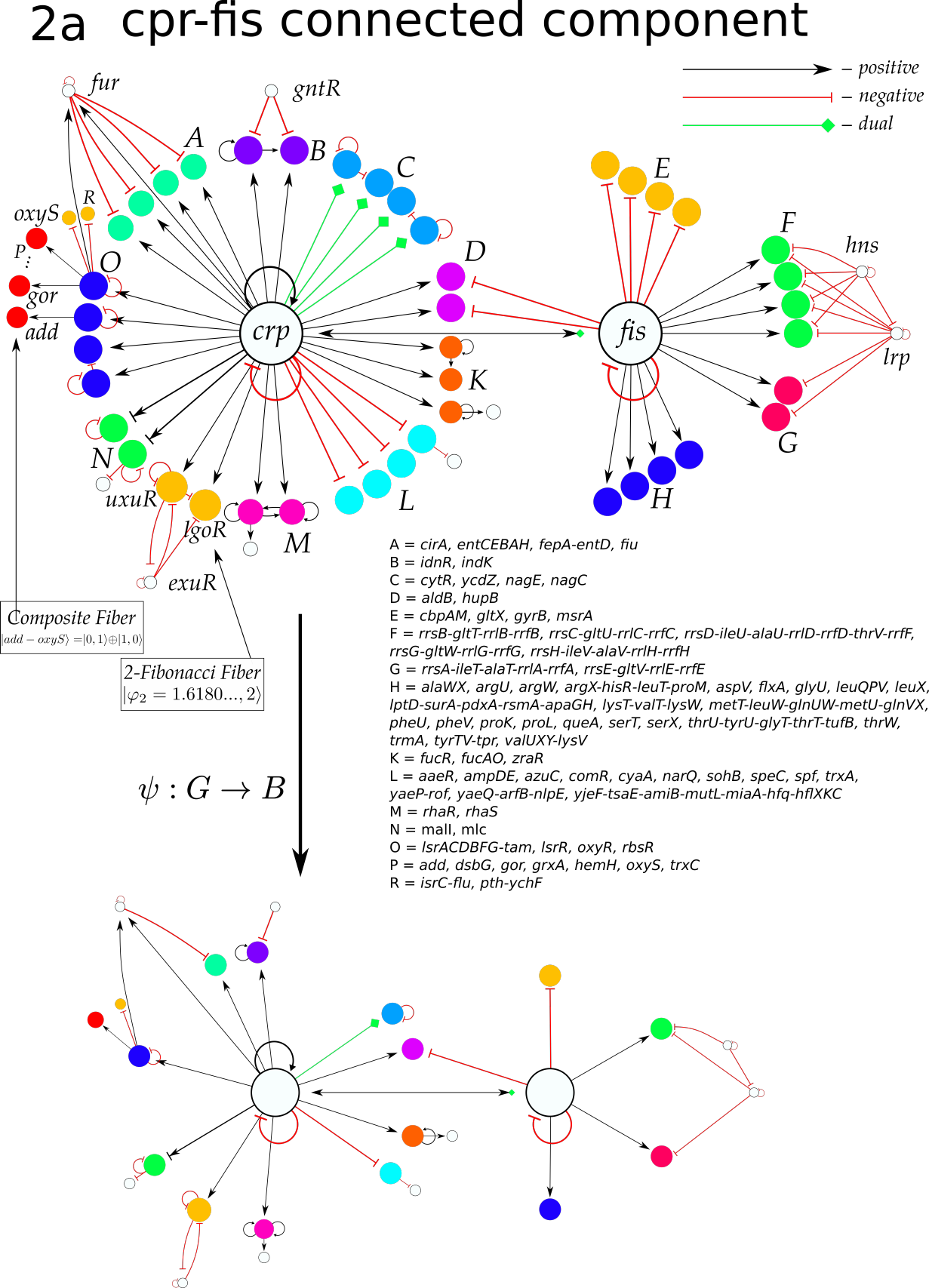}
  \centering
  \caption{\textbf{Fibers of the strongly connected components of
      sugar metabolism in {\it E. coli} TRN}.  Two-gene
    connected component of {\it crp-fis}. This component controls a
    rich set of fibers as shown. We also show the symmetry fibration
    collapsing the graph to the base. We highlight the fiber {\it
      uxuR-lgoR} which sends information to its regulator {\it exuR}
    and forms a 2-Fibonacci fiber $\rvert \varphi_2 = 1.6180.., l =
    2 \rangle$, as well as the double-layer composite
    $|add-oxyS\rangle = |0,1\rangle \oplus |1,1\rangle$. We denote the
    fibration symmetry transformation to its base by $\psi$. Figure reproduced
    from \citep{morone2020fibration}. Copyright \copyright ~2020, National Academy of
  Sciences U.S.A.}
\label{fig:componenta}
\commentAlt{Figure~\ref{fig:componenta}: 
No alt-text required.
}
\end{figure}

Going beyond these particular examples, \cite{morone2020fibration} performed a full fibration analysis of this biological network. The first step
is identifying the strongly connected components\index{strongly connected component } in the system, because
the classification of fibers\index{fiber }  is based on the cycles that the fibers
form.

Recall the concept of a strongly connected component, Definition \ref{def:SCC}: a
directed graph is strongly connected\index{strongly connected component } if there is a (directed) path between all
pairs of vertices; a strongly connected component\index{strongly connected } (SCC)\index{SCC } is a maximal induced subgraph that is strongly connected.
As stated earlier, we may omit the adjective `strongly'
when this does not cause confusion.

Beyond several single-gene strongly connected components like those
shown in Fig. \ref{cpxr}, the {\it E. coli} network has
three main SCCs,\index{SCC of E. coli@SCC of {\it E. coli} } which regulate a set of fibers:

\begin{itemize}
\item {\bf Sugar SCC:} A two-gene feedback loop strongly connected component composed of
  master regulators {\it crp-fis} involved in a myriad of functions
  related to carbon utilization and sugar metabolism
  (Fig. \ref{fig:componenta}).
\item {\bf pH regulation SCC}. The largest SCC at the core of the
network which is composed of genes involved in the pH-system that
regulate hydrogen concentration inside the cell, such  as
{\it GadX, GadW, Hns, GadE, GadF, YdeO, EvgA, FliZ}
(Fig. \ref{fig:componentb}).
\item {\bf Stress regulation SCC}.  A five-gene strongly connected
component involved in the stress response system (Fig. \ref{fig:componentc}).
\end{itemize}

Applying the fiber finding algorithm based on the minimal balanced
coloring of the network from Section \ref{sec:software} we find a rich
set of fibers that are regulated by each of the three components.
These fibers are then collapsed into the base by the symmetry
fibration $\psi: G \to B$, as shown in the figures.

Applying the symmetry fibration to the \emph{E.~coli} TRN leaves 
just 555 nodes in the base, 30\% the size of the original network, see
Table~\ref{tab:reduction}. Figure \ref{fig:fig3_assorted} show a
sample of the fibers in different functional circuits such as the
non-glucose carbon production circuit regulated by {\it crp}, and
amino-acid production fibers. A great variety of fibers is
observed. They are classified below into an hierarchy of fibration
building blocks.\index{building block !fiber }

\begin{figure}
  \centering
    \includegraphics[width=.7\textwidth]{fig14.4.pdf}
    \caption{ \textbf{Fibers of the strongly connected components of the
        pH regulation in {\it E. coli} TRN}. The core of the {\it
        E. coli} network is the strongly connected component formed by
      genes involved in the pH system as shown.  This component
      supports two Fibonacci fibers: 3-FF and 4-FF and fibers as
      shown.  Hollow colored circles indicate genes that are in fibers
      and also belong to the pH component. Figure reproduced from
      \citep{morone2020fibration}. Copyright \copyright ~2020, National Academy of
  Sciences U.S.A.}
    \label{fig:componentb}
\commentAlt{Figure~\ref{fig:componentb}: 
No alt-text required.
}
\end{figure}

As shown in Table \ref{tab:reduction}, the full \emph{E.~coli} genome
contains 4,690 genes, out which 1,843 genes express TFs that
are not isolated, i.e. they are TFs with at least a regulation to or
from another TF. Bacterial genomes contain operons,\index{operon } which are sets of
continuous genes that are transcribed together into mRNA by the same
RNA polymerase sharing the same promoter (the part of DNA that enables the
gene to be transcribed). As a result, operons are expressed together,
so they are trivially synchronized. But not always: some operons are
split into Transcriptional Units\index{transcriptional unit } (TU)\index{TU } which are operated by different
promoters. These promoters and end points are intercalated between the
genes of the operon. Thus sometimes the operon is partially expressed,
and sometimes it is fully expressed.  So synchronization in
operons is a bit complicated, but in general, pure operons (which do
not contain TUs) are considered to be fibers that synchronize trivially via
a common polymerase.

Interestingly, these operons are fibers, albeit of a trivial shape since
they are trivially synchronized (Section \ref{sec:operons}). To
simplify the analysis we just collapse them by a trivial fibration and
then analyze the remaining network, which in this case contains 879
TFs (Table \ref{tab:reduction}). In this network, 63\% of the genes are involved in fibers,
indicating a large coverage of fibers highlighting their importance for
the structure of the network. The remaining genes are mainly
dangling-ends, genes that send information to other parts of the
cellular network like the metabolism. When studying the TF, these genes
can be removed, and the remaining network contains just 42 genes from
the original 1,843.  This reduced network still preserves the same
dynamics as the original network, on account of the invariance of the dynamics under fibration, with all 
redundant information pathways collapsed into a single one.  This
remarkable reduction highlights the power of fibrations to tame
biological complexity. It is what we call the `minimal genome', studied
with more detail in Chapter \ref{chap:minimal}.

Most of the fibers regulated by these components do not belong to the
connected components, although they are regulated by them. This is
because they receive information from the connected component, but do not send information back
to it. This implies that we can classify these
fibers with simple integer dimensions, as we see next.  More
interestingly, fibers send information to the connected components,
creating cycles that produce fractal dimensions (see Section \ref{S:fractal_dim}) with more interesting
memory dynamics.

\begin{figure}[h!t]
  \centering
  \includegraphics[width=.7\textwidth]{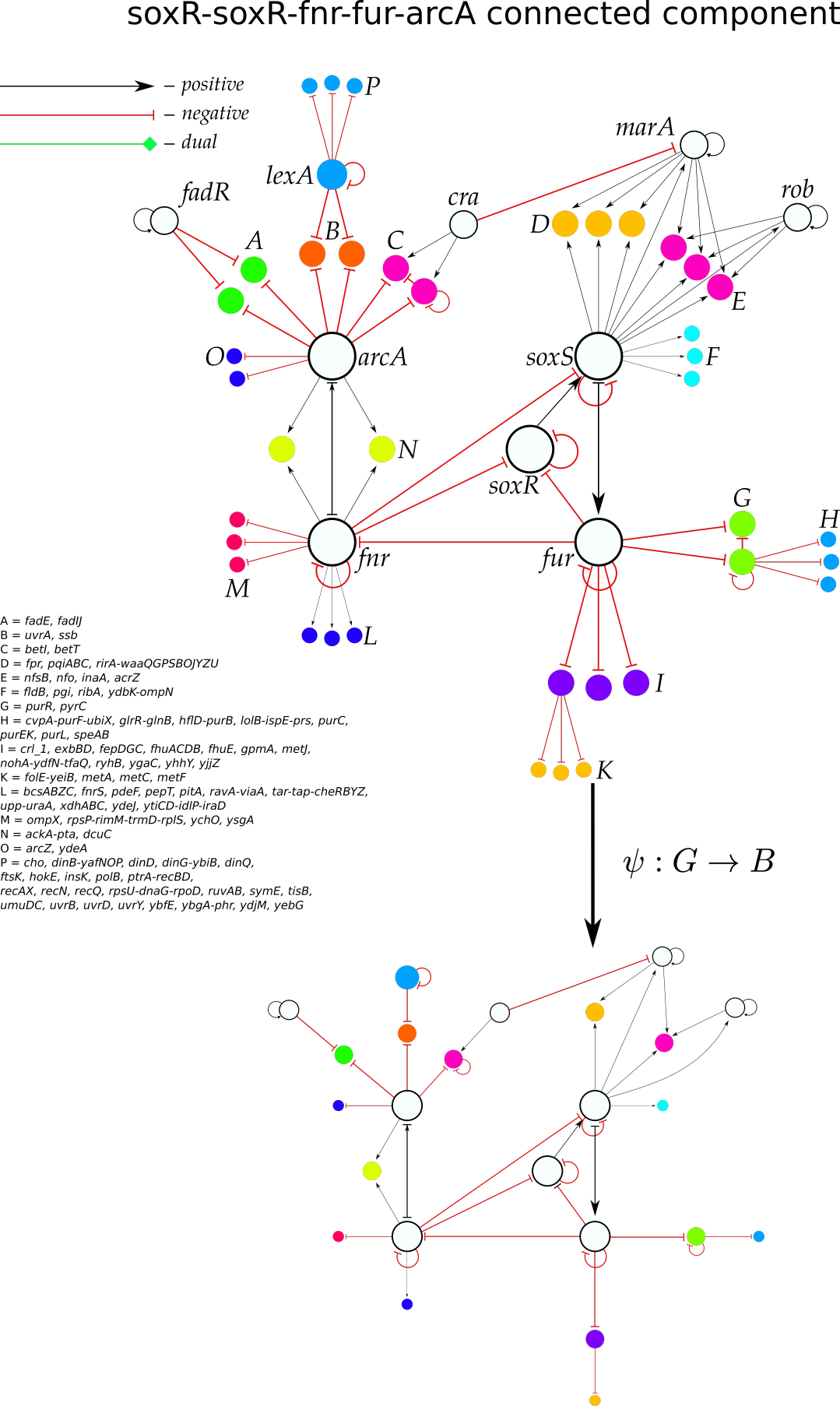} \centering
  \caption{ \textbf{Fibers of the strongly connected components of the
      stress response in {\it E. coli} TRN}. A five-gene connected
    component of {\it soxR, soxS, fnr, fur, } and {\it arcA} with its
    regulated fibers. Figure reproduced from
    \citep{morone2020fibration}. Copyright \copyright ~2020, National Academy of
  Sciences U.S.A.}
\label{fig:componentc}
\commentAlt{Figure~\ref{fig:componentc}: 
No alt-text required.
}
\end{figure}

\begin{table*}[ht]
  \centering
  \begin{tabular}{|l|r|r|}
    \hline
   \textbf{Reduction step}     & \textbf{Gene count} & \textbf{\%} \\
    \hline\hline
    Full Genome RegulonDB & 4,690 & -- \\
    \hline
    TRN (non isolated TFs)   & 1,843 & 100\%\\
    Operon-collapsed TRN    & 879 & 48\%\\
    \hline
    Base-TRN (collapsing fibers) & 555 & 30\%\\
    Minimal TRN (after trivial pruning) & 42 & 2\% \\
    \hline
 \end{tabular}
   \vspace{10pt}
  \caption{\textbf{Reduction count and fiber coverage.} The full \emph{E.~coli}
    genome contains 4,690 genes, of which
    1,843 genes express TFs that are not isolated, i.e. they are TFs
    with at least a regulation to or from another TF. The first
    reduction of operons is by a trivial fibration---which collapses
    the operons = trivial fibers, reducing the network to 879 TFs.
    The fibration further reduces the network to 30\% and after
    removing the trivial dangling-ends, and the final base is reduced to
    2\% of the original size conforming the minimal TRN. Table
    reproduced from \citep{alvarez2024fibration}.}
 \label{tab:reduction}
\end{table*}

\section{Searching for biologically meaningful
  building blocks}

In a gene network, how can we find (possibly overlapping) groups of
genes that form `meaningful circuits', i.e. circuits with a specific
biological function? The function might be performing
a specific signal processing task such as spike generation, filtering,
forming logical gates, etc.; or it might be regulating a specific biological
system, e.g. in response to external stimuli or via feedback
regulation. Circuits are likely to be meaningful if {\it (i)} they create a
meaningful dynamics or perform a meaningful regulation of signal
processing tasks, for example coexpression of target genes; {\it (ii)}
circuits of this shape occur many times in a network, much more often
than expected by chance (network motifs); {\it (iii)} they are biochemically
implemented through structures like graph fibers.

\begin{figure}[h]
 \centering \includegraphics[width=.95\textwidth]{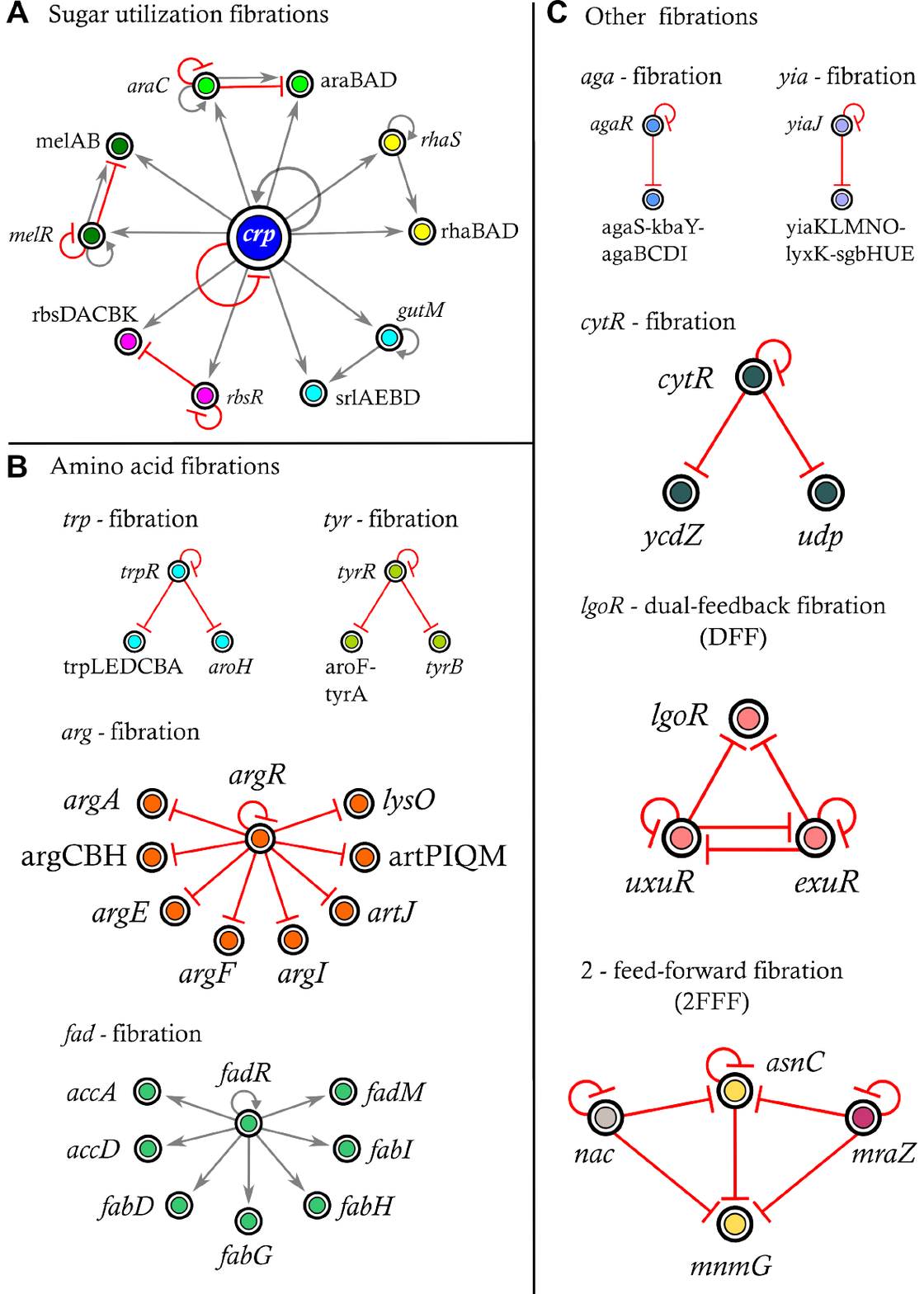}
  \caption {\textbf{Assorted sample of fibers in the TRN.}  (\textbf{a}) The
    non-glucose sugar utilization system contain many fibers with {\it
      crp} acting as the master regulator switch without synchronizing
    with any fiber. (\textbf{b}) Sample fibers responsible for amino-acid
    production in the cell. (\textbf{c}) More elaborate fibers.  }
\label{fig:fig3_assorted}
\commentAlt{Figure~\ref{fig:fig3_assorted}: 
A. Sugar utilization fibrations. 
B. Amino acid fibrations.
C. Other fibrations. 
}
\commentLongAlt{Figure~\ref{fig:fig3_assorted}: 
A. Sugar utilization fibrations. Central crp node sends arrows to
ten surrounding nodes, each forming pairs with same color.
B. Amino acid fibrations: Blue node trpR sends arrows to two other blue nodes.
Yellow node tyrR sends arrows to two other yellow nodes.
Red node argR sends arrows to 8 other red nodes.
Green node fadR sends arrows to 7 other green nodes.
C. Other fibrations. 
aga-fibration: Blue node agaR sends arrow to another blue node.
yia-fibration:  Blue node yiaJ sends arrow to another blue node.
cytR-fibration:  Dark green node cytR sends arrows to two other dark green nodes.
lgoR-dual feedback fibration: Red node lgoR sends arrows to two other red nodes,
which also send arrows to each other.
2-feed forward fibration (2FFF): yellow node receives one arrow
from buff node and one from red node. Second (lower) yellow node
receives one arrow
from buff node and one from red node, and also from yellow node. Top yellow
node sends arrow to lower yellow node. Top three nodes send arrows to themselves.
}
\end{figure}

Searching for these circuits as building blocks\index{building block } underlying the
structural organization of biological networks is one of the
cornerstones of systems biology\index{systems biology }
\citep{hartwell1999,alon2019,klipp2016book}.  In the framework of
network science, an enormous variety of network metrics have been
applied to biological networks, neural circuits, and structural and
functional brain networks in search of their fundamental building
blocks.  At the sub-circuit level the motifs\index{motif }
introduced by Alon\index{Alon, Uri }
and collaborators \citep{shen2002network,milo2002network,alon2019}, are the most widely used building blocks in biological
networks.  These are small subgraphs of the network, the most
prominent being the feed-forward loop, which are over-represented as
compared to a null model (defined by randomly rewiring the original
network while preserving the degree sequence). We define  motifs more
formally in Chapter \ref{chap:motif}.

This approach is blind to the fact that gene outputs depend on the
entire previous signal processing, which may involve
loops. Mathematically, the dynamics of the motif in isolation need not
correspond to its dynamics when embedded in the surrounding network,
because inputs to the motif affect its behavior. Over-representation
can be a useful indicator of potential functionality, but of itself it
carries no implications for dynamics. Thus it is questionable whether
network motifs would have functionality when embedded in the network.
Furthermore, genes may exhibit similar expression patterns, even if they
do not share the same inputs, as shown in Section
\ref{sec:carbon-utilization}.

Beyond motifs, there is a vast literature on frameworks use for finding
meaningful network structures based on ideas like small-world\index{network !small-world }
 \citep{watts1998}, scale-free\index{network !scale-free } \citep{barabasi1999}, modularity
\index{modularity } \citep{girvan2002}, community detection algorithms\index{community detection! algorithm }
\citep{blondel2008}, principal components,\index{principal component } clustering,\index{clustering } and centrality
analysis\index{centrality analysis } \citep{makse2024thescience} based on hubs\index{hub }
\citep{barabasi1999}, rich-clubs,\index{rich-club } Pagerank,\index{Pagerank } eigenvector centralities,
\index{eigenvector centrality }
$k$-core\index{k-core @$k$-core } \citep{kitsak2010identification}, or collective influence\index{collective influence}
\citep{morone2015influence,delferraro2018finding}, to name a few.
These network properties, while providing a summary description of the
structure of the networks, are too generic to illuminate the
functionality underlying biological networks, since they cannot be
traced to fundamental principles.

Even the statistical evidence has been criticized. \cite{fodor} argue
that common methods for searching for motifs involve testing a large
number of hypotheses in parallel, creating a high risk of false
positives. They consider two standard assumptions: a normal
distribution, and independence between candidates. They find cases
where both of these assumptions are violated, and suggest that this
casts doubt on some mainstream motif identification
techniques. \cite{stivala} suggest the use of exponential random
graph models to overcome such statistical shortcomings.

Furthermore, there is no experimental evidence for a link
between network motifs and dynamics
\citep{ingram2006}, and controversies remain on the functional role of
motifs in biological networks
\citep{ingram2006,payne2015,macia2009,ahnert2016}. These criticisms suggest that 
network motif structure does not correlate with function.

In contrast to network motifs, which are found by statistical
over-representation in the network, gene fibers\index{fiber, contrast with motif } reveal meaningful genetic
circuits with functional relevance using theory. These fibers are relevant even if they appear only once in a
network. Indeed, while motifs can be found on statistical grounds through
their number of occurrences, functional circuits that do not occur in
large numbers may easily go unnoticed. This remark implies, in particular, to
larger circuits, for which statistical over-representation as a motif
(by counting all subgraphs of that size) or as a module
\citep{hartwell1999} (by counting a `density' of edges within that
module) is hard to prove.

Fibrations \index{fibration } and their associated fibers\index{fiber }  provide a definition of
biological building blocks\index{building block } using first physical principles, rather
than by statistical significance. We use the principle of symmetry,
since it directly predicts synchronization, no matter how often these
structures are found in the network. The resulting circuits can
perform signal processing tasks, namely generating structurally
encoded, yet tunable patterns of coexpression.\index{coexpression }

In the coming sections, we apply fibration theory to gene regulatory
networks across species to find and classify the fibration building
blocks of biological networks. We then validate the predicted
coexpression of genes in the fibers of the gene regulatory networks
of bacteria. We also show that the fibration building blocks are
enriched by functional annotations and thus are biologically
significant in the cell. We compare the results with motifs in more
details in Chapter \ref{chap:motif}.

\section{Definition of fibration building blocks}
\label{sec:definition-building}

The dynamical state of a gene is encoded in the topology of the
input tree. In turn, this topology is encoded by a sequence, $a_i$,
defined as the number of genes in each $i-$th layer of the input tree,
see for instance Fig. \ref{cpxr}b. The sequence $a_i$ represents
the number of paths of length $i-1$ that reach the gene at the root.
This sequence is characterized by the branching ratio $r$ of the input
tree defined as:

\begin{definition} 
\label{def:BRIT}
{\bf Branching ratio of the input tree.} Given $a_i$ as the number of source genes with paths
of length $t - 1$ to the target gene in the input tree, the {\em branching ratio}\index{branching ratio } $r$ is defined as:
\begin{equation}
r=\lim_{i \to \infty}  \frac{a_{i+1}}{a_i},
\label{eq:branching}
\end{equation}
provided this limit exists, see Remark below. 

Another term for the branching ratio, which is common
in the literature and which we use on occasion
throughout the book, is `fractal dimension'.\index{fractal dimension }
\end{definition}

\begin{remark}
\label{r:branching_ratio}
The limit \eqref{eq:branching} need not exist, in which case
a more general definition is required. An example is 
Fig. \ref{fig:6node_input_tree}a, the input
tree of the 6-node network of Fig. \ref{fig:6node_orbital_quot}a.
In this case, the sequence $a_i$ is:
\[
1,1,2,2,2,4,4,4,8,8,8,\ldots
\]
The ratios $a_{i+1}/a_i$ are:
\[
1,1,2,1,1,2,1,1,2,\ldots
\]
which fails to converge.
Such networks are rare and unlikely to occur in biology.
When the limit in \ref{eq:branching} does not exist,
we can replace it by
\begin{equation}
r= \lim_{i \to \infty}  a_i^{1/i}
\label{eq:branching_new}
\end{equation}
The Perron--Frobenius Theorem\index{Perron--Frobenius Theorem } implies
that this limit always exists, and is equal to the spectral radius 
(here the maximal eigenvalue, which is real)
of the adjacency matrix of the `upstream subnetwork':\index{upstream subnetwork } the induced subgraph for all nodes connected to $i$ by a path. It takes the same value as \eqref{eq:branching}
when that limit exists. See
\citep{boldistewart2024}.
\end{remark}

The branching ratio represents the `average long-term' multiplicative growth rate of the number
of paths across the network reaching the gene at the root.  For
instance, the input trees of genes {\it baeR-spy} (Fig. \ref{cpxr}b)
encode a sequence $a_i=i$ with branching ratio $r=1$ representing the
single autoregulation loop inside the fiber.

The branching ratio organizes the topologies of the fibers in a neat fashion. For instance, the TRN of {\it E. coli} is organized into 91 different fibers.  The
complete list of fibers in {\it E. coli} appears in the supplementary
information section of \citep{morone2020fibration}.  We find a rich variety
of topologies of the input trees. Despite this diversity, the input
trees present common topological features that classify
all fibers into concise classes of fundamental {\it`fibration building
blocks'.}

The intuition for defining building blocks is as follows.\index{building block !intuition behind } The central
component of the building block is the fiber. Since the genes in the
fiber are synchronous, the building block captures the idea of
common functionality of the genes by associating function with
coherence. This is the most natural notion of a building block.
However, the fiber does not work in complete isolation since it
receives signals from the network, in particular from the
SCC to which it belongs. (It also sends signals to it.) Thus, it is common sense to add to the
central fiber of the building block  the external regulators of
the fiber. In the simplest case, a fiber just receives signals from
its regulator but does not send signals back to the regulators nor to
any other part of the network. In this case, the definition of
building block is unambiguous.  We associate a building block with this
fiber by considering the genes in the fiber plus the external
incoming regulators of the fiber.  

For instance, the fiber denoted by B
in the carbon SCC in Fig. \ref{fig:componenta}:
\begin{equation}
  \mbox{\it Fiber}  =  \{idnR, indK\}
\end{equation}
receives signals from
\begin{equation}
    \mbox{\it Regulators} = \{crp, gntR\}
\end{equation}
and do not send any signals outside the fiber. Then, the building
block is {\it simple} and it is formed by:
\begin{equation}
\begin{array}{rcl}
 \mbox{\it Simple Fibration Building Block Level 1} & = & \mbox{\it Fiber} + \mbox{\it Regulators}  \\
  & = & \{idnR, indK, crp, gntR\}.
\label{eq:simple-building} 
\end{array}
\end{equation}

We call it a simple level 1 building block since, from here, we can build more complex ones. The next level of complexity appears when the fiber sends signals back
to the regulators or to other genes in the SCC or to any other part of
the network through more than one path. This creates cycles that
are important to define the dynamics of the fiber. In principle, all
these cycles should be included in the definition of the building block
associated with the central fiber.

When the fiber sends signals back to the regulators through 
a {\it unique} path in the network from the fiber to the regulator,
that is when there is only a feedback loop between fiber and regulators,
then we consider the building block still {\it simple} but level 2. It is given by
(\ref{eq:simple-building}). For instance, the yellow fiber in Fig. \ref{fig:componenta}:
\begin{equation}
  \mbox{\it Fiber}  =  \{uxuR, lgoR\}
\end{equation}
is still a simple building block because it has a single cycle from fiber passing to one of its regulators {\it exuR}. We call this particular building block Simple Fibonacci building block and will be studied later:  
\begin{equation}
\begin{array}{rcl}
 \mbox{\it Simple Fibration Building Block Level 2} & = & \mbox{\it Fiber} + \mbox{\it Regulators} \\ 
& &  \,\mbox{\it (with Unique Path Through Regulators)} \\ & = & \{uxuR, lgoR, exuR, crp\} .
\label{eq:simple-building-fibonacci} 
\end{array}
\end{equation}

The majority of fibers in the TRN of
{\it E. coli} shown in Figs. \ref{fig:componenta},
\ref{fig:componentb} and \ref{fig:componentc} are simple Level 1 and 2 and
Eqs. (\ref{eq:simple-building}) and  (\ref{eq:simple-building-fibonacci}) suffices to define them.

However, the fiber
\begin{equation}
  \mbox{\it Fiber}  =  \{evgA, nhaR\}
\end{equation}
denoted in red in the pH SCC of Fig. \ref{fig:componentb} clearly
presents a more complex situation. This fiber (actually only the gene
{\it evgA}) is embedded in the SCC, implying that there could be more
than one cycle, starting from the fiber and passing through the
regulators.  In principle, the genes belonging to these paths
should be included in the building block of the fiber since all the
genes in these cycles are crucial to establish the dynamic
synchronization state of the fiber. These cycles can be complex and
long, and if the fiber is embedded in the SCC ,they may even include 
all cycles in the SCC.

At this point, we introduce a level of subjectivity into the definition
of a building block: we consider as part of the building block only the
shortest path. We call these building blocks {\it simple} level 3.\index{building block !simple }  We use
this definition of building block elaborated in
\citep{morone2020fibration,leifer2020circuits}.

For instance, the {\it evgA} fiber has a regulator:
\begin{equation}
  \mbox{\it Regulator}  =  \{hns\}
\end{equation}
that belongs to the SCC of {\it evgA} and {\it evgA} is connected to the  SCC.  Thus, there are pathways through the SCC that connect
the regulator and the fiber. Two of them are:
\begin{equation}
  {\rm Cycle \, 1}  =  evgA \to gadE \to gadX \to hns \to evgA
\end{equation}
\begin{equation}
  {\rm Cycle \,2}  =  evgA \to ydeO \to gadE \to gadX \to hns \to evgA.
\end{equation}
Since Cycle 1 is the shortest, only this path is included in the
building block of {\it evgA}. Not all genes in the SCC
participate in these paths. Where there are many shortest paths with
the same length, we either pick one at random or consider all of them.
We call this a:
\begin{equation}
\begin{array}{rcl}
\mbox{\it Simple Fibration Building Block Level 3}&=& \mbox{\it Fiber} + \mbox{\it Regulators}  \\
& & + \,\mbox{\it Shortest Path Through Regulators} \\ & = & \{evgA, ydeO, gadE, gadX, hns\}
\label{eq:complex-building} 
\end{array}
\end{equation}

All simple building blocks are regulated by isolated genes. The next level of complexity appears when the fiber is regulated by another fiber. This increases the complexity of the building block massively since there are many paths between the fibers, creating cycles of information that increase the fractal dimension of the input trees.
 We call these building blocks {\it complex}.\index{building block !complex }  They are abundant in the metabolic networks of Chapter \ref{chap:complex} and complex organisms.
\begin{equation}
\begin{array}{rcl}
\mbox{\it Complex Fibration Building Block}&=& \mbox{\it Fiber} + \mbox{\it Fiber Regulators}  \\
& & + \,\mbox{\it Shortest Path Through Fiber Regulators} 
\label{eq:complex-building2} 
\end{array}
\end{equation}

We formalize these intuitions in the following definition
\citep{morone2020fibration,leifer2022thesis}.

\begin{definition}
    \textbf{Fibration building block.} A {\it fibration building block}\index{fibration !building block }\index{building block !fibration } is a subgraph
    induced by the set of nodes consisting of:
    \begin{enumerate}
        \itemsep0em
        \item All the nodes in the fiber.
        \item All regulators that send inputs to the fiber.
        \item All nodes belonging to the shortest cycle, including a node in a fiber if any node in a fiber is a part of a cycle.
    \end{enumerate}
    \label{def:fibration-building}
\end{definition}

\begin{figure}[h!]
    \centering
    \includegraphics[width=.86\textwidth]{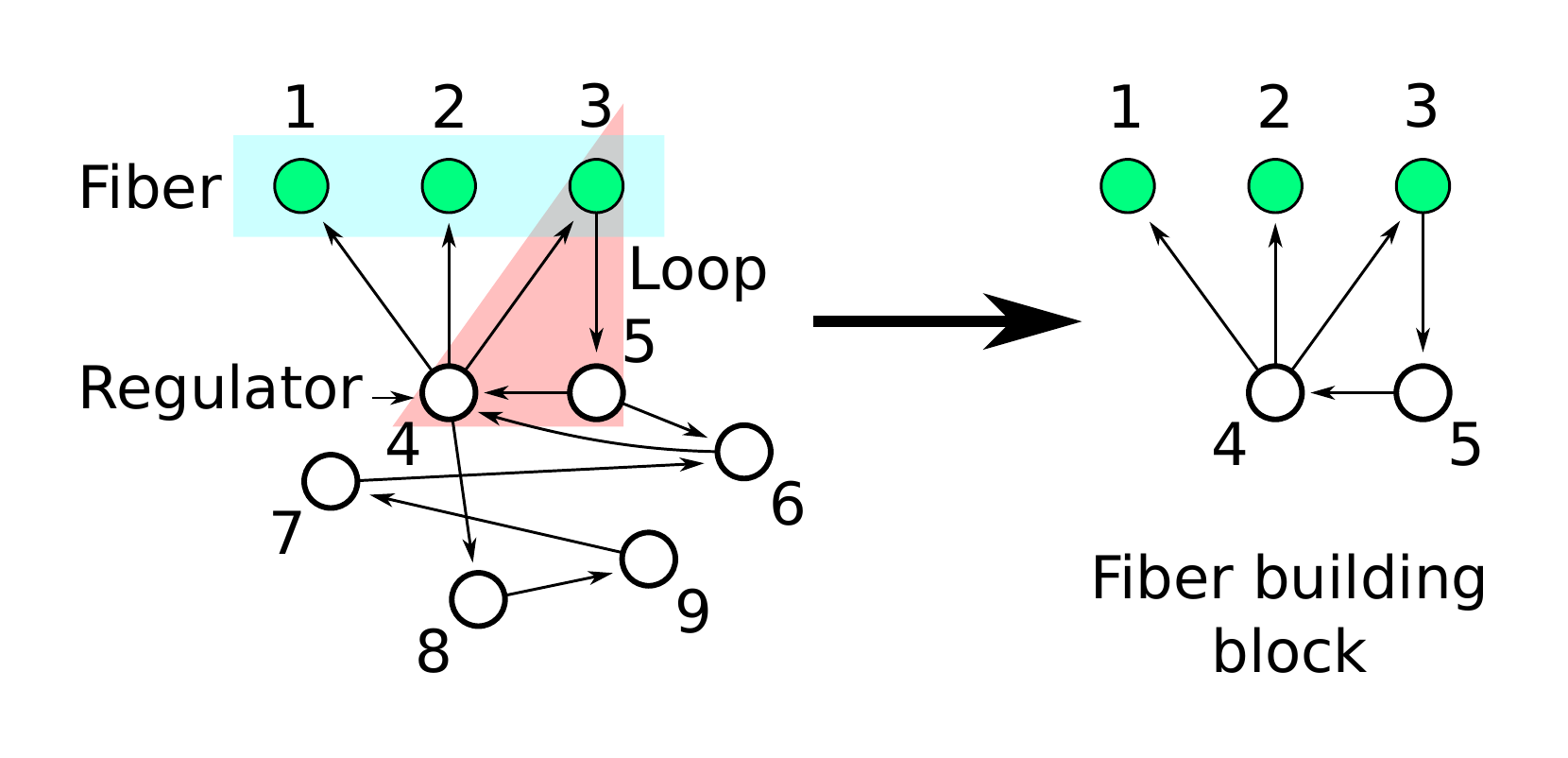} 
    \caption{\textbf{Example of a fibration building block} \citep{leifer2022thesis}. The building
      block corresponding to the fiber in green consists of $Fiber = \{1, 2, 3\}$,  fiber regulator $\{4\}$, and a
      node in the shortest cycle $\{5\}$ going through the regulator $4$. Longer cycles are neglected. }
    \label{Fig:BuildingBlockDefinition}
\commentAlt{Figure~\ref{Fig:BuildingBlockDefinition}: 
Left: Fiber:  green nodes 1,2,3. Regulator: white nodes 4, 5, 6, 7, 8, 9.
Arrow  35, 41, 42, 43, 48, 54, 56, 64, 76, 89, 97. Pink triangle labeled `loop'
round 3, 4, 5.
Right: Fiber building block: green nodes 1, 2, 3. White nodes 4, 5.
Arrows 35, 41, 42, 43, 54.
Horizontal black arrow between these, showing fibration.
}
\end{figure}

For example, consider the network in
Fig.~\ref{Fig:BuildingBlockDefinition} with a fiber in green. The building
block corresponding to the fiber consists of 5 nodes: (a) nodes $1$,
$2$, $3$ belong to the fiber, (b) node $4$ regulating this fiber, and
(c) node $5$ belonging to the shortest cycle that include a node in the
fiber. Despite the existence of the cycle $3 \to 5 \to 6 \to 4
\to 3$, node $6$ is not included in the building block because this
cycle is not the shortest.

A final distinction should be made between the input trees\index{input tree } of the
fiber and the input trees of the building block.  We exemplify
this case with the fiber G=\{ {\it purR , pyrC} \} in the stress SCC shown
in Fig. \ref{fig:componentc}.  The fiber and building block and input
trees of the building block are shown in Fig. \ref{fff}. 
This is the feed-forward fiber (FFF) already discussed in  Sections \ref{intuition},\ref{sec:ex_input_tree},
and \ref{S:6GC}.

This building block is embedded in the network, but the building block
ignores the paths reaching the regulator {\it fur}.  For instance,
{\it fur} receives signals from {\it soxS} and itself in the SCC.
Moreover, it also receives inputs from {\it oxyR}, although this is
not shown in Fig. \ref{fig:componentc} since {\it oxyR} does not
belong to the SCC.  However, all these regulators of {\it fur} (and
the paths that comprise its input tree) are not crucial to determine
whether {\it purR} and {\it pyrC} are synchronized, since the
regulators of {\it fur} are not connected to, nor receive any inputs
from, the fiber genes.  We do not consider the autoregulation loop at
{\it fur} to be in the building block, either.

Thus, the input trees of the building block are those of the circuit
composed of {\it fur-purR-pyrC} without considering any inputs of {\it
  fur} (including its own autoregulation).  The input tree of {\it fur} as
considered in the building block is empty, as shown in Fig. \ref{fff}b,
while the input tree of {\it fur} as embedded in the network is
obviously more complex.  The same consideration applies to the input
trees of the genes in the fibers.  The full input tree of {\it purR}
and {\it pyrC} in the network should be augmented by the paths
reaching {\it fur}. These larger input trees remain isomorphic,
demonstrating that the paths beyond {\it fur} do not affect
synchronization of the fiber. We conclude that the minimal circuit
that guarantees synchronization is what we call the  building block.

\begin{figure}[ht!]
    \centering
    \includegraphics[width=\textwidth]{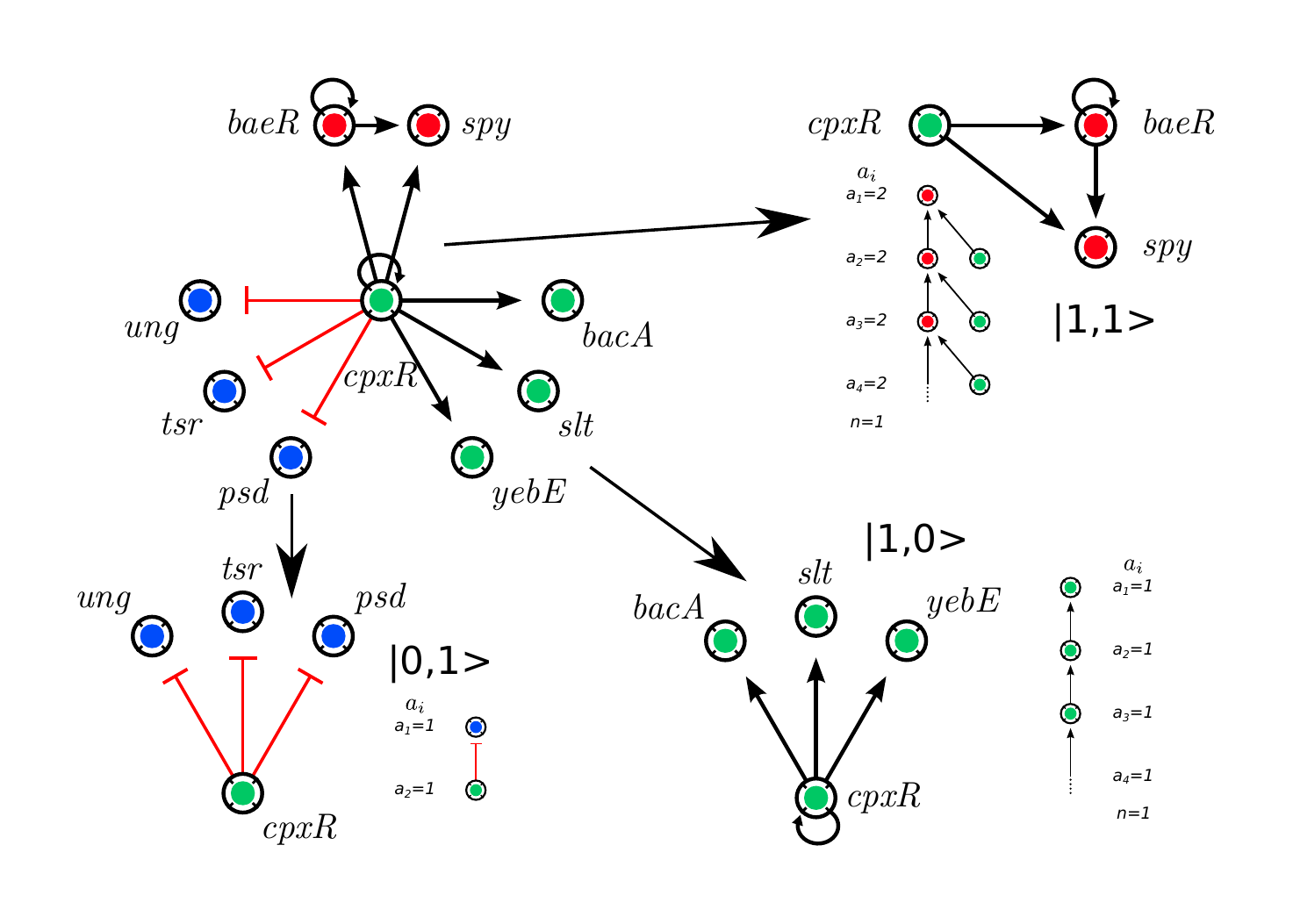}
    \caption{\textbf{Example of fiber building block decomposition
        and classification.} The presented network consists of three
      fibers and contains three corresponding building blocks \citep{leifer2022thesis}. The fiber
      building block corresponding to the nodes in green consists of
      nodes in the green fiber and no other nodes because
      this fiber is not regulated externally. The input tree of this fiber
      is an infinite chain with $a_i = 1$. Therefore
      this fiber building block is classified as $| n = 1, l = 0
      \rangle$. Similarly, the input tree of the fiber in red consists
      of an infinite chain with the addition of the external
      regulator, rendering this block as $| n = 1, l = 1
      \rangle$. The fiber building block in red has a finite input tree
      and one external regulator, and is therefore classified as $| n =
      0, l = 1 \rangle$.}
    \label{fig:buildingblockdecompositionexample}
\commentAlt{Figure~\ref{fig:buildingblockdecompositionexample}: 
Graph from Figure~\ref{cpxr}, with black arrows to three other graphs, as follows:
Green cpxR node sends arrows to three blue nodes.
Green cpxR node sends arrows to three green nodes.
Green cpxR node sends arrows to two red nodes. One red node
sends arrow to itself and the other red node.
}
\end{figure}

More subtle differences are exemplified in the building block
decomposition in Fig. \ref{fig:buildingblockdecompositionexample} of
the component regulated by {\it cpxR} that we encountered before in
Fig. \ref{cpxr}. This circuit is simple because it does not receive
any signal from other genes. Yet, there are some subtleties.

The central regulator {\it cpxR} plays the role of regulator for the
red fiber {\it baeR-spy} and the blue fiber containing {\it ung}, but
it belongs to the green fiber containing {\it bacA}.  While the fibers
form a non-overlapping partition, the building blocks can overlap. The
input tree of the building block of {\it baeR-spy} does not include
the autoregulation loop at {\it cpxR}. This encodes a sequence $a_i=2$
for all $i$, with branching ratio 1 representing the single
autoregulation loop inside the fiber and one external regulator. 

The input tree of the building block of genes {\it ung, tsr} and \textit{psd} is finite because the autoregulation loop
in the external regulator {\it cpxR}  is not
included in the building block.  This fiber is also an
orbit, as it arises from an automorphism permuting the genes. In a
sense, this is a trivial fiber, and the inputs suffice to determine
its synchronization either by fiber or by orbit. The other two fibers are
nontrivial, since the are not group symmetric. On the other hand,
{\it cpxR} appears as a part of the green fiber {\it bacA} and its loop
is included in the input tree of the genes in the building block.

In conclusion, we make a distinction between three different circuits:
\begin{itemize}
\item First, the actual fiber made of synchronized genes,
  like {\it purR-pyrC} in Fig. \ref{fff}a.
  \item Second, the building block associated with the fiber as in
    Fig. \ref{fff}a, which also includes the regulator: { \it fur,
      purR-pyrC}.
    \item Third, the fiber as embedded in the full network as in
      Fig. \ref{fig:componentc} with the added complexity of the input
      trees, i.e., with {\it fur} receiving signals from the rest,
      yet, representing information that is useless for understanding the dynamics
      of the fiber.
\end{itemize}

These considerations simplify the complexity of a building block to
its minimal description, permitting a systematic classification
of them all. It lets us obtain a concise topological classification
of all fibers in terms of the fiber numbers, as we show next.
Otherwise, if we were to include in the building blocks the paths
beyond the regulator or other longer paths, a simple classification
would not be possible.

\section{Classification of simple building blocks}
\label{sec:calculating}

We provide a hierarchical classification of building
blocks following \citep{morone2020fibration}.\index{building block !classification } Using symmetries, we implement a constructive procedure that
reveals a hierarchy of biological building blocks, ubiquitous across
species.  These conclusions introduce a theoretically principled
strategy to search for computational building blocks in biological
networks, as well as a route for conceiving synthetic biological
circuits. We apply the algorithms to find fibers explained in Section
\ref{sec:software} to find all fibers and then classify them into a
hierarchy of building blocks in the TRN of {\it E. coli}.  We start with
the {\it simple} building blocks that describe most of the {\it
  E. coli} TRN,\index{E.~coli TRN @{\it E.~coli} TRN } and then move on to describe a myriad of {\it complex}
building blocks that appear in the metabolism and more complex
species.

While we find a rich variety of topologies for input trees, we
observe that input trees present common topological features that
classify all fibers into concise classes of fibration
building blocks. The key to developing a building block classification
based on fibrations is to understand the hierarchical
organization of loops and cycles, inside the fibers and between fibers
and regulators.  At the simplest level this is captured by the lengths
of the cycles inside the fiber.  These cycles determine the
topology of input trees, hence the dynamics of the fiber.

The topological features of the fibers are encoded by the sequence
$a_i$ of the number of nodes in each $i-$th layer of the input
tree. The sequence $a_i$ represents the number of paths of length
$i-1$ that reach the gene at the root. The most basic input tree
topologies and corresponding sequences $a_i$ can be classified by
integer `fiber numbers'.

\vspace{20pt}
\begin{definition} {\bf Fiber number classification of simple fibration building blocks}.
  Simple building blocks are characterized by two {\it fiber numbers}:\index{fiber number }
\begin{equation}
  | n, l\rangle \,: \,\,\, \mbox{\rm simple building block
    classification }
\end{equation}
reflecting two features:
\begin{itemize}
\item The integer $n$ is the branching ratio\index{branching ratio } $r=n$ of the input tree defined in
   (\ref{eq:branching}), and represents the exponential growth of
  the number of paths through the network reaching the gene at the
  root.
\item The number of finite pathways starting at $l$ external
  regulators of the fiber. By convention, $n=0$ for finite input
  trees.
\end{itemize}
\label{def:fiber-numbers}
\end{definition}

For instance, the input trees of genes {\it baeR-spy} in
Fig.~\ref{fig:buildingblockdecompositionexample} have branching ratio
$n=1$ representing the single ($n=1$) autoregulation loop inside the
fiber and one external regulator with $l=1$. Therefore this
building block is classified as $| n=1, l=1\rangle$. On the other
hand, the input tree of genes {\it ung, tsr} and \textit{psd} is
finite with one external regulator, therefore this building block is
classified as $| n=0, l=1 \rangle$.

\begin{figure}[ht!]
    \centering
    \includegraphics[width=0.85\textwidth]{fig14.9.pdf} 
    \caption{\textbf{Classification of simple fibration building blocks with integer dimensions}. Basic
      fibration building blocks found in {\it E. coli}. These building
      blocks are characterized by a fiber that does not send back
      information to its regulator. They are characterized by two
      integer fiber numbers: $|n, l \rangle$. We show selected
      examples of circuits and input trees and bases. The statistical
      count of every class will be discussed later. The last example
      shows a generic building block for a general n-ary tree $|n,
      l\rangle$ with $l$ regulators. Figure reproduced from
      \citep{morone2020fibration}.  Copyright \copyright ~2020, National Academy of
  Sciences U.S.A.}
    \label{Fig:PNAS_Fig3a}
\commentAlt{Figure~\ref{Fig:PNAS_Fig3a}: 
Eight classes of genetic circuit, in four columns:   \vert  n,\ell \rangle , genetic circuit, 
input tree, base. Rows  \vert 0,1 \rangle ,  \vert  0,2 \rangle ,  \vert  0,3 \rangle ,  \vert  1,0 \rangle ,
  \vert  1,1 \rangle ,  \vert  1,2 \rangle ,  \vert  2,1 \rangle ,  \vert  n\geq 3, \ell \rangle .
Column 2 shows sample genetic circuits. Column 3 shows input trees.
Column 4 shows base. Further details in text and subsequent figures.
}
\end{figure}

\begin{figure}[ht!]
    \centering
    \includegraphics[width=0.7\textwidth]{fig14.10.pdf} 
    \caption{\textbf{Classification of simple building blocks with fractal dimensions (although
        more complex than Fig. \ref{Fig:PNAS_Fig3a})}.  Fibonacci and
      multilayer building blocks. These building blocks are more
      complex than the simpler ones shown in
      Fig. \ref{Fig:PNAS_Fig3a}. They are characterized by an
      autoregulated fiber that sends back information to its
      regulator. This creates a fractal input tree that encodes a
      Fibonacci sequence with golden branching ratio in the number of
      paths $a_i$. When the information is sent to the connected
      component that includes the regulator, then a cycle of length
      $d$ is formed and the topology is a generalized Fibonacci block
      with golden ratio $\varphi_d$ as indicated. Last panel shows a
      multilayer composite fiber with a feed-forward structure. Figure
      reproduced from \citep{morone2020fibration}. Copyright \copyright ~2020, National Academy of
  Sciences U.S.A.}
    \label{Fig:PNAS_Fig3b}
\commentAlt{Figure~\ref{Fig:PNAS_Fig3b}: 
Five classes of genetic circuit, in four columns:   \vert  n,\ell \rangle , genetic circuit, 
input tree, base. Rows   \vert  1.6180,2 \rangle  (2-FF),  \vert  1.4655 ,1\rangle  (3-FF),
  \vert  1.3802,1 \rangle  (4-FF),  \vert  phi_d,\ell \rangle  (d-FF),
  \vert  0,1 \rangle \oplus \vert  1,1 \rangle  (multilayer composite fiber).
Column 2 shows sample genetic circuits. Column 3 shows input trees.
Column 4 shows base. Further details in text and subsequent figures.
}
\end{figure}

These considerations lead to a systematic classification of all
building blocks.  The results for the {\it E. coli} TRN are
summarized in Fig.~\ref{Fig:PNAS_Fig3a} and Fig.~\ref{Fig:PNAS_Fig3b},
showing how dissimilar circuits can be concisely classified by
either integer dimensions $ |n, l\rangle$ or fractal dimensions.  The count for each building block is
shown in Table~\ref{tab:FiberBBstatistics}.  We explore this
classification one circuit at a time in the next chapter.

\begin{table*}[]
    \centering
    \begin{tabular}{| c | c |}
        \hline Building Block & Number in {\it E. coli} \\
        \hline\hline
        \textbf{$\rvert n = 0, l = 1 \rangle$} & 45 \\
        \textbf{$\rvert n = 0, l = 2 \rangle$} & 13 \\
        \textbf{$\rvert n = 0, l = 3 \rangle$} & 3 \\
        \textbf{$\rvert n = 1, l = 0 \rangle$} & 13 \\
        \textbf{$\rvert n = 1, l = 1 \rangle$} & 8 \\
        \textbf{$\rvert n = 1, l = 2 \rangle$} & 3 \\
        \textbf{$\rvert n = 2, l = 0 \rangle$} & 1 \\
        \textbf{$\rvert n = 2, l = 1 \rangle$} & 1 \\
        \textbf{$\rvert \varphi_d = 1.3802.., l = 1 \rangle$} & 1 \\
        \textbf{$\rvert \varphi_d = 1.4655.., l = 1 \rangle$} & 1 \\
        \textbf{$\rvert \varphi_d = 1.6180.., l = 2 \rangle$} & 1 \\
        Simple Multilayer Fiber & 1 \\
        \hline\hline
        \textbf{Total number of building blocks} & 91 \\
        \hline
    \end{tabular}
    \vspace{10pt}
    \caption{\textbf{Building block statistics.} We show the count of
      every building block in {\it E. coli} TRN defined by the fiber
      numbers. Table reproduced from \citep{morone2020fibration}. }
    \label{tab:FiberBBstatistics}
\end{table*}


\chapter[Simple Fibration Building Blocks]{\bf\textsf{Simple Fibration Building Blocks}}
\label{chap:hierarchy_2}

\begin{chapterquote}
  We use the hierarchy
  of fibration building blocks introduced in the previous chapter
  to analyze several small circuits that constitute such building blocks. They all occur widely in biology.
  We discuss their topology and investigate their synchronous dynamics
  using a variety of different models. We suggest that the
  resulting hierarchy of increasingly complex building blocks
 may have evolved through a sequence of changes in network
 topology, obtained by repeated application of
 the lifting property for fibrations of Section \ref{lifting}.
\end{chapterquote}

\section{The primordial building block: autoregulation (AR) loop}

\begin{figure*}[h!]
  \centering \includegraphics[width=0.7\linewidth]{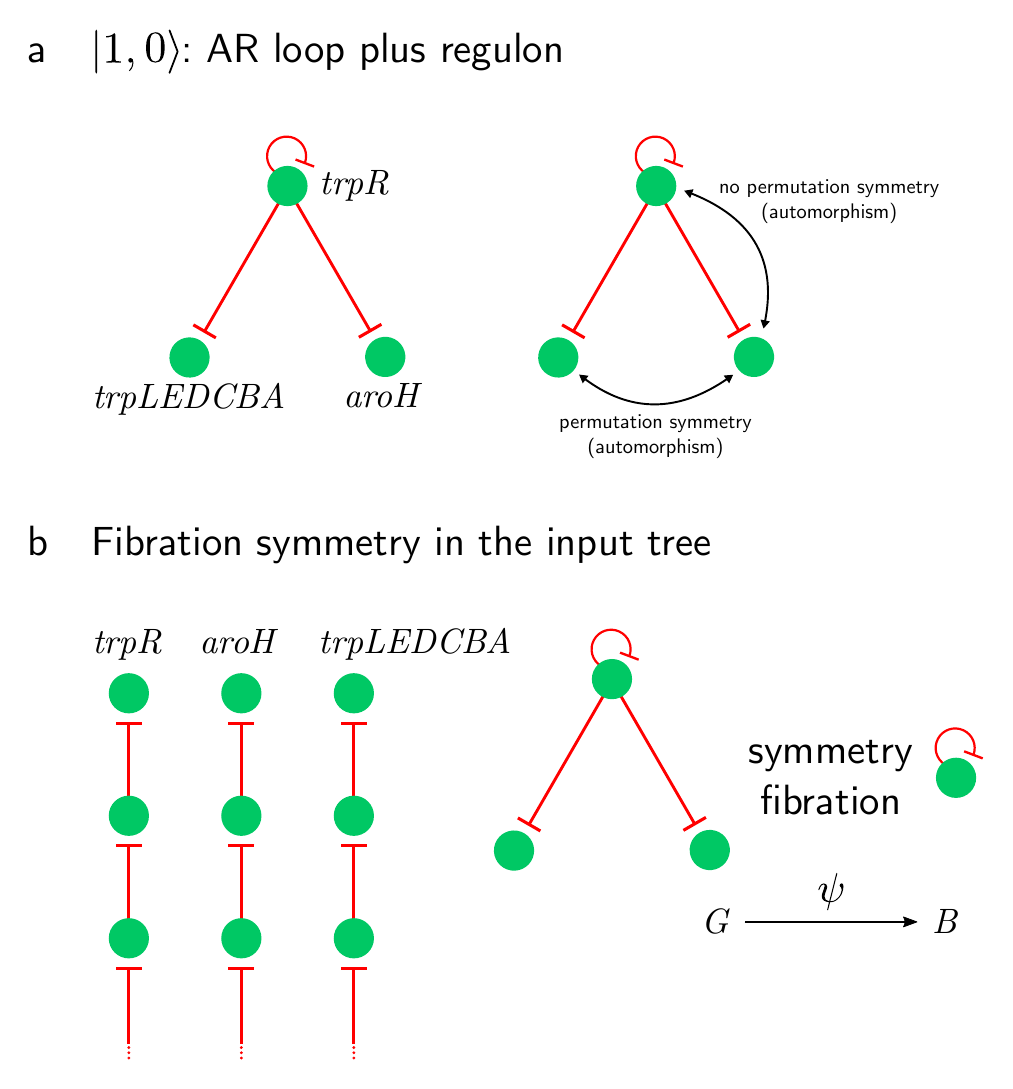}
  \caption{\textbf{Simple fibration building block 1: AR loop with
      regulon}. (\textbf{a}) Genes {\it aroH} and {\it trpLEDCBA} can
    be permuted by the $\mathbb{S}_2$ symmetry group, while
    \textit{trpR} cannot be permuted with them without changing the
    network. (\textbf{b}) \textit{trpR} receives an input only from
    itself, therefore its input tree is an infinite chain. {\it aroH}
    and {\it trpLEDCBA} receive an input from \textit{trpR}, which in
    turn receives an input from itself, turning these input trees into
    chains too. Therefore the input trees of all 3 genes are
    isomorphic to each other. Thus {\it aroH}, {\it trpLEDCBA} and
    \textit{trpR} belong to the same fiber and can synchronize their
    activity. The circuit has one loop and no external regulators, so
    it is classified as $\rvert n = 1, l = 0 \rangle$. Figure
    reproduced from \citep{leifer2021predicting}. Copyright  \copyright ~2021, The Author(s).}
  \label{ar}
\commentAlt{Figure~\ref{ar}: 
(a)   \vert  1,0 \rangle : AR loop plus regulon. Left: Green node trpR sends arrows to itself
and two other green nodes. Right: same figure with arrows top-lower right (no automorphism),
lower left/lower right (permutation symmetry, automorphism).
(b)  Fibration symmetry in the input tree. Left: Three vertical chains of three green nodes,
vertical upward arrows connecting them: trpR, aroH, trpLEDCBA. Right: Copy of (a left),
arrow to one green node connecting to itself. Label: symmetry fibration.
}
\end{figure*}

The first nontrivial form of synchronization in the hierarchy of
fibration building blocks occurs when a TF autonomously regulates its
own expression, establishing an autoregulation (AR) loop,\index{autoregulation loop }\index{AR loop } and
subsequently controls the expression of other genes. This phenomenon
is illustrated in Fig.~\ref{ar}a. This circuit is widely present in
\emph{E.~coli}.  One example is the tryptophan biosynthesis process,
where it is regulated by TrpR \citep{grove1987}. TrpR represses its
own expression, as well as the gene {\it aroH}
(2-Dehydro-3-deoxyphosphoheptonate aldolase) and the {\it trpLEDCBA}
operon, which encodes the enzymes involved in tryptophan
biosynthesis. The activation of this circuit is dependent on the
intracellular level of L-tryptophan \citep{ecocyc2018}. When the cell
contains sufficient tryptophan, the TF binds to the operon and
suppresses gene expression.

This circuit exhibits an automorphism
which corresponds to permutation of the genes regulated by {\it
  trpR} as shown in Fig. \ref{ar}a (right). It generates the symmetry group $\mathbb{S}_2$. This is a rather trivial
automorphism which appears many times in all regulatory networks. It
is called a {\it regulon}. A regulon is a group of genes that are regulated
as a unit by the same TF, and no other TF regulates any of the genes.

The interesting feature is in the AR of {\it trpR}.  This AR loop at
\textit{trpR} does not affect the automorphism permuting the regulated
genes. However, the AR loop introduces a
fibration symmetry between \textit{trpR}, \textit{trpLEDCBA}, and
\textit{aroH}. This symmetry cannot be captured by a
permutation of the nodes. For example, permuting the operon
\textit{trpLEDCBA} with \textit{aroH} preserves adjacency, but
permuting \textit{trpR} with either the operon or \textit{aroH} does
not preserve adjacency. Therefore, \textit{trpR} does not belong to
the symmetry group acting on \textit{trpLEDCBA} and
\textit{aroH}. However, \textit{trpR} is synchronized with
\textit{trpLEDCBA} and \textit{aroH} by the symmetry fibration, because they have isomorphic
input trees, as shown in Fig. \ref{ar}b (left).  The base of this
building block is simple: an AR loop as in Fig. \ref{ar}b
(right). This is the simplest form of fibration, reducing the
circuit to what could be considered as a primordial genetic circuit: a
single gene regulating itself.

Due to the AR loop, the input trees of this fiber satisfy $a_i=1$ for
all $i$, so the branching ratio is $n=1$.  This
circuit does not have any external regulators, so $l=0$.  We
therefore refer to this circuit as $ \rvert 1,0\rangle$.

While $n$ captures the loops, and $l$ captures the regulators, the
number of regulated genes is still unspecified.  Considering that 
many building blocks contain regulons, we extend the definition of
fiber numbers given in \ref{def:fiber-numbers} to include the number
$m$ of regulated genes of the fibers. This lead to the following
extended definition:
  
\begin{definition} {\bf Extended
    fiber number classification of simple fibration building blocks}.
Definition \ref{def:fiber-numbers} is extended to include the number $m$ of regulated genes in the fiber:
\begin{equation}
  | n, l, m \rangle \,: \,\,\, \mbox{\rm extended simple building block
    classification. }
\end{equation}
\label{def:extended}
\end{definition}

The building block in Fig. \ref{ar} becomes $| n, l, m \rangle = |
1, 0, 2 \rangle$, which, by the application of the fibration, becomes
$| 1, 0 \rangle$ when the 2-gene regulon\index{regulon } is collapsed. Since these two
circuits behave the same dynamically, we typically use the shorter
notation for the building block and ignore $m$.  While these are not a
complete set of invariants for circuit topologies, they provide a
useful coarse classification covering most circuits studied in this
book.

The AR building block is quite abundant in simple bacterial genomes\index{bacterial genome }
like the {\it E. coli} TRN\index{E.~coli TRN @{\it E.~coli} TRN } as shown in the statistics in Table
\ref{tab:FiberBBstatistics}.  However, AR loops are not present in
more complex genomes like eukaryotes.  The base of this circuit is
also one of the network motifs found by \cite{milo2002network} in the
same TRN.  In fact, \cite{milo2002network} already found that AR loops
are quite abundant in {\it E. coli}.  However, the fiber is more
complex than the AR motif in \citep{milo2002network}, since it
contains not only the AR base but also the regulon. Thus, the
fibration building block captures the correct synchrony of the genes,
which cannot be captured by the AR-motif building block.

Building upon this primordial circuit, the rest of the building blocks
appear in a well defined hierarchy of circuits, emerging from the
lifting property of the fibration, see Section \ref{lifting}.

\section{Dynamic repertoires in 
classes of fibration building blocks}

The simplest possible base for a fibration consists of one
autoregulated gene. By the lifting property (Definition
\ref{def:fibration}) this base can be lifted to produce its regulon,
conforming to the fiber observed in {\it E. coli}. We come back to
this idea in subsequent chapters: the evolution of the building blocks
into a hierarchical classification can be explained by the lifting
property applied to simple bases like the AR loop.  For now, we continue to
describe the simple circuits found in {\it E. coli}.\index{E. coli @{\em E. coli} }

\begin{figure}[!]
 \includegraphics[width=.8\textwidth]{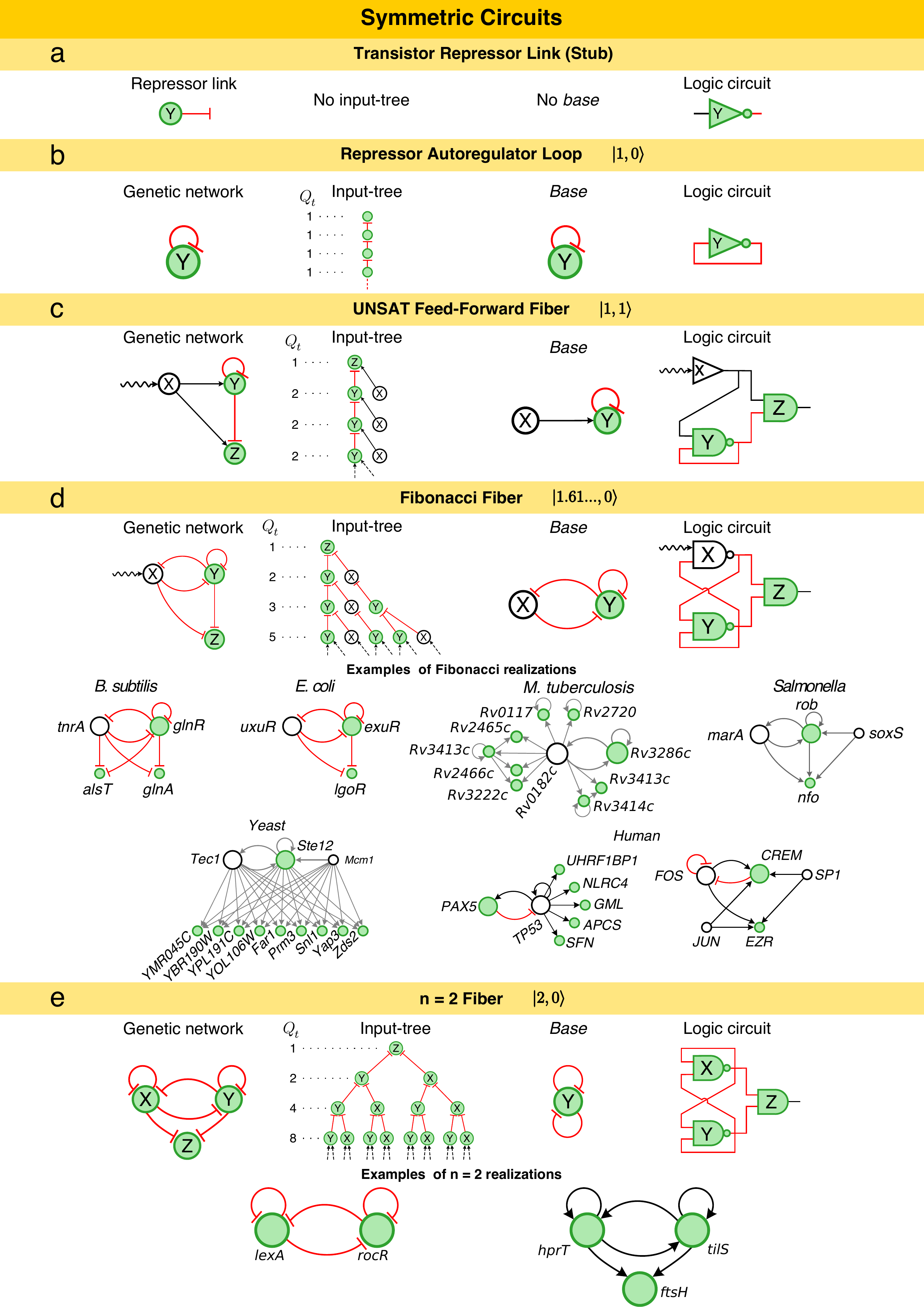} \caption{ \textbf{
 Hierarchy of simple fibration building blocks}.  Black arrows
 indicate activator links, red denotes repressors, and gray denotes
 unknown functionality. Examples are shown from all studied species.
 Symmetric circuits function as clocks. Adding autoregulation and
 feedback loops results in a hierarchy of increasingly complex
 circuits. Turning the repressor link (\textbf{a}) into repressor
 autoregulation (\textbf{b}) gives an input tree equivalent to its own
 base.  (\textbf{c}) Adding an external regulator creates the
 UNSAT-FFF, where genes Y and Z synchronize and
 oscillate. (\textbf{d}) Adding a second feedback loop gives an input
 tree that branches like the Fibonacci sequence.  (\textbf{e}) A
 second autoregulation at X results in a symmetric input tree with
 branching ratio $n=2$.  This fiber collapses into a base with two
 autoregulators. Figure reproduced
 from \citep{leifer2020circuits}. Copyright \copyright ~2020, Leifer {\it et al.} } \label{fig3_plos}
\commentAlt{Figure~\ref{fig3_plos}: A selection of circuits with
fibration symmetries. 
Illustrative purposes only. Any relevant ones discussed individually later.
No alt-text required.
}
\end{figure}

The procedure to build more complex fibers can be systematically
extended through an algebra of circuits that adds external regulators
and loops to grow the base of symmetric circuits
(Fig. \ref{fig3_plos}). In this space, the AR is the core loop unit
$|n=1, l=0\rangle$ (Fig.~\ref{fig3_plos}b).

From the AR base $| n=1, l=0\rangle$ we can increase the number of
regulator genes, $| 1, l\rangle$ with $l > 1$, as well as the
number of regulated genes, by lifting. If this is done by following the lifting
of the base, it does not affect the complexity (measured by the branching ratio\index{branching ratio } $n$ of the input tree\index{input tree })
because all the relevant dynamics remain constrained to the sole loop
in the fiber. This changes as soon as the fiber feeds information
back to the external regulator, as for the circuits in
Fig. \ref{fig3_plos}d, where the gene Y now regulates its own
regulator gene X.  
Addition of a second feedback loop
   results in an input tree that follows the Fibonacci sequence
   $a_t=1,2,3,5,8,...$. Here, gene X is not part of the 
     base. The branching structure of the input tree implies that the
   Fibonacci fiber can oscillate and synchronize.
   A second autoregulation at X results in a
   symmetric input tree with $a_t = 2a_{t-1}$ and branching ratio
   $n=2$.  This fiber collapses into a  base with two
   autoregulators.  Examples of the $n=2$ fiber are two fibers
   from the regulatory networks of {\it Bacillus subtilis}.  
    The important component of these circuits is the delay
in the feedback loop through the regulator from Y~$\rightarrow$~X and
back to Y.  We show in Chapter \ref{chap:breaking} that these
building blocks form the base of memory circuits that emerge via the
lifting property and by symmetry breaking, forming memory devices that mimic
the universal storage devices of computer memories. 

\begin{figure}
\centering \includegraphics[width=\textwidth]{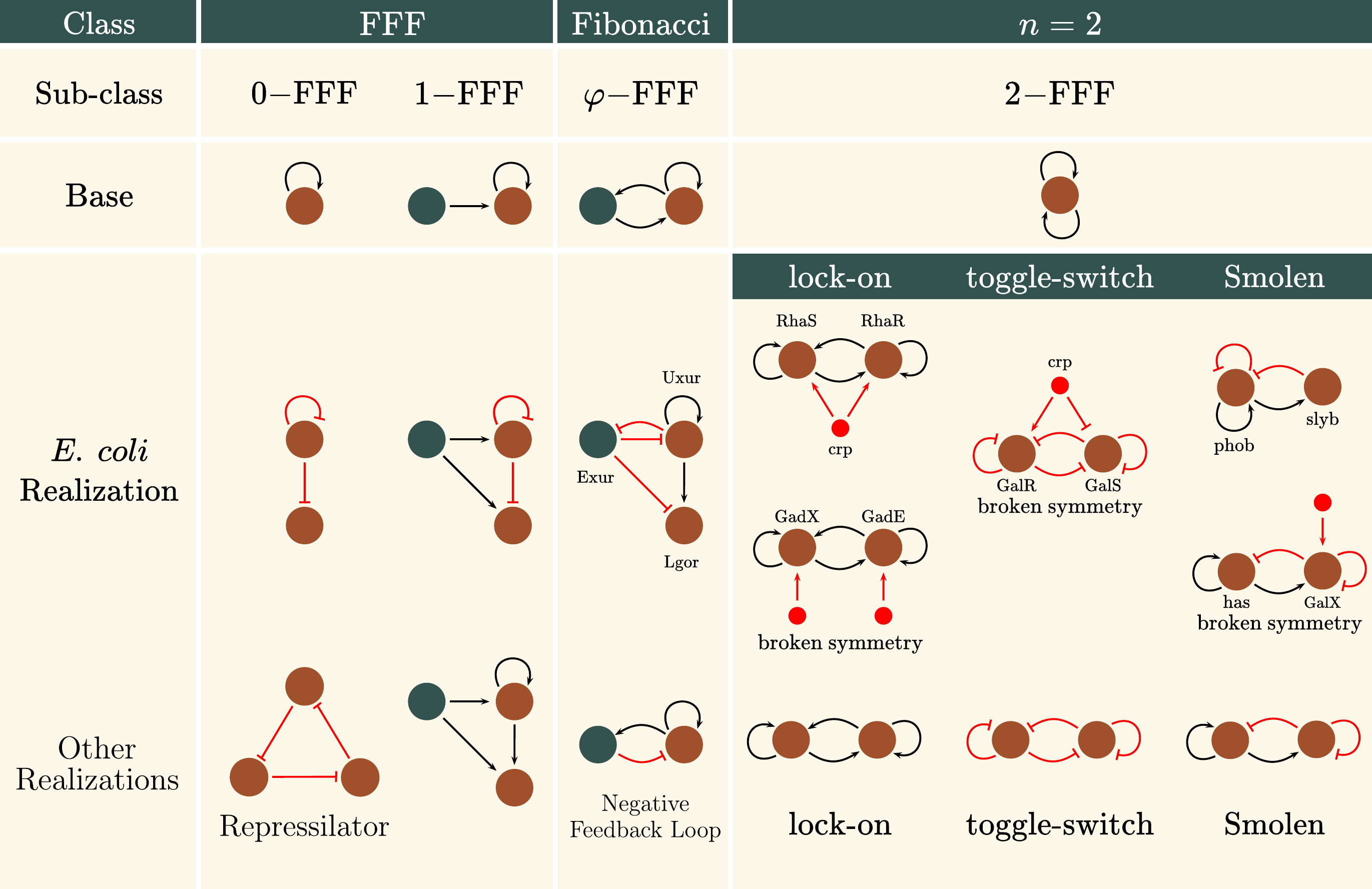}
\caption{\textbf{Evolution of simple fibration building blocks by lifting
    (duplication) and addition of edges.} Realizations of fibration
  building blocks and their broken symmetry version.  The
  consideration of external regulator nodes, feedback loops, and
  autoregulation loops results in circuits with an increasing level of
  complexity. {\em First column of graphs}: {\bf FFF} (feed-forward fiber). The base
  for the $0$-FFF and ($1$-FFF), namely, AR (autoregulation) loop and
  the AR loop with an external regulation unfold (or `lift') into different
  circuits found in the {\em E. coli} GRN, as the FFF, or synthetic
  ones, as the Repressilator. {\em Second column}: {\bf Fibonacci}. The
  Fibonacci ($\phi$-FF) base results in the Fibonacci circuit by
  unfolding the AR loop into two synchronized genes. An example present in the
  GRN of {\em E. coli} is the circuit made by the {\em Uxur}, {\em
    Exur}, and {\em Lgor} genes. {\em Third column}: {\bf {\em n}=2}. The
  presence of the second regulation loop allows diversity in
  combinations between gene regulations, resulting in a different subclass
  $n=2$ with different circuits and dynamics. The realizations
  presented are both symmetric and broken symmetric ones found in {\em
    E. coli} from \citep{leifer2020circuits}, which are related to
  previous GRNs found in the literature (lock-on, toggle-switch, and
  Smolen). Figure reproduced from \citep{stewart2024dynamics}. Copyright \copyright ~2024, The Author(s).  }
\label{fig:evolution}
\commentAlt{Figure~\ref{fig:evolution}: 
A selection of circuits with fibration symmetries. 
Illustrative purposes only. Any relevant ones discussed individually later.
No alt-text required.
}
\end{figure}

Lifting of genes mimics the
duplication mechanism in genetic evolution, see
Fig. \ref{fig:duplication}. Therefore, from
the starting point of an AR base, we can create a hierarchy of
circuits that mimics a biological evolutionary dynamic of growing
circuits, which increases the number of regulated genes
and adds edges in an orderly fashion. The simplest circuits involved are shown in Fig. \ref{fig:stability_circuits} in Chapter 8, where
we analyzed some features of their dynamics. Now we add
biological and dynamical detail.
The initial steps of this
hierarchical evolution of circuits of increasing complexity are
depicted in Fig. \ref{fig:evolution}, and will be discussed next,
circuit by circuit.

\section{The feed-forward fiber - FFF}
\label{sec:fff}
\index{feed-forward fiber }

We have seen that the AR base can be extended by lifting the base to
the regulon made of $m$ regulated genes. This process does not change
the dynamics of the resulting fiber since the regulon can be collapsed
back to the base by the fibration without changing its dynamics.  The
lifting can be done to any number $m$ of genes, increasing the
complexity from $m=0$ at the base to $m$ genes in the fiber.  Beyond
adding regulated genes, we can imagine that another way to increase
the complexity of the AR base is to add regulators by increasing
$l=0$ to $l=1$. When we also add an $m=1$ regulated gene we
create a new building block that is quite abundant across all species
and bio-networks.  We call it the Feed-Forward
Fiber (FFF) \citep{morone2020fibration} denoted by $\rvert 1, 1 \rangle$.

\begin{figure}
  \centering
  \includegraphics[width=0.7\linewidth]{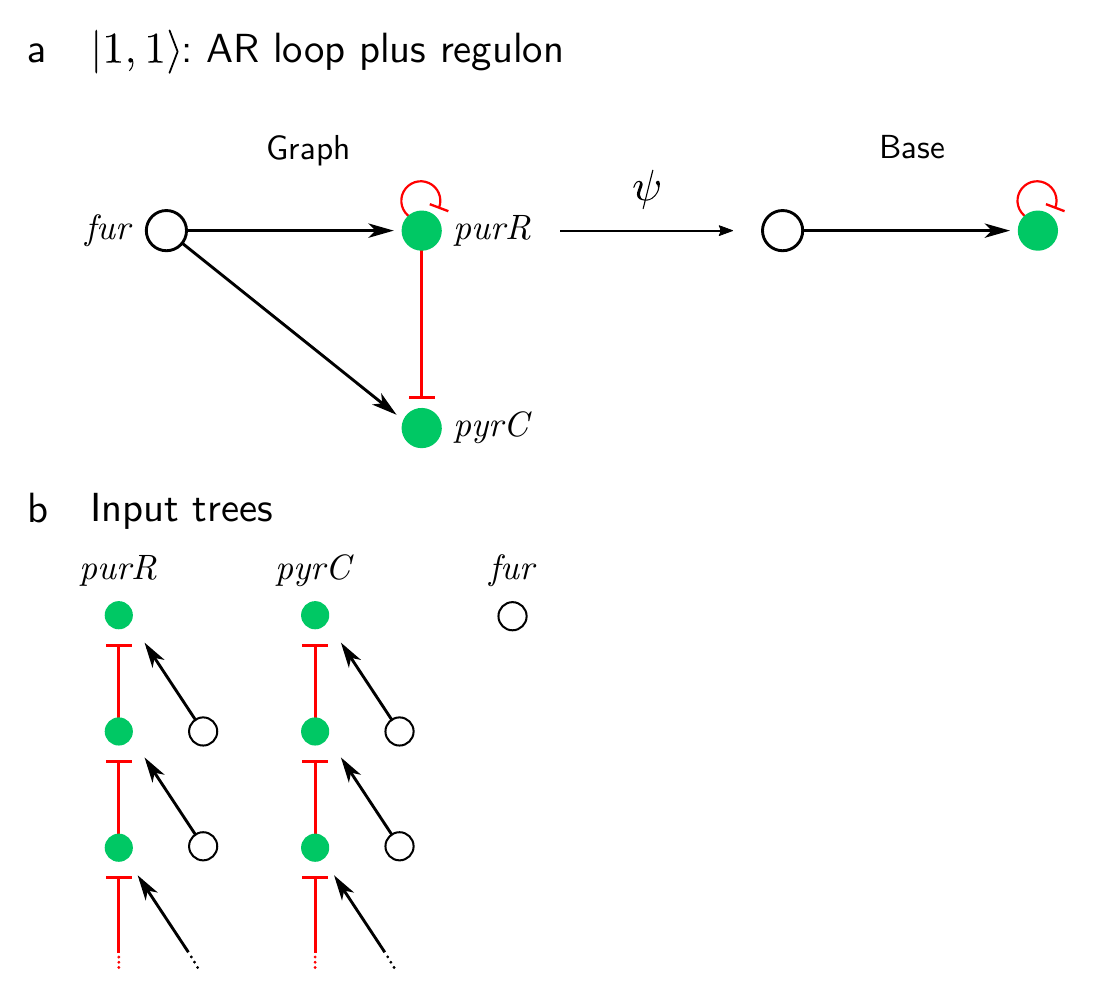}
  \caption{\textbf{Simple fibration building block 2: Feed-forward fiber
      (FFF)}.  It is composed of an AR loop with a regulon and
    external regulator. (\textbf{a)} \textit{purR} and its target gene
    \textit{pyrC} regulated by {\it fur} form a FFF. FFF has one loop
    and one external regulator and is classified as $\rvert n = 1,
    l = 1 \rangle$. \textit{purR} and \textit{pyrC} belong to the
    same fiber (shown in (\textbf{b)}) and are collapsed
    by the fibration $\psi$, while {\it fur} is left
    untouched. (\textbf{b)} \textit{purR} receives an input from
    itself, creating an infinite chain, together with regulator
    \textit{fur}, which does not have any inputs. Therefore the input
    tree of \textit{purR} is an infinite
    chain with an additional input on each layer. Similarly, \textit{pyrC} receives an input
    from \textit{purR} that leads to an infinite chain, and
    \textit{fur} creates an additional input. \textit{fur} does not
    receive any inputs, and therefore has an input tree of height
    0. Input trees of \textit{purR} and \textit{pyrC} are isomorphic,
    therefore \textit{purR} and \textit{pyrC} belong to the same fiber
    and synchronize their activity.  Figure reproduced from
    \citep{leifer2021predicting}.  Copyright  \copyright ~2021, The Author(s). }
  \label{fff}
\commentAlt{Figure~\ref{fff}: 
Described in caption/text. No alt-text required.
}
\end{figure}

An example of an FFF\index{FFF } is observed in the purine biosynthesis circuit in
\emph{E.~coli}, Fig.~\ref{fff}a. It is composed of the repressor TF
\textit{purR} and its target gene \textit{pyrC}, both regulated by the
master regulator {\it fur}.  The input trees of the genes in this FFF
are shown in Fig.~\ref{fff}a. We see that {\it pyrC} and {\it purR}
receive the same inputs from both {\it fur} and {\it purR}. On the
first layer of the input tree, we find that {\it purR} is an
autoregulator and also regulates the gene.  The second level of the
input tree contains exactly the same genes, and so on. The loop at {\it purR} creates an
input tree with infinitely many layers.

The FFF resembles the
feed-forward loop (FFL) network motif introduced by
\cite{milo2002network}, except for an additional AR at the
intermediate TF {\it purR}. This crucial addition transforms an FFL into an
FFF composed of two genes in the synchronized fiber.  This type of
building block is abundant in bacterial TRN
\citep{leifer2020circuits}, see Table \ref{tab:FiberBBstatistics}.

The dynamics of the $\rvert n = 1, l = 1 \rangle$ FFF building
block has been studied in detail in \citep{leifer2020circuits}.
The FFF has a synchronous oscillatory solution for
a certain type of regulation, which allows it to play the role of a
clock in the system. Therefore, the FFF is of high interest for
information processing networks. We now compare it with the analogous
FFL motif.

\subsection{Feed-forward loop (FFL) network motif}
\label{sec:comparison-ffl}
\index{feed-forward loop }\index{FFL }

\begin{figure*}[t!]
  \centering
  \includegraphics[width=0.7\linewidth]{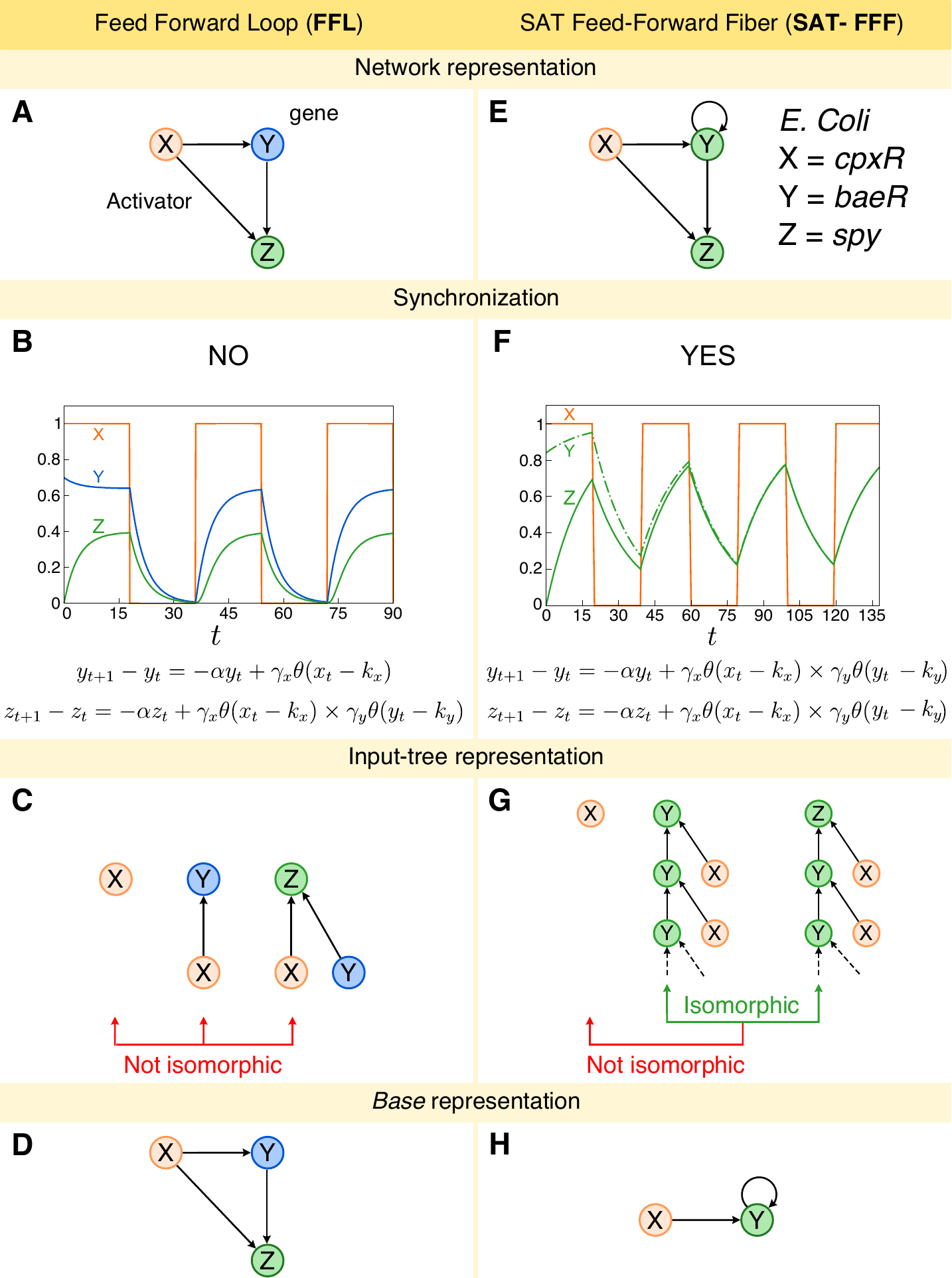}
  \caption{\textbf{Comparison between the Feed-Forward Loop (FFL) and
      feed-forward fiber (FFF).}  (\textbf{a}) FFL network
    representation.  (\textbf{b}) Numerical solution of FFL
    dynamics. The expression levels of genes Y and Z do not
    synchronize. The oscillation pattern presented is due to the
    square-wave behavior of gene X expression levels. (\textbf{c})
    Input trees of FFL. The input trees of genes X, Y, and Z are not
    isomorphic, so their expression levels do not
    synchronize.  (\textbf{d}) Base representation of FFL. The base is
    the same as the original circuits since there are no symmetries.
    (\textbf{e}) SAT-FFF network representation. Addition of
    autoregulation leads to a symmetry between the expression levels
    of genes Y and Z.  (\textbf{f}) The numerical solution of the
    SAT-FFF dynamics shows synchronization of the expression
    levels of genes Y and Z.  Again, the oscillation is due to the
    wave-like pattern of X.  (\textbf{g}) Genes Y and Z have
    isomorphic input trees and synchronize. The input tree of the
    external regulator X is not isomorphic to these, even though it
    directly regulates the fiber.  (\textbf{h}) Since Y and Z
    synchronize, gene Z can be collapsed into Y, resulting in a
    simpler base representation. Figure reproduced from
    \citep{leifer2020circuits}. Copyright \copyright ~2020, Leifer {\it et al.}}
  \label{fig1_plos}
\commentAlt{Figure~\ref{fig1_plos}: 
Described in caption/text. No alt-text required.
}
  \end{figure*}

We consider the dynamics of the most abundant network motif in
gene regulatory networks, the feed-forward loop (FFL)
\citep{shen2002network,alon2003a,alon2003b}. In particular
the coherent cFFL motif consists of three activator genes X, Y, and Z,
where the transcription factor expressed by gene X positively
regulates the transcription of Y and Z, and in turn Y regulates Z
(Fig. \ref{fig1_plos}a).  Figure ~\ref{fig1_plos}a shows an example of
the cFFL motif in {\it E. coli} with X={\it cpxR}, Y={\it baeR}, and
Z={\it spy}.  Numerical and analytic solutions for the expression
levels of the genes in the cFFL (Fig.~\ref{fig1_plos}b demonstrate
that the FFL cannot attain synchronization (unless for specific {\it ad hoc}
settings of parameters) nor oscillations in expression levels.  This is
consistent with previous research which has interpreted the
functionality of the FFL as a sign-sensitive signal delay in
transcriptional networks  \citep{alon2003b}.

\cite{leifer2020circuits} illustrate this result by presenting an
analytic solution of the FFL using a simple model of gene expression
dynamics that uses a discrete-time, continuous state variable model
with a logic Boolean interaction function in the spirit of the 
Glass--Kauffman model of biochemical
networks \citep{kauffman1973thelogical}. \cite{alon2019},
\cite{alon2003a} and \cite{alon2003b} studied the dynamics of the
expression levels $y_t$ and $z_t$ of genes Y and Z, respectively, as a
function of time $t$ in the cFFL. The discrete dynamical system is given by the following difference equations:
\begin{equation}
\begin{aligned}
y_{t+1} &= (1-\alpha) y_t + \gamma_x \theta(x_t-k_x),\\ z_{t+1} &=
(1-\alpha) z_t + \gamma_x \theta(x_t-k_x) \gamma_y
\theta(y_t-k_y)
\end{aligned}
\label{eq:fflbool}
\end{equation}
where $x_t$ is the expression level of gene X, $\alpha$ is the
degradation rate of the gene, $\gamma_x$ and $\gamma_y$ are the
strength of the interaction representing the maximum expression rate
of genes X and Y, respectively, and the thresholds $k_x$ and $k_y$ are
the dissociation constant between the transcription factor and biding
site. The expression level is measured in terms of abundance of gene
product, e.g., mRNA concentration.  The Heaviside step functions
$\theta(x_t-k_x)$ and $\theta(y_t-k_y)$ represent the activator
regulation from gene X and Y, respectively. They represent the Boolean
logic approximation of Hill input functions in the limit of strong
cooperativity \citep{alon2019}.  We consider an AND gate for the
combined interaction of transcription factors of genes X and Y onto
the binding sites of gene Z.  \cite{leifer2020circuits} show that
analogous results can be obtained with an OR gate and with ODE
continuum models.  Solutions for the FFL have been considered in the
literature \citep{alon2019}. Here, we adapt those results to the
particular models used in our studies to perform consistent
comparisons with the solutions of the FFF discussed next.

\cite{leifer2020circuits} showed that the expression levels of the
genes Y and Z do not synchronize; that is, $y_t$ and $z_t$ are not the same for all $t$. 
This behavior is exemplified in Fig.~\ref{fig1_plos}b, which shows the
solution for a particular set of parameters resulting in a
non-synchronized state. Such a state is obtained under the initial conditions
$y_0>k_y$ and $\alpha<\gamma_x$. Specifically, we use the following parameters:
$\alpha = 0.2$, $\gamma_x = 0.12$, $\gamma_y = 0.7$, $k_x = 0.5$ and
$k_y = 0.1$.  For this combination, $y_t$ and $z_t$ do not synchronize
since $y_t$ saturates at $y_t \to \gamma_x/\alpha = 0.6$ when $t \to
\infty$, and $z_t$ saturates at $z_t \to \gamma_x\gamma_y/\alpha =
0.42$, for $t \to \infty$.

In the figure, we set $x_t$ equal to a square wave and then monitor
the expression levels of $y_t$ and $x_t$. When $x<k_x$, both $y_t$ and
$z_t$ decay exponentially to zero. On the other hand, when $x>k_x$,
both variables evolve to saturate again at $y_t = \gamma_x/\alpha$ and
$z_t = \gamma_x\gamma_y/\alpha$, in agreement with the analytical
solution.

These results are consistent with a continuous variable approach
\citep{alon2019,alon2003a,alon2003b} using ODEs governing the dynamics
of expression levels $y(t)$ and $z(t)$:
\begin{equation}
    \begin{aligned}
    \dot{y}(t) &= -\alpha y(t) + \gamma_x \theta(x(t)-k_x),\\
    \dot{z}(t) &= -\alpha z(t) + \gamma_x \theta(x(t)-k_x)\ \gamma_y \theta(y(t)-k_y).
    \end{aligned}
    \label{eq_dynresults:fflode}
\end{equation}

This model contains many assumptions about the values of the different
parameters. For instance, $\alpha$ is the same for
both genes, and so is $\gamma_x$. In Section \ref{sec:more} we refine this
idealistic model of gene expression to discuss how these
approximations can be broken in a more realistic model.

This model shows that the expression levels from genes Y
and X do not synchronize.
In addition, $y(t)$ and $z(t)$ also do not reach
oscillatory states, in accordance with the results of the discrete
time model.

\subsection{Feed-forward fiber synchronization }

Fibers can predict which genes can synchronize, but they cannot
predict what kinds of synchronized states---e.g.~fixed point
synchronization or limit cycle oscillations---these genes may
reach. To determine the type of dynamical state achieved by the fiber,
we need to analyze its dynamics using systems of ODEs.  In general,
more complex input tree structures allow circuits to display
additional dynamical states. For instance, while a FFF fiber with all
activators leads to a simple fixed point, repressors and cycles in the
fiber can give rise to oscillatory limit cycles. Even a minimal loop, such as the AR in an
FFF, can produce oscillatory dynamics if the regulatory interactions
contain a delay \citep{leifer2020circuits}.  In reality, translation
and transcription might create such delays. Longer cycles, like those
observed in the Fibonacci fibers, allow for more complex functionality
beyond the simple FFF regulation of a single input function and can
yield oscillations with varying periods and more stability across the
parameter space, which extends the signal processing capabilities of
the circuits. We systematically describe the dynamics of the
simple building blocks, starting with the FFF.

The lack of synchronization in the FFL is easily explained by the lack
of symmetry in this circuit: gene Z receives input from X and Y, while
gene Y, instead, only from X, and therefore the inputs are not
symmetric (Fig.~\ref{fig1_plos}c).  However, a search of motifs in
biological networks show that the FFL circuit does not appear alone
in the bacterial TRN: instead, it often appears in conjunction with the AR
loop (like Y={\it baeR} in Fig.~\ref{fig1_plos}e) forming a synchronized FFF.  Because
 network motif algorithms search for these two circuits
separately, they were never found together, even though they
appear together, forming the FFF in the bacterial TRN. This
exemplifies the problem with network motifs or any searching
algorithm based on statistics alone: they have difficulties finding biologically meaningful circuits because statistical significance does not guarantee biological functionality.

Numerical simulations in Fig.~\ref{fig1_plos}f and analytical
solutions of the FFF done by \cite{leifer2020circuits} confirm 
the synchronization of genes Y and Z into coherent coexpression, as
predicted by the fibration.

The FFF with activator regulation in Fig.~\ref{fig1_plos}e has a
simple dynamic converging to a synchronized fixed point: all
interactions are satisfied, so the Heaviside step functions
evaluate to 1. We call this circuit SAT-FFF since its interactions
are always {\it satisfied} and positive.  Instead, when the
autoregulation is a repressor, the loop behaves as a logical NOT gate.
When expression is high, it inhibits itself, shifting to a low
state. Dually, if it is low, then it promotes itself to shift to a high
state. Hence, the activity of this gene oscillates
indefinitely \citep{atkinson2003development,kauffman1973thelogical}. We
call this circuit the UNSAT-FFF since the interactions are
{\it unsatisfied}. This is the simplest expression of {\it frustration}
 \citep{anderson1977more,kauffman1973thelogical}, a core concept in
complexity and spin glasses, which refers to a system that is always
in tension and thus never reaches a stable fixed configuration.

\subsection{Satisfied feed-forward fiber (SAT-FFF) synchronization }
\label{sec:satfff}
\index{satisfied feed-forward fiber }\index{SAT-FFF }

\begin{figure}[ht!]
    \centering \includegraphics[width=.4\textwidth]{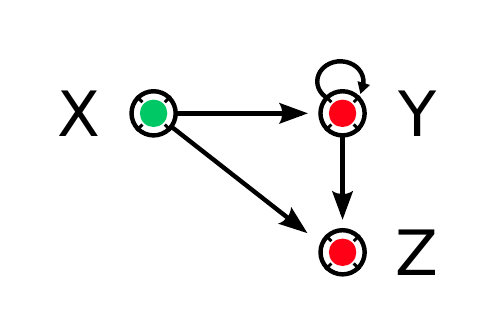}
    \caption{\textbf{SAT-FFF network representation.} SAT-FFF is an FFF
      in which all links between the nodes in the fiber are activation
      links, therefore, they are satisfied.  }
    \label{Fig:SAT-FFF}
\commentAlt{Figure~\ref{Fig:SAT-FFF}: 
Graph with nodes X(green), Y and Z (red). Sharp arrows XY, XZ, YY, YZ.
}
\end{figure}

The SAT-FFF is a feed-forward fiber with activator autoregulation
where all interactions are satisfied. That is, it does not present the
phenomenon of frustration, and the dynamics converge to a fixed
point. This can easily be seen by considering gene X to be high, which makes
genes Y and Z high, too. Finally, the configuration satisfies the AR
loop, so all bonds are satisfied. Below, we describe the solutions of
the SAT-FFF circuit in a discrete-time continuous variable model and
 ODE model.

\subsubsection{A. SAT-FFF discrete time model}
\label{Subsubsec:sat-discrete}

The discrete-time dynamics of the SAT-FFF with a logic interaction
term is given by:
\begin{equation}
    \begin{aligned}
    y_{t+1} &= (1-\alpha)y_t + \gamma_x\theta(x_t-k_x)\ \gamma_y\theta(y_t - k_y),\\
    z_{t+1} &= (1-\alpha)z_t + \gamma_x\theta(x_t-k_x)\ \gamma_y\theta(y_t - k_y).
    \end{aligned}
\end{equation}

The Heaviside function $\theta(y_t-k_y)$ represents 
activator feedback on the autoregulation of the Y gene. We consider an
AND gate for the interactions \citep{alon2019} (analogous results can
be obtained for OR gates). Writing down the set of equations for the
rescaled variables $\psi_t = y_t/k_y$ and $\zeta_t = z_t/k_y$, we get:
\begin{equation}
    \begin{aligned}
    \psi_{t+1} &= \beta\psi_t + \alpha\lambda\theta(x_t-k_x)\theta(\psi_t-1),\\
    \zeta_{t+1} &= \beta\zeta_t + \alpha\lambda\theta(x_t-k_x)\theta(\psi_t-1).
    \end{aligned}
\end{equation}

\begin{figure}[b!]
    \includegraphics[width=\textwidth]{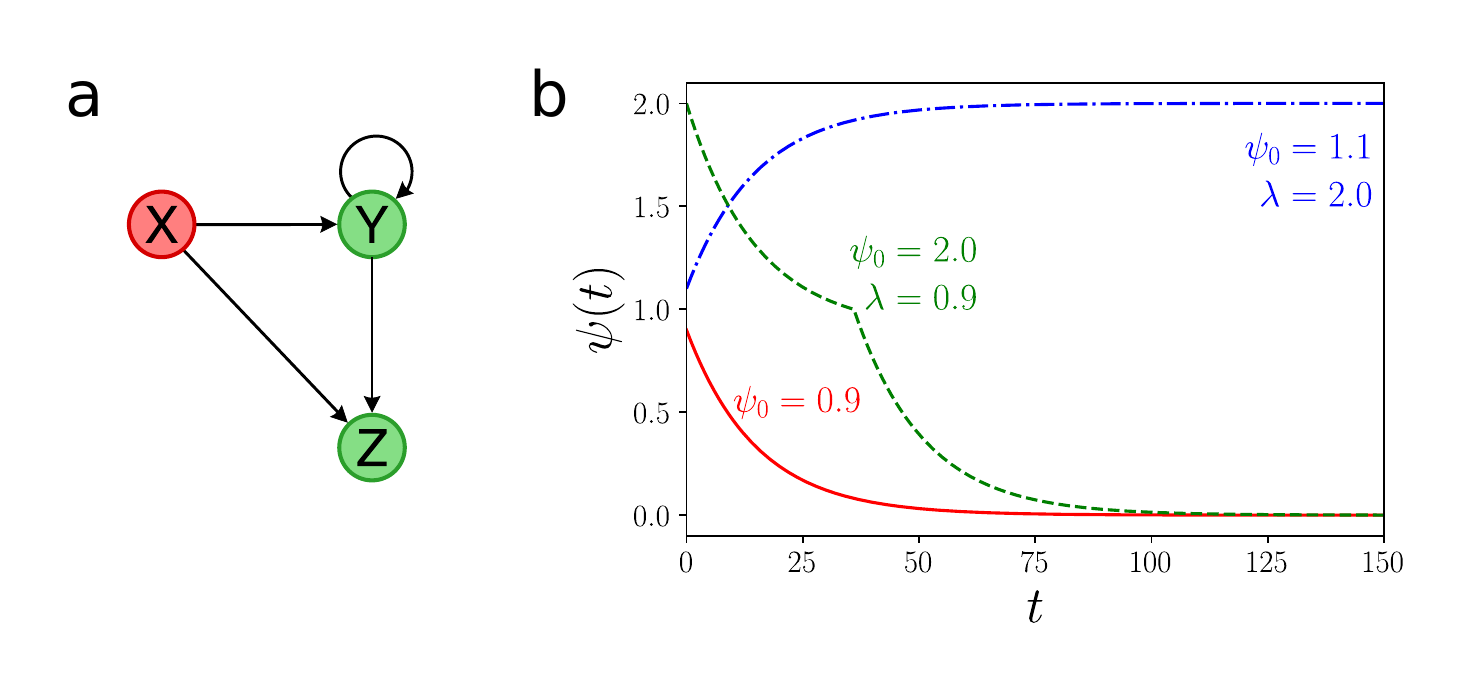}
    \centering
    \caption {\textbf{SAT-FFF}. (\textbf{a}) Network representation. (\textbf{b}) Different behaviors for the analytical solutions of
      $\psi_t$ depending on $\psi_0$ and $\lambda$. Figure reproduced
      from \citep{leifer2020circuits}. Copyright \copyright ~2020, Leifer {\it et al.}}
    \label{Fig:PLOS_smfig01}
\commentAlt{Figure~\ref{Fig:PLOS_smfig01}: 
(a) Graph with nodes X(green), Y and Z (red). Arrows XY, XZ, YY, YZ.
(b) Graph of analytic solutions of model ODE. One increases monotonically and saturates.
The others two decrease monotonically to zero.
}
\end{figure}

Here $\lambda = \gamma_x\gamma_y/\alpha k_y$ and
$\beta = (1-\alpha)$. Since the second terms on the
right-hand sides of both equations are equal, the dynamical variables
$\psi_t$ and $\zeta_t$ must synchronize, as well as $y_t$ and
$z_t$. Again, considering $x_t = x$ constant, for $x<k_x$, the
solutions for $\psi_t$ and $\zeta_t$ are trivial: both variables decay
exponentially as $\psi_t = \psi_0e^{-t/\tau}$ and $\zeta_t =
\zeta_0e^{-t/\tau}$, where $\tau^{-1} = -\log(1-\alpha)$. This
behavior is shown by the red solid line in
Fig.~\ref{Fig:PLOS_smfig01}b with $\psi_0 = 0.9$.

In terms of the iterative map, the dynamics of the SAT-FFF for the rescaled variable $\psi_t$ with $x>k_x$ is:
\begin{equation}
    \psi_{t+1} =  \beta\psi_t +  \alpha\lambda\theta(\psi_t-1) \equiv f(\psi_t),
    \label{eq_dynresults:fffpositiveY}
\end{equation}
so we find
\begin{equation}
    f^t(\psi) = f^{t-1}(\beta\psi)\theta(1-\psi) + f^{t-1}(\beta\psi+\lambda)\theta(\psi-1).
    \label{eq_dynresults:fffpositivemap}
\end{equation}
\noindent
This iterative map $\psi_t = f(\psi_t)$ provides different solutions
depending on $\psi_0$. Similarly to the case  $x<k_x$, if $\psi_0<1$
the solution decays to zero as $\psi_{t}= \psi_0e^{-t/\tau}$. However,
if $\psi_0>1$, there are two possibilities, depending on the values of
$\lambda=\gamma_x\gamma_y/\alpha k_y$.

First, if $\lambda>1$, the solution for both rescaled variables
converges to $\lambda$ as $\psi_t=\psi_0e^{-t/\tau} +
\lambda(1-e^{-t/\tau})$ and $\zeta_t=\zeta_0e^{-t/\tau} +
\lambda(1-e^{-t/\tau})$, such that
$\psi_{t\rightarrow\infty}\rightarrow \lambda$ and
$\zeta_{t\rightarrow\infty}\rightarrow \lambda$, represented by the
blue dash-dotted line in Fig.~\ref{Fig:PLOS_smfig01}b. For this case,
we use $\psi_0 = 1.1$ and $\lambda = 2$.

For $\lambda<1$, $\psi_t$ approaches $1$ at a time $t^\ast$ given by
\begin{equation}
    t^\ast = \left\lceil\frac{1}{\log(1 - \alpha)}\log\Bigg(\frac{1-\lambda}{\psi_0 - \lambda}\Bigg)\right\rceil .
\end{equation}
For $t>t^\ast$ the solutions decay to zero as $\psi_t=e^{-(t -
  t^\ast)/\tau}$ and $\zeta_t = \zeta_{t^\ast}e^{1(t -
  t^\ast)/\tau}$. This behavior is represented in
Fig.~\ref{Fig:PLOS_smfig01}b by the dashed green line, where we use
$\psi_0 = 2$ and $\lambda = 0.9$. The rescaled variables $\psi_t$ and
$\zeta_t$ always synchronize, and so do $y_t$ and $z_t$.  This can be
proved by finding the difference $\varepsilon_t = \psi_t - \zeta_t$. For
all the cases discussed above, $\varepsilon_t$ decays exponentially fast
as $\varepsilon_t = \left(\psi_0 - \zeta_0\right)e^{-t/\tau}$.

We can use the solution with $x_t$ constant to qualitatively
understand the SAT-FFF in general. An example of a SAT-FFF with
non-constant $x_t$ is depicted in Fig.~\ref{fig1_plos}. As shown,
the variables $y_t$ and $z_t$ synchronize, but with no internal
oscillations. We feed an external oscillatory pattern of $x_t$ as a
square wave. For $x_t<k_x$, both $y_t$ and $z_t$ decay
exponentially. When $x_t>k_x$, they tend to saturate at
$\gamma_x\gamma_y/\alpha$. The SAT-FFF synchronizes at a fixed point.

\subsubsection{B. SAT-FFF ODE model}
\label{Subsubsec:sat-ode}

Now we consider the ODE model of SAT-FFF to confirm results from the
discrete time continuous variable model in
\citep{leifer2020circuits}. The dynamics of gene X is driven by
outside sources, so we  consider only the dynamics of genes Y and Z,
which are described by:
\begin{equation}
    \begin{aligned}
        \dot{y} & = - \alpha y(t) + \gamma_x \theta(x(t)-k_x) \  \gamma_y \theta(y(t)-k_y), \\
        \dot{z} & = - \alpha z(t) + \gamma_x \theta(x(t)-k_x) \  \gamma_y \theta(y(t)-k_y).
    \end{aligned}
    \label{eq_dynresults:SAT-FFF-1}
\end{equation}
Taking $\psi(t) = y(t)/k_y$, $\zeta(t) = z(t)/k_y$ and $\delta = \gamma_x \gamma_y / k_y$ we transform \eqref{eq_dynresults:SAT-FFF-1} to:
\begin{equation}
    \begin{aligned}
        \dot{\psi} & = - \alpha \psi(t) + \delta \,\, \theta(x(t)-k_x) \  \theta(\psi(t)-1), \\
        \dot{\zeta} & = - \alpha \zeta(t) + \delta \,\, \theta(x(t)-k_x) \  \theta(\psi(t)-1).
    \end{aligned}
    \label{eq_dynresults:SAT-FFF-2}
\end{equation}
Without loss of generality, consider the case when $x(t)=x$ is constant over time. If $x<k_x$ then the solution of \eqref{eq_dynresults:SAT-FFF-2} is:
\begin{equation}
    \begin{aligned}
        \psi(t)_{x<k_x} & = \psi_{\rm 0}e^{-\alpha t},  \\
        \zeta(t)_{x<k_x} & = \zeta_{\rm 0}e^{-\alpha t},
    \end{aligned}
    \label{eq_dynresults:SAT-FFF-solution_1}
\end{equation}
where $\psi_{\rm 0}$ and $\zeta_{\rm 0}$ are the initial
conditions. 

Consider the case
$x>k_x$. Now \eqref{eq_dynresults:SAT-FFF-2} transforms into:
\begin{equation}
    \begin{aligned}
        \dot{\psi} & = - \alpha \psi(t) + \delta \,\, \theta(\psi(t)-1), \\
        \dot{\zeta} & = - \alpha \zeta(t) + \delta \,\, \theta(\psi(t)-1).
    \end{aligned}
    \label{eq_dynresults:SAT-FFF-3}
\end{equation}  
Because Y and Z belong to the same fiber, $\psi(t)$ and
$\zeta(t)$ synchronize and therefore $y(t)$ and $z(t)$ also synchronize, so we may consider only the dynamics of the first equation:
\begin{equation}
    \dot{\psi} = - \alpha \psi(t) + \delta  \,\, \theta(\psi(t)-1).
    \label{eq_dynresults:SAT-FFF-4}
\end{equation}
It is easy to see that for $\psi_{\rm 0} < 1$,
\eqref{eq_dynresults:SAT-FFF-solution_1} is the solution of
\eqref{eq_dynresults:SAT-FFF-4}. When $\psi_{\rm 0} > 1$ and
$\delta/\alpha > 1$, the solution is:
\begin{equation}
    \psi(t) = \delta/\alpha + (\psi_{\rm 0}-\delta/\alpha)e^{-\alpha t}.
    \label{eq_dynresults:SAT-FFF-solution_2}
\end{equation}

In the case $\psi_{\rm 0} > 1$ and $\delta/\alpha < 1$, the dynamics
of $\psi$ is split into two parts: one before $\psi$ decays to 1,
and the other one after $\psi$ crosses 1. The time when $\psi(t)$
crosses 1 is:
\begin{equation}
    t_c = \frac{1}{\alpha} ln(\frac{\psi_{\rm
        0}-\delta/\alpha}{1-\delta/\alpha}),
    \label{eq_dynresults:SAT-FFF-critical_time}
\end{equation}
and the dynamics can be written as:
\begin{equation}
    \begin{aligned}
        \psi(t) & = \delta/\alpha + (\psi_{\rm 0}-\delta/\alpha)e^{-\alpha t}         & \ \ {\rm for} \ \ t \in [0, t_c], \\
        \psi(t) & = \frac{\psi_{\rm 0}-\delta/\alpha}{1-\delta/\alpha} e^{-\alpha t}  & \ \ {\rm for} \ \ t \in [t_c, \infty].
    \end{aligned}
    \label{eq_dynresults:SAT-FFF-solution_3}
\end{equation}
To summarize, the solution is:
\begin{equation}
    \begin{aligned}
        \psi_{\rm 0} < 1                            & \ \ \to \ \ \psi(t) = \psi_{\rm 0}e^{-\alpha t} \\
        \psi_{\rm 0} > 1, \frac{\delta}{\alpha} > 1 & \ \ \to \ \ \psi(t) = \delta/\alpha + (\psi_{\rm 0}-\delta/\alpha)e^{-\alpha t} \\
        \psi_{\rm 0} > 1, \frac{\delta}{\alpha} < 1 & \ \ \to \ \
        \begin{aligned}
            \psi(t) & = \delta/\alpha + (\psi_{\rm 0}-\delta/\alpha)e^{-\alpha t}        & \ \ {\rm for} \ \ t \in [0, t_c] \\
            \psi(t) & = \frac{\psi_{\rm 0}-\delta/\alpha}{1-\delta/\alpha} e^{-\alpha t} & \ \ {\rm for} \ \ t \in [t_c, \infty].
        \end{aligned}
    \end{aligned}
    \label{eq_dynresults:SAT-FFF-solutions_summary}
\end{equation}
This solution is analogous to the one obtained for the discrete time
model above. That is, dynamics of the SAT-FFF is described by a
synchronous fixed point in the ODE model as well.

\subsection{Unsatisfied feed-forward fiber oscillates and synchronizes}
\label{Subsec:unsatfff}
\index{unsatisfied feed-forward fiber }\index{UNSAT-FFF }

\begin{figure}[ht!]
    \centering
    \includegraphics[width=0.4\textwidth]{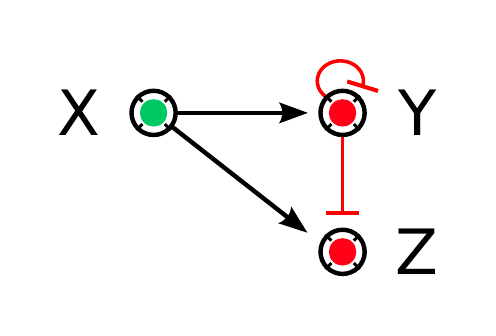} 
    \caption{\textbf{UNSAT-FFF network representation.} UNSAT-FFF is an
      FFF in which all links between the nodes in the fiber are
      repression links. }
    \label{Fig:UNSAT-FFF}
\commentAlt{Figure~\ref{Fig:UNSAT-FFF}: 
Graph with nodes X(green), Y and Z (red). Sharp arrows XY, XZ; barred arrows YY, YZ.
}
\end{figure}

\cite{leifer2020circuits} show analytically and numerically that the
UNSAT-FFF circuit has an oscillatory solution plus synchronization of
genes Y and Z (as expected from the fibration) using a discrete-time continuous variable model and an
ODE with time delay model. The time delay is related to the process of
transcription and translation. The conclusions made for SAT-FFF and FFL
hold in the framework of DDEs (Delay Differential Equations) but were
considered without the delay for simplicity.

\subsubsection{A. UNSAT-FFF discrete time model}
\label{Subsubsec:unsat-discrete}

The discrete-time dynamics of the expression levels of genes $y_t$ and
$z_t$ in the UNSAT-FFF are given by:
\begin{equation}
    \begin{aligned}
    y_{t+1} &= (1-\alpha) y_t + \gamma_x \theta(x_t-k_x)\ \gamma_y \theta(k_y-y_t),\\
    z_{t+1} &= (1-\alpha) z_t + \gamma_x \theta(x_t-k_x) \ \gamma_y \theta(k_y-y_t),
    \end{aligned}
    \label{eq_dynresults:fffnegative}
\end{equation}
where $\gamma_x$ and $\gamma_y$ are the strengths of the interaction
(maximum expression rate) of genes X and Y, respectively, and $k_x$
and $k_y$ are the respective dissociation constants of the same
genes. Similarly to the SAT-FFF case, synchronization between $y$ and
$z$ occurs because of the symmetry fibration that
collapses nodes Y and Z. However, the impact of the repressor
feedback loop on the dynamical behavior of this circuit is more
profound, since it leads to oscillations. Thus, while both SAT-FFF and
UNSAT-FFF lead to synchronization of Y and Z, the former synchronizes
into a fixed point and the later into an oscillatory limit cycle.

We set $\lambda=\gamma_x\gamma_y/k_y\alpha$, $\beta = 1-\alpha$,
and rewrite \eqref{eq_dynresults:fffnegative} for the
rescaled variables $\psi_t = y_t/k_y$ and $\zeta_t = z_t/k_y$ as:
\begin{equation}
    \begin{aligned}
    \psi_{t+1} &= \beta\psi_t + \alpha\lambda\theta(x_t-k_x)\theta(k_y-\psi_t),\\
    \zeta_{t+1} &= \beta\zeta_t + \alpha\lambda\theta(x_t-k_x)\theta(k_y-\psi_t).
    \end{aligned}
    \label{eq_dynresults:fffnegativepsi}
\end{equation}
We set $x_t = x$ constant in time for simplicity. For $x<k_x$,
the solutions decay exponentially as $\psi_t = \psi_0e^{-t/\tau}$ and
$\zeta_t = \zeta_0e^{-t/\tau}$, where $\psi_0$ is the initial
condition. For $x>k_x$, \eqref{eq_dynresults:fffnegativepsi}
defines an iterative map which satisfies the following recursive
equation:
\begin{equation}
    f^t(\psi) = f^{t-1}(\beta\psi)\theta(\psi-1) +
    f^{t-1}(\beta\psi+\alpha\lambda)\theta(1-\psi).
    \label{eq_dynresults:unsatfffmap}
\end{equation}
This iterative map results in different solutions depending on the
value of $\lambda$.  We first consider the case where the initial
condition is $\psi_0>1$. The solution of
\eqref{eq_dynresults:fffnegativepsi} is then $\psi_t =
\psi_0e^{-t/\tau}$, where $\tau^{-1}=-\log(1-\alpha)$. This solution
is correct as long as $\psi_t>1$, but ceases to be valid at a certain
time $t^*$ such that $\psi_t<1$, which is given by $t^*=\lceil\tau\log
\psi_0\rceil$.

\begin{figure}[ht!]
	\centering
        \includegraphics[width=\linewidth]{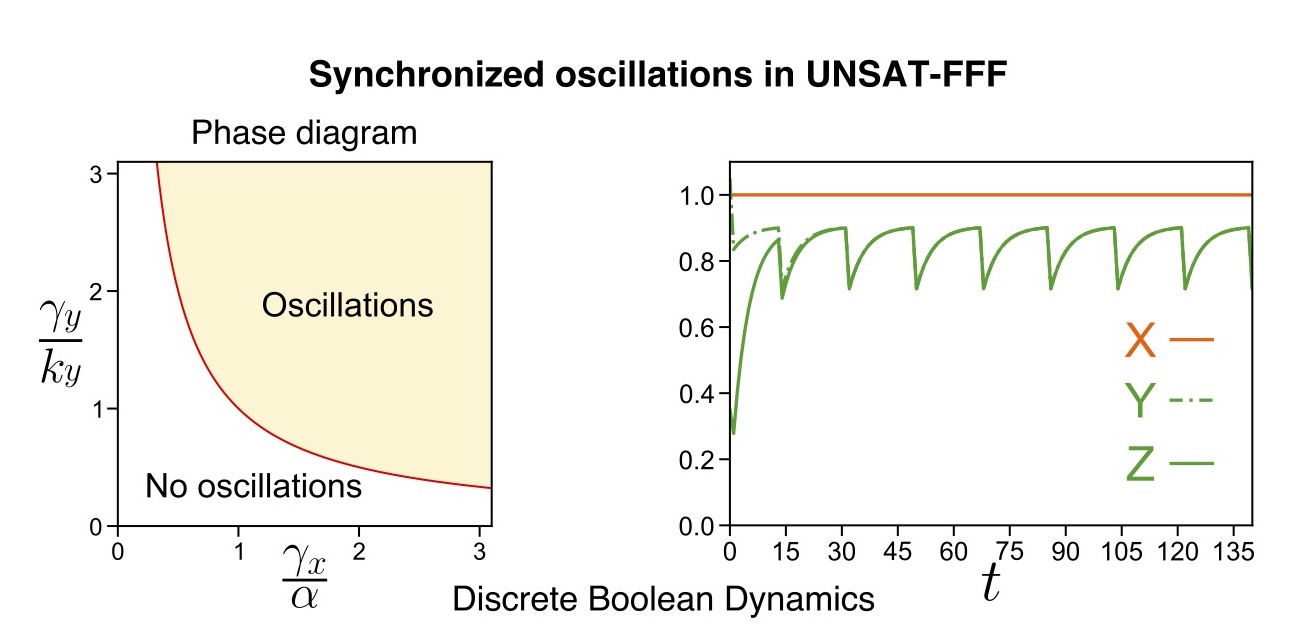}
    \caption{\textbf{Phase diagram of oscillations of the UNSAT-FFF.} An
      oscillatory phase is defined by the condition $\gamma_y/k_y >
      \left(\gamma_x/\alpha\right)^{-1}$. For instance, on the right
      side, we plot the solution of  \eqref{eq_dynresults:fffnegativepsi} for a set of
      parameters satisfying such condition. Specifically,
      $\alpha=0.205$, $\gamma_x = 0.454$, $\gamma_y=0.454$, $k_x=0.5$,
      and $k_y=1.0$. Figure reproduced from
      \citep{leifer2020circuits}.  Copyright \copyright ~2020, Leifer {\it et al.} }
    \label{Fig:PLOS_Fig2_g}
\commentAlt{Figure~\ref{Fig:PLOS_Fig2_g}: 
Left: Phase diagram, square with curve from top left to right bottom, decreasing rapidly at first and then slowing down. Above curve: oscillations. Below curve: no oscillations.
Right: Graph of solutions: X is constant, Y and Z synchronize in a sawtooth oscillation.
}
\end{figure}

Next, we consider the case $\psi_0<1$. In this case the solution is
given by $\psi_t = \psi_0e^{-t/\tau} + \lambda(1-e^{-t/\tau})$, which
is always valid for $\lambda<1$. Thus when $\lambda<1$ the system
does not oscillate, but converges monotonically to a fixed point
$\psi_\infty = \lambda$.  However, when $\lambda>1$ this solution
ceases to be valid at the time $t^* = \lceil \tau\log \frac{\lambda
  -\psi_0}{\lambda -1}\rceil$ such that $\psi_t>1$. Therefore the
solution $\psi_t$ oscillates in time for $\lambda>1$. For the case of
$\psi_0>1$, the explicit solution is given by the following general analytic
expression, which is plotted in Fig.~\ref{Fig:PLOS_Fig2_g} (right):
\begin{equation}
    \begin{aligned}
    \psi_t &= \psi_0e^{-t/\tau}\ \ \ &{\rm for}\ t\in\{0,1,
    \hdots,t_1\},\\ \psi_t
    &= \psi_1e^{-(t-t_1)/\tau}+\lambda(1 -
    e^{-(t-t_1)/\tau})\ \ \ &{\rm
      for}\ t\in\left\{t_1,\hdots,t_2\right\},\\ \psi_t &=
    \psi_2e^{-(t-t_2)/\tau}\ \ \ &{\rm
      for}\ t\in\Big\{t_2,\hdots,t_3\}.
    \end{aligned}
    \label{eq_dynresults:evol_unsatfff}
\end{equation}
Here
\[
t_1 = \left\lceil \tau\log \psi_0\right\rceil \qquad
t_2 = t_1+\left\lceil \tau\log
    \frac{\lambda-\psi_1}{\lambda-1}\right\rceil \qquad
t_3 = t_2+\lceil \tau\log
    \psi_2\rceil
\]
The general solution with initial condition $\psi(t_0)<1$ can be
written in a similar way.

Thus the main condition for oscillations in the circuits is
$\lambda>1$. If $\lambda<1$ there is no oscillatory behavior and
the solution $\psi_t$ converges monotonically to $\lambda$. Therefore,
the oscillatory phase is separated from the non-oscillatory phase by
the condition:
\begin{equation}
    \frac{\gamma_y}{k_y} = \left(\frac{\gamma_x}{\alpha}\right)^{-1},
\end{equation}
which is depicted in the phase diagram of 
Fig.~\ref{Fig:PLOS_Fig2_g} (left).

\subsubsection{B. UNSAT-FFF DDE model}
\label{Subsubsec:unsat-ode}

Now we elaborate on the solution of the ODE continuum model. Since
both genes Y and Z synchronize their behavior, the UNSAT-FFF circuit
can be reduced to study the base of the circuit consisting of a
negative autorregulation loop and an external regulator X
(Fig.~\ref{fig3_plos}c, right). This circuit has been
synthetically implemented by \cite{hasty2008} using a promoter that
drives the expression in the absence of LacI (and acts as a negative
feedback loop) or in the presence of IPTG, which acts as an
activator. It was shown experimentally that this circuit leads to
oscillatory behavior in the expression profiles. This result was
corroborated with a dynamical ODE model in \citep{hasty2008} which we
adapt to study the case of the UNSAT-FFF with ODE. See also the review
paper \citep{purcell2010}.

As before, we consider gene $x(t)=x$ constant in time and larger than
$x>k_x$, and rescale the expression of genes $y(t)$ and $z(t)$ as
$\psi(t) = y(t)/k_y$ and $\zeta(t) = z(t)/k_z$.  Since genes $y(t)$
and $z(t)$ synchronize their activities, then only one equation need
 be considered, that for $\psi(t)$.

The key to observing oscillations in a first-order ODE is to consider
the delay in the signal propagation in the circuit. Without the delay, the
dynamics converge to a fixed point; no oscillatory solution exists in
a first-order ODE continuous time model on $\mathbb{R}$. The situation is different in
the discrete-time model considered earlier. In this case, 
discrete-time plus a logic approximation leads to oscillations.

Negative feedback loop circuits with delays have been widely
investigated in the dynamical systems literature.  Here, we adapt the
negative feedback loop model with a delay of \cite[Supplementary Information Equation (6)]{hasty2008}. We
consider delays in the negative feedback loop, which is the key feature
explaining the experimentally observed robust oscillations in this
circuit \citep{hasty2008}.

Delays in a biological circuit arise from the combined processes of
intermediate steps like transcription, translation, folding,
multimerization and binding to DNA. This series of biological
processes is lumped into a single arrow between two genes in the
network representation of the circuit. In reality, this arrow
represents processes that should be modeled in a more detailed
manner. These biological processes can be approximately taken into
account by inserting a delay in the interaction term in the dynamical
equations. The interaction term can be written as $\delta \,\,
\theta(1-\psi(t - \tau))$, where $\tau$ represents the delay caused by
the process of self-repression not being
instantaneous. Therefore the dynamics of $\psi(t)$ is described by the
first-order delay-differential equation (DDE) :
\begin{equation}
    \dot{\psi} = - \alpha \psi(t) + \delta \,\, \theta(1-\psi(t - \tau)),
    \label{eq_dynresults:DDE}
\end{equation}
where $\tau$ represents the delay caused by the expression process.

We find analytical solutions to this equation following a procedure
outlined in \citep[Chapter V]{driver2012}. 
Initial conditions used for a DDE are not specified by the value of
the function at one point, but rather by a set of values of the
function on an interval of length $\tau$. The solution of a DDE cannot
be thought of as a sequence of values of $\psi(t)$ as in an ODE, but
rather as a set of functions $\{f_{0}(t), f_{1}(t), f_{2}(t),
\dots,\}$, defined over a set of contiguous time intervals $\{[-\tau,
  0], [0, \tau], [\tau, 2\tau], \dots,\}$.

Consider \eqref{eq_dynresults:DDE} with initial function
$f_0(t)$ for $t \in [-\tau, 0]$. Then for $t \in [0, \tau]$,
\eqref{eq_dynresults:DDE} looks like:
\begin{equation}
    \dot{\psi} = -\alpha \psi(t) + \delta \,\, \theta(1 - f_{0}(t-\tau)).
    \label{eq_dynresults:DDE_general_1}
\end{equation}
Moving the degradation term to the left and multiplying by $e^{\alpha
  t}$ we get:
\begin{equation}
    \dot{\psi}e^{\alpha t} + \alpha \psi(t)e^{\alpha t} = \delta \,\, e^{\alpha t} \theta(1 - f_{0}(t-\tau)).
    \label{eq_dynresults:DDE_general_2}
\end{equation}
Rewriting the left part, we obtain:
\begin{equation}
    \frac{d (\psi e^{\alpha t})}{dt} = \delta \,\, e^{\alpha t} \theta(1 - f_{0}(t-\tau)),
    \label{eq_dynresults:DDE_general_3}
\end{equation}
and integrating on the interval $\int_{0}^t$, we get:
\begin{equation}
    \psi e^{\alpha t} - \psi(0) = \delta \,\, \int_{0}^t e^{\alpha t'}
    \theta(1 - f_{0}(t'-\tau)) dt \,\,.
    \label{eq_dynresults:DDE_general_4}
\end{equation}
Since $\psi$ is continuous at 0 ($\psi(0) = f_{0}(0)$) and $\psi(t)$ for $t \in [0, \tau]$ is given by $f_{1}(t)$, we write:
\begin{equation}
    f_{1}(t) = f_{0}(0) e^{-\alpha t} + \delta \,\, \int_{0}^t e^{\alpha (t'-t)} \theta(1 - f_{0}(t'-\tau)) dt.
    \label{eq_dynresults:DDE_general_5}
\end{equation}

\begin{figure}[ht!]
    \includegraphics[width=\textwidth]{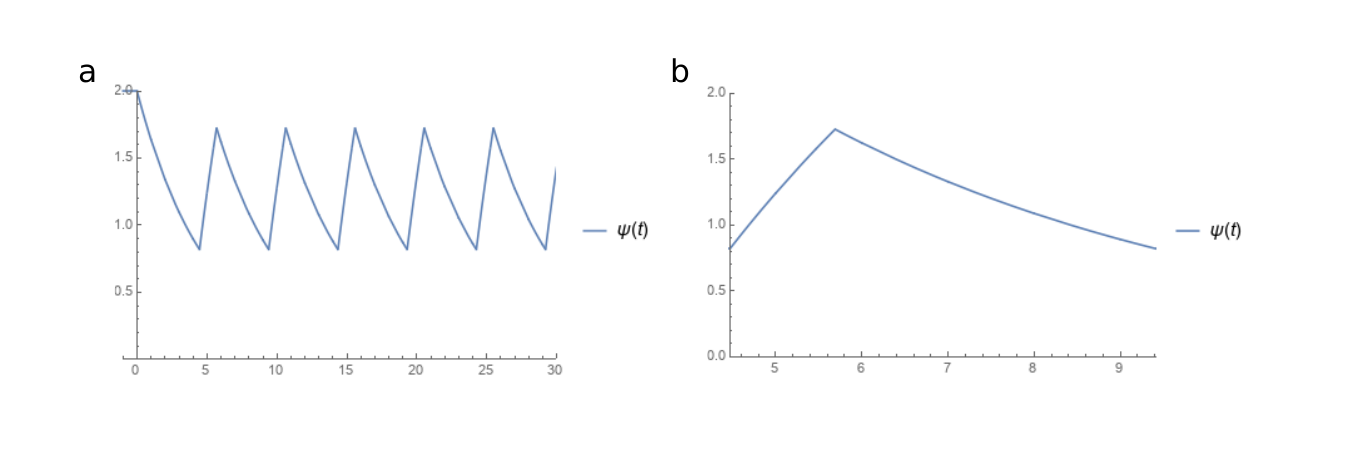}
    \caption {\textbf{UNSAT-FFF delay ODE model.}  (\textbf{a}) Solution
      of \eqref{eq_dynresults:DDE} using the recursion
      \eqref{eq_dynresults:DDE_general_recursive_equation} on $t
      \in [-\tau, 30\tau]$ for $f_{0}=2$, $\alpha = 0.2$, $\delta = 1$
      and $\tau=1$.  (\textbf{b}) One period of the oscillation of
      solution in {\bf (a)} consisting of two exponential
      pieces. Figure reproduced from \citep{leifer2020circuits}. Copyright \copyright ~2020, Leifer {\it et al.}}
    \label{Fig:PLOS_UNSAT-FFF_recursive}
\commentAlt{Figure~\ref{Fig:PLOS_UNSAT-FFF_recursive}: 
(a) sawtooth graph. (b) Piecewise smooth: increase first, then decrease.
}
\end{figure}

Following the same procedure we can derive the general formula for the
solution $\psi(t)$ on the interval $[k\tau, (k+1)\tau]$, assuming that
the solution on the previous interval $[(k-1)\tau, k\tau]$ is given by
$f_{k-1}(t)$. The solution is then given by the following iterative equation:
\begin{equation}
    \dot{\psi} = - \alpha \psi(t) + \delta \,\, \theta(1 - f_{k-1}(t-\tau)).
    \label{eq_dynresults:DDE_general_6}
\end{equation}
Applying the integrating factor method and integrating over
$\int_{k\tau}^t$ we obtain:
\begin{equation}
    \psi(t) = \psi(k\tau) * e^{\alpha(k\tau-t)} + \delta \,\, \int_{k\tau}^t e^{\alpha(t'-t)} \theta(1 - f_{k-1}(t'-\tau)) dt'.
    \label{eq_dynresults:DDE_general_recursive_equation}
\end{equation}
Using \eqref{eq_dynresults:DDE_general_recursive_equation} we 
recursively find functions $\{f_{0}(t), f_{1}(t), f_{2}(t), \dots,\}$
on the interval of interest, which provide the solution to
\eqref{eq_dynresults:DDE}. Using Mathematica we find functions on the interval $t \in
      [-\tau, 30\tau]$ for $f_{0}=2$, $\alpha = 0.2$, $\delta = 1$ and
      $\tau=1$ and put them together to find the solution plotted in
      Fig.~\ref{Fig:PLOS_UNSAT-FFF_recursive}a.

By\eqref{eq_dynresults:DDE_general_recursive_equation},
 all functions $f_{k}$ are the sum of an exponential
function and a constant. By \eqref{eq_dynresults:DDE},
when the Heaviside function is equal to zero we get a
solution that decays exponentially to zero. Likewise, when the
Heaviside function is equal to 1 we get a solution that grows exponentially
 to $\frac{\delta}{\alpha}$. In other words, the solution 
grows until $\theta(1 - \psi(t-\tau))$ changes to zero (i.e., when
$\psi(t-\tau)>1$), and  decays until $\theta(1 - \psi(t-\tau))$
changes to 1 (i.e., when $\psi(t-\tau)$  crosses 1 again, but from
the other side). Therefore we get oscillations consisting of two
exponential pieces. One period of the oscillation is shown in
Fig.~\ref{Fig:PLOS_UNSAT-FFF_recursive}b. The solution on this
interval is given by:
\begin{equation}
    \begin{aligned}
        \psi(t) = 5 - 10.2 * e^{-0.2t} & \ {\rm for} \ t \in [4.47, 5.69] \\
        \psi(t) = 5.4 * e^{-0.2t}      & \ {\rm for} \ t \in [5.69, 9.42],
    \end{aligned}
    \label{eq_dynresults:DDE_oscillation_solution}
\end{equation}
which is the predicted behavior. This circuit functions like a
capacitor, charging and discharging in an RC circuit.

\subsection{UNSAT-FFF clock functionality}
\label{sec:unsat-fff-functionality}

As shown above, the UNSAT-FFF has the
functionality of a clock,\index{clock } a fundamental unit in any computing system.
The solution of the discrete-time Boolean interaction
model for $\lambda>1$ oscillates in time, and so does the DDE
considered in the previous section. 

To compute the amplitude of oscillations $A_\psi$ for the rescaled
variable $\psi_t$, we recall that the iterative map $\psi = f(\psi)$
satisfies the recursive equation:
\begin{equation}
f^t(\psi) = f^{t-1}(\beta\psi)\theta(\psi - 1) + f^{t-1}(\beta\psi + \alpha\lambda)\theta(1 - \psi).
\end{equation}
The amplitude of oscillations $A_\psi$ is given by
\begin{equation}
A_\psi = \lim_{\psi\rightarrow 1^{-}}f(\psi) - \lim_{\psi\rightarrow
  1^+}f(\psi) = \alpha\lambda,
\label{eq:amp_psi}
\end{equation}
which implies that
\begin{equation}
A_\psi = \frac{\gamma_x\gamma_y}{k_y}.
\end{equation}

\begin{figure}[t!]
 \includegraphics[width=\textwidth]{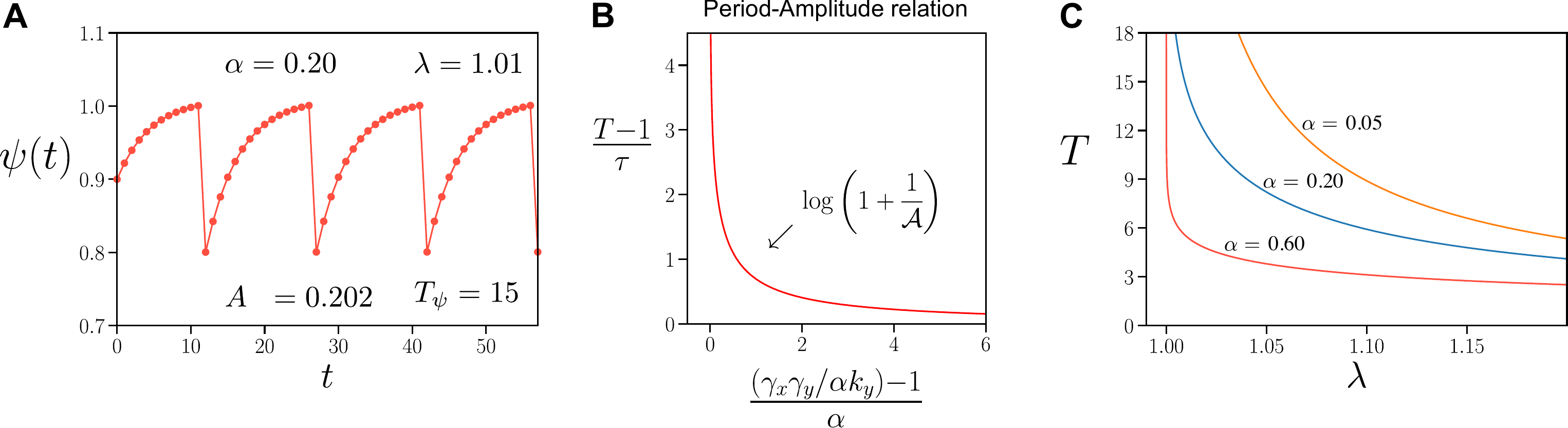}
 \centering
\caption {\textbf{Robust UNSAT-FFF clock functionality.} (\textbf{a})
  Solutions for $\alpha = 0.2$ and $\lambda = 1.01$. The values for
  $A_\psi = 0.202$ and $T = 15$ obtained from
 \eqref{eq:amp_psi} and \eqref{eq:period} agree perfectly 
  with those found by numerical simulation. (\textbf{b})
  Period-amplitude relationship in terms of the original set of
  parameters $\alpha$, $k_y$, $\gamma_x$, and $\gamma_y$. (\textbf{c})
  Period of oscillations as a function of $\lambda$ for different
  values of $\alpha$. Figure reproduced from
  \citep{leifer2020circuits}.  Copyright \copyright ~2020, Leifer {\it et al.}}
\label{sm:fig2}
\commentAlt{Figure~\ref{sm:fig2}: 
A. Sawtooth graph. B: Curve shaped like one component of a hyperbola. C. Three
such curves for alpha = 0.6, 0.2, 0.05.
}
\end{figure}

To find the period $T$ of the oscillations, we recall from
 \eqref{eq_dynresults:evol_unsatfff} that the solution for the
minimum value of $\psi$, $\psi_{\rm min}<1$, evolves to its maximum
value $\psi_{\rm max}$ in $T-1$ iterations as $\psi_{\rm max} =
e^{-(T-1)/\tau}\psi_{\rm min} + \lambda\left(1 -
e^{-(T-1)/\tau}\right)$.  Since $\psi_{\rm min} = (1-\alpha)\psi_{\rm
  max}$, because $\psi_{\rm max}>1$, we find:
\begin{equation}
T = \left\lceil1 + \tau\log\Bigg(1+\frac{\alpha}{\lambda-1}\Bigg)\right\rceil,
\label{eq:period}
\end{equation}
where we used $\psi_{max} = 1$. For example, using $\alpha = 0.2$ and
$\lambda=1.01$, we find $A_\psi = 2.02$ and $T = 15$, which agrees
with the numerical simulation in Fig.~\ref{sm:fig2}a.

Equation~(\ref{eq:period}) lets us define a rescaled amplitude
$\mathcal{A} = (\lambda - 1)/\alpha$ and a reduced period
$\mathcal{T} = (T-1)/\tau$ such that
\begin{equation}
\mathcal{T} = \log\left( 1 + \frac{1}{\mathcal{A}} \right),
\label{eq:period-amp}
\end{equation}
which corresponds to the {\it period-amplitude relationship} of the
UNSAT-FFF. A plot of this relationship is shown in 
Fig.~\ref{sm:fig2}b, where we plot $(T-1)/\tau$ as a function of
$\left[(\gamma_x\gamma_y/\alpha k_y)-1\right]/\alpha$.

Coming back to the original variable $y_t = k_y\psi_t$, 
the amplitude of oscillation of $y_t$, namely $A = k_yA_\psi$, is given by:
\begin{equation}
A = \gamma_x\gamma_y,
\label{eq:smamplitude}
\end{equation}
From ~(\ref{eq:period}) we can write the period of oscillation as
a function of the original parameters:
\begin{equation}
T = 1 -\frac{1}{\log(1-\alpha)}\log\left( 1 +
\frac{\alpha}{(\gamma_x\gamma_y/\alpha k_y) - 1} \right).
\label{eq:smperiod}
\end{equation}

Figure \ref{sm:fig2}c shows the period of oscillation $T$ as a
function $\lambda$ for $\alpha = 0.60$, $\alpha = 0.20$, and $\alpha =
0.05$.

\begin{figure}[t!]
 \includegraphics[width=0.6\textwidth]{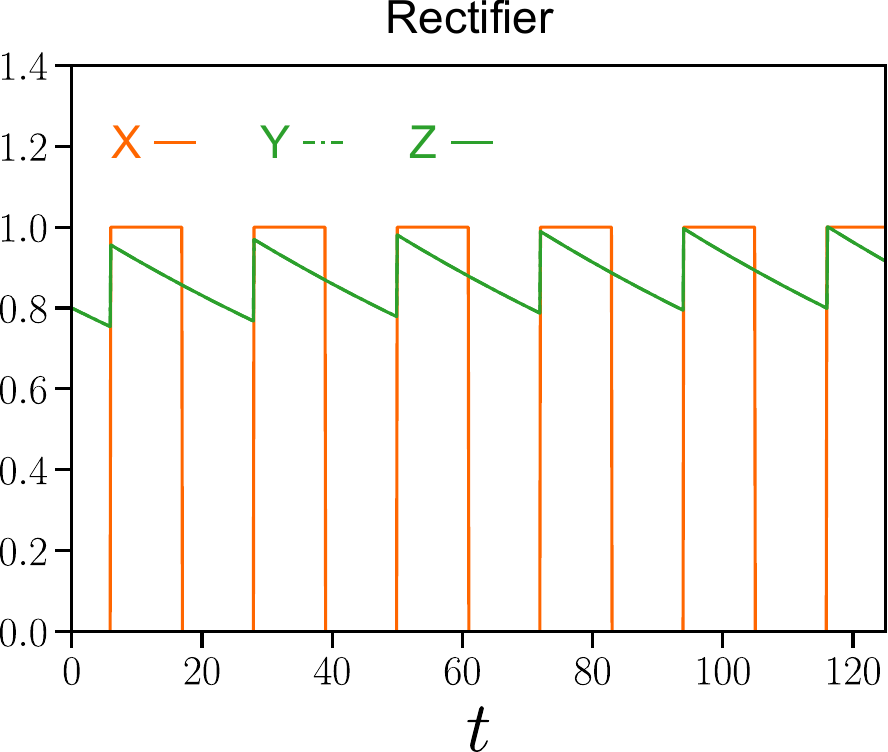} \centering
\caption {\textbf{UNSAT-FFF clock functionality.} The intrinsic relationship
between the period $T$ and amplitude of oscillations of the
UNSAT-FFF provides a stabilizing effect which results in predictable
and robust oscillations. Here, the UNSAT-FFF operates as a rectifier
for the set of parameters: $\alpha = 0.001$, $k_x = 0.5$, $k_y = 0.8$,
$\gamma_x = 0.45$, and $\gamma_y = 0.45$.}  
\label{sm:fig4}
\commentAlt{Figure~\ref{sm:fig4}: 
Graph in which X varies against time in a square wave, while Y and Z synchronize in
sawtooth oscillations.
}
\end{figure}

The clock functionality of the UNSAT-FFF can be understood by analyzing its
response function, i.e. the relation between oscillations at the input
and at the output of the circuit. The amplitude 
$A_y$, 
and period $T$ of the oscillations are not independent,
as in a harmonic oscillator, but are related through the
`period-amplitude' relation expressed by (\ref{eq:period-amp})
and Fig.~\ref{sm:fig2}b. From ~(\ref{eq:smperiod}), for $\alpha$
sufficiently small,
\begin{equation}
T - 1\sim \frac{k_y}{\gamma_x\gamma_y}= \frac{1}{A}\ ,
\label{eq:smpowerperiod}
\end{equation}
which constrains the `clock' ($T$) of the circuit to the power
($A_y$).  As a consequence, $A_y$ and $T$ cannot be controlled
arbitrarily, and this (A-T) constraint helps to stabilize the
UNSAT-FFF response against disturbance in the input X.

For example, for a given available power supply, the system is
constrained to dissipate this power, and when the UNSAT-FFF
oscillates, it is automatically set to operate on an extended time
window ($T$ large) at low amplitude $A$ when a small expression level
is required ($A$ small) and vice-versa.

If the signal from X has a frequency $f_x$ much higher than the
oscillator, the modulator $Y$ will transmit a signal oscillating with
very small amplitude $A\propto 1/T_x$. In other words, the UNSAT-FFF
can also operate like a rectifier for high-frequency signals coming
from X For example, Fig. \ref{sm:fig4} shows the UNSAT-FFF operating
as a rectifier for the set of parameters: $\alpha = 0.001$, $k_x =
0.5$, $k_y = 0.8$, $\gamma_x = 0.45$, and $\gamma_y = 0.45$.
Similarly, if the frequency $f_x$ of the baseband signal $x_t$ is
small compared to $f_y$, the modulator will transmit a signal with an
amplified bandwidth of order $f_y$.  

The stabilizing effect of the
negative feedback loop makes synchronization predictable and robust
through a symmetry fibration and endows the UNSAT-FFF circuit with a
reliable functionality independently of fine-tuning of
parameters. This is ideal for bacterial TRNs, which need to adapt to
large variation in environmental and growth conditions.  This
reliability, together with the simplicity of its feedback structure, may
explain why the FFF is ubiquitous in simpler genomes across the
bacterial domain like {\it B. subtilis}, {\it E. coli}, {\it
  Salmonella} and {\it M. tuberculosis}, although they are not present
in eukaryotes \citep{morone2020fibration}.

The findings related to clock functionality can be applied to other oscillatory building blocks, such as the Fibonacci fiber and the $n=2$ fiber, which we will explore next. The concept is that Fibonacci fibers with increasingly longer cycles can support more robust oscillatory patterns compared to the basic autoregulation negative feedback loop found in the FFF.

\section{Simple multilayer fiber}
\label{sec:multilayer}
\index{simple multilayer fiber }

Synchronization of the genes within the FFF building block is
guaranteed by their isomorphic input trees. These input trees contain
infinitely many layers. However, for this particular circuit
synchronization is guaranteed by the first layer of the input tree,
the input set, because all genes in the fiber have the same
in-neighbors. Thus, while the input tree is infinite, only the isomorphism
of the first layer is necessary to ensure synchronization. The next
level of complexity in the hierarchy of fibers is circuits, where
synchronization depends on deeper input layers of synchronized genes
and longer loops of information. This increases the complexity of the
circuits since the synchronization between genes connects coherently
distant areas in the network. We call this building block the {\em Simple
Multilayer Fiber}. 

Another way to increase the complexity is by the
appearance of longer cycles of information, as in the Simple
Fibonacci circuits explained in the next section.  We call these
circuits `simple' since the multilayers are composed of genes in
single fibers. Complex multilayer fibers made of multiple fibers are
treated in Chapter \ref{chap:complex}. Multilayer and Fibonacci
fibers are important in bacteria, although they appear in low
numbers. They are very abundant in more complex species, such as
eukaryotes, from yeast to humans. This compares to the simpler AR and FFF
(and operons), which are more prevalent in bacteria
\citep{leifer2020circuits}.

\begin{figure}
  \centering
  \includegraphics[width=0.8\linewidth]{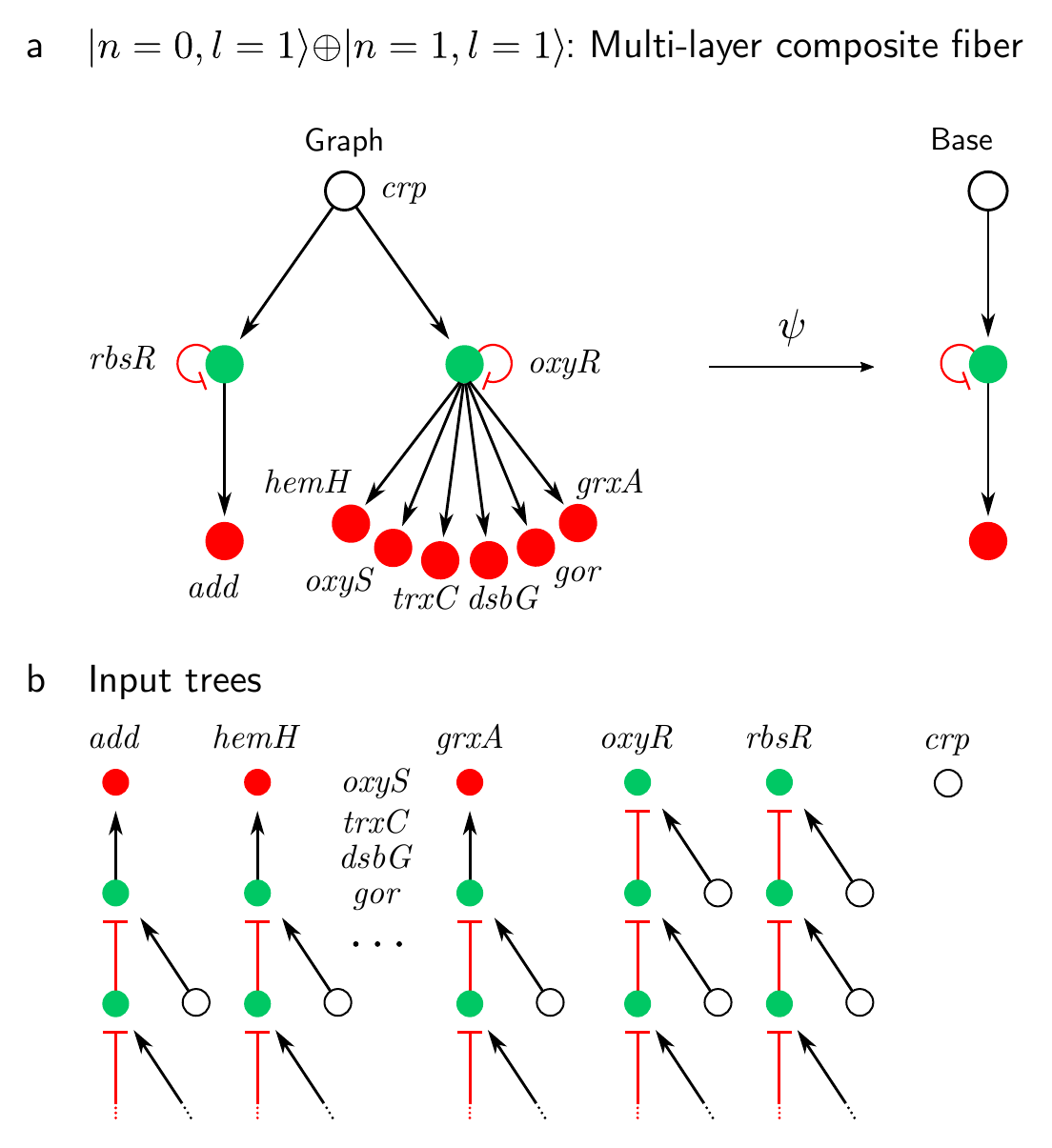}
  \caption{\textbf{Simple Multilayer Fiber.} (\textbf{a}) Circuit
    consists of two layers of fibers: {\it add, dsbG, gor, grxA, hemH,
      oxyS, trxC} classified as $\rvert n = 0, l = 1 \rangle$ and
    {\it rbsR, oxyR} classified as $\rvert n = 1, l = 1 \rangle$,
    therefore forming a multi-layer composite fiber $\rvert n = 0,
    l = 1 \rangle \oplus \rvert n = 1, l = 1 \rangle$. The fibration
    $\psi$ of this circuit collapses both fibers and leaves the
    regulator untouched. (\textbf{b}) Genes in the red fiber receive
    one input from the gene in the green fiber, which in turn receives
    an input from itself and the regulator. Therefore, input trees of
    genes in the red fiber resemble the sum of an input tree of
    $\rvert n = 0, l = 1 \rangle$, followed by the input tree of
    $\rvert n = 1, l = 1 \rangle$. Input trees of the green fiber
    are those of the FFF. The regulator node has no inputs. Thus,
    the multi-layer composite has two nontrivial fibers that can
    synchronize their activity. The gene {\it add} is separated from
    the rest of the red fiber by two steps, therefore allowing for 
    long-range synchronization in the network.  Figure reproduced from
    \citep{leifer2021predicting}. Copyright \copyright ~2021, The Author(s).}
  \label{composite}
\commentAlt{Figure~\ref{composite}: 
 Described in caption/text. No alt-text required.
}
\end{figure}

Figure \ref{composite}a shows an example of a multilayer composite fiber
in \emph{E.~coli} whose main regulator is {\it crp}. In this case,
{\it crp} is the inducer of a composite fiber, composed of {\it oxyR}
and {\it rbsR} and responsible for further downstream regulation of
several carbon utilization subsystems of genes.

The input tree of this fiber is isomorphic to that of
FFF $\rvert 1, 1 \rangle$, even though the building block has a very
different topology from that of the FFF, Fig.~\ref{fff}a. This first
layer of genes regulates, via {\it oxyR} and {\it rbsR}, a second fiber
composed of genes {\it add, dsbG, gor, grxA, hemH, oxyS, trxC}. If the
branch corresponding to {\it rbsR} is disregarded, the building block
of the fiber of genes {\it dsbG, gor, grxA, hemH, oxyS, trxC} is
classified as a single layer $\rvert 0, 1 \rangle$. Thus the building
block corresponding to the entire fiber in Fig.~\ref{composite}a in
red is a double-layer composite, which we denote by $\rvert add - oxyS
\rangle = \rvert 0, 1 \rangle + \rvert 1, 1 \rangle$, made of a series
of genes composing a single fiber of type $|0,1\rangle = | {\it add,
  dsbG, gor, grxA, hemH, oxyS, trxC} \rangle$, which is regulated by
two different transcription factors {\it rbsR} and {\it oxyR} that
form another fiber of type $|1,1\rangle = | rbsR, oxyR \rangle $.
This composite is important because it allows information to be
shared between two genes, {\it add} and {\it oxyS}, which are not
directly connected.

The gene {\it add} is two edges away from the other genes in its 
fiber, thus achieving synchronization at a distance of two in
the network. This is a clear example of global communication and
integration of unconnected genes in the network structure. The input
trees and base of this multilayer composite, depicted in
Fig.~\ref{composite}a, show that this composite fiber is the union
of two fundamental fibers.

Composite fibers satisfy a simple engineering `sum-rule': elementary
fibers are composed as a series of fibers in a predefined order where
the first layer is represented by an entry fiber (carrying
transcription factors), and the last layer is formed by a terminator
fiber of output genes (encoding enzymes).  This multi-layer composite fiber is
biologically significant because genes in the output layer synchronize
a genetic module that implements the same function even though the
genes in the module can be 
far apart in the network. Such functionally related modules could
not be identified by typical modularity algorithms, which cluster
nodes in modules of highly connected nodes.

We find that composite fibers are dominant in eukaryotes (yeast,
mouse, human). They resemble the building blocks of multilayered deep
neural networks where each subsequent gene in the layer synchronizes,
even though nodes can be distant in the network.  More
generally, composite fibers with multiple layers streamline the
construction of larger aggregates of fibration building blocks,
performing more complex functions in a coordinated fashion.

The biological relevance of these fibers results from the existence of
genes that can synchronize even though they do not share any common
inputs. So, what role does this composite fiber play in the regulatory
network?  Why has nature not chosen to give them the same regulatory
inputs in a single layer?  The main functionality of the multilayer
fibers is to communicate via synchronization of different fibers in
the network that are separated by
long distances.  
Multilayer fibers are the predominant method to achieve distant
synchronization, indicating the higher level of complexity in these
composite circuits.   This structure is critical to integrate
information in the large-scale network.

AR loops and FFF are commonly found in bacteria, while we observe only one type of multilayer fiber, as shown in Table \ref{tab:FiberBBstatistics}. Multilayer fibers are more prevalent in higher organisms, such as yeast and humans. This is supported by the analysis of fibers across various species in \citep{morone2020fibration,leifer2020circuits}, which highlights the increased complexity in the network structure and functionality of eukaryotes.

\section{Simple Fibonacci fibers}
\label{sec:ff}
\index{simple Fibonacci fiber }\index{Fibonacci fiber }

\begin{figure}
  \centering
  \includegraphics[width=0.75\linewidth]{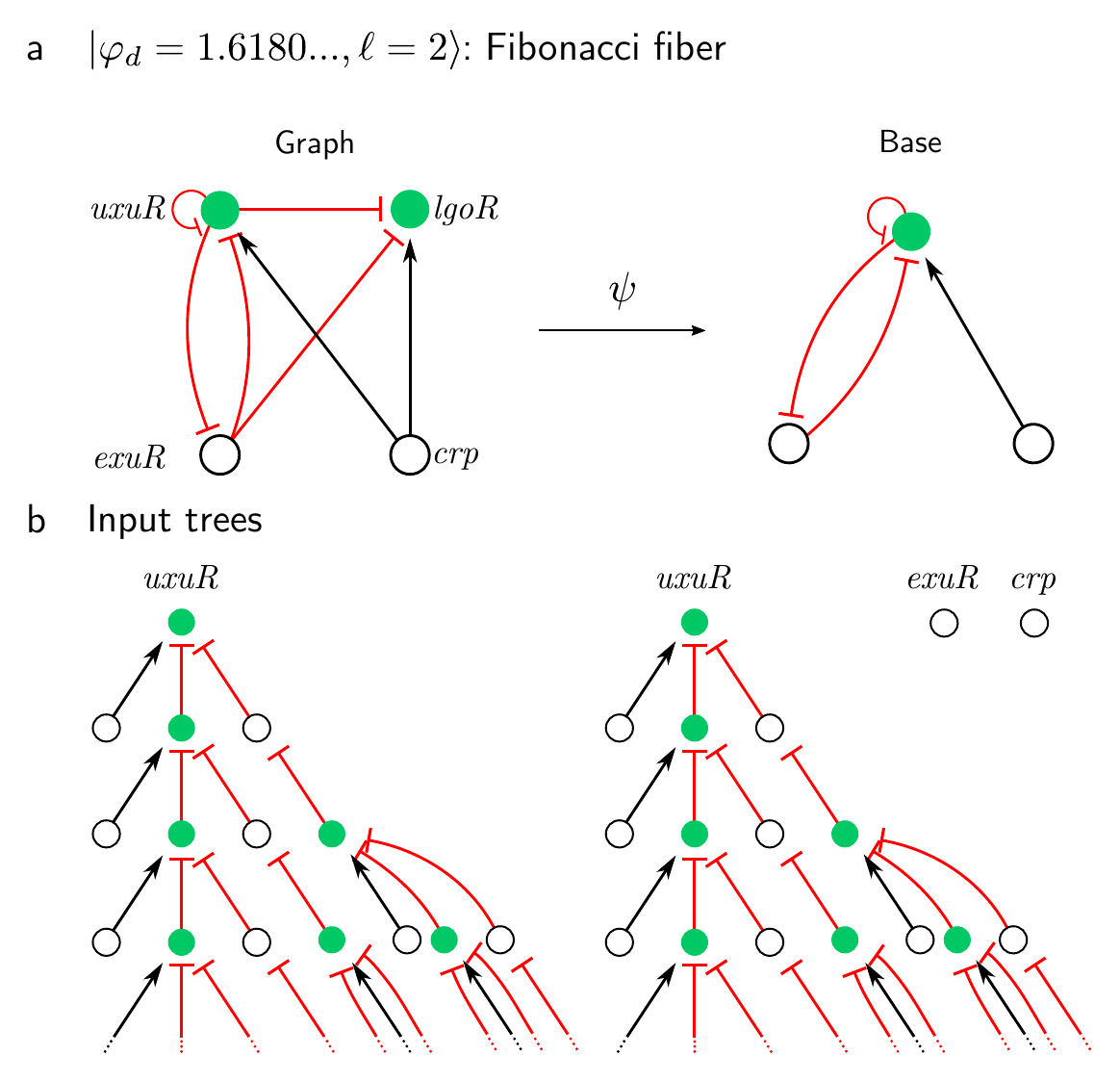}
  \caption{\textbf{Simple Fibonacci
    fiber (FF).} (\textbf{a}) (left) FF circuit is the FFF circuit with the
    additional edge from the fiber back to the regulator. In this
    example \textit{uxuR} sends back to \textit{exuR}, creating an
    extra loop in the circuit. The extra edge does not change the
    fiber. (\textbf{a)} (right) FF circuit is the FFF circuit with the
    fiber, so the fibration stays the same. (\textbf{b}) However, the
    extra loop changes the input tree of fiber nodes. \textit{uxuR}
    receives from itself and \textit{exuR}, which in turn receives
    from \textit{uxuR}, which creates an input tree with layer sizes
    following the Fibonacci sequence. The branching ratio then defines
    the first fiber number, so this FF is classified as $\rvert
    \varphi, l = 2 \rangle$, where $\varphi = 1.6180$.  Node
    \textit{lgoR} receives an input from \textit{exuR} and then from
    \textit{uxuR}, which means that even if there were no link from
    \textit{uxuR} to \textit{lgoR}, information would still be passed
    along through the regulator. This is another way for networks to
    process information.  Figure reproduced from
    \citep{leifer2021predicting}. Copyright \copyright ~2021, The Author(s).}
  \label{ff}
\commentAlt{Figure~\ref{ff}: 
 Described in caption/text. No alt-text required.
}
  \end{figure}

The next stage in the hierarchy is the Simple Fibonacci fiber (FF)
of Fig.~\ref{ff}a
\citep{morone2020fibration,leifer2020circuits}.  The FF shows a higher
level of complexity in the paths that regulate the fiber. To
understand the FF, and also the more complex FF that we discuss in
Chapter \ref{chap:complex}, we invoke the strongly connected
component of Definition \ref{def:SCC}. In general, a fiber may receive information from the entire
network through its input tree. When the fiber is not within an SCC,
information is processed only inside the fiber. This is the case for
all fibers described so far and characterized by integer fiber numbers
$n=0, 1$. The fiber may still receive signals from the
SCC, such as the case of the studied fibers so far. But, the fiber
does not send back signals, so the fiber, or any gene in the fiber, is
not part of the SCC.

However, if a fiber is connected to an SCC and sends back information
to its own regulators through the SCC, the level of complexity of the
fiber topology increases. In previous examples, the input trees are
infinite due to AR loops; here, the input tree becomes infinite due to
longer cycles in the SCC of the network. That is information cycles
around a longer loop before returning to the fiber. The input tree of
the fiber contains longer cycles of information arriving at the root
gene.  These cycles introduce extra terms in the sequence layers in
the input tree, leading to Fibonacci sequences in the number of paths.
There is an infinite number of possibilities for these cycles to
appear in a network. The simplest one is the Simple Fibonacci
fiber.

From the starting point of a $|1, 1\rangle$ FFF, many different
circuits can correspond to this base $|1, 1\rangle$. However,
only certain modifications conserve the topological class identified
by $|1, 1\rangle$. For instance, changing the sign of the edges is
allowed as long as the edges inputting to each gene are the same, but removing
the edge from the regulator to one of the genes in the fiber will
break the symmetry of the fiber, so it is not allowed.

Adding a second gene downstream of the AR in the FFF will conserve the
topological class, but only if it interacts with X following lifting.
This situation changes as soon as the fiber feeds information back to
the external world.  This is the case for the Fibonacci fiber in
Fig. \ref{ff}, where the gene {\it uxuR} now regulates its 
regulator gene {\it exurR}.  The inclusion of this second feedback
loop, the shortest of such a feedback loop from gene {\it uxuR} to its
regulator, forms a second cycle of length $d=2$, resulting in the
coexistence of two timescales in the network.  This, in turn,
increases the diversity of trajectories and delays in the network, a
dynamic complexity that is measured by the divergence of the branching
ratio of the input tree, which is captured by the sequence $a_t$.

The building block of the fiber {\it uxuR-lgoR} that is regulated by
the connected component {\it crp-fis} (Fig. \ref{ff}a) forms an
intricate input tree (Fig. \ref{ff}b) where the number of paths of
length $i-1$ is encoded in a Fibonacci-like sequence, the {\it Lucas numbers}:
\begin{equation}
a_i = 1, 3, 4, 7,
11, 18, 29, \ldots 
\end{equation}
This sequence leads to a non-integer branching ratio, the 
golden number $\varphi$:
\begin{equation}
\lim_{t\to\infty}\left(\frac{a_{t+1}}{a_t}\right) =\varphi =
\frac{1+\sqrt{5}}{2} = 1.6180...
\end{equation}
revealing that the input tree follows a Fibonacci recursion
\begin{equation}
  a_t= a_{t-1} + a_{t-2} ,
  \end{equation}
updating the current state two steps backwards.

The repressor interactions between genes X~=~{\it uxuR},
Y~=~{\it exuR}, and Z~=~{\it lgoR} 
function exactly as a Fibonacci fiber (Fig.~\ref{ff}a).  The important
component of these circuits is the delay in the feedback loop through
the regulator from Y~$\rightarrow$~X and back to Y captured by the
$a_{t-2}$ term in the Fibonacci sequence.

This topology arises in the genetic network due to the combination of
two cycles of information flow. First, the autoregulation loop inside
the fiber at {\it uxuR} creates a cycle of length $d=1$, which
contributes to the input tree an infinite chain with branching
ratio $n=1$. This sequence is reflected in the Fibonacci series by the
term $a_i = a_{i-1}$.  The important addition to the building block is
a second cycle of length $d=2$ between {\it uxuR} in the fiber and its
regulator {\it exuR}: {\it uxuR $\to$ exuR $\to$ uxuR}. This cycle
sends information from the fiber to the regulator and back to the
fiber by traversing a path of length $d=2$ that creates a delay of
$d=2$ steps in the information that arrives back at the fiber (Fig. \ref{ff}b). This short-term memory effect is captured
by the second term $a_i = a_{i-2}$ in the Fibonacci recursion, leading
to $a_i = a_{i-1} + a_{i-2}$ and the golden ratio limit. 

\cite{morone2020fibration} found three types of Fibonacci fibers in
the TRN of \emph{E.~coli}, as seen in Fig. \ref{Fig:PNAS_Fig3b}.
Eukaryotes like yeast and humans present a much richer variety of
Fibonacci fibers \citep{leifer2020circuits}. These circuits have been
synthetically implemented by 
\cite{hasty2008} using a hybrid promoter that drives the
transcription of genes {\it araC} and {\it lacI} forming a
dual-feedback circuit. The functionality of this circuit has been
demonstrated to be robust oscillations, due to the negative feedback
loop \citep{hasty2008}.  We show in Chapter \ref{chap:breaking}
that symmetry breaking in this Fibonacci circuit leads to the base of
the JK flip-flop, which is the universal storage device of computer
memories.

The above argument implies that an autoregulated fiber that further
regulates itself by connecting to its connected component via a cycle
of length $d$ encodes a generalized Fibonacci sequence of order $d$
defined as:
\begin{equation}
  a_i = a_{i-1} + a_{i-d}, \,\,\,\,\,\,\,\, \mbox{generalized Fibonacci sequence}
\end{equation}
with generalized golden ratio $\varphi_d$ (Fig. \ref{Fig:PNAS_Fig3b},
fourth row).

We find such a Fibonacci sequence in the {\it evgA-nhaR} fiber
building block (Fig. \ref{Fig:PNAS_Fig3b}, third row) linked to the pH
strongly connected components shown in Fig. \ref{fig:componentb}. This
fiber contains an autoregulation cycle inside the fiber and also an
external cycle of length $d=4$ through the pH strongly connected
component: {\it evgA $\to$ gad E $\to$ gadX $\to$ hns $\to$ evgA}
(Fig. \ref{Fig:PNAS_Fig3b}b, third row).

This topology forms a fractal input tree with sequence:
\begin{equation}
  a_i = a_{i-1} + a_{i-4} \,\,\,\,\,\,\,\, \mbox{generalized Fibonacci sequence of fiber {\it evgA}}
\end{equation}
and branching golden
ratio of a generalized Fibonacci sequence:
\begin{equation}
  \varphi_4=1.38028...
\end{equation}
Interestingly, this sequence has been indexed by the {\it The On-Line
Encyclopedia of Integer Sequences} as sequence A123456 in
\citep{sloane}.  We call this topology the {\em $\varphi_4$-Fibonacci fiber}.

Generalized Fibonaccis also appear inside strongly connected
components, like the {\it rcsB-adiY} $\varphi_3$-FF in the pH system
(Fig. \ref{Fig:PNAS_Fig3b}, second row).

As a final generalization, if the fiber participates in many
hierarchical cycles with the regulators of varying lengths up to a
maximum $d$, the Fibonacci sequence generalizes to:
\begin{equation}
  a_i = a_{i-1} + a_{i-2} + \dots + a_{i-1-d} + a_{i-d},
\end{equation}
  and the branching ratio $\varphi_d$ satisfies 
\begin{equation}
  d = - \frac{\log (2-\varphi_d)}{\log \varphi_d}.
  \end{equation}

In a Fibonacci fiber, genes connect to the SCC, and the SCC connects
back to the fiber via directed paths. If the SCC contains many paths
then, in principle, the building block associated with the FF should
contain all these paths through the SCC.  To define a building block
associated with a Fibonacci fiber, we consider only the shortest cycle
from the fiber through the SCC leading back to the fiber.  Longer
paths through the SCC are not included in the building block.  Consideration of these
neglected cycles in building blocks could substantially affect
the dynamics of gene expression in the fiber. However, they do not
affect the synchronization of genes in the FF.

\section{Binary tree fiber ($n=2$) and $n$-ary tree fibers}
\index{binary tree fiber }
\index{n-ary tree fiber@$n$-ary tree fiber }

The last type of simple building block in the hierarchy of fibers that
we find in the TRN of {\it E. coli} is the {\it binary tree fiber} with
$n=2$, shown in Fig.~\ref{fig3_plos}e with a core
$|2,l=0\rangle$.  It is characterized by two AR loops, leading to a
symmetric input tree; such a circuit is shown in
Fig.~\ref{fig3_plos}e.  This figure shows an $n=2$ binary tree fiber
resulting from the repressor regulations between genes X~=~{\it lexA}
and Y~=~{\it rocR}, and from the positive regulations between genes
X~=~{\it hprT}, Y~=~{\it tilS}, and Z~=~{\it ftsH}, both from the {\it
  B. subtilis} TRN studied in \citep{alvarez2024fibration}.

The sequence of information is coded in a sequence
  defined by
\begin{equation}
  a_i = 2 a_{i-1}, \,\,\,\,\,\,\,\, \mbox{binary tree fiber sequence}
\end{equation}
and we classify this fiber as $n=2$.  This procedure can be iterated
to any number of loops forming $n$-ary trees, as shown in the last row
of Fig. \ref{Fig:PNAS_Fig3a}, but the bacterial networks we have
studied do not contain any fiber with $n>2$, suggesting a practical
limit in complexity for these organisms.

\section{Evolutionary dynamics of fibration building blocks}
\label{sec:evolutionary}
\index{building block !evolutionary dynamics }

We have provided a constructive mechanism that reveals the hierarchy
of symmetric building blocks. Their broken symmetry counterparts,
which are also important, are
studied in Chapter \ref{chap:breaking}. The procedure mimics an
evolutionary growth procedure by recursively iterating the process
that expanded from a primordial AR loop\index{AR loop } to more complex circuits by
extensions to longer feedback loops producing functionalities like
synchronized clocks and oscillators. In parallel, a process of symmetry
breaking\index{symmetry breaking } of lifted circuits creates further functionalities beyond
synchronization and oscillations, providing logic computations and
memory storage through broken symmetry states, see 
Chapter \ref{chap:breaking}.

Summarizing, the starting point of an evolutionary process of growing circuits
 is the primordial AR loop with no external regulator and no
regulated genes:
\begin{equation}
  | 1, l=0, m=0\rangle, \,\,\,\,\,\,\,\,
  \mbox{primordial AR base}.
  \label{eq:bare}
  \end{equation}
This circuit can grow by lifting to $m$ regulated genes $| 1, l=0,
m=0\rangle$ following a transition:
\begin{equation}
  | 1, l=0, m=0\rangle \to | 1, l=0, m\ne0\rangle
  \,\,\,\,\,\,\,\, \mbox{lifting of the AR base.}
\end{equation}
  
Next, the bare AR base,  (\ref{eq:bare}), can receive a regulator
from another gene forming the base of a new circuit, the FFF,
following a transition:
\begin{equation}
  | 1, l=0, m = 0\rangle \to | 1, l=1, m = 0 \rangle \,\,\,\,\,\,\,\,
  \mbox{from AR base to FFF base.}
\end{equation}

This does not yet affect the complexity because all the relevant dynamics
remain constrained to the sole loop in the FFF circuit.
The next transition is again by lifting this FFF base to produce
the FFF with $m$ regulated genes:
\begin{equation}
  | 1, l=1, m = 0\rangle \to | 1, l=1, m \ne 0 \rangle \,\,\,\,\,\,\,\,
  \mbox{lifting of the FFF base.}
\end{equation}

These modifications conserve the topological class of the FFF,
providing its synchronization. Evolutionarily, the lifting property can
be thought of as a duplication mechanism (as discussed in Section
\ref{sec:duplication-lifting}) that conserves dynamic synchronization.
A duplication process produces more coherent genes that might be able
to reproduce a functionality or could be free to mutate and diverge
into other functions.

A transition to a new class of building blocks occurs as soon as the
autoregulated gene in the FFF feeds information back to the external
regulator, converting the FFF base into an FF base:
\begin{equation*}
  | 1, l=1, m = 0\rangle \to | \varphi_2 = 1.6180, l=1, m = 0 \rangle \,\,\,\,\,\,\,\,
  \mbox{from FFF base to FF base.}
\end{equation*}
This second feedback loop results in the coexistence of two
timescales in the network: one of the AR loop and another of length
two, resulting in the simple Fibonacci sequence of the input tree.
A new transition to the full FF is produced by lifting again:
\begin{equation*}
  | \varphi_2 = 1.6180, l=1, m = 0\rangle \to | \varphi_2 = 1.6180, l=1, m \ne 0 \rangle \,\,\,\,\,\,\,\,
  \mbox{lifting of the FF base.}
\end{equation*}

The AR loop and an
additional short cycle to the regulator still dominates the dynamics of this FF. This cycle can be enlarged
with the supposed improvement in functionality: oscillators like the
FF and FFF are known to be more stable when the feedback loops are
longer.  Besides, longer cycles to the regulator produce longer memory
in the dynamics. Thus the next steps include longer and longer cycles
to the regulator. These building blocks are generalized FFs with 
path length or regulatory cycles $d$. The first transition is from
$d=2$, the simple FF $\varphi_2-$FF to the generalized $\varphi_3-$FF
of length $d=4$, as seen in {\it E. coli} Fig. \ref{Fig:PNAS_Fig3b},
second row:
\begin{equation*}
  | \varphi_2 = 1.6180, l=1, m = 0\rangle \to | \varphi_3 = 1.4655,
  l=1, m = 0 \rangle \,\,\,\,\,\,\,\, \mbox{from  $\varphi_2-$FF
    to  $\varphi_3-$FF base.}
\end{equation*}
We consider the bases which can then be lifted to produce a variety of
regulated genes. The next natural transition observed in {\it E. coli} is
to the  $\varphi_4-$FF depicted in Fig. \ref{Fig:PNAS_Fig3b}, third row: 
\begin{equation*}
  | \varphi_3 = 1.4655, l=1, m = 0\rangle \to | \varphi_4 = 1.3802,
  l=1, m = 0 \rangle \,\,\,\,\,\,\,\, \mbox{from  $\varphi_3-$FF
    to  $\varphi_4-$FF base.}
\end{equation*}

While the evolution of the FF in bacteria ends here, we can imagine longer and
longer cycles generated by adding one intermediate gene at a time to
produce longer and longer memory effects.
\begin{equation*}
  | \varphi_{d} , l=1, m = 0\rangle \to | \varphi_{d+1},
  l=1, m = 0 \rangle \,\,\,\,\,\,\,\, \mbox{from  $\varphi_d-$FF
   to  $\varphi_{d+1}-$FF base.}
\end{equation*}
When $d\to \infty$ we find that $\varphi_\infty \to 1$. The feedback
loop effectively disappears and the resulting $\varphi_\infty-$FF
turns into a FFF with $n=1$.

The final transition in {\it E. coli} occurs when the regulator in the
FF acquires an AR loop, effectively turning it into a binary-tree fiber
with $n=2$:
\begin{equation*}
  | \varphi_{2} , l=1, m = 0\rangle \to | n=2,
  l=1, m = 0 \rangle \,\,\,\,\,\,\,\, \mbox{from  $\varphi_2-$FF
    to  binary-tree base,}
\end{equation*}
which can then be lifted to create more duplicated genes regulated by
the binary-tree base:
\begin{equation*}
  | n=2 , l=1, m = 0\rangle \to | n=2, l=1, m \ne 0 \rangle
  \,\,\,\,\,\,\,\, \mbox{lifting of the binary-tree base.}
\end{equation*}

While this systematic process can generate ever more complicated
circuits, the bacterial genome evolution stops here. We quantify their
complexity next.

\subsection{Complexity of the building blocks: fractal dimensions}
\label{S:fractal_dim}
The complexity\index{complexity } of the Fibonacci fiber\index{Fibonacci fiber } with feedback to the regulator
is the branching ratio\index{branching ratio } $1.6180...$, which tends back to 1 when $d\to
\infty$.  This golden ratio\index{golden ratio } can be thought of as the {\it fractal dimension}\index{fractal dimension } of
the input tree.\index{input tree }  The simplest
building blocks have integer dimensions $D=0$ (a regulon studied below
in Section \ref{sec:operons}), $D=1$: the AR loop fiber and the FFF
and $D=2$: the binary-tree fiber.  The complexity of the
AR loop and the FFF is the same; their only difference is the external
regulator.  The complexity class changes dramatically in the FF class,
which is now described by their fractal dimensions $\varphi_d$, where
$d=2$ for the simple FF generalized to any $d$ and tending to the FFF
as $d\to\infty$.  From $d=2$ to $n=2$, the next level from FF to
a binary tree, there is a discontinuity in fractal dimension values.

The fractal dimension of the FF is lower than the number of loops in
the circuit (two). The intuition of what this reveals is that, in this
circuit, the regulator is still not part of the base since it does not
receive input from itself. Hence, it is not within the fibration
symmetry of the fiber. This, in turn, indicates naturally that the
next element in the hierarchy of fibers results from the inclusion of
an AR loop leading to the binary-tree fiber, with complexity exactly 2.

The increasing complexity observed has an evolutionary basis rooted in the mechanism of gene duplication by lifting, which creates more connections, or edges, between existing genes. The fractal dimensions identified correspond to different classes of circuits, similar to how dimensions define universal classes of matter in physics. For instance, in the case of the simplest bacterial genome associated with the TRN, the fractal dimensions stop at 2. In Chapter \ref{chap:complex}, we explore the symmetries of metabolism, which is characterized by an extensive number of cycles. These cycles are organized in a manner that leads to progressively higher fractal dimensions, reaching an impressive maximum of 25.97, which indicates increasingly greater complexity.

\section{Biologically trivial fibers: operons and regulons}
\label{sec:operons}

The trivial building blocks leading to synchronization are operons\index{operon } and
regulons.\index{regulon } We recall these concepts. Operons are ubiquitous in bacteria \citep{jacob1961genetic}:
genes in an operon have a common promoter and are not transcribed into
individual mRNAs but as transcription units, yielding a single mRNA
strand containing several contiguous genes. Depending on the locations
of promoters and terminators, the transcription units can also
overlap.  The expression of genes in an operon is automatically
synchronized since they are translated together. In the case of
multi-promoter operons, we can group the genes into minimal
transcription units, each being controlled by the same combination of
promoters. The genes in such a minimal unit should be precisely
coexpressed (operons with several terminators can be subdivided
similarly). In our analysis, we take the operon as a single node in the
TRN. When two or more TFs belong to the operon, we leave one TF
associated with the operon and separate the remaining TFs from the
operon. While the synchronization in the operon is important
biologically, from the point of view of fibrations, they are trivial,
since their synchronization comes from a single promoter region and
polymerase.

\begin{figure}
  \centering
  \includegraphics[width=0.95\linewidth]{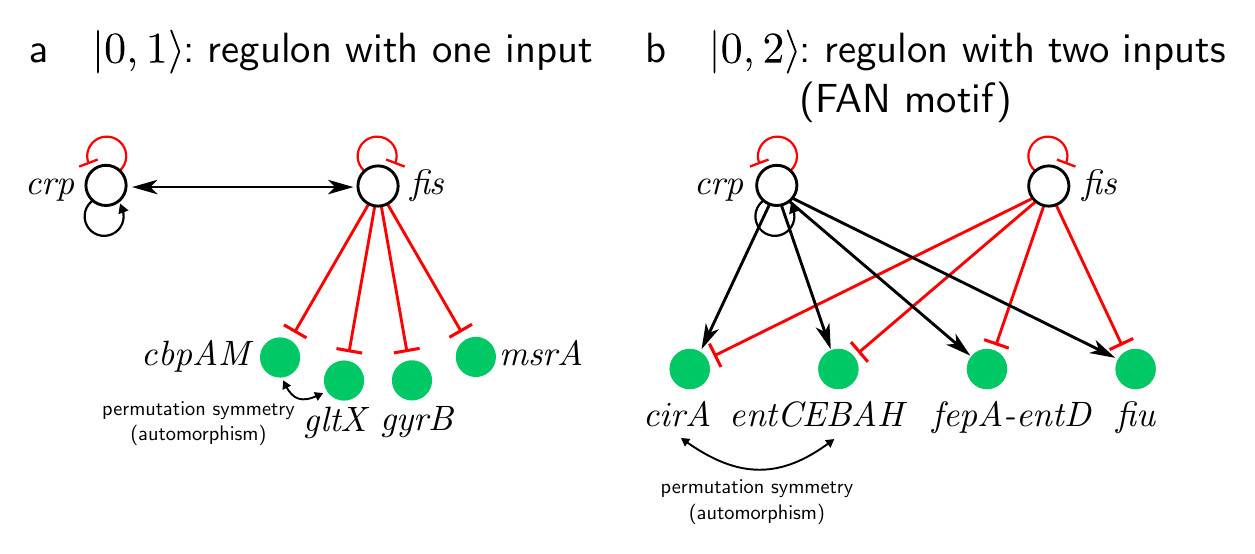}
  \caption{\textbf{Trivial fibration building blocks}: Regulons with
    co-regulation. (\textbf{a}) Genes \textit{cbpAM, gltX, gyrB} and
    \textit{msrA} are controlled by the same TF (\textit{fis}). Fiber
    numbers describing this circuit are $\rvert n = 0, l = 1
    \rangle$ since there are no loops and fiber has one regulator. Gene
    activity can synchronize because any two nodes can be permuted
    without any change in the network under the $\mathbb{S}_4$
    symmetry group. (\textbf{b}) Regulon circuit consisting of genes
    {\it clrA, fiu} and operons {\it entCEBAH, fepA-entD} controlled
    by two regulators \textit{crp} and \textit{fis} also synchronizes,
    because the symmetry group $S_4$ is preserved irrespective of the
    number of regulators. The fiber has two regulators and no cycles,
    and therefore is characterized by fiber numbers $\rvert n = 0,
    l = 2 \rangle$. Figure reproduced from
    \citep{leifer2021predicting}. Copyright \copyright ~2021, The Author(s). }
  \label{regulon}
\commentAlt{Figure~\ref{regulon}: 
 Described in caption/text. No alt-text required.
}
\end{figure}

Beyond operons, the next trivial network structure that can
synchronize genes is the regulon,\index{regulon } defined as the set of genes
regulated by a single TF. This kind of circuit is abundant in the {\it
  E. coli} TRN, see Table \ref{tab:FiberBBstatistics}.  Figure 
~\ref{regulon}a shows an example, the regulon of the transcription
factor Fis in \emph{E.~coli}. 
A note on notation: In bacteria, {\it fis} refers to the gene while Fis denotes the protein. Typically, there is one protein per gene, but eukaryotes exhibit variation due to splicing.

The regulon contains four units: the
operon \textit{cbpAM} and the genes \textit{gltX, gyrB, msrA}.  Gene
{\it fis} is, in turn, regulated by Crp. Crp and Fis are the two
master regulators of carbon utilization.  The regulon is an example
of the most trivial form of symmetry.  Assuming that the genes have no
other regulators, the symmetry can be captured by an automorphism of a
simple permutation in the regulon, for instance \textit{cbpAM}
$\leftrightarrow$ \textit{gltx}, or any permutation of the four genes
in Fig.~\ref{regulon}a.

These trivial fibration circuits are the same as some network motifs.
In the motif\index{motif } nomenclature
of \citep{alon2019} the $\rvert 0,2\rangle$ fiber shown in
Fig. \ref{regulon}b is a {\it FAN motif}, while the
$\rvert 0,1\rangle$ fiber depicted in Fig. \ref{regulon}a is a {\it star
  motif}.

The symmetry group of this building block formed by the two symmetric
genes is the symmetric group $\mathbb{S}_4$ (recall Definition \ref{def:symm_gp}).  Here, the symmetry group $\mathbb{S}_4$ of
the regulon in Fig. \ref{regulon}a contains all $4!=24$
permutations of the four genes.
The genes in the regulon are thus synchronized in orbits generated by
 automorphisms, so the regulon is also a fiber. 
The symmetry group $\mathbb{S}_4$
confirms the known fact that the genes in the regulon, namely \textit{cbpAM}, \textit{gltX}, \textit{gyrB}, and
\textit{msrA}, synchronize. However, these genes do not synchronize with the
regulator \textit{fis} since the regulator does not belong to the
orbit.  This is because it receives an edge from {\it crp}, thus
breaking the fibration symmetry with its regulon.

When the genes in a regulon are also under the control of other TFs,
there is a chance that the regulons might also preserve the
automorphism symmetries.  For instance, the single-regulon circuit
controlled by {\it fis} (Fig.~\ref{regulon}a) can be augmented by a
second regulator as in Fig.~\ref{regulon}b. The same symmetric group
$\mathbb{S}_4$ describes the symmetry among the genes {\it clrA, fiu} and
operons {\it entCEBAH, fepA-entD} since all of them can be
permuted. These genes are synchronous but not with the regulators
{\it crp}, {\it fis}.

These circuits are characterized by fiber numbers\index{fiber number } $\rvert n = 0, l
\rangle$ since they have no loops inside the fiber ($n=0$), and $l$
external regulators. These circuits do not need the input tree to be
characterized. This is because, in a circuit without cycles, $n=0$, only
the regulator is needed to characterize the dynamical state of
synchronization, so the input tree becomes the input set, reveling their triviality.

\section{Fibration building block landscape across networks and species}

In previous sections, we have shown that fibration building blocks are
organized in a hierarchy that can be classified using fiber numbers,
the branching ratio with integer and fractal dimensions, and the number
of regulators and regulated genes. This hierarchy is mainly based
on bacterial genetic networks.  \cite{morone2020fibration} extended these
results to complex networks of different domains and biological
networks of different species.  The fiber finding algorithm of Section
\ref{sec:software} was applied to 373 datasets across different
domains. Among biological networks, the analysis has been performed on
neural networks, gene regulatory networks, disease networks, signaling
pathways, and metabolic networks. Networks of different species have
been studied, including the bacteria \textit{E. coli},
\textit{B. subtilis}, \textit{Salmonella enterica} and
\textit{Mycobacterium tuberculosis}, along with networks of
\textit{Arabidopsis thaliana} and \textit{Drosophila melanogaster} and
of higher complexity like yeast \textit{Saccharomyces cerevisiae}, mouse
\textit{Mus musculus}, and human \textit{Homo sapiens} networks.

\begin{figure*}[t!]
	\centering
	\includegraphics[width=\linewidth]{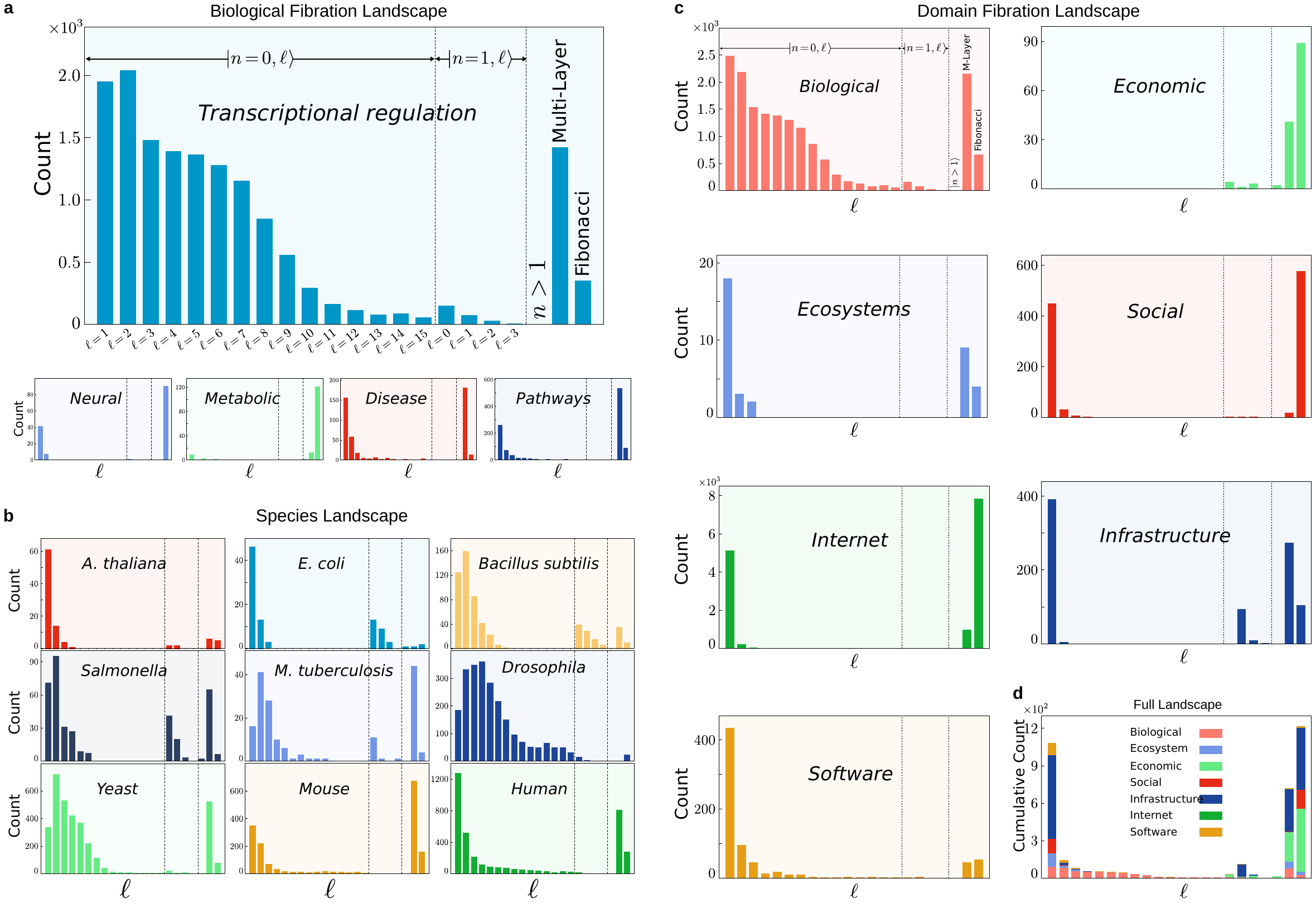}
	\caption{\textbf{Fibration building block
            landscape}. (\textbf{a}) Distributions of building block
          fiber numbers across different types of biological networks,
          summed over the species. (\textbf{b}) Distributions of
          building block fiber numbers across species, from bacteria to
          humans, summed over the type of the network. (\textbf{c})
          Distributions of building block fiber numbers across
          distinct domains of networks. (\textbf{d}) Distributions of
          building block fiber numbers for the union of all the domains
          in (\textbf{c}) normalized by the total number of nodes in the
          given domain. Results are scaled by multiplying by
          $10^4$. Figure reproduced from \citep{morone2020fibration}. Copyright \copyright ~ 2020, Leifer
{\it et al.}}
	\label{PNAS:Fig4}
\commentAlt{Figure~\ref{PNAS:Fig4}: 
Illustrative.  Described in caption/text. No alt-text required.
}
\end{figure*}

Figure \ref{PNAS:Fig4}a and b show that fibration building blocks\index{building block !for many species } are
 present in all the species mentioned above and types of
networks. Distributions of the building block fiber numbers (including
multi-layer and Fibonacci) across types of biological networks, summed
over species, are shown in Fig.~\ref{PNAS:Fig4}a. Figure \ref{PNAS:Fig4}b
shows distributions of the building block fiber numbers across
different species summed over the type of the network.
A common feature of these networks is that
the number of fibers decreases as $l$ increases, for both $n = 0$ and $n = 1$. Another interesting observation is that bacteria contain a
higher proportion of $n = 1$ building blocks compared to more complex
organisms like yeast, mice, and humans, which have a higher portion of
Fibonacci and multi-layered composite building blocks. One 
possible explanation might be that nodes in networks of
bacteria have more self-loops, while networks of mice and humans
contain longer cycles in their networks, creating more elaborated building blocks.

Beyond biological networks, \cite{morone2020fibration} examined networks in various domains, including ecosystems, the internet, software, economics, social interactions, and infrastructure. The distributions of fiber numbers for the building blocks studied are presented in Fig. \ref{PNAS:Fig4}c. Additionally, the normalized sum of all distributions across these domains and species is illustrated in Fig.~\ref{PNAS:Fig4}d, which shows the cumulative number of fibers for all domains and species per network size of 10,000 nodes. The results indicate the presence of symmetry fibrations, supporting the existence of fibration building blocks in these networks across multiple domains and species. This research serves as a foundational step toward applying the concept of symmetry fibration beyond biological contexts, helping to describe the building blocks of networks in any domain where synchronization plays a significant role.


\chapter[Fibration Analysis of Metabolic Networks: Complex Building Blocks]{\bf\textsf{Fibration Analysis of Metabolic Networks: Complex Building Blocks}}
\label{chap:complex}

\begin{chapterquote}
The complexity of fibration building blocks is intertwined with
the complexity of their cycles. These cycles can be within the fiber,
between the fiber and its regulators, or between fibers.  The simpler
 building blocks are found in the TRN of {\it E. coli}.  These
building blocks are characterized by integer branching ratios $r=0,
1, 2$. A few more complex (yet still simple) Fibonacci building blocks with
fractal branching ratios are also found in {\it E. coli}. TRNs from
more complex species show a high predominance of simple
Fibonaccis. The level of complexity explodes when we look at metabolic
networks. Thanks to intricate cycles, these networks are characterized
by highly complex fibration symmetries and building blocks.
\end{chapterquote}
  
\section{Metabolic networks}  

The fibration analysis of Chapter \ref{chap:hierarchy_2} studies
building blocks in the TRN, mainly in the bacterium {\it E. coli},
but also in more complex eukaryotes.  Since other network structures also exist, we focus next on metabolic networks. A {\em metabolic
network}\index{network !metabolic } is composed of the chemical reactions of metabolic pathways,
where enzymes transform metabolites that determine the physiological
and biochemical properties of a cell.  Figure ~\ref{Fig:1} shows a
typical construction of a metabolic network from its reactions,
enzymes, metabolic products, and educts. In the most typical case,
the metabolic network comprises metabolites as nodes and reactions as
edges (Fig.~\ref{Fig:1}a). This network representation does not capture the ODEs describing the dynamics of the metabolites. 
The hypergraph representation of Fig. \ref{F:metabolic_hypergraph}b in Section \ref{sec:hypergraph-metabolic} captures the correct representation of the ODEs for the fibration analysis of the metabolic network. A fibration analysis of these
networks in {\it E. coli} reveals a just simple structure of building
blocks.\index{building block }

However, by constructing a `dual' network where the enzymes are the
nodes and the edges are metabolites, a myriad of complex building blocks
are revealed.  These networks are called {\em enzyme networks}.\index{network !enzyme }
In these networks, directed links resemble the transmission of metabolic information from an enzyme source to target enzymes. Therefore, enzyme networks serve as natural extensions of the TRNs discussed in the previous chapter, with transcription factors acting as messengers.

Such a view follows a distributed
computational networked system captured by the fibration, where the
behavior of network nodes is fiberwise constant; that is, processors in
the same fiber are in the same state. Analogously, enzymes in fibers
of the enzyme network are synchronous in a way akin to
processing units that are synchronous in a computer. Such an
assumption suggests that enzyme-based fibers are elementary building
blocks that bundle metabolic information in metabolic pathways.

Below, we provide a fibration analysis of enzyme networks following
\citep{alvarez2024symmetries} that uncovers building blocks of much
greater complexity than those in TRNs. Compared to the {\it E. coli}
TRN---which is mostly dominated by simple fiber structures---more
complex structures exist in the metabolism, thanks to the intricate cycle
structures of their enzyme pathways. These cycles are made of
feed-forward and feedback structures that provide a systematic
classification in terms of the fractal dimensions of the input trees.

These building blocks are more functionally homogeneous than enzymes
in network motifs or network modules, pointing to a level of complexity
in the organization and architecture of biological networks that
topological motifs and modules found through statistical means simply
miss. As a consequence, such fibration building blocks harbor more
functional information compared to motifs and modules. Furthermore, since
fibers represent information flow that goes beyond the boundaries of
statistical over-representation of interactions between sets of nodes,
such groups of enzymes may well be a better entry point not only to
elucidate pathways from a different angle, but also to find novel
pathways. Along the same lines, such fibers may also be used 
to find new drug targets, since fibers capture information flow,
potentially providing a novel way to indicate points of therapeutic
intervention.

\section{Enzyme network of {\it E. coli}}
\index{network !enzyme }\index{E. coli @{\em E. coli} }

\begin{figure}[t!]
\centering
\includegraphics[width=0.95\textwidth]{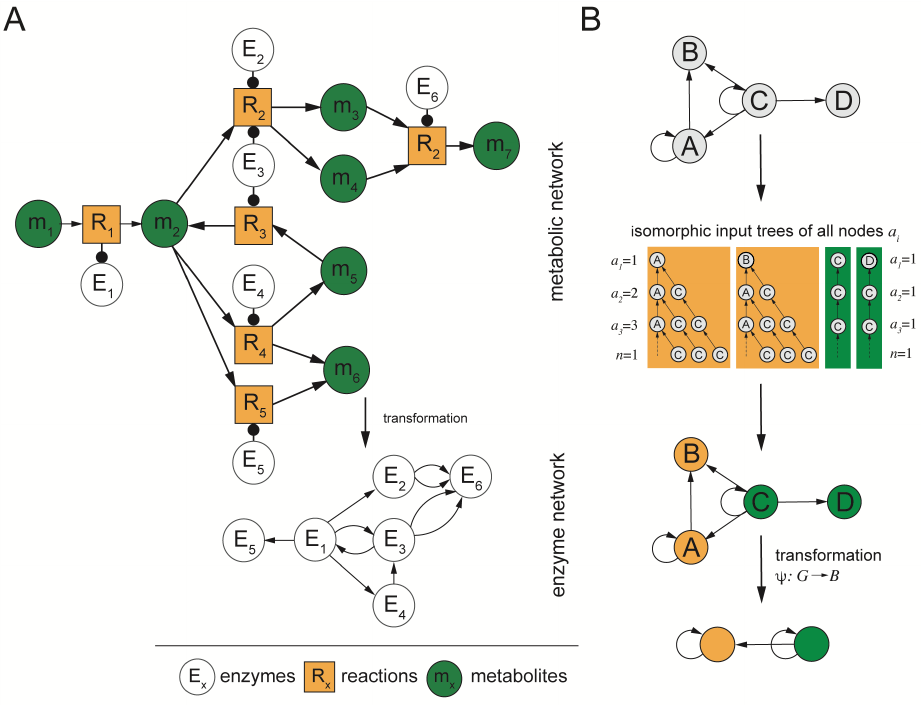}
\caption{\textbf{Construction of enzyme metabolic
    networks and determination of symmetric network characteristics.}
  (\textbf{a}) Enzyme network: Enzymes (E) are connected via a directed
  link when the products (m) of a preceding reaction (R) that is
  catalyzed by an enzyme are turned into a reactant in the subsequent
  reaction that is controlled through a different enzyme. (\textbf{b})
  Using such directed enzyme networks, we perform fibration
  analysis in search of the building blocks. Figure reproduced from
  \citep{alvarez2024symmetries}. }
  \label{Fig:1} 
\commentAlt{Figure~\ref{Fig:1}: 
A (top) Graph with nodes m1-m7 (green), R1-R5 (orange), E1-E6 (white).
Sharp arrows (activators) m1-R1, R1-m2, m2-R2, m2-R4, m2-R5, R2-m3, R2-m4,
R4-m5, R4-m6, m3-R2, m4-R2, R2-m7. Arrows with rounded ends (repressors): R1-E1, E2-R2, E3-R2, E3-R3, E4-R4, E5-R5, E6-R2.
A(bottom) graph with nodes E1-E6(white). Sharp arrows E1-E5, E1-E2, E1-E3(twice), E1-E4, E2-E6(twice), E3-E6(twice), E4-E3.
B (top) Graph with nodes A, B, C, D. Sharp arrows AA, AB, CA, CB, CC, CD.
B (second) Input trees of nodes show coloring with A=B, C=D is balanced.
B (third) Top graph with this coloring.
B(bottom) base: Orange node (for A, B) and green node (for C, D).
Arrows orange-orange, green-green, green-orange.
}
\end{figure}

\cite{alvarez2024symmetries} search for symmetries in the {\it
  E. coli} metabolic network built from the Ecocyc \citep{Keseler:21}
using the 'All Enzymes of E. coli K-12 Substrate MG1655' data set,
together with information on metabolic reactions \citep{Santos:18} that are
catalyzed by such enzymes from RegulonDB\index{RegulonDB } \citep{regulon2016}. These
datasets provide metabolic reactions and corresponding enzymes,
capturing 2,628 reactions between 2,093 metabolites.  To curb the
influence of ubiquitous metabolites such as ATP or H$_2$O, we sort
all metabolites according to their occurrence in the underlying
metabolic reactions and manually discard the most occurring metabolites.
After this manual curation, we obtain 2,628 reactions with 1,990
nontrivial metabolites, catalyzed through 1,930 enzymes. To construct
an enzyme-specific network (Fig.~\ref{Fig:1}a), we connect enzymes
when the product in a preceding reaction turns into a reactant in the
subsequent reaction. As a result, we obtain an enzyme-enzyme network
that captures 1,753 enzymes in a web of 22,011 directed edges.

\section{Fibration analysis of the {\it E. coli} enzyme network}

The first step in the analysis is to derive the SCCs\index{SCC } of the
network. The full enzyme network of 1,753 enzymes is
characterized by a large SCC comprising 70\% of the nodes. The other SCCs
are very small, composed of 4 enzymes on average. This is in contrast
to the TRN, which contains a few small SCCs of similar size.

Four function-specific metabolic subnetworks are then considered. The
relevance of metabolic reactions is established by the presence of
transcription factors that control the expression of the corresponding
enzymes, according to RegulonDB \citep{Santos:18}. In particular, we filter
metabolic reactions through enzymes that are regulated by
transcription factors expressed during oxidative stress 
\citep{Gama-Castro:14}. In the same way, we extract transcription
associated with carbon sources from the RegulonDB Sensor Units
Datasets \citep{Santos:18}.  We extract carbon metabolic pathways where
one-carbon moieties are transferred from donors to intermediate
carriers, ultimately used in methylation reactions or in the synthesis
of purine and thymidine, which are used in DNA building blocks.  We
also derive metabolic reactions of amino-acid metabolism from genes
related to a specific growth condition obtained from
\citep{Martinez:03}. We also account for oxidative stress as a
natural consequence of aerobic metabolism. Finally, a glycolysis
network is obtained by filtering all metabolic reactions catalyzed by
enzymes that appeared in the corresponding KEGG pathways
\citep{Kanehisha:23}. This is the major bacterial energy resource.  

\begin{figure}[h]
\centering
\includegraphics[width=\textwidth]{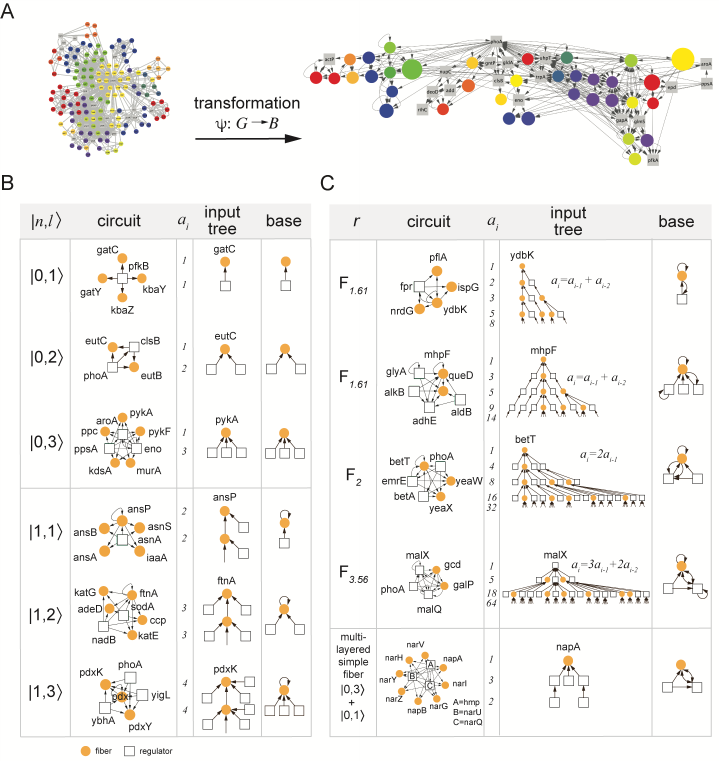}
\caption{\textbf{Fibers in the enzyme network.}  (\textbf{a}) Fibration applied to the glycolysis enzyme network in {\it
    E. coli}, where colors refer to fibers. (\textbf{b}) Simple
  fibration building blocks similar to those of the TRN are found in
  amino-acid synthesis, carbon metabolism, glycolysis, and oxidative
  stress enzyme networks.  (\textbf{c}) Examples of complex Fibonacci
  building blocks. Figure reproduced from
  \citep{alvarez2024symmetries}.}
  \label{Fig:2}
\commentAlt{Figure~\ref{Fig:2}: 
 Described in caption/text. No alt-text required.
}
\end{figure}

The fibers in these networks are obtained using the minimal balanced
coloring algorithm of Section \ref{sec:software}.  We find 44 fibers
in the amino-acid synthesis network, 32 fibers in the carbon
metabolism network, 35 fibers in the glycolysis network
(Fig.~\ref{Fig:2}a) and 30 fibers in the oxidative stress network. The lists of all fibers of each network can be found in
\citep{alvarez2024symmetries}. Fibers cover the networks
significantly: the majority of nodes appear in fibers (Table
~\ref{tab:breakdown-metabolic}). Moreover, at least roughly half of the
nodes belong to an SCC, while a large fraction of such nodes appear in
fibers as well.

\begin{table*}[h!]
  \centering
  \begin{tabular}{|l|rc|rc|}
    \toprule
     & \multicolumn{2}{c|}{\textbf{\emph{Amino Acid Synthesis}}} & \multicolumn{2}{c|}{\textbf{\emph{Carbon Metabolism}}}\\
     & \textbf{\emph{Nodes}} & \textbf{\emph{\%}} & \textbf{\emph{Nodes}} & \textbf{\emph{\%}}\\
    \textbf{\emph{Total nodes}}  & \textbf{261} & \textbf{100\%} & \textbf{149} & \textbf{100\%} \\
    Nodes in fibers    & 198 & 75.9\% & 95 & 63.8\%  \\
    \midrule
    In SCCs (in fibers) & 147 (107) & 56.3(41)\% & 93 (65) & 62.4(43.6)\% \\
    Connectors (in fibers) & 18 (5) & 6.9(1.9)\% & 22 (4) & 14.8(2.7)\% \\
    $k_{\rm out}$ shell (in fibers) & 96 (86) & 36.8(33)\% & 34 (26) & 22.8(17.4)\%\\
    \midrule    \midrule
    & \multicolumn{2}{c|}{\textbf{\emph{Glycolysis}}} & \multicolumn{2}{c|}{\textbf{\emph{Oxidative Stress}}}\\
     & \textbf{\emph{Nodes}} & \textbf{\emph{\%}} & \textbf{\emph{Nodes}} & \textbf{\emph{\%}}\\
    \textbf{\emph{Total nodes}}  & \textbf{153} & \textbf{100\%} & \textbf{168} & \textbf{100\%} \\
    Nodes in fibers    & 135 & 88.2\% & 137 & 81.5\%  \\
    \midrule
    In SCCs (in fibers) & 80 (66) & 52.3(43.1)\% & 80 (62) & 47.6(36.9)\% \\
    Connectors (in fibers) & 1 (0) & 0.7(0)\% & 9 (3) & 5.4(1.8)\% \\
    $k_{\rm out}$ shell (in fibers) & 72 (69) & 47.1(45.1)\% & 79 (72) & 47(42.9)\%\\
      \bottomrule
    
  \end{tabular}
  \vspace{10pt}
  \caption{ \textbf{Structural composition of the four enzyme
      subnetworks.}  We show the coverage of each network by
    fibers as well as the structural breakdown of the
    network. Typically, the networks consist of 4 SCCs, where one
    component captures the majority of nodes. Enzymes are also
    classified as connector nodes, or nodes in the $k_{\rm ou}$ shell, sending signals to other parts of the cell. Connector nodes send
    outputs to the SCCs, while the nodes in the $k_{\rm out}$ shell
     receive information only from nodes in the SCCs. Table reproduced
    from \citep{alvarez2024symmetries}.}
  \label{tab:breakdown-metabolic}
\end{table*}

\section{Complex fibration building blocks}

\index{building block !complex fibration }
A first analysis reveals that the simple
fibration building blocks found in the TRN in Chapter
\ref{chap:hierarchy_2} are still present in enzyme networks
(Fig.~\ref{Fig:2}b). These include all fibers with integer branching
ratio $n=0, 1, 2$, the simple Fibonaccis, and the multilayer simple
fibers. Their bases are sketched in the first three circuits in
Fig.~\ref{Fig:3}a.

However, these circuits form a minority of the fiber
structures encountered: most fiber structures in 
enzyme networks correspond to much more complex
structures. In particular, a pair of enzymes can
`communicate' through numerous different metabolites, so there are
multiple different edges between the enzyme nodes. Such links
lead to a more densely connected network structure, which produces a
high number of cycles between nodes. However, these cycles 
appear not only between nodes in the same fiber but also between nodes in different
fibers. Such a feature is ubiquitous in
metabolic networks. It produces much more complex structures, notably
the Composite Fibonacci Fibration building block.

\section{Composite Fibonacci fibration building blocks}

\begin{figure}[t!]
\centering
\includegraphics[width=0.95\textwidth]{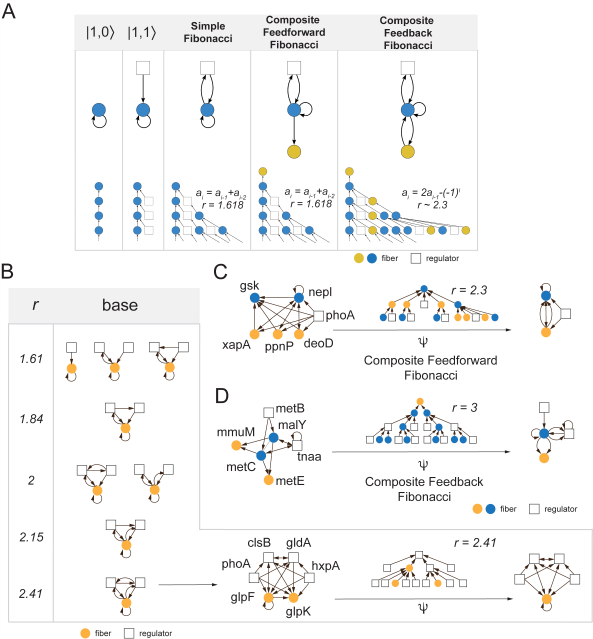}
\caption{\textbf{Complexity axes of Fibonacci
    fibers.}  (\textbf{a}) `Horizontal axis' shows transition from
  simple to complex types of building blocks through the addition of
  feedback loops. (\textbf{b}) `Vertical axis of
  complexity' corresponds to an increase in complexity, which increases the
  branching ratio. Such structures emerge when more
feedback loops between the underlying nodes are added. (\textbf{c}) In a feed-forward
  Fibonacci, we find a fiber \emph{mmuM-metE} with a non-zero branching ratio
  even though it does not send any feedback to its regulators because it
  is regulated by another fiber \emph{metC-malY}. (\textbf{d}) In a Feedback Fibonacci, two fibers (blue
  and yellow nodes) regulate each other, pointing to a branching input
  tree structure with $r = 2.3$. Figure reproduced from
  \citep{alvarez2024symmetries}. }
  \label{Fig:3} 
\commentAlt{Figure~\ref{Fig:3}: 
 Described in caption/text. No alt-text required.
}
\end{figure}

A key structure in enzyme networks is the \emph{Composite
  Fibonacci}  building block\index{building block !composite
  Fibonacci} composed of
more than one fiber (Fig.~\ref{Fig:3}a, c, d). These fibers are further split into
\emph{Composite feed-forward Fibonaccis}\index{composite feed-forward Fibonacci } and \emph{Composite Feedback
  Fibonaccis}.\index{composite Feedback
  Fibonacci } These building blocks are composed of either
through a fiber that regulates another one in a feed-forward manner or
by forming a feedback loop between the underlying fibers.
Composite feed-forward Fibonaccis are instances of
multi-layer fibers, in which a fiber regulates another in a
feed-forward manner, where the regulating fiber belongs to a Fibonacci
building block. In particular, the regulated fiber acquires a branching
input tree from a regulating Fibonacci fiber, allowing its own
input tree to branch. In Composite Feedback Fibonaccis, a
layer of complexity is added by forming a loop between the fibers, such
that each fiber acts as a regulator of the other.

These structures are abundant and can be systematically classified
through their complexity by two `complexity axes', facilitating
the structured organization of building blocks according to their
complexity levels. A `horizontal axis' relates the simple building blocks to more
complex types (Fig.~\ref{Fig:3}a). Traversing
different bases, we start with the base $\vert n = 1, l = 0\rangle$, a single fiber regulating itself. The addition of an
external regulator gives the structure $\vert n = 1, l =
1\rangle$.  Although both of these input trees are infinite as a
result of self-regulation, they do not branch
(Fig.~\ref{Fig:3}a). However, when we add an edge from the fiber back
to the regulator, we obtain the simplest Fibonacci fiber\index{Fibonacci fiber } with a
branching input tree\index{input tree } with a fractal dimension\index{fractal dimension } of $r = 1.618$.

Adding a feedback loop between a regulator and a
  fiber that points to a simple Fibonacci fiber lets the
  corresponding input tree branch, gives a \emph{Composite feed-forward Fibonacci},\index{composite feed-forward Fibonacci } since the
Fibonacci fiber is now acting as a regulator of another fiber
downstream. We show this case in Fig.~\ref{Fig:3}a, where we add a
yellow fiber that is regulated by a simple Fibonacci fiber in blue,
creating a Composite feed-forward Fibonacci fiber. The input tree of the 
yellow fiber is given by the input tree of the blue fiber plus 
the new yellow node that connects to the root of the
input tree of the blue fiber. Since the only difference between these
input trees is the extra yellow node, the branching ratio
remains unchanged from that of the simple Fibonacci and Composite
feed-forward Fibonacci fibers.  In turn, the addition of yet another
edge that connects the new fiber with the original Fibonacci fiber
creates a feedback loop, leading to a \emph{Composite Feedback
  Fiber}. In consequence, the underlying input tree changes,
leading to a branching ratio of 2.

A second `vertical axis of complexity' corresponds to an increase in
complexity through an increasing branching ratio. In Fig.~\ref{Fig:3}b
we increase the branching ratio\index{branching ratio } from the simplest Fibonacci ($r =
1.618$) to a building block with a branching ratio of $r = 2.41$ in a
Fibonacci building block that revolves around the regulation of {\it
  glpFK} through regulators that are connected through mutual
feedback loops.

In Fig.~\ref{Fig:3}c, since the blue
  regulator fiber is part of a Fibonacci structure, the input tree of
  fiber \emph{mmuM-metE} includes the input tree for \emph{malY-metC}. This
 is a branching tree, making its input tree branch. 
 
The type of Fibonacci structure in Fig.~\ref{Fig:3}d 
  features two (or even more) fibers representing a further step in
  complexity from previously observed fibers.

\begin{figure}[t]
\centering
\includegraphics[width=0.7\textwidth]{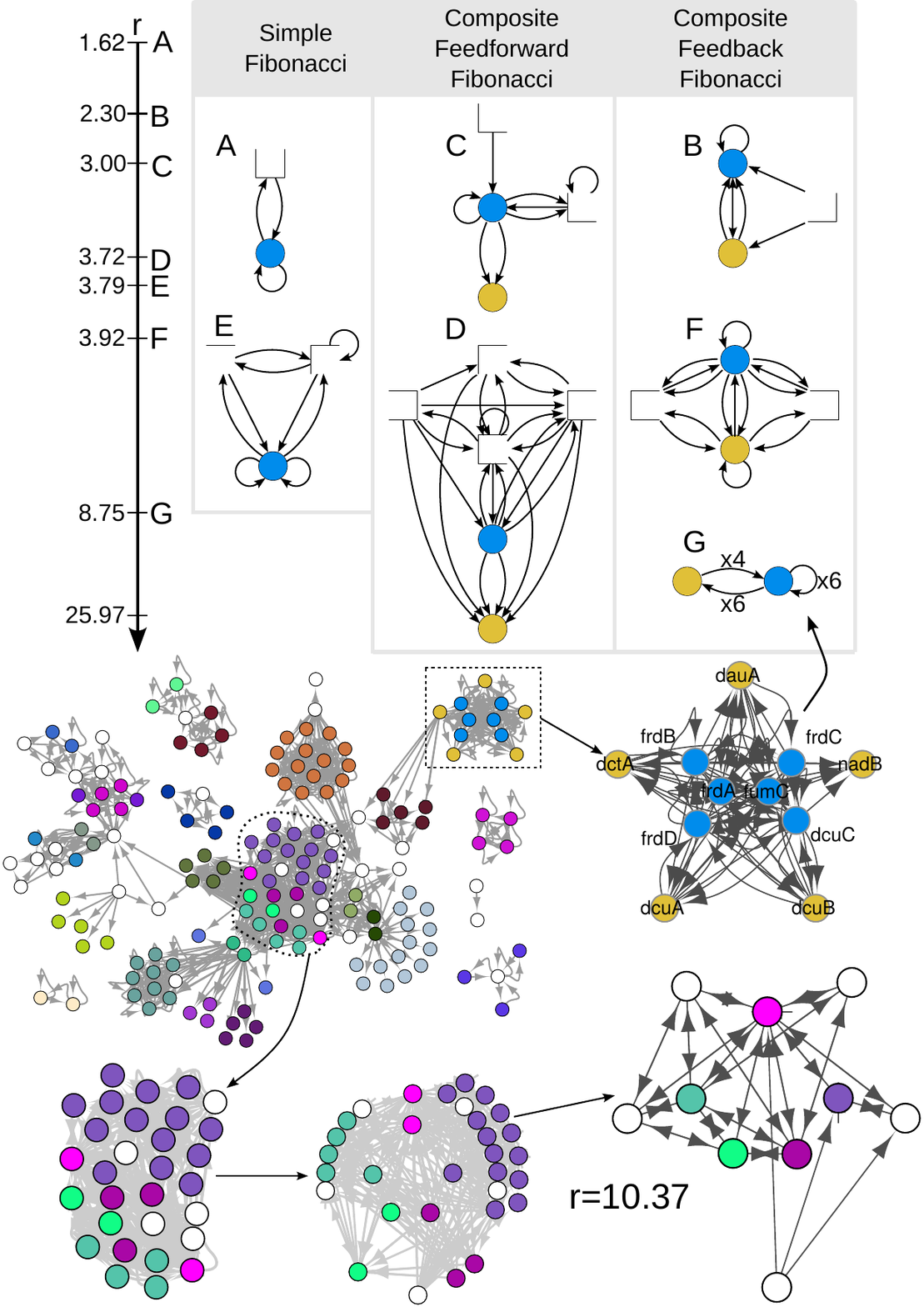}
\caption{\textbf{Complexity ladder of the branching
    ratio.} We show the bases of different building blocks
  corresponding to different types of \emph{Fibonaccis}. The axis on
  the left shows the range of branching ratios observed in the
  four different enzyme networks; labels correspond
  to the building blocks in the table. The first column depicts
  \emph{Simple} Fibonaccis. The second column shows three
  different \emph{Feedback Fibonaccis}. The third column depicts
  \emph{feed-forward Fibonaccis}, where the fiber in a Fibonacci
  structure (blue fiber) regulates a second fiber (yellow fiber) in a
  feed-forward manner. The corresponding
  complete building block of structure {\bf G } 
  illustrates how such a complex structure can be reduced to
  a much simpler, more comprehensible structure at its base.
  Figure reproduced from \citep{alvarez2024symmetries}.}
  \label{Fig:4}
\commentAlt{Figure~\ref{Fig:4}: 
 Described in caption/text. No alt-text required.
}
\end{figure}

All Fibonacci structures, such as simple Fibonaccis, Composite
feed-forward and Composite Feedback Fibonaccis are characterized by a
branching input tree. Specifying only the branching ratio\index{branching ratio }  of a
building block does not determine the class of the building
block. Two building blocks with the same branching ratio
can belong to different classes, while a simpler class can have a
higher branching ratio than a more complex class.  For example in
Fig.~\ref{Fig:4}, building block E corresponds to a simple Fibonacci
building block yet has a higher branching ratio $r = 3.79$
than blocks B and C, even though both are examples of Composite
Fibonacci building blocks.

One way to determine
the branching ratio\index{branching ratio }  of any input tree $r$, which is a simple practical option in
the cases we need, is to determine
the first few terms of the sequence $a_i$,
which is the number of nodes in the $i$th
layer of the input tree. This can be done by hand for some simple networks, using a recurrence relation \index{recurrence relation } such as the usual one for Fibonacci numbers\index{Fibonacci number }  if necessary. Then we compute the ratios $r_i=a_{i+1}/a_i$ and stop when 
the value of $r_i$, to the required accuracy,
doesn't change
\citep{alvarez2024symmetries}.

In general, however, there are two mathematical fine points. 
First, the sequence $a_{i+1}/a_i$ may converge slowly. 
Second---although it is unlikely to arise in
most applications---the sequence need not converge at all, as
discussed in Section \ref{sec:definition-building}, so \eqref{eq:branching_new}
must be used instead. In all cases the Perron--Frobenius Theorem\index{Perron--Frobenius Theorem } implies
that the branching ratio for a given node $k$, defined via
\eqref{eq:branching_new}, is the largest eigenvalue of the adjacency
matrix for the induced subnetwork defined by all nodes that occur in the input tree of $k$.  For a proof, which is not entirely straightforward, see \citep{boldistewart2024}. This eigenvalue can be computed rapidly by standard computer algebra packages.

In the next section, we analyze these complex building blocks in more detail.

\subsection{Composite Feed-Forward Fibonacci building blocks}
\index{building block !composite feed-forward Fibonacci }

A Composite Feed-Forward Fibonacci is a multi-layer building block with
a branching input tree, the key feature of Fibonacci fibers. Such a
structure occurs when a fiber in a Fibonacci structure regulates
(an)other fiber(s) downstream in a feed-forward manner. The Fibonacci
fiber-regulator, as part of a Fibonacci structure, has a branching input
tree that, in turn, gets `passed along' to the nodes it regulates. In
other words, the Fibonacci fiber `inherits' and passes its input tree
to the regulated fiber, causing the regulated fiber's input tree to
branch.
This is exemplified in ~Fig.~\ref{Fig:3}c where  yellow nodes
  \emph{mmuM} and \emph{metE} only receive input from both \emph{metC}
  and \emph{malY}. Focusing on \emph{metC} and \emph{malY} we observe
  that both nodes are embedded in a Fibonacci structure through their
  regulator \emph{tnaa}. As for the yellow fiber {\it mmuM} and {\it
    metE}, the input tree of the blue fiber is
  `inherited', thus prompting its own input tree to branch as
  well.

A Composite Feedback Fibonacci occurs when the feedback loop (or
loops) of a Fibonacci structure crosses two or more fibers, as
in Fig.~\ref{Fig:3}d. The input tree of the blue fiber \emph{gsk-nepI}
includes the nodes in the yellow fiber \emph{xapA-ppnP-deoD} as well,
since they act as regulators of the blue fiber. In turn, the nodes in
the blue fiber regulate the yellow fiber, thus creating a loop between
them.  As a consequence of the loops between these fibers, the blue
fiber also acts as a regulator to the yellow fiber. This loop causes
the input tree to branch as $a_i = 1, 5, 11, 26, 137, 314...$,
pointing to a branching ratio of $r = 2.3$.

As a consequence, one fiber can not be defined without the other(s),
since each fiber acts as a regulator for the other fiber(s), suggesting
that the separation of the regulator from the other fiber is
impossible. These structures can connect multiple fibers. In the simplest possible case, multiple
fibers are linked in a ring-like cycle
where each fiber is connected to only two others ({\it i.e.} one by an
input and one by an output). In turn, all the fibers can be directly
linked to each other, resembling an almost fully connected
graph. Distinguishing these two structures, only one feedback loop
({\it i.e.} one single cycle of length $n$) encompasses all fibers in
the first case, while a multitude of different feedback loops exists in
the other case. In more detail, in a base that connects $n$ fibers in
such a fully connected scheme the total number of cycles comprises
$\binom{n}{2}$ cycles of length 2 plus $\binom{n}{3}$
cycles of length 3, and so on up to one cycle of length $n$ (without
accounting for external regulators), pointing to a higher branching
ratio.

The oxidative enzyme network\index{network !oxidative enzyme } in Fig.~\ref{Fig:4} provides examples of
this kind of building block: the base of the structure shown at the
bottom right corner is almost a fully-connected network encompassing 5
different fibers and 4 regulators that participate in the feedback
loops (plus another regulator not included in any loop). We find a complex structure, pointing to an `entanglement' of numerous
cycles connecting several fibers with a high branching ratio.\index{branching ratio !high }  All
 enzyme networks\index{network !enzyme } exhibit this type of arrangement at the center
of the network, as shown in Fig.~\ref{Fig:4} in the oxidative
network. Such structures present the highest branching ratios.
The amino-acid network has one such structure with $r = 12.39$,
and another with $r = 25.97$, while the carbon network
with $r = 10.81$, the glycolysis network\index{network !glycolysis } has one with $r = 13.65$, and the
oxidative network has one with $r = 10.37$.

As a corollary, these two composite Fibonacci structures can be
combined, when a fiber in a Composite Feedback loop further
regulates---albeit in a feed-forward manner---other fiber(s), as shown
with the fibers regulated outwardly from the central Composite
  Feedback structure in the oxidative network at the bottom of
Fig.~\ref{Fig:4}.

Lastly, considering only \emph{shortest} cycles to
construct building blocks leads to a fiber that might be part of more
than one building block. Such a situation occurs when longer cycles
that cross (an)other fiber(s) are not included in a fiber's building
block, because they are longer than the shortest cycle between fibers and
regulators. For example, the addition of a second fiber to a simple
Fibonacci building block ($r = 1.618$) regulates the first fiber and
the only external regulator. This addition creates a new cycle of
length 3 that includes both fibers. When the building block of the
original fiber is constructed, this new cycle is ignored since
it is longer than the shortest cycle between the original fiber and
its external regulator. However, when constructing the building block
of the second fiber, the original fiber must be included since it
regulates this newly added fiber. This second building
block does include the longer cycle and both fibers, resulting in a
Composite Feedback Fibonacci building block. This is observed
numerous times in enzyme networks.

In such cases we consider the smaller building block to be the
`defining' one for the fibers included in both, since the
shortest cycles have a dominant effect on the dynamics of this fiber
or fibers. The bigger building block is naturally the
defining one for only the fiber(s) included in it, and it
can be considered a form of `higher order' correction on the dynamics
of the fiber(s) present in both building blocks.

\begin{figure}[h]
\centering
\includegraphics[width=.8\textwidth]{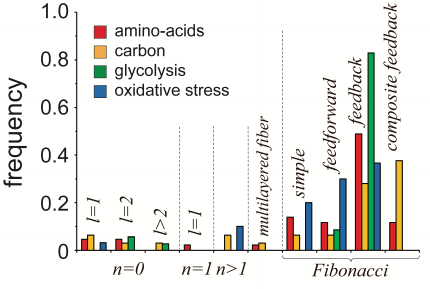}
\caption{\textbf{Fiber landscape in the enzyme
    networks.}  Distribution of fibers in the amino-acid synthesis,
  carbon metabolism, glycolysis and oxidative stress networks,
  suggesting that Fibonacci fibers are most abundant in all
  networks. Figure reproduced from \citep{alvarez2024symmetries}.}
  \label{Fig:5} 
\commentAlt{Figure~\ref{Fig:5}: 
 Described in caption/text. No alt-text required.
}
\end{figure}

\section{Fibration landscape of enzyme networks}

The large range of branching ratios found in  metabolic networks,
ranging from $r = 1.618$ to $r = 25.97$, prompts us to understand how adding or changing cycles influences the complexity of the
circuits. Figure ~\ref{Fig:4} showcases different examples of
observed single and complex composite Fibonacci building blocks and their
place in such a complexity landscape. The input tree is characterized
by the sequence $a_i$ of enzymes in the $i$th generation. As for
Fibonacci fibers,\index{Fibonacci fiber } this sequence can often be described by a closed form
recurrence relation,\index{recurrence relation } such as $a_i = a_{i-1} + a_{i-2}$, which is observed
with most basic Fibonacci fibers. The first term
($a_{i-1}$) represents a self-loop, a cycle of length 1, while the
second term ($a_{i-2}$) represents a cycle of length 2 between the
fiber and the regulator. Generally, a cycle\index{cycle } of length \emph{d}
contributes a term $a_{i-d}$ to the recurrence relation.

To increase the branching ratio of a building block, the number of
cycles needs to be increased, which can be achieved by changing the
multiplicity of the edges in the structure. For example, the block in
Fig.~\ref{Fig:4}g has essentially the same basic structure as the
simplest Fibonacci structure in Fig.~\ref{Fig:4}a. However, the
multiplicity of the edges is higher, indicated by $4$ directed edges
that connect the yellow fiber to the blue fiber. In turn, $6$ directed
edges link the blue fiber with the yellow fiber, while the self-loop also has
multiplicity 6, resulting in a structure with a highly elevated
branching ratio $r = 8.75$. Alternatively, we can add more nodes
and edges to increase the number of cycles, as
 in Fig.~\ref{Fig:3}b.  When the added nodes form
a part of a fiber, the added cycles create a Composite
Feedback Fibonacci, since they connect fibers through a feedback loop.

Many nodes in fibers belong to SCCs,\index{SCC } a natural consequence of
having both a high coverage of nodes that belong to fibers, as well as
a large percentage of nodes in the network belonging to SCCs. These
two facts create the perfect scenario for the existence of feedback
loops connecting multiple fibers, which point to very complex
Composite Feedback Fibonaccis with several fibers. We observe this
case in the building block at the center of the oxidative enzyme
network (Fig.~\ref{Fig:4} bottom), where we find 5
fibers and 4 regulators entangled in a myriad of cycles between them.
Such structures present the biggest branching ratios through
an elevated number of different cycles that connect multiple fibers
to a set of regulators. For example, the building block at
the center of the oxidative enzyme network (Fig.~\ref{Fig:4}) exhibits
a branching ratio $r = 10.37$.

The landscape of fibration building blocks in the studied metabolic
networks is dominated by highly complex Fibonacci structures
(Fig.~\ref{Fig:5}) rather than the simple blocks of the TRN.  Given
that these enzyme networks have many nodes in the SCCs and so many
more loops than in the TRN, we can expect
interactions between fibers, pointing to Fibonaccis as the most
predominant fiber structure in these networks.  In the TRN, we observe
only three Fibonaccis with a branching ratio\index{branching ratio } ranging between $1.38$
and $1.61$, while in enzyme networks, we observe values almost as
high as $\sim26$. Such an observation is mainly a consequence of the
increased number of cycles\index{cycle } in these networks, ensuring multiple
different possible ways to establish cycles between the fibers,
and drastically increasing their complexity.

What can we learn from this massive organization of biological complexity into fibration building blocks?
As fibers represent information flow that goes
beyond the boundaries of statistical over-representation of
interactions between sets of nodes, such groups of enzymes may well be
a better entry point to not only elucidate pathways from a different
angle, but also to find novel pathways. Along the same lines, such
fibers may also be used as a way to find new drug targets, as fibers
capture information flow, potentially providing a novel way to
indicate points of therapeutic intervention.


\chapter[Comparison Between Fibration Building Blocks, Motifs and Modules]{\bf\textsf{Comparison Between Fibration Building Blocks, Motifs and Modules}}
\label{chap:motif}

\begin{chapterquote}

This chapter explores the biological and statistical significance, as well as the functionality, of fibration building blocks through standard Gene Ontology analyses and $p$-value statistics. We compare the functionality of fibration building blocks to other methods of network decomposition, such as network motifs and modules. We discuss the insights that fibrations provide that are not captured by these alternative approaches, and we highlight the advantages of using fibrations to identify building blocks. Furthermore, we explain why alternative metrics based on statistical over-representation of circuits do not effectively capture the characteristics of functional building blocks.

\end{chapterquote}

\section{Biological significance of fibers}
  
Chapters \ref{chap:hierarchy_2} and \ref{chap:complex} have
characterized the topological features of fibration building
blocks as simple and complex.\index{building block } We have shown that increasingly larger cycle\index{cycle } arrangements
provide high complexity, but in a well-structured
way that leads to a systematic classification in terms of cycles and
branching ratios\index{branching ratio } (Figs.~\ref{Fig:3}, \ref{Fig:4}, \ref{Fig:5}). The
crux of the matter, however, is to prove that these
building blocks are significant for the functionality of the cell.

To this end, \cite{alvarez2024symmetries} consider the functional
similarity of enzymes in the fibers obtained in Chapter
\ref{chap:complex}, hypothesizing that participation in
synchronized groups of enzymes translate into functional
similarity. The mean similarity of Gene Ontology (GO) terms using
GoSemSim \citep{Yu:10} is determined over all pairs of enzymes in a
fiber.  The enrichment of a GO term is a guide to the biological
function of any group of genes.

\begin{figure}[h!]
\centering
\includegraphics[width=0.7\textwidth]{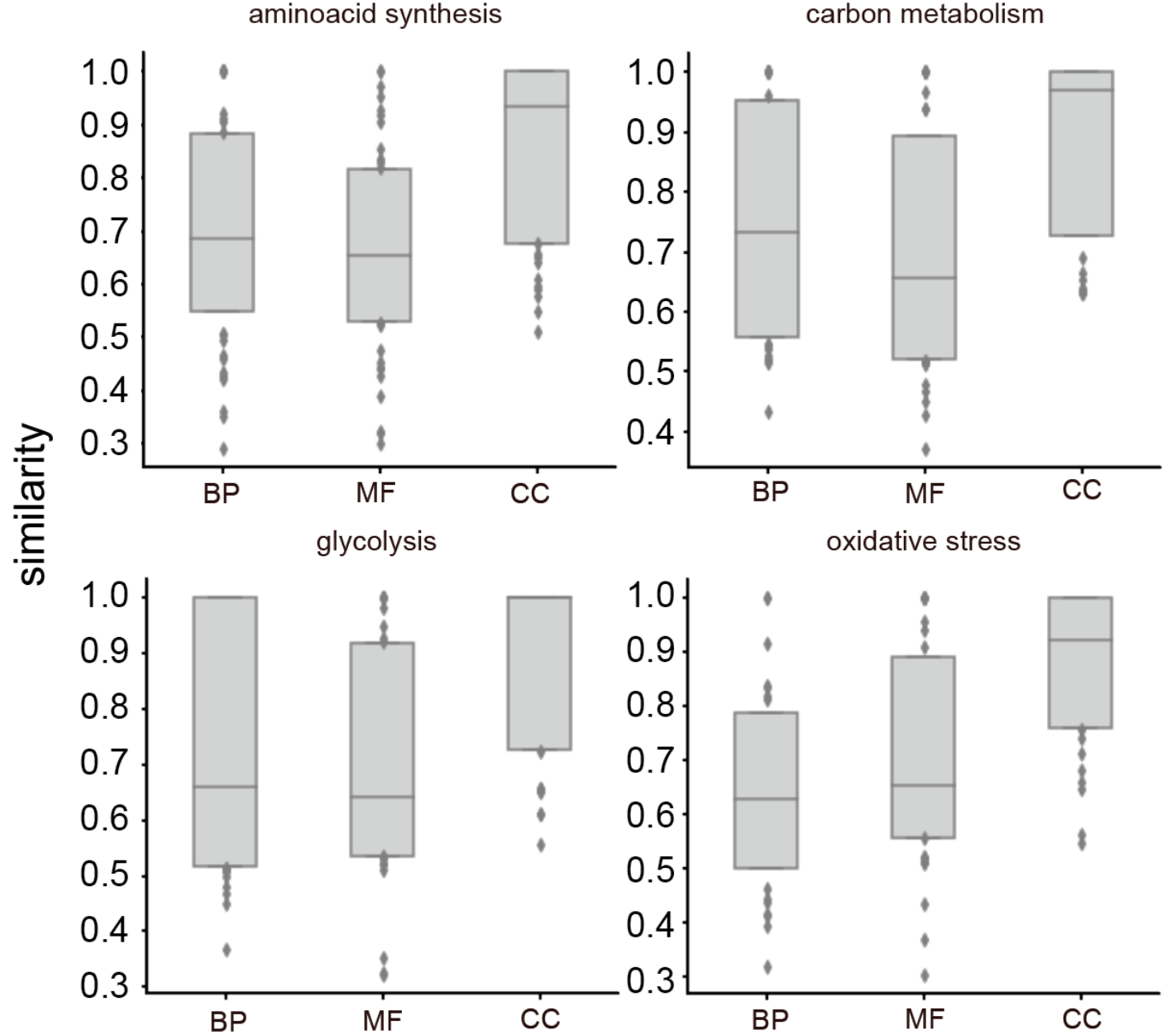}
\caption{\textbf{Biological significance of fibers.}
  Functional similarity of fibers in the amino-acid synthesis, carbon
  metabolism, glycolysis, and oxidative stress enzyme network. In each
  subnetwork, we calculated the mean similarities of all enzyme pairs
  in the fibers and their regulators.  Fibers have high biological
  significance. Enzymes in fibers show high average functional
  similarity in the underlying four enzyme networks. In all Gene
  Ontologies including the molecular function (MF), biological
  processes (BP), and cellular components (CC) ontologies and
  subnetworks, we observe that enzymes in fibers are functionally
  highly similar. Figure reproduced from
  \citep{alvarez2024symmetries}.}
  \label{Fig:S2} 
\commentAlt{Figure~\ref{Fig:S2}: 
Described in caption/text. No alt-text required.
}
\end{figure}

Fig. \ref{Fig:S2} shows the results for the enzyme networks studied
in Chapter\ref{chap:complex}: amino-acid synthesis, carbon metabolism,
glycolysis, and oxidative stress.  We observe that enzymes in fibers
are highly similar in their functions when we consider all three GO
categories: biological processes, molecular function, and cellular
components. As expected, when we include
the regulators of the fibers in the analysis, they are not as
functionally homogeneous as the enzymes within fibers are. This is because, most of
the time, the regulators are master regulators participating in different fibers and, therefore, in different functions.

To compare with other notions of elementary building blocks of biological networks,
we consider network motifs and modules, which are determined by
statistical over-representation and network partitioning.

\section{Network motifs}
\label{sec:comparison}

The search for building blocks of biological networks is often
motivated by the wish to find circuits that are easy to understand and 
occur frequently. They may also generate meaningful dynamics or perform
signal processing tasks. Evolutionary pressures on gene expression
dynamics may cause the evolution of suitable network structures: such
structures will evolve and be conserved. They tend to accumulate as
network motifs, or they may form symmetric structures such as fibers.

Are the symmetry-based building blocks the only way to
decompose the network into fundamental elements? The answer is no.
Meaningful patterns or building blocks may be defined in a number of
ways. As hard-coded patterns (such as operons),\index{operon } statistically
over-represented circuits (such as network motifs),\index{motif } circuits with
specific simple shapes or symmetries (such as fibers),\index{fiber } circuits forming modules, circuits
performing specific dynamic functions, or simply structures that are
found to be evolutionarily conserved. Of course, these properties may
also go hand in hand. 

The most widely used way to define building blocks of biological and
other networks are the motifs\index{motif }
introduced by Alon and collaborators \citep{milo2002network}, which 
identify circuits of high statistical significance.
These researchers pioneered the concept of a network motif as a
statistical feature of the local topology of a graph. They were
inspired by DNA sequence motif patterns, which are recurrent or
conserved sequences of nucleotides with biological significance for
the functioning of the cell, such as a binding site for a regulatory
transcription factor. They observed that in biological networks, some
simple motif network patterns appear more often than they would by
pure chance.

\cite{milo2002network} suggest looking for patterns of connectivity\index{pattern of connectivity } in
the network that occurs with a higher frequency than in random
networks, in the following sense. They identify all possible subgraphs of size $n$, calculate
the number of occurrences of each of them in the graph, and identify
ones that occur significantly more often than in random networks
generated by preserving the in- and out-degree for each node. They
interpret this increased frequency as evidence that these  motifs are
basic building blocks for biological mechanisms. This proposal has had
a major impact on systems biology \citep{alon2019,klipp2016book}, and
nowadays, finding  motifs is a popular tool for analyzing
network structure.

However, statistical abundance\index{statistical abundance } by itself does not imply that these
circuits are fundamental bricks of biological systems. Fundamental
circuits of biological networks can be underrepresented and appear
only once in the network, yet they can have overwhelming importance
for the functioning of the cell.

Motifs\index{motif } are good for finding basic building blocks that are
statistically repeated ubiquitously. However, the functional
significance of these network motifs has been debated, and results
indicate that the link between motif and function does not
exist \citep{ingram2006,payne2015,macia2009,ahnert2016}. Indeed,
except for the FAN motif, other motifs found by Alon and coworkers
\citep{milo2002network} do not lead to synchronization or other clear
functional roles. Thus, the mere statistical significance of the
network motif may not confer a definitive functionality within the
cell machinery.

Another popular way to decompose a network is community detection\index{community detection }
\citep{girvan2002}, also called network partitions or modularity\index{modularity }
\citep{hartwell1999,blondel2008}. However, a biological module\index{module } is not
a partition of the network that is usually identified with a
fundamental building block since many building blocks can form a
larger module.

Fibration building blocks\index{building block !fibration } present a different approach: they identify the
building blocks from a theoretical and principled approach through the
theory of symmetry, rather than by statistics.  Fibrations capture the topological properties, guaranteeing that
a circuit can be organized into minimal forms of coherent function and
logic computation.

By comparison, motifs cannot achieve these simple forms of
coherent functionality: they do not synchronize nor oscillate,
although admittedly, they were not designed to contribute toward these
functional forms.  The minimal canonical circuits that can achieve
such behavior is related to the simple topological principle of
symmetry invariance, which is necessary and sufficient to ensure that
a network configuration can result in a coordinated function of gene
dynamics, pointing towards a common functionality.

Below, we give a brief definition of network motif and modularity
(community detection) algorithms, and then compare the topological
decomposition and biological significance obtained by these methods to
the decomposition by fibration building blocks.  We explain why network
motifs and community detection fail to find the fundamental
bricks of biological networks.

\subsection{Calculating network motifs}
\label{S:network_motifs}

To find network motifs\index{motif !calculation of } of size $n$ in a graph $G$ we can,
in principle, take the following steps:

\begin{enumerate}
    \item Identify all distinct topological types of subgraphs of size
      $n$: $G^n_1 \dots G^n_m$ (we denote the total number of distinct
      graphs with $n$ nodes by $m$).
    \item Count the number of occurrences of $G^n_i$ in $G$.
    \item Assess the significance of each $G^n_i$.
\end{enumerate}

The first two steps of this process are illustrated using
Algorithms~\ref{algo:CountMotifs} and \ref{algo:FindMotifs} below. The
function CountMotifs($G, G_1, \dots, G_m$) counts the number of
occurrences of motifs $G_1, \dots, G_m$ in graph $G$. The function
FindMotifs($G, n$) finds all possible motifs of size $n$ in graph $G$
and calculates their frequency of occurrence.

However, as the last part of this section explains, these
two algorithms are seldom practical. On the one hand, there is a combinatorial explosion problem (line 1 of both algorithms); on the other hand, there is a graph-isomorphism test (line 4 of the first algorithm) for which there is no known polynomial-time algorithm.

\begin{algorithm}[H]
    \textbf{Input:} Graph $G$, $G_1, \dots, G_m$ ($m$ subgraphs to look for). \\
    \textbf{Output:} $l_1, \dots, l_m$ (appearance count of each graph in $\{G_i\}$).
    \begin{algorithmic}[1]
        \State Identify all subgraphs of graph $G$: $G^1, \dots, G^k$
        \State $l_i$ = array of m integers equal to 0
        \For{$i \in [1, m]$ and $j \in [1, k]$}
            \If{$G_i$ is isomorphic to $G^j$}
                \State $l_i$ = $l_i$ + 1
            \EndIf
        \EndFor
        \State \Return $\{l_i\}$
    \end{algorithmic}
    \caption{: CountMotifs($G, G_1, \dots, G_m$)}
    \label{algo:CountMotifs}
\end{algorithm}

\begin{algorithm}[H]
    \textbf{Input:} Graph $G=(N,E)$, $n$ - motif size of interest. \\
    \textbf{Output:} All unique graphs with $n$ nodes: $G^n_1, \dots, G^n_m$ ($m$ = number of unique graphs), $l_1, \dots, l_m$ (occurrence frequency for each graph $G^n_1, \dots, G^n_m$).
    \begin{algorithmic}[1]
        \State Identify all unique graphs of size $n$: $G^n_1, \dots, G^n_m$
        \State $l_i$ = CountMotifs($G, G^n_1, \dots, G^n_m$)
        \State \Return $\{G^n_i\}$, $\{l_i\}$
    \end{algorithmic}
    \caption{: FindMotifs($G, n$)}
    \label{algo:FindMotifs}
\end{algorithm}

\newpage
To describe the third step, we consider, without loss of generality, a
graph $G$ and a single motif $H$. We define a function $f(G) = {\rm
  CountMotifs}(G, H)$ that takes a graph $G$ and a motif $H$ and
returns the number of motifs in $G$ that are isomorphic to
$H$. Consider now a distribution $Y$ given by applying $f(G)$ to
random graphs from the set $\{G^R\}$. Then $Y$ can be found as $Y =
f(G^r), G^r \in G^R$. The problem of defining the statistical
significance of the motif splits into two steps: a) find the
distribution $Y$ of numbers of the motifs in random graphs, b) find
the significance of the measurement $x = f(G)$ given the distribution
$Y$.

The problem of finding the distribution of the occurrence frequency of
a motif in a random graph does not have an analytical
solution. Therefore, the distribution of $Y$ cannot be found exactly. In
practice, we can set up a computational experiment in which parameters
of this distribution can be found (the distribution is assumed to be
normal): $M$ random graphs are generated preserving the in- and
out-degree of each node in the original real network, and the number
of occurrences of $H$ is counted in each one of them ($\{l_i\}$).
The Central Limit Theorem\index{Central Limit Theorem } implies that for sufficiently large $M$ the
mean and standard deviation of $Y$ can be estimated as the mean and
standard deviation of $\{l_i\}$.

To discuss the significance of the measurement with the given
distributions, we use the $p$-value\index{p-value @$p$-value } and the $Z$-score.\index{Z-score @$Z$-score } Studies of
statistical significance are important not only to assess the
significance of motifs but for any significance study of any
building block. Explanations of how to calculate $p$-values and $Z$-scores
can be found in any book on statistics. However, due to their
importance for the present study, we now provide a succinct
introduction of the basic concepts involved in using $p$-values to assess significance.

Due to the combinatorial nature of the number of motifs for a fixed $n$,
i.e., the explosive factorial increase in the number of possible motifs
with $n$, only values of $n$ up to 5 are available. Then, for a fixed
$n$, the number of occurrences of each subgraph is recorded and
compared, using $p$-value
statistics, with the numbers of occurrences in random networks.

\subsection{Evaluating $p$-values and $Z$-scores}
\label{pvalue}

\index{p-value @$p$-value }\index{Z-score @$Z$-score }
In statistics, $p$-values are used in the framework of hypothesis
testing. The $p$-value is the probability of obtaining a measurement at
least as extreme as the observed one, given that the \textit{null
  hypothesis} is correct. The null hypothesis $H_0$ is a hypothesis to
be `nullified' in favor of an \textit{alternative hypothesis} $H_1$
that is complementary to the null hypothesis. That is, the alternative
hypothesis is considered valid and accepted if its complementary null
hypothesis is exceedingly unlikely. The $p$-value is the probability
of obtaining the observed measurement {\em given the null hypothesis},
which is different from the probability of the null hypothesis given the
observed measurement; that is, $P({\rm measurement} | H_0) \neq P(H_0
| {\rm measurement})$.

In a standard problem in which we are given the measurement $t$ and a
statistic $T$, there are three possible ways to find the $p$-value. If
we expect $t$ to be bigger than $T$ (one-sided right-tail test), the
$p$-value ($p_r$) is defined as
\begin{equation}
    p_r = P(T \geqslant t).
\end{equation}
If we expect $t$ to be smaller than $T$ (one-sided left-tail test), the $p$-value ($p_l$) is defined as
\begin{equation}
    p_l = P(T \leqslant t).
\end{equation}
And if we expect $t$ to be significantly different from $T$ (two-sided
test), then the $p$-value ($p$) is defined as
\begin{equation}
    p = 2 \min (p_r,p_l).
\end{equation}

The null hypothesis is rejected if the $p$-value is smaller than
a threshold significance value $\alpha$. A frequently used value of
$\alpha$ in biology is 0.05, or even a less adventurous value of 0.1. These values should be compared to the $p$-values required in physics for a Nobel Prize-bound significant result.  When the Higgs boson was discovered, the associated $p$-value was considered to be around $5\sigma$, which translates to a probability of roughly 1 in 3.5 million, meaning a very small $p$-value of approximately $3\times10^{-7}$.

As an  example, consider a coin tossing\index{coin tossing } experiment. The null hypothesis
$H_0$ is that the coin is fair ($q=0.5$). The alternative hypothesis
$H_1$ is that the coin is weighted towards heads ($q > 0.5$). Suppose
the desired significance level is $\alpha = 0.05$. Suppose that the
coin is tossed $n$ times and $m$ heads are observed. To find the $p$-value
($p$) of observing $m$ heads in $n$ tosses, assuming the coin is fair,
we sum the probabilities of obtaining $m$ or more heads using the
formula:
\begin{equation}
    p = \sum_{k = m}^{k=n}\frac{1}{2^n}\frac{n!}{k!(n-k)!}.
\end{equation}

For instance, if $n = 10$ and $m = 9$, the $p$-value is 
\begin{equation}
    p  =  \sum_{k = 9}^{k=10}\frac{1}{2^{10}}\frac{10!}{k!(10-k)!} 
     =  0.010 + 0.001 = 0.011 \leqslant \alpha = 0.05.
\end{equation}
This $p$-value is below the significance level of 0.05, so $H_0$ is deemed to be
unlikely and is rejected. Then $H_1$ is accepted with a significance
level of $0.05$ (or 95\%).

In this experiment, the probability distribution\index{probability distribution } assumed for the null
hypothesis\index{null hypothesis } is binomial\index{binomial distribution }\index{Gaussian distribution } (or its normal/Gaussian approximation), which
is well supported by the experiment.  It is important (but usually
under-emphasized) that the null hypothesis tacitly involves assuming a
specific probability distribution for the observations. Therefore, there can be two different reasons for `rejecting the null
hypothesis', which is the usual form of words used.  One is its low
probability, which in practice is seen as confirmation of the
alternative hypothesis despite the careful phraseology of `rejecting
the null hypothesis'. The other is that an inappropriate distribution
is assumed when calculating the probability of such an extreme
observation. Thus, it is vital to assume a plausible distribution. The
default assumption of a normal distribution may not be appropriate
despite the Central Limit Theorem.\index{Central Limit Theorem }

When the distribution of the statistic $T$ is normal, another metric
is often used in combination with the $p$-value. The {\em $Z$-score}\index{Z-score @$Z$-score } (also
referred to as standard score) counts how far the measured value is
from the mean of the distribution of the random value that the
measured value is being compared to. That is, if the null hypothesis
is that the measurement $x$ belongs to a normal distribution with mean
$\mu$ and standard deviation $\sigma$, then the $Z$-score ($Z$) of this
measurement is
\begin{equation}
z = \frac{x - \mu}{\sigma}.
\end{equation}

\begin{figure}
    \includegraphics[width=\linewidth]{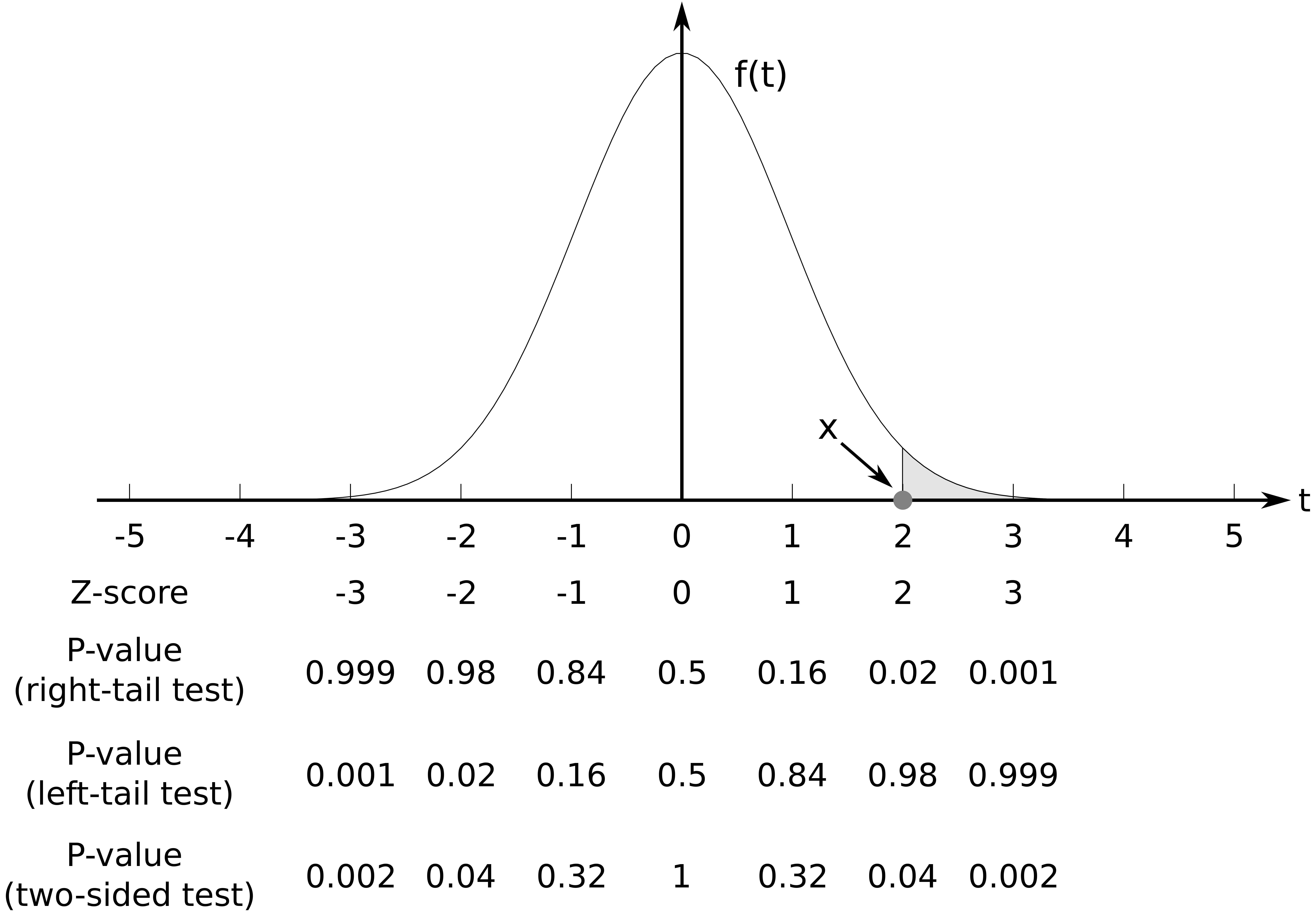}
    \caption{\textbf{$Z$-scores and $p$-values}. Example calculation of
      $Z$-scores and $p$-values given the observation $x$ (shown by the
      red dot) and the distribution $f(t) = N(0, 1)$. The $p$-value of
      $x$ in the right-tail test is the probability of observing value
      at least as low as $x$ and is given by $ \int_{x}^{\infty}
      f(t)\,dt$ (area in cyan). Since the standard deviation of $f(t)$
      is 1, $Z$-scores are equal to the value of $t$.}
    \label{Fig:motifs_zscore}
\commentAlt{Figure~\ref{Fig:motifs_zscore}: 
Graph of normal distribution. Point x=2 on x-axis
labeled x. Region under graph to right shaded.
Z-score -3, -2, -1, 0, 1, 2, 3.
P-value (right-tail test) 0.999, 0.98, 0.84, 0.5, 0.16, 0.02, 0.001.
P-value (left-tail test) 0.001, 0.02, 0.16, 0.5, 0.84, 0.98, 0.999.
P-value(two-sided test) 0.002, 0.04, 0.32, 1, 0.32, 0.04, 0.002.
}
\end{figure}

Figure \ref{Fig:motifs_zscore} illustrates the calculation of $Z$-scores
and $p$-values given the measurement $x$ and the normal distribution
$f(t)$ with zero mean and standard deviation one.

\subsection{Statistical significance of network motifs}
\label{sec:statistical}

To summarize, the significance of a  motif $H$ in a graph $G$ can
be found using the following steps:

\begin{enumerate}
    \item Count the number of occurrences of $H$ in $G$ as $x = f(G)$.
    \item Generate $M$ random graphs $G^1, \dots, G^M$ with the in- and out-degree of each node being equal to the in- and out-degree of the corresponding node in graph $G$.
    \item Find mean $\mu(Y) = \langle f(G_i)\rangle$ and standard deviation $\sigma(Y) = \sqrt{\langle f(G_i)^2 \rangle -\langle f(G_i)\rangle^2}$.
    \item Find $Z$-score as $z = \frac{x - \mu(Y)}{\sigma(Y)}$.
\end{enumerate}

Pseudocode for finding network motifs of size $n$ and
estimating their significance is given in
Algorithm~\ref{algo:GetMotifs}. The function ${\rm GetMotifs}(G,n)$ finds
all unique motifs of size $n$ in $G$, their number of occurrences in
$G$, and the significance of each motif.

\begin{algorithm}[H]
    \textbf{Input:} Graph $G=(N,E)$ ($n$ = motif size of interest). \\
    \textbf{Output:} All unique graphs with $n$ nodes: $G^n_1, \dots, G^n_m$ ($m$ = number of unique graphs), occurrence frequency for each graph $g^n_1, \dots, g^n_m$ and -score for each graph $z^n_1, \dots, z^n_m$.
    \begin{algorithmic}[1]
        \State Identify all unique graphs of size $n$: $G^n_1, \dots, G^n_m$
        \State $g^n$ = CountMotifs($G, G^n_1, \dots, G^n_m$)

        \State $l_{ij}$ = M by m array of integers equal to 0
        \For{$i \in [1, M]$}
            \State Generate $G^i = (N_H, E_H)$ with $|N_H| = |N|$ and $|E_H| = |E|$ such that
            $k_{in}(n^H_j) = k_{in}(n^G_j) \,\&\, k_{out}(n^H_j) = k_{out}(n^G_j)$ for $j \in [1, |N|]$
            \State $l_i$ = CountMotifs($G^i, G^n_1, \dots, G^n_m$)
        \EndFor
        
        \For{$i \in [1, M]$}
            \State $z^n_i = \frac{g^n_i - {\rm mean}(l_i)_j}{{\rm sd}(l_i)_j}$
        \EndFor
            
        \State \Return $\{G^n_i\}, \{g^n_i\}, \{z^n_i\}$
    \end{algorithmic}
Here \\
    $|N|$ = size of set $N$, \\
    $k_{in}$ and $k_{out}$ are the in- and out-degrees of a node, \\
    mean$(l_i)_j$ and sd$(l_i)_j$ are the mean and standard deviation of the sample $l_{ij}$ over the index $j$.
    \caption{: GetMotifs($G, n$)}
    \label{algo:GetMotifs}
\end{algorithm}

\subsection{Network motifs in complex networks}

\begin{figure*}
    \includegraphics[width=\linewidth]{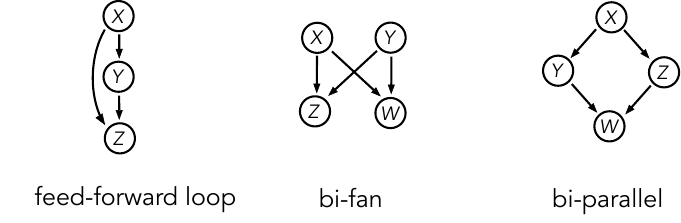}
    \caption{\textbf{Most common network motifs introduced by
        \cite{milo2002network}}. Feed-forward loop, bi-fan and bi-parallel.}
    \label{Fig:motifs_alon}
\commentAlt{Figure~\ref{Fig:motifs_alon}: 
Illustrative only.  Described in caption/text. No alt-text required.
}
\end{figure*}

\begin{figure*}
  \centerline{\includegraphics[width=.9\textwidth]{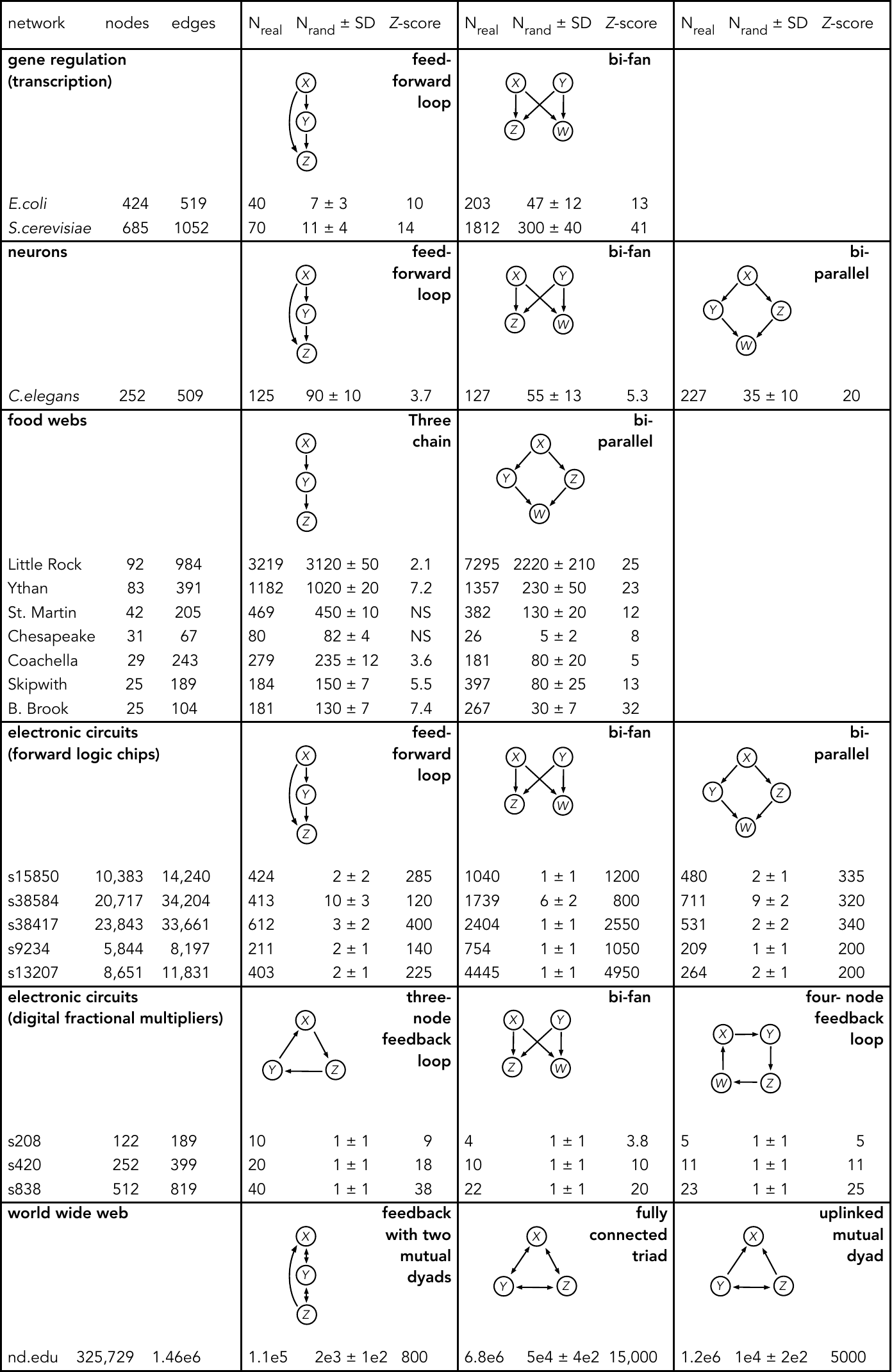}}
    \caption{ \textbf{Other network motifs in complex networks} after \citep{milo2002network}. Tree chain, 
three-node feedback loop, 
four-node feedback loop, 
feedback with two mutual dyads,
fully connected triad, and
uplinked mutual dyad.}
    \label{fig:motif}
\commentAlt{Figure~\ref{fig:motif}: 
Illustrative only.  Described in caption/text. No alt-text required.
}
\end{figure*}

\cite{milo2002network} apply the motif finding algorithm to a number
of networks, including gene regulatory, neural, food webs, electronic
circuits, and some others. Figure ~\ref{Fig:motifs_alon} shows some of
the most abundant motifs, and Fig. \ref{fig:motif} shows the remaining ones across networks. One of the most
abundant network motifs is the 
Feed Forward Loop (FFL)\index{feed-forward loop }\index{FFL }
\citep{alon2003a}, which we have already encountered in many discussions within
this book.  It is composed of a central Y gene that controls a Z gene,
with another X gene controlling both, all in a feed-forward
manner. Despite its name, there is no loop in this circuit.  The
functionality of the FFL is investigated in detail in Section
\ref{sec:fff}.  Briefly, it acts as a sign-sensitive delay element in
transcription networks, delaying the signal reaching gene Z, and
provides some protection from large short-lived fluctuations
\citep{alon2003b}.

As discussed in Section \ref{sec:fff}, the FFL is part of the
FFF, which is obtained from the FFL by the addition of an AR loop at
gene Y. The AR loop is another very common network motif.  Here, we see
one of the main problems with motif identification. Statistics alone will find these two circuits separately,
but it will miss the main point that almost all of the time, these circuits 
appear together, forming the FFF.  The FFF has the crucial property of
synchronization, while its separate pieces, FFL and AR, have no
coherent functionality, except that an FFL produces a
delay in the signal arriving at Z by the intermediate step at Y.
This has been proposed by \citep{alon2003b} to be useful since it
protects gene Z from large rapid fluctuations in the inputs. While
useful, this functionality does not seem to be of enough importance
to declare the FFL as the quintessential motif of biological
machineries.

The bi-fan motif\index{bi-fan motif } in Fig.~\ref{Fig:motifs_alon} is actually a
fiber, which is found by fibration analysis as the $|n=0, l=2, m=2
\rangle$ fiber.  It is a rather trivial fiber which produces
synchronization in the Z and W genes by their shared inputs and in
the TRN acts as a regulon, as discussed in Section
\ref{sec:operons}. Finally, the bi-parallel motif is also part of a
trivial fiber, in that the genes Y and Z are synchronized in a fiber
by sharing the same input, as discussed in Section
\ref{sec:operons}.

An important concept in this work is the distinction between
synchronization resulting from trivially sharing the same inputs (as
in operons and regulons) versus synchronization induced by a more
complex symmetry fibration. That is, the difference between
coregulation induced by a fiber (coexpression resulting from shared
input trees, which take into account extended paths in the network)
versus coregulation by a single input of a regulon. Both lead to
coexpression, but the former is more complex than the latter, which we
consider to be a trivial form of synchronization, unlike fiber
synchronization.

\subsection{Drawbacks of network motifs}
\label{sec:drawbacks}
\index{motif !drawbacks of }

The very high $Z$-scores obtained for the motifs \citep{milo2002network} point to the
conclusion that motifs do indeed appear more frequently in the
studied networks, and therefore, are `the basic building blocks of
complex networks'. While this deduction has had
considerable influence in systems biology
\citep{alon2019,klipp2016book}, the functional role of motifs is
still questionable \citep{ingram2006, macia2009, payne2015,
  ahnert2016}.  

\begin{figure}
    \includegraphics[width=.9\linewidth]{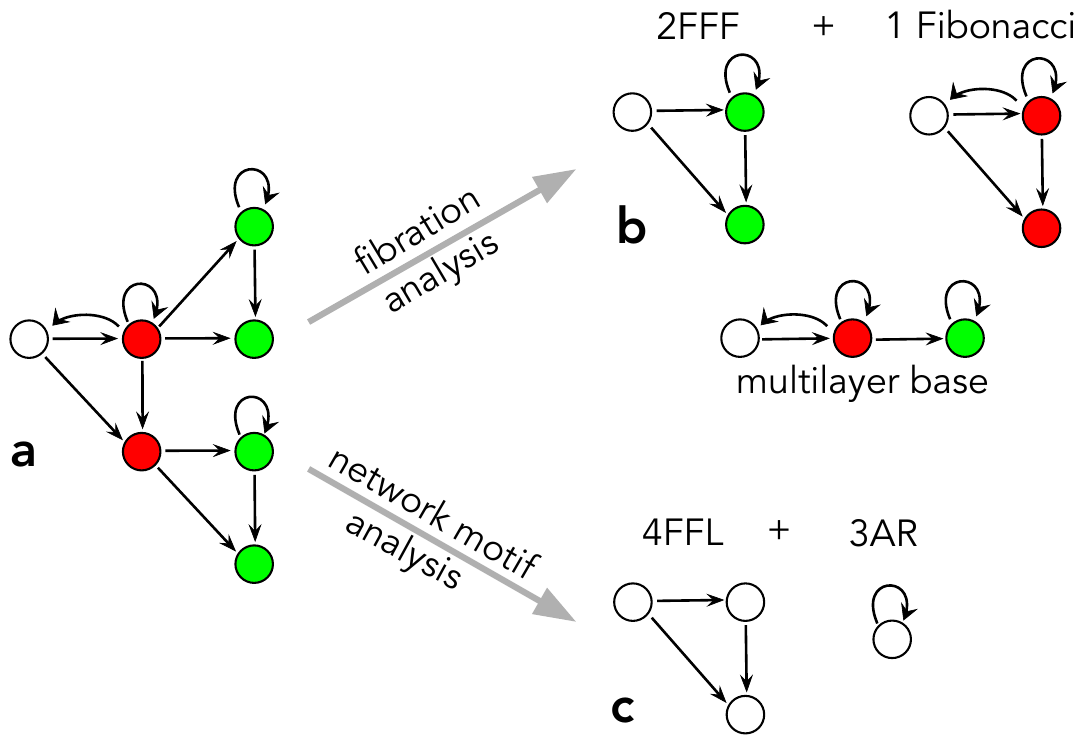}
    \caption{\textbf{Drawbacks of network motifs as building blocks.}  \textbf{(a)} Sample network with balanced coloring indicating fibration symmetry. \textbf{(b)} The sample network is composed of 2 FFFs and 1 Fibonacci fiber arranged into a multilayer base capturing the correct dynamics for the balanced coloring. \textbf{(c)} A motif analysis identifies 4 FFLs and 3 AR loops whose independent dynamics do not correspond to the dynamics embedded in the network.}
    \label{fig:isolation}
\commentAlt{Figure~\ref{fig:isolation}: 
(a) Network with one white node (call it W), two red (R1,R2), four green (G1-G4).
Arrows W-R1, W-R2, R1-W, R1-R1, R1-G1, R1-R2, R2-G3, R2-G4, G1-G1, G1-G2, G3-G3, G3-G4.
Arrow upward labeled `fibration analysis' points to:
(b)  2FFF + 1 Fibonacci
Network with nodes W, G1, G2; arrows W-G1, W-G2, G1-G1, G1-G2;
Network with nodes W, R1, R2; arrows W-R1, W-R2, R1-R1, R1-R2;
Multilayer base: Network with nodes W, R, G; arrows WR, RW, RR, RG, GG.
Arrow downward labeled `network motif analysis' to:
(b)  4FFL + 3AR
Network with nodes W1, W2, W3; arrows W1-W2, W1-W3, W2-W3;
Network with one node W and arrow WW.
}
\end{figure}

From the dynamical standpoint, there is an obvious
flaw in this logic. 
Motif dynamics studied in isolation can be very different from the dynamics of the same circuit when it is
      embedded in the network. This is because a motif is a
      subgraph, which ignores links to external
      nodes that can interfere with the motif dynamics. Thus, the
      isolated dynamics of the motif have nothing to do with the
      dynamics in the full network. This impedes any functional
      interpretation of the motif. Fibers, on the contrary, have
      the same synchronization properties
     when studied in isolation or in the network,
     as shown in the comparison between the FFL motif
      and the FFF in  Fig. \ref{fig:isolation}.

Thus, a network motif that is analyzed in isolation need
not behave in the same manner when it has inputs from the rest of the
network.  Therefore, any conclusion about the phenotype of a network
motif must be viewed with caution.  In contrast, fibration building blocks
are induced subgraphs that are designed in such a way that all the
inputs to the fiber are included in the analyzed network so that the
solution obtained for the fibers are the real solutions that would
be encountered when the fiber is embedded in the full network.

Another important limitation is the size of the motif. The problem of
finding all topologically distinct graphs of size $n$ becomes
computationally intractable even for fairly small $n$.  We could ask
whether, for instance, a complex fiber structure, such as any of the
Fibonacci fibers can be ever found by statistical analysis alone, \`a la
 motifs. The answer would be `no'. This circuit is too large to be
uncovered by statistical over-representation in a random search over all
possible circuits of a given node size.

If we believe that each gene should play a significant role and has
been evolutionarily optimized for that role, it makes sense to ask: How
is a given node regulated overall, and which nodes share the same coherent
patterns overall? Network fibers directly analyze the
`history' of input signals to a gene from its regulators, the
regulators of its regulators, etc., with an `infinite history' if the
regulation involves loops. They define an equivalence between genes
if their input histories look structurally the same and use this to
split the expression network exhaustively into gene groups, such
that---ideally---each group yields different expression patterns in
response to the network’s input signals, but all genes in the same
group shows the same expression pattern. The existence of fibers
addresses functionality directly through synchronization.

Some fibers are also motifs, like the $n=0$
class, which is analogous to a FAN motif or an autoregulation
loop, but fibers admit more general forms that are related to the
dynamical state of the genes and their function.
Thus, fibers can play the role of building blocks, as can network motifs,
but there are clear differences:

\begin{itemize}
  
\item Fibers are not defined (and revealed) by mere counting but by
  analyzing symmetry structures in the network.
  
\item Looking at the FFL from the viewpoint of fibrations brings an
  interesting insight. Consider an isolated, coherent positive FFL as
  an example (genes denoted by X, Y, Z in
  Fig. \ref{Fig:motifs_alon}). If the internal node Y contains a
  positive self-loop, then Y and Z form an FFF, so their
  expression is synchronized. As discussed, this internal
  self-regulation in FFLs is often observed and now acquires a natural
  explanation.

\item In analyzing the dynamics and function of network motifs, we
consider the motif in isolation. This works well for motifs in which
signals propagate in only one direction, as in the FFL. But if the motif
is embedded in a network, and the motif’s output signal feeds back
into its input, this is considered `out of scope'. This explains
why the motif dynamics need not persist when the motif is embedded in the
full network. In gene fibers, such infinite loops are not only handled
properly but are even part of the very definition of the fiber.
\end{itemize}

These differences are illustrated in Fig. \ref{fig:isolation}, which displays a sample network composed of FFFs and FFL motifs, along with an analysis of its fibration and network motifs building blocks.
In Fig. \ref{fig:isolation}b, we see that this network consists of 2 FFFs and 1 Fibonacci fiber building block, arranged into a multilayer fiber as depicted at the base. This analysis accurately represents the synchrony in the balanced coloring and shows no overlap among the fibers, except for the correct interactions between the regulators; one regulator can regulate multiple fibers. 
In contrast, the network motif analysis presented in Fig. \ref{fig:isolation}c reveals 4 FFLs and 3 AR loops. None of these motifs are functional, nor are they significant for the dynamics of the circuit. In particular, two FFLs overlap; the Fibonacci fiber is formed by two overlapping FFLs.

As remarked, the main problem with motifs is that edges connecting to the motif from the outside are not taken into account. These can change
the dynamics,\index{dynamics } compared to how the motif would behave on its own.
In contrast, fibers have, by definition, the same dynamical
behavior, whether they are analyzed in isolation or embedded in the
network. This might explain why the question of whether motifs have a
functional role remains controversial
\citep{ingram2006, macia2009, payne2015, ahnert2016}. Fiber building
blocks, on the contrary, are identified in the network structure in
such a way that their dynamics are much the same whether they are
studied in isolation or in the network.

Despite these drawbacks, motifs can still be a 
useful mathematical tool
to decompose a network. However, caution is required when
making any conclusion about the phenotype of a motif.

\section{Statistical significance of fibration building blocks}

Even though fibration building blocks\index{building block !fibration } are not found by statistical
over-representation, it is important to understand, after finding them by
a fibration analysis, whether they are statistically significant.
That is, whether these circuits could equally probably have been generated by
random addition of links to the network.  We can guess,
correctly, that the answer is `no'. Intuitively, intricate circuits such
as Fibonacci and FFF are unlikely to be the result of randomness. While
randomness plays an important role in evolution through mutations,
evolutionary pressure will eventually lead to very structured
circuits that cannot be compared to those obtained by random addition
of edges without selection for functionality.  Nevertheless, it is important to use the
statistical analysis applied to motifs in Section \ref{sec:statistical}
to confirm this intuition. 

\cite{leifer2020circuits} study the statistical significance of the
simple fibration building blocks discussed in Chapters
\ref{chap:hierarchy_1} and \ref{chap:hierarchy_2} using the regulatory networks of a wide range
of species spanning {\it A. thaliana}, {\it M. tuberculosis}, {\it
  B. subtilis}, {\it E. coli}, {\it salmonella}, yeast, mouse
and humans.
The datasets are described in Table \ref{Datasets_table}.  \cite{leifer2020circuits} also provide, in the
supplementary material sections, a plot of every single building block
found for all species and networks.  We show in Table~\ref{symm_table}
the count of building blocks across species and their associated
$Z$-scores;\index{Z-score @$Z$-score } the large $Z$-scores show that these circuits are statistically significant.

\begin{table*}[b!]
\scriptsize
 \resizebox{\textwidth}{!}{
\begin{tabular}{| l | l | l |}
\hline
{\bf Species} &  {\bf Additional information} \\
{\bf Database} & \\ 
\hline\hline
      
Arabidopsis Thaliana &  We use high-confidence functionally confirmed \\
ATRM \citep{ATRM} & transcriptional regulatory interactions from the \\
&ATRM database of the broadly used model plant \\
&Arabidopsis. \url{http://atrm.cbi.pku.edu.cn/} \\
\hline
Micobacterium Tuberculosis & Supplementary Information of  \citep{micobacterium_tuberculosis} \\
Research article \citep{micobacterium_tuberculosis} & \url{https://www.ncbi.nlm.nih.gov/pmc/articles/} \\
& \url{PMC2600667/bin/msb200863-s2.xls}\\								      
\hline
Bacillus subtilis &  We download the database from the SubtiWiki  website \\
SubtiWiki \citep{subtiwiki2018} & and consider all repressor and activation links \\
& as `Repression' and `Activation'. This database is \\
& considered to be the primary source of information for \\
& Bacillus. \url{http://subtiwiki.uni-goettingen.de/}\\
 \hline
 Escherichia coli &  We use the TF - operon interaction network \\
 RegulonDB \citep{regulon2016} & from \citep{regulon2016}. RegulonDB combines \\
 & transcriptional regulator interactions obtained by \\
 & curating literature and using NLP high-quality data \\
 & and partially confirmed experimentally \\
& and computationally predicted data. \\
& \url{http://regulondb.ccg.unam.mx/}\\
\hline
Salmonella SL1344 &  We use the regulatory layer of the strain \\
SalmoNet \citep{salmonet} &  Salmonella Typhimurium SL1344. SalmoNet \\
& consists of manually curated low-throughput \\
& and high-throughput experiments and predictions \\
& based on experimentally verified binding sites \\
& and TF-gene binding site data from \\
& RegulonDB. \url{http://salmonet.org/}\\
\hline
Yeast & We use the TF-gene regulatory and TF-gene \\
YTRP \citep{YTRP} & binding networks. Results of the TFPEs \\
& (Transcription Factor Perturbation Experiments) \\
& identify the regulatory targets of TFs. This is further \\
& refined by using literature-curated data. \\
& \url{http://cosbi3.ee.ncku.edu.tw/YTRP/Home}\\
\hline
Mouse &  Downloaded from TRRUST website. \\
TRRUST \citep{trrust} & TRRUST is constructed using sentence-based \\
& text mining of more than 20 million abstracts \\
& from research articles refined by manual curation. \\
& \url{https://www.grnpedia.org/trrust/}\\
 \hline
Human &   \\
TRRUST \citep{trrust} & Downloaded from TRRUST website.\\
TRRUST\_2 \citep{trrust} & Downloaded from TRRUST website and curated. \\
KEGG \citep{KEGG} &  We use KEGG API to download all pathways\\
&  of Human gene regulatory network. Then all \\
& networks are combined and duplicates are removed.\\
& \url{https://www.genome.jp/kegg/pathway.html} \\
\hline
\end{tabular}
}
\vspace{10pt}
\caption{\textbf{Description of biological dataset. } We use these datasets of biological networks to assess the statistical significance of fibration building blocks. All data are gathered from the stated sources. Table reproduced from \citep{leifer2020circuits}.}
\label{Datasets_table}
\end{table*}

  \begin{table*}[ht]
  \tiny
      \centering
      \begin{tabular}{| c | c c c | c c c |c c c |}
      \hline
        Species & Database & Nodes & Edges &  & AR Fiber & &  & FFF & \\
 & & & & $N_{\rm real}$ & $N_{\rm rand} \pm SD$ & $Z$-score & $N_{\rm real}$ & $N_{\rm rand} \pm SD$ & $Z$-score  \\
      \hline\hline
        Arabidopsis Thaliana & ATRM & 790 & 1431 & 2 & $0.2 \pm 0.5$ & 4 & 2 & $0 \pm 0$ & Inf \\      \hline
        Micobacterium tuberculosis & Research article & 1624 & 3212 & 11 & $0.7 \pm 0.8$ & 13.2 & 6 & $0.2 \pm 0.4$ & 14.6  \\
      \hline
        Bacillus subtilis & SubtiWiki & 1717 & 2609 & 35 & $0.3 \pm 0.5$ & 64.6 & 13 & $0.3 \pm 0.5$ & 23.4 \\
      \hline
        Escherichia coli & RegulonDB & 879 & 1835 & 14 & $0.2 \pm 0.5$ & 29.1 & 12 & $0.1 \pm 0.2$ & 49.4  \\
      \hline        Salmonella SL1344 & SalmoNet & 1622 & 2852 & 21 & $0.7 \pm 0.8$ & 25 & 14 & $0.2 \pm 0.4$ & 32 \\
      \hline        Yeast & & & & & 10 & & & 5 & \\
        & YTRP\_regulatory & 3192 & 10947 & 10 & $0.3 \pm 0.6$ & 17.3 & 4 & $0.2 \pm 0.4$ & 8.5 \\
& YTRP\_binding & 5123 & 38085 & 2 & $0.1 \pm 0.3$ & 6.3 & 0 & N/A & N/A  \\
      \hline
        Mouse & TRRUST & 2456 & 7057 & 1 & $0.1 \pm 0.4$ & 2.3 & 0 & N/A & N/A  \\
      \hline       
  Human & & & & & 1 & & & 1 &  \\
      
       & TRRUST & 2718 & 8215 & 0 & N/A & N/A & 0 & N/A & N/A  \\
      
        & TRRUST\_2 & 2862 & 9396 & 0 & N/A & N/A & 0 & N/A & N/A  \\
      
        & KEGG & 5164 & 59680 & 1 & $0.06 \pm 0.25$ & 3.76 & 1 & $0 \pm 0$ & $>3$  \\
      \hline
      \hline

        Species & Database & Nodes & Edges && Fibonacci fiber & & & $n = 2$ Fiber & \\
        & & & & $N_{\rm real}$ & $N_{\rm rand} \pm SD$ & $Z$-score & $N_{\rm real}$ & $N_{\rm rand} \pm SD$ & $Z$-score  \\ 
      \hline\hline
        Arabidopsis Thaliana & ATRM & 790 & 1431 &  5 & $0.3 \pm 0.6$ & 8.1 & 0 & N/A & N/A \\
      \hline
        Micobacterium tuberculosis & Research article & 1624 & 3212 &  4 & $1.7 \pm 1.4$ & 1.7 & 0 & N/A & N/A \\
      \hline
        Bacillus subtilis & SubtiWiki & 1717 & 2609 &  1 & $1.3 \pm 1.2$ & -0.2 & 2 & $0 \pm 0$ & 63.2 \\
      \hline
        Escherichia coli & RegulonDB & 879 & 1835 &  2 & $0.5 \pm 0.8$ & 1.9 & 1 & $0 \pm 0$ & $>3$ \\
      \hline
        Salmonella SL1344 & SalmoNet & 1622 & 2852 &  2 & $1.4 \pm 1.3$ & 0.5 & 3 & $0 \pm 0$ & $>3$ \\
      \hline
        Yeast & & & &  & 3 & & & 0 &  \\
        & YTRP\_regulatory & 3192 & 10947 &  2 & $1.8 \pm 1.3$ & 0.2 & 0 & N/A & N/A \\
        & YTRP\_binding & 5123 & 38085 &  0 & N/A & N/A & 0 & N/A & N/A \\
      \hline
        Mouse & TRRUST & 2456 & 7057 & 6 & $0.3 \pm 0.6$ & 9.3 & 0 & N/A & N/A \\
      \hline
        Human & & & & & 100 & & & 1 &  \\
        & TRRUST & 2718 & 8215 & 10 & $0.4 \pm 0.6$ & 16.3 & 0 & N/A & N/A \\
        & TRRUST\_2 & 2862 & 9396 & 11 & $0.4 \pm 0.7$ & 16 & 0 & N/A & N/A \\
        & KEGG & 5164 & 59680 &  79 & $0.6 \pm 0.7$ & 112 & 1 & $0 \pm 0$ & $>3$ \\
      \hline 
 \end{tabular}
    \vspace{10pt}
    \caption{\textbf{Fibers are statistically significant over many networks}. We
      report the $Z$-scores showing that all fibers found are
      statistically significant. We use a random null model with the
      same degree sequence (and sign of interaction) as the original
      network to calculate the random count $N_{\rm rand}$ and compare this
      with the real circuit count $N_{\rm real}$ to get the $Z$-score.
      Table reproduced from \citep{leifer2020circuits}.
      } 
    \label{symm_table}
  \end{table*}

\section{Modularity}
\label{sec:modularity}

Modularity\index{modularity } is a basic property of any network. It refers to the notion
that there exists modules or clusters of nodes in a network with
preferentially many links among themselves and with fewer links to nodes
in other modules. In social networks, modularity is synonymous with
community structure, and modularity and community detection algorithms
have primarily been designed to uncover coherent groups in social
networks. In these networks, it is expected that socially coherent
groups should have many links in common and be weakly connected to the
rest.  Many algorithms have been designed to capture this modular
structure in networks. There are excellent reviews covering the
extensive literature on community and modularity detection algorithms
\citep{fortunato2010community}; see also Chapter 9 of
\citep{barabasi2016network} for an introduction to community detection
and modularity. Here, we mention just a few of the most popular
methods. In Section \ref{sec:seaofmeasures}, we discuss their use for
identifying clusters. It should be noted that modularity applied to a functional network provided synchronized clusters, but applied to a structural network does not, more details in Section \ref{sec:synchrony-correlation}.

\subsubsection{Louvain modularity and hierarchical clustering}
\label{sec:louvain}

Louvain modularity\index{modularity !Louvain } is a popular algorithm designed to identify
communities or modules in networks \citep{blondel2008}. These
communities represent a partition of nodes into groups within which
the network connections are dense but sparser otherwise.

\begin{figure*}[b!]
  \includegraphics[width=\textwidth]{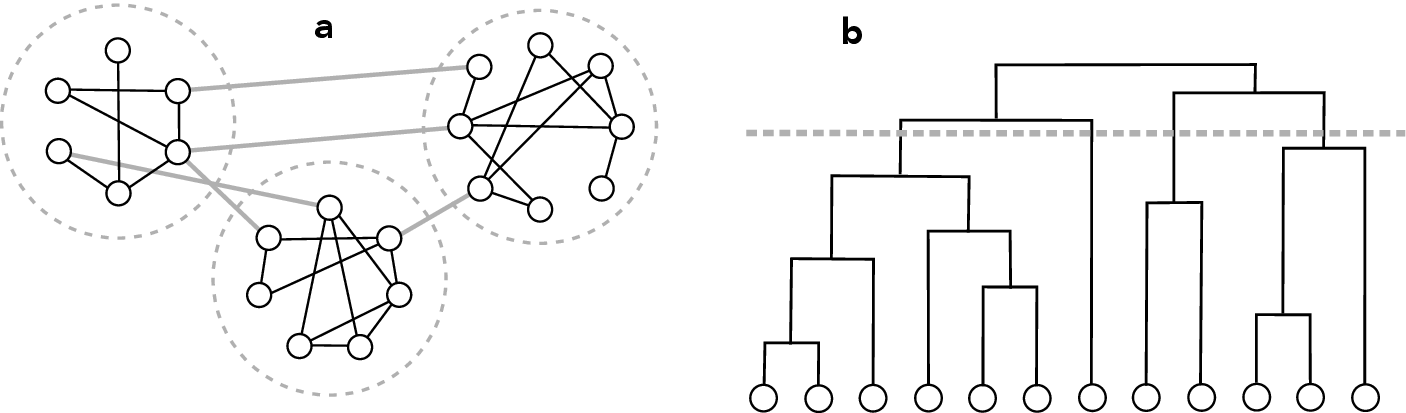}
  \caption{\textbf{Measurements of modularity.} \textbf{(a)} Louvain-type
      modularity. \textbf{(b)} Hierarchical modularity. }
    \label{fig:measurements}
\commentAlt{Figure~\ref{fig:measurements}: 
Illustrative, details irrelevant.
(a) Three small networks in dashed circles, some edges joining them
(b) Dendrogram (branching tree) with horizontal dashed line part way down.
}
\end{figure*}

Using a greedy agglomerative procedure, the Louvain clustering
algorithm\index{Louvain algorithm } maximizes the modularity $Q$ of a partition
$C={C_1,...,C_p}$ of a directed graph $G$
(Fig. \ref{fig:measurements}a):
\begin{equation}
\label{eq:Louvain}
    Q=\frac{1}{m}\sum_{u,v}\left[E_{uv}-\frac{d_u^{\rm in}d_v^{\rm out}}{m}\right]\delta(C_u,C_v),
\end{equation}
where $m=|E|$ is the number of edges of G, $E_{uv}$ represents the
existence (0 or 1) of an edge from community $u$ to community $v$,
$d_u^{\rm in/out}$ represent in/out degrees of community $u$. There are
many variations of this algorithm, including ones for undirected or weighted
networks \citep{fortunato2010community}.

Another popular algorithm was introduced by \cite{girvan2002} for
community detection in social networks. In the Girvan-Newman
algorithm,\index{Girvan--Newman algorithm } the \emph{edge betweenness centrality} (EBC) is defined to be the
number of shortest paths that pass through a given edge in a network.
Every edge is assigned an EBC score based on the shortest paths among
all the nodes in the graph.

Finding EBC scores is an iterative process: take one node at a time
and plot the shortest paths to the other nodes from the selected
node. Based on these shortest paths, compute the EBC scores for all the
edges. Repeat this process for every node in the graph. Now,
every edge gets $n$ scores corresponding to $n$ nodes in the
graph. These scores are added edgewise.

The Girvan--Newman algorithm uncovers communities in a graph by
iteratively removing the edges of the graph based on the EBC
score. Edges with the highest score are removed until the graph splits
into two. This constitutes one step of the algorithm.

Consider a particular division of the graph into $k$ clusters
(communities). Define a $k\times k$ symmetric matrix $e$, where the
element $e_{ij}$ is the fraction of all edges in the network that
link nodes in community $i$ to nodes in community $j$. Then the trace
${\rm Tr} \,e = \sum_i e_{ii}$ gives the fraction of edges in the
network that connects vertices in the same community. Clearly good
division in communities should have a high value for this trace.

Define $a_i = \sum_j e_{ij}$, which represents the fraction of edges
that connect to vertices in community $i$.  The modularity of a graph
is then defined as:
\begin{equation}
Q = \sum_i(e_{ii}-a_i^2)={\rm Tr} \, e - \| e^2 \| ,
\end{equation}
where $\| x \|$ indicates the sum of the elements of the matrix
$x$. The quantity $Q$ measures the fraction of the edges in the
network that connects vertices of the same type (i.e., within-community
edges) minus the expected value of the same quantity in a network with
the same community divisions but random connections between the
vertices. The sequence of splits determines a dendrogram,\index{dendrogram }
as in Fig.~\ref{fig:measurements}b.
Modularity is then calculated for every split of the
network as we move down the dendrogram and look for local peaks in
its value, which indicates particularly satisfactory splits.

Further popular methods to find communities and modules are
hierarchical clustering algorithms using agglomerative and divisive
methods \citep{fortunato2010community}. These methods also 
produce a
hierarchical tree or dendrogram, illustrating output generated by the
algorithm (Fig. \ref{fig:measurements}b). The circles at the bottom
represent individual nodes. These nodes are connected at a given level
of the tree according to different strategies measuring similarity
between nodes. As we move up the tree, nodes form larger communities,
culminating in a single community at the top. The dendrogram
illustrates an initially connected network splitting into smaller
communities. The resulting communities are subjective due to the
uncertainty of where to cut the dendrogram.

These methods just scratch the surface of the large number of algorithms. The interested reader is referred to extensive reviews in the literature \citep{fortunato2010community}.

\section{Comparison of fibrations,
motifs, and modules in enzyme networks}
  \label{sec:seaofmeasures}

Biological networks are modular
\citep{hartwell1999,brugere2018,pratapa2020,marbach2012,girvan2002}
and modularity and community detection algorithms have been used to
identify functional modules in biological networks \citep{girvan2002}.
Given the different ways to partition a network, it is of interest to
compare them not only at the topological level but, most importantly, on
their functional significance in the cell.  Below, we examine the
different partitions in TRNs and metabolic networks.

\subsection{Fibers and modules in TRNs}

\begin{figure}[t!]
  \includegraphics[width=.95\linewidth]{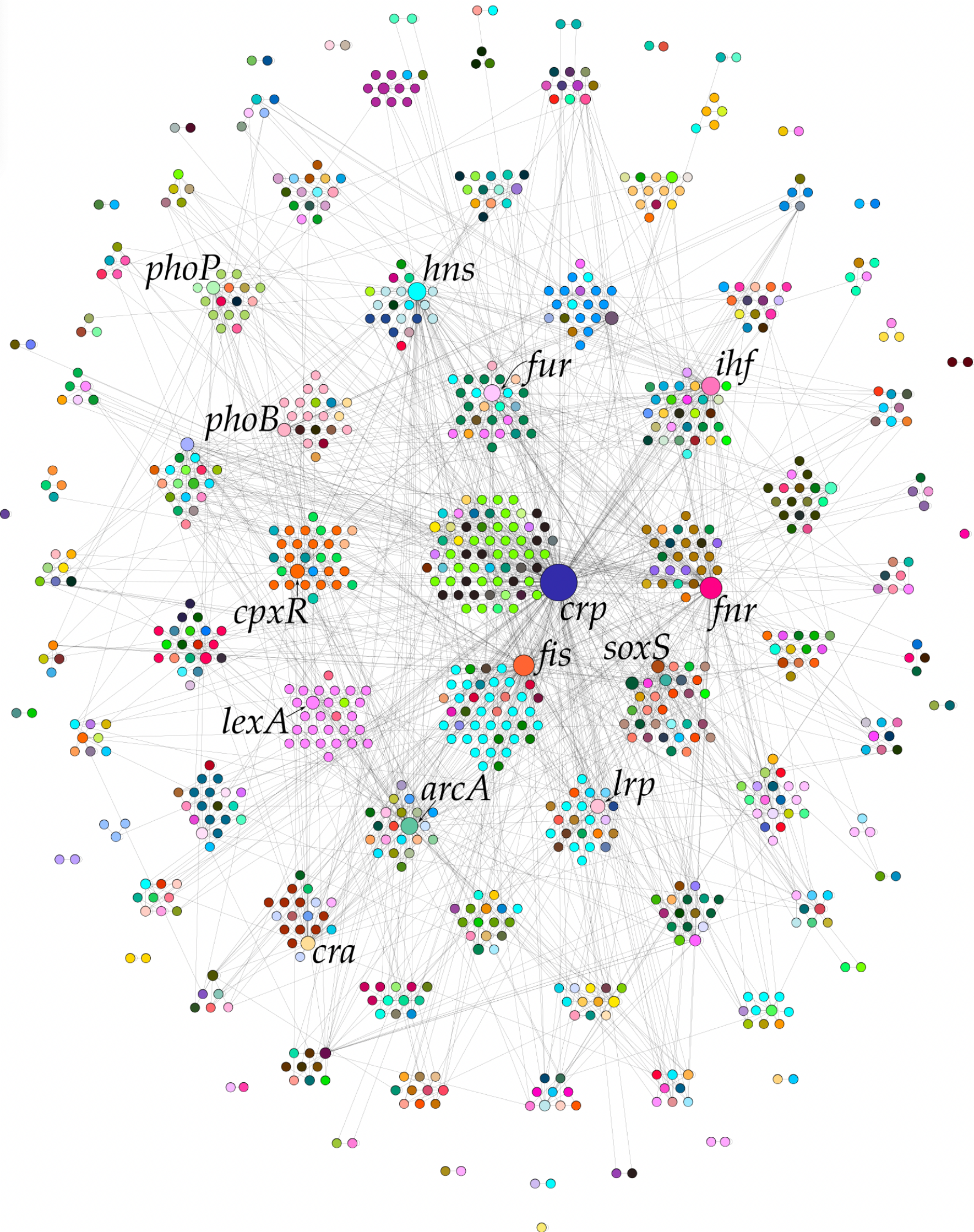}
    \caption{\textbf{Modularity versus fibers in TRN}. Louvain modularity
      detection algorithm applied to the TRN of \textit{E. coli}.  The modules obtained are clustered in
      space and the color of the genes indicates the fibers.
      Typically, modules are composed of many fibers, indicating that
      the fiber partition of synchronized genes is different than modularity.}
  \label{fig:modularity}
\commentAlt{Figure~\ref{fig:modularity}: 
Illustrative only.  Described in caption/text. No alt-text required.
}
  \end{figure}

We ran the Louvain community detection algorithm\index{Louvain algorithm } on the {\em E. coli}
TRN studied in Chapter \ref{chap:hierarchy_1} and found the modular
partition of Fig. \ref{fig:modularity}. We used the resolution parameter 0.16
      in the Louvain algorithm \citep{blondel2008}, which reaches the
      highest modularity (the density of links inside communities as
      compared to links between communities) at $Q=0.516$. Other
      parameters yield similar results. A visual comparison indicates
that the modules identified by the Louvain algorithm do not capture
the partition provided by the fibers. This is corroborated when we
calculate the performance of the Louvain modularity algorithm as the
fraction of genes classified correctly in the fiber, which is just
23\%. The results indicate that in the TRN, the functional partition
of the genetic network into synchronized fibers cannot be captured by
the modularity detection algorithm. 

The functional modules of synchronized genes that we identify using
fibrations in the TRN are difficult to detect with the community detection
algorithm. This is related to the role of hubs.\index{hub } A modularity algorithm
will put in the same module a hub like {\it crp} in the carbon SCC
shown in Fig. \ref{fig:componenta} together with all its regulated
genes since they are all connected. But dynamically this is not the
case since {\it crp} does not synchronize with its regulated genes
(they belong to different fibers).  Fibrations, on the other hand,
correctly classify the hub {\it crp} as an independent regulator not
synchronized with any gene in its regulon.  This is not an isolated
example. Figure \ref{fig:modularity} shows that the Louvain modularity
algorithm captures only 23\% of the synchronized functional fibers.

In general, modularity analysis may not capture functional
modularity when applied to a structural biological network like a
TRN, although in the next section, we see that the fibers may be
included within a module. However, modularity detection is applicable
to a biological functional network, as obtained by thresholding
a correlation matrix of fMRI BOLD signals in the brain or
coexpression. The difference between a structural network and a
functional network is investigated in Chapter
\ref{chap:synchronization}. In functional networks, similarity in
coexpression indicates synchronization and, therefore, functional
significance. Thus, a modularity algorithm works well to find
functional modules in functional networks but not in a structural
network like the TRN or a connectome (see Section \ref{sec:synchrony-correlation} for an application in the brain). In short, we have seen plenty
of examples where genes are not connected in the TRN but are still
synchronized, as in a multi-layer fiber.  In these cases,
fibration analysis should be used to identify the fibers associated with
function in the structural network. We develop this idea further in
Chapter \ref{chap:synchronization}.

\subsection{Fibers, modules and motifs in metabolic networks}

\begin{figure}[h]
\centering
\includegraphics[width=\textwidth]{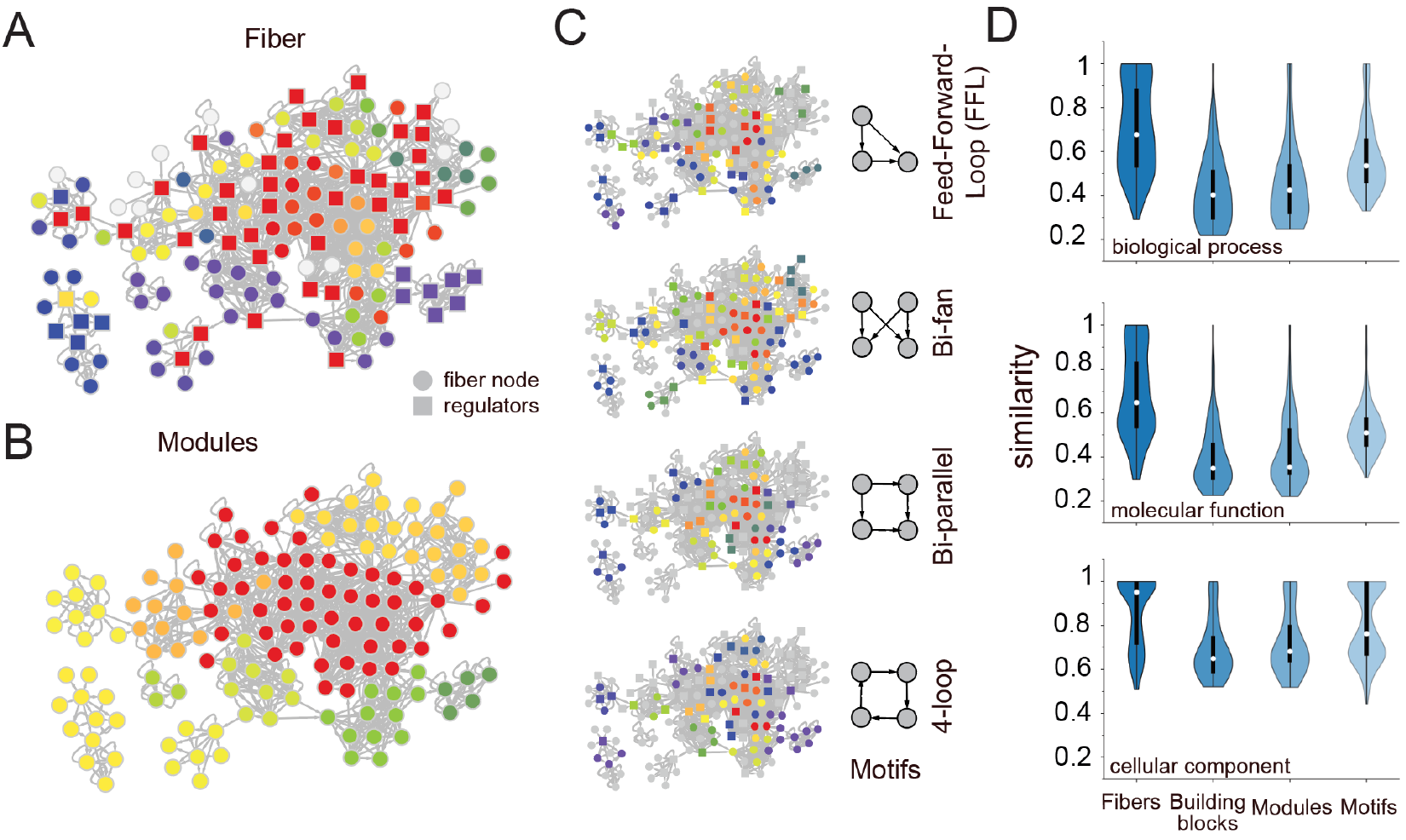}
\caption{\textbf{Biological significance of fibers,
    modules and motifs.} (\textbf{a}) Fibers in the carbon enzyme
  network, where colors refer to proteins in the same fiber. (\textbf{b}) Topological clusters were found using the Louvain
  algorithm. (\textbf{c}) Presence of significant
  motifs of different types. (\textbf{d}) Violin plots of the average
  functional similarity of enzyme pairs in fibers, clusters, motifs, and fibration building blocks using GO terms from the molecular
  function (MF), biological processes (BP), and cellular components
  (CC) ontologies. Enzymes in fibers show significantly 
  higher similarities compared to enzymes in cluster and motif
  building blocks ($P < 10^{-3}$, Mann-Whitney U-test). Figure
  reproduced from \citep{alvarez2024symmetries}. }
  \label{Fig:6} 
\commentAlt{Figure~\ref{Fig:6}: 
Illustrative only.  Described in caption/text. No alt-text required.
}
\end{figure}

The overwhelming presence of fibration building blocks and their
intricate cycle structure in the enzyme networks studied in Chapter
\ref{chap:complex} prompted \cite{alvarez2024symmetries} to compare
their characteristics and biological significance to network motifs
and modules (Fig.~\ref{Fig:6}).

As a consequence of the fundamentally different ways of detection, we
observe that motifs and modules are significantly less functionally
similar than fibers, suggesting that coherence between genes is a
better indicator of biological significance than the over-representation of topological patterns. Furthermore, motifs fail to
complement fibers, while network modules show the opposite. Such 
finding is probably rooted in the observation that modules partition
nodes in the network, increasing the chances that such groups of nodes
harbor fibers as well. Furthermore, the underlying fibers are bundles
of enzymes where metabolic information flows coherently,
suggesting that such patterns may also be reflected in the genomic
arrangements of genes in operons and fibers in the transcriptional
regulatory network. Indeed, we find that the overlap between fibers is
better established than motifs or modules, suggesting that fibers
point to biologically relevant, elementary building blocks.

Focusing on the carbon metabolism subnetwork of enzymes, we determine
modules using the Louvain algorithm and search for the most popular
network motifs: feed-forward-loops, Bi-Fan, Bi-parallel and 4-cycle
motifs \citep{shen2002network,milo2002network}.
A visual inspection of the coverage of fibers, modules, and motifs in
Fig. \ref{Fig:6}a-c indicates that modules partition nodes in the
underlying networks, while fibers and motifs, as expected, cover the
underlying network only partially. In particular, the determination of
modules and motifs is based on the connectivity of the underlying network,
while fibers account for similarities of input trees. As a
consequence, motifs and modules merely reflect local connectivity. In
contrast, fibers bundle together genes that share the same information flow,
suggesting that synchrony transcends simple connectivity. In other
words, nodes can be synchronized ({\it i.e.}  share the same input
tree) without being necessarily connected in contrast with motifs and
modules.

\begin{figure}[h]
\centering
\includegraphics[width=\textwidth]{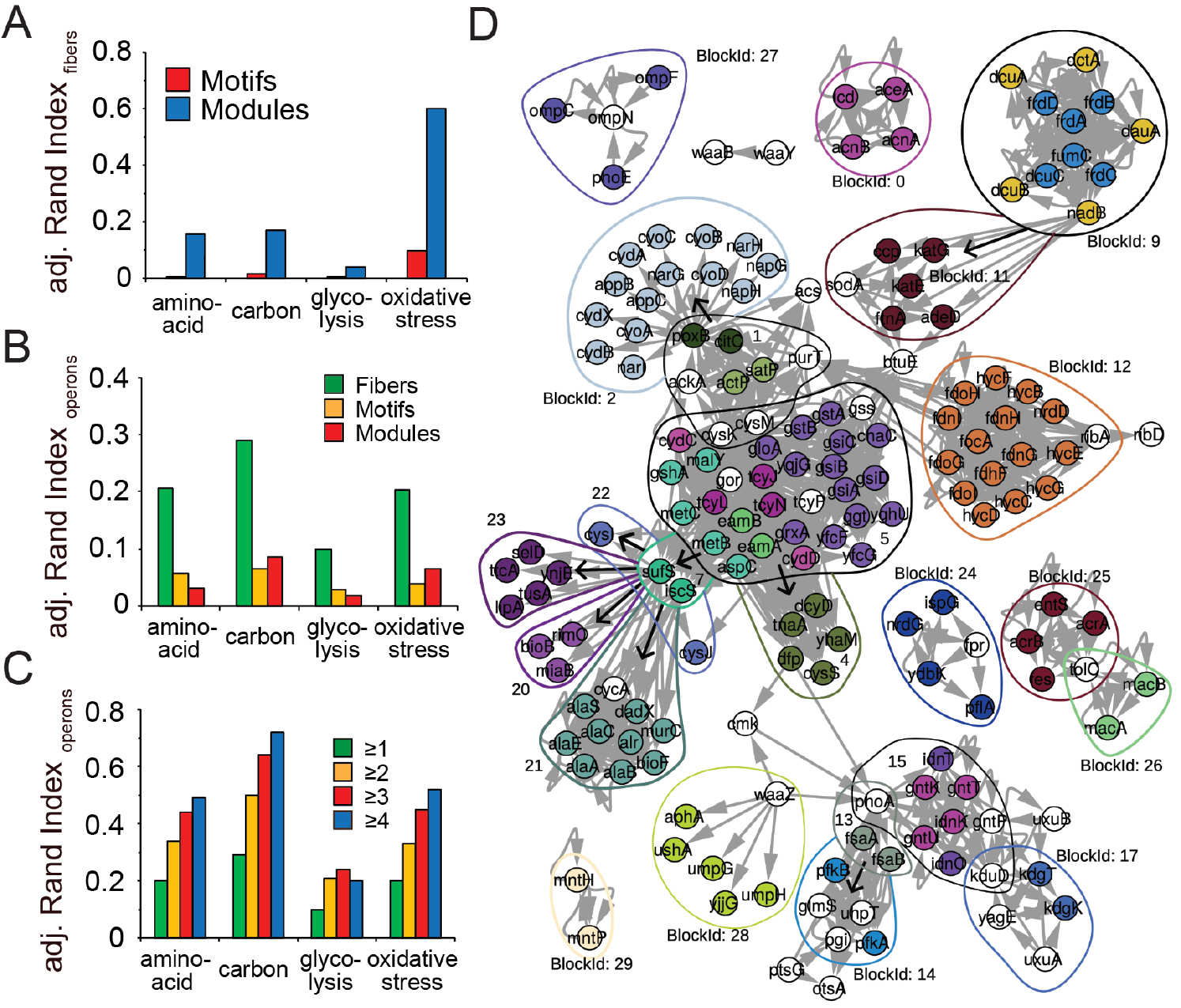}
\caption{\textbf{Comparison between fibers, modules, and motifs in the enzyme network.}  (\textbf{a}) Adjusted Ward index
  as a measure of the similarity of two partitions. Fibers overlap better with modules than with motifs in all enzyme
  networks. (\textbf{b}) Similarity between genes in
  operons and fibers, modules, and motifs indicate that fiber gene
  sets show the greatest overlap with operon gene sets. (\textbf{c}) The propensity of fibers in the underlying metabolic
  networks to overlap with genes in operons increases with elevated
  operon size. (\textbf{d}) Map of building blocks (enclosed by closed
  lines) and their constitutive fibers (nodes in color) in the
  oxidative stress network. Some regulators may belong to
  more than one building block, while the fibers are disjoint.  Figure
  reproduced from \citep{alvarez2024symmetries}. }
  \label{Fig:7} 
\commentAlt{Figure~\ref{Fig:7}: 
Illustrative only. Described in caption/text. No alt-text required.
}
\end{figure}

Given that fibers bundle information flow, as exemplified by their
input trees, we hypothesize that the functional roles of enzymes in the
same fiber should show a higher degree of functional similarity than motifs
and modules that are based on simple connectivity.  Furthermore,
motifs are  based only on statistical significance, which may
not be enough to guarantee homogeneous functionality.

As a benchmark, we compare the functional similarity of fibers to motifs
and modules, given the previous observations that strongly interacting
proteins have a heightened propensity to be involved in similar
biological functions and to appear in the same cellular components. In
Fig.~\ref{Fig:6}c, we observe that the average functional similarity
of enzyme pairs in fibers is higher than their module and motif
counterparts in all enzyme networks using GO terms from all three
ontologies. Similar results are found in all metabolic networks
separately studied in \citep{alvarez2024symmetries}. We further
consider the functional similarity of building blocks that, in the
most complex ways are composed of many fibers, such as composite
feed-forward and feedback Fibonaccis. As for their functional
similarity, we still find that building blocks have similarities that
compare to modules (Fig.~\ref{Fig:6}d).

Furthermore, we wonder to what extent fibers, modules and motifs
complement each other. Determining an adjusted Rand Index, we find
that modules overlap significantly better with fibers than motifs in
all four metabolic networks (Fig.~\ref{Fig:7}a). Such an observation
is rooted in the fact that modules arise from partitioning all nodes
in the underlying network, increasing the chance that a large
module harbors comparatively small fibers. Very few nodes
participate in more than one motif or fiber, which reduces the chance
that such structures substantially overlap.

As another indication of the biological relevance of genes that are
organized in fibers, we determine the overlap with genes that appear
in operons in {\it E. coli}, assuming that symmetries that dominate
the network topology are reflected in the regulatory organization of
gene sets. Utilizing 2,579 operon gene sets as of the RegulonDB
databases, we determine overlaps using an adjusted Rand Index, and find
that fibers overlap significantly better with operons than motifs and
modules in all four different metabolic networks
(Fig.~\ref{Fig:7}b). Furthermore, we refine our analysis by
considering operons that consist of at least a certain number of
genes. Fibers increasingly overlap with operons that
harbor more genes (Fig.~\ref{Fig:7}c).

These results suggest that fibers indeed bundle and potentially
coordinate enzymes that are involved in the same functions, pointing
to elementary building blocks in the underlying networks. In
particular, such fibers do not necessarily overlap with canonical
metabolic pathways, but offer a novel perspective for finding new
pathways.

Figure \ref{Fig:7}d maps the fibers and building blocks in the
oxidative stress network. As an example, we observe a fiber that is
composed of the \emph{fduABCD} operon of genes building the fumarate
reductase enzyme complex (BlockId 9, Fig. \ref{Fig:7}d). Such a fiber
is entangled with another fiber in a complex feedback loop that is
composed of subunits of the anaerobic C-4-dicarboxylate transporter
{\emph dcuABC} and {\emph dctA}, which is responsible for the uptake of
fumarate, succinate, L-aspartate and L- and D-malate under aerobic
conditions. While the underlying connections between the single
enzymes that are involved in the transport of metabolites and
corresponding reductase is highly complex, we break the underlying
complexity down to two fibers that not only bundle the underlying
functions effectively but also indicate which functions are connected
to each other on a dynamic basis. In other words, the complex
multiplicity links between enzymes are mapped to relatively simple
building blocks of enzymes that are dependent on each other, pointing
to the relevance of fibers as representatives of potential pathways.

The present evidence indicates that enzymes in fibers are functionally
more homogeneous than enzymes in network motifs or network
modules. This observation points to a hitherto unknown level
of complexity in the organization and architecture of biological
networks, which topological motifs and modules through statistical
means simply miss. As a consequence, such fibers may harbor more
functional information compared to motifs and modules. 

Additionally, fibers might offer a superior approach to uncovering pathways from a different perspective compared to motifs and modules. They could also help identify novel pathways that are obscured within existing representations of biological information. Furthermore, fibers could be utilized to discover new drug targets, providing an innovative means to pinpoint potential therapeutic interventions.


\chapter[Synthetic Biology Through Symmetries:
  the Cell as a Computer]{\bf\textsf{Synthetic Biology Through Broken Symmetries:
  the Cell as a Computer}}
\label{chap:breaking}

\begin{chapterquote}
We have seen how the functions of  biological networks can be
pictured as an orchestra of synchronized instruments captured by the
fibration symmetry. In this chapter we elaborate on an even stricter
and more taxing criterion for functionality of the network.  We ask
whether minimal biological circuits can perform core logic
computational programs. We show this through the mechanism of
`fibration symmetry breaking'. These circuits can then be used to
build computational molecular machineries. This helps in system
biology to design biological machines from the bottom up, following a systematic approach from
first principles of symmetry and symmetry breaking.
\end{chapterquote}

\section{Synthetic biology design through symmetry and broken symmetry}

Synthetic biology\index{synthetic biology } attempts to build circuits and systems 
to implement specific
functions. A long tradition at the interface of biology and
engineering proposes that the functional building blocks of these
systems should offer computational repertoires drawing parallels
between biological networks and electronic circuits.\index{circuit }  Indeed, the idea
of using electronic circuitry and devices to mimic aspects of gene
regulatory networks has been in circulation since the inception of
regulatory genetics by \cite{jacob1961genetic,jacob1977}. It
has been a driving force in synthetic biology, with several
demonstrations showing that engineered biological circuits can perform
computations analogous to electronic circuits and computers.

In 2000, the first two synthetic circuits were built in bacteria. These circuits
mimic two core functionalities of computer systems: timekeeping
clocks (oscillators), and memory.
A major breakthrough was the demonstration of a synthetic genetic
oscillator by \cite{elowitz2000}: the repressilator\index{repressilator } composed
of three genes in a ring interacting by repressors, discussed in
Section \ref{S:repressilator}.

Another basic circuit with oscillatory behavior is a purely negative
feedback loop, comprising two genes $A$ and $B$, where $A$ inhibits $B$
and $B$ activates $A$. This circuit by itself
could function as a pulse generator since it is capable only of
 damped oscillations \citep{alon2019}. However,
in the presence of noise it generates sustained oscillations with a reliable
frequency but ill-defined amplitude. Some simple additions can be made
to this basic structure to improve its oscillatory behavior. For
example, \cite{purcell2010} show that from an analytical
standpoint the addition of an activation self-loop in gene $B$
provides an amplified negative feedback loop; however, only damped
oscillations have been seen \emph{in
  vivo} \citep{purcell2010}. Furthermore, addition of an
inhibiting self-loop in gene $A$ gives the Smolen oscillator of Section \ref{S:metab+smol},\index{Smolen oscillator }
which exhibits sustained oscillations \citep{hasty2008} in a
robust and highly tunable manner.

A second major advance was made by \cite{gardner2000construction}, who
demonstrated a genetic toggle-switch\index{toggle-switch } allowing memory storage in the
form of a bit (discussed in Section \ref{S:bifurcations}).  The
toggle-switch is a bistable two-way switch, analogous to an
electronic flip-flop that functions as a binary memory in a computer.  Bistability\index{bistability } is the key feature of
this circuit: it can be switched between two stable equilibria by
different inputs. \cite{gardner2000construction} showed the property
of bistability in this circuit using the concentrations of two
repressors and their effective synthesis rates.

Another prominent design to store memory is the lock-on circuit, 
discussed in Section \ref{S:lock-on}, which,
unlike the toggle-switch, stores memory in a `permanent' form. This is
a genetic circuit consisting of a positive autoregulation (PAR)
feedback loop, which works as a bistable one-way switch circuit
sustaining two stable states: both genes inactive or both genes active
after either one has been activated \citep{tyson2003sniffers}.  In
contrast to the toggle-switch, the lock-on is a one-way switch: once
the circuit switches to the activated case it cannot return to
its previous state. This type of lock-on circuit take a role in
developmental processes characterized by a state transition, such as
apoptosis (programmed cellular death) \citep{tyson2003sniffers}.

Since the introduction of the repressilator and toggle-switch two
decades ago, an explosion of activity \citep{cameron2014brief,
  khalil2010synthetic} has demonstrated a myriad of genetic circuits
that are able to perform basic logical operations necessary for a
computational device \citep{dalchau2018computing,
  tanenbaum2016structured, tyson2003sniffers,
  khalil2010synthetic,fussenegger2010synchronized}. Such circuits are
constructed using feedback loops, both positive and
negative, \citep{tyson2003sniffers, dalchau2018computing} and are
executed by synthetic switches and oscillators designed from simple
components such as interacting genes or protein-protein interactions.
They includes various toggle-switches for memory storage
\citep{gardner2000construction,atkinson2003development,kramer2004anengineered,kramer2005hysteresis,ajofranklin2007rational,ham2008design}
and logic operations
\citep{anderson2007environmental,guet2002combinatorial,rackham2005anetwork},
oscillators
\citep{elowitz2000,atkinson2003development,fung2005asynthetic,hasty2008,tigges2009atunable},
pulse generators \citep{basu2004spatiotemporal}, and edge
detectors \citep{tabor2009asynthetic}.

Remarkably, \cite{leifer2020circuits} and \cite{alvarez2024fibration}
find the presence of circuits closely resembling of all these synthetic
circuits at the core of the {\it E. coli} TRN. This suggests a
view of the minimal TRN core as a logical computational machine \citep{kondev2014bacterial}, to
be discussed in Chapter \ref{chap:minimal}. 
In this chapter we perform a systematic
analysis of circuit capable of logical computations in TRNs and
present algorithms to seek them in any network through fibration
symmetry breaking. This chapter thus extends Chapter \ref{chap:hierarchy_2} by considering electronic analogs.

Since these circuits can be constructed artificially to perform
computations, it is reasonable to expect to observe them, or some
close variation, in the core computational subset of the network. We
expect to find memory storage circuits, as well as oscillating circuits
for timekeeping. In the case of the TRN of simple model bacteria like
{\it E. coli}, we expect to observe the simpler forms of these known genetic
circuits from synthetic biology, which is indeed the case, as will be
shown next. These circuits can be understood through the concept of a
broken-symmetry fibration, introduced by \cite{leifer2020circuits}.

We formalize this quest mathematically, by showing that this principle
of symmetry breaking, when applied to circuits
with symmetries, reveals a hierarchy of genetic circuits across species
from bacteria to humans. These circuits map to the fundamental building blocks
of electronic architectures \citep{horowitz2015thearts}: starting with the transistor and
progressing to mirror circuits, ring oscillators, and complex logic
integrated circuits involving flip-flops. The functionality of these
circuits is analogous to electronic operations: they act like clocks
and counters in their symmetric states, and as toggles, latches and
memory storage units (flip-flops) in their symmetry breaking states.

So far we have seen that a large part of the regulatory network is
made of synchronized fibers. In the case of {\it E. coli}, 66\% of the
genes can be characterized by symmetries. In this chapter we show that
the remaining genes can also be accounted for by symmetry breaking
(and a few more regulatory functions), so we can classify {\it every
  single gene} in the bacterium's genome by its function.  Beyond the
fibers identified by surjective symmetry fibrations, the network can be
further reduced by the application of an injective fibration, which
identifies a reduced driver networks elaborated in Chapter
\ref{chap:minimal}. The remaining network is organized into three
giant strongly connected components, which control the fibers. We
further decompose these components into circuits identified by broken
fibration symmetries. The circuits play a crucial functional role in
the network, acting as memory storage flip-flops analogous to building
blocks of memory in digital computers.  These logical genetic circuits
are mainly composed of feedback loops \citep{alvarez2024fibration} or
frustrated topologies \citep{leifer2020circuits}. Negative feedback
loops (or {\it frustrated} circuits) give rise to oscillating
circuits, while positive feedback loops allow for memory storage.

\section{A biological transistor at the core of genetic circuits } 

To understand the computational rationale for symmetric and
symmetry-broken circuits made of repressors, we map them to electronic
analogs.  We start our analogy with electronic circuits with the
simple fibration building blocks found in the {\it E. coli} TRN
displayed in Fig. \ref{fig3_plos}, spanning from the AR loop to the
FFF.

A first realization is that a repressor link between a source and target gene in a TRN and can be mapped to a
NOT gate, which is a primary function of the transistor
(Fig. \ref{fig2-transistor}a). A repressor or inhibitory interaction
is, in turn, the primary interaction of all biological systems, from
genetic networks to the brain.

The mappings between electronic\index{circuit !electronic } and gene regulatory circuits\index{circuit !gene regulatory } are
straightforward. Even though nothing is circulating in a genetic
circuit, the information flow between genes represented by the TF can
be thought of as circulating charges in electrical circuits. The
expression level of a gene is analogous to the voltage in an
electrical circuit.  A battery does work to produce a given voltage
level; likewise a gene does work to produce a desired expression
level.  Connecting a battery to a device through a wire is like
connecting a gene to another one through a regulatory link; the
electric current represents the rate of change of the expression
levels in the genetic circuit.

\subsection{Repressor interaction maps to a transistor}
\index{transistor }

\begin{figure}[!h]
    \centering
    \includegraphics[width=.9\textwidth]{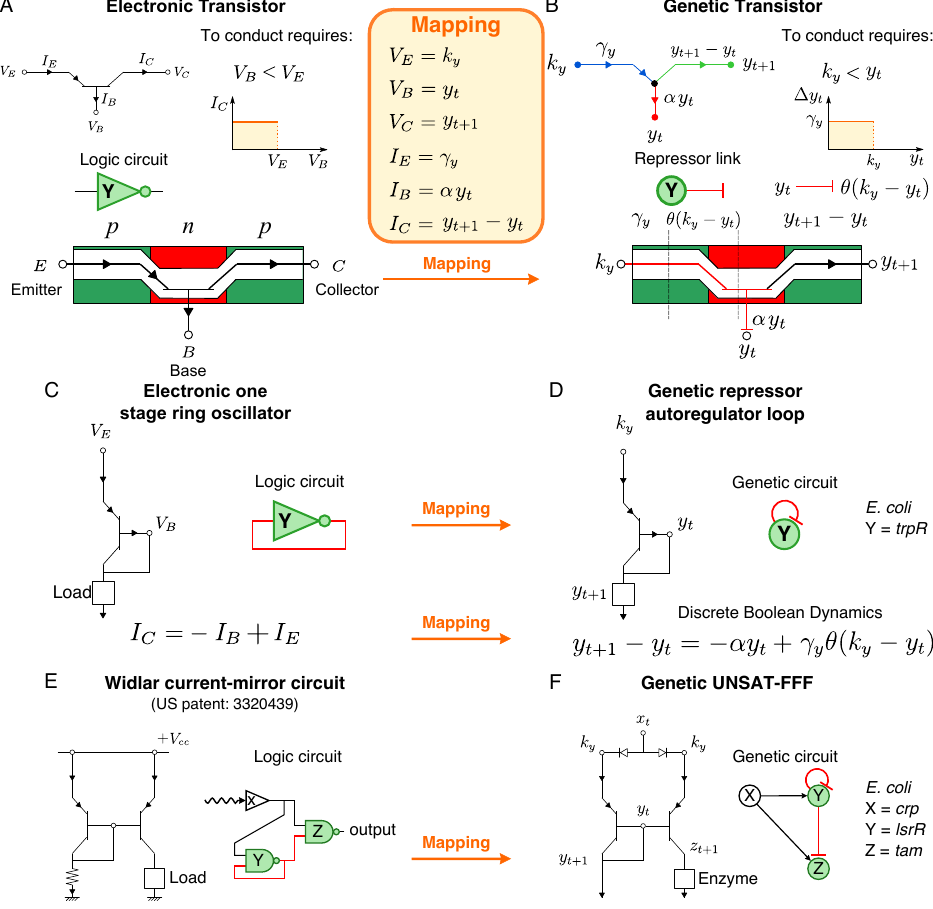}
\caption{ \textbf{Mapping between electronic and biological symmetry
    fibration circuits.}  (\textbf{a}) A pnp transistor allows current
  flow if the voltage applied to its base is lower than the voltage at
  its emitter ($V_B<V_E$). Since it has a high (low) output for a
  lower (high) input, it is logically represented by a NOT gate.  The
  yellow box shows the mapping between the pnp transistor and the
  biological repressor. (\textbf{b}) A repressor regulation link plays
  the role of the pnp transistor since the rate of expression of a
  gene is repressed by gene Y if $k_y<y_t$. (\textbf{c}) Connecting
  the base of the transistor to its collector forms a one stage
  ring oscillator. (\textbf{d}) This connection is translated to the
  biological analog as a repressor autoregulation at gene Y.  In
  this way, the rate expression of gene Y is able to oscillate,
  depending on the parameters $\alpha$, $k_y$ and
  $\gamma_y$. (\textbf{e}) Widlar current-mirror circuit and (\textbf{f})
  its biological analog (UNSAT-FFF). By mirroring the ring
  oscillator, the Widlar mirror circuit allows synchronization and
  oscillations. \textbf{(g)} Phase diagram of oscillations of the
  UNSAT-FFF. An oscillatory phase is defined by the condition
  $\gamma_y/k_y > \left(\gamma_x/\alpha\right)^{-1}$. For instance, on
  the right side we plot the solution of the discrete dynamics for a
  set of parameters satisfying such a condition. Specifically,
  $\alpha=0.205$, $\gamma_x = 0.454$, $\gamma_y=0.454$, $k_x=0.5$, and
  $k_y=1.0$. Figure reproduced from \citep{leifer2020circuits}. Copyright \copyright ~ 2020, Leifer {\it et al.}}
\label{fig2-transistor}
\commentAlt{Figure~\ref{fig2-transistor}: 
Described in caption/text. No alt-text required.
}
\end{figure}

We begin the analogy with the simplest circuit: a single gene with a
feedback loop with repression (AR loop,
Fig.~\ref{fig2-transistor}d). At the simplest Boolean level, the
dynamics for the expression level $y_t$ is described by the discrete
time model with Boolean interaction \citep{kauffman1973thelogical}:
\begin{equation} \Delta y_t = y_{t+1} - y_t = -\alpha y_t
  + \gamma_y\theta(k_y-y_t).
\label{eq:analog}
\end{equation}
Here, $\alpha$ is the degradation rate of the Y gene product,
$\gamma_y$ is the maximum expression rate of gene Y, and $k_y$ is the
dissociation constant.  The Heaviside step function $\theta(k_y-y_t)$
reflects the repressor autoregulation in the Boolean logic
approximation.

\cite{leifer2020circuits} show that the genetic repressor interaction,
shown as the stub in Fig.~\ref{fig2-transistor}b, is the genetic
analog of the solid-state transistor, shown in
Fig.~\ref{fig2-transistor}a.

The formal (mathematical) analogy mediated by the charge transport
process in semiconductor materials indicates that the repressor
interaction, symbolized by the $\theta(k_y-y)$ term in
 (\ref{eq:analog}), works like a logic NOT gate.  At the
foundations of electronics, it was understood that a NOT gate has to
convert a low input into a high output, and thus it requires an
amplifier, which cannot be made out of a combination of diodes. The
device that does the job is the transistor.\index{transistor }

A transistor is typically made up of three semiconductors, a base
sandwiched between an emitter and a collector
(Fig.~\ref{fig2-transistor}a).  The current flows between the emitter
and collector only if voltage applied to the base is lower than at the
emitter ($V_B<V_E$) and thus the transistor acts as a switch and
inverter. In the genetic circuit, the expression $y_t$ drives the rate
of expression of gene Y, like the voltage drives current around an
electric circuit.  Comparing (\ref{eq:analog}) to the pnp
transistor in Fig. \ref{fig2-transistor}a leads to an analogy in
which the expression $y_t$ is an analog for the base potential $V_B$
of a transistor, $k_y$ is an analog for $V_E$, $\gamma_y$ is an analog
for the emitter current $I_E$, $\alpha y_t$ is an analog for $I_B$, and
$\Delta y_t$ is an analog for $I_C$. Then \eqref{eq:analog}
provides the genetic equivalent of the equation for a transistor's
collector current $I_C = I_E - I_B$ (Fig. \ref{fig2-transistor}cd).

The mapping between a transistor and a repressor is formalized in
(Fig. \ref{fig2-transistor}a, b):
\begin{equation}
  V_E  = k_y \quad
  V_B  = y_t \quad
  V_C  = y_{t+1} \quad
  I_E  = \gamma_y \quad
  I_B  = \alpha y_t \quad
    I_C  = y_{t+1} - y_t.
\label{eq:mapping-transistor}
\end{equation}

\subsection{AR loop maps to a ring oscillator}

In the AR loop shown in Fig. \ref{fig2-transistor}d the repressor
link connects to its own gene. Analogously, the collector of the
transistor connects to the base. Thus, the repressor AR genetic
circuit of Fig.~\ref{fig2-transistor}d becomes a one-stage ring
oscillator\index{ring oscillator } (Fig.~\ref{fig2-transistor}c) where the collector of the
transistor connects to the base forming the minimal signal feedback
loop. As shown in Fig.~\ref{fig2-transistor}d, an example of the AR is
the gene Y~=~{\it trpR} from the {\it E. coli} transcriptional
network.

\subsection{FFF maps to a Widlar current-mirror electronic circuit}
\index{Widlar current-mirror circuit }

The repressor AR loop can be extended to the FFF by symmetrizing it,
adding gene Z, such that it synchronizes with Y to express an enzyme
that catalyzes a biochemical reaction
(Fig.~\ref{fig2-transistor}f). The circuit is completed by an
external regulator X that maintains the symmetry between Y and Z. The
resulting UNSAT-FFF\index{UNSAT-FFF } is shown in Fig.~\ref{fig2-transistor}f as {\it
  E. coli}'s X~=~{\it crp}, Y~=~{\it lsrR}, Z~=~{\it tam}.
\cite{leifer2020circuits} show that the frustrated topology UNSAT-FFF
$\rvert n= 1, l = 1 \rangle$ give rise to oscillatory behavior
(Section \ref{Subsec:unsatfff}).

The UNSAT-FFF maps to the so-called Widlar current-mirror electronic
circuit shown in Fig. \ref{fig2-transistor}e, a popular
building block of integrated circuits used since the foundation of
the semiconductor industry---1967 US patent
\citep{widlar1969some,widlar1967patent,widlar1969design,horowitz2015thearts}. It
serves two key functions as we show below: mirror synchronization of
$y_t=z_t$ and oscillatory activity (Fig. \ref{fig2-transistor}g) which
could be used as a timekeeping clock.

Interestingly, a single self-inhibiting gene can also
exhibit oscillatory behavior if the self-inhibition signal exhibits a
delay as shown in Section \ref{Subsec:unsatfff}, allowing for its expression levels to increase before its
self-inhibition decreases them. This is actually a very sensible
prediction, given that there is some natural delay in the system while
the gene is transcribed, then translated into a TF and finally
binds to the gene's promoter region, which requires the concentration of
the TF to be high enough for binding to take place.

\section{Gene duplication in evolution mimics lifting in a fibration}
\label{sec:duplication-lifting}

Chapter \ref{chap:robustness} discussed the robustness of a genetic
network evolving by gene duplication\index{gene duplication } which generates rich fibrations
as compared to one with only automorphisms.  Figure \ref{summary}
exemplifies, anecdotally, how gene duplication and divergence can lead
to a robust fiber-rich gene network.

\begin{figure}
  \includegraphics[width=.5\linewidth]{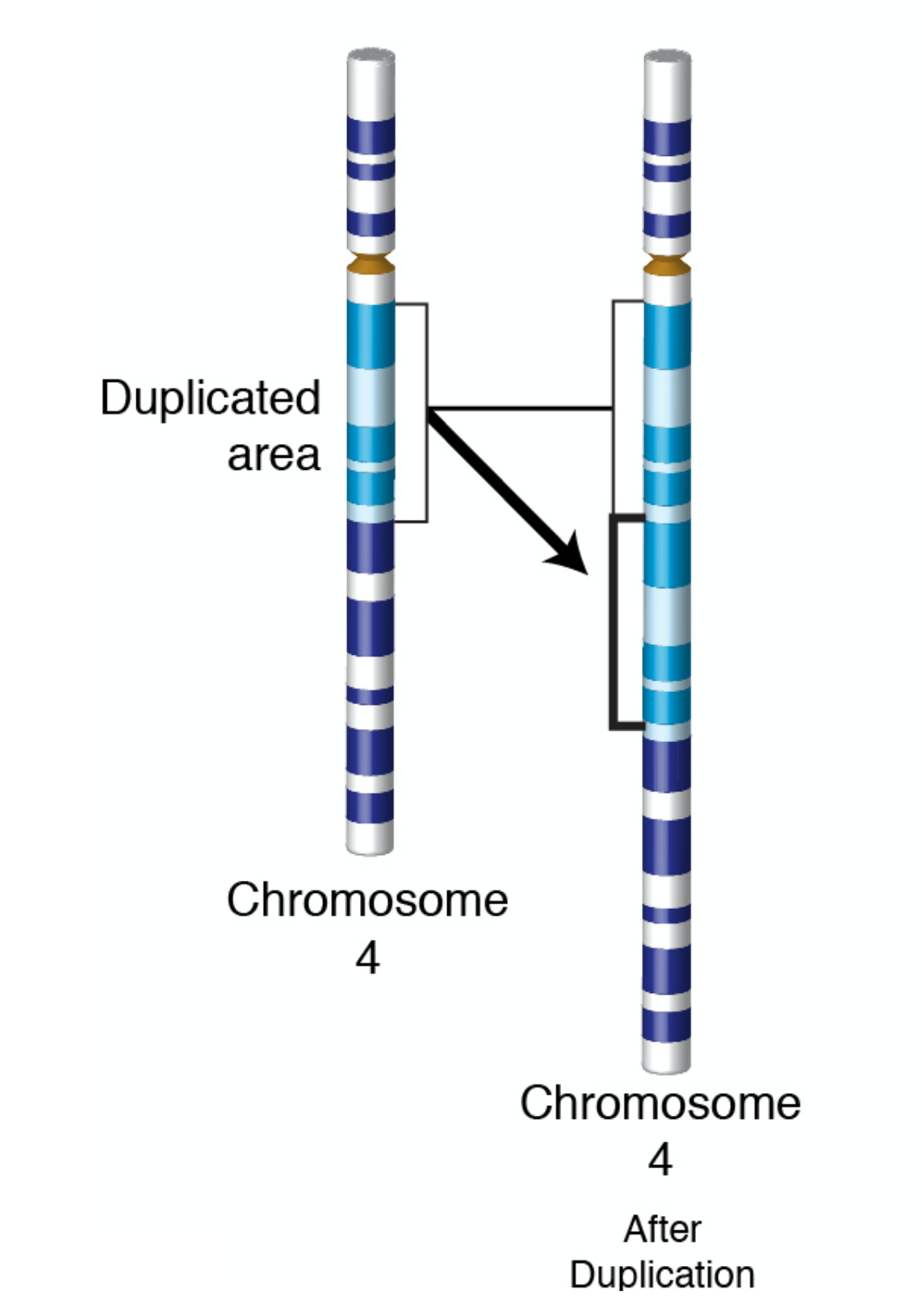} \caption{
    \textbf{Duplication event in the genome}.  A region of the chromosome
    is duplicated copying a number of genes.  When the duplication
    follows the lifting property, the event increases the number of
    genes in a fiber without affecting their synchronization. Thus, it
    is analogous to a gauge transformation in the genome as discussed
    in Section \ref{genetic-connection}.  Reprinted from Wikipedia
    \url{https://en.wikipedia.org/wiki/Gene_duplication}.}
  \label{fig:duplication}
\commentAlt{Figure~\ref{fig:duplication}: 
Left: Chromosome with one segment marked `duplicated area'.
Right: Chromosome with two neighboring copies of the
segment marked `duplicated area' .
}
\end{figure}

Mathematically, fibrations are built by the lifting property, and this
mechanism can be mapped to the duplication of genes and edges in the
biological graph.  This suggests that the emergence of fibration
symmetry in the TRN is a major design principle of genetic networks,
consistent with their dynamical evolution as fibrations are created.
Gene duplication (see Fig. \ref{fig:duplication}) is a crucial mechanism in the evolution of genomes and genetic systems. This process generates new genes by duplicating existing ones when a segment of the chromosome  that contains a gene is replicated, resulting in the cell possessing two paralogous copies of a set of genes.

Gene duplications typically occur due to errors in DNA replication and repair mechanisms. This process not only duplicates the gene itself but also its promoter region—the DNA sequence adjacent to the gene that contains the binding sites for transcription factors (TFs). This duplication creates redundancy within the cell.

Below we show that gene duplication is analogous to the lifting
operation that defines the fibration (see Section \ref{lifting}). 
This strongly suggest an evolutionary drive towards fibrations symmetries in the genome.
It also 
provides a natural model for the evolution of fibration symmetry and
broken symmetry circuits in genetic networks. It defines a hierarchy of broken
symmetry circuits with toggle-switch memory functionalities like
the electronic flip-flop. At the same time, the fibration building blocks
of the previous section work as clocks with varying levels of
sophistication.  Symmetry and broken symmetry circuits provide
computing functionalities to the cell composing all building blocks of
TRNs.

\begin{figure*}[t!] 
   \begin{center}
    \includegraphics[width=\textwidth]{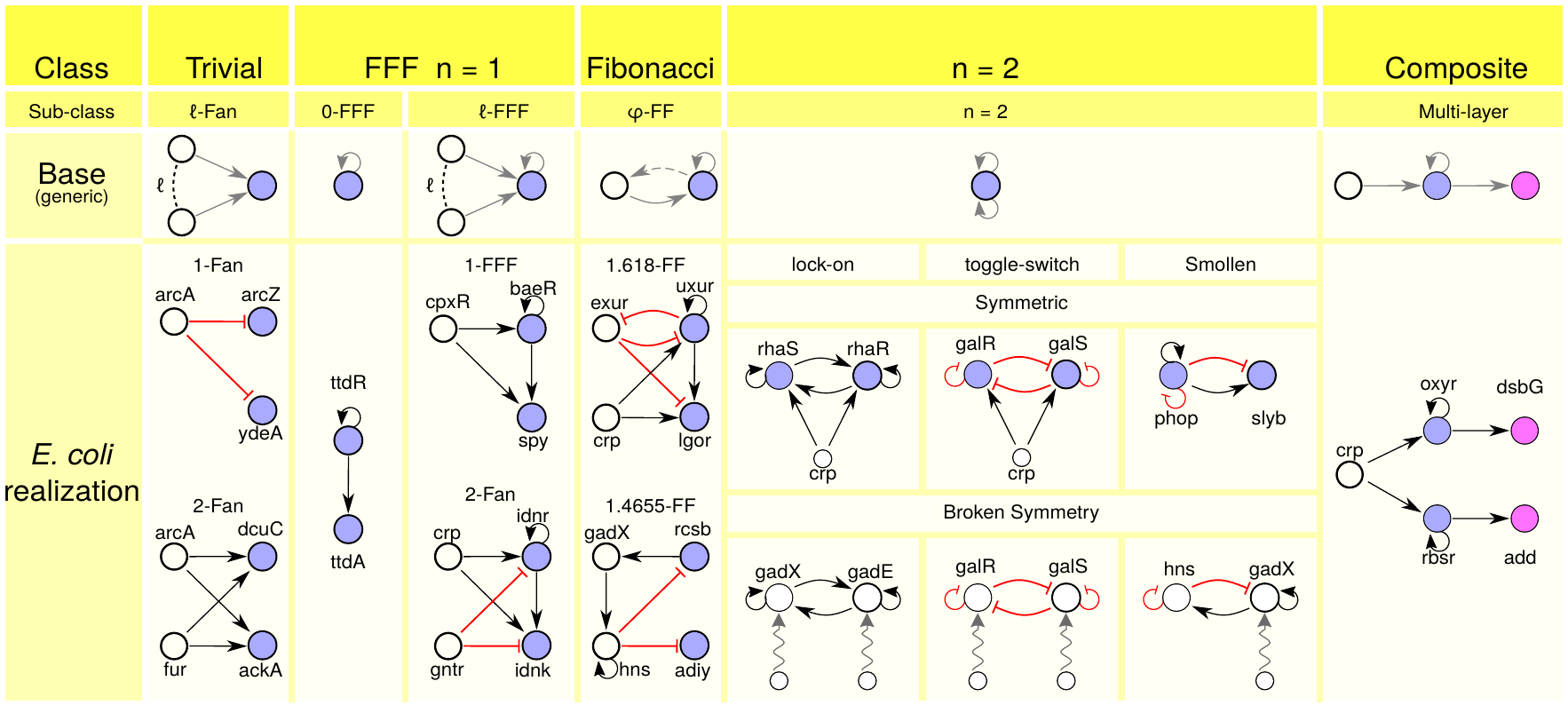}
    \end{center}
    \caption{ \textbf{Bases of simple fibration building blocks in
        TRNs}. These circuits serve as bases for duplicated circuits
      leading to a broken symmetry circuit performing as a flip-flop. TRNs can be seen as assemblies of five basic fibration
      building blocks (top row): (i) Trivial fibers with $|n=0,
      l\ne 0 \rangle$.  (ii) The AR loop $|n=1, l =0\rangle$ and
      FFF $|n=1, l\ne 0 \rangle$. (iii) The Fibonacci fiber,
      $\varphi$-FF. (iv) the $|n=2,l\rangle$ fibers.  When this
      symmetry is broken it forms the memory and oscillatory circuits.
      (v) Multilayer fibers of the previous ones. By adding different
      types of the previous four building blocks in a sequential manner,
      a composite fiber is obtained. Figure reproduced from
      \citep{alvarez2024fibration}. }
    \label{fig:circuitsOverview}
\commentAlt{Figure~\ref{fig:circuitsOverview}: 
Described in caption/text. No alt-text required.
}
\end{figure*}

The evolutionary design principle of the TRN follows the lifting of
genes and edges from a primordial set of fibration bases depicted in
Fig. \ref{fig:circuitsOverview}.  At the most basic level this leads
to the simple building blocks described in Chapter
\ref{chap:hierarchy_2}.  Starting from the base of these building blocks,
by duplication we obtain symmetric fibers, and then by breaking their
symmetries we obtain logic circuits for memory storage and timekeeping.  This lifting-driven dynamical evolution creates the core of
the computational decision-making of the cell, which constitutes the
apparatus of its genetic collective intelligent device.

We explain this new hierarchy next.

\begin{figure}[!ht]
    \centering
    \includegraphics[width=.8\linewidth]{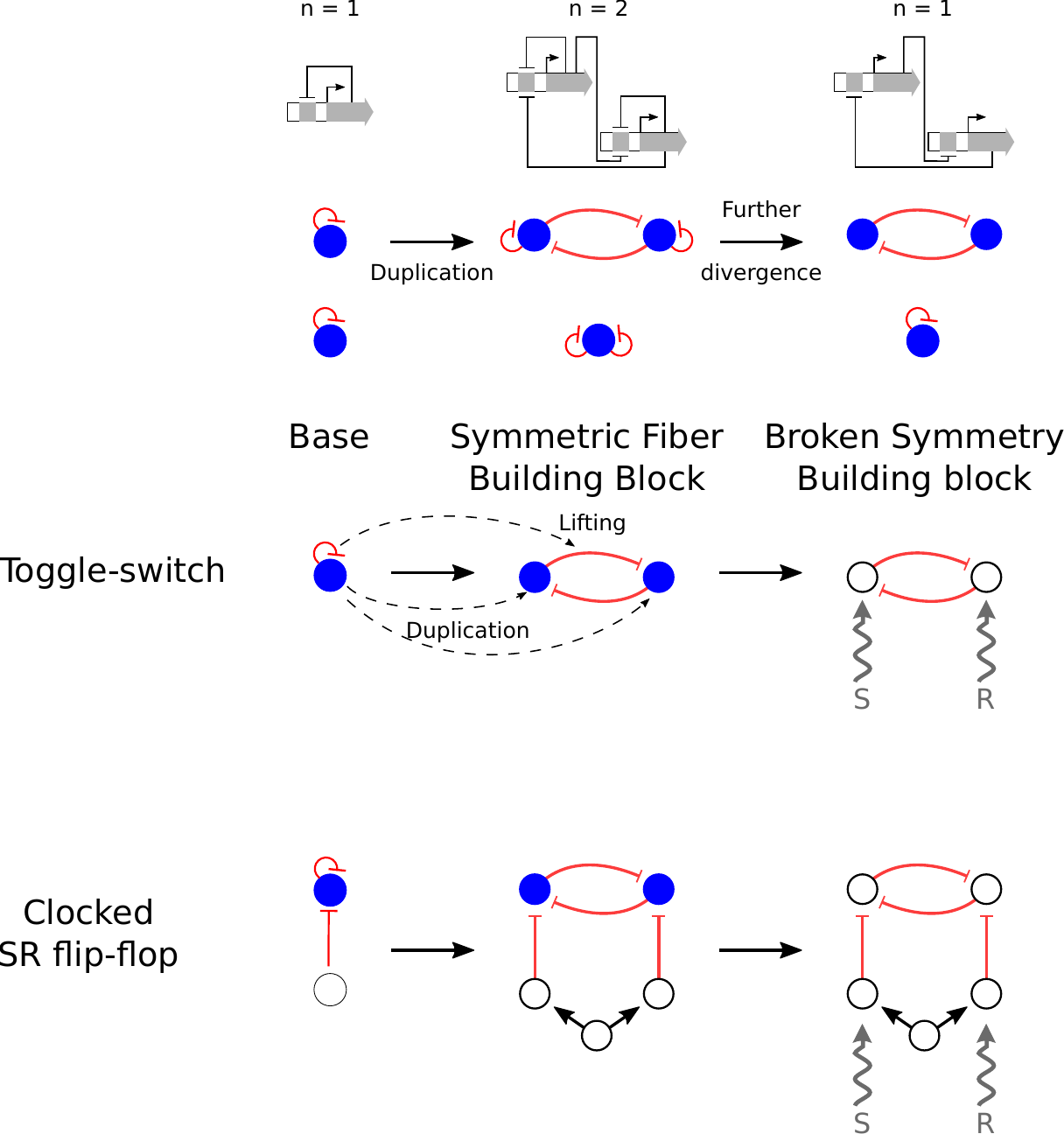}
    \caption{\textbf{Broken symmetry circuits and duplication/lifting process.}
      How flip-flops emerged by the lifting operation analogous to duplication and broken symmetries. Starting with an initial simple structure a gene duplication
      process can create a symmetric fiber structure, analogous to a
      lifting process. The structure resulting from the
      duplication may  undergo further modifications
    (top). The circuit formed depends on the initial
      structure that is replicated (bottom).  For each
      circuit, the symmetric form is at the center, while the
      broken symmetry process is on the right. The {\it
        set-reset (S-R)} inputs break the symmetry by sending
      different regulations to an otherwise synchronous pair of
      nodes. The {\it S-R} inputs can be any different nodes sending
      different signals or can be the same node sending different
      signals to the nodes in the fiber.  Figure reproduced from
      \citep{alvarez2024fibration}.}
    \label{fig:circuits_hierarchy}
\commentAlt{Figure~\ref{fig:circuits_hierarchy}: 
Described in caption/text. No alt-text required.
}
\end{figure}

Gene duplication\index{gene duplication } has been proposed as the mechanism behind the
emergence of 
motifs \citep{milo2002network,shen2002network,alon2003a}.  Fibers
could emerge in an evolutionary setting of random rewiring (mutation)
and selection for functional structures. However, such evolution
would probably be slow. In contrast, gene duplication generates
fiber structures fast and `for free', with subsequent minor
modifications of the duplicated arrows (mutations in the regulatory
region modifying incoming edges, or in the coding region modifying
outgoing edges).

In evolution, multiple paralogue copies of a gene may emerge via gene
duplication events, and subsequent evolution may lead to differences
in their coding or regulatory regions. If the gene codes for a
transcription factor, the duplication will correspond to a duplication
of a node in the TRN, and the subsequent changes may cause a
subsequent rewiring of outgoing or incoming edges, respectively.  The
first step, the process of taking a gene and `opening it up'
into one or multiple other identical genes, is akin to the lifting
operation between a collapsed base graph and its original graph. The
lifting operation can be understood as opening up a collapsed
fiber node in the base into the full set of original synchronous nodes
that share the same color. The key feature of the lifting operation
is that the nodes obtained by this opening up have the exact same
inputs and input trees as the node that is being opened
  up. This ensures that the new network's dynamics is the same as
the original collapsed base graph. This is a great property to have if
the duplication event is really duplicating the dynamics as well. That
is, this is a change without a change to the cell; a symmetry. 

The cell is then free to diverge.
Further mutations in the coding region then cause the
two original copies of the gene to perform different functions, and
thus diversify the bacteria's functions while creating bigger fibers.
This process has been described also in Section
\ref{genetic-connection} in terms of curvature and connection in the
genetic network.

This process does not only duplicate existing genes. The most important
part is the subsequent divergence. By virtue of creating a duplicate
of the previous genes with identical dynamics, one of the genes is
ready to mutate and diverge. In particular, the symmetry now can be
broken.  If the symmetry of some of the symmetric building blocks
obtained by duplication is broken, logical circuits are obtained as we
explain next.  By taking the basic fiber building blocks in
Fig.~\ref{fig:circuitsOverview} and duplicating their bases we obtain
different replica circuits still preserving the same dynamics and
synchrony. Nevertheless, if there are added regulations that break
 symmetry, the result is a circuit that performs logical
functions: a symmetry-broken circuit.

It is important to point out, however, that there exist duplication
processes that do not preserve the inputs. For example, a gene with 
self-regulation, once duplicated, would still have  self-regulation,
but would also have cross-regulation with its copy,
meaning it has gone from having one input to two, which changes the original
input tree; see Fig.~\ref{fig:circuits_hierarchy}. Therefore,
evolution or some other mechanism, is needed to ensure at least some
duplication events preserve the input tree of the original gene. This
implies that evolution is selecting for duplications that preserve the
dynamics of the network, like lifting-driven duplication.

\section{Symmetry breaking in physics}
\label{sec:breaking-physics}

While symmetry principles stand as crucial elements within natural
laws, much of the world's complexity emerges from mechanisms of
symmetry breaking,\index{symmetry breaking !in physics } which encompasses various ways in which nature's symmetry
can be veiled or disrupted \citep{anderson1977more}.  Any situation in
physics, in which the ground state (i.e., the state of minimum energy)
of a system has less symmetry than the system itself, exhibits the
phenomenon of symmetry breaking.  For instance, the phases of
matter are characterized by different symmetries. At higher
temperatures, matter takes on a `higher symmetry' phase (e.g., liquid,
paramagnetism, normal conductivity, and fluidity), while at lower
temperatures, the symmetries of the phases are broken to `lower
symmetry' (e.g., solid, ferromagnetism, superconductivity, and
superfluidity).

Most symmetry laws in physics are broken in one way or another. One
such mechanism is spontaneous symmetry breaking, where the laws of
physics remain symmetric, but the system's ground state exhibits a
lower symmetry than the full system, as in a
paramagnetic-to-ferromagnetic phase transition. For temperatures below
the critical value $T_{c}$, the magnetic moments of the atoms of a
ferromagnetic material are partially aligned within magnetic domains,
producing a net magnetic moment even though the atoms interact through
a spin-spin interaction, which is invariant under rotation. Thus the
rotationally invariant symmetry of the system is broken to this ground
state, which has nonzero magnetic moment. As the temperature increases,
this alignment is destroyed by thermal fluctuations and the net
magnetization progressively reduces until it vanishes at $T_{c}$. The
orientation of the magnetization is then random. Each possible direction is
equally likely to occur, but only one is chosen at random, resulting
in a zero net magnetic moment. So the rotational symmetry of the
ferromagnet is manifest for $T > T_{c}$ with zero magnetic moment, but
is broken by the arbitrary selection of a particular
(less symmetric) ground state with non-zero magnetic moment for $T <
T_{c}$.

Another type is explicit symmetry breaking, where the dynamics is
only approximately symmetric, yet the deviation caused by the breaking
forces is minimal. Now we can consider the symmetry violation as a
small correction to the system. An example is the spectral line
splitting in the Zeeman effect\index{Zeeman effect } due to a magnetic interaction
perturbing the Hamiltonian of the atoms involved.

The type of symmetry breaking we find in genetic circuits is more of an
explicit type. The lifting of the base produces a symmetry that is
explicitly broken by external regulators, for instance {\it S} and
{\it R} in a flip-flop.

\section{Hierarchy of broken symmetry circuits: duplication and memory}

\cite{alvarez2024fibration} and \cite{leifer2020circuits} find that
circuits with broken fibration symmetry act as circuits performing
basic logic computations of memory storage in the TRN like toggle-switches and flip-flops in electronics (Fig. \ref{fig:circuits_hierarchy}).  These circuits are identified
by starting with a symmetric circuit, which originates from a fiber
building block, and breaking its symmetry by adding extra edges acting
as regulators that break the symmetry in the fiber structure.  The
breaking of symmetry is the result of a process of gene duplication
that starts with the symmetric base of a fiber building block, which
is duplicated to a (still) symmetric structure. This new structure may
then incorporate new regulations that result in symmetry breaking
of this duplicated circuit, as seen in
Fig.~\ref{fig:circuits_hierarchy}.  Crucially, if the result of a
duplication process respects the lifting property and preserves the
fibration symmetry (Fig.~\ref{fig:circuitsOverview}), this new
 symmetric structure is prone to obtain new regulations that
 result in symmetry breaking.

Strictly speaking, the duplication process of a base does not
necessarily produce a symmetric structure or a structure with the same
fiber classification. For example, if an AR repressor gene is
duplicated into two genes (Fig.~\ref{fig:circuits_hierarchy}, top), a
precise duplication of the auto-repressor should produce a structure
in which the two paralogue genes mutually repress each other as well
as also auto-repress themselves. The original structure corresponds to
a $|n = 1, l=0\rangle$ AR loop, while the new duplicated structure
would in fact correspond to a $|n = 2,l=0\rangle$ binary-tree
instead, which accounts for the two interactions that the new
paralogue genes receive.

However, the initially duplicated structure may undergo further
divergent modifications, like the removal of the self-inhibitions in
this case, which result in a structure that preserves the same
fiber classification since the input relations are the same as in the
original structure.  This means that the duplicated circuit can be
lifted from its AR base, or, analogously, application of the
fibration to the duplicated circuit results in the base.

This leads to a structure with the same dynamics as the base, by
respecting the lifting process. This new structure, when collapsed via
a symmetry fibration, gives the original structure
that was replicated. The further process of modification after
duplication can be thought of as a divergence process of the
duplicated structure. In this case, there are now multiple synchronous
genes where before there was only one. We call this process a {\it
  lifting-driven duplication event.}

This mechanism provides a concise explanation for the presence of
genetic fibrations, but the key point is that breaking the symmetry of  duplicated
circuits gives rise to a series of logic circuits. The process of
duplication as lifting and eventually symmetry breaking also provides
a suitable model for artificially generating an ensemble of networks
with a prescribed number of fibers and broken symmetry logic circuits.

\section{Lifting the AR base: toggle-switch or SR flip-flop circuit}

By duplicating a self-repressing gene in such a way that the lifting
property is respected, the duplicated circuit results in two mutually
repressed genes forming a two-node negatively autoregulated (NAR)
fiber (a fiber that autoregulates itself in an inhibitory manner) as
shown in the first duplication event in the upper part of
Fig.~\ref{fig:circuits_hierarchy}. This duplicated circuit is still 
symmetric. However, its dynamics is not the same as the
negative autoregulated loop in the base, since each gene receives two
inputs in the duplicated circuit, and the base receives only one. So,
this duplication event does not conserve lifting, and the duplicating
genes cannot be a fiber. If the genes further `lose' the
autoregulation, the resulting circuits is a lift of the
base, the fibration is restored, and the dynamics is
conserved.

\begin{figure}[h]
    \centering
    \includegraphics[width=.5\textwidth]{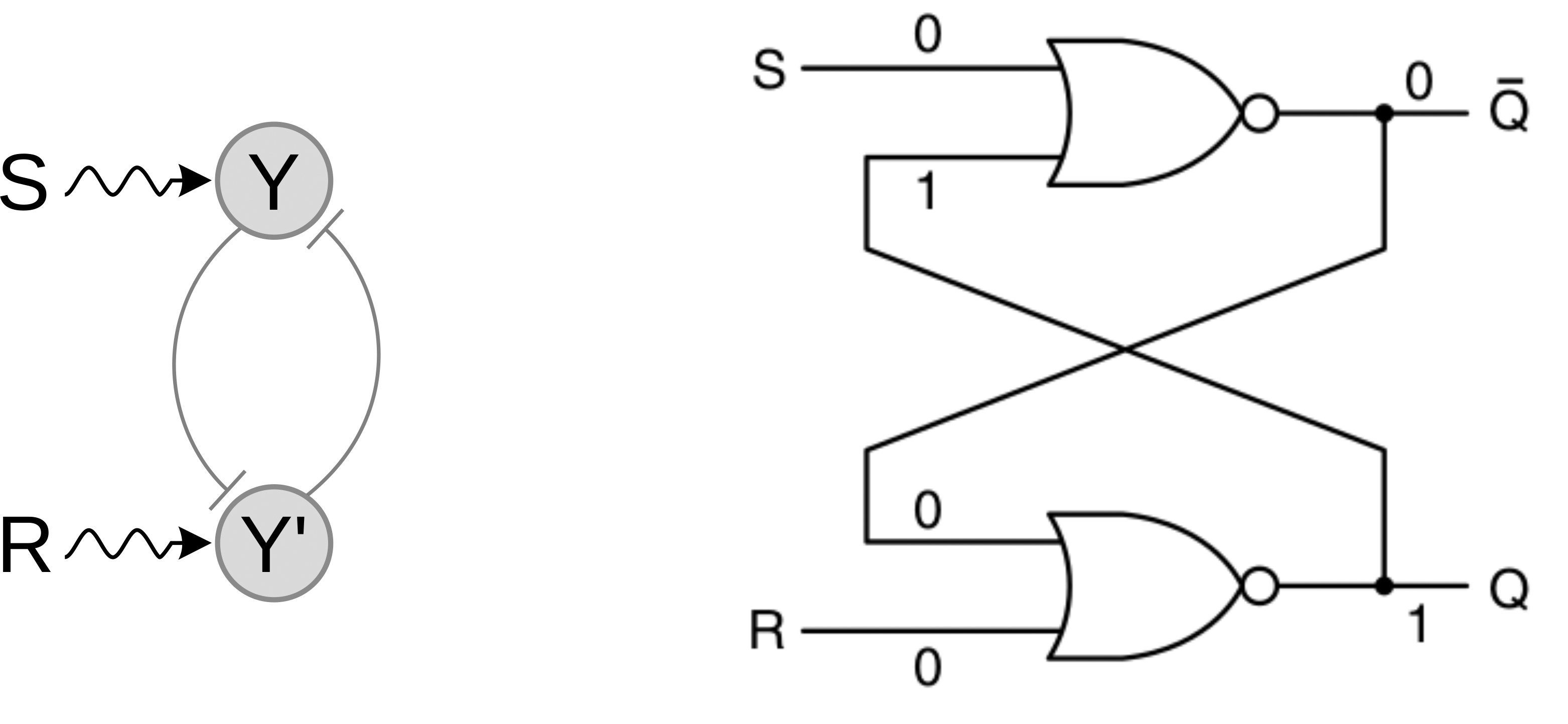}
    \caption{\textbf{SR flip-flop}. In a genetic network, this circuit is
      obtained by lifting the AR base and breaking symmetry with
      two regulators S and R.}
    \label{fig:NAR-osc}
\commentAlt{Figure~\ref{fig:NAR-osc}: 
Left: Nodes Y and Y', with repressor arrows XY, YX. Inputs S and R;
wavy arrows SY, RY'.
Right: Circuit diagram for flip-flop. Two NOR gates. 
Inputs to top gate: S on input line 0; input line 1. Output 0 connects to input line 0 of bottom gate.
Output line Q bar.
Inputs to bottom gate:  input line 0; R on input line 1.Output 1 connects to input line 1 of top gate.
Output line Q.
}
\end{figure}

When different regulators are added to each gene, the symmetry is
broken, and the resulting circuit is the bistable switch or
toggle-switch\index{toggle-switch } shown in Fig.~\ref{fig:circuits_hierarchy} center.  As we have seen, this
circuit is analogous to a flip-flop in
electronics \citep{tanenbaum2016structured}, as shown in Fig.
\ref{fig:NAR-osc} with the different inputs for each gene being the
\emph{set (S)} and \emph{reset (R)} switches.

This circuit is the bistable\index{bistable }
toggle-switch (Section \ref{S:bi_tog}) \citep{gardner2000construction,leifer2020circuits} that
stores one bit of information, since it has
two possible stable and reciprocal states.  One of these stable
  states corresponds to one gene being expressed and the other 
inhibited; its reciprocal case is the other stable
  state, and the genes switch roles. Even if the symmetry is restored and both inputs return
to being identical, the state of the circuit remains unchanged, 
  storing the previous state as one bit of information. It is
possible to switch between the two states by toggling the \emph{S-R}
inputs to the circuit, which is why this  circuit is referred to as a {\em bistable}
two-way switch.

The simplest form of a bistable switch corresponds to this
flip-flop:\index{flip-flop } the two genes inhibit each other. If other forms of
self-regulation are added to this basic structure, the resulting
structure still posses two different stable states, so it 
remains a bistable switch, albeit with different dynamics.

Such a circuit of positive feedback loops constitutes a memory storage
device (Fig. \ref{fig:NAR-osc}).  It consists of two genes mutually
repressing each other (which makes it a double-negative loop). The
toggle-switch exhibits two stable complementary steady states: either
gene Y is expressed and represses gene Y$'$, or vice versa. This  stores one bit of binary information depending
on which of the two states is present. Since these states are
complementary, we are free to label one state as $Q$ (say $Q=0$) and
the other as its logical opposite $\bar{Q}$ ($\bar{Q}=1$).

In the case of the flip-flop, what select the state of the circuit are
the \emph{set (S)} and \emph{reset (R)} switches shown in
Fig. \ref{fig:NAR-osc}. The analogous \emph{set/reset} switches
for the genetic toggle-switch are the broken symmetry inputs: external
regulators of the feedback loop that break their input symmetry
isomorphisms, meaning regulators that do not regulate both $A$ and $B$
equally, as exemplified in Fig. \ref{fig:circuits_hierarchy}. Such unequal external
regulators break symmetry, causing the system to select
one of the two possible states.

When the symmetry is first broken it causes the circuit to express one
of the two genes. The external regulator \emph{`set'} drives the
circuit to express gene Y. However, if this regulator is absent,
making both \emph{S} and \emph{R} absent, the input tree isomorphism
is restored, and the circuit remains in the same state (expressing
Y). This allows the circuit to retain the memory of its
previous state. Thus symmetry breaking of the circuit endows it with
the ability to store memory. The state of the circuit can be toggled
to the other state if instead the \emph{R} external regulator is
present, so that it causes the system to express Y$'$ instead of
Y. Again, if the symmetry is then restored, the circuit still
preserves the memory of its previous state.

The SR flip-flop does not provide synchronized outputs. After the
input signals arrive at the logic gates, each gate provides its output
without waiting for the output of the other one. This results in fast
oscillations which, in the  application to integrated
circuits in digital electronics, are undesired. In digital
electronics the input $S = 0$ and $R=0$ is therefore `forbidden',
since the NAND gates set both $Q=1$ and $\overline{Q}=1$, which
violates the logical state $\overline{Q} =$ {\bf not} $Q$.  

Biological realizations of the AR symmetry breaking class are shown in
Fig. \ref{fig:broken-hierarchy}, both from human regulatory networks
with genes {\it NFKB1} and {\it HOXA9} (upper), and the regulatory
network of genes {\it IRF4} and {\it BCL6} (bottom). Gene {\it NFKB1}
further regulates two genes, {\it BST1} and {\it HAX1} as its outputs,
but this regulation does not affect the functionality of the
flip-flop. The SCCs of \textit{E. coli} are also composed of many flip-flops, more on this in next chapter.

\section{Lifting the FFF base: clocked SR flip-flop circuit}
\index{clocked SR flip-flop }

The process just described corresponds to breaking symmetry in
an $\rvert n = 1, l = 0 \rangle$ circuit with inhibition loops.  An
analogous duplicating-by-lifting symmetry breaking process for the
circuit generated from an FFF building block base, shown at the bottom of Fig.~\ref{fig:circuits_hierarchy}, gives a circuit resembling
a clocked \emph{SR
  flip-flop} \citep{leifer2020circuits,horowitz2015thearts,
  tanenbaum2016structured}, which basically acts as a more complex
flip-flop or toggle-switch.  This process can be repeated with
more complex fiber building blocks, following the hierarchy of
complexity depicted in Fig. \ref{fig:broken-hierarchy}.

As shown in the middle column of Fig. \ref{fig:broken-hierarchy}, the
X and Y genes from the FFF base are lifted to create a circuit
isomorphic to the Clocked SR flip-flop.  The base contains one
external regulator X that is also lifted to X and X$'$ so that the
\emph{S-R} inputs are received by these additional regulators along
with the signal from an extra \emph{`clock'} gene.  The symmetry is
broken by regulators or inducers of genes X and X$'$ acting as S (set)
and R (reset) of memory, and it is restored when S~=~R, leaving the
system `magnetized'.  This new circuit is the clocked SR flip-flop in
which the effect of the \emph{S-R} inputs is now conditional on the
presence or not of the \emph{clock}.

Lifting the FFF adds a second level
of NAND logic gates to the SR flip-flop via gene X. In order to have
consistent logic operations, an input clock gene CLK is added which
symbolizes the activation of gene X, since gene X needs to receive
input for its activation.

The second level flip-flop inverts the outputs of the previous SR
flip-flop logic circuit, meaning that when $S=1$ and $R=0$ ($S=0$ and
$R=1$), the circuit outputs $Q=1$ and $\overline{Q}=0$ ($Q=0$ and
$\overline{Q}=1$). The input $S=0$ and $R=0$ results in an unchanged
state. When the input of gene CLK is low, $CLK = 0$, the output of the
second level of both NAND gates outputs high signals, independently of
the values of S and R, assuring that the outputs $Q$ and
$\overline{Q}$ remain unchanged. However, when the clock input is
$CLK=1$, it allows the first level of NAND gates to change the outputs
for $S\neq R$. The clocked SR flip-flop also has a forbidden state
when $S=1$ and $R=1$ in digital electronics.

In Fig. \ref{fig:broken-hierarchy}, we show two biological
realizations of the Clocked SR flip-flop: the set of
genes $\{${\it CLOCK, NR0B2, NR3C1, E2F1} and {\it TP53}$\}$ and the
set $\{${\it CEBPB, DDIT3, PRDM1, CEBPA} and {\it MYC}$\}$. Interestingly, there is a gene called CLOCK exactly at the clock position in this flip-flop. This flip-flop turns on and off the activity of TP53, a major tumor protein that is mutated in most of human cancers. Both
examples are from human  genetic regulatory networks. The outputs of
the flip-flops, like {\it E2F1} and {\it TP53}, further regulate a set
of genes each as indicated by the red genes in the figure. These circuits are also abundant in \textit{E. coli}, see next chapter.

\section{Lifting the Fibonacci base: JK flip-flop circuit}

Adding now an extra edge feeding information back into the regulator
of the FFF gives a \emph{Fibonacci} building block, described in
detail in Section \ref{sec:ff}, an additional step in the complexity
hierarchy of broken symmetry circuits. As for the FFF broken symmetry
circuit, the lifting of the Fibonacci base 
results in a circuit analog to the \emph{JK
  flip-flop},\index{JK flip-flop } the most widely used of all flip-flop designs
\citep{horowitz2015thearts}.  It is a universal flip-flop since it
can be configured to perform as any other flip-flop.  The JK flip-flop
is basically an SR flip-flop with a longer feedback loop.  In its
symmetric state it acts as a toggle command that changes the output
to a logical current value.

\begin{figure*} 
    \centering
    \includegraphics[width=\textwidth]{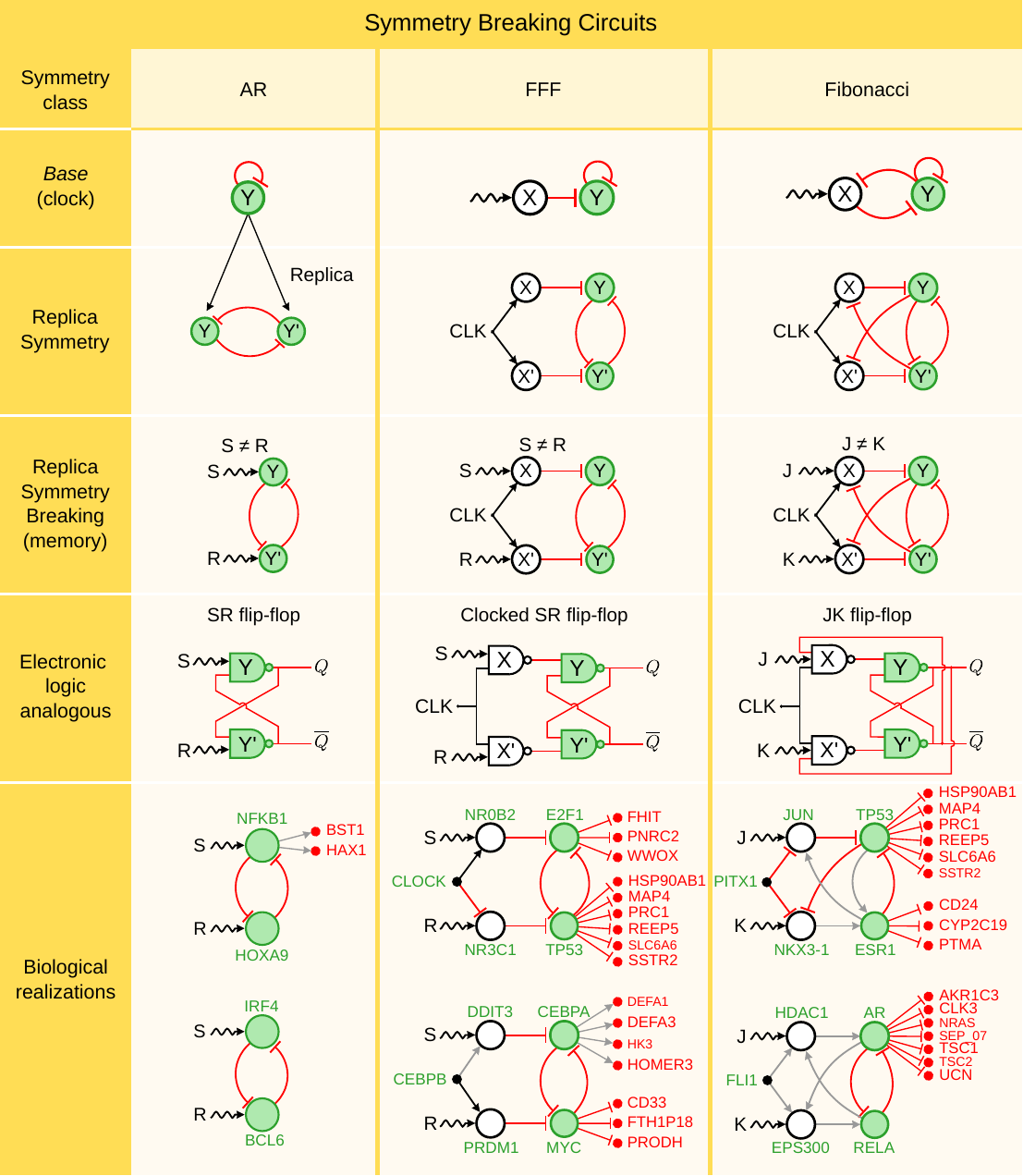}
    \caption{\textbf{Broken symmetry hierarchy.}  {\em First column}:  \textbf{AR}. The lifting-driven duplication of the AR 
      circuit results in a circuit analogous to the SR flip-flop. The symmetry between Y and Y$'$ is broken by the inputs S
      and R, such that $\textrm{S} \neq \textrm{R}$, resulting in a
      pair of logical outputs $Q$ and $\overline{Q} = $ {\bf not}($Q$).
    {\em Second column}: \textbf{FFF}.
      Following the same strategy, we lift the FFF base through 
       duplication. This operation adds a
      second level of logic gates to the SR flip-flop. {\em Third column:} \textbf{Fibonacci}.  The
      lifting-duplication of the Fibonacci base results in a
      logic circuit  analogous to the JK flip-flop.
       Figure reproduced from
      \citep{leifer2020circuits}. Copyright \copyright ~2020, Leifer {\it et al.} }
    \label{fig:broken-hierarchy}
\commentAlt{Figure~\ref{fig:broken-hierarchy}: 
Described in caption/text. No alt-text required.
}
\end{figure*}

In its symmetric state, the JK flip-flop is isomorphic to the
symmetric Fibonacci base. In the symmetry broken phase, it acts as a
memory device which presents two possibilities
(Fig. \ref{fig:broken-hierarchy} third column). A `chiral' symmetry
(where Y feeds X and Y$'$ feeds X$'$) and a `parity' symmetry (left-right
reflection, where Y feeds X$'$ and Y$'$ feeds X). This last one is the one
realized in biological circuits, see Fig.~\ref{fig:broken-hierarchy},
third column. Examples of JK flip-flops are abundant in gene
regulatory networks in humans and mice. 

These more complex circuits were not observed in the bacterial TRN
studied in this work but in more complex species like yeast and
human. Two examples of the Fibonacci broken symmetry circuits are
shown in Fig. \ref{fig:broken-hierarchy} for the sets of genes
$\{${\it PITX1, JUN, NKX3-1, TP53} and {\it ESR1}$\}$ and $\{${\it
  FLI1, HDAC1, EPS300, AR} and {\it RELA}$\}$, both from human genetic
networks. We also show the set of genes regulated by these flip-flops.

      The external regulator genes, J and K, provide inputs
      that are logically processed by the circuit according to the
      type of interaction links between the genes: activators (black
      arrows) or repressors (red flat links).  The outputs of the
      circuits (green genes) regulate the expression levels of other
      genes (in red) without affecting the circuit
      functionality. Gray arrows correspond to interactions with
      unknown functionality. 
      
      These are wonderful examples of the TRN
      computational core. 
Crucial to the dynamics of these circuits are feedback loops between
different genes. Their presence implies that the circuits are always embedded
into the SCCs of the network. It means that we can interpret the SCCs of
the network as the modules where the logical computations are
performed.
This process can be extended to higher complexity fiber building
blocks to obtain ever more complex broken symmetry circuits.

\section{The TRN as a computer}
\index{TRN !as a computer }

This notion of genetic circuits performing basic logic computations
becomes important when trying to understand the TRN in bacteria,
discussed at greater detail in Chapter \ref{chap:minimal}. The TRN can be
understood as a computational machine in charge of computing the
decision-making of the bacterium. \cite{kondev2014bacterial} has put this
idea very clearly: for bacteria (and us), it is all about `To eat or
not to eat sugar'. This binary decision is made by an internal
digital computer embedded in the TRN.

The core driver of the TRN is an ensemble of logic circuits
containing both memory storage devices resembling toggle-switches, and
timekeeping devices in the form of genetic oscillators. This is of
relevance, given that from the perspective of modern computational
devices any computer needs at the most basic level these two modules:
a memory device component and a timekeeping component. This motivates
the question of how we can understand the bacterium's decision-making
process as a logical computational process, and to understand a
bacterium as a computational device. In the particular case of
\emph{E. coli}, we find negative feedback loops as oscillators, and
both toggle-switches and FFF broken symmetry circuits.  However, we
have not found Fibonacci broken symmetry circuits which appear in more complex species like humans. More in-depth
discussion is given in Chapter \ref{chap:minimal}.

In order to fully understand the dynamics of this circuit it is crucial
to understand their external inputs and how are they embedded in or
connected to the rest of the TRN. This complements what has been done
in their implementation and design by the synthetic biology
community. On the one hand, external inputs produce the
symmetry breaking that allows some of the circuits to store memory;
on the other hand, understanding how these circuits are embedded in
the full network, and which functional modules of the TRN communicate
with each other, and how, are central to understanding bacterial
processes as logical computational ones.

\section{Algorithm to find fibration symmetry breaking circuits}
\index{algorithm !broken fibration symmetry }

To find the circuits mentioned above, we run the algorithm developed
by \cite{leifer2020circuits} looking for
the induced subgraphs of the  network whose connectivity is
identical to the logic circuits we are looking for. The algorithm is
explained in more detail in the Supplementary Information of
\citep{leifer2020circuits}. Corresponding codes can be found at
{\small\url{https://github.com/makselab/CircuitFinder}}.

Briefly, the algorithm consists of two steps: (1) Find the subgraphs of
the minimal network isomorphic to a broken symmetry building block,
and (2) Keep only those subgraphs that are induced. Some of the
observed circuits, however, have a slightly different topology due to
self-regulation of some genes in these circuits, so their dynamical
behavior might be different.

\subsection{Software to find broken symmetry circuits}
\label{Sec:AlgoBrokenSymm}

\cite{leifer2020circuits}, supplementary section `Algorithm to find broken
symmetry circuits', explains the algorithm to find broken symmetry
circuits.\index{algorithm !broken fibration symmetry }
The internal dynamics of each node in the network is defined by its
inputs, therefore the internal dynamics of the circuit is defined by
its topology. We believe that external inputs to all the nodes of the
circuit can switch the state of the circuit, but cannot change the
function that it performs. Therefore, the circuit we are looking for
is a circuit that has a given structure, with the possibility of external
inputs. Hence, the required algorithm is one that can find all the
circuits with given topology `embedded' in a network. Mathematically,
this is expressed as the search for an {\em induced subgraph} of a graph, see Definition \ref{induced}.

To count the number of occurrences of broken symmetry circuits in a
given graph we count the number of appearances of induced subgraphs with
the given topology. The problem of finding broken symmetry circuits
consists of three steps:
\begin{itemize}
\item Create a duplicate by lifting the symmetry circuit.
  \item Find subgraphs isomorphic to the replica symmetry
    circuit.
  \item
    Remove subgraphs that are not induced.
\end{itemize}

\begin{figure}[b!]
    \centering
    \includegraphics[width=\textwidth]{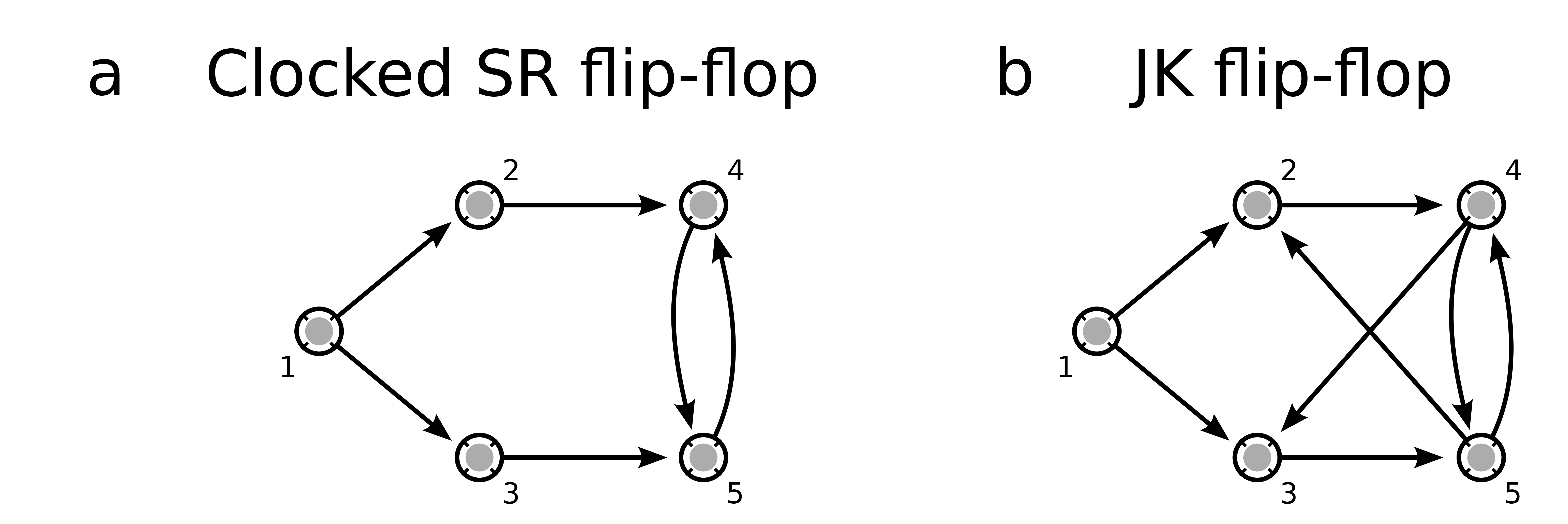}
    \caption{\textbf{Examples of the broken symmetry circuits.} (\textbf{a})
      Clocked SR flip-flop. (\textbf{b}) JK flip-flop.}
    \label{Fig:AlgoBrokenSymm}
\commentAlt{Figure~\ref{Fig:AlgoBrokenSymm}: 
(a) Clocked SR flip-flop. Nodes 1, 2, 3, 4, 5; arrows 12, 13, 24, 35, 45, 54.
(b) JK flip-flop. Nodes 1, 2, 3, 4, 5; arrows 12, 13, 24, 35, 43, 45, 52, 54.
}
\end{figure}

The first step is described in detail in Section \ref{sec:duplication-lifting}.
The second step is to find subgraphs isomorphic to the lifted
symmetry circuit. The general idea is to choose a subgraph and check
whether it is isomorphic to the circuit concerned, and to continue doing this for all
possible subgraphs of the network. However, this straightforward
approach is computationally expensive. For example,
two of the broken symmetry circuits in
Fig.~\ref{Fig:AlgoBrokenSymm} consist of 5 nodes. The 
computational time required to check this circuit is $~N^5$, where $N$
is the total number of nodes in the graph, which means that for big
enough graphs the problem becomes computationally
infeasible. The same problem occurs in the search for motifs.

Different approaches to this problem and applications to graph
matching and pattern recognition have been widely studied and are
 reviewed in \citep{conte2004}. Time costs can be cut if
unprofitable paths are identified and skipped in the search space. One
of the recent works in the field is the VF2 algorithm developed by
\cite{cordella2004}. It is designed to deal with large graphs and uses
state-of-the-art techniques in order to reduce computational time. In
our code, we use the algorithm implemented in a popular R package
igraph \citep{csardi2006} as a function {\em subgraph}$\_${\em isomorphisms(...)}.

The third step is to remove all subgraphs that are not induced or,
simply speaking, have extra links between the genes in the broken
symmetry circuit. Our procedure is: take a node set identified
above, find the induced subgraph of the complete graph with this node
set, and compare the adjacency matrix of the induced subgraph with the
adjacency matrix of the circuit (modulo permutation). If the matrices
are different, then the circuit is removed. All remaining circuits are
the broken symmetry circuits. By agreement, multi-links and self-loops
are removed from the network prior to consideration.

By applying the steps described above, we get the full list of induced
subgraphs that are isomorphic to the given circuit, which we take as
broken symmetry circuits. The implementation of this code  can be
found at {\small\url{https://github.com/makselab/CircuitFinder}}.

A few biological realizations of this symmetry-breaking process are
shown in Fig.~\ref{fig:broken-hierarchy}, last
row. \cite{leifer2020circuits} perform a systematic search for
these circuits using the algorithm above. The full list of symmetry broken
circuits in gene regulatory networks across species appears in the
Supplementary Materials of \citep{leifer2020circuits}.
Table~\ref{asymm_table} shows the $Z$-scores of these circuits
indicating their high significance.

\begin{table*}[ht]
\scriptsize
\centering
      \begin{tabular}{| c | c c c | c c c |}
        \hline
     Species & Database & Nodes & Edges & & SR flip-flop &  \\
& & & & $N_{\rm real}$ & $N_{\rm rand} \pm SD$ & $Z$-score \\
 \hline\hline
        Arabidopsis thaliana & ATRM & 790 & 1431 & 47 & $1.6 \pm 1.2$ & 36.40 \\
\hline
        Micobacterium tuberculosis & Research article & 1624 & 3212 & 6 & $1.7 \pm 1.4$ & 3.20\\
 \hline
 Bacillus subtilis & SubtiWiki & 1717 & 2609 & 3 & $2.1 \pm 1.4$ & 0.6 \\
 \hline
  Escherichia coli & RegulonDB & 879 & 1835 & 14 & $2.1 \pm 1.4$ & 8.40 \\
  \hline
  Salmonella SL1344 & SalmoNet & 1622 & 2852 & 6 & $1.4 \pm 1.2$ & 3.80 \\
  \hline
  Yeast & & & & & 27 & \\
   & YTRP\_regulatory & 3192 & 10947 & 9 & $5 \pm 2.5$ & 1.60 \\
   & YTRP\_binding & 5123 & 38085 & 31 & $21.6 \pm 5.8$ & 1.60 \\
   \hline
        Mouse & TRRUST & 2456 & 7057 & 82 & $4 \pm 2.1$ & 37.70 \\
\hline
        Human & & & & & 192 & \\
        & TRRUST & 2718 & 8215 & 89 & $4.3 \pm 2.1$ & 40.50 \\
     & TRRUST\_2 & 2862 & 9396 & 103 & $5 \pm 2.3$ & 43  \\ 
 \hline\hline

      Species &  & Clocked SR flip-flop & & & JK flip-flop & \\ 
      & $N_{\rm real}$ & $N_{\rm rand} \pm SD$ & $Z$-score & $N_{\rm real}$ & $N_{\rm rand} \pm SD$ & $Z$-score \\
      \hline\hline
      Arabidopsis thaliana &   3 & $0.2 \pm 0.5$ & 5.80 & 2 & $0 \pm 0$ & $>3$ \\ 
 \hline
 Micobacterium tuberculosis & 0 & N/A & N/A & 0 & N/A & N/A \\
 \hline
 Bacillus subtilis & 0 & N/A & N/A & 0 & N/A & N/A \\
 \hline
 Escherichia coli & 3 & $0.3 \pm 0.8$ & 3.30 & 0 & N/A & N/A \\
  \hline
  Salmonella SL1344 &0 & N/A & N/A & 0 & N/A & N/A \\
  \hline
  Yeast && 58 & & & 1 & \\
YTRP\_regulatory  & 3 & $3 \pm 3.6$ & 0 & 0 & N/A & N/A \\
YTRP\_binding  & 192 & $103.3 \pm 45.6$ & 1.90 & 2 & $6.8 \pm 6.1$ & -0.8 \\
 \hline
        Mouse & 216 & $1.9 \pm 2.7$ & 79.50 & 25 & $0.004 \pm 0.06$ & 417 \\
\hline
  Human  & & 566 & & & 90 & \\
TRRUST  & 247 & $3.5 \pm 4.8$ & 50.60 & 45 & $0.02 \pm 0.2$ & 225 \\ 
TRRUST\_2  & 319 & $5.9 \pm 7.2$ & 43.20 & 45 & $0.02 \pm 0.3$ & 150 \\ 
  \hline
  \end{tabular}
  \vspace{10pt}
   \caption{\textbf{Lifting-induced duplication circuits by fibration symmetry breaking are statistically significant over many networks and species.} We report the corresponding $Z$-score statistics as computed in Table~\ref{symm_table} for the symmetry circuits.}
    \label{asymm_table}
\end{table*}


\chapter[Taming Biological Complexity with Symmetries: the Minimal Genome]{\bf\textsf{
Taming Biological Complexity with Symmetries: the Minimal Genome}}
\label{chap:minimal}

\begin{chapterquote}
We 
show that the TRN can be reduced to its minimal components, or `minimal
genome'---an effective minimal base network obtained by the application
of fibrations.  The central maps used for dimensional reduction of
the TRN are two fibrations: the minimal surjective
graph fibration and the inverse injective fibration.  The entire
biological network is distilled into a machine that performs computations over
time.
This computation is executed by a set of flip-flops, which are symmetry-breaking circuits discussed in Chapter \ref{chap:breaking}, along with clocks and synchronized fibers, which are symmetric circuits explored in Chapters \ref{chap:hierarchy_1} and \ref{chap:complex}. 
In this system, part of the network functions as a memory device, while another part operates as synchronized clocks. Consequently, each gene is assigned a specific function within the regulatory machine, allowing the network to be reduced to its minimal functional form.
\end{chapterquote}

\section{Fibration complexity reduction}

The sheer complexity of biological networks can often make it
challenging to decipher the underlying mechanisms that govern their
behavior. Traditional methods for analyzing biological networks, such
as graph theory and differential equations, often result in
large, unwieldy models that are difficult to interpret and
analyze. These models can also suffer from issues such as over-fitting,
where the model is too complex and thus unable to generalize to new
data, or under-fitting, where the model is too simple and thus unable
to capture the nuances of the underlying biology.

To overcome these challenges, researchers have developed a variety of
methods for complexity reduction\index{complexity reduction } in biological networks. Such
techniques aim to simplify the network while retaining its essential
features, making it easier to analyze and understand. These methods
range from simple heuristics such as filtering out low-confidence
interactions, to more sophisticated techniques such as clustering and
dimensional reduction through machine learning techniques.

One of the most interesting applications of our new notion of
symmetry is to the simplification of biological systems, particularly
networked biological systems, by reducing the complexity of
the system while still preserving its key dynamical features. For
instance, describing the set of ODEs underlying gene expression
dynamics in a TRN requires
knowing a large number of microscopic parameters governing gene
input functions. These parameters capture the entire process from
transcription, translation, protein folding to mRNA and protein
degradation, including details such as binding and unbinding to DNA,
ribosome and polymerase binding, mRNA and protein lifetimes, etc. 
\citep{klipp2016book}. This complexity inherent to the large
dimensionality of the multiparameter space is common to all processes
in biology. Finding low-dimensional effective models to describe the
dynamics is crucial for understanding how function and behavior in
biological systems emerge from their complex underlying microscopic
dynamics.

In physical systems, very often we are also confronted with
high-dimensional models in order to fit experimental data. However, in physics,
well defined theoretical models have often been developed, which allow
the definition of theoretical descriptions of complex phenomena with
few or no free parameters to tune, using the underlying symmetries of
the system. For example, the Standard Model\index{Standard Model } of particle physics, with
only (!) 18 parameters \citep{cahn1996eighteen} can describe all
fundamental particle interactions (with the exception of gravity) with
extremely high precision. Renormalization group\index{renormalization group } techniques are used to
`smooth out’ the details of the system, focusing on the effective
`larger picture’ models that arise at critical points (liquid-gases
transition, superconductivity, superfluidity, phases, etc.). These
results led to the idea of universality in critical phenomena where
simple macroscopic properties of the system emerge from their
microscopic parameters while being independent of them. With this, the
dimensionality of complex systems is effectively reduced by the
underlying symmetries of the system, while still preserving the
fundamental properties. This makes symmetry principles the fundamental
cornerstone of most physical laws.

In the biological sciences, however, a {\em systematic} way to perform
dimensional reduction is still lacking, despite a large amount of work
\citep{stephens2011searching}. \cite{alvarez2024fibration} address the
issue of complexity reduction in biology for the TRN of \emph{E. coli}
by reducing the network to its minimal computational apparatus. This
dimensional reduction is derived in two main steps: first collapse
genes with isomorphic input trees using surjective fibrations, which we
detail below; then identify the core driver subnetwork which drives
the dynamics and logic computations of the entire system with
injective fibrations and $k$-core decomposition.

\section{CoReSym: reducing the TRN to its minimal computational core}
\label{Sec:EcoliReduction}

\begin{figure*}[!t] 
    \centering
    \includegraphics[width=.7\textwidth]{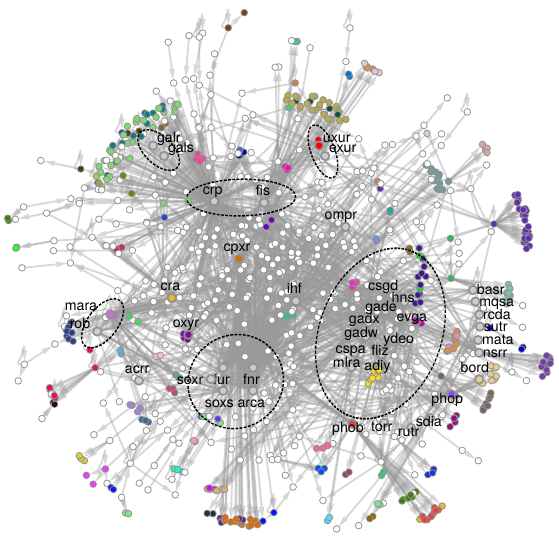}
    \includegraphics[width=.7\textwidth]{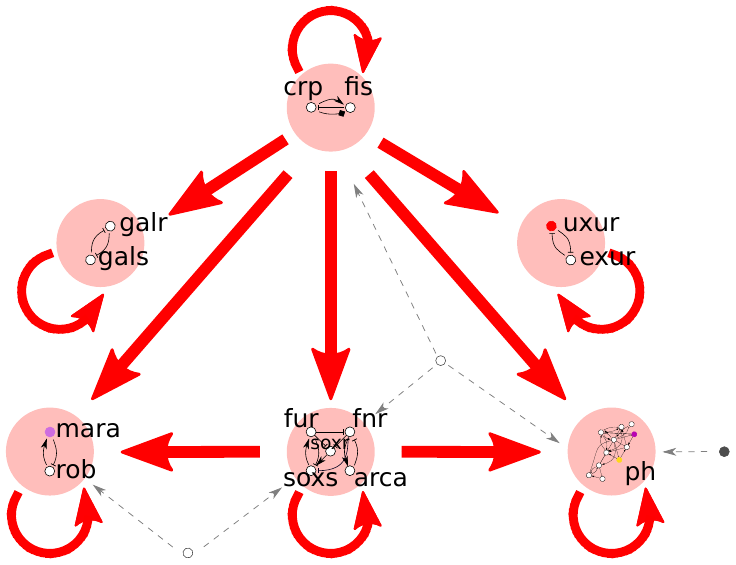}
\caption{ \textbf{The effective minimal TRN obtained through CoReSym}. Top: The weakly connected component of the 879-node
  operon-TRN for \emph{E.~coli}: sizes of nodes and font size are
  proportional to the out-degree of the node.  Bottom: The
  effective minimal TRN obtained after the application of CoReSym,
   illustrating the signal flow between its different components
  (bigger red nodes), which are the strongly connected components (SCCs) of the
  network. Figure reproduced from \citep{alvarez2024fibration}.}
\label{fig:main-reduced}
\commentAlt{Figure~\ref{fig:main-reduced}: 
Illustrative. Described in caption/text. No alt-text required.
}
\end{figure*}

\cite{alvarez2024fibration} introduce a scheme, called Complexity
Reduction by Symmetry or {\it CoReSym},\index{CoReSym } to reduce a gene regulatory
network to its {\it minimal genome} in a way that preserves its
dynamics and uncovers its computational capabilities.
Figure ~\ref{fig:main-reduced} shows the reduction procedure for the
case of \emph{E.~coli}: the TRN on the top is reduced to the
simpler, more comprehensible structure on the bottom, which elucidates
how the network works.

Below we elaborate on how to use the fibration machinery to obtain
such a minimal genome in bacteria. The dimension reduction process
 consists of five steps:

\begin{itemize}
\item Step 1. Collapsing: graph reduction via symmetry fibrations, by
  collapsing all the redundant symmetric pathways. This represents a
  reduction of 70\% of the original \emph{E.~coli} TRN.
\item Step 2. Pruning: reduction via $k$-core decomposition, analogous
  to an inverse injective fibration. This step removes nodes like
  enzymes that do not regulate other genes directly.  This represents
  a further reduction to only 2\% of the original \emph{E.~coli}
  TRN.
\item Step 3. Identification of the Minimal TRN: breaking down the core
  minimal network into strongly connected components (SCC).
\item Step 4. Computational core: Identification of symmetry breaking
  building blocks in the Minimal TRN.
\item Step 5. Simple cycles: Identification of simple cycles in the
  Minimal TRN's SCCs.
\end{itemize}
Computationally, the steps in CoReSym consist in searching for fibration
building blocks \citep{morone2020fibration}, symmetry breaking
building blocks \citep{leifer2020circuits} and cycles in the core
network \citep{purcell2010}.

\begin{figure}[t!] 
    \centering
    \includegraphics[width=.4\linewidth]{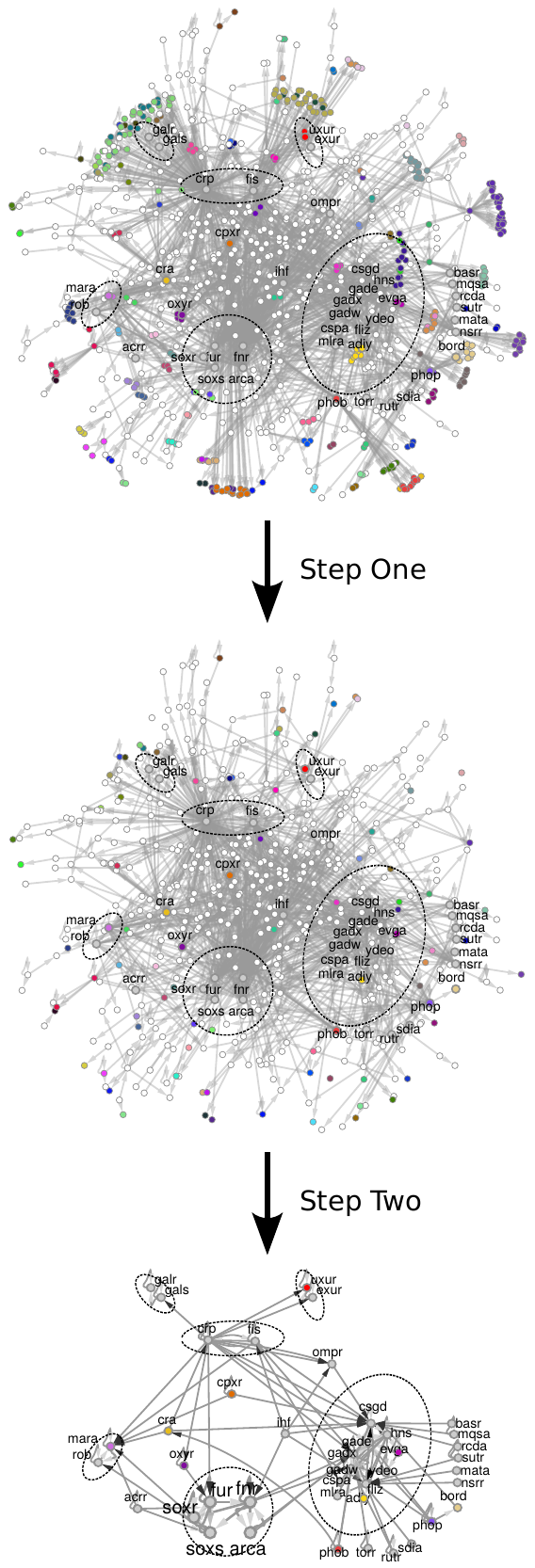}
    \caption{\textbf{Steps of CoReSym.} We start with the full network.
      SCCs are marked by ellipses; names shown are only for the genes
      belonging to the core of the network, obtained after 
      reduction.  Various nodes with notably high out-degree do not belong to the core network. Step 1 of
      our procedure collapses all fibers into one representative node,
      resulting in the base network of the symmetry fibration.  The
      remaining nodes maintain their positions.
      Step 2 removes all null-paths ending at
      nodes with no output via the $k$-core decomposition, resulting
      in the core of the network with only 42 nodes. Figure reproduced
      from \citep{alvarez2024fibration}. }
          \label{fig:steps-reduced}
\commentAlt{Figure~\ref{fig:steps-reduced}: 
Illustrative. Described in caption/text. No alt-text required.
}

\end{figure}

Steps 1 and 2 are concerned with the reduction of the network,
removing all elements in it that do not contribute to its
computational capabilities (Fig. \ref{fig:steps-reduced}). This reduces
the TRN to the core network at the heart of the decision-making
processes: the minimal TRN. Step 1 eliminates all information
pathway redundancies. Step 2 removes the nodes that only
receive signals or pass through it without contributing to the
decision-making process; these nodes are responsible for communicating
the output from the minimal TRN to the periphery. Steps (3)-(5)
analyze this minimal TRN. In step (3) we focus on the large-scale
structure of the minimal TRN: how the components of this core network
are connected to each other. The last two steps `zoom in' to look at
the small-scale, or local, structures within the minimal TRN's
components, by looking at the logical circuits (step 4) and how these
are connected with each other, as well as providing  information on the
connectivity structure within the different components (step 5).

Step 1 and 4 decompose the genetic network into its building
blocks via fibration symmetries and broken symmetries,
respectively. Together, these steps lead to the identification of the
function for every single gene in the TRN as belonging to
three general classes of genes:

\begin{itemize}
\item  (1) A set of synchronized symmetric fibers following the hierarchies of Chapters \ref{chap:hierarchy_1} and \ref{chap:complex} 
\item  (2) Regulators of the SCCs
\item  (3) Broken fibration symmetry circuits within the SCCs following the hierarchy of Chapter
\ref{chap:breaking}.
These
  can be further classified into:
  
\quad(3.1) memory devices (flip-flops or toggle-switches): broken
      symmetries of the AR, FFF, Fibonacci and $n=2$ fibers.

  \quad(3.2) oscillators.  
\end{itemize}

The synchronized nodes obtained by symmetry fibration analysis are characterized by five basic type of fibers, Fig.~\ref{fig:circuitsOverview}:

\begin{itemize}
\item (1) Trivial fibers made of regulons, operons, etc., with no
  loops within the fibers; they are $n=0$ fully
  feed-forward. They make up 67\% of the building blocks in
  \emph{E.~coli}.
\item (2) Feed-forward fibers making up 26.4\% of \emph{E.~coli}'s building
  blocks. They are subdivided in sub-classes according to the
  number of external regulators:

    \quad (2.1) 0-FFF, 14\%,
    
    \quad (2.2) 1-FFF, 9\%, 
    
    \quad (2.3) 2-FFF, 3\%.

\item (3) Fibonacci fibers $\varphi$-FF with fractal dimensions
  $\varphi = 1.61, 1.31,$ and 1.15, representing 3.3\% of the building
  blocks. 
\item (4) $n=2$ fiber (2.2\%). 
\item (5) Composite multilayer fibers: a composition made of the previous ones
  (1.1\%).
\end{itemize}

These building blocks are arranged in a large-scale
structure made of SCCs with a rich cycle structure, assembled into an
effective feed-forward structure that we call the {\em minimal TRN}. This
structure contrasts with previous representations of the bacterial TRN
as a purely feed-forward network  \citep{ma2004hierarchical,
  martinez2008functional,dobrin2004aggregation,shen2002network}.

\begin{table*}[h!]
  \centering
  \begin{tabular}{|c|c|}
    \hline
    Name     & Description \\
    \hline
    \hline
    Full Network & 879 \\
    In Fibers & 416\\
    \hline
    Base     &  555\\
    \hline
    Minimal ($1-$(out)core)    & 42\\
    In SCCs & 24\\
    1st type regulators & 11\\
    2nd type regulators & 4\\
    3rd type regulators & 2\\
    \hline
  \end{tabular}
  \vspace{10pt}
   \caption{\textbf{Count of genes in the dimensional reduction of
     \emph{E. coli} TRN.} Table reproduced from
     \citep{alvarez2024fibration}.}
  \label{tab:eco}
\end{table*}

We use the TRN of \emph{E. coli}\index{E. coli @{\em E. coli} } from RegulonDB\index{RegulonDB }
\citep{regulon2016} and we treat it under the idealized assumptions
stated before in Sections \ref{sec:introfib}. Most important for this
purpose is the `uniform broadcast' idealization explained in detail in
Section \ref{ideal}: for genes within a fiber, a) their input
functions depend only on the transcription factor and not on the
target gene's binding site, which is mostly satisfied, and b) that
their constants for degradation and maximal synthesis rates are equal
across the fiber's genes. As explained in
Section \ref{sec:operons}, we take an operon to be a single
node in the TRN, which removes this set of trivial fibers from the
analysis. This means that our initial global TRN is made up of 879
nodes, see Table \ref{tab:eco}.

\section{Symmetry fibrations reduce the network yet preserve information flow}

By considering `message passing' between genes by means of transcription
factors in the TRN, we understand this network as an
information-processing network.\index{network !information-processing } We simplify this network using
fibration symmetries.  In terms of the flow of information, fibers
define redundant pathways of information, since all genes in a fiber receive exactly
the same information. Collapsing the fibers then reduces the
network considerably, while still preserving the dynamics in the TRN,
given that we are eliminating only redundancies in the flow of
information without eliminating any information pathway. The
information flow is the feature we wish to preserve, in order to
understand the decision-making processes of bacteria. The maximum
reduction corresponds to the symmetry fibration that reduces the network to its minimal base, a part of the minimal genome. This step
accounts for a great reduction in the number of genes present, given
that out of the 879 genes, 416 of them belong to 92 fibers, see Table
\ref{tab:eco}.

\section{Finding the driving core}

The second step in the reduction finds the core of the network that is
responsible for driving the dynamics of the global TRN. This
corresponds to performing an inverse injective fibration.

\subsection{Injective fibrations}

\begin{figure*}[!ht]
    \centering
    \includegraphics[width=.9\textwidth]{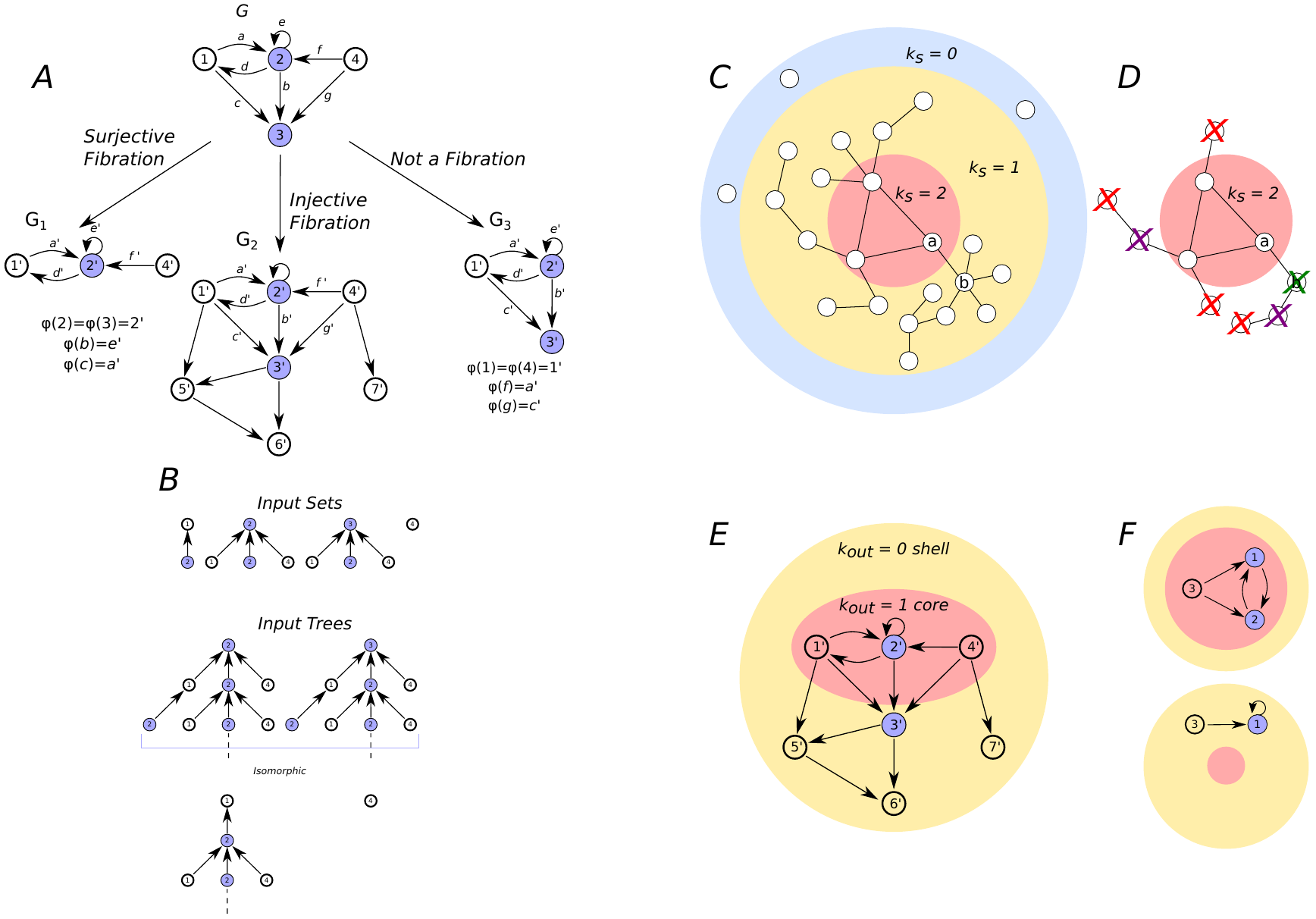}
    \caption{\textbf{Injective fibration and $k$-core decomposition.}
      (\textbf{a}) Graph $G$, a subgraph of the TRN of \emph{E.~coli},
      shows a Fibonacci building block. The map from graph $G$ to
      graph $G_1$, on the left, corresponds to a surjective fibration:
      all nodes with isomorphic input trees are collapsed to one
      (nodes $2$ and $3$ collapsed to $2'$), and all input trees are
      preserved. The map from graph $G$ to graph $G_2$ is an injective
      fibration: all input trees are preserved and  extra
      nodes are included. The map from the graph $G$ to graph $G_3$, which
      maps node $4$ to $1'$ does not correspond to a fibration, since the input tree of $4$ (seen in (\textbf{b})) is not
      preserved in its image node $1'$ in graph $G_3$. The same
      problem occurs with the images of nodes $2$ and $3$ ($2'$ and
      $3'$ respectively). (\textbf{b}) Input sets and input trees of
      nodes in graph $G$.  (\textbf{c}) Schematic drawing of the
      $k$-core decomposition of an undirected network. Even though
      node $b$ has a higher degree than node $a$, it is connected to
      nodes with smaller degree and has therefore a smaller coreness
      than node $a$. (\textbf{d}) Example of how to obtain the $k = 2$
      core of the network in (\textbf{c}). First, the nodes with
      degree less than $2$ are removed. The remaining nodes are those
       shown; successive iterations remove the nodes crossed with
      colored \emph{X}. Different colors stand for successive
      iterations: red is the first iteration, the second is purple,
      and the last is green.  (\textbf{e}) Example of the $k$-core
      decomposition of graph C from (\textbf{a}). Even though node
      $5'$ on the outer $k_{\rm out} = 0$ shell (in red) has one
      output, once nodes $6'$ and $7'$ in the shell are removed, it
      will then be left with no output, and will be removed as
      well. All the remaining nodes in the $k_{\rm out} = 1$ core have
      at least one output after performing this process. (\textbf{f}) The
      bottom network corresponds to the symmetry fibration of the
      network at the top. The SCC of nodes $1$ and $2$ is also a fiber, and
      after collapsing the fiber the SCC is lost, so the collapsed
      network becomes an acyclic graph, which does not have a $k_{\rm
        out} = 1$ core. Figure reproduced from
      \citep{alvarez2024fibration}.}
    \label{fig:fibration}
\commentAlt{Figure~\ref{fig:fibration}: 
Illustrative. Described in caption/text. No alt-text required.
}
\end{figure*}

An injective fibration\index{fibration !injective } formalizes the intuitive notion that
under certain conditions the dynamics of an entire system may be
driven by only a subset of its constitutive elements. This is
formalized by \cite[Lemma 5.2.1]{lerman2015b}, where
it is shown that for an injective (one-to-one) fibration $\varphi: G
\rightarrow G_2$ the dynamics of the bigger graph $G_2$ is driven by
the dynamics of the smaller graph $G$. In fact $G$ is (isomorphic to) a subgraph of
$G_2$. This can be seen in Fig.~\ref{fig:fibration} for the injective
fibration from $G$ to $G_2$.

The emphasis here is on the injective nature of the map $\varphi$. For
a mapping to a bigger graph to be a fibration, like the one shown in
Fig.~\ref{fig:fibration}, it must satisfy the lifting property. This
requires that each edge in $G_2$ whose target node is an image
from a node in $G$ (i.e. $1',2',3',4'$), can be lifted
to a unique edge in $G$. This implies that no new edges are allowed to
target any of the nodes of the original graph (i.e. $1',2',3',4'$),
satisfying the lifting property.

As a consequence, all the added nodes in $G_2$ (i.e. $5', 6', 7'$)
must strictly be targets only of the image nodes. Hence signals flow
only outwards from $G$, and so the dynamical state of the outer nodes
is driven by the dynamics of the original smaller graph $G$. In other
words, the subgraph $G$ of $G_2$ drives or controls the entire
dynamics of $G_2$.  This, in turn, guarantees that the dynamics of the
original graph is preserved in the image graph $G_2$.

\cite{lerman2015b} show that when a fibration $\varphi: G
\rightarrow B$ is injective, the domain $G$ corresponds to the
subnetwork of the system driving the dynamics of the entire system
$B$, the codomain. Since $\varphi$ is a fibration,
all nodes $i'$ in $B$ that
are not image nodes must be targets of image nodes and must
not send information back to $G$; so the information flows only
outwards of $G$. In our case, the problem we face is going in the
opposite direction to the map: from the entire system we want to
find the subsystem responsible for its dynamics, meaning we want to
construct an inverse of the injective fibration.

\subsection{$k$-core decomposition}

Therefore we need to determine the nodes that only receive
information, or that only pass information along, since these are
clearly not responsible for driving the dynamics. This is done by the
{\em $k$-core decomposition}\index{k-core decomposition @$k$-core decomposition } of the network  \citep{kitsak2010identification}. This decomposition `peels off’ layers, formally called {\em
  $k$-shells},\index{k-shell @$k$-shell } from the network, Fig. \ref{fig:fibration}c, d.  This is
achieved by assigning a {\em coreness index}\index{coreness index } $k_s$ to each node,
corresponding to the respective shell each node belongs to. The
$k$-shell corresponds to the set of nodes with coreness $k_s$.

The smaller $k_s$ is, the more peripheral the node is, as can be seen in
Fig.~\ref{fig:fibration}c.  The coreness of a node captures the degree
of the nodes it is connected to. For example, node $b$ in
Fig.~\ref{fig:fibration}c has a smaller coreness ($k_s = 1$) than node
$a$ ($k_s = 2$), even though $b$ has a higher degree of $5$ compared
to $a$'s degree of $3$. This is because $b$ is connected mostly to
nodes with degree $1$, in contrast to $a$, which connects to nodes
with higher degree, making it more
\emph{influential} \citep{kitsak2010identification} or central.

\begin{definition} {\bf $k$-core and $k$-shell of the network.}
The {\em $k$-core}\index{k-core @$k$-core } of a network is the maximal induced subgraph, subject to the condition that the
number of edges of every node within the $k$-core is at least $k$. The
coreness of each node is then the $k$-shell\index{k-shell @$k$-shell }  number $k_s$ for which the node
belongs to the $k_s$-core but not to the $(k_s + 1)$-core
(Fig. \ref{fig:fibration}c, d).
\label{kcore}
\end{definition}

Since the TRN corresponds to a directed network, every node has an
in-degree\index{in-degree } and an out-degree.\index{out-degree } However, the out-degree is not relevant here,
since the network formalism we are using 
represents all variables involved in the dynamics of a node by 
the source nodes of its
input set.
Thus we iteratively eliminate the nodes with no out-degree,
which corresponds to the $k_{\rm out} = 0$ shell. This means that
all paths that end at a node with no output are removed,
so what remains are the paths that loop back to some previous node in
the path. Because of this, all the strongly connected components of the
network are still present. The result is the $k_{\rm out}= 1$ core of the
network.

The order in which these two steps are applied
matters. If we find the $k_{\rm out} = 1$ core before eliminating
redundancies in information pathways, the resulting network might
still have redundant pathways that are not actually computing anything
relevant, only transmitting information outward. These pathways should
not be part of the core driver of the network. In particular,
given a 2-node SCC whose nodes are also symmetric (belong to the same
fiber), if both nodes have out-degree at least 1 they 
still belong to the $k_{\rm out} = 1$ core. However, if we apply the
symmetry fibration first, these nodes collapse to a single node with
only a self-loop, meaning that they may no longer be part of the
$k_{\rm out}= 1$ core.  By definition the 1-shell of the TRN is made of
enzymes and other proteins except for TFs. A TF cannot be part of the
out-going 1-shell, since a TF always regulates at least one other
gene, so it has out-degree $1$ or more.

Most of the 1-shell then comprises enzymes that catalyze metabolic
reactions. When we consider a TRN, these enzymes send information to
the metabolic network. This information can then return to the TRN
since metabolites and small molecules can bind to the TF to activate
or inhibit their activity. We do not consider this important type of
regulation here, so the outgoing 1-shell can be removed from the
network for now. An integrated biological network should include this loop of information by considering couplings between the TRN and the
metabolic networks, as well as the protein-protein interaction
network.

After this decomposition, the network is reduced from 555 nodes to
only 42 (see table \ref{tab:eco}), which correspond to the driver core
of the TRN. The dynamics of the global TRN then propagates outward
from this minimal TRN.  Table~\ref{tab:reduction2} shows how the number
of nodes decreases at each step of CoReSym. Table~\ref{tab:breakdown}
shows the coverage.

\begin{table*}[b]
  \centering
  \begin{tabular}{lrr}
    \toprule
    \textbf{Reduction step}     & 
    \emph{Gene count} & \textbf{\emph{\%}}\\
    \midrule
    Step 0.0: Full Genome RegulonDB & 4,690 & -- \\
    \hline
    Step 0.1: TRN (non isolated TFs)   & 1,843 & 100\%\\
    Step 0.2: operon-collapsed TRN    & 879 & 48\%\\
    \hline
    Step 1: Base-TRN (collapsing fibers) & 555 & 30\%\\
    Step 2: Minimal TRN (after trivial pruning) & 42 & 2\%\\
    \bottomrule
  \end{tabular}
  \vspace{10pt}
  \caption{\textbf{Reduction count and fiber coverage.} The full \emph{E.~coli}
    genome contains 4,690 genes, of which 1,843 genes express TFs
    that are not isolated, i.e. they are TFs with at least one regulation
    to or from another TF. The first reduction of operons is
    performed by a trivial fibration that collapses the operons
    that are trivial fibers. This is in reality just a part of Step 1
    and reduces the network to the nontrivial TRN with 879 TFs, which
    is then used as the initial step of the nontrivial reduction
    process. The fibration reduces the network to 30\%, and after
    removing the trivial dangling ends the final base is reduced to
    2\% of the original size, constituting the minimal TRN. Table
    reproduced from \citep{alvarez2024fibration}.}
  \label{tab:reduction2}
\end{table*}

\begin{table*}[b]
  \centering
  \begin{tabular}{lcc}
    \toprule
    \textbf{operon-TRN breakdown}     &  
    \emph{Gene count}& \textbf{\emph{\%}}\\
    \midrule
    operon-TRN  & 879 & 100\%\\
    \hline
    Nodes in fibers    & 416 & 47.3\%\\
    $k_{\rm out}$ shell (not in fibers) & 431 & 49\%\\
    Nodes in SCCs (not in fibers) & 20 & 2.3\%\\
    Connectors (not in fibers) & 12 & 1.4\%\\
    \bottomrule
  \end{tabular}
  \vspace{10pt}
  \caption{\textbf{Breakdown of the operon-TRN network}. Step 1 collapses
    the 416 nodes in fibers into just 92 fibers, to give the
    Collapsed-fibers 555-node network. Step 2, removes all the Shell
    nodes and 82 of the fiber-collapsed nodes,  leaving only the 42
    nodes in the Minimal TRN of \emph{E.~coli}. Table
    reproduced from \citep{alvarez2024fibration}.}
  \label{tab:breakdown}
\end{table*}

\section{Structure and composition of the minimal TRN: a simple computer}

The next step in the reduction of the TRN is to calculate the
strongly connected components.\index{strongly connected component }

The overall structure of the reduced TRN, the computational minimal
core of \emph{E. coli} TRN corresponds to a feed-forward structure
between the SCCs\index{SCC } (with autoregulation loops within them), with the
information flowing outward from the \emph{crp-fis} SCC, the most
central master regulators genes in \emph{E. coli}, illustrated in
Fig. \ref{fig:main-reduced} bottom. In addition to this
feed-forward structure, the reduced TRN is completed by regulator
nodes that connect the SCCs to each other, and also feed information
directly into them. We obtain 6 SCCs (Fig. \ref{fig:surjmin}, three of
them already discussed in Section \ref{sec:fibration-analysis}); one
with 11 nodes which is in charge mostly of the \emph{pH} and stress
response of the cell (pH SCC); one with 5 nodes (the \emph{soxS} SCC);
and the rest with only two nodes (\emph{crp-fis, marA-rob, galR-galS}
and \emph{uxuR-exuR}). The \emph{crp-fis} SCC works as a type of
master regulator, controlling the rest of the SCCs and being regulated
by only two external outputs: the \emph{cra-fiber} and the
\emph{ihfAB} TF. The reduced TRN can also be understood as an ensemble
of simple genetic logic circuits that perform basic logical
computations, like those explained in the previous section.

\subsection{The minimal TRN}

The minimal TRN network is defined as follows:
\begin{definition}
    \textbf{Minimal TRN network} \citep{alvarez2024fibration}. The {\em minimal TRN}\index{minimal TRN } is an
    effective network in which each SCC is a super-node;
    that is, a subnetwork that can be considered to be
    a node in some related network. A directed
    edge between two super-nodes means that there is at least one
    directed edge between a node of the corresponding SCC and a node in the SCC at the target of the edge. (This is a standard construction in graph theory, called the {\em condensation} of the graph.)
    \label{def:minimal-trn}
    \end{definition}

\begin{figure*}[!ht] 
  \centering
  \includegraphics[width=.85\linewidth]{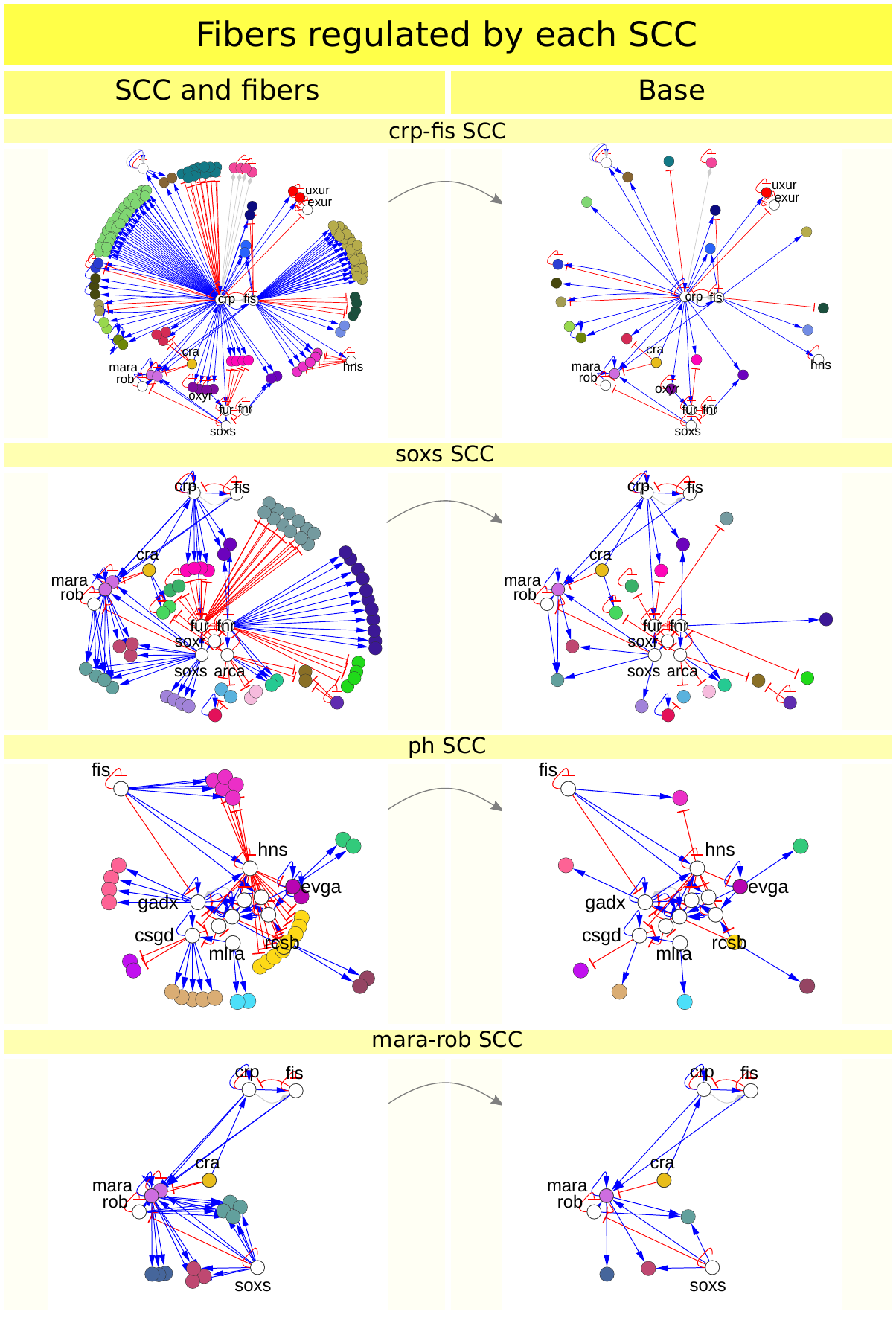}
  \caption{ \textbf{Symmetry fibrations applied to the main SCCs in the {\it
        E. coli} TRN.}  (Left) Genes in the same fiber share
    the same color. (Right) All fibers have been collapsed.
    Labeled nodes belong to SCCs or regulators to them.  Blue edges
    with arrows represent activation, red edges with bars represent
    inhibition, and gray edges with rhombuses represent dual regulation
    (dual means that the TF can act as a activator or repressor
    depending on how it is activated by ligands). Figure reproduced
    from \citep{alvarez2024fibration}. }
    \label{fig:surjmin}
\commentAlt{Figure~\ref{fig:surjmin}: 
Illustrative. Described in caption/text. No alt-text required.
}
\end{figure*}

Figure \ref{fig:surjmin} shows the application of the fibration to each of the SCCs  (see also
Figs. \ref{fig:componenta}, \ref{fig:componentb} and
\ref{fig:componentc}) and the minimal TRN of {\it E. coli} is shown in Fig. \ref{fig:main-reduced} (bottom).

The minimal TRN is a feed-forward
structure without cycles, only loops at the super-nodes. At the root is the carbon SCC,
which works as a type of master regulator, controlling the rest of the
SCCs. All these SCCs are regulated by different genes. For instance
the \emph{cra}-fiber and \emph{ihfAB} regulates the carbon SCC.

The importance of the carbon \emph{crp-fis} SCC is exemplified by its
central position in the minimal network.  This two-node SCC also
regulates the biggest fibers, since the main function of the bacterial
cell is to process sugars.  Also of interest is that there are some
fibers that are jointly regulated by two SCCs, and that the
\emph{uxuR-exuR} and \emph{galR-galS} do not regulate any fibers of
their own, although they do regulate other genes and operons.

The \emph{galR-galS} SCC and the \emph{uxuR-exuR} SCC receive signals
from \emph{crp-fis}, but do not receive or send signals to the rest of
the SCCs. They compute their state solely based on \emph{crp-fis}
SCC's input and then send their corresponding outputs to the genes
that they regulate. The \emph{crp-fis} pair acts as a toggle-switch
that turns on and off, depending on environmental clues.

The other SCCs are arranged in the shape of a feed-forward motif:
\emph{crp-fis} SCC feeding the \emph{soxS} SCC and the \emph{pH} SCC,
with the \emph{pH} SCC also receiving from the \emph{soxS} SCC and
another similar structure with the \emph{marA-rob} SCC in the same
role as the \emph{pH} SCC: receiving from both \emph{crp-fis} and
\emph{soxS}. The overall structure of the computational minimal core
of the \emph{E.~coli} TRN corresponds to two feed-forward
structures between the SCCs.

Inside each SCC there is a bunch of toggle-switch\index{toggle-switch } circuits which can
be seen as broken symmetry circuits from fibration symmetries that
provide the network with its computational capacity. Most of the
oscillators and synchronized fibers, which
are purely symmetric circuits, are controlled by these switches.

\begin{figure*}[!t] 
    \centering
    \includegraphics[width=.65\linewidth]{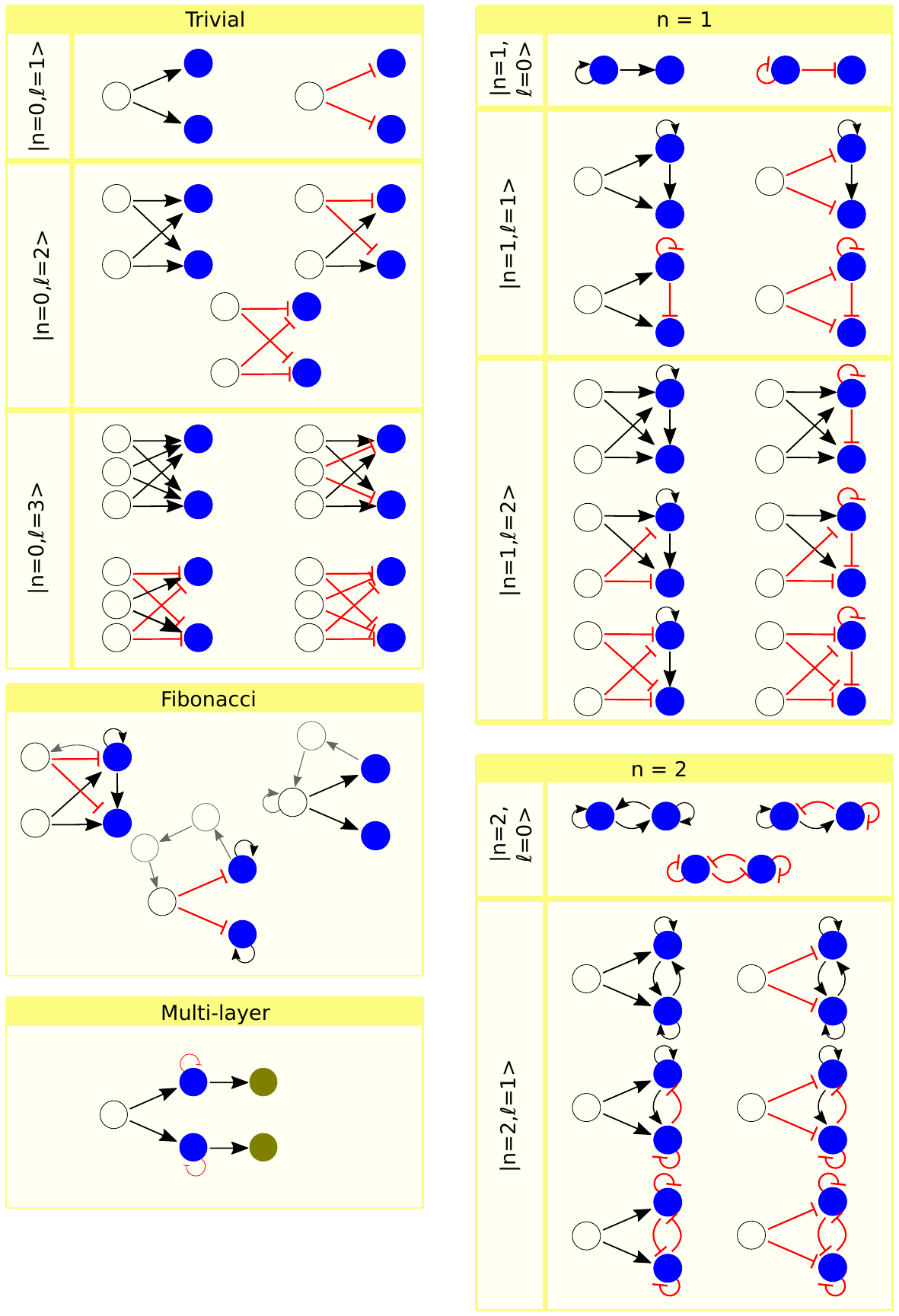}
    \caption{\textbf{Possible circuits of the observed building block
        structures from \emph{E.~coli} and \emph{B. subtilis}}. These circuits are all the
      possibilities for building blocks that can be used to assemble
      a computational TRN from the bottom up.  We take the observed $|n,
      l\rangle$ classes in both bacteria, and show all the possible
      combinations of activation and inhibition regulations for each
      class, i.e. all possible configurations that would still have a
      symmetric pair of nodes. Figure reproduced from
      \citep{alvarez2024fibration}.}
    \label{fig:poss-fibs}
\commentAlt{Figure~\ref{fig:poss-fibs}: 
Illustrative. Described in caption/text. No alt-text required.
}
\end{figure*}

Figure~\ref{fig:poss-fibs} shows all the different possible
configurations of building blocks observed in both
\emph{E.~coli} and \emph{B.~subtilis} by \cite{alvarez2024fibration}.
A computational TRN can be now assembled from such a circuits.

\subsection{Toggle-switches (flip-flops) in the TRN}
\index{toggle-switch }

\begin{figure*}[t!]
    \centering
    \includegraphics[width=.72\linewidth]{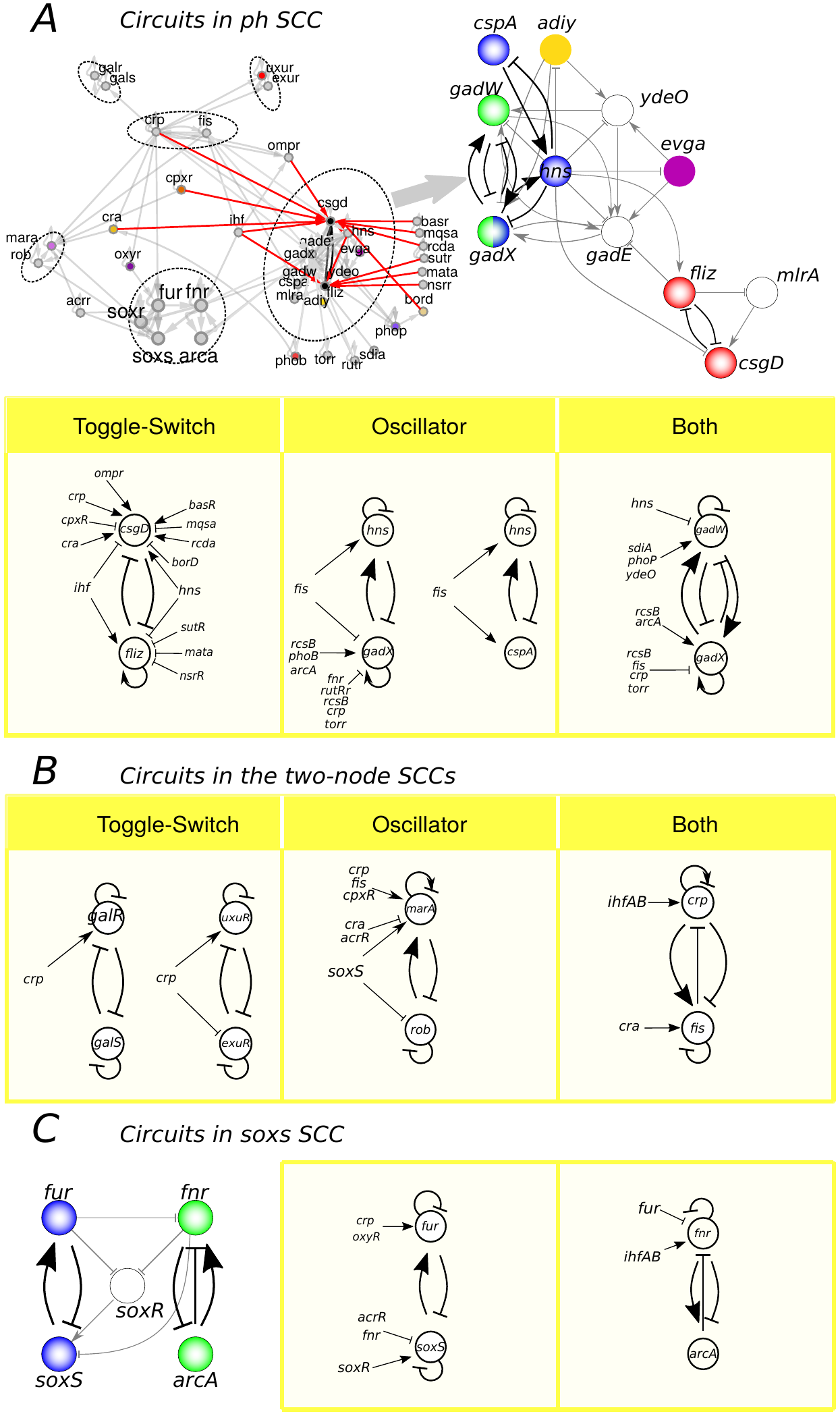}
    \caption{\textbf{Electronic circuits found in \emph{E.~coli} TRN.}
      (\textbf{a}) The largest SCC in charge of mostly pH responses,
      and the electronic circuits embedded in it. \emph{E.~coli}'s
      minimal TRN is shown with red links for the symmetry breaking
      inputs to the toggle-switch \emph{fliz-csgd}. (\textbf{b})
      Electronic circuits forming two-node SCCs. (\textbf{c})
      Electronic circuits in the \emph{soxS} SCC. Nodes involved in
      toggle-switches are colored in red in the SCC depictions,
      oscillating nodes in blue, and nodes in circuits that can be
      either in green. For every circuit, the incoming signals that
      break the symmetry are shown. Figure reproduced from
      \citep{alvarez2024fibration}.}
    \label{fig:all_circuits}
\commentAlt{Figure~\ref{fig:all_circuits}: 
Illustrative. Described in caption/text. No alt-text required.
}
\end{figure*}

Inside the SCCs of the TRN we found a variety of genetic circuits with
broken symmetries. In
total 12 different pairs of genes were found to be involved in a
number of  logic circuits; see Fig.~\ref{fig:all_circuits} for
a depiction of the full symmetry breaking for all electronic circuits
in~\emph{E.~coli}.

The \emph{galR-galS} and the \emph{uxuR-exuR} SCCs correspond to
circuits almost identical to toggle-switches 
except for their self-inhibitory loops which compute their state based
on the \emph{crp-fis} SCC’s state, and then send this information
outward to the fibers that they regulate. The \emph{uxuR-exuR} SCC
forms a nicely regulated `almost toggle-switch': when \emph{crp} is
active it seems to induce the SCC to the state of \emph{uxuR} being
expressed, and a very similar mechanism applies to the
\emph{galR-galS} SCC, with \emph{galR} being active.  Both of these
circuits present additional negative autoregulations in each gene
(see Fig.~\ref{fig:all_circuits}), so their dynamics varies compared
to the classic toggle-switch design, but their switching function
remains the same.

The pair \emph{uxuR-exuR} cycle with the regulations from \emph{crp}
is the \emph{Mutual Repression Network with Negative Autoregulation}
studied by \cite{hasan2019improvement}. This circuit can show two
distinct stable states, and therefore serves as a memory; when
\emph{crp} is active, it can induce a state in which \emph{uxuR} is
expressed while \emph{exuR} is repressed.

The third toggle-switch-like circuit\index{toggle-switch } is between
\emph{csgD-fliz} in the \emph{ph} SCC, with numerous possible ways for its
symmetry breaking to occur. This circuit contains a positive
autoregulation. As shown in \citep{leon2016computational}, this allows
for two stable states, making it possible for it to function as a
memory device.

The remaining 3 SCCs are arranged in the effective minimal TRN
(Fig. \ref{fig:main-reduced}) as an FFF-like effective building block
as an overarching structure: \emph{crp-fis} SCC feeding the
\emph{soxS} SCC and the \emph{pH} SCC, with the \emph{pH} SCC also
receiving from the \emph{soxS} SCC and another similar structure with
the \emph{marA-rob} SCC in the same role as the \emph{pH} SCC,
receiving from both the \emph{crp-fis} and the \emph{soxS} SCC. If we add
an effective AR loop to each SCC, symbolizing that the
information also travels inside the SCC, the result somewhat resembles an FFF
at the global level of the whole network.

Overall, we observe six circuits almost identical to toggle-switches,\index{toggle-switch } three
being an SCC (\emph{galR-galS},
\emph{uxuR-exuR} and \emph{crp-fis}), while the other three form part of the
\emph{pH} SCC (\emph{fliz-csgD, fliz-gadW} and
\emph{gadW-gadX}). Based on their inputs, \emph{galR-galS} and
\emph{uxuR-exuR} may possibly develop only one state: \emph{galS} and
\emph{uxuR} active, respectively. The remaining ones seem 
capable of both stable states, given the numerous possible 
ways for their symmetry to break. Most of these states would be only
transient, because of self-inhibitory loops for most genes, with the
exception of the states where \emph{crp}, \emph{csgD} or \emph{gadX}
are active in their respective circuits, given their self-activation.

\subsection{Oscillators (clocks) in the TRN}
\index{clock }

For oscillator-type circuits we observe eight circuits:
\emph{crp$\mapsto$fis, rob$\mapsto$marA, soxS$\mapsto$fur,
  fnr$\mapsto$arcA, gadX$\mapsto$hns, cspA$\mapsto$hns,
  gadW$\mapsto$gadX, gadX$\mapsto$gadW}. Most of the observed
oscillators are known to have damped oscillations; however,
\emph{gadX$\mapsto$gadW, csgD$\mapsto$gadW} and \emph{crp$\mapsto$fis}
may have the same topology as the more robust Smolen oscillator.

Four of them are purely negative feedback loops (NFBL, oscillatory
type): \emph{rob $\mapsto$ marA} a SCC by itself; \emph{soxS $\mapsto$
  fur} in the \emph{soxS} SCC; and \emph{gadX $\mapsto$ hns} and
\emph{cspA $\mapsto$ hns} in the \emph{ph} SCC.  All of these are
autoregulated, but autoregulation in \emph{gadX~$\mapsto$~hns}
(Fig.~\ref{fig:steps-reduced}) makes it the Smolen oscillator,
a more robust type of oscillator \citep{hasty2008}.

\subsection{Additional electronic circuits: toggle-switch and clock-type}

There are also three pairs of nodes that can exhibit various behaviors,
given that they can send various types of regulation messages between
them.  For example, \emph{crp} can send either an activation or
repression signal to \emph{fis}, which means that \emph{crp-fis} can
be an mutually repressed circuit (toggle-switch type)\index{toggle-switch }
 or an NFBL circuit (oscillating
type), possibly even a Smolen oscillator\index{Smolen oscillator }
 given its autoregulations. A similar situation occurs for \emph{fnr-arcA} in the \emph{soxS}
SCC, which can be either mutually repressed or Smolen. Lastly, \emph{gadW-gadX} in the \emph{ph}
SCC, in which both genes send both activating and repressing signals,
can be a mutually repressed, an NFBL, or a positive autoregulation (PAR) feedback loop such as in a lock-on
circuit. On top of this, one of the possible NFBL configurations
includes a Smolen oscillator.

\begin{figure}
    \centering
    \includegraphics[width=0.7\linewidth]{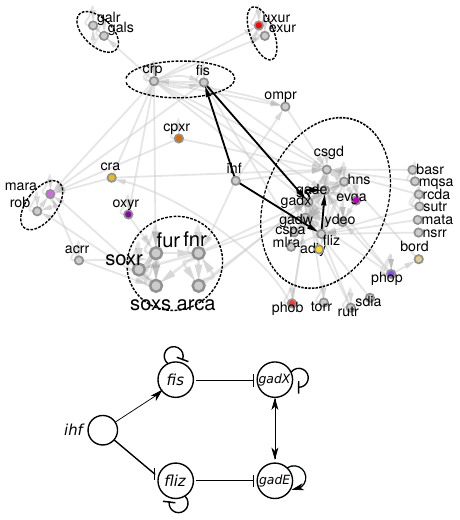}
    \caption{\textbf{FFF electronic circuits}. {\em Top}: These circuits connect the main SCC
      \emph{crp-fis} to the \emph{pH}-SCC. {\em Bottom}: The
      isolated circuit. Figure reproduced from
      \citep{alvarez2024fibration}.}
    \label{fig:fff-circuits}
\commentAlt{Figure~\ref{fig:fff-circuits}: 
Top: Illustrative subset of genes from E.coli, forming network.
Bottom: One circuit isolated from this network. Five nodes 
ihf, fis, gadX, fliz, gadE. Activator arrows ihf-fis, gadX-gadE, gadE-gadE, gadE-gadX.
Repressor arrows fis-fis, fis-gadX, fliz-fliz, flz-gadE, gadX-gadX, gadE-gadE.
}
\end{figure}

There are also three FFF\index{FFF } circuits, shown in
Fig.~\ref{fig:fff-circuits}, connecting to the \emph{pH} SCC: one to
the master regulator \emph{crp-fis} SCC, and two others that connect through
different paths to the \emph{soxS} SCC. Remarkably, the three FFFs
are regulated by the same clock: transcription factor \emph{ihfAB}. For the three FFF circuits, the underlying
feedback loop is a double-positive autoregulation (double-PAR) feedback
loop, which works as a bistable lock-on circuit, which is less
dynamic.

Only one \emph{lock-on} feedback loop is present, and it is part
of the FFF type circuit. Table \ref{tab:circuits} shows the rich
variety of electronic circuits found in {\it E. coli}, as well as in
another well-studied bacterium {\it B. subtilis}, as calculated in
\citep{alvarez2024fibration}.

\begin{table*}[h]
\small\sf\centering
  \begin{tabular}{lll}
    \toprule
  \bf  Circuit type &   \bf \emph{E.~coli}  &  \bf \emph{B. subtilis} \\
    \midrule
    Toggle-switch type & \emph{galS-galR}, \emph{uxuR-exuR}, & \emph{lexa-rocr}, \emph{glnr-tnra}\\
    & \emph{csgD-fliz} \\
    \hline
    Oscillator type & \emph{rob $\mapsto$ marA}, \emph{soxS $\mapsto$ fur}, & \emph{sigk $\mapsto$ gere}, \\
    & \emph{cspA $\mapsto$ hns}, \emph{gadX $\mapsto$ hns}, & \emph{siga $\mapsto$ spo0a} \\
    \hline
    Lock-on types & & \emph{sigf-sigg}, \emph{siga-sigh},\\ & & \emph{siga-sigd}, \emph{sigd-swra} \\
    \hline
    Capable of various types & \emph{crp-fis}, \emph{gadW-gadX},  \\
     & \emph{fnr-arcA} \\
    \hline
    FFF type & \emph{ihf}$\mapsto$ \{\emph{fis,fliz}\} $\mapsto$ \emph{gadX-gadE} \\
     & \emph{ihf}$\mapsto$ \{\emph{fnr,fliz}\} $\mapsto$ \emph{gadX-gadE} \\
     & \emph{ihf}$\mapsto$ \{\emph{fnr,fliz}\} $\mapsto$ \emph{gadW-gadE} \\
    \bottomrule
  \end{tabular}
  \vspace{10pt}
  \caption{\textbf{List of gene circuits in \emph{E.~coli} and
      \emph{B. subtilis}}. The circuits found in \emph{E.~coli} are
    described in detail in Fig.~\ref{fig:all_circuits}. Table
    reproduced from \citep{alvarez2024fibration}. }
  \label{tab:circuits}
\end{table*}

\subsection{Structure of cycles}

Cycles\index{cycle } are very important, as we have seen before. They are a way to create
longer memory\index{memory } between logic circuits,\index{logic circuit } and correspond to
different ways of regulating them.

By definition, an SCC\index{SCC } is basically a complicated arrangement of
feedback loops between its constituent nodes. As such, we want to probe
its structure by studying the independent simple cycles present in the
minimal TRN. Each of these cycles represents longer-term memory
where signals loop around, as well as being responsible for the
interconnectedness of the logic circuits.

\begin{definition} {\bf Simple cycle.}
\label{def:simple_cycle}
  A {\it simple cycle} is a cycle such that every node is traversed
  only once, and the list of independent cycles corresponds to those
  that cannot be constructed from other cycles.
\end{definition}

SCCs can be thought of as `information vortices'\index{information vortex } where the
signals can cycle. This is represented as the AR loop in every SCC in
the effective minimal TRN in Fig. \ref{fig:main-reduced}. These vortices
are relevant because they are constituted from feedback loops, without
which the TRN's computational capacity would be drastically reduced to
a `combinatorial' nature only. Indeed, feedback loops are what allow
for the more complex dynamics of the logic circuits, and this is why they are
embedded in the SCCs. The SCCs, being information vortices, correspond to a very intricate cobweb of feedback loops.

To study this phenomenon, \cite{alvarez2024fibration} look for all the independent
simple directed cycles in the network. They look for a list of the independent simple
cycles, the {\it cycle base}, since all other cycles can be constructed as a
sum of these, given that they span the cycle vector space
 \citep{kavitha2009cycle, gruber2012digraph}.  This is related to the
concept of Betti numbers\index{Betti number } in simplicial
homology\index{simplicial homology } \citep{munkres2018elements}, where the $k$-th Betti number
is the number of $k$-dimensional holes in a topological space. A cycle in
this sense describes a type of topological hole. In particular, for undirected graphs the first Betti number $b_0 = |CC|$, where
$|CC|$ is the number of connected components, while the
next Betti number $b_1 = |E| - |N| - |CC|$ is the number of
independent simple cycles \citep{berge2001theory}.

To construct this cycle base, cycles must be allowed
to traverse a directed edge in the opposite direction, represented by a
negative contribution. For example, in a feed-forward loop motif we
can construct a cycle by traversing at least one of the edges in
a `negative' direction  \citep{kavitha2009cycle}. This does not make
biological sense, however, so we work with the `incomplete' base of
cycles allowed only by existing pathways in the network, whose
combinations span only the biologically existing pathways.

Under this mathematical restriction, in directed graphs, determining
the number of cycles is an $NP$-complete
problem \citep{gruber2012digraph}. However, the networks studied are small
enough to be analyzed using state-of-the-art techniques, producing output
in a matter of minutes. The  algorithm employed is loosely based on
\emph{Johnson's algorithm}\index{Johnson's algorithm } \citep{johnson1977efficient}. It starts by
breaking the network into SCCs and searches for cycles within each. Each component is analyzed by  (i) enumerating the nodes in
  the SCC,  (ii) choosing one node (the initial/final node of the
  cycle), (iii) looking for the outgoing neighbors of this node,
  and (iv) looking for all the simple paths (paths without
  repeating nodes) back from each neighbor to the initial/final node.

\begin{figure}[t!] 
    \centering
    \includegraphics[width=.7\linewidth]{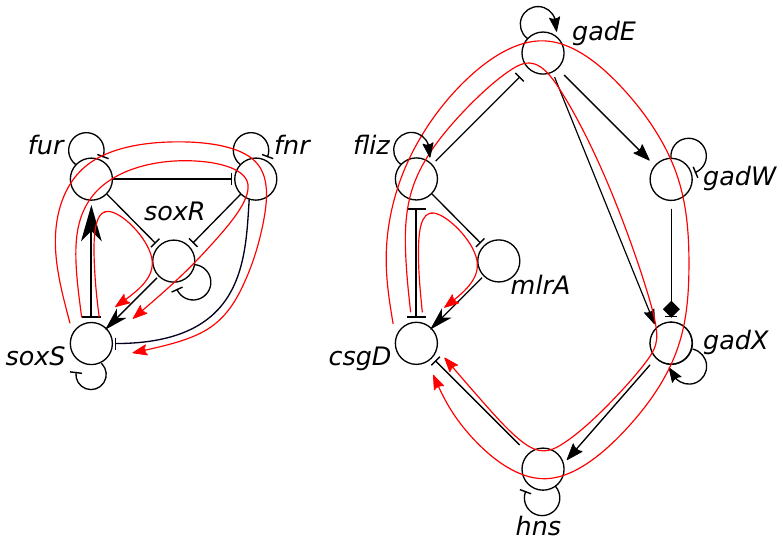}
    \caption{\textbf{Example of simple directed cycles in
        \emph{E.~coli}.} All are different cycles that cross through
      the logic circuits \emph{soxS}-\emph{fur} on top and
      \emph{csgD-fliz} on the bottom. Figure reproduced from
      \citep{alvarez2024fibration}.}
    \label{fig:cycles}
\commentAlt{Figure~\ref{fig:cycles}: 
Left: Network with nodes fur, fnr, soxR, soxS. Arrows showing three cycles
(details in text). 
Right: Network with nodes gdE, fliz, gadW, mlrA, csgD, gadX, hns.
Arrows showing three cycles (details in text). 
}
\end{figure}

In total, we find 41 simply directed cycles in the minimal core of the
\emph{E.~coli} TRN (Fig. \ref{fig:cycles}).  Only one does not
cross any logic  circuit, which supports the
point that connectivity between the different cycles is an
important feature of these networks.

Four of them are the two-node SCCs, five are located in the \emph{soxS}
SCC, and the remaining 32 in the \emph{ph} SCC. Each cycle
contains a pair of nodes in a logic circuit, so all cycles
longer than two nodes pass through two nodes that are also connected by
a logic circuit. All logic circuits are, of course, two-node
cycles.

For example, in the case of the \emph{soxS} SCC we observe 2 two-node
cycles (circuits \emph{soxS $\mapsto$ fur} and \emph{fnr-arcA}). The
remaining three cycles in this component all pass through
\emph{soxS~$\mapsto$~fur}, and as such can be considered to be longer loops
from \emph{soxS} to \emph{fur}, as in
Fig.~\ref{fig:cycles}. There are numerous cases where the
cycles even cross through multiple nodes that are connected by a logic
circuit ; this can also be seen in
Fig. \ref{fig:cycles} where the longer loops cross
through \emph{gadE-gadW}, \emph{gadW-gadX}, and \emph{gadX-hns}, all of
which are parts of different logic circuits. In a way, all the loops
shown can be considered to be loops of various lengths between the circuits
\emph{soxS $\mapsto$ fur} and \emph{csgD$-$fliz}. A full list of all
the cycles and code is provided at the repository
{\small\url{https://github.com/makselab/MinimalTRNCodes}}.  In this analysis
we ignore loops since they connect a node to itself in a cycle of
length 1, so they do not really contribute to the computational machinery.

\begin{figure}[t!] 
    \centering
    \includegraphics[width=.5\linewidth]{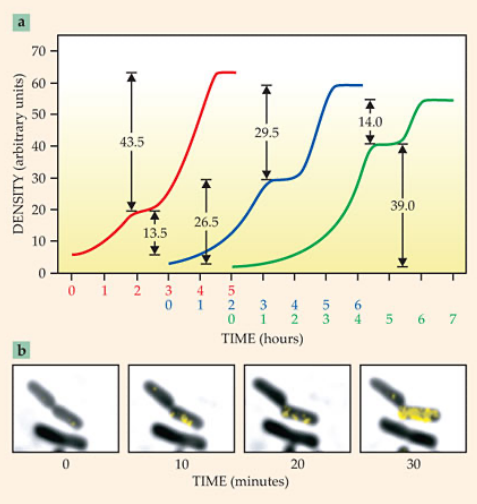}
    \caption{\textbf{Bacterial decision making in action.} {\bf (a)} Growth of {\it
        E. coli} bacterial culture in an environment of glucose
      and lactose mixed in proportions 1:3, 1:1, and 3:1 (curves from
      left to right, respectively). The observed amount of growth in
      each phase (see the plateau) is proportional to the amount of
      glucose and lactose in the cell's environment. Figure reproduced
      from \citep{mullerhill1996}. {\bf (b)} In an environment with
      lactose, cells switch to a cellular state that can digest
      lactose as indicated by the appearance of yellow-labeled lactose
      permease from left to right. Figure reproduced from
      \citep{choi2008astochastic}. The bacterial cell controls these
      states like a computer' using the flip-flop formed by the {\it
        crp-fis} genes. They are the master regulators of the sugar
      utilization circuits catabolizing the different sugars. Figure
      reproduced from \citep{kondev2014bacterial} with the permission of AIP Publishing.}
    \label{fig:kondev}
\commentAlt{Figure~\ref{fig:kondev}: 
Described in caption/text. No alt-text required.
}
\end{figure}

\section{Assembling the TRN from its building blocks}

Recall the statement in \citep{kondev2014bacterial} that the most important decision
for bacteria (and most humans) is `To eat or not to eat sugar'
(Fig. \ref{fig:kondev}). Crucial to understanding these
decision-making routines are the embedded logical gene circuits in the
TRN with their respective inputs. As explained in Chapter
\ref{chap:breaking}, it is symmetry breaking that allows these circuits
to make binary decisions---as for the toggle-switch,
whose inputs break symmetry and force the circuit into one of two
possible states. Besides this, it is necessary to understand how these
circuits are connected and embedded in real networks, to figure out how
these circuits are utilized in living cells and how their dynamics
differ. This approach complements what has been done so far by the synthetic
biology community, implementing such circuits in vivo and working with
them independently of the rest of the TRN.

In Section \ref{sec:introfib} we discussed how to conceive of such a
networked system as some form of von Neumann cellular automaton:\index{von Neumann cellular automaton } every
node can be thought of as a finite-state machine, which has a defined
state at every moment, and whose future state is determined
predictably by its current state and the inputs from its
neighbors. We have also shown that this network is composed in part
of (binary) memory devices in the form of toggle-switches and
timekeeping devices, or synchronized clocks in the form of genetic
circuit oscillators. Therefore this network can be interpreted as
 a von Neumann cellular automaton, hence as a simple type of
computer.
From a modern perspective any computer, at the
most basic level, is composed of memory devices and timekeeping
devices \citep{horowitz2015thearts, tanenbaum2016structured}. Therefore
the reduced TRN, hence the global TRN, can be regarded as
a simple logical computer. Indeed, Sidney Brenner\index{Brenner, Sidney } has argued that
cells are good examples of Turing\index{Turing machine } and von Neumann machines.\index{von Neumann machine } This
motivates the question of how these circuits are integrated into real
genetic networks, and how understanding this would help to illuminate
the process of decision-making by the cell, viewed as a computational
process.

\begin{figure}[t!]
    \centering \includegraphics[width=.6\textwidth]{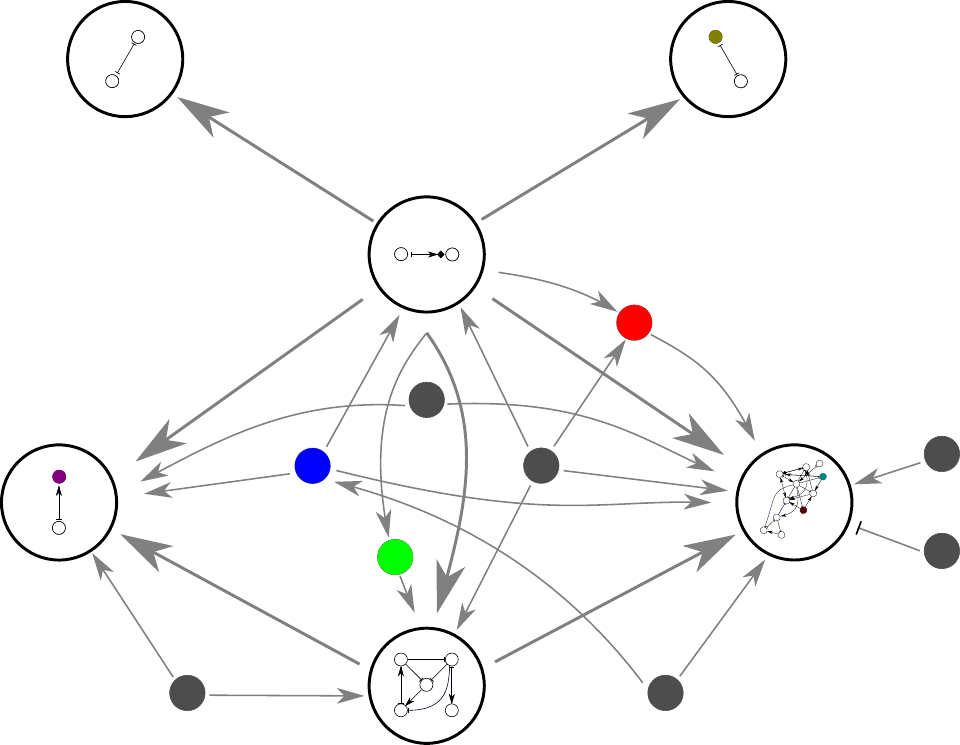}
    \caption{\textbf{Effective representation of the minimal TRN of {\em
          E. coli}}.  The structure is feed-forward with the carbon
      SCC at the center, with autoregulation loops inside the SCCs
      and single gene regulators that regulate 1, 2 and 3 SCCs.}
    \label{fig:rep}
\commentAlt{Figure~\ref{fig:rep}: 
Described in caption/text. No alt-text required.
}
\end{figure}

We can understand the bacterium as a simple logical computer: the cell
takes signals from the surrounding environment as inputs, and computes a
proper response as an output (a change in the expression levels of
genes) which then propagates outward from the core driver network.

Ultimately, every single gene in the bacterial
TRN is classified as belonging to one of the following classes (Fig. \ref{fig:rep}):

\begin{itemize}
  \item $\rvert r, l\rangle$: a synchronized symmetric fiber
    characterized by a building block made of either $r=n=0, 1, 2$
    loops and $l$ `external regulators', or a fractal dimension $r$
    of Fibonacci building blocks, and multilayer composites of these
    basic circuits.

\item Combinatorial logic circuits belonging to a few strongly connected
  components made of broken fibration symmetries, which play the roles
  of digital electronic circuits for memory storage flip-flops or
  oscillatory clocks.

\item A series of external 1-, 2-, and 3-regulators connecting the
  SCCs of the structure.
  
  \end{itemize}


\chapter[From Structure to Function: Cluster Synchronization in Genetic Networks]{\bf\textsf{From Structure to Function: Cluster Synchronization in Genetic Networks}}
\label{chap:synchronization}

\begin{chapterquote}
What does the existence of fibrations mean from a biological
standpoint? In genetics, fibration synchrony means coexpression. In
this chapter we consider experimental expression profiles predicted
from gene fibers.  We show that gene expression levels of the genes
inside each fiber synchronize in the manner predicted by fibration
theory. Genetic synchronization provides the functional relation, and
symmetry is the link between the structure of the underlying TRN and
its function via synchronization of its constitutive units. This
closes the gap from structure to function.
\end{chapterquote}

\section{Structure $\rightsquigarrow$ function in genetic networks}
\index{structure-function relation }

Systems biology has developed the capacity to look at the activity of
the entire set of proteins of an organism, such as the $100,000+$
proteins expressed by the $\sim$20,000 human genes. This expression
can be monitored simultaneously for all genes using transcriptomic
techniques like microarrays or RNA-seq.  When the patterns of gene
expression are observed through correlation functions, it is found
that the genes separate into groups. These groups are sets
of genes with a similar activity or gene
coexpression,\index{coexpression } in a statistical sense, implying cluster synchronization
\citep{klipp2016book,aguiar2022}.

The subject of this chapter is whether this form of synchronization
can be predicted from fibrations\index{fibration } in the underlying TRN. Gene fibrations
provide a way to ensure gene coexpression by a specific network
wiring, without necessarily placing the genes in the same
transcription unit (i.e., under the control of the same promoter), and
even without necessarily placing the genes in the same regulon\index{regulon } under
the control of the same transcription factor.

\begin{figure}
  \centering
  \includegraphics[width=\linewidth]{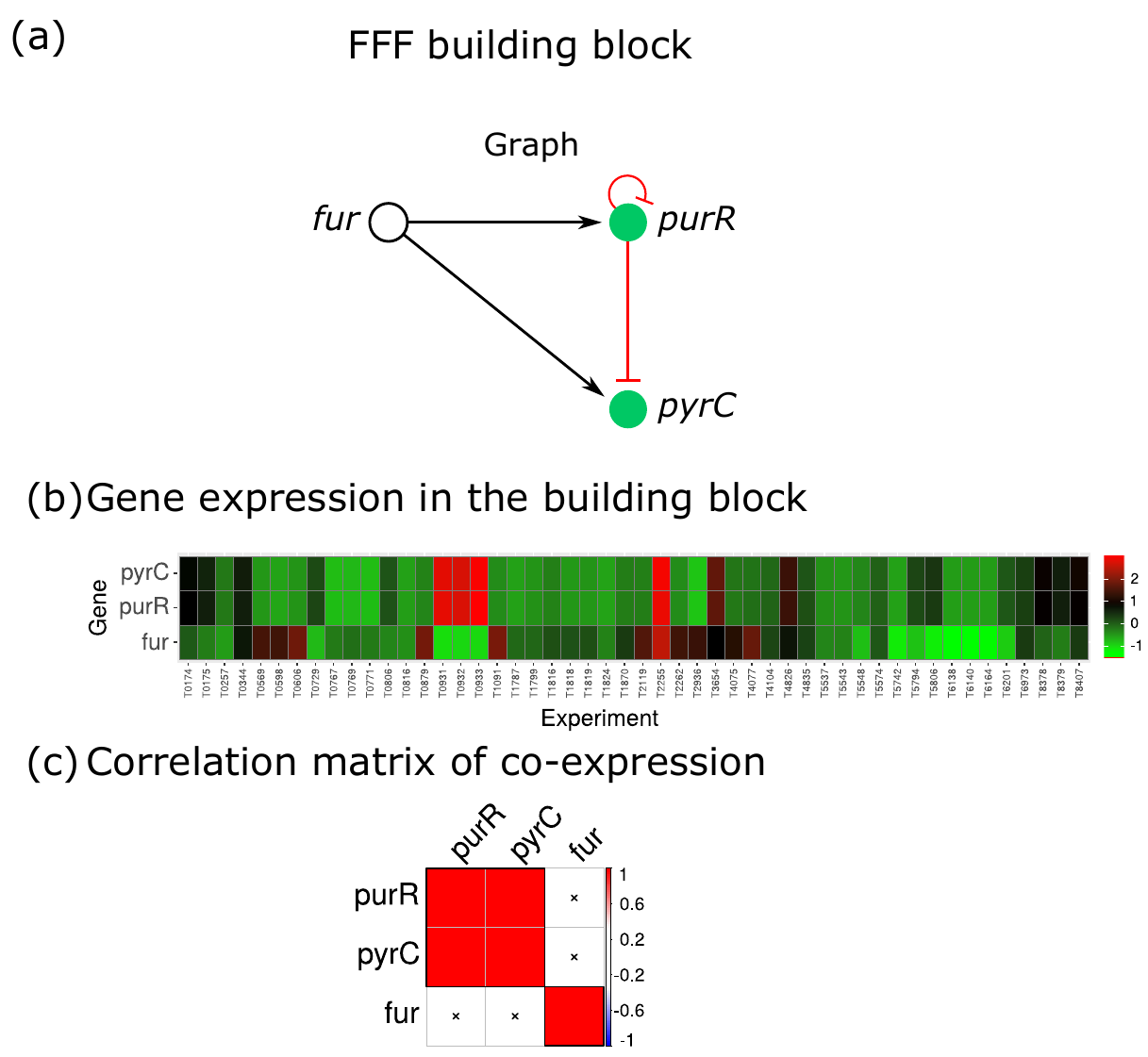}
  \caption{\textbf{Heat map and correlation matrix of an FFF in {\it
        E. coli}}.  Genes are rows, and experiments are columns. The
    color bar indicates the differential expression level in AU
    with respect to the wild-type condition of a given gene and
    experiment.}
  \label{fig:heatmap-fff}
\commentAlt{Figure~\ref{fig:heatmap-fff}: 
Top: FFF building block. Graph with nodes fur, purR, pyrC.
Activator arrows fur-purR, fur-pyrC. Repressor arrows purR-purR, purR-pyrC.
Middle: Heat map of gene expression.
Bottom: Correlation matrix of co-expression splits into 3x3 blocks, rows and columns labeled
by fur, purR, pyrC. High correlations between purR and pyrC and between
fur and itself. Other correlations close to zero.
}
\end{figure}

\begin{figure}
  \centering
    \includegraphics[width=\linewidth]{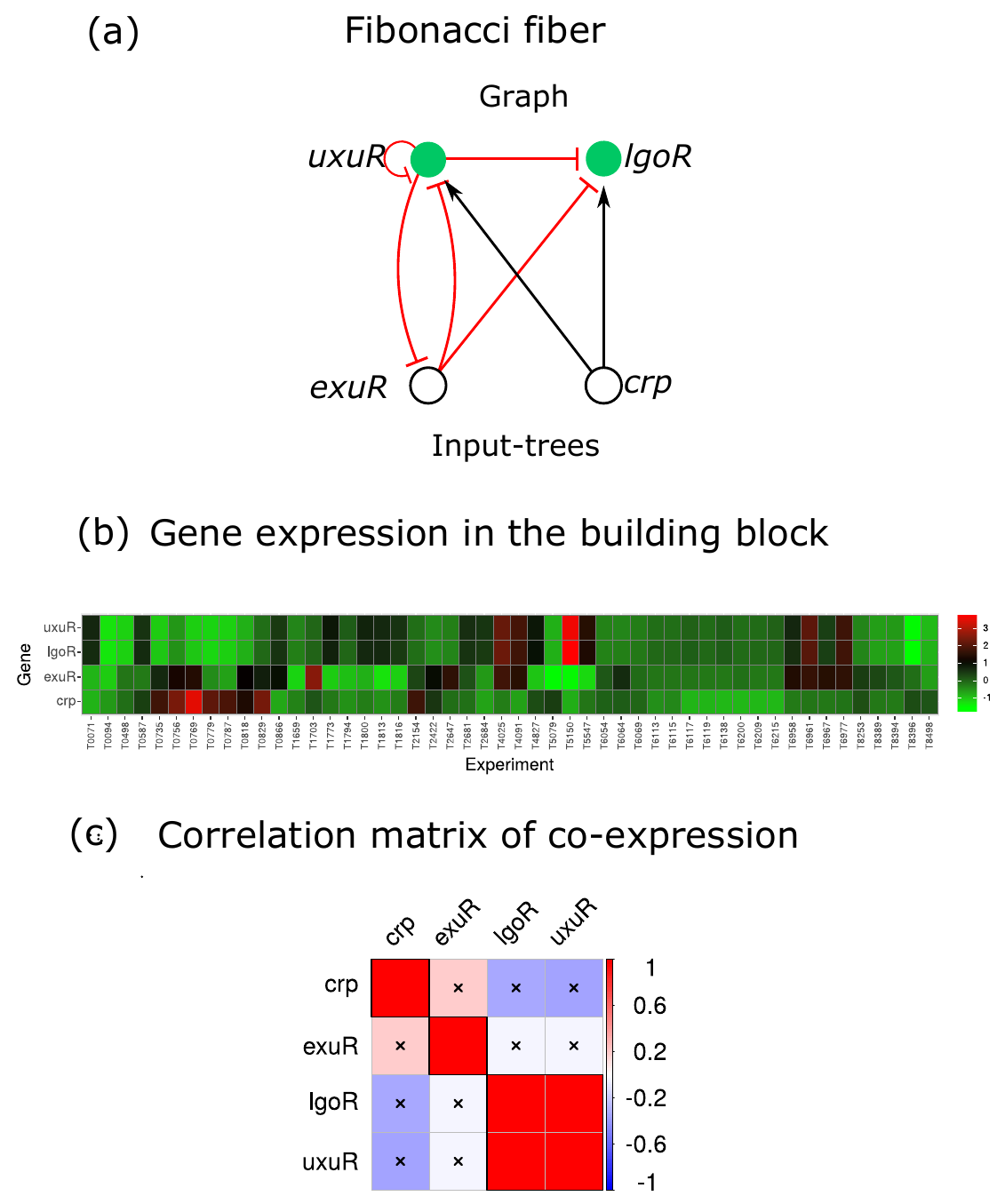}
    \caption{\textbf{Heat map and correlation matrix of a Fibonacci fiber
        in {\it E. coli}}.  Genes are rows and experiments, are
      columns. The color bar indicates the differential expression
      level in AU with respect to the wild-type condition of a given gene
      and experiment.}
  \label{fig:heatmap-ff}
\commentAlt{Figure~\ref{fig:heatmap-ff}: 
Top: FF building block. Graph with nodes uxuR, lgoR, exuR, crp.
Activator arrows crp-uxuR, crp-lgoR. Repressor arrows uxuR-uxuR, uxuR-lgorR, uxuR-exuR, exuR-uxuR, exuR-lgorR.
Middle: Heat map of gene expression.
Bottom: Correlation matrix of co-expression splits  into 4x4 blocks, rows and columns labeled
by uxuR, lgoR, exuR, crp. High correlations between crp and itself,
exuR and itself, lgorR and uxuR. Other correlations close to zero.
}
\end{figure}

In general, we seek groups of genes that have similar
gene expression activity since these similarities are
examples of cluster synchronization,\index{synchronization !cluster } revealing common biological
functions. Gene annotation based on patterns of gene expression have
been widely used in the literature \citep{klipp2016book}. Typically, a
heat map measures the expression of genes under different
conditions, such as those in Figs. \ref{fig:heatmap-fff} and 
\ref{fig:heatmap-ff} for {\it
  E. coli}. These heat maps are obtained from databases that compile a
large number of experimental conditions over many laboratories in the
world and over many species.  In a heat map, genes are rows, while different
experiments, conditions, or subjects appear in columns. Clusters across
the rows produce sets of genes with highly correlated gene expression,
which can be interpreted as working in synchrony, see
 (\ref{eq:time2}) below, across different experiments, and
therefore potentially suggest similar functional annotations for the
genes. Clustering across the columns, instead, leads to sets of
similar conditions.

\begin{remark}\rm
In a heat map,\index{heat map } genes cluster over conditions/subjects, which contrasts with the
definition of synchronization as the similarity of gene expression of a
single condition over time.  It is possible to identify the
correlation over conditions/subjects with the synchrony over time if we assume
ergodicity in the data. We elaborate on this in Section \ref{sec:gene-expression}, see Eqs. (\ref{eq:time}) and (\ref{eq:time2}).
\end{remark}

Under this assumption, we can study correlation profiles of
gene expression in search of synchronized genes, and then test whether
this notion of synchrony emerges from the fibers in the underlying TRN, bridging the gap between function and structure.
We test these ideas with transcriptome\index{transcriptome } data.  Based on idealized ODE
models of gene expression, we discuss whether the predicted
synchronization in expression profiles for genes in fibers can, in
reality be measured from the experimental correlated expression
profiles where the idealized symmetry is only approximate.

\section{Structural network and functional network}

Relating the structural network\index{network !structural } to the functional\index{network !functional } one is a
longstanding question for all biological systems, from genes to the
brain
\citep{park2013structural,friston2011functional,hartwell1999,tononi1994}.

As claimed in Section \ref{sec:structure_to_function}, symmetry\index{symmetry } is the missing link to uncover how structure determines
function in biological networks, from the TRN to the brain.  Chapters
\ref{chap:brain1}, \ref{chap:brain2}, and \ref{chap:brain3} deal with
structure-function in the brain\index{brain }
\citep{park2013structural,friston2011functional}, where we 
consider structural networks made from axons and bundles of axons of
white-matter fiber tracts.  In this chapter, we distinguish between
the structural network and functional networks in gene regulatory
networks.

The structural network or structural graph\index{network !structural } is the TRN: the set of
genes and edges that form the underlying network of physical binding
between the TF expressed by the source gene and the DNA binding
promoter site of the target gene.

When the structural graph is equipped with state variables and
dynamical equations, it becomes a network ODE. When the
ODE represents a lumped model of gene activity such as those studied
in Chapter \ref{chap:stability}, the state variables are the
concentration of proteins expressed by the gene, and the dynamical
equations are those described in the lumped models used in Chapter
\ref{chap:stability_examples}. We are interested in cluster synchronization of
these state variables.

The functional network\index{network !functional } is obtained from the correlations between gene
activity. A link in the functional networks implies some common
functionality between the genes. This functionality can be measured by
many means; the most common is to calculate a Pearson correlation coefficient\index{correlation coefficient! Pearson }
between the activity of the genes and to consider a link
to exist when the correlation is above a certain threshold. 
Once the functional network is
defined, clustering\index{clustering } techniques can be applied to find clusters of
genes. These are identified as genes of cluster synchronization.

Although the function is usually given a broader meaning, here, we
identify `function' with synchronization of genetic activity as
measured by any transcriptomic technique.

The `structure $\rightsquigarrow$ function' concept is applied throughout the
book to other examples, such as brain synchronization.  This turns out
to be an important concept since, somehow, the role of the functional
network is sometimes confused with that of the structural network. In
these examples, we clearly see that the functional network is expected
to be a complete subgraph for genes that are synchronized and
belong to a common fiber. In contrast, the structure network of the fiber
does not need to be fully connected, and indeed, never is. A fully
connected functional network implies that all the genes can
synchronize, but this does not mean that the structural network must
reflect the same fully connected nature. In fact, this is never the
case. Thus, structural connectivity cannot be directly inferred from
 functional connectivity despite many efforts in the field to
prove the opposite. 

\section{Cluster synchronization from gene coexpression}
\label{sec:gene-expression}

Examples of synchronization are abundant in biology, where the
continuous expression of genes is routinely measured.  Gene
coexpression\index{coexpression } indicates that proteins act as a synchronous unit in
time.  This represents a direct measure of synchronization via the
expression activity of each gene as a function of time:
\begin{equation}
  \lim_{t\to\infty} (x_i(t)- x_j(t)) = 0,
  \label{eq:time}
\end{equation}
where $x_i(t)$ and $x_j(t)$ are the expression levels of genes $i$ and $j$ in the
synchrony cluster and the equation holds for all pairs $ i,j$ in the cluster.

This direct measure of synchronization contrasts with what is usually
measured in the cell by gene coexpression using transcriptomic
techniques such as microarray\index{microarray } and RNA-seq\index{RNA-seq } experiments
\citep{barrett2012}. These are indirect measures of synchronization,
because gene activity is not measured as a function of time but  is
measured across different experimental conditions or across
subjects of a given organism. These gene expression profiles are snapshots of
the expression levels (usually mRNA concentration) of all the genes
measured at once and then repeated over experimental conditions or
different subjects. In this case, the time series of
 (\ref{eq:time}) are replaced by $T$ data points of gene activity
$x_i(t_k)$ with $k=1, \dots, T$, over different experimental conditions and/or subjects, creating a discrete series of $T$
events.  Perfect cluster 'synchrony' over all conditions
then implies:
\begin{equation}
  (x_i(t_k)- x_j(t_k)) = 0, \,\,\,\, \forall \, k=1, \dots, T .
  \label{eq:time2}
\end{equation}

Clearly, the two conditions, (\ref{eq:time}) and  (\ref{eq:time2}), need not be the same. Thus, replacing the condition for synchronization over time in Eq. (\ref{eq:time}) with a synchronization over conditions/subjects in Eq. (\ref{eq:time2}) is not always possible and should be considered an approximation.
When these conditions are the same, the system is said to be ergodic. Our assumption in the rest of the book is that the 
synchronization over experimental conditions is a good proxy of time synchronization.

Synchronization\index{synchronization !imperfect } is never perfect, as expected from  (\ref{eq:time2}). Instead, it occurs in real systems as a degree
of synchronization in the data that can be captured by
statistical techniques.

The most common and simpler (linear) statistical method is the correlation
function, which defines a correlation matrix\index{correlation matrix } $C_{ij}$ from the Pearson
correlation coefficient.\index{correlation coefficient! Pearson } It quantifies synchrony between
genes $i$ and $j$ over $T$ conditions as:
\begin{equation}
  C_{ij}=\frac{1}{T}\sum_{k=1}^T\left(\frac{x_i(t_k)-\mu_i}{\sigma_i}\right) \left(\frac{x_j(t_k)-\mu_j}{\sigma_j}\right) ,
  \label{eq:pearson}
\end{equation}
where $x_i(t_k)$ is the mRNA concentration of gene $i$ in condition
$t_k$, and $\mu_i$ and $\sigma_i$ are the mean and standard deviation
of $x_i(t_k)$ averaged over all conditions:
\begin{equation}
    \begin{array}{ccl}
    \mu_i & = & \langle x_i(t_k)\rangle  \\ 
    \sigma_i & = &
        \sqrt{\langle x_i(t_k)^2 \rangle - \langle x_j(t_k)
        \rangle^2}.
    \end{array}
\end{equation}
Here, $\langle \cdot \rangle$ represents a sample average taken over
the series of different conditions (or over different
individuals): $\langle x_i(t_k) \rangle =
(1/T)\sum_{k=1}^T = x_i(t_k)$, when $T$ measurements are taken, each
one at time $t_k$. Here, $t_k$ is just an index and nothing
depends on time.

The correlation $C_{ij}$ ranges from $-1$ to 1.  $C_{ij}>0$ corresponds
to positive correlation\index{correlation } between genes $i$ and $j$, $C_{ij}<0$ corresponds to 
anticorrelation,\index{anticorrelation } and $C_{ij}=0$ indicates lack of correlation.

The correlation matrix\index{correlation matrix } can be calculated not only from Pearson
correlation coefficients\index{correlation coefficient! Pearson } like (\ref{eq:pearson}), but from many 
other measures of synchrony.  In fact, there is a zoo of
strategies to interrogate synchrony between any given dynamical
variables, not only for gene expression but also for any other biological time
series, such as neural activity.  The most commonly used and simplest measure is
the linear correlation function of (\ref{eq:pearson}) and covariance.

However, this correlation measure may not capture the correct
synchrony in the signal. For instance, two signals may have different
amplitudes yet be synchronous in phase. In some applications,
it is desirable to capture this kind of synchronization, particularly
in the brain.  The Phase Locking Value\index{phase locking value }
\citep{bruna2018phase,lachaux1999measuring} captures this kind of
synchronization, which is relevant, for instance, for synchronization in the
brain. We discuss this in Chapters \ref{chap:brain1} and
\ref{chap:brain3}, where we also employ a more direct measure of
synchrony, the {\it Level of Synchronicity} (LoS), which directly
measures  (\ref{eq:time}), and provides a stronger condition of
synchrony. To be considered `fully' synchronous, the neurons
need not only be active at the same time (phase synchrony), but also
to have the same value (amplitude synchrony).

There are also nonlinear measures, such as mutual information\index{mutual information } and
covariograms,\index{covariogram } that focus on the statistical dependence of two signals.
Maximum entropy\index{maximum entropy } methods and Bayesian networks\index{network !Bayesian } are also very popular
ways to obtain effective models and functional networks from data.

Together, these methods let us derive a correlation matrix
from the data. Once the correlation matrix $C_{ij}$ is known, different methods
can be used to obtain synchrony clusters from it. In general
there are two methods:
\begin{itemize}
\item Threshold this matrix to obtain the functional network, which is
  then used to obtain the synchronous clusters by network analysis
  (Section \ref{sec:thresholding}).
    \item
      Analyze $C_{ij}$ directly by any hierarchical clustering
      algorithm (Section \ref{sec:hierarchical-clustering}) to obtain
      the synchrony clusters.
\end{itemize}

\subsection{Thresholding method to find cluster synchronization}
\label{sec:thresholding}

The simplest way to obtain a functional network is by thresholding\index{thresholding } the
correlation function. This creates undirected links between nodes $i$
and $j$ if the correlation exceeds a given threshold value $p$
\citep{bullmore2009complex,rubinov2010complex,gallos2012asmall}.
Thresholding, in principle, produces the adjacency matrix\index{adjacency matrix } of the
functional network $A_{ij}$ by including an undirected
edge when the correlation between $i$ and $j$ is higher than the
threshold $p$.  However, $C_{ij}$ can be positive, signaling
correlations, or negative, signaling anticorrelations. This
introduces a dilemma that has to be solved for the particular
analysis. In general, correlation and anticorrelation are two types
of synchrony. They are two extreme cases in a continuum of
different types of oscillating data, which arise from a phase shift
that can take any value between zero (correlation) to 180$^\circ$
(anticorrelation). Unfortunately,  (\ref{eq:pearson}) ignores all
these intermediate values, while the phase locking value captures this
more refined level of synchrony.

Different choices of the threshold $p$ can produce different adjacency matrices, leading to different types of functional networks.  They can
be unweighted ($A_{ij}=\{0, 1\}$), signaling the absence or presence of
an edge, yet neglecting anticorrelations:
\begin{align}
    A_{ij} = \left\{
    \begin{array} {lcc}
      {1} \quad \mbox{if} \quad { C_{ij} > p ,}  \\
      {0}  \quad \mbox{otherwise.}
    \end{array}
    \right.
\label{eq:aij}
\end{align}
Alternatively, we can consider not only correlations but also anticorrelations, as
proxies of synchrony:
\begin{align}
    A_{ij} = \left\{
    \begin{array} {lcc}
      {1} \quad \mbox{if} \quad { | C_{ij} | > p ,}  \\
      {0}  \quad \mbox{otherwise.}
    \end{array}
    \right.
\end{align}
While this last definition includes anticorrelations, it
places them on the same footing as positive correlations. That is,
the method does not distinguish between both, which is in general, not
advisable.

If we assign the weight information to the edges, we obtain a
weighted functional network ($A_{ij}\in \mathbb{R}$) by
considering the strength of the correlation interaction. Here there are also two
alternatives.  We can ignore negative correlations:
\begin{align}
    A_{ij} = \left\{
    \begin{array} {lcc}
      {C_{ij}} \quad \mbox{if} \quad { C_{ij} > p}  \\
      {0}  \quad \mbox{otherwise.}
    \end{array}
    \right.
\end{align}
or we can consider anticorrelations also to be a form of
synchronization:
\begin{align}
    A_{ij} = \left\{
    \begin{array} {lcc}
      {C_{ij}} \quad \mbox{if} \quad { |C_{ij}| > p ,}  \\
      {0}  \quad \mbox{otherwise.}
    \end{array}
    \right.
    \label{eq:aij-w}
\end{align}

Below, we  use (\ref{eq:aij}) to analyze gene
expression data, being the simplest method of all, but more sophisticated
analysis can be done with more sensitive methods.

\begin{figure}[t!]
  \centering 
  \includegraphics[width=.5\linewidth]{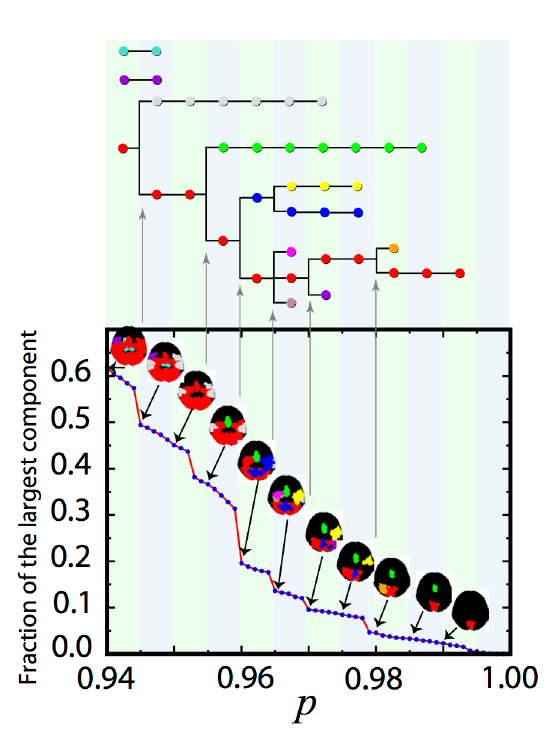}
\caption{\textbf{Hierarchical formation of synchrony clusters in a
    brain network.} An example of a brain functional network from
  human fMRI in a dual task experiment from \citep{sigman2008brain}, as
  analyzed in \citep{gallos2012asmall}. The size of the largest connected
  component of nodes (as measured by the fraction of the total system
  size) is a function of the percolation threshold $p$. As we lower $p$,
  the size of the largest component increases at jumps when new
  modules emerge, grow, and finally are absorbed by the largest
  component. We display the resulting clusters for the indicated $p$.
  The jumps make the identification of clusters clear.  The tree in
  the upper figure shows the hierarchical merging of clusters into a
  hierarchical network of networks. The topography of these clusters
  reflects coherent patterns of cluster synchrony for a
  visual/auditory activity. We observe clusters covering the anterior
  cingulate (AC) region, a cluster covering the medial part of the
  posterior parietal cortex (PPC), and a cluster covering the medial
  part of the posterior occipital cortex (area V1/V2) along the calcarine
  fissure. }
\label{perco-brain}
\commentAlt{Figure~\ref{perco-brain}: 
Described in caption/text. No alt-text required.
}
\end{figure}

The main question is how to choose the threshold $p$.\index{threshold, choice of }  A large
 $p$ reveals clusters of strongly correlated
genes, which in principle, should be disconnected from each
other.  As we lower the value of $p$, these clusters start to connect
(percolate) to each other through weak links.  The choice of the
threshold $p$ is an art in itself. It appears in any analysis of
clustering in diverse ways.

In principle, we could pick a unique threshold.  This leads to a well
defined functional network from where the clusters can be obtained by
modularity, such as the Louvain algorithm \citep{blondel2008} of Section \ref{sec:louvain}.  Section \ref{sec:carbon-utilization}
employs this simple technique to analyze gene expression in
{\it E. coli}.

There are more elaborate ways to pick the threshold, which can reveal
more structure in the data.  \cite{gallos2012asmall} and
\cite{reis2014avoiding} demonstrate that by applying percolation
analysis to $C_{ij}$, a hierarchical structure emerges with critical
values $p_c$ that reveals modules of cluster
synchrony.  Such an analysis has been applied to fMRI brain signals in
\citep{gallos2012asmall,gili2024fibration} and is discussed in
more detail in Chapter \ref{chap:brain3}.

We monitor the size of the largest connected component of the
functional network as we lower $p$ from 1 to 0, as in
Fig.~\ref{perco-brain}. This is interpreted as a percolation process
\citep{bunde-havlin}. The jump in the largest connected component at
$p_c$ signals the merging of two clusters, A and B, formed by strong
links for $p>p_c$ and interconnected by weak links with $p<p_c$.  The
clusters A and B are composed of nodes with incoming links inside
their clusters, defined by their strong links. Each node in A or B
has a number of outgoing weak links connecting to nodes in the other
clusters. Within each cluster, the nodes are
synchronous. Another question is whether these clusters are also
synchronous between themselves. We investigate this question in
Chapter \ref{chap:brain3}.

Finer modules can be obtained by lowering the threshold. Figure
\ref{perco-brain} shows the detailed behavior of a modular network
obtained from an fMRI BOLD signal in an individual performing a dual
visual and auditive task obtained in \citep{sigman2008brain} and analyzed
in \citep{gallos2012asmall}. The hallmark of hierarchical cluster
synchronization is the jumps in the percolation process
seen in Fig.~\ref{perco-brain}.  The figure shows the size of the
largest connected component during percolation for the correlation
function as we lower a threshold $p$. At larger values than $p=0.98$,
three large clusters are formed localized to the medial occipital
cortex (red), the lateral occipital cortex (orange), and the anterior
cingulate (green).  Critical values occur at the jumps. At
$p_c=0.978$ the orange and red clusters merge, as revealed by the
first jump in the percolation plot.  As $p$ continues to decrease, this
mechanism of network formation and absorption repeats, defining a
hierarchical process depicted in the tree in the top panel of
Fig.~\ref{perco-brain}.

This mechanism represents the iteration of a process by which clusters
form at a given $p$ value and merged via comparably weaker links. This
process is recursive. The weak links of a given transition become the
strong links of the next transition, in a hierarchical fashion,
resulting in a hierarchical percolating process where clusters merge
with each other, and sharp transitions are observed in the size of the
largest network (red in Fig.~\ref{perco-brain}) as it absorbs smaller
ones.  The modules have consistent topographic projections on the map
of the brain \citep{gallos2012asmall,gallos2012conundrum}.  The
signature of hierarchical synchronization in a brain structure is,
then, the hierarchical sequence of clusters as a function of the
threshold $p$, as shown by the jumps in Fig. \ref{perco-brain}.
\cite{bardella2016} applied a modified percolation method, where rather
than monitoring the jumps in the largest connected component, the
number of components is followed.

This percolation analysis\index{percolation analysis } is not restricted to the brain but can
be performed on any data that is expected to cluster; in particular,
gene coexpression data. In Chapter \ref{chap:brain3} it will be
applied to gene coexpression patterns in the human brain. This
chapter also develops a more sophisticated thresholding method,
called clique synchronization  (\ref{eq:1}), which  captures 
synchrony within the data better than the simple percolation thresholding
of $C_{ij}$ does.

Thresholding $C_{ij}$ is a way to
regularize the data, and this is the domain of machine learning. A  widely used thresholding method in machine learning is the
Graphical Lasso (Glasso).  There
are countless methods in machine learning to obtain sparse
representations of data, such as Ridge regression or Lasso (least
absolute shrinkage and selection operator), a regularization technique
introduced by \cite{tibshirani1996regression}.  Here we
elaborate on  Glasso (implementation available at
{\small\url{http://www-stat.stanford.edu/~tibs/glasso}}), which uses Lasso to
obtain a sparse representation of the data and then provide a
graphical representation by thresholding spurious correlations. This is completely analogous to thresholding $C_{ij}$
\citep{sojoudi2016}.

To avoid spurious correlations introduced by random covariates, and to
reduce the dimensionality of the problem, graphical lasso\index{graphical lasso }
\citep{friedman2008} infers a sparse
representation $\hat{J}$ from the correlation matrix $C_{ij}$. Graphical
\index{functional network !graphical lasso } lasso assumes a
multivariate Gaussian distribution\index{multivariate Gaussian distribution } for the dynamical variables
$x_i(t)$, and then infers the model's parameters from experimental
data. The estimation of this matrix is made by minimizing the
log-likelihood of a multivariate normal distribution:
\begin{equation}
  \log {\cal L}(\hat{J}) = \log \det [ \hat{J} ] - {\rm Tr}[ \hat{C} \hat{J} ] -
  \lambda|\hat{J}| ,
  \label{lasso}
\end{equation}
where the penalty parameter\index{penalty parameter } $\lambda$ controls the sparsity of
the resulting network as estimated by the number of total
connections. Here $\lambda$ plays the
same role that $p$ does in the thresholding method \citep{sojoudi2016}. After
obtaining $\hat{J}$ for a given $\lambda$,  a graph of the
functional network is obtained by applying a small threshold
$\varepsilon$ to $\hat{J}$, such that spurious interactions are avoided. That is, an edge in the functional network exists if
$\hat{J}_{ij}>\varepsilon$. This produces a sparse representation of
the functional network.
  
As in the thresholding\index{thresholding } method, the critical step is to choose
$\lambda$. To select the appropriate value, two
criteria can be applied, which are valid across all biological data: they
are clustered, yet integrated \citep{tononi1994}.  This means that
the existence of clusters does not imply that these clusters are
completely isolated. They must still be integrated into the system as a
whole.  In neuroscience, this is referred to as the `integration versus
segregation problem',\index{integration vs.
segregation problem} and it is a longstanding issue to understand this problem for 
the brain \citep{tononi1994}. Brain networks are integrated, i.e.,
signals are exchanged through the whole architecture within and
between different brain clusters.  Therefore, the functional network
must guarantee global connectivity. At the same time, the functional
brain network is segregated into clusters.

Based on these considerations, the value of $\lambda$ was fixed by
\cite{delferraro2018finding} so that all nodes are connected through
a path (to allow for brain integration) and, at the same time, are
segregated into clusters. Their procedure follows a  percolation
process similar to thresholding the correlation matrix, as in
Fig. \ref{perco-brain}.  We decrease the value of $\lambda$ to
regularize the functional network less and less until all relevant
clusters are integrated into a global network, meaning that they are all
connected.  For this, we monitor the jumps in the size of the largest
connected component of the functional network to identify the weak and
strong links defining the clusters.  

What is interesting is that it has been mathematically proved by
\cite{sojoudi2016} that the functional network obtained by
thresholding $C_{ij}$ is exactly the same as the one obtained by 
the Glasso $J_{ij}$. That is, if we build a functional network by
thresholding $C_{ij}$ by $p$ and a network by Glasso\index{Glasso } with penalty
$\lambda$, the networks are exactly the same when $p=\lambda$.  This
remarkable result brings some sort of validity to the whole method.
It also facilitates obtaining a sparse representation of data by just
thresholding $C_{ij}$, which is a much faster operation than inverting
Glasso to obtain $J_{ij}$. In fact, Glasso cannot be applied to very
large network sizes, while thresholding can.

After the functional network is obtained, either by thresholding
$C_{ij}$ or by Glasso or any other method, the clusters of
synchrony can be obtained by applying any clustering method to the
resulting graph.  The most common method is to apply modularity\index{modularity } or
community detection algorithms\index{community detection } (like Louvain, Section
\ref{sec:modularity}). These can identify clusters that are highly
connected internally, but have few connections to other clusters.  Be reminded: modularity algorithms should be applied to the functional
network, but not to the structural one.

Modularity produces modules that are just an approximation to cluster
synchrony. In principle, a perfectly synchronous cluster is a clique
of nodes, fully connected among themselves but with no connections
outside the cluster. However, this ideal case never occurs in nature, so
modularity is just a first approximation used to find them in real data. In Section \ref{sec:functional_clusters}
 we discuss a more refined method based on 
clique synchronization algorithm that better captures these clusters.

\subsection{Hierarchical clustering to find cluster synchronization}
\label{sec:hierarchical-clustering}

There are also alternatives to obtain clusters directly from the
correlation matrix $C_{ij}$ without going through the functional network. Clustering\index{clustering } is a well-developed branch of machine
learning,\index{machine
learning } and a large number of techniques exist.

Unsupervised machine learning clustering methods   are hierarchical
clustering and agglomerative and divisive class, determined by
Euclidean distance and others. Some of these methods have been discussed in
Section \ref{sec:modularity}.  In hierarchical modularity
(Fig. \ref{fig:measurements}), the dendrogram\index{dendrogram } can be cut at any level
to partition the data into clusters. Again, this permits a
degree of liberty to choose the thresholding\index{thresholding } level, and again, this
parameter should be determined in advance. In principle, thresholding
the dendrogram should be analogous to thresholding $C_{ij}$, although
the resulting clusters are not the same.

Other unsupervised methods include $K$-means, which
requires specifying the number of clusters in advance to find the set
of clusters that minimizes the distance from each data point to the
clusters. The drawback of this method is that there is no a priori
information on how many clusters there are in the data; also, there is
no information on the centroids of the clusters, so random
initialization of the centroids is needed. Thus, there is no
guarantee of convergence with the right answer. Many other
approaches improve on these methods (fuzzy $k$-means, etc.).

\section{Gene coexpression synchronization analysis from massive datasets}

Many specialized databases contain information on the
transcriptome\index{transcriptome } across species, such as Colombos\index{Colombos } \citep{colombos2016},
Ecomics\index{Ecomics } \citep{kim2016}, and SubtiWiki\index{SubtiWiki } \citep{subtiwiki2018}. On
another level, there are databases of gene annotation that are
manually curated to reflect the interplay between different genes in
their respective processes and functions. Ecomics contains microarray
and RNA-seq experiments gathered from NCBI GEO \citep{barrett2012},
including several \textit{E. coli} strains in 3,579 experimental growth
conditions for 4,096 genes.  The SubtiWiki dataset contains experiments
in the GEO database that give 104 experimental conditions for {\it Bacillus}
genes.  Raw data on gene expression among different experiments and
platforms is pre-processed to obtain normalized expression levels by
using noise reduction and bias correction normalized data across
different platforms.

\cite{leifer2021predicting} use these datasets to test gene
synchronization via fibrations in {\it E. coli} and {\it
  B. subtilis}.  The Ecomics portal \citep{kim2016} collects
microarray and RNA-seq experiments from different sources including
the NCBI Gene Expression Omnibus (GEO)\index{Gene Expression Omnibus } public database \citep{barrett2012}
and ArrayExpress\index{ArrayExpress } \citep{kolesnikov2015}.  The data is also compiled at
the Colombos\index{Colombos } web portal \citep{colombos2016}.  We use Ecomics\index{Ecomics } since it
provides the data in  wild-type (WT) conditions, rather by
providing data based on the fold-change that compares a mutation or
perturbation to the WT. Using Ecomics, we obtain a
set of experimental conditions where the particular genes in a given
fiber have been significantly expressed. For this task we follow
standard gene expression analysis, as developed in {\small\url{colombos.net}} and
\citep{colombos2016} for expression levels in \emph{E. coli}:
first to identify the set of growth conditions where the genes in a
given fiber are significantly expressed with respect to random noise, and
then to test synchronization through correlations in gene activity
using these conditions.  We then repeat the scheme using the
conditions in Subtiwiki for {\it B. subtilis}\index{B. subtilis @{\it B. subtilis} } \citep{subtiwiki2018}.

For a given set of genes in a fiber, we find the experimental
conditions for which the genes have been significantly expressed by
comparing the expression samples over different biological
conditions. To filter conditions where the genes are expressed we use
the Inverse Coefficient of Variation (ICV), similar to that applied
by Colombos.

The experimental conditions are ranked by:
\begin{equation}
  \mbox{ICV}_i=|\mu_i|/\sigma_i ,
  \label{eq:icv}
\end{equation}
where $\mu_i$ is the average expression level of the genes in the
condition $i$ and $\sigma_i$ is the standard deviation.
\cite{colombos2016} select those conditions with ICV$_i>1$, i.e.,
where the average expression levels in the particular condition $i$
are higher than the standard deviation.  This score reflects
that, in a relevant condition, the genes show an increment of their
expression level, above the individual variations caused by random noise.
The expression profiles obtained are organized by the experimental
conditions, which are labeled according to the GEO database
\citep{barrett2012}.  A heat map such as shown in
Fig. \ref{fig:heatmap-ff} is obtained. For each set of genes, there is
a particular set of conditions where they have been activated, as determined
by the ICV.   The conditions obtained depend on the
fiber and, generally speaking, any pair of fibers may have
correlations calculated under suitable conditions. To deal with this
issue, the union of conditions in which both fibers are activated is
considered when determining correlations between nodes in different
fibers.  From these data, we calculate the coexpression matrix using
the Pearson correlation coefficient\index{correlation coefficient! Pearson } between the expression levels of
two genes under the relevant conditions. Then the functional network
is obtained by thresholding, and the synchrony clusters are obtained
by modularity.

\begin{figure*}
  \centering \includegraphics[width=.8\linewidth]{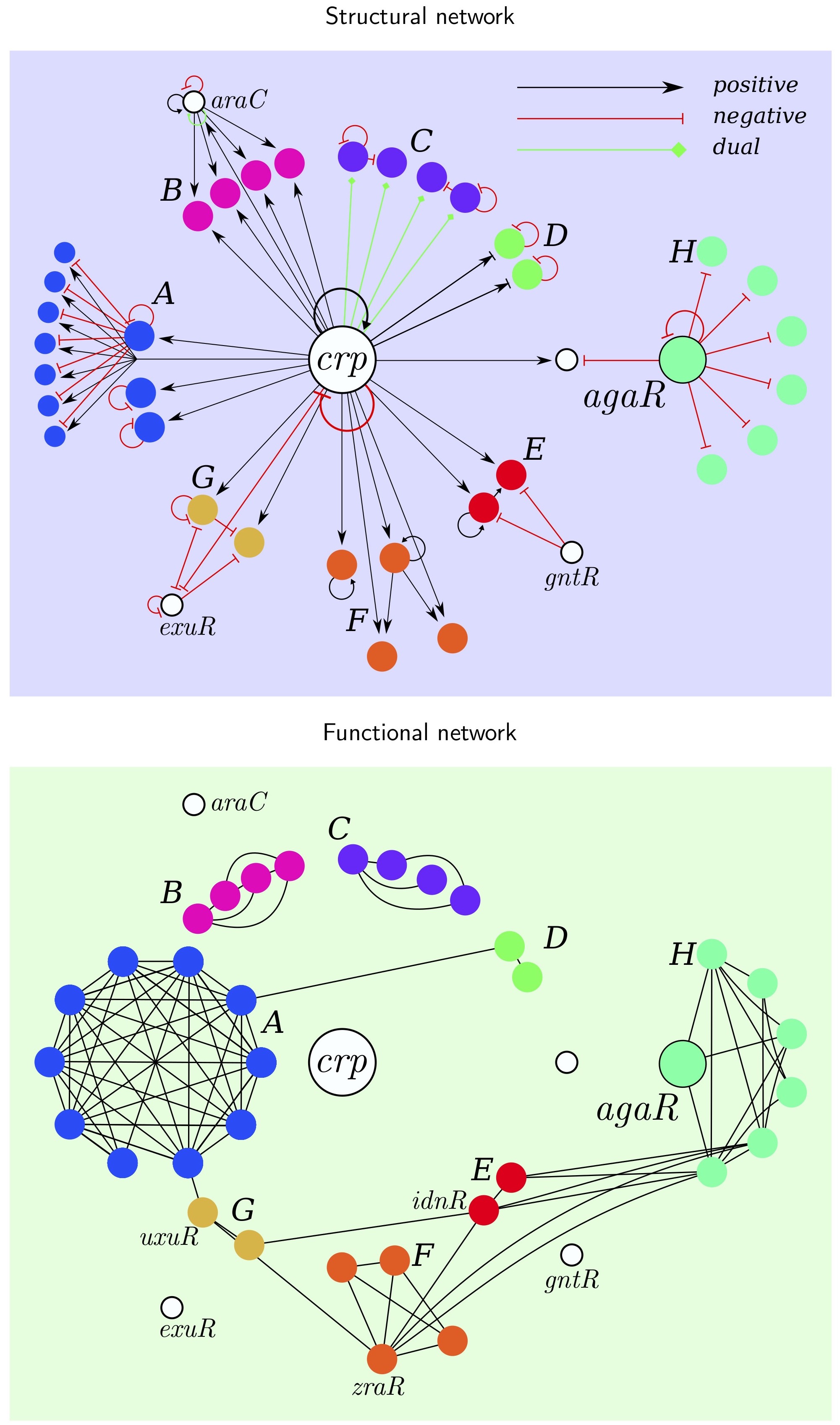}
\caption{\textbf{Structure $\rightsquigarrow$ function in the carbon utilization
    network of {\it E. coli}}.  {\em Top}: The structural network represents the
  topology of the gene regulatory network of alternative carbon
  sources. 
  {\em Bottom}: Functional
  network obtained by thresholding the correlation matrix $C(i,j)$ in
  Fig.~\ref{fig:carbonc} at $0.6$. That is, any two genes with 
  correlation above $0.6$ are connected, and any two genes with
  correlation below $0.6$ are disconnected.}
\label{fig:carbona}
\commentAlt{Figure~\ref{fig:carbona}: 
Illustrative. Described in caption/text. Top: structural network.
Main point is that genes crp and agaR
are highly connected to other genes in the structural network.
Bottom: functional network. Main point is that genes crp and agaR have similar connectivity
in the functional network.
}
\end{figure*}

\subsection{Gene synchrony in fibration circuits}

Figs. \ref{fig:heatmap-fff} and \ref{fig:heatmap-ff} show the
coexpression analysis in the FFF\index{FFF } and Fibonacci circuits in {\it
  E. coli}.  The expression profiles of the genes in these circuits
are filtered by ICV over all experimental conditions in Ecomics to
obtain the heat maps. Then the Pearson correlation matrix is
calculated over these conditions, as shown.

The expression profiles for the FFF fiber are composed of the main TF
regulator {\it fur} and the fiber genes {\it purR} and {\it pyrC}, which
should be synchronized. The fiber predicts no synchronization between
{\it purR} and {\it pyrC} and the regulator {\it fur}.  
In the FFF, {\it
  fur} should neither synchronize with {\it purR} nor with {\it pyrC}.
This is corroborated by the correlation matrix in
Fig. \ref{fig:heatmap-fff}c.

The lack of synchronization between the fiber genes {\it purR-pyrC}
and its regulator occurs even though the fiber is directly
regulated by {\it fur}, that is, direct regulation does not
necessarily lead to synchronization. This is because, beyond this
direct regulation, other TF regulations inside the fiber synchronize the genes within the fiber but not outside. As
predicted, genes are highly coexpressed within fibers but not
significantly correlated with the regulator genes.

This is corroborated by the analysis of the FF circuit in
Fig. \ref{fig:heatmap-ff}.  Figure \ref{fig:heatmap-ff}c shows strong
correlations between the genes {\it lgoR} and {\it uxuR} within the
Fibonacci fiber, yet the regulators {\it exuR} and {\it crp} (which
also regulate many other fibers) remain unsynchronized with the fiber.

A functional network obtained by thresholding the correlation
matrix is shown in Fig.~\ref{fig:carbona}, bottom. We use a threshold
$C_{ij} > 0.8$ to produce a functional network where a link exists if
the Pearson correlation coefficient\index{correlation coefficient! Pearson } between gene $i$ and $j$, $C_{ij}$
is above the threshold. Links in this functional network exemplify
significant synchronization, with possible functional relations
between genes in the fiber indicated by significant levels of
synchronization.

\subsection{Structure $\rightsquigarrow$ function in the carbon utilization circuit of {\it E. coli}}
\label{sec:carbon-utilization}

\cite{leifer2021predicting} perform a large-scale analysis of gene
synchronization expression patterns in the carbon utilization  subgraph\index{network! carbon utilization }
of {\it E. coli}
controlled by the {\it crp} gene shown in Fig. \ref{fig:componenta}. This subgraph controls genes
involved in the metabolism of alternative carbon sources, i.e.,
all sugars except glucose. It contains genes involved in
pathways used to catabolize sugars like fucose, galactosamine, ribose,
N-acetylglucosamine, maltose, ribose, galacturonate and L-idonate,
which are used as an alternative to glucose, the primary source of
sugar intake. The resulting structural network of the TRN can be seen
in Fig.~\ref{fig:carbona}a. We apply the fiber finding algorithm and
obtain the colored fibers shown in the figure. This subgraph is part
of the carbon SCC in Fig. \ref{fig:componenta} and analyzed in
detail in Section \ref{sec:fibration-analysis}.

Although the genes have a related general biological function of
carbon utilization, each pathway is activated by the presence of
different sugars in the system. For example, the presence of arabinose
activates the arabinose catabolism pathway and the presence of other
sugars like ribose or maltose activates their respective
pathways. Therefore, the entire carbon utilization system does not
have to be coexpressed at the same time.

\begin{figure*}
  \centering{ \includegraphics[width=\linewidth]{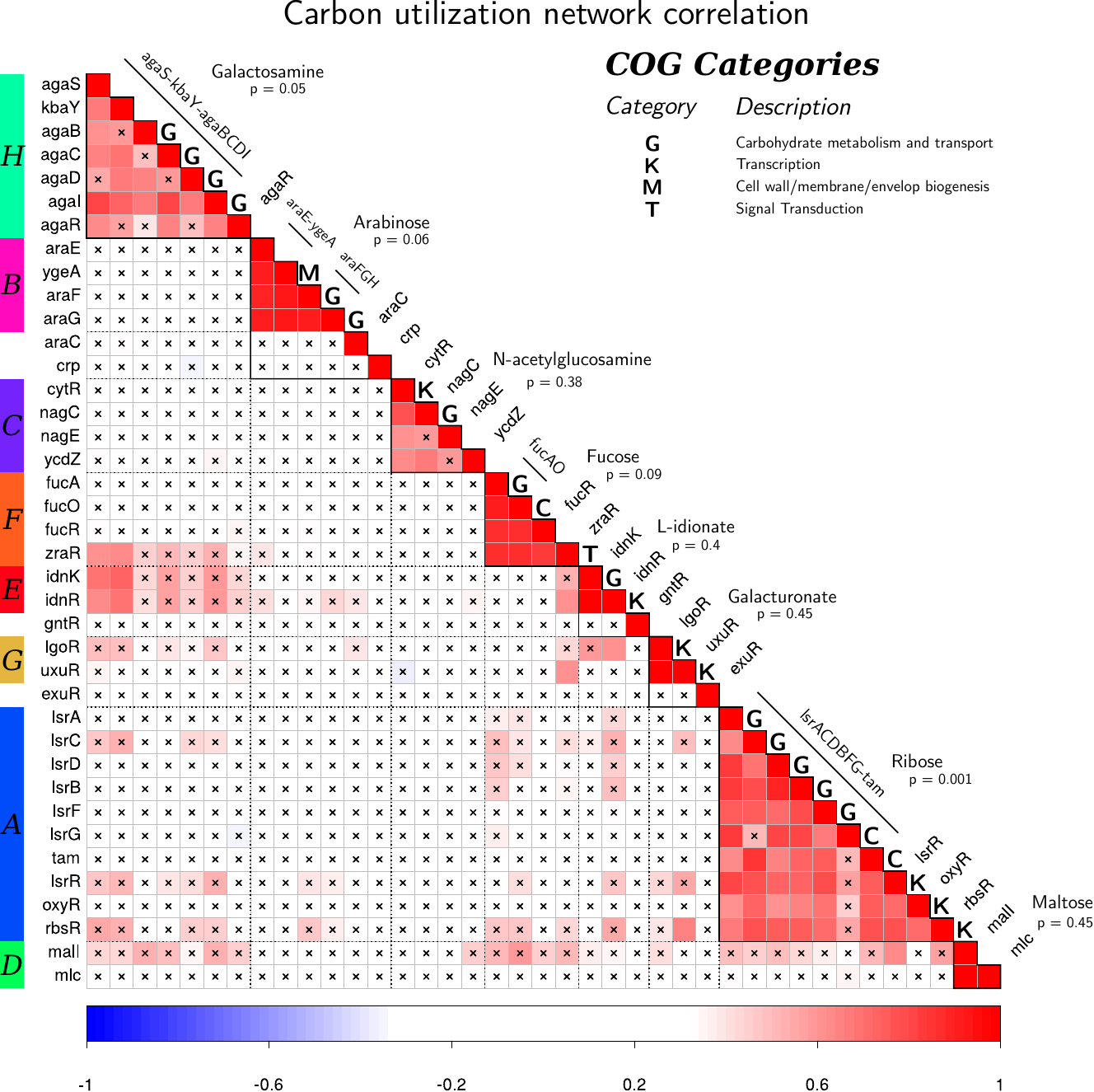}}
  \caption{\textbf{Correlation matrix of carbon utilization network of
      {\it E. coli}.}  This is obtained by a filtering method based on
    the ICV. Operons are shown with a black line underneath the gene
    name along the diagonal. Correlations below 0.6 are marked with
    black crosses. UniProt database \citep{uniprot2022} provides COG
    categories. The type of a fiber building block regulator defines the
    function of each block and is obtained from RegulonDB
    \citep{regulon2016}. Figure reproduced from
    \citep{leifer2021predicting}. Copyright \copyright ~2021, The Author(s). }
\label{fig:carbonc}
\commentAlt{Figure~\ref{fig:carbonc}: 
Illustrative. Described in caption/text. No alt-text required.
}
\end{figure*}

Investigation of this network shows a measurable coexpression within
fibers and low coexpression across fibers, as evidenced by the
correlation matrix $C_{ij}$ in Fig.~\ref{fig:carbonc}.

The activities of the genes in operons are reported individually in
Ecomics, so we plot the activity of the individual genes. Genes in one
operon are marked in the plot; for instance, the operon {\it fucAO} is
unpacked as the two genes {\it fucA} and {\it fucO}.  Synchrony
within operons is given by the common polymerase reading, so it is
trivial, hence not a proof of nontrivial fiber synchrony,
although unpacking the operon creates a fiber. The test of
fiber synchronization is then to compare the activity of any gene in
the operon with that of the genes outside the operon.  For instance {\it
  fucAO} with {\it fucR} and {\it zraR}.

Figure \ref{fig:carbona} (bottom) shows the associated functional
network, obtained by thresholding\index{thresholding } the correlation matrix\index{correlation matrix } of
Fig. \ref{fig:carbonc}.  A threshold $C_{ij} > 0.6$ was used to
produce a functional network where a link between genes $i$ and $j$
exists if the Pearson correlation coefficient\index{correlation coefficient! Pearson } $C_{ij}$ from
Fig. \ref{fig:carbonc} is above the threshold $0.6$. Links in this
functional network exemplify synchronization between genes in the
fiber, according to their high pair correlation. The resulting
functional network corresponding to the carbon circuit is shown in
Fig.~\ref{fig:carbona} (bottom).

There is a lot to unpack here. In the ideal case of perfect synchrony
in the fibers, we would expect to see a functional network
made of a clique for  the genes in each
fiber, and no connections between genes in different fibers or
between fibers and regulators. This pattern would indicate perfect
synchrony. We see this approximately, but not perfectly, in
Fig.~\ref{fig:carbona} (bottom). For instance, the
fiber A working on ribose catabolism is almost a clique: fully
connected internally, but it has a connection to a fiber D responsible for
maltose catabolism. There is still some synchronization between
fibers, but it is very weak.

The fibers B and C involved in arabinose and N-acetylglucosamine
utilization are indeed perfectly synchronized. They form a clique in
the functional network with no external links.  We also observe that
the regulators {\it crp}, {\it araC}, {\it exuR} and {\it gntR} are fully
disconnected in the functional network, yet connect to the set of
fibers in the structural network, as predicted.

Modularity algorithm provides the colors in
Fig.~\ref{fig:carbona} (bottom), which correspond exactly to the
balanced colors obtained by fiber analysis of the TRN in
Fig.~\ref{fig:carbona} (top).  This demonstrates the structure $\rightsquigarrow$
function relation in the {\it E. coli} TRN.

The functional network\index{network !functional } shows a high correlation (synchrony) for
genes in the same fiber, and a lack of correlation between
genes in different fibers. The synchrony and lack of
it observed in and outside fibers does not depend on whether the
genes in the fiber/outside fiber are connected in the structural
network.

A link in the functional network does not imply one in the structural
network, that is, in the TRN, and vice versa. The most clear example of
how different the functional network is from the structural TRN is the
role of the master regulator {\it crp}.  This regulator is clearly the
hub of the entire carbon utilization circuit in the TRN of
Fig.~\ref{fig:carbona} (top), since it regulates all the genes in this
circuit, together with many others that are not considered part of the carbon
utilization. Yet in the functional network\index{network !functional } of Fig.~\ref{fig:carbona} (bottom), {\it crp} is an isolated gene, not functionally connected with
any of its regulated genes. That is, its regulon does not predict its
functionality via synchronization. Indeed, while {\it crp} regulates a
myriad of fibers, it does not synchronize with any of them.

Guilt by association techniques to annotate genes should be applied
only to the functional network. Applying this approach to the
structural TRN would give the wrong result, as shown for
\textit{crp}. In the multi-layer composite, fiber A in
Fig. \ref{fig:carbona} (top), the genes \textit{hemH-oxyS} are not
linked with \textit{add}; in fact, they are quite far apart in the
network, separated by a distance of two steps, but  they
are still highly correlated and functionally related by synchronization,
which is indicated by their connections in the functional network.

We conclude that the functional network\index{network !functional } does not directly reflect the
structural network. These two networks measure different aspects of
gene activity. The structural network is the network of physical
interactions between genes via a TF, expressed by the source gene,
binding to the promoter region of the source gene. A direct link here
means direct physical interaction. On the other hand, the functional
network measures the functional association of two genes emanating
from their synchronization and high coexpression.

These differences seem not to be clear in the literature, especially
when the structural links are inferred directly from the functional ones, as
is routinely done
\citep{brugere2018,horvath2011,zhang2005,langfelder2008,bansal2007,tegner2003,liao2003,wang2014,marbach2012,butte1999,maertens2018,chen2008}.

The only way to infer structural links from functional links or viceversa is using theory. In Chapter \ref{chap:function} we develop a method to infer the
structural network from the functional one using fibrations, which resolves this issue. We
will develop a symmetry-driven reconstruction of the structural network
that guarantees that the resulting network reproduces the
synchronization observed dynamically in the functional network. The
method reconstructs links in the structural network independently of whether
they are connected or not in the functional network.

\begin{figure*}[ht]
	\centering
	\includegraphics[width=\linewidth]{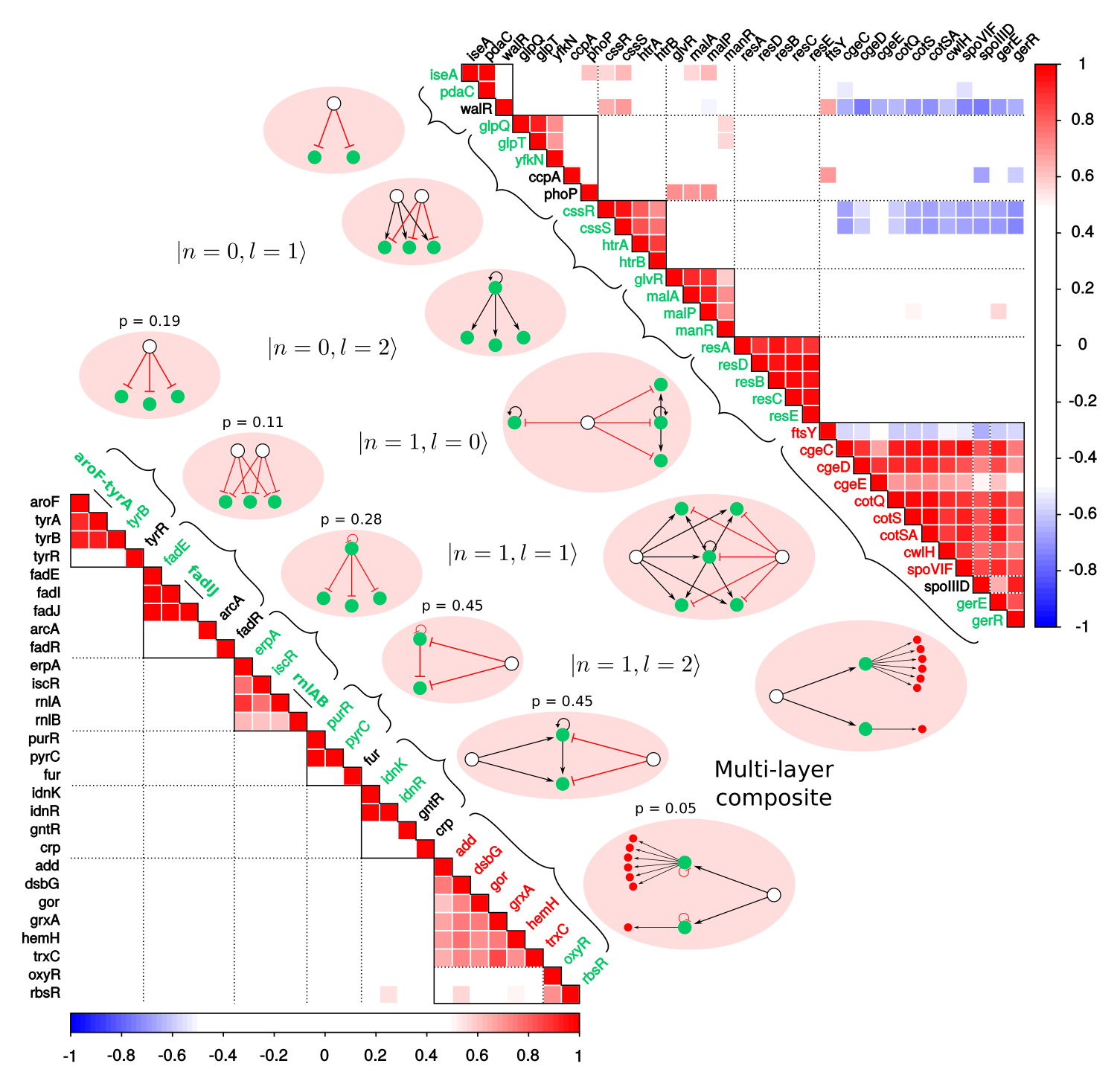}
	\caption{\textbf{Correlations in the hierarchy of fiber building
            blocks in \textit{E. coli} and \textit{B. subtilis}.}
          Fiber building blocks are represented as pink ovals. White
          nodes represent regulators and green and red nodes represent
          fibers. Black links represent activation and red links
          represent repression. Cross-correlations are shown with the
          lower diagonal (\textit{E. coli}) and upper diagonal
          (\textit{B. subtilis}) correlation matrices. Black lines
          indicate building blocks and black dotted lines indicate
          cross-correlations between different fibers. Operons are
          shown by a line underneath the gene names along the
          diagonal. Figure reproduced from
          \citep{leifer2021predicting}. Copyright \copyright ~2021, The Author(s).}
	\label{BMC:Fig10}
\commentAlt{Figure~\ref{BMC:Fig10}: 
Illustrative. Described in caption/text. No alt-text required.
}
\end{figure*}

\subsection{Synchronization in the hierarchy of fibers in {\it E.coli} and {\it B.Subtilis}}

\cite{leifer2021predicting} also present a study of coexpression
patterns in \textit{B. subtilis}, Fig. \ref{BMC:Fig10}. The circuits presented include simple fibers:
$\rvert n = 0, l = 1\rangle$, $\rvert n = 0, l = 2\rangle$, $\rvert n
= 1, l = 0 \rangle$, $\rvert n = 1, l = 1\rangle$, and $\rvert n = 1,
l = 2\rangle$; and multi-layer composite fibers in both species.

The observed correlations largely confirm synchrony within
fibers, and lack of correlation between fibers. However, there are some interesting exceptions. For
instance, the genes {\it cssR} and {\it cssS} in $\rvert n = 0, l =
2\rangle$ in {\it B. subtilis} present large anticorrelations with a
fiber (the multi-layer containing the gene {\it cgeC}). This
 anticorrelation may indicate extra transcriptional
regulations between these fibers. These types of correlations can be
used to guide the search for missing regulation edges, which are
ubiquitous in genetic network reconstructions. 

The existence of unexpected extra synchronization is also an indication of
symmetries that are undetected in the TRN due to incomplete
data. It is also observed that the multi-layer composite circuit
example in Fig.~\ref{BMC:Fig10} is overly synchronized. That is,
rather than exhibiting no synchronization between the two layers of
the building block (in the figure, the two fibers in the multilayer
are indicated by names in red and green), the circuit exhibits
complete synchronization among all genes. We hypothesize that this can
be explained by the self-activation link at the gene \textit{spoIIID},
which would have turned this building block into $\rvert n = 1, l =
0\rangle$, which is supposed to exhibit complete synchronization.
Chapter \ref{chap:function} discusses an algorithm to perform this
synchrony-based reconstruction of the TRN, guided by symmetries.


\chapter[All Biology Conspires Against Synchrony, so How Can We Be Alive?]{\bf\textsf{All Biology Conspires Against Synchrony, so How Can We Be Alive?}}
\label{chap:alive}

\begin{chapterquote}

Genes fibers of an idealized model exhibit perfect symmetry
and exact synchrony. This is never realized in biology, as nothing is
ideal in living systems. In fact, we expect that perfect
synchrony and symmetry can easily be broken by the multitude of
parameters and interactions, all different, of real systems. 
This leads to the question: how can a predictable pattern of synchrony, which is essential for survival, be achieved in biology? 
How can we survive in the presence of noise and disorder competing against symmetry? This question was vividly posed by \cite{schrodinger1974what} in his seminal book, {\it What is Life?} Decades later, we still lack a definitive answer. While we do not claim to provide one here, we hope that the following speculations may contribute to the understanding of the issue.

\end{chapterquote}

\section{Modeling in Biology}

First, some remarks on modeling assumptions are in order. In this
chapter---in fact, for much of the book---we start by working with idealized
models that ignore many aspects of real biology.  For example, when we
say that two nodes are `synchronized', we mean that their dynamic
states are {\it identical}, not merely very similar. This, of course, is
never realized exactly in practice, but it leads to `clean'
mathematics.  It is also a standard modeling assumption in much of the
synchronization community. Likewise, we work with 'idealized' ODEs based on uniform parameters and many simplifying assumptions that allow for their mathematical treatment.

In return for these simplifications, it becomes easier to understand
the underlying mathematical principles.  In later sections,  we discuss more realistic modeling assumptions. It is worth bearing in
mind that all models involve a trade-off. Simple models are easier to
understand, but oversimplified models can be misleading. On the other
hand, complex models may be more realistic, but they can be much more
difficult to analyze and to understand. Moreover, complex models are
often very sensitive to the exact assumptions being made: realistic
ingredients are no guarantor of accuracy.

Ultimately, what makes a model worthwhile is not so much how realistic
its ingredients are but how much insight it gives and how well its
{\it predictions} corresponds to reality---what matters is not so much
what is put in, but what we get out.  
For instance, astronomers
regularly model the solar system as a collection of point masses. No
actual planet is a point, and point masses cannot exist in reality;
yet these models are accurate enough to get a space probe close to its
destination with high precision. Only when it has to land is it
necessary to model the complex gravitational field and planetary
terrain in greater detail.

\section{Synchronization in biology is not ideal}
 \label{sec:not-ideal}

One of the main advantages of using symmetries to characterize 
dynamics is that the synchrony patterns that we predict are
combinatorial/topological properties of the network, thus largely
independent of the explicit form of the dynamical equations.
Therefore, we do not need to specify the equations that
describe the dynamics of the system---be it a neural system, a genetic
network, or any other physical, chemical, or biological system---beyond
the general form from Definition
\ref{D:admissible_ODE_def}, as long as the ODEs are
admissible for the graph. We, therefore, derive all our
results for a general type of dynamics, which can then be specialized
to particular systems.  However, particular systems are
characterized by a large number of parameters and most of them are
required to be the same for symmetry to exist. Unfortunately, this
`uniformity condition' is never strictly realized in biology. How is the coherent synchronization necessary for life to emerge?

We treat this question below. We first elaborate on an ideal model of
gene expression that predicts perfect synchrony.\index{synchrony !perfect } Then we show that
the conditions for this model are not realized in biology due to the
broad parameter space of interactions among genes, proteins, and small molecules.  We discuss how to reconcile this
result with the observed (quasi-perfect) synchrony\index{synchrony !quasi-perfect } in real systems.

\section{An ideal ODE model of gene expression dynamics}

Gene expression\index{gene expression } dynamics in cells depend on a variety of processes
including transcription, translation, protein folding, mRNA and
protein degradation, and mRNA and protein dilution in growing cells
\citep{karlebach2008modelling}. All of these processes are stochastic
and gene-specific, which prevents complete and precise
synchronization of expression profiles. However, strictly speaking,
synchrony via symmetries requires an idealized deterministic model of
gene expression with not too many parameters, where, in general, all
interactions and parameters are the same.

We first define this ideal model, where fibers and synchrony are perfect.\index{synchrony !perfect }
Each gene has a variable describing
the cellular concentration of the gene product, i.e.~the protein
encoded by the gene. Assuming short mRNA half-lives (and linear,
unregulated mRNA degradation), transcription and translation can be
effectively described by a single production rate, with the protein
product concentration as the output variable. However, for reasons of
data availability, we instead consider mRNA concentrations to be the
relevant gene product.  If the finite duration of transcription and
translation is fast as compared to cell growth and protein lifetime
(or when describing cell steady states) it can be ignored. This
justifies an ODE without time delay.  The expression dynamics of the
given gene depends on two contributions: degradation and synthesis
(the arguments are independent of this assumption):
\begin{equation}
  \label{eq:generalODEmain}
    \frac{dx(t)}{dt}= -\alpha_x \,x(t) + f_{x}(y,z,\dots) .
\end{equation}
The input function $f$ describes the transcription rate as a function
of the active TF concentrations of the incoming genes. It can be
modeled, for simplicity ,by a Hill function of
other's TF activities $y$, $z$, $\dots$.
In this model, the internal dynamics is given by the first term and the interaction by the second.

By the Rigid Equilibrium Theorem\index{Rigid Equilibrium Theorem } \citep[Theorem 14.9]{GS2023}, a crucial ingredient for synchronization of the
genes in a fiber is that the input functions are the same within
the fiber. This is realized if the Hill functions\index{Hill function } mediated by a given
TF are the same, i.e., determined by the same kinetic constants, which
must be independent of the target gene. We elaborate on this modeling
assumption and many others in detail below, using the FFF as a
prototypic example.

\subsection{A more realistic ODE model of the FFF}
\label{sec:more}

To understand the role of the parameters, we describe a quite general
model where all the constants and parameters defining the dynamics are
different. We use the FFF. As in \citep{karlebach2008modelling}, the
amount of product (measured by mRNA concentration) of the three genes
in the FFF---namely X = external regulator to the fiber, Y =
autoregulated TF in the fiber, Z = enzyme in the fiber regulated by
Y---are described by dynamical variables $(x(t), y(t), z(t))$
obeying the system of ODEs:

\begin{eqnarray}
  \label{eq:generalODE}
    \frac{dx(t)}{dt}&=&-\alpha_x \,x + f_{\rm ext}\nonumber\\
    \frac{dy(t)}{dt}&=&-\alpha_y \,y + \gamma_{y} \, f_{y}(x, y) \nonumber
    \\
    \frac{dz(t) }{dt}&=&-\alpha_z \,z + \gamma_{z}\, f_{z}(x, y) .
  \label{Equations:FFF}
\end{eqnarray}

Here $x(t), y(t)$ and $z(t)$ are the time-dependent concentrations of
gene products $i = $ X, Y, Z measured by the mRNA concentration in the
cell, $\alpha_i$ is the mRNA degradation rate constant for gene $i$, and
$\gamma_{i}$ is the maximal synthesis rate of the mRNA product of gene
$i$.

The input functions $f_{y}$ and $f_z$ of the genes Y and Z can be
modeled by different forms, e.g. motivated by statistical mechanics
models of TF binding based on binding states and their energies. In
our simplified model we follow 
\cite{kaplan2008}, who have experimentally determined gene input functions
for the carbon utilization system of {\it E. coli}, showing that the
contributions are approximately multiplicative (AND gates) in many
cases. Other gates can also be assumed: using OR gates
instead of AND does not break the symmetry as long as the same (AND/OR)
model is used for all genes in the fiber. Therefore we assume that
the input functions of gene X and Z are a multiplicative function of
the inputs from gene X and gene Y:
\begin{equation}
\begin{array}{lll}
  f_{y}(x, y) &=& f_{x\to y}(x)\, f_{y\to y}(y) \\
  f_{z}(x, y) &=& f_{x\to z}(x)\, f_{y\to z}(y)\\
\end{array}
  \label{Equations:FFF0}
\end{equation}
where $f_{i\to j}$ is the input function of gene $j$ characterizing
the binding probability of the TF expressed by gene $i$ on the DNA
binding site of gene $j$. Thus, for instance, the input function of
gene Y is a logical AND function of the input function from gene X,
$f_{x\to y}(x)$, and from gene Y, $f_{y\to y}(y)$ through the AR loop.

The existence of synchrony in this model of FFF requires a 
uniformity assumption:\index{uniformity assumption } the functional forms of the input functions in the
fiber must be of the same type. For instance, we cannot mix a Hill
function in one link with a Boolean function in another. Typically
\citep{karlebach2008modelling,alon2019,klipp2016book,kaplan2008}, 
the functional form of the single input functions is fitted by
Hill functions.  Then we model all input functions assuming Hill
interactions of the following form \citep{alon2019}:
\begin{equation}
\begin{array}{lll}
  f_{x\to y}(x) = \frac{x^{n_{xy}}}  {K_{xy}^{n_{xy}} + x^{n_{xy}}}, &\quad &
   f_{y\to y}(y) = \frac{ y^{n_{yy}}} {K_{yy}^{n_{yy}} +y^{n_{yy}}}, \\
  f_{x\to z}(x) = \frac{x^{n_{xz}}}  {K_{xz}^{n_{xz}}+x^{n_{xz}}},   & \quad &
  f_{y\to z}(y) = \frac{y^{n_{yz}}}  {K_{yz}^{n_{yz}} + y^{n_{yz}}}.
\end{array}
  \label{Equations:FFFa}
\end{equation}
These functions are determined by a set of kinetic parameters. In general, they can be written as:
\begin{eqnarray}
  f_{i\to j}(x_i) = \frac{x_i^{n_{ij}}}  {K_{ij}^{n_{ij}} + x_i^{n_{ij}}},
    \label{Equations:FFFc}
\end{eqnarray}
where $x_i$ represents gene $i$, $K_{ij}$ is the dissociation constant
between TF $i$ and the promoter DNA region of gene $j$ (that is, the
TF concentration at which half of the binding sites are occupied), and
$n_{ij}$ is the cooperative Hill coefficient of the input function of
genes $j$ representing the cooperative binding of the TF $i$ at the
DNA binding site of gene $j$. Finally, in (\ref{eq:generalODE}),
$f_{\rm ext}$ represents an external input to gene X that is
irrelevant for the existence of synchronization in the FFF, and
exemplifies the input regulation of gene X by the rest of the
network. Genes Y and Z may also have outgoing links, but they do not
appear in the dynamical equations because their effect shows up only
in the dynamics of other nodes.

\section{The non-idealized ODE model does not lead to synchronization}

The dynamical system represented by
(\ref{Equations:FFF})--(\ref{Equations:FFFa}) with all the kinetic
constants different has no symmetries and no synchrony subspaces. That
is, the graphical representation of the dynamical system
 (\ref{Equations:FFF}) does not have any fibration symmetry; the
input trees of distinct genes are not isomorphic. This is because an admissible graph representing these equations will have all
the edges/arrows in the graph different since they represent
different interaction terms in the ODE.

In the graph fibration formalism,\index{fibration !formalism } two arrows are `the same' if and
only if the propagated state of the source node is processed in the
same way by the respective target nodes. Different types of arrows can
occur, so the `output' effectively depends on the target. Indeed, in
this formalism, the only output from a node is propagation of its state
to the target. However, different targets {\em process} this state in different ways, specified by the arrow types that determine the
components of the ODE. Possibly different signals from a node $c$,
depending on its state $x_c$, are packaged into different component functions $f_c$ of the admissible ODE. Thus,
arrows indicate the `influence network'\index{network !influence }: which nodes
affect which, and how they do so. See Section \ref{SS:SMEq}. 

Therefore, in principle, this dynamical system cannot sustain any
synchronized activity of its genes (except in extremely special
`non-generic' circumstances).  In the underlying FFF graph, all
links have different strengths $\gamma_i$, and the same goes for the other kinetic
constants that define the input functions. Different nodes would
receive different input signals, destroying synchrony.  Thus, it is
clear that there can be no isomorphic input trees between the genes in
such a graph. The Rigid Equilibrium Theorem \citep[Theorem 14.9]{GS2023} makes this statement precise, with a rigorous proof.

\section{The uniformity assumption: idealized FFF model leads to synchronization}
\label{ideal}

Recall that the condition for the existence of isomorphisms between
input trees is a local isomorphism in which nodes are mapped to nodes
and edges are mapped to edges, one-to-one.  In contrast, in the
ODE ~(\ref{Equations:FFF})-(\ref{Equations:FFFa}), all parameters
and input functions are different. In a network representation of the
FFF circuit representing
(\ref{Equations:FFF})--(\ref{Equations:FFFa}), each edge
represents a different class of interaction given by their different
weights $\gamma_{ij}$ or different input functions $f_{i \to
  j}$. Therefore, in (\ref{Equations:FFF})-(\ref{Equations:FFFa})
an isomorphism cannot exist between Y and Z since the links cannot be
mapped between the input trees representing the activity of the genes.

To obtain an isomorphism between input trees, corresponding links
mapped by the isomorphism must be of the same type.  In the case of
TRNs, these links must represent the same strength of interaction and
the same input function. We thus arrive at the {\it `uniformity
  assumption'}\index{uniformity assumption }
 required for the existence of perfect symmetry
fibrations and perfect synchronization. To satisfy this condition we
require an idealized system of equations where the single gene input
functions and parameters (binding constants, synthesis strengths and
degradation strengths) are the same for all genes that belong to a
fiber.  We do not require further equalities between parameters for
distinct fibers.  Specifically, we require:

\begin{itemize}
  \item The strengths of the maximal synthesis rates of genes in the
    fiber are the same. For the FFF,
    \begin{equation} \gamma_y =
      \gamma_z \equiv \gamma.
    \end{equation}
    The degradation constants are the same as well to guarantee the same internal dynamics, otherwise, nodes will have different types, breaking the symmetry:
\begin{equation}
  \alpha_y = \alpha_z \equiv \alpha,
\end{equation}
but they can be absorbed into the definition of $\gamma_i$.
    
  \item The Hill functions $f_{i\to j}$ producing the outputs of gene
    $i$ are of the same type and are characterized by the same
    parameters. Thus, the single Hill function parameters in $f_{i\to
      j}$ depend only on the properties of the source $i$th TF, but
    not on the properties of the target binding site at gene $j$. In
    the FFF, for example, for the input trees of genes Y and Z to be
    isomorphic, we require the input functions $f_{y \to y}$ and $f_{y
      \to z}$ to have the same shape; a condition that is achieved by
\begin{equation}
    K_{yy} = K_{yz} \equiv
    K_y,
\end{equation}
and Hill cooperative coefficients
\begin{equation}
n_{yy} = n_{yz} \equiv
n_y.
\end{equation}
Additionally, the outputs of gene X must be the same:
\begin{equation}
K_{xy} = K_{xz} \equiv K_x \,\,\,\, \mbox{\rm and} \,\,\,\, n_{xy} =
n_{xz} \equiv n_x.
\end{equation}
The outputs of X and Y need not to be the same, though, so in general
$K_x\neq K_y$ and $n_x\neq n_y$.
 \end{itemize}
    
Under these uniformity conditions, the AR link Y $\to $ Y has the same
strength and type (input function) as the link Y $\to$ Z, and X also
regulates Y and Z in the same way. Only under these conditions can an
isomorphism between Y and Z occur, and perfect synchronization between
Y and Z arise in the FFF:
\begin{equation}
  y(t)=z(t).
  \end{equation}
In general, for an isomorphism in a fiber to exist and to guarantee
synchronization, it is required that for each link received by one
gene in a fiber, there exists a link received by other genes in a
fiber of the same strength and interaction type.

While this condition may seem very restrictive, we argue below that it
is not difficult to be satisfied in TRNs if the interaction strengths
between a TF and a DNA binding domain depend mainly on the source TF
and less on the DNA structure of the promoter region. If so, we can
drop the target-dependence on the coefficients in the gene input
functions and use: $K_y$ and $n_y$, and $K_x$ and $n_x$.

Assuming this uniformity within fibers, we now consider an idealized
ODE model with symmetrized gene input functions.  The ODE
~(\ref{eq:generalODE}) is then replaced by:
\begin{equation}
\begin{array}{lcl}
  \frac{dx}{dt}&=&-\alpha_x \,x + f_{\rm ext}, \\
  \frac{dy}{dt}&=&-\alpha \,y + \gamma \,\, f_x(x)\,  f_y(y), \\
  \frac{dz}{dt}&=&-\alpha \,z + \gamma \,\, f_x(x)\,  f_y(y), 
  \label{Equations:FFF2}
\end{array}
\end{equation}
with idealized input functions:
\begin{eqnarray}
  f_{x}(x) = \frac{x^{n_x}} {K_{x}^{n_x} + x^{n_x}} ,&\,\,\,\,\,\,\,&
   f_{y}(y) = \frac{y^{n_y}} {K_{y}^{n_y}+ y^{n_y}}.
  \label{Equations:FFFb}
\end{eqnarray}

As discussed in Chapter~\ref{chap:fibration_1}, under
these assumptions, genes Y and Z
belong to the same fiber, and there is a synchronous solution $z(t) =
y(t)$.  It can be shown that this solution is more stable and has a
larger basin of attraction for the system than other solutions
\citep{lerman2015b,nijholt2016}, so it is highly probable that
\begin{equation}
\lim_{t\to\infty} (z(t)- y(t)) = 0.
\end{equation}

So, under these uniform assumptions, gene synchronization is achieved
via a symmetry fibration. It is still reasonable to ask: where on Earth
could we find a system that satisfies all these conditions on the
parameters so precisely?

\section{Breaking of uniformity}
\label{breaking1}

In reality, the input functions will not be identical, leading to
symmetry breaking\index{symmetry breaking } in the synchronization of the expression profiles.
There are many possible reasons for this. Below, we discuss some of
them, and in Section \ref{sec:necessary}, we argue how this uniformity
assumption\index{uniformity assumption } might be realized. This is fundamentally a modeling
assumption, and like all such, it should be judged by whether its
predictions can be verified, not by how closely it mimics reality. A
map that is the same size as the territory is usually useless.

First, we enumerate what could go wrong for perfect symmetries in
biological networks.
\begin{enumerate}
\item The parameters of the gene input functions depend, among other
  things, on promoter gene sequences and ribosome binding sites and
  can be expected to be similar between genes within a transcription
  unit (e.g., in an operon with a single promoter). Differences can be expected between genes in
  different transcription units (but belonging to the same gene
  fiber). Also, different AND/OR gates
  \citep{alon2019,kaplan2008} and combinatorial gates can affect the
  existence of symmetries.
  \item Our simplified model ignores many biochemical details
    (e.g. gene-specific mRNA lifetimes) that may alter the expression
    profiles of individual genes, and hence reduce correlations
    between different profiles.
\item Since gene expression is inherently stochastic, even larger
  deviations from coexpression are expected to arise at the
  single-cell level. Again, this particularly concerns genes that
  reside in one fiber, but in different transcription units.

\item We have assumed that mRNA production rates depend on TF
  activities via simple Hill functions. For each gene, the effects of
  different TFs are assumed to be multiplicative. The product of Hill
  functions represents a simple TF binding kinetics, and can be seen
  as a simplified version of the gene input functions
  \citep{bintu2005}.  Multiplicative gene input functions were
  experimentally observed in \emph{E.~coli} \citep{kaplan2008}.
  However, the combinatorial nature of these interactions
  \citep{buchler2003} implies that many other gates, for instance OR
  gates, are possible \citep{alon2006}. Again, the main assumption to
  keep the symmetry fibration and consequent synchronization is that
  these gates are the same for genes inside the fiber, which may not
  be true in general.
\item Degradation was assumed to be linear and non-regulated, with
  identical degradation constants for all genes in the fiber. For
  concentration variables, degradation can effectively include
  dilution effects in growing cells. In fast-growing bacteria, and
  assuming long-lived gene products, the degradation constant is
  approximately given by the cell's specific growth rate. The
  assumption of linear, non-regulated degradation of mRNA and protein
  has often been made in models, including NCA and the ODE models used
  to study the dynamics of network motifs
  \citep{alon2019,liao2003}. Typically, bacterial protein lifetimes
  are much longer than the cell cycle period, while mRNA lifetimes are
  still considerably smaller. Long protein lifetimes have been shown
  experimentally in \emph{E.~coli}, \citep{klipp2016book}.
  
\item Importantly, the activity of TFs can be modulated by ligands
  whose concentration carries information about the state of the
  cell. In the TRNs studied here, these additional regulations are
  ignored. Considering them may lead to a further subdivision of gene
  fibers, and to a predicted desynchronization of genes that are
  currently predicted to be synchronized.  These effectors are
  obtained by searching for bacterial growth experimental conditions
  where fibers are activated, as explained below.

\item Our network reconstructions are certainly incomplete, leading to
  errors in fibrations and to mispredictions about
  coexpression. Adding another TF to the network may destroy the
  symmetry of a fiber and lead to its subdivision, entailing a change
  in predicted coexpression patterns.
\item We consider the TRN in isolation. In reality, it is part of a
  larger network consisting of several layers (including
  transcription, translation, protein interactions, and metabolism,
  with different types of regulation arrows connecting them).
  Fibration analysis of such networks, acknowledging the tight
  feedback regulation between transcriptional regulation and other
  cellular systems, might bring additional insights about the coexpression
  of genes in the context of an entire cell. An extended fibration
  theory of integrated biological networks is still in its infancy.
\end{enumerate}

For all these reasons, synchronization of gene expression can only be
partial, if it exists at all.  It can be asked: Why do we start by
assuming idealized symmetries across all the Part I of this book, just
to later conclude that these symmetries may not exist in
reality? Our answer is that this procedure---deriving basic, general laws under
simplicity assumptions and treating deviations from them as necessary
but small corrections---has been a very successful strategy in physics
and has some mathematical justification.  A biological example occurs in the
repressilator \citep{elowitz2000}, a synthetic genetic
circuit. Here, the idealized model has perfect $\Z_3$ symmetry, leading
to a rotating wave state in which all nodes have the same periodic
waveform, with each node being one-third of a period out of phase with
the previous one. If the symmetry is broken by the different
parameters of the ODE---even strongly---the waveforms become
different. Still, their peaks remain very close to one-third of a period
out of phase with each other. On the whole, phase relations seem less sensitive to perturbations than amplitudes are.

Since detailed models of gene expression (including precise gene input
functions and covering all possible effectors) are lacking, we take
idealized symmetric input functions as a working hypothesis to be
tested experimentally {\em a posteriori}. If this hypothesis is accepted, we
can model TRNs where different genes in a fiber are modeled with the
same gene input function. This theory predicts the exact coexpression of
genes in each fiber. Deviations from this behavior can then be
interpreted as `weak symmetry breaking', which could be treated by
perturbation theory,\index{perturbation !theory } allowing for additional variation in gene
expression.

In this picture, we first imagine a hypothetical---biochemically
possible---cell in which all assumptions are justified and in which
perfect coexpression is realized. We then consider the real cell to be
a `weakly symmetry-broken' version, possibly evolved to achieve
specific patterns of non-coexpression for genes that otherwise would
be symmetric. Such symmetry breaking occurs, for example, if two
transcription factors in a fiber are differently regulated by
different interaction strengths. The resulting subdivided fibers allow
for more diverse and adapted expression profiles while retaining
coexpression under certain conditions.

\subsection{How physicists deal with this problem}
\label{perturbation}

Variations from ideal symmetry\index{symmetry !variations from } are ubiquitous in nature.  As 
discussed in Section \ref{sec:breaking-physics}, there are two major
types of symmetry breaking in physics: spontaneous and forced (or
induced). Spontaneous symmetry breaking\index{symmetry breaking !spontaneous } occurs when the model remains
symmetric, but a fully symmetric state becomes unstable so that a less
symmetric state bifurcates. There is an extensive mathematical theory
of pattern formation via spontaneous symmetry breaking; see, for
example \citep{stewart2003book}. Forced symmetry breaking\index{symmetry breaking !forced } occurs when
the model itself becomes asymmetric, for example, by parameters ceasing
to be exactly equal. This type of symmetry breaking has also been
widely studied.

The kind of symmetry breaking we are dealing with in this
chapter is of this forced kind, and it is not the same as that
discussed in the electronic circuit analogy in Chapter
\ref{chap:breaking}. That symmetry breaking is discrete and breaks
completely the functionality of the circuit by adding different
regulators to a symmetric circuit.  The present breaking is dynamical
and occurs smoothly as the different parameters of the ODE model are
moved away from the uniform assumption of all being equal. This
dynamical loss of synchronization by different parameters can be
treated by perturbation theory,\index{perturbation !theory } and it does not represent a new
principle of functionality of the cell. It is just that cells are noisy
and we must take this into account.

Variations from exact symmetry can be found
in all theories that are computable in physics \citep{landau1977quantum},
including the theory of the Higgs boson and the discovery of new
hadrons by Murray Gell-Mann \citep{weinberg1995,gellmann1995,georgi2018}.
Assuming symmetries but acknowledging that they are broken has been a
typical step in many theoretical physics discoveries.  In a
theoretical model, we first identify the exact symmetry and then
proceed by perturbation expansion to describe the real system.  This
is a successful theoretical procedure in physics, and we believe that
it could bring fruitful results to biology.

Indeed, we could argue that no symmetry is exactly realized in nature,
or is not broken. Even in the SU(3) flavor symmetry of chromodynamics
in the Standard Model\index{Standard Model } of particle physics, the constituent masses of
quarks up, down, and strange are not perfectly symmetric ($m_u = 336$
MeV and $m_d=340$MeV, $m_s=486$Mev). Despite this asymmetry between
masses, by assuming perfect symmetry (all masses are the same), we can put
the three quarks into a triplet (which assumes that they can be
interchanged by an SU(3) symmetry transformation) and predict the
existence of all the composite hadrons in the octet of\index{Gell-Mann, Murray }
Gell-Mann---such as protons, neutrons, and the heavier baryons---with
high precision \citep{gellmann1995,weinberg1995,georgi2018}. The
remaining small physical differences in quantum behavior arising from
the asymmetries are then treated with perturbation theory\index{perturbation !theory } from the
ideal symmetric case.  In biology, this perturbative theoretical
approach can be readily translated to describe the cell's biological
network's departure from ideal symmetry, which can be quantified and
made computable by perturbation theory, and therefore, point to the
efficacy of the perturbation regime.

We elaborate below on how this symmetry breaking in gene fibers can be
handled, at the level of perturbation theory,\index{perturbation !theory } in the `weak symmetry
breaking regime'.  If the heterogeneities between gene input functions
are small, they can be handled mathematically by a perturbation
expansion around the symmetric state.

Variability is an intrinsic property of biological systems and a main
driver for phenotypic change among organisms and species.  Biological
variations by symmetry breaking may explain phenotypic variations
since ideal symmetries must be broken for evolution to exist.

If a biological system is close enough to the idealized symmetry, we
can invoke conditions of hyperbolicity\index{hyperbolicity } \citep{guckenheimer1983},
discussed in Chapter \ref{chap:stability}, to infer that robust
approximately synchronous patterns of expression exist. A theory of
perturbation applied for approximate fibrations has been developed by
\cite{coram1977}.  This theory shows that a perturbative expansion can
be rigorously controlled, and the `loop' corrections (referring to
theoretical methods used to deal with perturbation theory)\index{perturbation !theory } can be
analytically computed, in principle, to any order.

This is stated in Theorem 2.6, page 282 in \citep{coram1977}: `If $p:
E \to B$ is an approximate fibration, and if $E$ and $B$ are
manifolds, then for every $b \in B$, $F_b$ satisfies the small loops
condition.'  Thus, approximate fibrations are computable and are
backed up by rigorous mathematical results in the framework of
category theory.  Experimental evidence, e.g. Fig. \ref{fig:carbonc},
suggests that symmetry breaking must be weak because weak symmetry
breaking from the input functions implies weak breaking of synchrony.
An extended theory of quasi-fibrations developed by \cite{boldi2021},
and pseudo-balanced colorings developed by
\cite{leifer2022symmetry-driven} and discussed in Section
\ref{algo-pseudo}, can characterize biological networks beyond
the idealized cases presented here. We have proposed an analogous
theory of pseudosymmetry\index{pseudosymmetry } groups to explain pseudosymmetries in the
{\it C.elegans} connectome in  \citep{morone2019symmetry}.

It should be recalled, though, that fibrations\index{fibration } form groupoids,\index{groupoid } not
groups.  Such a theory would complement and augment the validity of
the present approach to networks that may not be as completely
characterized as the bacterial networks studied here. A
pseudofibration\index{pseudofibration } theory would be especially handy to describe
symmetries in human biological networks that are notoriously less
complete and, of course, more complex than bacterial
networks. Mathematical efforts in this direction are welcome. The next
chapter discusses some of them.

\section{Is symmetry necessary in biological networks?}
\label{sec:necessary}

\index{symmetry !necessity for } 
From a mathematical point of view, the answer is `yes'. If there is
robust and stable synchronization in a system with a hyperbolic
equilibrium, then the only way to obtain such synchronization is via
the fibration symmetries in the underlying network. The Rigid
Equilibrium Theorem \citep[Theorem 14.9]{GS2023} proves that there is no
other way to obtain robust synchronization other than with (fibration) symmetries;
see  Chapter~\ref{chap:fibration_1}.

In general, dynamical states of synchronization without any underlying
network symmetry is `fragile' in the sense that the states are
destroyed by small perturbations of the ODE that respect network
structure \citep{stewart2006}. 
Approximate synchrony---as is typically observed in reality---need not be destroyed. However, approximate synchrony is different from a complete lack of synchrony, and it requires explanation. The simplest explanation is the approximate symmetry of the network,
which takes us back to idealized models.

More robust synchrony can be
imposed by making additional modeling assumptions. A common one is to
assume there is a distinguished rest state $0$ and impose the
condition that the function $f$ in the ODE $\dot x = f(x)$ satisfies
$f(0)=0$---which happens, for instance, in \ref{E:MSF} because the coupling assumed is generalized diffusive. In such cases, all nodes can synchronize at $0$. However, in
the absence of such special constraints, robust synchrony can only be achieved through symmetries in the underlying network.  Armed
with this mathematical result, we now elaborate on the experimental
evidence for symmetries and synchronization in biological networks.

We address how to overcome the limitations of the application of the
fibration model to biological network, as described in Section
\ref{breaking1}.  The following discussion is very limited and should
be seen as an initial speculation, definitively not definite, about the
problem.

Our ODE model of gene regulation is a simplified one, as discussed in
Section \ref{breaking1}, many biochemical details are missing and
identical gene input functions are assumed for all the outgoing links
from a given gene. Since none of these assumptions is valid in
reality, can we still claim that symmetries in network structures
entail symmetries in gene expression?  Instead of making this claim,
we assume that these idealized symmetries will always be broken in
reality, which precludes exact synchronization and introduces
additional variation in expression.  The exact symmetries of an
idealized network is interpreted as predicting {\em approximate}
symmetry/synchrony in the real biological system modeled by that
network.

What are the main factors that lead to such symmetry breaking?\index{symmetry breaking } Do
observed gene expression patterns, despite these effects, still
display a vestige of the unbroken symmetry? And, since synchronized
expression patterns are beneficial and widespread in biological
systems, could there be evolutionary pressures that counter symmetry
breaking?

\subsection{Weak symmetry breaking by differing gene input functions}

To be coexpressed precisely, genes in a fiber must share the same
regulatory input functions as in the idealized models
(\ref{Equations:FFF2}) and (\ref{Equations:FFFb}). In contrast,
differences in these functions or their quantitative parameters, as in (\ref{Equations:FFF}) and (\ref{Equations:FFFa}), lead to
weak symmetry breaking, i. e.,  incomplete coexpression.  The simple
input functions used in our model describe TF binding, but if we were
to fit them to expression data, they would actually capture the entire
process of transcriptional and translational regulation, including
details such as mRNA stability, ribosome binding, and protein
stability. Given all these details, our effective `gene input
functions' would certainly differ between genes (with the exception of
genes in the same operon in bacteria) even when the remaining
differences occurring post-transcriptionally may be negligible. In
all other cases, there is no biochemical reason why gene input
functions should be identical.

Thus, on the one hand, in reality, we can expect symmetry breaking\index{symmetry breaking } to
cause genes in a fiber to become desynchronized. On the other hand, we
observe measurably higher coexpression within fibers as in
Figs. \ref{fig:carbonc} and \ref{BMC:Fig10}.  Hence, if gene input
functions within fibers are found to be significantly similar; this
may indicate selection pressures opposing potential symmetry breaking,
and would, at the same time justify gene fibers as a theoretical tool
of analysis.

\subsection{Similarities between measured gene input functions}
\label{breaking4}

We first recall that perfect synchronization\index{synchronization !perfect } in a fiber requires only
conditions on the input links to the genes in the fiber.  This means
that a given input function's shape and parameters should depend on
the source gene.

This point is discussed in Section \ref{ideal} from the conditions
leading to symmetrized equations
(\ref{Equations:FFF2})--(\ref{Equations:FFFb}).  Biological
evidence for this kind of shared input function among genes can be
found in the work of \cite{kaplan2008}. They
measure the input functions of genes in the well-characterized gene
regulatory system of sugar catabolism in \emph{E. coli}. Direct
measurements of gene input functions, describing the effects of
effector molecules (acting on a transcription factor) on subsequent
gene expression, have been performed.

The observed input functions were quite diverse, but a careful
analysis of the data in \citep{kaplan2008} reveals that the reported
diversity does not affect the existence of fibration symmetries. On
the contrary, the data support the existence of symmetrical input
functions in precisely those genes needed for the fibrations to
occur. As we saw in Section \ref{ideal}, the requirement of uniform
input functions for fiber synchronization is that the input function
of a given TF, or a pair of TFs that regulate a particular set of
genes, depends on parameters of the input functions of the source TFs
and not on parameters specified by the binding sites of the target
genes.  That is, the input functions over a set of genes that are
regulated by the same TF are expected to be approximately similar,
even though other input functions of other TFs might be
different. This may not be satisfied in real systems and may lead to
weak symmetry breaking from the predicted fibers.

\begin{figure}
  \centering \includegraphics[width=0.6\linewidth]{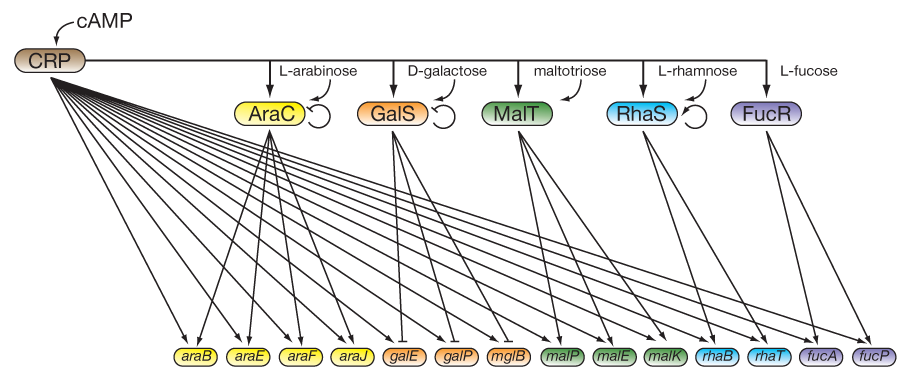}
  \includegraphics[width=\linewidth]{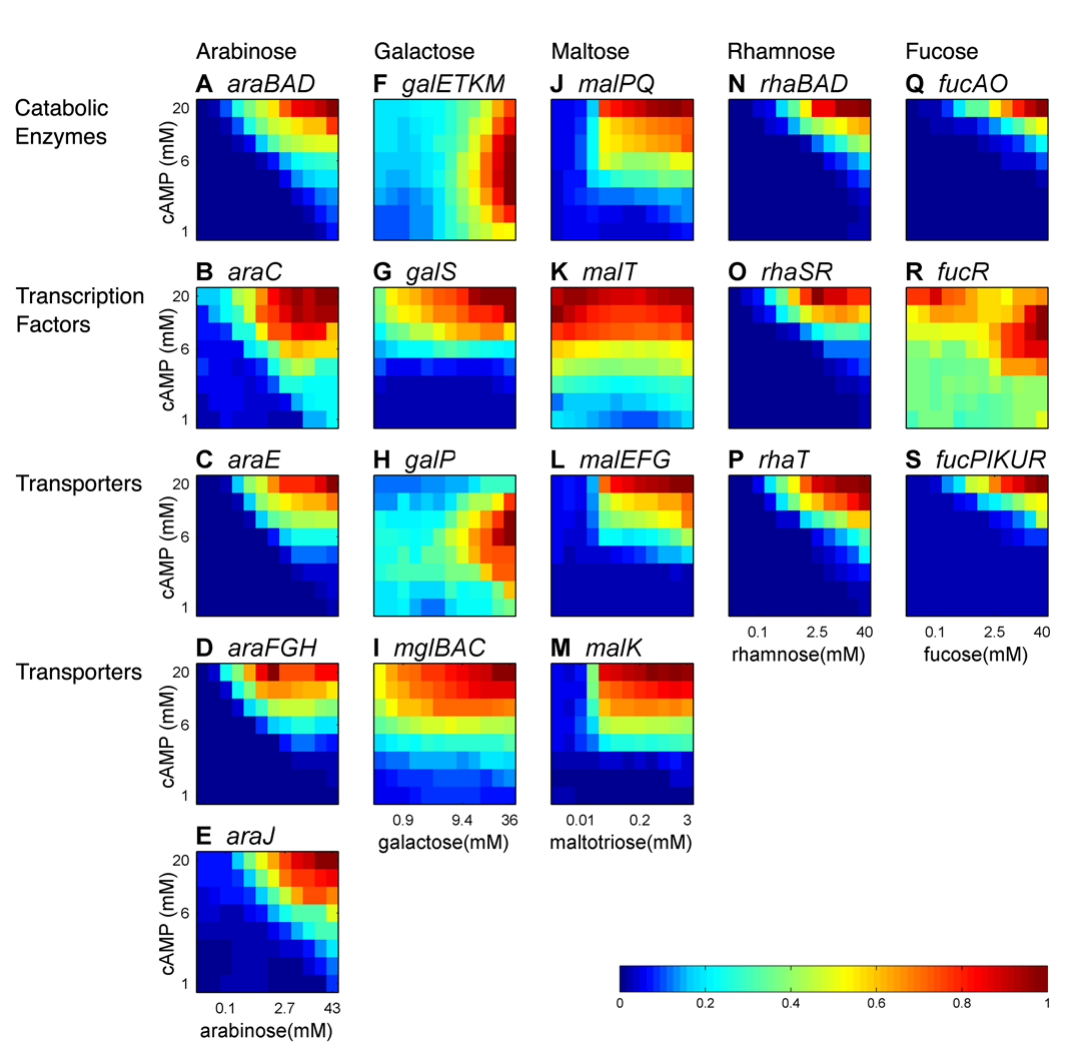}
  \caption{\textbf{Study of input functions by \citep{kaplan2008}}. {\em Top}:
    Regulon of {\it crp} in the {\it E. coli} sugar consumption network.  {\em Bottom}: Input functions of {\it E. coli} sugar
    genes. {\bf (A–E)} arabinose system; {\bf (F–I)} galactose system;
    {\bf (J–M)} maltose system; {\bf (N–P)} rhamnose system; {\bf
      (Q–S)} fucose system. Input functions are defined as the
    promoter activity at each of the 96 combinations of the two input
    signals, cAMP and the sugar. The x- and y-axes correspond
    respectively to sugar and cAMP concentrations in mM.  
    Figure and caption reproduced from \citep{kaplan2008}. Copyright \copyright ~ 2008, Elsevier Inc.}
  \label{fig:kaplan}
\commentAlt{Figure~\ref{fig:kaplan}: 
Described in caption/text. No alt-text required.
}
\end{figure}

This situation is studied in \citep[Fig. 2]{kaplan2008}, discussed
in \citep[Fig.~9.18]{klipp2016book} and reproduced here in
Fig. \ref{fig:kaplan}. In this figure, the same
    cAMP levels are used in all input functions, and the same sugar
    levels are used in each column. Promoter activity is the rate of
    GFP fluorescence accumulation per OD unit in exponential
    phase. The figure shows promoter activity normalized to its
    maximal value for each promoter. Rows arranged by biological role.  
    This figure shows that the input functions of
\emph{E. coli} sugar genes, regulated by the same TFs, are similar,
even though the claim in \citep{kaplan2008} is that they are diverse
and quite different.  This claim is correct when we compare all of
the input functions across different TFs. But when we look at them
per TF to see whether the input functions are similar for a given TF,
i.e., column by column in Fig. \ref{fig:kaplan} (bottom), then the
similarities are evident.

\begin{figure}
  \centering
  \includegraphics[width=\linewidth]{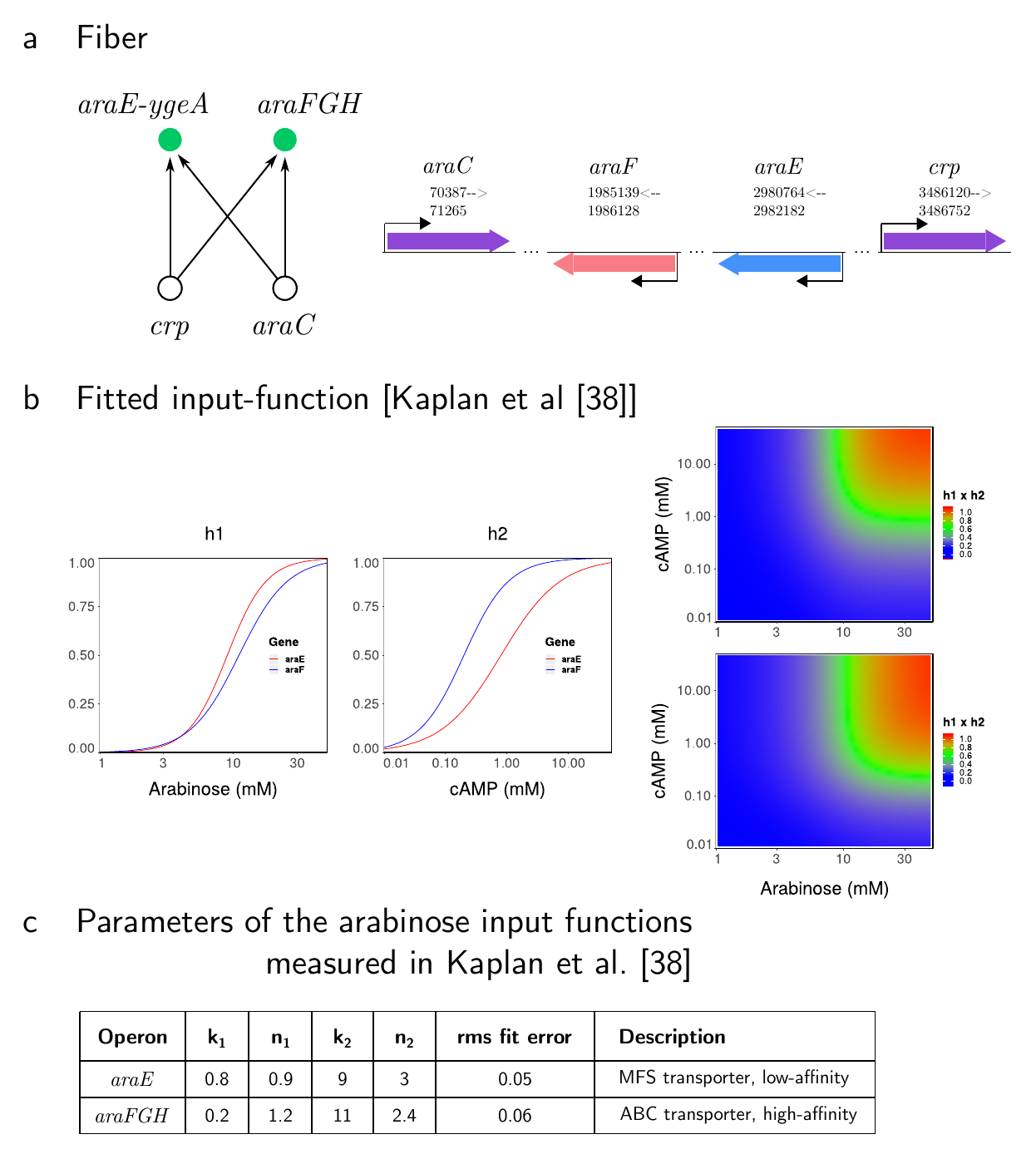}
  \caption{\textbf{Input function study of the arabinose consumption
      circuit from \citep{kaplan2008}}.  (\textbf{a}) Circuit producing
    arabinose with its fiber in green.  (\textbf{b}) Input functions
    measured experimentally and modeling in \citep{kaplan2008}. (\textbf{c}) Coefficients measured for the Hill input functions of {\it
      araE-ygeA} and {\it araFGH} promoters. Data 
    reproduced from \citep{kaplan2008}. Copyright \copyright ~ 2008, Elsevier Inc.}
  \label{fig:input}
\commentAlt{Figure~\ref{fig:input}: 
(a) Fiber has four nodes: araE-ygeA and araFGH (green);
crp and araC (white). Arrows from each white node to each green node.
(b) Fitted curves to the observed input function are sigmoidal increasing.
(c) Measured parameters of arabinose input functions have very low
rms fit error (0.05, 0.06).
}
\end{figure}

We can observe this behavior in the arabinose\index{arabinose } circuit of
Fig. \ref{fig:input}.  Arabinose is the molecular effector that
activates AraC. AraC is also induced by \emph{crp} (which is activated
by a small molecule cAMP) and regulates the catabolic enzymes {\it
  araBAD} and the transporters {\it araE-ygeA, araFGH, araJ} that
bring the sugar inside the cell. The first column of panels in
Fig. \ref{fig:kaplan} shows the set of input functions measured
experimentally as a function of the activator effectors cAMP and
arabinose.  The input functions for the genes \emph{araBAD, araE-ygeA,
  araFGH, araJ} and \emph{araJ} are similar, as noted in
\citep{kaplan2008}. That is, the input functions of the target genes
depend mainly on the source genes, {\it araC} and {\it crp}, but are
the same across the target genes: \emph{araBAD, araE-ygeA, araFGH,
  araJ} and \emph{araJ}. The input functions can be factorized into
simple functions of Crp and AraC (\ref{fig:input}.

Figure \ref{fig:input} reproduces some of the results of
\citep{kaplan2008}.  The circuit of the fiber \emph{araE-ygeA, araFGH}
with external regulators, Crp and AraC is shown in
Fig. \ref{fig:input}a. The genes in the fiber and regulators are not
contiguous in the genome:
{\it araE-ygeA} is located at position 2,980,764 in the genome
and {\it araFGH} at
1,985,179; their regulators are at 70,387 ({\it araC}) and 3,486,120
(Crp), so the similarities of input functions are not due to shared
promoters or short-distance effects in the genome \citep{junier2016}.

The parameters defining the Hill functions,\index{Hill function } obtained in
\citep{kaplan2008}, are shown in the table of Fig.~\ref{fig:input}c
with the shapes of the Hill functions plotted in the figures.  The
Hill input functions of {\it araE-ygeA} and {\it araFGH} are almost
identical, and the coefficients for both Hill functions are equal
within error bars. For instance, the cooperative coefficients
$n_1=3\pm 1$ and $n_2=2.4\pm 0.4$ are within error bars, and so are the
other coefficients---except for the dissociation constants $K_1 = 0.8
\pm 0.6$ (mN) and $K_2 = 0.2 \pm 0.1$ (nM), which are close to but
slightly beyond the error bars.

The Hill functions fitted with these parameters are shown in
Fig. \ref{fig:input}b, supporting the conjecture of symmetrically related Hill
functions in the fiber.  The input functions can be factorized into
simple functions of Crp and AraC (Fig. \ref{fig:input}b), as assumed
in (\ref{Equations:FFF2})--(\ref{Equations:FFFb}).  The effects of
CRP and AraC are multiplicative on each regulated gene, and each
multiplicative contribution is well described by a Hill function of
the type used in the ideal model.

This result is corroborated by other gene input functions. For
example, the input functions in the galactose system show strong
pairwise similarities (see the second column in
Fig. \ref{fig:kaplan}). The input functions of the operon {\it
  galETKM} and the gene {\it galP} are similar to each
other. Likewise, the input function of {\it galS} is similar to
that of {\it
  mglBAC}, yet they are quite different from the {\it galP} functions.
These similarities are conditions needed for synchronization in the
galactose fiber, which is split in two. Likewise, the remaining carbon
circuits in Fig. \ref{fig:kaplan} suggest the isomorphisms in the
input functions needed for the symmetry fibration. In the maltose
circuit, the input functions of {\it malPQ, malEFG, malK} are similar,
yet different from {\it makT}; in the rhamnose circuit, {\it rhaBAD,
  rhaSR, rhaT} are all similar; and in the fucose circuit, {\it fucAO}
and {\it fucR} are relatively similar, explaining the synchronization
observed in this carbon circuit in Fig.~\ref{fig:carbonc}.

Thus, the observed Hill input functions in this circuit are consistent
with uniformity conditions for a symmetry fibration.  These experiments
support the existence of local equivalences between genes in a fiber
through uniform input functions.  Eventually, the effectiveness of the
fibration in capturing synchronization in gene coexpression should be
validated experimentally by coexpression pattern analysis.

\subsection{TF binding sites}

What could be the molecular origin of the similarity in the Hill input
functions observed above? Below, we offer a tentative answer to this
question, which could lead to ideas for experimental and analytical
tests of the molecular origins of the fibration symmetries.  We
elaborate on whether the mere existence of binding motifs\index{motif !binding } in the DNA
sequences of binding sites\index{binding site } (these motifs are in the DNA sequence and
should not be confused with network motifs) could lead to similar
gene input functions, as required by symmetry.  In principle,
\cite{alon2006} have shown that a gene input function can be changed
very flexibly by only small changes in the sequence. In principle,
these results could already rule out a molecular origin for symmetric
input functions such as binding motifs.

Indeed, the observed similarity is surprising because gene input
functions can depend on minute changes in the binding site sequences
\citep{alon2006}.  What determines a gene input function is a number
of factors that control the binding and unbinding properties of the
regulators to DNA \citep{klipp2016book}.  Likewise, promoter strengths
(and ribosome binding site strength) differ over orders of magnitude
\citep{zelcbuch2013}.  Furthermore, different TF binding sequences
allow for a range of binding strengths, corresponding to a wide range
of effective parameters in the gene input functions.

However, it seems likely that there should be a molecular explanation
of the symmetries as well. Confirming or denying this view would require
an analysis that compares TF binding site sequences or promoter
sequences across genes, within and between fibers, and that shows
significant differences between the two cases. Such an analysis is
beyond the scope of the present book. So, as it stands, the arguments
below are merely conjectural. The molecular basis of Hill functions
has been described in \citep{alon2006}, and it would be an interesting
problem to reconcile these results with the uniformity
assumptions\index{uniformity assumption } required for symmetries and synchronization that we impose here.

Among the different factors that determine a Hill function,\index{Hill function } below we
analyze existing data on the binding DNA motifs of the TF at the
promoter's nucleotide sequence \citep{klipp2016book} that may lead to
similar gene input functions. A given TF will bind preferentially to
promoter sequences that contain well-defined TF-binding motifs.  These
motifs are used in bioinformatics analysis to determine the binding
probability of a TF to DNA through position-specific scoring matrices
(PSSM). Such matrices are constructed using statistics on binding sites' DNA
sequences, characterized experimentally for a given TF
\citep{hertz1990,medina2011}. Evidence is seen in experimental studies
of DNA motif analysis of TFs, showing enriched sequence-logo binding
site DNA sequences. See, for instance, the TrpR repressor for
tryptophan biosynthesis in {\it E. coli} logo analysis 
\citep{medina2011}.

\begin{figure}
  \centering
  \includegraphics[width=\linewidth]{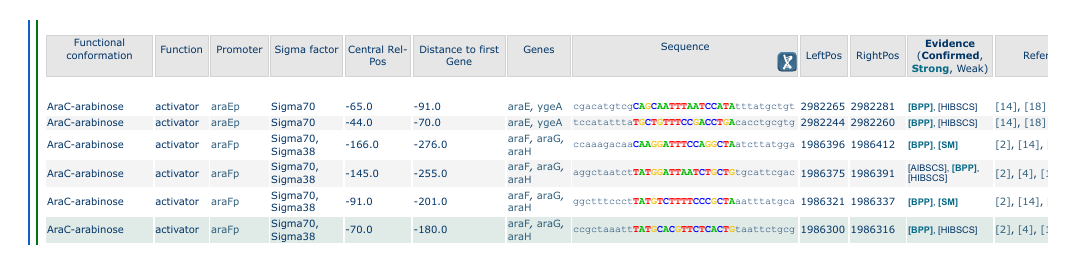}
\includegraphics[width=\linewidth]{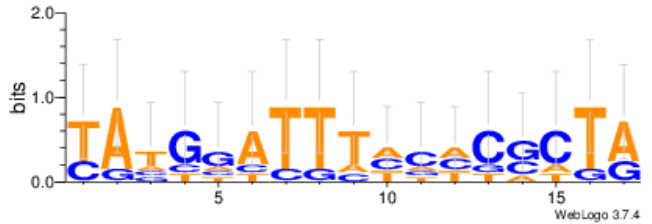}
\caption{\textbf{Logo analysis of the binding sites.} We perform a logo
  analysis using \citep{crooks2004} at promoters araEp and araFp for
  the TF AraC. (\textbf{a}) Sequences of binding sites of the promoters
  of {\it arae-ygeA} and {\it araFGH}. (\textbf{b}) Sequence logo
  analysis of these binding sites {\it araE-ygeA} and {\it araFGH}
  shows strong similarities between the two promoters, as demonstrated by the
  enrichment at several DNA sites. }
  \label{input2}
\commentAlt{Figure~\ref{input2}: 
Described in caption/text. No alt-text required.
}
\end{figure}

We analyze the sequences of the binding sites for the arabinose
circuit controlled by the TF AraC, whose gene input functions are shown
in the first column of Fig. \ref{fig:kaplan}. The TF AraC binds to two
promoters (it binds to more but we study only two).  One binding site
is promoter araEp, which controls the expression of the gene {\it
  araE-ygeA}, which expresses the proteins AraE and YgeA. The other
binding site that we study is promoter araFp which expresses the
operon {\it araFGH} producing proteins AraF, AraG, and AraH.  The
sequence, positions, and features of these two promoters are shown in
Fig.~\ref{input2} (top). There are two promoter sequences
for araEp and four for araFp.
Figure \ref{input2} (bottom) shows the sequence logo analysis from
\citep{crooks2004} of these sequences, revealing enrichment at
positions 1, 2, 4, 6, 7, 8, 9, 13, 14, 15, 16, 17.

Is the similarity in these DNA sequence motifs evidence for the
similarity of the input functions of AraC at these promoter sites?
First, we see whether this enrichment similarity is consistent with the
similarity in input functions across the first column in
Fig. \ref{fig:kaplan} obtained in \citep{kaplan2008}.  Here we just
point out that the existence of these well-defined DNA binding motifs
is consistent with \citep{kaplan2008} since the input functions are
quite similar; for instance, compare panel C ({\it araE}) and panel D
({\it araFGH}) in Fig. \ref{fig:kaplan}.

Are all these results also consistent with the conditions needed for
the validity of symmetry fibrations and the concomitant gene
coexpression observed experimentally?  This is not definitive. Many
other factors affect these binding strengths. But this partial
evidence points to an affirmative answer.  More studies are needed to
determine the molecular underpinning of these symmetries. We conjecture
that, eventually, the synchronization and fibration symmetries we
observe at the TRN level might be explained by symmetries underlying
the DNA sequences that control the binding and unbinding of molecules
to DNA for their regulation. This could shed light on the structure of
the gene regulatory code as discussed next.

\section{Selection pressures for symmetry}
\label{sec:selection_pressures}
\index{symmetry !selection pressure for }

Despite all the possible molecular biological reasons for symmetry
breaking, our study demonstrates gene coexpression within fibers as
long as meaningful sets of biological samples are considered (i.e.,
samples in which the genes are significantly expressed).  For gene
pairs within a transcriptional unit, this coexpression would not be
surprising. However, we found coexpression also in genes that are
controlled by different transcription factors. Thus, despite
inevitable asymmetries among gene input functions, the observed
symmetry breaking is not very strong.  There are two possible
explanations.  First, gene input functions (when normalized to the
typical TF levels) may just not be very diverse, so network structure
alone may determine coexpression\index{coexpression } patterns.  However, there may also be a
second, functional explanation, by which gene input function would be
specifically similar within gene fibers.

While some gene input functions in \emph{E.~coli} mapped by
\citep{kaplan2008} show diversity ranging between approximate AND and
OR gates, depending on relatively small variations of the promoter
sequence, the diversity can be grouped as shown in Section
\ref{breaking4}, and this grouping agrees with the conditions needed
for the existence of a fibration.

What might the evolutionary role of this similarity be? Evolutionary
pressure for robust gene coexpression would lead not only to the
emergence of fibers (which support partial coexpression), but also to
convergence of gene input functions within gene fibers during
evolution, to make this coexpression more precise. The gene input
functions of sugar genes reported by \cite{kaplan2008} point in this
direction: the genes activated by the regulator AraC (although
belonging to different operons) show consistently similar responses to
arabinose and cAMP, indicating similar gene input functions inside the
fiber. The same holds for the gene sets activated by TFs in the other
fibers in the circuits considered, GalS, MalT, RhaS, and
FucR. Strikingly, the self-activator RhaS shows the same expression
response as its target genes, exactly in line with the fact that they
belong to one fiber.

Finally, since edges in a graph are merely proxies for gene input
functions, and since the effects of TFs may gradually change during
evolution, the appearance and disappearance of edges, and the
evolution of gene input functions themselves are just two
descriptions of the same process. We may even extend this
conclusion. If selection pressures can make gene input functions
within a gene fiber converge for synchronization and function, we may
speculate that the same pressures that adjust gene input functions may
also act towards adding more genes to a fiber by adding or removing
input arrows.  The evolution of network structures and input functions
(and even of gene function itself) may go hand in hand. For example,
if gene products are involved in the same process, selection for
coexpression may first put them into one fiber and then make their
gene input functions converge (even before these genes may happen to
be placed in one operon).

\section{Towards a gene regulatory code}
\label{sec:isthere}
\index{gene regulatory code }

The {\it genetic code}\index{genetic code } is a set of rules engraved on the DNA sequence used
by the cell to translate information from DNA into proteins.  The
`letters' of this code are the codons\index{codon } that consist of triplets of base
pairs.

While the genetic code determines the amino acid sequence of the
protein from the DNA of genes, other parts of DNA, like the binding
sites of the TFs, determine where and when these TFs bind to produce
more proteins. The machinery that expresses these genes is controlled
by the {\it gene regulatory code}.\index{gene regulatory code } In turn, this regulatory code 
must be engraved in the structure of the gene regulatory network,
which has fibration symmetries.

The full information flow goes from DNA to DNA.  The genetic code
transmits information from the DNA sequence of the gene expressing the
protein to the amino acid sequence of the protein.  This protein is a
TF that binds to another DNA sequence at the promoter site of the
target gene, expressing another protein.  The information is then
transferred from the TF to the new protein expressed at the target
site. 

The transfer of information is then between two DNA sequences, that of
the source gene and the binding site of the regulated gene.  This
information flow is represented by an arrow in the regulatory graph,
and all the arrows together build the fibration-rich regulatory code.
It would not be surprising if we were to find symmetries in these
sequences as well, which would reveal the regulatory code underlying
gene regulation.\index{gene regulatory code }


\chapter[From Function to Structure: Network Inference Reconstruction]{\bf\textsf{From Function to Structure: Network Inference Reconstruction}}
\label{chap:function}

\begin{chapterquote}
Chapter \ref{chap:synchronization} deals with the structure $\rightsquigarrow $
function relation by testing how the fibration symmetries of the
structural transcriptional network determine the functional
synchronization observed in gene coexpression. This validation
largely relies on the completeness of the structural network. But real
structural networks in biology are rarely complete. On the other hand,
the functional network can be obtained for a fairly large number of
biological nodes and, in many cases, for all of them, e.g., gene
coexpression patterns of entire organisms. We use this information to
infer the structural network of connections using a symmetry-driven
reconstruction algorithm guided by balanced colorings obtained from
cluster synchronization of the biological units. Thus, the present
chapter deals with the inverse problem: how to use function $\rightsquigarrow $
infer structure. The algorithm reconstructs incomplete gene regulatory
networks from coexpression patterns as well as incomplete connectomes
from neural synchronization data. This represents a crucial
application of fibration theory to understanding biological pathways,
with potential applications for reconstructing pathways for drug development and the understanding
of disease.
\end{chapterquote}

\section{Biological networks are never complete}
\label{sec:incomplete}

The concept of symmetry is at the basis of physics, but we have seen
that it is not (yet) as pervasively adopted in biology.
Symmetries in biological networks have been very hard
to find since the time of Monod's first formulation of the symmetry
problem in regulatory networks \citep{monod1970symmetry}.
In our
opinion, three crucial aspects have contributed to determining the reduced
attention to the role played by symmetries in biological systems:
\begin{itemize}
\item Historically, the symmetries of a network were identified with
  automorphisms, a very rigid form of symmetry, rarely
  observed in biology. Furthermore, automorphisms are not the right
  tool to describe the notion of synchronization since they miss
  synchronies that only fibrations can identify.
\item Even playing with the more general symmetry of fibration,
  biological systems are never exactly symmetrical: the high
  redundancy of biological systems leaves space for imperfections and
  small asymmetries. These do not affect the overall working of the
  system, but fail to be captured by a rigorous mathematical definition
  of exact fibration.
\item Experimental observations providing the basis for constructing
  biological networks are inherently imperfect and incomplete, because
  biological systems are noisy and disordered. Therefore perfect
  symmetries are never realized in biology, reducing the chances of
  identifying the symmetries captured by either fibrations or
  automorphisms.

\end{itemize} 

For instance, Chapters \ref{chap:hierarchy_2} and \ref{chap:complex}
have shown that graph fibrations uncover symmetries in  gene
regulatory networks that are validated by synchronization patterns in
gene coexpression in Chapter \ref{chap:synchronization}. However,
these results were shown for one particular organism, the bacterium {\it
  E. coli}, that is well-studied with a gene regulatory network and
metabolism that has been almost completely mapped.
In general, this is seldom the case.  Structural networks are known only
partially because only some pathways can be studied and mapped.
This is particularly true for higher-level organisms like humans, where
only certain gene interactions are studied. The inherent
incompleteness and disordered nature of biological data preclude the
application of the fibration formalism \emph{as it is}. As a
consequence, the algorithms to identify fibrations discussed so far
may not uncover all the symmetries of the network.

\cite{boldi2021} and \cite{leifer2022symmetry-driven}  address
this problem by proposing an extended theory of quasifibrations and
pseudobalanced colorings, respectively (see Section
\ref{algo-pseudo}).  These theories attempt to capture more realistic
patterns of almost-synchronization of units in biological networks.

Based on the theory of pseudo-balanced coloring, the network reconstruction algorithm developed by \cite{leifer2022symmetry-driven} (described in Section \ref{sec:repair}) aims to identify the optimal modifications needed for a network to achieve a symmetrically balanced coloring. This algorithm does not require prior knowledge of the balanced coloring. However, as we discuss next, it is often the case that the balanced coloring is known {\it a priori} through cluster synchronization. This information can be invaluable for reconstructing missing links in an incomplete network. Consequently, this algorithm is simpler and more practical than the alternative presented in 
 Section \ref{sec:repair} \citep{leifer2022symmetry-driven}.

Indeed, while determining complete structural networks is a challenge,
obtaining complete functional networks for all genes in an organism
can be done routinely.  Current advances in systems biology make it possible to
measure the activity of the entire transcriptome of a given species.
For instance, the mRNA expressed by the $\sim$20,000 human genes can
be monitored simultaneously in cells using transcriptomic techniques
like microarrays or RNA-seq, even at the single-cell level 
\citep{klipp2016book}. In the brain, single neuronal calcium activity can reveal whole-body synchronization in small connectomes, while fMRI can reveal similar synchronization at the mesoscopic level of larger brains. These patterns reveal the cluster
synchronization emerging from the symmetries of the underlying
structural network, as seen in Chapter \ref{chap:synchronization}.

\begin{figure}[ht!]
	\centering
	\includegraphics[width=\linewidth]{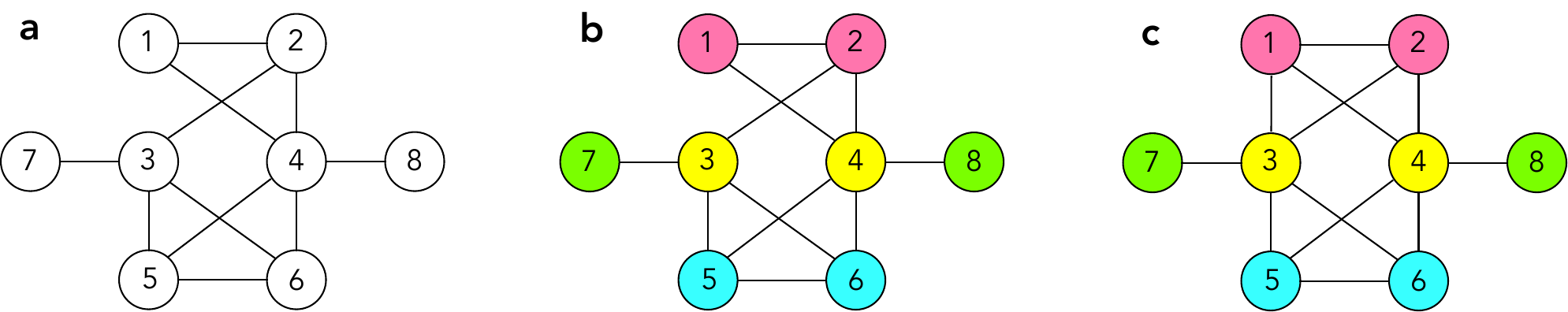}
	\caption{\textbf{Symmetry-driven inference algorithm informed by
            balanced coloring synchronization.} (\textbf{a}) 
            Uncolored structured network. This network is incomplete. (\textbf{b}) A balanced coloring is obtained experimentally, measuring the cluster synchrony.
            It fails to be balanced at nodes 1 and 3. (\textbf{c})
            Repairing the network by adding the minimal number of edges: an edge from 1 to 3
            makes the coloring balanced. 
            This example shows how a `missing' link can
          break a balanced coloring. If we know the
          coloring in advance, it should be easy to infer this
          missing link. Thus, an optimization algorithm that 
          looks for the minimal number of links to be added (or
          removed) from a network to satisfy a previously known
          balanced coloring would be an effective way to reconstruct the network.
          Such an algorithm makes use of fibration theory and 
          balanced coloring to solve the following optimization problem: find the
          minimal number of changes to edges.}
        \label{fig:Fig1-pseudo}
\commentAlt{Figure~\ref{fig:Fig1-pseudo}: 
Left: Graph with nodes 1-6. Edges 12, 14, 23, 24, 35, 36, 37, 45, 46, 48, 56.
Middle: same graph with nodes colored: 1,2 (red), 3,4 (yellow);
5, 6 (blue); 7, 8 (green). right same graph with one extra edge 13.
Same coloring as middle graph.
}
\end{figure}

We take advantage of these experimental advances to solve the inverse
problem, from function $\rightsquigarrow$ structure.\index{function-structure relation } Given an (almost) complete
functional network from which cluster synchronization\index{synchronization !cluster } can be extracted,
at least for a (large) set of nodes, we identify these synchrony clusters with the balanced colorings of the structural network. We are interested in inferring
the underlying structural network supporting this synchrony
(Fig. \ref{fig:Fig1-pseudo}). Two main systems come to mind: (1)
inferring the TRN from gene coexpression\index{coexpression } synchrony, and (2) inferring
a connectome\index{connectome !inferring from synchronization } from neural synchronization, which we treat in Chapters
\ref{chap:brain1}, \ref{chap:brain2} and \ref{chap:brain3}.

We wish to determine the missing links\index{missing link } to complete these networks to
fully characterize pathways responsible for gene expression and
pathways in the brain.  The existence of missing links and variability
across samples preclude the identification of perfect symmetries in
the connectivity structure. However, a symmetry-driven inference algorithm
informed by cluster synchronization is able to reconstruct these
missing links and remove erroneous links in the biological network.

\begin{figure}[ht!]
	\centering
\textbf{(a)} \includegraphics[width=\linewidth]{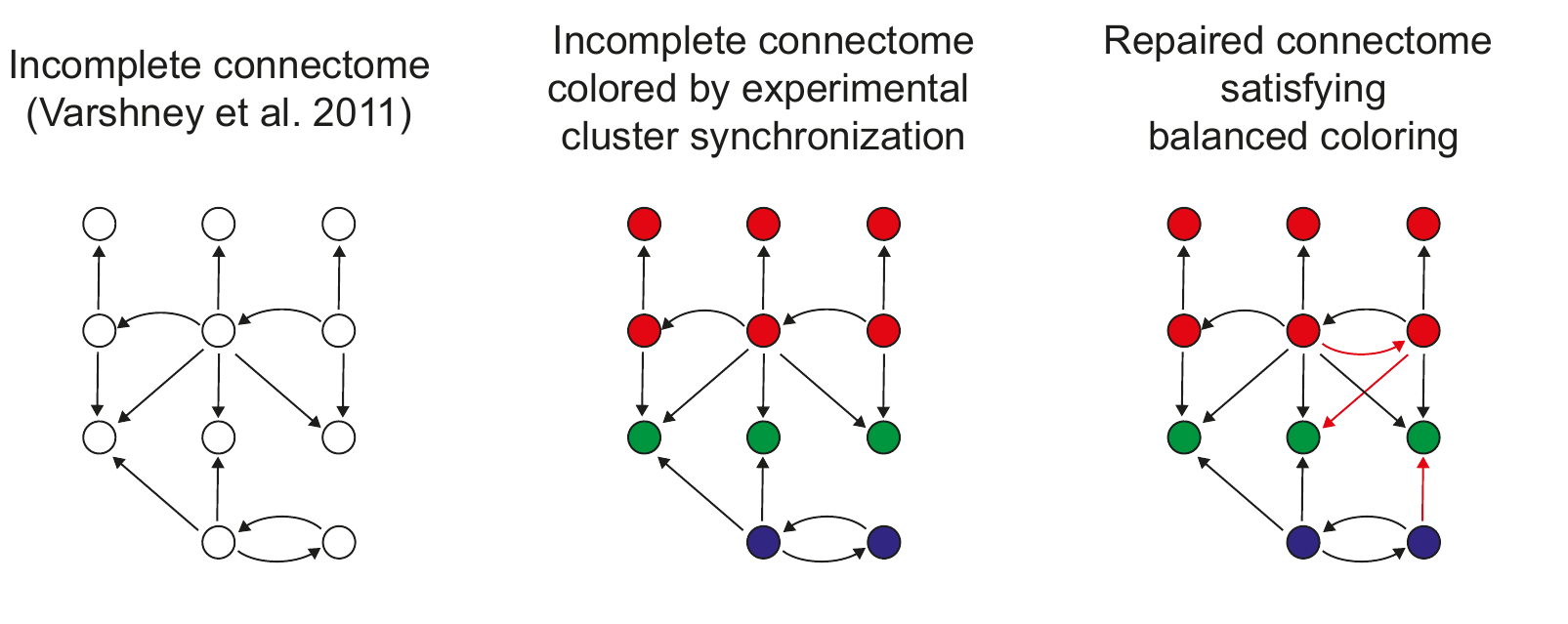}
\textbf{(b)} \includegraphics[width=\linewidth]{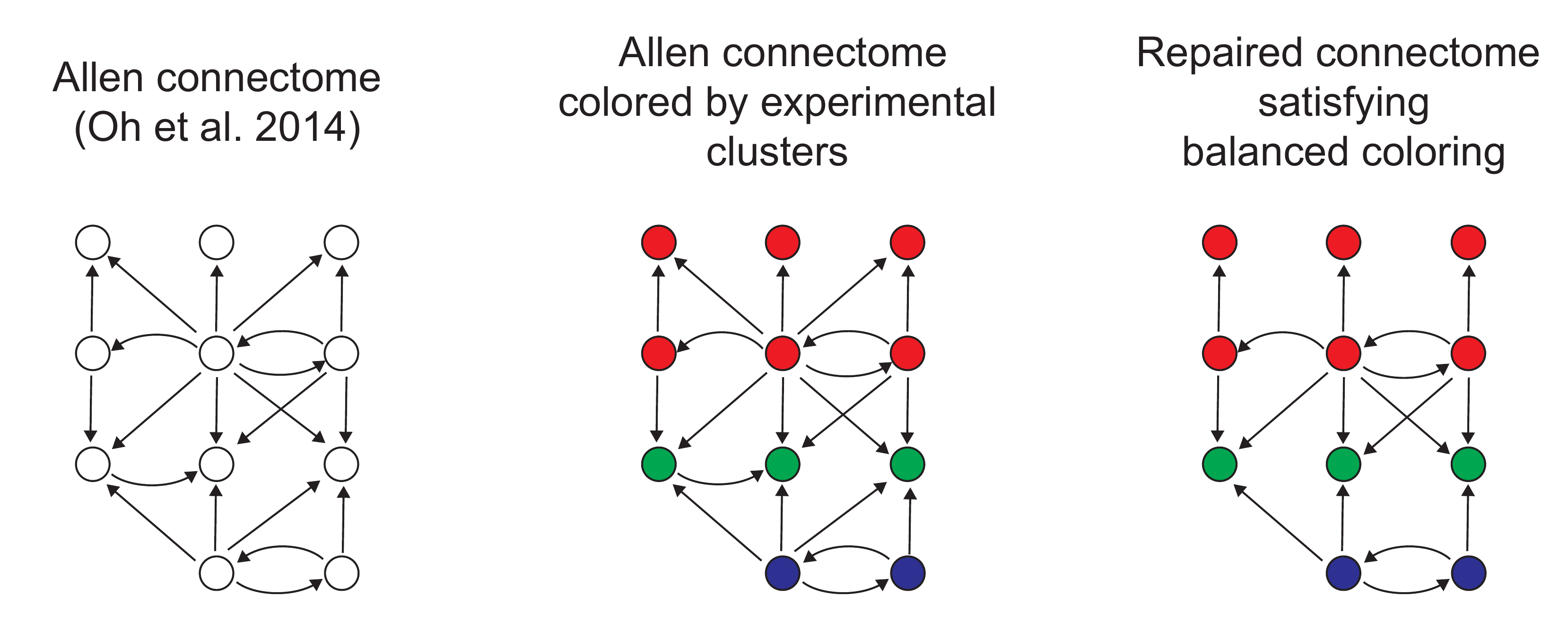}
	\caption{\textbf{Applications of symmetry-driven inference algorithm of structural networks.} (\textbf{a}) Link prediction. Section \ref{sec:link-prediction} and Chapter \ref{chap:brain1}.
             (\textbf{b}) Pathway reconstruction. Section \ref{sec:inferring} and Chapters \ref{chap:brain2}, \ref{chap:brain3}.
            }
        \label{fig:two-applications}
\commentAlt{Figure~\ref{fig:two-applications}: 
Top left: incomplete connectome with 11 nodes (in rows of three, three, three, two) and
14 arrows. Details irrelevant. Top middle:  incomplete connectome
colored by experimental cluster synchronization. Same graph with top two rows colored red, third 
row green, bottom row blue. Top right: repaired connectome satisfying balanced coloring.
Like top left but with three extra arrows added to obtain balanced coloring.
Bottom left: Allen connectome. Incomplete connectome with 11 nodes (in rows of three, three, three, two) and
14 arrows. Details irrelevant. Bottom middle: Allen connectome
colored by experimental cluster synchronization. Same graph with top two rows colored red, third 
row green, bottom row blue. Bottom right: repaired connectome satisfying balanced coloring.
Like bottom left but with one arrow deleted to obtain balanced coloring.
}
\end{figure}

As it turns out, the reconstruction algorithm will solve two problems that we treat in turn (Fig. \ref{fig:two-applications}):

\begin{itemize}
\item Link prediction (Fig. \ref{fig:two-applications}a): reconstructing an incomplete structural network (Section \ref{sec:link-prediction}). We will apply this algorithm to reconstructing the connectome of {\it C. elegans} in Chapter \ref{chap:brain1}.
\item Pathway reconstruction (Fig. \ref{fig:two-applications}b): from a (quasi) complete structural network, infer the pathways that are active in a given task (Section \ref{sec:inferring}). We will apply this algorithm to infer the pathways active in the memory engram in mice in Chapter \ref{chap:brain2} and in the human brain in Chapter \ref{chap:brain3}. 
\end{itemize}

\section{Link prediction in biological networks}
\label{sec:link-prediction}

Modeling biological systems as networks with symmetries poses
 challenges. Firstly, experimental data on networks
collected by different techniques are never complete due to
experimental errors and the impossibility of measuring every possible
link. 
Secondly, natural variability\index{variability } across individuals of the same species
results in different networks making interpretations of the data across
specimens more difficult.  This is particularly important in the
brain.  The neuronal network of {\it C. elegans}
contains 302 individually identifiable neurons, and the
wiring diagram includes 890 gap junctions and 6,393 chemical
synapses \citep{white1986thestructure,varshney2011structural}.  The
number of neurons across these animals is very consistent
\citep{white1986thestructure}, while the gap-junctions and chemical
synapses are reproducible from animal to
animal within 25\% variability  \citep{varshney2011structural,hall1991posterior}.  Due to its small size and
relative completeness, the neural network of {\it C. elegans} has been
a formidable model system in which to search for design principles underpinning
the structural organization and functionality of neural networks.

Despite the fact that 302 neurons are very constant across worms,  no two worms will ever have the same connectome. The concept of pseudo-balanced colorings arises
naturally from the observation that connectomes vary from animal to
animal. As just stated, \cite{varshney2011structural} have estimated
this variation experimentally 
to be $25\%$ of the total connections.  We consider this variability across individual
connectomes to be an intrinsic property consistent with biological
diversity and evolution. Furthermore, the number of connections is
subject to change from animal to animal through plasticity, learning,
and memory, so it cannot be ignored.

On the other hand, while connectomes\index{connectome } vary from animal to animal,
functions developed from them, such as locomotion, are barely
distinguishable between different worms, and still show some vestige of
an ideal symmetry. In fact, the function is preserved despite the 25\%
variation in the connectomes.  Despite this inherent disorder, the function
and neural activity of {\it C. elegans} exhibits strong regularities,
as observed, for instance, in its oscillatory locomotion patterns \citep{stephens2008dimensionality} and
the neural synchronization observed in neuronal activity
\citep{kato2015}.  

This means, for instance, that a single balanced coloring in Fig. \ref{sec:link-prediction}a center specifying a function could correspond to different slightly noisy connectomes for every animal shown on the left. Yet, upon the reconstruction of a few links, the ideal common average connectome can be revealed, shown on the right with the new links in red. Consequently, we expect deviations from exact
symmetries to be relatively `small'. Exact symmetries of the
connectome should be considered as an idealization, and we do not
expect them to be realized exactly. 

Thirdly, organisms adjust to a changing habitat using neural
plasticity, epigenetics, and other adaptations. For example, even two
organisms that are identical at a given moment may become
different in the future. Organisms survive and benefit from these
small variations, so evolution through natural selection
occurs. Therefore, networks that model these systems exhibit not only
differences across individuals but also have missing experimental
links that can mask their underlying regularities. In particular, the
missing experimental interactions and natural variations make the
existence of perfect symmetries difficult to find in biological
networks.  This
difficulty has persisted despite the ubiquitous existence of
synchronization across all biological networks, which is a
manifestation of (approximate) symmetries in the underlying networks that support 
biological activity.

What is the anatomical substrate for this synchrony? Figure \ref{fig:two-applications}a shows an example where, upon
reconstruction of a few missing links, an incomplete network will
reveal an underlying ideal perfect symmetry that is the unique
`blueprint' of the symmetry of the species.  Each network
can be thought of as a small variation on this perfect blueprint. The experimentally observed networks
are all different and are always pseudosymmetric; they can be 
modified slightly to exhibit the ideal blueprint of the species.

Reconstructing a network to reveal ideal symmetries is an instance of
link prediction\index{link !prediction } problem in complex networks \citep{Lu11}. Link
prediction is the problem of determining the probability of the
existence of a link between a pair of nodes in the network that is
missing from the data \citep{Lu11, Getoor05, Liben07, Chen05,
  Newman08}. It can also be generalized to removing spurious links from the network. Previous work has proposed several approaches to find missing
links based on different statistical and Markov chain models; see the
review in \citep{Lu11}. However, functional and dynamical features of
incomplete networks, such as synchronization,  are not taken
into consideration in the reconstruction. The approach presented here employs network
symmetry and synchronization as an organizing principle to guide
 link prediction\index{link !prediction } and to reconstruct the network from its incomplete
state.

\subsection{Previous reconstruction algorithms for missing data}
\label{sec:inferring}\index{reconstruction algorithm }

Existing approaches to network reconstruction \citep{brugere2018} 
infer TRN structures from coexpression based on pairwise correlations
between the expression of genes; examples are `Weighted Gene Correlation
Network Analysis'\index{Weighted Gene Correlation
Network Analysis } (WGCNA)\index{WGCNA } \citep{horvath2011,zhang2005,langfelder2008}. Such approaches
typically assume that shared expression profiles reflect control by
the same TFs
\citep{klipp2016book,horvath2011,bansal2007,brugere2018,tegner2003,liao2003,wang2014}.
Reconstructing TRN direct interactions from pairwise correlations of
gene expression in high-throughput data has been treated by many 
\citep{marbach2012,bansal2007,brugere2018,butte1999,maertens2018,chen2008}.

Statistical methods for TRN reconstruction using expression data
\citep{horvath2011,zhang2005,marbach2012,bansal2007,brugere2018,butte1999,maertens2018,chen2008}
typically do not distinguish between TRN and functional
coexpression network \citep{roy2014,klipp2016book}, that is, between
the TRN and the network obtained typically by thresholding\index{thresholding } a
coexpression matrix\index{coexpression matrix } or correlation matrix.\index{correlation matrix }

This logic is also used to parametrize quantitative models
\citep{kaplan2008,liao2003,klipp2016book} and it underlies many
attempts to infer regulation mechanisms from the observed expression
profiles. However, as is clear from our analysis of, e.g.,  multilayer
composite fibers in Section \ref{sec:multilayer}, this approach has
problems since genes can show coexpression without sharing 
transcriptional inputs. They might just as well be regulated by
different TFs that, by the symmetry of the network structure, tend to
be coexpressed.

Thus, the difference between the functional coexpression network and
the TRN is clearly demonstrated in multilayer composite fibers. In
theory, this coregulation could be described by models, given the
(usually unknown) activity profiles of different TFs. If the TRN
structure implies such indirect coregulation, how can we infer this by
network analysis alone? Comparing genes by their direct in-going
links are not enough; instead, we need to go back and compare the
entire chain of regulations behind these inputs.  This process
is achieved by the input tree and the symmetry fibration.

Basically, this implies that the typical approach to network
reconstruction, which assumes that coexpression means direct
co-regulation needs to be revised. It indicates the need to search
for indirect interactions beyond co-regulated genes, including chains
of regulations behind coexpressed genes.

\subsection{Missing link algorithms using topological measures of the network}\index{Missing link !algorithm }

The link prediction\index{link !prediction } problem is a problem of predicting the probability
of the existence of an edge between two unconnected nodes, given the
observed data. Typically, this probability is defined as the
similarity between the nodes concerned, obtained using a topological
measure of the graph \citep{Newman08, Liben07, Lu11}. To repair a graph
using a link prediction approach, an empirically chosen number of 
top-ranked edges is predicted to exist \citep{Chen05}. Common
measures of similarity are divided into local, global, and quasi-local
classes. We review some of them 
and propose a way to estimate their performance following \cite{Lu11}.

First, we introduce a few indices characterizing the local topology of
the graph. Let $S_{xy}$ be the distance between nodes $x$ and $y$, let
$\Gamma(x)$ be the neighborhood of node $x$ (the set of nodes connected to
$x$ by a directed edge), let $\mid X \mid$ be the cardinality of set $X$, and let $k_x$ be the
degree of node $x$.

The \textit{preferential attachment index}\index{preferential attachment index } calculates the similarity
between two nodes based on their degree:
\begin{equation}
    S_{xy}^{PA} = k_x\ k_y.
\end{equation}
This measure is widely used in scale-free networks,\index{network !scale-free } and is generalized
by the {\it Randic index}\index{Randic index } \citep{randic75,KP10}. In particular,
\begin{equation}
\label{E:scale-free metric}
s(G) =
\sum_{x, y \in V} S_{xy}
\end{equation}
is called a {\it scale-free metric} and is used as a measure of
scale-freeness of the graph \citep{Li05}.

A degree-based metric is also used while generating a random
scale-free network with the Barabasi--Albert model\index{Barabasi--Albert model } \citep{Albert02}. In
this model, preferential attachment implies that each newly added node
in a generated graph is connected to other nodes with a probability
proportionate to their degrees.  The scale-free metric\index{scale-free metric } \eqref{E:scale-free metric} is used as a measure of
scale-freeness of the graph.

The {\it Common Neighbors Similarity metric}\index{common neighbors similarity metric } is based on the assumption
that nodes that have a lot of common neighbors are likely to be
neighbors themselves. This assumption has had success in social
networks\index{network !social } \citep{Newman01, Kossinets06}; however, it does not increase
the symmetry of the network. \textit{Common Neighbors (CN)} is defined
by:
\begin{equation}
    S_{xy}^{CN} = \,\,\mid \Gamma(x) \cap \Gamma(y) \mid.
\end{equation}
There are two differently normalized versions of the Common Neighbors
metric: the \textit{Salton Index}\index{Salton index } normalized by the degree of the nodes
\begin{equation}
    S_{xy}^{Salton} = \frac{S_{xy}^{CN}}{\sqrt{k_x \times k_y}} = \frac{\mid \Gamma(x) \cap \Gamma(y) \mid}{\sqrt{k_x \times k_y}},
\end{equation}
and \textit{Jaccard similarity}\index{Jaccard similarity } normalized by the total size of the
node neighborhood:
\begin{equation}
    S_{xy}^{Jaccard} = \frac{S_{xy}^{CN}}{\mid \Gamma(x) \cup \Gamma(y) \mid} = \frac{\mid \Gamma(x) \cap \Gamma(y) \mid}{\mid \Gamma(x) \cup \Gamma(y) \mid}.
\end{equation}

A popular measure of similarity based on the global topology of
the graph is the \textit{Katz Index}.\index{Katz index } This index counts the number of
paths between
specified nodes (in the sense of this book; more technically, a `walk', which can visit nodes and edges more than once) of increasing length, weighted by the coefficient
$\beta < 1$. Here $(A)^k_{xy}$ is the number of paths of length $k$ between
nodes $x$ and $y$ \citep{Newman18}. The Katz index is:
\begin{equation}
    S_{xy}^{Katz} = \beta A_{xy} + \beta^2 (A)^2_{xy} + \beta^3 (A)^3_{xy} + \dots.
\end{equation}
This series converges when $\beta$ is less than the inverse of the largest eigenvalue of $A$, and the sum is
\begin{equation}
    S_{xy}^{Katz} = ((I - \beta A)^{-1}-I)_{xy},
\end{equation}
where $I$ is the appropriate identity matrix.

These methods are used in typical missing link\index{missing link } studies. Since all
methods are heuristic, the accuracy of each method needs to be
assessed with traditional hypothesis testing performance metrics. The
confusion matrix\index{confusion matrix } in Table~\ref{tbl:ConfusionMatrix} introduces four
basic metrics: true positive (TP), false negative (FN), false positive
(FP) and true negative (TN), which compares edges identified by the
manual `ground truth' solution (true/false) with the edges obtained by
a method (positive/negative). Other important metrics include
Precision, Recall, Miss Rate, Accuracy, and F-measure, which are
calculated as functions of TP, FN, FP, and TN as:
\begin{equation}
    \begin{array}{lcc} 
        {\rm Precision} & = & \displaystyle \frac{TP}{TP + FP}, \\
        \vspace{1pt}\\
       {\rm Recall} & = & \displaystyle \frac{TP}{TP + FN}, \\
        \vspace{1pt}\\
        {\rm Miss Rate }  & = & \displaystyle \frac{FN}{FN + TP}, \\
         \vspace{1pt}\\
        {\rm Accuracy} & = & \displaystyle \frac{TP + TN}{TP + FP + FN + TN}, \\
         \vspace{1pt}\\
        {\rm F-measure} & =& \displaystyle \frac{TP}{TP + \frac{1}{2}(FP + FN)}.
    \end{array}
\end{equation}

Using the above methods, we can choose the edges
to be repaired with the highest priority.
When the network is sparse, the
TN value is very high for all methods. Therefore, the best metric
for overall performance assessment is the F-measure,\index{F-measure } which is
proportional to the number of edges guessed correctly and inversely
proportional to the sum of missed and falsely identified edges.

\begin{table*}[ht]
\footnotesize
    \begin{center}
        \begin{tabular}{ c  c  c | c |}
            & & \multicolumn{2}{c}{\textbf{Prediction by a method}} \\
            \cline{3-4}
            & & \multicolumn{1}{|c|}{Predicted} & Not predicted \\
            \cline{2-4}
            \multirow{3}{*}{\rotatebox[origin=c]{90}{\textbf{truth}}} & \multicolumn{1}{|c|}{ } & True Positive (TP). Correct prediction. & False Negative (FN). Type II error. \\
            & \multicolumn{1}{|c|}{Predicted} & Edge repaired by ground truth & Edge repaired by a ground truth, \\
            & \multicolumn{1}{|c|}{ } & and a method & but not repaired by a method \\
            \cline{2-4}
            \multirow{3}{*}{\rotatebox[origin=c]{90}{\textbf{Ground}}} & \multicolumn{1}{|c|}{ } & False Positive (FP). Type I error & True Negative (TN). Correct rejection. \\
            & \multicolumn{1}{|c|}{Not predicted} & Edge repaired by a method, & Edge not repaired by either \\
            & \multicolumn{1}{|c|}{ } & but not repaired in ground truth & ground truth or a method \\
            \cline{2-4}
        \end{tabular}
    \end{center}
    \vspace{10pt}
    \caption{\textbf{Confusion matrix.} The confusion matrix identifies
      four variables, TP, TN, FP, and FN, depending on the positive or
      the negative outcome of the predicted result and the ground
      truth. The comparison assumes that we know the ground truth. For
      instance, we start with a perfectly symmetric network and then
      remove a few links. Then, we test the methods to find these
      missing links. The ground truth is known since we removed the
      links. True and false correspond to the ground truth and
      positive and negative correspond to the prediction.}
    \label{tbl:ConfusionMatrix}
\end{table*}

\section{Inferring the pathways guided by cluster synchronization}
\label{sec:inferring2}

A second application of the reconstruction algorithm is to determine the pathways in the structure network associated with a given synchronization task that produces a balanced coloring in the network, see Fig. \ref{fig:two-applications}b. Here, it is assumed that the connectome is well-known.

The main application is in the brain, although the method can be applied to any network. It is based on an allegory of a highway network and the one-to-many degeneracy into many functions elaborated by \cite{park2013structural}.
A baseline connectome serves as a detailed map of the intricate highway system that underlies the brain's various information pathways. 
This is illustrated in Fig. \ref{fig:two-applications}b left by the Allen connectome for the mouse (see Chapter \ref{chap:brain2}).
These pathways are the substrate for all functional tasks that the brain performs. It is important to note that the specific routes along this neuroanatomical highway vary depending on the nature of the task at hand. According to the hypothesis proposed by \cite{park2013structural}, the brain's functional activity selectively engages a subset of the available connections within this expansive 'highway' connectome. This selective engagement allows the brain to adaptively operate in diverse functional states, whether at rest, during memory retrieval, or while executing motor functions. The phenomenon described as a 'one-to-many' degenerate structure-function relationship highlights how a single, static architecture of neural connections can give rise to a multitude of distinct functional states, showcasing the remarkable flexibility and complexity of the brain's operations.\index{connectome !symmetry of }

Thus, active pathways in the brain adjust according to the task at hand \citep{friston2011functional, park2013structural}. Algorithmically, starting with a baseline structural network (shown on the left of Fig. \ref{fig:two-applications}b) and a balanced coloring of the network (depicted in the center, which is obtained from a specific task), the algorithm will identify the minimal number of removals (and possibly additions) to the connectome. This process aims to discover a substrate network that reproduces the balanced coloring, as illustrated on the left in the figure.
This algorithm will be applied in Chapter \ref{chap:brain2} to the mouse brain and Chapter \ref{chap:brain3} to the human brain.

\section{Link prediction in biological networks via 
  symmetry and synchronization}
\index{link !prediction }

We now develop the algorithm to infer the structural network from a balanced coloring of the
functional network. \cite{avila2024link,avila2024symmetries,gili2024fibration} use the pseudobalanced coloring theory of
\citep{leifer2022symmetry-driven} to develop a mixed
integer linear programming (MILP) \citep{bertsimas1997introduction}
 algorithm
to optimize a minimal link removal (and addition) from a baseline
structural network to satisfy a predefined balanced coloring of the network.

The inference algorithm can be summarized in the
following steps:
\begin{itemize}
    \item First, 
      identify the baseline structural network  (left of Fig. \ref{fig:two-applications}a or b) that forms the graph of all known structural
      connections between the nodes.  This
      is the baseline structural network that will be repaired by the
      algorithm.
    \item Using any synchronization measure, find the functional
      network of the system from experimental activity data. This
      can be the gene coexpression matrix or synchronous neural
      data. Extract the cluster synchronization of the nodes from the
      functional network.  Assign a color to each cluster. Assign to each node in the baseline
      structural network a color according to the cluster
       to which they belong (center of Fig. \ref{fig:two-applications}a or b). This is the balanced
      coloring that guides the
      network reconstruction.
    \item Apply the  MILP algorithm to reconstruct the baseline
      network by adding/removing the minimal number of edges
      until the balanced coloring of the decimated graph matches the
      coloring obtained from the functional network
      (right of Fig. \ref{fig:two-applications}a or b).
      According to the application, adding,
      removing, or both should be applied.
\end{itemize}

\section{Integer linear program to infer the structural
  network from cluster synchronization}
\label{sec:integer}  

The crucial step in this scheme is finding minimal additions/removals that satisfy the
coloring condition. This problem is formulated as a mixed integer program that
is solvable for modestly sized instances. \cite{avila2024link}
formulate the problem of finding the minimum perturbations to induce a
minimal balanced coloring as an integer linear program, i.e., an
optimization problem where the decision variables are all integers.
\cite{avila2024symmetries} have applied this algorithm to reconstruct
the {\it C. elegans} connectome (Chapter \ref{chap:brain1}).
\cite{garcia2024minimal} used it to unravel the minimal engram
structure during memory formation in mice (Chapter \ref{chap:brain2}), and
\cite{gili2024fibration} applied this algorithm to the human brain
(Chapter \ref{chap:brain3}). The most updated code appears in \url{GitHub.com/MakseLab}.

We consider a directed graph, $G = (V,E)$, where $V$ denotes the set of
nodes, and $E$ denotes the set of directed edges (recall that an undirected edge is
considered as two directed edges). Let  
$n=|V|$ be the number of nodes and let $m=|E|$  be the number of directed edges. We also
define
\begin{equation}
E^C = \{ ij : i,j \in V, ij \not\in E\}
\end{equation}
 as the set of ordered pairs of nodes for which no directed edge
 exists in $G$, which we refer to as {\it non-edges}. These ordered
 pairs represent potential edges that could be added to the graph $G$.
 We define $\alpha, \beta$ as constant parameters that govern the
 objective function's relative importance between edge removal and edge
 addition.

\begin{definition} {\bf Symmetry-driven inference algorithm guided by balanced coloring synchronization.}
Let $\mathcal{S}$ denote a coloring of $G$.  This coloring is provided by the cluster synchronies from the
functional network synchronization.  We wish to determine the minimum
number of edges to add or remove so that $\mathcal{S}$ becomes a balanced
coloring of $G$.
\end{definition}

The following integer programs are guaranteed to find a balanced
coloring, but are not guaranteed to find a minimal balanced coloring.

The model's three families of binary decision variables are defined as follows. \\
For $ij \in E$,
\begin{equation}
    \label{eq:remove_vars}
r_{ij} = 1 \mbox{ if edge $ij$ is removed}, 0 \mbox{ otherwise.}
\end{equation}
For $ij \in E^C$,
\begin{equation}
    \label{eq:add_vars}
a_{ij} = 1 \mbox{ if non-edge $ij$ is added}, 0 \mbox{ otherwise.}
\end{equation}
For $P, Q, R \in \cS$ with $P\not=Q$ and for $i \in P, j \in Q$
\begin{equation}
    \label{eq:imbalance_vars}
s_{ijR} = 1 \mbox{ if $i$ and $j$ are imbalanced on $R$}, 0 \mbox{ otherwise.}
\end{equation}
The role of the linear constraints below is to set
up a system of linear equalities and inequalities that, if satisfied by
these decision variables cause the given coloring to be a balanced coloring of the resulting repaired graph.

The objective function is to minimize the weighted sum of edges
removed and edges added.
The function is then defined as:
\begin{equation}
\label{eq:mip_obj}
f_{\alpha,\beta}(r,a) = \alpha \sum_{ij \in E} r_{ij} + \beta \sum_{ij \in E^C} a_{ij}.
\end{equation}
This equation represents a simplified version of unweighted
networks (which include multigraphs if we replace
positive integer weights by edges of suitable multiplicity). Below, we generalize the approach to include weighted networks.  The main
constraint ensures that $\cS$ is a balanced coloring of the perturbed
graph $G$.
\begin{equation}
\begin{array}{l}
    \displaystyle\sum_{ip \in E: i \in S} (1 - r_{ip}) + \sum_{ip \in E^C:i \in S} a_{ip}  = \\
    \displaystyle\sum_{iq \in E: i \in S} (1 - r_{iq}) + \sum_{iq \in E^C: i \in S} a_{iq}; 
   \quad p,q \in T; S,T \in \cS.
\end{array}
       \label{eq:balancing}
\end{equation}

Constraints \eqref{eq:balancing} are imposed for every pair of nodes $p,q$
that have the same color and for every color set. for a given
edge $ij \in E$, the quantity $1-r_{ij}$ is 1 if the edge is not
removed and 0 if it is removed. Also, for $ij \in E^C$, the quantity
$a_{ij}$ is 1 if $ij$ is a newly created edge and 0 otherwise. Thus,
the left-hand side of \eqref{eq:balancing} represents the edges that
enter into a given node $p$ from the color set $\cS$, and the right-hand
side represents the edges entering node $q$ from
$\cS$. Using the same sums, \eqref{eq:atleastone} ensure that the
in-degree is at least 1 for every node:
\begin{equation}
    \label{eq:atleastone}
    \sum_{ip \in E} (1 - r_{ip}) + \sum_{ip} a_{ip}
    \geq 1, \quad \quad \quad p \in V.
\end{equation}

The following constraints are valid for minimal balanced colorings;
i.e., they are necessary but not sufficient.
\begin{equation}
\label{eq:minimal_nec1}
\begin{array}{l}
 \displaystyle\sum_{ip \in E: i \in R} (1 - r_{ip}) + \sum_{ip:i \in R} a_{ip} -  
\displaystyle\left(\sum_{iq \in E: i \in R} (1 - r_{iq}) + \sum_{iq: i \in R} a_{iq}\right)\\
     \qquad \geq s_{pqR} - ns_{qpR}; 
     \qquad p\in S; q\in T; R,S \not= T \in \cS, 
\end{array}
\end{equation}

\begin{equation}
\label{eq:minimal_nec2}
\begin{array}{l}
 \displaystyle\sum_{iq \in E: i \in R} (1 - r_{iq}) + \sum_{iq:i \in R} a_{iq} -  
  \displaystyle   \left(\sum_{ip \in E: i \in R} (1 - r_{ip}) + \sum_{ip: i \in R} a_{ip}\right) \\
     \geq  s_{qpR} - ns_{pqR} ;
   \qquad  p \in S; q\in T; R,S \not=T \in \cS,     
\end{array}
\end{equation}

\begin{equation}
\label{eq:minimal_nec3}
\begin{array}{l}
    s_{pqR} + s_{qpR} \leq 1; \qquad
    p\in S; q\in T; R,S \not= T \in \cS;      
\end{array}
\end{equation}
\begin{equation}
\label{eq:minimal_nec4}
 \displaystyle    \sum_{R \in \cS} (s_{pqR} + s_{qpR})  \geq 1; \qquad
    p\in S; q\in T; S,T \in \cS.
\end{equation} 

The inequalities \eqref{eq:minimal_nec3} keep at most one of the two
binary variables $s_{pqR}$ equal to 1 for every color $R$. If
both are zero, then the inequalities \eqref{eq:minimal_nec1} and
\eqref{eq:minimal_nec2} force $p$ and $q$ to be balanced for the
color $R$. If one is zero, the total in-adjacent nodes of color $R$
for $p$ and $q$ differ by at least 1. In particular, for
color $R$, if $s_{pqR}=1$ and $s_{qpR}=0$, then the number of
in-adjacent nodes to $p$ is at least 1 greater than that to color
$q$. The converse is also true.

The inequalities \eqref{eq:minimal_nec4} force one of $s_{pqR}$
or $s_{qpR}$ to equal 1 for at least one color $R$. This is a
necessary but not sufficient condition for the coloring to be {\it
  minimal}. For example, if two color partitions have no edges between
them, the same number of edges to all other colors, and the same
positive number of internal edges, then \eqref{eq:minimal_nec4} is
satisfied since different colors would register as an
imbalance. However, the union of these two color partitions is
balanced and has one color fewer; i.e., the coloring is no longer
minimal. That being said, experiments by \cite{gili2024fibration}
yield strong evidence that in practice, the necessary condition 
enforces the minimal balanced condition, since a 
minimal balanced coloring\index{coloring !minimal balanced } was found for all test cases.

The complete model is then:
\begin{equation}
\begin{array}{ll}
\min f_{\alpha,\beta}(r,a) & \\[1mm]
\mbox{subject to} & \eqref{eq:balancing},
\eqref{eq:atleastone},
\eqref{eq:minimal_nec2}, \eqref{eq:minimal_nec2},\eqref{eq:minimal_nec3},
\eqref{eq:minimal_nec4}, \\[1mm]
& r_{ij}, a_{k\ell}, s_{pqR} \in \{0,1\}, ij
\in E,\\[1mm]
& k\ell \in E^C, p \in P, q \in Q, P\not=Q,R \in \cS,
\end{array}
\label{eq:complete}
\end{equation}
where (\ref{eq:minimal_nec4}) within the equation above is a reference
to select only one of its sub-equations.

The objective and constraint functions are linear. We then solve the
integer linear programs with the solver \cite{gurobi}. \cite{avila2024link} have benchmarked
and validated this algorithm.  The uniqueness of the solution is tested by
developing an independent solver based on the quasifibration
framework of \citep{boldi2021}.  In all cases considered, we find the
same solution using the quasifibration formalism and MILP.

The drawback of this technique is that MILPs are, in general,
NP-Hard. However, decades of research from the 1990s and
computational advances have increased our ability to solve MILPs
by several orders of magnitude. Since 2000, advances in hardware alone have increased the speed
by a factor of 180 to 1000, and algorithmic progress has rendered many
previously intractable problem classes solvable in
seconds \citep{koch2022progress}.  Future directions are to
investigate algorithmic improvements to solve larger instances.

The weighted case requires new variables and a modification of the
constraints \eqref{eq:balancing}. Let $w_{ij}$ denote the weight of
edge $ij$ for $ij \in E$. Let $\varphi \in [0,1]$ denote the maximum
percentage by which a preexisting edge weight can be adjusted, and let
$\tau$ denotes the maximum weight that a new edge can possess.
Let $\mu_{ij}$ denote the modification of weights for preexisting
weights for $ij \in E$. Let $\sigma_{ij}$ denote the weight of a new
edge if one is added for $ij \in E^C$.

The main constraint to ensure that $\cS$ is a balanced coloring
of the weighted graph, $G$ is:
\begin{equation}
    \label{eq:balancing-weighted}
    \begin{array}{l}
\displaystyle
    \sum_{ip \in E: i \in S} \left(w_{ip}(1 - r_{ip}) + \mu_{ip} \right)
    + \sum_{ip:i \in S} \sigma_{ip} \\
\displaystyle
    \quad =
    \sum_{iq \in E: i \in S} \left(w_{iq}(1 - r_{iq}) + \mu_{iq}\right) + \sum_{iq \in E: i \in S} \sigma_{iq}, 
    \quad  p,q \in T, S,T \in \cS.
    \end{array}
\end{equation}
The following constraints ensure that the new weight variables are
restricted to a percentage of the current weight or an absolute number
for edge weight modifications or new edge weights, respectively: $
-\varphi w_{ij} \leq \mu_{ij} \leq \varphi w_{ij}.$ 

As mentioned, the algorithm is not just about repairing
missing links,\index{missing link } but identifying the routes that sustain the functional
networks in different dynamical states.  Even within a single connectome, the
brain can never have exact symmetries due to its high
complexity. Structural brains are all different, but a certain level
of ideal symmetry must be common to all of them in order to guarantee
the performance of the same function, even though not all
structural brains are identical.  We elaborate on these applications
in the next three chapters.


\chapter[Fibration Theory of the Brain I: \hbox{\it C. elegans} Locomotion]{\bf\textsf{Fibration Theory of the Brain I: \hbox{\it C. elegans} Locomotion }}
\label{chap:brain1}

\begin{chapterquote}
Physics hopes for a simplified vision of nature, and this wisdom was
clearly realized in the fiber bundle picture of the fundamental forces of nature (Chapter \ref{chap:bundles}).  The hope for a
simplified vision in biology encounters its biggest challenge in the
search for an organizing theory of the brain.  The brain works by
processing signals traveling mainly through the connectome of synaptic
connections and the information flow in these networks is related to
the fibration; a rigorous extension of the fiber bundle.  The question
arises: Can a theory based on fibration symmetries help 
uncover the organizing principles of the brain by shedding light on the
structure of the connectome and its relation to function? This chapter
starts by tackling this question for one of the simplest neural systems
mapped as of today: the 302-neuron system of the nematode {\it
C. elegans}. We discuss synchronization in the organism's locomotion connectome. Succeeding chapters treat similar questions for rodents and
the human brain.
\end{chapterquote}

\section{Can we hope for an organizing theory of the brain?}

\begin{figure}
  \centering
  \includegraphics[width=\linewidth]{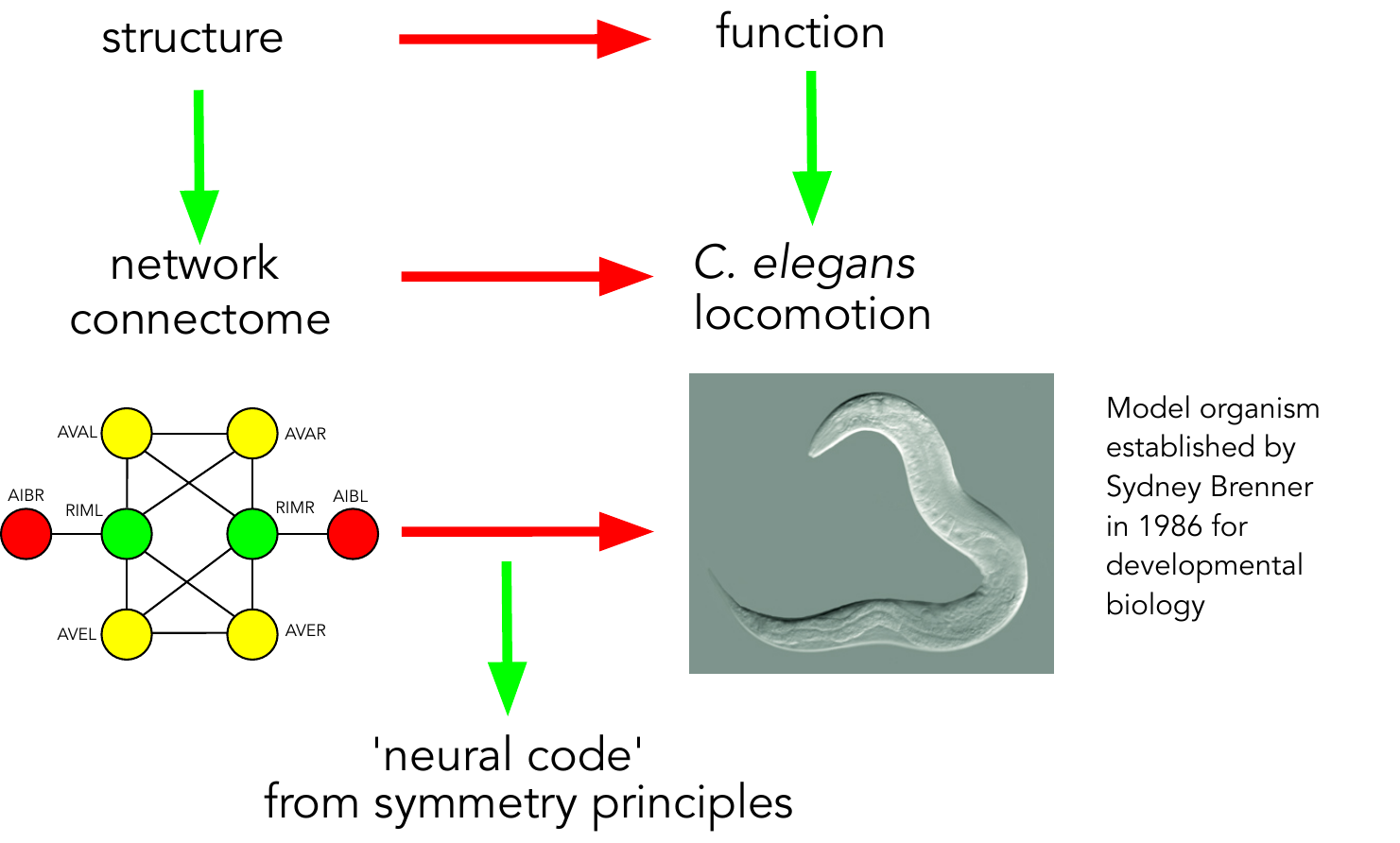}
   \caption{\textbf{Structure $\rightsquigarrow$ function relation in the
       nematode connectome.} The structure is characterized by
     its symmetries, which determine the patterns of locomotion of the
     phenotype.}
\label{structure-function}
\commentAlt{Figure~\ref{structure-function}: 
Diagram with labels structure, function, network connectome, C elegans locomotion,
neural code. Also small gene circuit, details irrelevant and picture of nematode.
Horizontal red arrows from structure to function; network connectome to C elegans locomotion;
gene circuit to picture of nematode.
Vertical green arrows from structure to  network connectome; function to C elegans locomotion;
bottom red arrow to `neural code from symmetry principles'.
}
\end{figure}

The central hypothesis of the present book applied to the brain, is that the
structure of the connectome---that is, the pattern of interconnections
between neurons---is not random, and the regularities observed in the
structure of connectomes\index{connectome } can be explained by symmetries of the
biological network, which is then manifested in neuronal synchrony and
function. This hypothesis is exemplified in
Fig. \ref{structure-function}.
It addresses a core problem in systems neuroscience
\citep{kandel,yuste,alivasatos,hartwell,tononi1994}:
how the function of the brain emerges from the structure of the
connectome through the neural code. Searching for such organizing
principles in a biological system is a goal of systems science in
general, but this question becomes particularly difficult for the brain because of the enormous complexity of the brain
connectivity.

\begin{figure*}
  \includegraphics[width=\textwidth]{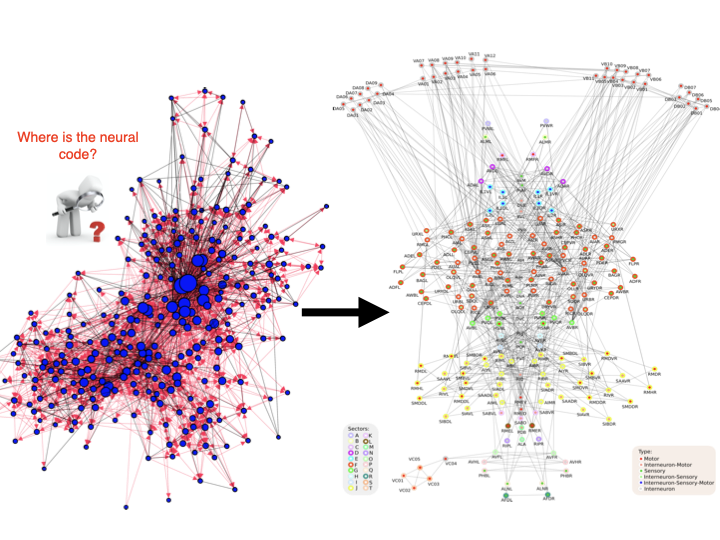}
  \caption{ \textbf{Where on Earth is the neural code? Symmetries bring
      order out of biological disorder.} {\em Left}: The worm connectome, as
    represented in typical plotting software like Gephi, shows
    pictorially random characteristics like a small-world and
    scale-free dominated by hubs. Node size is proportional to the
    degree. Where is the neural code represented in such a
    connectome? {\em Right}: For comparison, this plot represents the organization of the
    same connectome by its symmetries. This organization agrees in
    broad terms with the classification into classes: interneurons,
    sensory and motor neurons, and their combinations. Might this figure
    be a more faithful representation of the structure of brains than
    random models? 
    }
  \label{full}
\commentAlt{Figure~\ref{full}: 
Illustrative. Described in caption/text. No alt-text required.
}
\end{figure*}

Typically, the structure of connectomes has been modeled by random
models \citep{bullmore_perc,van2013network,watts1998,barabasi2009scale}. A
typical configuration of a random model resembles Fig. \ref{full}
left, which depicts the entire {\em C. elegans} connectome.\index{C. elegans connectome @ {\em C. elegans} connectome } Features
of randomness, such as the scale-free nature \citep{barabasi2009scale},
are suggested by the broad range of sizes of the nodes (which are
plotted in proportion to the degree of the node), and the random
disorder of a small-world network \citep{watts1998}, are apparent in the long-range links.

In contrast, we can plot the same network by placing each node in a
two-dimensional plot that respects its symmetries. This plot results
in the ordered structure on the right of Fig. \ref{full}. This figure
depicts the same connectome as that on the left but with the
neurons in harmony with their symmetries. While this is just a
pictorial representation, it makes the point that an organizing
principle for the brain could extract order from the biological
disorder inherent in small-world and scale-free random models of
the brain.

In Sections \ref{genetic-connection} and \ref{sec:duplication-lifting}
we developed the idea that gene duplication provides a plausible
mechanism for the emergence of symmetry structures in genetic
networks. In Section \ref{sec:necessary}, we also speculated on whether
symmetries are a consequence of biological evolution.\index{evolution }  Evolution, with
its strong selective pressure toward coherent structures, might
 be sufficient to produce connectomes with symmetries, structured
hubs, and small-world networks, all at the same time.

Similarly, we can argue that synchronized fibers in biology cannot be
discovered only through statistical comparisons with random models.
Fibers determined by symmetries are not network motifs. Important symmetry
structures might appear only once in the network and would not then
be statistically over-represented.  In the tradition of physics,
synchronized fibers are uncovered only with guidance from theory, not
through statistical analysis. Fibers may not be over-represented, as network motifs are (by definition), but
they are significant for function.

Since symmetries require strong structural patterns of connections, which can easily be broken by
a single missing or extra link, we 
wonder whether symmetries can survive the natural variability and
complexity of the brain.  Indeed, even for the simplest of all
connectomes, that of {\it C. elegans},\index{C. elegans connectome @ {\em C. elegans} connectome } no two connectomes of
different worms are ever exactly the same.  As discussed in Section
\ref{sec:link-prediction}, there is a natural variability of around
25\% between connectomes of different worms, and this variability is
sufficient to break all symmetries. Thus, to capture symmetries in
real connectomes, with missing links and variability from animal to
animal, we introduced the concepts of pseudosymmetry\index{pseudosymmetry } and
quasifibration\index{quasifibration } in Section \ref{algo-pseudo}, and in Chapter \ref{chap:function} we presented algorithmic
versions that can reconstruct imperfect connectomes and realize their inherent
underlying ideal symmetries. These
`almost-symmetries' can capture natural differences between
connectomes of organisms in the same species, and can tell us how far
away from ideal symmetry, the connectome of a given worm is.

Even for the smallest known connectome it is
already difficult to find symmetries, as evidenced by the need to use
complex NP-hard algorithms for pseudosymmetries and quasifibrations.
We might, therefore, ask whether it is hopeless to search for symmetries in more
complex brains, and in particular, the human brain.  As soon as we
consider other model organisms beyond {\it C. elegans}, the complexity
of brain connectivity becomes intractable.  The sizes of known
connectomes of model organisms span 11 orders of magnitude from the
302 neurons of {\it C. elegans} \citep{white1986thestructure} to the
100,000 neurons of the larval zebrafish \citep{ahrens} and the adult
fruit fly \citep{block},  
the 70 million of the house rat, the 6.4 billion of the rhesus macaque,
and ending with the 86 billion neurons in the human brain, which is,
alas, not the largest in the animal kingdom. Certainly, we cannot
expect such complex connectomes to respect the strict rules leading to
exact symmetries.

In this chapter and the next two, we explore the symmetries of connectomes across
all these scales and species. We devote some time to {\it
  C. elegans}, showing that the entire connectome at the single-cell
level can be ordered in terms of symmetries.  We then analyse more
complex species, rodent, and human brains at a coarse-grained level
with known partially reconstructed mesoscopic connectomes and exhibit
further symmetries in those.  In these more complex connectomes, our
approach is to use the reconstruction algorithms of Chapter
\ref{chap:function} to invert the arrow from function $\rightsquigarrow$ structure,
and to infer a meaningful mesoscale connectome\index{mesoscale connectome } that provides
predictions for the structure and response function of the brain.
These predictions 
are then validated experimentally.

We argue that synchronization is widespread in these systems, so
symmetry at a given scale (perhaps not the cellular scale, but larger
axonal fiber scales) should support these coherent states. We analyze
in detail synchronization of groups of the order of thousands of
neurons, as measured by c-Fos activity in rodents under a memory
consolidation tasks and in fMRI signals obtained in humans performing
language-related tasks. We argue that symmetries in the underlying
network connectomes of large-scale axonal fibers could support brain
synchronization at large scales in the human brain. Network symmetries
lead directly to the ensemble of neurons that synchronize their
activity and can thus be associated with an ensemble of functionally
related neurons.

A note on nomenclature: to avoid confusion, we henceforth refer to the
anatomical connections between mesoscopic areas in the human brain
across white matter as `axonal fibers'\index{axonal fiber } or `white matter tracts'.\index{white matter tract } We
reserve the term `fiber' for a fiber of a fibration.

\section{Dynamics and structure $\rightsquigarrow$ function in the brain}

The Hebbian learning\index{Hebbian learning } rule \citep{hebb1949} has been informally
characterized as: 
\begin{floatingbox}[h!]
\label{hebb1}
  \processfloatingbox{Hebbian rule \citep{hebb1949}}
                     {\,\,\,\,\,\,\,\,\,\, \,\,\,\,\,\,\,\,\,\,\,\,\,\,\,\,\,\,\,\,\,\,\,\,\,\,\,\,\,\,\,\,\,\,\,\,\,\,\,\,\,\,\,\,\,\,\,\, `cells that fire together wire
                       together'}
\end{floatingbox}

This is not quite what Hebb\index{Hebb, Donald } intended; his idea was that if one
cell should {\em cause} another to fire, the strength of the connection between them should increase. Thus `fire together' should not be
interpreted as synchrony.

Our basic principle is similar but differs in one vital respect: a {\it direct} connection between the cells is not required. It is:

\begin{floatingbox}[h!]
\label{hebb2}
\processfloatingbox{ Our hypothesis} {\,\,\,\,\,\,\,\,\,\,\,\,\,\,\,\,\,\,\,\,\,\,\,\,\,\,\,\,\,\,\,\,\,\,\,\,\,\,\,\,\,\,\,\,\,\,\,\, `cells that
  fire together have symmetries together'}
\end{floatingbox}

If a connectome possesses symmetries, we expect the symmetries to be
reflected in the neural synchronization and then in the functionality
of the corresponding circuit.  Moreover, we expect symmetry measures
to outperform other popular connectivity measures when identifying functional building blocks.

Finding organizing principles for the wiring of the brain from
symmetries also lead to interesting analogies with the wiring of
artificial devices, and in Chapter \ref{chap:ai} we discuss how this
information might inspire a new generation of AI architectures-based
on fibrations.

The theoretical framework of symmetries does not directly take into
account the dynamical system of equations describing neuronal activity.
To be precise, it assumes that the structure of the connectome can
determine the functional classification of the neurons independently of
the dynamics.  The question arises: Can the dynamics break
symmetry, even when the network topology is symmetric?

The answer is `yes'.  Just as solutions to nonlinear differential
equations can spontaneously break the symmetry of these equations; it
is possible for certain dynamic solutions to a network ODE to break
(at least some) symmetries of the network. Such behavior is the {\em
  synchrony breaking} discussed in Chapter \ref{chap:stability}.
Recall that we assume that the dynamics are specified by a set of
admissible equations\index{admissible !equation } (Definition \ref{admissible})---equations that
respect the fibration symmetry of the network.  The existence of symmetries
imposes a hard constraint on the dynamics, and the induced synchronization
is largely independent of the type of dynamical equation used to
describe the system---as long as the equations are admissible.

However, the dynamics could break the symmetries imposed by the
structure.  Imagine, bursting neurons following a Hodgkin--Huxley model\index{Hodgkin--Huxley model }
\citep{izhikevich} or similar.  There could be a set of parameters and
initial conditions leading to solutions that break the symmetries of
the network.  Indeed, if the fully symmetric state becomes unstable as
some parameter varies, the system can (and usually will) bifurcate to
a less symmetric state. This causes synchronized clusters to break
into smaller ones. Section \ref{S:metab+smol} illustrates such possibilities for the
metabolator and Smolen networks, with a mere two nodes. Even in this simple case,
there are many different combinations of stable states.

Moreover, the dynamics of neurons are not only regulated by 
interactions via the connectome but also by other interactions, most
notably by neuromodulators\index{neuromodulator }, which can regulate a population of neurons
\citep{bargmann1993}; some of them are discussed in terms of the multiplex models of Section \ref{sec:heterogeneous}. These interactions could induce a plausible breaking of
symmetry that is not taken into account by the network symmetry.  What
part of the functionality of the neural circuit can be captured by the
structure of the connectome itself?  Results below indicate that
classifications, such as command interneurons in different locomotions\index{locomotion }
and motor neurons are related to the building blocks of {\it
  C. elegans}. This approach explores how much of the functionality of
the connectome can be captured by its network topology.

\section{The 
{\it C. elegans} neural system}
\label{sec:celegans}

The nematode {\it C. elegans}\index{C. elegans@{\it C. elegans} } is a primitive organism, over 600
million years old, with a simple neural system made of a network of
gap junctions and another network of chemical synapses. This
connectome has been studied over the last four decades and fully
mapped, neuron by neuron and axon by axon, using electron microscopy
to scan every single synaptic connection. The smallness and
completeness of this connectome make it a perfect system to start
searching for symmetries in neural networks.

The {\it C. elegans} connectome\index{C. elegans connectome @ {\em C. elegans} connectome } is paradigmatic in neuroscience. {\it
  C. elegans} was first established as a model organism in genetics
by Sydney Brenner in 1963 to study developmental biology
\citep{brenner1974}. By 1986, its connectome was mapped by
\cite{white1986thestructure}. Today it remains the only connectome
reconstructed in its entirety. Despite it being the simplest of the
known connectomes and having been fully mapped four decades ago, we
still do not know how to simulate the mind of the worm. We do not even
know how to simulate in a computer its elegant sinusoidal locomotion
(reflected by its name) using a model based on the relevant part of the connectome, 
although principal component analysis yields a simple
dynamical system based on `eigenworms'\index{eigenworm } \citep{stephens2008dimensionality}.
A fair question is whether a complete
knowledge of the synapse-level wiring diagram of a brain, as is
available for the worm, is enough to understand the neural code, and,
from there, to predict motion, function, and behavior. The search for
an organizing principle for the brain is wide open.

Fibration symmetry allows the study of building blocks\index{building block } of connectomes
and their relation to synchronization and function.  The symmetry
hypothesis can be tested rigorously in {\it C. elegans} using the
synapse-resolution connectome in conjunction with dynamical data from
the nervous system at single-cell resolution.  This approach can then be
 extended to analyze available connectomes across more complex
species, which are yet not complete. However, partial
reconstructions and dynamical data exist, so the symmetry hypothesis
can be tested across numerous species, from annelids, fruit flies,
zebrafish, rodents, macaque, to humans.  These results in turn, can be
used to develop a unifying symmetry theory of the brain to uncover
synchronized neural populations that could be functionally related at
different scales, from the single cell in {\it C. elegans} to the
mesoscopic scales of the human brain captured by fMRI activation.

As stated above, the neuronal network of  {\it C. elegans} is made of
302 neurons, which are individually identifiable. Its wiring diagram
includes 890 gap junctions and 6,393 chemical synapses, according to the
reconstruction curated by \cite{varshney2011structural}.  
Because of its small size and relative completeness, the neural
network of {\it C. elegans} has been a formidable model system in the
search for design principles underpinning the structural organization
and functionality of neural networks
\citep{white1986thestructure,chalfie1985,bargmann1993,gray2005,chen2006,varshney2011structural,kato2015,yan2017}.

Symmetries of connectomes are either permutations of neurons in the
neural circuit (orbits) or fibration symmetries of the input trees
that define the synchronized fibers.  \cite{morone2019symmetry} analyze the symmetries of the {\it C. elegans} connectome
by looking at the circuits involved in the forward and backward
locomotion tasks.  Backward locomotion of the animal is supported
by the activation of neural classes of interneurons AVA, AVE, AVD,
AIB and RIM, and two classes of motor neurons, VA and DA (ventral
and dorsal motor neurons of class A).  Similarly, forward locomotion
is supported by the activation of motor neuron classes VB and DB
through the interneuron classes AVB, PVC, and RIB. These
neurons are used, together with the connectome of gap-junctions and chemical
synapses from  \citep{varshney2011structural}, to construct the neural
circuits for forward and backward locomotion.

\begin{figure}[t!]
\centering \includegraphics[width=.75\textwidth]{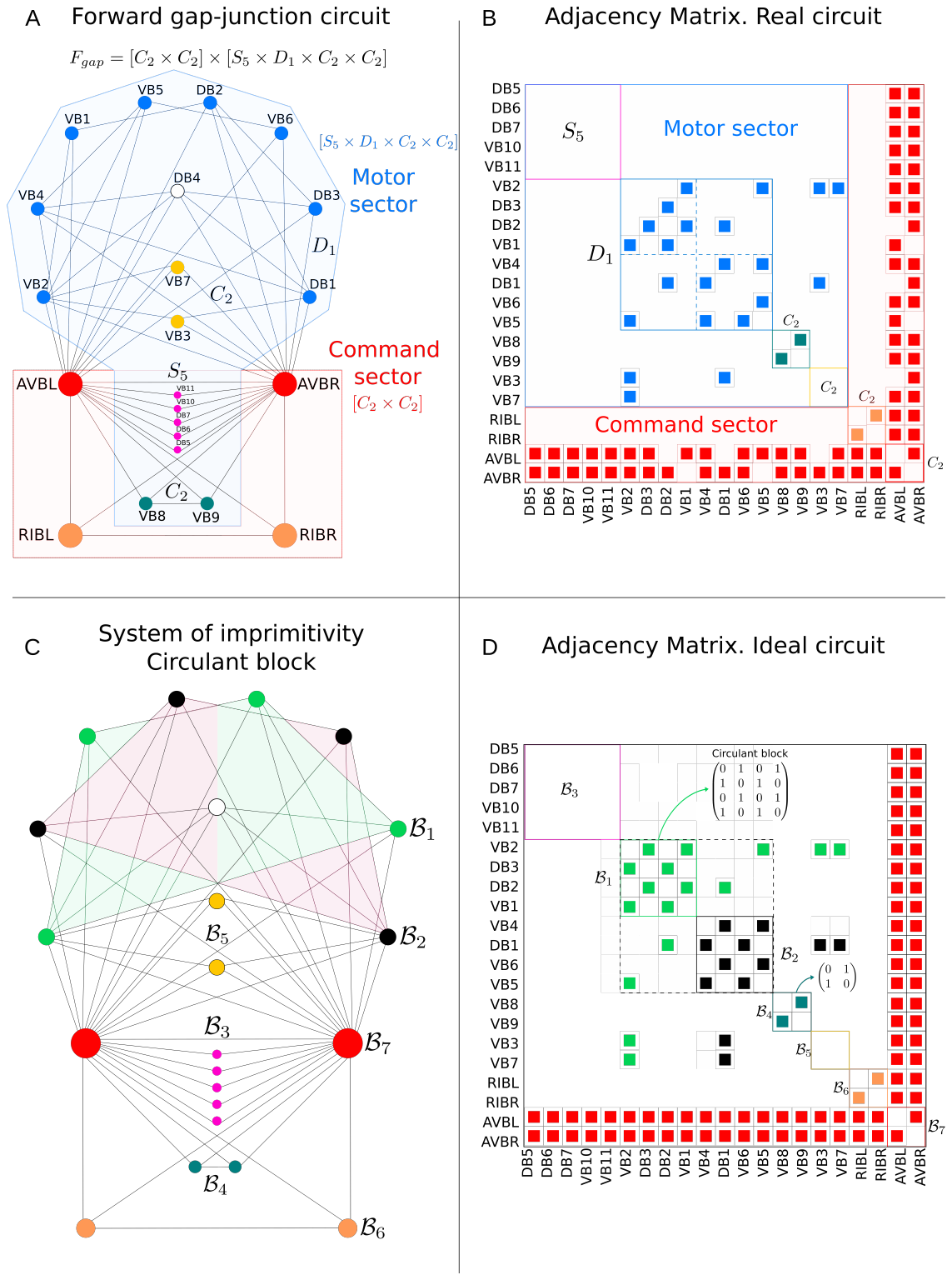}
\caption{\textbf{Symmetry group ${\bf F}_{\rm gap}$ of the forward
    gap-junction circuit of {\it C. elegans}.}  (\textbf{a}) Real circuit
  with connections from \citep{varshney2011structural} showing the
  normal subgroup factorization and its match with the broad
  classification of command interneurons and motor neurons. This
  circuit has pseudosymmetries since it is not complete. (\textbf{b})
  The adjacency matrix of (\textbf{a}) shows the subgroup structure and its
  matching with broad neuronal classes.  The incompleteness of the
  dataset is apparent in the matrix. (\textbf{c}) Ideal circuit of (\textbf{a}). This circuit is a hand-crafted reconstruction of the
  original circuit in (\textbf{a}) obtained by adding the smallest set of
  edges to obtain symmetry below the variability threshold of 25\%. An
  internal structure made of blocks of imprimitivity
  \citep{morone2019symmetry} is visible in the cycles shown in the
  circuit. (\textbf{d}) Ideal adjacency matrix showing how the subgroups
  are broken down into blocks of imprimitivity. Figure reproduced from
  \citep{morone2019symmetry}. Copyright \copyright ~2019, The Author(s).}
\label{fig:forwardgap}
\commentAlt{Figure~\ref{fig:forwardgap}: 
Described in caption/text. No alt-text required.
}
\end{figure}

Figure \ref{fig:forwardgap} shows a hand-crafted reconstruction of the
forward locomotion gap-junction circuit\index{forward locomotion gap-junction circuit } in {\it C. elegans} from
\citep{morone2019symmetry}. It shows that the
symmetries of this locomotion circuit are determined by automorphisms.  The symmetry group can be
factorized into a direct product of normal subgroups, as shown in the
figure:
\begin{equation}
{\bf F}_{\rm gap} = \mathbf{C}_2 \times \mathbf{C}_2 \times \mathbf{S}_5 \times
\mathbf{D}_1 \times \mathbf{C}_2\times \mathbf{C}_2 .
\end{equation}

We have already explained that automorphisms are more restricted
symmetries than fibrations and are less common in biology. Despite this,
automorphisms appear in the gap-junction connectome.  They may be
present because the {\it C. elegans} gap-junction circuit is quite
elementary as an evolving brain. However, we have not
found any other biological network described by automorphisms. More
complex organisms exhibit almost no automorphisms; most of their
symmetries are fibrations.

Automorphisms constitute a special class of fibration symmetries: the
orbits are the fibers, and the fibration is a projection from orbits to
a set of representatives. Gap-junctions are undirected links, but the
chemical synapse network is made from directed links and also has many
automorphisms \citep{morone2019symmetry}.  Again, this may
be due to the simplicity of the neural system.  However, in a later
study, \cite{avila2024symmetries} also found extra fibration
symmetries not captured by automorphisms. So, in this
network, we find both orbits (of the symmetry group of a subnetwork)
and fibers that are not orbits of any such group.  We discuss this
more general framework in succeeding sections.

\section{Synchronization in  
{\it C. elegans} neural system}
\label{sec:celegans-synchronization}

Experimental evidence for synchronization in brain activity is
abundant and has been observed at all scales, from single neurons to
large-scale EEG waves and fMRI activity. There are excellent books and
reviews that elaborate on all these examples
\citep{strogatz2018,arenas2008,pikovsky2001}.

Evidence of synchronization in neural populations in {\it C. elegans}
can be obtained by probing brain dynamics at single-cell resolution
and in real-time, i.e., sub-second scale.  Many experimental groups are
working on full-body single-cell resolution.\index{single cell resolution } Below, we review the work of
Manuel Zimmer and collaborators at the University of Vienna
\citep{kato2015,zimmer2017,Skora2018}, which is particularly suitable
for testing the relation between symmetries and synchronization under
locomotion; see also \citep{leifer2016}. These groups have pioneered
experiments using Ca$++$ imaging data that encompass the whole nervous
system, including head and tail ganglia as well as the ventral nerve
cord containing all motor neurons. This permits a study of the
relationship between synchronization and network structure for the
particular function of locomotion, which has also been extensively
studied before \citep{bargmann1993}.

Data have been generated under various experimental conditions,
including variation in feeding status, developmental state, sensory
stimulation, and genotypes
\citep{Schrodel2013,kato2015,zimmer2017,Skora2018}.  Results for
real-time and single-cell resolution whole-brain calcium imaging in
{\it C. elegans} have been published in \citep{Schrodel2013}, and
subsequent descriptions of internal brain dynamics that correspond to
behavioral action commands, as well as arousal and sleep states, are
published in \citep{kato2015,zimmer2017,Skora2018}. These dynamics
involve a large proportion of neurons in the animals' brains, 
providing sufficiently large datasets to analyze thousands of
potential neuronal pairwise interactions.

\begin{figure}
  \centering
  \includegraphics[width=\linewidth]{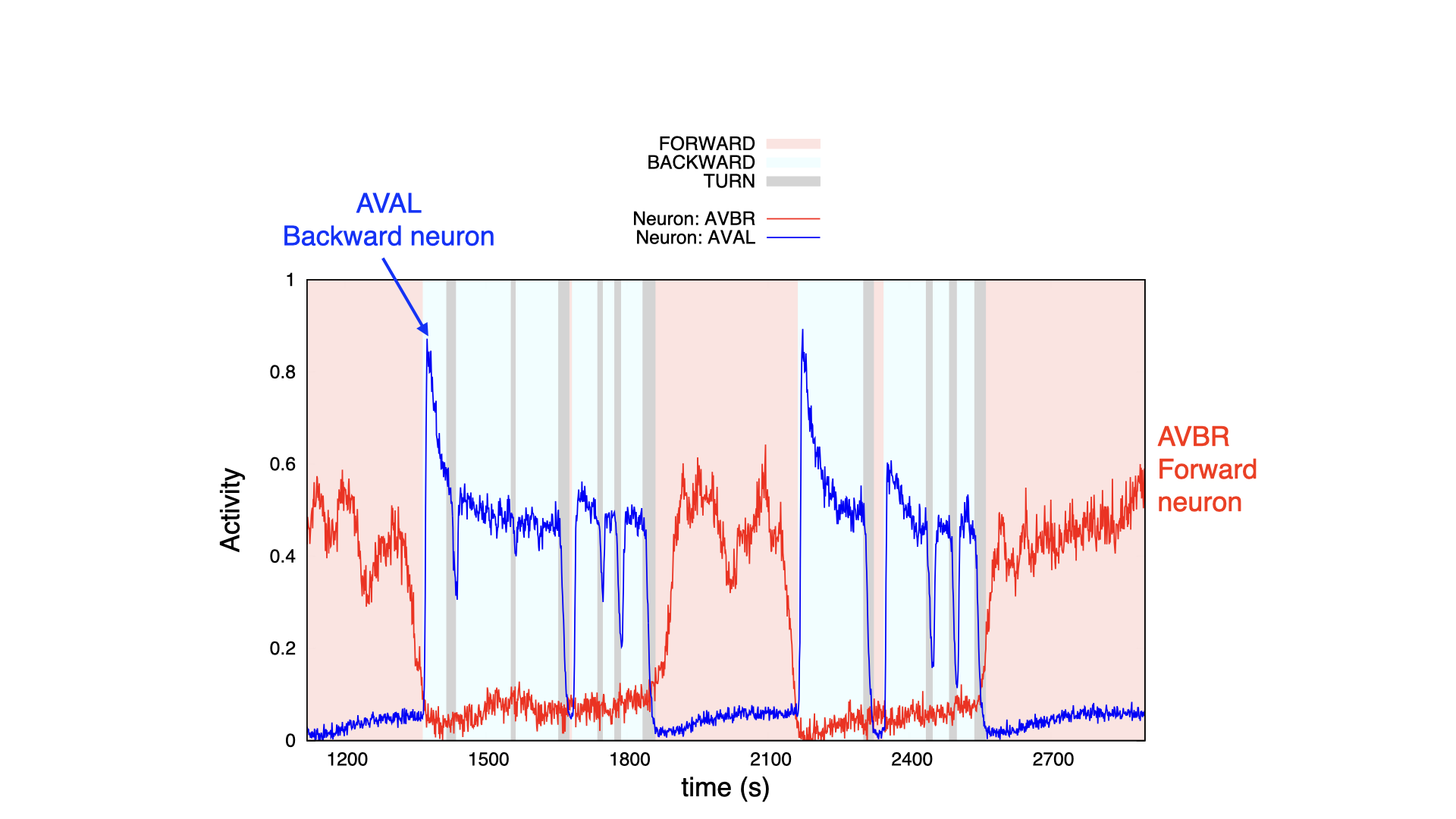}
   \caption{\textbf{Neuronal activity at single-cell resolution.} Calcium image
     activity of two neurons, AVAL and AVBR, involved in forward and backward {\it
       C. elegans} locomotion, respectively, are shown by their anticorrelated behavior. Data collected by the Zimmer lab.}
\label{two-neurons}
\commentAlt{Figure~\ref{two-neurons}: 
Described in caption/text. No alt-text required.
}
\end{figure}
     
\begin{figure}
  \centering
  \includegraphics[width=\linewidth]{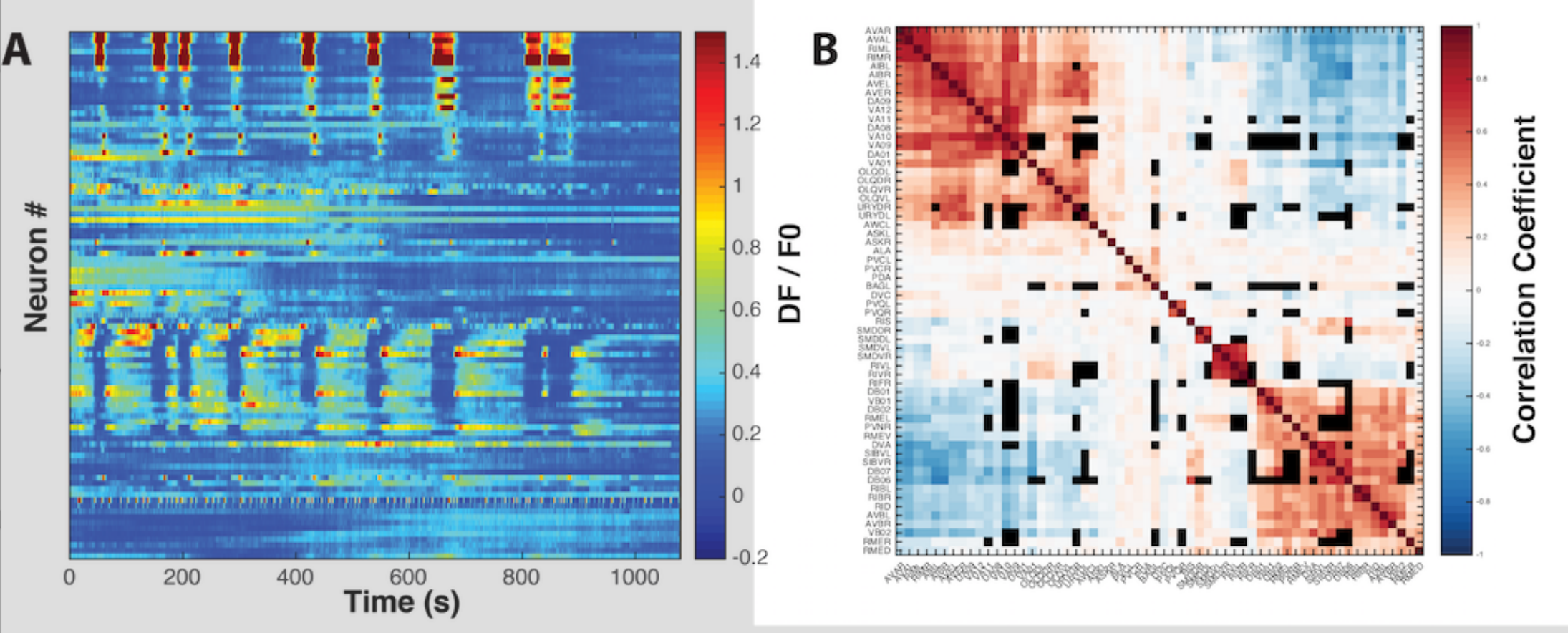}
   \caption{\textbf{Synchronization in worm locomotion in the experiments
       of \cite{kato2015}. } {\bf (a)} Heat map of whole brain Ca$++$
     imaging data of brain activity in a typical worm. {\bf (b)}
     Matrix of pairwise correlations between the activities of activated
     neurons under locomotion. Spontaneous activity is dominated by
     synchronies as shown in the heat map. The correlation matrix
     shows the Pearson coefficient between the neurons, which reveals
     clusters of synchronized neurons. Figure reproduced from
     \citep{kato2015}. Copyright \copyright ~ 2015,  Elsevier Inc.  }
\label{1A}
\commentAlt{Figure~\ref{1A}: 
Described in caption/text. No alt-text required.
}
\end{figure}

The functional imaging assays permit a first-level analysis of
synchronization in the connectome.  In these experiments, transgenic
worms expressing a nuclear-localized Ca$++$ -sensor (GCaMP) in a
pan-neuronal fashion are immobilized in microfluidic devices
\citep{Schrodel2013,kato2015}. This enables volumetric
fluorescence imaging at sufficient speed and optical resolution to
capture the activity of most neurons with single-cell resolution in
the animal's head ganglia \citep{Schrodel2013,kato2015}. Figure \ref{two-neurons}
shows an example of
calcium image activity in two neurons involved in forward locomotion
(AVBR)\index{AVBR } and backward locomotion (AVAL).\index{AVAL }  Neural synchronization is shown in
Fig. \ref{1A}, obtained from \citep{kato2015}.

Figure \ref{1A} shows that spontaneous network activity in the {\it
  C. elegans} brain is dominated by the synchronization of the neurons.
Figure \ref{1A}a shows the heat map of an exemplary whole brain Ca$++$
imaging dataset. Each row corresponds to a neuron, the color bar represents
relative fluorescence change, and Fig. \ref{1A}b shows the matrix of
pairwise correlation coefficients averaged across four such worms.
The correlation matrix in Fig. \ref{1A}b is sorted by
agglomerative clustering. Evident are two large synchronous neuronal
ensembles of backward and forward neurons, respectively. The backward
set of neurons is formed by interneurons AVAL, AVAR, AVEL, AVER, RIML,
RIMR, AIBL, and AIBR, while the forward set of interneurons comprises
AVBL, AVBR, RIBL, RIBR, among others.  Each set of neurons supports
the corresponding motor neurons: classes VA and DA for backward
locomotion and motor neuron classes VB and DB for forward
locomotion. Inside these large groups of neurons, smaller clusters are
further synchronized, indicating partitions into
fibers. These fibers require a more refined analysis of synchronization developed next.

Notably, brain activity under these conditions occurs spontaneously,
in the sense that dynamics arise in the absence of any acutely
delivered external and time-varying sensory stimulus.  Clusters of
synchronously active neurons, shown in Fig. \ref{1A}b, generate
activity patterns corresponding to behaviors such as forward crawl,
reverse crawl, ventral turn, and dorsal turn.  Importantly, fibers
of synchronous network activity can be robustly observed across all of
these conditions.  Thus, synchronous neuronal ensembles are
functionally related.  Since these dynamics and synchronies are not
acutely triggered or bound by fluctuating external inputs, they must
emerge as a property of the network itself.

Taken together, these features make this an ideal test case to
scrutinize the symmetry hypothesis.

\section{Synchrony and symmetry in the worm}
\label{sec:sync-symm}

Given that the structural connectome and neural synchronization in the
whole body worm is readily available, we would expect that the forward
arrow from structure $\rightsquigarrow $ function can easily be
validated in these animals, as was done for the TRN of {\it E. coli}
in Chapter \ref{chap:synchronization}. However, this is not the
case. Despite being fully mapped, no available connectome is complete,
and the connectomes of different animals are different. For instance,
if we analyze any available {\it C. elegans} connectome
\citep{varshney2011structural} and calculate either automorphisms or
fibrations, we will find just a few. Except in
Fig. \ref{fig:forwardgap} where \cite{morone2019symmetry} have curated
the data with a few hand-crafted modifications. However this is not
feasibly in general.  Yet there is an obvious synchronization of their
neurons, as seen in Fig. \ref{1A}. This means that this is an instance
of the inverse problem of reconstruction, discussed in Chapter
\ref{chap:function}.

\cite{avila2024symmetries} have developed a pipeline that utilizes the
algorithm developed in Section \ref{sec:integer} to repair the
connectome\index{connectome !repair } of {\it C. elegans}, guided by the synchronization dynamics
measured experimentally. Once this connectome is obtained, we can
perform further perturbations to the neurons, such as ablation or
inhibition, guided by fibration theory, to improve the understanding
of the neural system.

The method repairs the connectivity structure of a baseline
connectome, taken as the one curated by \cite{varshney2011structural},
and aims for a fibration symmetric connectome that reflects the
observed synchronization. This method provides a means to idealize
connectomes conceptually. The threshold for connectome modifications
is set to a strict 25\% variation based on inter-animal partial
connectome differences \citep{hall1991posterior}, to ensure behavioral
consistency despite neural network discrepancies. That is, the method
modifies a baseline connectome to reproduce the experimentally
obtained balanced coloring and it accepts the new connectome if the
changes to the connectivity is below the 25\% natural variation of the
connectome. This conceptual framework proposes a novel inference
scheme for reconstructing connectomes (or networks in general) while
retaining accuracy in modeling dynamical neural behavior.

\begin{figure}
  \centering \includegraphics[width=\textwidth]{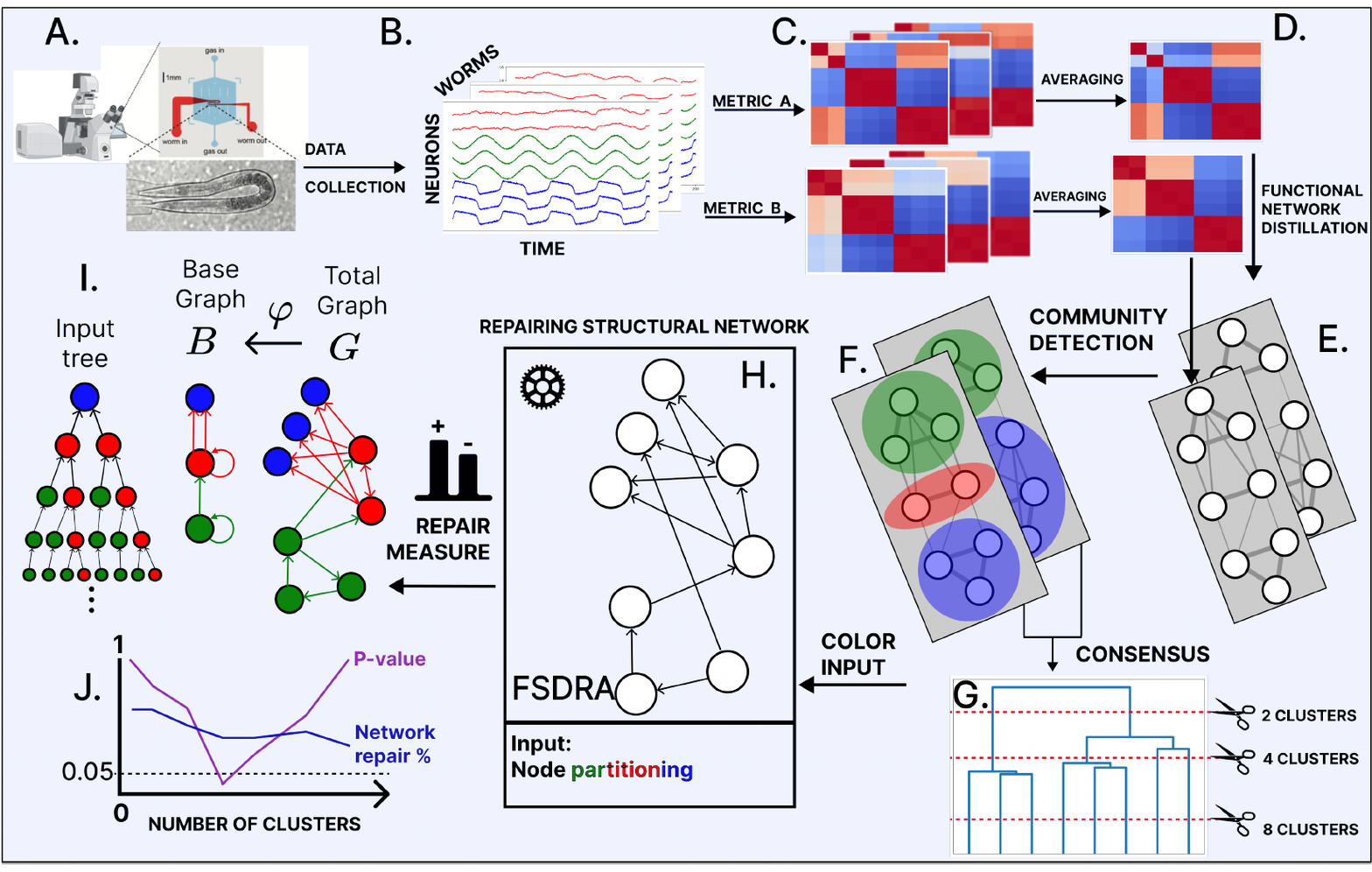}
\caption{\textbf{Pipeline for the structural network repair of the
    backward locomotion of {\it C. elegans} based on neural recordings.}
  This pipeline is quite general and can be applied to any dataset to
  infer the network from the synchronization. Figure reproduced from
  \citep{avila2024symmetries}. }
\label{fig:pipeline}
\commentAlt{Figure~\ref{fig:pipeline}: 
Described in caption/text. No alt-text required.
}
\end{figure}

Figure \ref{fig:pipeline} describes the pipeline of the method
developed by \cite{avila2024symmetries}:

\begin{enumerate}
\item The process begins at Fig. \ref{fig:pipeline}a with a
  microfluidic device that maintains constant gas and pressure,
  creating non-disruptive conditions for proper fluorescent Ca$++$
  imaging in {\it C. elegans} via a microscope and camera setup.

\item Time series data for activity traces of multiple neurons are
  recorded simultaneously, as shown in Fig. \ref{fig:pipeline}b. This
  procedure is performed on multiple worms under same
  conditions.
  
\item Various metrics that capture synchrony are applied to the
  recorded time series dynamics, producing correlation matrices that
  capture functional synchrony (Fig. \ref{fig:pipeline}c).

\item The correlation matrices shown in Fig. \ref{fig:pipeline}d are
  averaged across worms to obtain an average synchrony matrix.
  
  \item A percolation process is then applied to obtain the functional
    network (Fig. \ref{fig:pipeline}e) using standard methods
    \citep{gallos2012asmall,gili2024fibration}. Starting from a
    disconnected graph, links between nodes are progressively added
    in decreasing order of weight of the averaged correlation
    matrix.

\item Averaging the functional networks leads to a consensus matrix\index{consensus matrix }
  across different methods of synchrony (Fig. \ref{fig:pipeline}f).

\item Clustering algorithm\index{hierarchical clustering
  algorithm } is implemented to find a partition of synchrony clusters
  (Fig. \ref{fig:pipeline}g). Each neuron is assigned a color
  according to its cluster of synchrony. These colors are used as
  inputs to the repair algorithm in the next step.

\item We implement the mixed integer linear programming (MILP)
  algorithm \index{MILP algorithm } of Section \ref{sec:integer} to
  add/remove a minimal number of edges from the baseline incomplete
  connectome to produce an ideal fiber-symmetric network that
  reproduces the coloring in the consensus partition obtained in the
  previous step (Fig. \ref{fig:pipeline}h).

\item The algorithm produces a fiber-symmetric network with cluster
  synchronization that reproduces the experimental data. This network
  can be collapsed into a smaller representation, the base
  graph, where nodes belonging to the same fiber have isomorphic input
  trees (Fig. \ref{fig:pipeline}i).

\item For each consensus partition, a permutation $p$-value test is
  performed by permuting node labels and repairing the structural
  network 1,000 times. The frequency with which the initial partition
  outperforms permuted versions is measured by the modifications
  needed to create a fiber-symmetric network. The partition with the
  lowest $p$-value is chosen as the optimal solution
  (Fig. \ref{fig:pipeline}j).
\end{enumerate}

\subsection{Whole nervous system recording and motor neuron characterization}
\label{ca2plus}

\begin{figure}
 \centering \includegraphics[width=\textwidth]{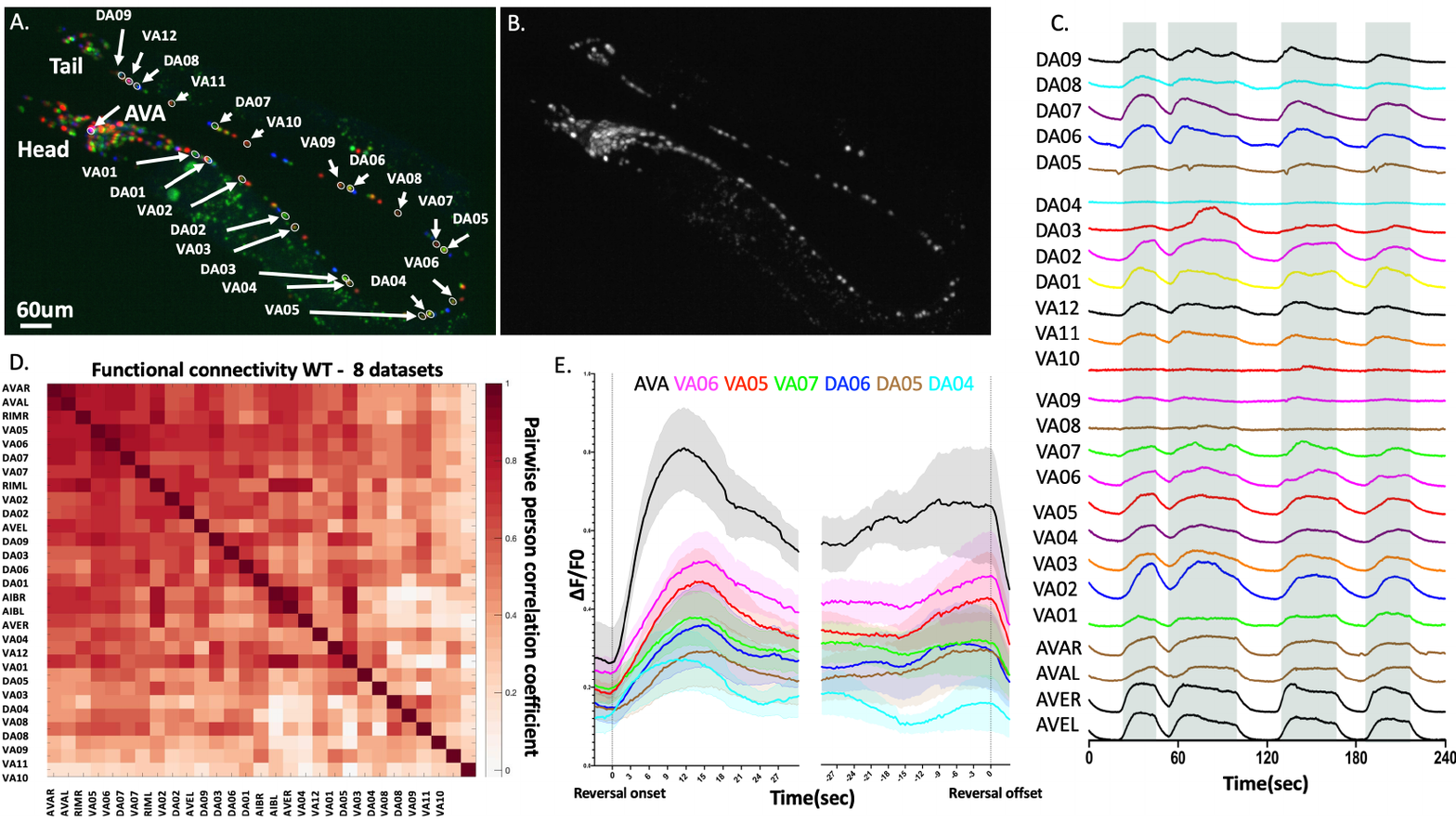}
\caption{ \textbf{Whole nervous system recording reveals synchronous
    motor neuron activity.} (\textbf{a}) NeuroPAL labeled worm with
  selected neuronal cell class identities indicated.  (\textbf{b}) Same
  worm as in (\textbf{a}) showing NLS-GCaMP6f labeling.  (\textbf{c})
  Activity time series (DF/F0) of selected neurons; gray shading indicates
  reversal command states defined by AVAL activity.  (\textbf{d})
  Pairwise Level of Synchronization ($LoS$) among selected
  neurons. Each matrix entry is the average of $n=3-8$ pairwise
  observations.  (\textbf{e}), (\textbf{f}) Triggered average ($\pm$ SEM)
  to reversal command onset (\textbf{e}) and offset (\textbf{f}), both
  defined by reference neuron AVAL. Shown are 3 example neurons of
  each VA and DA motorneuron class. Averages were calculated across
  $N=3-8$ recordings including $n=x-y$ events.  Figure reproduced from
  \citep{avila2024symmetries}.}
\label{fig:neuroPAL}
\commentAlt{Figure~\ref{fig:neuroPAL}: 
Described in caption/text. No alt-text required.
}
\end{figure}

The setup of \citep{avila2024symmetries} records Ca$++$ activity of
motorneurons\index{motorneuron } involved in backward locomotion along
the worm's length. Identification of each neuron is done by relying on
a neuron dictionary such as NeuroPAL that is able to capture almost
all nervous system activity with single-cell resolution
\citep{yemini2021neuropal} (Fig. \ref{fig:neuroPAL}). This involves
almost 300 neurons per recording, including the head ganglia, the
complete ventral cord, and the tail ganglia
(Fig. \ref{fig:neuroPAL}a). Figure \ref{fig:neuroPAL}c show a typical
multi-neuron time series, with discernible calcium activity patterns
and known neurons \citep{kato2015,Uzel2022neuronhubs} participating in
the reverse motion. These include reversal
interneurons\index{interneuron } such as AVAL/R and AVEL/R, which are
major coordinators of the motor activity of dorsal and ventral
reversal neurons, DA, and VA respectively.

\subsection{How to measure synchronization}
\label{sec:synchronization_measures}

The task of determining synchronization between two or more
simultaneously recorded signals has been carried out in many different
ways, depending on the characteristics that we are interested in
measuring and determining their level of similarity. This results in a
diverse array of techniques aiming to capture synchronization through
different approaches \citep{schoffelen2009source}.  Naturally, all
this leads to a zoo of methods that try to capture the level of
synchronization between signals through different aspects
\citep{schoffelen2009source}. Many of the techniques have been
discussed already in Section \ref{sec:gene-expression}. Below, we
expand on some methods more suitable for brain dynamics.

These techniques are generally classified into four primary groups
based on their focus: time domain versus frequency domain, and methods
that account for directional dependencies (like measuring causality in
the signals) versus those that do not (like a correlation function).
Some of these methods can be observed in Table
\ref{table:synchronization_measures}.

\begin{table*}
\centering
\begin{tabular}{|c|c|c|} 
\hline
 & \textbf{Time Domain} & \textbf{Frequency Domain} \\ \hline
\textbf{With Directional} & Cross-correlation  & Phase Locking Value  \\ 
\textbf{Dependency} & Granger Causality  &  Directed Transfer Function  \\
 & Transfer Entropy &  Partial Directed Coherence \\
\hline
\textbf{No Directional} & Pearson Correlation  & Coherence  \\ 
\textbf{Dependency} &  Spearman's Rank Correlation  & Spectral Correlation  \\
 & Mutual Information & Phase Coherence \\
\hline
\end{tabular}
\vspace{10pt}
\caption{\textbf{Synchronization measures in time and frequency domains.}
  There is a sea of measures used in the literature
  \citep{schoffelen2009source}. Each captures different aspects
  of synchronization.}
\label{table:synchronization_measures}
\end{table*}

Selecting the appropriate metrics can sometimes be difficult.  It is
The better approach is to use the information that these methods
provide in a consensus manner to decipher the synchronization in the
system.

\subsection{Measures of synchrony}
\index{synchrony !measure of }

In the neuroscience community, correlation is the de facto synchrony
measurement for neuronal data
\citep{kato2015,hallinen2021decoding,creamer2022correcting}. The
Pearson correlation coefficient\index{correlation coefficient }
(\ref{eq:pearson}) is one of the more popular metrics to capture
synchrony.  Its main lure is that it captures whether two signals have
the same distribution or whether they are monotonically related.
Different correlation measures can be used, such as Spearman, Kendall,
or Distance Correlation plus Covariance to aid in casting a wider net
to capture linear or nonlinear correlation dependencies among signals.

Another popular measure of synchronization is the Phase-Locking Value
(PLV) \citep{bruna2018phase,lachaux1999measuring}, which measures
synchrony in the phase rather than the amplitude of the signal. This
metric is used in Chapter \ref{chap:brain3} in the human brain.

We also use the Level of Synchrony\index{level of synchrony measure }
($LoS$) introduced in \citep{phillips2011gentle,avila2024fibration} to
determine how closely two signals are to perfect synchrony, as in
(\ref{eq:time}). This is done for $V_i$, which measures the dynamic of
a neuron using its membrane potential:
\begin{equation}\label{eq:cluster_synchrony}
  V_{i}(t) = V_{j}(t)  \,\,\, \forall i,j\in C_k.
\end{equation}
Here, $C_k$ is the synchrony cluster.

$LoS$ evaluates how closely synchronized two signals are over time, by
considering their instantaneous differences and scaling them through a
parameter $\sigma$:
\begin{equation}\label{eq:LoS}
        LoS_{ij} = \frac{1}{N+1}\sum_{t=0}^{N}
        \exp\left(-\frac{[V_{i}(tT/N) - V_{j}(tT/N)]^2}{2\sigma^2}\right).
\end{equation}
Here, $N+1$ is the total number of time steps in which $LoS$ is
applied between the signals from neurons $i$ and $j$, and $T$ is the
total time interval considered. The range of the $LoS$ function is $[0,1]$,
where a value of 1 indicates full synchrony, akin to
\eqref{eq:cluster_synchrony} as exemplified by identically colored
coded signals in Fig. \ref{fig:pipeline}b. A value of 0 indicates no
synchronization. The parametric term $\sigma$ serves as a scale to
define a benchmark for closeness between two points in time.

This parameter deals with natural variations in biological systems. No
two neurons are perfect copies of each other, so if two similar
neurons at rest are stimulated by identical signals, their outputs
(membrane potential) will naturally vary in intensity and phase by
small amounts. With this concept in mind, we can implement multiple
versions of the $LoS$, each with a different value for $\sigma$. When
the value of $\sigma$ is zero, each signal will be synchronous only
with its exact copy, as the value of $\sigma$ increases, it reaches a
state in which all signals are synchronous with all others. Thus, an
ideal value of $\sigma$ lies somewhere between zero and an upper
limit; the largest difference between signals at some time $t$ should
suffice.

\subsection{Synchrony and correlation matrix}
\label{sec:synchrony-correlation}

Once the particular metric of synchrony is chosen, either a Pearson
correlation (\ref{eq:pearson}), a {\em LoS} (\ref{eq:LoS}), or any
similar measure, an average correlation matrix\index{correlation
  matrix } is calculated. These matrices are first computed for each
of the $n$ subjects, capturing the synchronous activities observed
within each individual as depicted in Fig. \ref{fig:pipeline}c. These
individual matrices are then averaged in groups, as in
Fig. \ref{fig:pipeline}d.

The simplest procedure to build the functional network\index{network
  !functional } follows a percolation process introduced by
\cite{gallos2012asmall}, like that explained in Section
\ref{sec:thresholding}.  The functional entries are taken one at a
time, starting with the largest value and continuing in descending
order. The procedure halts when it incorporates the final neuron into
the network, as represented by Fig. \ref{fig:pipeline}e, leading to a
functional network as seen in Fig. \ref{fig:pipeline}e and exemplified
in Fig. \ref{perco-brain}. More sophisticated methods of percolation
can be used as well; see the analogous problem for gene coexpression
data in Sections \ref{sec:thresholding} and
\ref{sec:hierarchical-clustering}.

\subsection{Clique synchronization}
\label{sec:functional_clusters}

The functional network is used to find the synchrony clusters.
Various methods, such as cluster and community detection algorithms,
can be applied to extract these functionally synchronous clusters of
neurons from a functional network, reviewed before. Two additional
methods are briefly explained below.

The clique synchronization\index{clique synchronization } method introduced by
\cite{gili2024fibration} decides that a node belongs to a clique if the
average of its internal edges is bigger than any of its external
edges. Any non-assigned node is forced into the clique from which it
receives the highest average weight.

Ideally, perfect cluster synchronization is a non-overlapping, fully
connected induced subgraph (clique)\index{clique } embedded in the functional
network. Since ideal synchrony cannot be expected in real
data we relax this condition, allowing the fully connected subgraph
to be connected by weak inter-clique links. We define a {\em cluster
synchronization $N$-clique} as an induced, fully connected subgraph of
the functional network composed of $N$ nodes that satisfies the
following conditions:
\begin{equation}
\begin{array}{l}
\displaystyle\sum_{i<j}^{1,N} \sigma(x_i (t),x_j(t)) \geq {N(N-1)\over 2}\sigma(x_k
(t),x_{k'}(t)) \\ \qquad k= 1,..., N\mbox{ and } k' \in \mathcal{M}_k
\end{array},
\label{eq:1}
\end{equation}
where $\mathcal{M}_k$ is the set of nearest neighbors of node
$k=1,...,N$ not belonging to the clique concerned and $\sigma(x_i
(t),x_j(t))$ is the chosen metric of synchrony between time series
$x_i(t)$ and $x_j(t)$ of nodes $i$ and $j$, respectively.

The clusters of synchronized nodes are again obtained by applying a
percolation threshold procedure to the clique synchronization
matrix. Starting from a disconnected graph, links between nodes are
progressively added in decreasing order of weight in the correlation
matrix (i.e., degree of synchronization), starting from the largest. A
synchronized clique is found as soon as \eqref{eq:1} is satisfied.

The largest clique is chosen when two partially overlapping cliques
satisfy these conditions. If the size of these two cliques is the
same, the clique with the largest edge weight average is chosen.  The
process stops when the weight of the links to be added does not allow
further cliques to form. This process defines a hierarchy of cluster
synchronization according to the order of clique appearance in the
percolation process.

\subsection{Modularity and community detection}
\label{sec:mod_comm}

Another method to find clusters of synchrony are Louvain community
detection and any hierarchical clustering such as those discussed in
Section \ref{sec:louvain}. This kind of method decides the number of
clusters and the number of nodes in each cluster by optimizing the
modularity function \eqref{eq:Louvain}, which is solely
cluster-dependent.

The Louvain measurement at every execution assigns every neuron to its
synchrony cluster. It proceeds by randomly grouping neurons into
bigger clusters, accepting mergers only if the modularity value
\eqref{eq:Louvain} increases, and halting once this value can no
longer increase. Due to the stochastic nature of this process, each
execution may lead to a slightly different result. For this reason,
the Louvain method is executed 1,000 times for a distilled network
that retains the partition with the highest measured modularity.

\subsection{Consensus cluster synchronization}

A consensus is created to leverage the information obtained from all
the unique partitions across all methods, from Louvain to clique synchronization and metrics, from $LoS$ to different correlations measures,  and across parameters of these methods, following \citep{tian2022scmelody}. All these
matrices can be summed and normalized.  The consensus matrix is shown
in Fig. \ref{fig:consensus_solutions}a.

From this consensus, we can obtain various partitions depending on
where the dendrogram is sliced. In Fig. \ref{fig:consensus_solutions}a
the dendrogram is sliced at a value of 3  to obtain
its three major groups observed in the consensus matrix. Each block is further divided into
smaller diagonal blocks, indicating the solution with the least amount
of modification to convert the collapsed Varshney
connectome\index{Varshney connectome } into a fiber-symmetric solution
(5 clusters for a cutoff at 2).

\section{Symmetry-driven reconstruction of the {\it C. elegans} locomotion connectome}

\begin{figure}[t!]
\centering \includegraphics[width=\textwidth]{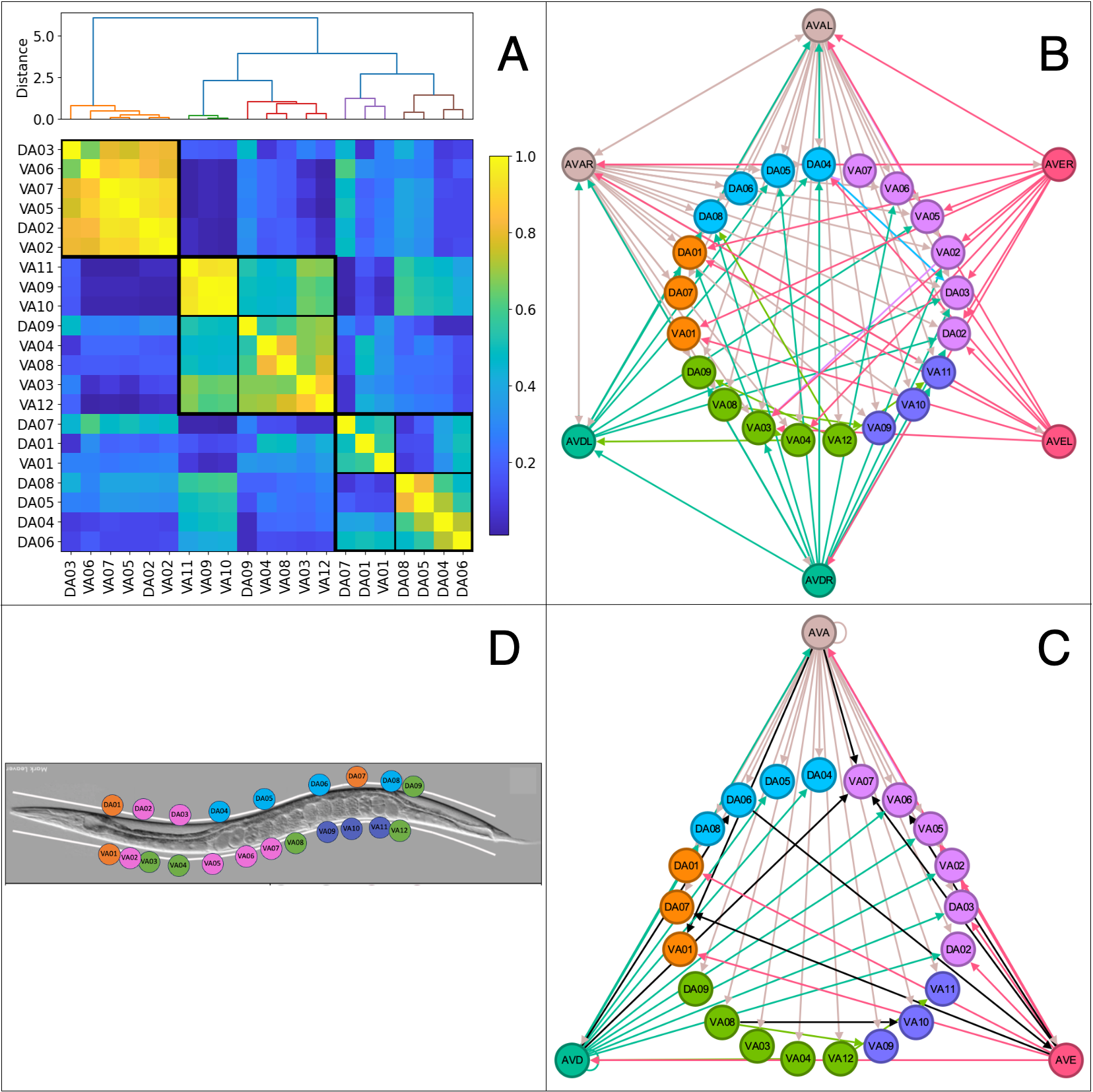}
\caption{\textbf{Fibration reconstruction of the locomotion connectome.}
  (\textbf{a}) 
  The averaged consensus co-occurrence matrix shows (five) cluster synchronization of the neuron activity. (\textbf{b}) Original
  binary Varshney chemical synaptic network among backward
  motor-neurons and their three primary interneuron pairs AVAL/R, AVEL/R
  and AVDL/R. (\textbf{c}) Inferred locomotion connectome obtained by applying the inference algorithm to a collapsed version of 
  (\textbf{b}). (\textbf{d})
    Anterior-posterior distribution of motor neurons (DAn,
  VAn) involved in the backward locomotion in 
  \textit{C.elegans} colored by the clusters of synchrony.   Figure reproduced from
  \citep{avila2024symmetries}.}
\label{fig:consensus_solutions}
\commentAlt{Figure~\ref{fig:consensus_solutions}: 
Described in caption/text. No alt-text required.
}
\end{figure}

The resulting clusters of neurons, found to be synchronous via the
above procedures are used to repair the raw chemical synapse
connectome of \textit{C.elegans} for the backward locomotion gait shown in Fig. \ref{fig:consensus_solutions}b. We
use the linear integer program described in Section \ref{sec:integer}
to modify this binary network by adding and removing the fewest edges.

The goal is to obtain a network with a fiber partition satisfying the
balanced coloring from cluster synchronization.  Using the clusters
found in the previous section, including consensus clustering, several
repairs can be produced, one for each method used to calculate the
clusters.  The optimal solution for each type of clustering is chosen
to be the one requiring the smallest number of modifications of the
raw connectome by addition and removal of edges.

Figure \ref{fig:consensus_solutions}c shows the final repaired network for the
optimal reconstruction of the connectome that corresponds to consensus
clustering. The final repairs are around 20\% of
the original edges, below the 25\% threshold for acceptance of the
solution.  Figure \ref{fig:consensus_solutions}d shows the neurons of
the {\it C. elegans} connectome, with the balanced coloring obtained
from the cluster analysis. 

\section{Perturbations and symmetry breaking by ablation and inhibition}
\label{ablation}

Perturbing a network to determine its function is a central problem in
biology and network science, encompassing brain/genetic network
theory. The development of specific experiments to measure network
dynamics by intervening with certain nodes is an experimental paradigm
with vast consequences for understanding the functionality of the
connectome.

Symmetries predict that symmetry-breaking by, for instance, ablation
of a specific neuron or a group of neurons or any other perturbation
such as optogenetics or transgenic inhibition, should lead to
desynchronization of nodes in the affected fibers.  These experiments
are difficult to implement since it is desirable to manipulate more
than one neuronal class to break left-right symmetry.
\cite{yan2017} used control theory to guide the manipulation of
the connectome and obtained new functionalities. However, a network
with strong symmetries is uncontrollable
\citep{russo2011symmetries}. On the other hand, symmetry
considerations can guide ablation experiments to break or create new
synchrony patterns in the network.

Many of the observed network symmetries stem from the left-right
symmetric body plan of nematodes. Permutations of left/right
pairs (e.g., AVAL/AVAR) make up a large component of the symmetry
operations. Thus, manipulating a neuronal class, e.g., AVA, will, in
almost all cases, represent a symmetric interrogation.
The vast majority of genetic drivers that are available for
optogenetics or transgenic inhibition affects both the left and the
right member of each class. In contrast, laser ablation is the
only reliable method that allows loss of function of individual
neurons, i.e., single members of a class.

Ablation experiments\index{ablation } can be guided by theory so that
they systematically search for the minimal ablations (ideally just one
or a few neurons) that have a maximal effect on breaking symmetry in
the network. Fibrations and symmetry groups predict that breaking
symmetries by laser ablation should lead to specific asynchronies in
neuronal activity patterns, while control ablations that do not affect
symmetries should not affect synchronies but could, perhaps, affect
other activity patterns.

In the case of control ablations, there can still be other effects on
network activity, such as altered signal amplitudes and frequencies,
and the theory predicts specific effects on synchronies. An example of
this kind is given in \citep{kato2015}. When both AVAL and AVAR are
inhibited, animals are largely unable to execute reverse crawling
actions due to a disconnect of the brain from the motor
sector. However, network synchronies across AVA connections remain
largely intact.

The theoretical reconstruction of the connectome suggests performing
systematic {\it in silico} searches, providing examples to apply laser
ablation procedures, and then to perform whole nervous system activity
recordings in worms, which could lead to a complete characterization
of synchronies in the connectome. The results of these experiments
would provide a rigorous test of the symmetry framework.

\subsection{Future work}

The use of integer linear programming to adjust connectomics data
based on functional imaging, demonstrate the use of a novel computational
approach to neuroscientific research. This technique refines existing
neural network models and provides a quantifiable method for aligning
theoretical predictions with empirical observations.
As shown in Fig. \ref{fig:consensus_solutions}, this points
towards an anatomical organization of the chemical network that
sustains the smooth sinusoidal movement of the worm while moving
backward. Our results suggest that there could be at least three major
groups of activity (anterior, midbody, and posterior) that relate to
the shape of this sinusoidal movement. Ultimately, our results also
suggest that  this activity  can be further subdivided into five
groups, which could be related to how the different motor neurons
connect to different muscle segments. 
Future ongoing experiments
imaging muscle activity in immobilized and freely moving conditions
will surely help to clarify this hypothesis.  

The results of \cite{avila2024symmetries} have substantiated the
pivotal role of connectome structure, specifically fibration
symmetries, in orchestrating neuronal synchronization in {\it
  C. elegans}. The refined understanding that structural connectomics
provides to underpin significant aspects of neural functionality, furthers our
grasp of the physical basis of neural synchronization. This
synchronization is crucial since it forms the foundation for coordinated
motor outputs and behavioral responses in organisms. The use of
advanced calcium imaging and graph theory to correlate these
structural motifs with dynamic patterns of neural activity offer a
compelling model for predicting neuronal behavior based on underlying
anatomical data.

Thus, the implications are not restricted to \textit{C.elegans}. The
principles of neuronal synchronization, facilitated by connectomic
structures have analogs across bilaterian species, including mice and
humans \citep{gili2024fibration} as investigated in the next two
chapters.  Investigating these principles in more complex nervous
systems reveals  insights into how brain-wide synchronization
patterns contribute to complex behaviors and cognitive
functions.


\chapter[Fibration Theory of the Brain II: the Minimal Engram in Mice]{\bf\textsf{Fibration Theory of the Brain II: \hbox{the Minimal} Engram in Mice}}
\label{chap:brain2}

\begin{chapterquote}
In 1949, Donald Hebb\index{Hebb, Donald } postulated that the substrate of memory resides
in the strengthening of synaptic connections, which occurs when the
neurons concerned are active at very similar times.  Hebb's principle
states that if one neuron firing consistently causes another to fire,
then the strength of the connection increases.  In this causal form,
Hebb's postulate has been largely confirmed experimentally. Important
though the postulate has been, it left unanswered the fundamental
question of how the memory engram, as a whole, is engraved in the
structure of the network connectome when new information is
encoded. What are the fundamental building blocks in the brain network
that form memory engrams to assimilate new information into our
knowledge base?  This chapter shows how engram formation is reflected
in the fibration symmetries of synaptic connections.
\end{chapterquote}

\section{One-to-many relation between structure and function in the engram}

It has been proposed by \cite{park2013structural} that the brain's functional activity selectively engages a subset of connections within the connectome 'highway' to operate in various functional states. This 'one-to-many' degeneracy in the structure-function relationship allows for the emergence of diverse functional states (e.g., resting, memory, movement) from a single, consistent connectome architecture. Although the postulate has been important, it left unanswered what the fundamental theoretical framework that links the connectome's structure to its function is. 
In this context, we focus on the formation of 'memory engram networks,' which serve as the neural substrate for storing and retrieving information related to experiences. We first show that mesoscopic brain areas exhibit covariance in the expression of immediate-early genes (c-Fos) across different animals, which is, as discussed in Section \ref{sec:gene-expression}, a form of cluster synchronization, not in time, but across animals. 

We utilize this synchronization to infer the substrate engram network by applying the inference algorithm developed in Section \ref{sec:integer}.
The theory yields falsifiable predictions that can be tested experimentally, thereby assessing the applicability of the entire symmetry framework in describing the brain.

The fibration reconstruction predicts that the Agranular insula, both its dorsal and ventral aspects, along with the basolateral amygdala, play a crucial role in the particular engram network. We experimentally validate this prediction by using targeted pharmacogenetic inactivation of these areas, demonstrating significant impairment in memory formation, retrieval, and performance on behavioral tasks compared to control experiments. 

Thus, the symmetry framework may serve as the missing link to connect function and structure and could also be a valuable tool for assessing the role of essential nodes in the broader context of synchronous systems, extending from other brain connectomes to genetic networks in health and disease.

\section{Where in the connectome are memories stored?}

In 1921, evolutionary biologist \cite{semon1921themneme} proposed the
idea of an {\it engram}\index{engram } as {\it `the enduring though primarily latent
modification in the irritable substance produced by stimulus'.} His
work inspired generations of neuroscientists to join the search for a
memory fingerprint in the brain.

A large body of experimental evidence supports the idea that memories
are made, or encoded, through the strengthening of synapses in
distributed associative circuits
\citep{rolls1998neural,roy2022brain,poo2016what}. These circuits
consist of interconnected excitatory and inhibitory neurons in
different brain sub-regions or network nodes.
However, memory\index{memory } is not a single unitary process, and it is neither
located in one particular place in the brain nor static in time.
Engrams are known to be distributed across different brain regions, defining separate neuronal assemblies (which represent multimodal
aspects of a given memory). How these distant assemblies are
coordinated to form a specific engram is still unknown.

Memory is, rather, a distributed and dynamic phenomenon that enriches
our knowledge base with new experiences. As we explore an unknown
place, for instance, a new memory of the context is formed in the
brain. The spatial landmarks, the pathways, the presence of
potentially useful resources (e.g., food), and the timing of the
exploration episode are stored in memory. After some time, navigating
that environment to collect food from the known resources and even
planning the time of the excursion and the pathway ahead are aided by
recollecting the stored memory traces.  These later excursions do not
require significant encoding of new information until, for instance,
one food resource is extinguished. At that moment, memory needs to be
updated to adapt to the new environmental configuration.

How do these memory engrams evolve across such contingencies? What is
the network structure of the memory engram at the moment of encoding,
during retrieval or updating? More importantly, which network building
blocks critically define information routing in the memory engram, 
how is the engram engraved in the structure of synaptic connections,
and how do engrams evolve during different learning cycles?  The
integrated, distributed, and specialized network activity leads to
synchronization in the information processing system, and, like any
synchronization phenomenon, it is expected to be supported by the
symmetries of the underlying engram connectome.\index{engram }\index{connectome }

\section{How to measure engram synchronization}

While memory
is thought to be encoded through modifications of the weights of
synaptic connections \citep{hebb1949},\index{synaptic connection } it is not known how the
multiple and dispersed cell assemblies activated by a particular
experience are engraved in the structure of the connectome. Time is
important for establishing associations at the cognitive level, as is
the timing of inputs for stabilizing synaptic modifications. Thus,
synchronization between experience-relevant neuronal populations is
fundamental for building memory engrams.

If the formation of the engram is a synchronization\index{synchronization } phenomenon among
neural populations supported by the modification of the weights of the
synaptic connections, it is natural to enquire whether the underlying
symmetries of the connectivity of the neurons in the engram could
determine the building blocks of the memory networks. This leads to
the synchronization of experience-relevant neural populations and
their binding into coherent engrams,\index{engram } thereby supporting new memory
formation.  Finding the building blocks\index{building block } of engram synchronization
reveals the distributed nature of memory.  It could also reveal the
mechanisms for segregation of synchronized function in the presence of integration of information in the entire brain, which is a
fundamental problem in neuroscience
\citep{tononi1994,gallos2012asmall}.

Synchrony\index{synchrony } in the engram\index{engram } can be tested in a real memory\index{memory } engram network
recorded from awake and freely behaving rodents involved in learning
tasks.  Animals acquire ‘knowledge’ no less than humans, so it should
be possible to use animal models of memory formation to secure a
mechanistic understanding through causal experiments.

Modern memory engram technologies, aided by molecular biology, make it
possible to tag activated neurons\index{neuron !activated } in the engram network in time
windows that can be experimentally controlled. It has been shown that
activation of neurons is accompanied by the expression of plasticity
related immediate-early genes (IEGs) such as c-Fos
\citep{vann2000using,vanelzakker2008environmental,josselyn2015finding}. In
this way, for instance, the engram cells associated with the formation
of a contextual fear memory can be genetically tagged
\citep{lei2007genome,dheeraj2016memory} and revealed (with
fluorophores),\index{fluorophore } and the activity can be manipulated (with c-Fos-driven
expression of optogenetic\index{optogenetics } tools). Later reactivation of c-Fos positive
neurons tagged during a particular experience evoke the memory
encoded during that experience, demonstrating that c-Fos labeled
neuronal assemblies constitute a memory engram \citep{ryan2015engram}.

\begin{figure}
  \centering{
  \includegraphics[width=.55\linewidth]{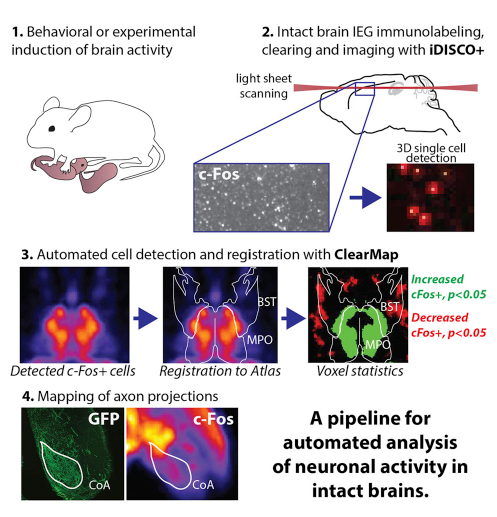}
  \includegraphics[width=.44\linewidth]{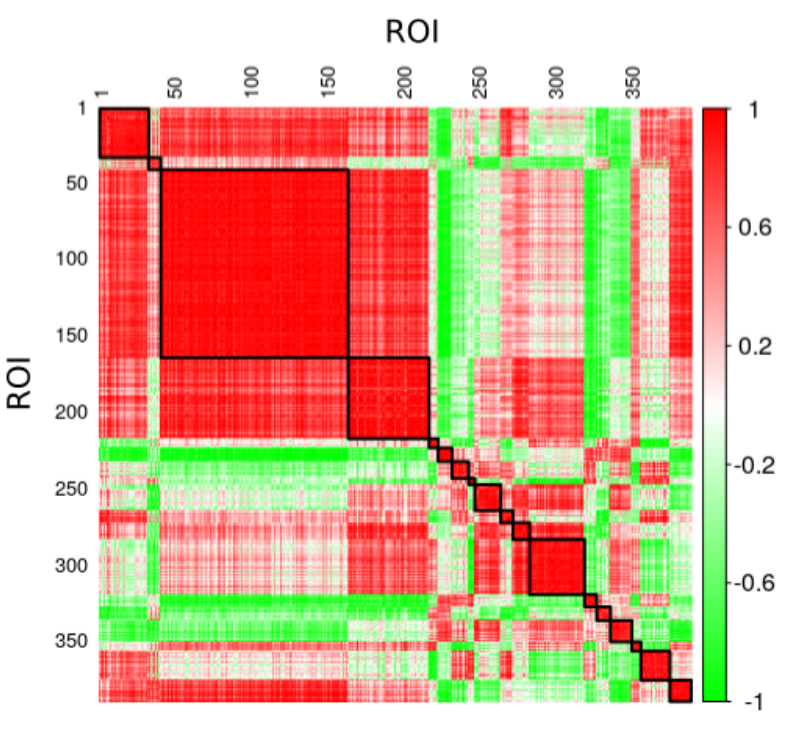}}
\caption{\textbf{Synchronization in engram formation in mice.}
  {\em Left}: Pipeline designed by \cite{renier2016mapping} for the
  automated analysis of neuronal activity in intact brains.  {\em Right}:  Using c-Fos
  expression\index{c-Fos } data obtained from \citep{renier2016mapping}, we show the
  synchronization of ROIs\index{ROI } in the brain engram network made of 380 ROIs
  in mice. Similar synchrony is observed at the global level in the
  memory network under LTP in \citep{delferraro2018finding}. The
  structural origin of the synchronization in the memory engram can be
  thought of in terms of symmetries of the underlying connectome. Figure reproduced from \citep{renier2016mapping}. Copyright  \copyright ~ 2016,  Elsevier Inc.}
\label{fig:renier}
\commentAlt{Figure~\ref{fig:renier}: 
Described in caption/text. No alt-text required.
}
\end{figure}

Using this technology, \cite{renier2016mapping} established an
experimental pipeline to
measure engram formation in mice,  shown on the left of Fig. \ref{fig:renier}.  The right of Fig. \ref{fig:renier}
shows the existence of synchronized neurons in the engram using these
data \citep{renier2016mapping}.  In each of these experiments, a mouse was placed in a cage to explore a new environment ($n$=4
mice). Then, mice
were sacrificed and processed histologically to find the specific set
of neurons that were active during the task. This was done by staining
the c-Fos\index{c-Fos } immediate-early gene, which correlates with neuronal
activity.  High-resolution optical imaging with light-sheet microscopy
and immediate-early genes (IEGs)\index{immediate-early gene }\index{IEG } expression (c-Fos) allows the
quantitative assessment of neuronal assemblies in the complete brain,
formed in response to a particular behavior.
Following brain clarification\index{brain clarification } (making tissue transparent), the active
c-Fos+ cells were imaged and quantified using available tools (e.g.,
ClearMap).\index{ClearMap }

Next, the number of cells in different regions of interest (ROIs) was
computed, based on the\index{Allen Brain Atlas } Allen Brain Atlas---an open-source parcellation
atlas for the mouse brain\index{mouse brain } \citep{oh2014amesoscale}. Using the sample
data, a set of 380 ROIs was considered to cover the complete brain to
compute a matrix of Pearson correlations\index{correlation } across all ROIs. Figure
\ref{fig:renier} (right) shows the resulting correlation matrix with
hierarchical clustering across the 380 ROIs. The correlation matrix,
analyzed by typical clustering methods, shows evidence for a set of
highly correlated synchronized ROIs supporting distributed
synchronization at the mesoscale.

Similarly, \cite{delferraro2018finding} observed high correlations in
long-term potentiation (LTP) experiments in rodents.  These
synchronies are manifestations of symmetries at the meso- and macro-scale analogous to those observed in the synchronized locomotion of
{\it C. elegans} discussed in Chapter \ref{chap:brain1} at the neural
level.  \cite{delferraro2018finding} confirmed synchronization in
distributed memory networks using fMRI data and pharmacogenetic
interventions in anesthetized rodents based on the synaptic plasticity
paradigm of LTP as a laboratory model of memory formation in rats. 

\cite{garcia2024minimal} study how engram
synchronization\index{synchronization } is manifested in the
symmetries of the mesoscopic connectome in mice.  They show that the
binding of dispersed cell assemblies into memory engrams is controlled
by a set of building blocks for network synchronization that can be
identified by fibration\index{fibration } theory.  An extrinsic value
of this analysis is the application of the mathematical tools for
neuronal network\index{network !neuronal } synchronization and
integration to understand different cognitive functions and
pathological alterations.

\begin{figure}
  \centering
  \includegraphics[width=\linewidth]{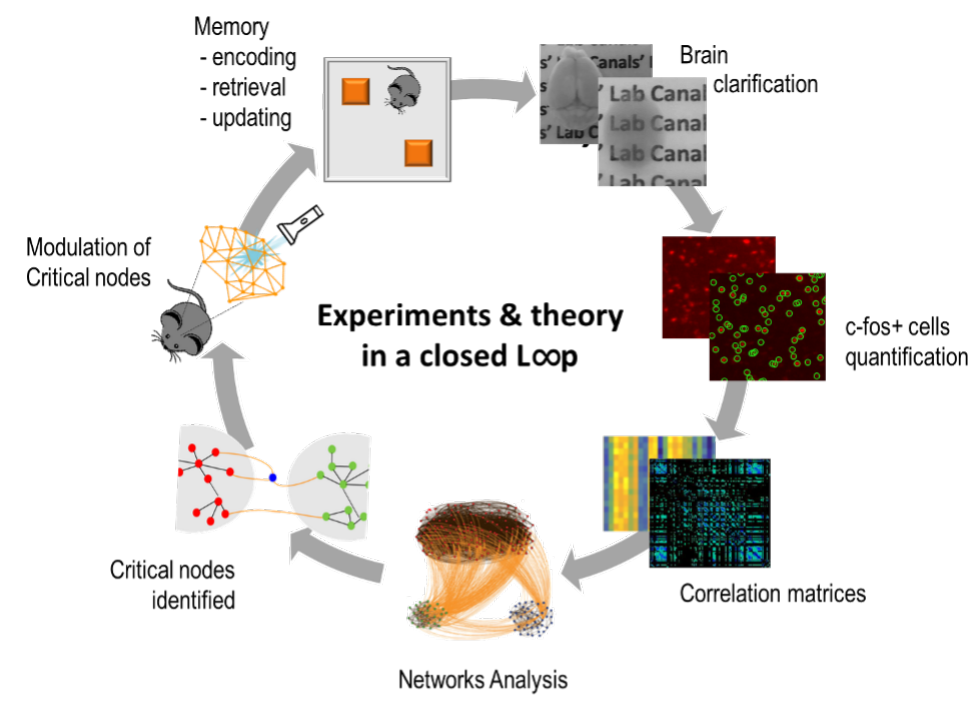}
\caption{\textbf{Closed-loop design to study memory engram
    synchronization and symmetry} \citep{garcia2024minimal}.
  Genetically and histologically labeled memory engram networks are
  reconstructed from clarified brains, critical nodes are identified, and
  their activity is modulated with optogenetic tools. Closing the loop,
  the outputs of these manipulations are read out at the
  cognitive and network levels.}
\label{fig:engram-design}
\commentAlt{Figure~\ref{fig:engram-design}: 
Described in caption/text. No alt-text required.
}
\end{figure}

\begin{figure}
  \centering
  \includegraphics[width=\linewidth]{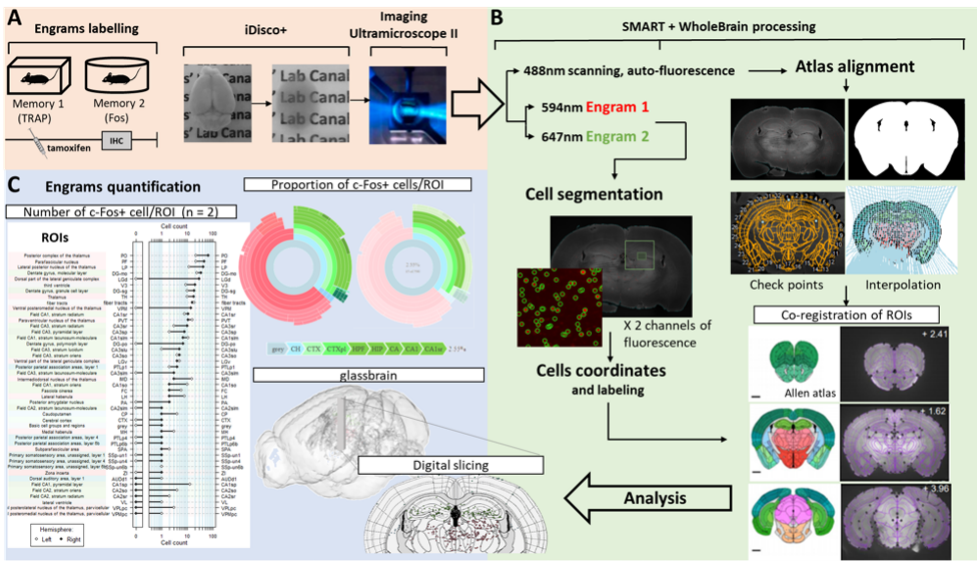}
\caption{\textbf{Whole-brain c-Fos+ cell quantification and memory
    engram network construction} \citep{garcia2024minimal}. (\textbf{a}) After labeling engram cells at two time points of memory
  formation, brains undergo iDisco+ procedures for tissue clarification
  and light sheet microscopy. (\textbf{b}) Imaging is done at three
  different wavelengths. At 488 nm, the autofluorescence of the tissue
  provides a complete 3D anatomical image optimal for atlas alignment
  (Allen Brain Atlas). At 594 nm and 647 nm, we image the c-Fos+ cells
  (genetically or immunohistologically labeled) of the two memory
  engrams, respectively. (\textbf{c}) The number of positive cells in
  each region of interest (ROI) defined in the atlas is counted per
  animal ($n = 7$). The size of the ROIs used to build the network is
  dynamically adjusted (circular hierarchical plot) to define the
  spatial resolution and to test the impact of coarse-graining in the
  final results. ROIs differentiate left and right hemispheres (dot
  plot). Finally, this information is used to build correlation
  matrices and perform network analysis to obtain the clusters of
  ROIs synchronized by the engram.}
\label{fig:construction}
\commentAlt{Figure~\ref{fig:construction}: 
Described in caption/text. No alt-text required.
}
\end{figure}

\cite{garcia2024minimal} recorded memory engram networks from awake
and freely behaving rodents involved in learning tasks of (1) memory
encoding\index{memory !encoding } (engram formation), (2) memory
retrieval\index{memory !retrieval } (engram reactivation), and (3)
memory updating\index{memory !updating } (engram
remodeling). Following a closed-loop design between theory and
experiment (Fig. \ref{fig:engram-design}), they used models of memory
formation to secure a mechanistic understanding with causal
experiments. They implemented the brain clarification\index{brain
  clarification } technique together with dynamic c-Fos quantification
(Fig. \ref{fig:construction}). They performed $n=7$ mice experiments
in which engram cells were recorded at two-time points, one during
memory encoding (genetically labeling c-Fos expression) and the other
during memory updating and recall (staining c-Fos with
immunolabeling). This method, developed in
\citep{denardo2019temporal}, provides dynamic information on engrams
since each animal provides two temporal instances of the engram
map. The first is at the inception of the memory; the second is when
it is recalled and updated, and the original information is enriched.

High-resolution optical imaging with light-sheet microscopy and
IEGs\index{IEG } expression\index{c-Fos } allows the
quantitative measurement of activated neuronal assemblies in the
complete brain of the mouse in response to a particular behavior
\citep{renier2016mapping,denardo2019temporal,roy2022brain}. c-Fos
measurements allow for studies in freely behaving animals and provide high
sensitivity to neuronal activity changes. After performing the
behavioral experiment, animals are sacrificed, and the perfused brains
are extracted and processed for tissue clarification.  Then, genetically
labeled or immunohistochemically stained c-Fos+ nuclei are imaged
under a light-sheet microscope. The 3D brain images of each 
subject are co-registered to a common anatomical template. This step
is critical since it makes it possible to identify and label the
same ROIs consistently in the population of subjects used to construct the
network.

\section{Inferring the engram network}

\begin{figure}
  \centering
  \includegraphics[width=\linewidth]{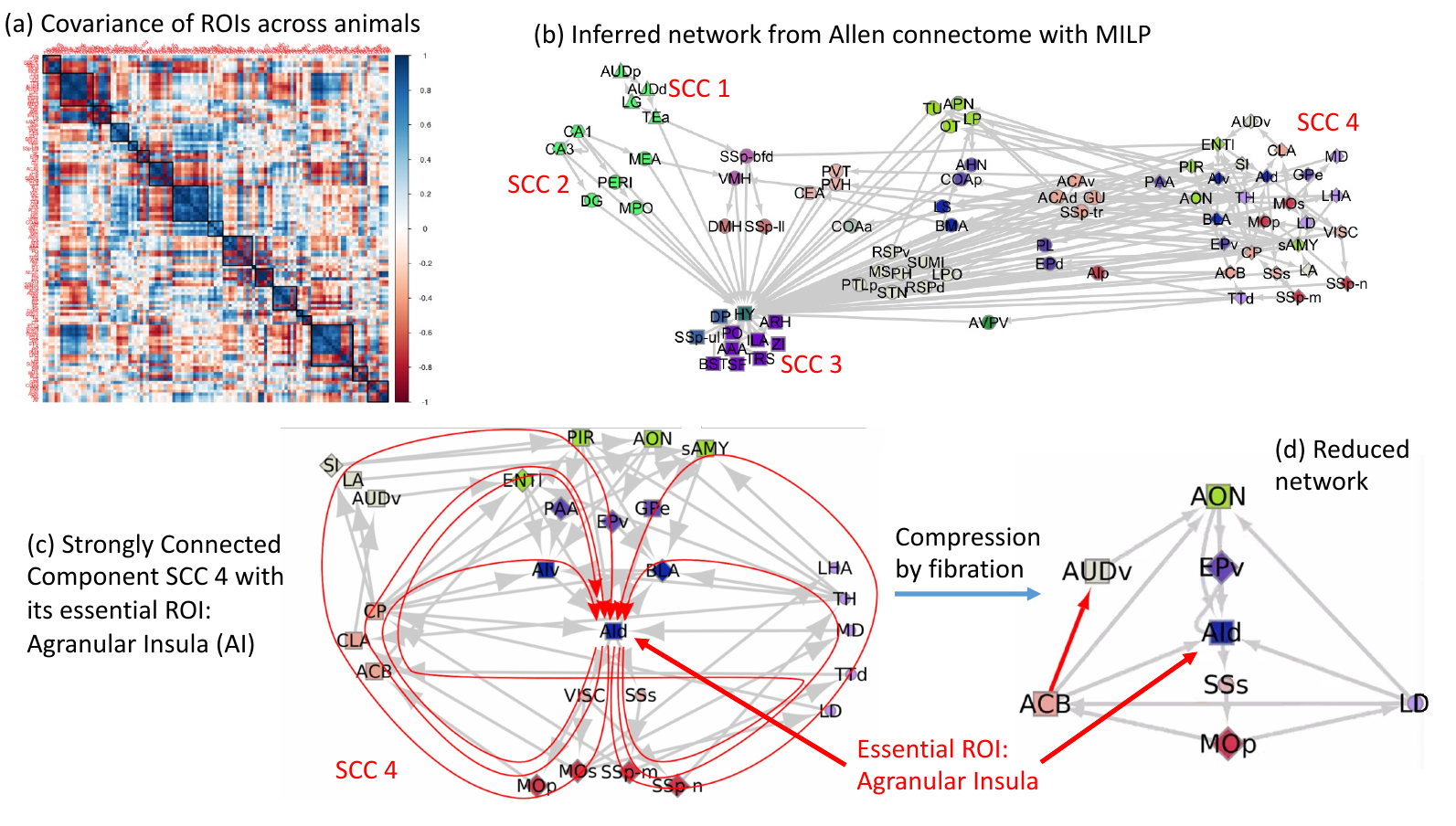}
\caption{\textbf{Inferring the engram connectome.}  \textbf{(a)} Covariance matrix between activated ROIs showing the hierarchical clustering into 18 clusters of coherence across animals. \textbf{(b)} The memory engram: Using the clusters from \textbf{(a)}, MILP then infers the substrate connectome to which graph tools are applied to find essential nodes. Four SCCs were identified; SCC 4 was the most unexpected. \textbf{(c)} SCC4 is analyzed, and the Agranular Insula is identified as the central node of this SCC and a primary candidate for disrupting its integrity. \textbf{(d)} By applying a fibration, the network can be compressed by collapsing the colored clusters into one representative node of the cluster for further graph processing. Results obtained by \citep{garcia2024minimal}. }
\label{fig:inferring-engram}
\commentAlt{Figure~\ref{fig:inferring-engram}: 
Described in caption/text. No alt-text required.
}
\end{figure}

\cite{garcia2024minimal} compute a covariance matrix across 380 ROIs
as defined in \citep{renier2016mapping} by averaging over all animals
considered. The resulting covariance matrix is shown in
Fig. \ref{fig:inferring-engram}a, which provides evidence of coherence
in the engram formation across animals.  The analysis is very similar
to the gene coexpression correlation\index{correlation } analysis in
Chapter \ref{chap:synchronization} and the
synchronization\index{synchronization } analysis in Chapter
\ref{chap:brain1} in {\it C. elegans}. However, the small number of
animals used in this study ($n=7$) requires massive regularization
techniques to make the correlations to be significant. It must be said
that $n=7$ is the largest c-Fos study so far.

The bundles of axonal tracts\index{axonal tract } between ROIs are
obtained from the Allen Brain Atlas connectome
\citep{oh2014amesoscale}.\index{connectome !mouse } These tracts
constitute the baseline connectome of the whole mouse brain.
It\index{connectome } represents the available highway system
composing the underlying information routes of the brain involved not
only in memory but in any functional task.  However, which routes of
this highway are utilized depends on the type of task for which the
brain is responding. The main hypothesis is that the brain's
functional activity utilizes a subset of the links available in the
highway connectome to operate in each functional state.

The `one-to-many' degenerate structure-function postulate of
\cite{park2013structural} allows the emergence of diverse functional
states from a unique static connectome architecture.  In the present
case, it means that, given the Allen baseline connectome (one),
different subsets of this connectome mediate different engrams (many).

To uncover this engram network, we use the inference algorithm
developed in Section \ref{sec:integer} by assigning a color to each
cluster of synchrony obtained from the covariance matrix.  The
baseline connectome is considered fairly complete, although some
missing links might exist. Therefore, the algorithm allows for the
removal of links as well as addition (with a lower probability).

\section{The minimal engram network}

Figure \ref{fig:inferring-engram}b shows the resulting inferred engram
network. Several steps of reduction have been applied here. First, we
identify 72 ROIs (out of 380) significantly activated in the
engram. Then, we group these ROIs in 18 clusters obtained through the
clustering of the covariance matrix in
Fig. \ref{fig:inferring-engram}a. Each cluster is assigned a color. We
then reconstruct the baseline Allen connectome
Fig. \ref{fig:inferring-engram}b to infer a network that satisfies the
balanced coloring based on the 18 clusters. After obtaining this
connectome, graph analysis follows.

The first step is to remove (trivial) ROIs that do not contribute to
the integrity of the network. These are nodes that have no
out-degree. They only receive information but do not transmit it.  The
next step is to obtain the graph's SCCs, where information travels
across cycles.  We find that there are 4 SCCs in the network
(Fig. \ref{fig:inferring-engram}b). Two of them are expected: one is
dominated by the hypothalamus (HY in SSC 3), which is a super hub
receiving signals from all SCCs, and another by hippocampal CA1 and
CA3, which is central for spatial memory formation. All of the SCCs
send information to SCC 3, both directly and indirectly, through paths
that cross other 'bridge' nodes. In fact, SSC 3 acts as an information
sink, receiving input from many ROIs.  The rest of the ROIs are not in
SCC, but they are feed-forward connectors of the SCCs.

\begin{figure}
  \centering
  \includegraphics[width=\linewidth]{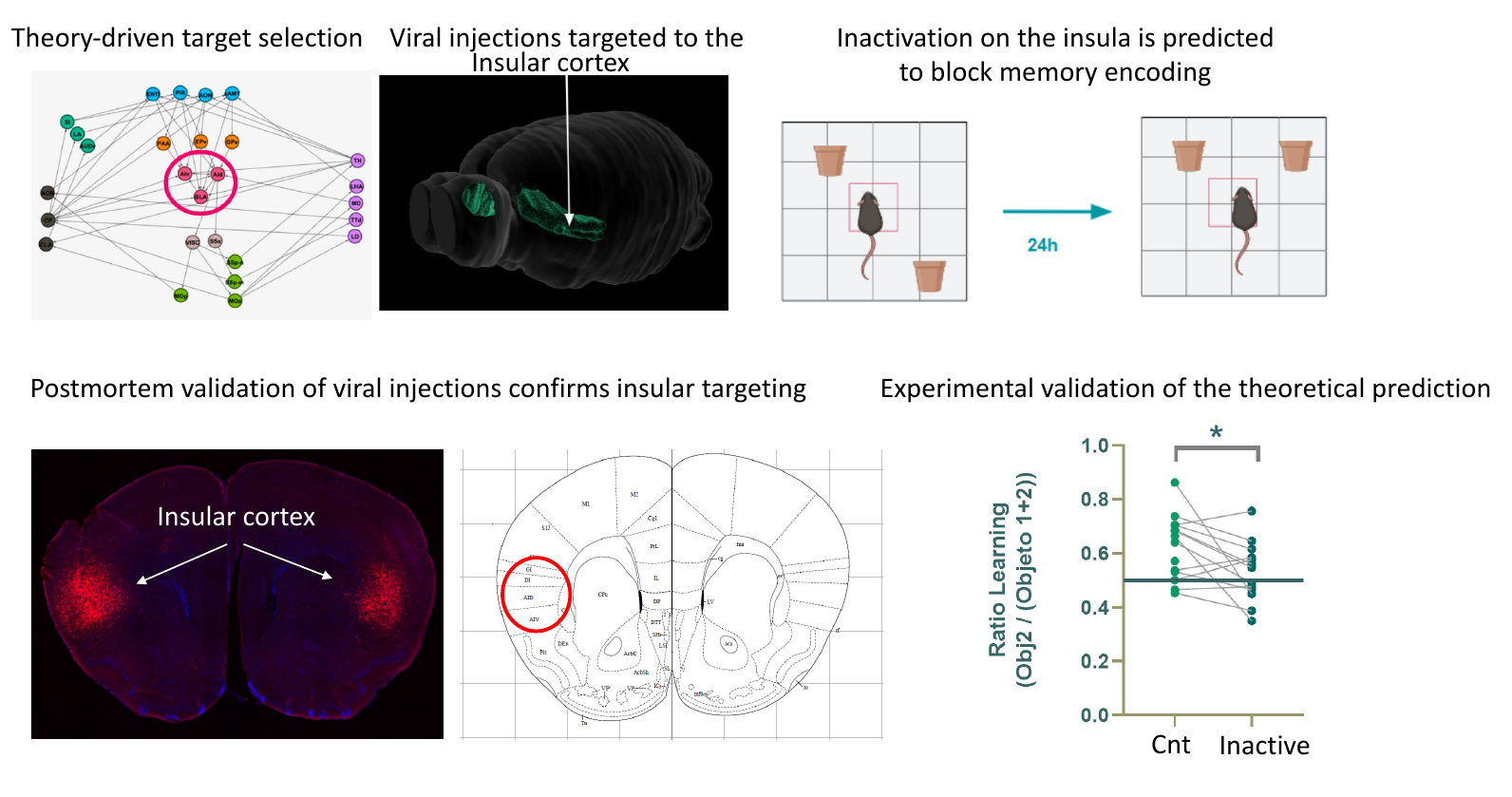}
\caption{\textbf{Disrupting the minimal engram.} Inactivation of the insula using 13 mice. Results obtained by \citep{garcia2024minimal}. After determining the target coordinates based on mouse anatomy, DREADD-expressing viruses were injected to inhibit the insula. Postmortem verification (red staining in the histology and red circle in the anatomical atlas) confirmed the accuracy of the procedure. In these animals, memory formation occurs normally when they encode contextual information with an intact insula (Control, light green dots). However, when the insula is inactivated ('inactive'), the animals lose the ability to encode new memories (dark green dots), as shown in the bottom-right figure.  The learning ratio on the right quantifies the animal's exploration of the two objects in the arena, with higher ratios demonstrating a better memory representation of the two objects' location. The asterisk denotes statistical significance ($p>0.05$ paired t-test).}
\label{fig:disrupting}
\commentAlt{Figure~\ref{fig:disrupting}: 
Described in caption/text. No alt-text required.
}
\end{figure}

We further do a network analysis of SCC 4 using different centralities
to find the essential nodes of this component
\citep{makse2024thescience}. The essential nodes that hold this SCC
together are part of the insular cortex, more specifically, the
Agranular Insula (AI) on its dorsal (AId) and ventral (AIv) aspects,
which, together with the basolateral amygdala (BLA) form a fiber at
the center of SCC 4. This is also a hub of SCC 4 and has high
betweenness centrality since there are many paths, the two red cycles
in Fig. \ref{fig:inferring-engram}c, that go through this fiber.  Upon
inactivation of the AI, the whole SCC is disintegrated, revealing the
essential character of this fiber.  The network can be reduced for
further processing by collapsing each of the fibers into the base, as
seen in Fig. \ref{fig:inferring-engram}d.

Using the knowledge amassed in the entire book, we have obtained the
{\it 'minimal engram'} network. In this network, each supernode is a
SCC. SCC 1, 2, and 4 send information to the sink hypothalamus-SCC
3. Connector nodes control and act as intermediates between these
SCCs. Inside the SCCs, a cycling structure is found composed of memory
toggle-switch building blocks.  Amazingly, the structure of this
minimal engram is quite similar to the minimal TRN uncovered in
Fig. \ref{fig:main-reduced}.

The analysis predicts that upon inactivating the AI, the SCC 4 would
disintegrate, and memory formation would be dramatically impaired, in
comparison to the inactivation of other nodes in the same SCC or
others nodes, which are defined as non-essential. Inactivation of
non-essential nodes may not produce a large effect since the network
contains other pathways to transmit the information (for instance,
other cycles in SCC 4).

\cite{garcia2024minimal} confirmed experimentally this prediction by
pharmacological inactivation ($n=13$) of the insular cortex neurons
identified by the graph model.  To achieve inhibition of this region,
they use parvalbumin (PV)-cre transgenic mouse, enabling
cell-type-specific protein expression via adeno-associated viruses
stereotactically injected into the target area, as done in the previous
studies \citep{delferraro2018finding}. Parvalbumin-expressing neurons
are inhibitory interneurons, so activating this population induces
strong inhibition in the insula. The results, presented in
Fig. \ref{fig:disrupting}, demonstrate that insula inhibition
negatively impacts memory formation. Figure \ref{fig:disrupting}
bottom right shows that when the insula is inactivated ('Inactive'),
the animals lose the ability to encode new memories (dark green dots)
in comparison with control experiments ('Cnt').

This result offers promising evidence for the symmetry theory
inference of the network and the identification of critical nodes within it. It
represents an encouraging first step in demonstrating that the
symmetry approach can yield falsifiable predictions that can be tested
experimentally. Numerous future analyses are expected. For example, it
would be valuable to clarify whether this type of manipulation causes
a disconnection of a significant component, such as the SCC 4 of the
memory engram.

Overall, fibration appears to be an effective theory for bridging the
gap between structure and function, helping to identify essential
nodes in the network.


\chapter[Fibration Theory of the Brain III: 
the Human Language Network]{\bf\textsf{Fibration Theory of the Brain III: 
\hbox{the Human} Language Network}}
\label{chap:brain3}

\begin{chapterquote} 
Understanding the human brain organization is the final frontier for the
applicability of fibration theory.  This chapter explores the
relationship between structural connectivity (connectome) and the
emergent synchronization of mesoscopic regions of interest in the
human brain network. Among the allowed patterns of structural
connectivity, synchronization elicits different fibration symmetry
subsets according to the functional engagement of the brain. The
resting state of the brain is a particular realization of the cerebral
synchronization pattern characterized by a fibration symmetry that is
broken in the transition from rest to a task engaged in language.
\end{chapterquote}

\begin{figure*}[b!]
\centering \includegraphics[width=\textwidth]{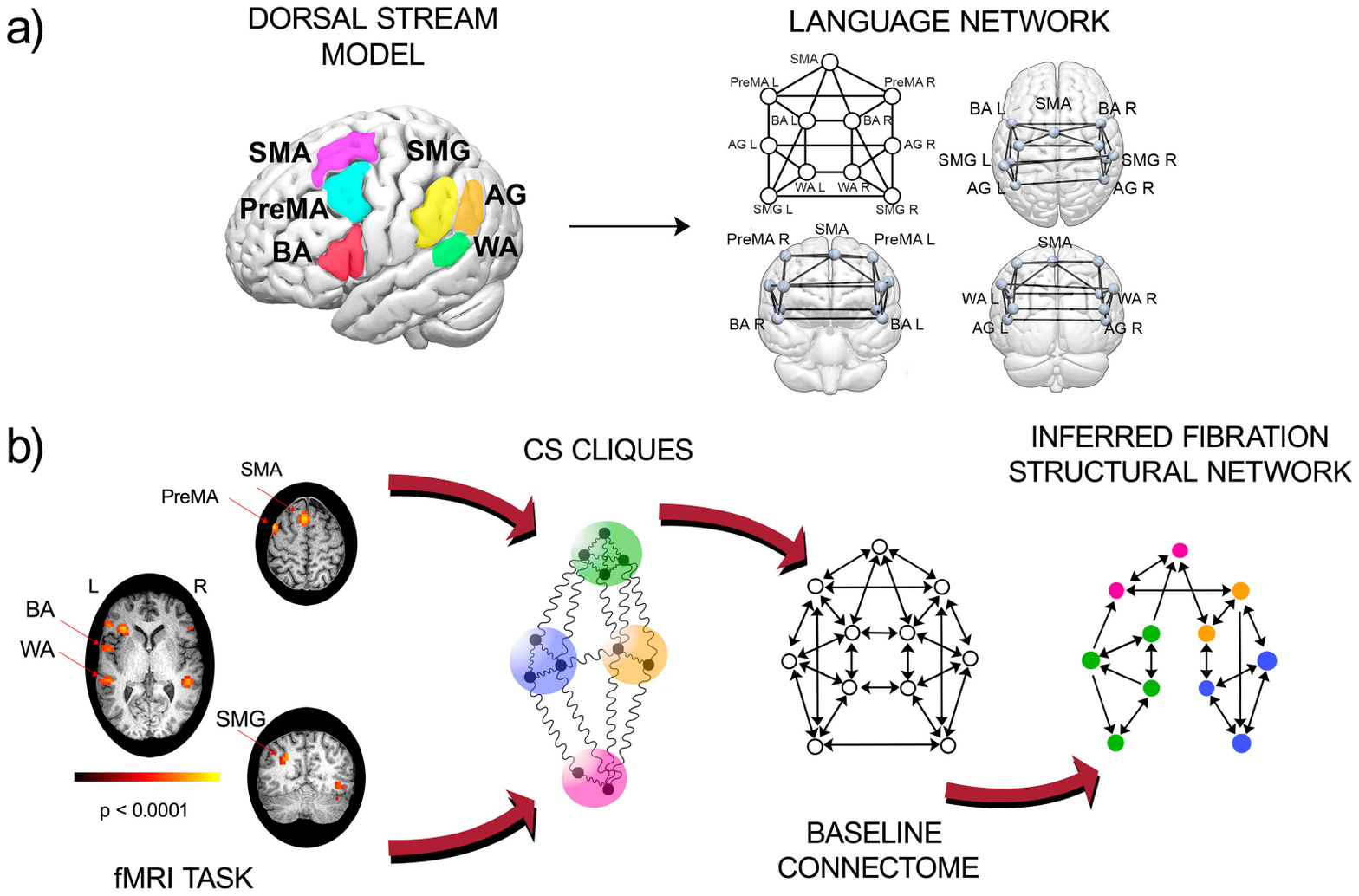}
	\caption{\textbf{Synchronization-guided inference
            algorithm.} We exemplify the algorithm in the case of a
          brain network. (\textbf{a}) Left: Nodes of the primary language
          network are given by the dorsal stream of the dual-stream
          model localized in the 3D brain. Right: dual (dorsal) stream
          baseline connectome showing the fiber tracts between the
          ROIs in (\textbf{a}). (\textbf{b}) Pipeline for inference of the
          structural network from dynamical data. Left: fMRI images
          for a task or resting state over many subjects are taken as
          inputs to calculate the group-average cluster synchronization
          cliques among the nodes. The clusters are identified with
          the balanced colors or fibers in the baseline connectome. A
          mixed integer programming algorithm is employed to optimally
          infer the structural network (right) that sustains the
          coloring cluster pattern obtained from the dynamics. Figure
          reproduced from \citep{gili2024fibration}.}
	\label{fig:intro-gili}
\commentAlt{Figure~\ref{fig:intro-gili}: 
Described in caption/text. No alt-text required.
}
\end{figure*}

\section{Structure and function in the human brain}

The structure $\rightsquigarrow$ function relation has been deeply studied in the human
brain at the mesoscopic level \citep{park2013structural,friston2011functional}. Studies use diffusion
and functional MRI to correlate white matter tracts\index{white matter tract } to the functional
coupling between the appropriate region of interests\index{region of interest } (ROIs)\index{ROI }
\citep{honey2009predicting,koch2002an,greicius2009resting,vandenheuvel2009functionally,hagmann2008mapping}.
Previous studies have looked at correlations between structural
connectivity (SC, e.g., from diffusion tensor imaging, DTI) and
resting-state functional connectivity (rsFC, from fMRI) for
anatomically defined ROIs
\citep{honey2009predicting,koch2002an,greicius2009resting,vandenheuvel2009functionally,hagmann2008mapping}.
While the SC resembles the rsFC, two ROIs can be structurally
connected but not functionally related, and vice versa.
Thus, these fMRI studies have demonstrated that anatomical
connectivity does not always translate into functional connectivity.

\cite{6deco2013} studied the structure-function relation via theory and
modeling.
A theoretical framework to study the link between anatomical
structure and resting-state dynamics were proposed based on DTI tractography
and parcellation to provide the neuroanatomical structural network
averaged across subjects.
A neurodynamical model is used to model BOLD signals in resting and
task-based cognitive states. The model is validated by comparing
predicted spatiotemporal patterns with empirical functional
connectivity data.

When the graph has no structure---think, for instance, of an
Erd\H{o}s-Rényi graph wherein nodes are connected
at random---nothing can be said about its function. Away from
randomness, many structures have been found in graphs. Neural graphs display modules, motifs, hubs, small worlds, fractality,
and others \citep{bullmore2009complex,rubinov2010complex}. However, most of these structures say little about the
function of the network.

However, when symmetries determine the structure of the graph, the
relationship between structure and synchronization becomes
clear. If we identify synchronization with function, then a structure-function relation can be studied from a theoretical point of view. In this chapter, we explore this hypothesis at the mesoscopic
scales measured by fMRI activity in the human brain.  We show how
symmetries of the mesoscale connectome\index{mesoscale connectome } (white matter tracts)\index{white matter tract } can
explain the synchronization of brain activity at scales measured by
fMRI in the human brain.

This chapter develops the symmetry viewpoint of the human brain,
letting empirical activity synchronization drive reconstructions of
the connectome based on symmetry fibrations,\index{fibration !reconstruction of
connectome } and inferring how
communicability shapes communication patterns among regions of the
human brain using the reconstruction algorithms of Chapter
\ref{chap:function}.

Our contention is not that symmetries should describe human connectomes
exactly. Instead, these symmetries must be realized in biology in an
approximate way as quasifibrations or pseudobalanced colorings, as
described in Section \ref{algo-pseudo}.  Synchronization in the
fMRI activity drives the inference of an ideal connectome that has
vestiges of an underlying symmetry common to all subjects, shedding
light on the engagement of the brain in different functional states.

\section{Human language network}
\label{sec:human}

Activity in the human brain\index{brain !human } occurs at many scales. Spatial
scales range from nanoscales at synaptic connections to systems of
neurons and their organizations, to large-scale network structures
such as cortical and sensory-motor networks. Temporal activity also occurs on
widespread timescales, ranging from the milliseconds of action
potential transmission to the minutes to a lifetime of large-scale
plasticity changes triggered by experience and memory.

There are many experimental ways to interrogate brain activity 
at these different scales.  Functional magnetic resonance imaging\index{functional magnetic resonance imaging } (fMRI)\index{fMRI }
lumps all these scales together into a blood-oxygen-level-dependent\index{blood-oxygen-level-dependent signal }
(BOLD)\index{BOLD } signal.  The BOLD signal is not a direct measure of neuronal
activity, as given, for instance, by action potentials. It is an
indirect measure of neural activity that measures changes in
deoxyhemoglobin induced by alterations in blood flow and levels of
oxygen in the brain. These changes are coupled to neuronal activity by
neurovascular coupling.

BOLD signals can be used to map the functional connectivity of  ROIs
in the brain at scales of a few millimeters and a few seconds.
In the human brain, activity is measured in an fMRI voxel,
which is typically around one mm$^3$ and contains about a few hundred thousand neurons per voxel in the cerebral cortex.  Whole
brain fMRI imaging interrogates the brain at these intermediate
temporal and spatial resolutions, but it is not sensitive to the msec
temporal scales of neural activity.

To understand these data, we can examine the correlation\index{correlation } functions of BOLD activity, as discussed in Chapter \ref{chap:synchronization}. This approach provides evidence of synchronization in the brain at broader scales. In this chapter, we investigate the phenomenon of synchronization in the human brain during a language task and demonstrate its connection to the underlying symmetries in the structural connections of white matter tracts.\index{white matter tract }

\cite{gili2024fibration} have analyzed ($n=20$) fMRI networks and DTI
from healthy individuals and brain tumor patients at Memorial Sloan
Kettering Cancer Center (MSKCC) in New York City.  These subjects
performed a language task and resting state (RS), allowing the construction of
networks associated with expressive language. Language function has
been widely studied using both resting state and task-based fMRI. In
task-based fMRI, we can be reasonably sure of the functional
activation in the brain concerned
\citep{holodny2002translocation,li2019functional,li2020core,li2021monolingual}.
The pipeline of the analysis is shown in Fig. \ref{fig:intro-gili}.

The subjects perform two different language-associated tasks,\index{language } as well
as the typical resting state protocol. The task-based protocol
emphasizes a language function known to be associated with the left
frontal lobe. Phonemic fluency\index{phonemic fluency } (letter) and verb generation\index{verb generation } tasks are
applied as part of the presurgical language task panel for
preoperative planning for tumor surgery at MSKCC. During the phonemic
fluency task, subjects generate words that begin with a presented
letter (for example: a patient presented with the letter `A' may generate
words such as ‘apple’, ‘apron’, or ‘ashtray’).
Subjects are also instructed to lie in the scanner with their eyes
open, to try to think of nothing in particular, and to keep fixating
on a central cross on a screen during RS.

According to current dual-stream models of language, different ROIs are
 involved in the language task
\citep{hickok2007cortical,hickok2022dual}. The areas that are most
responsive in language task paradigms have been studied in
\citep{li2019functional,li2020core,li2021monolingual}.  In tasks that
are focused on language production, only regions of the dorsal stream
are part of the analysis.
Specifically, the most current dual-stream model of
\cite{hickok2022dual} asserts that the following  anatomically defined ROIs
of the dorsal stream are involved in these tasks: supplementary motor
area (SMA), premotor area (premotor, left and right), supramarginal
gyrus (supramarginal, left and right), Broca's Area\index{Broca's area } (BA, left and
right), angular Gyrus\index{angular Gyrus } (angular, left and right), Wernicke's Area\index{Wernicke's area } (WA,
left and right).

The BA and WA are the primary areas responsible for language
expression and comprehension. The SMA has
largely been considered to be involved in controlling speech-motor
functions. It has also been shown by \cite{hertrich2016role} that the
SMA performs several higher-level control tasks during a speech
communication and language comprehension. The angular gyrus is
associated with complex language functions (i.e., reading, writing, and
interpreting what is written), and the supramarginal gyrus is
involved in the phonological processing of highly cognitive
tasks. Processing an action verb depends in part on activity in a
motor region that contributes to planning and executing the action
named by the verb, and the premotor cortex is known to be functionally
involved in the understanding of action language
\citep{chang2023representing}.

\section{Synchrony in the human brain at the mesoscale scale}

Cluster synchronization\index{synchronization !cluster } can be determined using the procedures
discussed in previous chapters. In addition,  \cite{gili2024fibration} has
developed a method specifically tailored to fMRI data, as follows.

\begin{figure*}[t!]
\centering
\includegraphics[width=\textwidth]{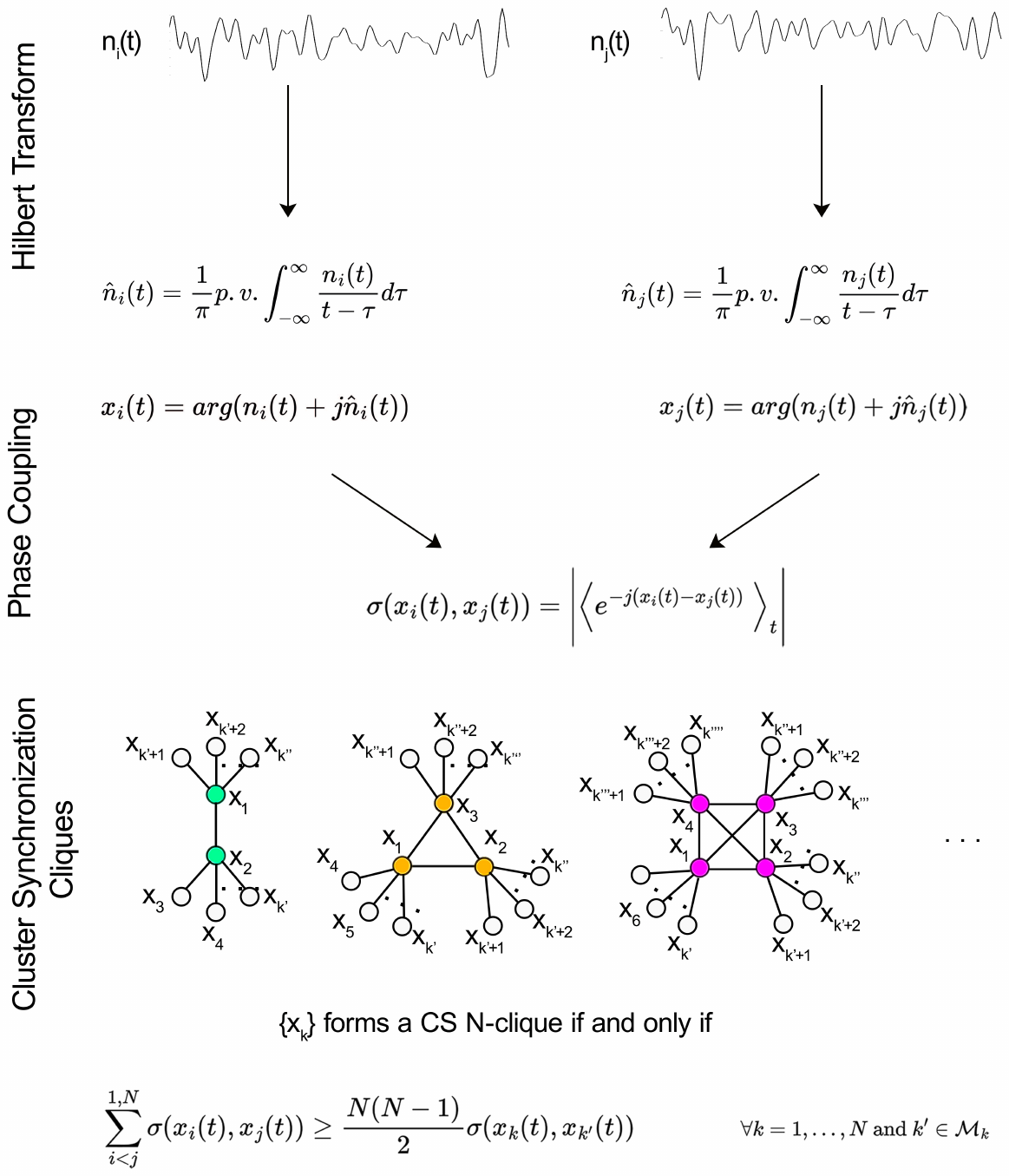}
\caption{ \textbf{Schematics of the synchronization clustering algorithm
    for fMRI}. Pairs of time series coming from pairs of cerebral ROIs
  are Hilbert transformed and entered into the phase-locking value
  calculation. Once all pairs of regions of interest are included
  in the calculation, the synchronization clustering algorithm is
  implemented. Figure reproduced from \citep{gili2024fibration}.}
	\label{fig:plv}
\commentAlt{Figure~\ref{fig:plv}: 
Described in caption/text. No alt-text required.
}
\end{figure*}

First, we use standard methods to build the functional network from
the time-dependent, fMRI-measured BOLD signal. For a single subject,
we measure synchronization via the Phase
Locking Value\index{phase locking value } (PLV) \citep{bruna2018phase,lachaux1999measuring} for
the BOLD time series from each ROI pair.\index{BOLD time series }
This metric captures phase synchronization, ignoring the amplitude. Thus, two signals with different amplitudes can still be synchronized by their phases.
BOLD time series were extracted from all voxels in a sphere of radius 6
mm centered on the coordinates of each ROI. Each ROI was composed of
123 voxels. Once time series had been obtained for the eleven ROIs included
in the study (by spatial averaging the BOLD signal within each ROI at
each time point), the synchronization was calculated as follows.

Given the BOLD signals $n_u(t)$ and $n_v(t)$ coming from regions $u,v
= 1, ..., N$ ($N = 11$), their instantaneous phases $\phi_{n_u}(t)$
and $\phi_{n_v}(t)$ can be obtained by means of their Hilbert
transforms (Fig. \ref{fig:plv}).  The PLV
$\sigma(\phi_{n_u}(t),\phi_{n_v}(t))$ is then given by
\begin{equation}
    \sigma(\phi_{n_u}(t),\phi_{n_v}(t))=|{\langle e^{-j(\phi_{n_u}(t)-\phi_{n_v}(t))} \rangle}_t| ,
\end{equation}
where $j$ is the imaginary unit.

To test the statistical significance of the PLVs, a non-parametric
permutation test was run by generating surrogate ROI signals randomly
rearranged and eventually time-reversed (1,000 permutations). This
procedure generates a null distribution that has
the same parameters (mean and standard deviation) as the
original data and has similar (but not identical) temporal dynamics.  This
process produced a null distribution of t-statistics that provided the
one-tailed $p$-value estimated using a generalized Pareto
distribution for the tail of the permutation distribution
\citep{winkler2016faster}. Correction for multiple comparisons was
provided by thresholding\index{thresholding } statistical maps at the 95th percentile
(P$<$0.05, FDR) of the maximum t distribution from the permutation
\citep{winkler2014permutation}.

The PLVs were then entered in a $N\times N$ correlation matrix,
representing the correlation/synchronization or PLV
matrix. Finally, the PLV matrices were averaged across subjects in
each experimental condition (resting state, phonemic fluency task,
verb generation task).  The functional network was then obtained by
thresholding the group-averaged correlation matrix.

The cluster synchronization was then obtained from the functional
networks by applying the clique synchronization algorithm of Section
\ref{sec:functional_clusters} defined in  (\ref{eq:1}).

The cliques were then detected by applying the standard percolation
threshold procedure \citep{gallos2012asmall, mastrandrea2023information}
to define a hierarchy of synchrony clusters according to the order
of clique appearance in the percolation process.  These clusters are
the balanced colorings or fibers that form the input of the MILP of
Section \ref{sec:integer}, which is used to infer the connectome.

\section{Patterns of synchrony during resting state and language task}
   \label{sec:patterns}

\begin{figure}[t!]
  \centering
\includegraphics[width=\textwidth]{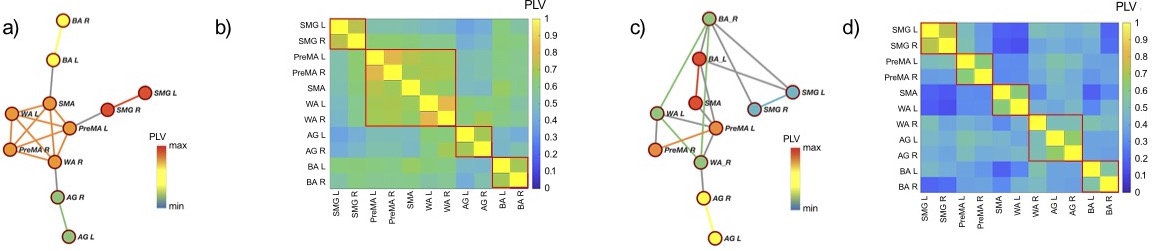}
\caption{\textbf{Resting state and language functional network.} (\textbf{a}) Cluster
  synchronization of the functional network obtained under the RS
  condition. Each color represents a cluster. (\textbf{b}) PLV matrix
  under RS. The red-lined boxes are visual indicators for the clusters
  of ROIs. (\textbf{c}) Cluster synchronization of the functional network
  obtained under task-based language. (\textbf{d}) PLV matrix for the 11
  ROIs. Figure reproduced from \citep{gili2024fibration}.}
	\label{fig:figure3}
\commentAlt{Figure~\ref{fig:figure3}: 
Described in caption/text. No alt-text required.
}
\end{figure}

 The results of the RS functional
network are shown in Fig. \ref{fig:figure3}a, in which the ROIs are
colored according to the clusters obtained from the PLV correlation
matrix shown in Fig. \ref{fig:figure3}b. It is known from the
literature that the RS-fMRI functional network is approximately
left-right symmetric; i.e., Broca left, and Broca right are activated
and synchronized in RS
\citep{teghipco2016disrupted,seitzman2019state}. These results confirm
this evidence by demonstrating left-right symmetry in the
synchronization of the language ROIs under RS. We find a central CS
composed of a pentagonal clique
\citep{li2019functional,li2020core,li2021monolingual} of two bilateral
pairs of regions (premotor and Wernicke's area)\index{Wernicke's area } and the supplementary
motor area, and three CS, each composed of a bilateral pair of regions
(supramarginal gyrus, angular gyrus, and Broca's area,\index{Broca's area }
Fig. \ref{fig:figure3}a).

The same analysis is applied under the phonemic fluency task
(Fig. \ref{fig:figure3}c, d), showing the clusters formed by SMA and
Broca's left to be the most synchronized. The second most synchronized
cliques were angular left and angular right. The third most synchronized
were premotor left and premotor right), followed by Wernicke's left,
Wernicke's right, Broca's right, and finally, the supramarginal left,
supramarginal right clique. This clear synchrony pattern is then used
as the input of the symmetry-based algorithm of Section \ref{sec:integer} to infer the pathways of white
matter tracts that are responsible for the synchronization.

\section{From synchronization to the structural network}

The active pathways of the brain adjust according to the requirements
of the task \citep{friston2011functional,park2013structural}. In the
present case, given our network of mesoscopic brain regions and all
the possible structural connections among them that are identified as
the bundles of axons forming the white matter fiber tract connectome,
different subsets of these structural connections mediate different
functional conditions. Consistent with this idea, a given functional
activity (whether at rest or task-specific) is sustained by a specific
pathway of the brain structure, which strictly depends on
the activity itself. We use this idea to match the
patterns of balanced colorings\index{coloring !balanced } from cluster synchronization\index{synchronization !cluster } with the
symmetries of the underlying connectome.\index{connectome !symmetry of }

\begin{figure}
    \centering
    \includegraphics[width=\textwidth]{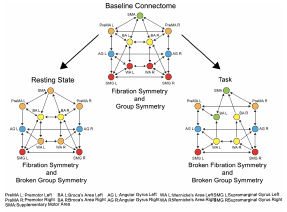}
\caption{\textbf{Fibration symmetry breaking in the human brain.}  (\textbf{a}) Baseline connectome. Structural network between anatomically defined ROIs
  from the literature. This network has the largest symmetry, which is
  a global automorphism group. Here, the symmetry group is the same as the
  symmetry fibration and the five orbits are the same as the five
  fibers (balanced coloring with five different colors). (\textbf{b}) Rest.
  Reconstruction of the structural network for the group average of
  subjects under RS, using the balanced coloring from
  Fig. \ref{fig:figure3}a and the MILP algorithm. The network has only
  local symmetry, given by fibrations with four fibers (balanced coloring with four different colors), but no global symmetry. (\textbf{c}) Task. The
  symmetry of (\textbf{b}) is broken by lateralization under the
  language task. This structural network (reconstructed using the
  coloring from Fig. \ref{fig:figure3}c) shows less symmetry (i.e.,
  broken symmetry with more fibers/five balanced colors shown in the
  figure) than in (\textbf{b}). Figure reproduced from
  \citep{gili2024fibration}.}
  \label{fig:figure8}
\commentAlt{Figure~\ref{fig:figure8}: 
Described in caption/text. No alt-text required.
}
\end{figure}

The pipeline to infer the active pathways in the brain is exemplified in Figs. \ref{fig:two-applications}b and \ref{fig:intro-gili}, and was used to infer the memory engrams in the mice connectome in Chapter \ref{chap:brain2}. The algorithm receives the
initial baseline connectome as input, which represents the communication highway. This connectome can be complete or incomplete. In the
case of the language function studied here, years of brain analysis
have determined the main white matter tracts\index{white matter tract } between the main ROIs
considered. The ROIs are anatomically defined in Section
\ref{sec:human}. The baseline structural
connectome is obtained from the literature, and it is shown in
Fig. \ref{fig:figure8}a.

From this initial connectome (highway system), only a subset of routes
is needed to guarantee the existence of the synchronous clusters
(fibers/balanced coloring) that satisfy the synchrony pattern observed
in Fig. \ref{fig:figure3}. The working hypothesis is that the
brain connectome is stable over the timescales of the experiments,
i.e., no plasticity reorganization of the white matter tracts\index{white matter tract }
 occur. If the
knowledge of the connectome is complete, then any modification we want
to induce on it cannot include the addition of new links, so only
decimation processes are admitted to lead to the structural network that
matches the coloring of the brain regions to the synchrony
coloring. If the knowledge of the connectome is incomplete, then the
algorithm also allows the addition of new links. The algorithm from
Section \ref{sec:integer} accounts for both situations.

The symmetry-driven MILP algorithm\index{MILP algorithm } (Section \ref{sec:integer}) applied to the brain is as follows.
It finds the minimal number of links to be removed from the connectome to match the
balanced coloring of the graph with the one obtained experimentally
from the functional network. The scheme for
synchronization-to-symmetry matching in the brain network can be
summarized in the following steps, Fig. \ref{fig:intro-gili}:
\begin{itemize}
    \item For a given set of ROIs, identify the baseline connectome
      that form the graph of all the permitted structural connections
      among them;
    \item By means of the PLV synchronization measure, find the
      synchrony clusters according to clique synchronization,
       (\ref{eq:1}). Each cluster is assigned a color
      such that nodes with the same symmetry have the same
      color;
    \item Partition and color nodes according to the
      fibration/group symmetries,
    \item Decimate the initial connectome with MILP by removing the
      minimal number of links until the symmetry coloring of the
      connectome matches the synchronization coloring.
\end{itemize}
This optimization problem is solvable with MILP for modestly sized
instances.

The algorithm starts with the known structural network containing all
the pathways (highway system)  shown in Fig. \ref{fig:figure8}a. We see that this baseline connectome already
displays a remarkably symmetric structure.  This network represents
the underlying communication highways of the brain involved in
language.

\section{Symmetry breaking in the human language connectome}
\index{symmetry breaking !in human language connectome }

In general, only a subset of routes is needed to guarantee the
existence of synchronous ROIs that satisfy the observed fMRI
synchronization.  Applying the pipeline in Fig. \ref{fig:intro-gili}
to the baseline connectome of Fig. \ref{fig:figure8}a, informed by the
fMRI PLV correlation matrix from Fig. \ref{fig:figure3}ac, we
obtain two reconstructed connectomes for resting and language tasks,
shown in Fig. \ref{fig:figure8}b,c respectively. That
the algorithm can return a solution with minimal removals
from the baseline connectome indicates the feasibility of the
solution. Otherwise, the algorithm would return either no solution or
remove all links since any coloring is balanced when there are no
links.

We find that the resting state is a particular realization of the
cerebral synchronization pattern, characterized by a defined symmetry
that is broken in the rest-to-task transition, determining
differential recruitment of structural connections according to its
functional state. Consequently, we show that (among the allowed
patterns of structural connectivity) synchronization elicits different
subsets according to the different functional engagements of the brain,
shaping the necessary communication routes.

We find progressive and different symmetry-breaking processes. The
baseline language connectome (highway system) is characterized by the
most symmetric configuration. As shown in Fig. \ref{fig:figure8}a,
the numbers of orbits and fibers are equal, so the symmetry is a
group symmetry---and, by definition, also a fibration. That is, the symmetry is a global
automorphism.\index{automorphism } Once the
brain dynamics enters its resting state (Fig. \ref{fig:figure8}b), this
global symmetry is broken, and only the local fibration symmetry
survives. Synchronization emerges as a functional condition, being the
resting state characterized by only fibration symmetry and a total
loss of global group symmetry. This means that the small perturbation
represented by brain synchronization, overlapping the static network,
suppresses the automorphism but also enhances the biological
configuration that allows for activity stability.

Finally, during the execution of the language task, the activity is
largely polarized in recruiting areas devoted to language
function (Fig. \ref{fig:figure8}c). The known lateralization of
brain activity during language execution induces a further fibration
symmetry breaking observed in the lateralization or breaking of
left-right symmetry between Broca's left and Broca's right. This
induces five synchronized clusters in the task, as compared to 4 in the
resting state, with a concomitant loss of symmetry together
with a total loss of global automorphism (one orbit for each ROI). (A
larger number of synchronized clusters corresponds to less symmetry.)

The result is significant for understanding the brain in terms of
complex systems. Indeed, much of the physical world's complexity
emerges from mechanisms of symmetry breaking \citep{anderson1977more},
and nature's symmetry can be broken in various ways, as discussed in many situations in Part I.  One such
mechanism is spontaneous symmetry-breaking
\citep{anderson1977more,wilkseck,wilczek2016beautiful,weinberg1995},
which explains all matter and interactions from phase transitions
(ferromagnets, superconductivity, etc) to the standard model (Higgs
boson, etc). As in the theory of phase transitions in physical
systems, we find that the brain switches from a dynamical state (the
resting state) to another (the execution of a task), inducing a
fibration symmetry breaking of the structural connectivity pathways that
sustains synchronization of the brain regions' activity in the
different conditions.

\section{Application of symmetry theory to find essential areas in the brain}

\begin{figure}
    \includegraphics[width=\linewidth]{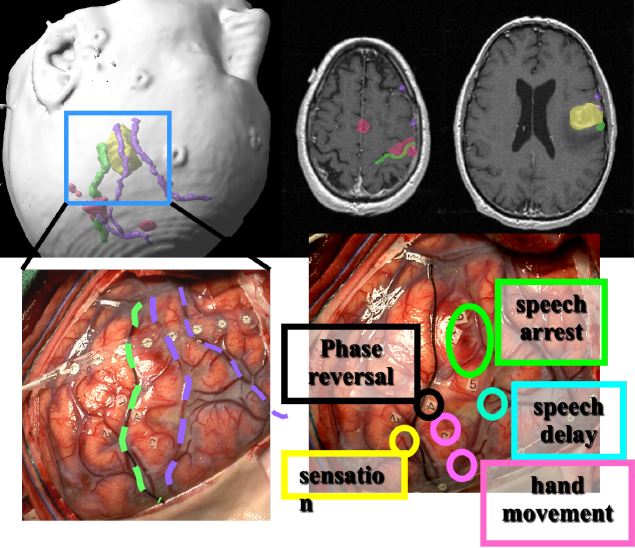}
    \caption{\textbf{Application of inference graph theory to the brain.}  Example
      of how fMRI data is presented to the neurosurgeon for  glioma
      resection. A 3D rendering (top left) in the projection the
      neurosurgeon will see during the craniotomy portrays the central
      sulcus (green), the hand motor homunculus (red), the tumor
      (yellow), and cortical veins (purple). The same information is
      seen in the 2D images (top right). The fMRI data correlates
      well with the intraoperative findings. The bottom left
      shows the exposed brain during the operation, with the green and
      purple lines corresponding to their counterparts on the 3D MR
      image. The results of the direct cortical stimulation (bottom
       right ) correspond to the locations predicted by
      fMRI-based theory. A fibration theory of the brain can identify
      the essential areas in the brain in advance, which helps the
      neurosurgeon perform optimized tumor resections with minimal
      consequences for the patient. Figure courtesy of Dr. Andrei
      Holodny, MSKCC. }
  \label{fig:operating}
\commentAlt{Figure~\ref{fig:operating}: 
Described in caption/text. No alt-text required.
}
\end{figure}

A fibration theory of the brain\index{fibration !theory of brain } improves our understanding of the
design principles of neural circuits, as well as how their structure
influences their function. This theory has immediate medical relevance in
neurosurgery.  For instance, we are working with neurosurgeons and neuroradiologists at Memorial Sloan Kettering
Cancer Center in New York City. They routinely perform awake brain surgery during craniotomy
for tumor resection on patients with brain tumors\index{fibration !application in neurosurgery }. During the operation, neurosurgeons perform direct cortical
stimulation of different areas in the brain to search for
essential areas for expressive language function (Fig. \ref{fig:operating}).

In the clinical setting, Dr. Holodny and his team of neuroradiologists at MSKCC
present the neurosurgeons with fMRI activation maps, which depict the
locations of eloquent cortices (such as language and motor areas)
adjacent to tumors, which neurosurgeons use to plan and guide the
resection of gliomas and other intracranial masses
(Fig. \ref{fig:operating}).  Notwithstanding its advantages, fMRI
clearly has limitations. One of the main problems facing the
pre-operative evaluation of brain tumors by fMRI is that this
technology depicts activations of both essential and non-essential
functional areas. For a neurosurgeon, this
distinction is of paramount importance. In this case, essential
language areas are those whose direct, intraoperative stimulation by
electrodes leads to speech arrest. Resection of such areas leads to
severe language deficits. In contradistinction, non-essential language
areas are defined as those whose direct, intraoperative stimulation
by electrodes does not lead to speech arrest. Resection of such areas
is thought to be safe.

While fMRI technology alone does not allow one to differentiate
between core areas, the inference network method used in this chapter and tested in the mouse brain in Chapter \ref{chap:brain2} allows for this
distinction. A fibration theory that can predict these essential areas in the brain
contributes to optimizing the decision-making process
through which neurosurgeons select which parts of a tumor to
resect.\index{fibration !application for essential brain areas } Currently, routinely performed preoperative fMRI scans do not
discern which areas of fMRI activation represent `essential' language
areas. The neurosurgeon, therefore, does not know, in advance of the
operation, which areas of the tumor can safely be resected.

A graph theory of the brain has the potential to predict those areas
that are essential for language function.  Such a result will be
unprecedented: prediction/testing of brain ROIs that are `essential'
for function via direct cortical stimulation has never before been
attempted for the human brain.

Development and testing (in the setting of direct intraoperative
cortical stimulation) of theories of organization and response to
brain network perturbations should lead to the inference of general
principles regarding the network organization of brain circuits and
the changes they undergo in pathological states.

Finally, the development of theories of the organization of the
connectome should lead to  general principles
regarding network organization, applicable to areas outside
neuroscience, including general information-processing complex systems.

\chapter[Overview and Outlook]{\bf\textsf{Overview and Outlook}}
\label{chap:outlook}

\begin{chapterquote}
Fundamental laws of physics are conceptually simple based on a high
degree of symmetry, yet their consequences are enormously
rich. 
 We have explored whether the same
reasoning can be applied to understand biological systems. Although
the group symmetries of physics have some
biological applications, the more
relaxed concept of fibration symmetry is more widely applicable,
and provides insights into fundamental issues in biology.
In the first section of this final chapter we summarize some of the key messages of the book, and in the second section we sketch possible topics for future applications. 

\end{chapterquote}

\section{Overview}

In this section we summarize some of the more important conclusions to be drawn from the ideas presented in this book.

Global symmetries,
formalized in terms of groups, have been enormously
successful as a basis for fundamental physics. 
They also
have some applications to biology. However,  more general
and more flexible notions of symmetry are better suited 
to biology, especially in the age of Big Data Omics.

We learned that these symmetries are organized in a hierarchy, progressing from less strict to more strict:

\begin{floatingbox}[h!]
\label{hyerarchy}
  \processfloatingbox{Hierarchy of Symmetries}
                     {\,\,\,\,\,\,\,\, homomorphisms $\to$ fibrations  or  opfibrations $\to$ coverings $\to$ automorphisms}    
\end{floatingbox}
describing successively different domains of nature:
\begin{floatingbox}[h!]
\label{hyerarchy2}
  \processfloatingbox{Application Domains of Symmetries}
                     {\,\,\,\,\,\,\,\, biology (genes/brain) $\to$ artificial intelligence $\to$ physics and geometry}    
\end{floatingbox}

There is an elegant and substantial body of mathematical theory
connected with fibration symmetries. It includes fundamental concepts
such as equitable partitions, also called balanced colorings, which
are determined by the fibers of a fibration.  It provides general
methods for analyzing models, computing stabilities, and studying
bifurcations. These methods fit into a coherent framework rather than
just being the result of one-off numerical calculations on some
specific model.

Fibration symmetries in biological networks provide information about
biological function and its relation to the form (topology) of the
network.  However, addressing this question is challenging because
biological data are incomplete.  We have discussed the problem
of incomplete/missing data in biological networks by developing a
reconstruction optimization algorithm for the network that reproduces
the observed synchronization.

Fibration symmetries permit simple routes to evolutionary
change. Under selection pressure for coexpression, fibration
symmetries may come into existence when genes are duplicated by
lifting processes and are later conserved. This is just a possibility,
but we studied it empirically.

Through evolutionary rewiring and selection, regulatory networks can
be modified to perform signal processing tasks.  Genetic circuits that
are performant, cheap, and robust may evolve many times. Some can be
identified statistically as network motifs.  However, motifs
determined by statistical over-representation should be treated with
caution because the dynamics of a motif considered in isolation may
differ from its dynamics when embedded in the full network.
In contrast, fibration building blocks can be determined directly,
without the need for a null model, even if they appear only once in a
network.  Dynamics arising from fibration symmetries---in particular
synchrony---occur in the full network, not just in the motif.
Genetic fibers provide insights into coexpression that could not have
been achieved by alternative theories, e.g., network motifs, modules,
etc.
  
Our approach bridges the gap between structure and synchronization and
can be used to further assess the functionality of the biological
network via perturbations to its structure. The concept of fibration
symmetry suggests a number of potential experiments. For example, the
predictions of the theory can readily be checked by controlled gene
knockout experiments\index{gene
knockout experiment } designed to break the symmetries of the
fiber. Rather than the `trial and error' protocols currently employed,
we envisage theoretically guided protocols of gene mutations designed
to break specific symmetries of the fibers. This idea also
suggests numerous
investigations in synthetic genetic circuits
which can be designed in a systematic way to build a living GRN made of
fibration symmetries and their broken symmetries.

Going beyond the symmetries found in TRNs, the fibration
framework can be adapted to include any type of regulation, or any
type of interaction among biomolecules. This includes protein-protein
interactions,\index{protein-protein
interaction } metabolic reactions, and metabolic effects on
transcription, as well as signaling pathways that control cell
behavior.  We envisage an integrative model of gene regulations and
metabolism that includes all interactions and molecules that
contribute to cellular processes that could be valid across species
from bacteria to eukaryotes.

\begin{figure}[h]
  \includegraphics[width=\textwidth]{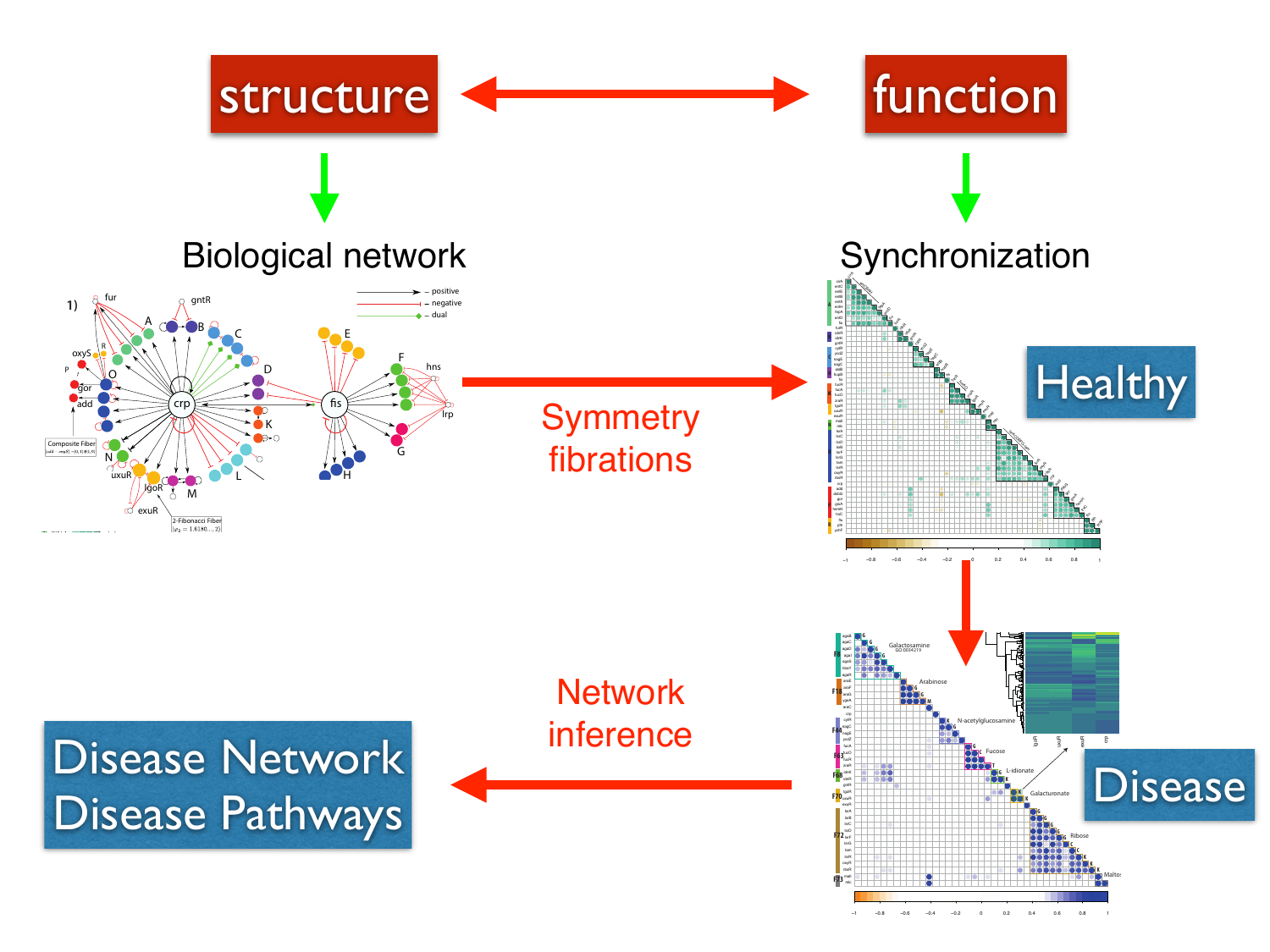}
  \caption{ \textbf{Finding disease pathways with fibrations.}
    Inferring disease pathways\index{disease !pathway } using fibrations involves analyzing
    gene/neural network synchronization. A healthy genetic or neural
    network can be established through synchronization experiments and
    inference optimization algorithms developed in Section
    \ref{sec:integer}. This healthy network serves as a baseline to
    infer a disease network by feeding the differential
    synchronization patterns in the disease state into the inference
    algorithm.  }
  \label{fig:disease}
\commentAlt{Figure~\ref{fig:disease}: 
Components are--- 
Top row left to right: box (structure); box (function);
Middle row left to right:  image (biological network); image of correlation matrix (synchronization) labeled `healthy';
Bottom row left to right:  box (disease network/disease pathways);  image of 
correlation matrix (synchronization) labeled `disease'.
Red arrows link structure to function in both directions,
biological network image to synchronization; healthy to disease;
disease to box (disease network/disease pathways).
}
\end{figure}

 \section{Outlook}

To round off the book, we focus on two of the most promising
applications of fibration theory: the understanding of
disease\index{disease } and AI. We discuss how
fibration theory could help to predict disease pathways in the regulatory networks, to assist designing 
therapeutics, and to design novel artificial neural architectures that
might inspire further work in these directions. We make no attempt at
complete coverage, and anticipate that unexpected new developments
will introduce new directions, not mentioned here. The aim is to try
to justify the view---which, of course we hold, rightly or
wrongly---that this book is just a beginning.

{\bf Building the AI of the future.}\index{artificial intelligence }
The application of fibrations to artificial neural networks opens the door to developing more efficient architectures and novel training algorithms. Symmetries facilitate the dimensional reduction of parameter space, leading to faster convergence and lower resource consumption. Furthermore, these symmetries can enhance the robustness and interpretability of neural networks, as they often reflect underlying regularities in the data associated with the problem being solved. Along the path of geometric deep learning, fibration symmetries and covering spaces represent an extended blueprint for the development of the novel neural architectures of the future.

{\bf Uncovering disease pathways.}\index{disease !pathway } Symmetry theory is also
significant for the understanding, diagnosis, and development of
treatment strategies for disorders thought to be due to damage to
connectivity. In the case of brain diseases, this includes, e.g.,
Alzheimer’s disease,\index{Alzheimer’s disease } ADHD,\index{ADHD } strokes,\index{stroke } traumatic brain injury,\index{brain injury } and
schizophrenia\index{schizophrenia }
\citep{van2019cross}.

A primary application of the synchronization-driven inference method
of Section \ref{sec:integer} is the understanding of disease
pathways.\index{disease !pathway } The inference of pathways from
dynamical data on healthy subjects can be extended to neurological or
psychiatric conditions as well as genetic ones, allowing the
identification of differential disease pathways, leading to an
understanding of the disease, establishing the diagnosis, and
ameliorating the consequences (see Fig. \ref{fig:disease}).

Moreover, the method is beneficial for drug development by targeting
the inferred structural network of a specific disease onto a healthy
one. The treatment of neurological and psychiatric diseases through
invasive intervention (surgery) or non-invasive intervention
(electric/magnetic stimulation) will benefit from the identification
of the patterns of symmetry and synchronization, and their breaking
processes, to reduce side effects or to optimize the effectiveness of
the application.

Understanding the organization and response to perturbations of
genetic and brain networks leads to general principles regarding
network organization of biological circuits and the changes they
undergo in pathological states. This occurs because
desynchronization/gain of synchronization can be seen as a
pathological state, e.g., cancer. The analyses have the potential to
predict the circuits and pathways responsible for pathologies, which
can subsequently be targeted to restore the healthy state.

Such advances are crucial for developing more accurate and
dynamically responsive neural and genetic models, which can be pivotal
in both basic research and clinical applications, such as designing
targeted treatments for neurological disorders involving dysregulated
neural synchronization in complex neural systems.

Furthermore, exploring the impact of structural variations
on the synchronization in disease models could open new therapeutic
avenues, particularly for neurological disorders where dysregulation\index{dysregulation }
of synchronized activity is evident.

We hope this book will inspire these exciting developments in the future.



  \backmatter

  \appendix
%
%
%

\appendix
\chapter{Software, code, and data}
\label{sec:list-software}

We describe a list of software packages used in each 
chapter. This is handy for those readers who want to delve into the
application of fibrations to the study networks in various fields.
The general repositories to reproduce all the results in the book and to perform new fibration analyses are available at
{\small\url{https://GitHub.com/MakseLab}} and
{\small\url{https://osf.io/4ern8}}. All URLs were accessible as of February 2025.

\section{List of software implementations, codes, and datasets used in this book}
\index{software implementation }

\begin{enumerate}
\item {\bf Codes and data for Chapters \ref{chap:hierarchy_1} and  \ref{chap:hierarchy_2}}

The software for Chapters \ref{chap:hierarchy_1} and \ref{chap:hierarchy_2} contains the primary codes used to conduct a fibration symmetry analysis of networks. This analysis involves calculating balanced colorings and constructing simple building blocks of fibers. These codes are essential for any fibration analysis, as they systematically take a network, identify all fibers, and categorize the building blocks into distinct groups of simple components. The software addresses the forward problem by starting with a graph and identifying and classifying all building blocks based on their symmetries. Additionally, the codes in Chapter \ref{chap:breaking} will enhance this study by focusing on symmetry-breaking building blocks. 
Ultimately, the codes for Chapter \ref{chap:minimal} will integrate all these elements to perform a dimensional reduction, resulting in the minimal fibration base of the network, akin to a 'minimal genome'.

\begin{itemize}

\item Fibration symmetries and fibration building blocks--- Algorithms \index{building block !fiber !code } developed by  \cite{morone2020fibration} based on balanced coloring algorithms by  \cite{kamei2013} and  \cite{cardon1982partitioning}. Available
at {\small\url{https://GitHub.com/MakseLab/FibrationSymmetries}}, and 
{\small\url{https://osf.io/4z2hb}}.\index{fibration !algorithms}

\item  Fibration symmetries and fibration building block--- Data \index{building block !fiber !dataset } and analyzed networks developed by \cite{morone2020fibration}. Available at  {\small\url{https://osf.io/z793h}} and  {\small\url{https://osf.io/pwe56}}, respectively. This includes a list of all simple building blocks found in the TRN networks across species.

\item    Fast Fiber Partition---Julia implementation of the Fast Fibration algorithm developed by   \cite{monteiroAlgorithm}. Available at {\small\url{https://GitHub.com/MakseLab/FastFibration}}.

\item  Automorphisms---  McKay's {\it Nauty}\index{Nauty algorithm } algorithm~\citep{nauty}
to find automorphisms is available at {\small\url{https://pallini.di.uniroma1.it}}. Pynauty, a Python library based on Nauty to work with graph
    automorphisms, is available at {\small\url{https://GitHub.com/pdobsan/pynauty}}.
    
\end{itemize}

\item{\bf Codes and data for Chapter \ref{chap:complex}}

These codes systematically classify the complex building blocks described in Chapter \ref{chap:complex}.

\begin{itemize}
\item The codes and data to find the complex building blocks defined in Chapter \ref{chap:complex} developed by \cite{alvarez2024symmetries} are available at {\small\url{https://github.com/MakseLab/Symmetries-in-Metabolic-Networks-of-E.-coli}} and {\small\url{https://osf.io/2ntr3}}. They are mirrored from {\small\url{https://github.com/luisalvarez96/Symmetries-in-Metabolic-Networks-of-E.-coli}}.
This includes a list of all complex building blocks found in the metabolic network of {\it E. coli}.
\end{itemize}

\item {\bf Codes and data for Chapters \ref{chap:motif} and \ref{chap:breaking}}

Codes and data to find the symmetry-breaking circuits leading to various forms of flip-flops and oscillators used in Chapter \ref{chap:breaking} as well as motif  finders  developed by 
\cite{leifer2020circuits}
are available at:

\begin{itemize}

\item Symmetry breaking Building blocks Finder Circuit--- Algorithm for symmetry breaking circuits developed by \cite{leifer2020circuits} is available at {\small\url{https://GitHub.com/MakseLab/CircuitFinder}}.

\item  DDE solver--- Delayed ODE (DDE) solver from Mathematica  developed by \cite{leifer2020circuits} to study dynamics of building blocks at
{\small\url{https://GitHub.com/MakseLab/CircuitFinder/blob/master/delayODE.nb}}.

\item Data and analyzed networks for this chapter developed by \cite{leifer2020circuits} are available at {\small\url{https://osf.io/jhmau}}.

\end{itemize}

\item{\bf Codes and data for Chapter
\ref{chap:minimal}}

This code enhances the fibration analysis by generating the minimal effective network, similar to a minimal genome or minimal TRN, for any given network. 

\begin{itemize}
\item Codes and data to perform CoReSym, the complexity reduction analysis based on symmetries used in Chapter \ref{chap:minimal}, and developed by \cite{alvarez2024fibration}, are available at
{\small\url{https://github.com/MakseLab/MinimalTRNCodes}} mirrored from 
{\small\url{https://github.com/luisalvarez96/MinimalTRN}}. Further data is available at {\small\url{https://osf.io/eh6ps}}.

\end{itemize}

\item{\bf Codes and data for Chapter \ref{chap:synchronization}}

These codes analyze the functional network of any system and produce the cluster synchronization of the network nodes. The analysis focuses on gene co-expression as a
form of cluster synchronization.

\begin{itemize}
\item Correlations and synchronization measures and analysis.\index{cluster synchronization !code } 
Code and gene co-expression data developed
 by \cite{leifer2021predicting} are available at
 ~{\small\url{https://github.com/MakseLab/geneCoexpressionFibration}}.
 
 \item Data,
 co-expression and building blocks of genetic networks developed by \cite{leifer2021predicting} are available
 at {\small\url{https://osf.io/kbauj}}.

\end{itemize}

\item{\bf Codes and data for Chapter \ref{chap:function}}

Perhaps the most important algorithms for practical applications are here. These codes deal with the problem of reconstructing an incomplete network guided by the observed cluster synchronization, as described in Section \ref{sec:link-prediction}. The codes also solve the `one-to-many' network inference problem of finding the activated pathways for a given synchrony pattern as explained in Section \ref{sec:inferring}. These codes are then used in the brain analysis in Chapters \ref{chap:brain1},  \ref{chap:brain2} and \ref{chap:brain3}.
We also include code and data to deal with similar problems: quasi-fibrations (Section \ref{sec:quasifibrations}) and pseudo-balanced colorings and network reconstructions using the symmetry-driven algorithm described in Section \ref{sec:repair}. This last algorithm is more general since it reconstructs a network without knowing the balanced coloring. 

\begin{itemize}

\item  The network reconstruction algorithm is based on mixed integer linear programming (MILP) and is guided by experimental cluster synchrony, as discussed in Section \ref{sec:integer}. This algorithm takes a balanced coloring and a baseline network to reconstruct or infer a network that satisfies the balanced coloring. This code addresses the inverse problem: given a functional network, it infers the corresponding structural network guided by cluster synchrony obtained experimentally.
It was developed in \citep{avila2024symmetries,garcia2024minimal, gili2024fibration}, and it is the basis for brain studies in Chapters \ref{chap:brain1},  \ref{chap:brain2} and \ref{chap:brain3}, respectively. It is available at {\small\url{https://github.com/MakseLab/genes_coloring}}.

\item Pseudo-balanced coloring and network reconstruction via symmetries without knowing the balanced coloring. This integer
            programming 
            developed by \cite{leifer2022symmetry-driven} \index{pseudo-balanced coloring !code } solves a more difficult problem than in Section \ref{sec:integer}, since the coloring is not known {\it a priori}. It is available 
            at ~{\small\url{https://github.com/MakseLab/PseudoBalancedColoring}}. An analogous problem is the
link prediction code \index{network reconstruction !code }
            from \citep{leifer2022symmetry-driven} available at {\small\url{https://osf.io/26u3g}}.  Data and analyzed networks
            at {\small\url{https://osf.io/prt5g}}.
           
\item Yet, another formulation for incomplete networks. Quasi-fibration finder and network reconstruction - \index{graph fibration !quasi-fibration !code } Code for building quasi-fibration symmetries of graphs developed by
\cite{boldi2021} at {\small\url{https://github.com/MakseLab/QuasiFibrations}}. Data and analyzed networks at
{\small\url{https://osf.io/amswe}}.
\end{itemize}

\item{\bf Codes and data for brain analysis for Chapters \ref{chap:brain1}, \ref{chap:brain2}, and \ref{chap:brain3}}

The main application of the inference and reconstruction algorithms from Chapter \ref{chap:function} is to infer brain networks. 
These algorithms are applied to three cases.
More algorithms include functional brain network
reconstructions and network analysis.

\begin{itemize}

\item {\it C. elegans} locomotion inference algorithm--- Chapter \ref{chap:brain1}. Code and data to reconstruct the connectome using synchronization data developed by \cite{avila2024symmetries} are available at {\small\url{https://github.com/makselab/c_elegans_neural_network}} and 
{\small\url{https://osf.io/9zvtq}}.

\item Minimal engram in the mouse brain--- Chapter \ref{chap:brain2}.
Code and data for mouse connectome inference based on Allen connectome and engram data developed by \cite{garcia2024minimal} are available at
{\small\url{https://github.com/MakseLab/MouseConnectome}}.

\item Inferring the human language brain network--- Chapter \ref{chap:brain3}. Code and data to infer a human connectome developed by \cite{gili2024fibration} is available at {\small\url{https://github.com/MakseLab}}.

\item Functional brain network construction from functional MRI signal--- Series of codes and scripts perform a functional network inference construction from fMRI tasked-based data by \cite{delferraro2018finding} at  {\small\url{https://github.com/gdelfe/Functional-network-construction-from-fMRI-signals}}. \index{functional brain network !percolation !code } Tools to construct functional networks and graph theoretical analysis at {\small\url{http://kcorebrain.com}}.

\item  Software to analyze brain images--- \index{neural influencer code } \index{functional brain network influencer code }  Calculate influential areas, centralities, and $k$-core analysis developed by \cite{li2020core} is available    at 
{\small\url{https://github.com/MakseLab/Healthy_brain_networks}}.
A {\it C. elegans} locomotion simulator and network analysis developed by \cite{avila2024symmetries} is available at
{\small\url{https://github.com/MakseLab/C.-elegans_locomotive_simulator}}.

\end{itemize}

\item{\bf Codes and data for Chapter \ref{chap:ai}}

\begin{itemize}

\item Code and data for the fibration analysis of deep neural networks developed by \cite{velarde2024therole} are available at {\small\url{https://osf.io/m4cxg}}.

\end{itemize}

\end{enumerate}

  \endappendix

  \addtocontents{toc}{\vspace{\baselineskip}}

  \bibliography{thebib,paolo,osvaldo}
  \label{refs}
  \bibliographystyle{cambridgeauthordate}

  \cleardoublepage

\printindex




\end{document}